%% file: fpwg_yellow_report.tex
\documentclass[a4paper,11pt]{cernrep}

\usepackage{epic,eepic,pspicture,mathptmx,times,graphpap}

\usepackage{xspace}
\usepackage{color}
\usepackage{pdflscape}
\usepackage{atlasphysics}
\usepackage{graphicx}
\usepackage{caption}
\usepackage{subcaption}
\usepackage{booktabs}
\usepackage{multirow}
\usepackage{epstopdf}
\usepackage{placeins}
\usepackage{appendix}
\usepackage{chngcntr}
\usepackage{etoolbox}
\AtBeginEnvironment{subappendices}{%
\chapter*{Appendix}
\addcontentsline{toc}{chapter}{Appendices}
\counterwithin{figure}{section}
\counterwithin{table}{section}
}

\input{softdiffraction/softdiffraction_macros.tex}

\input{forward/forward_macros.tex}

\begin{document}
\title{Forward Physics at the LHC}

\input{title/titlepage}

\cleardoublepage

\tableofcontents
\cleardoublepage

\setcounter{page}{1}
\pagenumbering{arabic}

\chapter*{Foreword}
\input{fore/foreword.tex}

\chapter*{Introduction}
 \label{chap:introduction}
\addcontentsline{toc}{chapter}{Introduction}

\large {\bf Editors: J. Bartels, N. Cartiglia, C. Royon} \\
\large {\bf Internal Reviewers: M. Arneodo, V. Khoze} \\

\input{introduction/introduction.tex}

\chapter{Running conditions and beam induced backgrounds}
\label{chap:bema}

\large {\bf Editors: V. Avati, C. Royon} \\

\input{introduction/acceptance.tex}

\input{introduction/background.tex}

\input{introduction/beamcond.tex}

\input{introduction/introduction.bbl}
\chapter{Monte Carlo}
\label{chap:montecarlo}
\large {\bf Convener and Editor: L. Harland-Lang} \\
\large {\bf Internal Reviewers: H. Jung, M. Ruspa} \\

\input{montecarlo/montecarlo.tex}

\chapter{Soft Diffraction and Total Cross section}
\label{chap:softdiffraction}
\large{\bf Conveners and Editors: V. Avati, T. Martin} \\
\large{\bf Internal Reviewers: P. Grafstrom, V. Khoze} \\

\input{softdiffraction/softdiffraction.tex}

\input{softdiffraction/softdiffraction.bbl}
\chapter{Inclusive Hard Diffraction}
\label{chap:harddiffraction}
\large{\bf Conveners and Editors: M. Ruspa, M. Trzebi\'nski} \\
\large{\bf Internal Reviewers: P. Bussey, T. Martin, M. Obertino} \\

\input{harddiffraction/harddiffraction.tex}

\input{harddiffraction/harddiffraction.bbl}
\chapter{Central Exclusive Production}
\label{chap:cep}
\large{\bf Conveners and Editors: L. Harland-Lang, V. Khoze, M. Saimpert} \\
\large{\bf Internal Reviewers: R. Mc Nulty, M. Rangel} \\

\input{cep/cep.tex}

\chapter{Cosmic Ray Physics, Particle multiplicities, correlations and spectra}
\label{chap:cosmic}
\large{\bf Convener and Editor: T. Pierog} \\
\large{\bf Internal Reviewers:  D. D'Enterria, T. Sako, D. Salek} \\

\input{cosmic/cosmic.tex}

\input{cosmic/cosmic.bbl}
\chapter{Heavy Ion Physics}
\label{chap:ion}
\large{\bf Convener and Editor: D. Takaki} \\
\large{\bf Internal Reviewers: V. Avati, M. Murray} \\

\input{heavyion/heavyion.tex}

\input{heavyion/heavyion.bbl}
\chapter{BFKL and saturation}
\label{chap:bfkl}
\large{\bf Conveners and Editors: J. Bartels, H. Jung, C. Marquet} \\
\large{\bf Internal Reviewers: A. Sabio Vera, S. Wallon} \\

\input{forward/forward.tex}

\input{forward/forward.bbl}
\chapter{Detectors}
\label{chap:detectors}
\large{\bf Conveners and Editors: V. Avati, J. Baechler} \\
\large {\bf Internal Reviewers: M. Bruschi} \\

\large {\bf LHC Forward Physics Working Conveners and Editors:  N. Cartiglia, C. Royon} \\

\input{detectors/detectors.tex}

\input{detectors/detectors.bbl}
\chapter*{Summary and Conclusion}
\label{chap:conclusion}
\input{summary/summary.tex}

\chapter*{Acknowledgements}
\input{acknowledgements/acknowledgements.tex}

\end{document}

%% file: softdiffraction/softdiffraction_macros.tex
\newcommand{\pyeight}{\texttt{Pythia 8}\xspace}
\newcommand{\py}{\texttt{Pythia}\xspace}
\newcommand{\epos}{\texttt{EPOS}\xspace}
\newcommand{\hpp}{\texttt{Herwig++}\xspace}
\newcommand{\logxix}{\ensuremath{\log_{10}\left(\xi_{\textrm{X}}\right)}\xspace}
\newcommand{\lnxix}{\ensuremath{\ln\left(\xi_{\textrm{X}}\right)}\xspace}
\newcommand{\logxiy}{\ensuremath{\log_{10}\left(\xi_{\textrm{Y}}\right)}\xspace}
\newcommand{\MX}{\ensuremath{M_{\textrm X}}\xspace}
\newcommand{\MY}{\ensuremath{M_{\textrm Y}}\xspace}
\newcommand{\Nch}{\ensuremath{N_{\textrm{Ch}}}\xspace}
\newcommand{\detaf}{\ensuremath{\Delta\eta^{\textrm{F}}}\xspace}
\newcommand{\ptcut}{\ensuremath{p_{\texttt{T}}^{\texttt{Cut}}}\xspace}

%% file: forward/forward_macros.tex
\newcommand{\be}{\begin{equation}}
\newcommand{\bea}{\begin{eqnarray}}
\newcommand{\ena}{\end{eqnarray}}

\newcommand{\slfrac}[2]{\left.#1\middle/#2\right.}

%% file: title/titlepage.tex
%%%%%%%%%%%%%%%%%%%%%%%%%
%%%%%  TITLE PAGE  %%%%%%
%%%%%%%%%%%%%%%%%%%%%%%%%
\begin{titlepage}

% Header ---------------------------------------------------
\vspace*{-1.5cm}

\hspace*{-0.5cm}
\begin{tabular*}{\linewidth}{lc@{\extracolsep{\fill}}r}
\vspace*{-1.4cm}\mbox{\!\!\!\includegraphics[width=.09\textwidth]{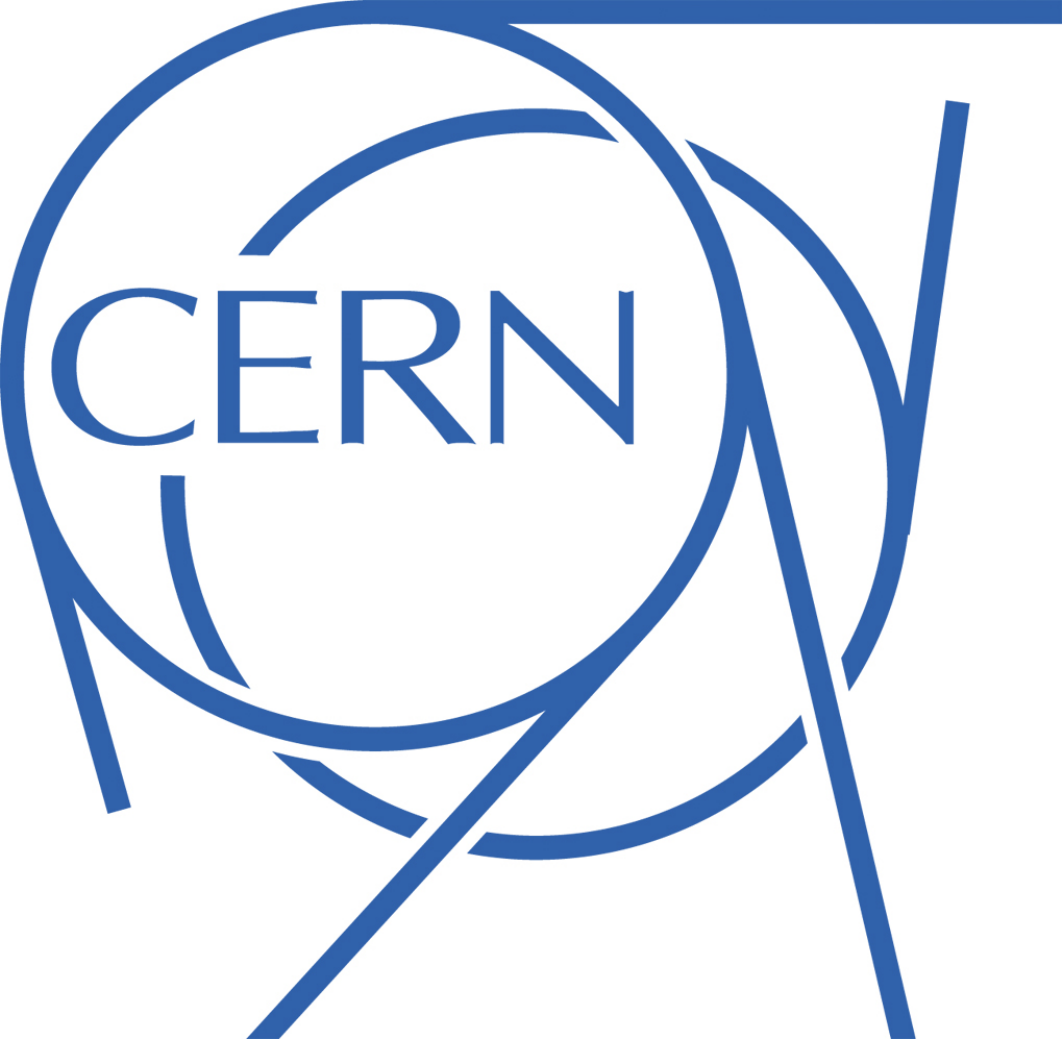}} & & \\
 & & CERN-PH-LPCC-2015-001 \\
 && SLAC-PUB-16364 \\
&& DESY 15-167 \\
   September 3 2015
% & & \today\\
 & & \\
\hline
\end{tabular*}

\vspace*{2cm}

% Title --------------------------------------------------
{\bf\boldmath\Large
\begin{center}
LHC Forward Physics\\
%\textcolor{red}{\textbf{INCOMPLETE INTERNAL DRAFT DOCUMENT}}
\end{center}
}

\vspace*{1cm}

% Authors -------------------------------------------------
%\begin{center}
%The LHC Forward Physics Working Group
%\end{center}
\input{authors/fpwg_authors.tex}

\vspace{\fill}

% Abstract -----------------------------------------------
\begin{abstract}
%\vspace{0.5cm}

 % \noindent

\noindent 
The goal of this report is to give a comprehensive overview of the rich field of forward physics, with a special attention to the  topics that can be studied at the LHC.  The report  starts presenting a selection of the Monte Carlo simulation tools currently available, chapter 2, then  enters  the rich phenomenology of  QCD at low, chapter 3, and high, chapter 4,  momentum transfer,
while the unique scattering conditions of central exclusive production are analyzed in chapter 5. The last two experimental topics, Cosmic Ray and Heavy Ion physics are presented in the chapter 6 and 7 respectively.  Chapter 8 is dedicated to the BFKL dynamics, multiparton interactions, and saturation. The report  ends with an overview of the forward detectors at LHC. Each chapter is correlated with a comprehensive bibliography, attempting to provide to the interested reader with a wide opportunity for further studies. 
\end{abstract}

\vspace*{2.0cm}
\vspace{\fill}

\end{titlepage}

\pagestyle{empty}  % no page number for the title

%%%%%%%%%%%%%%%%%%%%%%%%%%%%%%%%
%%%%%  EOD OF TITLE PAGE  %%%%%%
%%%%%%%%%%%%%%%%%%%%%%%%%%%%%%%%

%  empty page follows the title page ----
\newpage
\pagestyle{plain}
\pagenumbering{roman}
\setcounter{page}{2}
\mbox{~}

\cleardoublepage

%% file: authors/fpwg_authors.tex
 %\vspace{1cm}
\centerline{\large \bf Editors: N.~Cartiglia, C.~Royon} 
%\vspace{1cm}
\centerline{\large\bf The LHC Forward Physics Working Group}

\begin{flushleft}

\small
K.~Akiba$^{21}$,
M.~Akbiyik$^1$,
M.~Albrow$^2$,
M.~Arneodo$^{3,4}$,
V.~Avati$^{5,6}$,
J.~Baechler$^6$,
O. Villalobos Baillie$^{87}$,
P.~Bartalini$^7$,
J.~Bartels$^8$,
S.~Baur$^1$
% ...

%\\

%\bigskip\\
{\it\footnotesize
$ ^1$ Karlsruhe Institute of Technology (KIT), Karlsruhe, Germany \\
$ ^2$ Fermilab, Batavia, USA \\
$ ^3$ INFN Sezione di Torino, Italy \\
$ ^4$ Universit\'a del Piemonte Orientale, Novara, Italy \\
$ ^5$ AGH University of Science and Technology, Krakow, Poland  \\
$ ^6$ CERN, Geneva, Switzerland \\
$ ^7$ Central China Normal University (CCNU), Wuhan, Hubei, China \\
$ ^8$ University of Hamburg, Germany \\
% ...
$ ^{21}$ Universidade do Estado do Rio de Janeiro (UERJ), Rio de Janeiro, Brazil \\
% ...
$^{87}$ University of Birmingham, Birmingham, UK\\
% ...
}

\end{flushleft}

%% file: fore/foreword.tex
In early 2013 the LHC Forward Physics and Diffraction Working Group was
formed, as part of the activities of common interest to the LHC experiments
organized by the LHC Physics Centre at CERN (LPCC, http://cern.ch/lpcc). The
primary goal of the WG was to coordinate, across the experiments and with
the theoretical community, the discussion of the physics opportunities,
experimental challenges and accelerator requirements arising from the study
of forward phenomena and diffraction at the LHC. The mandate of the group
included the preparation of a Report, to outline a coherent picture of the
forward physics programme at the LHC, taking into account the potential of
the existing experiments -- including possible detector upgrades --, the
possible beam configurations and performance of the accelerator, and the
optimization of the LHC availability for these measurements, in view of the
priority need to maximize the LHC total integrated luminosity.

The WG was set up by the LPCC in coordination with the management of the
ALICE, ATLAS, CMS, LHCb, LHCf and TOTEM experiments, which nominated their
representatives in the WG steering group and the WG co-chairs. The steering
group identified theory conveners, to oversee the relevant sections of the
Report, and created three subgroups to focus the WG activity, reflecting the
physics goals appropriate to different LHC running conditions: 
\begin{itemize}
\item low pileup and luminosity (few 10 pb$^{-1}$), 
\item medium luminosity (few 100 pb$^{-1}$), 
\item high luminosity (100 fb$^{-1}$).  
\end{itemize}

All interested physicists were then invited to attend the 16 WG meetings
held so far, and to contribute to the writing of this Report, which
hopefully represents the unanimous views of the broad forward-physics
community. The detailed information about the WG, including the composition
of the steering committee and of the subgroups' conveners, the list of
meetings, the link to the WG material and to its mailing list subscription,
can be found in the WG web page at 

http://cern.ch/LPCC/index.php?page=fwd\_wg

As requested by the LHC experiments committee (LHCC), and following the
several presentations delivered to the committee in the course of the WG
activity, this final Report will be submitted to the LHCC, and will form the
basis for its internal discussions and recommendations on the requests by
the experiments for beam time and detector upgrades, related to forward
physics, during Run 2 of the LHC and beyond. More in general, we trust that
this Report will promote the deeper understanding and appreciation of the
value of this component of the LHC physics programme, and will encourage
further progress and the development of new ideas, both on the theoretical
and experimental fronts. 

The chairs of the LHC Forward Physics WG

%% file: introduction/introduction.tex
%\section{Introduction}

For a successful run of the LHC it is essential to have a full understanding
of the complete final states.
This includes, besides the central region, also the kinematic region as close as possible to the forward direction. 
New physics is mainly searched for in the central region where factorization theorems for inclusive cross sections allow 
the use of  parton densities and hard subprocesses whose cross sections can be calculated by using 
perturbative theory.  However, there is a rich physics content outside this kinematic region, in particular close to  the forward directions. 
Prominent examples include the final states with high forward 
multiplicities, as well as those with rapidity gaps, notably
in elastic, diffractive, and central exclusive processes. 
Some of these configurations originate from purely nonperturbative
reactions, while others can be explained in terms of multiparton chains 
or other extensions of the perturbative QCD parton picture. Future 
progress in this field requires the combination of thorough
experimental measurements and extensive theoretical work.

Monte Carlo generators are indispensable for analyzing experimental data 
and comparing them with theoretical predictions. Their further 
development requires detailed studies of the forward region. The
most successful and most frequently used Monte Carlo event generators
(Madgraph, Pythia, Herwig) were initially written with focus on the central
region, considered as the most promising for discovering new physics.
Higher order QCD calculations have been implemented, and corrections due 
to multiparton interactions are now being included. Nevertheless, there 
remain important aspects that 
require further attention. Most importantly, when extending these 
event generators to the forward direction, it becomes necessary to 
include diffractive (elastic and inelastic) final states.  The importance 
of this rapidity gap physics has been demonstrated, in particular, by the 
HERA data.
At the LHC, final states with rapidity gaps are ascribed to rescattering 
effects (multiparton chains) that reduce the probability of finding 
kinematic regions devoid of jets or particles. This suppression 
(commonly referred to as 'suppression due to survival  factors') has to 
be taken into account by the event generators, a task that
still presents both conceptual and practical difficulties. 
On the other hand, there are event generators specifically developed
for the forward direction (EPOS, PHOJET, QGSJET) that have proven 
to be particularly successful in predicting, for example, forward
energy flow and multiplicities. A third class of specialised
Monte Carlo generators  
has been developed: CASCADE and HEJ for small-x and BFKL physics, 
POMWIG, FPMC and SuperCHIC for central exclusive production. 
What is still missing are Monte Carlo  
generators that simulate final states dominated by saturated parton distributions. 
In summary, the most ambitious goal in the field of Monte Carlo simulation
is the development of generators that include precision QCD calculations,
and simulate multiparton interactions as well as final states with rapidity
gaps. Clearly, the study of forward physics plays a central role in making
progress along these lines. 

The measurement of elastic $pp$ scattering at the highest available
energies is a 'must' at the LHC. This includes the measurement of 
$\sigma_{tot}$, $d\sigma_{el}/dt$ over the largest possible $t$ region 
(specially at small $t$-values), and, more generally, the 
study of the composition of the total cross section in terms of 
elastic, diffractive and inelastic contributions.  
These measurements represent a textbook example of forward physics. 
The observed rise of the total cross section at the ISR, the Tevatron  
and at HERA and its compatibility with unitarity has always been  a 
topic of central interest in particle physics.  One of the goals is 
the connection of  $pp$ scattering at collider energies with cosmic 
ray physics: we are now in the novel situation in which the LHC 
energies are within the cosmic ray energy domain and it is thus possible
to connect these two branches  of particle physics. 
The high energy run of LHC will allow to provide new data points in the cosmic ray spectrum. 
On the theoretical side the rise of the total cross section raises the question of unitarity, one of the basic principles of particle physics. How do $\sigma_{tot}$ and $\sigma_{el}$ reach their respective 
unitarity bounds? Is there an Odderon, as predicted by perturbative QCD?  
Theoretical answers cannot be obtained from perturbative calculations 
alone: there are important nonperturbative aspects in high energy 
forward 
scattering that reside in the region of large impact parameter. 
In contrast to the static potential of low energy QCD, in the high 
energy scattering of two hadrons both the profile function and the transverse energy composition are energy 
dependent, and their understanding within QCD therefore requires new tools. 
In this situation, experimental measurements are most important.            

The appearance of rapidity gaps in $pp$ scattering as well as the presence
of intact protons in the final state that can be measured - in particular when accompanied by a hard scale (jets or heavy particles) - 
is part of the complicated structure of multiple interactions. In contrast 
to deep inelastic scattering where, for diffractive final states, 
multiple interactions are strongly suppressed, in $pp$ scattering a rapidity gap 
in a single parton chain is likely to be filled by production from another chain. 
This leads to the suppression of rapidity gaps and the destruction of the 
scattered protons, leading to a suppression of the
visible diffractive cross section encoded in the survival factor $S^2<1$.      
The thorough measurement of final states with rapidity gaps and intact 
protons therefore serves as a valuable tool for understanding the 
event structure in $pp$ scattering. The most promiment examples 
include the single diffractive production of jets, $Z$ and $W$ bosons, 
as well as the central exclusive production reactions (double Pomeron exchange).  
These events allow to further constrain the Pomeron structure in terms
of quarks and gluons, as initially investigated at HERA, in the completely new kinematical domain reached at the LHC.
Diffractive final states originating from double Pomeron exchange attract
attention also from another perspective.
Double Pomeron exchange allows the formation of new states from pure 
gluons: the glueball states that have been under discussion for many 
years, heavy flavor states as well as beyond-standard-model objects. 
Tagging of the intact protons allows for a clean spin-parity analysis of the produced states.  
The presence of rapidity gaps between the protons and the centrally produced
system along with that of intact protons can also be due to photon exchange, i.e.
in such final states LHC serves as a $\gamma\gamma$-collider.
This opens the door to the electroweak sector, e.g. to the search for
anomalous couplings of vector bosons and photons.

Forward physics allows to address specific aspects of QCD dynamics 
that go beyond the collinear approximation, notably BFKL 
and small-x physics.  The BFKL Pomeron has been derived for the high 
energy scattering of partons, but its theoretical interest has 
become much broader, and now includes aspects of integrability and 
the connection with gravity and string theory.  Consequently there is a strong 
motivation to establish its existence in the real world of strong 
interactions. Already at HERA and at the Tevatron special final states 
were identified as providing potentially clear signals, most 
notably the Mueller-Navelet jets with a large rapidity separation 
between between two jets of comparable transverse momenta, 
and the so-called jet gap jet events, where two jets are separated
by a gap devoid of particles.
Such configurations have already been investigated in previous runs 
of the LHC (in particular, angular decorrelations),  but it has become clear that further evidence has to be searched for.    
Both the increase in energy and the recent theoretical developments strongly motivate new efforts.
For example, with the higher machine energy it will be possible to address, 
apart from the celebrated angular decorrelation between the jets, also 
the BFKL intercept: a comparison of Mueller-Navelet jets at different 
machine energies (7 TeV and 13 TeV) with fixed momentum fractions of the 
parton densities allows a direct measurement of the cross section as a 
function of the rapidity separation, i.e. the BFKL interecpt. 
Another BFKL related measurement that has not been carried out yet consists
of varying the transverse momenta of the two Mueller-Navelet jets: when the
momenta are of the same order, the BFKL Pomeron should dominate, whereas for
very different transverse momenta the DGLAP description applies.
BFKL dynamics can be tested also in another way.
With modern calculational tools it has become possible to address multiparton final states within the collinear factorization. In the region of large rapidities, these subprocesses 
generate logarithms of energy and thus can directly be compared with LO or NLO BFKL.
A new Monte Carlo (BFKLex) has been designed and developed specially for probing BFKL dynamics.
Interest in small-$x$ corrections to parton densities has been 
stimulated by deep inleastic $ep$ scattering at HERA and by heavy 
ion collisison, studied both at RHIC and in the previous LHC run. 
One of the prominent ideas is the saturation of gluon densities at 
small $x$ and low $Q^2$ that arises from multi-parton chains and 
their recombination.  At the LHC one of the most promising places for 
searching signals for saturation is the kinematic region very close to the forward directions, in particular the Drell-Yan production of lepton pairs. Here LHC energies allow to access a much larger
kinematic region than previous machines. More information on saturation is expected from the 
measurements of two-particle correlations: here it is mandatory to have rapidity intervals as large as possible.                

Understanding the sources and the propagation of cosmic rays are central questions of
astroparticle physics. While there is increasing evidence that supernova remnants accelerate 
cosmic rays up to energies of $\sim Z\times 10^{14}$ \UeV\ (with $Z$ the charge of
the cosmic ray nucleus), the sources of the particles of energies up to $10^{20}$ \UeV\ are not known. 
Cosmic ray physics needs a good understanding of $p$-air collisions in 
the forward directions.  Indeed, air shower simulations represent a 
key ingedient needed to analyze cosmic ray data. Monte Carlo generators 
developed for cosmic ray physics  (EPOS, PHOJET, QGSJET)  are already 
quite successful in describing $pp$ collisions in the forward direction and,
for further improvements, it will be useful to study proton-oxygen 
collisions at the LHC. LHC energies have 
now reached regions that are close to cosmic ray physics and thus will allow 
to understand and to fine-tune hadronic models used for air shower simulation.

Finally, we describe the relevance of forward physics for heavy
ion physics. Ultraperipheral collisions (UPC)  of 
nuclei (protons and nuclei) at the LHC provide a tool complementary 
to $pp, pA$ collisions for testing high energy QCD dynamics. For example, 
studies of UPC allow to measure nucleon and nucleus PDFs in a 
wide range of $x$ down to  $x\ge 10^{-5}$  for  much smaller virtualities 
than in $pp$ and  $pA$ collisions. Photon induced processes can also be probed in ion-ion and p-ion interactions
given the fact that the intensity of the photon flux grows as the square 
of the charge of the accelerated particle, implying that heavy ions 
are a more efficient source of photons than protons. 

We finish the document by describing the new detectors that are being or
will be installed at the LHC by the ALICE, ATLAS, CMS, LHCb, LHCf and TOTEM
collaborations. They will allow fulfilling the rich program of forward
physics mentioned above and described in detail in the document.

%% file: introduction/acceptance.tex
In this chapter, we describe briefly the acceptance of the forward
detectors in the ATLAS and CMS/TOTEM experiments, as well as the induced
backgrounds and the different running conditions at the LHC, that will
be used in the next chapters of this document.

\section{Acceptance of Forward Detectors}
\label{sec:soft:forwardacceptance}

In this chapter, the proton impact position at forward detector locations for various optics settings and at an energy of $\sqrt{s} = 14$ TeV is discussed in vicinity of the ATLAS Interaction Point (IP1). The detailed studies of the proton behaviour for other energies can be found in \cite{Trzebinski_optics}. Similar results are expected for the CMS-TOTEM Interaction Point.

The amount of data delivered by a collider is described by its instantaneous luminosity, which can be calculated as:
$$L = \frac{n \cdot N_1 \cdot N_2 \cdot f \cdot \gamma}{4 \cdot \pi \cdot \varepsilon \cdot \beta^*} F,$$
where $N_1$ and $N_2$ are the number of particles per bunch in beam 1 and 2, correspondingly, $n$ is the number of colliding bunches (beam pairs), $f$ is the beam revolution frequency, $\varepsilon$ is its emittance, $\beta^*$ is the betatron function at the Interaction Point (IP), $\gamma$ is beam Lorentz factor, and $F$ is the geometric luminosity reduction factor due to the crossing angle at the Interaction Point:
$$F = \left( 1 + \left( \frac{\theta_c^*\,\sigma_z^*}{2\,\sigma^*} \right)^2 \right)^{-1/2},$$
where $\theta_c^*$ is the crossing angle, $\sigma_z^*$ -- the bunch length, and $\sigma^*$ -- the transverse beam size\footnote{In this section, asterisk denotes values at the Interaction Point.}. The crossing angle is introduced in order to avoid unwanted parasitic interactions, \textit{i.e.} when the bunches collide with each other away from the IP.

In terms of the accelerator optics, the value of the betatron function, $\beta$, at a point is the distance from this point to the next at which the beam is twice as wide. The lower the value of the betatron function at the IP ($\beta^{*}$), the smaller the beam size is ($\sigma \sim 1/\sqrt{\beta}$), and thus the larger the instantaneous luminosity is. During standard data taking at the LHC, one tries to decrease the value of $\beta^{*}$ in order to maximise the delivered luminosity. These settings are commonly referred as the collision optics. Such an approach introduces large pile-up, which, as will be shown later in this report, makes diffractive measurements very difficult, if not impossible. Therefore one would like to have a few runs dedicated to the studies of diffraction. In such runs the magnets settings may be unchanged, but the proton population in bunches should be decreased, in order to keep the pile-up at reasonably low level.

Processes at very small $|t|$ such as elastic scattering require a dedicated machine configuration, known as \textit{high-}$\beta^*$ optics. The modifications include:
\begin{itemize}
  \item a high value of the betatron function, which implies a very low beam angular divergence (angular momentum spread) at the IP,
  \item low intensity bunches, needed to minimise the intra-beam scattering effects and to avoid an additional proton transverse momentum smearing,
  \item small number of bunches, to operate without a crossing angle,
  \item parallel-to-point focusing -- a special feature obtained with a phase advance of $\psi = \pi/2$ to the forward detectors that causes the protons scattered at the same angle to be focused at the same point in the forward detector (in case of the discussed ALFA detectors such focusing occurs in the vertical, $y$, coordinate),
\end{itemize}

Another important parameter is the beam emittance, $\varepsilon$, which is a measure of the average spread in the position-momentum phase space. The LHC has been designed to obtain $\varepsilon = 3.75\,\mu$m$\cdot$rad, but due to its outstanding performance this value was about $2\ \mu$m$\cdot$rad the average during Run 1. In the following, the design value of the emittance is used in the calculations of the beam properties around the forward detectors, whereas the actual one is employed when the beam behaviour at the IP is computed. Such an approach is consistent with the one of the LHC machine group and the real experimental conditions.

The beam sizes and the beam divergence (momentum angular spread) at the ATLAS Interaction Point for various LHC optics are listed in \Tref{tab_beam_size_IP}. 
These results were obtained using the \textsc{MAD-X} program \cite{MADX, PTC}, input with the relevant LHC optics files {\cite{optic_files}}.

\begin{table}[!htbp]
  \caption{LHC beam transverse size and beam divergence at the ATLAS IP for $\sqrt{s} = 14$ TeV, various $\beta^{*}$ optics modes and emittance values.}
  \label{tab_beam_size_IP}  
  \begin{center}  
    \begin{tabular}{c | c c | c c }
    \toprule
    \multirow{2}{*}{$\beta^*$ [m]} & \multicolumn{2}{c}{beam transverse size [mm]} &  \multicolumn{2}{c}{beam divergence [MeV]} \\
    & $\varepsilon = 2\ \mu$m$\cdot$rad & $\varepsilon = 3.75\ \mu$m$\cdot$rad & $\varepsilon = 2\ \mu$m$\cdot$rad & $\varepsilon = 3.75\ \mu$m$\cdot$rad\\
    \midrule
0.55  & 0.012 & 0.017 & 150   & 210\\
90  & 0.16  & 0.21 & 12  & 17 \\
1000  & 0.52  & 0.71 & 3.6 & 5.0 \\
    \bottomrule
  \end{tabular}
  \end{center}
\end{table}

The beam size at the forward detector's location determines the minimum distance from the beam to which the detectors can be safely inserted. Its knowledge is important for both the event simulations and data analysis, as it defines the kinematic regions that are accessible for a given optics settings. The results for the AFP and ALFA cases are listed in \Tref{tab_beam1_size}. It is worth recalling that although beam 1 and beam 2 are not identical, the differences in their transverse size at the location of forward detectors are negligible. For the ALFA(AFP) detectors only $y$($x$) width is meaningful since they approach the beam in the vertical(horizontal) plane.

\begin{table}[htbp]
  \caption{LHC beam size in $x$ at AFP and in $y$ at ALFA stations for different $\beta^{*}$ optics modes for nominal and low emittance. Calculations were done for $\sqrt{s} = 14$ TeV.}
  \label{tab_beam1_size}  
  \begin{center}
    \begin{tabular}{c | c c c c c c c c c}
     \toprule
    $\beta^*$ & \multicolumn{2}{c}{$\sigma^{204}_{x}$ [mm]} & \multicolumn{2}{c}{$\sigma^{212}_{x}$ [mm]} & \multicolumn{2}{c}{$\sigma^{237}_{y}$ [mm]} & \multicolumn{2}{c}{$\sigma^{245}_{y}$ [mm]}\\
    \mbox{}[m] & $\varepsilon = 2$ & $\varepsilon = 3.75$ & $\varepsilon = 2$ & $\varepsilon = 3.75$ & $\varepsilon = 2$ & $\varepsilon = 3.75$ & $\varepsilon = 2$ & $\varepsilon = 3.75$\\
\midrule
0.55    & 0.14  & 0.19& 0.10  & 0.14  & 0.21  & 0.28  & 0.17  & 0.23  \\
90    & 0.43  & 0.59  & 0.36  & 0.49  & 0.48  & 0.66  & 0.44  & 0.60  \\
1000    & 0.56  & 0.76  & 0.48  & 0.65  & 0.17  & 0.23  & 0.16  & 0.22  \\
\bottomrule 
    \end{tabular}
  \end{center}
\end{table}

For all the measurements that are possible with forward detectors, it is crucial to understand the connection between the scattered proton momentum and the position in the detector. This is usually expressed in terms of the geometric acceptance, which is defined as the ratio of the number of protons with a given relative energy loss ($\xi$) and transverse momentum (\pt) that reached the detector to the total number of the scattered protons. Obviously, not all scattered protons can be measured in forward detectors as they can be too close to the beam to be detected or can hit some LHC element (a collimator, the beam pipe, a magnet) upstream of the detector. In the calculations presented below, the following factors were taken into account:
\begin{itemize}
  \item beam properties at the IP,
  \item beam pipe aperture,
  \item properties of the LHC magnetic lattice,
  \item detector geometry,
  \item distance between the beam centre and the active detector edge.
\end{itemize}

\begin{figure}[!htbp]
\centering
  \includegraphics[width=0.32\textwidth]{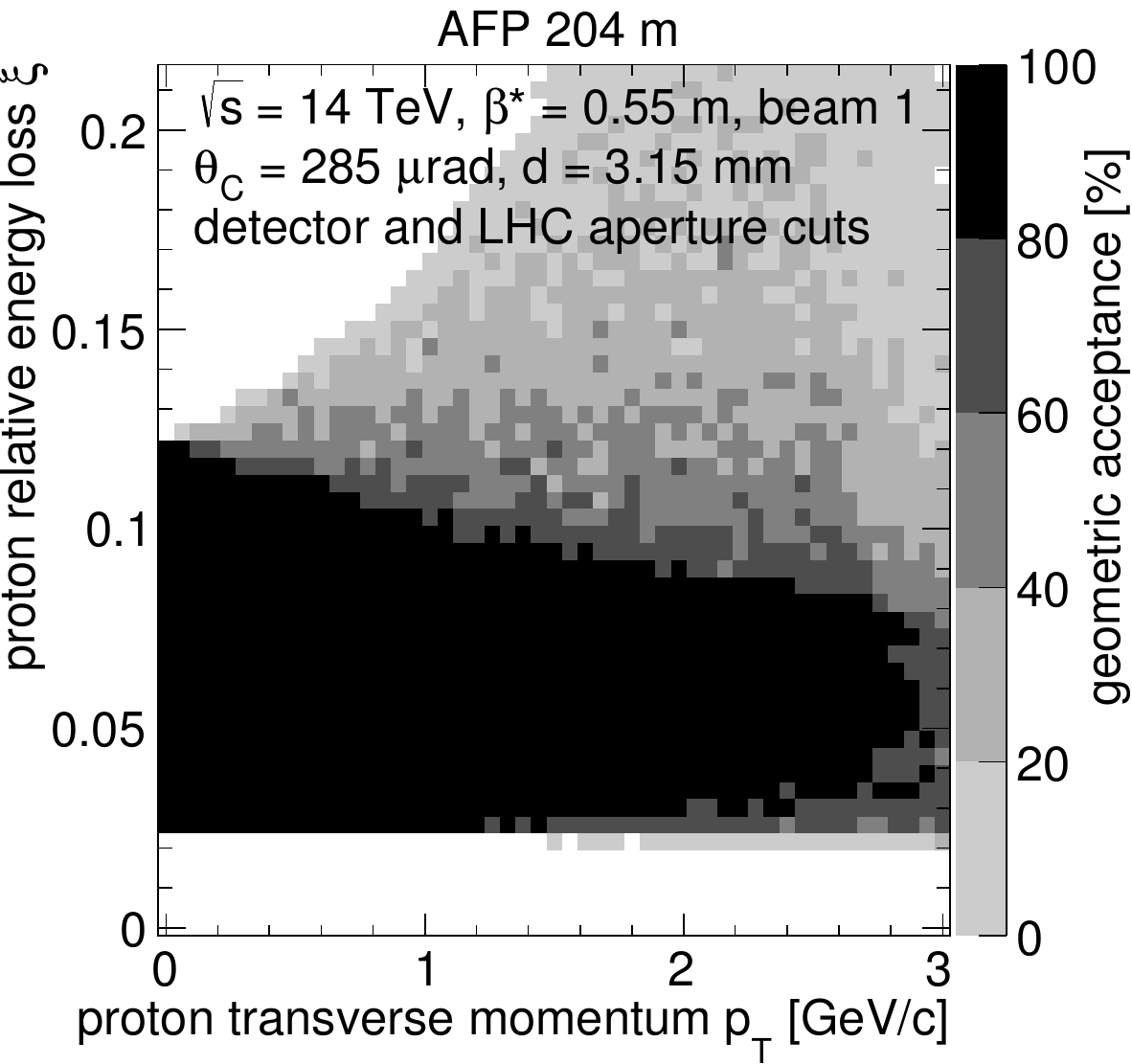}
  \includegraphics[width=0.32\textwidth]{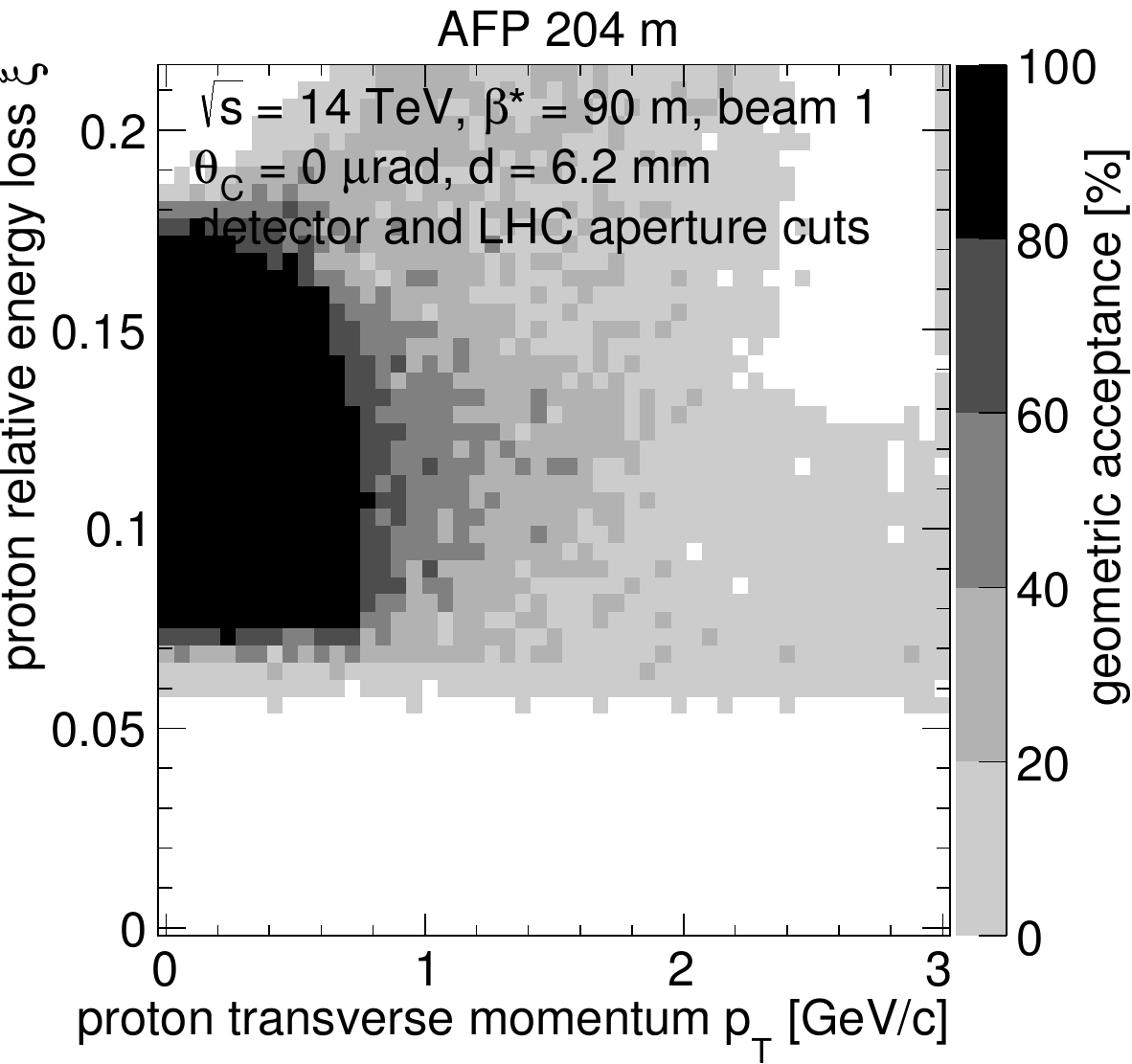}
  \includegraphics[width=0.32\textwidth]{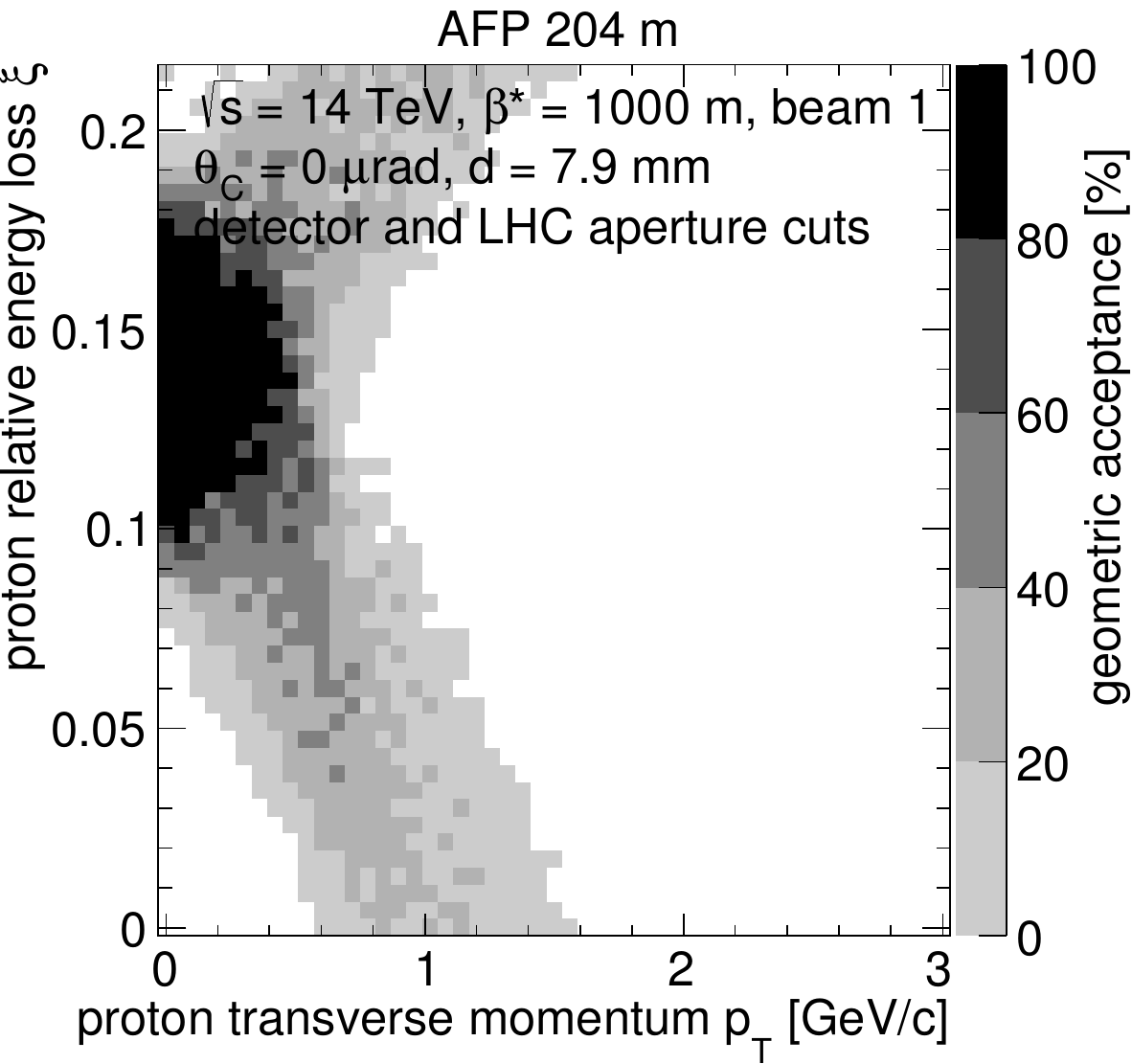}
\caption{Geometric acceptance of the AFP detector as a function of the proton relative energy loss ($\xi$) and its transverse momentum ($\pt$) for different LHC optics settings. The beam properties at the IP, the beam chamber and the detector geometries, the distance between the detector edge and the beam centre were taken into account. The beam energy was set to 7 TeV and the distance from the beam is calculated taking into account the nominal emittance value of 3.75~$\mu$m$\cdot$rad and 0.3 mm of dead material.}
\label{fig_acceptance_AFP}
\end{figure}

The geometric acceptance of the first AFP station (planned to be installed 204 m from the IP1) for $\sqrt{s} = 14$ TeV is shown in \Fref{fig_acceptance_AFP}. The distance from the beam centre was set to 15 $\sigma$ for the collision optics, to 10 $\sigma$ for the \textit{high-}$\beta^*$ ones (\textit{cf.} \Tref{tab_beam1_size}). In order to account for the dead material of the Roman Pot window a 0.3 mm distance was added in all cases.

For collision optics, the region of high acceptance is limited to $\pt < 3$ GeV and $0.02 < \xi < 0.12$. These limits change to $\pt < 1$ GeV and $0.07 < \xi < 0.17$ and $0.1 < \xi < 0.17$ for $\beta^* = 90$ and $1000$ m optics, correspondingly.

\begin{figure}[!htbp]
\centering
  \includegraphics[width=0.32\textwidth]{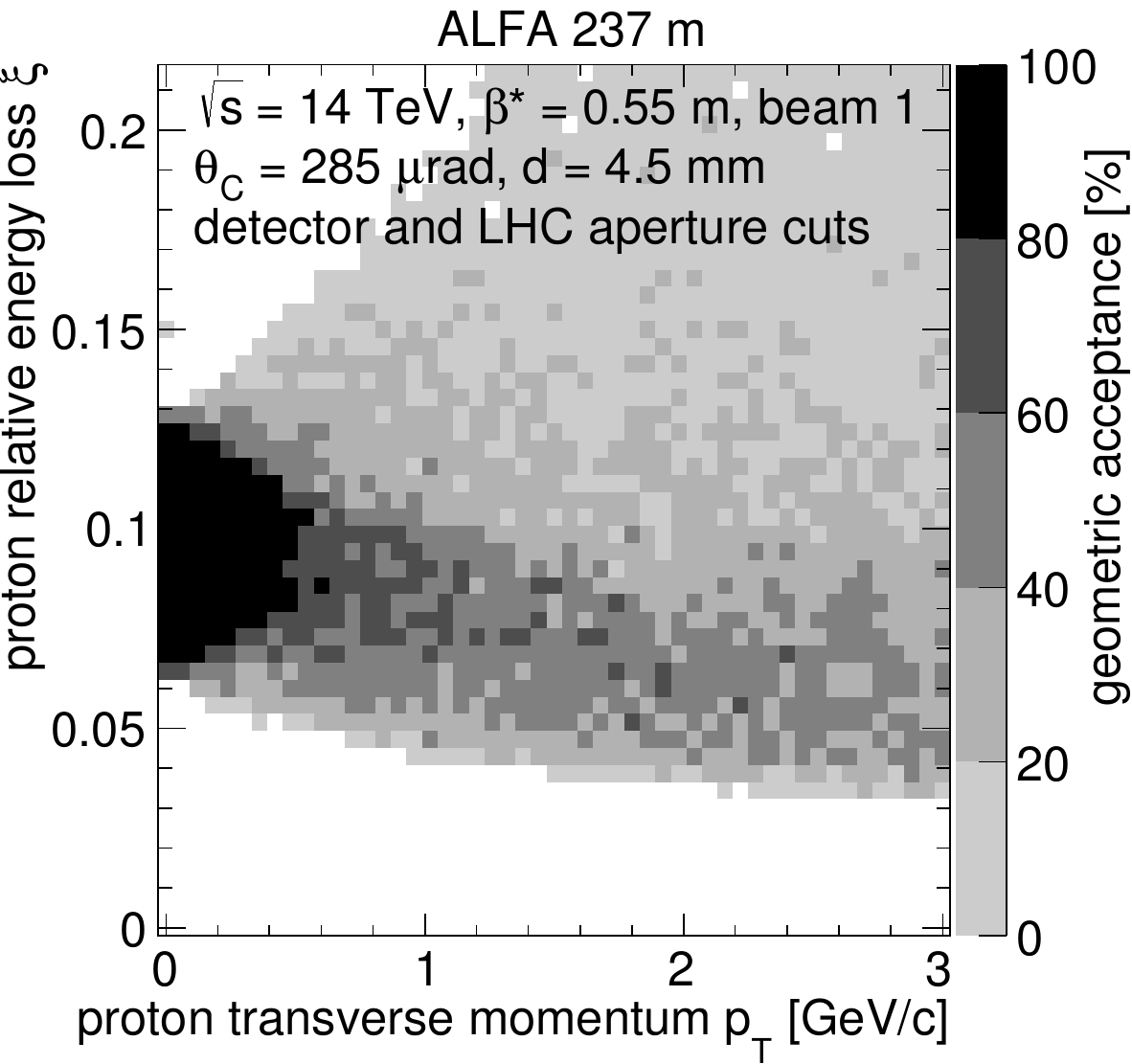}
  \includegraphics[width=0.32\textwidth]{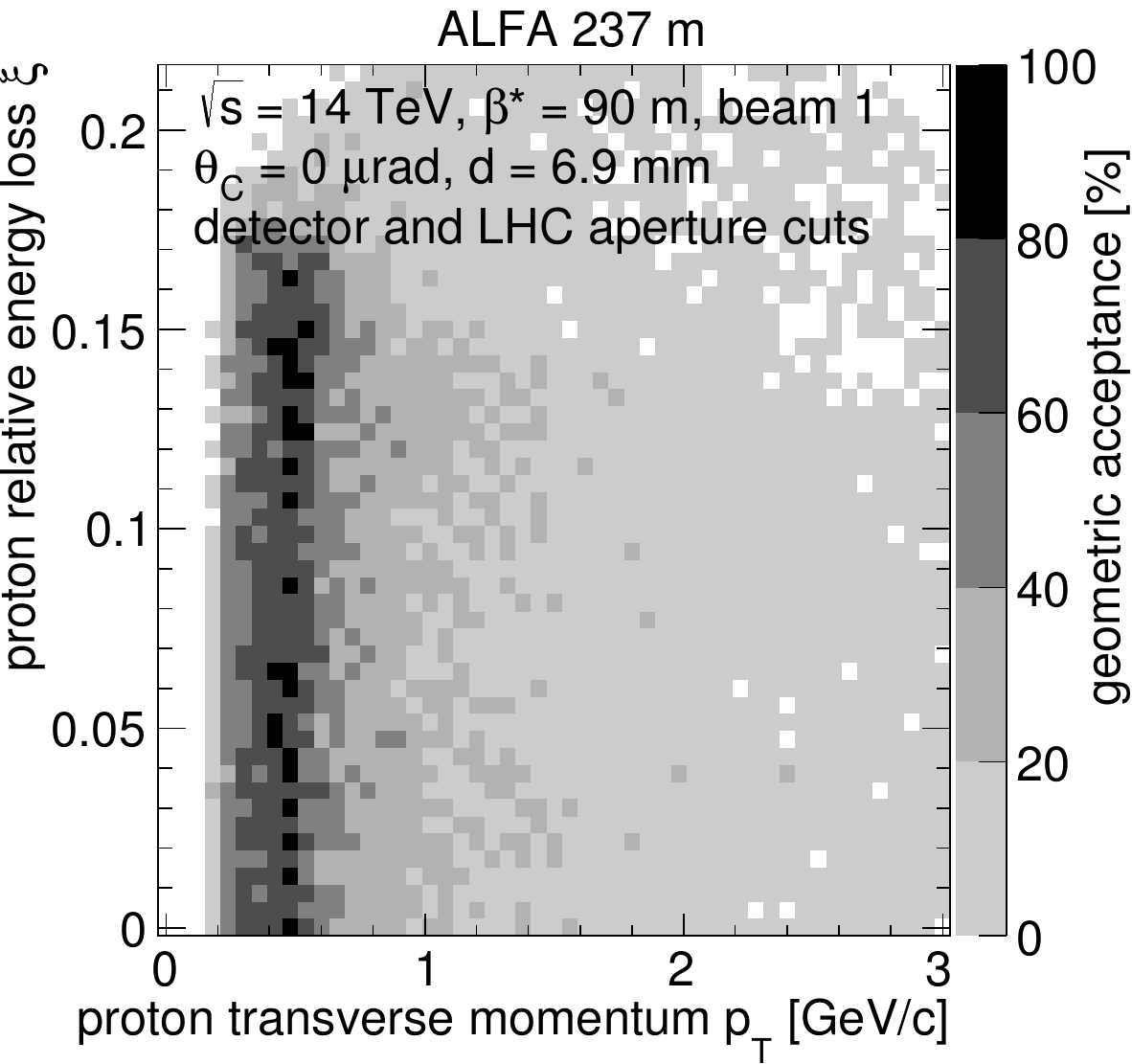}
  \includegraphics[width=0.32\textwidth]{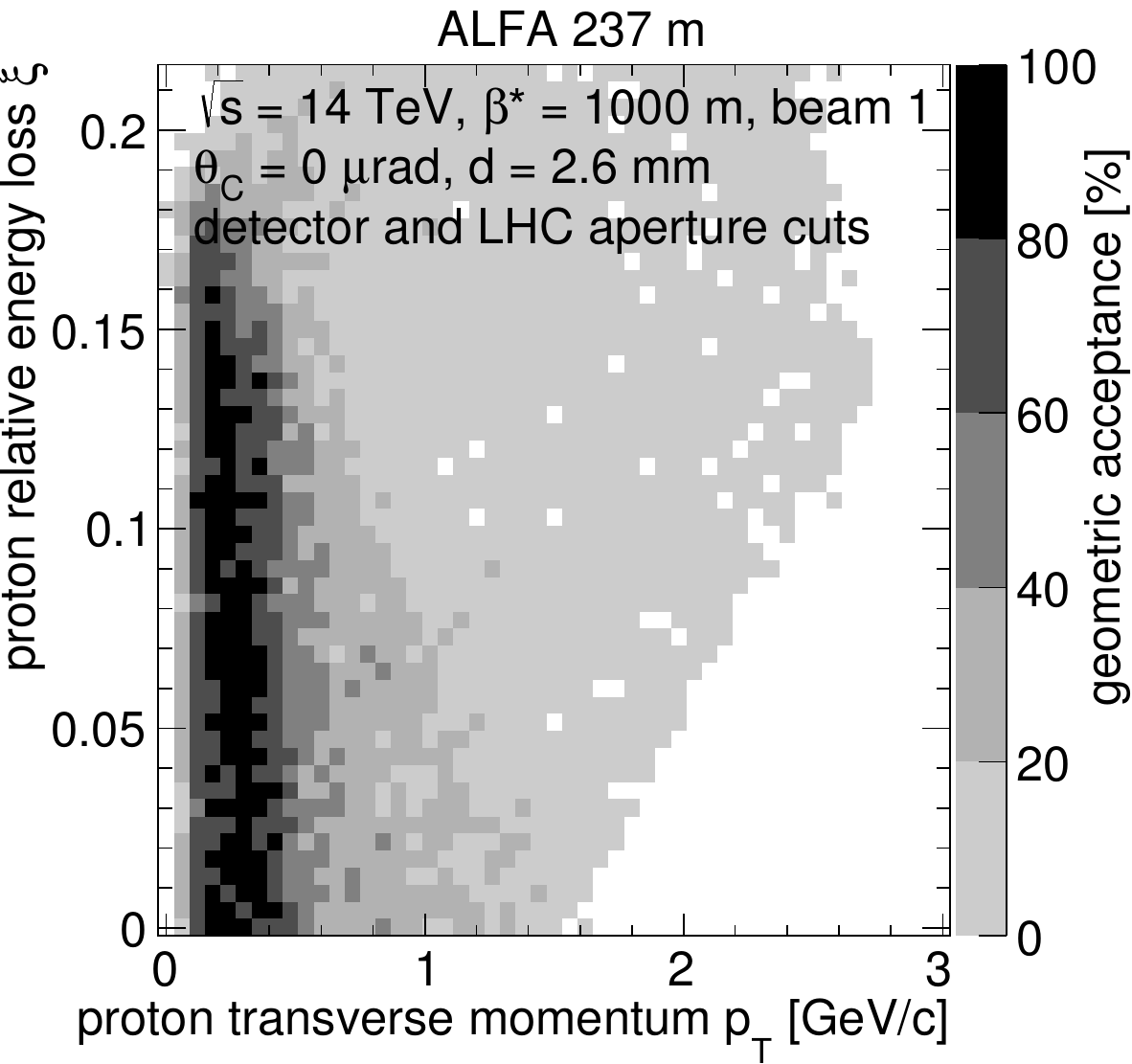}
\caption{Geometric acceptance of the ALFA detector as a function of the proton relative energy loss ($\xi$) and its transverse momentum ($\pt$) for different LHC optics settings. The beam properties at the IP, the beam chamber and the detector geometries, the distance between the detector edge and the beam centre were taken into account. The beam energy was set to 7 TeV and the distance from the beam is calculated taking into account the nominal emittance value of 3.75~$\mu$m$\cdot$rad and 0.3 mm of dead material.}
\label{fig_acceptance_ALFA}
\end{figure}

The results for the first ALFA station (located 237 m from the IP1) are shown in \Fref{fig_acceptance_ALFA}. For this case the distance from the beam centre was set to 15 $\sigma$ for collision optics and to 10 $\sigma$ for the \textit{high-}$\beta^*$ ones and a 0.3 mm of dead material was added.

For collision optics the region of high acceptance ($>80\%$) is limited by $\pt < 0.5$ GeV and $0.06 < \xi < 0.12$, which is significantly smaller than in case of the AFP detectors. The picture changes drastically when \textit{high-}$\beta^*$ optics is considered, as these settings are optimised for the elastic scattering measurement in which access to low $\pt$ values for $\xi = 0$ is crucial. One should also note that the limit on the minimum value of the proton $\pt$ decreases with the increase of the $\beta^*$ value. In other words, the higher the $\beta^*$ is, the smaller $t$ values are accessible. It is worth mentioning that the lower value of accessible $\pt$ depends on the distance between the beam centre and the detector edge as was shown \cite{Trzebinski_transport}.

\begin{figure}[htb]
\centering
\includegraphics[height=2.7in, width=0.49\linewidth]{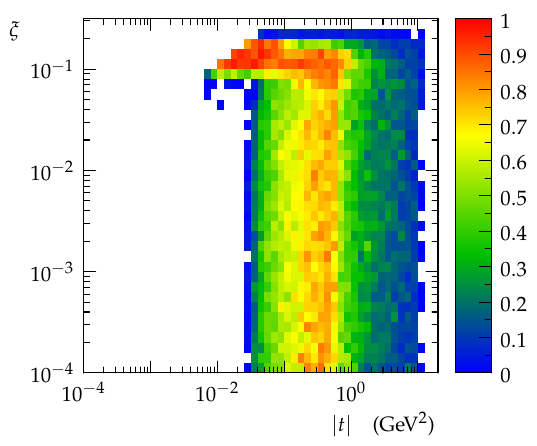}
\includegraphics[height=2.7in, width=0.49\linewidth]{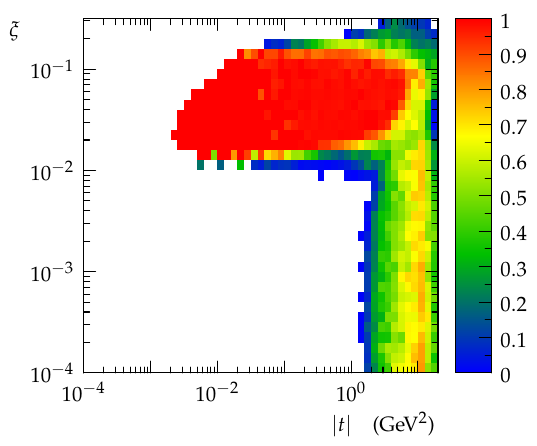}
\caption{Geometric acceptance of the TOTEM-RP detectors (vertical and horizontal) as a function of the proton relative energy loss ($\xi$) and its squared four-momentum transfer ($t$) 
for different LHC optics settings (left, $\beta^*$=90\,m; right, low-$\beta^*$),
at the beam energy of 7 TeV and at detector distance from the beam corresponding to 10$\sigma$+0.5\,mm.}
\label{fig:totemacc}
\end{figure}

Similar considerations can be done for the acceptance in the leading proton detectors in IP5.
The detailed acceptance studies of the TOTEM and CT-PPS detectors have been published already elesewhere (\cite{totem-upgrade-tdr,ctpps-tdr}).
As example in Fig.~\ref{fig:totemacc} is shown the acceptance for high (90\,m) and low beta optics, for both vertical and horizontal detectors.

%% file: introduction/background.tex
\section{Background: pp induced background}
\label{sec:machinebackground}

In addition to the genuine physics processes from the hard interaction or from pileup events,
``machine-induced'' backgrounds mainly due to beam halo or secondary particles must be taken into account, as
more than one track per bunch crossing can arrive to the RP.
The electronics associated to the timing detector can measure without ambiguity only the traversing time of one particle
per bunch crossing, hence the detector plane must be properly segmented
as the deterioration of the timing detector resolution has a direct impact on
the background suppression.

The contribution of the
beam-related background has been added to the background from the physics interactions in many studies presented in this report.

The machine-induced background contribution at z=220\,m is estimated 
by extrapolating the TOTEM measurements at $\sqrt{s}=8$~TeV during Run-I to the Run-II data-taking conditions.  

\noindent
The beam-related background has two components: the ``collision debris" and the ``beam
halo" background. The ``collision debris"  contains particles from showers generated in the
vacuum pipe aperture limitations that eventually produce a signal in the RPs. This fraction
of the background scales with $\mu$ (defined as the mean number of inelastic interaction per bunch crossing)
as the number of vertices (pile-up) generated in the bunch crossing.
The ``beam halo" contribution is due to beam protons travelling far from the central
beam orbit and hitting the RPs; this contribution is expected to scale with the beam current
($\approx \sqrt\mu$). 
\footnote{In fact $I_{beam} \propto n_{bunch} N_{proton}$, where $n_{bunch}$ is the number of bunches in the LHC ring and $N_{proton}$ is the number of protons in a bunch,  while the pile-up is proportional to $N^2_{proton}$.}. 
In this study the background is calculated per bunch crossing and the effective scaling is done based on the parameter $\mu$.

Different approaches have been used to understand how to extract the background component from the
data and how to extrapolate it to higher pile-up conditions.
The detailed procedure is described in~\cite{Mirko:2014}.
It can be summarized as in the following:
the background probability per bunch-crossing is estimated from a zero-bias data sample (random trigger on bunch crossing). 
which includes all events, from both background and physics processes.
In order to subtract the contribution from physics processes, the multiplicity of the leading protons
reconstructed in the pots is estimated with a dedicated sample of simulated events (without pileup) for the very high pile-up case (low-$\beta^*$) or 
by using the information of the T2 telescope for moderate pile-up (high-$\beta^*$).
By comparing the multiplicity of the primary tracks with the average cluster multiplicity per detector plane from
the data (zero-bias data sample) it is then possible to subtract this contribution, and to extract 
the probability distribution of the background per bunch-crossing  as well as its spacial distribution in the detector.

%90m
In the  high-$\beta^*$ scenario the beam-beam background has been estimated by selecting events with tracks
in both arms of T2: in this sub-sample the probability to have at least a cluster in the RPs for
events without elastic candidates was found to be 1.5\%. In this estimate the contribution of the
high-mass diffraction is already subtracted (about 0.5\%).
The beam halo contribution was calculated as the probability to have a proton track reconstructed in the vertical RPs when both T2 arms are empty and no elastic signature is present (i.e. no collinear protons on the other arm).
The estimate is conservative and probably overestimates the beam-halo, as the selection includes contributions from low mass SD (no signal in T2 with possibly a single proton in the RPs acceptance) and a small fraction of elastic events with a proton on 
one arm escaping the detection (due to smearing and edge effects).
This background, assumed to scale with $\sqrt{\mu}$, is $\approx 2-3\%$ for each vertical RP in condition with $\mu=$0.5.

In conclusion, the beam-beam background probability estimated for a scenario with high-$\beta^{*}$ and $\mu=$0.5 is about 3\% per BX.

In the  low-$\beta^*$ scenario, the probability per BX to have an additional track due to the beam-beam background has been found to be 80\% at $\mu$=50.

\begin{figure}[htb]
\centering
\includegraphics[height=2.7in, width=0.99\linewidth]{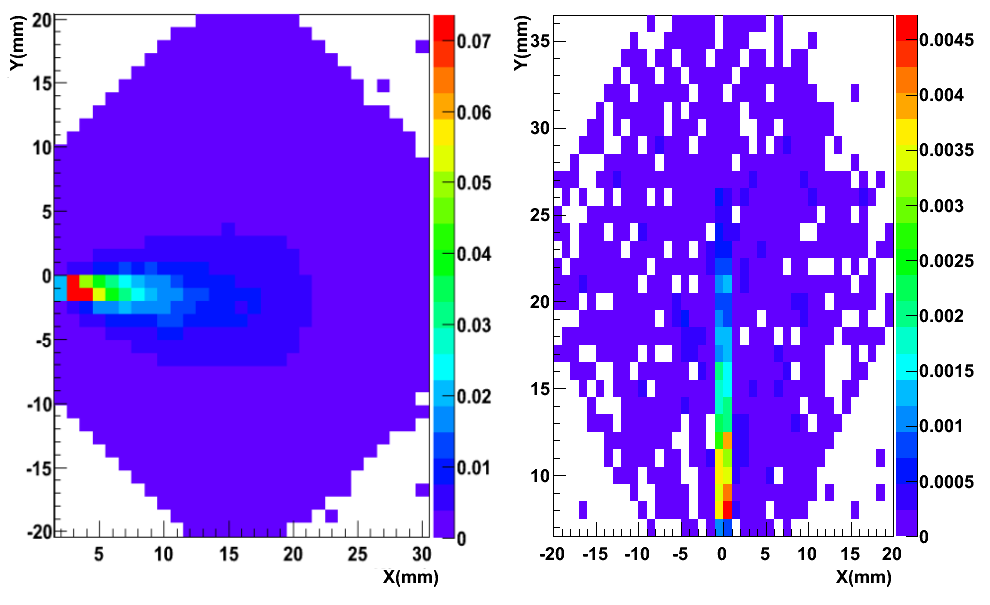}
\caption{Left: occupancy/BX*mm$^2$ in the horizontal RP (low-$\beta^{*}$, $\mu=50$, 6$\sigma$ approach). Right: occupancy/BX*mm$^2$ in the vertical RP ($\beta^{*}=$90m, $\mu=0.5$, 9.5$\sigma$ approach).
Not included in the plot the corrections factor 2 (1.2) which accounts for multiple tracks inefficiency (see text).  
}
\label{OccupLowHighB}
\end{figure}

The extrapolated occupancies are shown in Figure~\ref{OccupLowHighB} for $\mu=50, 0.5$. 
The occupancy values reported in Figure~\ref{OccupLowHighB} are not corrected by a factor 2-1.2 to account for the limited multi-track capability. 

The beam background estimation is necessary for a proper optimization of the timing detector design: this extrapolation has been used for detectors development in the TOTEM Timing Upgrade
TDR~\cite{totem-upgrade-tdr} and in the CT-PPS TDR~\cite{ctpps-tdr}.
Moreover several studies have been performed to understand the impact of such background on the physics process selection (see Chapter~\ref{chap:cep})
and on the trigger optimization for high luminosity runs~\cite{Mirko:2014,ctpps-tdr}.

%% file: introduction/beamcond.tex
\section{Different running conditions}

We briefly describe in this section the different running scenarii at the LHC,
namely the low, medium and high luminosity runs (Table~\ref{tab:runcond}.

First we would like to stress that there is a complementarity between the low
and high $\beta^*$ measurements. At high $\beta^*$ in the forward detector
acceptance of ATLAS/ALFA and CMS-TOTEM it is possible to be sensitive to very
low $\xi$ values and thus to small diffractive masses, which corresponds to much
higher cross section with respect to high diffractive masses. The small amount
of luminosity available at high $\beta^*$ will this be enough to fulfill the
diffractive program at low masses. \\
On the contrary, at low $\beta^*$, the
horizontal roman pots of ATLAS/AFP and CT-PPS will be needed and the acceptance
is better at high $\xi$ and thus high diffractive masses. The cross section for
such processes is much smaller but high luminosity of the order of 100s of
fb$^{-1}$ will be available, allowing even searching for beyond standard model
effects. In such high luminosity and high pile up conditions, the rapidity gap
method to detect diffractive events in ATLAS/CMS is impossible. \\
An intermediary
case is with LHCb that can acculumate a reasonable amount of luminosity
(typically a few fb$^{-1}$) with little pile up and can use the rapidity gap
method to measure diffraction since the beam are partially defocused close to
LHCb. The Alice collaboration concentrates more on heavy ion and p-ion runs and
will measure diffraction in those runs where pile up is negligible and the
rapidity gap method can also be used.

The low luminosity runs (without pile up) allow performing multiplicity and
energy flow measurements useful to tune MC as well as to measure the total and
soft diffractive cross sections in the ATLAS/ALFA and TOTEM
experiments for a typical $\beta^*$ up to 1000 m (see chapter 3). 
Additional measurements such as single diffraction, low mass
resonances, glueballs, jet production in double Pomeron  are possible at 
non-zero but little pile up, for $\beta^*$ between 20 and 90 m.
The default configuration studied in this document for these runs is 
at high $\beta^* \sim 90$m and the vertical
forward detectors of ATLAS/ALFA and TOTEM can be used, together with the forward
detectors such as T1/T2 from TOTEM.
In a few days of data taking at a pile
up of $\sim$0.1, typically  5 to 10
pb$^{-1}$ can be acculmulated..

Medium luminosity runs are set-up specifically for the different 
LHC experiments. 
The CMS-TOTEM and ATLAS (ALFA and AFP) can accumulate low pile up
data in low and high $\beta^*$ special runs at low luminosity. 
It is then  possible to accumulate
10 to 100 pb$^{-1}$ at high $\beta^*$ with a pile up $\mu \sim$1 with a couple
of weeks
of data taking and about the same amount of luminosity at low $\beta^*$ with 
the same time scale at $\mu \sim$2 to 5. Let us mention again that both running
conditions are usueful since they access different kinematical domains, namely
small and large diffractive masses  The LHCb experiment can accumulate 
a few fb$^{-1}$ of data at small pile up..

High pile up data taking conditions means working at the same luminosity 
delivered to the experiments during standard data taking periods. The 
conditions in ATLAS and CMS are such to have a pile up $\mu$ between 20 and 
100. It is also possible to
collect data at a lower pile up $\mu \sim$25 by restricting data
taking to end of store 
(we estimated that up to 40\% of the total luminosity can be collected in 
this way) or to use events  originating only from the tails of the vertex 
distribution, where the pile-up conditions are less severe. Typical luminosities
will be 100s of fb$^{-1}$ in such conditions.

\begin{table}[ht]
\centering 
\small
 \caption{Summary of the machine parameters for the different running conditions. } 
\begin{tabular}{|c|c|c|c|c|c|c|p{4cm}|}
\hline
\hline
Conditions & $\beta^*$  &  N  & N$_b$ & $\mu$  & L & L$_{int}$ &  Physics \\
 & [m] &  [10$^{11}$ p] & &  (pileup) &[cm$^{-2}$s$^{-1}$] &  [24h] &   \\
\hline
\multirow{5}{*}{LOW}    & & & & & & & \\
                        & $\geq$ 1000 & 0.7 & 2        & 0.004 & 10$^{27}$          & 0.1/nb    &  $\sigma_{tot}$; Coulumb region \\
                        & 19          & 0.1 & 40       & 0.01  &  5$\cdot$10$^{28}$ & 4.8/nb    & Lhcf Run; Multiplicity;  Energy flow; Inelastic cross section \\
\hline
\multirow{6}{*}{MEDIUM} & & & & & & & \\
                        & 19          & 0.7 & 40       & 0.4   &  2$\cdot$10$^{30}$ & 0.17/pb      & High cross section diffraction \\
                        & 90          & 0.7 & 156--700 & 0.1   &10$^{30}$--10$^{31}$& 0.2--1/pb & $\sigma_{tot}$; low mass diffraction; Hard diffraction\\
                        & 90          & 1.5 &      700 & 0.6   &5$\cdot$10$^{31}$   & 4.4/pb    & Glueball searches; CEP \\                        
\hline
& & & & & & & \\
\multirow{2}{*}{HIGH}   & 0.5         & 1.15& 2800     &       &                    &           &    LHCb programme   \\
                        & 0.5         & 1.15& 2800     & 30    &         10$^{34}$  & 1/fb      &  Exclusive dijets, anomalous coupling      \\
 \hline
 \end{tabular} 
  \label{tab:runcond} 
\end{table}

%% file: montecarlo/montecarlo.tex
    \newcommand{\Shrimps}{S\protect\scalebox{0.8}{HRiMPS}\xspace}
    \newcommand{\NGW}{N_{\mathrm{GW}}}
    \newcommand{\dtwo}{{\rm d}^2} 
    \newcommand{\done}{{\rm d}}
    \newcommand{\Wabs}{\mathcal{W}_{\mathrm{abs}}}
    \newcommand{\smin}{s_{\mathrm{min}}}
    \newcommand{\qcut}{Q_{\mathrm{cut}}}
    \newcommand{\oforder}[1]{\mathcal{O}\left(#1\right)}
    \newcommand{\Sherpa}{S\protect\scalebox{0.8}{HERPA}\xspace}

    \renewcommand{\d}{\mathrm{d}}
    \newcommand{\units}[1]{\,\hbox{#1}}

    \newcommand{\egamma}{E_{\gamma}}
    \newcommand{\pb}{\units{pb}}
    
        \newcommand{\phojet}{\texttt{PHOJET}\xspace}
    \newcommand{\dpmjet}{\texttt{DPMJET~III}\xspace}
    \newcommand{\pythia}{\texttt{PYTHIA~6}\xspace}
    
            \def\pomwig{{\sc Pomwig} \hspace{.05cm}}
    \def\herwig{{\sc Herwig} \hspace{.05cm}}
    \newcommand{\PO}{I\!\!P}
    \newcommand{\RO}{I\!\!R}
\newcommand{\xpom}{x_{\PO}}
    
    \section{Introduction}
    
Monte Carlo (MC) simulations of high energy physics are an essential part of the LHC forward physics programme. Such simulations are important as a means to compare the available models of diffractive physics with LHC measurements, as well as a tool to tune to hadronic data and hence provide a phenomenological description of soft QCD effects, an understanding of which is essential for a wide range of high energy physics analyses, including searches for BSM physics. In addition these are crucial in the modelling of cosmic ray physics, as described in Chapter~\ref{chap:ion}. A large range of MC generators that deal with diffractive processes explicitly are available, many of which have been used in the experimental analyses described in this report, while conversely, these MC generators rely on future diffractive measurements at the LHC to constrain and improve the theoretical models contained within them. In Sections \ref{sec:epos}--\ref{sec:shrimps} some of the most widely used and up--to--date such MC generators for diffractive physics are described: the basics of the underlying theoretical models are summarised, and the outlook for the future is discussed, in particular in terms of the possibilities for and importance of future LHC measurements. Central exclusive production, discussed in Chapter~\ref{chap:cep}, requires a different theoretical approach to standard inclusive processes and is not currently included in the available general purpose MC event generators. A selection of MC generators that deal dominantly with this exclusive process are on the other hand available, which are discussed in Sections~\ref{sec:exhume}--\ref{sec:superchic}. This (non--exhaustive) list of MC generators for diffraction and CEP is intended to serve as a reference point for some results in this report, where these MC generators are used. Finally, in Section~\ref{sec:mctuning} a selection of comparison plots between LHC Run I diffractive measurements and MC predictions are shown: this serves as an indication of the way in which, already, such measurements can be of great use for MC tuning, with future data increasingly allowing differentiation between the model inputs.
    
%%%%%%%%%%%%%%%%%

%%%%%%   EPOS

%%%%%%%%%%%%%%%%

   \section{EPOS LHC}\label{sec:epos}
      
EPOS LHC~\cite{Pierog:2013ria:ch2} is a minimum bias MC hadronic
generator used for both heavy ion interactions and cosmic ray air shower 
simulations. It is based on \epos~1.99~\cite{Pierog:2009zt:ch2,Werner:2005jf:ch2}
retuned to reproduce LHC data on a higher precision level. \epos is based on a hadronic model which provides a consistent treatment of the cross section
calculation and particle production, taking into account energy conservation,
in both cases according to parton-based Gribov-Regge 
theory~\cite{Drescher:2000ha}. In this approach, the basic ingredient is
the purely imaginary amplitude of a single pomeron exchange, which is the sum of a
(parameterized, Regge--like) soft contribution  
$G_0(\hat{s},b)=\alpha_0(b)\hat{s}^{\beta_0}$
 and a semi-hard contribution based on the convolution of a soft pre-evolution (the part of the amplitude corresponding to a Regge--like soft evolution, from an arbitrary low virtuality Q$^2$ to the minimum hard scale Q$_0^2$ necessary to start the hard evolution),
 a DGLAP based hard evolution and a standard LO QCD 2$\to$2 cross section. The latter needs complex calculations but can be fitted to a simple
Regge-like term: $G_1(\hat{s},b)=\alpha_1(b)\hat{s}^{\beta_1}$. $\hat{s}=sx^+x^-$ 
is the fraction of energy (mass) carried by the pomeron and $b$ the impact
parameter of the nucleon-nucleon collision. Further details can be found 
in~\cite{Drescher:2000ha}.

Both cross sections and particle production are based on the total amplitude 
$G=\sum_iG_i$ via a complex Markov-Chain MC. The particle production process
has two main components. Firstly, there are the strings composed from pomerons (2 strings per 
pomeron, with ISR and FSR and the soft contribution from the non-perturbative pre-evolution, below the fixed scale $Q^2_0$, included); at high energy many pomerons can be exchanged in parallel in each event (MPI), covering the mid-rapidity part. Secondly, there are the remnants, which carry the remaining energy and quarks and mostly cover the fragmentation region. A remnant 
can be as simple as a resonance or a string elongated along the beam axis if 
its mass is high enough and is treated the same way for both diffractive and
non-diffractive events. Another particularity of \epos is that on an event-by-event basis,
% if the energy density
%is too high, string segments are merged to have a collective hadronization,
%with particle spectra generated according to statistical decay with an additional 
%flow~\cite{Pierog:2013ria:ch2}.
if the particle density of the secondaries produced by the string fragmentation is too high (more than about 3 or 4 hadrons per fm$^3$), then string segments are merged to form clusters. Clusters are subsequently decayed following the microcanonical ensemble with additional flow to mimic the particle spectra obtained after  hydrodynamical evolution and freezout hadronization (statistical collective hadronization).

    \subsection{Diffractive contribution}\label{epos:diff}

To generate inelastic events where new particles are produced, following standard AGK (Abramovski, Gribov, Kancheli~\cite{Abramovski73-e}) cutting rules configurations of cut (inelastic) pomerons (with amplitude G) and uncut (elastic) pomerons (with amplitude -G) are generated. Configurations having the same number of cut pomerons, and any number of uncut ones, belong to the same class of inelastic events. As a consequence a class of inelastic event is defined by its number of cut pomerons and the sum of all possible elastic (uncut) pomeron exchanges. 

A low mass diffractive event will be produced if only the 
remnants are excited and no inelastic (i.e. cut) pomeron is exchanged. To have such a
contribution consistently produced by the MC, a third term $G_2$ is added to the
total amplitude. Unlike the pomeron exchange discussed above, this diffractive exchange
will not produce central strings (except in the case of central diffraction) but
will allow the remnant to gain a heavier mass as some excited state. It can be 
defined as
\begin{equation}\label{g2}
G_{2}(x,s,b)=\alpha_{2}x^{-\alpha_{diff}}\exp\left\{ -\frac{b^{2}}{\delta_{2}(s)}\right\} 
\end{equation}
where $\alpha_{2}$ is a free parameter depending on the remnant type.
To use the same form as in the case of the soft pomeron, we have
\begin{equation}
\delta_{2}=4\cdot0.0389\cdot\left(R_{diff}^{pro}+R_{diff}^{tar}+\alpha'_{diff}\ln s\right)
\end{equation}
with 2 free parameters $R_{diff}^{rem}$ and $\alpha'_{diff}$.  Since $\hat{s}=M^2$, $\alpha_{diff}$
is fixed at 1 to have a mass distribution following the usual $1/M^2=\hat{s}^{-\alpha_{diff}}$. 
$\alpha_{2}$, $R_{diff}^{rem}$ and
$\alpha'_{diff}$ can be fixed by fitting all cross sections (total,
elastic, inelastic, single diffractive and elastic slope).\\

With $G$ defined as
\begin{equation}
G(x^{+},x^{-},s,b)=\sum_{i=0}^{2}G_{i}(x^{+},x^{-},s,b)
\end{equation}
it is possible to have a soft diffractive interaction if only $G_2$ is 
exchanged, while in the case of multiple interactions, $G_2$ can be produced together
with $G_0$ and/or $G_1$. In future developments, $G_2$ will be used to get the 
mass of the remnant in all cases (including non-diffractive events), while in 
\epos~LHC an independent $1/M^{2\alpha_{remn}}$ distribution is currently used to 
fix the mass of the excited remnants.

As a consequence, a high mass diffractive event will occur if a hard pomeron
is exchanged without remnant excitation. A free parameter is introduced to fix this probability.

    \subsection{Inclusive Cross Sections}

One fundamental quantity is the function $\Phi$, due to the contribution of all elastic 
pomeron (virtual) exchanges, which can be written as :
\begin{eqnarray}
\Phi\left(x^{\mathrm{proj}},x^{\mathrm{targ}},s,b\right) & = & \sum_{l=0}^{\infty}\int\, \prod_{\lambda=1}^{l}dx_{\lambda}^{+}dx_{\lambda}^{-} \;\frac{1}{l!}\,\prod_{\lambda=1}^{l}-G(x_{\lambda}^{+},x_{\lambda}^{-},s,b)\,\label{eq:pierog:phi}\\
 & \times & F_{\mathrm{remn}}\left(x^{\mathrm{proj}}-\sum_{\lambda}x_{\lambda}^{+}\right)F_{\mathrm{remn}}\left(x^{\mathrm{targ}}-\sum_{\lambda}x_{\lambda}^{-}\right).\nonumber \;,
\end{eqnarray}
where $x^{\mathrm{proj}}$ and $x^{\mathrm{targ}}$ are the momentum fractions not used in
inelastic pomeron exchange and $ F_{\mathrm{remn}}$ is a vertex function with 
the remnant to guarantee energy conservation ($\sum_{\mathrm{ine}}x_{\mathrm{ine}}+\sum_{\mathrm{ela}}x_{\mathrm{ela}}<1$).

\begin{figure}
    \centering\includegraphics[width=.5\linewidth,angle=-90]{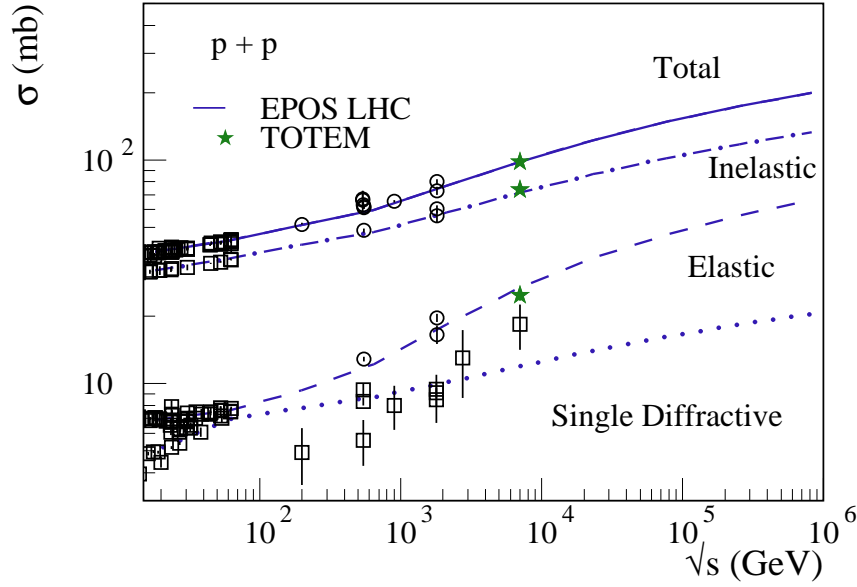}
    \caption{Total, inelastic, elastic and single diffractive {\it p-p} cross section calculated with EPOS~LHC. Points are data from~\protect\cite{Caso:1998tx} and the stars are the LHC measurements by the TOTEM experiment~\protect\cite{Csorgo:2012dm:ch2}.}
    \label{fig:pierof:sigma}
    \end{figure}

For $x^{\mathrm{proj}}$=$x^{\mathrm{targ}}=1$, the $\Phi$ function can be seen as 
the probability to have only elastic pomeron exchange without any new particles
produced, for a given impact parameter $b$. This then leads to the inelastic 
cross section definition
\begin{equation}
\sigma_{\mathrm{ine}}(s)=\int d^{2}b\,\left(1-\Phi(1,1,s,b)\right)\;.
\end{equation}
An elastic scattering corresponds to the sum of 
elastic pomeron exchanges, with at least one exchange, which
can be written as
\begin{equation}
\sigma_{\mathrm{ela}}(s)=\int d^{2}b\,\left(1-\sqrt{\Phi(1,1,s,b)}\right)^2\;,
\end{equation}
and then
\begin{equation}
\sigma_{\mathrm{tot}}(s)=\sigma_{\mathrm{ine}}(s)+\sigma_{\mathrm{ela}}(s)=2\int d^{2}b\,\left(1-\sqrt{\Phi(1,1,s,b)}\right)\;.
\end{equation}
The elastic slope $B$ ($d\sigma_{\mathrm{ela}}/dt=A\exp(+Bt)$ for $t \to 0^-$)  is then
\begin{equation}
B=\frac{1}{2}\frac{\int db^{2}b^{2}\left(1-\sqrt{\Phi(1,1,s,b)}\right)}{\int db^{2}\left(1-\sqrt{\Phi(1,1,s,b)}\right)}\;.
\end{equation}
All free parameters entering in the definition of $G$ and $ F_{\mathrm{remn}}$ can 
be tuned by a combined fit of all hadronic cross sections 
(Fig.~\ref{fig:pierof:sigma}), particle multiplicity and the proton
structure function $F_2$ (including the $Q^2$ independent correction at high 
energy/high mass needed to reduce the rise of the cross sections). For this reason minimum bias measurements
are important for more exclusive channels, where special configurations
of the amplitudes $G_i$ are tested. 
%The use
%of a different saturation scale $Q_s^2$ for each pomeron is subject of
%the current development of \epos.

    \subsection{Diffractive Cross Sections}

The diffractive cross section is now defined as that due to at least one inelastic exchange $G_{2}$,
but with no other inelastic contribution; this can not be calculated 
analytically. Since $G_{2}$ is only dominant for $\sum x<\!\!<1$, we can
write
\begin{eqnarray}
\sigma_{\mathrm{diff}}(s,b) & \sim & \Phi\left(1,1,s,b\right)\\
 & \times & \left[\sum_{m=0}^{\infty}\frac{1}{m!}\prod_{\mu=1}^{m}\int dx_{\mu}^{+}dx_{\mu}^{-}G_{2}(x_{\mu}^{+},x_{\mu}^{-},s,b)-1\right]\nonumber \\
 & \sim & \Phi\left(1,1,s,b\right)\left[\exp\left\{ \int dx^{+}dx^{-}G_{2}(x^{+},x^{-},s,b)\right\} -1\right]\;.
\end{eqnarray}
%or in practice, comparing the formula with MC results to fix
%$MCorr$
In practice, a parameter $MCorr$ is introduced to evaluate the diffractive cross section without making such an approximation. The numerical value of $MCorr$ can be fixed by a fit to the cross section obtained from exact MC simulations (equivalent to numerical integration) using
\begin{equation}
\sigma_{\mathrm{diff}}(s)=\int d^{2}b\,\left(\Phi\left(1,1,s,b\right)\left[\exp\left\{ \frac{\chi}{MCorr}\right\} -1\right]\right)\;,
\end{equation}
with $\chi=\int dx^{+}dx^{-}G_{2}(x^{+},x^{-},s,b)=\frac{\alpha_{2}}{(1-\alpha_{diff})^{2}}\exp\left\{ -\frac{b^{2}}{\delta_{2}(s)}\right\} $.

Then defining the probability $R_{\mathrm{pro}}$ and $R_{\mathrm{tar}}$  to have 
projectile or target excitation respectively, the single diffractive cross
section can be written as
\begin{equation}
\sigma_{\mathrm{sd}}(s)=R_{\mathrm{pro}}.(1-R_{\mathrm{tar}}).\sigma_{\mathrm{diff}}(s)+(1-R_{\mathrm{pro}}).R_{\mathrm{tar}}.\sigma_{\mathrm{diff}}(s)\;,
\end{equation}
and as a consequence the double diffractive cross section is simply
\begin{equation}
\sigma_{\mathrm{dd}}(s)=R_{\mathrm{pro}}.R_{\mathrm{tar}}.\sigma_{\mathrm{diff}}(s)\;,
\end{equation}
and the low mass (soft) central diffraction cross section is
\begin{equation}
\sigma_{\mathrm{cd}}(s)=(1-R_{\mathrm{pro}}).(1-R_{\mathrm{tar}}).\sigma_{\mathrm{diff}}(s)\;.
\end{equation}
For central diffraction, since none of the remnants are excited, two strings without remnant
connections are used to produce particles at mid rapidity (with two rapidity 
gaps). This is very similar to the method used to treat high mass diffraction
but without any hard contribution (only soft strings). A better treatment of 
central diffraction with resonance production is under development.

Future LHC measurements of diffractive mass and rapidity gap distributions
are extremely important to further constrain the parameters of the model for
both low mass (soft) diffraction and high mass (hard) diffraction.

%%%%%%%%%%%%%%%%%%%

%%%%%%  PHOJET

%%%%%%%%%%%

     \section{PHOJET}

\phojet is a MC event generator \cite{Engel:1994vs,Engel:1995yd} designed for simulating soft
and semi--hard hadronic
interactions, suited for describing accelerator events selected with minimum bias triggers.
Special care is taken to have a self--consistent model for all partial cross sections, including
the interplay of soft, hard, as well as diffractive and non--diffractive interactions~\cite{Engel:1994vs}.
Each inelastic
interaction configuration is related through unitarity to a contribution to the elastic amplitude.

While \phojet was originally developed for hadron--hadron, photon--hadron, and photon--photon interactions
(hadron = p/$\pi$/K)~\cite{Engel:1995yd},
it has later been extended and included as a building block in the \dpmjet MC package~\cite{Roesler00a}
to also apply it to
hadron--nucleus\cite{Bopp:2005cr}, nucleus--nucleus~\cite{Bopp:2004ip},
and photon--nucleus interactions~\cite{Engel:1996yb,Roesler:1998wy}.
The description of hadronic interactions
of photons is limited to real and weakly--virtual photons, and no attempt is made to model deep--inelastic
scattering.

The theoretical framework of the model is the Dual Parton Model~\cite{Capella:1992yb}
in which color flow topologies derived from the expansion of QCD for large numbers of color and
flavour~\cite{tHooft:1973jz,Veneziano:1974fa} are unitarized in an eikonal--like model.
The Dual Parton Model is closely related to the the Quark--Gluon--String Model~\cite{Kaidalov:1999zb},
although there are differences in the practical implementations.

    \subsection{Inclusive and total cross sections}

A detailed description of all partial cross sections can be found in~\cite{PhD-RE}. In the following only 
a very brief summary is given.

Applying the optical theorem an elastic scattering amplitude is constructed from the sum of soft and hard
interactions. All interactions leading to transverse momenta of partons smaller than $p_\perp^{\rm cutoff}$
are attributed to soft interactions, for which the parton interpretation is only valid in
analogy to the topological expansion of QCD. The Born cross section for soft interactions is parameterized 
by $\sigma_s = g^2 s^\Delta_{\rm eff}$.
Interactions with large momentum transfer, corresponding
to partonic final states with $p_\perp > p_\perp^{\rm cutoff}$, are called hard (or semi--hard) interactions and
described by leading--order perturbative QCD
\begin{equation}
\frac{{\rm d}\sigma_{\rm hard}}{{\rm d}^2 p_\perp} =
\int {\rm d}x_1 {\rm d}x_2
\sum_{i,j,k,l} \frac{1}{1+\delta_{kl}} f_{i|A}(x_1,\mu^2)
f_{j|B}(x_2,\mu^2) \frac{{\rm d}\hat{\sigma}_{i,j\rightarrow
k,l}(\hat{s})}{{\rm d}^2 p_\perp} .
\end{equation}
The transverse momentum cutoff is increased with the collision energy~\cite{Bopp:1994cg} to obtain a good 
description throughout the collider energy range~\cite{Moraes:2007rq}. 
Charm quarks are treated as massless and heavier quarks are not included in the calculation.
The parameters of the amplitude for
soft interactions are fitted to obtain a good description of the total, elastic, and diffractive cross
sections and the 
forward slope of the differential elastic cross section at collider energies. Therefore,
the soft parameters, and in particular $\Delta_{\rm eff}$, depend on the set of parton densities
and the $p_\perp^{\rm cutoff}$
used for the fit.

The sum of the amplitudes of soft and hard interactions form the pomeron amplitude, the basic building
block of \phojet. Pomeron--pomeron interactions are only explicitly included at lowest order for a number of
graphs of interest (triple--pomeron for single diffractive dissociation, loop--pomeron for
double--diffraction dissociation, and two combined triple--pomeron graphs, sometimes called
double--pomeron scattering, for central diffraction~\cite{Engel:1996aa}).
All unitarity cuts of these graphs are accounted for following the AGK cutting rules~\cite{Abramovski73-e}.

The partial (soft, hard, triple--pomeron, loop--pomeron, double--pomeron) amplitudes are unitarized
in a two--channel eikonal model~\cite{Kaidalov:1979jz}. The two channels are the ground states of the scattering
particles and 
effective low--mass excitations of the ground states, that are used to describe low--mass diffraction dissociation,
similar to the Good--Walker model~\cite{Good:1960ba}. Low--mass excitations are limited to $M_D^2 < 5$\,GeV$^2$.

Photon interactions are described using the Vector Dominance Model (VDM) for soft (resolved) photon processes
and QCD/QED matrix elements are used for hard processes and point--like photon interactions.
VDM form factors are taken to extend the description from real
photons to photons of virtuality up to $Q^2 \sim 1-2$\,GeV$^2$.

    \subsection{Modelling of inelastic final states}

As a first step the cross sections for different inelastic final states (diffractive and
non--diffractive topologies) are calculated. Thanks to the two--channel 
unitarization of the amplitudes the sizes of the diffractive cross sections are directly linked to, for example, the
multiplicity distribution in non--diffractive interactions, leading to strong model constraints. A high--energy event can
be built up of a superposition of unitarity cuts of all the aforementioned amplitudes and exhibits, in general, a very
complex topology. Hard interactions are sampled first without considering any phase space constraints.
In the next step,  working from 
the highest $p_\perp$ downward, the generated
hard interactions are completed with angular ordered initial state radiation and, if needed, soft partons.
The algorithm for generating initial state radiation is very similar to that
described in \cite{Sjostrand85}. Sometimes, depending on the number of interactions and 
available phase space, it may not be possible for all of the soft and hard interactions
to be realized: in this case, priority is given to those with the highest $p_\perp$.

The partonic color flow of each event is sampled explicitly in the large $N_c$ limit~\cite{tHooft:1973jz}. An option for 
soft color reconnection is implemented but currently not activated as it would not be compatible with the underlying
ideas of the topological expansion of QCD. Partons are combined to color--neutral strings according to their color charges
and \pythia~\cite{Sjostrand:2006za}
is used to generate final state radiation for hard interactions. String fragmentation and hadronization
is also done with \pythia using an optimized set of fragmentation parameters.

One special feature of \phojet is the generation of multiple soft and hard interactions
in single and double diffractive dissociation, and in double pomeron scattering.
A description of the single interaction scenario is given in~\cite{Engel:1995sb} and 
the extension to multiple interactions is discussed in~\cite{Engel:1997zf}.
Inspired by the Ingelman--Schlein approach~\cite{Ingelman:1984ns}
the implementation of multiple interactions is analogous to that 
in non--diffractive interactions except that a virtual pomeron state is used to replace one or
two of the scattering hadrons. Correspondingly, hard interactions are generated with parton densities for the pomeron
(i.e.\ diffractive parton densities). The suppression of hard interactions with large rapidity gaps, 
due to the gap survival probability, is accounted for by generating 
multiple--interaction graphs. A prediction of this model is that the increase with the mass/energy of the pseudorapidity plateau
of charged particles in diffractive interactions is similar to or faster than that observed for
non--diffractive interactions~\cite{Engel:1997zf}.

    \subsection{Plans and future developments}

Work is ongoing to implement new parton densities in \phojet and to carry out the corresponding cross section
fits and fragmentation parameter optimization within a timescale of one year.
On a somewhat longer time scale the implementation of a microscopic model of parton density saturation,
which is currently accounted for only
in a rather crude way and independent of the impact parameter of the collision, is foreseen.

        %%%%%%%%%%%%%%%%%%%%%%%%%%
    
    %%%%%%%%%   POMWIG
    
    %%%%%%%%%%%%%%%%%%%%%%%%%%%

     \section{\pomwig}\label{sec:pomwigmc}
     
\begin{figure}
\begin{center}
\includegraphics[width=0.45\textwidth]{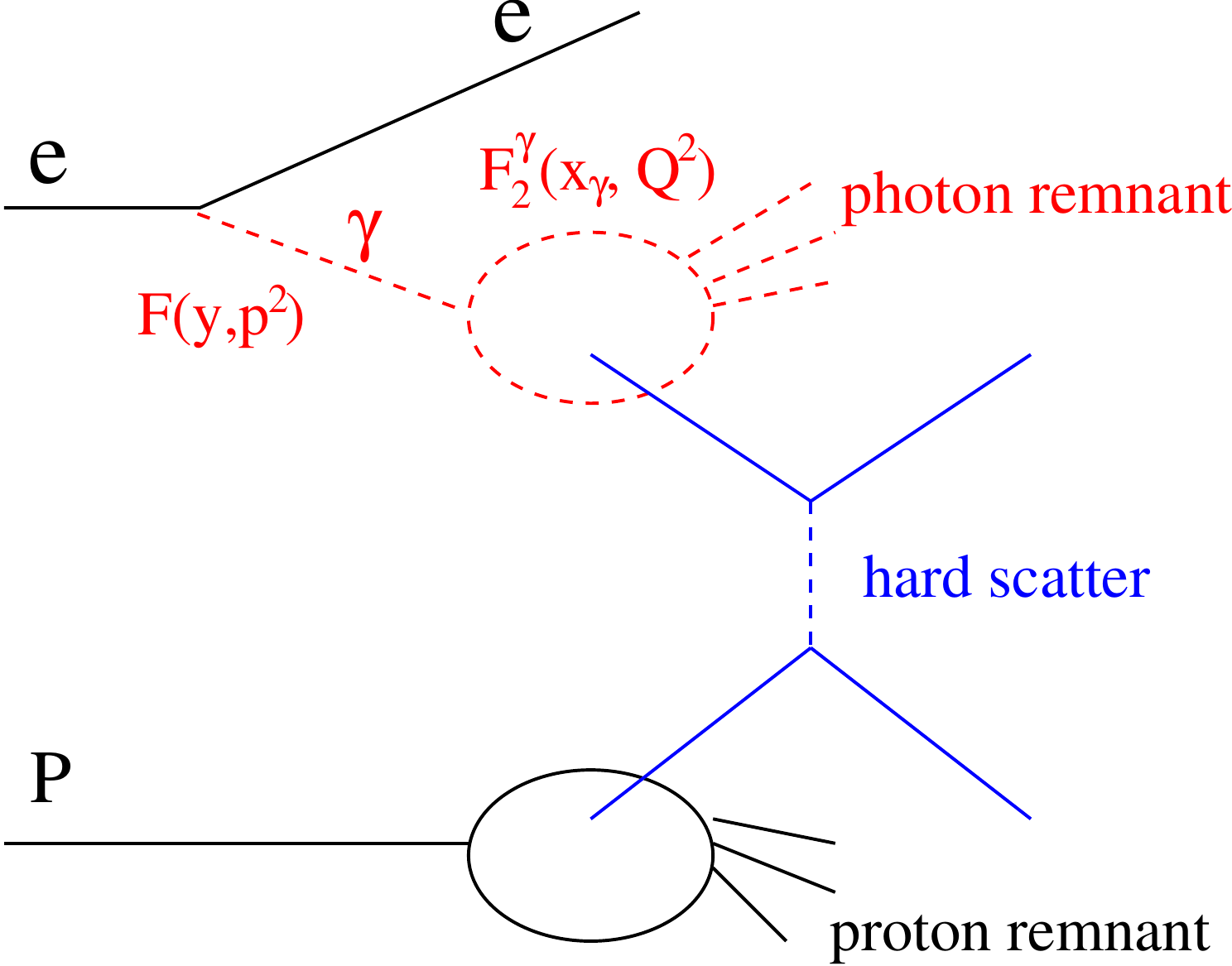}
\includegraphics[width=0.47\textwidth]{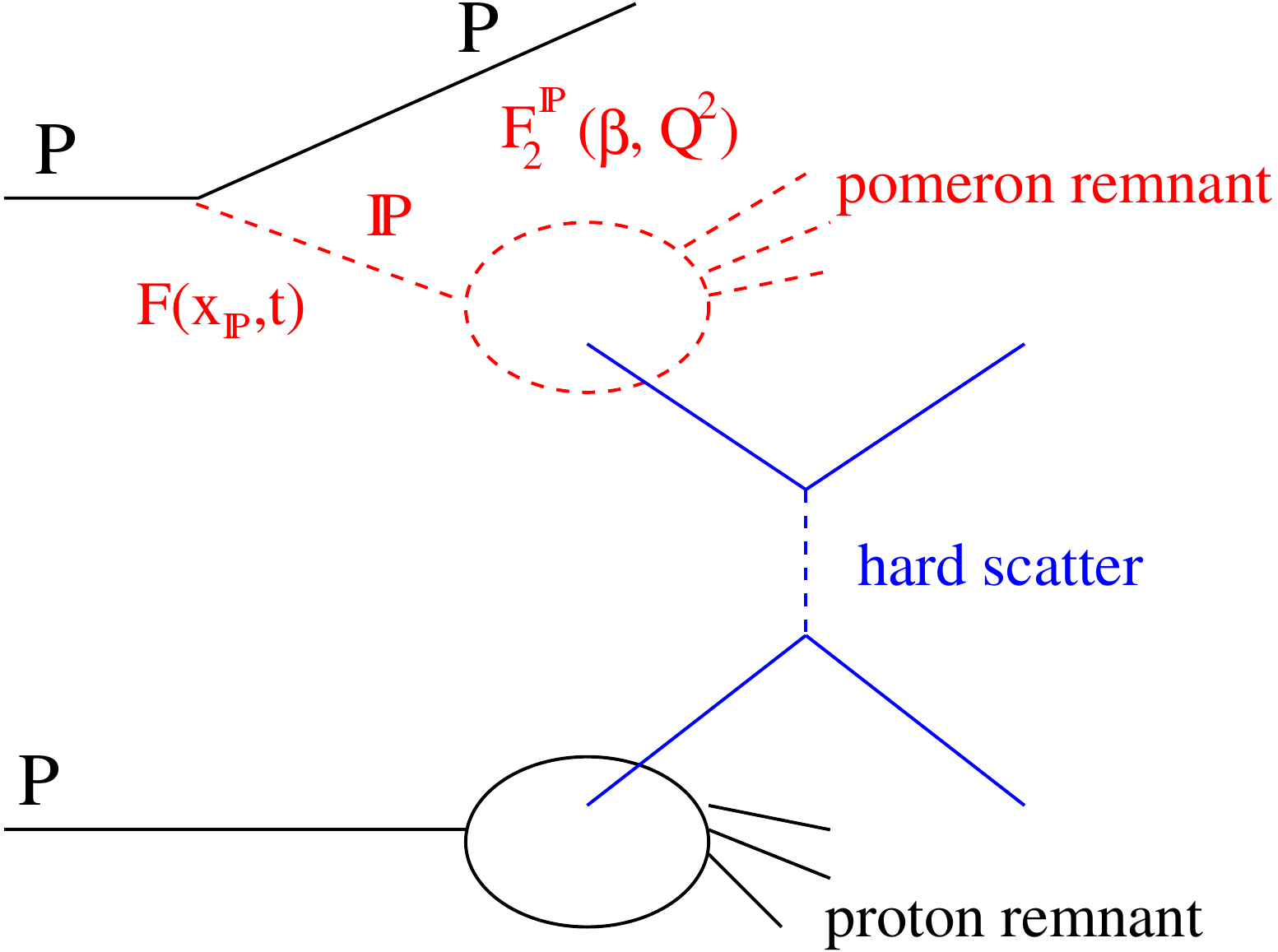}
\caption{The \pomwig model. Photoproduction in ep (or ee) collisions is replaced with pomeron or reggeon exchange in $pp$ (or $ep$) collisions.}
\label{photopom}
\end{center}
\end{figure}

     \pomwig is a modification to the \herwig event generator that allows for the simulation of diffractive interactions. The modifications are simple once it is noticed that pomeron exchange events in hadron--hadron collisions look very much like resolved photoproduction events in lepton--hadron collisions \cite{Cox:2000jt:ch2}. In resolved photoproduction in electron--proton collisions, for example, the process is modelled by the incoming electron radiating a quasi--real photon according to a flux formula. The photon is then treated as a hadronic object with a structure function, which undergoes a collision with the beam proton. Similarly, single diffractive interactions in proton--proton collisions may be modelled by assuming that one of the beam protons emits a pomeron, again according to some flux formula, which subsequently undergoes an interaction with the other beam proton (see Fig.~\ref{photopom}). \herwig will automatically choose to radiate a photon from a beam lepton if a hard subprocess is selected which requires a hadronic structure for the beam lepton. An example would be choosing \herwig subprocess 1500 (QCD $2 \rightarrow 2$ scattering) in an electron--proton collision. 

To simulate a single diffractive interaction in $pp$ collisions, therefore, the photon flux should be replaced with a suitable pomeron flux factor, and the photon structure function with a pomeron structure function, and  \herwig should be run in $ep$ mode rather than $pp$ mode. The electron is identified with the proton which remains intact after the diffractive scattering, and POMWIG replaces the final--state electron by an intact, forward going proton in the event record. This process may be generalised to include sub-leading Regge exchanges, and to simulate double pomeron collisions. 

%\section{Installing the Code}
%\label{sec:code}

The code can be obtained from \cite{pomwigcode}. The routines supplied function with all currently available Fortran versions of \herwig from 5.9 onwards. Full installation details can be found at \cite{pomwigcode}. The example main program provided generates double pomeron $\rightarrow$ Higgs events at the LHC, for a Higgs mass of 115 GeV, and using H1 2006 pomeron Fit A parton densities.

\begin{table}
\begin{center}
\begin{tabular}{|l|l|l|l|l|l|} \hline

  Quantity  & Value  \\ \hline
  $\alpha_{\PO}$ & 1.203 \\ \hline
  $\alpha_{\RO}$ & 0.50 \\ \hline   
  $\alpha^{'}_{\PO}$ & 0.26  \\ \hline
  $\alpha^{'}_{\RO}$ & 0.90  \\ \hline
  $B_{\PO}$ & 4.6 \\ \hline
  $B_{\RO}$ & 2.0 \\ \hline
  $C_{\RO}$ & 16 \\ \hline
  
\end{tabular}
\caption{The default parameters in \pomwig}\label{pomtable}
\end{center}
\end{table}

The default parameters for the pomeron and reggeon fluxes are those found by the H1 Collaboration in \cite{Aktas:2006hy:ch2}, for the case in which no interference is assumed between the pomeron and reggeon contributions to $F_2^{D(3)}$, as shown in Table~\ref{pomtable}. The reggeon contribution is not well constrained by the H1 data, but is only important at high $\xpom$ and low-$\beta$. In hadron--hadron collisions, the variable $\xpom$ is commonly referred to as $\xi$. The fluxes are parameterised as
\begin{equation}
f_{\PO /p}(\xpom)=N\int^{t_{min}}_{t_{max}}\frac{e^{\beta_{\PO}(t)}}{\xpom^{2\alpha_{\PO}(t)-1}}dt
\end{equation}
\begin{equation}
f_{\RO /p}(\xpom)=C_{\RO}\int^{t_{min}}_{t_{max}}\frac{e^{\beta_{\RO}(t)}}{\xpom^{2\alpha_{\RO}(t)-1}}dt
\end{equation}
where $\alpha_{\PO}(t) = \alpha_{\PO}(0)+\alpha^{'}_{\PO}t$ and $\alpha_{\RO}(t) = \alpha_{\RO}(0)+\alpha^{'}_{\RO}t$. The normalisation of the flux is arbitrary in the case of the H1 pomeron structure function. The  \texttt{H1QCD} routine is implemented such that the generated cross section will always match $F_2^{D(3)}$ as measured by H1 at $\xpom = 0.003$, irrespective of the parameters chosen for the flux. The normalisation of POMWIG diffractive cross sections is not expected to match LHC data. In particular, rapidity gap survival effects are not taken into account in POMWIG. The CMS Collaboration estimated a gap survival probability in single diffractive dijet production of approximately 0.1 in the range $0.0003 < \xpom <  0.002$  \cite{Chatrchyan:2012vc:ch2}.

Finally, details of the \texttt{POMPYT} MC for diffractive interactions, based on a similar approach to that described above, can be found in~\cite{Bruni:1993is}.

    %%%%%%%%%%%%%%%%%%%%%%%%%%
    
    %%%%%%%%%   PYTHIA
    
    %%%%%%%%%%%%%%%%%%%%%%%%%%%

\section{PYTHIA 6 \& 8}\label{sec:pythiamc}
The starting point for the 
modelling of soft--inclusive QCD processes is common to both PYTHIA
6~\cite{Sjostrand:2006za} 
and PYTHIA~8~\cite{Sjostrand:2007gs:ch2}. Both generators are therefore discussed
together here, with the new features that are only
available in PYTHIA 8 being pointed out where they occur. (Note that PYTHIA 6 has been in
a legacy state since 2013, and is now no longer officially
maintained.)

\subsection{Inclusive Cross Sections}
The default total, elastic, and inelastic cross
sections are obtained from Regge fits. For $pp$, the 1992
Donnachie--Landshoff 
parametrization~\cite{Donnachie:1992ny} is used,  with one pomeron and one
reggeon term, 
\begin{equation}
\sigma^{pp}_\mathrm{tot}(s) \ = \ 
   21.70\, s^{0.0808}
 + 56.08\, s^{-0.4525}~\mathrm{mb},
\end{equation}
with the $pp$ CM energy squared, $s$, in units of
$(\mathrm{GeV})^2$. For $p\bar{p}$ collisions, the reggeon coefficient 
changes to
$98.39$. (See~\cite{Donnachie:1992ny,Schuler:1993td,Sjostrand:2006za}
for other beam types.) The elastic cross section is approximated by a simple
exponential falloff with momentum transfer, $t$ (valid at small $t$),
related to the total cross section via the optical theorem, 
\begin{equation}
\frac{\mathrm{d}\sigma^{pp}_\mathrm{el}(s)}
     {\mathrm{d}t} \ = \ \frac{(\sigma^{pp}_\mathrm{tot})^2}{16\pi}
 \exp\left(B^{pp}_\mathrm{el}(s)\, t\right)
\hspace*{0.5cm}\to\hspace*{0.5cm}~
\sigma^{pp}_\mathrm{el}(s) \ =
\ \frac{(\sigma^{pp}_\mathrm{tot})^2}{16\pi
  B^{pp}_\mathrm{el}(s)} 
~,
\end{equation}
with $B_\mathrm{el}^{pp} = 5 + 4s^{0.0808}$ being the $pp$ elastic slope (in
$\mathrm{GeV}^{-2}$), defined using the same power of $s$ as the pomeron
term in $\sigma_\mathrm{tot}$, to maintain sensible
asymptotics at high energies. The inelastic cross section is defined
by 
\begin{equation}
\sigma_\mathrm{inel}(s) = \sigma_\mathrm{tot}(s) -
\sigma_\mathrm{el}(s)~.
\end{equation}
The relative breakdown of the inelastic cross
section into 
single--diffractive (SD), double--diffractive (DD), and non--diffractive
(ND) components is given by the following parametrizations~\cite{Schuler:1993td,Schuler:1993wr}:
\begin{eqnarray}
\frac{\mathrm{d}\sigma^{pp\to Xp}_{\mathrm{SD}}(s)}{\mathrm{d}t\,\mathrm{d}M_X^2} 
 & = & F_\mathrm{SD}\, 
\frac{g_{3\mathbb{P}}\beta_{p\mathbb{P}}^3}{16\pi
  M_X^2}\,\exp\left(B_{\mathrm{SD}}^{Xp}\,t\right) 
\\[2mm]
\frac{\mathrm{d}\sigma_\mathrm{DD}^{pp}(s)}{\mathrm{d}t\,\mathrm{d}\,M_1^2\mathrm{d}M_2^2} 
 & = & F_\mathrm{DD}\,
\frac{g_{3\mathbb{P}}^2\beta_{p\mathbb{P}}^2}{16\pi M_1^2M_2^2}\,\exp\left(B_{\mathrm{DD}}\,t\right)~, 
\\[2mm]
\sigma^{pp}_\mathrm{ND}(s) & = & \sigma^{pp}_\mathrm{INEL}(s) - \int \left(
  \mathrm{d}\sigma_{\mathrm{SD}}^{pp\to Xp}(s) 
+ \mathrm{d}\sigma_{\mathrm{SD}}^{pp\to pX}(s) 
+ \mathrm{d}\sigma_\mathrm{DD}^{pp}(s)
\right)~,\label{eq:sigND}
\end{eqnarray}
with $M_X$, $M_1$, $M_2$ being the diffractive masses, and 
the pomeron couplings ($g_{3\mathbb{P}}$, $\beta_{p\mathbb{P}}$),
diffractive slopes ($B_\mathrm{SD}$, $B_\mathrm{DD}$), and ``Fudge Factors''
($F_\mathrm{FD}$, $F_\mathrm{DD}$) given
in~\cite{Schuler:1993td,Schuler:1993wr,Sjostrand:2006za,Navin:2010kk}.  
Note in particular that the ND cross section is
only defined implicitly, via eq.\eqref{eq:sigND}.  
Note also that, in PYTHIA 8, a central--diffractive (CD) component has
recently been added as well, with a cross section $\sigma_\mathrm{CD}
\sim 2\,\mathrm{mb}$. 

Precision measurements at high energies, in particular by
TOTEM~\cite{Antchev:2013iaa:ch2,Antchev:2013paa}, have highlighted that
$\sigma_\mathrm{tot}(s)$ actually grows a bit faster at large $s$, 
and more recent
fits~\cite{Cudell:1996sh,Donnachie:2013xia} are consistent with
using a power $s^{0.096}$ for the pomeron term. Updating the
total cross section formulae in PYTHIA 8 is planned for a
future revision. 
Alternatively, PYTHIA 8 optionally allows a Minimum Bias
Rockefeller (MBR) model to be used, which comes with its own parametrizations of
all $pp$ and $p\bar{p}$ cross sections~\cite{Ciesielski:2012mc}. As a
last resort, it is also possible to set your own user--defined cross
sections (values only, not functional forms), see the HTML manual's
section on ``Total Cross Sections''.

Cross sections for hard (parton--initiated) 
processes are obtained from perturbative $2\to 1$ and $2\to 2$ matrix elements
folded with parton distribution functions (PDFs). 
There are also
extensive (and automated) facilities to interface higher--order processes 
and/or matrix--element corrections from external matrix--element
generators such as ALPGEN~\cite{Mangano:2002ea} or
MADGRAPH~\cite{Alwall:2011uj}.  For inclusive QCD  samples, internal cross 
sections are defined in such a way that the high--$p_\perp$ tail of the inclusive
QCD cross sections (above) is correctly normalized to the perturbative $2\to 2$
result~\cite{Sjostrand:1987su}.  

\subsection{Dynamical modelling}

In PYTHIA, the modelling of hard (parton--initiated) physics processes
is based on a factorized  picture of perturbative 
matrix elements, combined with 
the standard machinery of initial-- and final--state parton showers,
interfaced with the Lund string hadronization model~\cite{Andersson:1998tv}. 
In the context of multi--parton--interaction (MPI) 
models, this picture can be extended to cover all
$p_\perp$ scales (including soft
ones)~\cite{Sjostrand:1987su}, via the introduction of an infrared
regularization scale, $p_{\perp 0}$, 
which is a main tuning parameter
of such models. Physically, $p_{\perp 0}$ expresses a colour screening / 
saturation scale, which is assumed to modify 
the naive LO QCD $2\to2$ cross sections in the following way,
\begin{equation}
\frac{\mathrm{d}\sigma_{2\to 2}}{\mathrm{d}p_\perp^2} \ \propto \
 \frac{\alpha^2_s(p_\perp^2)}{p_\perp^4} \ \to \  
 \frac{\alpha^2_s(p_\perp^2+p_{\perp 0}^2)}{(p_\perp^2+p_{\perp 0}^2)^2} ~,
\end{equation}
such that the divergence for $p_\perp \to 0$ is regulated. 
In practice, the optimal value for $p_{\perp 0}$ (and its scaling with
the hadron--hadron CM energy) depends on several
factors: the PDFs at low $x$~\cite{Schulz:2011qy,Skands:2014pea}, the
IR behaviour of 
$\alpha_s$, the IR regularization of the parton showers, and the possible
existence of other significant IR physics effects, such as colour
(re)connections~\cite{Sjostrand:1987su,Rathsman:1998tp,Skands:2007zg,Sjostrand:2013cya,christiansen}.   
There is also 
an implicit dependence on the 
assumed transverse mass--density of the proton~\cite{Corke:2011yy}. Accepting
the presence of these caveats and dependencies, MPI is the basic
concept driving the modelling of all inelastic non--diffractive
events in both PYTHIA 6 and 8, with the latter using a more 
recent formulation~\cite{Corke:2010yf:ch2} with more advanced options. 
(The modelling of diffraction differs more significantly between the
generators, and will be discussed below.) 

In PYTHIA 6, two explicit 
MPI models are available, an ``old'' one based on 
virtuality--ordered
showers~\cite{Sjostrand:1985xi,Bengtsson:1986et,Bengtsson:1986hr} with
no showers off the additional MPI interactions and a comparatively
simple beam--remnant treatment~\cite{Sjostrand:1987su}, and a ``new''
one based on (interleaved) $p_\perp$--ordered showers~\cite{Sjostrand:2004ef},
including MPI showers and  
a more advanced beam--remnant treatment~\cite{Sjostrand:2004pf}. In
both cases, only partonic QCD $2\to 2$ processes are included among
the MPI (hence no multiple--$J/\psi$, multiple--$Z$, etc type
MPI processes). Most
LHC tunes (e.g., the ``Perugia'' ones~\cite{Skands:2010ak}) 
use the ``new'' $p_\perp$--ordered framework. 
Diffractive events are treated as purely non--perturbative, 
with no partonic substructure: a diffractive mass, $M$, is 
selected according to the above formulae, and the final state produced
by the diffractively excited system is modeled as a single 
hadronizing string with invariant mass $M$, stretched along the beam
axis (with two strings in the case of double diffraction). 

In PYTHIA 8, there is (so far) only one MPI model, extending and
improving the $p_\perp$--ordered one from PYTHIA 6. 
The main differences are: full interleaving of final--state showers with
ISR and MPI~\cite{Corke:2010yf:ch2}; a richer mix of MPI processes,
including electroweak processes and multiple--$J/\psi$ and --$\Upsilon$
production (see the HTML manual under ``Multiparton
Interactions:processLevel''); an option to select the second MPI ``by
hand'' (see the HTML manual under ``A Second Hard Process''); 
an option for final--state 
parton--parton rescattering~\cite{Corke:2009tk} (mimicking a mild 
collective--flow effect in the context of a dilute parton system, 
see the HTML manual under
``Multiparton Interactions: Rescattering''); colour reconnections are
handled somewhat differently (see the HTML manual
and~\cite{Sjostrand:2013cya,christiansen}); and an option for an  
$x$--dependent transverse proton size~\cite{Corke:2011yy}. 
Furthermore,
future development of PYTHIA will only occur in the context of PYTHIA
8, so more advanced models are likely to only be available there, and not
in PYTHIA 6. 
An example where the treatment in PYTHIA 8 already far
surpasses the one in PYTHIA 6 is hard diffraction (for soft diffraction,
the modelling is the same between 6 and 8, though the diffractive and 
string--fragmentation tuning parameters may of course differ). The
default modelling of hard diffraction in PYTHIA 8 is described
in~\cite{Navin:2010kk} and follows an Ingelman--Schlein
approach~\cite{Ingelman:1984ns} to introduce partonic substructure in
high--mass diffractive scattering. (``High--mass'' is defined as
corresponding to diffractive masses greater than about 10 GeV, though
this can be modified by the user, see the HTML manual under
``Diffraction''.)
This gives rise to harder $p_\perp$
spectra and diffractive jets. A novel feature of the PYTHIA 8
implementation is that hard diffractive interactions can include MPI 
(inside the pomeron--proton system such that the rapidity gap is not
destroyed), with a rate governed by the (user--specifiable)
pomeron--proton total cross section, $\sigma_{p\mathbb{P}}$. This
predicts that there should be an ``underlying event'' also in hard
diffractive events, which could be searched for, say, in the region
``transverse'' to diffractive jets, and/or in association with
diffractive $Z$ production, which is currently being implemented in
PYTHIA 8.
Finally, as mentioned above, an
alternative treatment relying on the min--bias Rockefeller (MBR) model
is also available in PYTHIA 8~\cite{Ciesielski:2012mc}. 

The most recent PYTHIA 8 tune is currently the Monash
2013 tune~\cite{Skands:2014pea}, which however did not explicitly
attempt to retune the diffractive components. 
Important remaining open questions include dedicated tuning studies in
the context of diffraction, for instance to constrain the total
pomeron--proton cross section, $\sigma_{p\mathbb{P}}$, which controls
the amount of MPI in hard diffractive processes, the sensitivity to
the diffractive PDFs, and dedicated tests of string--fragmentation
parameters in the specific context of diffractive final states, as
compared with LEP--tuned parameters. The question of colour
reconnections (CR) is likewise
pressing~\cite{Rathsman:1998tp,Skands:2007zg,Sjostrand:2013cya,christiansen},
and disentangling its causes and effects is likely to be a crucial
topic for soft--QCD studies to unravel during the coming years. This
will require the definition and study of CR--sensitive observables and a
detailed consideration of the interplay between PDFs, MPI, and diffractive
physics, with MPI possibly contributing to destroying rapidity gaps
in ``originally'' diffractive events, and CR possibly creating them in
``originally'' non--diffractive
ones~\cite{Edin:1995gi:ch2,Rathsman:1998tp}.

%  \bibliography{mybib}{}
%\bibliographystyle{h-physrev}

%%%%%%%%%%%%%%%

%%%%   QGSJET

%%%%%%%%%%%%%%

\section{QGSJET-II}\label{sec:qgsjet}

The QGSJET-II model \cite{Ostapchenko:2007qb,Ostapchenko:2013pia} has been developed
within the Reggeon Field Theory (RFT) \cite{Gribov:1968fc} framework. The underlying
physics picture  is  one of multiple
scattering processes: the interaction is mediated by multiple parton
cascades which develop between the projectile and the target. Using
the RFT language, those cascades are represented by exchanges of composite
objects characterized by vacuum quantum numbers -- pomerons. The properties
of the underlying ``elementary'' parton cascades thus define the behavior
of the pomeron amplitude. In order to  match with perturbative
QCD, one applies the ``semihard pomeron'' scheme: describing
the parton evolution in the region of relatively high virtualities
$|q^{2}|>Q_{0}^{2}$ using the DGLAP formalism and using a phenomenological
soft pomeron amplitude for non--perturbative ($|q^{2}|<Q_{0}^{2}$)
parton cascades \cite{Drescher:1999zy,Liu:2001hz}. The respective RFT scheme is thus
based on the amplitude of the ``general pomeron'' which is the sum
of the soft and the semihard ones.%
\footnote{It is worth stressing that the respective amplitude is no
longer that of the pomeron pole, being characterized by more complicated
$s$-- and $t$--dependences \cite{Drescher:2000ha}.}
 The $Q_{0}^{2}$ scale has no fundamental meaning here, being
just a border between the two treatments applied to otherwise smooth
parton dynamics.

The beauty of the RFT scheme is that it allows one to develop
a coherent framework for calculating total and elastic cross sections
for hadron--hadron (hadron--nucleus, nucleus--nucleus) scattering and
for deriving partial cross sections for various configurations of
inelastic final states, including diffractive ones \cite{Baker:1976cv}.
This is based on the optical theorem and on the 
AGK cutting rules \cite{Abramovski73-e}. While the former states that the total
cross section, being the sum of all the respective partial cross sections,
including the elastic one, is equal to the $s$--channel discontinuity of
the elastic scattering amplitude, the latter, expressed qualitatively, states that in the high energy limit there
is no interference between final states of different topologies. This
allows one to calculate partial cross sections \textsl{for
all possible configurations of final states} by considering
unitarity cuts of various elastic scattering diagrams and identifying
the contributions of cuts of certain topologies with the desired
cross sections.

  A particular configuration for an inelastic collision thus
contains a number of ``elementary'' production processes described
by cut pomeron contributions and  an arbitrary number of elastic
(virtual) re--scattering processes described by uncut pomerons (see also the discussion in Section~\ref{epos:diff}). To obtain
the respective partial cross section, one has to sum over all the relevant
contributions, i.e.\ ones which have the desirable cut pomeron topology
and \textsl{any number of uncut pomerons}. This is
rather easy to do within the eikonal framework: considering independent
pomeron exchanges between the projectile and the target. In this way
one arrives at the usual simple expressions for the inelastic cross
section, and for relative probabilities of multiple inelastic interactions,
which are employed in most MC generators. The scheme
can be further generalized to include a treatment of \textsl{low
mass diffraction}  by applying the Good--Walker formalism \cite{Good:1960ba}:
 considering
the projectile and target hadron states to be superpositions of a
number of elastic scattering eigenstates characterized by different
vertices for their coupling to the pomeron \cite{Kaidalov:1979jz}. However, to
treat nonlinear processes, like the splitting/fusion of parton cascades
\textsl{or high mass diffraction}, one has to consider
so--called enhanced pomeron diagrams, which describe pomeron--pomeron
interactions \cite{Kancheli:1973vc,Cardy:1974yp,Kaidalov:1986tu}.\footnote{In principle, high mass 
diffraction may be treated within
the Good--Walker framework. However, such a scheme would have a weak
predictive power as one has to parameterize empirically the energy--dependence
of Good--Walker eigenstates.}% 

  An explicit treatment of nonlinear contributions to the interaction
dynamics, based on an all--order re--summation of enhanced pomeron diagrams
\cite{Ostapchenko:2006vr:ch2,Ostapchenko:2006nh,Ostapchenko:2010gt:ch2}, is the distinctive feature of the QGSJET-II
model. Various (generally complicated) final states,
including diffractive ones, for inelastic collisions are generated
by the   MC procedure in an iterative fashion, based
on the respective partial cross sections \cite{Ostapchenko:2013pia}.
 It is noteworthy that the
positive--definiteness of the latter is a very nontrivial fact; it
is only achieved after a full resummation of all the contributions
for a particular final state of interest, i.e. summing over any number
(and topology) of virtual rescatterings described by uncut pomerons
\cite{Ostapchenko:2013pia,Ostapchenko:2010gt:ch2}. In the particular case of diffractive production,
this generates important absorptive corrections (the rapidity gap `survival factor' discussed throughout this report) which, on the one hand, assure
$s$--channel unitarity of the scheme and on the other result in a
nontrivial dependence of the respective cross sections on the masses
$M_{X}$ of the diffractive states produced \cite{Ostapchenko:2010gt:ch2,Ostapchenko:2011nk}.
\begin{figure}
    \centering  \includegraphics[height=4.5cm,width=5.5cm]{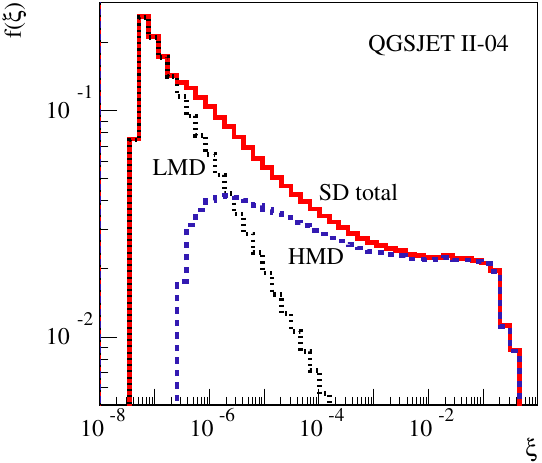}
    \caption{$f_{{\rm SD}}(\xi)=\frac{\xi}{\sigma_{{\rm SD}}}\,\frac{d\sigma_{{\rm SD}}}{d\xi}$ for single diffractive $pp$ collisions at $\sqrt{s}=7$ TeV (solid); partial contributions of low (LMD) and high (HMD) mass diffraction are shown by dotted and dot--dashed
lines respectively.}
    \label{fig:SD}
    \end{figure}

As an illustration, in Fig.~\ref{fig:SD}  the calculated 
$\xi=M_{X}^{2}/s$
distribution for single diffraction is shown.
 Apart from the sharp peak at small $M_{X}$, which is due to
the contribution from low mass diffraction, with decreasing $M_{X}$ one
observes a strong steepening of the $M_{X}$--dependence of high mass
diffraction. This effect is produced by a strong impact parameter $b$
dependence of the  absorptive corrections discussed above: at small
$b$, strong absorptive effects lead to an approximate 
$\propto dM_{X}^{2}/M_{X}^{2}$
shape of the mass spectrum, while in peripheral (large $b$) collisions,
the $M_{X}$--dependence approaches the triple--pomeron asymptotics
\cite{Ostapchenko:2010gt:ch2,Ostapchenko:2011nk}. Such a behavior has indeed been
observed by CMS and TOTEM, as discussed in \cite{Ostapchenko:2014mna} and illustrated
in Table \ref{tab: SD-totem}.% 
\begin{table*}[t]
\begin{tabular*}{1\textwidth}{@{\extracolsep{\fill}}lccccc}
\hline 
$M_{X}$ range & $<3.4$ GeV & $3.4-1100$ GeV & $3.4-7$ GeV & $7-350$ GeV &
 $350-1100$ GeV\tabularnewline
\hline 
\hline 
TOTEM \cite{Antchev:2013haa,Oljemark} & $2.62\pm2.17$ & $6.5\pm1.3$ & $\simeq 1.8$
 & $\simeq 3.3$ & $\simeq 1.4$\tabularnewline
QGSJET-II-04 & 3.9 & 7.2 & 1.9 & 3.9 & 1.5\tabularnewline
\hline 
\end{tabular*}\caption{$\sigma_{pp}^{{\rm SD}}$ (mb) at $\sqrt{s}=7$ TeV for different
ranges of mass $M_{X}$ of diffractive states produced.\label{tab: SD-totem}}
\end{table*}%

\begin{figure}[t]
\centering
\includegraphics[width=0.8\textwidth,height=2cm]{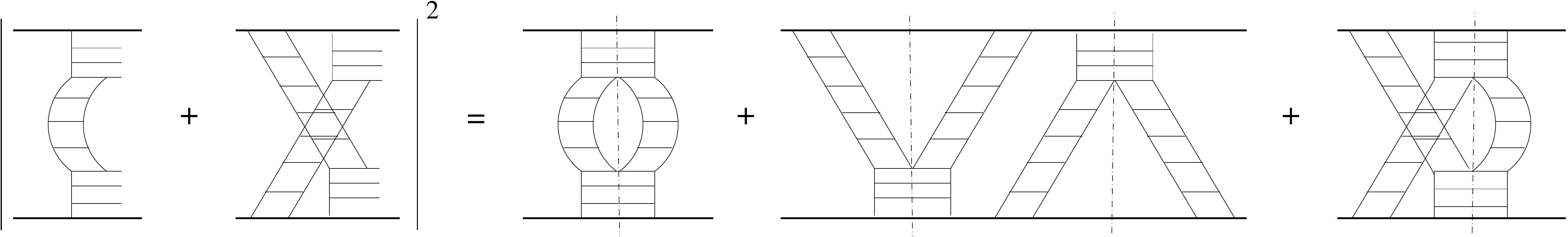}
\caption{Lowest order contributions to $\sigma _{\rm DD}$: squared sum of the
respective amplitudes (lhs) and the corresponding cut diagrams (rhs);
 the cut plane is indicated by dot--dashed lines.\label{fig:DD-nonGW}}
\end{figure}

 Another nontrivial predicted effect is the interference between
different contributions to the  double diffractive cross section 
$\sigma _{\rm DD}$ \cite{Ostapchenko:2010gt:ch2}, which
is illustrated for the lowest order (with respect to the triple--pomeron
coupling) graphs in Fig.\ \ref{fig:DD-nonGW}.
 Because of the finite pomeron slope, at large
$b$  the process is dominated by the usual ``pomeron
loop'' contribution - 1st graph in the right--hand side (rhs)
of the figure. On the other hand, moving to smaller $b$,
one obtains a significant contribution from a superposition of two
(projectile and target) single diffraction processes characterized
by overlapping rapidity gaps - 2nd graph in the rhs. 
In addition, the interference
between the two contributions produces a (negative) contribution
corresponding to the 3rd graph in the rhs.

\begin{figure}[t]
\centering
\includegraphics[width=0.55\textwidth,height=3cm,clip]{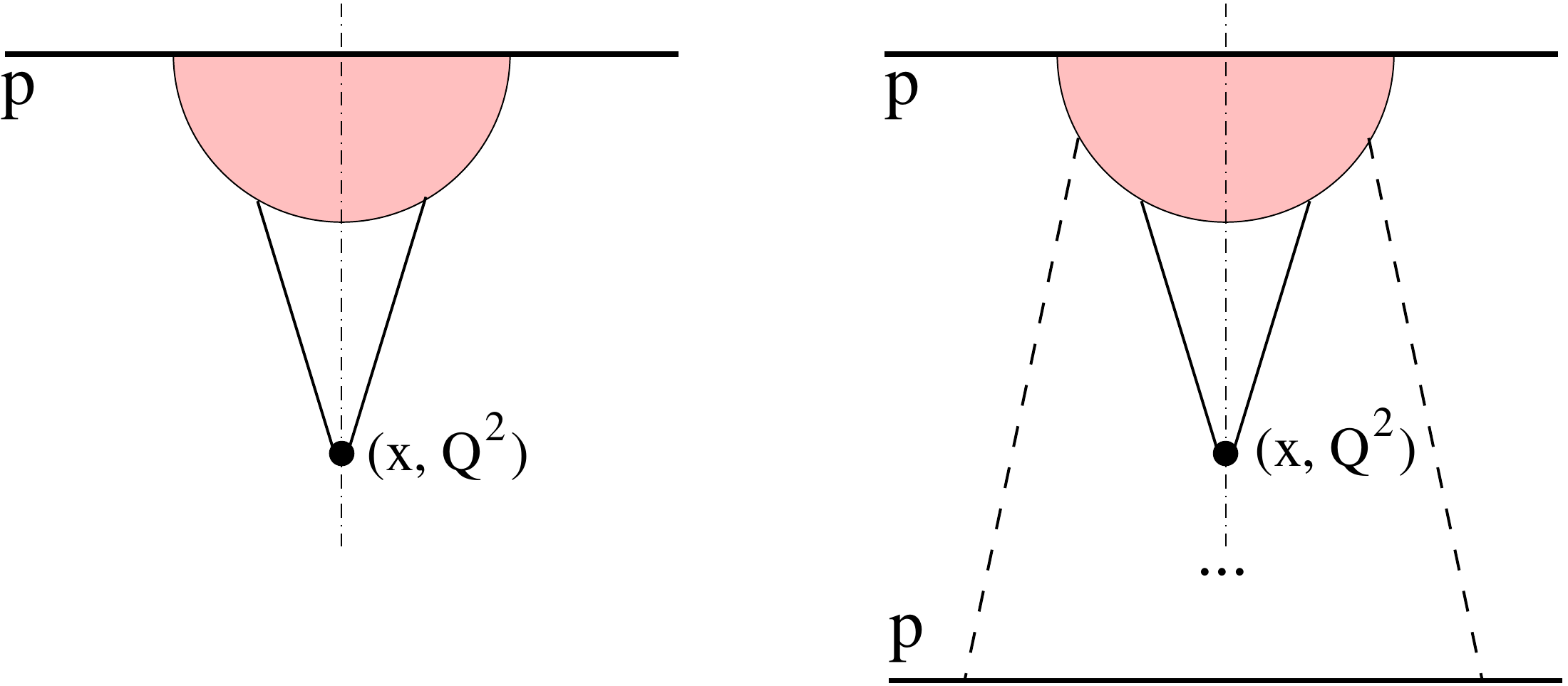}
\caption{Schematic view of parton distributions as ``seen'' in 
DIS (left)  and in $pp$ collisions (right). A low $x$ parton (sea quark or
gluon) originates from the initial state ``blob'' and interacts with a virtual
``probe''. The universal PDFs measured in DIS are affected by the rescattering
of intermediate partons from the initial state cascade 
(hidden in the  ``blob'')
 off the parent proton. In $pp$ interactions the initial ``blob''
 is affected itself by the collision process: due to the rescattering of 
 intermediate partons off the partner (here, target) proton, as indicated by
   dashed lines.}
\label{pdfs}       % Give a unique label
\end{figure}

 Finally, it is worth recalling the relationship between absorptive
corrections due to enhanced pomeron graphs and the breakdown of 
collinear QCD factorization for quantities that are not fully inclusive,
e.g.\ for jet production in specially triggered events or for diffractive
dijet production. Unlike the universal
parton distribution functions (PDFs) measured in deep inelastic scattering (DIS),
cross sections for particular inelastic final states depend on non--universal
PDFs which are influenced by absorptive corrections due to intermediate
parton rescattering off the partner hadron and hence depend on the properties
of the particular final state of interest \cite{Ostapchenko:2007qb}, as depicted in Fig.\
\ref{pdfs}.

    %%%%%%%%%%%%%%%%%%%%%%%%%%%%%%%
    
    %%%%%%%%%%%%%%%%%% Shrimps
    
    %%%%%%%%%%%%%%%%%%%%%%%%%%%%%%%%

 \section{\Shrimps}\label{sec:shrimps}
\subsection{Inclusive properties and the KMR model}
The \Shrimps MC generator aims at a complete description of Minimum Bias QCD events at hadron
colliders and, most notably, at the LHC.  It is based on the model by Khoze,
Martin, and Ryskin (KMR)~\cite{Ryskin:2009tj,Ryskin:2009tk:ch2}.  The model 
rests on the description of the incident hadrons through the superposition of 
$\NGW$ diffractive or Good--\-Walker states~\cite{Good:1960ba,Gustafson:2012hg},
typical for models of soft interactions.  Interactions between these
states are described through single eikonal factors, related to the incident 
eigenstates, which emerge from effective parton densities, $\Omega_{ik}$, associated with them:
\begin{equation}\label{Eq:SingleChannel_WithoutKT}
  \begin{split}
    \Omega_{ik}(Y,\,B_\perp) \;=\;& 
    \frac{1}{\beta_0^2}\,
    \int\dtwo b_\perp^{(1)}\dtwo b_\perp^{(2)}
    \delta^2\left(\vec B_\perp-\vec b_\perp^{(1)}+\vec b_\perp^{(2)}\right)
    \Omega_{i(k)}\left(y,\,b_\perp^{(1)}\right)
    \Omega_{(i)k}\left(y,\,b_\perp^{(2)}\right)\,.
  \end{split}
\end{equation}
Here $B_\perp$ is the impact parameter of the two hadrons, while the individual
$\vec{b}_\perp^{(1,2)}$ label the partonic position with respect to the two
incident hadrons in the transverse plane; $\beta_0^2$ is a normalisation
factor with units of area (or cross section) and of the typical size of a
hadronic total cross section and $Y$ is the beam rapidity. With this form of the eikonal, the
total, elastic and inelastic cross sections are for instance given by
\begin{equation}
  \label{Eq:TotalXSecs_SumGoodWalker}
  \begin{split}
    \sigma^{pp}_{\rm tot}(Y)  \;=\; & 
    2\int\dtwo B_\perp\;\left\{\sum\limits_{i,k=1}^{\NGW}\,|a_i|^2|a_k|^2\,
    \left\{1-\exp\left[-\frac{\Omega_{ik}(Y,B_\perp)}{2}\right]\right\}
    \vphantom{\sum\limits_{i,k=1}^S}\right\}
    \\
    \sigma^{pp}_{\rm el}(Y)   \;=\; &   
    \int\dtwo B_\perp\;\left\{\sum\limits_{i,k=1}^{\NGW}\,|a_i|^2|a_k|^2\,
      \left\{1-\exp\left[-\frac{\Omega_{ik}(Y,B_\perp)}{2}\right]\right\}
      \right\}^2\\
    \sigma^{pp}_{\rm inel}(Y) \;=\; &   
    \int\dtwo B_\perp\;\left\{\sum\limits_{i,k=1}^{\NGW}\,|a_i|^2|a_k|^2\,
    \left\{1-\exp\left[-\Omega_{ik}(Y,B_\perp)
      \vphantom{\frac{\Omega_{ik}(Y,B_\perp)}{2}}\right]\right\}
    \vphantom{\sum\limits_{i,k=1}^S}\right\} \equiv
    \sum\limits_{i,k=1}^{\NGW}\,\sigma^{(ik)}_{\rm inel}(Y)\;.
  \end{split}
\end{equation}
Here, the $a_i$ are the coefficients in the expansion of the proton wave function in terms of Good-\--Walker states.
Low-mass diffractive dissociation can proceed in three ways, namely by the
transition of either one of the two hadrons or of both of them into excited 
states.  They can be labeled as single diffraction of hadron 1 or 2, $SD1$
and $SD2$, respectively, or by double diffraction, $DD$.  For instance, the 
differential cross section with respect to the momentum transfer $t=-Q^2$ 
for the sum of elastic scattering and single diffraction of hadron 1 can be
written as 
\begin{equation}
  \begin{split}
    \frac{\done\sigma_{el+SD1}(Y)}{\done t} \;=\; &
    \frac{1}{4\pi}\sum\limits_{i,j,k=1}^{\NGW}\left\{
    \vphantom{\sum\limits_{i,j,k=1}^{\NGW}}
    |a_i|^2|a_k|^2|a_j|^2
    \int\dtwo B_\perp\,
    \exp\left[i\vec Q_\perp\cdot\vec B_\perp
      \vphantom{\frac{\Omega_{ij}(Y,B'_\perp)}{2}}\right]\,
    \left\{1-\exp\left[-\frac{\Omega_{ik}(Y,B_\perp)}{2}\right]\right\}
    \right.\\
    &\hspace*{34mm}\times\left.
    \int\dtwo B'_\perp\,
    \exp\left[i\vec Q_\perp\cdot\vec B'_\perp
      \vphantom{\frac{\Omega_{ij}(Y,B'_\perp)}{2}}\right]\,
    \left\{1-\exp\left[-\frac{\Omega_{ij}(Y,B'_\perp)}{2}\right]\right\}
    \vphantom{\sum\limits_{i,j,k=1}^{\NGW}}      \right\}\,.
    \label{Eq:DiffSigmaSD1}
  \end{split}
\end{equation}
The incident parton densities $\Omega_{i(k)}\left(y,\,b_\perp^{(1)}\right)$ and 
$\Omega_{(i)k}\left(y,\,b_\perp^{(2)}\right)$ of Good--\-Walker state $i$ or $k$ 
in the presence of $k$ or $i$ are the solutions of coupled differential 
equations, describing their evolution in rapidity.  Their boundary values, the
initial parton densities at the incident hadronic rapidities of
\begin{equation}
  Y = \pm\log\frac{\Ecm}{m_{\mathrm{had}}}
\end{equation}
with $\Ecm$ the centre-of-mass energy of the hadron collision and
$m_{\mathrm{had}} = 1\UGeV$ a typical hadronic scale, are fixed through form factors.
In the \Shrimps implementation of the KMR model, $\NGW = 2$ with initial parton 
densities given by dipole-like form factors modified with an exponential to 
guarantee numerical stability,
\begin{equation}\label{Eq:GoodWalkerStatesFormFactors}
  {\cal F}_{1,2}(q_\perp) \;=\; 
  \beta_0^2\;(1\pm \kappa)\frac{\exp\left(-
    \displaystyle\frac{\xi(1\pm \kappa)q_\perp^2}{\Lambda^2}\right)}
       {\left(1+\displaystyle\frac{(1\pm\kappa)q_\perp^2}{\Lambda^2}
         \right)^2}\,,
\end{equation}
and therefore
\begin{equation}\label{Eq:BoundaryConditionsWithoutKT}
  \begin{split}
    \Omega_{i(k)}(-Y/2,\,b_\perp^{(1)}) \;=\;& F_i(b_\perp^{(1)})\\
    \Omega_{(i)k}(+Y/2,\,b_\perp^{(2)}) \;=\;& F_k(b_\perp^{(2)})\,,
  \end{split}
\end{equation}
where $F_j(b_\perp)$ are the Fourier transforms of the
form factors ${\cal F}_j(q_\perp)$. The parton densities increase with increasing rapidity distance from the 
hadron, driven by a parameter $\Delta$, which could be identified as a 
reggeon, and in particular a pomeron intercept.  This exponential ``gain'' 
term is counteracted by an absorptive correction, $\Wabs$ which is interpreted 
as parton recombination.  It is parametrised by a constant $\lambda$, which 
could consequently be identified as being connected to the triple-pomeron 
vertex, and reads
\begin{equation}\label{Eq:Wabs}
  \begin{split}
    \Wabs^{(ik)}(y,\,b_\perp^{(1)},\,b_\perp^{(2)})
    \;=\;&
    \left\{\frac
        {1-\exp\left[-\frac{\lambda}{2}
            \Omega_{i(k)}(y,\,b_\perp^{(1)},\,b_\perp^{(2)})\right]}
        {\frac{\lambda}{2}\Omega_{i(k)}(y,\,b_\perp^{(1)},\,b_\perp^{(2)})}
        \right\}
    \left\{\frac
        {1-\exp\left[-\frac{\lambda}{2}
            \Omega_{(i)k}(y,\,b_\perp^{(1)},\,b_\perp^{(2)})\right]}
        {\frac{\lambda}{2}\Omega_{(i)k}(y,\,b_\perp^{(1)},\,b_\perp^{(2)})}
        \right\}\,.
  \end{split}
\end{equation}
Together, therefore
\begin{equation}\label{Eq:DEQs_WithoutKTDep}
  \begin{split}
    \frac{\done\Omega_{i(k)}(y,\,b_\perp^{(1)},\,b_\perp^{(2)})}{\done y} \;=\;& 
    +\Wabs^{(ik)}(y,\,b_\perp^{(1)},\,b_\perp^{(2)})
    \;\cdot\Delta\cdot\;\Omega_{i(k)}(y,\,b_\perp^{(1)},\,b_\perp^{(2)})\\
    \frac{\done\Omega_{(i)k}(y,\,b_\perp^{(1)},\,b_\perp^{(2)})}{\done y} \;=\;& 
    -\Wabs^{(ik)}(y,\,b_\perp^{(1)},\,b_\perp^{(2)})
    \;\cdot\Delta\cdot\;\Omega_{(i)k}(y,\,b_\perp^{(1)},\,b_\perp^{(2)})\,,
  \end{split}
\end{equation}
Taken together this yields a reasonably good description of total, elastic,
inelastic and diffractive cross sections in $pp$ and $p\bar{p}$ collisions
at various centre--of--mass energies, see Fig.~\ref{Fig::xsecs}.

\subsection{Exclusive properties}

\subsubsection{Parton--parton interactions}
%In order to link the KMR model with a truly exclusive partonic language, the 
%\Shrimps model assumes, in a first step, that the specific Good--\-Walker 
%eigenstates taking part in the interaction are fixed, thereby also fixing 
%the number and positions in transverse space of the individual parton--\-parton
%scatters.  The former, i.e. the number of partonic interactions, is given by a 
%Poissonian in the eikonal $\Omega_{ik}(B_\perp)$, after the eigenstates $i$ 
%and $k$ and their distance in impact parameter has been probabilistically 
%determined.  In a similar way the individual positions of the 
%parton--\-parton scatters is then also fixed.  

In order to link the KMR model with a truly exclusive partonic
language, the \Shrimps model assumes, in a first step, that while the
proton is a superposition of Good--Walker eigenstates, the interaction
projects onto one of these states. This happens for both colliding
hadrons, and the corresponding contributions for each possible combination of
Good--Walker states to the inelastic cross section can be
read off from (\ref{Eq:TotalXSecs_SumGoodWalker}). After choosing
the channel $ik$ in which the interaction is taking place, the impact
parameter distribution is given by $\done \sigma^{(ik)}_{\rm
inel}(Y)/\done B_\perp$. The number of partonic interactions is given
by a Poissonian in the eikonal $\Omega_{ik}(B_\perp)$ and the positions
of the individual parton--parton scatters is determined
probabilistically according to the parton densities
$\Omega_{i(k)}(y,\,b_\perp^{(1)},\,b_\perp^{(2)})$ and
$\Omega_{(i)k}(y,\,b_\perp^{(1)},\,b_\perp^{(2)})$.

The individual interactions between partons are interpreted as being given
by cut pomerons, effectively multiple gluon emissions along a gluon-ladder
ordered in rapidity.  Since the KMR model has no notion of energies or 
light--cone momenta of the partons constituting the incident hadron states,
suitable PDFs must be constructed, which are then convoluted with a total 
parton--\-parton cross section mediated through pomeron exchange, or a 
reggeised $t$--channel gluon, to yield the inelastic cross section,  
\begin{equation}
  \label{Eq::Parton2to2xsec}
  \sigma_{\mathrm{inel}}^{(ik)}(\smin) =
  \frac{1}{2S}\sum\limits_{\tilde{i},\tilde{k}}\;
  \int\limits_{\smin}^{s_{\mathrm{max}}}\,\done\hat s
  \int\,\done\hat y\,\left[f_{\tilde{i}/h_1}(x_1,\mu_F^2)
    f_{\tilde{k}/h_2}(x_2,\mu_F^2)
    \left(\frac{\hat{s}}{\smin}\right)^{\eta_{ik}}\right]\,.
\end{equation}
Here, the lower limit for the centre-\-of-\-mass energy squared of the 
partonic $2\to 2$ scattering, $\smin$ is a parameter, and the corresponding
upper limit can be conveniently set to the hadronic centre-\-of-\-mass energy 
squared, $s_{\mathrm{max}} = S = \Ecm^2$.  The parameter $\smin$ is fixed by
equating this cross section with the one obtained from the eikonal,
\begin{equation}
  \sigma^{(ik)}_{\rm inel}(Y)\,=\,\sigma_{\mathrm{inel}}^{ik}(\smin)\,,
\end{equation}
and therefore depends on the Good--\-Walker eigenstates.  The exponent 
$\eta_{ik}$ is given by the product of $\Delta$ and $\Wabs$ for the given 
combination of $i$ and $k$.  It is interesting to note that this typically 
reduces the bare pomeron intercept $\Delta\approx 0.3$ to $\eta\approx 0.1$, 
in remarking agreement with parametrisations of the pomeron, e.g.\ 
in~\cite{Donnachie:1992ny}.  

\subsubsection{Infrared--continued parton density functions and strong coupling}
In order to also capture the dominant 
non--perturbative parts of the cross section, the PDFs in the \Shrimps model
are continued into the infrared region, allowing $\mu_F = 0$ to be set in
the calculation of the cross section above.  In the \Shrimps model, the 
basic assumption is that at $\mu_F = 0$ only valence components of the
proton exist, where the valence gluon distribution follows in shape the 
valence quarks.  The transition between the perturbative regime and the
non--perturbative extension is smooth: starting from an IR--cut parameter 
$\qcut\approx\rm{2}\,{GeV}$, the sea components of the PDFs are phased out 
linearly with $\mu_F^2$ such that
\begin{equation}
  f_{\mathrm{sea}/h_1}(x,\,\mu_F^2) = \left\{
  \begin{array}{lcl}
    f_{\mathrm{sea}/h_1}(x,\,\mu_F^2) & \mathrm{for} & \mu_F\ge \qcut\\
    \frac{\mu_F^2}{\qcut^2}\;f_{\mathrm{sea}/h_1}(x,\,\qcut^2) & 
    \mathrm{for} & \mu_F<\qcut\,,
  \end{array}
  \right.      
\end{equation}
while the quark valence distributions behave as
\begin{equation}
  f_{q_\mathrm{val}/h_1}(x,\,\mu_F^2) \,=\,
    f_{q_\mathrm{val}/h_1}(x,\,\mathrm{max}\{\mu_F^2,\,\qcut^2\})\,.
\end{equation}
The valence gluon component is normalised such that the momentum sum rule
is satisfied.

For each individual partonic $2\to 2$--scattering, a gluon $t$--channel 
exchange is assumed; incoming flavours and kinematics are selected according 
to (\ref{Eq::Parton2to2xsec}), and the outgoing partons are supplemented with 
a transverse momentum according to the form factors of 
(\ref{Eq:GoodWalkerStatesFormFactors}).  These initial configurations serve 
as starting points for further gluon emissions off the $t$--channel gluon.  The 
strong coupling which appears in the additional radiation off the $t$--channel 
gluon is infrared continued, as
\begin{equation}
  \bar\alpha_S(\mu^2)\;=\;\alpha_S(\mu^2+q_0^2)
\end{equation}
with $q_0^2$ being $\oforder{1\UGeV^2}$.  

\subsubsection{Building gluon ladders}
The $2\to 2$--scattering provides the starting point of further emissions
and defines an active rapidity interval $[y_{i-1},\,y_{i+1}]$ for them.  
However, at this point the model includes a diffractive component by
allowing the $t$--channel gluon to either be in an octet state or to
be re--\-interpreted as a colour singlet, a pomeron.  Phrased differently,
a decision is to be made as to whether the exchange corresponds to a cut pomeron or 
not, which is thereby related to the absorption part of the evolution equation.
This choice is achieved probabilistically, based on the parton densities.  
The corresponding weights for singlet or octet exchange along the active 
interval is given by
\begin{equation}
  \begin{split}
    \mathcal{W}_1\;=\;& 
    \left[1-\exp\left(-\frac{1}{2}\,\lambda^2\,
      \frac{\Omega_{i(k)}(y_{i+1})-\Omega_{i(k)}(y_{i-1})}
           {\Omega_{i(k)}(y_{i-1})}\right)\right]^2\\[2mm]
    \mathcal{W}_8\;=\;& 1-\exp\left(-\lambda^2\,
    \frac{\Omega_{i(k)}(y_{i+1})-\Omega_{i(k)}(y_{i-1})}
         {\Omega_{i(k)}(y_{i-1})}\right)\,,
  \end{split}
\end{equation}
following the logic already encoded in the expressions for the elastic
and inelastic cross sections, c.f.~(\ref{Eq:TotalXSecs_SumGoodWalker}).  
Of course, if the active rapidity interval is associated with a singlet 
exchange, no further emissions will happen off this part of the ladder.

\begin{figure}[h!]
  \begin{center}
    \includegraphics[width=0.25\textwidth]{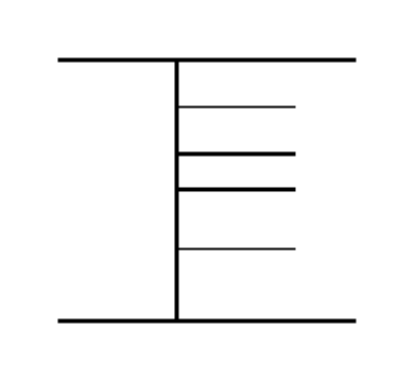} 
    \includegraphics[width=0.25\textwidth]{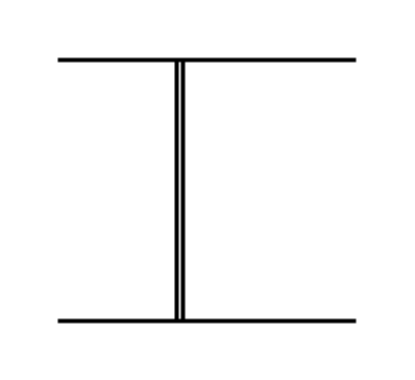} 
    \includegraphics[width=0.25\textwidth]{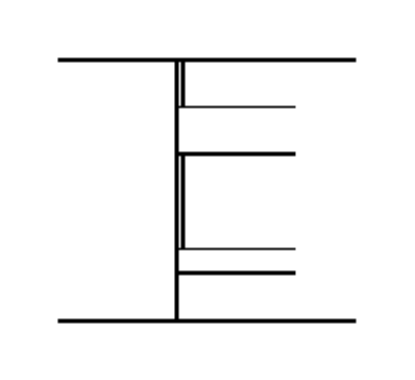} 
    \caption{
      Ladders with different colour topologies: one with only octet 
      propagators (left), a pure singlet exchange (middle), and a ladder with 
      both singlet and octet exchanges (right). Octet propagators are denoted 
      by single, singlets by double lines.
      \label{Fig::ColouredLadder}
    }
  \end{center}
\end{figure}

The additional emission off the $t$--channel are driven by a Sudakov form 
factor--like structure, $\Delta(y_{i-1},\,y_{i+1})$, which yields the probability 
for no emission in the active rapidity interval,
\begin{equation}
  \Delta(y_{i-1},\,y_{i+1}) \;=\;
  \exp\left\{-\int\limits_{y_{i-1}}^{y_{i+1}}\,\done y_i
  \int\frac{\done k_{\perp,i}^2}{k_{\perp,i}^2+Q_0^2(y_i)}\,
  \frac{C_A\bar\alpha_S(k_{\perp,i}^2)}{\pi}\,\Wabs(y_i)
  \right\}\,.
\end{equation}
The IR regulator $Q_0^2$ appearing 
in the equation above, guaranteeing the convergence of the integration over 
the transverse momentum of the emitted gluon, $k_{\perp, i}$ scales with the
parton densities as
\begin{equation}
  Q_0^2(y_i) = \frac{\lambda q_0^2}{%\left[
    \left(\frac{\Omega_{i(k)}(-Y,b_\perp^{(1)},b_\perp^{(2)})}
         {\Omega_{i(k)}(y,b_\perp^{(1)},b_\perp^{(2)})} \right)^2
         +\left(\frac{\Omega_{(i)k}(Y,b_\perp^{(1)},b_\perp^{(2)})}
         {\Omega_{(i)k}(y,b_\perp^{(1)},b_\perp^{(2)})} \right)^2
         %\right]^{-1} 
  }\,,
\end{equation}
where $\lambda$ is introduced in (\ref{Eq:Wabs}). In other words, the denser the parton soup at the emission rapidity, the more
transverse momentum the emitted gluon must have in order not to be absorbed.
However, after each emission, the available rapidity interval shrinks, with
$y_i$ replacing $y_{i-1}$.  Some example ladder types are exhibited in Figure
\ref{Fig::ColouredLadder}.
Each ladder is finally reweighted in such a way that its hardest interaction 
follows a rough estimate of perturbative QCD cross sections.

\subsubsection{Rescattering}
\begin{figure}[h!]
  \begin{center}
    \includegraphics[width=0.25\textwidth]{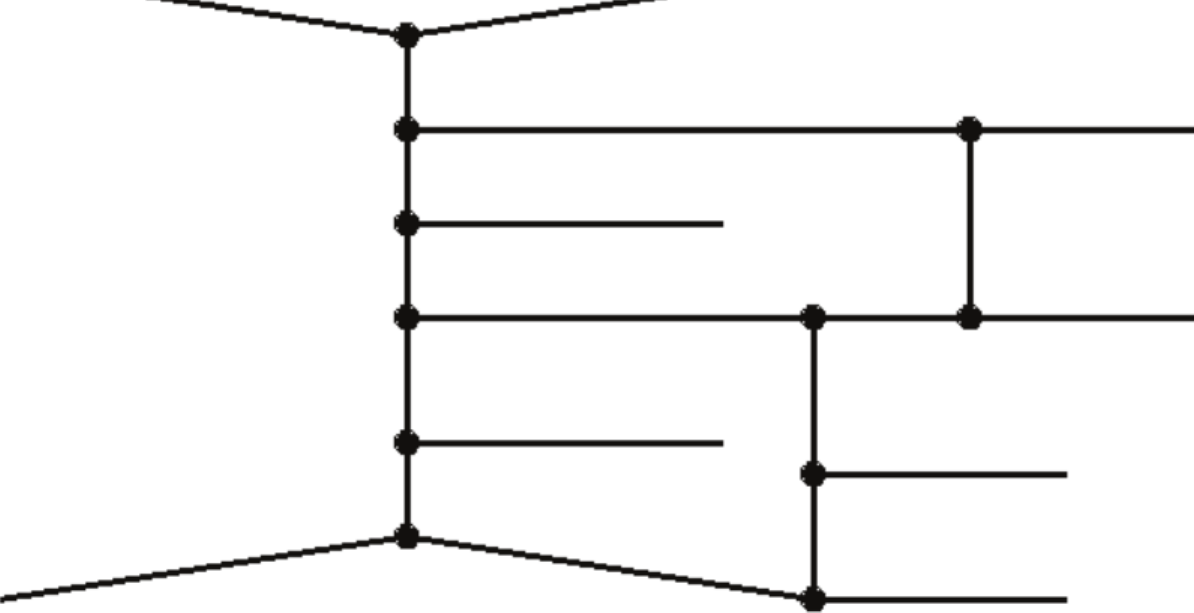}
    \caption{Rescattering of partons off a primary ladder and the
      subsequent secondaries.  In the \Shrimps model they are all located
      at the same position in impact parameter space, and the rescatter
      probability is given by the parton densities.
      \label{Fig::Rescatter}
    }
  \end{center}
\end{figure}

Another important aspect of the model is that it allows the rescattering of
partons produced at the same position in impact parameter space, 
giving rise to a cascade of further ladders and potentially mixing the offsprings
of different such ladders, as exemplified in Figure \ref{Fig::Rescatter}.
In the \Shrimps model, the rescattering appears probabilistically, with a 
rescattering probability between two partons $i$ and $j$ given by
\begin{equation}
  \mathcal{P}_{\mathrm{resc}}(y_i,\,y_j)\;=\;
  \frac{1}{N_{\mathrm{resc}}!}\,\left(
  \frac{s_{ij}}{\mathrm{max}\{s_{ij},\,\smin\}}\right)^\eta\,
  \mathcal{W}_8\,,
\end{equation}
where $N_{\mathrm{resc}}$ counts the number of rescatters that already happened
before arriving at this pair of partons.
\subsection{The link to hadrons}
The emerging parton ensemble undergoes further (collinear) parton showering; in the 
\Sherpa event generator~\cite{Gleisberg:2003xi,Gleisberg:2008ta:ch2} this is 
achieved through the native parton shower based on Catani--Seymour subtraction 
kernels~\cite{Nagy:2006kb,Schumann:2007mg} with suitably defined starting 
conditions (avoiding double counting), typically given by the relative transverse momentum the partons 
have w.r.t.\ their colour partner. After the generation of all ladders and parton showering but before hadronization, colour is re--arranged through a colour reconnection model. The transition to hadrons is facilitated 
through \Sherpa's cluster fragmentation model, in the spirit 
of~\cite{Winter:2003tt}, and supplemented with the intrinsic modelling
of hadron decays, QED final state radiation etc.

\subsection{Selected predictions}

\begin{figure}[h!]
 \centering
  \includegraphics[angle=-90,width=0.48\textwidth]{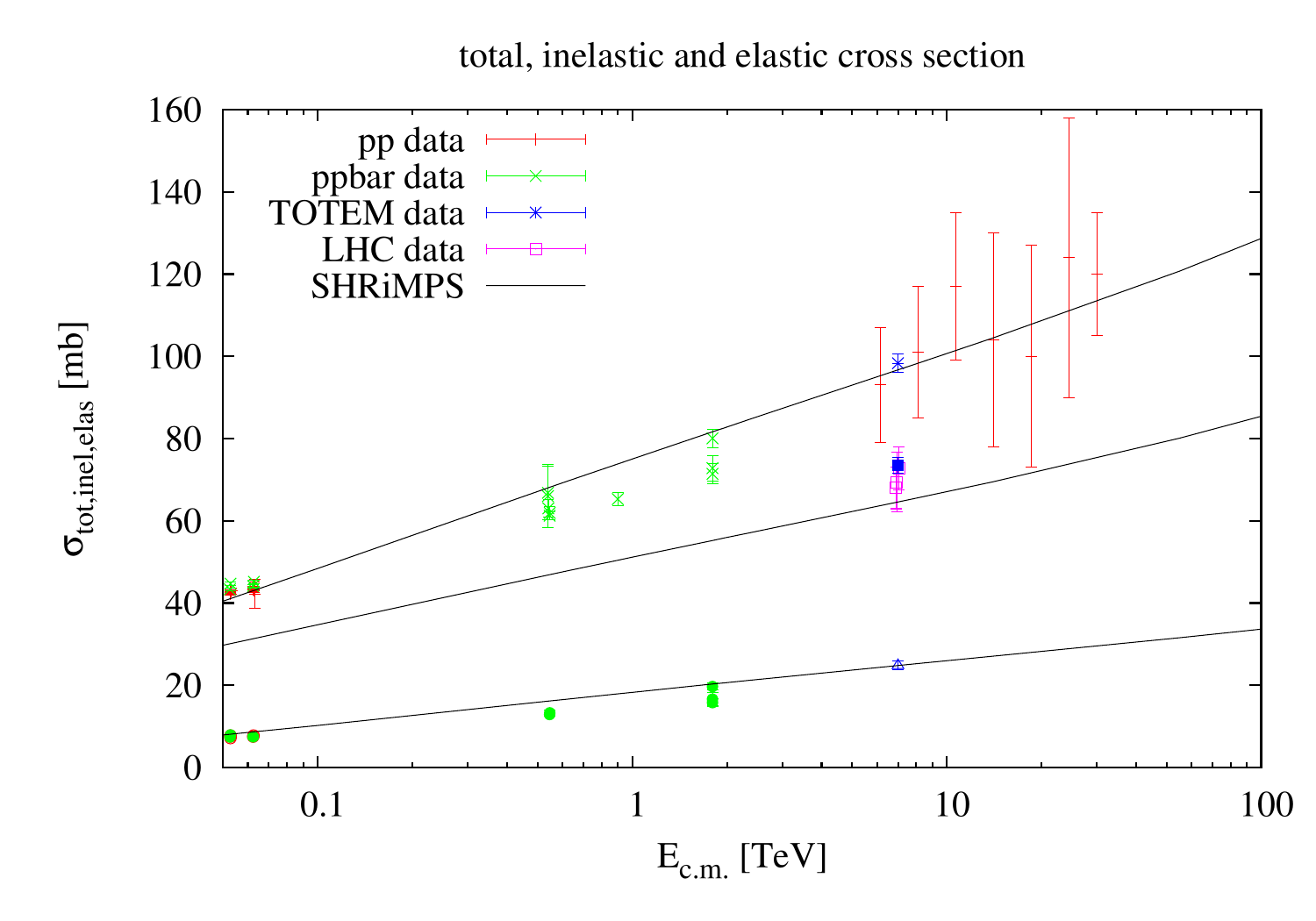}
  \includegraphics[angle=-90,width=0.48\textwidth]{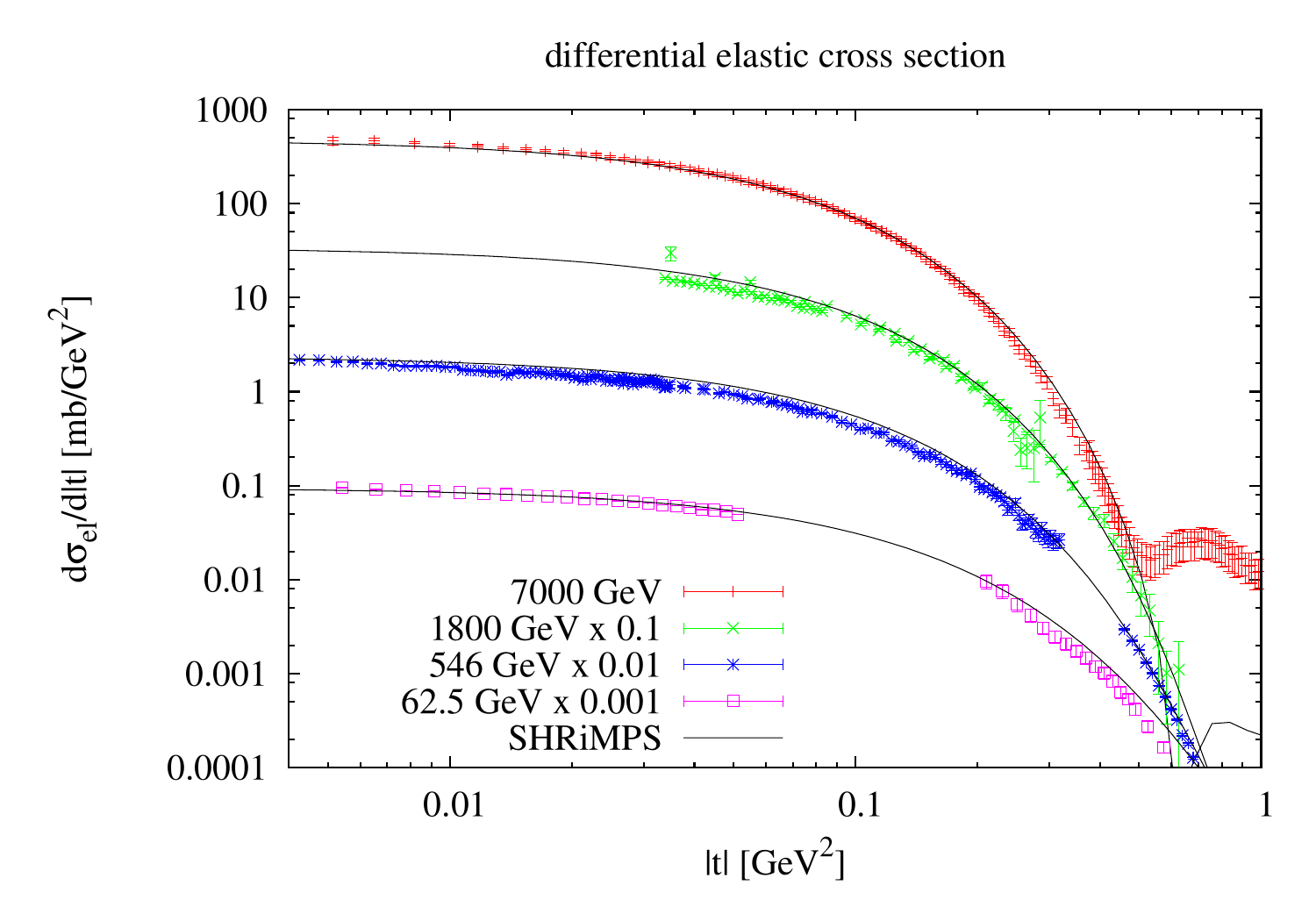}
  \caption{\textbf{LHS:} Total, inelastic and elastic cross section 
   compared data from $p+p$ and $p+\bar p$ 
   collisions~\protect\cite{Nakamura:2010zzi} and 
   LHC data from TOTEM~\protect\cite{Antchev:2011vs},
   ATLAS~\protect\cite{Aad:2011eu},
   CMS~\protect\cite{Chatrchyan:2012nj} and 
   ALICE~\protect\cite{Poghosyan:2011sp}. \textbf{RHS:} Differential elastic cross section 
       compared to data from the LHC~\protect \cite{Antchev:2011zz,Antchev:2013gaa}, 
       the ISR~\protect \cite{Amos:1985wx,Kwak:1975yq}, the
       SPS~\protect \cite{Bernard:1987vq,Bozzo:1985th,Battiston:1983gp}
       and Tevatron~\protect \cite{Abe:1993xx,Amos:1990fw}.}
       \label{Fig::xsecs}
\end{figure}
The parameters entering the eikonal (Eq.~\ref{Eq:BoundaryConditionsWithoutKT} and \ref{Eq:DEQs_WithoutKTDep}) are constrained by the total, inelastic and elastic as well as the differential elastic cross sections. The version of the KMR model forming the basis of the \Shrimps model can be seen to yield a decent description of these quantities at various beam energies (Fig.~\ref{Fig::xsecs}).

\begin{figure}[h!]
 \centering
  \includegraphics[width=0.48\textwidth]{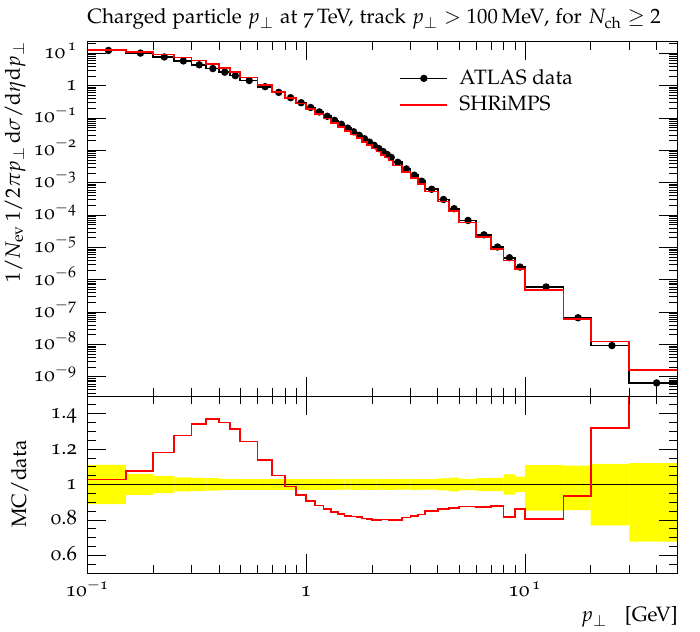}
  \includegraphics[width=0.48\textwidth]{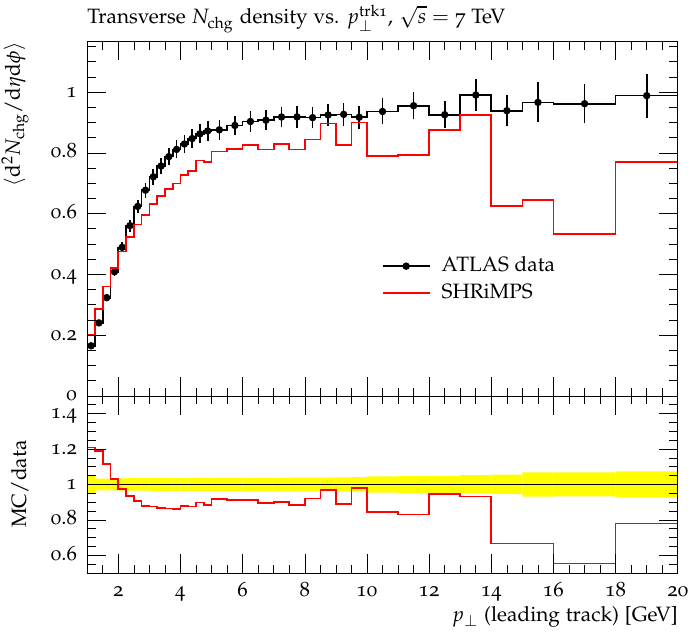}
  \caption{\textbf{LHS:} Charged particle transverse momentum spectrum
as an example for minimum bias observables measured by
ATLAS~\protect\cite{Aad:2010ac:ch2} at $\sqrt{s} = 7\,\rm{TeV}$ requiring at least
two charged particles with $p_\perp > 100\,\rm{MeV}$ within the acceptance;
\textbf{RHS:} Charged particle density in the transverse region as example for
underlying event observables measured by ATLAS~\protect\cite{Aad:2010fh} at
$\sqrt{s} = 7\,\rm{TeV}$ for $p_\perp > 500\,\rm{MeV}$.}
 \label{Fig::minbias+ue}
\end{figure}

The charged hadron transverse momentum distribution in minimum bias events
(i.e.\ requiring at least two charged particles with $p_\perp > 100\,\rm{MeV}$
within the detector acceptance) in \Shrimps is compared to experimental data in
the left panel of Fig.~\ref{Fig::minbias+ue}. In contrast to the very global
observables of the minimum bias measurements, the underlying event observables
study the correlation between soft and hard components of the event. An 
example can be seen in the right panel of Fig.~\ref{Fig::minbias+ue}, namely the
charged particle density in the region transverse (in azimuthal angle,
$60^\degree < |\Delta \phi| < 120^\degree$) to the hardest track in the event.
This region is considered to be mainly populated by the interactions of
partons not involved in the hard process.

Overall, \Shrimps is in reasonable agreement with a variety of different
measurements.

\subsection{Summary and outlook}
The \Shrimps model aims at describing minimum bias QCD interactions in hadronic
collisions, and in particular at the LHC.  Starting from the inclusive model
of Khoze, Martin, and Ryskin~\cite{Ryskin:2009tj,Ryskin:2009tk:ch2}, it adds
an interpretation of this fairly inclusive interaction picture in terms 
of an exclusive partonic language, fit for implementation in a MC
event generator, which in turn would take care of subsequent parton showering,
i.e.\ a DGLAP evolution for the fragmentation of the hard partons, and the
hadronization.  This has been realised through an implementation as a new
module, the \Shrimps module, of the multi--purpose MC event generator 
\Sherpa; somewhat in parallel the ideas of this implementation also form the 
base of the more theory--driven considerations of~\cite{Ryskin:2011qe}
% 
% \subsection*{Acknowledgments}
% We are grateful to our colleagues in the \Sherpa team for fruitful discussions
% and a very pleasant collaboration over many years.  
% This work has been supported in part by the European Commission through 
% the ``MCnet'' Initial Training Network PITN-GA-2012-315877.

%%%%%%%%%

%%%  Dime

%%%%%%%%%%

\section{\texttt{Dime}}

The \texttt{Dime} MC is an event generator~\cite{Harland-Lang:2013dia:ch2} for meson pair CEP, proceeding via the double reggeon exchange mechanism of the type show in Fig.~\ref{fig:central_double_diffraction_diagrams}. In this `one--meson exchange' model (see for instance~\cite{Azimov:1974fa,Pumplin:1976dm,Lebiedowicz:2011nb}) the mesons are produced via pomeron--pomeron fusion, with an intermediate off--shell meson exchanged in the $t$--channel. The CEP cross section is given by
\begin{equation}\label{ncross}
\sigma^{CEP}=\frac {1}{16\pi(16\pi^2)^2}\int dp^2_{1\perp}dp^2_{2\perp}dy_3dy_4dk^2_{\perp}\frac{|\mathcal{M}|^2}{s^2}\;,
\end{equation}
where $\sqrt{s}$ is the c.m.s. energy, $p_{1\perp}, p_{2\perp}$ are transverse momenta of the outgoing protons, $k_\perp$ is the meson transverse momentum and $y_{3,4}$ are the meson rapidities. Ignoring secondary reggeon contributions and soft survival effects for simplicity (these will be discussed below), the production amplitude, $\mathcal{M}$, is given by the sum $\mathcal{M}=\mathcal{M}_{\hat{t}}+\mathcal{M}_{\hat{u}}$ of the $t$ and $u$--channel contributions, with $\hat{t}=(P_1-k_3)^2$, $\hat{u}=(P_1-k_4)^2$, where $P_i$ is the momentum transfer through pomeron $i$, and $k_{3,4}$ are the meson momenta. We have
\begin{equation}\label{namp}
\mathcal{M}_{\hat{t}}=\frac 1{M^2-\hat{t}} F_p(p^2_{1\perp})F_p(p^2_{2\perp})F^2_M(\hat{t})\sigma_0^2
\bigg(\frac{\hat{s}_{13}}{s_0}\bigg)^{\alpha_{I\!\!P}(p^2_{1\perp})}\bigg(\frac{\hat{s}_{24}}{s_0}\bigg)^{\alpha_{I\!\!P}(p^2_{2\perp})}\;,
\end{equation}
where $M$ is the meson mass, $s_0=1\,{\rm GeV}^2$ and $\alpha_{I\!\!P}(p^2_{i\perp})=1.08-0.25\, p^2_{i\perp}$, for $p^2_{i\perp}$ measured in ${\rm GeV}^2$~\cite{Donnachie:1992ny}, and $s_{ij}=(p_i'+k_j)^2$ is the c.m.s. energy squared of the final--state proton--meson system $(ij)$.  The proton form factors are often taken for simplicity to have an exponential form, $F_p(t_i)=\exp(B_it_i/2)$, although as in~\cite{Khoze:2013dha:ch2} a slightly different parameterization is taken here.

We can see from (\ref{namp}) that the cross section normalisation is set by the total meson--proton cross section $\sigma_{\rm tot}(M p)=\sigma_0 (s_{ij}/s_0)^{\alpha(0)-1}$ at the relevant sub--energy; the factor $\sigma_0$ can be extracted for example from the fits of~\cite{Donnachie:1992ny}. While this is therefore well constrained for the cases of $\pi\pi$ and $KK$ production, there remain other elements and possible additions to the model, which are in general less constrained by the available data. These are: the form factor $F_M(\hat{t})$ in (\ref{namp}) of the pomeron coupling to the off--shell meson, the possibility to produce additional particles in the pomeron fusion subprocess, and the effect of Reggeization of the meson exchange in the $t$--channel. 

In \texttt{Dime} three different choices for the form factor $F_M(\hat{t})$ can be chosen, an exponential ($\sim \exp(b_{\rm exp}t)$, an `Orear--like' form ($\sim \exp(b_{\rm or}\sqrt{-t})$) and a power--like form ($\sim 1/(1-t/b_{\rm pow})$), with the parameters fitted to ISR data on $\pi^+\pi^-$ CEP~\cite{Breakstone:1990at}. Any possible effect of meson Reggeization is currently omitted from the MC, as it is not clear that this effect will be important in the relevant kinematic regime, when the mesons are produced relatively centrally, without a large separation in rapidity between them. A simple phenomenological model is used for the possibility to produce additional particles in the pomeron fusion subprocess that would ruin the exclusivity of the event; this may be turned off or on in the MC. Finally soft survival effects are included using the approach of~\cite{Khoze:2013dha:ch2}; all four model implementations described there are included in the MC. It is important to emphasise that a full treatment of the survival factor is given in the MC: it is included at the amplitude level, accounting for the differential dependence of the survival factor on the particle kinematics, rather than simply applying an overall multiplicative factor. For further discussion of these issues and description of the MC, see~\cite{Harland-Lang:2013dia:ch2}.

Currently, the \texttt{Dime} MC implements $\pi^+\pi^-$, $K^+K^-$, $\pi^0\pi^0$, $K^0K^0$ and $\rho_0(770)\rho_0(770)$ production. In the $\rho_0\rho_0$ case the mesons are decayed via $\rho_0 \to \pi^+\pi^-$, including the finite $\rho_0$ width, according to phase space only\footnote{A more complete treatment should account for the different $\rho$ polarization states, which may in general have distinct form factors $F_M(\hat{t})$, however given the lack of information about these such possible polarization effects are omitted in the current version of the MC.}, while the factor $\sigma_0$ in (\ref{namp}) is set by default to the reasonable estimate $\sigma_0^{I\!\! P}=10 \, {\rm mb}$, i.e. of order the $\pi^+\pi^-$ cross section, but taking a lower value due the larger $\rho_0$ mass. This somewhat arbitrary input is necessary due to the lack of $\rho_0 p$ scattering data with which to set the normalization (another reasonable choice may be to take $\sigma_0^{I\!\! P}=$13.63 mb as in $\pi^{\pm} p$ 
scattering~\cite{Donnachie:1992ny}). For 
$\rho_0\rho_0$ production, secondary reggeons are not included and any spin effects are currently ignored in the production subprocess. Given the relative uncertainty in the $\rho_0\rho_0$ cross section normalisation,  any effect from additional particle production is currently omitted, although this could in principle be included in the future.

%%%%%%%%%%%

%%%% Exhume

%%%%%%%%%%%

\section{\texttt{ExHuME}}\label{sec:exhume}

The Exclusive Hadronic MC Event (\texttt{ExHuME}) generator~\cite{Monk:2005ji:ch2} produces events for CEP processes. It is based on the `Durham' model described in Section~\ref{CEP:gluonex:intro} but with some simplifying assumptions. The starting point is to write the CEP cross section for the production of system $X$ of invariant mass $M_X$ and rapidity $y_X$ in the factorized form
\begin{equation}\label{fact}
\sigma=\mathcal{L}\left(M^2_X,y_X\right)\hat{\sigma}\left(M^2\right)\;
\end{equation}
  where $\hat{\sigma}$ is the subprocess cross section, which is written in terms of a colour averaged amplitude (\ref{Vnorm}), see e.g.~\cite{Khoze:2001xm,HarlandLang:2010ep:ch2} for more details. The factor $\mathcal{L}$ corresponds to the effective luminosity for producing the system $X$, and is written as (see also (\ref{bt}))
  \begin{equation}
  M_X^2\frac{\partial \mathcal{L}}{\partial y_X \partial M_X^2  \partial t_1  \partial t_2}=\left\langle S^2\right\rangle F_p(t_1)F_p(t_2) \left(\frac{\pi}{N_C^2-1}\int \frac{{\rm d}{\bf Q}_\perp^2}{{\bf Q}_\perp^4}\,f_g(x_1,x_1', Q_1^2,\mu^2)f_g(x_2,x_2',Q_2^2,\mu^2)\right)^2\;,
  \end{equation}
  where the $F_p(t_i)$ are the elastic proton form factors, for momentum transfer $t_i\approx-{\bf p}_{i\perp}^2$,  and are taken to have a simple exponential form. The $f_g$'s in (\ref{bt}) are the skewed unintegrated gluon densities of the proton, described in Section~\ref{CEP:gluonex:intro}. The factor $\left\langle S^2\right\rangle$ is the \emph{average} survival factor, which is taken to have a constant value. 
  
 \texttt{ExHuME} generates events for the CEP of a Standard Model Higgs boson, via the $gg \to H$ subprocess, and dijet and diphoton production, via the $gg\to gg$, $gg\to q\overline{q}$ and $gg\to \gamma\gamma$ subprocesses, respectively. 
% This MC implementation played a crucial role in the comparison of the `Durham' model with the CDF measurement of exclusive jet production~\cite{Aaltonen:2007hs}, which leant strong support to this perturbative approach. It has also been used in the CMS search~\cite{CMS:2012amw} for exclusive $\gamma\gamma$ production, with the corresponding limits found to be quite close to the MC predictions.
However, it should be noted that certain simplifying assumptions that have been made in this MC, and in the FPMC generator discussed in Section~\ref{sec:fpmc} which uses a similar framework, are not always reliable. In particular, as discussed in Section~\ref{sec:tagcepmotglu} the soft survival factor is not constant, but rather will depend on and effect the distribution in the proton transverse momenta ${\bf p}_\perp$. Moreover, the factorization of (\ref{fact}) only holds if the effect of any non--zero proton ${\bf p}_\perp$ inside the hard process matrix element is neglected. That is, it only includes a $J_z^P=0^+$ component, with $q_{1_\perp}=q_{2_\perp}=-Q_\perp$ taken when calculating $\hat{\sigma}$. For some processes, such as $\chi_{c(1,2)}$ production~\cite{HarlandLang:2009qe:ch2} this can be a very bad approximation. Thus, in such situations as when the $J_z^P=0^+$ component is not necessarily dominant and/or the protons are tagged, these approximations may be very bad indeed. Conversely, if the $J_z^P=0^+$ component is indeed strongly dominant, and/or the proton transverse momenta are simply integrated 
over, these simplifications are more reliable.

%%%%%%%%%%%%%%%%
%
%%%%%%  FPMC
%
%%%%%%%%%%%%%%%
%
\section{FPMC}\label{sec:fpmc}
  \subsection{Introduction}

The idea of FPMC is to produce single diffraction, double pomeron exchange, exclusive diffraction and photon--induced processes within the same framework. The diffractive and exclusive processes are implemented by modifying the \hbox{HERWIG} routine for the $e^+e^-\rightarrow(\gamma\gamma)\rightarrow X$ process. In case of the two--photon $pp$ events, the Weizs\"{a}cker--Williams (WWA) formula describing the photon emission off point--like electrons is substituted for the photon flux which properly describes the coupling of the photon to the proton, taking into account the proton electromagnetic structure. For central exclusive production, a look--up table of the effective gluon--gluon luminosity computed by \texttt{ExHuME}~\cite{Monk:2005ji:ch2}, see Section~\ref{sec:exhume}, is implemented. In case of pomeron/reggeon exchange, the WWA photon fluxes are replaced by the pomeron/reggeon fluxes multiplied by the diffractive parton density functions. 

 For processes in which the partonic structure of the pomeron is probed, the existing HERWIG matrix elements for non--diffractive production are used to calculate the production cross sections. The list of particles is corrected at the end of each event to change the type of particles from initial--state electrons to hadrons and from the exchanged photons to pomerons/reggeons, or gluons, depending on the process. 

 All these fluxes are implemented in the {\verb FLUX } routine. The user
selects the desired production mechanism by selecting the {\verb NFLUX } 
parameter. Their overview is shown in Table~\ref{fpmc:fluxes}. The energy which is carried by the exchanged object (photon/pomeron/reggeon/gluon) from the colliding particles is driven by the 
parameters {\verb WWMIN } and {\verb WWMAX }, representing the minimal and maximal momentum fraction loss $\xi$ of the collided hadron.
\begin{table}  
\centering
%     \caption{Available fluxes}
%   	\vspace{0.25cm}
  	\begin{tabular*}{0.7\textwidth}{@{\extracolsep{\fill}}|r|l|} 
    	\hline
			\textbf{NFLUX} & \textbf{Flux} \\ \hline  %\hline
			9  &  QCD factorized model, Pomeron flux \\ %\hline
			10 &  QCD factorized model, Reggeon flux \\ %\hline
%			11 & 	QCD Bialas--Landshoff \\ %\hline
			12 &	QED flux from Cahn, Jackson; $R\sim1.2A^\frac{1}{3}$ \\ %\hline
			13 & 	QED flux from Drees et al., valid for heavy ions only\\ %\hline
			14 & 	QED flux in pp collisions, from Papageorgiou\\ %\hline
			15 & 	QED flux in pp collisions, from Budnev et al.\\ %\hline
			16 & 	QCD KMR flux \\
			17 & QCD factorized model, Pomeron--Reggon flux \\
			19 & QCD factorized model, Pomeron Reggeon fluxes \\
			20 & QED flux Budnev -- QCD factorized model,
			Photon--Pomeron \\
			21 & QCD factorized model, Reggeon--Pomeron fluxes \\
			22 & QED flux Budnev -- QCD factorized model, 
			Pomeron--Photon \\
	\hline
		\end{tabular*}
\caption{Overview of available fluxes which are implemented in the FPMC generator. The QED flux corresponds to the photon exchange. 
The QCD flux corresponds to the pomeron/reggeon exchange, or to the gluon exchange in the case of the CEP predicted by the KMR calculation.}
\label{fpmc:fluxes}
\end{table}

\subsection{Two--photon interactions}

Two--photon production in $pp$ collisions is described in the framework of the Equivalent Photon Approximation 
(EPA)~\cite{Budnev}. The almost real photons (with low photon virtuality $Q^2=-q^2$) are 
emitted by the incoming protons, producing an object $X$ in the $pp\rightarrow pXp$ process, 
through two--photon exchange $\gamma\gamma\rightarrow X$. The precise form for the photon spectrum is given by (\ref{flux}).
\begin{figure}
\begin{center}
\includegraphics[width=7.5cm]{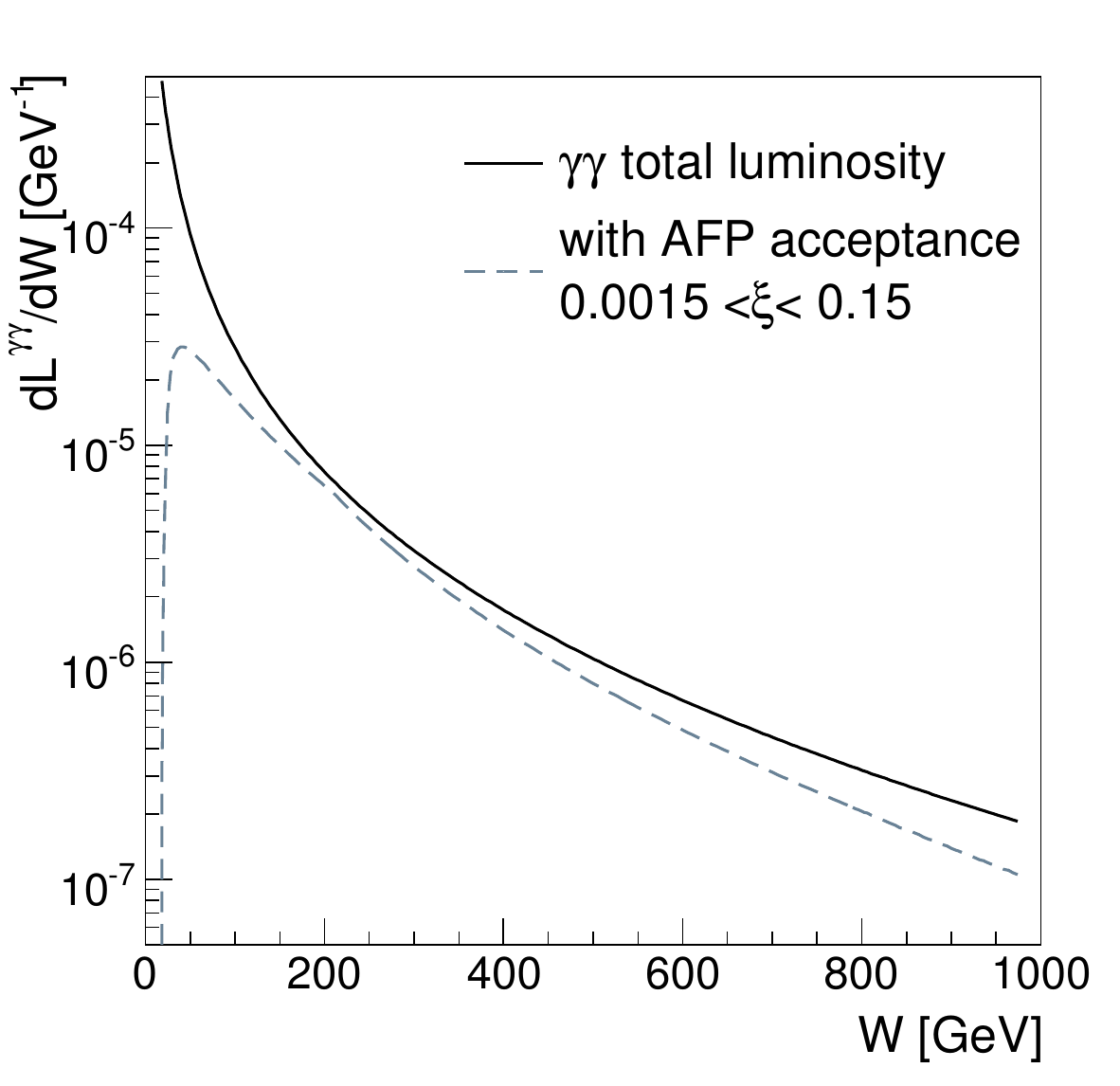}
\caption{Relative effective $\gamma\gamma$ luminosity in $pp$ collisions at $14\units{TeV}$
as a function of the  two--photon invariant mass. The maximal virtualities of the
emitted photons are set to $Q^2_{max}=2\,\GeV^2$. The dashed curve
shows the photon spectrum within the ATLAS or CMS forward detector acceptance.}
\label{luminosity}
\end{center}
\end{figure}
%
%The contribution to the integral above $Q^2_{max}\thickapprox2\GeV^2$  is very small. 
%The $Q^2$--integrated photon flux also falls rapidly as a function of the photon energy $E_{\gamma}$, 
%which implies that the two--photon production process prefers small masses $W\approx2\sqrt{E_{\gamma1}E_{\gamma2}}$. 
Integrating the product of the
photon fluxes $f(E_{\gamma1})\cdot f(E_{\gamma2})\cdot \d E_{\gamma1}\cdot \d E_{\gamma2}$ from both protons over the photon  energies while keeping
the two--photon invariant mass fixed to $W$, one obtains the two--photon effective
 luminosity spectrum $\d L^{\gamma\gamma}/\d W$.

The effective $\gamma \gamma$ luminosity is shown in Fig.~\ref{luminosity} as a function of the mass $W$. The production of heavy objects is particularly interesting at the LHC, where new particles could be produced in a very clean environment. The production rate of massive objects is however limited by the photon luminosity at high invariant masses. The integrated two--photon luminosity above $W>W_0$ for $W_0=23\GeV,\ 2\times m_W\thickapprox160\GeV$, and $1\TeV$ is $1\%$, $0.15\%$ and $0.007\%$, respectively, of the luminosity integrated over the whole mass spectrum. The luminosity spectrum for $0.0015<\xi<0.15$ is also shown in the figure (calculated in the limit of low
$Q^2$, thus setting $E_{\gamma}=\xi E$). 
 
Using the effective relative photon
luminosity $\d L^{\gamma\gamma} \slash \d W$, the total cross section reads 
\begin{equation}
      \frac{\d\sigma}{\d\Omega}=\int\frac{\d \sigma_{\gamma\gamma\rightarrow X}(W)}{\d\Omega}\frac{\d L^{\gamma\gamma}}{\d W}\d W 
\label{eq:sm:totcross}
\end{equation}
where $\d \sigma_{\gamma\gamma\rightarrow X}/\d\Omega$ denotes the differential 
cross section of the sub--process $\gamma\gamma\rightarrow X$, dependent on the invariant mass of the two--photon system.

In FPMC, the formula (\ref{eq:sm:totcross}) is implemented in the routine 
FLUX. It is normalized by the beam energy and is actually dimensionless, parameterized by the momentum fraction loss of the proton $\xi=E_\gamma/E$.

Many photon--induced processes have been implemented in FPMC, namely:
\begin{itemize}
\item dilepton production
\item standard model $\gamma \gamma$ production including lepton, quark and $W$
loops
\item anomalous quartic coupling production of $WW$, $ZZ$ and $\gamma \gamma$
\item anomalous trilinear production of $WW$ and $ZZ$
\item SM Higgs boson production
\end{itemize}

\subsection{Implementation of pomeron and reggeon exchanges in inclusive
diffraction}

Diffractive parton density functions (DPDF) were measured at HERA. The 
outcome of the fits are the values of the pomeron and reggeon trajectories 
$\alpha_{I\!\!P}(t)=\alpha_{I\!\!P}(0)+\alpha'_{I\!\!P}$t, $\alpha_{I\!\!R}(t)=\alpha_{I\!\!R}(0)+\alpha'_{I\!\!R}$t 
governing the corresponding flux energy 
dependence, and the pomeron/reggeon parton distribution
functions $f_{P/p}(\beta,\,Q^2)$, $f_{R/p}(\beta,\,Q^2)$. Only the 
normalization of the product of the diffractive structure function 
$f^D_i(x, Q^2, x_P, t)$ and of the 
pomeron/reggeon flux is fixed by the QCD fits. 
Therefore, the normalization of the fluxes is conventionally fixed at 
$x_P=x_R=0.003$ such that
\begin{equation}
x_P \int_{t_\mathrm{cut}}^{t_\mathrm{min}}f_{P/p}\,\d t =1 
\end{equation}
where $|t_\mathrm{min}|\simeq m_p^2x_P^2/(1-x_P)$ is the minimum 
kinematically accessible value of $|t|$, $m_p$ is the proton mass and 
$|t_\mathrm{cut}|=1.0\GeV^2$. The normalization of the reggeon flux is 
defined in  the same way.

The pomeron and reggeon parameters obtained in the most recent 
H1 QCD fits are shown in Table~\ref{parameters}.
\begin{table}
\centering
\begin{tabular}{|c|c|c|}
\hline
Fit parameter &  Fit A &  Fit B\\
\hline\hline
$ \alpha_P(0)$ & $1.118\pm0.008 $  & $1.111\pm0.007 $  \\
$n_R$ & $(1.7\pm0.4)\times10^{-3}$ & $(1.4\pm0.4)\times10^{-3}$ \\
\hline
$\alpha'_P$ &  \multicolumn{2}{l|}{\hspace{2cm}$0.06^{+0.19}_{-0.06}\GeV^{-2}$}\\
$B_P$ & \multicolumn{2}{l|}{\hspace{2cm}$5.5_{+0.7}^{-2.0}\GeV^{-2}$}\\ 
$ \alpha_R(0)$  &\multicolumn{2}{l|}{\hspace{2cm}$0.5\pm0.10$} \\
$\alpha'_R$ &\multicolumn{2}{l|}{\hspace{2cm}$0.3^{+0.6}_{-0.3}\GeV^{-2}$} \\
$B_R$ & \multicolumn{2}{l|}{\hspace{2cm}$1.6_{+0.4}^{-1.6}\GeV^{-2}$}\\ 
\hline
\end{tabular}
\caption{Diffractive structure function parameters of QCD Fit A and Fit B 
fits \cite{Aktas:2006hy:ch2}. These structure functions are used as defaults in FPMC.}
\label{parameters}
\end{table}
The implemented diffractive parton densities are summarized in Table~\ref{fpmc:pdfs} and can be selected with the IFIT parameter. The flux parameters are fixed in the routine HWMODINI where the initial parameters are 
set. The parton densities are used in the routine HWSFUN where the call to the 
H1 tables (the source code can be found at \cite{Aktas:2006hy:ch2}) is made.

\begin{table}
\centering
  	\begin{tabular}{|r|l|c|} 
    	\hline
			\textbf{IFIT} & \textbf{PDF set} & Source \\ \hline  \hline
			10 & H1 (old) & \cite{Royon:2006by} \\  %\hline
			20 & Zeus  (old) &\cite{Royon:2006by} \\  %\hline
			30 & combined H1 and Zeus (old) & \cite{Royon:2006by}\\ \hline
         100& H1 Fit B &\cite{Aktas:2006hy:ch2} \\ 
         101& H1 Fit A &\cite{Aktas:2006hy:ch2} \\ \hline 
		\end{tabular}
\caption{Implemented diffractive parton density functions in FPMC. The most recent are the H1 Fits A and Fit B  IFIT=101, 100.}
\label{fpmc:pdfs}
\end{table}

Predictions for the single diffractive and double pomeron exchange 
dijet cross sections for various jet $p_T$ thresholds are summarized in Table~\ref{fpmc:dijets}. They are given assuming pomeron exchange only, since 
the contribution from sub--leading exchanges is found to be negligible at the LHC.
Similarly, the single diffractive $W$ and $Z$ production cross sections 
are shown in Table~\ref{fpmc:wz}. All numbers are calculated with the H1 Fit B 
parton density functions, with a cut on the maximum momentum fraction 
loss of the proton $\xi_{max}=0.1$. The rates are not corrected for the 
survival probability which is expected to be 0.06 at the 
LHC \cite{Khoze:2000wk_4}.  

\begin{table}
\centering
\begin{tabular}{|ccc|}
  \multicolumn{3}{c}{$\sqrt{s}=1.96\TeV$}  \\
\hline
 PTMIN [\GeV] &   SD dijets [pb]  &  DPE dijets [pb]  \\
\hline
\hline
10\GeV &   $180\cdot10^{5} $& $429\cdot 10^{3}$\\
15\GeV &   $29\cdot10^5$& $42\cdot10^3$\\
25\GeV &   $23\cdot10^5$& $1.3\cdot 10^3$\\
\hline
\end{tabular}
\begin{tabular}{|ccc|}
  \multicolumn{3}{c}{$\sqrt{s}=14\TeV$}  \\
\hline
 PTMIN [\GeV] &   SD dijets [pb] &  DPE dijets [pb]   \\
\hline
\hline
15\GeV &   $107\cdot10^{6}$ & $5.2\cdot10^6$\\
25\GeV &   $14\cdot10^6$ &  $5.4\cdot10^5$\\
35\GeV &   $3.5\cdot10^6$ & $1.1\cdot10^5$\\
\hline
\end{tabular}
\caption{Single diffractive and double pomeron exchange dijet cross sections for various thresholds at the Tevatron and the LHC. No survival probability factor, which is expected to be approximately 0.06, is applied.}
\label{fpmc:dijets}
\end{table}

\begin{table}
\centering
\begin{tabular}{|cccc|}
\hline
process  & $\sqrt{s}=1.96\TeV$    &  $\sqrt{s}=14\TeV$  &\\
\hline
 $W\rightarrow$ anything+Gap  &   468\pb  & 9570\pb &  IPROC=11499  \\
 $Z/\gamma\rightarrow$ anything+Gap  &   640\pb  & 6292\pb  &  IPROC=11399\\
\hline
\multicolumn{4}{c}{ Flags: TYPEPR='INC', TYPINT='QCD', PART1='P', PART2='E+', 
WWMAX=.1 } \\
\hline
\end{tabular}
\caption{Total single diffractive production cross section of $W$ and $Z/\gamma$ bosons at $\sqrt{s}=14\TeV$. No survival probability factor, which is expected to be 0.06, was applied.}
\label{fpmc:wz}
\end{table}

Recently, jet--gap--jet events were also implemented in DPE  following
the Mueller--Tang formalism~\cite{jgjpap}.

\subsection{Implementation of exclusive production}
The implementation of central exclusive Higgs and dijet productions is not done in terms of a flux, as in the cases discussed above, but rather in terms of the effective gluon--gluon luminosity. The calculation of the effective gluon--gluon luminosity in exclusive events \cite{Khoze:2001xm} is available in the ExHuME generator, see Section~\ref{sec:exhume}. It is convenient to study the forward processes in the same framework with the same hadronization model. We therefore adopted the ExHuME calculation of the gluon--gluon luminosity in FPMC.

CEP is implemented by means of look--up tables of the 
gluon--gluon luminosity calculated by ExHuME (Lumi()routine)
as a function of the momentum fraction losses of the scattered 
protons  $\xi_1,\,\xi_2$. It is evaluated and added to the event weight after 
generation of both of $\xi_1,\,\xi_2$. The rest of the event is then 
generated with the  $gg\rightarrow q\bar{q},gg, H$  matrix elements 
respecting the $J_z=0$ selection rule. The effective 
gluon--gluon luminosity included in FPMC and the one calculated by 
ExHuME (v1.3.3) are in good agreement. 
%

%%%%%%%%%

%%%   STARLIGHT

%%%%%%%%%

\section{STARLIGHT}

STARLIGHT is a MC event generator for electromagnetic interactions in nucleus-nucleus, proton-nucleus, and proton-proton collisions~\cite{starlight}. Simulations are performed for ultra-peripheral collisions, where the nuclei/protons are separated by impact parameters larger than the sum of their radii. In these collisions, purely hadronic interactions are strongly suppressed while the cross sections for electromagnetic interactions remain large~\cite{Bertulani:2005ru,Baltz:2007kq:ch2}. Two-photon and photonuclear/photon-proton interactions are included in the model. The main focus is on exclusive particle production where the nuclei remain intact, $A+A \rightarrow A+A+X$, but general photonuclear interactions $\gamma + A \rightarrow X$ are included through an interface to the DPMJET MC. The model is primarily developed for interactions at high energy colliders such as RHIC, the Tevatron, and the LHC. 

The electromagnetic field is treated as an equivalent flux of photons, and the photon spectrum is calculated in impact parameter space. Working in impact parameter space is preferable when dealing with hadronic beams, since it provides the clearest way to suppress interactions where the beams interact hadronically. In simple terms, this means that interactions with impact parameters $b < R_1 + R_2$ have to be excluded ($R_{1,2}$ are the nuclear radii). In STARLIGHT, the exclusion of hadronic interactions is done through a calculation of the  hadronic interaction probability using the Glauber model.

The dominating exclusive particle production mechanism in high-energy nucleus-nucleus collisions is photonuclear vector meson production~\cite{Klein:1999qj}. In these interactions, the photon fluctuates to a vector meson, which becomes real by scattering ``elastically'' off the target nucleus. For momentum transfers $|t| < (1/R)^2$, the vector meson couples coherently to all nucleons in the target and the cross section is enhanced. For larger momentum transfers, the vector meson may scatter quasi-elastically off a single nucleon. Coherent and incoherent photonuclear production of the $\rho^0$, $\omega$, $\phi$, $J/\psi$, $\psi(2S)$, and $\Upsilon(1S,2S,3S)$ vector mesons are included in STARLIGHT. In all cases, including asymmetric systems such as proton-nucleus collisions, either projectile can act as photon emitter or target. 

The photonuclear vector meson production cross section is calculated from the corresponding $\gamma + p \rightarrow V + p$ cross section using the Glauber model. The photon-proton cross section is obtained from phenomenological fits to data, mostly from the electron-proton collider HERA. Interference between the two photon emitter and target configurations will modify the transverse momentum spectrum at low momenta~\cite{Klein:1999gv}. This interference may be optionally included. 

STARLIGHT also includes two-photon production of single pseudo-scalar and tensor mesons as well as dilepton pairs~\cite{Baltz:2009jk:ch2}. The total cross section is obtained by convoluting the photon spectra from the two beams with the two-photon cross section, $\sigma(\gamma \gamma \rightarrow X)$, under the requirement that there should be no accompanying hadronic interaction in the same event. For single meson production, $\sigma(\gamma \gamma \rightarrow M)$ is proportional to the two-photon decay width, $\Gamma_{\gamma \gamma}$, while for dilepton pair production the Breit--Wheeler cross section $\sigma(\gamma \gamma \rightarrow l^+ l^-)$ is calculable from lowest order QED. Both for two-photon and photonuclear production of single mesons, the decay into two charged daughter particles is simulated taking into account the effects of polarization on the decay angle for mesons with $J > 0$. 

In collisions of truly heavy ions (e.g. Au at RHIC or Pb at the LHC), the probabilities of exchanging multiple photons in a single event is high~\cite{Baltz:2002pp}. These additional photons typically have low energy but can lead to the breakup of one or both nuclei. Two-photon and photonuclear particle production can be simulated in STARLIGHT for various breakup scenarios of one or both beam nuclei. 

General photonuclear interactions $\gamma+A \rightarrow X$ can be simulated with the DPMJET model~\cite{Engel:1996yb}. STARLIGHT includes an interface to run DPMJET with photon spectra appropriate for heavy-ion beams. Emission of a photon from one or both nuclei in the same event can be simulated, with photon spectra calculated as described in~\cite{Djuvsland:2010qs}. 

STARLIGHT has been found to give a good description of two-photon production of dilepton pairs in heavy-ion collisions at RHIC~\cite{Adams:2004rz,Afanasiev:2009hy} and the LHC~\cite{Abbas:2013oua:ch2,Nystrand:2014vra}. Exclusive photonuclear production of $\rho^0$ mesons in heavy-ion collisions~\cite{Nystrand:2014vra,Abelev:2007nb,Adler:2002sc} and $J/\psi$ mesons in $pp$/$p \overline{p}$ collisions~\cite{Aaltonen:2009kg:ch2,Aaij:2014iea:ch2} are also well reproduced. The cross section for photonuclear $J/\psi$ production at the LHC is found to be overestimated, presumably because nuclear gluon shadowing is not included in the model~\cite{Abbas:2013oua:ch2}.

%%%%%%%%%%%%%

%%%%%  Superchic

%%%%%%%%%%%%%

\section{\texttt{SuperChic}}\label{sec:superchic}

\subsection{Version 1}

The original version of \texttt{SuperChic} was designed to generate events for the CEP of $\chi_{c,b}$ and $\eta_{c,b}$ quarkonia, as described in~\cite{HarlandLang:2009qe:ch2,HarlandLang:2010ep:ch2}. Rather than integrating (\ref{bt}) directly, with the dependence on the outgoing proton $\bf{p}_\perp$ included inside the integral, an expansion was performed, so that for small proton $\bf{p}_\perp$, the $\chi_{c0}$ amplitude may for example be written as 
\begin{align}\nonumber
T_0 &\propto \int \frac{{\rm d}^2 Q_\perp ({\bf q}_{1_{\perp}}\!\cdot\!{\bf q}_{2_{\perp}})}{Q_\perp^2 {\bf q}_1^2 {\bf q}_2^2}\,f_g(x_1,x_1',Q_1^2,\mu^2)f_g(x_2,x_2',Q_2^2,\mu^2)\\ \label{taylor0}
&\approx C_0+C_1({\mathbf p^2_{1_{\perp}}}+{\mathbf p^2_{2_{\perp}}})+C_{12}({\mathbf p_{1_{\perp}}}\cdot{\mathbf p_{2_{\perp}}})+\cdots \; .
\end{align}
Squaring (\ref{taylor0}) and keeping only the leading terms in $p_{i_\perp}^2$, we can see that this expansion is equivalent to making the replacement (at lowest order in $p_{i_\perp}^2$)
\begin{equation}\label{exp}
e^{-b{\mathbf p^2_{i_{\perp}}}} \to e^{-(b-2\frac{C_1}{C_0}){\mathbf p^2_{i_{\perp}}}} \; .
\end{equation}
Thus to a first approximation we expect the inclusion of non--zero $p_\perp$ in the amplitude calculation to simply result in a change in the effective slope of the proton form factor. This approach may be readily extended to the higher spin $\chi_{1,2}$ and odd-parity $\eta$ states, see~\cite{HarlandLang:2009qe:ch2}. This allows a more precise inclusion of non--zero proton $\bf{p}_\perp$ effects than is given by simply assuming the forward proton limit when calculating the subprocess matrix element (an assumption that would moreover give a vanishing cross section for the $\chi_{1,2}$ and $\eta$ states). The decay $\chi_c \to J/\psi\gamma\to \mu^+\mu^-\gamma$ is also included in the MC, including full spin correlations, see~\cite{HarlandLang:2009qe:ch2}. This MC treatment was subsequently extended to include $\gamma\gamma$~\cite{HarlandLang:2010ep:ch2} and meson pair ($\pi\pi$, $KK$, $\rho\rho$, $\eta(')\eta(')$) CEP within the approach of~\cite{HarlandLang:2011qd:ch2,Harland-Lang:2013bya}.

In addition, \texttt{SuperChic} models the photoproduction process of C--odd vector mesons ($J/\psi$, $\Upsilon(1S)$, $\psi(2S)$). The cross sections are normalised using a fit to HERA data~\cite{Aktas:2005xu}
\begin{equation}\label{herafit}
\frac{{\rm d}\sigma(\gamma^*p\to X\,+\,p)}{{\rm d}{\bf p}_{2\perp}^2}=\frac{N}{b}\left(\frac{W}{1 {\rm GeV}}\right)^\delta e^{-b{\bf p}_{2\perp}^2}\;,
\end{equation}
with $\delta= 0.72$, $N = 3 $ nb in the case of $J/\psi$ production, while for $\Upsilon(1S)$ production the fit of~\cite{Motyka:2008ac:ch2} is taken, which gives $\delta=1.63$ and $N=0.12$ pb. In the case of $\psi(2S)$, the same value of  $\delta= 0.72$ as for the $J/\psi$ is taken, with $N=0.498$ nb: any difference in the energy scaling cannot be reliably determined from the limited statistics HERA data~\cite{Adloff:2002re}. The photon flux is given as in~\cite{Albrow:2008pn}, while a simplified form for the survival factor, as in~\cite{Schafer:2007mm:ch2}, is used. A Regge scaling behaviour is taken for the slope $b$, with
\begin{equation}
b=b_0+4\alpha'\log\left(\frac{W}{W_0}\right)\;,
\end{equation}
with $b_0=4.6\,{\rm GeV}^{-2}$, $W_0=90$ GeV, and $\alpha'=0.16\,{\rm GeV}^{-2}$. For the $J/\psi$ and $\Upsilon(1S)$ the decay to $\mu^+\mu^-$ is included, with full spin correlations, while the $\psi(2S)\to J\psi \pi^+\pi^-\to \mu^+\mu^-\pi^+\pi^-$ is also included, distributed according to phase space. In addition, in the case of $J/\psi$ and $\Upsilon$ production, the simple leading order QCD cross section is also included, as in~\cite{Ryskin:1992ui}, with both MSTW08~\cite{Martin:2009iq:ch2} and CTEQ6~\cite{Pumplin:2002vw} PDFs.

\subsection{Version 2}

An update of  \texttt{SuperChic} is currently close to release, which will address some of the limitations present in the previous version. In particular, the approximation (\ref{exp}) is no longer applied: instead, the exact $p_\perp$ dependence of the matrix element is used and, significantly, soft survival effects are included at the amplitude level, that is differentially and not as an overall constant factor. In this way the influence of the survival factor on the distribution of the outgoing protons, which as discussed in Section~\ref{sec:tagcepmotglu} can be quite significant, is included. In addition to this, the code has been re-written to allow all elements of the Durham model (PDF choice, skewness effects, model of soft survival) to be adjusted by the user in a relatively straightforward way. Finally, the range of processes generated is increased to include 2 and 3 (quark and gluon) jet, Higgs boson and double $J/\psi$ production, in the first instance. This project is ongoing, and further developments are planned for the future.

%%%%%%%%%%%%%%

%%%% MC plots

%%%%%%%%%%%%%%

\section{LHC forward measurements and MC tuning}\label{sec:mctuning}

In this section a small selection of comparison plots between LHC diffractive measurements and MC predictions are shown, in all cases made using the MCPLOTS repository~\cite{Karneyeu:2013aha}. These are intended to serve as an indication of the way in which already such measurements can be of great use in tuning the available MCs, with further data to come increasingly allowing a differentiation between the model inputs.

\begin{figure}[h]
    \centering  \includegraphics[scale=0.28]{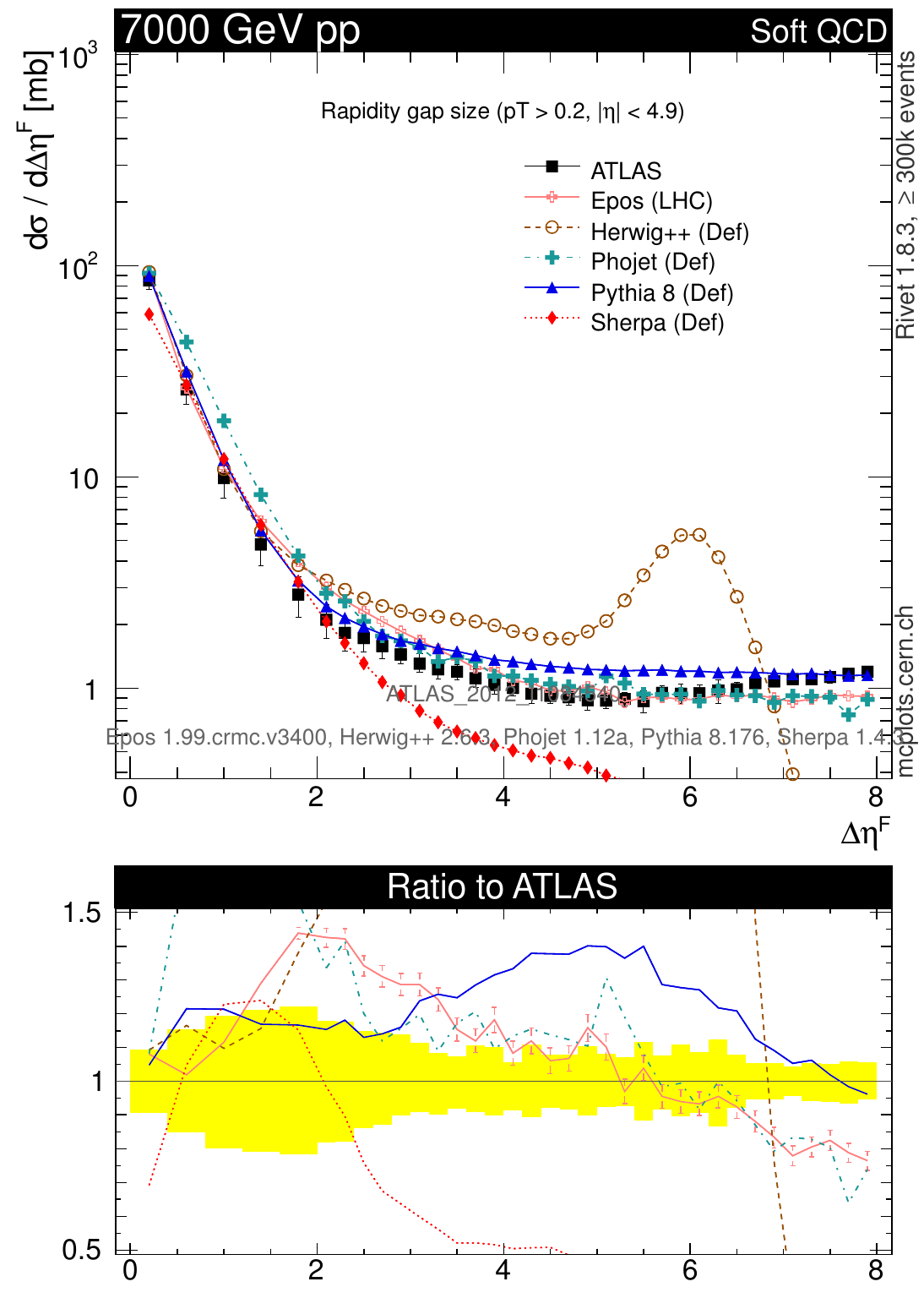}
    \includegraphics[scale=0.28]{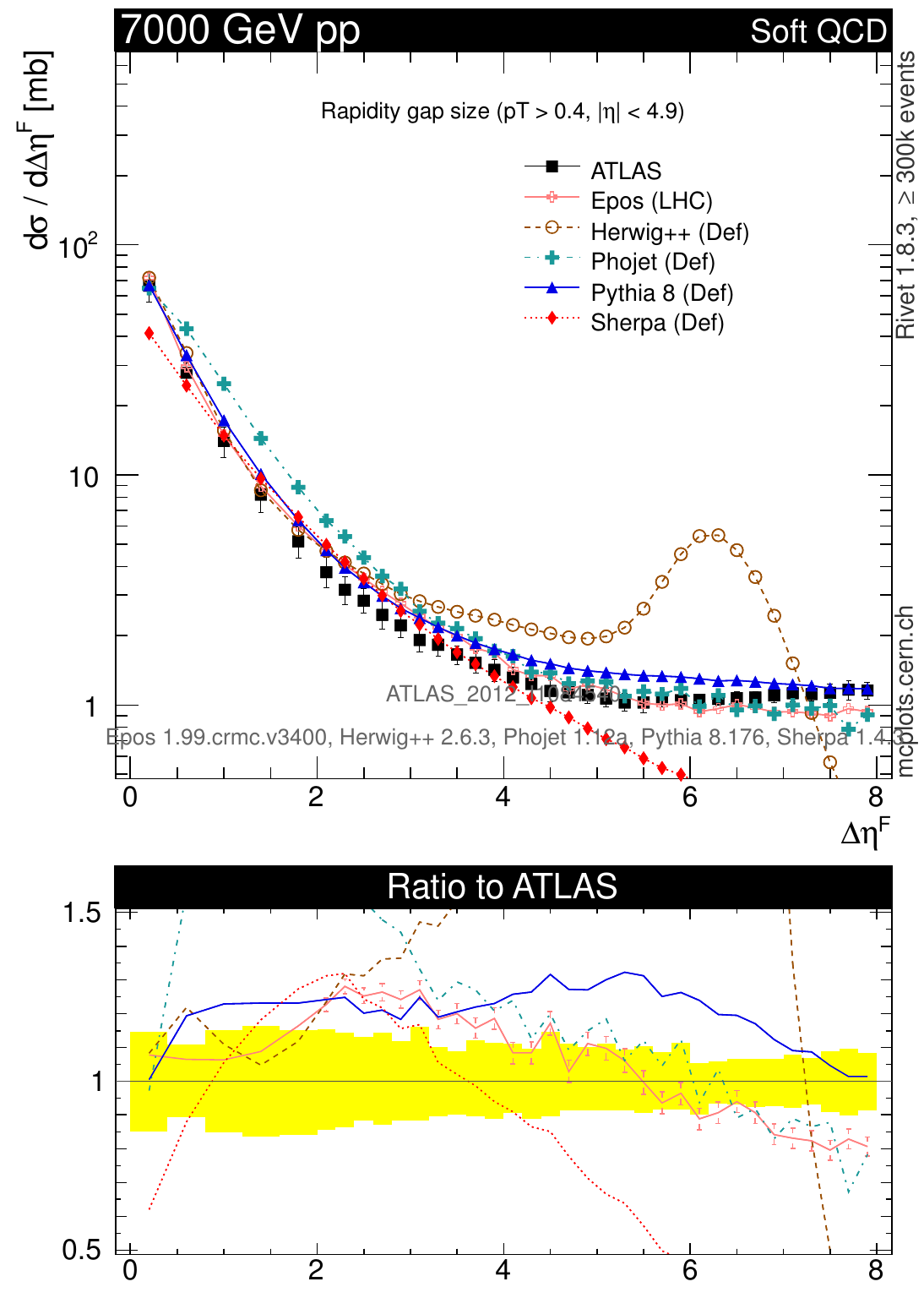}
    \caption{Comparison of MC distributions with ATLAS forward rapidity gap cross section~\cite{Aad:2012pw:ch2} at $\sqrt{s}=7$ TeV, as a function of the gap size $\Delta \eta_F$ in which no final state particles are produced above some threshold $p_\perp^{\rm cut}$.}
    \label{mcplots:gapsize}
    \end{figure}

In Fig.~\ref{mcplots:gapsize} a comparison with the ATLAS forward rapidity gap cross section~\cite{Aad:2012pw:ch2} at $\sqrt{s}=7$ TeV, as a function of gap size, $\Delta \eta_F$, in which no final state particles are produced above a transverse momentum threshold $p_\perp^{\rm cut}$, is shown. Such data is invaluable for tuning the various input parameters of the MCs, such as the form of the pomeron flux. Overall it is clear that there is a large spread in MC predictions: Herwig++ has no explicit diffractive model, although it nevertheless generates quite large rapidity gap events, but clearly fails to describe the shape or magnitude of the data; PYTHIA tends to overestimate the data, a result which remains true for other available tunes; Sherpa clearly struggles to describe the shape and size of the data at higher $\eta_F$; EPOS (LHC re--tune) gives the best overall agreement.

  \begin{figure}[h]
    \centering  \includegraphics[scale=0.28]{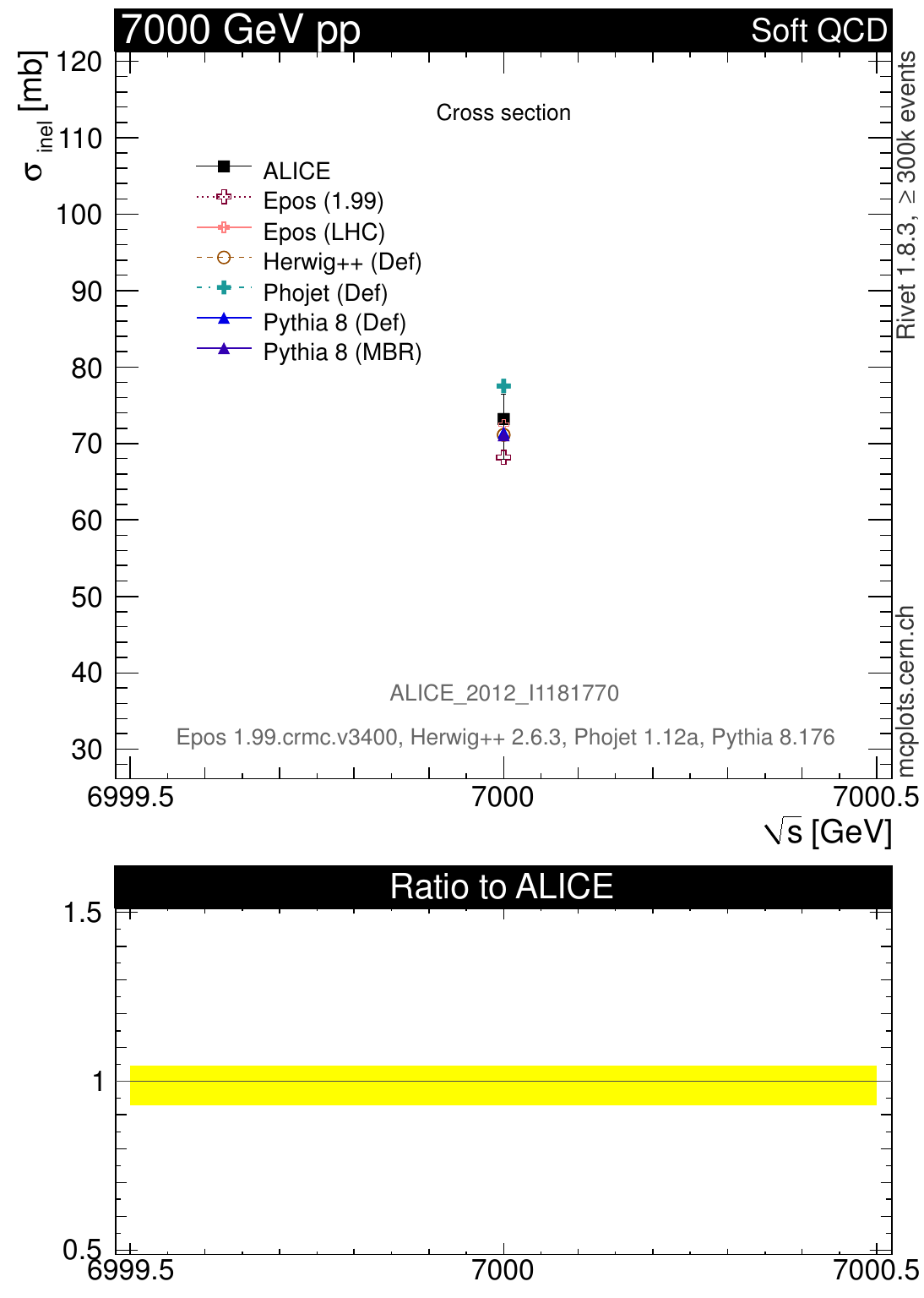}
    \includegraphics[scale=0.28]{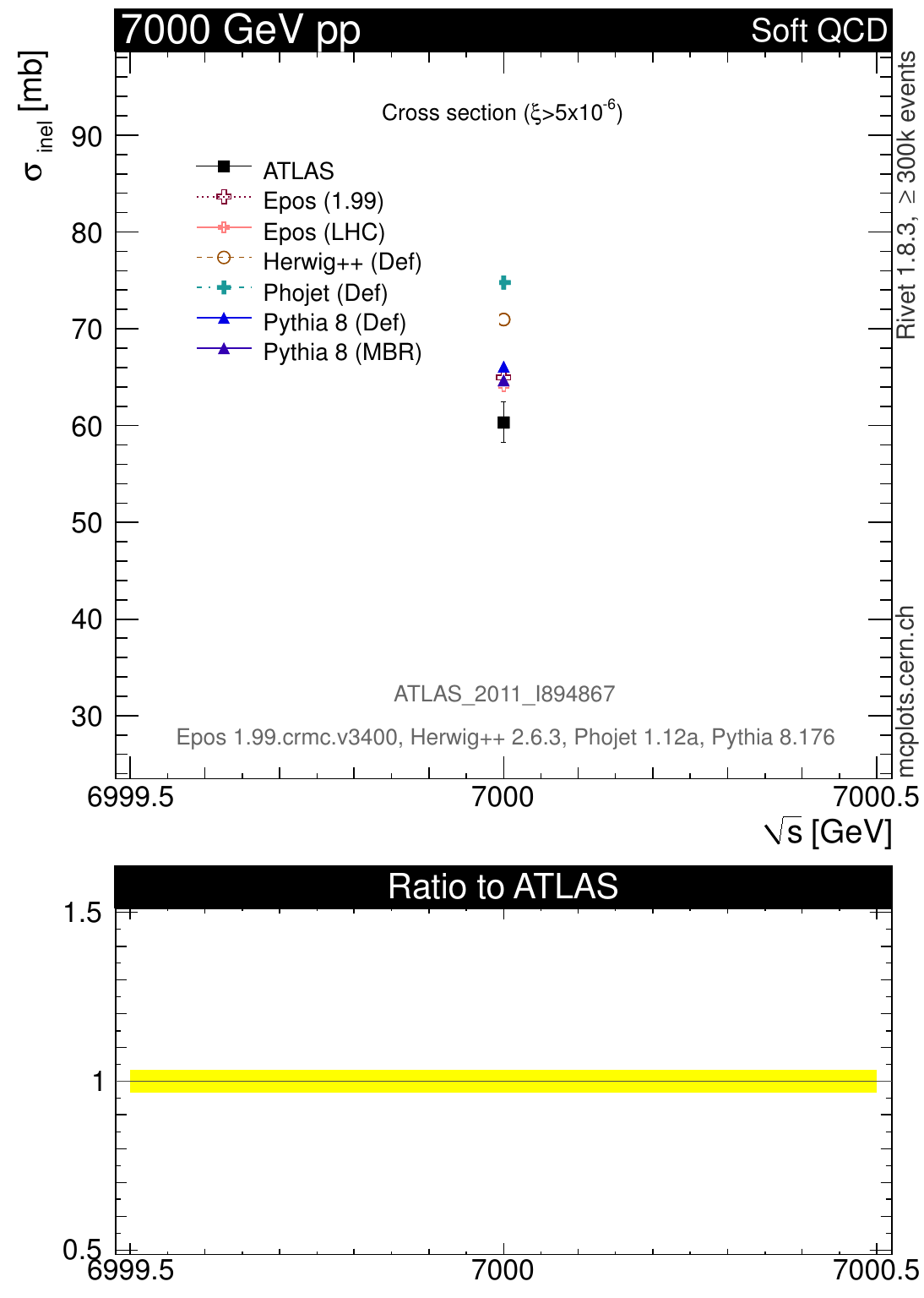}
    \includegraphics[scale=0.28]{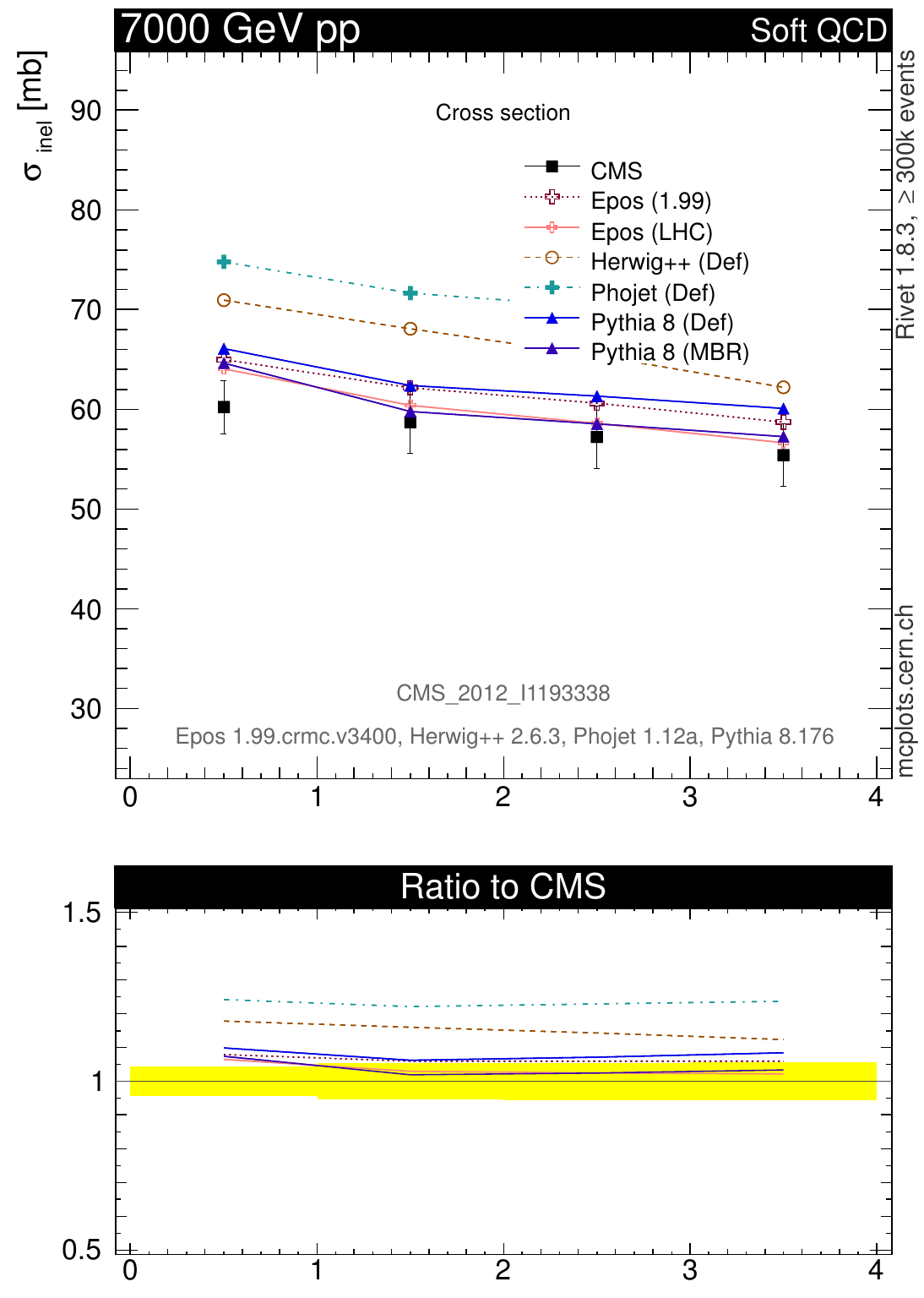}
    \caption{Comparison of MC predictions with ALICE~\cite{Abelev:2012sea}, ATLAS~\cite{Aad:2011eu} and CMS~\cite{Chatrchyan:2012nj} measurements of the inelastic cross section at $\sqrt{s}=7$ TeV.}
    \label{mcplots:inel}
    \end{figure}

         \begin{figure}[h]
    \centering  \includegraphics[scale=0.28]{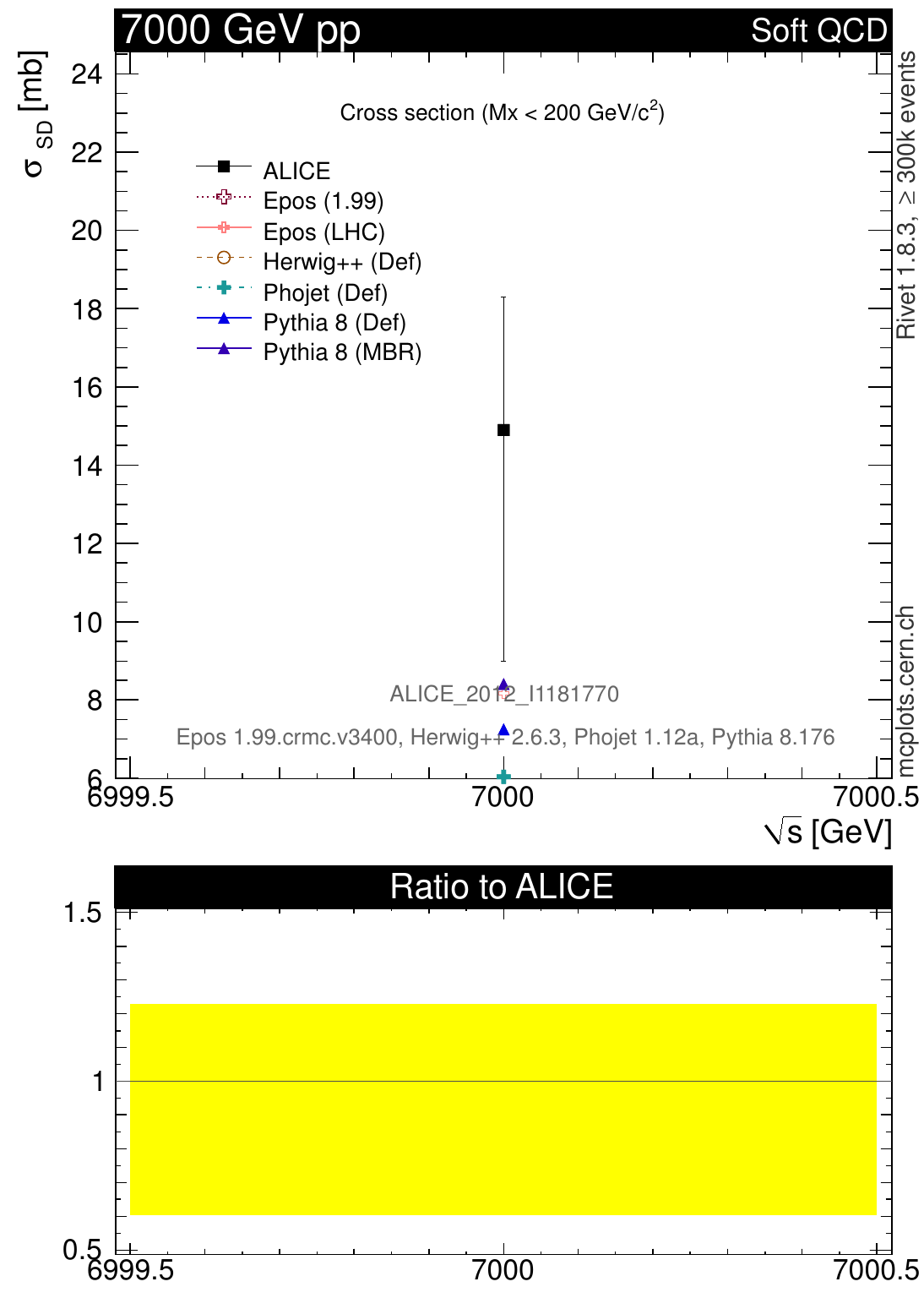}
    \includegraphics[scale=0.28]{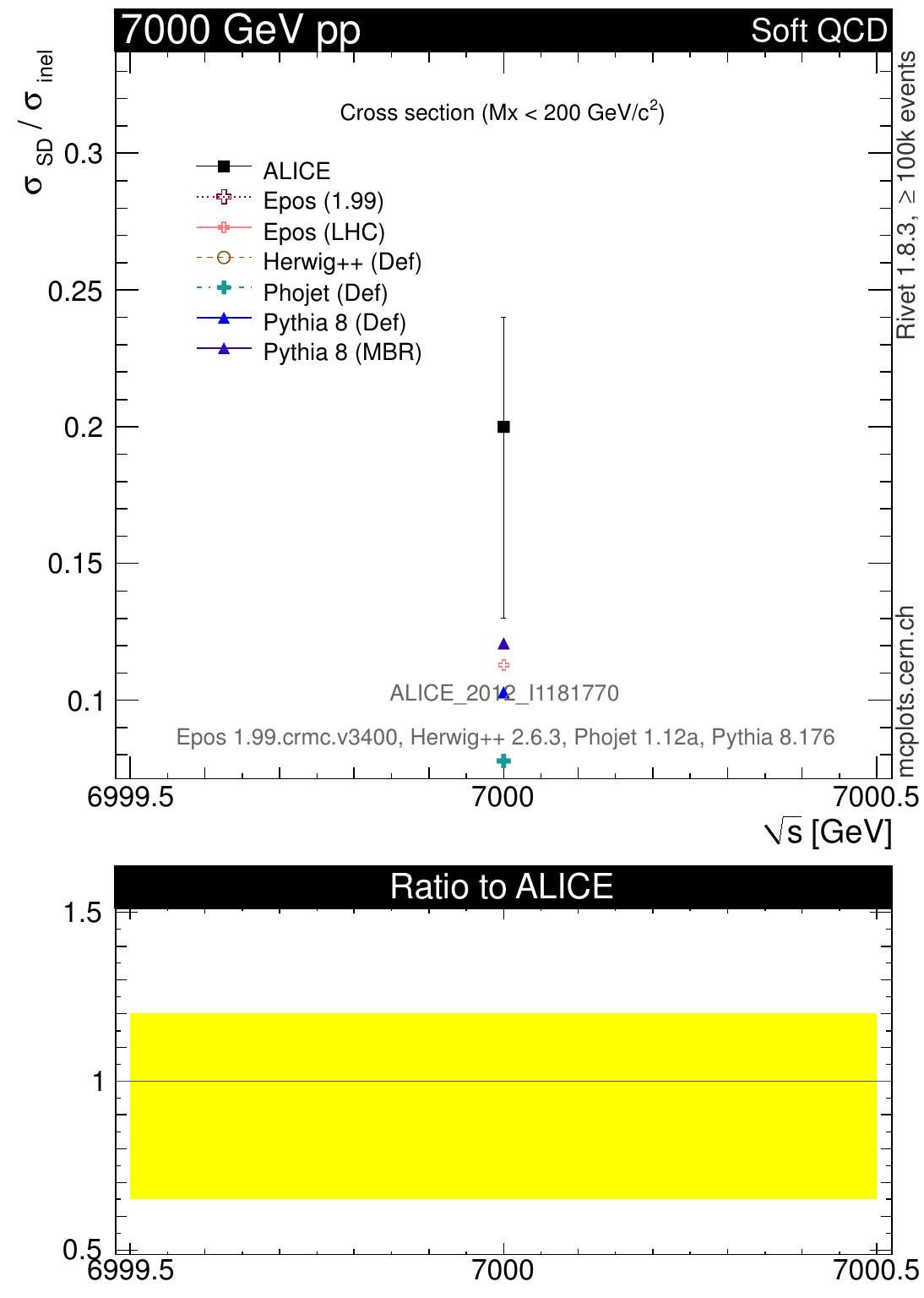}
    \caption{Comparison of MC predictions to the ALICE~\cite{Abelev:2012sea} measurements of the ratios of the single--diffractive (for a diffractive mass $M_X<200$ MeV) and double--diffractive cross sections (for a gap $\Delta \eta > 3$) to the inelastic cross section.}
    \label{mcplots:aliceddsd}
    \end{figure}
    
In Fig.~\ref{mcplots:inel} ALICE~\cite{Abelev:2012sea}, ATLAS~\cite{Aad:2011eu} and CMS~\cite{Chatrchyan:2012nj} measurements of the inelastic cross section at $\sqrt{s}=7$ TeV are shown. The CMS data correspond to two independent methods, one using the forward calorimeters (first bin in figure), and another using the central tracker, with requirements of there being more than one, two or three tracks with $p_\perp>200$ MeV and $|\eta|<2.4$; these measurements are therefore sensitive to different phase space regions. The ATLAS and CMS (calorimetry) measurements corresponds to the region $\xi=M_X^2/s>5\times 10^{-6}$, below which low mass diffraction is not seen within the detector acceptance, while the ALICE measurement corresponds to an extrapolation to this low $M_X$ region. We can see that all MCs tend to overestimate the ATLAS and CMS ($\xi>5\times 10^{-6}$) measurements, but the agreement is better for the CMS central track--based measurement. Altogether there is broad agreement between the MCs and data, but with higher precision measurements in the future, it will be possible to place more stringent constraints on these predictions. 

Finally in Fig.~\ref{mcplots:aliceddsd} the ALICE~\cite{Abelev:2012sea} measurements of the ratios of the single--diffractive (for a diffractive mass $M_X<200$ MeV) and double--diffractive cross sections (for a gap $\Delta \eta > 3$) to the inelastic cross section are shown. The MC predictions agree broadly within the quite large experimental uncertainties, but clearly the higher precision that will come from CMS, TOTEM and CMS+TOTEM measurements, combined with this data, will be of great use in tuning the MCs.

MC simulations are an essential part of the LHC forward physics programme,
both as a means to compare the available models of diffractive physics with
LHC measurements, as well as a tool to tune to hadronic data and hence
provide a phenomenological description of soft QCD effects, an understanding
of which is essential for wide range of high energy physics analyses,
including searches for BSM physics. In addition such models are crucial in
the modelling of cosmic ray physics. In this chapter a range of MC
generators for diffractive and exclusive processes have been considered. A
description of these MCs, which are widely used in the analyses presented in
this report, as well as discussion of the possibilities for further
constraints from future LHC data, have been presented.

%= bibliography ===================================
%\bibliographystyle{h-physrev}
%\bibliography{refs}

%% file: softdiffraction/softdiffraction.tex
\section{Introduction}

%BEGIN A MARTIN INTRODUCTIOn
High energy elastic proton-proton scattering is an important fundamental reaction,
which provides information on the $pp\to pp$ strong interaction
amplitude, and  - via unitarity, about the sum of all inelastic processes as well.  The LHC reaches sufficiently 
high energies that data should be able to distinguish between the different asymptotic 
scenarios for high energy interactions.

The $t$-slope of the elastic amplitude determines the value of the
interaction radius. Moreover, after transformation into the impact parameter 
($b_t$) representation, the elastic scattering amplitude (together with the
total cross section) allows us to trace how the strong interaction
at high energies approaches the black disk limit.
In turn, proton diffractive dissociation is driven by the
probability of parton elastic scattering. Therefore its
mass- and $t$-dependences provide (integrated) information about
the proton's partonic wave function; that is, about the $k_t$ and rapidity
distributions of the partons inside a proton.

On the other hand, the survival probability of Large Rapidity Gaps
(which are an essential feature of diffractive dissociation events, 
and arise from the exchange of a colour singlet)
reflects the probability of an additional inelastic soft interaction
in the multi-particle process.

In high energy $pp$ collisions about 40$\%$ of the total cross section comes 
from diffractive processes, like elastic scattering and single and double diffractive 
dissociation. We need to study these soft interactions to understand the structure 
of the total cross section, and the nature of the underlying events which accompany 
the rare hard sub-processes. Indeed, the hope is that a detailed study of these elastic 
and quasi-elastic soft processes will allow the construction of a Monte Carlo simulation which merges 
the soft and hard high energy interactions in a reliable and consistent way.
%END A MARTING INTRODUCITON

This chapter outlines the probability of detecting a proton in the forward detectors for high cross 
section elastic and inelastic proton interactions, and investigates details of the modelling in MC.

Past studies of large rapidity gaps in soft events are summarised and future prospects are listed, 
including the use of forward and very forward detectors
%the CASTOR detector at CMS \cite{SOFT_CMS:2013mda} and forward shower  counters 
to increase the acceptance of the LHC experiments to diffractive signatures. 

Results on the total, elastic and inelastic cross section measurements are also summarised along with the outlook for Run-II.

\section{Detecting soft diffraction with Forward Detectors}
%%%%%%%%%%%%%%%%%%%%%%%%%%%%%%%%%%%%%%%%%%%%%%%%%%%%%%%%%%%%%%%%%%%%%%%%%%%%%%%%%%%%%%%
%begin maciej 2 
%%%%%%%%%%%%%%%%%%%%%%%%%%%%%%%%%%%%%%%%%%%%%%%%%%%%%%%%%%%%%%%%%%%%%%%%%%%%%%%%%%%%%%%
%\section{Soft Diffraction with Forward Detectors}
Forward detectors offer a unique opportunity to combine information about the centrally produced system and the intact protons. This additional information will be used 
to significantly increase the purity of diffractive samples and, in some cases,
%for e.g. some exclusive production channels, 
make the measurement possible, e.g. for some exclusive production channels. One can also construct dedicated diffractive triggers utilising the coincidences between forward and central detectors.

%In this section two problems are addressed: registering an intact proton in the detector and probability of reconstructing a soft vertex for given experimental conditions. The former studies are crucial to calculate \textit{e.g.} rates, whereas the later ones are needed as one vertex requirement will be an important selection criterion in hard diffractive measurements.

Protons were generated using \textsc{Pythia 8} \cite{SOFT_Pythia8} with MBR tune\footnote{It should be noted that the differences between various MC generators are known to be significant and even a factor of 2 in the predicted cross sections can be expected.} \cite{SOFT_MBR} and assuming $\sqrt{s} = 14$ TeV. The following processes were taken into account: minimum-bias (\textsc{Pythia} process code = 101), elastic scattering (102), single diffraction (103 and 104), double diffraction (105) and central diffraction (106). Generated protons were then transported using \textsc{FPTrack} \cite{SOFT_FPTrack} to the forward detector position. 
The vertex position was smeared accordingly to values from \Tref{tab_beam_size_IP} and the momentum spread (\textit{cf.} \Tref{tab_beam_size_IP}) was applied, tables are in Section~\ref{sec:soft:forwardacceptance}.

All probabilities are for single interactions, including both elastic scattering and inelastic collisions. The interpretation of these data is in terms of tagging probabilities for high cross section processes. By multiplying this probability by an average pile-up value, the probability of a tag from a soft interaction forming a background to other hard processes is obtained. 

Under real experimental conditions, a single-sided horizontal detector
such as AFP will not be able to reconstruct elastically scattered protons
unless their \pt\ is large enough. Moreover, such detector will never reconstruct both elastic
protons since one of them will be deflected in the un-instrumented
direction.

\subsection{Per Interaction Probability of Single and Double Tag}

\begin{figure}[!htbp]
\centering
  \includegraphics[width=0.49\textwidth]{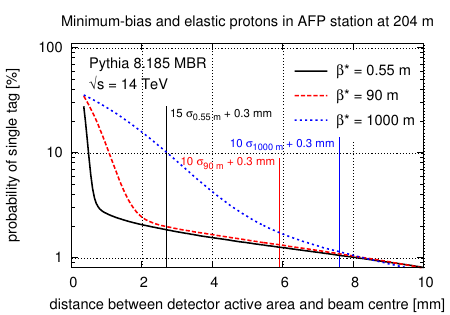}
  \includegraphics[width=0.49\textwidth]{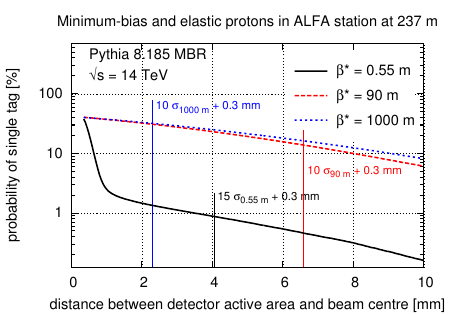}
\caption{Probability per soft interaction of observing elastic or minimum-bias proton in one of the forward detector for the AFP (\textbf{left}) and ALFA (\textbf{right}) detectors. 
The solid black lines are for the $\beta^* = 0.55$ m optics, dashed red ones for $\beta^* = 90$ m and the dotted blue -- for $\beta^* = 1000$ m. The vertical lines  mark the distance for each setting.}
\label{fig_prob_ST}
\end{figure}

The probability of observing a scattered proton in a forward detector depends on the distance between the detector active area and the beam centre -- \textit{cf.} \Fref{fig_prob_ST}. In this figure the solid black lines mark the results for the $\beta^* = 0.55$ m optics, dashed red ones for $\beta^* = 90$ m and the dotted blue -- for $\beta^* = 1000$ m. Due to the fact that the beam size depends on the optics used (see \Tref{tab_beam1_size}), for each setting the distance in $\sigma$ is marked with vertical lines. The additional distance of 0.3 mm represents the so-called `dead edge' -- the area between the edge and active part of the detector.

The AFP detectors are expected\footnote{The exact value depends obviously on the real beam intensity and will be fixed during the run.} 
to operate at 15 $\sigma$ during the runs with the collision optics and at 10 $\sigma$ during the high $\beta^*$ ones. As can be seen from \Fref{fig_prob_ST} (left) 
this translates to $1 - 2\%$ chance of observing scattered proton in the detector. These protons originate mainly from single diffractive events. 
There is also a contribution from double diffraction and non-diffractive events, which starts to be important at larger distances (higher $\xi$). 
For the $\beta^* = 90$ m and, especially, $\beta^* = 1000$ m a contribution of the elastics scattering is also visible, 
but as it decreases rapidly with increasing detector distance from the beam. %, it would not be measurable at 10 $\sigma$. 
Note that the elastic contribution also cannot be reconstructed in AFP as the other scattered proton is lost due to the single-sided horizontal acceptance.

For the ALFA detectors and $\beta^* = 0.55$ m, the situation is similar to one for AFP, except that the expected probability of observing a scattered proton at 15 $\sigma$ distance is about two times smaller. For the \textit{high-}$\beta^*$ optics the situation changes drastically, as the contribution of the elastic scattering is dominant for all considered distances.

\begin{figure}[!htbp]
\centering
  \includegraphics[width=0.49\textwidth]{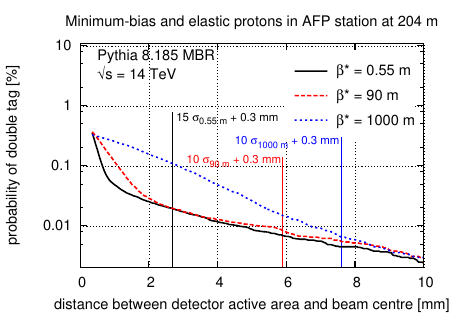}
  \includegraphics[width=0.49\textwidth]{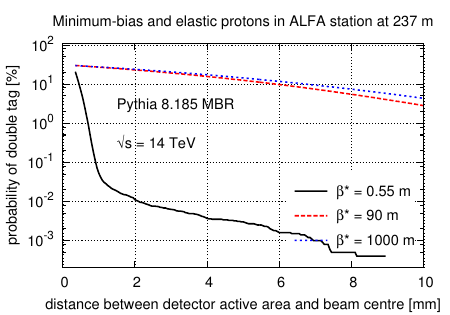}
\caption{Probability per soft interaction of observing double tagged event in the forward detector for the AFP (\textbf{left}) and ALFA (\textbf{right}) detectors. The solid black lines are for the $\beta^* = 0.55$ m optics, dashed red ones for $\beta^* = 90$ m and the dotted blue -- for $\beta^* = 1000$ m. The vertical lines on the left figure mark the distance for each setting, they are at the same distances for the right hand figure.}
\label{fig_prob_DT}
\end{figure}

The probability of registering protons on both sides of the IP (so-called double tag) is shown in \Fref{fig_prob_DT}. Similarly to the single tag case the solid black lines marks results for the $\beta^* = 0.55$ m optics, dashed red ones for $\beta^* = 90$ m, the dotted blue -- for $\beta^* = 1000$~m and the relevant beam smearing were considered.

For $\beta^* = 0.55$ m and 15 $\sigma$ distance between the AFP detector and the beam centre the probability of observing a double tag event is about $2 \cdot 10^{-4}$. The main contribution comes from central diffraction. For \textit{high-}$\beta^*$ optics and 10 $\sigma$ distance this probability is about $8 \cdot 10^{-5}$. For the distances larger than 10 $\sigma$ the main contribution comes from double and central diffractive processes. The single diffraction process plays a secondary role.

In the case of the ALFA detectors and collision optics, the probability of observing a double tagged event at 10 $\sigma$ distance is about $3 \cdot 10^{-5}$. These protons come mainly from central diffraction. For the \textit{high-}$\beta^*$ optics and 10 $\sigma$ distance the probability of observing a double tag event is very high: 0.1 for $\beta^* = 90$ m and 0.2 for $\beta^* = 1000$ m. This is not surprising, since these events are in $\sim 95\%$ of cases due to the elastic scattering.

\subsection{Soft Vertex Reconstruction}

In hard diffractive analyses the background is mainly due to hard  non-diffractive events. Proton tagging allows us to eliminate some of these events. However, due to pile-up, there could be the situation where a hard event is produced together with a soft one which contains forward proton(s). Requiring exactly one vertex reconstructed in the central detector allows further background reduction. Apart from knowing how often the vertex originating from soft event (hereafter referred to as the `soft vertex') is reconstructed when there is a diffractive proton in the forward detector, one needs also take into account that there are cases in which soft vertices are not visible. There are two main sources of vertex reconstruction inefficiency:
\begin{itemize}
  \item the soft event is produced too close to a hard one; due to finite detector resolution and reconstruction algorithms the vertices are merged,
  \item there are not enough tracks pointing to the soft vertex.
\end{itemize}

In the presented studies, the vertex was assumed to be reconstructed if there are at least four charged particles within the ATLAS tracker ($|\eta| < 2.5$). In order to account for the detector efficiency, each particle had a certain probability of being registered. The thresholds were set to:
\begin{itemize}
  \item 50\% for particles with $100 < \pt < 500$ MeV and
  \item 90\% for particles with $\pt > 500$ MeV.
\end{itemize}
These values are reflecting the behaviour of ATLAS inner detector \cite{SOFT_ATLAS_track_reco}, but are also similar for the CMS experiment. The minimal distance below which vertices are merged was set to 1.5 mm.

The probabilities should be multiplied by an average pile-up to yield the prediction of soft vertex reconstruction for the running conditions in question.

\begin{figure}[!htbp]
\centering
  \includegraphics[width=0.49\textwidth]{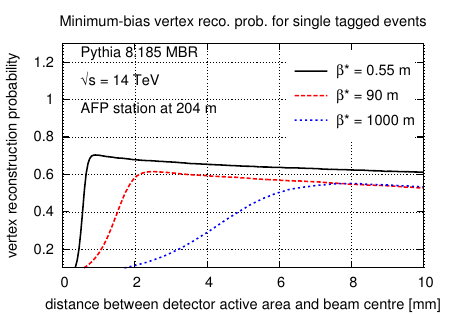}
  \includegraphics[width=0.49\textwidth]{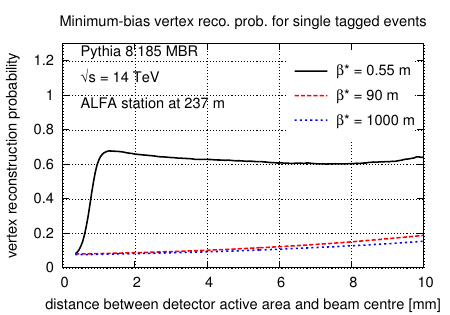}
\caption{Soft vertex reconstruction probability for single tagged events tagged in the AFP (\textbf{left}) and ALFA (\textbf{right}) detectors. The solid black lines are for the $\beta^* = 0.55$ m optics, dashed red ones for $\beta^* = 90$ m and the dotted blue -- for $\beta^* = 1000$ m. The vertex is assumed to be reconstructed if there are at least four charged particles in the ATLAS tracker ($|\eta| < 2.5$). Particles with $100 < \pt < 500$ MeV have 0.5 chance to be detected whereas the probability for the ones with $\pt > 500$ MeV was set to 0.9.}
\label{fig_prob_ST_vertex}
\end{figure}

The vertex reconstruction probability under the condition that the proton is tagged in the forward detector as a function of the distance between detector active area and the beam centre is shown in \Fref{fig_prob_ST_vertex}. For $\beta^* = 0.55$ m it ranges between 0.6 and 0.7. In the case of the AFP (left) and \textit{high-}$\beta^*$ optics, this probability at the 10 $\sigma$ distance is on average smaller by 0.1 than that for the collision optics. This situation is a bit different for the ALFA detectors (right) -- due to the fact that the elastic scattering plays an important role in the wide range of distances, the probability to have a vertex in the event is much smaller. The shapes of the presented distributions are a consequence of a non-trivial interplay between the kinematics of forward proton and central system multiplicity.

\begin{figure}[!htbp]
\centering
  \includegraphics[width=0.49\textwidth]{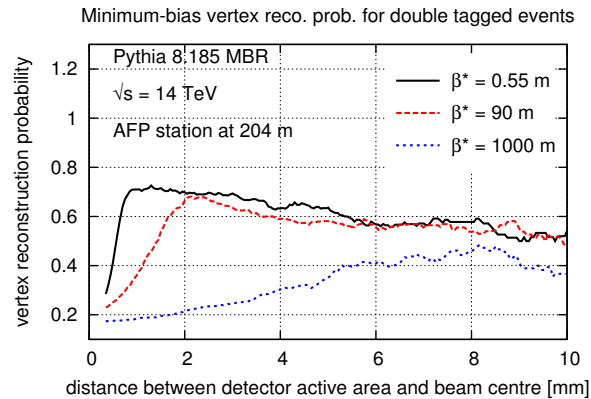}
\caption{Soft vertex reconstruction probability for double tagged events tagged in the AFP detector. The solid black lines are for the $\beta^* = 0.55$ m optics, dashed red ones for $\beta^* = 90$ m and the dotted blue -- for $\beta^* = 1000$ m. The vertex is assumed to be reconstructed if there are at least four charged particles in the ATLAS tracker ($|\eta| < 2.5$). Particles with $100 < \pt < 500$ MeV have 0.5 chance to be detected whereas the probability for the ones with $\pt > 500$ MeV was set to 0.9.}
\label{fig_prob_DT_vertex}
\end{figure}

The probability to reconstruct the soft vertex under the condition that there is a double tag in the AFP detector is shown in \Fref{fig_prob_DT_vertex}. One concludes that the shapes of the presented dependences are qualitatively very similar to those in the single tag case.

\subsection{Proton and Vertex reconstruction Conclusion}

Studies are presented which quantify the probability of reconstructing single and double proton-tagged events using the AFP and ALFA forward detectors (CT-PPS, TOTEM results are expected to be broadly similar). The results are presented for combined elastic, inelastic and MBR diffractive interactions at different LHC optics settings as a function of the detector-beam separation. 

The probability of also reconstructing the soft vertex within the central tracking volume is subsequently investigated in conjunction with the forward proton tag. These studies together indicate an approximate (MC dependent) rate of proton tags from soft interactions which may be used in one of two ways. It quantifies the expected statistics to be gained from a minimum bias data taking campaign for studies on soft interactions with forward proton tags. Or by multiplying the probability by an average pileup value, it yields the chance that a hard interaction will overlap with a pileup event with forward proton tags - a possible source of background to smaller cross section hard diffractive processes, as investigated in Section~\ref{ch4_sec_backgrounds}.

\section{Physics sources and properties of forward protons from inelastic interactions}
\label{sec:timm:forwardprotonsources}
%
%\TODO{WHEN PRESENT - THIS SECTION SHOULD MOVE UP ONE INTO THE MC CHAPTER} ??? Or should it?? No - keeping here
Forward proton detectors such as AFP, ALFA and TOTEM reconstruct protons from elastic scattering, single-diffractive interactions and double-Pomeron exchange. Other physics mechanisms may exist however which produce protons within the acceptance of forward proton detectors, these too are explored in this section and with the aim of allowing for greater discrimination between soft physics models.

Soft pseudorapidity gaps, devoid of any final state particles above a low threshold (typically a few hundred \MeV) have long been used as a probe of diffractive interactions \cite{SOFT_Ansorge:172602,SOFT_Bernard:1986yh,SOFT_Affolder:2001vx,SOFT_Aad:2012pw,SOFT_CMS:2013mda}. At the LHC, the diffractive kinematics are such that for a single diffractive interaction to leave a rapidity gap within the acceptance of the main detectors (here taken as $|\eta| < 5$), the scattered proton must be in the interval $-6 < \logxix < -2$. Smaller diffractive masses escape down the beam-line, where specialist forward detectors are needed to measure them, while larger diffractive masses span the full $\eta$ range of the central detector and hence do not leave a reconstructible pseudorapidity gap within.

The lowest $\xi$ for which a single diffractive event will start to leave a pseudorapidity gap in the central detector is two times smaller than the lowest $\xi$ reconstructible by AFP. Therefore if we wish to probe pseudorapidity gaps in a soft diffractive enriched data sample, we must explore other mechanisms by which a forward proton tag may be obtained.

Diffraction in the \pyeight generator is described in Section~\ref{sec:pythiamc}, of key interest to this study is the modelling of double dissociative interactions where the mass of the smaller diffractive system (denoted \MY) is of the order of the proton mass. For $\MY < m_p + 1 \GeV$ (where $m_p$ is the proton mass), the system is decayed isotropically by \py into a two body system while for masses up to 10 \GeV\ the system is hadronised from a string with the quantum numbers of the originating proton. Only higher mass resolved systems (with a probabilistic turn on, starting from a diffractive mass of 10 \GeV) are subjected to a perturbative modelling with \cite{SOFT_Capella:1995pn} proton-Pomeron PDF from H1 and the full \py machinery for parton showers and MPI employed within the diffractive system\footnote{Note that this is fully contained within the proton-Pomeron interaction and cannot interfere with or destroy the rapidity gap generated via the Pomeron exchange.}.

\begin{landscape}
\begin{figure}[p]
\centering
\begin{subfigure}[b]{0.47\textwidth}
\includegraphics[width=\textwidth]{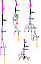}
\caption{}
\label{fig:timm:Py8_DD:a}
\end{subfigure}
\hspace{5mm}
\begin{subfigure}[b]{0.65\textwidth}
\includegraphics[width=\textwidth]{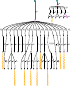}
\caption{}
\label{fig:timm:Py8_Epos_DDLike:b}
\end{subfigure}
\caption{Example low mass dissociative systems from four independent \pyeight interactions (a) which result in a forward proton within AFP acceptance. 
Equivalent example for decay products from two independent inelastic proton interactions simulated by \epos (b) which result in a high and low mass system. In all cases the forward protons are highlighted in magenta. }
\label{fig:timm:Py8_DD}
\end{figure}
\end{landscape}

These low mass diffractive excitations are predicted to often result in the beam baryon number being retained by a proton, but one with a significantly lower energy due to the two body decay or low mass string fragmentation. This is illustrated in \Fref{fig:timm:Py8_DD:a} where the hadronisation of four independent low mass diffractive systems in \pyeight are visualised with \texttt{MCViz} \cite{SOFT_mcviz}. In each case, the forward proton (highlighted in magenta) was found to be produced within the kinematic acceptance of the AFP detector, assuming collision optics. The cross section prediction by \pyeight of double diffractive interactions which produce a forward proton within AFP acceptance is 0.3 mb (3.3 \% of the total double dissociative prediction, see \Tref{tab:timm:crosssections}). For all but the highest mass diffractive systems, there is very little correlation between the size of the two diffractive dissociations \MX and \MY.
%, see \Fref{fig:timm:MC_XiMax_Vs_XiMin}.

This high cross section process will allow for these protons, produced through low mass forward \MY systems, to be used as independent tags to study the dynamics of the larger \MX system in 
minimum bias interactions and will be exploited in generator feasibility studies reported in this document (see Sections~\ref{sec:timm:rapgaps}, \ref{sec7:EnergyFlow} and \ref{sec7:Multiplicity}).

Other MC generators considered are \epos and \hpp. \epos uses a parton based Gribov-Regge model and is described in Section~\ref{sec:epos}, \epos interactions resulting in 
a forward proton within AFP acceptance are illustrated in \Fref{fig:timm:Py8_Epos_DDLike:b} for two independent systems, one of low mass and one of high mass. 

\begin{figure}[ht]
\centering
\includegraphics[width=\textwidth]{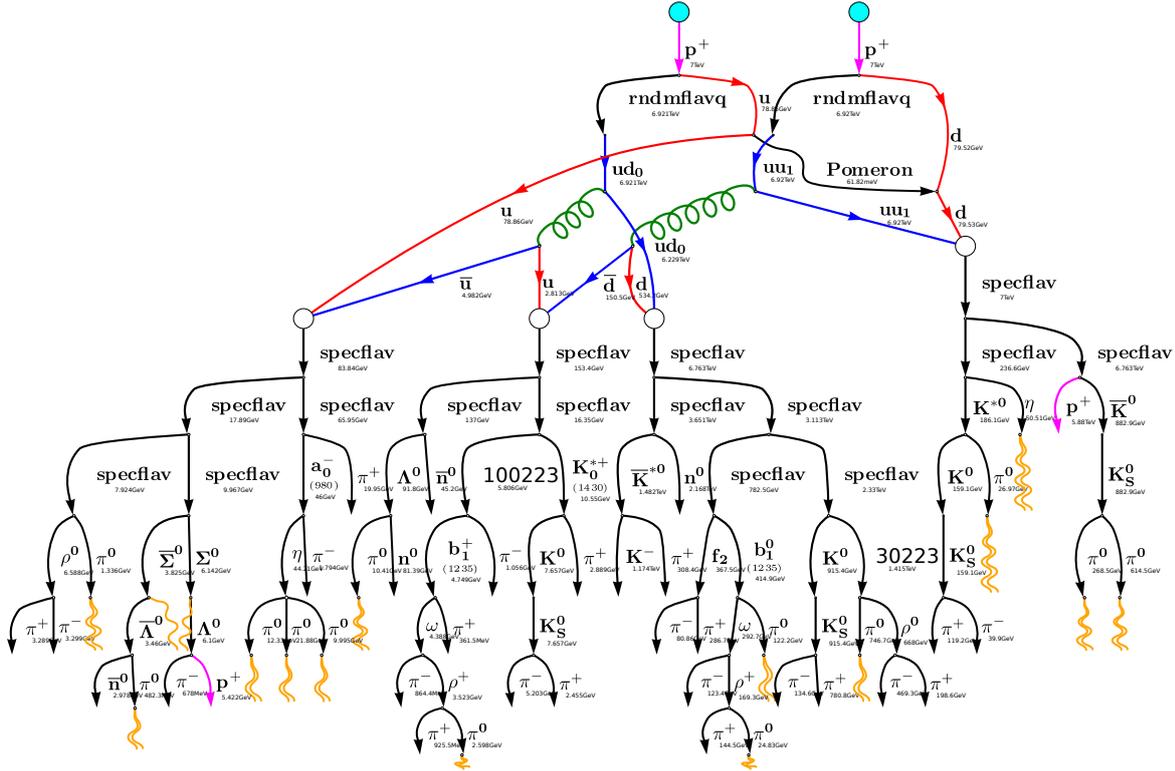}
\caption{Example inelastic interaction from the \hpp underlying event model. Two protons are produced in the final state (highlighted in magenta). The proton on the right hand side of the diagram is within AFP acceptance.}
\label{fig:timm:Hpp_DDLike}
\end{figure}

\hpp is discussed in Section~\ref{sec:pomwigmc}, events are generated using the \hpp underlying event model where the hard scatter matrix element is set to the unit matrix 
and particle production is generated solely from the simulation of $h$ semi-hard (containing object with $\pt > 3.36 \GeV$) and $n$ soft scatters where $h$ and $n$ are each 
chosen per event via the sampling of Poisson distributions. For the case $h=0$ and $n=0$, only the beam remnants are present. Although not explicitly modelling soft diffractive 
interaction here, \hpp is also capable of generating protons within the acceptance of AFP and additionally uses the cluster hadronisation model. An example of a \hpp event with a 
forward proton tag is illustrated in \Fref{fig:timm:Hpp_DDLike}.
%
%\begin{figure}
%\centering
%\includegraphics[width=0.5\textwidth]{figs/softdiffraction/MC_XiMax_Vs_XiMin}
%\caption{Correlation of the larger (\logxix) and smaller (\logxiy) diffractive masses for double dissociative diffraction in \pyeight. 71\% of the predicted cross section satisfy the condition $\logxiy < -6$ resulting in events where the smaller diffractive mass is fully contained at $|\eta| > 5$. Note, this figure was simulated at $\sqrt{s} = 7 \TeV$ however its message holds too at $\sqrt{s} = 14 \TeV$ where the low mass kinematic limit extends out further to $\logxiy > -8.3$.}
%\label{fig:timm:MC_XiMax_Vs_XiMin}
%\end{figure}
%

\subsection{MC versions and tagged proton selection for inelastic studies}
\label{sec:timm:protontagging}
The subsequent studies on the rapidity gap spectra, Section~\ref{sec:timm:rapgaps}, transverse energy density, Section~\ref{sec7:EnergyFlow} and charged particle multiplicities, Section~\ref{sec7:Multiplicity} use the following generators. \pyeight version 8.176 with the option \texttt{Tune:pp=8} (ATLAS MB Tune A2 with the MSTW 2008 LO PDF) \cite{SOFT_ATL-PHYS-PUB-2012-003}, \epos version \texttt{1.99.crmc.v3200} with the tune \texttt{EPOS-LHC} and \hpp version 2.7.0 with the \linebreak\texttt{LHC-UE-EE-4.in} run card using the \texttt{CTEQ6L1} PDF, modified to increase $\sqrt{s}$ to 14 \TeV.
For event selections including a forward proton tag, acceptance efficiency and resolution maps for AFP and ALFA were provided by \cite{SOFT_Trzebinski:1616661} binned in $p_{\textrm{T,}p}$ and $\xi_p = E_{\textrm{beam}} - E_{p} / E_{\textrm{beam}}$, see Section~\ref{sec:soft:forwardacceptance}. Final state protons produced by the MC are checked against these maps and a forward proton tag is generated if $P(p_{\textrm{T,}p}, \xi_p) \geq r$ where $P(p_{\textrm{T,}p}, \xi_p)$ is the probability of detection for the given proton kinematics and $r$ is a uniformly distributed random number over the range 0--1.

Primary considered scenarios were AFP tracking detectors at a distance of 2.0 mm (10$\sigma$) from the beam and collision optics, $\beta^* = 0.55$ m, this gives access to a large range of $p_{\textrm{T,}p}$ for $\xi_p > 0.01$. For ALFA a distance of 4.5 mm (5$\sigma$) was used with $\beta^* = 90$ m optics which accesses $\xi_p < 0.2$ for $p_{\textrm{T,}p} > 0.1 \GeV$. See Section~\ref{sec:forward_detector_systems} for additional details on the current and future forward detectors at ATLAS and CMS. These settings are optimistic regarding how close the detectors will be able to approach the beam, however increasing the distance to 20(10)$\sigma$ for AFP(ALFA) results only in a small increase in the minimum reconstructible $\xi_p$($p_{\textrm{T,}p}$) in AFP(ALFA) and does not change the conclusions due to the large (millibarn) cross sections.

$\beta^* = 0.55$ collision optics were also investigated for the ALFA detector. It was found that the MC predictions were in good agreement with the AFP predictions in terms of shape, however due to the reduced acceptance (ALFA does not posses horizontal detectors) the predicted number of diffractive events is further reduced from the AFP estimates by around a factor of 4 ($\sim 0.4$ mb for single and double diffraction combined).

The cross sections and overall probabilities of acquiring exactly one forward proton tag per inelastic interaction are listed for the considered MC and forward detector arrangements in \Tref{tab:timm:crosssections}.
\begin{table}[h]
\caption{MC predictions for the inelastic cross section at $\sqrt{s} = 14 \TeV$, including a breakdown of \pyeight into the diffractive and non-diffractive sub components. Also listed is the probability per event that exactly one proton is reconstructed in the forward detectors, based on the probabilistic acceptance as a function of $\xi_p$ and $p_{\textrm{T,}p}$.}
\begin{tabular}{lcccc}
\hline\hline
               & Cross Section             & AFP Tag Prob.      & ALFA Tag Prob.   & ALFA Tag Prob.     \\
               & $\sqrt{s} = 14 \TeV$ (mb) & $\beta^* = 0.55$ m (\%) & $\beta^* = 90$ m (\%) & $\beta^* = 0.55$ m (\%) \\ \hline
\hpp UE-EE-4   & 78.0                      & 0.7                     & 0.3                   & 0.2                     \\
\epos LHC      & 80.1                      & 4.6                     & 1.7                   & 1.1                     \\
\pyeight A2    & 79.3                      & 2.5                     & 0.8                   & 0.6                     \\
\phantom{--}\pyeight A2 SD & 12.9 (16\%)   & 11.5                    & 3.9                   & 2.7                     \\
\phantom{--}\pyeight A2 DD & 8.9  (11\%)   & 3.3                     & 0.6                   & 0.8                     \\
\phantom{--}\pyeight A2 ND & 57.5 (73\%)   & 0.4                     & 0.2                   & 0.1                     \\
\hline\hline
\end{tabular}
\label{tab:timm:crosssections}
\end{table}

Sources of background, such as discussed in Section~\ref{sec:machinebackground}, and the effects of pileup are not included in these studies.

\subsection{Kinematics of tagged proton samples}
Of the \pyeight, \epos and \hpp interactions with exactly one forward proton tag in AFP for the $\beta^* = 0.55$ collision optics, a double-diffractive enhanced sample is selected by requiring a pseudorapidity gap $\Delta\eta > 4$ between any pair of neighbouring final state particles in the event. Single diffractive events are rejected by requiring the smaller system's mass to be greater than the proton mass.

\begin{figure}
\centering
\begin{subfigure}[b]{0.47\textwidth}
\includegraphics[width=\textwidth]{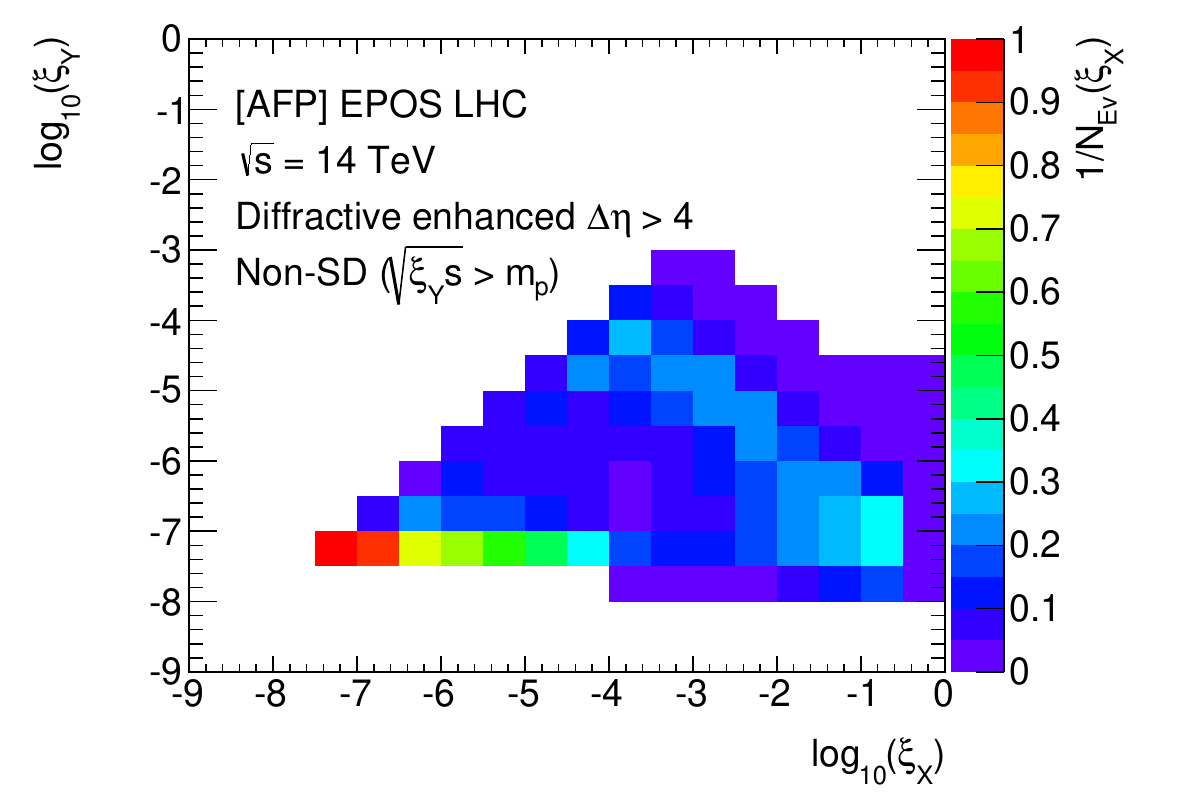}
\caption{}
\label{fig:timm:2d_datapoints_EPOS}
\end{subfigure}
\hspace{5mm}
\begin{subfigure}[b]{0.47\textwidth}
\includegraphics[width=\textwidth]{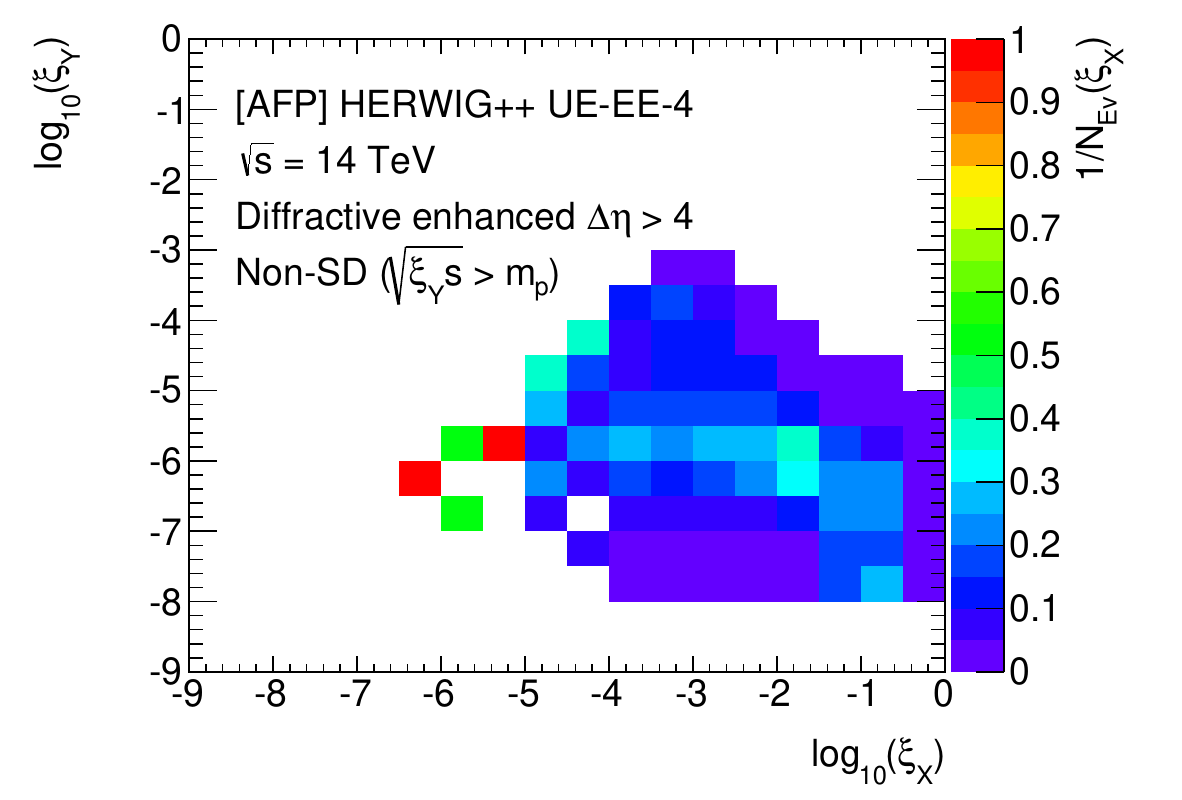}
\caption{}
\label{fig:timm:2d_datapoints_HPP}
\end{subfigure}
\begin{subfigure}[b]{0.47\textwidth}
\includegraphics[width=\textwidth]{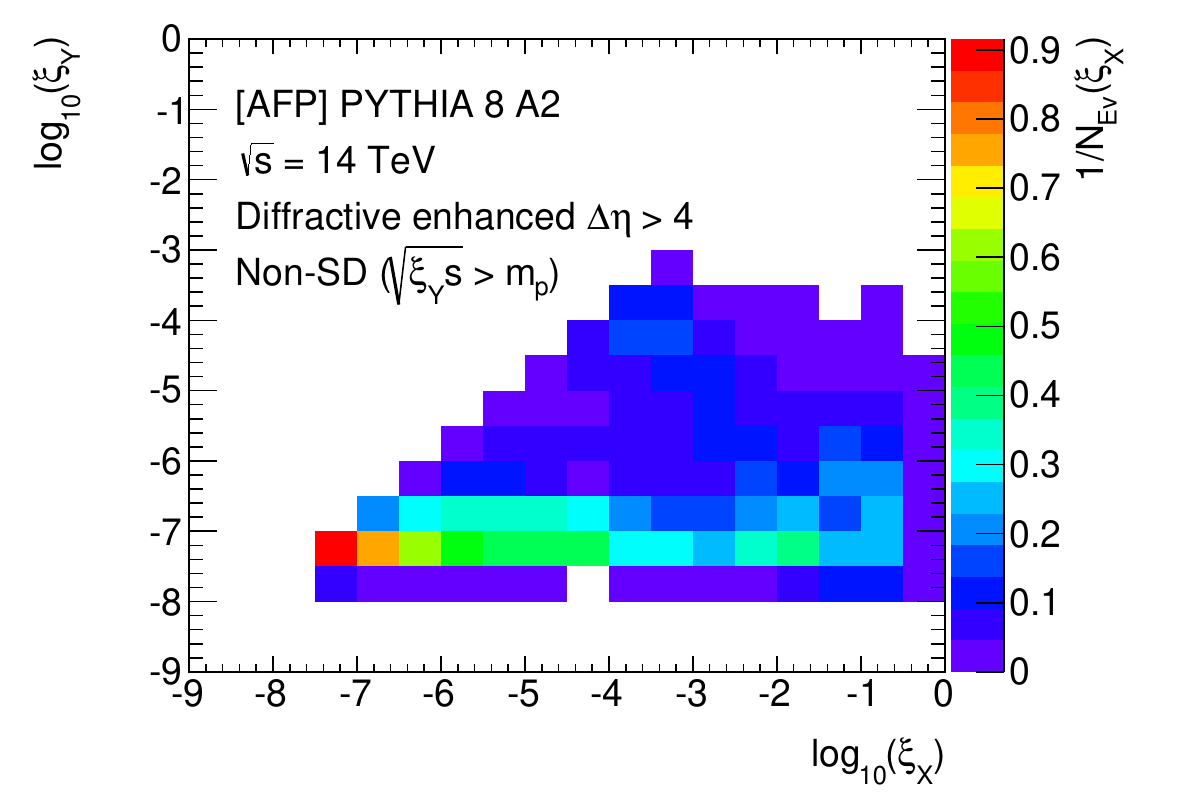}
\caption{}
\label{fig:timm:2d_datapoints_PY8}
\end{subfigure}
\caption{Correlations of invariant masses of systems either side of the largest pseudorapidity gap for \epos (a), \hpp (b) and \pyeight (c) generators using a double-diffractive enhancing selection of $\Delta\eta > 4$ and $\MY > m_p$. Distributions are normalised to unity in columns of \logxix.}
\label{fig:timm:2d_datapoints}
\end{figure}

All final state particles from either side of the largest pseudorapidity gap in the event are combined into two systems which are identified as \MX and \MY (where $\MX > \MY$). The correlations of the two systems are plotted in \Fref{fig:timm:2d_datapoints}. The mechanism of forward protons generated through low mass dissociation in \pyeight from Section~\ref{sec:timm:forwardprotonsources} is observed as the excess of events with $\logxiy = -7$. \epos also shows this independence of the variables, but it only holds for $\logxix < -4$. For larger \MX, \epos and \hpp display a less prominent anti-correlation between \MX and \MY. %XXX TODO - comment on this when 100% sure of cause!

\section{Soft pseudorapidity gaps}
\label{sec:timm:rapgaps}
\subsection{Previous measurements}
\label{sec:timm:rapgaps:prev}
Large pseudorapidity gaps devoid of all final state particles above a lower experimental cut off (typically of order a few hundreds of \MeV) are a characteristic signature of diffractive interactions.

The soft pseudorapidity gap cross section was measured at $\sqrt{s} = 7 \TeV$ by ATLAS \cite{SOFT_Aad:2012pw} and CMS \cite{SOFT_CMS:2013mda}. Gaps size are expressed here in terms of the event variable \detaf, 
this is defined as the largest of the two forward rapidity regions extending to (at least) $\eta = \pm4.9$ which contain no final state particles above a threshold $\pt > \ptcut$. The value of $\ptcut$ is 
varied in the range\footnote{Note that for \ptcut $\geq 400 \MeV$, the starting hemisphere of the rapidity gap as calculated at \ptcut = 200 \MeV\ is used to fix the side of the detector from which \detaf is measured.} 200--800 \MeV\ and as \ptcut increases the  
modelling of hadronisation in the MC is tested, especially for small values of \detaf where gap fluctuations from hadronisation effects dominate.

CMS in addition measures the cross section for single and double dissociation as a function of $\xi = M_X^2/s$, where $M_X^2$ is the mass of the larger (in the case of double diffraction) diffractive system~\cite{SOFT_cms_dd_sd}.

TOTEM has measured the DD cross section in the forward rapidity range~\cite{SOFT_si_dd_7} using the T1 and T2 telescopes, and the SD cross section 
using the T1 and T2 telescopes and the forward proton~\cite{SOFT_si_sd_7}. 
An estimate of the low-mass diffractive events ($M_X$ < 3.4 GeV/c$^2$) with no charged particles in the $|\eta| <$ 6.5 range has been obtained 
by estimating the difference between the total inelastic cross section determined using elastic scattering and the optical theorem and the inelastic cross section measured using the T1 and T2 telescopes (see Section~\ref{s:inel}).

\begin{figure}
\centering
\includegraphics[width=0.4\textwidth]{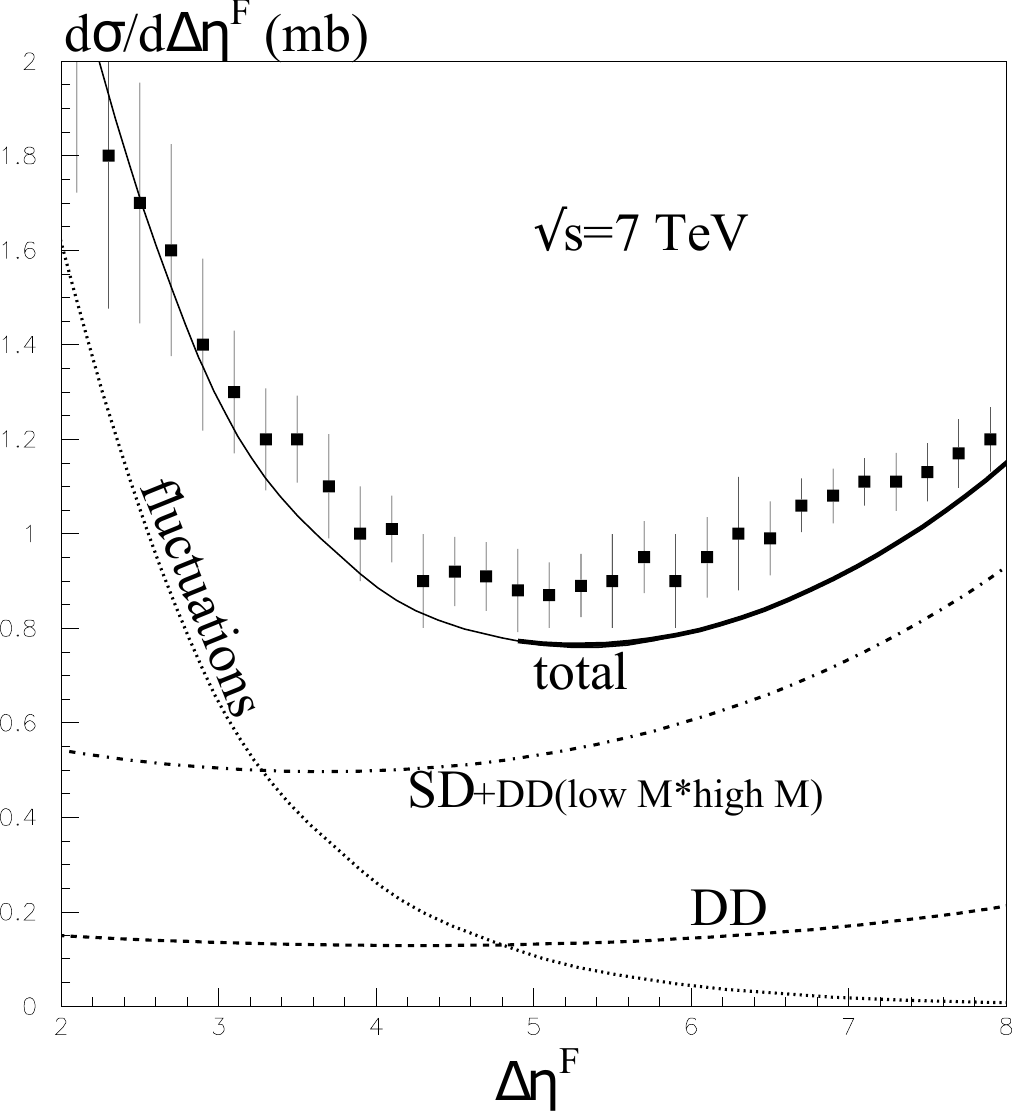}
\caption{Figure from \cite{SOFT_Khoze:2013jsa} of the forward pseudorapidity gap size as measured by ATLAS \cite{SOFT_Aad:2012pw} versus the proposed global description. }
\label{fig:soft:durhamModel}
\end{figure}

These data, along with other measurements including data from TOTEM and CDF, are being employed to test theoretical modelling which aims to globally describe LHC elastic and diffractive data. 
An example is shown in \Fref{fig:soft:durhamModel}. See Section~\ref{sec:shrimps} and \cite{SOFT_Khoze:2014aca,SOFT_Khoze:2013jsa,SOFT_Martin:2014gfa,SOFT_Ryskin:2011qe} for additional details.

\begin{figure}
\centering
\includegraphics[width=.47\textwidth]{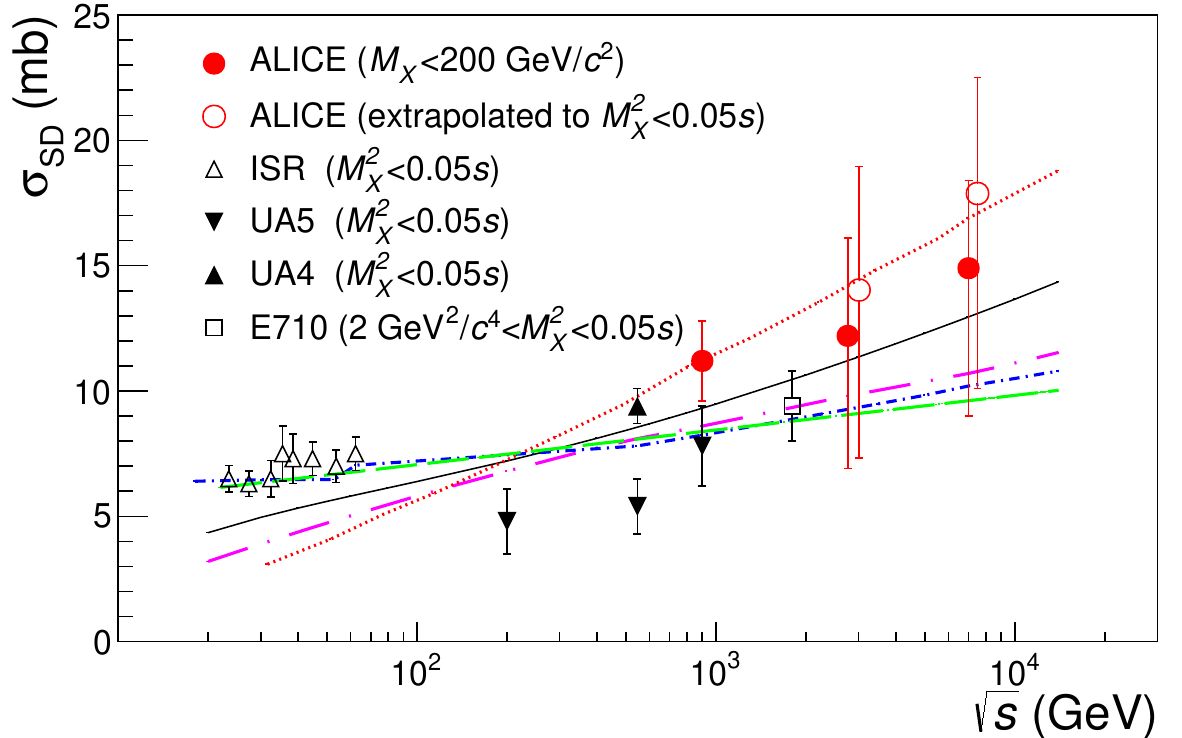}
\includegraphics[width=.47\textwidth]{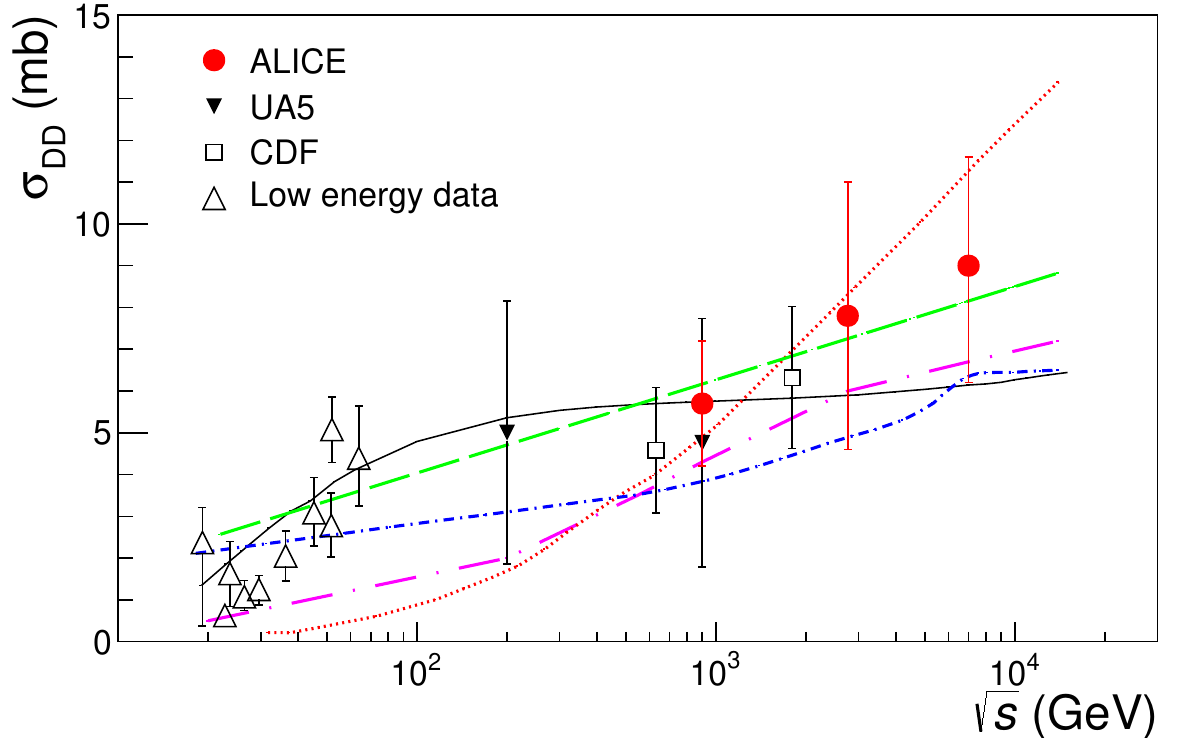}
\caption{
Single diffractive cross section as a function of centre-of-mass energy (left) and for the double diffractive cross section (right) where $\Delta\eta > 3$. SD data from other experiments are for $M_X^2 < 0.05s$.
ALICE measured SD points are shown with full red circles, in order to compare with data from other experiments these were extrapolated to
$M_X^2 < 0.05s$ (open red circles), when needed. Theoretical model predictions, shown as lines, all correspond to $M_X^2 < 0.05s$.}
\label{fig:ALICE_SD_DD}
\end{figure}

%Other experiments have made use of the kinematics of diffractive signatures to measure the inelastic cross section. 
The ALICE collaboration used minimum bias data at 0.9, 2.76 and 7 TeV to calculate the single (SD) and double (DD) diffraction cross sections at these three energies \cite{SOFT_Abelev:2012sea}. 
In Run-I the data were triggered using the VZERO detector and the two innermost layers of the Inner Tracking System, with a combined pseudorapidity coverage $-3.7\le \eta \le 5.1$. 
Offline selections were based on the largest forward and central pseudorapidity gaps between tracks at forward and central pseudorapidity, and on the ratio between these two gaps. 
These are used to define two samples which are strongly enhanced in SD and DD events. These were compared to distributions from two different event generators (PYTHIA 6 [Perugia-0, tune 320] \cite{SOFT_ALICEPy6WTune} and PHOJET \cite{SOFT_PHOJETRef}) 
and the diffractive fractions were obtained from an adjustment of the fractions assumed in these two generators (see paper for details). For SD, an extrapolation was performed to estimate the rate for unobserved low-mass 
diffractive events according to the parametrisation of Kaidalov and Poghosyan \cite{SOFT_Kaidalov}.
The resulting SD and DD cross sections are shown as a function of energy in \Fref{fig:ALICE_SD_DD}. The inelastic cross-sections derived from these measurements are discussed in Section~\ref{sec:soft:sigmatot:prev}. % i.e. section 3.5
In Run-II the measurement will be repeated using essentially the same method, but with increased pseudorapidity coverage (to $-7 \le \eta \le 6.3$) using the new AD counters described in Section~\ref{sec:detecor:lhcb:herschel}. %TODO - replace this with the ALICE descr.

\subsection{Future soft pseudorapidity gaps studies with a proton tag}
%This does not affect the AFP or ALFA triggered samples where the direction is taken from the tagged proton.
By correlating pseudorapidity gaps with forward proton tags, we present an analysis which has the potential to offer enhanced sensitivity to the modelling of ${\rm d}\sigma/{\rm d}\xi$ over four orders of magnitude in $\xi$.

For the proton tagged event selection, the near side of the detector to the proton tag is defined as where the gap starts. 

\begin{figure}
\centering
\begin{subfigure}[b]{0.47\textwidth}
\includegraphics[width=\textwidth]{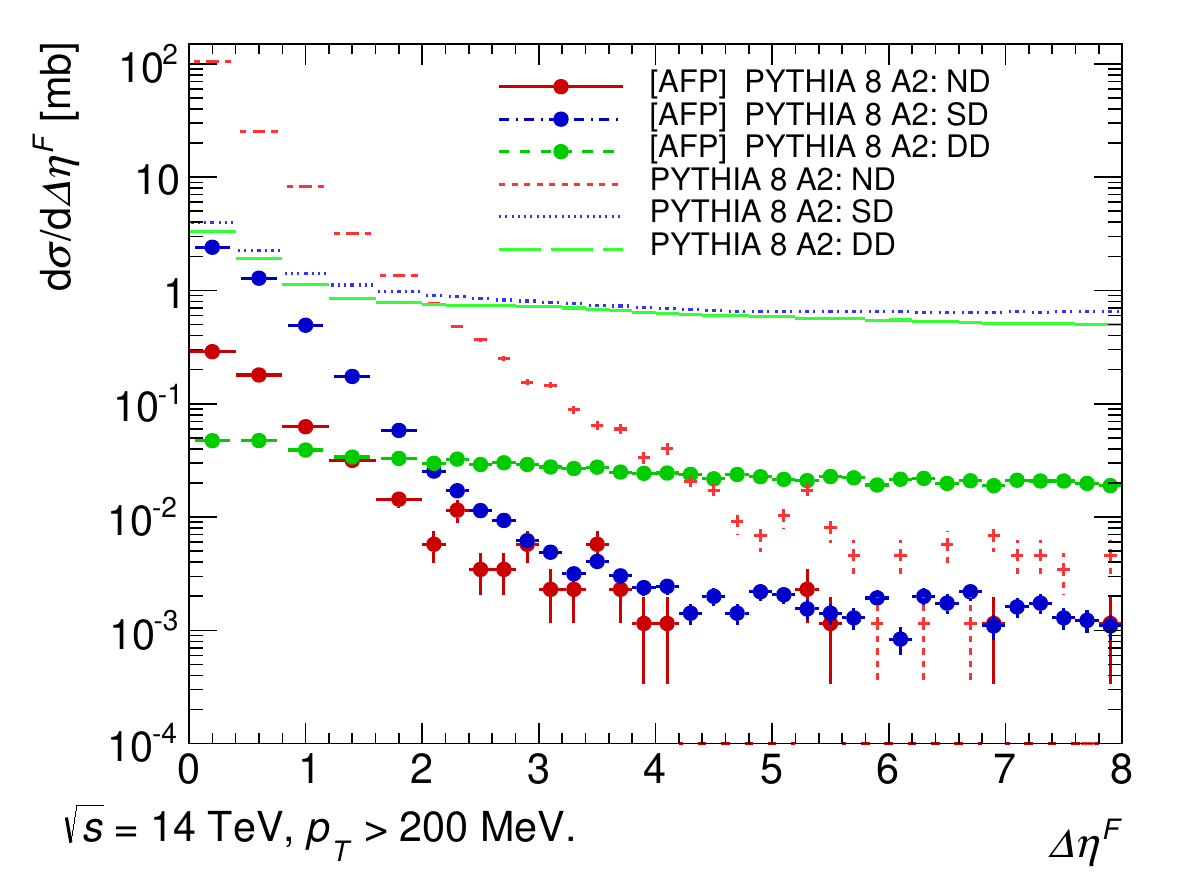}
\caption{}
\label{fig:timm:diffractiveGaps_1:a}
\end{subfigure}
\hspace{5mm}
\begin{subfigure}[b]{0.47\textwidth}
\includegraphics[width=\textwidth]{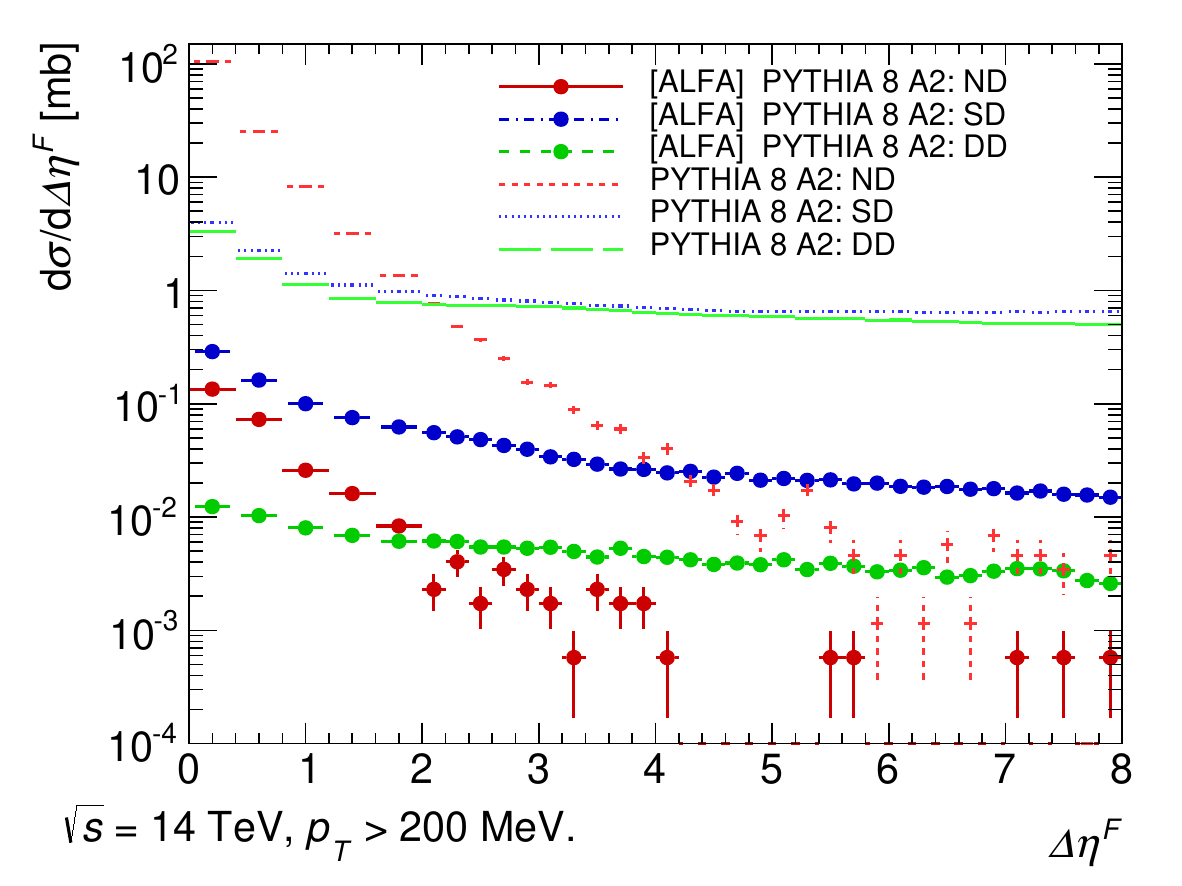}
\caption{}
\label{fig:timm:diffractiveGaps_11:b}
\end{subfigure}
\caption{Forward rapidity gap cross sections in the range \detaf < 8 compared between single-diffractive, double-diffractive and non-diffractive components of the inelastic cross section from \pyeight, (a) for the inclusive and AFP selections and (b) for the inclusive and ALFA selections (see text). }
\label{fig:timm:diffractiveGaps_1}
\end{figure}

Diffractive topologies are isolated at large gap sizes, the distribution is truncated at \detaf = 8 due to experimental trigger inefficiencies for larger gaps (CMS are able to trigger events up to \detaf = 8.4). The \pyeight decomposition of the inelastic cross section is explored in \Fref{fig:timm:diffractiveGaps_1}  (see Section~\ref{sec:timm:protontagging} for forward proton selection and MC details). Here we observe an exponential fall both for the strongly  suppressed non-diffractive events  and the  single diffractive events. The exponential fall of the single diffractive events  is expected because only high mass single diffractive events result in a large enough fractional momentum loss to enter the acceptance of AFP. These high mass systems span all of ATLAS and only contain rapidity gaps from hadronisation fluctuations.

%There is a small tail of single diffractive events out to large gap sizes and this comes from events with large $|t|$ which is a much smaller cross section and is suppressed further by the AFP geometric acceptance.
The hypothesis that for double dissociation in \pyeight, the low mass system decaying to a forward proton provides an independent tag is illustrated in the flatter behaviour of the double diffractive cross section, this follows from the relation $\lnxix \propto \detaf$. The small exponential slope is likely due to the residual effect of hadronisation from large diffractive masses.

The key conclusions is that a high purity diffractive sample is predicted where single diffraction is dominant at small gap sizes and double diffraction is dominant at large gap sizes.

When requiring an 90 m optics ALFA tag as in \Fref{fig:timm:diffractiveGaps_11:b}, the non-diffractive component is observed to be even more suppressed than for the 0.55 m optics AFP case. However the lack of any large acceptance at high $\xi_p$ also results in the large suppression of the diffractive components, with the double diffractive being much more heavily suppressed than the single diffractive.

\begin{figure}
\centering
\begin{subfigure}[b]{0.47\textwidth}
\includegraphics[width=\textwidth]{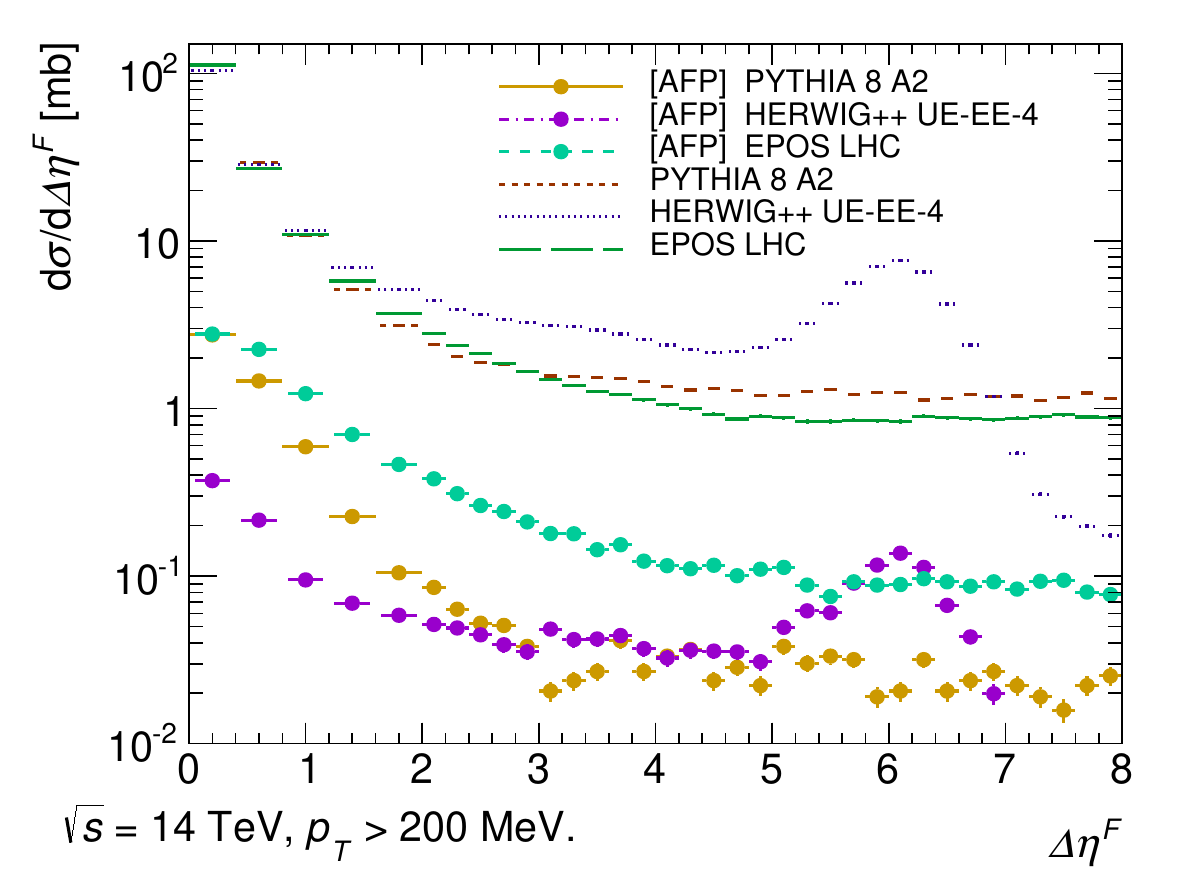}
\caption{}
\label{fig:timm:diffractiveGaps_2:a2}
\end{subfigure}
\hspace{5mm}
\begin{subfigure}[b]{0.47\textwidth}
\includegraphics[width=\textwidth]{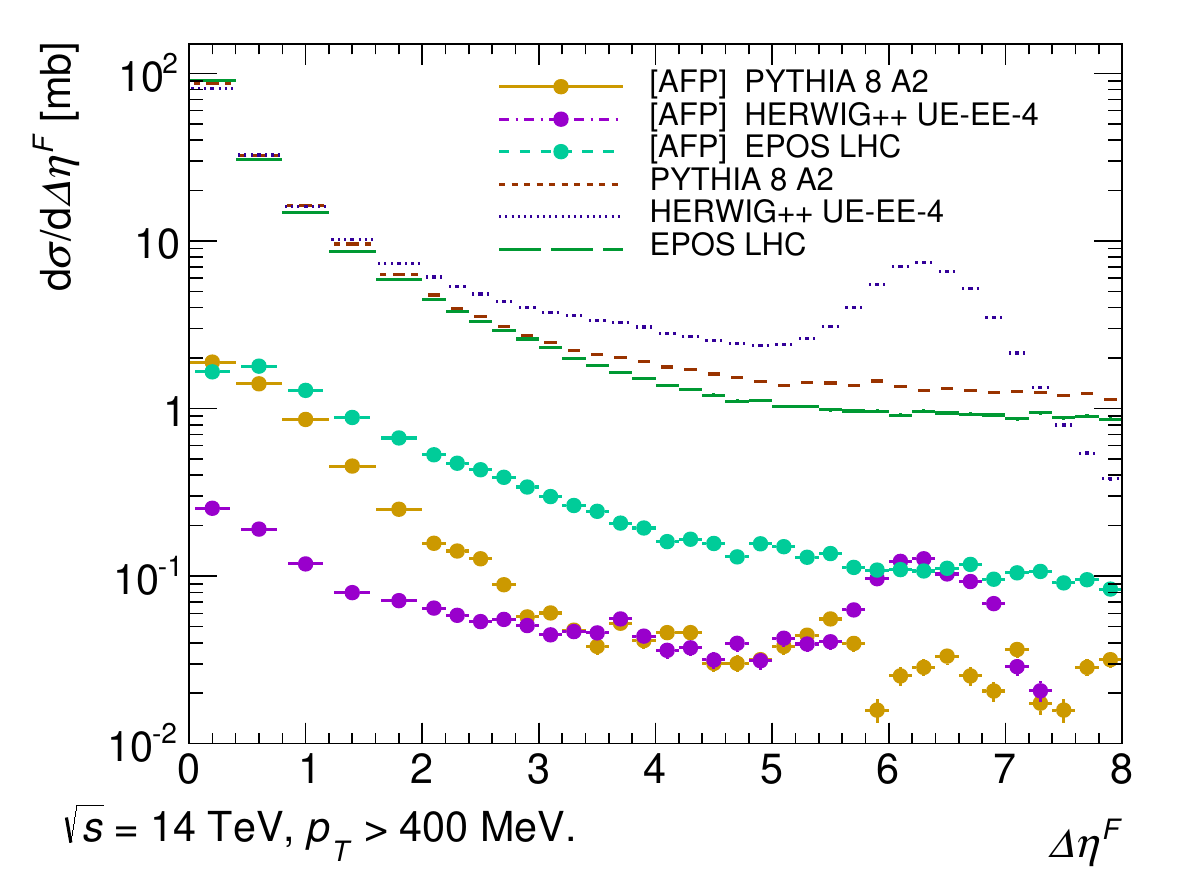}
\caption{}
\label{fig:timm:diffractiveGaps_4:b}
\end{subfigure}
\begin{subfigure}[b]{0.47\textwidth}
\includegraphics[width=\textwidth]{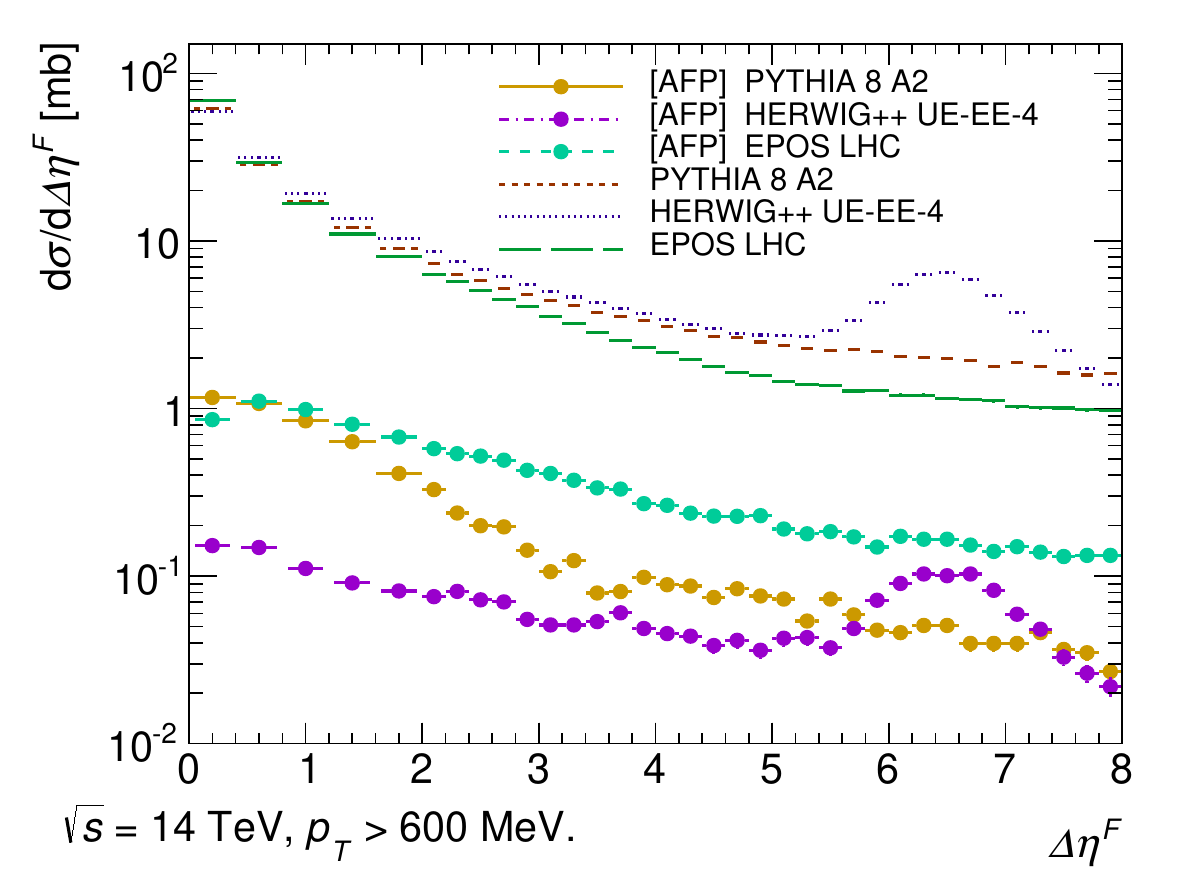}
\caption{}
\label{fig:timm:diffractiveGaps_6:c}
\end{subfigure}
\hspace{5mm}
\begin{subfigure}[b]{0.47\textwidth}
\includegraphics[width=\textwidth]{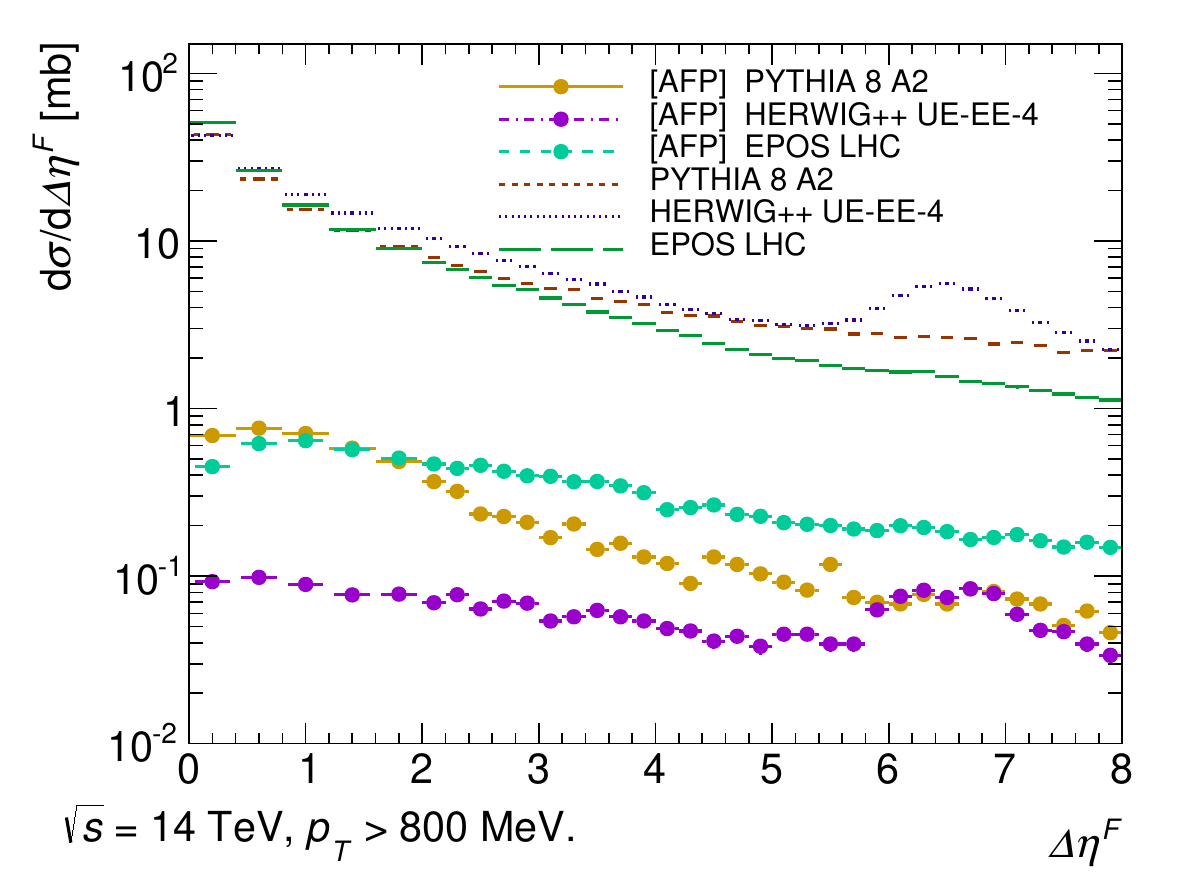}
\caption{}
\label{fig:timm:diffractiveGaps_8:d}
\end{subfigure}
\caption{Forward rapidity gap cross sections in the range \detaf < 8 compared between MC models for the inclusive and AFP selections for \ptcut = 200 (a), 400 (b), 600 (c) and 800 (d) \MeV (see text). The rise in cross section with \hpp\ at $\detaf=6$ is a issue known to the authors which is due to clustering of beam remnants.}
\label{fig:timm:diffractiveGaps_4}
\end{figure}

In \Fref{fig:timm:diffractiveGaps_2:a2}, the inelastic cross section predictions of \pyeight, \hpp and \epos are plotted differential in \detaf at $\sqrt{s} = 14 \TeV$ for an inclusive sample and for a sample requiring exactly one forward proton tag from AFP.
For the inclusive selection, \epos and \pyeight are in rough agreement regarding the relative flatness of the diffractive tail, disagreeing at the 30\% level regarding the normalisation. \hpp generates an excess of events with \detaf = 6 which is a known by-product of the cluster hadronisation of beam remnants.

Upon requiring a forward proton tag from AFP, the overall cross section predictions fall significantly and inline with the acceptances from \Tref{tab:timm:crosssections}. All three generators do however still predict a long tail, with the difference in normalisation between \epos and \pyeight now around a factor of 7.5. \hpp also retains its excess of events at \detaf = 6. 
%For the ALFA selection, the cross sections are predicted to be yet further suppressed, with a significant change of slope in \pyeight for small gaps
The MC all remain sufficiently separated to allow for good model discrimination power given sufficiently precise data.

It is concluded from \Fref{fig:timm:diffractiveGaps_1:a} that the AFP selection greatly suppresses non-diffractive interactions in \pyeight allowing for a higher purity probe of the fragmentation of the $p$--$I\!P$ system.

One method used is to gradually increase \ptcut, allowing for hadronisation fluctuations to create larger pseudorapidity gaps and hence studying the \pt\ and $\eta$ dependences of soft particle production. This is presented in \Fref{fig:timm:diffractiveGaps_4}, where the \ptcut cut is varied over 200--800 \MeV. 

\begin{figure}
\centering
\includegraphics[width=0.45\textwidth]{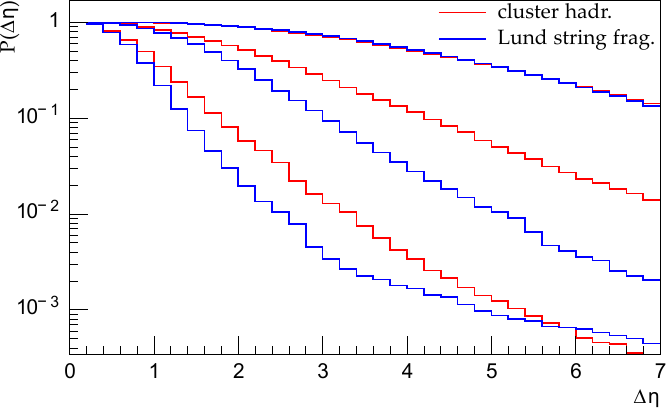}
\caption{Plot from \cite{SOFT_Khoze:2010by} illustrating the probability of finding a hadron level pseudorapidity gap within $|\eta| < 5$ for different choices of \ptcut, shown for Sherpa 2.2.1 with the default cluster hadronisation and when instead using Lund string fragmentation from Pythia. From top to bottom the values of \ptcut\ are 1.0, 0.5 and 0.1 \GeV.}
\label{fig:KKMRZGapFluc}
\end{figure}

This scan in the \ptcut defining the gap was originally motivated by \cite{SOFT_Khoze:2010by} to study the differences in pseudorapidity gap fluctuations possible between different hadronisation models, see \Fref{fig:KKMRZGapFluc}.
\subsection{Soft pseudorapidity gap studies with CASTOR}
The CASTOR calorimeter provides a unique detector at LHC to measure charged and neutral
energy deposits in the very forward phase space. The detector extends the CMS acceptance to a
pseudorapidity of -6.6. In this location of phase space, CASTOR is very sensitive to the production
of medium- and low-mass excited states and can thus be used to study diffractive dissociation. 

It was shown that CASTOR is well suited to distinguish double diffraction from single diffraction,
and that it can contribute to studies of soft diffraction \cite{SOFT_CMS:2013mda}. 
The RMS noise level per calorimeter cell is of the order of 100--300 \MeV, which provides a very
good environment to search for rapidity gaps under the condition that the luminosity and subsequently pileup levels are not
high. Since CASTOR has no segmentation in pseudorapidity, only gaps larger than the acceptance
of CASTOR can be observed.

\begin{figure}
\centering
\begin{subfigure}[b]{0.47\textwidth}
\includegraphics[width=\textwidth]{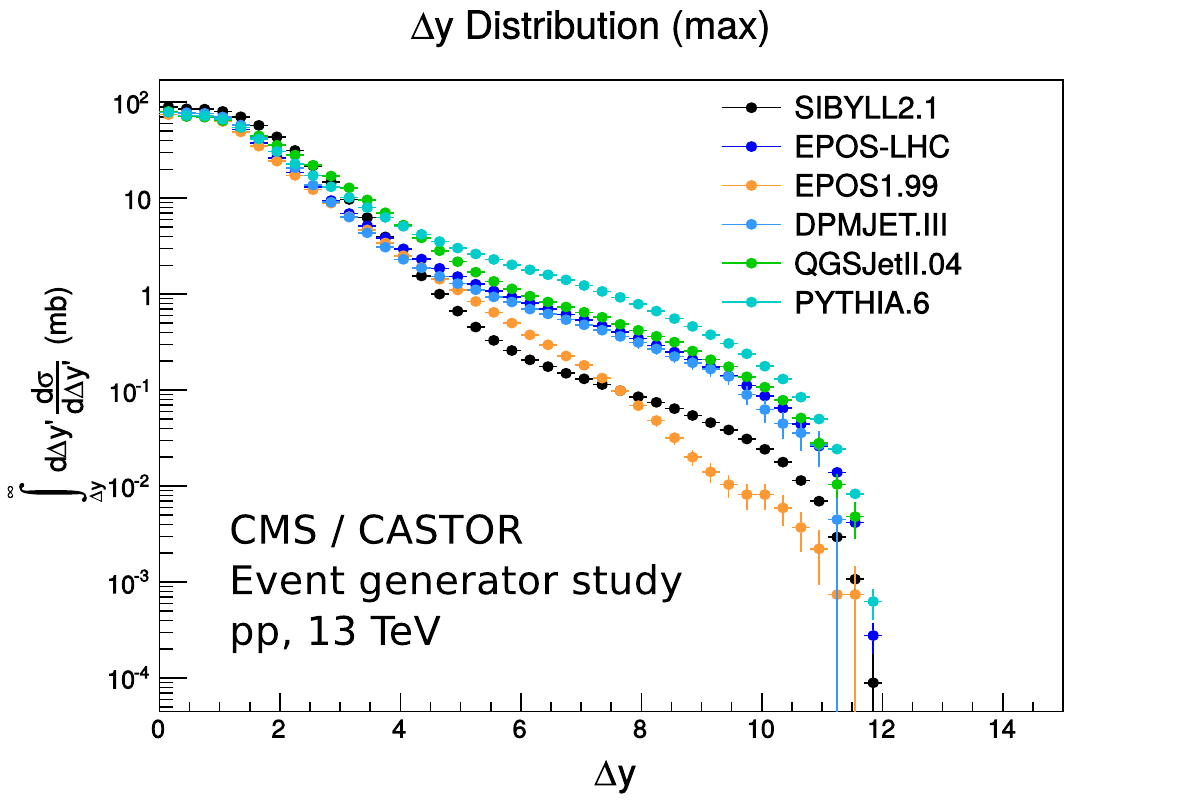}
\caption{}
\label{fig:castorgaps:a}
\end{subfigure}
\hspace{5mm}
\begin{subfigure}[b]{0.47\textwidth}
\includegraphics[width=\textwidth]{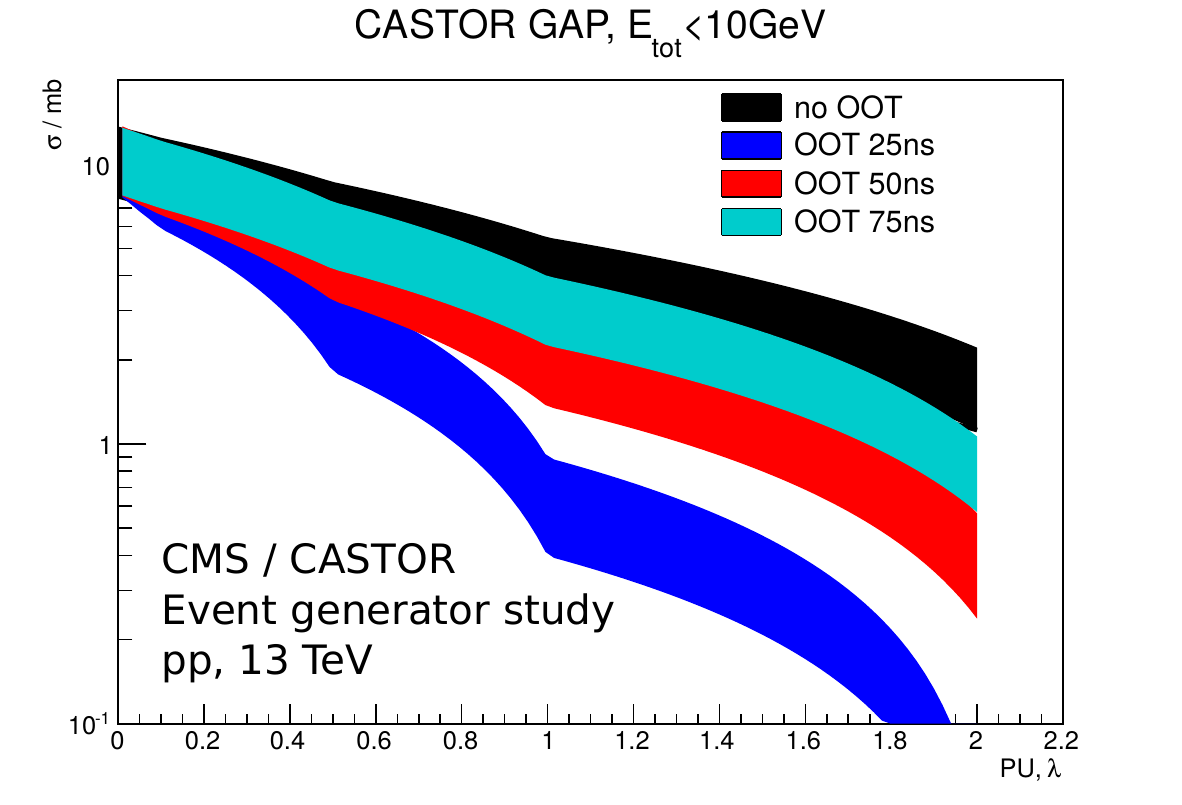}
\caption{}
\label{fig:castorgaps:b}
\end{subfigure}
\caption{CASTOR/CMS event generator level distributions at  $\sqrt{s}=13 \TeV$ of (a) the distribution of rapidity gaps sizes in 13 TeV proton-proton collisions. The transition to low-mass diffraction for rapidity gaps larger then 10 is only accessible with the CASTOR detector and T2. (b) The dependence of detection efficiency of rapidity gaps ($E_{\rm CASTOR} < 10 \GeV$) on pileup for different bunch spacing scenarios. }
\label{fig:castorgaps}
\end{figure}

The use of CASTOR allows for the soft pseudorapidity gap spectrum to be investigated over a larger range.
As seen in \Fref{fig:castorgaps:a}, very little integrated luminosity is required, around 10 nb$^{-1}$ with low pileup ($\mu < 0.05$) is sufficient for many
studies. Here rapidity gaps in CASTOR are defined as being events for which there was less than 10 \GeV\ total energy
deposit in the acceptance of CASTOR. The impact of out-of-time pileup events destroying any gap is highlighted for different bunch configurations in \Fref{fig:castorgaps:b}.

Such data will contribute complementary information to other soft diffractive
measurements that will be performed e.g. by the TOTEM Collaboration. The CASTOR data can
be studied together with the TOTEM T2 tracking station data. In this way, rapidity gaps within
the acceptance of CASTOR are resolved and correlation measurements of \MX vs. $\Delta\eta$ in
this very forward phase space interval can be performed. These are unique measurement
opportunities at LHC in respect to soft diffraction. Firstly, to measure soft diffractive cross sections
in differential bins of mass as well as rapidity gaps and, secondly, to measure single diffractive mass
distributions, if possible correlated to the momentum loss of the surviving proton.

\subsection{Extending pseudorapidity gaps with forward shower counters}

As described in Section~\ref{sec:TheAliceDiffractiveDetector}, the ALICE, CMS and LHCb collaborations have installed forward scintillator shower counters (FSC) which surround the beam pipe at large distances from the central detectors and provide a veto on very forward activity such as is found with low mass diffractive dissociations.

Although with lower granularity in $\eta$, such detectors will allow rapidity gap measurements to extend out yet further from the coverage of the central detectors to (in the case of CMS) $|\eta| = 9$ \cite{SOFT_Albrow:2008az,SOFT_Albrow:2014lta} hence allowing rapidity gaps up to $|\Delta\eta| = 18$ to be measured.

Another primary use of FSC is motivated by LHCb. As the LHC beam is defocused at the LHCb IP to provide low-pileup conditions. The LHCb FSC will be of use in determining the exclusivity of central production events by vetoing on low mass dissociation. %For more details on rapidity gaps in central exclusive production see Section~\ref{subsec:rapiditygap}. % no longer a good single place to ref

Signals in forward scintialltor counters may also be used to measure low mass diffractive states and the data will also be of use for luminosity and beam condition motoring.

\subsection{Soft rapidity gap conclusion}
Rapidity gaps in soft events have been shown in Run-I to be sensitive probes of the physics of diffraction and hadronisation. Giving us a greater phenomenological understanding of the proton. This is made possible with small samples of data take at very low pileup in combination with the forward detectors.

In Run-II, the addition of proton tags is shown in MC to provide an event sample which is strongly diffractive-enhanced. Allowing for greater sensitivity to differences between proton-proton and Pomeron-proton interactions. Moreover it is predicted that the soft pseudorapidity gap spectrum will isolate the contributions from single diffractive dissociation at small gap sizes and double diffractive dissociation at large gap sizes. 

The continued use of the CMS CASTOR detector, along with the TOTEM T1 and T2 telescopes and the range of Forward Shower Counters being installed at LHC experiments will fill the forward aperture, allowing for a much larger span of reconstructible pseudorapidity gap sizes and hence access to lower diffractive masses.

\input{softdiffraction/totem_sigmatot.tex}

\section{Conclusions \& Running Conditions}
The current Roman pot based detectors at the LHC are well located to make precision measurements of elastically scattered protons and will continue their program at higher energies which will be obtained in future runs. 
In addition, we have shown how joint analysis of forward tagged protons correlated with activity in the central detectors will allow for much greater soft physics model discrimination with very little integrated luminosity. Requiring a 
forward proton is expected to be a very good way of suppressing the non-diffractive component of the cross section while different LHC optics configurations will be used to enhance either the single or double diffractive 
components of the diffractive proton cross section. This will allow for a better handle on soft hadronisation effects originating from a Pomeron-proton vertex.   

%Such studies require a minimum bias data sample of the order of ten million events to be obtained at very low pileup conditions ($\mu \sim 0.01$) with combined data taking in the forward and central detectors at 
%$\beta^*$ optics of both 0.55 and 90 m. For the latter, this can be combined with runs to collect elastic scattering data and in ATLAS, the AFP project will have much greater acceptance for the collision optics of $\beta^* = 0.55$ m. 

%% file: softdiffraction/totem_sigmatot.tex
%\documentclass[11pt]{cernrep}

%\usepackage{epic,eepic,pspicture,mathptmx,times,graphpap}

%------------------------------------

\def\un#1{\,{\rm #1}}
\def\unt#1{\,{\rm (#1)}}
\def\der{{\rm d}}

%------------------------------------

\section{Measurements of the Inelastic Cross-Section}
\label{sec:soft:sigmatot:prev} 
%TimM - addint LHCb and other bits and bobs

Measures of the total inelastic cross section during Run-I were performed by ALICE, ATLAS, CMS and LHCb central detectors and by TOTEM. 
Such studies rely on measuring the cross section from charged particle production within the kinematic and fiducial acceptance of the detectors and subsequently performing a model 
dependent extrapolation to the total inelastic cross section.

\begin{figure}[htp]
\centering
\includegraphics[width=0.7\textwidth]{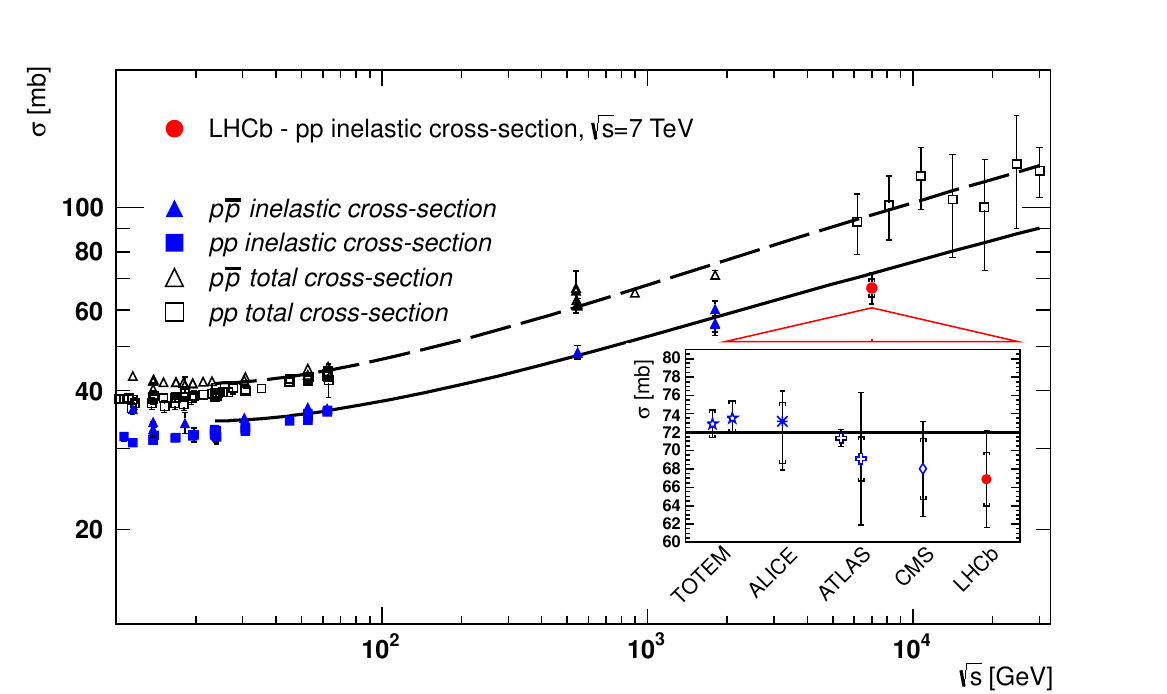}
\caption{%
Comparison of model-extrapolated total inelastic cross section measurements by ALICE, ATLAS, CMS and LHCb (discussed above) along with precision measurements by ATLAS (ALFA) and TOTEM (discussed below). Figure from LHCb \cite{SOFT_Aaij:2014vfa}.
}
\label{fig:lhcb_sigmaIn}
\end{figure}

%New ALICE contribution
The ALICE collaboration made a measurement of $\sigma_{\mathrm inel}$ which makes use of the determination of the single diffractive and double diffractive cross sections 
(see Section~\ref{sec:timm:rapgaps:prev}) to determine the inelastic cross section. 
Measurements were made at three energies (0.9, 2.76 and 7 \TeV). The 7 \TeV\ measurement is shown in \Fref{fig:lhcb_sigmaIn}, and all the values can be found in reference \cite{SOFT_Abelev:2012sea}.
%Old for ALICE
%The ALICE collaboration measurement \cite{SOFT_Abelev:2012sea} of $\sigma_{\rm inel}$ was based on an event sample triggered by the ALICE VZERO hodoscopes or the the two innermost layers of the ALICE Inner Tracking System with overall coverage of $-3.7 < \eta < 5.1$.

ATLAS measured the cross section \cite{SOFT_Aad:2011eu} corrected to the acceptance of the Minimum Bias Trigger Scintillators, this corresponds to all non-diffractive events along with diffractive events 
with mass $M_X^2/s > 5\times10^{-6}$. 

CMS also corrects to this mass range using data from the forward calorimeters ($3 < |\eta| < 5$) and, as an independent method, 
CMS measures the cross section for events containing two or more charged particles with $\pt > 0.2 \GeV$ within $|\eta| < 2.4$ \cite{SOFT_Chatrchyan:2012nj}. 

LHCb measures \cite{SOFT_Aaij:2014vfa} the cross section for one or 
more charged particles with $\pt > 0.2 \GeV$ within $2 < \eta < 4.5$.

TOTEM measured directly the inelastic cross sections using the T1 and T2 telescopes with a 3.1 < $|\eta|$ < 4.7 and
5.3 < $|\eta|$ < 6.5 coverage at $\sqrt{s}$ = 7 TeV (see Section~\ref{s:inel})~\cite{SOFT_si_inel_7}.
In addition higher precision on the inelastic measurement has been achieved by deriving $\sigma_{\mathrm inel}$ from the total cross section measurements based either on the luminosity independent method (TOTEM) or on 
the elastic scattering measurement (TOTEM and ALFA). See next Sections for details.

%TOTEM and ALFA obtained higher precision through directly measuring the cross section for elastic scattering at small angles. 
%The inelastic and hence total cross sections are then assessable via the optical theorem. The remainder of this chapter focuses on this method of determination of the total cross section.   

All results are in agreement to within experimental uncertainty, see \Fref{fig:lhcb_sigmaIn}. 
However the measurement based on larger pseudorapidity coverage like ALICE and TOTEM tend to give larger inelastic cross section values that in addition are better in agreement with the precision measurements 
of the inelastic cross-section using elastic scattering and the optical theorem. This points to a lack of the models to predict the cross section of low-mass diffraction in the 3.4 -- 26 GeV/c$^2$ mass range. 
After model uncertainties due to extrapolation are included, the typical precision of these measurements is 5 -- 11\%. 

%\begin{document}
\section{Measurements of the Total, Elastic and Inelastic Cross-Section with the TOTEM detectors}
\label{sec:soft:sigmatot:totem}

%\section{The TOTEM experiment}

%The TOTEM experiment ~\cite{SOFT_totem_jinst} is dedicated to forward hadronic phenomena at the LHC. The three pillars of its physics programme are: an accurate determination of the total cross-section, a measurement of elastic scattering in a wide kinematic range and studies of diffractive processes. This paper is focused on the first two. 

%The physics programme brings special requirements for the detector apparatus. In particular, large pseudorapidity coverage (to detect most fragments from inelastic collisions) and excellent acceptance for outgoing diffractive and elastic protons. To accomplish this task, TOTEM comprises three subdetectors: the inelastic telescopes T1 and T2 and a system of Roman Pots (RP) for leading proton detection.

%The RP system consists of two stations placed at $+220\un{m}$ (right arm) and $-220\un{m}$ (left arm) from the LHC interaction point (IP) 5. Each of the stations is composed of two units (far and near with respect to the IP) separated by about $5\un{m}$, which is beneficial for reconstructing proton kinematics and discrimination from background. Each unit includes two vertical (top and bottom) and one horizontal RP. The RPs are movable beam-pipe insertions that can bring sensitive detectors to sub-millimetre distance from the beam once it is stable. Each RP hosts 5 back-to-back mounted pairs of silicon strip sensors with reduced ($\approx 50\un{\mu m}$) insensitive margin on the edge facing the beam, in order not to loose protons scattered to very low angles.

\begin{figure}[htp]
\flushleft
\includegraphics[width=0.7\textwidth]{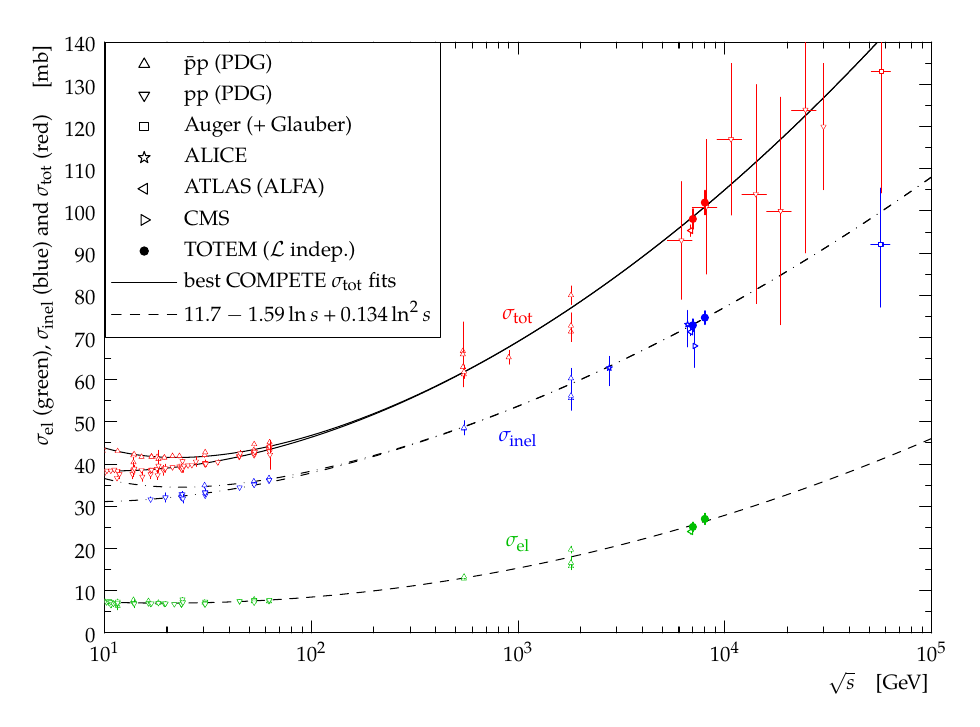}
\includegraphics[width=0.27\textwidth]{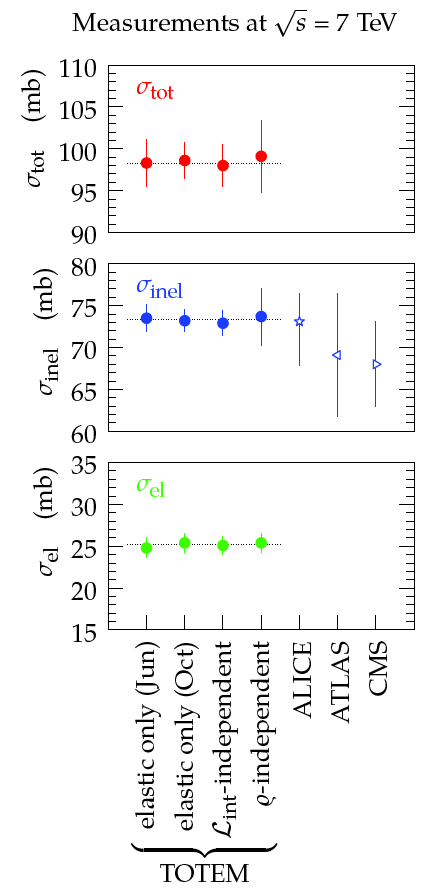}
\caption{%
{\em Left}: compilation ~\cite{SOFT_si_tot_7,SOFT_si_tot_8,SOFT_pdg,SOFT_alice_inel,SOFT_atlas_el_7,SOFT_cms_inel_7,SOFT_auger} of total, inelastic and elastic cross-sections plotted as a function of the centre-of-mass energy $\sqrt s$. 
The continuous black lines (lower for $\rm pp$, upper for $\rm \bar pp$) represent the best fits of the total cross-section data by the COMPETE collaboration ~\cite{SOFT_compete}. 
The dashed line results from a fit of the elastic scattering data. The dash-dotted curves correspond to the inelastic cross-section and is obtained as the difference between the continuous and dashed fits.
{\em Right}: detail of the measurements of total, inelastic and elastic cross-sections at $\sqrt s = 7\un{TeV}$. 
The circles represent the four TOTEM measurements, the other points show the measurements of other LHC collaborations.
}
\label{fig:sigmas}
\end{figure}

The TOTEM experiment~\cite{SOFT_totem_jinst} has measured the total, elastic and inelastic proton-proton cross-section
during LHC Run-I, at $\sqrt{s}$ = 7 and 8 TeV.
The data samples collected at different centre-of-mass energy have all been obtained in dedicated
runs (most with special beam optics) with Roman Pots approaching the beam close
enough to detect elastic events with squared four-momentum transfer $|t|$ as low as possible. 
The available data samples, $|t|$ ranges, 
event statistics and analysis status/publication reference are listed in Table~\ref{tab:samples}.
All the published total, elastic and inelastic cross-section results by TOTEM are summarised and put in context of earlier measurements in \Fref{fig:sigmas}.

\begin{table}[htp]
\caption{List of available data samples. The LHC optics is characterised by the betatron function value at the IP, $\beta^*$. 
The RP approach to the beam is given in multiples of the transverse beam size, $\sigma$. The number of elastic events corresponds to both diagonals after the proton tagging.
}
\label{tab:samples}
\begin{center}
\begin{tabular}{|c|c|c|c|c|c|}\hline
$\beta^*\unt{m}$ & RP approach & $|t|$ range & elastic  & inelastic   & Results \cr 
  &   & $\unt{GeV^2}$ &   events &  events &   \cr\hline\hline
\multicolumn{6}{|c|}{$\sqrt{s}$ = 7 TeV}\\
\hline
$90$  & $10\un{\sigma}$					& $0.02\hbox{ to } 0.4$ 		& $15\un{k}$	& & $\sigma_{\rm tot}$ ~\cite{SOFT_si_el_7_90a}\cr\hline
$90$  & $4.8 \hbox{ to }6.5\un{\sigma}$	& $0.005\hbox{ to } 0.4$ 		                & $1\un{M}$	& $5.54\un{M}$ & $\sigma_{\rm tot}$, $\sigma_{\rm inel}$, $\sigma_{\rm el}$, $(d\sigma/dt)_{\rm el}$  ~\cite{SOFT_si_el_7_90b} ~\cite{SOFT_si_inel_7}
~\cite{SOFT_si_tot_7}\cr\hline
$3.5$ & $7\un{\sigma}$					& $0.4\hbox{ to } 2.5$			& $66\un{k}$	& &  $(d\sigma/dt)_{\rm el}$  ~\cite{SOFT_si_el_7_3p5}\cr\hline
$3.5$ & $18\un{\sigma}$					& $2\hbox{ to } 3.5$	        & $10\un{k}$	& &  $(d\sigma/dt)_{\rm el}$ in progress\cr\hline\hline
\multicolumn{6}{|c|}{$\sqrt{s}$ = 8 TeV}\\
\hline
$1000$	& $3\hbox{ -- }10\un{\sigma}$	& $0.0006\hbox{ to }0.2$	& $352\un{k}$	&           & $\rho$, $(d\sigma/dt)_{\rm el}$ in progress \cr\hline
$90$	& $6\hbox{ -- }9.5\un{\sigma}$	& $0.01\hbox{ to }0.3$		& $0.65\un{M}$	& $4\un{M}$ & $\sigma_{\rm tot}$, $\sigma_{\rm inel}$,$\sigma_{\rm el}$ ~\cite{SOFT_si_tot_8}\cr\hline
$90$	& $9.5\un{\sigma}$			& $0.03\hbox{ to }1.4$	& $7.2\un{M}$	&           & $(d\sigma/dt)_{\rm el}$ in progress \cr\hline\hline
\multicolumn{6}{|c|}{$\sqrt{s}$=  2.76 TeV}\\
\hline
$11$	& $5\hbox{ -- }13\un{\sigma}$	& $\approx 0.06\hbox{ to }0.5$	& $45\un{k}$	&       1.5M    & $\sigma_{\rm tot}$, $\sigma_{\rm inel}$, $\sigma_{\rm el}$, $(d\sigma/dt)_{\rm el}$  in progress \cr\hline
\end{tabular}
\end{center}
\end{table}

\subsection{Total cross-section}\label{s:tot}

Three complementary  methods have been used to determine the total cross-section. The methods, having very different systematic dependences,
give results in excellent agreement. 

\begin{itemize}

\item The first method exploits only elastic scattering measurement. By applying the optical theorem the following formula for the total cross-section $\sigma_{\rm tot}$ is obtained:

\begin{equation}\label{eq:sitotm1}
\sigma_{\rm tot}^2 = {16\pi\over 1+\rho^2} {1\over {\cal L}} \left. {\der N_{\rm el}\over \der t}\right |_{t = 0}\ ,
\end{equation}
where ${\cal L}$ stands for the integrated luminosity and $\der N_{\rm el}/\der t|_{t = 0}$ is the elastic differential rate extrapolated to $t = 0$.
For the $\rho$ parameter the COMPETE~\cite{SOFT_compete} preferred-model extrapolation has been used (0.141 $\pm$ 0.007 at $7\un{TeV}$).

\item The second method relies on summing the elastic event rate $N_{\rm el}$ (obtained by integrating and extrapolating the differential rate) and the inelastic event rate $N_{\rm inel}$ (measured by the inelastic telescopes T1 and T2):
\begin{equation}\label{eq:sitotm2}
\sigma_{\rm tot} = {1\over {\cal L}} (N_{\rm el} + N_{\rm inel})\ .
\end{equation}
This method does not require the value of $\rho$ as input from external sources and it doesn't rely on the Optical Theorem, but in addition it proofs its validity (at 3.5\% level).

\item The third method is luminosity-independent. 
The method requires the
simultaneous measurements of the inelastic and elastic
rates, as well as the extrapolation of the latter in the
invisible region down to  $t = 0$.
Moreover, the value of the integrated luminosity can be determined as well:

\begin{equation}\label{eq:sitotm3}
\sigma_{\rm tot} = {16\pi\over 1+\rho^2} {\der N_{\rm el}/ \der t |_{t = 0}\over N_{\rm el} + N_{\rm inel} }\ ,\qquad
{\cal L} = {1+\rho^2\over 16\pi} { (N_{\rm el} + N_{\rm inel})^2 \over \der N_{\rm el}/ \der t |_{t = 0}}\ .
\end{equation}

\end{itemize}

At $\sqrt s =7\un{TeV}$, all three methods have been used, all exploiting the $\beta^*=90\un{m}$ optics. 
The method based on elastic inputs only, Eq.~(~\ref{eq:sitotm1}), is described in ~\cite{SOFT_si_el_7_90a} while
in ~\cite{SOFT_si_tot_7} all the three methods are described (second row from top in Table~\ref{tab:samples}). 
\Fref{fig:sigmas} (right) shows the consistency of all four total cross-section results.

At  $\sqrt s =8\un{TeV}$ only the luminosity-independent results on elastic, inelastic and total cross-section have been published ~\cite{SOFT_si_tot_8}. 
Moreover, the analysis of the $\beta^*=1000\un{m}$ data is in progress: the separation of Coulomb and nuclear effects is at reach, thus yielding methodically more accurate results.

\subsection{Elastic scattering}\label{s:el}

\begin{figure}[htp]
\begin{center}
\includegraphics[width=\textwidth]{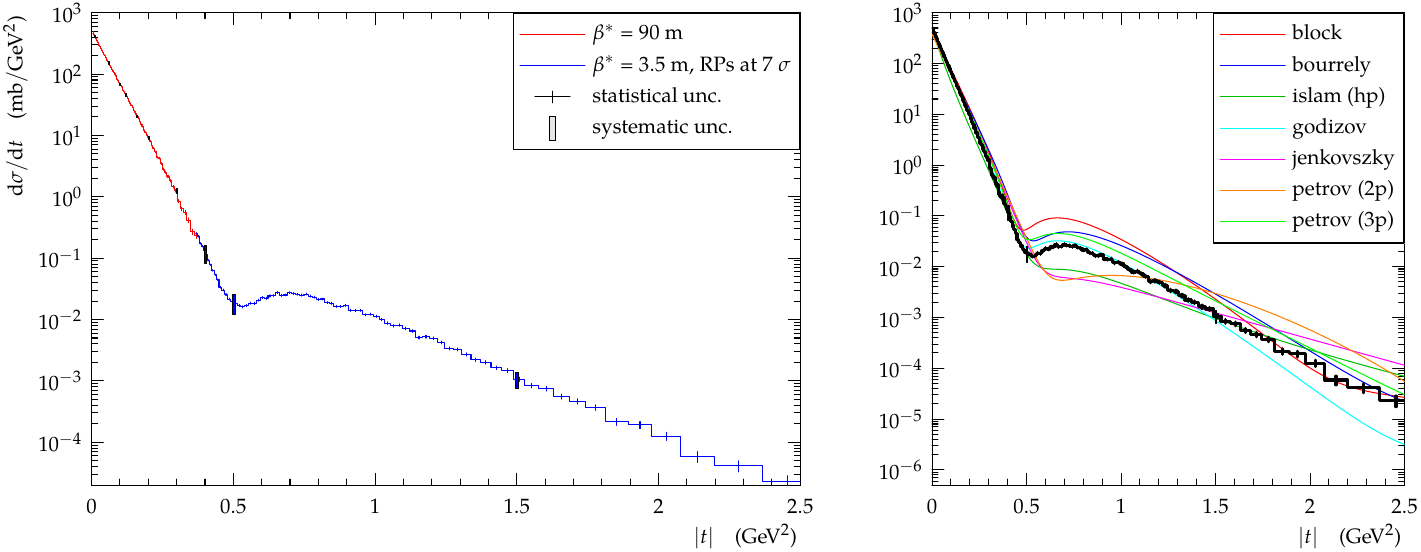}
\caption{Left: The elastic differential cross-section measurements by TOTEM at $\sqrt s =7\un{TeV}$. Right: The measured d$\sigma$/dt compared to the predictions of several models~\cite{SOFT_block,SOFT_bourrely,SOFT_islam,SOFT_jenkovszy,SOFT_petrov}}
\label{fig:ela7}
\end{center}
\end{figure}

At the centre-of-mass energy $\sqrt s =7\un{TeV}$  the differential cross-section of elastic scattering,
d$\sigma/dt$, has been measured by TOTEM in the range 0.005 $<|t|<$ 2.5 GeV$^2$  (see \Fref{fig:ela7}), 
extending from the almost exponential forward peak (d$\sigma/dt$ $\propto \exp(-Bt)$ with B = 19.9 $\pm$ 0.3 GeV$^{-2}$) ~\cite{SOFT_si_el_7_90b} through the dip-bump region (with the minimum observed at 0.53 $\pm$ 0.01 GeV$^2$) to the
large-$|t|$ domain exhibiting a power-law behaviour, $\propto |t|^{-7.8}$ ~\cite{SOFT_si_el_7_3p5} . The $|t|$-range analysed so far has
been covered by two data sets and will be extended at its upper bound to about 3.5 GeV$^2$ with a
third data set already under analysis (see Table~\ref{tab:samples}).

The direct measurement of the cross section, based on the observation of 91\% of the elastic events, gives
$\sigma_{el}= (25.43 \pm 1.07)$ mb.
The direct and indirect\footnote{From the luminosity and $\rho$ independent method} evaluation of the elastic cross section is summarised in Table~\ref{tab:samples} and \Fref{fig:sigmas}.

At the centre-of-mass energy $\sqrt s =8\un{TeV}$ a first data set ($0.01<|t|<0.3 \un{GeV^2}$) has been analysed and used for the luminosity independent measurement of the total cross section~\cite{SOFT_si_tot_8}.
The elastic cross-section has been derived independently from the luminosity ($\sigma_{el}= (27.1 \pm 1.4)$ mb).
Details of the analyses can be found in ~\cite{SOFT_si_el_7_90b,SOFT_si_tot_8}.

More analyses  with two different machine optics are in progress:
\begin{itemize}

\item With $\beta^*$ = 1000m optics, $|t|$ values from  $6\cdot10^{-4}\un{GeV^2}$ to 0.2 $\un{GeV^2}$ have been reached,
and the interference between electromagnetic (Coulomb) and strong (nuclear) interactions has been observed for the first time at the LHC. 
  
This interference gives some sensitivity to the phase of the nuclear amplitude mainly at $t$ = 0  
and allows to separate the Coulomb and nuclear effects (beneficial for determinations of the {\em nuclear} total cross-section).
However, the precise functional form of the scattering amplitude in the interference
region is not known from first principles and thus is model dependent.

The preliminary results for $\rho$ 
are conditional to the functional form of modulus and phase of the nuclear amplitude, and to the choice of the interference formula.

\item
The second set at $\beta^*$= 90m is characterised by very high statistics (7 M events) in
the range  $0.03<|t|<1.4 \un{GeV^2}$ . Its strong statistical power enables an in depth
analysis of the exponential slope $B(t)$.

\end{itemize}

\begin{figure}[htp]
\begin{center}
\includegraphics[width=0.6\textwidth]{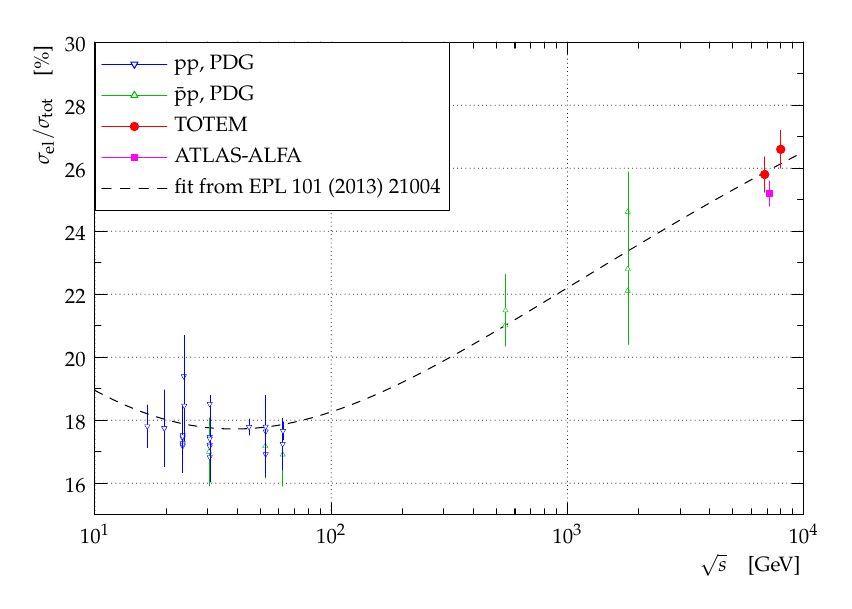}
\caption{The ratio of the elastic to total cross-section as a function
of energy~\cite{SOFT_pdg,SOFT_si_tot_7,SOFT_si_tot_8}. The dashed line shows the
ratio of the $\sigma_{\rm el},\sigma_{\rm tot}$ fits from \Fref{fig:sigmas}}
\label{fig:ratio}
\end{center}
\end{figure}

The ratio $\sigma_{\rm el}/\sigma_{\rm tot}$ can give some insights into the shape and the opacity of the proton, subject to model-dependent
theoretical interpretations. The steady rise of this ratio
with energy (\Fref{fig:ratio}) is often interpreted as the increase of proton size and opacity with energy.

\subsection{Inelastic scattering}\label{s:inel}

At the centre-of-mass energy $\sqrt s =7\un{TeV}$  the inelastic cross-section has been directly measured by TOTEM using inelastic events triggered by the T2 telescope~\cite{SOFT_si_inel_7}.
The T2-visible inelastic cross section, in the pseudorapidity range 5.3$<|\eta|<$6.5, has been measured to be (69.73 $\pm$ 2.88) mb.
 
After including the contributions of events with tracks measured only in T1 telescope (3.1$<|\eta|<$4.7), the contribution of diffractive events with no tracks in T1 and a rapidity gap covering T2, 
and the contribution of low mass diffraction with all final particles at $|\eta| >6.5$, the total inelastic
cross-section has been determined to be (73.7 $\pm$ 3.4) mb. 
The T1+T2 telescopes are sensitive to diffractive masses larger than 3.4 GeV. Although the extrapolation range is very small compared to other LHC experiments, low mass contribution is the second largest uncertainty of the
inelastic cross-section measurement (after the luminosity).

An estimate of the contribution of low mass diffraction can be obtained by comparing the inelastic cross section
measurement obtained from
elastic scattering ~\cite{SOFT_si_el_7_90b} with
with the direct measurements as described before.
From their difference,
$$\sigma_{\rm tot}^{\rm RP} - \sigma_{\rm el}^{\rm RP} - \sigma_{\rm inel,|\eta| <6.5}^{\rm T2} = 2.62 \pm 2.17 \,{\rm mb}$$,
an upper limit of 6.31 mb at 95\% confidence
level on the cross-section for events with diffractive masses below 3.4 GeV has been deduced.
 
At the centre-of-mass energy $\sqrt s =8\un{TeV}$, the inelastic cross-section has been derived independently from the luminosity, $\sigma_{inel}=(74.7 \pm 1.7$) mb.

\subsection{TOTEM Plans at $\sqrt s =13\un{TeV}$}

During LHC Run-II, TOTEM plans to perform the measurement of the total, elastic and inelastic cross-section at the  energy  $\sqrt s =13\un{TeV}$ based on the methods described above.
The high-beta optics ($\beta^* \le 90\un{m}$) is expected to have the same performance as at lower energies: the lower acceptance limit in $|t|$ is roughly 
$|t|_{min,13TeV} \approx 2 |t|_{min,7TeV}$, still allowing a good extrapolation of the differential elastic cross section to $t=0$ at $\beta^*=90\un{m}$ but not enough to acces the Coulomb interference region.
A higher beta optics ($\beta^*=2.5\un{km}$) is foreseen in order to access the required $|t|$-range.
Moreover the differential elastic cross section will be measured, up to the high $|t|$ values, and further studies of the exponential behaviour at low-$|t|$ are envisaged.
The full menu of diffractive measurements is described elsewhere in this document.

\section{Measurement Total and Elastic cross-section with ALFA}\label{s:tot_alfa}
The ATLAS precision measure of the total $pp$ cross section was performed with the ALFA detector using the luminosity dependent parametrisation of the optical theorem in Equation \ref{eq:sitotm1} 
during a dedicated $\sqrt{s} = 7 \TeV$ run with $\beta^* = 90$ m optics.

ALFA uses parallel-to-point focusing optics in the vertical plane to translate the scattering angle $\theta$ at the interaction point to a vertical displacement at the detector. 
This angle was reconstructed from the impact points and beam transport matrix using the so called `subtraction method' by exploiting that elastic protons will be reconstructed back-to-back in the forward and backward instrumentation.

\begin{figure}[htp]
\begin{center}
\includegraphics[width=0.47\textwidth]{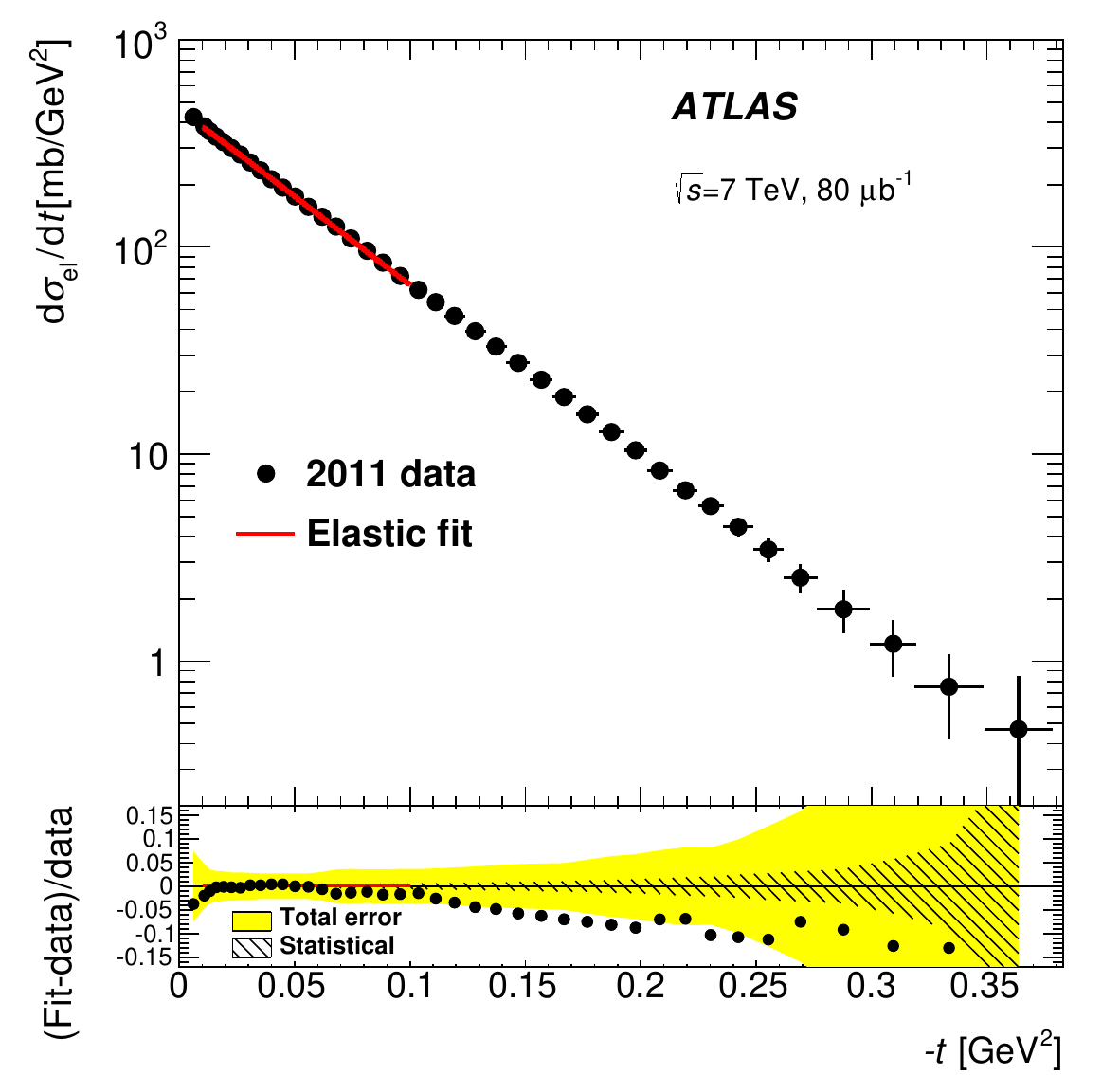}
\caption{Left: A fit of a parametrised form of the differential elastic cross section, reconstructed with the subtraction method.}
\label{fig:atlas:tslope}
\end{center}
\end{figure}

Both background and efficiency determination were data driven and ALFA fits the $-t$ spectrum (\Fref{fig:atlas:tslope}) in the range of $> 10\%$ acceptance, $0.01 < -t < 0.1 \GeV^2$, 
which yields fits to the $B$ slope $B = 19.73\pm0.14 {\rm(stat.)} \pm 0.26 {\rm(syst.)} \GeV^{-2}$ and total cross section of $\sigma_{\rm tot} = 95.35\pm0.38 {\rm(stat.)} \pm 1.25 {\rm(exp.)} \pm 0.37 {\rm(extr.)}$ mb \cite{SOFT_atlas_el_7}. 
The `extra.' uncertainty covers the extrapolation to the optical point $|t|\rightarrow 0$ and the dominant uncertainty comes from the luminosity and beam energy.

\subsection{ALFA Plans at $\sqrt s =13\un{TeV}$}

For the upcoming Run-II period of the LHC, dedicated periods for special fills with high $\beta^*$ are envisaged. The ALFA approved physics programme at the energy $\sqrt{s}$ = 13 \TeV\ will initially 
focus on total and elastic cross-section and diffractive measurements at $\beta^*$ = 90 m.

During the Long Shutdown 1 (2014) several upgrades have been performed on the ALFA detector to ensure good performance during Run-II. Most notably, a RF-protection system has been installed and the 
distance between stations has been increased to 8 m improving the local angle resolution by factor of 2 which will improve the final precision.

For the shutdown 2015--2016 it is foreseen to install a new set of cables to power separately Q4 and Q7, allowing higher values in $\beta^*$ and therefore giving access to lower $|t|$ values. 
The ultimate long term goal of ALFA is to run at very high $\beta^*$ (2.5 km) optics to study the Coulomb interference region and
to obtain a calibration of the absolute luminosity.

%% file: harddiffraction/harddiffraction.tex
%\documentclass[11pt,oneside,a4paper]{article}
%
%%\usepackage{atlasphysics} % Usefull HEP things
%
%\usepackage[a4paper,vscale=0.8, hscale=0.8]{geometry}
%
%%\input{softdiffraction_macros.tex}
%%\input{forward_macros.tex}
%
%\usepackage{graphicx}
%\usepackage{booktabs}
%\usepackage{amsmath}
%\usepackage{color}
%\usepackage{multirow}
%
%\usepackage[T1]{fontenc}
%
%\usepackage{hyperref}
%
%\graphicspath{{../}}
%
%\usepackage{setspace}
%
%\onehalfspacing
%
%\begin{document}

%%%%%%%%%%

\section{Introduction}
\label{ch4_sec_introduction}
Hard diffractive processes are important part of the studies performed in a high energy physics since their discovery in the UA8 experiment~\cite{UA8_1, UA8_2}. Data collected by HERA and Tevatron detectors allowed to significantly increase this knowledge. Nevertheless, at the LHC era, many questions still remain open.

The definition of diffraction is connected to the exchange of a colourless object (the Pomeron). Such exchange does not only leave the interacting proton intact, but also creates a region in rapidity devoid of particles -- a large rapidity gap (LRG). Such signature is often a requirement in the diffractive event selection. However, as the gap is created between the scattered proton and the Pomeron remnants, \textit{i.e.} in the very forward direction, it is sometimes beyond the detector acceptance. Moreover, a gap could be created also in the non-diffractive events as a result of a fluctuation of the final state particles. Alternatively to requiring the LRG, the intact proton can be directly tagged, provided the adequate instrumentation is available. 

During Run I a number of diffractive measurements were done at the LHC by the ATLAS \cite{ATL_gen_1, ATL_gen_2, ATL_gen_3, ATL_gen_4}, CMS \cite{CMS_gen_1, CMS_gen_2, CMS_gen_3, Chatrchyan:2012vc, CMS_gen_5}, TOTEM \cite{TOTEM_gen_1, TOTEM_gen_2, TOTEM_gen_3, TOTEM_gen_4, TOTEM_gen_5}, ALICE \cite{ALICE_gen_1, ALICE_gen_2} and LHCb \cite{LHCb_gen_1, LHCb_gen_2} experiments. These analyses, preformed at $\sqrt{s}$ of 7 or 8 TeV will be continued also at Run II, when the centre-of-mass energy will be increased to 13 TeV.

This chapter describes the foreseen hard diffractive programme of the CMS/TOTEM and ATLAS experiments, with special attention paid to measurements utilising forward proton tagging techniques (see Chapter~\ref{chap:bema} for details on the instrumentation). These results include experimental motivation as well as an estimation of the obtainable significances along with the technical requirements for the collection of suitable datasets. In particular, the possibilities of measuring various single diffractive and double Pomeron exchange events will be discussed.

\section{Backgrounds}
\label{ch4_sec_backgrounds}
\begin{figure}[!htbp]
\centering
    \includegraphics[width=0.45\textwidth]{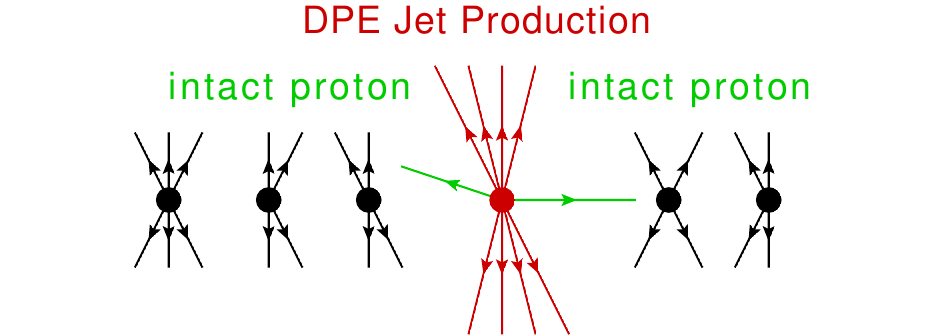}\hfill
    \includegraphics[width=0.45\textwidth]{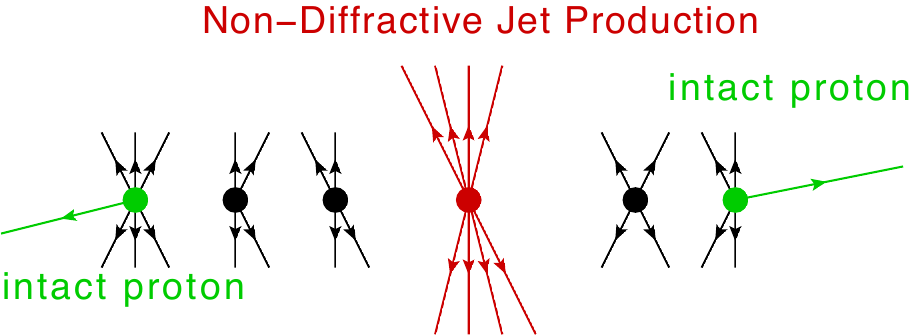}
    \caption{An example of diffractive signal (\textbf{left}) and non-diffractive background (\textbf{right}) in non-zero pile-up environment. The non-diffractive event is overlaid with minimum-bias protons visible in the forward detectors.}
    \label{ch4_background}
\end{figure}

In measurements using a forward tagging technique in a non-zero pile-up environment a variety of backgrounds are present. They may originate from events in which a non-diffractive system is produced together with a forward proton coming from a different interaction in the same bunch crossing. For example, a hard single diffractive production process might be mimicked by a hard non-diffractive interaction overlaid with an intact proton coming from the minimum-bias events. In case of double Pomeron exchange (DPE) production, the background may come from non-diffractive or single diffractive events (see Fig.~\ref{ch4_background}).

In addition to pile-up events, other particles circulating with the LHC beam (the so-called beam halo) can be detected in the forward detectors and act as a background. The realistic way of simulating such events is not yet known. Fortunately, at low luminosities (low pile-up values) they can be sufficiently suppressed by data driven approaches based on the correlations between the central system and the forward protons. In order to reduce these backgrounds, the following selection criteria can be applied:
\begin{itemize}
  \item proton tag,
  \item one vertex reconstructed in central detector,
  \item correlation between kinematics of proton(s) and central system.
\end{itemize}
Since the mentioned backgrounds are similar for all measurements described in this Chapter, their treatment is discussed commonly in this Section.

\subsection{Proton Tag}
The presence of the forward proton(s) is a natural requirement in the following analyses. In order to mimic a diffractive event, there has to be a proton visible in the forward detector.

The probability of having intact protons originating from soft events depends on various factors, such as the LHC optics settings and the acceptance of the forward detectors. These issues are discussed in details in Section~\ref{sec:soft:forwardacceptance}. 

\subsection{One Vertex Requirement}
Another constraint that may suppress pile-up backgrounds comes from the single vertex requirement. Unfortunately, this rejection is not 100\% effective due to:
\begin{itemize} 
  \item finite resolution of the central trackers which may result in the merging of nearby vertices,
  \item too few tracks originating from the soft pile-up vertex.
\end{itemize}

In the ATLAS feasibility studies presented in this Chapter, the vertex is assumed to be reconstructed if there are at least four charged particles in the tracker ($|\eta| < 2.5$). In order to account for the detector efficiency, each particle was assigned a probability of being reconstructed. The thresholds were set to:
\begin{itemize}
  \item 50\% for the particles with $100 < p_T < 500$ MeV and
  \item 90\% for the ones with $p_T > 500$ MeV.
\end{itemize}
These values reflect the performance of the ATLAS inner detector~\cite{minbias_tracks}, but are also similar for the CMS experiment~\cite{Khachatryan:2010nk:ch4}. The minimal distance below which vertices are merged was set to 1.5 mm.

\subsection{Relative Energy Loss Difference}
\begin{figure}[!htbp]
\centering
\includegraphics[width=0.49\textwidth]{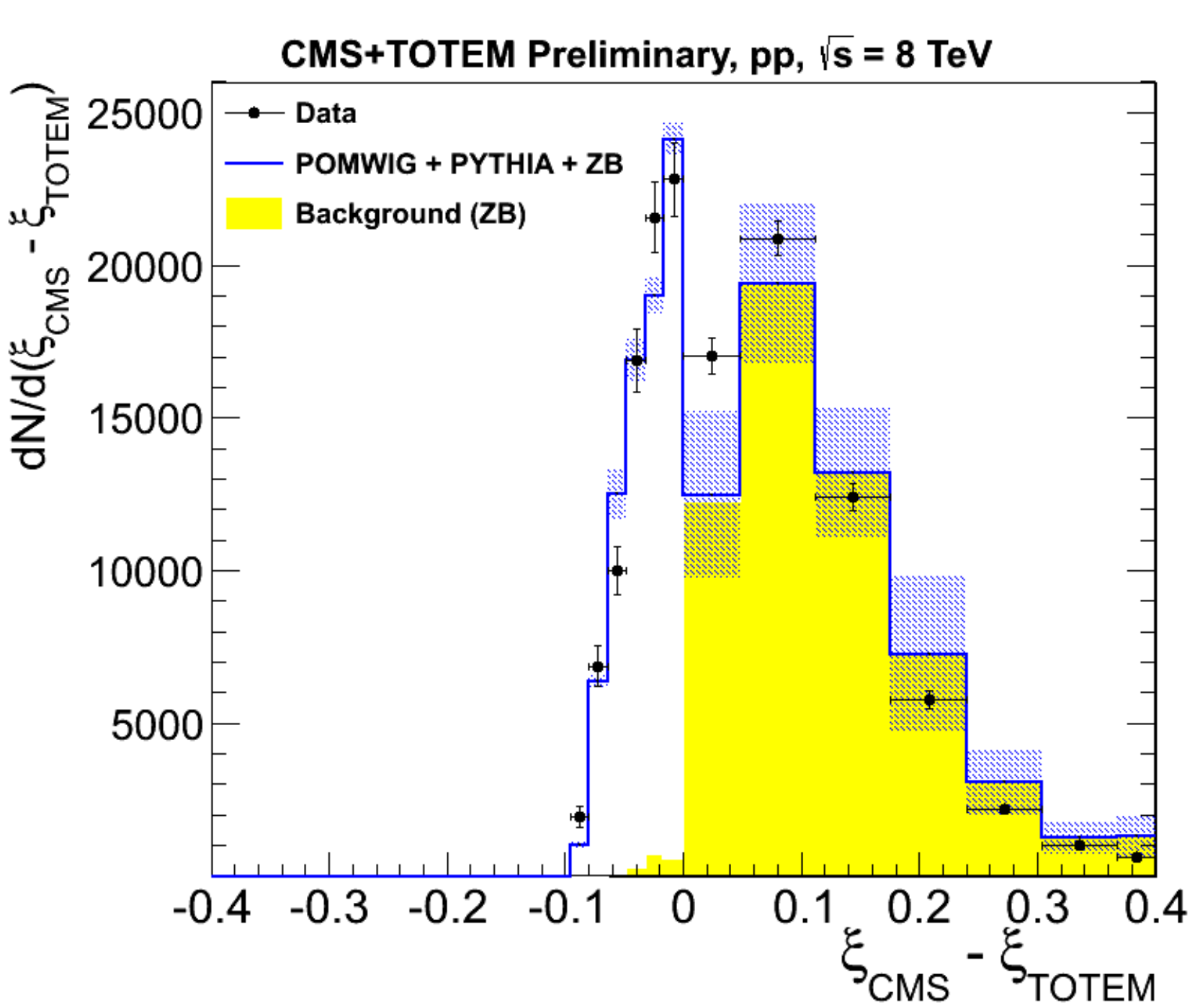}\hfill
\includegraphics[width=0.49\textwidth]{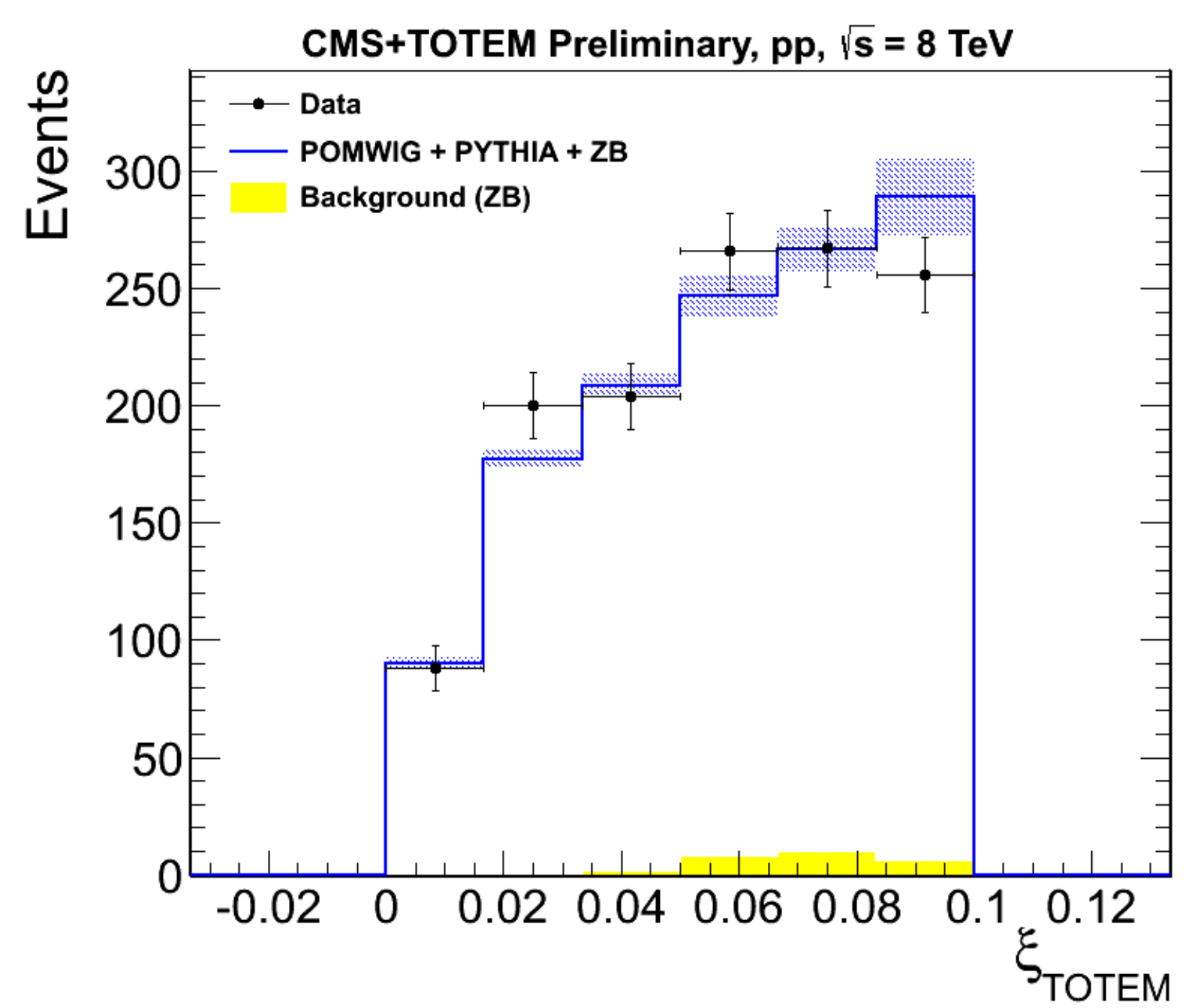}
\caption{\textbf{Left:} difference between the longitudinal momentum loss of the proton reconstructed with CMS and with TOTEM in the single diffractive dijet production process. The data points (full circles) are compared to a mixture of MC and ZB events. \textbf{Right:} longitudinal momentum loss of the proton reconstructed with TOTEM after application of the background subtraction condition $\xi_{\rm CMS}-{\xi}_{\rm TOTEM} > 0$.}
\label{ch4_backgrounds_xi}
\end{figure}

In order to suppress pile-up and beam-halo backgrounds a data driven approach based on the correlations between the central system and the forward protons can be used. For example, in a CMS-TOTEM analysis of single diffractive dijet production with Run I data, beam halo and pile-up backgrounds were subtracted by comparing the longitudinal momentum loss of the proton reconstructed with CMS ($\xi_{\rm CMS}$, obtained summing up the energies and longitudinal momenta of all final state particles) and that reconstructed with the TOTEM Roman Pots (RP) ($\xi_{\rm TOTEM}$)~\cite{CMS-DP-2015-005}. The difference $\xi_{\rm CMS}-{\xi}_{\rm TOTEM}$ is shown in Fig.~\ref{ch4_backgrounds_xi} (left). 

The data in Fig.~\ref{ch4_backgrounds_xi} are compared to a mixture of MC (containing signal and non-diffractive background) and zero bias data events. These data were collected with the CMS and TOTEM detectors in proton-proton collisions at $\sqrt{s}=8\;{\rm TeV}$ during a dedicated run with $\beta^{*}=90\;{\rm m}$ (and therefore include events with protons originating from pile-up and particles from beam halo). It is worth stressing that such conditions are similar to those expected for the low-luminosity, low-pile-up scenarios during the LHC Run II. 

Background events populate the kinematically forbidden region of $\xi_{\rm CMS}-{\xi}_{\rm TOTEM} > 0$. The requirement $\xi_{\rm CMS}-{\xi}_{\rm TOTEM} < 0$, applied in Fig.~\ref{ch4_backgrounds_xi} (right) where ${\xi}_{\rm TOTEM}$ is shown, selects mostly signal events. The remaining contamination of background was found to be $\sim 4$\%.

\subsection{Running Conditions}
\label{ch4_sec_run_conditions}
As run conditions are not fixed and the optimal data-taking conditions differ process by process studied, it is useful to discuss the measurements as a function of the average number of interactions per bunch crossing $\mu$ (so-called pile-up).

\begin{figure}[htbp]
  \centering
  \includegraphics[width=0.49\textwidth]{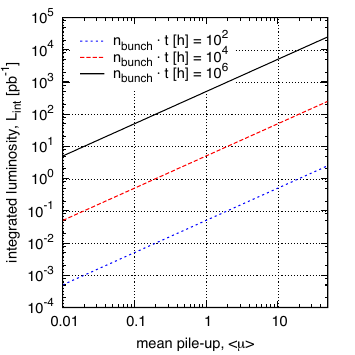}\hfill
  \includegraphics[width=0.49\textwidth]{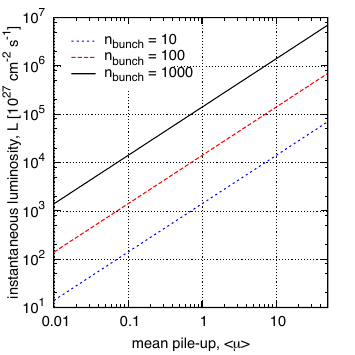}
  \caption{Integrated (\textbf{left}) and instantaneous (\textbf{right}) luminosity as a function of pile-up. The lines represent three different bunch configurations. The tick marks on the vertical scale are at multipliers 2, 5, and 8.}
  \label{fig_lumi}
\end{figure}

The integrated luminosity as a function of $\mu$ is shown in Fig.~\ref{fig_lumi}. The lines represent the product of the number of colliding bunches $n_{bunch}$ and run time, $t$, in hours. For example, the collection of a 5 pb$^{-1}$ data sample at $\mu$ = 0.1, requires $n_{\rm bunch}\times t({\rm h})=10^5$ equivalent to 100~h (1 week) of running with $n_{\rm bunch}=1000$ colliding bunches.

\section{Factorisation Tests}
\label{ch4_sec_factorisation_tests}
A key physics issue in diffractive processes is whether the diffractive Parton Distributions Functions (DPDFs) are universal, \textit{i.e.} whether the collinear factorisation~\cite{Collins:1997sr} holds or not. DPDFs were extracted from high precision HERA data by performing perturbative QCD fits at next-to-leading order accuracy and include a full experimental and theoretical error estimation~\cite{H1LRG06_fit,ZEUS10_fit}. Support to the factorisation theorem was provided by analyses of diffractive dijet cross sections in DIS. These results, despite large theoretical errors, are well described by next-to-leading order predictions based on DPDFs extracted from the inclusive diffractive DIS data~\cite{Aktas:2007bv,Aaron:2010su,Chekanov:2007aa}. However, the hard scattering factorisation was proven to fail in $p\bar{p}$ collisions at the Tevatron~\cite{Aaltonen:2012tha,Aaltonen:2010qe}, where the single diffractive production cross sections of dijet and the electro-weak bosons were overestimated by an order of magnitude w.r.t. predictions based on HERA DPDFs. In Ref.~\cite{Kaidalov:2001iz}, it was shown that this breakdown can be explained by screening effects quantified by the so-called rapidity gap survival probability. In photoproduction at HERA ($Q^2 \simeq 0$), the exchanged photon, which is real or quasi-real, can either interact directly with the proton (so-called direct photoproduction) or behave like a hadron, first dissolving into partonic constituents that then scatter off the target (so-called resolved photoproduction). For the latter process, the factorisation is expected to fail like in the hadron-hadron case. Whether H1 and ZEUS dijet phoproduction data show a suppression, as predicted by theory~\cite{Kaidalov:2001iz,Kaidalov:2003xf}, has been a dilemma for the last decade~\cite{Aaltonen:2012tha,Aktas:2007bv, Chekanov:2007aa}.

The factorisation theorem is at the heart of modern QCD phenomenology at hadron colliders. It provides a crucial predictivity to the theory and, so far, has been tested and verified by all phenomenological analyses. Understanding the mechanism, responsible for the striking breaking of factorisation in hard diffraction, would unveil the non-perturbative phenomena behind it. Single diffractive production processes, like Drell-Yan and vector boson production, are among the best tools to look for such effects in proton-proton collisions at the LHC energies. Moreover the concept of photoproduction of diffractive dijets at HERA can be revisited at the LHC with the flux of quasi-real photons in ultraperipheral collisions (UPS)~\cite{Baltz:2007kq:ch4, Baur:2001jj:ch4}, relying on the notation of equivalent photon approximation. 
%It should be noted that getting a full quantitative first principles analysis requires an eveluation of a virtual corrections to the $\gamma^* \rightarrow q\bar{q}$ impact factor, which are presently under study~\cite{Boussarie:prep}. 
In order to get a full quantitative prediction for the diffractive photoproduction of two and three jets, one needs both the $\gamma \to q \bar{q} g$ impact factor at leading order   \cite{Boussarie:R1} and the virtual corrections to the $\gamma \to q \bar{q}$ impact factor \cite{Boussarie:prep}.

\subsection{Predictions }
\begin{figure*}[!htbp]
\begin{minipage}{0.49\textwidth}
 \centerline{\includegraphics[width=1.0\textwidth]{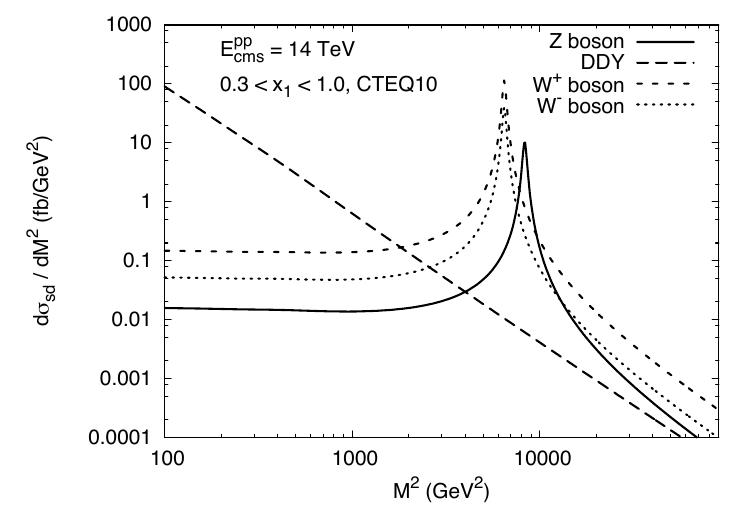}}
\end{minipage}
\hspace{0.5cm}
\begin{minipage}{0.46\textwidth}
 \centerline{\includegraphics[width=1.0\textwidth]{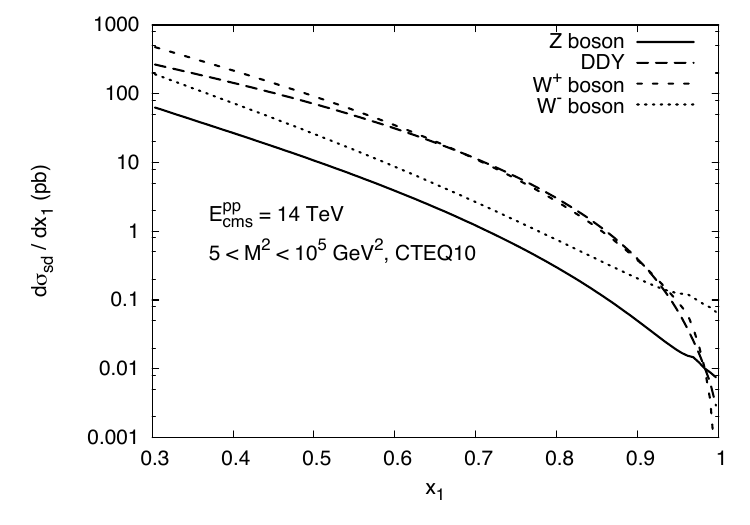}}
\end{minipage}
   \caption{
Diffractive gauge boson production cross section as a function of boson invariant mass squared $M^2$ (\textbf{left}) and boson fractional light-cone momentum $x_1$ (\textbf{right}) in $pp$ collisions at $\sqrt{s}=14$ TeV. Solid, long-dashed, dashed and dotted curves correspond to $Z$, $\gamma^*$, $W^+$ and $W^-$ bosons, respectively. The CTEQ10 PDF parametrization \cite{Lai:2010vv} was used.}
 \label{fig:CS-LHC}
\end{figure*}

The single diffractive cross sections for $Z^0,\,\gamma^*$
(diffractive DY) and $W^{\pm}$ bosons production, calculated for 
$\sqrt{s}=14$ TeV according to the model~\cite{Kopeliovich:2006tk,Pasechnik:2011nw,Pasechnik:2012ac}, are presented in Fig.~\ref{fig:CS-LHC}. The cross section is shown differentially in the dilepton mass squared (left plot) and in the longitudinal momentum fraction (right plot). These plots  do not reflect particular detector constraints. The $M^2$ distributions were integrated over the {\it ad hoc} interval of fractional boson momentum of $0.3<x_1<1$, corresponding to the forward rapidity region (at not extremely large masses). Then the mass distribution is integrated over the
potentially interesting invariant mass interval of $5<M^2<10^5$~GeV$^2$.

The $M^2$ distributions of the $Z^0$ and $W^{\pm}$ bosons clearly demonstrate their resonant behaviour: in the resonant region it significantly exceeds the corresponding diffractive Drell-Yan component. Only for very low masses the $\gamma^*$ contribution becomes important. For $x_1$ distribution, when integrated over the low mass and resonant regions, the diffractive $W^+$ and $\gamma^*$ components become comparable to each other, both in shapes and values, whereas the $W^-$ and, especially, $Z$-boson production cross section are noticeably lower. The $W^-$ cross section is smaller than the $W^+$ one due to differences in valence $u$- and $d$-quark densities (dominating over sea quarks at large $x_q$) in the proton. The precise measurement of differences in the forward diffractive $W^+$ and $W^-$ rates would allow to constrain the quark content of the proton at large values of $x_q\equiv x_1/\alpha$.

\begin{figure}[!htbp]
\begin{minipage}{0.49\textwidth}
 \centerline{\includegraphics[width=1.0\textwidth]{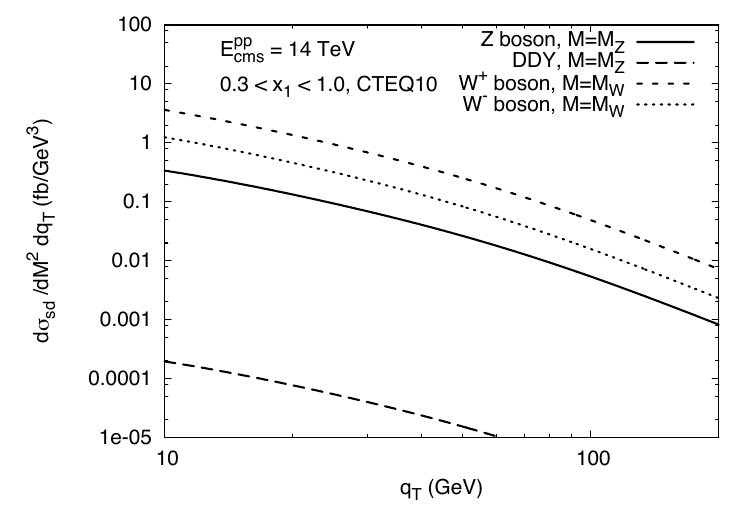}}
\end{minipage}
\hspace{0.5cm}
\begin{minipage}{0.46\textwidth}
 \centerline{\includegraphics[width=1.0\textwidth]{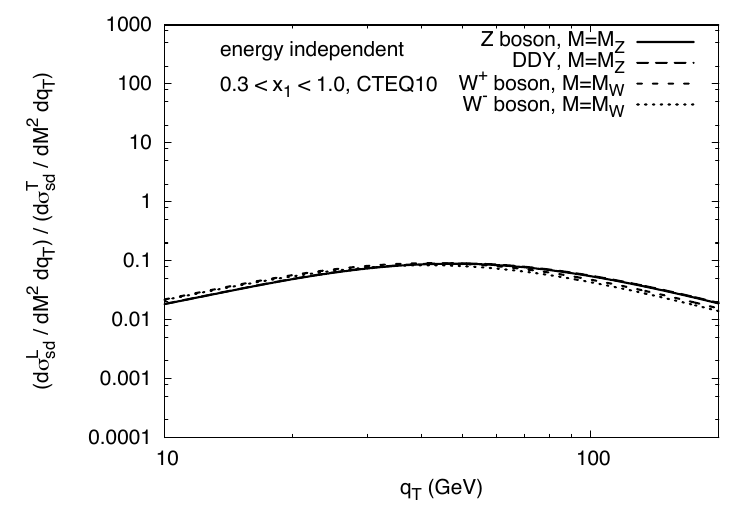}}
\end{minipage}
   \caption{
\textbf{Left}: double-differential diffractive gauge boson production cross section as function of the di-lepton transverse momentum $q_{\perp}$ in $pp$ collisions at $\sqrt{s}=14$ TeV.  \textbf{Right}: longitudinal-to-transverse gauge bosons polarisations ratio as a function of the di-lepton transverse momentum $q_{\perp}$. In both plots, the invariant mass is fixed as $M=M_Z$ in the $Z^0,\gamma^*$ production case and as $M=M_W$ in the $W^{\pm}$ production case.}
 \label{fig:dqt}
\end{figure}

From the phenomenological point of view, the distribution of the forward diffractive cross section in the dilepton transverse momentum $q_{\perp}$ could also be of major importance. In Fig.~\ref{fig:dqt} (left) the dilepton transverse momentum $q_{\perp}$ distribution of the double-differential diffractive cross section at  $\sqrt{s}=14$ TeV is shown for the dilepton invariant mass fixed at a corresponding resonance value -- the $Z$ or $W$ mass. The shapes turned out to be smooth and the same for different gauge bosons, whereas the normalisation is different. The longitudinal-to-transverse gauge boson polarisations ratio, $\sigma^L/\sigma^T$, shown in Fig.~\ref{fig:dqt} (right), does not strongly vary for different bosons. It is peaked at about the half of the resonance mass, and uniformly decreases to smaller/larger $q_{\perp}$ values.

\begin{figure*} [!htbp]
\begin{minipage}{0.49\textwidth}
 \centerline{\includegraphics[width=1.0\textwidth]{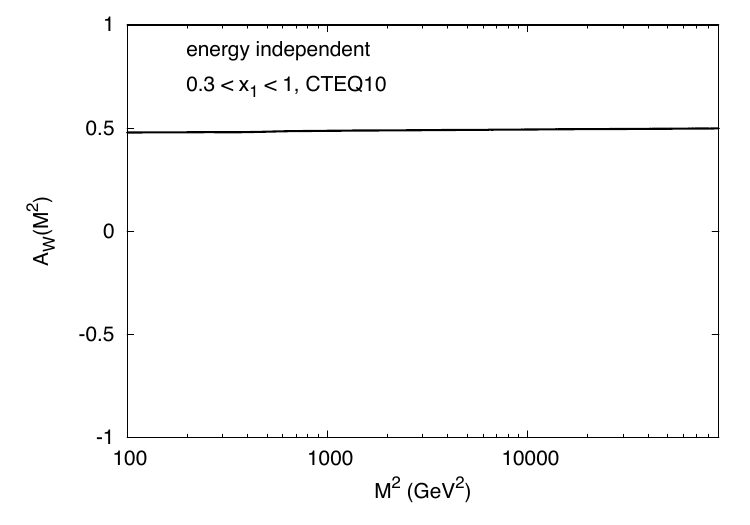}}
\end{minipage}
\hspace{0.2cm}
\begin{minipage}{0.48\textwidth}
 \centerline{\includegraphics[width=1.0\textwidth]{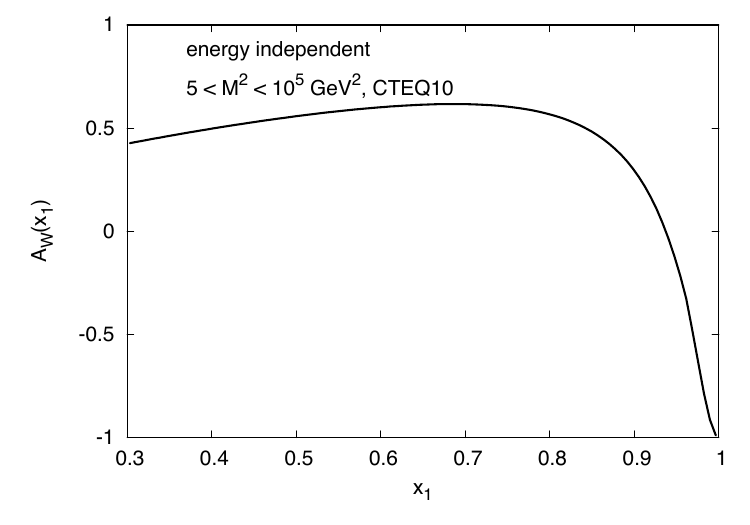}}
\end{minipage}
   \caption{
\small Charge asymmetry in the single diffractive $W^+$ and $W^-$ cross sections as a function of $M^2$, at fixed $x_1=0.5$ (\textbf{left}),  and as a function of $x_1$, at fixed $M^2=M_W^2$ (\textbf{right}). The solid lines correspond to $\sqrt{s}=14$ TeV, the dashed lines to the RHIC energy $\sqrt{s}=500$ GeV.}
 \label{fig:AW}
\end{figure*}

Due to its sensitivity to the difference between $u$- and $d$-quark PDFs at large $x$, the $W^{\pm}$ charge asymmetry, $A_W$, is a crucial observable.  It is shown in  differentially as a function of the dilepton invariant mass squared $M^2$ and integrated over the $0.3<x_1<1.0$ interval in Fig.~\ref{fig:AW} (left) and as a function of the boson momentum fraction $x_1$ and integrated over the $5<M^2<10^5$ GeV$^2$ interval in Fig.~\ref{fig:AW} (right). $A_W$ turns out to be independent on both the hard scale $M^2$ and the center of mass energy. One concludes that, due to different $x$-shapes of valence $u,\,d$ quark PDFs, at small $x_1$ the diffractive $W^+$ bosons' rate dominates over $W^-$ one. However, when $x_1\to 1$ the $W^-$ boson cross section becomes increasingly important and strongly dominates over the $W^+$ one.

\begin{figure*} [!htbp]
\begin{minipage}{0.49\textwidth}
 \centerline{\includegraphics[width=1.0\textwidth]{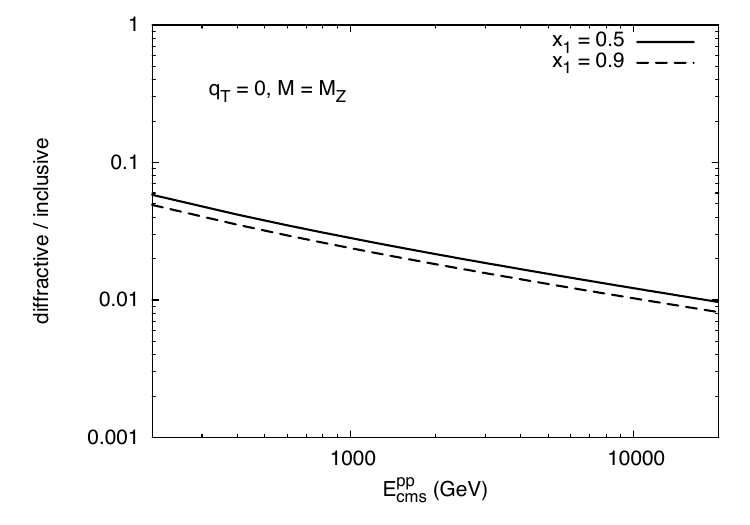}}
\end{minipage}
\hspace{0.2cm}
\begin{minipage}{0.48\textwidth}
 \centerline{\includegraphics[width=1.0\textwidth]{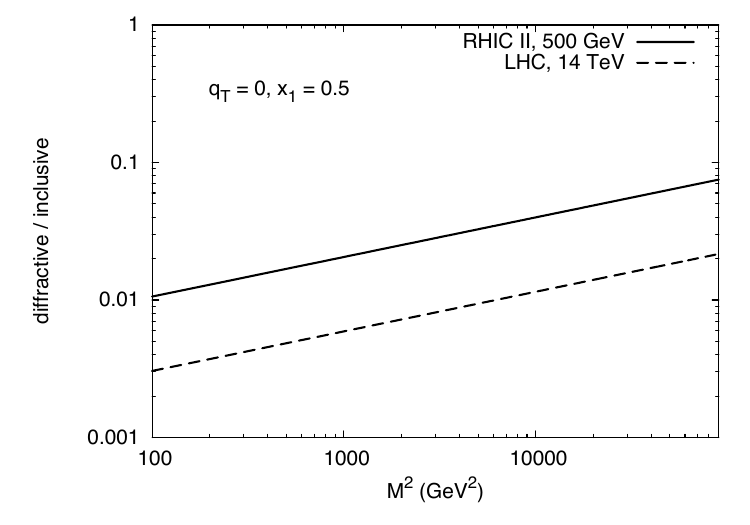}}
\end{minipage}
   \caption{
\small The diffractive-to-inclusive ratio of the gauge bosons production cross sections in $pp$ collisions as a function of the center of mass energy (\textbf{left}) and the dilepton invariant mass $M^2$ (\textbf{right}). It does not depend on the type of the gauge boson and quark PDFs.}
 \label{fig:ratio1}
\end{figure*}

The diffractive to inclusive ratio, $\sigma_{sd}/\sigma_{incl}$, shown in Fig.~\ref{fig:ratio1}, is independent on the type of the gauge boson, its polarisation or quark PDFs. In this respect, it is the most convenient and model independent observable. The ratio decreases with energy, but increases with the hard scale, thus it behaves opposite to what is expected according to the diffractive factorisation-based approaches. Therefore, measurements of the single diffractive gauge boson production cross section, at least at two different energies, would provide an important information about the interplay between soft and hard interactions in QCD, and their role in the formation of diffractive excitations and colour-screening effects.

\section{Single Diffractive Jet Production}
\label{ch4_sec_SD_JJ}
\begin{figure}[!htbp]
\centering
\includegraphics[width=0.25\textwidth]{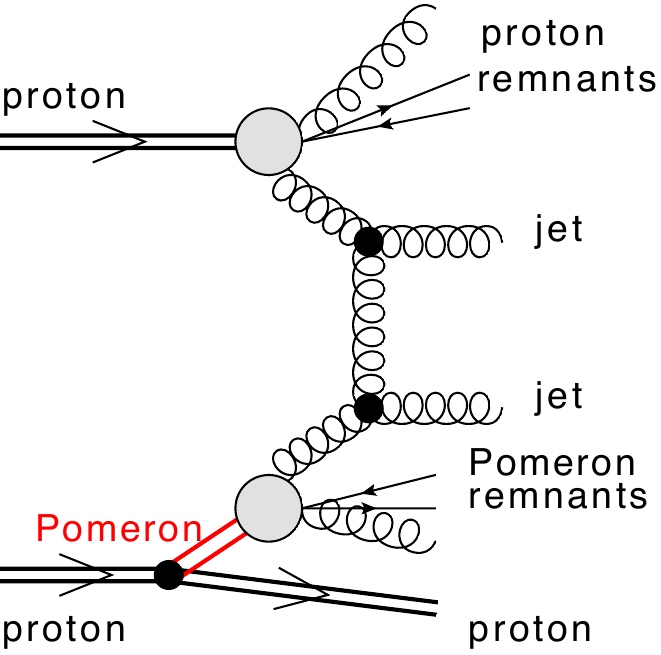}\hspace{0.2\textwidth}
\includegraphics[width=0.25\textwidth]{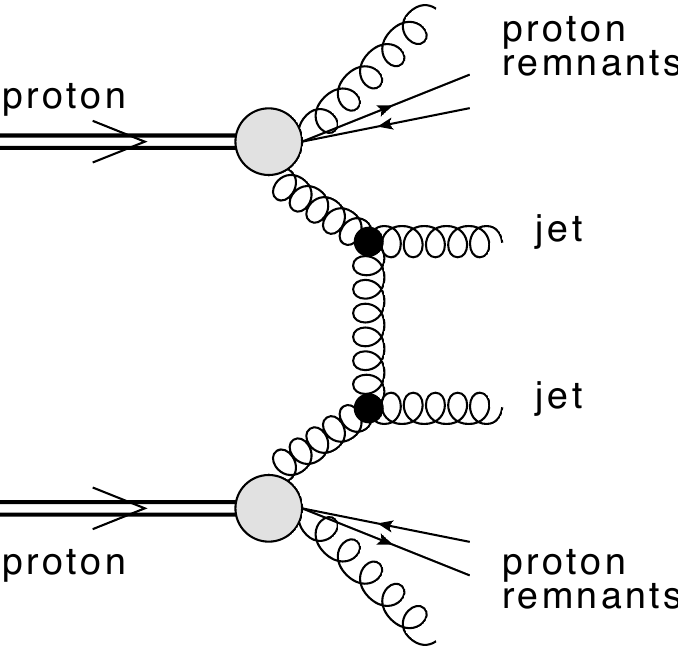}
\\[0.2cm]
\includegraphics[width=0.4\textwidth]{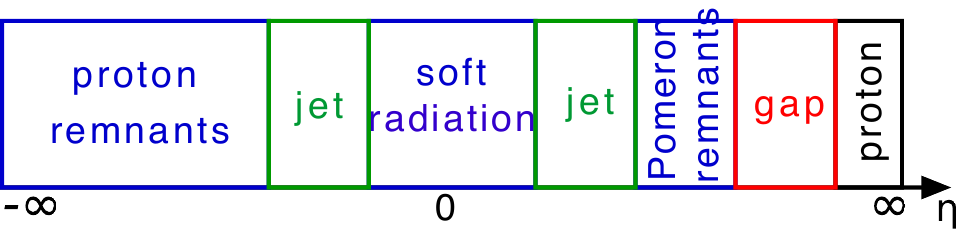}\hspace{0.1\textwidth}
\includegraphics[width=0.4\textwidth]{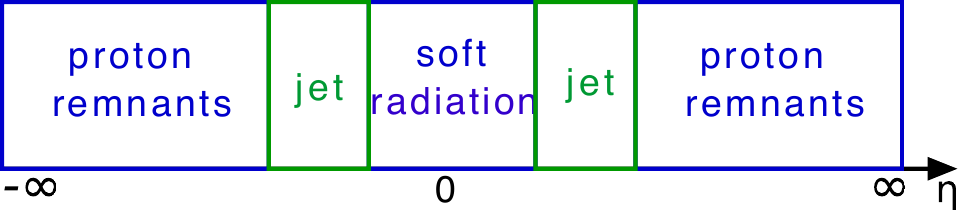}
\caption{\textbf{Left:} single diffractive jet production -- one interacting proton stays intact, the second one is destroyed and two jets are produced. \textbf{Right:} non-diffractive jet production -- the interacting protons are destroyed and two jets are produced.}
\label{ch4_SD_JJ_diag}
\end{figure}

In the single diffractive jet production process (Fig.~\ref{ch4_SD_JJ_diag} left) a jet system is produced in the central region and one of the protons emits a Pomeron, stays intact and is scattered at very high pseudorapidity. Depending on the momentum lost in the interaction, the intact proton may be detected by proton taggers. Unfortunately, not all such protons will survive. This is due to the additional soft interactions between the diffracted proton and the rest of the final state. Such effect will be hereafter quantified by the so called gap survival probability factor. For hard single diffractive processes at $\sqrt{s} = 13$ TeV such probability is estimated to be of about 0.1~\cite{KMR_surv}.

It is informative to compare the single diffractive jet production to the non-diffractive one (Fig.~\ref{ch4_SD_JJ_diag} right). In the latter process, both interacting protons are destroyed and two jets are produced; low-$p_T$ particles populate the pseudorapidity region between the two jets and the proton remnants.

By studying single diffractive jet production, the universality of the Pomeron in $ep$ and $pp$ collision can be probed~\cite{Royon_DPE_JJ}. Moreover, the gap survival probability can be quantified: a good experimental precision will allow for comparison to theoretical predictions and differential measurements of the dependence of the survival factor on (for example) the mass of the central system. Finally, the QCD evolution of the gluon and quark densities in the Pomeron can be tested and compared with the HERA measurements.

\begin{figure}[hbtp]
  \begin{center}
    \includegraphics[width=0.49\textwidth]{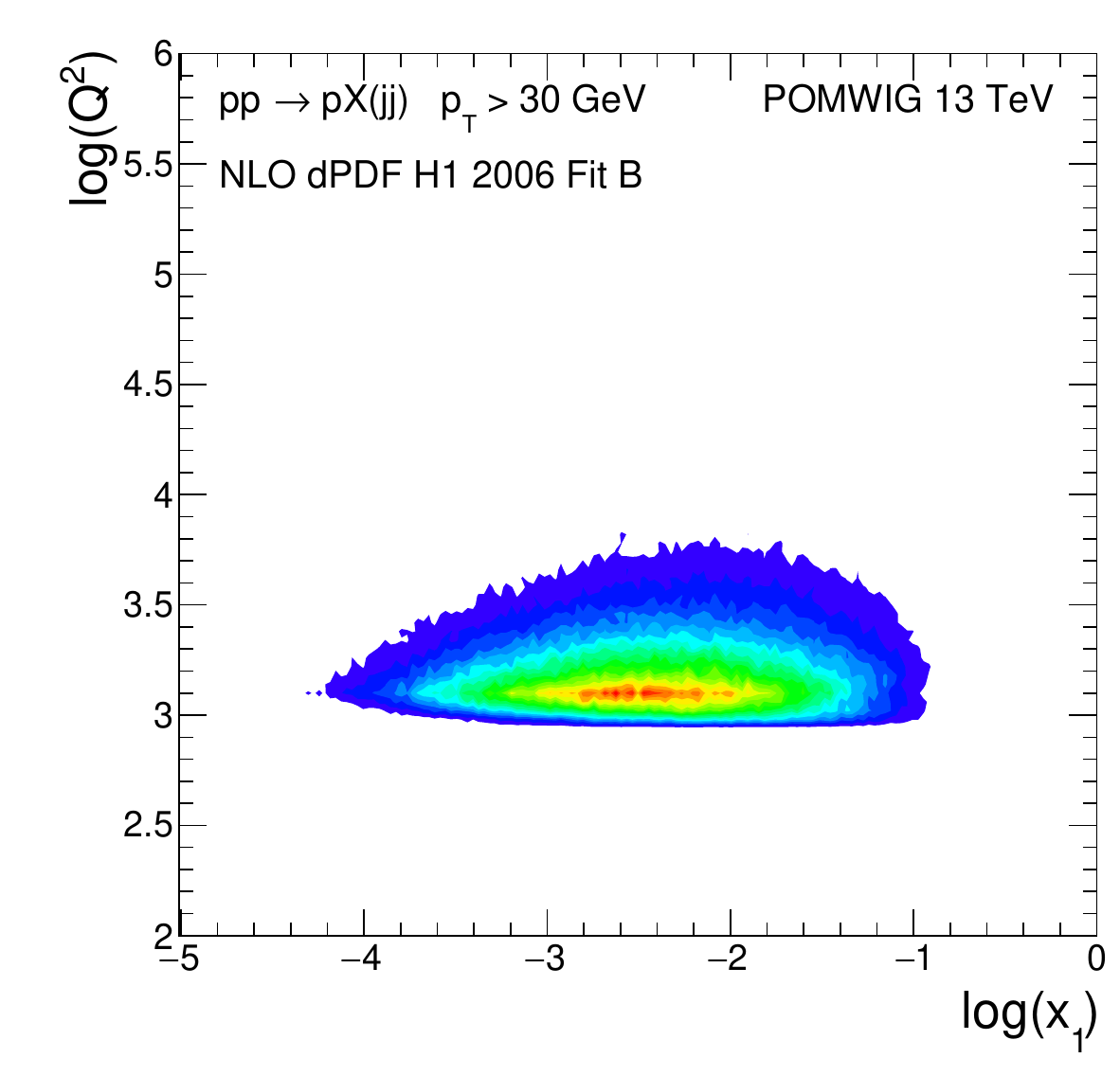}
    \includegraphics[width=0.49\textwidth]{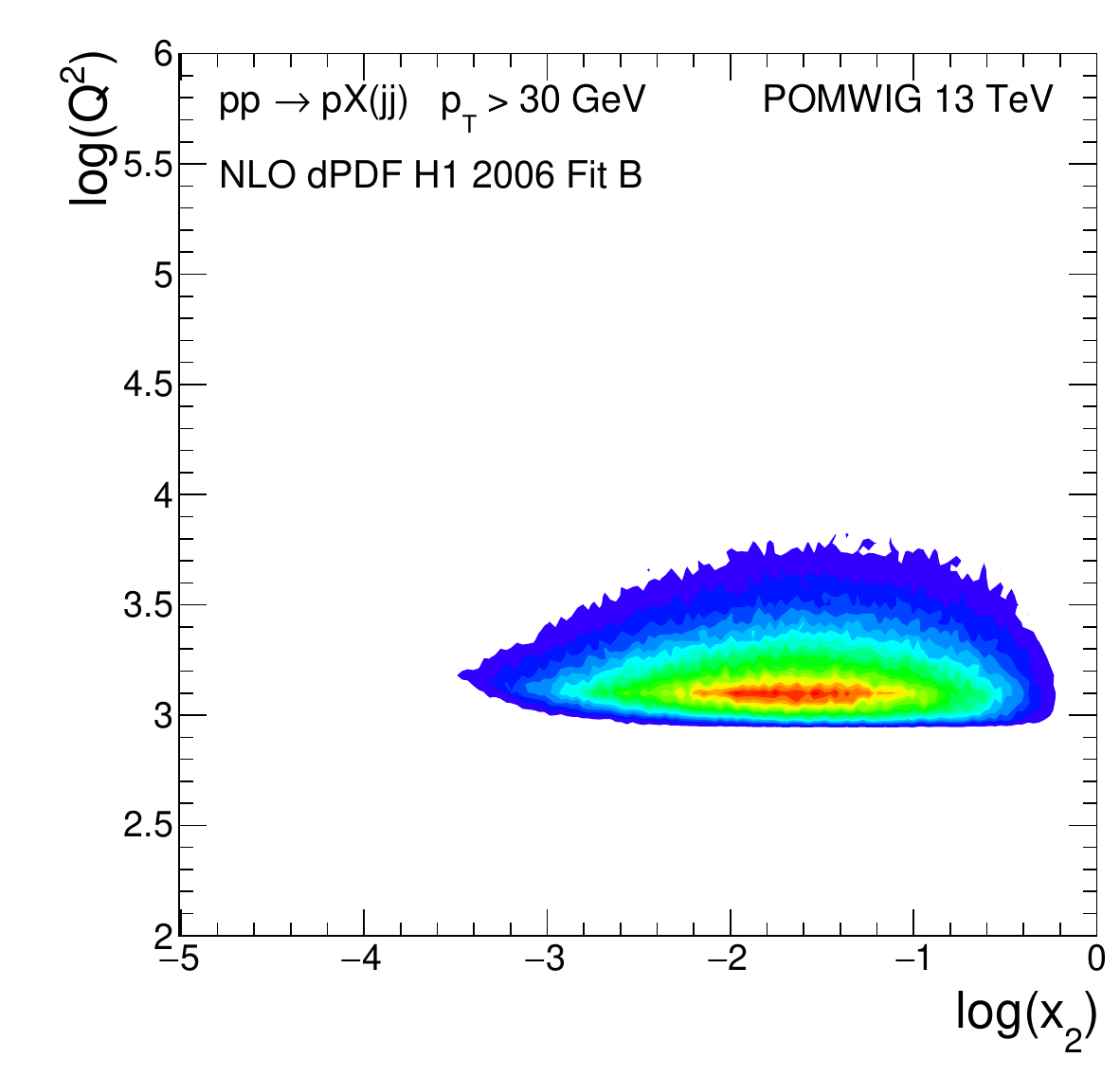}
    \caption{
    %Distribution of the hard scale squared $Q^2$ versus the momentum fraction of the parton initiating the hard scattering in single diffractive dijet production, $pp \rightarrow p{\rm jj}X$. Events were simulated with {\sc POMWIG}, version $2.0$ beta~\cite{Cox:2000jt}. The outgoing proton is scattered in the positive $z$ direction. 
    %The NLO H1 2006 Fit B~\cite{Aktas:2006hy} is used for the diffractive dPDF and the pomeron flux calculation. The following values were used: $\alpha_{\mathbb{P}}(0)=1.111$, $\alpha{\rm '}=0.06~{\rm GeV}^{-2}$ and $\beta_{\mathbb{P}}=5.5~{\rm GeV}^{-2}$. 
    %The scale is defined as $Q = \sqrt{2\,\hat{s}\,\hat{t}\,\hat{u}/(\hat{s}^2 + \hat{t}^2 + \hat{u}^2)}$ and corresponds roughly to the transverse momentum of the outgoing parton. 
    %Left: Momentum fraction of the incoming parton in the positive $z$ direction. Right: Momentum fraction of the incoming parton in the negative $z$ direction.
    Distribution of the hard scale $Q^2$ for single diffractive dijet production simulated with {\sc Pomwig} and the momentum fraction of the parton initiating the hard scattering in the outgoing proton direction (\textbf{left}), and in the opposite direction (\textbf{right}).
    }
    \label{fig:Q2_vs_x-SD-POMWIG}
  \end{center}
\end{figure}

\begin{figure}[hbtp]
  \begin{center}
    \includegraphics[width=0.49\textwidth]{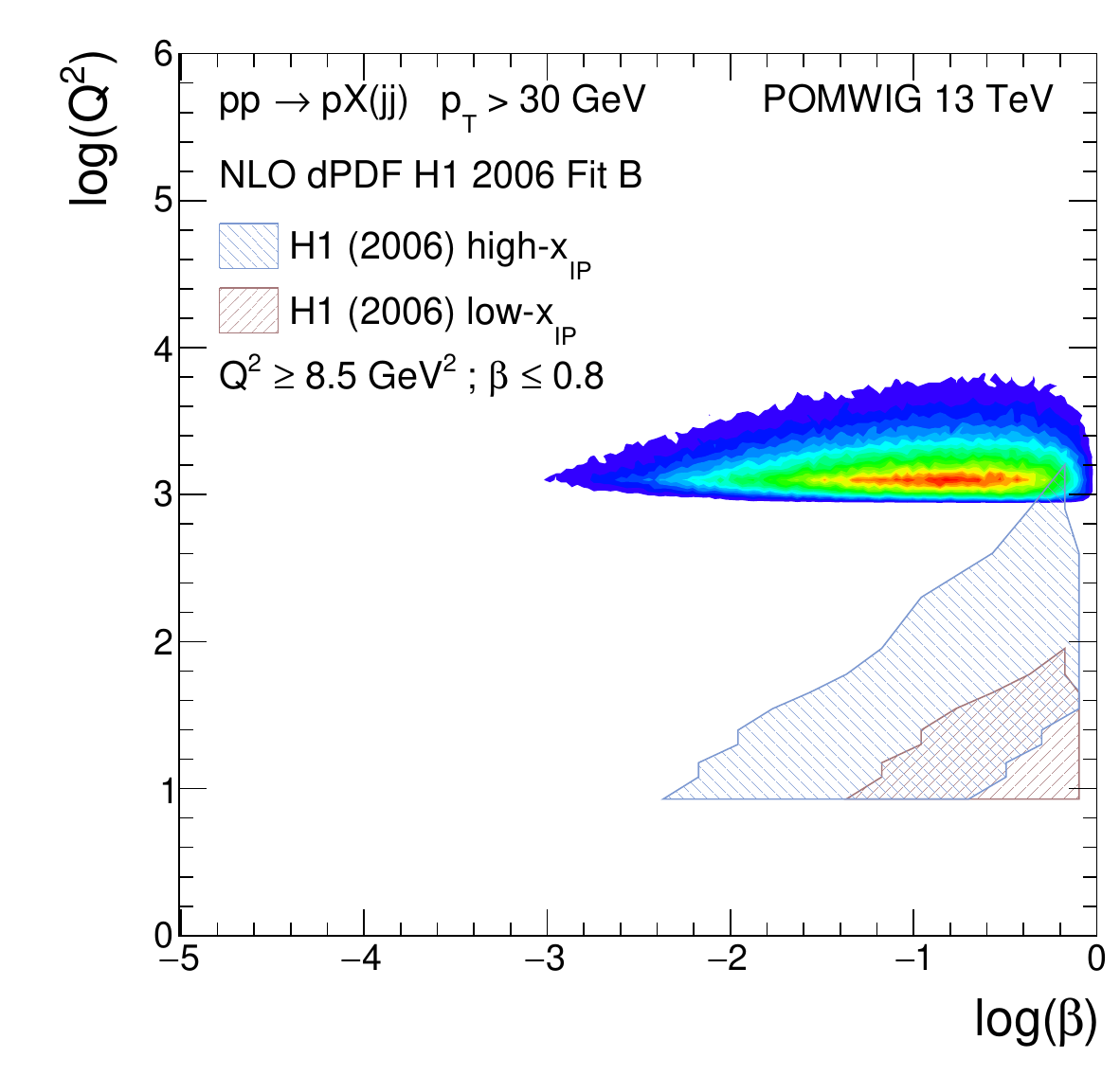}
    \caption{
    %Distribution of the hard scale squared $Q^2$ versus $\beta$ in single diffractive dijet production, $pp \rightarrow p{\rm jj}X$, where $\beta$ is the momentum fraction at which the diffractive PDF is probed. Hence, $x = \beta\,\xi$, where $\xi$ is the longitudinal momentum loss of the proton in such events and $x$ the momentum fraction of the parton initiating the hard scattering. Events were simulated with {\sc POMWIG}, version $2.0$ beta~\cite{Cox:2000jt}. 
    %The NLO H1 2006 Fit B~\cite{Aktas:2006hy} is used for the diffractive dPDF and the pomeron flux calculation. The following values were used: $\alpha_{\mathbb{P}}(0)=1.111$, $\alpha{\rm '}=0.06~{\rm GeV}^{-2}$ and $\beta_{\mathbb{P}}=5.5~{\rm GeV}^{-2}$. 
    %The scale is defined as $Q = \sqrt{2\,\hat{s}\,\hat{t}\,\hat{u}/(\hat{s}^2 + \hat{t}^2 + \hat{u}^2)}$ and corresponds roughly to the transverse momentum of the outgoing parton. 
    Distribution of the hard scale $Q^2$ and $\beta$, the momentum fraction at which the diffractive PDF is probed, for single diffractive dijet production simulated with {\sc Pomwig}. The region covered by the H1 data used to extract the NLO H1 2006 Fit B~\cite{H1LRG06_fit} is also shown.
    }
    \label{fig:Q2_vs_beta-SD-POMWIG}
  \end{center}
\end{figure}

It must be pointed out that going from the HERA to the LHC kinematics means extrapolating the diffractive parton distribution functions well beyond the region in which they have been measured. 
The HERA coverage in photon virtualities, $Q^2$, reaches typically several hundred GeV$^2$, at least one order of magnitude 
below that of the LHC data, where in single diffractive dijet production the scale corresponds roughly to the transverse momentum of the outgoing parton. 
Figure~\ref{fig:Q2_vs_x-SD-POMWIG} shows the distribution of $Q^2$ and the momentum fraction of the parton initiating the hard scattering in single diffractive dijet events. The simulation was performed with {\sc Pomwig}, version $2.0$ beta~\cite{pomwig}. The outgoing proton is 
scattered in the positive $z$ direction and the outgoing parton has transverse momentum $p_T > 30$~GeV. 
The NLO H1 2006 Fit B~\cite{H1LRG06_fit} was used for the diffractive PDF (DPDF) and the Pomeron flux calculation. This is one of several DPDF fits performed with the HERA data (see Sect.~\ref{ch4_sec_factorisation_tests}). 
Figure~\ref{fig:Q2_vs_beta-SD-POMWIG} shows the distribution of $Q^2$ and $\beta$, the fractional momentum of the diffractive exchange carried by the struck quark. The coverage in $\beta$ extends at the LHC down to $10^{-3}$, below that of the HERA data. One may note that $x= \beta\xi$, where $\xi$ (or $x_{I\!P}$) is the longitudinal momentum loss of the proton in such events and $x$ the momentum fraction of the parton initiating the hard scattering. 
Also shown is the HERA measurement region used to extract the diffractive parton distribution functions in Ref.~\cite{H1LRG06_fit}; the regions corresponding to values of $x_{I\!P}$ from $0.0003$ to $0.003$ (low-$x_{I\!P}$) and from $0.01$ to $0.03$ (high-$x_{I\!P}$) are shown separately. The $Q^2$ values range from $Q^2 = 8.5$~GeV$^2$ up to a maximum of $Q^2 = 1600$~GeV$^2$, while $\beta$ has a maximum value of $0.8$. 
It should be noted that decreasing the transverse momentum requirement to values of $20$~GeV or lower could substantially reduce the extrapolation with respect to the HERA measurement region.

\subsection{ATLAS Feasibility Studies for $\sqrt{s} = 13$ TeV}
\label{ch4_sec_SD_JJ_ATLAS}
In the following studies the FPMC generator~\cite{FPMC} was used to generate diffractive jet samples. Non-diffractive jets were generated by \textsc{Pythia8}~\cite{Pythia8}. Pile-up was generated using \textsc{Pythia8} with the MBR tune~\cite{MBR} and the following processes were included: non-diffractive production, elastic scattering, single diffraction, double diffraction and central diffraction. The vertex position was smeared according to values from tables in Section~\ref{sec:soft:forwardacceptance}. 

In order to calculate the proton transport through the LHC structures between the ATLAS Interaction Point (IP) and the forward detectors, the \textsc{FPTrack}~\cite{FPTrack} program was used. For a given distance between the forward detector and the beam, diffractive protons were checked to be within the detector acceptance. The proton energy was reconstructed based on the procedure described in~\cite{forward_unfolding}.

Jets were reconstructed using the \textsc{FastJet} package with the anti-$k_T$ ($R = 0.4$) algorithm~\cite{FastJet}. A particle was considered visible in the tracker if the criteria described in Section~\ref{ch4_sec_backgrounds} were fulfilled. Three thresholds for the transverse momentum of the leading jet were considered: $p_T^{\rm jet1}$ of 20~GeV, 50~GeV and 100~GeV. The sub-leading jet was required to have $p_T^{\rm jet2} > 20$ GeV. In order to assure reconstruction of a hard vertex, both leading jets were required to be inside the acceptance of the ATLAS tracker ($|\eta| < 2.5$).

\begin{figure}[!htbp]
\centering
\includegraphics[width=0.49\textwidth]{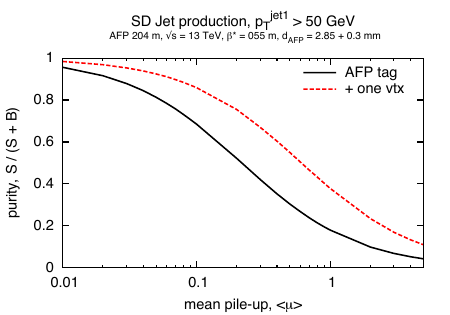}\hfill
\includegraphics[width=0.49\textwidth]{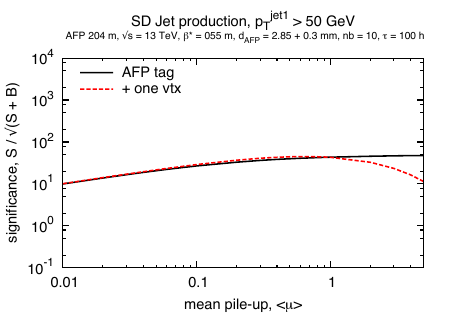}
\caption{Single diffractive jet production with protons tagged in the AFP detectors for $\sqrt{s} = 13$ TeV and $\beta^* = 0.55$~m: purity (\textbf{left}) and significance (\textbf{right}) for jets with $p_T > 50$ GeV  as a function of average pile-up. The number of bunches multiplied by data collecting time (in hours) was assumed to be 1000. In the analysis the following cuts were considered: proton tag in the AFP detector (black solid line) and tag + exactly one reconstructed vertex (red dashed line).}
\label{ch4_SD_JJ_AFP_beta055}
\end{figure}

The results for the AFP detector and $\beta^* = 0.55$~m optics are shown in Fig.~\ref{ch4_SD_JJ_AFP_beta055}. In the left plot the purity, hereafter defined as the ratio of the signal to the sum of signal and background events, is presented as a function of the mean pile-up, $\mu$. In this figure, the black solid line is for events with proton tag in the AFP detector whereas the red dashed line is for those with a tag and exactly one reconstructed vertex. Purity greater than 50\% was obtained for $\mu \sim 0.5$. Moreover, it grows rapidly to values greater than 80\% for $\mu < 0.1$. This plot was done for jets with $p_T > 50$~GeV,  but the purity is not significantly different for the other considered $p_T$ thresholds (\textit{cf.} Tab.~\ref{ch4_tab_SD_JJ}).

In Fig.~\ref{ch4_SD_JJ_AFP_beta055} (right) the statistical significance, hereafter defined as the number of collected signal events over the square root of the sum of the accepted signal and background events, is presented as a function of the mean pile-up for jets with $p_T > 50$~GeV. To compute the statistical significance, a certain integrated luminosity must be assumed. For the presented results this was done by setting the number of bunches ($n_b$) multiplied by the data collecting time ($t_{data}$, in hours) to be $n_b \cdot t_{data} = 1000$. This can be interpreted as one hour of data-taking with 1000 bunches, or 10 hours with 100 bunches, (\textit{cf.} Fig.~\ref{fig_lumi}).

\begin{figure}[!htbp]
\centering
\includegraphics[width=0.49\textwidth]{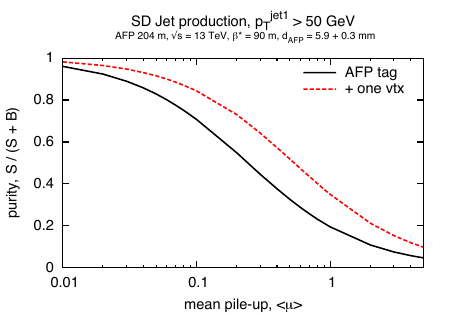}\hfill
\includegraphics[width=0.49\textwidth]{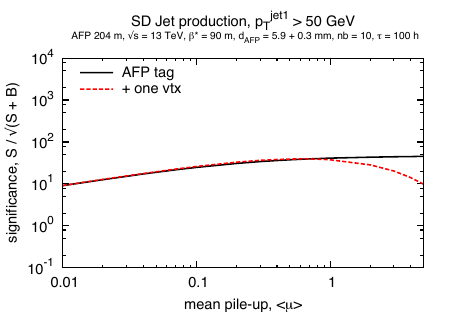}
\caption{Single diffractive jet production with protons tagged in the AFP detectors for $\sqrt{s} = 13$ TeV and $\beta^* = 90$~m: purity (\textbf{left}) and significance (\textbf{right}) for jets with $p_T > 50$ GeV as a function of average pile-up. The number of bunches multiplied by the data collecting time (in hours) was assumed to be 1000. In the analysis the following cuts were considered: proton tag in the AFP detector (black solid line) and tag + exactly one reconstructed vertex (red dashed line).}
\label{ch4_SD_JJ_AFP_beta90}
\end{figure}

\begin{figure}[!htbp]
\centering
\includegraphics[width=0.49\textwidth]{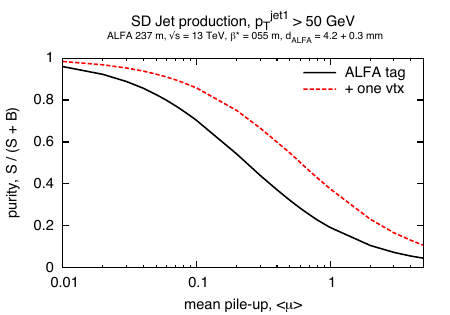}\hfill
\includegraphics[width=0.49\textwidth]{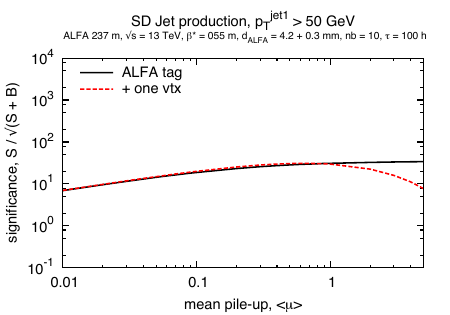}
\caption{Single diffractive jet production with protons tagged in the ALFA detectors for $\sqrt{s} = 13$ TeV and $\beta^* = 0.55$~m: purity (\textbf{left}) and significance (\textbf{right}) for jets with $p_T > 50$ GeV  as a function of average pile-up. The number of bunches multiplied by the data collecting time (in hours) was assumed to be 1000. In the analysis the following cuts were considered: proton tag in the ALFA detector (black solid line) and tag + exactly one reconstructed vertex (red dashed line).}
\label{ch4_SD_JJ_ALFA_beta055}
\end{figure}

Studies for the AFP detector and $\beta^* = 90$~m optics are shown in Fig.~\ref{ch4_SD_JJ_AFP_beta90}. For such an optics configuration and for a 10 $\sigma$ distance from the beam, the purity and significance are similar to the case discussed above. The same conclusions were obtained in the case of the ALFA detector and $\beta^* = 0.55$~m optics (\textit{cf.} Fig.~\ref{ch4_SD_JJ_ALFA_beta055}).

\begin{figure}[!htbp]
\centering
\includegraphics[width=0.49\textwidth]{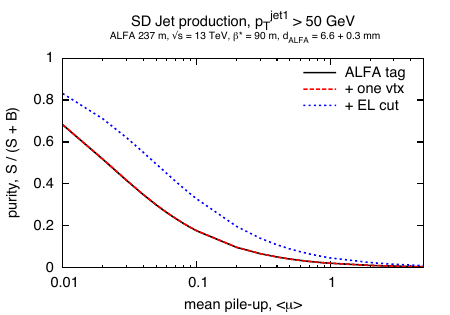}\hfill
\includegraphics[width=0.49\textwidth]{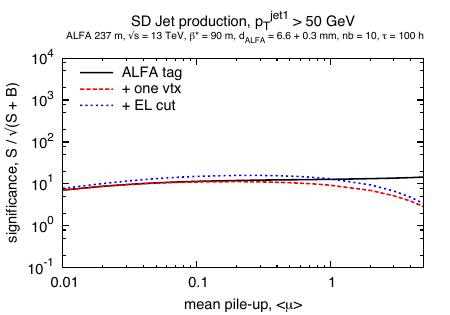}
\caption{Single diffractive jet production with protons tagged in the ALFA detectors for $\sqrt{s} = 13$ TeV and $\beta^* = 90$~m: purity (\textbf{left}) and significance (\textbf{right}) for jets with $p_T > 50$ GeV  as a function of average pile-up. The number of bunches multiplied by the data collecting time (in hours) was assumed to be 1000. In the analysis the following cuts were considered: proton tag in the ALFA detector (black solid line), tag + exactly one reconstructed vertex (red dashed line) and tag + one vertex + elastic veto (blue dotted line).}
\label{ch4_SD_JJ_ALFA_beta90}
\end{figure}

The situation changes dramatically when ALFA detectors and $\beta^{*} = 90$~m optics are considered (see Fig.~\ref{ch4_SD_JJ_ALFA_beta90}). Due to the acceptance for the elastic scattering, the purity is only higher than 50\% for the mean pile-up smaller than 0.02. Filtering out the elastic events (blue dotted line) increases the purity. However, an average pile-up of less than 0.05 is still needed. In conclusion, the optimal data taking conditions for such configuration are for $\mu \sim 0.01$.

The summary of these feasibility studies is presented in Table~\ref{ch4_tab_SD_JJ}. The rate was calculated for 100 bunches.

\section{Single Diffractive Z, W and J/$\Psi$ Production}
\label{ch4_sec_SD_WZ}
\begin{figure}[!htbp]
  \begin{center}
    \includegraphics[width=7.0cm]{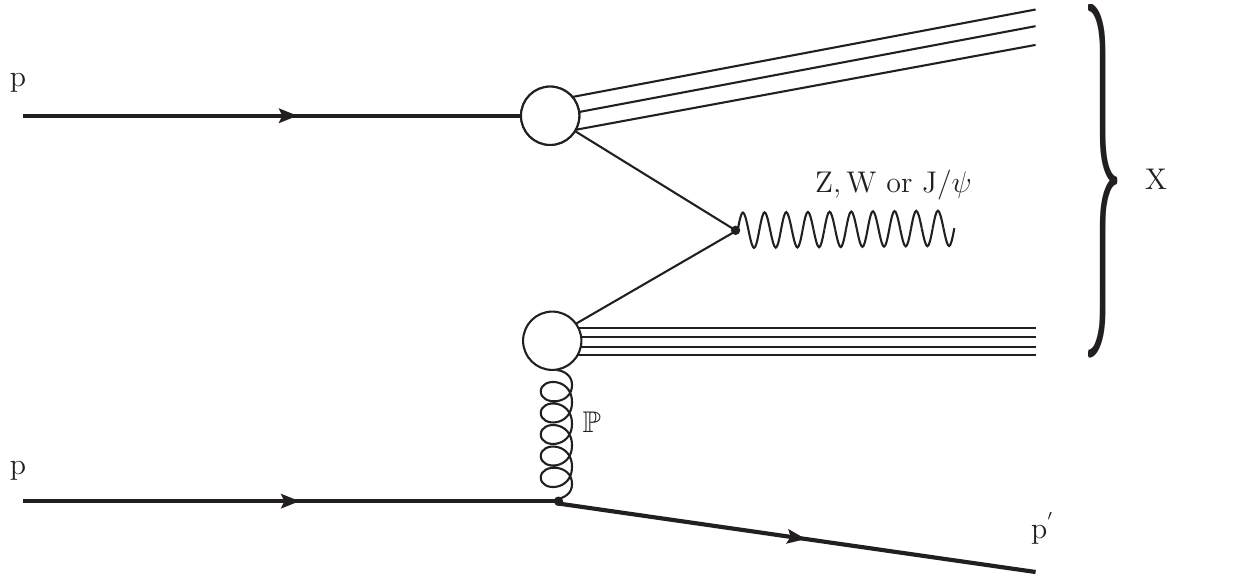}
    \caption{Single diffractive Z, W and ${\rm J}/\psi$ production.}
    \label{ch4_SDWZ_diag}
  \end{center}
\end{figure}

The leading order diagram for single-diffractive Z, W boson, or ${\rm J}/\psi$ meson production is shown in Fig.~\ref{ch4_SDWZ_diag}. The two final-state particles originating on either side of the colour-singlet are, in general, well separated by a rapidity gap.

The single-diffractive production of $W/Z$ bosons and $J/\psi$ meson is sensitive to the diffractive structure function of the proton, notably to its quark component, since many of the observed production modes can originate from quark fusion. Moreover, as in case of the single diffractive dijet production, the diffractive boson production can help quantifying the gap survival probability. 

\subsection{CMS-TOTEM Feasibility Studies for $\sqrt{s} = 13$ TeV}
\label{ch4_sec_SD_WZ_CMS}
The results presented in this Section are based on a Monte Carlo study presented in Ref.~\cite{CMS-PAS-FSQ-14-001}. These studies were done for the low-luminosity, low-pile-up LHC runs and illustrate the potential for physics measurements with the CMS and TOTEM experiments at the beginning of Run II.

\subsubsection{Signal and Background Simulation}
\label{ch4_sec_SD_WZ_CMS_simulation}
Single-diffractive Z and W boson production was simulated with {\sc Pomwig}~\cite{pomwig} in the electron and muon decay channels. Single-diffractive J$/\psi$ production was generated with {\sc Pompyt}~\cite{pompyt}. Pile-up events were simulated using the {\sc Pythia}8 Monte Carlo event generator~\cite{Pythia8} with A2 tune. Pile-up events were added to the signal with a probability $p(n;\mu)$, where $n$ is the number of pile-up events given by a Poisson distribution with an average of $\mu = 1$.

The presented predictions include a rapidity gap survival probability of 0.1, which provides a good description of CMS diffractive dijet data~\cite{Chatrchyan:2012vc}. According to {\sc Pomwig}, the cross-section for the lepton decay of the $Z$ boson is $12.1$~pb. For the $W$ boson decaying into lepton and (anti)neutrino, the predicted cross-section is of about $131$~pb. The {\sc Pompyt} cross-section for diffractive J$/\psi \rightarrow \mu\mu$ production is $2.5$~nb.

\subsubsection{Trigger Strategy}
The trigger strategy in Run II will be similar to the one in Run I: the signal accepted by CMS (TOTEM) will be sent to TOTEM (CMS) to trigger the readout. In Run I, with 112 bunches and a pile-up of $\mu \sim0.07$, the detector trigger selections and corresponding rates were:
\begin{itemize}
  \item for single-diffractive Z or W selection: 
  \begin{itemize}
    \item at least one muon (electron) with $p_{\rm T}> 7 $~GeV ($57$~Hz) or
    \item at least two muons (electrons) with $p_{\rm T} > 3$~GeV ($22$~Hz),
  \end{itemize}
  \item for single-diffractive J$/\psi$ selection at least two muons with non-zero $p_T$ and $\vert\eta\vert < 2.45$ ($45$~Hz).
\end{itemize}

\subsubsection{Event Selection}
\label{ch4_sec_SD_WZ_CMS_selection}
The simulation and reconstruction software used in this study did not include the description of the forward proton detectors. Instead, an acceptance table was used to quantify the probability that a proton is measured. This table was determined on the basis of a parametrisation of the proton propagation in the LHC beam line~\cite{totemsoftware}.

The Z or W boson and J$/\psi$ meson final states were selected using the central CMS detector in the range $|\eta| \ < \ 2.5$. The detailed MC simulation of the CMS detector response was based on GEANT4~\cite{geant}. All events were required to be within the acceptance of the TOTEM Roman pots on either side of the interaction point and to have exactly one reconstructed vertex.

\noindent For the specific samples the following additional criteria were applied:
\begin{itemize}
  \item ${\rm Z}\rightarrow {\rm e}^{+}{\rm e}^{-}$ and ${\rm Z}\rightarrow {\mu}^{+}{\mu}^{-}$:
  \begin{itemize}
  \item both leading leptons were required to have $p_{\rm T} > 20$~GeV and to fulfil the isolation criteria described in Ref.~\cite{anboson};
  \item the dilepton system invariant mass was required to be within the range of $60 < {\rm M}_{ll} < 110~{\rm GeV}$;
  \end{itemize}
  \item ${\rm W}^{\pm}\rightarrow {\rm e}^{\pm}{\nu}_{e}$ and ${\rm W}^{\pm}\rightarrow {\mu}^{\pm}{\nu}_{\mu}$:
  \begin{itemize}
    \item the leading lepton was required to have $p_{\rm T} > 20$~GeV and to fulfil the isolation criteria~\cite{anboson};
    \item events with an additional lepton with $p_{\rm T} > 10$~GeV were rejected;
    \item the transverse mass of the lepton-neutrino system, ${\rm M}_{\rm T} = \sqrt{2{\rm E}_{\rm T,\rm l}\cdot{\rm E}_{{\rm T},\nu}[1-\cos(\phi_{l}-\phi_{\nu})]}$, was required to be in the range $60 < {\rm M}_{\rm T} < 110~{\rm GeV}$;
  \end{itemize}
  \item J$/\psi \rightarrow \mu^{+}\mu^{-}$:
  \begin{itemize}
    \item at least two muons were required with opposite charge, with $|\eta|<2.45$;
    \item the dimuon system invariant mass was required to be in the range $3.05 < {\rm M}_{\mu^{-}\mu^{+}} < 3.15~{\rm GeV}$.
   \end{itemize}
\end{itemize}

\begin{figure}[!htbp]
  \begin{center}
    \includegraphics[width=7.0cm]{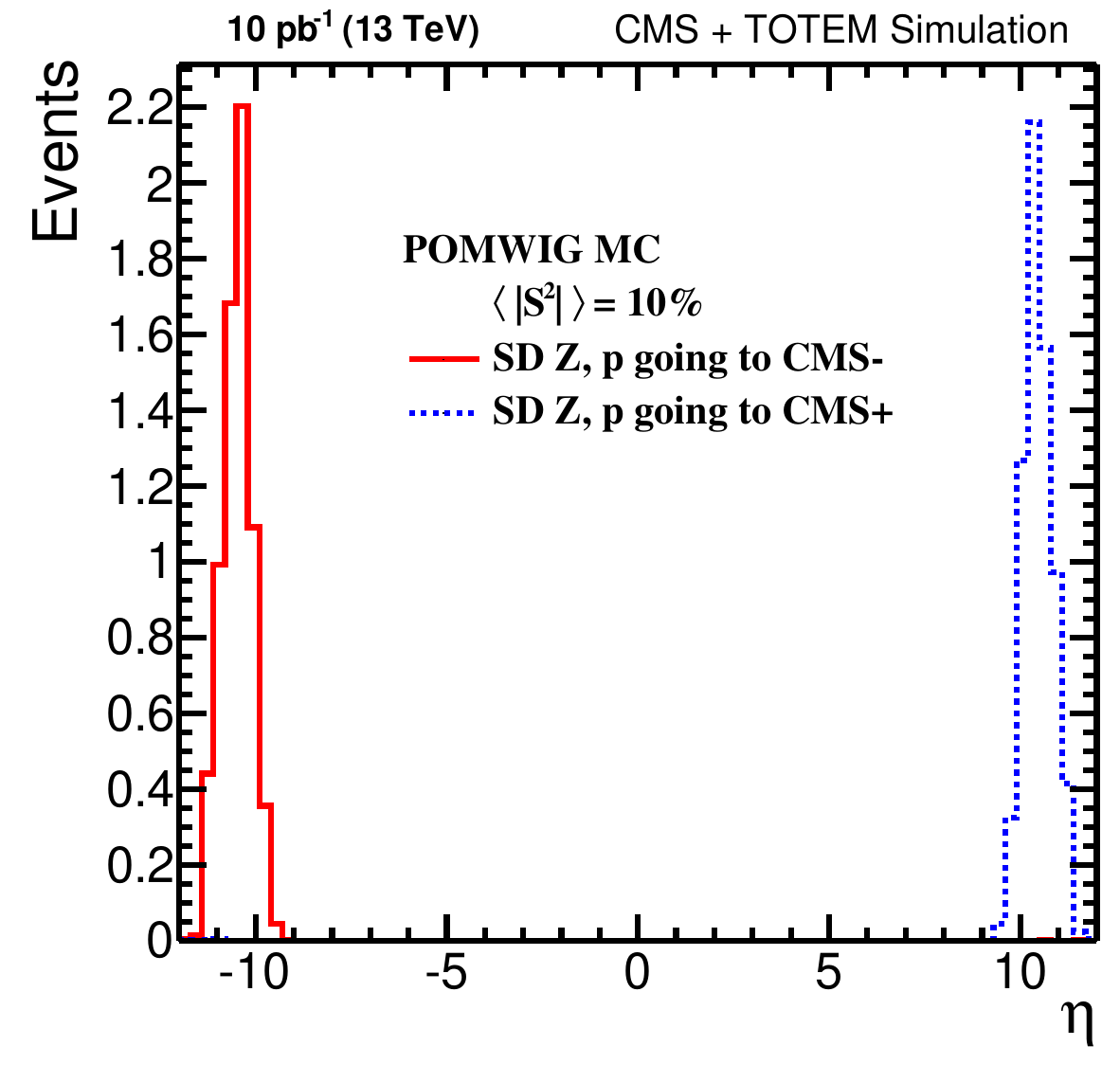}
    \includegraphics[width=7.0cm]{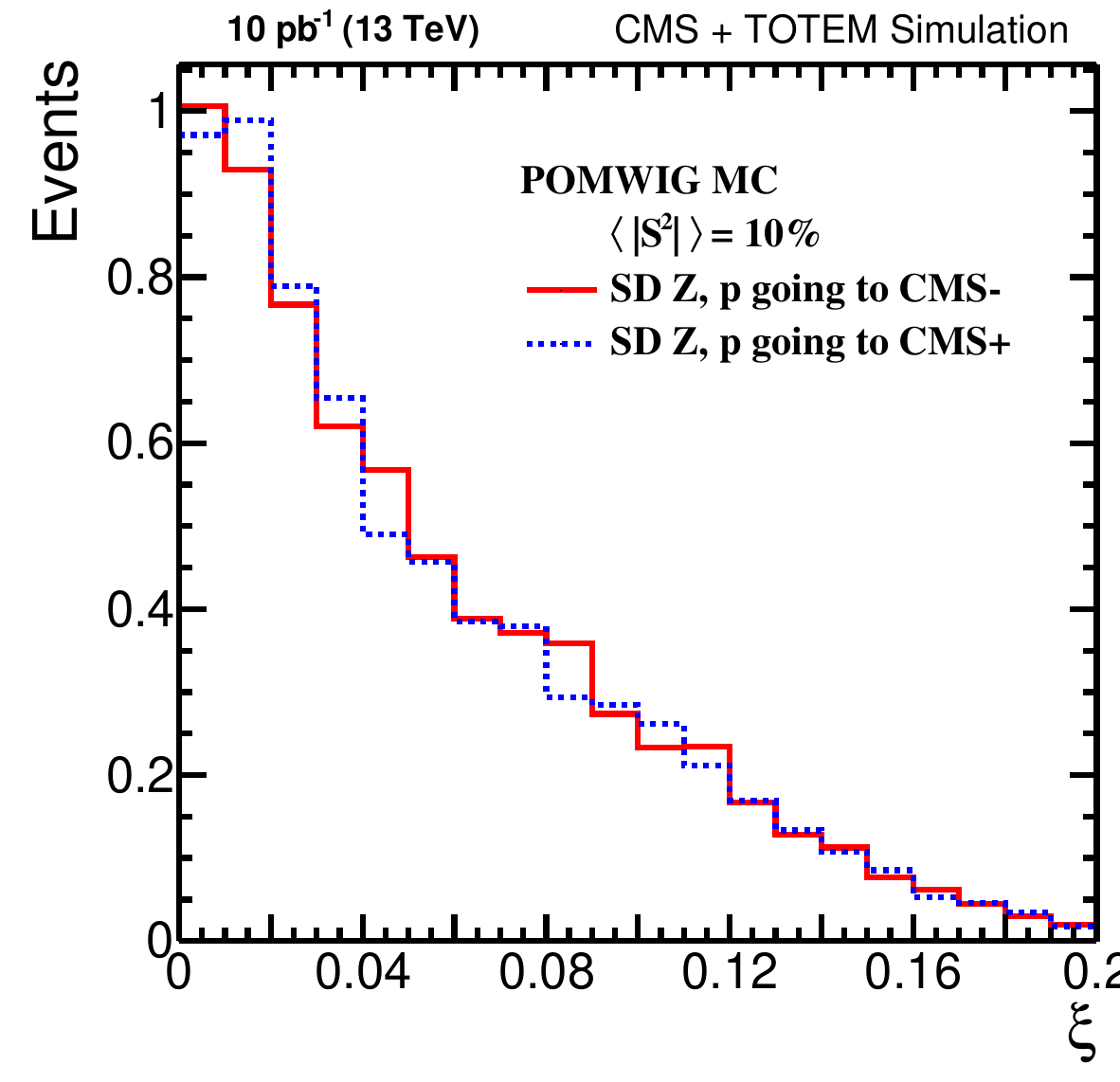}
    \caption{Distributions of the pseudorapidity $\eta$ (\textbf{left}) and of $\xi$ (\textbf{right}) of the protons tagged in the TOTEM RP detector stations in ${\rm Z}\rightarrow {\rm e}^{+}{\rm e}^{-}$ events. Outgoing protons in the CMS $z$-negative direction are shown in red (solid line) and protons in the $z$-positive direction in blue (dashed line). Events were simulated with the {\sc Pomwig} MC and normalized to an integrated luminosity of $10~{\rm pb}^{-1}$. A gap survival probability of 0.1 was used.}
    \label{ch4_sec_SD_WZ_CMS_zeproton}
  \end{center}
\end{figure}

\begin{figure}[!htbp]
  \begin{center}
    \includegraphics[width=7.0cm]{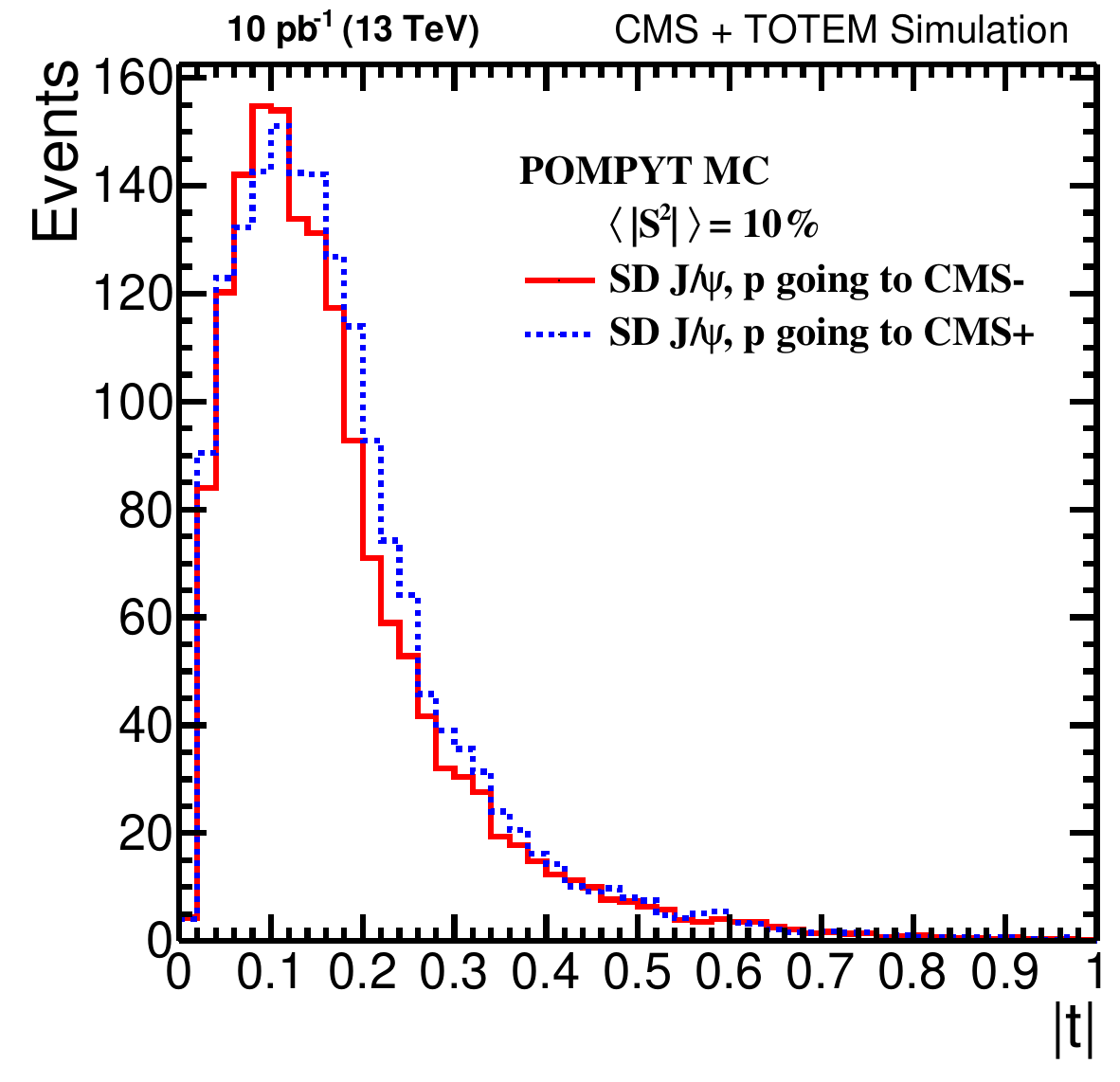}
    \includegraphics[width=7.0cm]{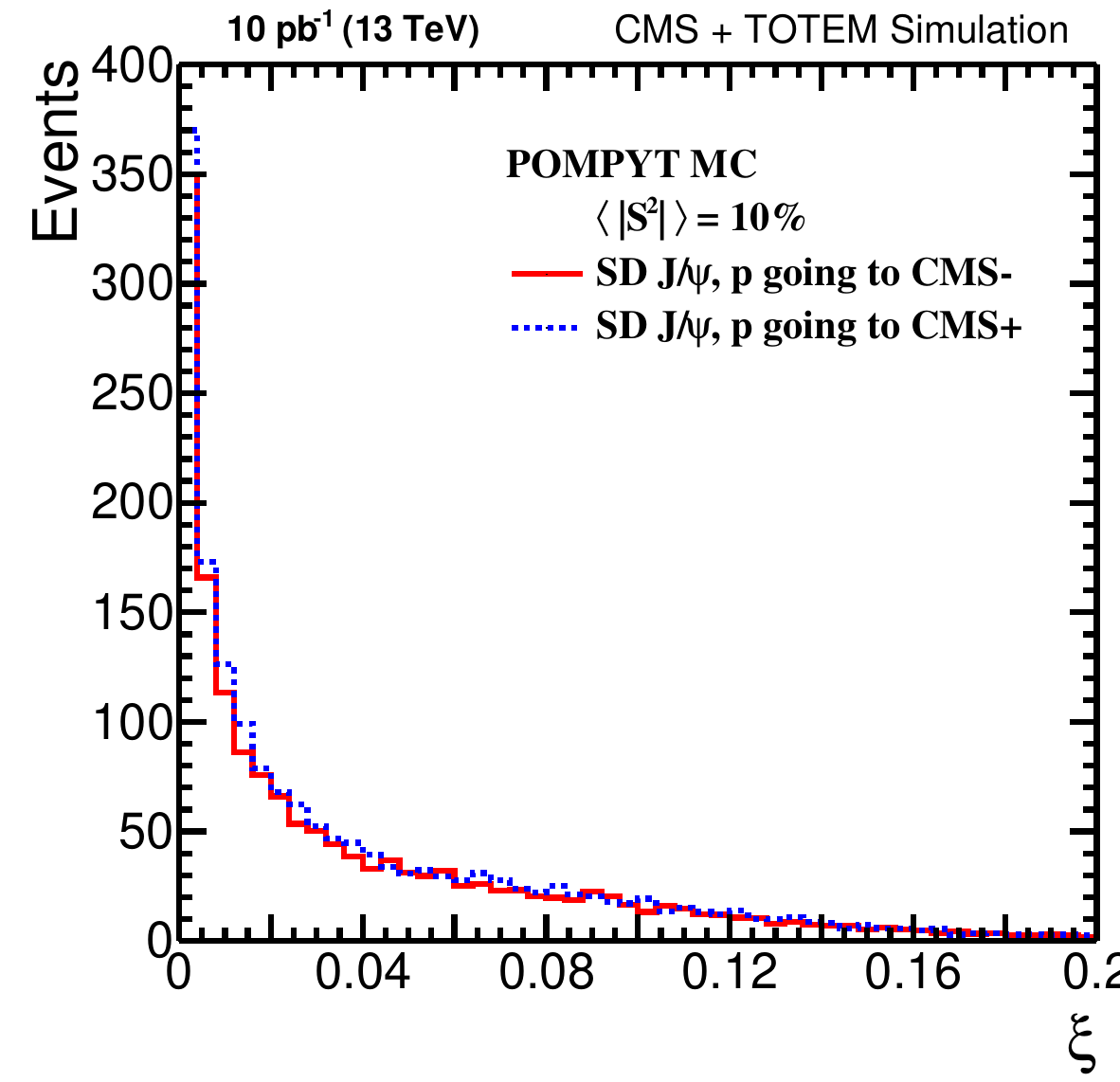}
    \caption{Distributions of $t$ (\textbf{left}) and $\xi$ (\textbf{right}) of protons tagged in the TOTEM RP detector stations in J$/\psi (\mu^{+}\mu^{-})$ events. Outgoing protons in the CMS $z$-negative direction are shown in red (solid line) and protons in the CMS $z$-positive direction in blue (dashed line). Events were simulated with the {\sc Pompyt} MC and normalized to an integrated luminosity of $10~{\rm pb}^{-1}$. A gap survival probability of 0.1 was used.}
    \label{ch4_sec_SD_WZ_CMS_jproton}
  \end{center}
\end{figure}

\begin{figure}[!htbp]
  \begin{center}
   \includegraphics[width=7.0cm]{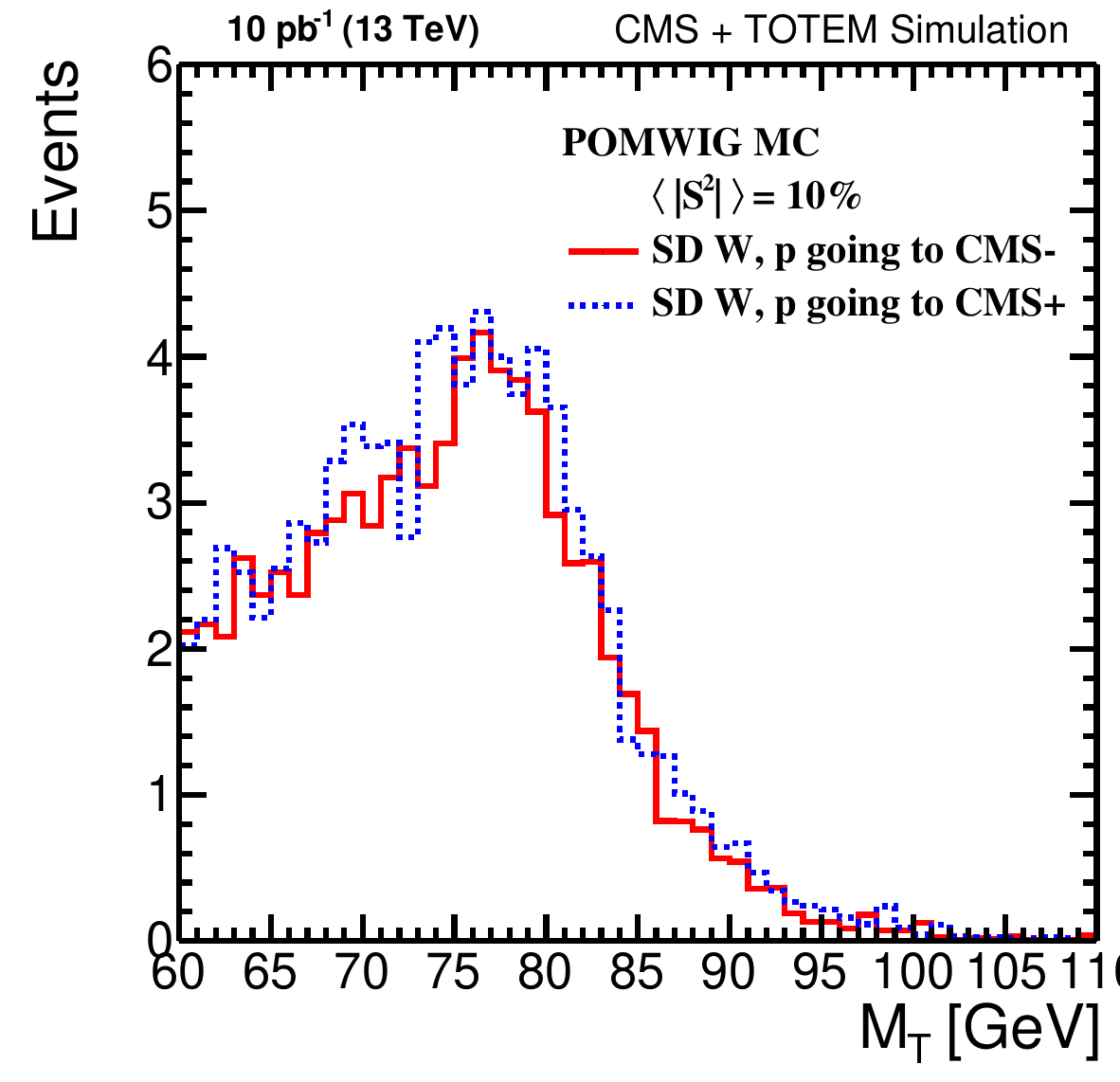}
    \includegraphics[width=7.0cm]{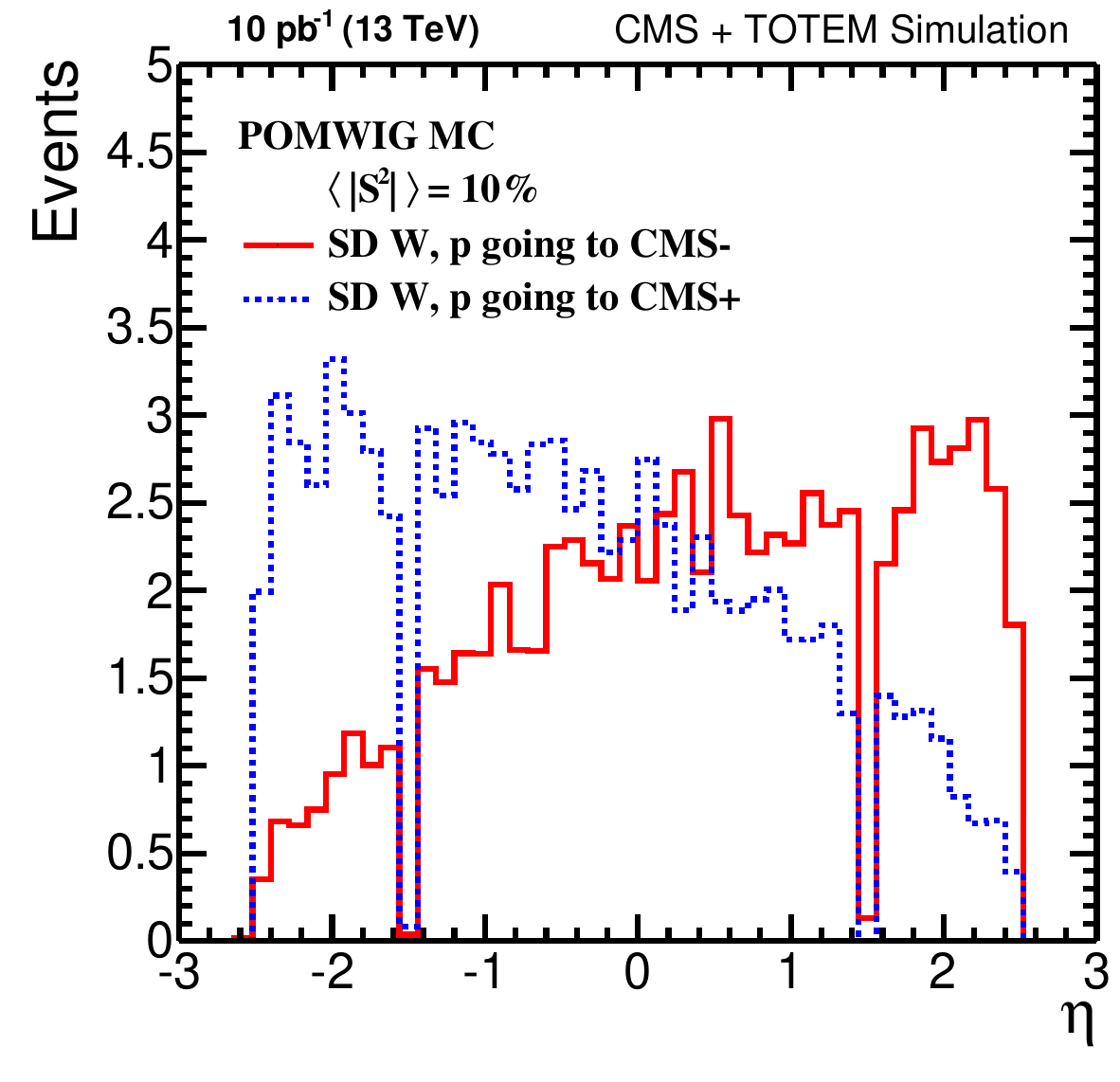}
    \caption{Distributions of the transverse mass ${\rm M}_{\rm T}$ (\textbf{left}) and the leading lepton pseudorapidity (\textbf{right}) in ${\rm W}^{\pm}\rightarrow {\rm e}^{\pm}{\nu}_{e}$ events. Events with a proton detected by TOTEM RP detector stations in the CMS $z$-negative (positive) direction are shown in solid red (dashed blue) line. Events were simulated with the {\sc Pomwig} MC and normalized to an integrated luminosity of $10~{\rm pb}^{-1}$. A gap survival probability of 0.1 was used.}
    \label{ch4_sec_SD_WZ_CMS_wekin}
  \end{center}
\end{figure}

The distributions of pseudorapidity, $\eta$, and relative energy loss, $\xi$, of the protons tagged in the forward detectors in ${\rm Z}\rightarrow {\rm e}^{+}{\rm e}^{-}$ events are shown in Fig.~\ref{ch4_sec_SD_WZ_CMS_zeproton}. Figure~\ref{ch4_sec_SD_WZ_CMS_jproton} shows the distributions of $t$ and $\xi$ of protons tagged in J$/\psi \rightarrow \mu^{+}\mu^{-}$ events. Figure~\ref{ch4_sec_SD_WZ_CMS_wekin} shows the distributions of the transverse mass and the leading lepton pseudorapidity in ${\rm W}^{\pm}\rightarrow {\rm e}^{\pm}{\nu}_{e}$ events. 

\subsubsection{Results}
\label{ch4_sec_SD_WZ_CMS_results}

Table~\ref{sigmatable} summarizes the results of the visible cross-section ($\sigma_{\rm vis}$) for the electron and muon channels for the considered production channels. The obtained visible cross-section summed over all considered channels are:
\begin{itemize}
  \item $3.38 \pm0.03~{\rm pb}$ for SD Z,
  \item $36.7 \pm0.3~{\rm pb}$ for SD W,
  \item $332.5 \pm 2.9~{\rm pb}$ for J$/\psi$.
\end{itemize}

Table~\ref{eventnumbers} gives an overview of the expected event yields assuming an integrated luminosity of $10~{\rm pb}^{-1}$. The results shown are further corrected by the TOTEM proton reconstruction efficiency of $92.5 \pm 2.5\%$~\cite{totemreco}.

\begin{table*}[!htbp]
\begin{center}
\caption{Overview of the visible cross-section values obtained in the SD ${\rm Z}\rightarrow {\rm e}^{+}{\rm e}^{-}$, ${\rm Z}\rightarrow {\mu}^{+}{\mu}^{-}$, ${\rm W}\rightarrow {\rm e}{\nu}_{e}$, ${\rm W}\rightarrow {\mu}{\nu}_{\mu}$ and J$/\psi \rightarrow \mu^{+}\mu^{-}$ production channels, shown for events with a proton detected in the CMS $z$-negative or $z$-positive directions. The uncertainties shown are statistical.
\label{sigmatable}}
\vspace{0.3cm}
\begin{tabular}{cccccc}
\hline
 & {}  ${\rm Z}\rightarrow {\rm e}^{+}{\rm e}^{-}$ & ${\rm Z}\rightarrow {\mu}^{+}{\mu}^{-}$ & ${\rm W}^{+}\rightarrow {\rm e}^{+}{\nu}_{e}$ & ${\rm W}^{+}\rightarrow {\mu}^{+}{\nu}_{\mu}$ & J$/\psi(\mu^{+}\mu^{-})$ \\
                          &  &  & ${\rm W}^{-}\rightarrow {\rm e}^{-}\bar{\nu}_{e}$ & ${\rm W}^{-}\rightarrow {\mu}^{-}\bar{\nu}_{\mu}$ & \\
\hline
$\sigma_{\rm vis}$~[pb] & $1.34\pm0.02$ & $2.04\pm0.02$ & $16.37\pm0.21$ & $20.30\pm0.23$ & $332.5\pm2.9$\\ \hline
\end{tabular}
\end{center}
\end{table*}

\begin{table*}[!htbp]
\begin{center}
\caption{
Overview of the expected event yields with the statistical uncertainty, for an integrated luminosity of $10~{\rm pb}^{-1}$ in the SD Z or W and J$/\psi$ production channels.
\label{eventnumbers}}
\vspace{0.3cm}
\begin{tabular}{cccc}
  \hline
   LHC Scenario & SD Boson Z & SD Boson W & SD J$/\psi$ \\
\hline
  $10~{\rm pb}^{-1}$ & $30 \pm 1$ & $340 \pm 10$ & $3080 \pm 90$\\
\hline
\end{tabular}
\end{center}
\end{table*}

\subsection{ATLAS Feasibility Studies for $\sqrt{s} = 13$ TeV}
\label{ch4_sec_SD_WZ_ATLAS}
In this Section the possibility of observing single diffractive W and Z events using proton tagging technique and the ATLAS detector is discussed. The single diffractive $W/Z$ events were generated by FPMC with a gap survival factor of 0.1~\cite{KMR_surv1}, whereas the non-diffractive samples were obtained using \textsc{Pythia8}. The visible cross-section, distance between forward detector and LHC beam and probability of having a minimum-bias tag were taken analogously as in Section~\ref{ch4_sec_SD_JJ}. In the following, only results for $W \rightarrow l\nu$ (where $l$ means an electron or muon) are shown, since the shapes of the distributions in the case of $Z \rightarrow ll$ are similar.

\begin{figure}[!htbp]
\centering
\hfill
\includegraphics[width=0.49\textwidth]{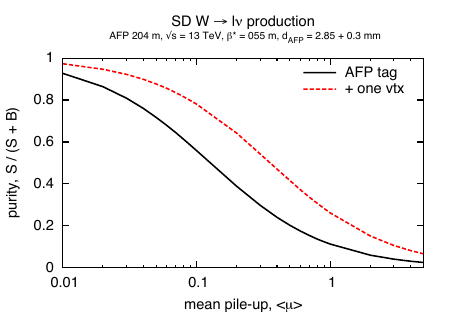}\hfill
\includegraphics[width=0.49\textwidth]{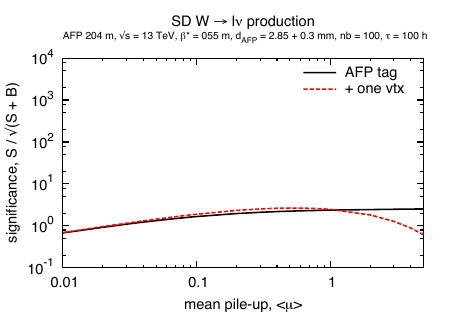}\hfill\hfill
\caption{Single diffractive $W \rightarrow l \nu$ production with protons tagged in the AFP detectors for $\sqrt{s} = 13$ TeV and $\beta^* = 0.55$~m: purity (\textbf{left}) and significance (\textbf{right}) as a function of average pile-up. The number of bunches multiplied by the data collecting time (in hours) was assumed to be 10000. In the analysis the following cuts were considered: proton tag in the AFP detector (black solid line) and tag + exactly one reconstructed vertex (red dashed line).}
\label{ch4_SD_W_lnu_AFP_beta055}
\end{figure}

The purity and significance for the AFP detector and $\beta^* = 0.55$~m optics is shown in Fig.~\ref{ch4_SD_W_lnu_AFP_beta055}. In this figure the black solid line is for events with a proton tag in the AFP detector whereas the red dashed line is for events with a proton tag and exactly one reconstructed vertex. A purity greater that 50\% is obtained for $\mu \sim 0.2$ and grows to values greater than 80\% for $\mu < 0.06$. The number of bunches multiplied by the data collecting time\footnote{For the reference see Section~\ref{ch4_sec_SD_JJ_ATLAS} and Fig.~\ref{fig_lumi}.} was assumed to be 10000, which is the minimal value for observing such events. Similar conclusions can be driven for the cases of AFP runs with $\beta^* = 90$~m and ALFA with $\beta^* = 0.55$~m optics.

%\begin{figure}[!htbp]
%\centering
%\hfill
%\includegraphics[width=0.32\textwidth]{purity_sd_W_lnu_z204_beta90_d5_9}\hfill
%\includegraphics[width=0.32\textwidth]{significance_sd_W_lnu_z204_beta90_d5_9_L2}\hfill\hfill
%\caption{Single diffractive $W \rightarrow l \nu$ production with protons tagged in the AFP detectors for $\sqrt{s} = 13$ TeV and $\beta^* = 90$~m: purity (\textbf{left}) and significance (\textbf{right}) as a function of average pile-up. The number of bunches multiplied by data collecting time was assumed to be 1000. In the analysis the following cuts were considered: proton tag in the AFP (black solid line) and tag + one reconstructed vertex (red dashed line).}
%\label{ch4_SD_W_lnu_AFP_beta90}
%\end{figure}
%
%\begin{figure}[!htbp]
%\centering
%\hfill
%\includegraphics[width=0.32\textwidth]{purity_sd_W_lnu_z237_beta055_d4_2}\hfill
%\includegraphics[width=0.32\textwidth]{significance_sd_W_lnu_z237_beta055_d4_2_L2}\hfill\hfill
%\caption{Single diffractive $W \rightarrow l \nu$ production with protons tagged in the ALFA detectors for $\sqrt{s} = 13$ TeV and $\beta^* = 0.55$~m: purity (\textbf{left}) and significance (\textbf{right}) as a function of average pile-up. The number of bunches multiplied by data collecting time was assumed to be 1000. In the analysis the following cuts were considered: proton tag in the ALFA (black solid line) and tag + one reconstructed vertex (red dashed line).}
%\label{ch4_SD_W_lnu_ALFA_beta055}
%\end{figure}

\begin{figure}[!htbp]
\centering
\hfill
\includegraphics[width=0.49\textwidth]{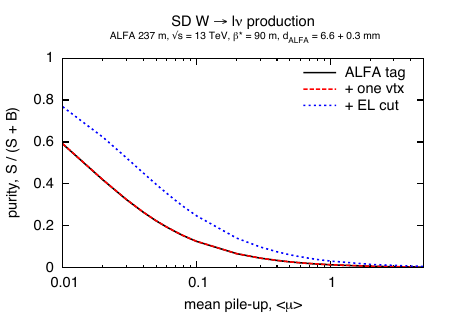}\hfill
\includegraphics[width=0.49\textwidth]{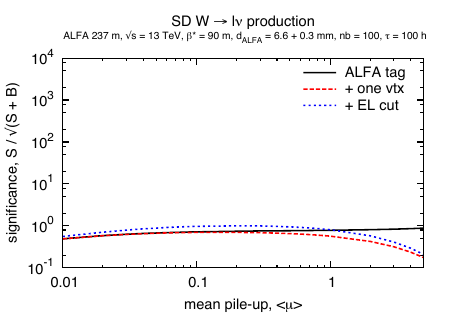}\hfill\hfill
\caption{Single diffractive $W \rightarrow l \nu$ production with protons tagged in the ALFA detectors for $\sqrt{s} = 13$ TeV and $\beta^* = 90$~m: purity (\textbf{left}) and significance (\textbf{right}) as a function of average pile-up. The number of bunches multiplied by the data collecting time (in hours) was assumed to be 10000. In the analysis the following cuts were considered: proton tag in the ALFA detector (black solid line), tag + exactly one reconstructed vertex (red dashed line) and tag + one vertex + elastic veto (blue dotted line).}
\label{ch4_SD_W_lnu_ALFA_beta90}
\end{figure}

The situation worsens for the ALFA run with $\beta^* = 90$~m optics. As can be observed in Fig.~\ref{ch4_SD_W_lnu_ALFA_beta90}, even after the rejection of double-tagged events from the elastic background, the purity is greater than 50\% only for $\mu < 0.02$. 

From the presented studies it is clear that a significant measurement of diffractive $W$ and $Z$ boson production cannot be achieved unless the data-taking conditions are as follows:
\begin{itemize}
  \item pile-up not larger than 0.1,
  \item number of bunches greater than couple of hundreds,
  \item data collecting time of at least 100 hours.
\end{itemize}

\section{Double Pomeron Exchange Jet Production}
\label{ch4_sec_DPE_JJ}
\begin{figure}[!htbp]
\centering
\includegraphics[width=0.25\textwidth]{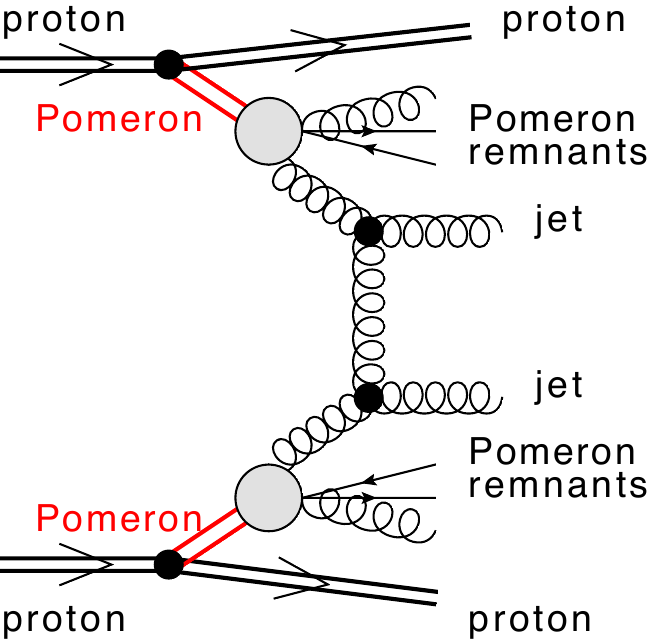}
\caption{Double Pomeron exchange jet production -- both interacting protons stays intact and two jets are produced centrally.}
\label{ch4_DPE_JJ_diag}
\end{figure}

In double Pomeron exchange (DPE) jet production, shown in Fig.~\ref{ch4_DPE_JJ_diag}, two jets are created and a colourless object is emitted from both interacting protons. As discussed in Section~\ref{ch4_sec_SD_JJ}, the additional soft interactions can break the two protons. At $\sqrt{s} = 13$ TeV the rapidity gap survival probability is estimated to be 0.03~\cite{KMR_surv1}. In case of double Pomeron exchange this factors is expected to be smaller than in SD cases. Nevertheless, it is worth stressing that since the theoretical uncertainties are quite large, the determination of the gap survival probability using data is a very important measurement.

\begin{figure}[!htbp]
\centering
\hfill
\includegraphics[width=0.45\textwidth]{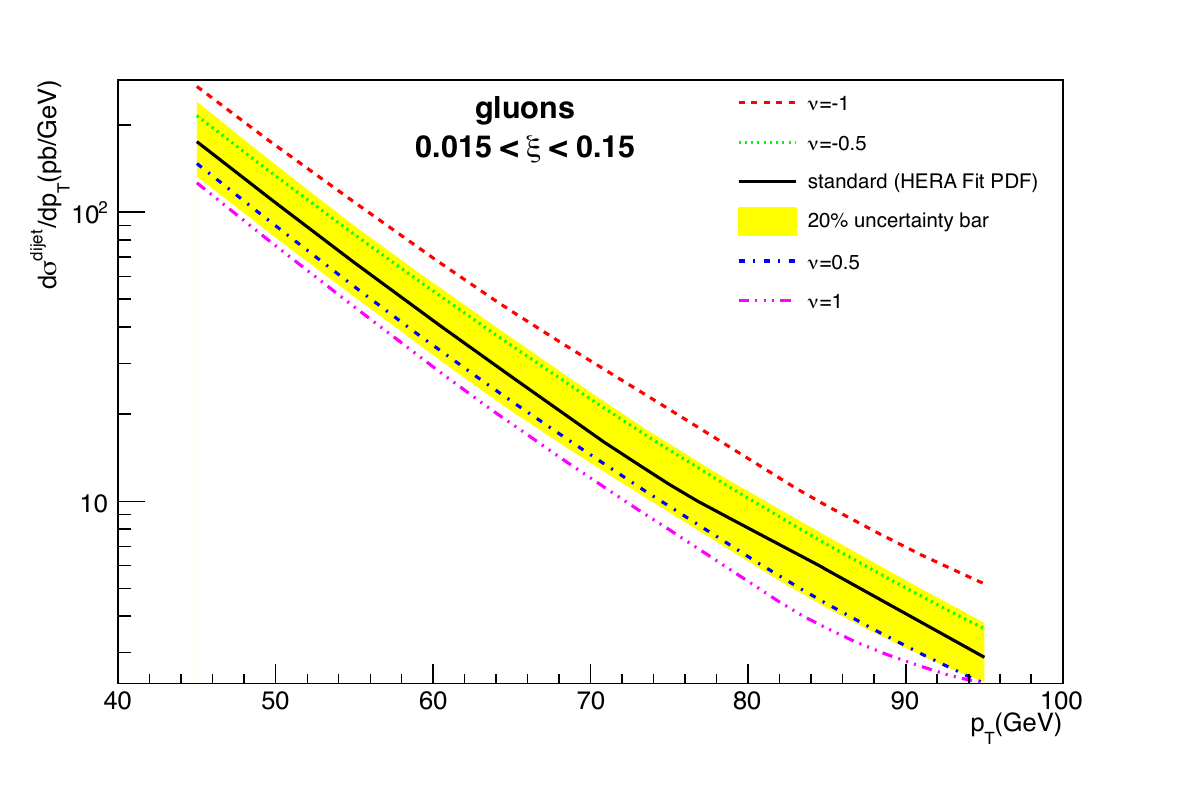}\hfill
\includegraphics[width=0.45\textwidth]{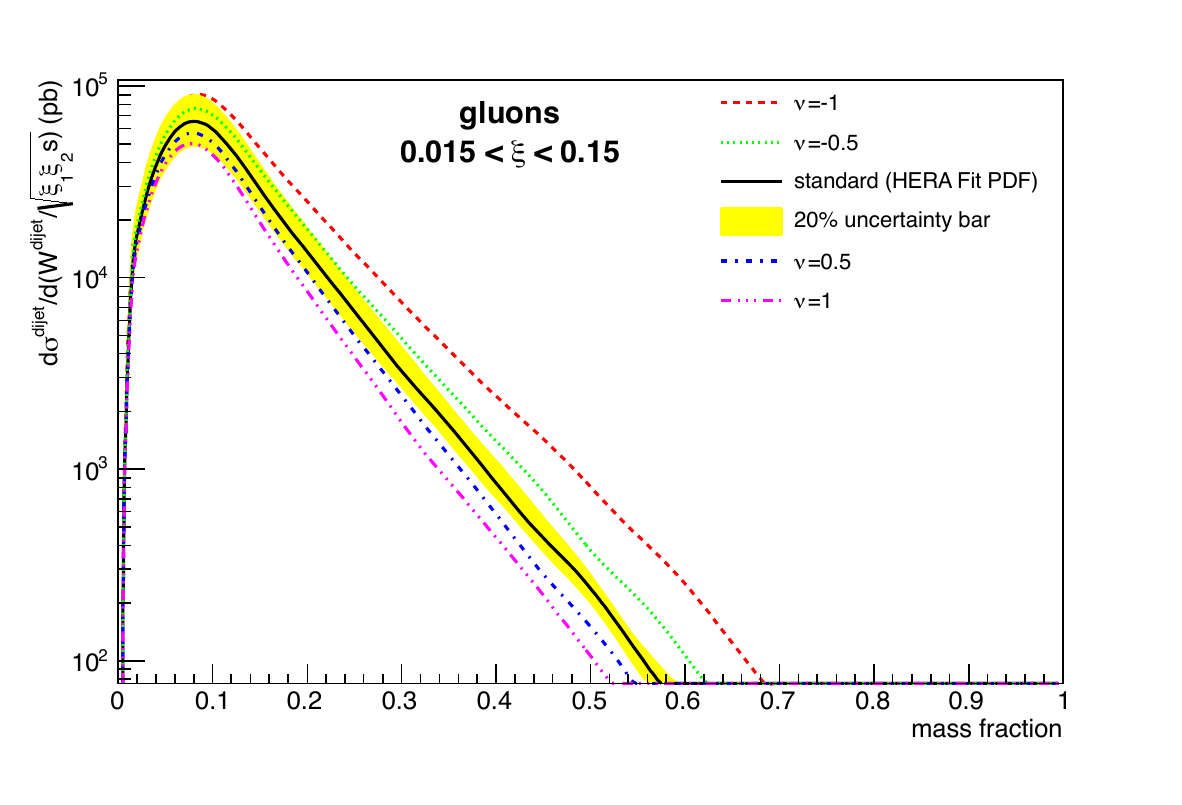}\hfill\hfill
\caption{Cross-section of DPE jet production as a function of leading jet $p_T$ (\textbf{left}) and mass fraction (\textbf{right}). The different curves correspond to different modifications of the Pomeron gluon density extracted from HERA data (see text).}
\label{ch4_DPE_JJ_theory}
\end{figure}

The DPE jet production is sensitive to the gluon density in the Pomeron~\cite{Royon_DPE_JJ}. This is shown in Fig.~\ref{ch4_DPE_JJ_theory}, where the gluon density is modified by $(1 - x)^\nu$. The central black line displays the cross-section value for the gluon density of the Pomeron as measured at HERA and including an additional survival probability of 0.03~\cite{KMR_surv1}. The yellow band shows the effect of a 20\% error on the gluon density, taking into account the normalisation uncertainties. The dashed curves display the expected cross-section sensitivities at the LHC to the gluon density distribution.

Unfortunately, due to the constant ratio between the curves for various gluons densities, it will be difficult to distinguish if observed changes in the absolute gluonic parton cross-section are due to the gluon density or to the survival probability. Hence the so-called mass fraction, defined as the ratio of the dijet mass to the total diffractive mass\footnote{The diffractive mass was computed as $\sqrt{s \xi_1 \xi_2}$, where $\xi_{1,2}$ are the proton fractional momentum carried by each Pomeron and $\sqrt{s}$ the center-of-mass energy.}, is introduced. As can be observed in Fig.~\ref{ch4_DPE_JJ_theory} (right), the curves corresponding to the different values of $\nu$ diverge faster at high values of the dijet mass fraction.

\subsection{ATLAS Feasibility Studies for $\sqrt{s} = 13$ TeV}
\label{ch4_sec_DPE_JJ_ATLAS}
The results presented in this Section were obtained analogously as the ones in Section~\ref{ch4_sec_SD_JJ}: after the event generation the protons were transported to the forward detector location, their energy was reconstructed taking into account various experimental effects, the jets were obtained using the anti-$k_t$ algorithm and tracks were required to fulfil the reconstruction criteria. The distance between forward detectors and beam centre was set to 15~$\sigma$ and 10~$\sigma$ in case of $\beta^*=0.55$~m and $\beta^*=90$~m optics, respectively.

\begin{figure}[!htbp]
\centering
\includegraphics[width=0.49\textwidth]{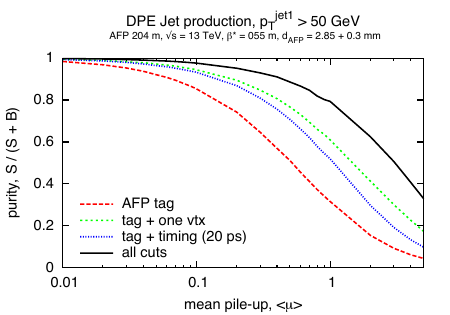}\hfill
\includegraphics[width=0.49\textwidth]{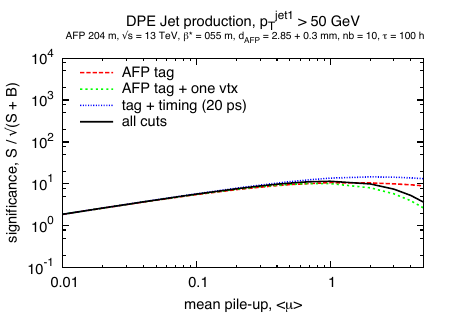}
\caption{Double Pomeron exchange jet production with both protons tagged in the AFP detectors for $\sqrt{s} = 13$ TeV and $\beta^* = 0.55$~m: purity (\textbf{left}) and significance for jets with $p_T > 50$ GeV (\textbf{right}) as a function of average pile-up. The number of bunches multiplied by the data collecting time (in hours) was assumed to be 1000. In the analysis the following cuts were considered: double proton tag in the AFP detector (red line), tag + exactly one reconstructed vertex (green line), tag + timing requirement (20 ps, blue line) and all cuts (black line).}
\label{ch4_DPE_JJ_AFP_beta055}
\end{figure}

The results for the the AFP detector and $\beta^* = 0.55$~m optics are shown in Fig.~\ref{ch4_DPE_JJ_AFP_beta055}. In these figures, the red line shows events with a double proton tag in the AFP detectors, the green line events with a double tag and exactly one reconstructed vertex, the blue line events with a double tag and finally the black line represents all these cuts. Since in DPE events there are two outgoing protons, their time-of-flight can be calculated and compared to the position of the hard vertex. The resolution of the AFP timing detectors was assumed to be 20 ps and the cut was done at 2$\sigma$. These values were taken similarly as in exclusive jet analyses (see the next Chapter for details).

\begin{figure}[!htbp]
\centering
\includegraphics[width=0.49\textwidth]{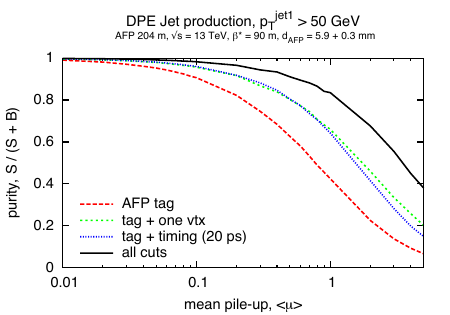}\hfill
\includegraphics[width=0.49\textwidth]{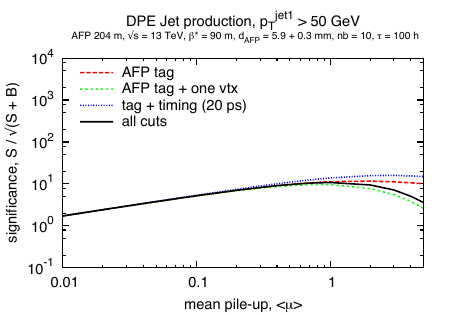}
\caption{Double Pomeron exchange jet production with both protons tagged in the AFP detectors for $\sqrt{s} = 13$ TeV and $\beta^* = 90$~m: purity (\textbf{left}) and significance for jets with $p_T > 50$ GeV (\textbf{right}) as a function of average pile-up. The number of bunches multiplied by the data collecting time (in hours) was assumed to be 1000. In the analysis the following cuts were considered: double proton tag in the AFP (red line), tag + exactly one reconstructed vertex (green line), tag + timing requirement (20 ps, blue line) and all cuts (black line).}
\label{ch4_DPE_JJ_AFP_beta90}
\end{figure}

The statistical significance is maximised for $\mu \sim 1$. For such pile-up values the purity is of about 80\%. Data taken at smaller pile-up result in a smaller statistical significance, however with a very high purity ($>95$\%). Assuming the number of bunches multiplied by the data collecting time of 1000 and $\mu \sim 1$, jets with $p_T$ up to 100 GeV could be measured. Similar conclusions are drawn in case of the AFP detector and $\beta^* = 90$~m optics (see Fig~\ref{ch4_DPE_JJ_AFP_beta90}).

\begin{figure}[!htbp]
\centering
\includegraphics[width=0.49\textwidth]{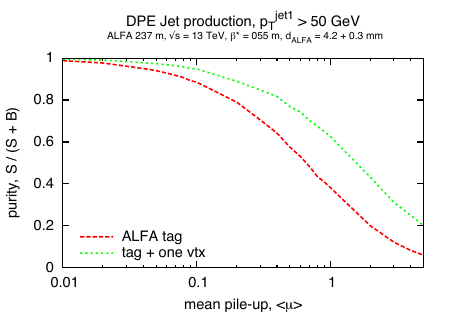}\hfill
\includegraphics[width=0.49\textwidth]{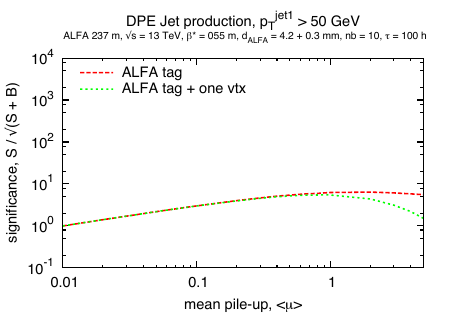}
\caption{Double Pomeron exchange jet production with both protons tagged in the ALFA detectors for $\sqrt{s} = 13$ TeV and $\beta^* = 0.55$~m: purity (\textbf{left}) and significance for jets with $p_T > 50$ GeV (\textbf{right}) as a function of average pile-up. The number of bunches multiplied by the data collecting time (in hours) was assumed to be 1000. In the analysis the following cuts were considered: double proton tag in the ALFA detector (red line) and tag + exactly one reconstructed vertex (green line).}
\label{ch4_DPE_JJ_ALFA_beta055}
\end{figure}

The results for the ALFA detector and $\beta^* = 0.55$~m optics are shown in Fig.~\ref{ch4_DPE_JJ_ALFA_beta055}. Since at present there is no plan to install timing detectors in ALFA, only two constraints were considered: double proton tag (red line) an exactly one reconstructed vertex (green line). At the maximal significance ($\mu \sim 1$) the purity is of about 60\%. Going to smaller pile-up values increases the purity, but reduces the statistical significance.

\begin{figure}[!htbp]
\centering
\includegraphics[width=0.49\textwidth]{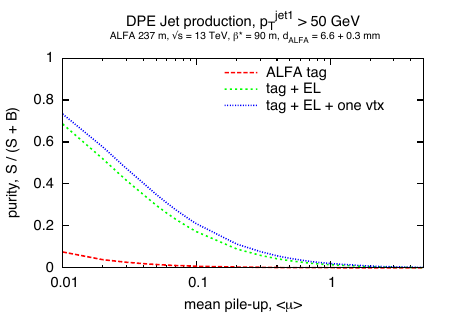}\hfill
\includegraphics[width=0.49\textwidth]{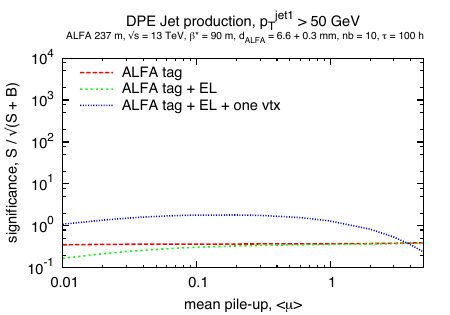}
\caption{Double Pomeron exchange jet production with both protons tagged in the ALFA detectors for $\sqrt{s} = 13$ TeV and $\beta^* = 90$~m: purity (\textbf{left}) and significance for jets with $p_T > 50$ GeV (\textbf{right}) as a function of average pile-up. The number of bunches multiplied by the data collecting time (in hours) was assumed to be 1000. In the analysis the following cuts were considered: double proton tag in the ALFA detector (red line), tag + elastic veto (green line) and tag + elastic veto + exactly one reconstructed vertex (blue line).}
\label{ch4_DPE_JJ_ALFA_beta90}
\end{figure}

Similarly as in the case of SD jet production, the measurement worsens dramatically when $\beta^* = 90$~m optics is considered (see Fig.~\ref{ch4_DPE_JJ_ALFA_beta90}). This is due to the fact that this optics was designed to measure the elastic scattering with ALFA, thus such events are well within the acceptance and contribute as a background. Fortunately, the elastic signature is relatively easy to filter out by using kinematic constraints. In the following studies, the filtering efficiency was assumed to be 100\%, \textit{i.e.} all generated elastic events were removed. Unfortunately, even after such selection the purity is greater than 60\% only for $\mu < 0.02$. This is due to high acceptance for the soft central exclusive processes. This means that even for jets with $p_T \sim 50$~GeV a significant measurement is possible only in long runs ($\sim 100$ h) with hundreds of bunches.

A summary of the studies presented above is presented in Tab.~\ref{ch4_tab_DPE_JJ}. Moreover, in this table the purity and statistical significance for other considered jet $p_T$ thresholds are given. The rate was calculated assuming 100 colliding bunches.

\section{Double Pomeron Exchange Photon+Jet Production}
\label{ch4_sec_DPE_GammaJet}
\begin{figure}[!htbp]
\centering
\includegraphics[width=0.25\textwidth]{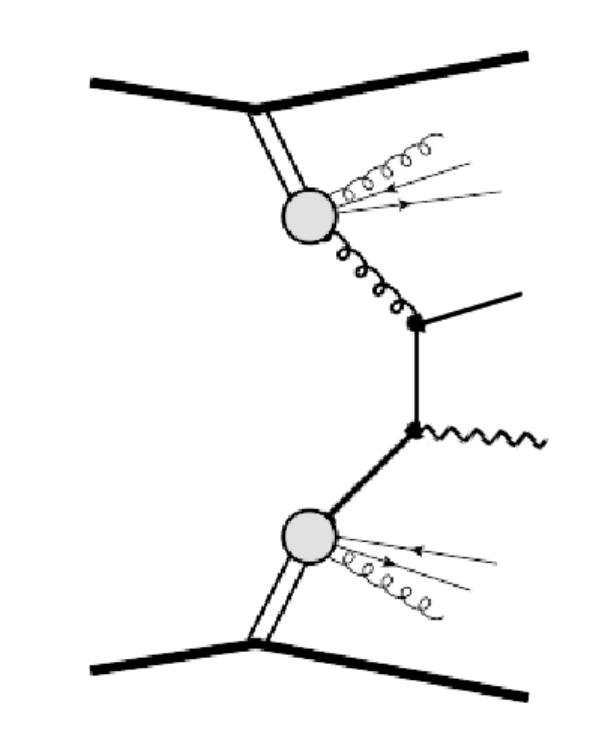}
\caption{Double Pomeron exchange photon+jet production -- both interacting protons stays intact and a (quark) jet and a photon are produced centrally.}
\label{ch4_DPE_GammaJet_diag}
\end{figure}

In double Pomeron exchange mode also events containing a (quark) jet and a photon could be produced. In such case one Pomeron emits a gluon, whereas from the other one a quark is taken. A diagram for such production is presented in Fig.~\ref{ch4_DPE_GammaJet_diag}.

A measurement of photon+jet events produced in DPE mode can be used to test the Pomeron universality between HERA and LHC. Moreover, the Pomeron quark content can be probed: the QCD diffractive fits performed at HERA assumed that $u = d = s = \bar{u} = \bar{d} = \bar{s}$, since data were not sensitive to the difference between the different quark component in the Pomeron. As will be shown, the LHC data would allow to check if this assumption was correct. For example, if a value of $d/u \ne 1$ will be favoured by data, the HERA QCD diffractive fits will have to be modified.

\begin{figure}[htb!]
\centering
\includegraphics[width=0.49\textwidth]{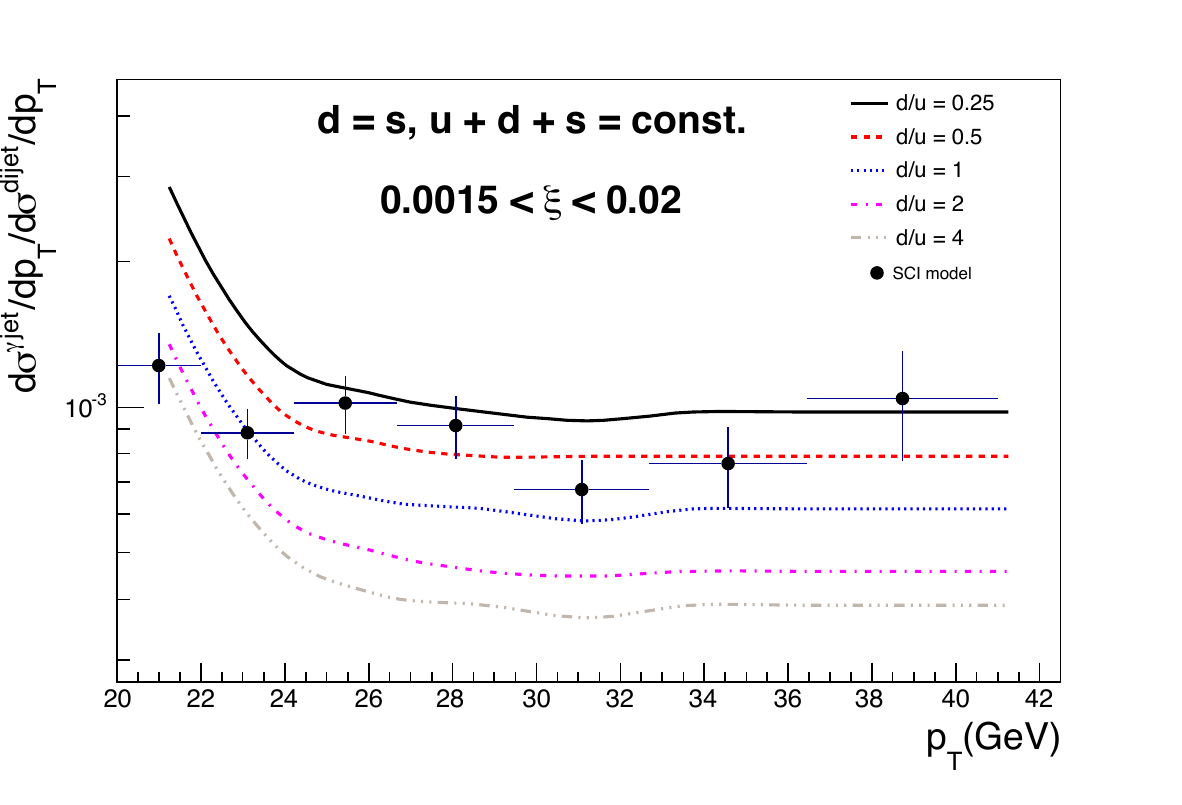}\hfill
\includegraphics[width=0.49\textwidth]{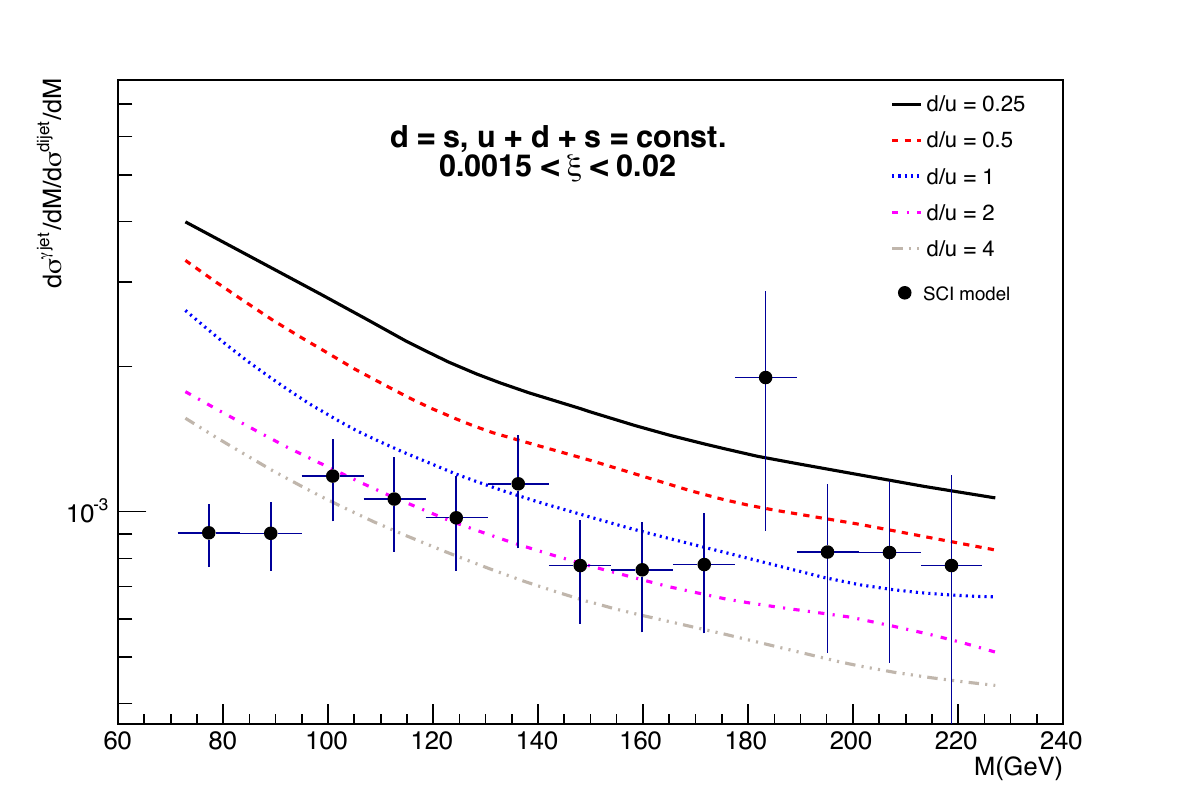}
\caption{\textbf{Left:} ratio of $\gamma$+jet over dijet differential cross-section as a function of the leading jet $p_T$. \textbf{Right:} ratio of $\gamma$+jet over dijet differential cross-section as a function of the diffractive mass $M=\sqrt{s \xi_1 \xi_2}$. The different curves correspond to different ratios $d/u$ inside the Pomeron. Proton-proton collisions at $\sqrt{s}=14$~TeV are assumed.}
\label{ch4_DPE_GammaJet_motivation}
\end{figure}

Observables that can probe the quark content in the Pomeron at the LHC are the transverse momentum of the leading jet ($p_T$) and the diffractive (missing) mass $M = \sqrt{s \xi_1 \xi_2}$. They are shown in Fig.~\ref{ch4_DPE_GammaJet_motivation} for different assumptions, namely $d/u$ varying between 0.25 and 4. For comparison the predictions of the Soft Colour Interaction (SCI) model~\cite{Edin:1995gi} are presented.

\subsection{ATLAS Feasibility Studies for $\sqrt{s} = 14$ TeV}
\label{ch4_sec_DPE_GammaJet_ATLAS}
\begin{figure}[htb!]
\centering
\includegraphics[width=0.49\textwidth]{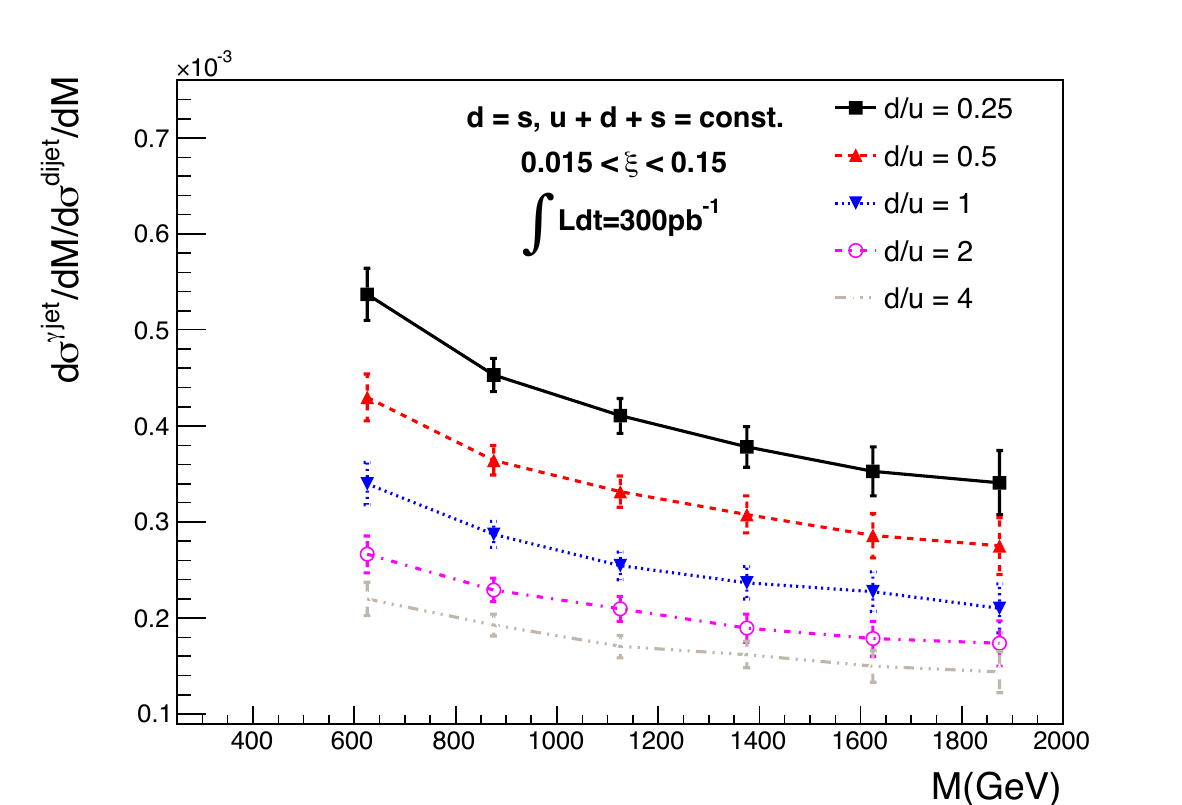}
\caption{DPE $\gamma$+jet to dijet differential cross-section ratio as a function of the diffractive mass $M=\sqrt{\xi_1 \xi_2 s}$ for different values of d/u within the acceptance of the 210 m proton detectors.}
\label{ch4_DPE_GammaJet_results}
\end{figure}

Results, based on Ref.~\cite{gammajet}, for the AFP detectors and $\beta^* = 0.55$~m are shown in Fig.~\ref{ch4_DPE_GammaJet_results}. In this plot the differential cross-section ratio of the DPE $\gamma$+jet events to the non-diffractive dijets is presented as a function of the diffractive mass. Most of the systematic uncertainties cancel since the mass distributions for $\gamma$+jet and dijet events are similar. Taking into account that the typical mass resolution is of the order of 2 to 3$\%$, a significant measurement can be done with an integrated luminosity of 300~pb$^{-1}$.

\section{Double Pomeron Exchange Jet-Gap-Jet Production}
\label{ch4_sec_DPE_JGJ}
\begin{figure}[!htbp]
\centering
\includegraphics[width=0.25\textwidth]{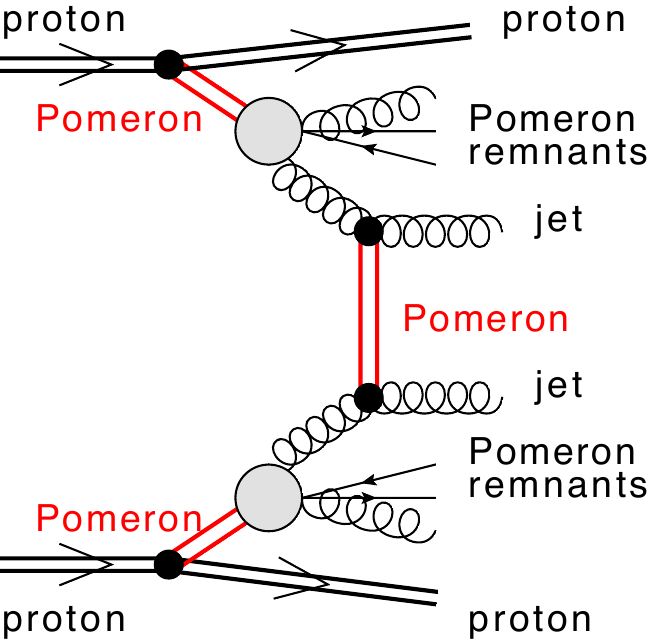}\\[0.2cm]
\includegraphics[width=0.4\textwidth]{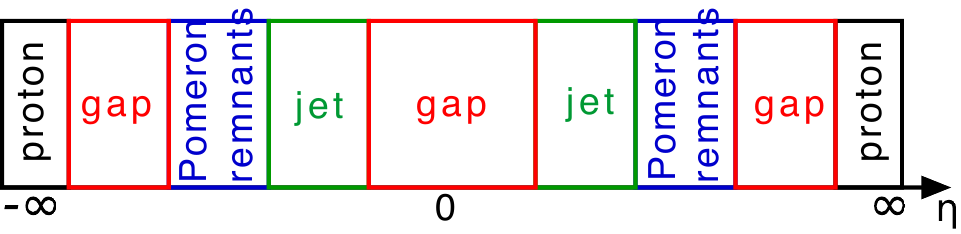}
\caption{Double Pomeron exchange jet-gap-jet production: both interacting protons stay intact and two jets are produced. In both cases the object exchanged in the $t$ channel is colour singlet and there is a gap in rapidity between the two jets.}
\label{ch4_DPE_JGJ_diag}
\end{figure}

A jet-gap-jet event features a large rapidity gap with a high-$p_T$ jet on each side. Across the gap, the object exchanged in the $t$-channel is a colour singlet and carries a large momentum transfer. When the rapidity gap is sufficiently large, the perturbative QCD description of jet-gap-jet events is performed in terms of a Balitsky-Fadin-Kuraev-Lipatov (BFKL) Pomeron~\cite{JGJ}. The jet-gap-jet topology is also produced in the single diffractive and the double Pomeron exchange processes. In such events, a colour singlet is exchanged between the protons and in the $t$-channel between the jets. The signature, shown in Fig.~\ref{ch4_DPE_JGJ_diag}, is two intact protons scattered in the forward regions and a gap in rapidity between the jets.

The process of double Pomeron exchange jet-gap-jet production was never measured experimentally. By studying its properties, the BFKL model can be tested, \textit{e.g.} by comparing the fraction of DPE JGJ to all DPE jet events. In case of DPE such ratio is larger than the corresponding fraction in ``standard'' JGJ production, since in DPE events the penalty of the gap survival probability applies to both the DPE JGJ and the total DPE cross-sections~\cite{dpe_jgj}.

\subsection{ATLAS Feasibility Studies for $\sqrt{s} = 14$ TeV}
\label{ch4_sec_DPE_JGJ_ATLAS}
In order to simulate the DPE jet-gap-jet production the FPMC program was employed. The used version contained an implementation for summing over non-conformal spins in the leading logarithm (LL) and next-to-leading logarithm (NLL) approximations~\cite{dpe_jgj}. 

A crucial element of the DPE jet-gap-jet measurement is the probability to tag the protons with forward detectors. In the presented analysis, ATLAS and AFP were taken as central and forward detectors, respectively. The leading jet was required to have a transverse momentum greater than 40~GeV. The transverse momentum of the sub-leading jet was required to be greater than 20~GeV. 

In the following studies, a gap is defined as a rapidity interval devoid of final state particles with a transverse momentum greater than 200~MeV. The two leading jets were required to be on the opposite pseudo-rapidity hemispheres and the rapidity gap ($\eta_g$) was requested to be symmetric around zero. These requirements are somewhat arbitrary and were introduced due to the simplicity -- the central tracker region has the highest efficiency for reconstructing low-$p_T$ tracks. Obviously, this analysis can be extended to events with non-symmetric gaps which will increase the visible cross-section.

\begin{figure}[!htbp]
\centering
\includegraphics[width=0.4\textwidth]{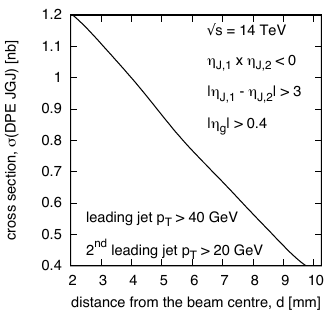}\hfill
\includegraphics[width=0.4\textwidth]{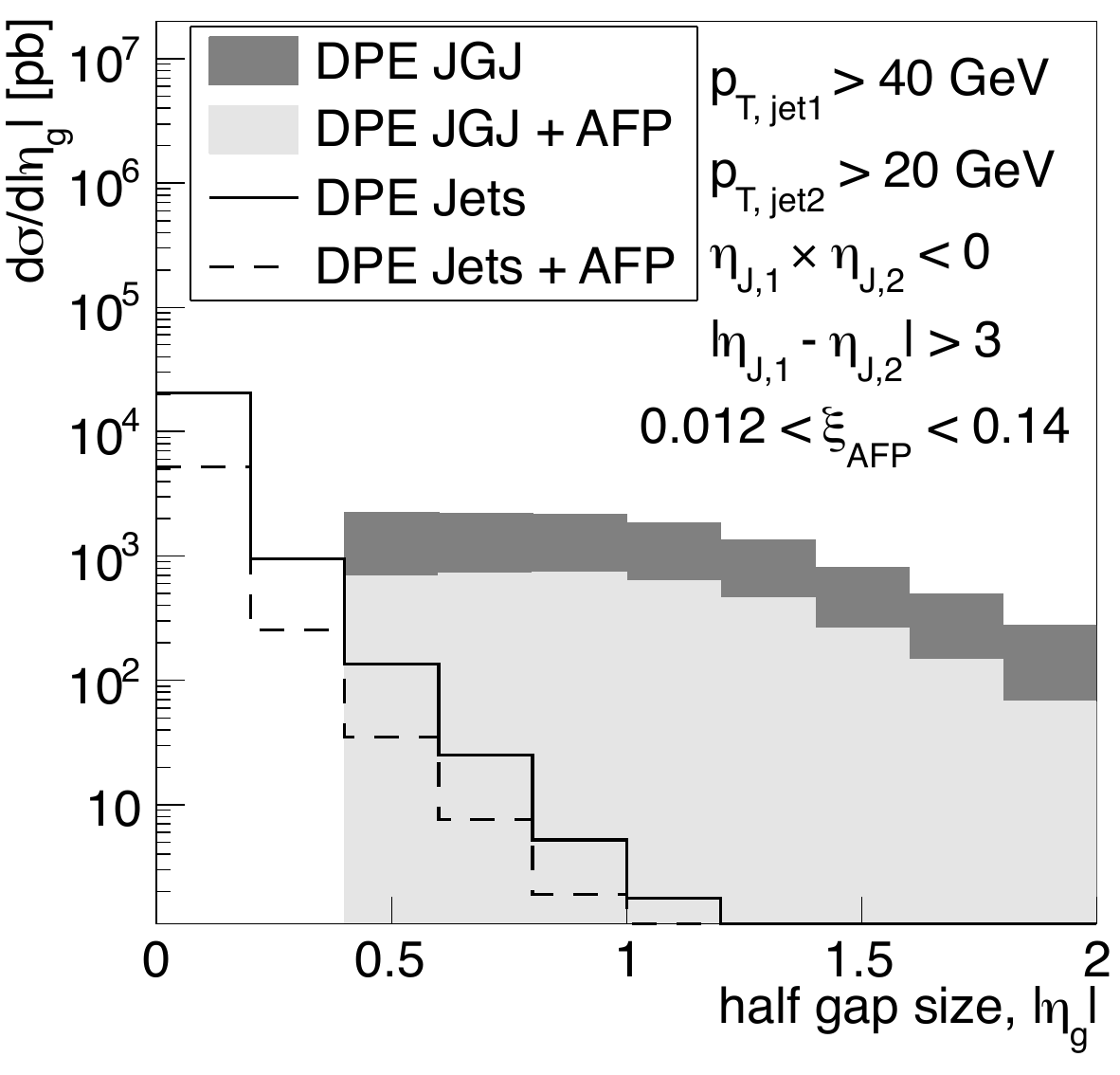}
\caption{\textbf{Left:} visible cross-section as a function of the distance between the detector and the beam centre. \textbf{Right:} gap size distribution for DPE jets and DPE jet-gap-jet events with and without the AFP tag requirement. For large enough gaps $\Delta\eta_g>0.5$, the gap-between-jets events are not dominated by fluctuations of dijets events.}
\label{ch4_DPE_JGJ_distance_momentum}
\end{figure}

Since both protons need to be tagged in the AFP stations, not all events can be recorded. As illustrated in Fig.~\ref{ch4_DPE_JGJ_distance_momentum} (left), the visible cross-section depends on the distance between the AFP active detector edge and the beam centre. For this analysis, a distance of 3.5~mm was assumed. This results in a visible cross-section of about 1~nb.

The main background to the DPE jet-gap-jet production will be the DPE jet production. For such processes a gap between the jets may be due to the fluctuations in hadronisation, but this background is significantly reduced by requiring large gap sizes. This is illustrated in Fig.~\ref{ch4_DPE_JGJ_distance_momentum} (right), where the DPE background is shown as a continuous and dashed lines whereas the DPE JGJ signal is plotted as a grey area. The probability of having a gap due to a fluctuation falls exponentially with the increase of the gap size. For example, if $|\eta_g| > 0.5$, the background will mimic the signal in less than $5\%$ of cases.

Larger gap sizes are increasingly dominated by the jet-gap-jet process. However, the cross-section also falls steeply with an increase of the gap size. Assuming both protons tagged in AFP, a good balance between the signal to background ratio and the visible cross-section was found for a gap of $|\eta_g| \sim 0.5$.

\begin{figure}[!htbp]
\centering
\includegraphics[width=.49\textwidth]{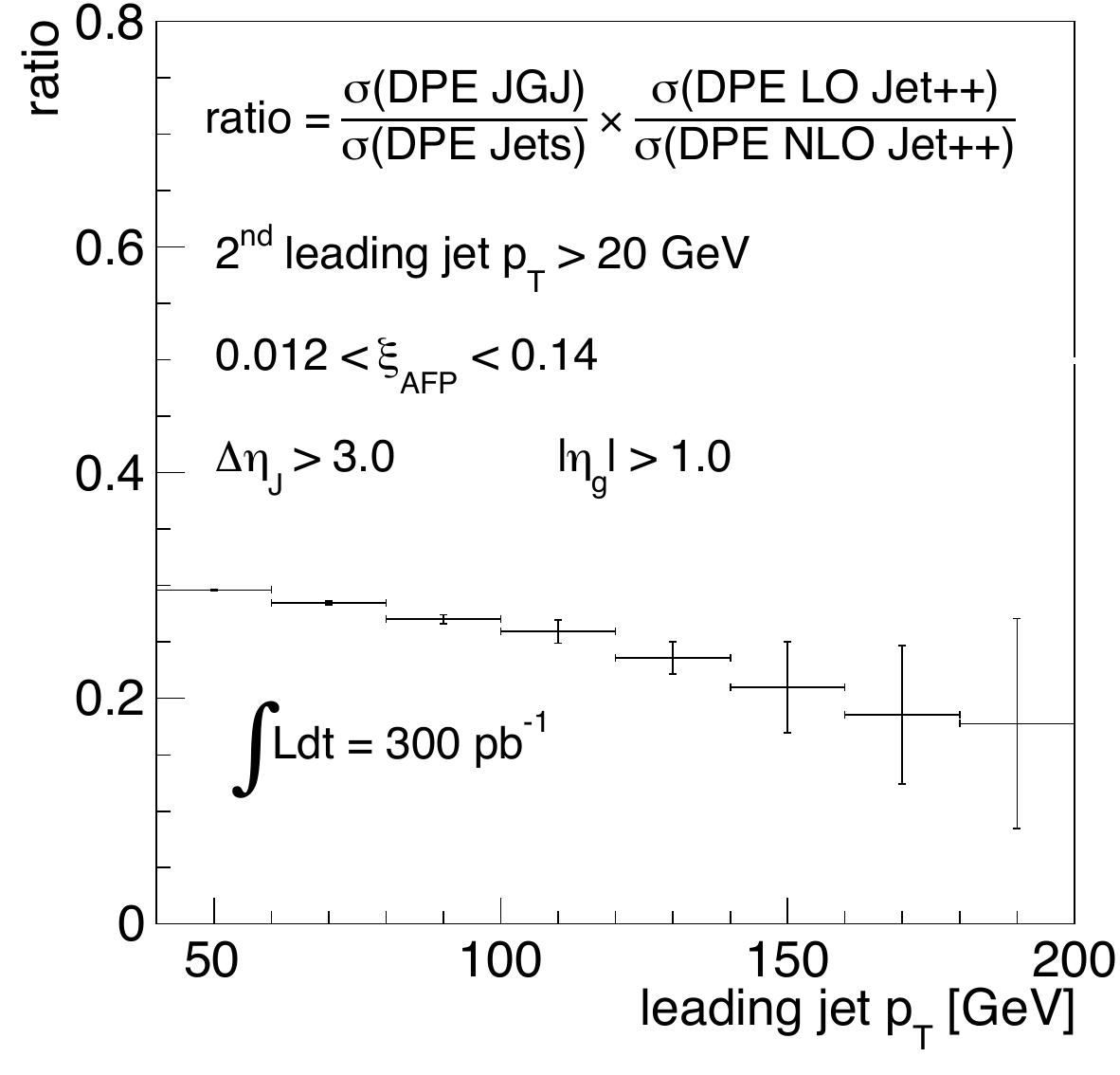}
\hfill
\includegraphics[width=.49\textwidth]{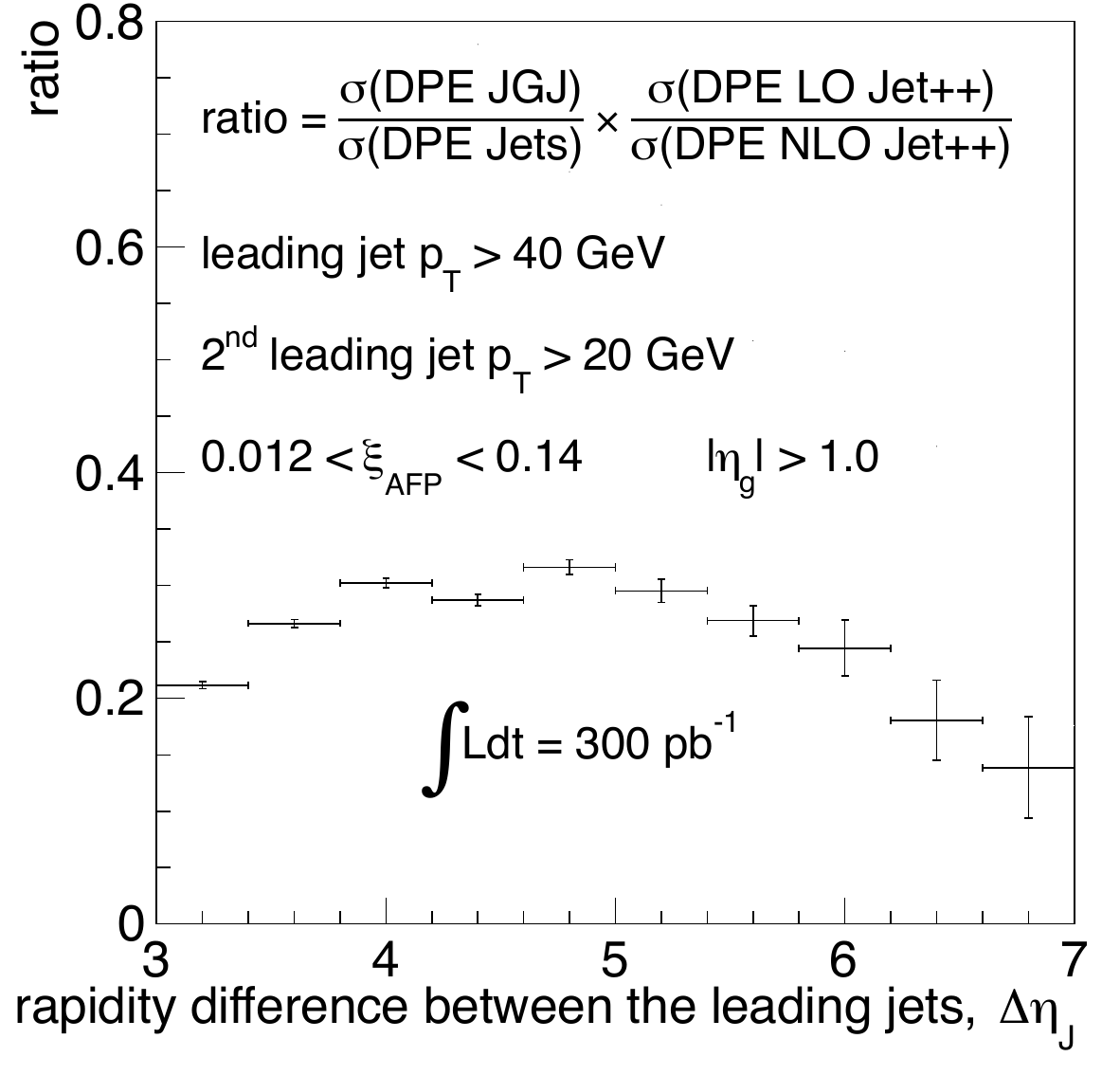}
\caption{Predictions for the DPE jet-gap-jet to DPE jet cross-section ratio at the LHC, as a function of the leading jet transverse momentum $p_{T}$ (\textbf{left}), and of the rapidity difference between the two leading jets $\Delta\eta_J$ (\textbf{right}). For both plots, an integrated luminosity of 300~pb$^{-1}$ was assumed.}
\label{ch4_dpe_jgj_ratio}
\end{figure}

The DPE jet-gap-jet event ratio is plotted in Fig.~\ref{ch4_dpe_jgj_ratio} as a function of the transverse momentum of the leading jet and as a function of the pseudorapidity difference between the two jets with the highest transverse momentum, $\Delta\eta_{J}$. To take into account the NLO QCD effects, absent in the FPMC program, the LO ratio was corrected by the cross-section ratio $\sigma$(DPE LO Jet++)/$\sigma$(DPE NLO Jet++) obtained with the \textsc{NLO Jet++} program~\cite{NLOJET}. The detailed description of this procedure can be found in Ref.~\cite{dpe_jgj}. To verify the statistical power of this measurement, statistical errors corresponding to 300~pb$^{-1}$ of integrated luminosity were computed.

\section{Conclusions}
CMS and ATLAS have a dense program of hard diffractive measurements. Several of them are feasible with moderate integrated luminosities and/or benefitting from the special low luminosity runs foreseen in 2015. Studies described in this Chapter were done under the assumption that protons are tagged in forward detectors: AFP or ALFA in case of the ATLAS experiment or TOTEM Roman pots for CMS. The analyses assumed either collision ($\beta^*$ = 0.55~m) or special, high-$\beta^*$, optics.

The measurements of single diffractive $J/\psi$, $W$ and $Z$ bosons were shown to be possible with with 10 pb$^{-1}$ of integrated luminosity. The amount of data needed to be collected for single diffractive or double Pomeron exchange studies depend strongly on the jet $p_T$ threshold. Nevertheless, the low $p_T$ jets could be measured already with few inverse picobarns collected. In order to study the double Pomeron exchange $\gamma$+jet and jet-gap-jet productions more data is needed. In particular, it was shown that the measurement of later two processes is possible with O(300) pb$^{-1}$.

By studying diffractive production a number of QCD tests can be performed. For example, such measurements may shed a light on the problem of the Pomeron universality in $ep$ and $pp$ collision. Moreover, the gap survival probability can be quantified and the QCD evolution of the gluon and quark densities in the Pomeron can be tested and compared with the HERA measurements. Some of the diffractive processes, like double Pomeron exchange jet-gap-jet production, were never measured experimentally. Moreover, by studying the properties of JGJ events QCD models, like BFKL, can be tested.

\section*{Acknowledgments}
    
Pieces of this chapter has been supported in part by Polish National Science Centre grant number 2012/05/B/ST2/02480.

\begin{subappendices}
\section{Expected Statistics of Single Diffractive Jet Measurement}

\begin{table}[!htbp]
  \caption{Purity, statistical significance, number of events and rate for single diffractive jet production for various optics settings and ATLAS forward detectors. The sub-leading jet was required to have $p_T > 20$~GeV. The number of bunches multiplied by the data collecting time (in hours) was assumed to be 1000. The rate was calculated for 100 bunches.}
  \label{ch4_tab_SD_JJ}
\begin{center}
{\small
  \begin{tabular}{l | c | c | c | c}
\toprule
\multicolumn{1}{c|}{\textbf{Leading jet}} & \multirow{2}{*}{\textbf{Purity}} & \multirow{2}{*}{\textbf{Significance}} & \textbf{Number of} & \textbf{rate}\\ 
\multicolumn{1}{c|}{\textbf{$\mathbf{p_T}$ threshold}} & & & \textbf{events} & \textbf{[Hz]}\\ 
\midrule
\multicolumn{5}{c}{\textbf{AFP 204 m, $\boldsymbol{\beta^*}$ = 0.55 m, $\boldsymbol{\mu}$ = 0.1}}\\
\midrule
$p_T > 20$ GeV	&		0.7	&		64	&		10000	&	5	\\
$p_T > 50$ GeV	&		0.85	&		21	&		1000	&	0.3	\\
$p_T > 100$ GeV	&		0.8	&		7	&		100	&	0.02	\\
\midrule
\multicolumn{5}{c}{\textbf{AFP 204 m, $\boldsymbol{\beta^*}$ = 90 m, $\boldsymbol{\mu}$ = 0.1}}\\
\midrule
$p_T > 20$ GeV	&		0.65	&		56	&		8000	&	3	\\
$p_T > 50$ GeV	&		0.85	&		19	&		800	&	0.3	\\
$p_T > 100$ GeV	&		0.8	&		5	&		50	&	0.01	\\
\midrule
\multicolumn{5}{c}{\textbf{ALFA 237 m, $\boldsymbol{\beta^*}$ = 0.55 m, $\boldsymbol{\mu}$ = 0.1}}\\
\midrule
$p_T > 20$ GeV	&		0.7	&		45	&		5000	&	3	\\
$p_T > 50$ GeV	&		0.85	&		15	&		500	&	0.2	\\
$p_T > 100$ GeV	&		0.85	&		3	&		20	&	0.01	\\
\midrule
\multicolumn{5}{c}{\textbf{ALFA 237 m, $\boldsymbol{\beta^*}$ = 90 m, $\boldsymbol{\mu}$ = 0.01}}\\
\midrule
$p_T > 20$ GeV	&		0.65	&		20	&		1000	&	5	\\
$p_T > 50$ GeV	&		0.8	&		7	&		100	&	0.3	\\
$p_T > 100$ GeV	&		0.8	&		1	&		5	&	0.02	\\
\bottomrule
  \end{tabular}
}
\end{center}
\end{table}

\section{Expected Statistics of Double Pomeron Exchange Jet Measurement}

\begin{table}[!htbp]
  \caption{Purity, statistical significance, number of events and rate for double Pomeron exchange jet production for various optics settings and ATLAS forward detectors. The sub-leading jet was required to have $p_T > 20$~GeV. The number of bunches multiplied by the data collecting time (in hours) was assumed to be 1000. The rate was calculated for 100 bunches.}
  \label{ch4_tab_DPE_JJ}
\begin{center}
{\small
  \begin{tabular}{l | c | c | c | c}
\toprule
\multicolumn{1}{c|}{\textbf{Leading jet}} & \multirow{2}{*}{\textbf{Purity}} & \multirow{2}{*}{\textbf{Significance}} & \textbf{Number of} & \textbf{rate}\\ 
\multicolumn{1}{c|}{\textbf{$\mathbf{p_T}$ threshold}} & & & \textbf{events} & \textbf{[Hz]}\\ 
\midrule
\multicolumn{5}{c}{\textbf{AFP 204 m, $\boldsymbol{\beta^*}$ = 0.55 m, $\boldsymbol{\mu}$ = 0.1}}\\
\midrule
$p_T > 20$ GeV	&		0.95	&		49	&		5000	&	0.2	\\
$p_T > 50$ GeV	&		0.95	&		16	&		500	&	0.01	\\
$p_T > 100$ GeV	&		0.95	&		2	&		10	&	0.0004	\\
\midrule
\multicolumn{5}{c}{\textbf{AFP 204 m, $\boldsymbol{\beta^*}$ = 90 m, $\boldsymbol{\mu}$ = 0.1}}\\
\midrule
$p_T > 20$ GeV	&		0.95	&		38	&		3000	&	0.1	\\
$p_T > 50$ GeV	&		0.95	&		12	&		300	&	0.01	\\
$p_T > 100$ GeV	&		0.95	&		3	&		20	&	0.0005	\\
\midrule
\multicolumn{5}{c}{\textbf{ALFA 237 m, $\boldsymbol{\beta^*}$ = 0.55 m, $\boldsymbol{\mu}$ = 0.1}}\\
\midrule
$p_T > 20$ GeV	&		0.9	&		22	&		1000	&	0.05	\\
$p_T > 50$ GeV	&		0.95	&		7	&		100	&	0.003	\\
$p_T > 100$ GeV	&		0.95	&		2	&		5	&	0.0001	\\
\midrule
\multicolumn{5}{c}{\textbf{ALFA 237 m, $\boldsymbol{\beta^*}$ = 90 m, $\boldsymbol{\mu}$ = 0.01}}\\
\midrule
$p_T > 20$ GeV	&		0.6	&		11	&		300	&	2	\\
$p_T > 50$ GeV	&		0.7	&		1	&		2	&	0.05	\\
$p_T > 100$ GeV	&		0.6	&		0	&		0.2	&	0.003	\\
\bottomrule
  \end{tabular}
}
\end{center}
\end{table}

\end{subappendices}

\FloatBarrier

%\bibliographystyle{unsrt}
%\bibliography{harddiffraction}
%%
%
%
%\end{document}

%% file: cep/cep.tex
\newcommand{\br}{\mbox{\boldmath $r$}}
\newcommand{\bkappa}{\mbox{\boldmath $\kappa$}}
\newcommand{\bk}{\mbox{\boldmath $k$}}
\newcommand{\bDelta}{\mbox{\boldmath $\Delta$}}
\def\Pom{{\bf I\!P}}
\newcommand{\dpe}{D$\, \mathrm{I\!\!\!P}$E }
\DeclareRobustCommand{\fbi}{\ensuremath{\mathrm{fb}^{-1}}}
\newcommand{\lln}{<\!\!\!\!<}
\newcommand{\rr}{\mbox{\boldmath $r$}}
\newcommand{\rb}{\mbox{\boldmath $b$}}
\newcommand{\rd}{\mbox{\boldmath $\Delta$}}
\newcommand{\dif}{\mathrm{d}}
    \renewcommand{\d}{\mathrm{d}}

\section{Introduction}
\label{sec:intro}

Central Exclusive Production (CEP) is the reaction
\begin{equation}\nonumber
pp({\bar p}) \to p+X+p({\bar p})\;,
\end{equation}
where `$+$' signs are used to denote the presence of large rapidity gaps, separating the system $X$ from the intact outgoing protons (anti--protons). Over the last decade there has been a steady rise of theoretical and experimental interest in studies of this process in high--energy hadronic collisions, see~\cite{Albrow:2010yb,Harland-Lang:2014lxa,Harland-Lang:2014dta} for reviews. On the theoretical side, the study of CEP requires the development of a framework which is quite different from that used to describe the inclusive processes more commonly considered at hadron colliders. This requires an explicit account of both soft and hard QCD, and is therefore sensitive to both of these regimes. Moreover, the dynamics of the CEP process leads to unique predictions and effects which are not seen in the inclusive mode. Experimentally, CEP represents a very clean signal, with just the object $X$ and no other hadronic activity seen in the central detector (in the absence of pile up). 

In addition, in such reactions the outgoing hadrons can be measured by installing proton tagging detectors, situated far down the beam line from the central detector, which can provide information about the mass and quantum numbers of the centrally produced state; this is the aim of the CT-PPS and CMS-TOTEM detectors, and the ALFA and AFP detectors for ATLAS. This chapter will discuss the motivation and possibilities for performing measurements of exclusive processes both with and without tagged protons at low to medium luminosity, as part of special high $\beta^*$ runs with ATLAS and CMS, or during general LHCb and ALICE running, and at higher luminosity with tagged protons, where tools such as precision timing detectors will be fundamental to control pile up effects.

The CEP process requires the $t$--channel exchange of a color--singlet object, so that the outgoing protons can remain intact. More generally, in order for the cross section not to vanish with rising rapidity gaps between the final state particles, the $t$--channel exchanges cannot transfer charge, isospin, or color. One possibility to achieve this is the two--photon fusion process $\gamma\gamma \to X$, where the radiated quasi--real photons couple to the electromagnetic charge of the whole protons. Another possibility is to consider so--called Double Pomeron Exchange (DPE), where both protons interact strongly, `emitting' pomerons, which then `fuse' to create the object $I\!\!P I\!\!P \to X$. Provided the object X mass is large enough, this process can be considered in the framework of pQCD, that is by considering gluon rather than pomeron interactions. Finally it is possible for `photoproduction'  reactions to occur, where both photon and pomeron (gluon) emission take place, i.e. $I\!\!P \gamma \to X$. All three processes will be considered in this chapter; the label `CEP' will be used in all cases.

The outline of this chapter is as follows. In Section~\ref{sec:analysis}, the methods of selecting exclusive events, namely through proton tagging or rapidity gap based techniques, are discussed, and the relevant features of the LHC experiments for performing CEP measurements are summarised. Theoretical details, as well as the experimental results and outlook are presented in Sections~\ref{CEP:gluonex} and~\ref{CEP:phoex} for a range of QCD and photon exchange/photoproduction exclusive processes, respectively. Finally, in Section~\ref{sec:exploratory}, the possibility for performing exploratory searches, and probing BSM physics are discussed.

\clearpage

\section{Analysis techniques and detectors to study exclusive processes at the LHC}\label{sec:analysis}

In this Section, we start by presenting the two methods for selecting exclusive events in collision data applied so far in analyses: proton tagging and rapidity gap detection, see Section~\ref{subsec:anatech}. Then, we summarise briefly the advantages and challenges of the different detectors at the LHC when measuring CEP processes: in Section~\ref{subsec:anatech:LHCbALICE} the LHCb and ALICE detectors and in Section~\ref{subsec:anatech:CMSATLAS}, the CMS and ATLAS detectors, are discussed. The reader is referred to Chapter~\ref{chap:detectors} for further details of the relevant detectors.

\subsection{Analysis techniques}
\label{subsec:anatech}

Proton tagging is both a very challenging and powerful technique. For a CEP process, by detecting the intact protons in addition to the system produced in the central detector the full event kinematics are reconstructed, which is generally not possible in a hadron--hadron collider. Detecting the outgoing protons is the only way to get a pure sample of exclusive events experimentally: indeed, the kinematic variables reconstructed from the forward and the central detectors can be compared in order to reject a very large fraction of the traditional backgrounds encountered in CEP measurements (e.g. quasi-exclusive, dissociated proton events). That is, a comparison between the central mass, $M_{\rm central}$, computed from the centrally produced particles, and $M_X$, computed from the outgoing protons, can be performed. More specifically, the transverse ($p_T$) and longitudinal ($p_z$) momentum of the central state and the two protons may also be compared, and the rapidity gaps predicted by the proton fractional momentum loss $\xi$ measurements can be verified. In all cases consistency between the central and proton systems is expected in the case of CEP, but not in general for background events. Furthermore, the full kinematics reconstruction has recently been shown to make various CEP measurements competitive in searches for BSM physics in the nominal LHC high luminosity running, compared to standard LHC searches using the central detector only (see Section~\ref{sec:exploratory}). This presents completely new possibilities for forward physics and CEP measurements. 
 As well as serving as the most effective way to select exclusive events experimentally, proton tagging is also of great interest theoretically, as a means to measure the momenta of the outgoing intact protons in the CEP reactions, see Sections~\ref{sec:tagcepmotglu} and~\ref{sec:tagcepmotphot}.

Without tagging the outgoing protons it is still possible to select events which are dominantly due to CEP,  by using rapidity gap methods, i.e. demanding that there is no additional hadronic activity associated with the event in a large enough region of rapidity. In order to veto quasi-exclusive interactions containing low mass proton dissociation, a wide rapidity coverage of veto detectors is advantageous. This technique is most readily applied to higher cross section processes, for example exclusive $J/\psi$ photoproduction where one may require (excluding the decay of the central system) no reconstructed tracks, no central calorimeter energy deposits above noise, and no activity in forward scintillators or calorimeters. The exact requirements vary between experiments due to different detector technologies covering differing solid angles. Such techniques require the probability of more than one proton interaction per bunch crossing to be small. However, for smaller cross section processes other techniques to select dominantly exclusive events, in the presence of pile--up, can be still applied, for example vetoing on any additional tracks associated with the interaction vertex.

\subsection{Central exclusive production at LHCb and ALICE}\label{subsec:anatech:LHCbALICE}

No proton tagging detectors are currently installed at LHCb, and so the determination of the exclusivity of an event depends on no activity being seen in an active detection region that extends over as large a pseudorapidity range as possible. The LHCb detector is fully instrumented for pseudorapidities 2 $< \eta <$ 4.5, and includes a high-precision tracking system consisting of a silicon-strip vertex detector (VELO) surrounding the $pp$ interaction region, which has sensitivity to charged particles in the backwards region ($-3.5<\eta<-1.5$), as well as extending the sensitivity in the forward region to $1.5<\eta <5$. Thus the Run--I rapidity coverage sums to roughly $5.5$ units in rapidity. During Run--II, the newly installed HERSCHEL forward scintillation detectors will allow vetoes on additional particle production up to $|\eta|\approx 8$ extending the detection of a rapidity gap to up to about 12 units, see Section~\ref{sec:detecor:lhcb:herschel}. The triggering capability of the LHCb detector, being designed for low mass objects, is well suited to CEP. It consists of a two-stage system, a fast hardware trigger followed by a software trigger that applies a full event reconstruction. For CEP, the hardware stage triggers on muons with transverse momentum above \mbox{400 MeV},  or electromagnetic or hadronic energy  above 1000 MeV, all of which are placed in coincidence with a charged multiplicity of less than 10 deposits in the scintillating-pads (SPD). The software trigger is configured to select concidence with a low charged particle multiplicity requirement.  The data-taking conditions at LHCb are advantageous for the selection of CEP events. Unlike ATLAS and CMS, where there were typically 20 interactions per beam-crossing in the 2012 data-taking, the beams are defocused at LHCb, resulting in an average of about 1.5 proton-proton interactions per beam-crossing.  Consequently, about 20\% of the total luminosity has a single interaction and is usable for CEP. During high--luminosity Run--II conditions, LHCb intend to run with an average of 1.1 proton--proton interactions per collision.

Due to specific detector restrictions, ALICE requires a reduced luminosity in $pp$ interactions at IP2,
therefore the instantaneous luminosity delivered to ALICE for $pp$ collisions
is adjusted accordingly. The experiment aims to collect $pp$ data for CEP studies during dedicated
runs for diffractive studies, as well as during runs dedicated
for minimum-bias studies. Another advantage of the ALICE experiment for performing
CEP and diffractive studies is its low-$p_T$ reach. It has very good track
reconstruction and particle identification efficiency starting from
$p_T \simeq$ 150 MeV. MC simulations show that the invariant mass and
transverse momentum resolution for the two-track system is better than 0.5$\%$
and that the systematic shifts are negligible. In order to enhance the ALICE capabilities for diffractive studies, two new scintillator counters have been installed during LS1 in both sides of the interaction region, covering pseudorapidities up to $|\eta| \simeq 7$, see Section~\ref{sec:TheAliceDiffractiveDetector} for further details. In addition, ALICE is currently introducing a dedicated online trigger
to be used during Run-II for selecting a sample with an enhanced CEP contribution. It will require an activity
in the central barrel and no activity up to pseudorapidities $|\eta| \simeq 7$.

\subsection{Central exclusive production at CMS and ATLAS}\label{subsec:anatech:CMSATLAS}
\label{subsec:cepatcmsatlas}

In the case of ATLAS or CMS, the tracker is of particular importance. In both cases it is designed for efficient and precise reconstruction of the trajectories of charged particles with transverse momentum above $1\,$GeV in the pseudorapidity range $|\eta|<2.5$. Special tracking algorithms extend the tracking capabilities down to $p_T \sim$ 0.1$\,$GeV, essential for CEP studies. 

The CMS detector already includes proton taggers associated with the Interaction Point (IP), with both the existing TOTEM experiment and the CMS-TOTEM Precision Proton Spectrometer (CT-PPS) upgrade soon to start data taking, see Section~\ref{sec:totemdetector} and ~\ref{sec:ctppsdetector}. For the ATLAS collaboration, the existing ALFA detectors allow intact protons to be detected at high $\beta^{*}$, similarly to TOTEM, and the properties of the future AFP detectors are similar to the CT-PPS upgrade, see Sections~\ref{sec:exp} and~\ref{sec:afp}. In the following some further details are given in the case of CMS-TOTEM, but it should be emphasised that much of this discussion also applies in the ATLAS-ALFA-AFP case. 

The combination of the CMS and TOTEM experiments gives an exceptionally large pseudorapidity coverage for tracking and calorimetry that is well suited for studies of diffractive processes like CEP. The addition of proton timing detectors with $\sim$ 50 ps timing resolution in the vertical Roman Pots (RPs) allows access to CEP  processes with $O$(pb) cross sections. The Forward Shower Counters (FSC), covering 6 $\lesssim |\eta| \lesssim$ 8 in rapidity, can be used to veto proton diffractive dissociation and extend rapidity gap measurements. Since protons with any fractional momentum loss $\xi$ can be detected in the vertical RPs with low pile--up $\beta^*$ = 90 m optics (see Chapter~\ref{chap:bema}), the mass coverage in CEP and photon exchange reactions extends to any central system mass $M_X$, as long as the $|t|$ of both scattered protons is larger than $\sim$ 0.04 GeV$^{2}$. This is therefore complementary to the reach of the CMS-TOTEM precision proton spectrometer (CT-PPS) discussed below. Already, data has been taken during a common CMS-TOTEM $\beta^*$ = 90 m run at $\sqrt{s}$ = 8 TeV in July 2012, showing the feasibility of CEP measurements. The available double-arm RP jet (jet or lepton) triggered sample corresponds to an integrated luminosity of $\sim$ 0.003 ($\sim$ 0.1) pb$^{-1}$. 

The CT-PPS (equivalent in ATLAS to the AFP project) adds precision proton tracking and timing detectors in the very forward region on both sides of CMS at about 220--240m from the IP, to study CEP in proton-proton collisions. At $\sqrt{s}$ = 13 TeV and in normal high-luminosity conditions, with the
CT-PPS detectors at 15$\sigma$ from the beam,
values of the central system mass $M_X \gtrsim 300$ GeV will be accessible.
Even with an average of 50 pile up events, the backgrounds can be suppressed by matching the reconstructed values of $M_{central}$ (in the central detector) and $M_{X}$ (in the CT-PPS), by requiring small charged multiplicity associated to the di-lepton vertex for the case of leptonic final states (i.e. $X = e^+e^-, \mu^+\mu^-, \tau^+\tau^-$ and $W^+W^-$), and by exploiting the proton timing constraint on the $z$-vertex position.

Finally, it should be noted that complementary measurements of rapidity gaps during special runs at low luminosity without proton tagging are also possible with ATLAS and CMS detectors alone.
 
 \section{QCD processes}\label{CEP:gluonex}
 
 In this Section, theoretical discussion of CEP processes that proceed via the strong interaction, and motivations for future measurements, are presented.
 
  \subsection{Introduction}\label{CEP:gluonex:intro}

\begin{figure}[ht]
\begin{center}
\includegraphics[scale=1.2]{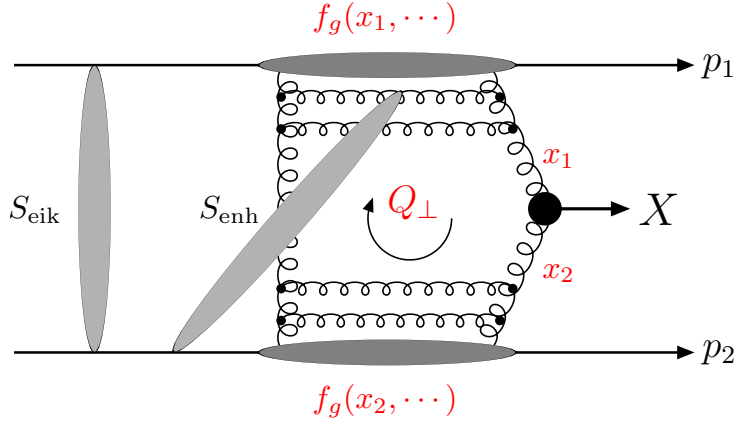}
\caption{The perturbative mechanism for the exclusive process $pp \to p\,+\, X \, +\, p$, with the eikonal and enhanced survival factors 
shown symbolically.}
\label{fig:pCp}
\end{center}
\end{figure}

The CEP process may be mediated purely by the strong interaction, in the language of Regge theory proceeding via double Pomeron exchange. In this case, and when the mass of the system, $X$, produced in the CEP reaction is sufficiently large, a perturbative QCD approach becomes applicable~\cite{Khoze:2001xm,Albrow:2010yb,Harland-Lang:2014lxa,Harland-Lang:2015rev}, and we may consider the two--gluon exchange diagram shown in Fig.~\ref{fig:pCp}. This approach, often referred to as the `Durham model', was developed in papers such as~\cite{Khoze:2000jm,Khoze:2001xm} and has undergone much development   in subsequent years; see~\cite{Harland-Lang:2014lxa} for a review and~\cite{Heinemeyer:2007tu,Pasechnik:2007hm,Coughlin:2009tr,Pasechnik:2009qc,d'Enterria:2013yra,Harland-Lang:2013xba} for some examples of further theoretical and phenomenological work. It represents a novel application of perturbative QCD, as well as requiring an account of soft diffractive physics. For such processes it is found that a dynamical selection rule operates~\cite{Harland-Lang:2014lxa,Kaidalov:2003fw}, where $J_z^{PC}=0^{++}$ quantum number states (here $J_z$ is the projection of the produced object angular momentum on the beam axis) are dominantly produced; this simple fact leads to many interesting and non--trivial implications for CEP processes, which are not seen in the inclusive case.
     
     Within this approach, the perturbative CEP amplitude is written as~\cite{Khoze:2001xm,Harland-Lang:2014lxa}
     \begin{equation}\label{bt}
  T=\pi^2 \int \frac{d^2 {\bf Q}_\perp\, \mathcal{M}}{{\bf Q}_\perp^2 ({\bf Q}_\perp-{\bf p}_{1_\perp})^2({\bf Q}_\perp+{\bf p}_{2_\perp})^2}\,f_g(x_1,x_1', Q_1^2,\mu^2; t_1)f_g(x_2,x_2',Q_2^2,\mu^2; t_2) \; ,
    \end{equation}
    where $\mathcal{M}$ is the color-averaged, normalised sub-amplitude for the $gg \to X$ process:
\begin{equation}\label{Vnorm}
\mathcal{M}\equiv \frac{2}{M_X^2}\frac{1}{N_C^2-1}\sum_{a,b}\delta^{ab}q_{1_\perp}^\mu q_{2_\perp}^\nu V_{\mu\nu}^{ab} \; .
\end{equation}
    Here $a$ and $b$ are color indices, $M_X$ is the central object mass, $V_{\mu\nu}^{ab}$ is the $gg \to X$ vertex, $q_{i_\perp}$ and $x_i$ are the transverse momenta and momentum fractions of the incoming gluons, respectively, and $x'_{i}$ are the momentum fractions of the screening gluon, which does not couple to the hard subprocess. The $f_g$'s in (\ref{bt}) are the skewed unintegrated gluon densities of the proton. These correspond to the distribution of gluons in transverse momentum $Q_\perp$, which are evolved in energy up to the hard scale $\mu \sim M_X$, such that they are accompanied by no additional radiation, as is essential for exclusive production. In the $x' \sim Q_\perp^2/s \lln x \sim M_X/\sqrt{s}$ region relevant to CEP, these can be expressed in terms of the conventional gluon PDFs, and a `Sudakov factor', $T_g$, which resums the logarithmically enhanced higher--order corrections, and corresponds to the (Poissonian) probability of no extra parton emission from each fusing gluon. This factor is essential in ensuring a perturbatively stable result~\cite{Coughlin:2009tr,Harland-Lang:2014lxa}.
    
    In addition to this amplitude (\ref{bt}) for the exclusive production of an object $X$ in a short--distance interaction, it is also necessary to include the probability that extra particles are not produced in additional soft proton--proton interactions (`rescatterings'), independent of the hard process, i.e. as a result of underlying event activity. This probability is encoded in the so--called `eikonal survival factor'~\cite{Bjorken:1992er,Ryskin:2009tk,Ostapchenko:2010gt,Khoze:2014aca,Gotsman:2014pwa}, $S^2_{\rm elk}$: while this is a soft quantity which cannot be calculated using pQCD, it may be extracted from hadronic data. Although there is some uncertainty in the precise level of suppression (in particular in its dependence on the c.m.s. energy $\sqrt{s}$), it is found to be a sizeable effect, reducing the CEP cross section by about two orders of magnitude. It is in addition expected that there may be some suppression due to rescatterings of the protons with the intermediate partons in the hard process. This is encoded in the so--called `enhanced' survival factor~\cite{Ryskin:2009tk,Martin:2009ku,Gotsman:2014pwa}: while this is expected to have a much less significant effect than the eikonal survival factor, the precise level of suppression remains uncertain and may be clarified by future CEP measurements.

     It is in principle possible to consider the CEP of any $C$--even particle which couples to gluons within this mechanism, and an important advantage of these reactions is that they provide an especially clean environment in which to investigate in detail the properties of a wide range of SM and BSM states~\cite{Khoze:2001xm,Kaidalov:2003ys,Heinemeyer:2007tu,HarlandLang:2010ep,Harland-Lang:2014lxa}. In addition, as described above, the theoretical framework is sensitive to both hard and, through the survival factors, soft aspects of QCD, as well as depending sensitively on the gluon PDF in the low $x$ and $Q^2$ region, where it is currently quite poorly determined from global fits. This process therefore provides a very promising framework within which to study various aspects of QCD, both perturbative and non--perturbative, as well as new physics at the LHC. 
     
Finally, we note that in the past other approaches have been taken to modelling the QCD--mediated CEP process discussed above. Most notably, the `Saclay' model~\cite{Boonekamp:2003wm}, an implementation of the calculation of~\cite{Bialas:1991wj}, did not include a Sudakov factor in the amplitude (\ref{bt}) but rather the low--$Q_\perp$ infrared--unsafe region was suppressed by the introduction of `non--perturbative' gluon propagators, the parameters of which were fitted to total and elastic cross section data (see e.g.~\cite{Forshaw:2005qp} for further discussion). This favours much lower average gluon $Q_\perp$ than in the Durham approach so that, it turns out, the $J_z=0$ selection rule discussed above would not necessarily hold, and in addition leads to a much gentler fall in the cross section with $M_X$, as well as generally much larger predicted cross sections. These latter predictions were found to be in strong disagreement with the CDF measurement of exclusive jet production~\cite{Aaltonen:2007hs} (which was on the other hand in good agreement with the Durham model predictions);  moreover, from a theoretical point of view the omission of the Sudakov factor, a crucial element in the perturbative calculation, lacks clear justification. For these reasons, such models are generally less used in current phenomenological work.

 \subsection{Forward proton tagging: phenomenological insight and advantages}\label{sec:tagcepmotglu}

In an exclusive reaction any transverse momentum $p_\perp=|\mathbf{p}_\perp|$ of the outgoing protons is transferred to the central object $X$. For this reason a measurement of the distributions with respect to such variables as the magnitude of the proton $p_\perp$ and the angle $\phi$ between the proton $p_\perp$ vectors (which is only possible with proton tagging detectors) is sensitive to the structure of the $gg\to X$ vertex, and the spin--parity of the produced object. Moreover, it is found that additional soft interactions, which generate the soft survival factor $S^2_{\rm eik}$, can have a very strong and model--dependent effect on these distributions.
     
\begin{figure}[t]
\begin{center}
\includegraphics[scale=0.75]{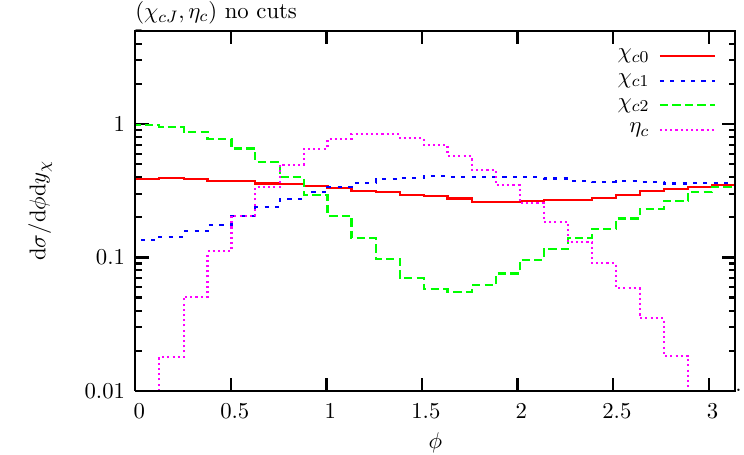}
\caption{Distribution~\cite{HarlandLang:2010ys} (in arbitrary units) within the perturbative framework of the difference in azimuthal angle of the outgoing protons for the CEP of different $J^P$ $c\overline{c}$ states at $\sqrt{s}=14$ TeV and rapidity $y_X=0$.}\label{surv1}
\end{center}
\end{figure}

     In general, we can write down an expression for the CEP cross section at $X$ rapidity $y_X$ as~\cite{HarlandLang:2010ep}
\begin{equation}\label{ampnew}
\frac{{\rm d}\sigma}{{\rm d} y_X}=\langle S^2_{\rm enh}\rangle\int{\rm d}^2\mathbf{p}_{1_\perp} {\rm d}^2\mathbf{p}_{2_\perp} \frac{|T(\mathbf{p}_{1_\perp},\mathbf{p}_{2_\perp})|^2}{16^2 \pi^5} S_{\rm eik}^2(\mathbf{p}_{1_\perp},\mathbf{p}_{2_\perp})\; ,
\end{equation}
where $T$ is given by (\ref{bt}), and $\langle S^2_{\rm enh}\rangle$ is the averaged `enhanced' survival factor discussed above, which is expected to depend very weakly on the proton $p_\perp$, and is therefore not relevant to the current considerations~\cite{Ryskin:2009tk,Martin:2009ku}. As mentioned above, the $\mathbf{p}_{\perp}$ dependence of the hard amplitude $T$ is strongly sensitive to the quantum numbers of the produced state. For example, for small $p_\perp$ it can be shown that we expect the squared amplitudes for the CEP of an object of spin--parity $J^P$ to behave as~\cite{Kaidalov:2003fw}
\begin{align}\label{R0p}
|{T}_{0^+}|^2 &\sim {\rm const.}\;,\\ \label{R1p} 
|{T}_{1^+}|^2 &\sim ({\bf p}_{1_\perp}-{\bf p}_{2_\perp})^2\;,\\ \label{R0m}
|T_{0^-}|^2 &\sim {\bf p}_{1_\perp}^2{\bf p}_{2_\perp}^2\sin^2{\phi}\;.
\end{align}
Such a behaviour is seen in Fig.~\ref{surv1}, which shows distributions with respect to the azimuthal angle $\phi$ at $\sqrt{s}=14$ TeV, for the case of $\chi_c$ and $\eta_c$ CEP within the Durham approach (very similar distributions are expected for the higher mass $\chi_b$, $\eta_b$): this effect is driven by the different Lorentz forms of the $gg\to X$ couplings, depending on the $J^P$ of the state $X$. A measurement of this distribution is therefore directly sensitive to the nature of the produced state, as well as more generally the structure of the production subprocess. It is moreover the case that the $J^{PC}=0^{++}$ selection rule discussed in Section~\ref{CEP:gluonex} is exact in the limit of exactly forward protons (i.e. $p_\perp=0$), and becomes weaker as the proton $p_\perp$ is increased. Within the Durham approach, it is found that 
\begin{equation}\label{simjz2}
\frac{|T(|J_z|=2)|^2}{|T(J_z=0)|^2}\sim \frac{\langle p_\perp^2 \rangle^2}{\langle Q_\perp^2\rangle^2}\;,
\end{equation}
where $\langle p_\perp^2 \rangle$ is the average squared proton transverse momentum, and $\langle Q_\perp^2\rangle \sim $ a few ${\rm GeV}^2$ is the average squared transverse momentum going round the gluon loop. Thus by selecting events with higher or lower proton $p_\perp$, the relative fraction of non--$J_z^P=0^+$ states can be enhanced or suppressed, respectively.

In addition, it can be seen from (\ref{ampnew}) that the eikonal survival factor depends on the proton ${\bf p}_\perp$ vectors. Physically, this is to be expected, as the survival factor cannot be a simple multiplicative constant, but will rather depend on the impact parameter of the colliding protons. Loosely speaking, as the protons become more separated in impact parameter, we should expect there to be less additional particle production, and so for the survival factor to be larger (consequently, the average survival factor is much larger in the case of photon--mediated processes, where larger impact parameters are favoured, when compared to QCD processes). As the transverse momenta ${\bf p}_{i_\perp}$ of the scattered protons are nothing other than the Fourier conjugates of the proton impact parameters, ${\bf b}_{it}$, this leads to the ${\bf p}_{i_\perp}$ dependence seen in (\ref{ampnew}).

\begin{figure}
\begin{center}
\includegraphics[scale=0.58]{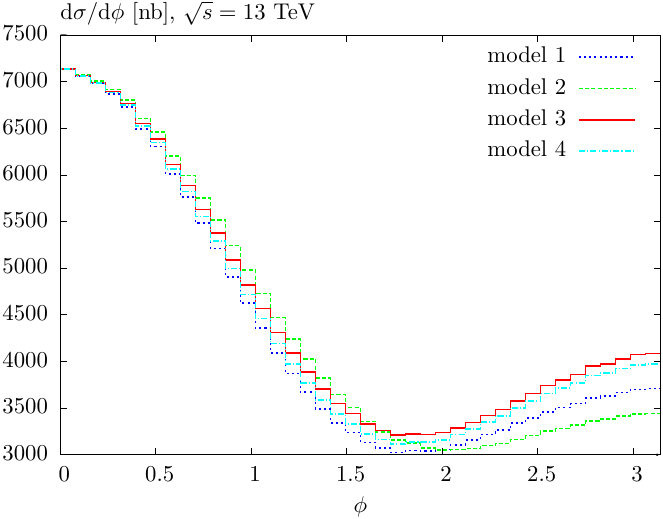}
\includegraphics[scale=0.58]{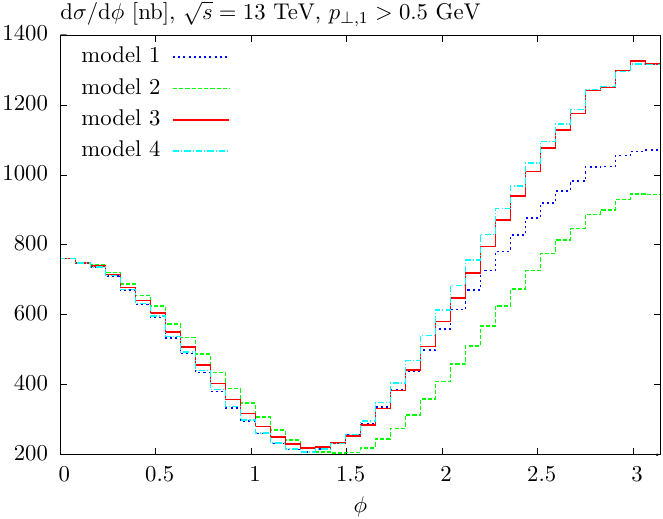}
\caption{Differential cross section ${\rm d}\sigma/{\rm d}\phi$ for the process $pp\rightarrow p + \pi^{+}\pi^{-} + p$, where $\phi$ is the azimuthal angle between the outgoing proton $p_\perp$ vectors, at the $\sqrt{s}=13$ TeV LHC, for the four soft models of~\cite{Khoze:2013dha}. In the left plot the proton $p_\perp$ is unconstrained, while in the right plot an additional cut of $p_\perp>0.5$ GeV is placed on one proton. In both cases, a cut of $|y_\pi|<2$ on the centrally produced pions is placed. For display purposes the predictions are normalized in the first $\phi$ bin, to the model 1 predictions. Plots from~\cite{Harland-Lang:2013dia} are made using \texttt{Dime} MC~\cite{Dime}.}\label{LHCphi}
\end{center}
\end{figure}
 
 In Fig.~\ref{LHCphi}  the $\phi$ distribution at the LHC ($\sqrt{s}=13$ TeV) for $\pi^+\pi^-$ CEP is shown, with four different models for the eikonal survival factor, as described in~\cite{Khoze:2013dha}. A very distinct `diffractive' dip structure is observed, with the distributions reaching a minimum at a particular value of $\phi$. This destructive interference is completely driven by the effect of these additional `screening' corrections which generate the soft survival factor. In particular, to account for soft survival effects   the CEP amplitude including rescattering effects, $T^{\rm res}$, should be calculated, by integrating over the transverse momentum ${\bf k}_\perp$ carried round the Pomeron loop (represented by the grey oval labeled `$S_{\rm eik}^2$' in Fig~\ref{fig:pCp}). The amplitude including rescattering corrections is given in the simplest approach by
\begin{equation}\label{skt}
\mathcal{M}^{\rm res}(s,\mathbf{p}_{1_\perp},\mathbf{p}_{2_\perp}) = \frac{i}{s} \int\frac{{\rm d}^2 \mathbf {k}_\perp}{8\pi^2} \;\mathcal{M}_{\rm el}(s,{\bf k}_\perp^2) \;\mathcal{M}(s,\mathbf{p'}_{1_\perp},\mathbf{p'}_{2_\perp})\;,
\end{equation}
where $\mathbf{p'}_{1_\perp}=({\bf p}_{1_\perp}-{\bf k}_\perp)$ and $\mathbf{p'}_{2_\perp}=({\bf p}_{2_\perp}+{\bf k}_\perp)$, while $\mathcal{M}^{\rm el}(s,{\bf k}_\perp^2)$ is the elastic $pp$ scattering amplitude in transverse momentum space, see for example~\cite{Khoze:2002nf,Harland-Lang:2014lxa} for more details. The `bare' amplitude excluding rescattering effects must be added to this to give the full result: it is the interference between this screened and the unscreened amplitude which generates these clear diffractive dips seen in Fig.~\ref{LHCphi}. For a particular value of $\phi$ this interference is strongest, resulting in the observed minimum in the $\phi$ distribution. As the form of the screened amplitude depends on the particular soft model, we may expect the position and depth of this minimum to be sensitive to this, as well as depending on the particular cuts imposed on the proton $p_\perp$. In fact, it appears from Figs.~\ref{LHCphi} that the position of the minimum does not depend too strongly on the choice of model, but nonetheless the overall shape of the $\phi$ distribution does show some variation. Thus a measurement of distributions with respect to $\phi$ (or the magnitude of the proton $p_\perp$, where similar dipping structure may be seen) could help differentiate between the available models of soft physics which are needed to calculate the survival factors. Although the case of $\pi^+\pi^-$ CEP is considered here, such diffractive dip structure is expected to be observable in any CEP process, such as exclusive jet or quarkonium production, see e.g.~\cite{HarlandLang:2010ys}.

  \subsection{Conventional quarkonium production}
  
  \subsubsection*{Motivation and theory}
  
The exclusive production of heavy quarkonium~\cite{HarlandLang:2009qe,HarlandLang:2010ep,Khoze:2004yb,Pasechnik:2009bq,Pasechnik:2009qc,Yuan:2001nu,Petrov:2004nx,Bzdak:2005rp,Rangel:2006mm} provides a valuable test of the QCD physics of bound states, with the predictions for the range of available $J^{PC}$ states exhibiting distinct features in the exclusive mode. The direct production channel can be easily selected, that is without feed--down contributions, and only the `core' color--singlet component of the state is probed, due to the requirement that no additional hadronic particles are present. 
    
Exclusive $\chi_{cJ}$ production  has been observed by both CDF at the Tevatron~\cite{Aaltonen:2009kg} and LHCb~\cite{LHCb:2011dra} at the LHC (see the following section for further details), and quite high production cross sections are expected: the Durham framework predicts total cross sections for the $\chi_{c0}$, $\chi_{c1}$ and $\chi_{c2}$ at $\sqrt{s}$ = 13 TeV of $\sim$ 340 nb, $\sim$ 8.0 nb and $\sim$ 4.4 nb, respectively~\cite{HarlandLang:2010ep}, with an uncertainty of about a factor 2--3. It is clear from these results that the cross sections for the three different spin states are predicted to follow a strong hierarchy: due to the $J^{PC}=0^{++}$ selection rule described above for the $\chi_{c2}$ (within the non--relativistic approximation), and due to the Landau--Yang theorem~\cite{Landau:1948kw,Yang:1950rg} for the $\chi_{c1}$, the cross sections for these higher spin states are expected to be at the level of a few percent of the $\chi_{c0}$ cross section. Such a suppression is not expected or seen in the inclusive mode~\cite{Aaij:2013dja}, where all three spin states are observed to give comparable contributions before branching. In the $\chi_c \to J/\psi\gamma$ decay channel, for which the $\chi_{c(1,2)}$ branching ratios are much higher, we should expect to see non--negligible contributions from all three states. Crucially, in the case of the LHCb data, it was possible to distinguish between the three different spin states, with results that were found qualitatively to support this expectation.

 However, there remain some open questions related to the $\chi_{c2}$, for which an apparent enhancement relative to theory expectations is seen by LHCb. As discussed in more detail in~\cite{Harland-Lang:2014lxa}, this may be due to proton dissociation not seen in the LHCb detector acceptance; while this is quite poorly understood theoretically, it is expected to preferentially enhance the higher spin states, in particular the $\chi_{c2}$.  Alternatively, such an enhancement may be due to additional `non--perturbative' corrections in the theory calculation, as the mass scale of the $\chi_c$ may be too low to allow a purely perturbative approach, which assumes that $M_X \sim M_\chi \gg Q_\perp$, as well as to relativistic corrections to the $\chi_c$ wave function. This issue can only be fully clarified with further higher statistics data from the LHC, with for example the HERSCHEL system (see Chapter~\ref{chap:detectors}), not used in any existing measurements, being a particularly effectively way to reduce the effect of proton dissociation at LHCb. Measurements with the CMS-TOTEM and ALFA in runs at high $\beta^{*}$, for which the proton tagging detectors can effectively eliminate the effect of proton dissociation, would also be very useful. Such data would give a much cleaner comparison with theory, and could be sensitive to any transition to a non--perturbative regime for these lower $\chi_c$ masses, where a Regge theory based approach can be taken.
      
An observation of the higher mass bottomonium $\chi_b$ states, for which the mass scale is safely in the perturbative regime, would provide a more stringent test of the theory. The predicted $\chi_{b0}$ cross is $\sim 100$ pb at $\sqrt{s}=14$ TeV~\cite{HarlandLang:2010ep}, and a similar hierarchy in spin states to the $\chi_c$ case is predicted, but with a negligible $\chi_{b1}$ cross section due to the higher mass. It is also worth noting that the spin assignments of the $P$--wave $\chi_{bJ}$ states still need experimental confirmation~\cite{Agashe:2014kda}, and so this is an issue which the spin--parity selecting properties of CEP could shed light on.
      
 Other observables of interest include the $\chi_c$ states via two body ($\pi\pi$, $KK$...) decays, for which the exclusive continuum background is expected to be manageable~\cite{HarlandLang:2012qz,Lebiedowicz:2011cw}. The CEP of the odd--parity $\eta_{c,b}$ states, for which the cross sections are predicted to be similarly suppressed to the higher spin $\chi_{c,b}$ states, would also represent a further potential observable. As discussed in Section~\ref{sec:tagcepmotglu}, the distributions of the outgoing protons are expected to be highly sensitive to the spin--parity of the produced quarkonium state, as well as to the soft survival factors. Finally, exclusive photoproduction of $C$--odd quarkonia ($J/\psi$, $\psi(2S)$, $\Upsilon$...) is of much interest; this is discussed in more detail in Section~\ref{CEP:phoex}.
     
     \subsubsection*{Experimental results and outlook}

     A favourable decay mode of the $\chi_c$ meson is to $J/\psi\gamma$, with the
only significant experimental background being contamination from $\psi(2S)\rightarrow J/\psi \pi^0\pi^0$
where only one photon is identified from the subsequent pion decays.

\begin{figure}[ht]
\centerline{\includegraphics[width=0.49\linewidth]{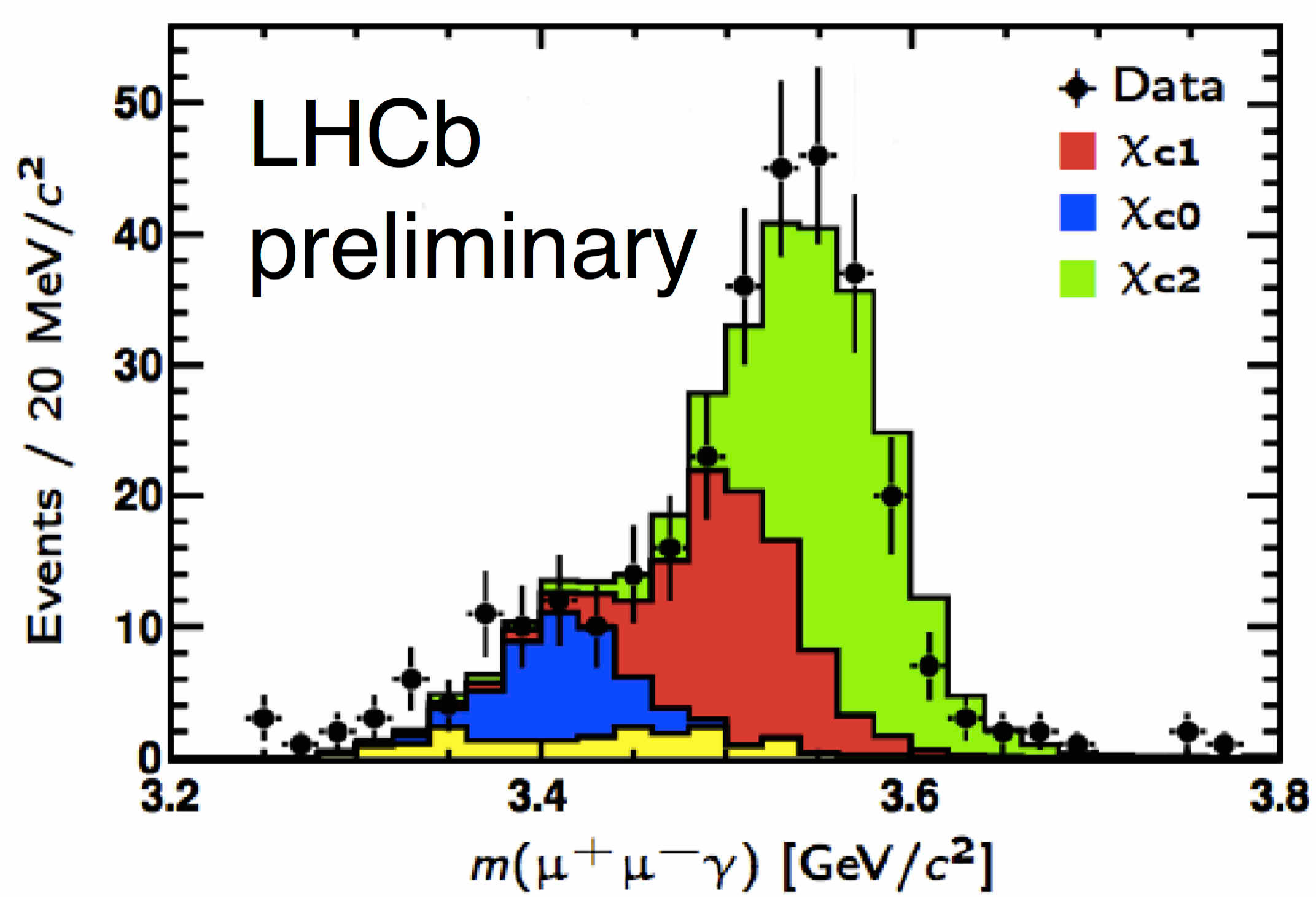}}
\caption{
Invariant mass of the di-muon plus photon system in events having no other activity
inside LHCb.
\label{fig:chic}}
\end{figure}

LHCb has made preliminary measurements~\cite{LHCb:2011dra} of the
production of $\chi_c$ mesons with $37 {\rm\ pb^{-1}}$ of data.
The selection of events proceeds as for the $J/\psi$ selection in Sec.~\ref{sec:photooutlook}
but now one (rather than no) photon candidate is required.  The invariant mass of the di-muon plus photon system
is shown in Fig.~\ref{fig:chic}  fitted to expectations from the SuperChic simulation~\cite{superchic,HarlandLang:2009qe} for 
$\chi_{c0},\chi_{c1}.\chi_{c2}$ signal contributions and the $\psi(2S)$ background.
The CDF collaboration made the first observation~\cite{Aaltonen:2009kg} of CEP of $\chi_c$ mesons but because
of the limited mass resolution, assumed it all to consist of $\chi_{c0}$ mesons.
The mass resolution of LHCb is sufficiently good to distinguish the three states.
In this decay mode, the contribution from
$\chi_{c2}$ dominates although much of that is due to the higher branching fraction
for this state to decay to $J/\psi\gamma$.
Unfortunately, the resolution is not
good enough to separate the three states completely and so
the fraction of the sample that is exclusively produced is determined for the
whole sample and is estimated to be
$0.39\pm0.13$ using the $p_T$ of the reconstructed meson.
The cross sections times branching fractions are measured to be
$9\pm5,16\pm9, 28\pm 12$\ pb for $\chi_{c0},\chi_{c1},\chi_{c2}$, respectively,
slightly higher but in reasonable agreement with the 
theoretical predictions of 4, 10, 3 pb for purely exclusive production.  
Only the relative cross sections for $\chi_{c2}$ to $\chi_{c0}$ of $3\pm 1$ appears to be somewhat 
higher
in the data than the theory expectation that they are roughly equal. This is consistent with the CDF 
 measurement of $\pi^+\pi^-$ CEP~\cite{Aaltonen:2015uva}, where a limit on the $\chi_{c0}\to \pi^+\pi^-$ cross section is set which indicates that less than $\sim 50\%$ of the previously observed $\chi_c \to J/\psi \gamma$ events at the Tevatron~\cite{Aaltonen:2009kg} are due to the $\chi_{c0}$. As discussed above, one possible reason for this discrepancy is that the fraction of elastic exclusive events in the sample
differs for each of the three resonances.
With greater statistics, a more sophisticated fit can be performed in order to estimate the
fraction of exclusive events separately for each $\chi_c$ state.

Further discrimination of the $\chi_c$ states is possible by considering different decay modes.
Of particular interest are the decays to two pions or two kaons, which are not possible for
$\chi_{c1}$ and are about four times higher for $\chi_{c0}$ than for $\chi_{c2}$.
In addition, the mass resolution in this channel is about a factor of three better than in the
$\mu\mu\gamma$ channel.
Making use of their ability to trigger on hadronic objects with low transverse momentum
during the $\sqrt{s}=$7 and 8 TeV running,
LHCb has collected a large sample of low-multiplicity data in single proton-proton interactions
(without pile-up) corresponding to an integrated luminosity of about 600 pb$^{-1}$.
Consequently, the observation of $\chi_c$ states in the $\pi\pi$ and $KK$ modes ought to be possible, so long as the backgrounds from the double pomeron exchange production of pairs of pseudo-scalar mesons is not too large.

 In addition to this rapidity gap based analysis there are also possibilities for future LHC measurements with tagged forward protons.  With CMS-TOTEM and ALFA, the different $\chi_c$ states can be easily separated in charged-particle-only final states, and the proton dissociation background can be eliminated using proton tagging. A preliminary analysis has been performed on the data of the common CMS-TOTEM $\beta^*$ = 90 m run at $\sqrt{s}$ = 8 TeV in July 2012. The available data set contains a few $\chi_{c}$ exclusive 
candidates, consistent with the CDF and LHCb measurements. 
In the case that the $\chi_c$ decays to 
two- or three-$\pi^+\pi^-$ pairs or to $K^+K^-\pi^+\pi^-$, the tracker dE/dx can be used to confirm the pion or kaon hypothesis. This combined with higher branching ratios compared to the $\pi^+\pi^-$ and $K^+K^-$ final states, where no particle identification is possible, makes these three decay modes the most promising in terms 
of signal-to-background ratio. In 5 pb$^{-1}$ 
of integrated luminosity, more than 1000 $\chi_{c0}$ candidates are expected in each of the decay modes ($2(\pi^+\pi^-)$, $3(\pi^+\pi^-)$, $K^+K^-\pi^+\pi^-$). This will allow a good determination of the cross section$\times$branching ratio as well as a detailed study of the 
azimuthal angular difference $\Delta\phi$ of the outgoing protons for each decay mode separately. According to the predictions of~\cite{HarlandLang:2010ep}, 5 pb$^{-1}$ does not seem be sufficient to be able to study exclusive $\chi_{c1}$  and $\chi_{c2}$ production. However, given the possible discrepancy regarding exclusive $\chi_{c2}$ production between the LHCb measurement discussed above, it might well be that the exclusive $\chi_{c2}$ production cross section at $\sqrt{s}$ = 13 TeV is almost an order of magnitude higher than predicted and hence feasible to study even with an integrated $\beta^*$ = 90 m luminosity of $\sim$ 5 pb$^{-1}$.  Such an observation (or the absence of it) would provide valuable information about the model of~\cite{Harland-Lang:2014lxa}, and its application to $c\overline{c}$ systems.
     
     \subsection{`Exotic' quarkonium production}
     
      \subsubsection*{Motivation and theory}
     
     In addition to conventional quarkonia states, there are possibilities for the observation and study of `exotic' charmonium--like states, which have been discovered over the past 10 years~\cite{Brambilla:2010cs}. In some cases, the $J^{PC}$ quantum numbers of these states have not been determined experimentally and often a range of interpretations are available: a $D^0\overline{D}^{*0}$ molecule, tetraquarks, $c\overline{c}g$ hybrids, the conventional $c\overline{c}$ charmonium assignment, and more generally a mixture of these different possibilities. Considering the CEP of such objects, then the effect of the $J_z^{P}=0^{+}$ selection rule, as well as a measurement of the distribution of the outgoing proton momenta, may help to fix the quantum numbers of the centrally produced system. 
       
     One possibility is the  CEP of the $Y(3940)$, in particular via the $J/\psi \omega$ channel, which could help to resolve current uncertainties~\cite{Albuquerque:2013owa,Sreethawong:2013qua} in the interpretation of this state.  Another particularly topical example is the $X(3872)$, for which the quantum numbers are determined~\cite{Aaij:2013zoa} to be $J^{PC}=1^{++}$, but a concrete interpretation remains elusive. In the case of a dominantly $D^0\overline{D}^{*0}$ interpretation, the hadroproduction of such a state with the size of cross section observed~\cite{Chatrchyan:2013cld} in the $X(3872)$ case, if possible at all, should in general take place in an environment where additional particles are emitted~\cite{Artoisenet:2009wk,Bignamini:2009sk}, so that the initially produced short--distance $c\overline{c}$ pair can form the loosely--bound $D^0\overline{D}^{*0}$ state. The observation of the $X(3872)$ in the exclusive mode, via for example the $X(3872)\rightarrow J/\psi\pi^+\pi^-$ decay channel, where any additional hadronic activity is vetoed on, would therefore strongly disfavor such a dominantly molecular $D^0\overline{D}^{*0}$ interpretation. The $X(3872)$ may instead be dominantly a  conventional $\chi_{c1}(2{}^3 P_1)$ state, in which case the cross section is expected to be of a comparable size to the ground--state $\chi_{c1}$, which has already been observed by LHCb. If, as may be more realistic~\cite{Voloshin:2007dx}, it is a mixture of a $\chi_{c1}(2P)$ and a molecular $D^0\overline{D}^{*0}$ state, then the size of this ratio will also  be driven by the probability weight of the purely $c\overline{c}$ component; if this is small, that is the molecular component is dominant, then the $X(3872)$ cross section will be suppressed relative to the $\chi_{c1}(1P)$.
       
 \subsubsection*{Experimental results and outlook}
 
LHCb is well positioned to observe exotic states with $J/\psi$  mesons in the final state.
About 600 pb$^{-1}$ of data without pile-up was taken in Run--I, and 1.5 ${\rm fb}^{-1}$ is expected in Run--II.  High efficiency triggering and reconstruction of $J/\psi$ mesons
has been demonstrated~\cite{Aaij:2014iea}, and the reconstruction of other final state particles, including $\pi^0$,  can be
performed provided the transverse momentum of the final state charged and neutral objects
is greater than about 100 MeV. This measurement will be complementary to the ones performed at low diffractive masses by the CMS-TOTEM and ALFA experiments.
       
 \subsection{Photon pair production}\label{sec:gamgam}

 \subsubsection*{Motivation and theory}

The CEP of a pair of photons ($\gamma\gamma$) produced via an intermediate quark loop represents an experimentally clean test of the perturbative CEP mechanism and is less sensitive to some of the theoretical uncertainties which are present, for example, in the case of $\chi_c$ CEP, due to the higher invariant masses, $M_X=M_{\gamma\gamma}$, which are accessible here. A measurement of the differential cross section with respect to $M_{\gamma\gamma}$ is a particularly useful observable, sensitive to the effect of the Sudakov factor, as well as the gluon PDF and the theoretically challenging `enhanced' survival factor $S^2_{\rm enh}$. 

\begin{figure}[t]
\begin{center}
\includegraphics[scale=0.65]{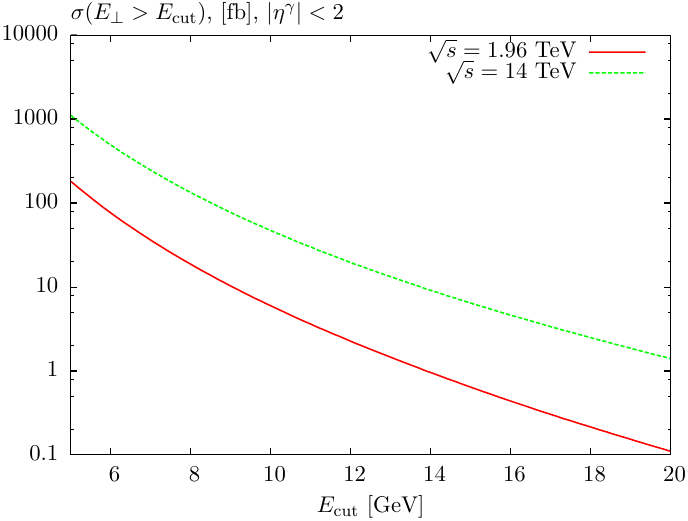}
\includegraphics[scale=0.65]{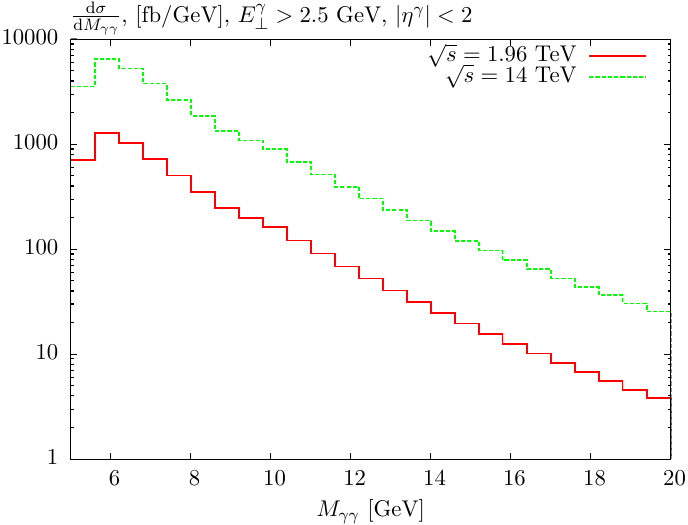}
\caption{Cross sections for $\gamma\gamma$ CEP at $\sqrt{s}=1.96$ and $14$ TeV~\cite{Harland-Lang:2014lxa}, calculated using MSTW08LO PDFs~\cite{Martin:2009iq} as a function of the cut on the photon transverse energy $E_\perp>E_{\rm cut}$ and invariant mass distribution ${\rm d}\sigma/{\rm d}M_{\gamma\gamma}$ for $E_\perp>2.5$ GeV. In both cases the photon pseudorapidity is required to lie within $|\eta^\gamma|<2$. Predictions made using the \texttt{SuperChic}~\cite{superchic} MC.}\label{gam1}
\end{center}
\end{figure}

In Fig.~\ref{gam1} some representative predictions for $\gamma\gamma$ CEP at the LHC and Tevatron are shown. Already exclusive $\gamma\gamma$ data have been taken: the 2011 CDF observation~\cite{Aaltonen:2011hi} of 43 $\gamma\gamma$ events in $|\eta(\gamma)|<1.0$ with no other particles detected in $-7.4<\eta<7.4$ was found to be in reasonable agreement with the Durham prediction. As can be seen from Fig.~\ref{gam1} the predicted cross sections are somewhat larger at the LHC, due to the larger gluon density at the lower $x$ values probed. Further tests at the higher $\sqrt{s}$ of the LHC would be very useful: by considering for example the ratio of cross sections at different $\sqrt{s}$ values various theoretical uncertainties decrease, and the energy dependence of the cross section, driven primarily by the survival factors and the $x$ dependence of the gluon density, are probed.

   \subsubsection*{Experimental results and outlook}

As discussed above,  exclusive $\gamma\gamma$ data has been taken by the CDF collaboration at the Tevatron. More recently, at the LHC, CMS~\cite{Chatrchyan:2012tv} has presented a search for exclusive $\gamma\gamma$ events using 36 pb$^{-1}$ of data collected at $\sqrt{s}=7$ TeV. The trigger and analysis methods were identical to the CMS $\gamma\gamma \rightarrow ee$ measurement, apart from the selection of two photons instead of two electrons. While no candidate diphoton events were observed, the corresponding limits were close to the theoretical predictions, providing a strong motivation for further searches at the LHC in the future. LHCb will be able to measure $\gamma\gamma$ CEP for $E(\gamma)>2$ GeV
with about 1.5 ${\rm fb}^{-1}$ of Run--II data,
using a new di-photon trigger that builds on the experience of the low multiplicity triggers
used in Run--I. 

Such a measurement could also be performed during Run--II during dedicated high $\beta^{*}$ runs by ALFA or CMS-TOTEM~\cite{Fichet:2014uka}. Due to the low instantaneous luminosity of those special runs
it should be possible to implement a dedicated di-photon trigger with
$p_{T}$ thresholds as low as
$p_{T1,2}>$ 5 GeV, in which case we can expect $\sim 10$s of events, for a typical integrated luminosity of
0.1 fb$^{-1}$.

     \subsection{Light meson pair production}\label{sec:lightmeson}
     
  \subsubsection*{Motivation and theory}

     Another interesting CEP process is the production of light meson pairs~\cite{HarlandLang:2011qd,Harland-Lang:2013qia} \linebreak[4]($X=\pi\pi, KK, \rho\rho, \eta(')\eta(')$). At sufficiently high meson transverse momentum $k_\perp$, a perturbative approach combining the Durham model with the `hard exclusive' formalism~\cite{Brodsky:1981rp,Benayoun:1989ng} to evaluate the meson production subprocess may be taken. The basic idea of the latter approach is that the hadron--level amplitude can be written as a convolution of a (perturbatively calculable) parton--level amplitude, $T$, and a `distribution amplitude' $\phi$, which contains all the (non--perturbative) information about the binding of the partons in the meson. It has been shown that within this approach the distinct features of the relevant parton--level helicity amplitudes $gg \to q\overline{q} q\overline{q}$, $q\overline{q}gg$, $gggg$ lead to some highly non--trivial predictions. 
     
\begin{figure}
  \begin{center}
  \includegraphics[scale=0.65]{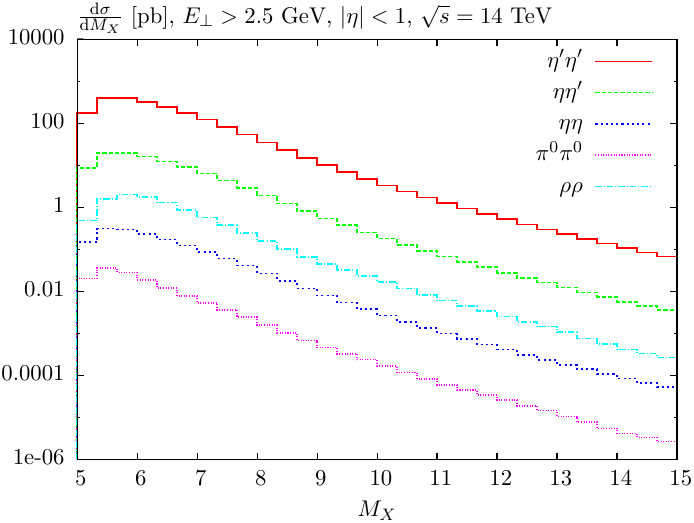}
  \caption{Differential cross section ${\rm d}\sigma/{\rm d} M_X$ for the CEP of meson pairs, for meson transverse energy $E_\perp>2.5$ GeV and pseudorapidity $|\eta|<1$. Predictions made using \texttt{SuperChic} MC~\cite{superchic}.}\label{vsm}
  \end{center}
  \end{figure}

     In particular, in the case of flavour non--singlet mesons ($\pi\pi,KK$...), it is found that there is a strong suppression in the CEP cross section, due to the vanishing of the parton--level production amplitudes for $J_z=0$ incoming gluons and the $J_z=0$ CEP selection rule~\cite{HarlandLang:2011qd}. In the case of flavour--singlet mesons ($\eta(')\eta(')$...) a different configuration of the outgoing partons can enter, with the effect that the $J_z=0$ amplitudes do not vanish, and so the corresponding CEP cross sections are expected to be much larger. The flavour--singlet cross sections may even be sensitive to the size of the gluonic component of the $\eta'$ (and, through mixing, $\eta$), via the valence $gg$ contributions to the $gg \to \eta(')\eta(')$ amplitudes~\cite{Harland-Lang:2013ncy}. Currently the long--standing issue concerning the extraction of the gluon content of the $\eta'$ (and $\eta$) remains uncertain, in particular due to non--trivial theory assumptions and approximations that must be made, as well as the current experimental uncertainties and limitations~\cite{Thomas:2007uy,DiDonato:2011kr}. It has in particular been shown that even a small gluonic component could lead to a sizeable increase in the predicted $\eta(')\eta(')$ CEP cross sections~\cite{Harland-Lang:2013ncy}. A representative invariant mass distribution, for typical central LHC cuts (the results are similar for forward production, as in the case of LHCb), are shown in Fig.~\ref{vsm}, and the expected enhancement in the $\eta'\eta'$ cross section is clear. Also shown are predictions for the vector $\rho\rho$ case.
     
\begin{figure}[htbp]
      \centering
      \includegraphics[width=0.35\textwidth]{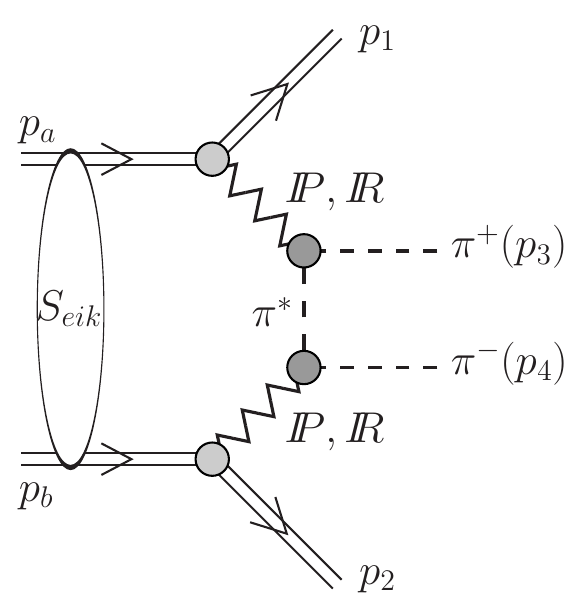}
      \hspace{0.1\textwidth}
      \includegraphics[width=0.35\textwidth]{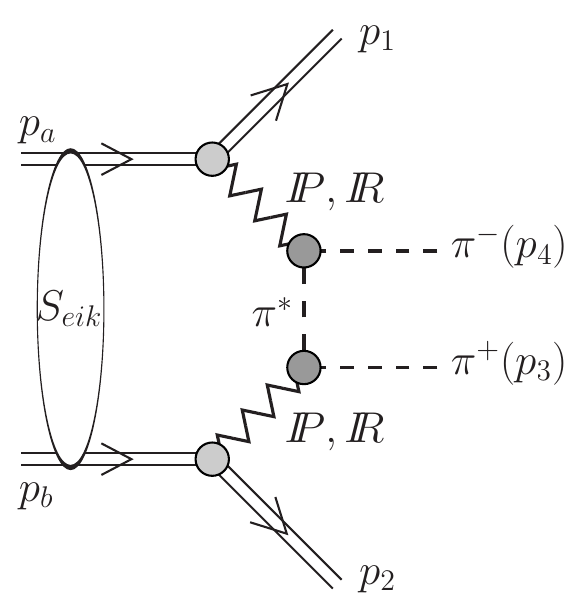}
      \caption{\label{fig:central_double_diffraction_diagrams}
      The `non--perturbative' mechanism for the CEP of
      $\pi^{+}\pi^{-}$ pairs. }
    \end{figure}

     As the meson transverse momentum $k_\perp$ and/or meson pair invariant mass decreases, we will expect to enter a regime where a Regge--theory inspired approach is more applicable~\cite{Lebiedowicz:2012nk,Harland-Lang:2013dia}. This mechanism of production is shown in Fig.~\ref{fig:central_double_diffraction_diagrams} for the case of non-resonant $\pi^{+}\pi^{-}$ production. In this case, it has been shown that for example $\pi^+\pi^-$ CEP can serve as a probe of the tools of Regge theory, and of the uncertain question of the transition to the perturbative regime discussed above. In addition, the observation of light meson pairs with tagged protons can serve as a detailed probe of the models of soft physics which are used to calculate the soft survival factors (see Section~\ref{sec:tagcepmotglu}).
     
       \subsubsection*{Experimental results and outlook}

An important possible background to the $\gamma\gamma$ CEP process discussed in Section~\ref{sec:gamgam} is the exclusive production of a pair of $\pi^0$ mesons, with one photon from each $\pi^0$ decay undetected or the two photons merging. At first sight it would appear that the cross section for this purely QCD process may be much larger than the $\gamma\gamma$ cross section and so would constitute an appreciable background. In the CDF measurement~\cite{Aaltonen:2011hi} of $\gamma\gamma$ CEP, despite previous hints of a non--negligible $\pi^0\pi^0$ contribution in earlier data~\cite{Aaltonen:2007am}, of the 43 candidate $\gamma\gamma$ events, the contamination caused by $\pi^0\pi^0$ CEP was observed to be very small ($< 15$ events, corresponding to a ratio $N(\pi^0\pi^0)/N(\gamma\gamma)<0.35$, at 95\% CL). This supports the prediction discussed in the previous section, namely that there is a strong dynamical suppression in the production cross section for pairs of flavour--non--singlet mesons such as $\pi^0\pi^0$; without this dynamical suppression the expected cross section would be $\sim$ 2 orders of magnitude higher. Nonetheless, this prediction, and in particular the expected hierarchy in production cross sections between flavour--non--singlet and singlet meson pairs has yet to be confirmed or disproved experimentally, by a direct observation safely in the perturbative region; this would represent a very interesting observation to be made at the LHC. 

On the other hand, in the region of lower system invariant masses and (more significantly) meson transverse momenta, where the tools of Regge theory should be applied, a variety of data exist at RHIC, the Tevatron and LHC. CDF data on $\pi^+\pi^-$ production at $\sqrt{s} = 900$ and 1960 GeV have been presented in~\cite{Aaltonen:2015uva},  and a preliminary measurement of $\pi^+\pi^-$ and $K^+K^-$ production has been performed by CMS~\cite{denterria}.

ALICE has taken data on central diffractive pion and kaon pair production~\cite{Schicker:2014wvk}. During Run--I, candidate CEP events were selected offline from data taken with
a minimum-bias trigger (at least one charged particle anywhere in 8 units of
pseudorapidity) by requiring an activity in the central barrel ($|\eta| < 0.9$)
surrounded with large gaps (up to $|\eta| \simeq 4.5$) on both sides.  While the statistics were quite low, qualitatively
quite pronounced resonant structures were observed for masses below 2~GeV in both the $\pi\pi$
and $KK$ channels. Some preliminary results can be found in~\cite{ALICE_CP_thesis}. In addition, in the four--track event sample an indication of interesting features is observed, relating to the fragmentation of
produced system into the $4\pi$ and $2\pi 2 K$ final states.

 LHCb has triggered on low multiplicity $\pi\pi$ and $KK$ systems when the transverse momentum of the hadrons was greater than about 500 MeV.  In Run--II, the trigger threshold will be lowered, allowing an investigation of the region around the $f_0(980)$.  Studies are ongoing on $\rho\rho$ and $\eta(')\eta(')$ states using Run--I data, and these will allow dedicated triggers to be installed during Run--II in order to collect high statistics samples.

    \begin{figure}[htbp]
      \centering
      \includegraphics[width=0.5\textwidth]{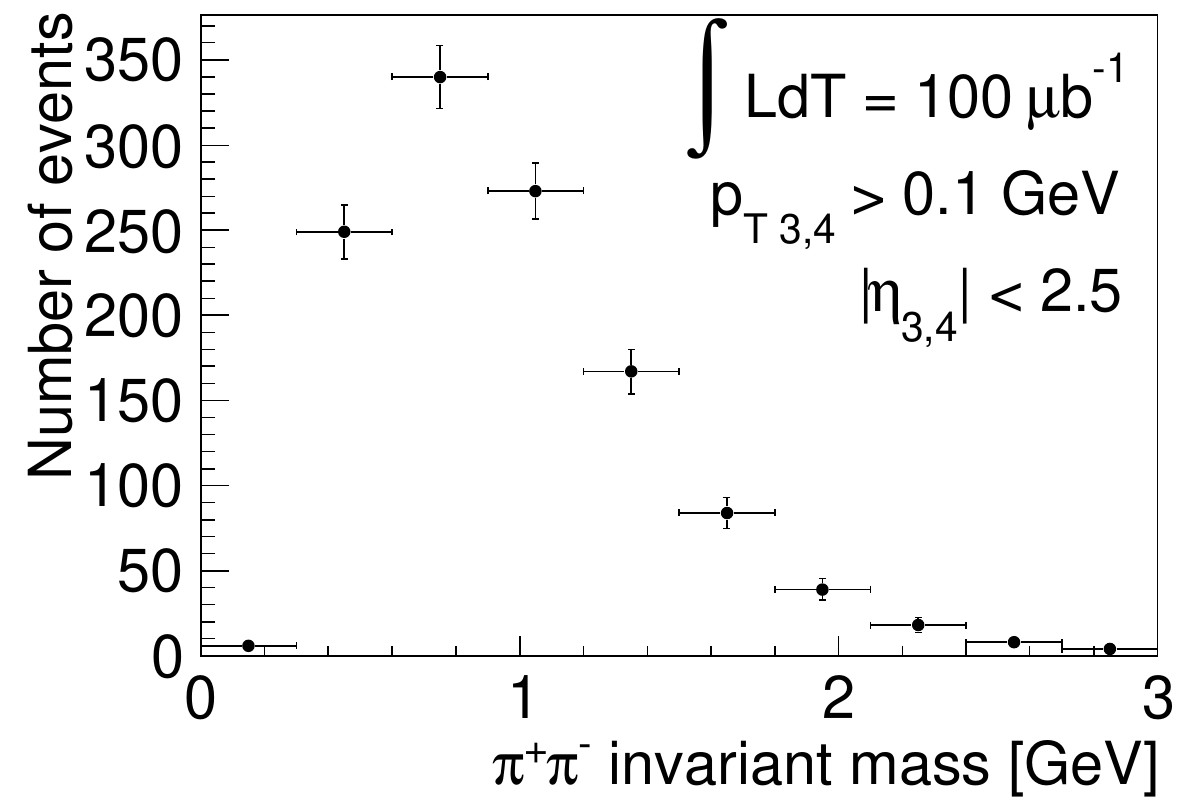}
      \caption{Simulated measurement of the $\pi^+\pi^-$ invariant mass
        distribution. The pions are measured with the ATLAS tracking detector and
        protons with the ALFA stations. An integrated luminosity of 100~$\mu$b$^{-1}$
      is assumed and only the statistical errors are plotted.}
      \label{fig:M}
    \end{figure}
    
In addition to these measurements, selected with rapidity gap based techniques, the possibility of observing exclusive meson pair production with tagged protons is of much interest and has been the subject of detailed studies. The STAR collaboration at RHIC~\cite{Adamczyk:2014ofa} have reported a preliminary measurement of $\pi^+\pi^-$ CEP with tagged protons, in broad agreement with Regge expectations. 
At the LHC, in~\cite{Staszewski:2011bg} the case of exclusive $\pi^+\pi^-$ is considered within the context of the ATLAS + ALFA detectors (similar possibilities exist with the CMS + TOTEM detectors). While a value of $\sqrt{s}=7$~TeV is taken in this study, these results give a very good indication of the measurement possibilities at the higher $\sqrt{s}$ values relevant to Run-II.  Fig.~\ref{fig:M}, taken from~\cite{Staszewski:2011bg},  presents a possible measurement of the $\pi^+\pi^-$
    invariant mass that could be performed with 100~$\mu$b$^{-1}$ of integrated
    luminosity (30~hours of data acquisition time assuming the luminosity value of
    $10^{27}$~cm$^{-2}$s$^{-1}$) for the experimental conditions assumed above. This corresponds to the non--resonant model depicted in Fig.~\ref{fig:central_double_diffraction_diagrams}: there will be a range of resonance ($\rho_0$, $f_0$, $f_2$) structure in addition to this continuum contribution.
         
     \subsection{Production of low mass resonances and glueballs}

\subsubsection*{Motivation and theory}

The CEP process effectively turns the LHC into a gluon-gluon collider and provides an excellent opportunity to study gluon systems  
with a longitudinal momentum fraction $x \sim$ 10$^{-4}$ and, in particular, to search for glueball candidates\footnote{It should be noted that the final--states for the processes considered in this section are often the same as those discussed in Section~\ref{sec:lightmeson}, where the case of non--resonant light meson pair production is considered.}. Glueballs are 
predicted by QCD as gluon bound states with no valence quark content.  The absence of valence quarks, in 
combination with the $J_z^{PC}$ = 0$^{++}$ selection rule~\cite{Kaidalov:2003fw}, makes CEP reactions an ideal place to search for them. QCD lattice calculations foresee a $J^{PC}$ = 0$^{++}$ ground state and a 2$^{++}$ state followed by a spectrum of excited states \cite{Mathieu:2008me, Chen:2005mg}. The $f_0(1500)$ or the $f_0(1710)$ are generally regarded as potential glueball $0^{++}$ states since one of them is in excess to the meson $SU(3)$ multiplet and both are compatible with a glueball in terms of mass, spin, parity, and decay channels (having e.g. a suppressed $\gamma\gamma$ mode). Recent unified lattice calculations~\cite{Morningstar:1999rf, Chen:2005mg} predict the $0^{++}$ glueball at $\sim$~1700 MeV  within $\sim100\,$MeV overall uncertainty (statistical and systematic), thus favouring the $f_0(1710)$ as a glueball candidate. Formerly proposed meson--glueball mixings~\cite{Kirk:2000ws,Kirk:1999fy,Kirk:1999df} relied on an incorrect mass hierarchy ($u\bar{u}$, $d\bar{d}$, $gg$, $s\bar{s}$) and have been made obsolete by further calculations~\cite{Lee:1999kv, Cheng:2006hu} based on the correct ($u\bar{u}$, $d\bar{d}$, $s\bar{s}$, $gg$) mass hierarchy giving a $\gtrsim95\,\%$  glueball purity at 
 $\sim$~1700 MeV. Measuring the CEP of a resonance as well as its decay 
branching ratios ~\cite{Mathieu:2008me, Ochs:2013gi} could help to establish its glueball nature.
  
The WA102 experiment~\cite{Kirk:2000ws,Kirk:1999fy,Kirk:1999df} reported that the $f_0(1710)$ branching ratio into $K^+K^-$ exceeded its branching ratio to $\pi^+\pi^-$, contrary to the case of the $f_0(1500)$, disfavouring a glueball interpretation of the $f_0(1710)$. This measurement leads to the conclusion that the coupling of the  $f_0(1710)$ to $s$--quarks must be larger than the $u$,$d$-quark couplings, a result which is not expected for a glueball state (although in~\cite{Chanowitz:2005du} a possible coupling to quark mass for the decays of pure gluonic states is noted). Moreover, the predicted decay mode into $\rho\rho$ has not been observed so far. An observation of the $f_0(1710)\rightarrow\rho\rho$ decay at the LHC would, in addition to being the first observation of this mode, alter the  $K \bar{K}$ vs `pionic' branching ratios and therefore change the expected couplings to $u$,$d$-quarks vs $s$-quark to values more consistent with those expected for glueballs. In relation to the measurement of the decay to $K^+K^-$, it would also bring additional knowledge about the coupling to quark masses.

The $f_0(1710)$ mass measurements (consistently pointing to a 1700--1710~MeV mean value within uncertainties) do not allow the Particle Data Group (PDG) to do a reliable average due to the systematically shifted measurements by BELLE and BES. Currently, the most precise existing measurement gives 1701~MeV from ZEUS~\cite{Chekanov:2008ad}. A high precision measurement at the LHC could give the decisive word about the $f_0(1710)$ mass. The $f_0(1710)$ has been interpreted in the past as an $f_2$ by several experiments, although it has been consistently found to be an $f_0$ by modern experiments and the issue is considered solved~\cite{Agashe:2014kda}. However a thorough spin analysis is mandatory for any $f_0(1710)$ measurement to confirm the quantum numbers of the measured resonance, as well as to cross-calibrate the purity of the event selection with the mass measurement. While for the $f_0(1500)$ the yields, decay channels and branching ratios have been extensively measured, the $f_0(1710)$ branching ratios are controversial in the literature, being largely unknown, and the main decay channels are described as `seen' by the PDG~\cite{Agashe:2014kda}. As already mentioned, allowed decay modes such as $\rho\rho$ have never been observed. A systematic and quantitative study of the decay modes and other properties of the $f_0(1710)$ can be performed at the LHC via CEP.

Former experiments (ISR, SPS, WestArea,\dots) did not have sensitivity to the $\rho\rho$ decay, due to their limited reach in invariant mass ($\lesssim$ 1.5~GeV) and/or to the $4\pi$ final state, and those few that had the sensitivity e.g.~\cite{Breakstone:1993ku}, had the analysis faked by the old assumption that the $f(1710)$ was an $f_2$ (as wrongly measured by several previous experiments at that time and also reported by the PDG). Attempts from modern experiments (FNAL, LHC, RHIC) lacked either the purity due to the absence of double proton tagging or the mass resolution for the two charged particles final states. The unique characteristics of the LHC are a $\sqrt{s}$ such that $\sim1$--$10\,$~GeV invariant masses can be produced diffractively with $\xi_{1,2}\sim10^{-3}$--$10^{-4}$, ensuring purely gluonic exchanges. 

 While the quantum numbers of glueballs that can be \emph{singly} produced in exclusive double pomeron exchange are dominantly $J^{PC}$ = (even)$^{++}$, many other quantum numbers are allowed in theory, see e.g.~\cite{Morningstar:1999rf}. Any glueballs can be pair produced in CEP, and this is a promising channel since the pomeron is dominated by gluons, especially at the low--$Q^2$ values relevant here. One potential strategy is to select single interaction events with central mass $M_X\sim$  4 - 10 GeV between two large rapidity gaps (detecting one or both protons would be beneficial, but may not be essential), and then to require the state $X$ to have zero charge, strangeness, etc., and finally, dividing the central event into two neutral clusters, to select cases where their masses are consistent with being equal and plot their average mass.  The decays of the two clusters may be different, e.g. $(\pi^+\pi^-\pi^+\pi^-)$ and ($\pi^+\pi^-K^+K^-)$, possibly containing meson resonances such as $\rho^0$ and $\phi$, etc. Unfortunately the branching fractions of these states are not well predicted, so this would largely be a data--driven search.  It would probably require large statistics, e.g. $10^{5}$--$10^{6}$ events with no pile up in that $M_X$ range, but the CEP cross sections are not expected to be small ($\sim$ nb)  and with an optimised trigger high statistics could be collected in a dedicated period of a few days.
     
Especially interesting would be the discovery of  ``oddballs'' ~\cite{Qiao:2014vva}, which have quantum numbers such as $J^{PC} = 0^{--}$ (a $G_{0-}$), and which cannot be composed of two gluons (by C--parity) or $q\bar{q}$, so that they do not mix with conventional mesons. The mass of $G_{0-}$ has been estimated to be about 3.8 GeV, and it should be relatively narrow; a possible decay mode is $G_{0-} \rightarrow \gamma + f_1(1285) \rightarrow \gamma + 4\pi$. Exclusive production of $G_{0-}$ pairs at the LHC could be an excellent discovery mechanism, and it would open a new chapter in non--perturbative QCD. A more speculative glueball topology in the string model is a barred-loop with two 3-string junctions (there are no quarks, which would be the ends of open strings). The most likely decay would be through triple-string-breaking to two baryons. 
     
\subsubsection*{Experimental results and outlook}
%Figure 1
\begin{figure}[htbp]
\includegraphics[width=\linewidth]{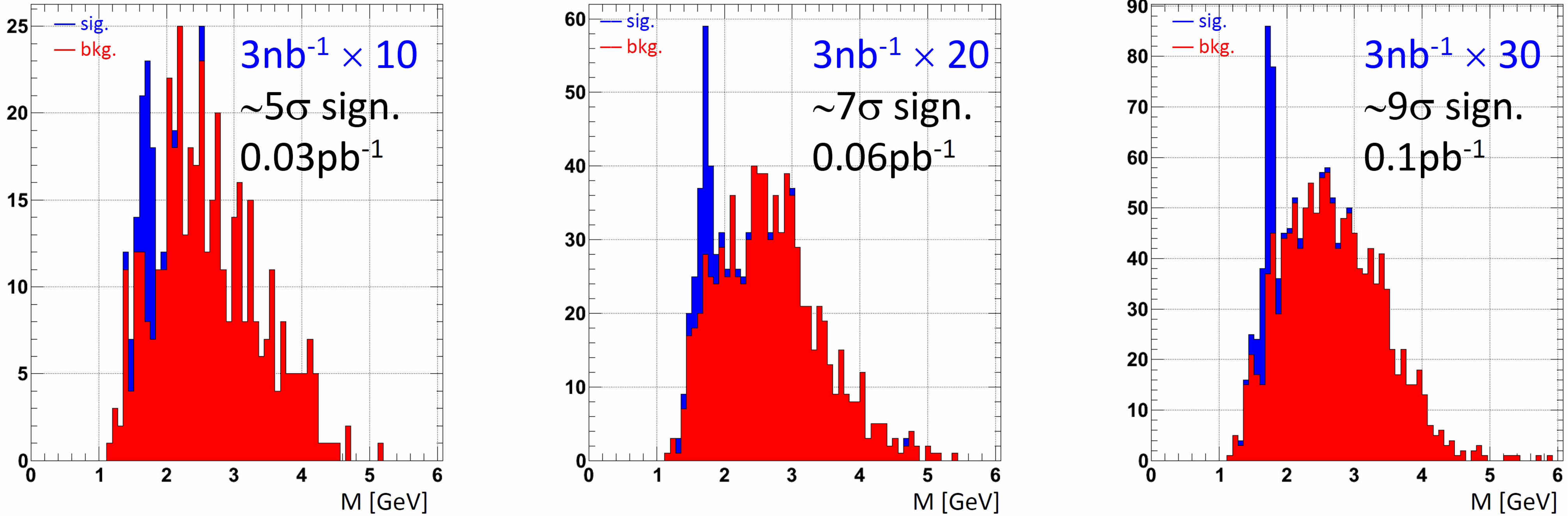}     
 \caption{Simulated signal and background mass distributions for exclusive $f_0(1710)\rightarrow\rho^0\rho^0\rightarrow2(\pi^+\pi^-)$ production in CMS-TOTEM. Three different integrated luminosities are given: $0.03\,$pb$^{-1}$, $0.06\,$pb$^{-1}$ and $0.1\,$pb$^{-1}$ together with the local peak significance. The background estimate from non-resonant exclusive $\rho^0\rho^0$ production is based on the \texttt{Dime MC}~\cite{Harland-Lang:2013dia}. Other backgrounds like exclusive $2(\pi^+\pi^-)$ and $\pi^+\pi^-\rho^0$ production as well as  exclusive production of adjacent $f_2$ states are not taken into account.} 
\label{f1710_plot_01}
\end{figure}
A preliminary analysis of low mass resonance production has been performed on the data of the common CMS-TOTEM $\beta^*$ = 90 m run at $\sqrt{s}$ = 8 TeV in July 2012. There is excellent mass resolution ($\sim$ 20-30 MeV) with the tracker for charged-particle-only final states, allowing the produced resonances to be clearly identified, without further steps like model- and parameter-dependent partial-wave analysis. 
Events with two RP protons and only two or four charged particles in the tracker with zero total charge are selected in the double arm RP triggered sample~\cite{timing_TDR}. The background has been shown to be low by selecting events with the same criteria but with a non-zero net charge for the charged particles. In the analyzed data sample of  $\sim$ 3 nb$^{-1}$, $\sim$ 1000 $\pi^+\pi^-$ and a few tens of $\rho^0\rho^0$ exclusive candidates were found, where for the latter two $\pi^+\pi^-$ combinations are required to be compatible with the $\rho^0$. 

A preliminary analysis of the common CMS-TOTEM data reveals sensitivity to events showing the possible decay of
$f_0(1710)\rightarrow\rho^0\rho^0\rightarrow2(\pi^+\pi^-)$. Due to the limited amount of data, the background due to non-resonant exclusive $\rho^0\rho^0$ production is estimated with the DIME Monte Carlo~\cite{Harland-Lang:2013dia} event generator. Fig.~\ref{f1710_plot_01} shows simulated signal distributions of $f_0(1710)\rightarrow\rho^0\rho^0\rightarrow 2(\pi^+\pi^-)$ together with background due to non-resonant exclusive $\rho^0\rho^0$ production with their local significance, for three different integrated luminosity scenarios. According to the simulation, at least $0.06\,$pb$^{-1}$ is required to observe the resonance. A similar integrated luminosity is needed for the measurement of $f_0(1500)\rightarrow K^+K^-$.

A precise measurement of the branching ratios of the $f_0$ resonances is essential in the context of identifying the resonances as glueball candidates. As the branching ratios for low mass resonances may easily differ by an order of magnitude (e.g. for $f_0(1500)$: Br~($K\bar{K}$) $\approx$ 9$\,\%$, Br~(${\eta\eta}$) $\approx$ 5$\,\%$ and Br~(4$\pi$) $\approx$ 50$\,\%$)  and assuming a similar range for $f_0 (1710)$, a factor of ten of integrated luminosity higher than that estimated for observing $f_0(1710)\rightarrow\rho^0\rho^0\rightarrow2(\pi^+\pi^-)$, would be required in order to precisely measure the  $\pi^+\pi^-$, K$^+$K$^-$ and $2(\pi^+\pi^-)$ decay modes. As the backgrounds from exclusive $2(\pi^+\pi^-)$ and $\pi^+\pi^-\rho^0$ production as well as the adjacent $f_2$ states were not taken into account in the above analysis, a detailed measurement of the $f_0(1500)$ and the $f_0(1710)$ branching ratios will in reality require slightly more i.e. an integrated $\beta^*$ = 90 m luminosity of $\sim1\,$pb$^{-1}$. 

\begin{figure}[h]
\begin{center}
\includegraphics[width=0.32\textwidth]{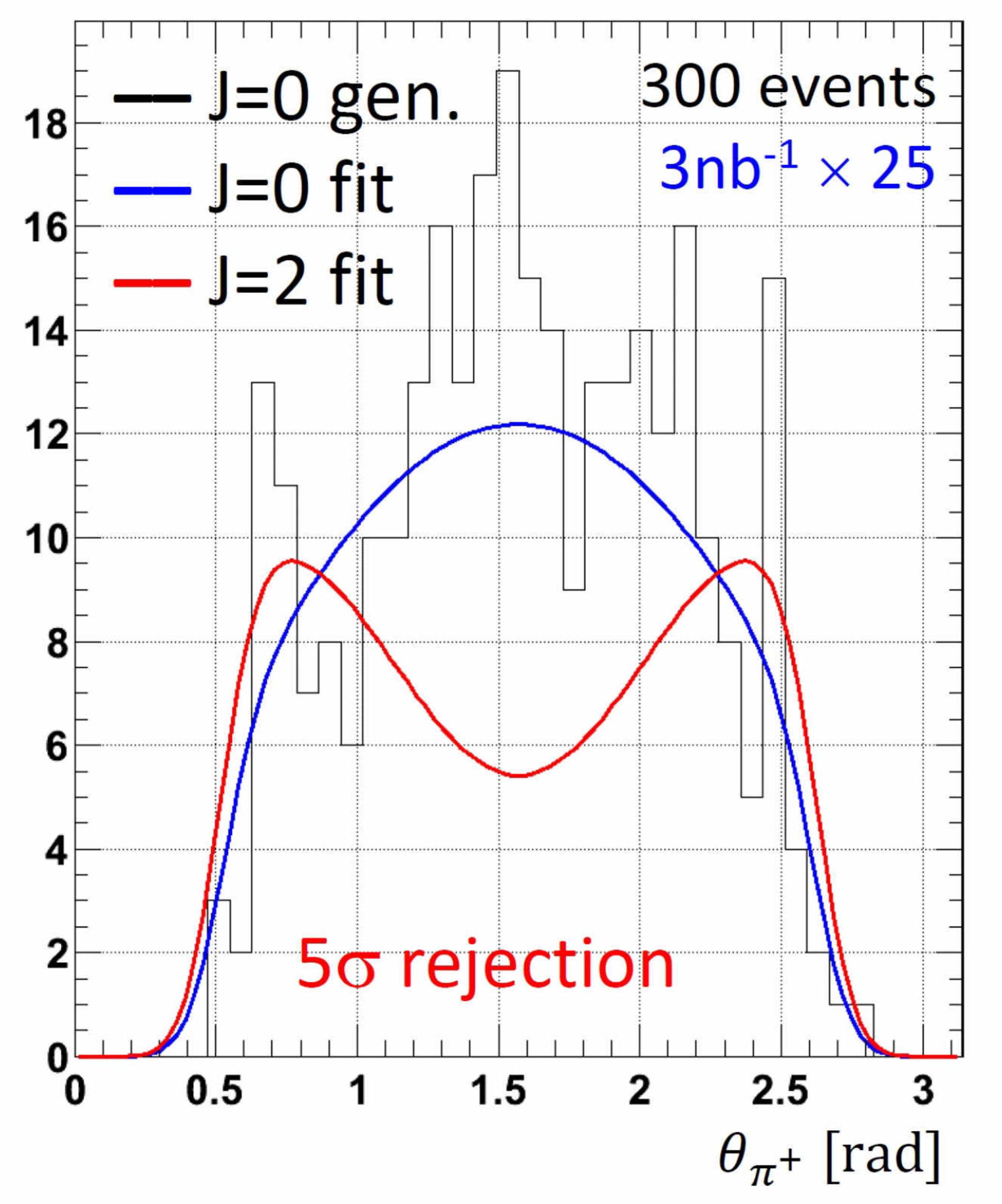}
\includegraphics[width=0.327\textwidth]{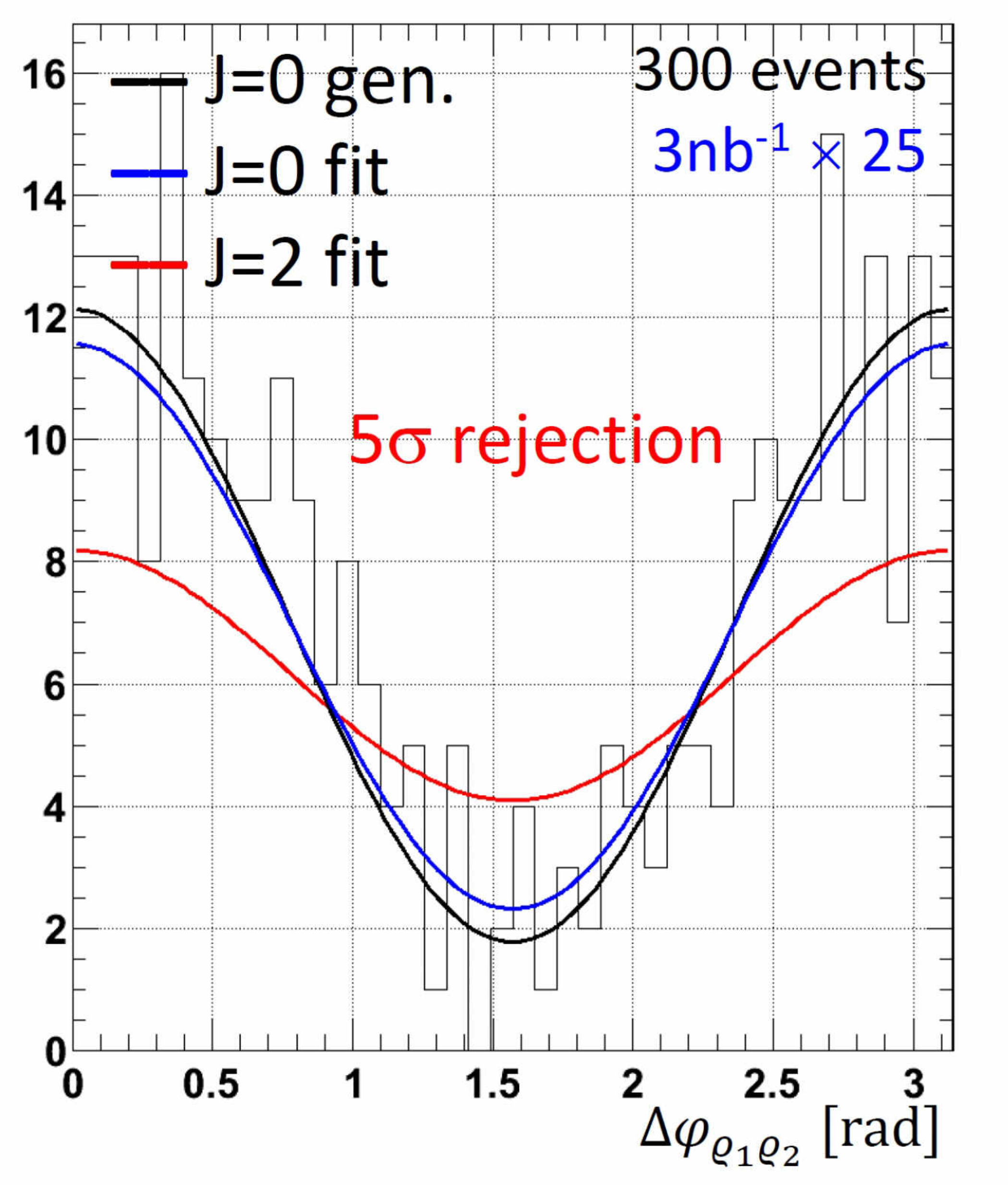}
\includegraphics[width=0.325\textwidth]{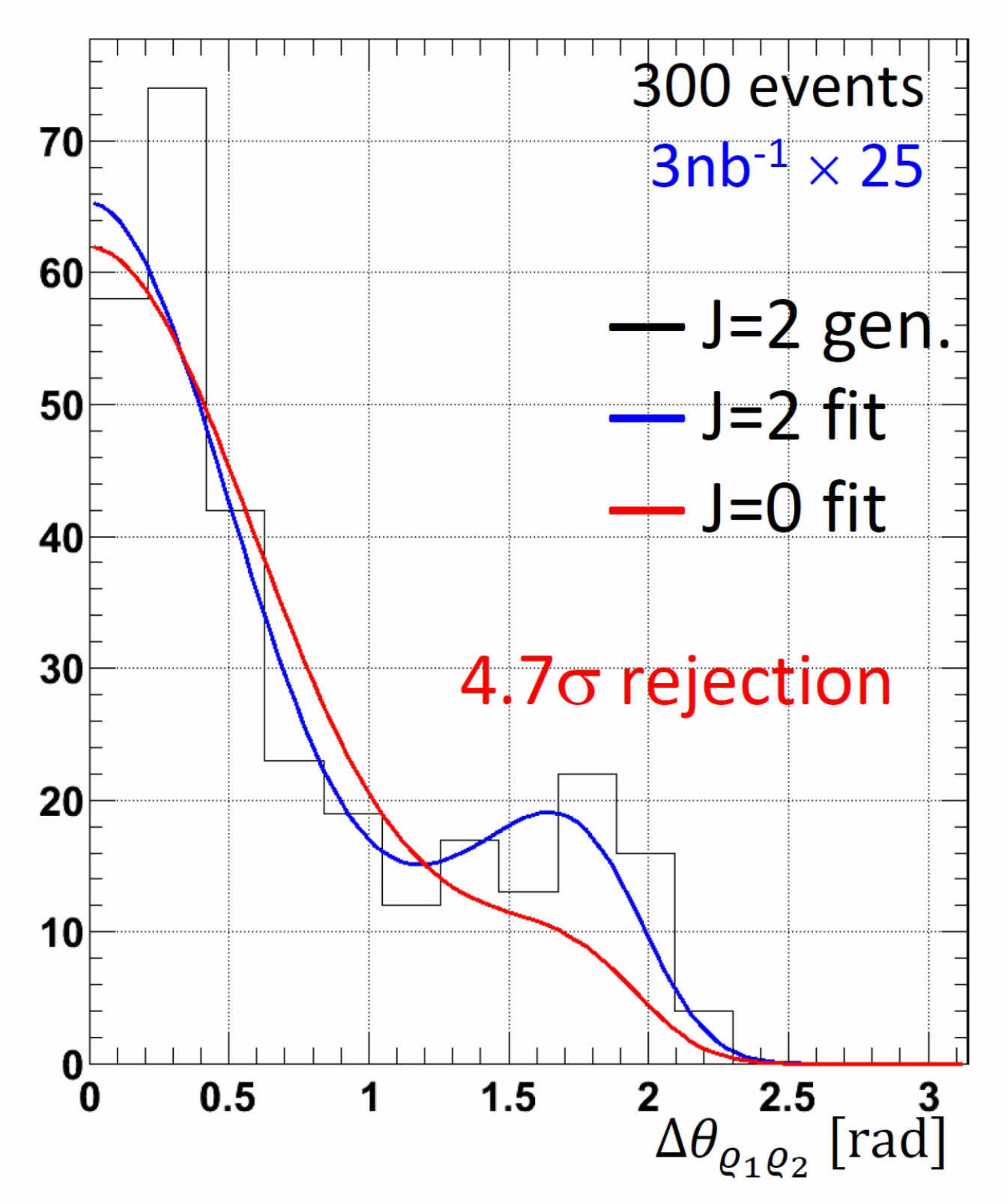}
 \caption{Simulated spin analysis in exclusive $f_J \rightarrow \rho^0 \rho^0 \rightarrow 2(\pi^+\pi^-)$ production. Left: Distribution of the polar angle $\theta_{\pi^+}$ of the $\pi^+$ from the decay of the $\rho^0$ with $\eta >0$. Center: Distribution of the azimuthal angle difference $\Delta \varphi_{\rho_1\rho_2}$ between the two $\pi^+\pi^-$ pairs. Right: Distribution of the polar angle difference $\Delta \theta_{\rho_1\rho_2}$ between the two $\pi^+\pi^-$ pairs produced. All plots correspond to an integrated luminosity of $\sim$ 75$\,$nb$^{-1} $. The histograms represent the simulated samples assuming either $J=0$ (left and center) or $J=2$ (right).  The blue (red) curves are the fits to the simulated data for the (in)correct hypothesis. The significance quoted in the histograms refers to the incompatibility to data of the fit with the incorrect hypothesis. No contributions from backgrounds have been included in the simulated samples.}
\label{f1710_plot_02}
\end{center}
\end{figure}

Finally, a study to estimate the required luminosity for a detailed angular momentum analysis, based on the $f_{J} \rightarrow \rho^0\rho^0 \rightarrow 2(\pi^+\pi^-)$ decay is summarised below; such an analysis is of high importance to give full confidence that the measured branching ratios of the potential low mass glueball candidates are correct. The study~\cite{timing_TDR} is carried out with a simplified detector acceptance model and in addition no background is assumed. The integrated luminosity requirements for an angular momentum study are illustrated in Fig.~\ref{f1710_plot_02} (left) that shows the sensitivity of the spin $J$ determination allowed by the distribution of the polar angle $\theta_{\pi^+}$ of the $\pi^+$ from the decay of the $\rho^0$ with $\eta >0$, produced in the reaction $f_J\rightarrow \rho^0 \rho^0\rightarrow2(\pi^+\pi^-)$. The rejection of an incorrect $J=2$ hypothesis is possible with at least 300 events, corresponding to an integrated luminosity of $\sim$ 75$\,$nb$^{-1}$. A similar integrated luminosity requirement is imposed by the spin determination from the azimuthal and polar angle difference ($\Delta \varphi_{\rho_1\rho_2}$, $\Delta \theta_{\rho_1\rho_2}$) between the $\pi^+\pi^-$ pairs, as illustrated in Figs.~\ref{f1710_plot_02} (middle) and~\ref{f1710_plot_02} (right). The angular correlations between the leading protons will also have sensitivity to the spin of the centrally produced exclusive state, see Section~\ref{sec:tagcepmotglu}. 

However, the considerations illustrated by Fig.~\ref{f1710_plot_02} can be considered only as rough estimates since they do not take into account the background from exclusive $2(\pi^+\pi^-)$, $\pi^+\pi^-\rho^0$ and $\rho^0\rho^0$ production. Moreover, in the vicinity of the $f_0(1710)$ there are other resonances, such as the $f_2(1640)$ or $f_2(1810)$, which partially overlap in the invariant mass spectrum. The decay amplitude coupling constants of a given resonance may differ as a function of the invariant mass $M$. Finally the $\rho^0\rho^0$ angular momentum $L_{\rho^0\rho^0}$ needs to be properly determined. 
A realistic spin-parity analysis therefore requires a study of the angular amplitudes as a function of the invariant mass in a wider interval than the resonance width itself, to make the deconvolution of the overlapping contributions coming from adjacent resonances and background possible. Similar approaches were already successfully employed in low mass resonance studies, see e.g.~\cite{Barberis:1999cq}.

The spin-parity analysis therefore has to be performed in mass steps $\Delta M$. The smallest step size $\Delta M$ is limited by the mass reconstruction resolution $\sigma(M)\approx$ 20-30 MeV. The largest possible step size could be a fraction of the resonance width ($\sim$ 100 MeV) but nevertheless should not exceed $\sim$ 40 MeV. Taking all the above into account, a full spin-parity analysis of the exclusive production of $f_0$ states will need to be made in mass bins with a size $\Delta M=$ 30-40 MeV and when requiring sufficient statistics in each $\Delta M$ bin, it is strictly only fully feasible with a integrated $\beta^*$ = 90 m luminosity of $\sim$ 4-5 pb$^{-1}$. A similar analysis will be possible in ATLAS-ALFA but will suffer more from contamination at large rapidities since the forward coverage in ATLAS is worse than in CMS-TOTEM. 
                
     \subsection{Quarkonium Pair Production}\label{sec:gluonex:djpsi}
     
     \subsubsection*{Motivation and theory}
     
     Motivated by the LHCb measurement of exclusive double $J/\psi$ and $J/\psi \psi(2S)$ production~\cite{Aaij:2014rms}, see below, the first calculation of exclusive double $J/\psi$ production in hadronic collisions was presented in~\cite{Harland-Lang:2014efa}. After an analysis of the Born--level $gg \to J/\psi J/\psi$ amplitudes within the non--relativistic quarkonium approximation, the predicted exclusive cross sections were found to be suppressed by the CEP dynamical $J_z=0$ selection rule, although still of a reasonable size: depending on the choice of gluon PDF and model of soft survival factor the predicted cross sections were found to lie in the $\sim 2-7$ pb range, in reasonable agreement (considering other theoretical uncertainties, such as the variation of the renormalization/factorization scale) although somewhat lower than the LHCb measurement of $24\pm 9$\ pb. It is worth emphasising that without the effect of the $J_z=0$ selection rule, the predicted cross section would be $\sim 2$ orders of magnitude higher. 
     
\begin{figure}
\begin{center}
\includegraphics[scale=0.8]{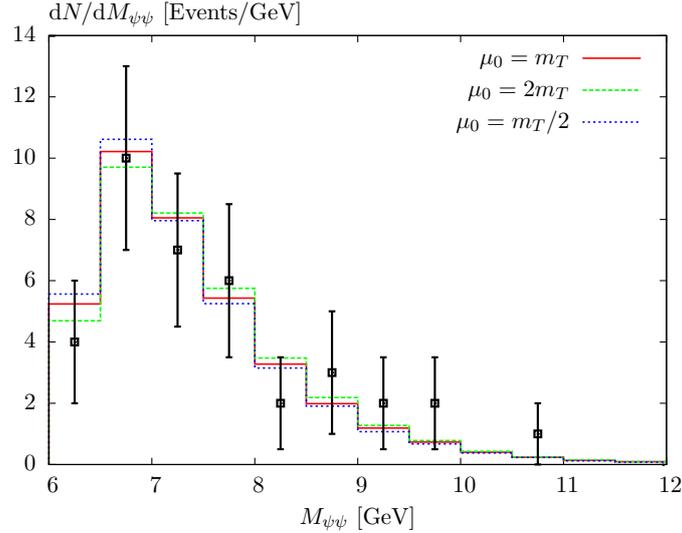}
\caption{Comparison of LHCb measurement~\cite{Aaij:2014rms} of $J/\psi J/\psi$ invariant mass distribution with theory prediction, calculated as described in the text. In all cases the result is normalized to the data.}\label{lhcbm}
\end{center}
\end{figure}

The shape of the invariant mass distribution, which has a much smaller theoretical uncertainty than the absolute cross section, is shown in Fig.~\ref{lhcbm}, normalised to the data, and is seen to describe the LHCb measurement well, within the (quite large) experimental errors. Further higher statistics measurements would clearly allow a closer comparison with theory. Other observables such as the $J/\psi$ transverse momentum and the rapidity difference $\Delta(y)$ between the mesons, are also of interest; in the latter case, the predicted distribution is found to be broader than in the inclusive process, an effect which is driven by the CEP selection rule, which enhances the amplitudes with $J_z=0$ incoming gluons, for which the  $\Delta(y)$ distribution is much broader. In inclusive production, some broadening is also expected (and observed~\cite{Abazov:2014qba}), but here it is generated by the `double parton scattering' production mechanism~\cite{Kom:2011bd}, where each $J/\psi$ is produced in independent scatters. In~\cite{Harland-Lang:2014efa}, this contribution was found to be very small in exclusive production, and thus the pure, single--parton scattering (and colour--singlet) contribution is probed. These reactions are also in principle quite sensitive to additional particles which might be produced in decay chains that involve exotic particles.
     
Finally, it is possible to estimate the expected cross sections for $J/\psi \psi(2S)$ and $\psi(2S) \psi(2S)$ production, giving
\begin{equation}
 \sigma^{J/\psi J/\psi} \, :\, \sigma^{J/\psi \psi(2S)}\, : \, \sigma^{\psi(2S) \psi(2S)} = 1\,:\,0.40\,:\,0.044\;,
\end{equation}
to be compared with the LHCb measurement of
\begin{equation}\label{lhcbdatr}
\frac{\sigma(J/\psi \psi(2S))}{\sigma(J/\psi J/\psi)}=1.1^{+0.5}_{-0.4}\;,
\end{equation}
assuming the same exclusive fraction in both cases.  There is clearly reasonable agreement, but with further data it will be possible to make a more precise statement about this. The cross sections involving $\chi_c$ mesons are estimated to be much smaller, although this needs to be confirmed by a full calculation.
On the other hand, as the formation amplitude of the pseudoscalar $\eta_c$ meson is proportional to the same value of the wave function at the origin as in the $J/\psi$ case, and the CEP pair production mechanism may also be produced by additional so--called `ladder' diagram (similar to the case of $\eta(')\eta(')$ production discussed in Section~\ref{sec:lightmeson}), we may expect the cross section for double $\eta_c$ production to be of the same size or even bigger than for the $J/\psi$.

It is worth emphasising that the work of~\cite{Harland-Lang:2014efa} represents the first calculation of exclusive double $J/\psi$ production, while the LHCb data~\cite{Aaij:2014rms} is the first ever measurement of this process. This therefore represents a very new topic of investigation, and there is a great deal of further theory work to be pursued: for example, the effect of relativistic corrections, in particular in the case of the $\psi(2S)$, the calculation of higher--order contributions and a full calculation for the case of $\chi_c$ and $\eta_c$ final states has not yet been fully considered. Further measurements of these processes, and more differential tests of the theory, will be essential in pursuing such a programme.

\subsubsection*{Experimental results and outlook}

\begin{figure}[ht]
{\includegraphics[width=0.49\linewidth]{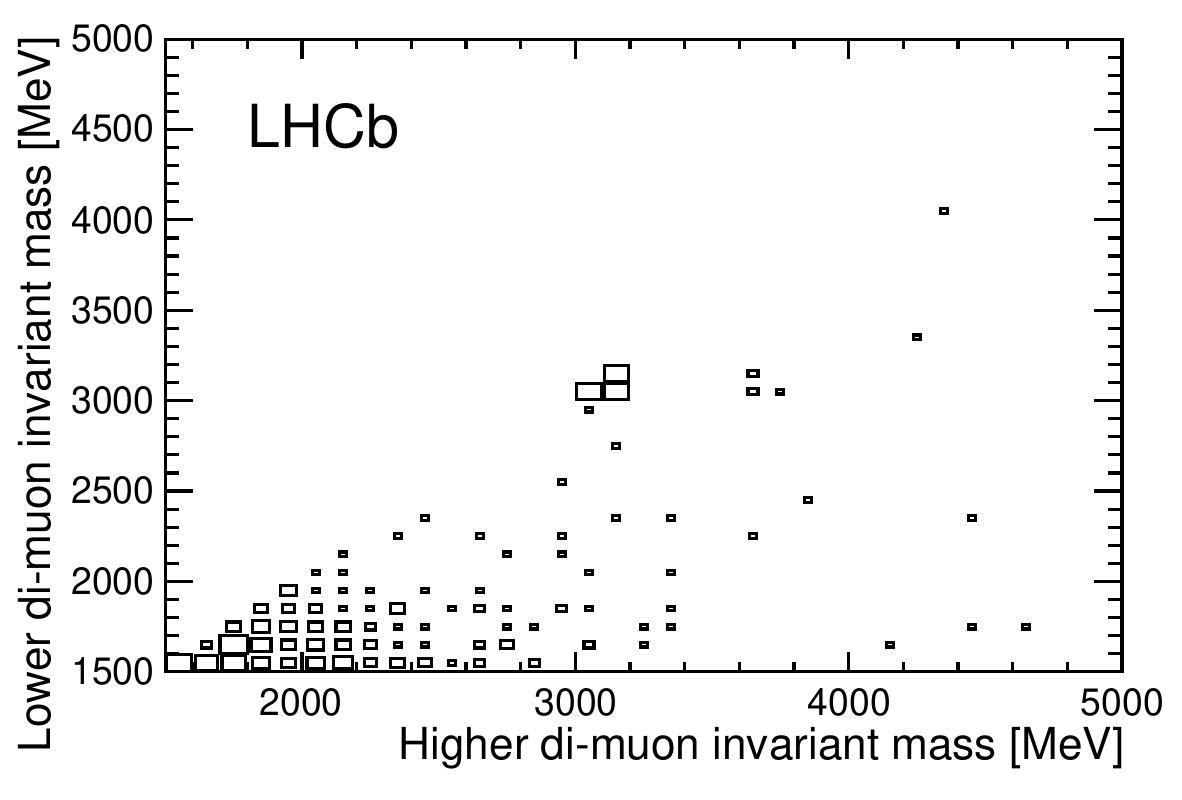}}
{\includegraphics[width=0.49\linewidth]{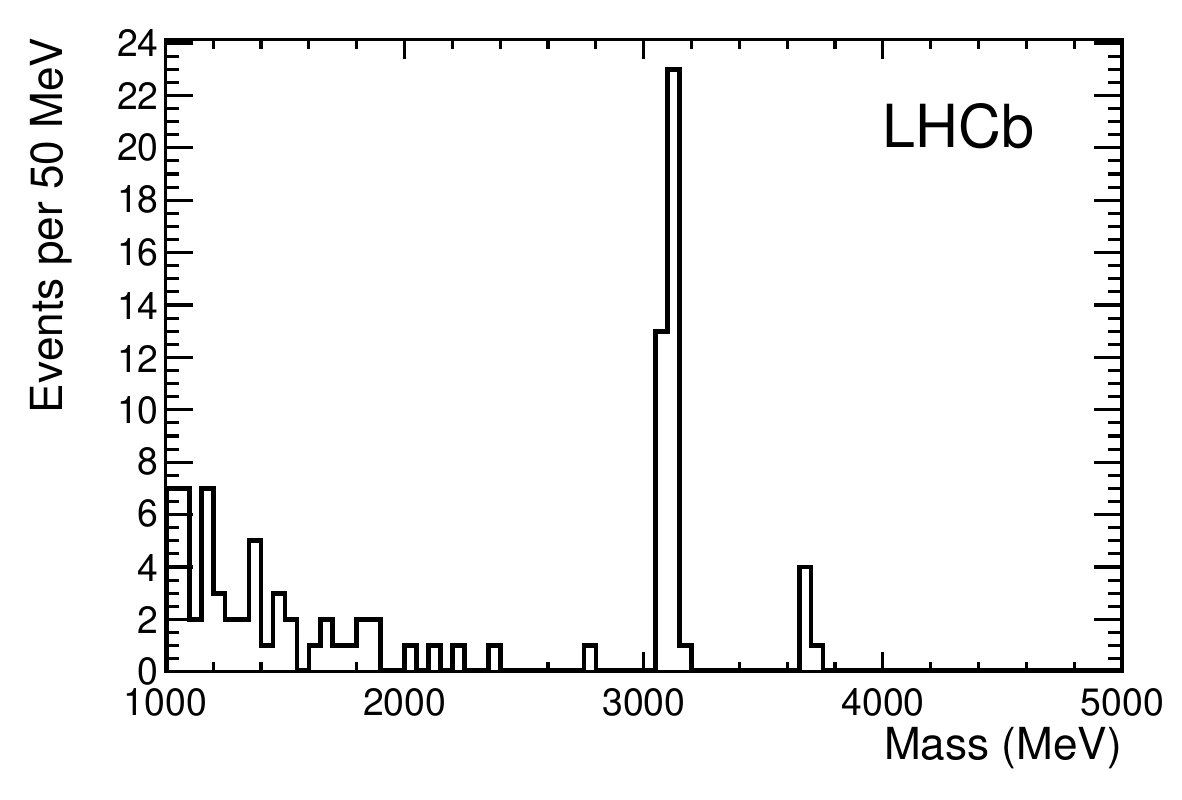}}
\caption{
(Left) Invariant masses of the two di-muon candidates.
(Right) The higher mass di-muon candidate having required the lower mass candidate to
be consistent with the $J/\psi$ mass.
\label{fig:2dmu}}
\end{figure}

LHCb has recently made measurements,  using a data sample corresponding to $3{\rm\ pb}^{-1}$, of double charmonia~\cite{Aaij:2014rms}, $J/\psi J/\psi, J\psi\psi(2S), \psi(2S)\psi(2S)$, $\chi_{c0}\chi_{c0}$,
$\chi_{c1}\chi_{c1},\chi_{c2}\chi_{c2}$. The selection proceeds in a similar fashion to that described in Sec.~\ref{sec:photooutlook},
although now four charged tracks (at least three of which are identified muons) and no other activity are required to select pairs of S-wave states, 
while one or more photons are required to select pairs of P-wave states.
Very few low multiplicity events have three or more identified muons.  The invariant mass
distribution of the two pairwise combinations is given in the left plot of 
Fig.~\ref{fig:2dmu} and shows
an accumulation of events at the $J/\psi$ and $\psi(2S)$ masses in a region of phase space that is
otherwise empty.  
The right plot in Fig.~\ref{fig:2dmu} shows the higher mass combination
when asking that the lower mass combination is consistent with the $J/\psi$ meson.
There are 37 $J/\psi J/\psi$,
5 $J/\psi\psi(2S)$ and no $\psi(2S)\psi(2S)$ candidates.
The only substantial background to the $J/\psi J/\psi$ signal comes from 
$J/\psi\psi(2S)$ where $\psi(2S)\rightarrow J/\psi X$ with $X$ unreconstructed.
After correcting for detector acceptance and efficiencies, 
the measured cross sections for pairs of S-wave mesons with $2<y<4.5$,
which are exclusive {\it within the LHCb acceptance},
are
$
\sigma^{J/\psi J/\psi} = 58\pm 10 \pm 6 {\rm\ pb}$, 
$\sigma^{J\psi\psi(2S)} = 63 ^{+27}_{-18} \pm 10 {\rm \ pb} $, and
$\sigma^{\psi(2S)\psi(2S)} < 237 {\rm\ pb}$ at the 90\% confidence level.
The search for P-wave pairs has a single candidate for $\chi_{c0}\chi_{c0}$ 
that is also consistent with
$J/\psi\psi(2S)$ production, and so upper limits at the 90\% confidence level
are set on the production of 
$\chi_{c0},\chi_{c1}$ and $\chi_{c2}$ pairs at 69, 45 and 141 pb, respectively.

The numbers quoted above are for di-mesons detected in the absence of any other
activity inside the LHCb acceptance.  In order to compare with theory predictions,
a correction needs to be made for events which are not truly exclusive.
This is determined
to be $(42\pm13)\%$ with a large uncertainty due to the low number of $J/\psi J/\psi$ events observed, and
leads to a measurement of elastic CEP $J/\psi J/\psi$ with $2<y<4.5$,
at an average $\sqrt{s}=7.6$ TeV,
of $24\pm 9$\ pb.
This is in fair agreement with the predictions of~\cite{Harland-Lang:2014efa}, see Section~\ref{sec:gluonex:djpsi}.
There is a sizeable uncertainty on the theoretical prediction, due in large part to the poorly
understood low-$x$ gluon PDF that enters with the fourth power in the theoretical calculation.
More data, both to pin down the
gluon PDF (as described in Sec.~\ref{sec:photooutlook}) and to improve the $J/\psi J/\psi$ CEP measurement will enable a more precise 
comparison.  

\begin{figure}[ht]
{\includegraphics[width=0.49\linewidth]{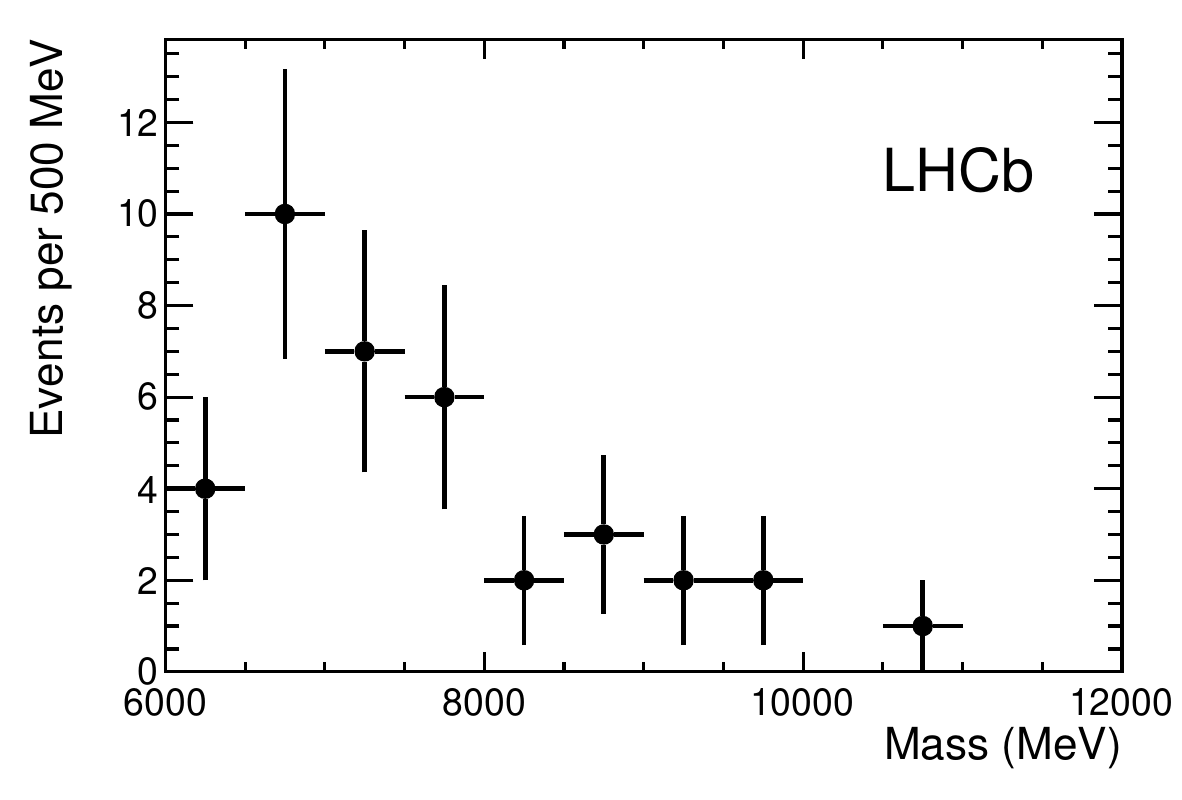}}
{\includegraphics[width=0.49\linewidth]{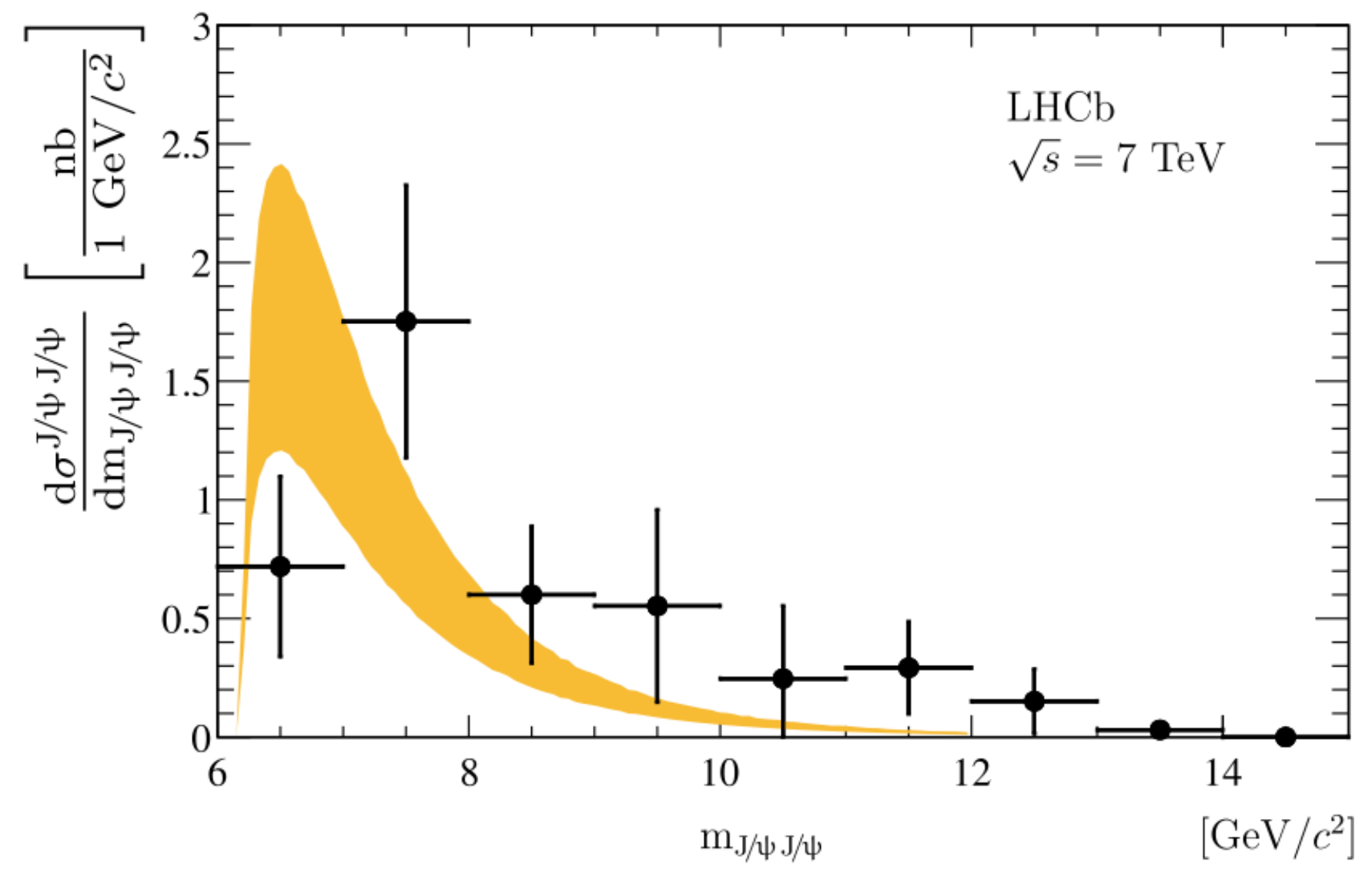}}
\caption{
Invariant mass of the $J/\psi J/\psi$ system in (left) exclusive
and (right) inclusive events.  The shaded area is the theoretical 
prediction of~\cite{Berezhnoy:2011xy}.
\label{fig:cfdjj}}
\end{figure}

In Fig.~\ref{fig:cfdjj},
the invariant mass of the exclusive signal is compared to that of an inclusive measurement
of double $J/\psi$ production, performed by LHCb~\cite{Aaij:2011yc}; both have a similar shape.
The data in the inclusive measurement are shifted to slightly higher masses than the theory,
and this has been discussed as possible evidence for
double parton scattering~\cite{Kom:2011bd}
or tetraquark states~\cite{Berezhnoy:2011xy}.  The former is negligible in CEP due to the ultra-peripheral
nature of the collision (see~\cite{Harland-Lang:2014efa} for further discussion), and thus with more statistics, the exclusive measurement 
will become sensitive to the presence of
higher mass resonances. Similar complementary measurements will be possible with proton tagging at high $\beta^{*}$ in CMS-TOTEM and ATLAS-ALFA.

        \subsection{Jet production}\label{CEP:gluonex:jets}
     
  \subsubsection*{Motivation and theory}
     
    Exclusive jet production~\cite{Martin:1997kv,Khoze:2006iw}, in particular of a 2--jet system ($jj$), has been of great importance in testing the underlying perturbative CEP formalism. Moreover, as discussed below, there is much potential to measure this process at the LHC, in particular with both protons tagged using the installed and proposed forward proton spectrometers: the expected production cross sections can be as high as the nanobarn level, depending on the precise event selection and in particular the $M_X$ range probed. Indeed, as discussed below, already a sample of `exclusive--like' 2--jet and 3--jet events has been collected in a combined CMS+TOTEM run at 8 TeV. Most events with two scattered protons and central jets will correspond to central diffractive (CD) jet production, i.e they will not be exclusive, but will have additional particle production from the Pomeron remnants. Exclusive production may be regarded as a particular case of CD jet production with only the jets in the final state, and no Pomeron remnants. It proceeds through the mechanism shown in Fig.~\ref{fig:pCp}, via the $gg\to gg, q\overline{q}$ and $gg \to ggg,gq\overline{q}$ subprocesses for 2-- and 3--jet production, respectively.
  
     The different behaviour of the parton--level helicity amplitudes relevant to exclusive jet production leads to some highly non--trivial predictions. For example, considering quark jets, the $gg\to q\overline{q}$ amplitudes are given by
\begin{align}\label{qqexc0}
\mathcal{M}\left((g(\pm)g(\pm)\to q_h \overline{q}_{\bar{h}}\right) &=\frac{\delta^{cd}}{N_c}\frac{16\pi\alpha_s}{(1-\beta^2\cos^2\theta)}\frac{m_q}{M_X}(\beta h \pm 1)\delta_{h,\bar{h}}\;,\\ \label{qqexc2}
\mathcal{M}\left((g(\pm)g(\mp)\to q_h \overline{q}_{\bar{h}}\right) &=\pm h\frac{\delta^{cd}}{2N_c}8\pi \alpha_s\left(\frac{1\pm h \cos\theta}{1\mp h\cos\theta}\right)^{1/2}\delta_{h,-\bar{h}}\;,
\end{align}
for gluons of `$\pm$'  helicity and quarks of helicity $h$, while $c,d$ are the outgoing quark color labels, and $\beta=(1-4m_q^2/M_X^2)^{1/2}$. We can see that the $J_z=0$ amplitude involves a helicity flip along the quark line, and vanishes as the quark mass $m_q \to 0$. Thus we expect a strong suppression in the CEP cross section for quark di-jets, relative to the $gg$ case, for which the $gg\to gg$ amplitudes with $J_z=0$ incoming gluons display no such suppression.  In this way the exclusive mode offers the possibility to study almost purely (over $99\%$ for typical event selections) gluonic and, crucially, isolated jets~\cite{Khoze:2000jm} (produced by the collision of a color--singlet $gg$ state), shedding light on the underlying properties of these jets (such as multiplicity, particle correlations etc) in a well--defined and comparatively clean exclusive environment. In Table~\ref{table:exjets}, some representative predictions for exclusive two and three jet production are shown and this $gg/q\overline{q}$ hierarchy is clear. The quite large predicted cross sections are also evident.     
      
%\begin{table}[h]
%\begin{center}
%\begin{tabular}{ccccc}
%\hline
%$M_X({\rm min})$ & $gg$ & $q\overline{q}$ & $ggg$ & $gq\overline{q}$ \\ \hline
%75 & 110 & 0.53 & 2.8 & 0.27\\
%150 & 3.7 & $8.9 \times 10^{-3}$ & 0.50 & 0.049 \\
%250 & 0.13 & $2.6 \times 10^{-4}$ & 0.012 & $7.1 \times 10^{-4}$ \\
%\hline
%\end{tabular}
%\caption{Parton--level predictions for exclusive two and three jet production cross sections (in pb) at the LHC for different cuts on the minimum central system invariant mass $M_X$. The jets are required to have transverse momentum $p_\perp>$ 20 GeV for $M_X({\rm min})=75,100$ GeV and $p_\perp>$ 40 GeV for $M_X({\rm min})=250$ GeV and pseudorapidity $|\eta|<2.5$.  The Anti--$k_t$ algorithm with $R=0.6$  is used in the three jet case. The quark jet cross sections are summed over five flavours, so that the purely $b$ quark cross sections are approximately a factor of five smaller than those displayed here. Predictions made using the \texttt{SuperCHIC} 2 MC~\cite{harlandlandfut}.}
%\label{table:exjets}
%\end{center}
%\end{table}

\begin{table}[h]
\begin{center}
\begin{tabular}{|c|c|c|c|c|c|}
\hline
$M_X({\rm min})$ & $gg$ & $q\overline{q}$ & $b\overline{b}$ & $ggg$ & $gq\overline{q}$ \\ \hline
75 & 120 & 0.073 & 0.12 &6.0 &0.14\\
150 &4.0  & $1.4 \times 10^{-3}$ &  $1.7 \times 10^{-3}$ &0.78 & 0.02\\
250 & 0.13 & $5.2 \times 10^{-5}$ &$5.2 \times 10^{-5}$   &0.018 &$5.0 \times 10^{-4}$ \\
\hline
\end{tabular}
\caption{Parton--level predictions for exclusive two and three jet production cross sections (in pb) at the LHC for different cuts on the minimum central system invariant mass $M_X$ at $\sqrt{s}=13$ TeV. The jets are required to have transverse momentum $p_\perp>$ 20 GeV for $M_X({\rm min})=75,100$ GeV and $p_\perp>$ 40 GeV for $M_X({\rm min})=250$ GeV and pseudorapidity $|\eta|<2.5$.  The Anti--$k_t$ algorithm with $R=0.6$  is used in the three jet case and the $q\overline{q}$ cross sections correspond to one quark flavour.  Predictions made using the \texttt{SuperChic} 2 MC~\cite{harlandlandfut}.}
\label{table:exjets}
\end{center}
\end{table}

     In the case of three jet production, that is $q\overline{q}g$ and $ggg$ jets, this suppression in the $q\overline{q}$ exclusive di-jet cross section also leads to some interesting predictions~\cite{Khoze:2006um,Khoze:2009er}. In particular, we expect the behaviour of the $q\overline{q}g$ amplitude as the radiated gluon becomes soft to be governed by the corresponding Born--level, $q\overline{q}$, amplitude. This is expected to lead to an enhancement of `Mercedes--like' configurations for the $q\overline{q}g$ case, where all three partons carry roughly equal energies and are well separated. More generally, it would be of much interest to investigate the difference in the predicted event shape variables, which may be quite different between the experimentally distinguishable $b\overline{b}g$\footnote{Predictions are presented in Table~\ref{table:exjets} for $gq\overline{q}$ production with massless quarks, however the corresponding cross sections with $b$--quarks are expected to be similar.} and $ggg$ cases, as well as to the corresponding inclusive cases. 
     
     In addition, in~\cite{Harland-Lang:2015faa} so--called `planar radiation zeros'  were shown in~\cite{Harland-Lang:2015faa} to be present in 5--parton QCD amplitudes, that is, a complete vanishing of the Born--level amplitudes, independent of the particle polarisations, when their momenta lie in a plane and satisfy certain additional conditions on their rapidity differences. These were seen in particular to occur in the $gg \to ggg$ and, in certain cases, the $gg\to q\overline{q} g$ amplitudes, when the initial--state gluons were in a colour--singlet configuration. This is precisely the situation for exclusive 3--jet production, and so the presence of such zeros may be detectable in the CEP process. 
     
       \subsubsection*{Experimental results and outlook}
\begin{figure}[htbp]
      \centering
      \includegraphics[width=0.7\textwidth]{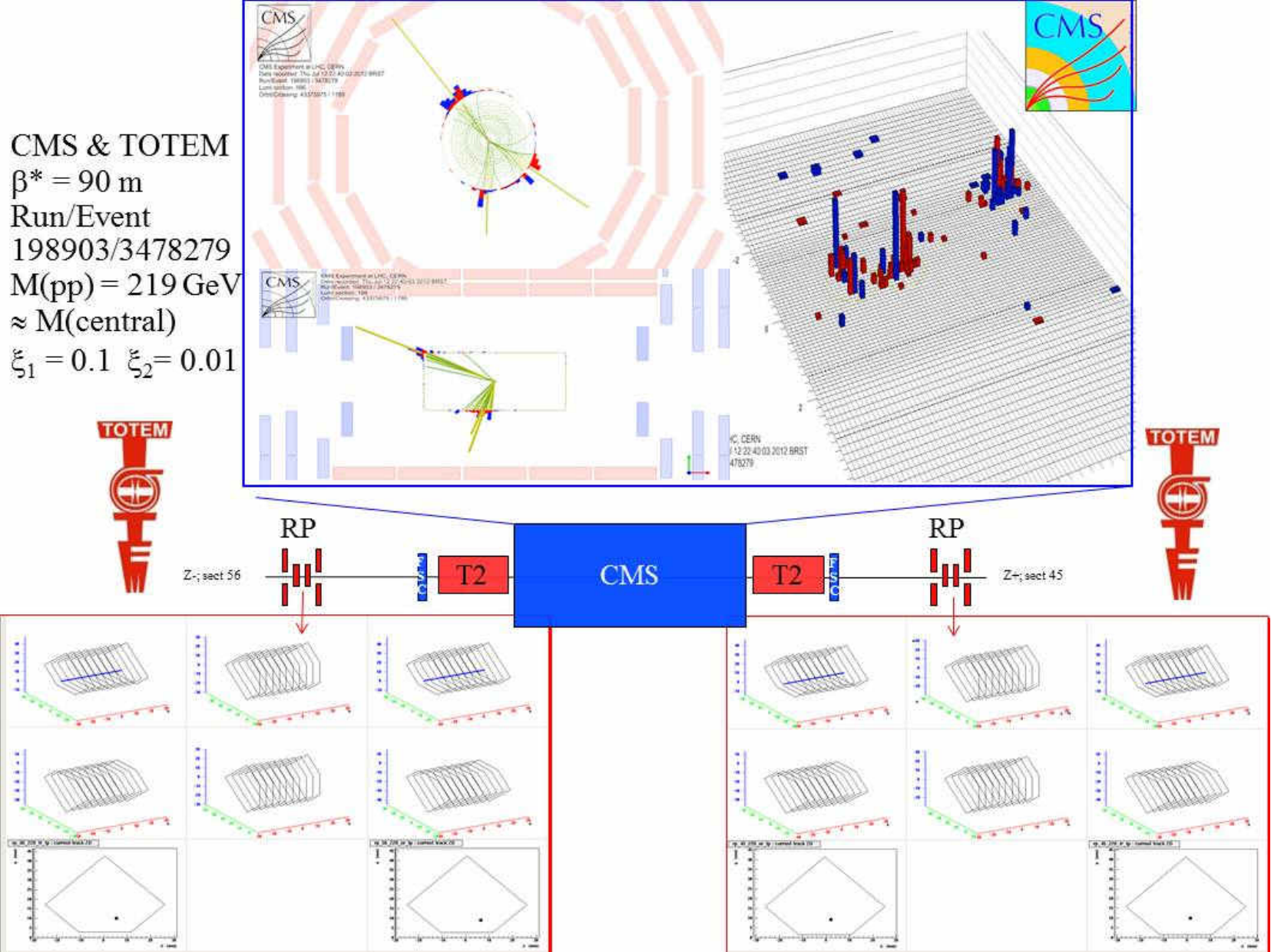}
      \caption{A central diffractive three-jet event recorded by TOTEM and CMS in a $\beta^*$ = 90m run at $\sqrt{s}$ = 8 TeV.
 The upper part of the figure displays the central part of the event, as seen in CMS; the 
 lower part displays the proton information in the 
TOTEM Roman Pots.}
      \label{fig:trijet}
    \end{figure}
  In 2008, the CDF collaboration reported~\cite{Aaltonen:2007hs} the observation and cross section measurement of exclusive jet production using a data sample of $310$ ${\rm pb}^{-1}$, at $\sqrt{s}=1.96$ TeV and for $E_\perp^{\rm jet}>10$ GeV, selected by tagging the outgoing anti--proton and requiring a rapidity gap in the proton direction (which was not tagged). They presented both dijet invariant mass $M_{jj}$ and jet transverse momenta $E_\perp^{\rm jet}$ distributions, out to quite high $M_{jj} \sim$ 130 GeV, and $E_\perp^{\rm jet} \sim 35$ GeV, and it was found that the perturbative approach of the Durham model described the data well. This observation was later supported by the measurement of the D0 collaboration~\cite{Abazov:2010bk}, which found evidence for exclusive dijet production with $M_{jj}>100$ GeV. The first study of di-jet production at $\sqrt{s}=7$~TeV is presented in~\cite{Chatrchyan:2012vc:ch5}, which however is limited to single-diffractive (SD) di-jet production and has no measurement of the scattered proton. An older study also exists with Tevatron data, presented in~\cite{Kepka:2007nr}.

In high $\beta^*$ runs, CMS-TOTEM and ATLAS-ALFA can study CD dijets with $E_T >$ 20 GeV at any $M_X$.  Some two- and three-jet events, though not truly exclusive since $M(jj,jjj) <\!\!< M(pp)$, were already seen  by CMS and TOTEM during the short high-$\beta^*$ run in July 2012. Common data were recorded with a CMS trigger on two jets with $E_T >$ 20 GeV. Selecting events with a proton in each direction in the  TOTEM RPs, extremely clean events with jets were found, as shown in Fig.~\ref{fig:trijet}. With 100 pb$^{-1}$ of high $\beta^*$ running, a sample of about 10,000 CEP jet events with $M_X >$ 60 GeV is expected, since the expected visible cross section for CMS-TOTEM is about 100 pb \cite{Harland-Lang:2014lxa}. The expected number of background events is significantly lower\cite{timing_TDR}. The obtained sample will enable studies of the azimuthal difference $\phi$ between the  scattered protons,  the shape of the proton $t$-distribution and the overall cross section behaviour with $M_X$, providing a good test of different models  \cite{Harland-Lang:2014lxa, Petrov:2007kn,Ryutin:2012np}. 

Such  high $\beta^*$  measurements are complementary to the possibilities with the CT-PPS and AFP detectors, in standard LHC running, which only have access to $M_X \gtrsim 300$ GeV , but with much higher integrated luminosities. Measurement feasibilities in this scenario have been the subject of detailed studies, in the case of both detector set--ups: these are summarised in the following sections.

%%%

 \subsection{Jet production: ATLAS feasibility study}

In this section, a feasibility study for exclusive jet production, performed for $\sqrt{s} = 14$~TeV, and using the ATLAS detector equipped with the AFP stations is summarised. The full analysis is described in~\cite{ATLAS_EXC_JJ}.

Exclusive dijet events were generated using the \textsc{FPMC} generator~\cite{FPMC:ch5}. Further details of the  tools used for generating non--diffrative dijet and single/double Pomeron exchange events, as well as additional proton--proton interactions are given in~\cite{ATLAS_EXC_JJ}. In order to simulate the detector response, all events were reconstructed using the ATLAS full simulation chain \cite{ATLAS_reco}.

 Due to the distance from the beam assumed in this analysis, the minimal energy loss visible in the AFP detectors is $\xi_{min} \approx 0.02$, which translates to a jet momentum of about 140--150 GeV. One proton is required in each AFP station, reducing the ND background by an order of magnitude. The difference between the primary vertex $z$ position reconstructed by the ATLAS main detector and the one reconstructed from the AFP time measurement, $\Delta z$, is required to be less than 3.5 mm.

\begin{figure}
\centering
\includegraphics[width=.40\textwidth]{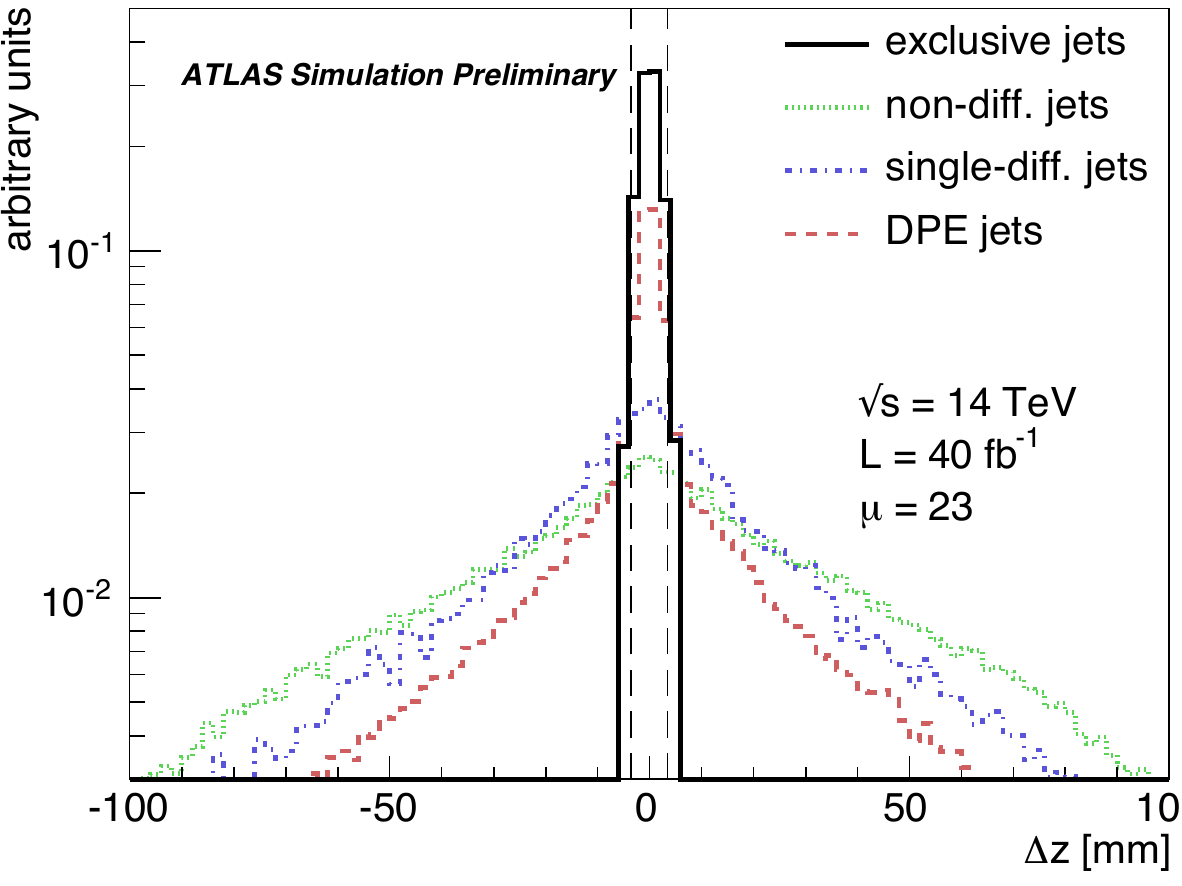}
\caption{The difference of the vertex $z$-coordinate as reconstructed by the ATLAS main detector and the one reconstructed from the AFP time measurement. The integral of the distribution is normalised to 1. The AFP time resolution of 10 ps has been assumed for background rejection. The exclusive signal is plotted as a solid black line, whereas the backgrounds are a dotted green (non-diffractive jet production), dashed-dotted blue (single diffractive jet production) and dashed red (double Pomeron exchange jet production) lines. The black dashed line represents the value of the applied cut.}
\label{fig_timing_cut}
\end{figure}

\begin{figure}
\centering
\includegraphics[width=.40\textwidth]{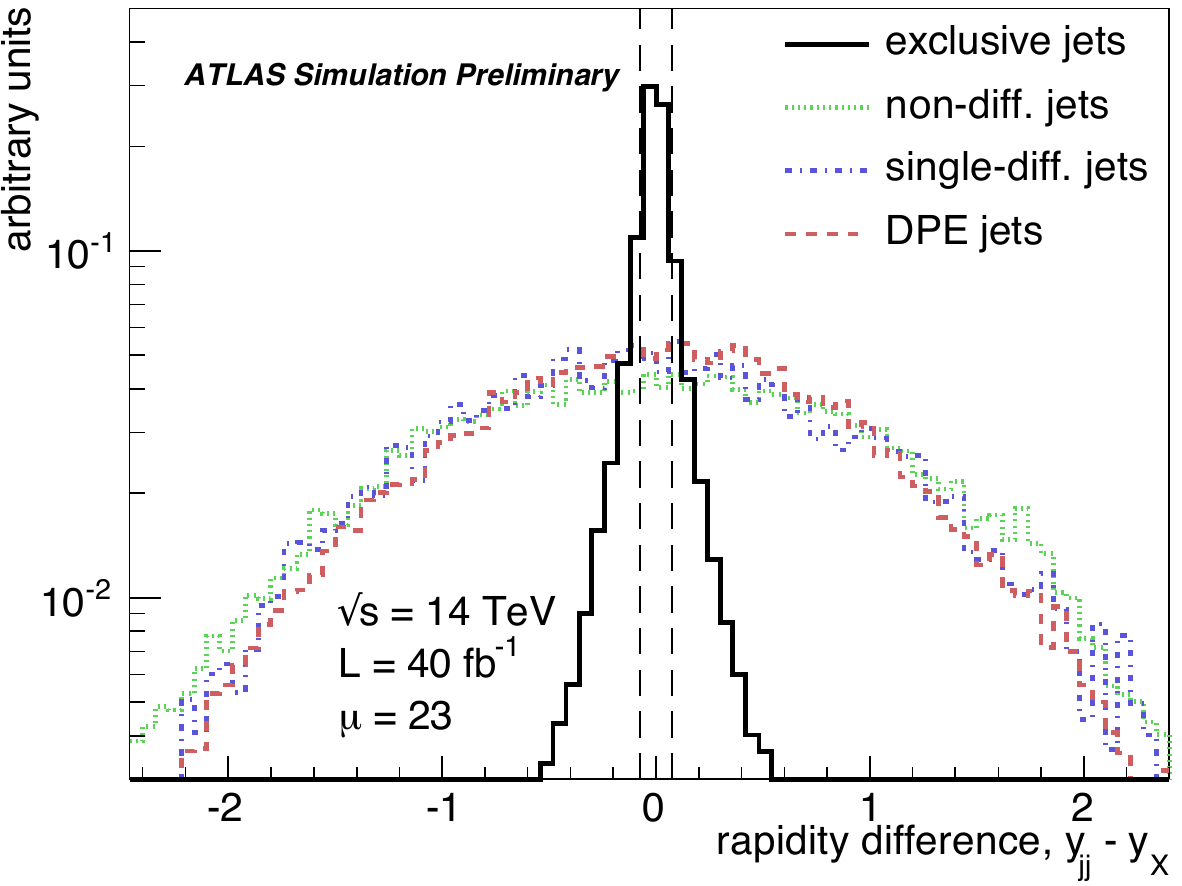}\qquad\qquad
\includegraphics[width=.40\textwidth]{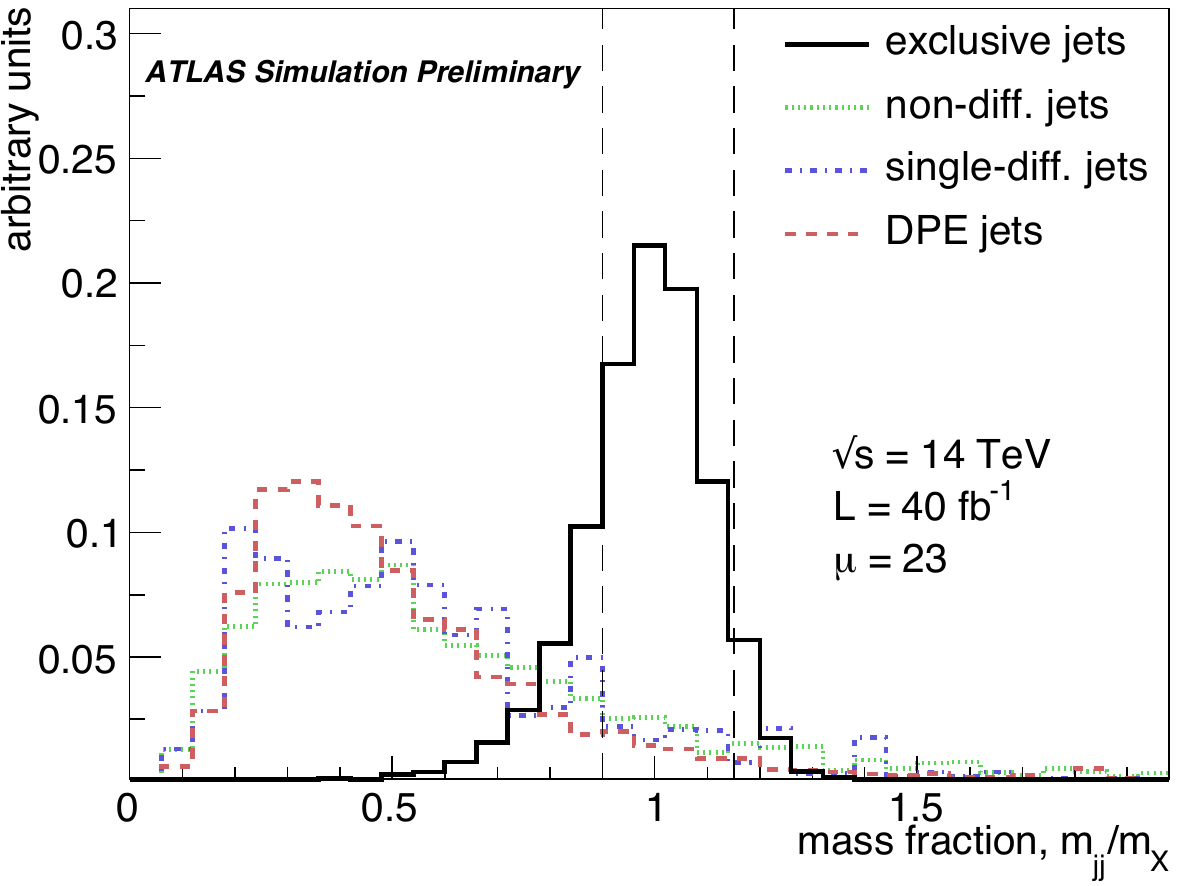}
\caption{\textbf{Left:} Difference, $y_{jj} - y_{X}$, of the rapidity of the jet system ($y_{jj}$) and the rapidity of the proton system $y_{X} = 0.5 \cdot \ln \left( \frac{\xi_{1}}{\xi_{2}} \right)$. \textbf{Right:} Ratio of the jet system mass to the missing mass $m_{X} = \sqrt{s\cdot\xi_{1}\cdot\xi_{2}}$. The $\xi_{1}$ and $\xi_{2}$ are relative energy losses of protons tagged in the AFP stations. The integral of the distribution is normalised to 1. The black dashed line represents the value of the applied cut.}
\label{fig_jet_rapidity_and_mass_fract_cut}
\end{figure}

The $\Delta z$ distribution for signal and background events is shown in Figure \ref{fig_timing_cut}. The broad distributions for ND and SD jet production are due to the size of the beamspot, as in these cases protons are coming from pile-up interactions. The tails in the case of DPE jet production are due to the events in which one `hard' proton was not seen in the AFP, but there was an additional pile-up proton. For DPE jets with $p_T > 150$ GeV such a situation is quite probable as protons are expected to lose a lot of their initial energy. The exclusive signal was generated with both protons in the AFP acceptance; here, the width is mainly due to the AFP timing resolution. When more than one proton was observed in a given station, all combinatoric possibilities were considered and the one with the smallest $\Delta z$ was taken.

For signal events, the kinematics of the central dijet system can be estimated from the forward proton measurements, and correlated with the kinematics reconstructed from the jet four-momenta.  Fig.~\ref{fig_jet_rapidity_and_mass_fract_cut} (left) shows the ratio of the dijet mass reconstructed from the jet four-vectors to that obtained from the proton kinematics. The exclusive signal can be enhanced with respect to the background by the following cuts: $|y_{jj} - y_{X}| < 0.075$ and $0.9 < \frac{m_{jj}}{m_{X}} < 1.15$. These requirements provide further background reduction by about three orders of magnitude.

\begin{figure}
\centering
\includegraphics[width=.75\textwidth]{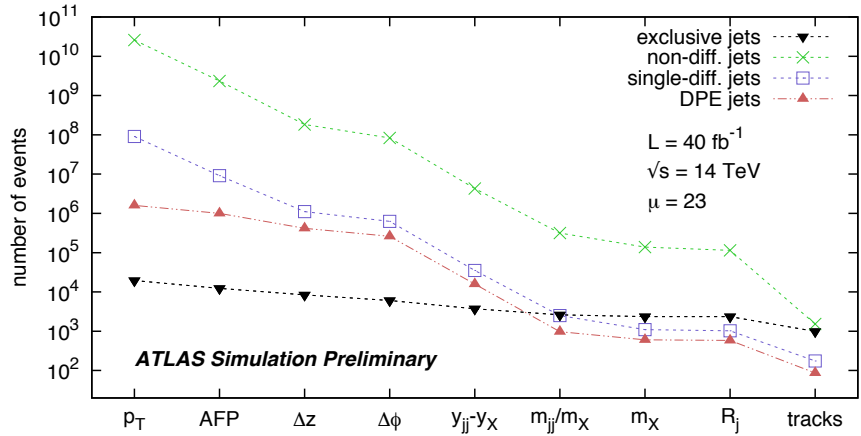}
\caption{Number of events accepted after a given cut for signal (exclusive jet production) and background (double Pomeron exchange (DPE), single diffractive (SD) and non-diffractive (ND) jet production) processes for an integrated luminosity of 40 fb$^{-1}$ and an average number of interactions of $\mu = 23$ as a function of the applied consecutive cuts. The AFP time resolution of 10 ps has been assumed for background rejection.}
\label{fig_cut_summary_mu23}
\end{figure}

\begin{figure}
\centering
\includegraphics[width=.4\textwidth]{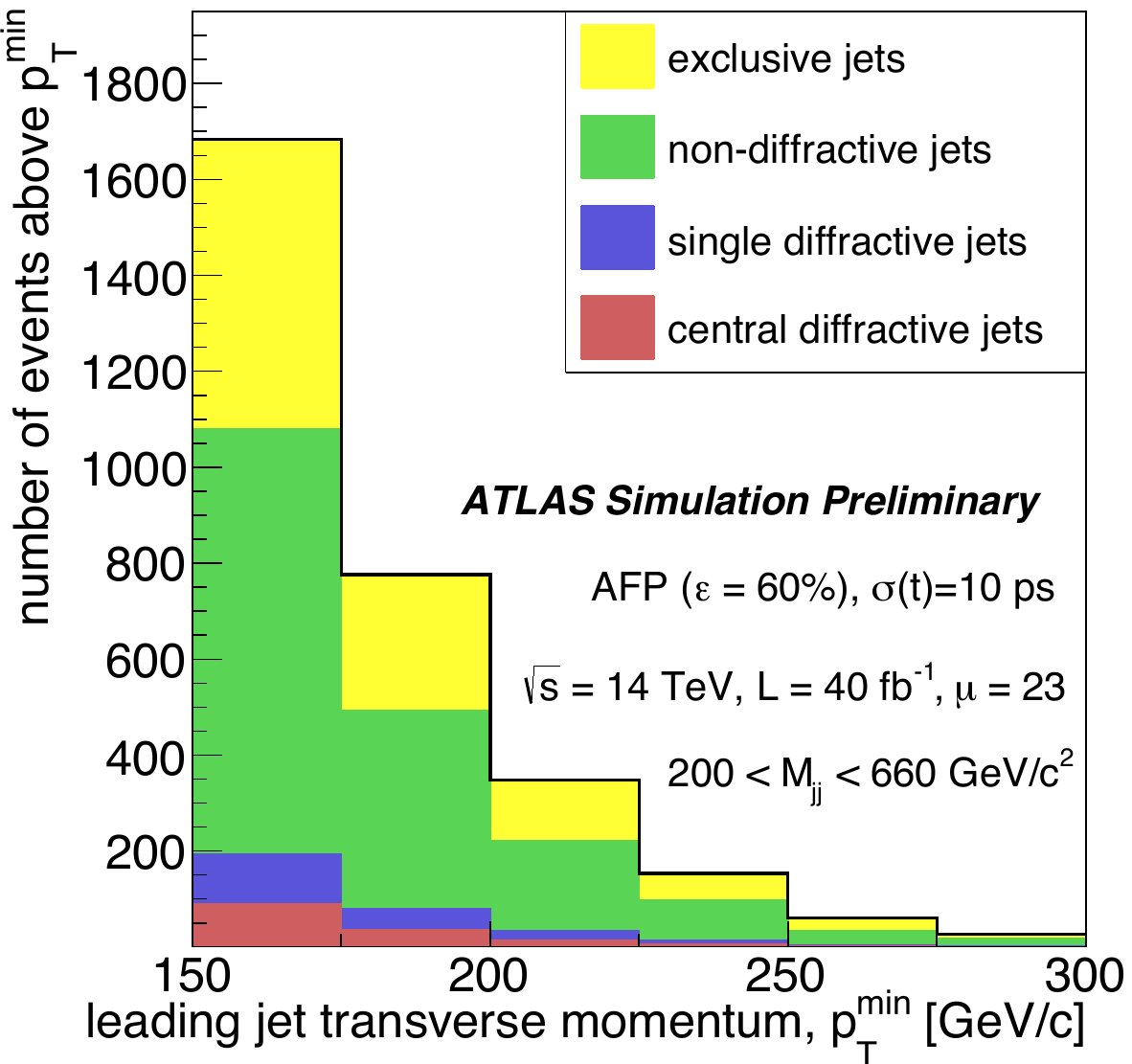}\qquad\qquad
\includegraphics[width=.4\textwidth]{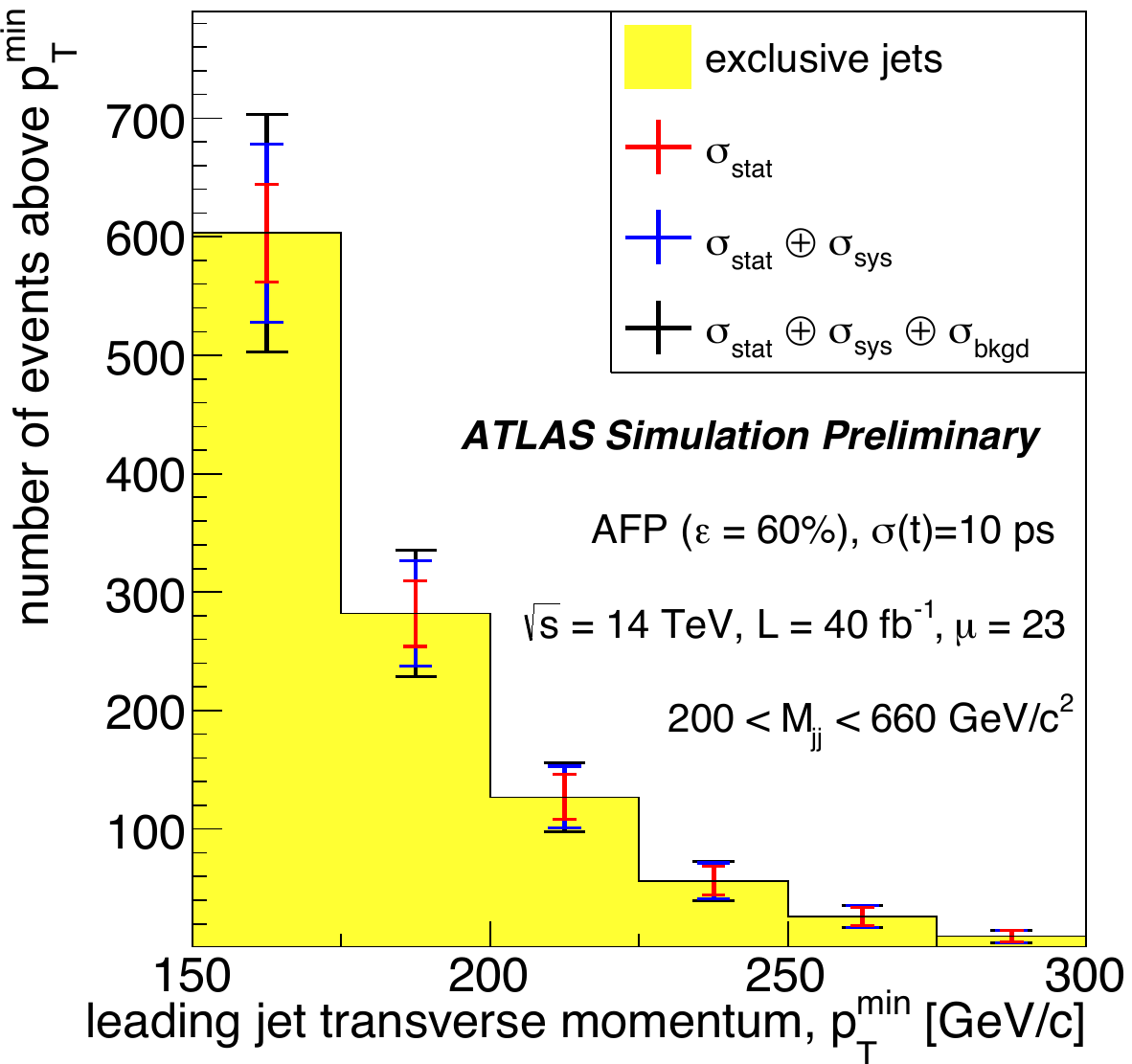}\\ \vspace{0.5 cm}
\includegraphics[width=.4\textwidth]{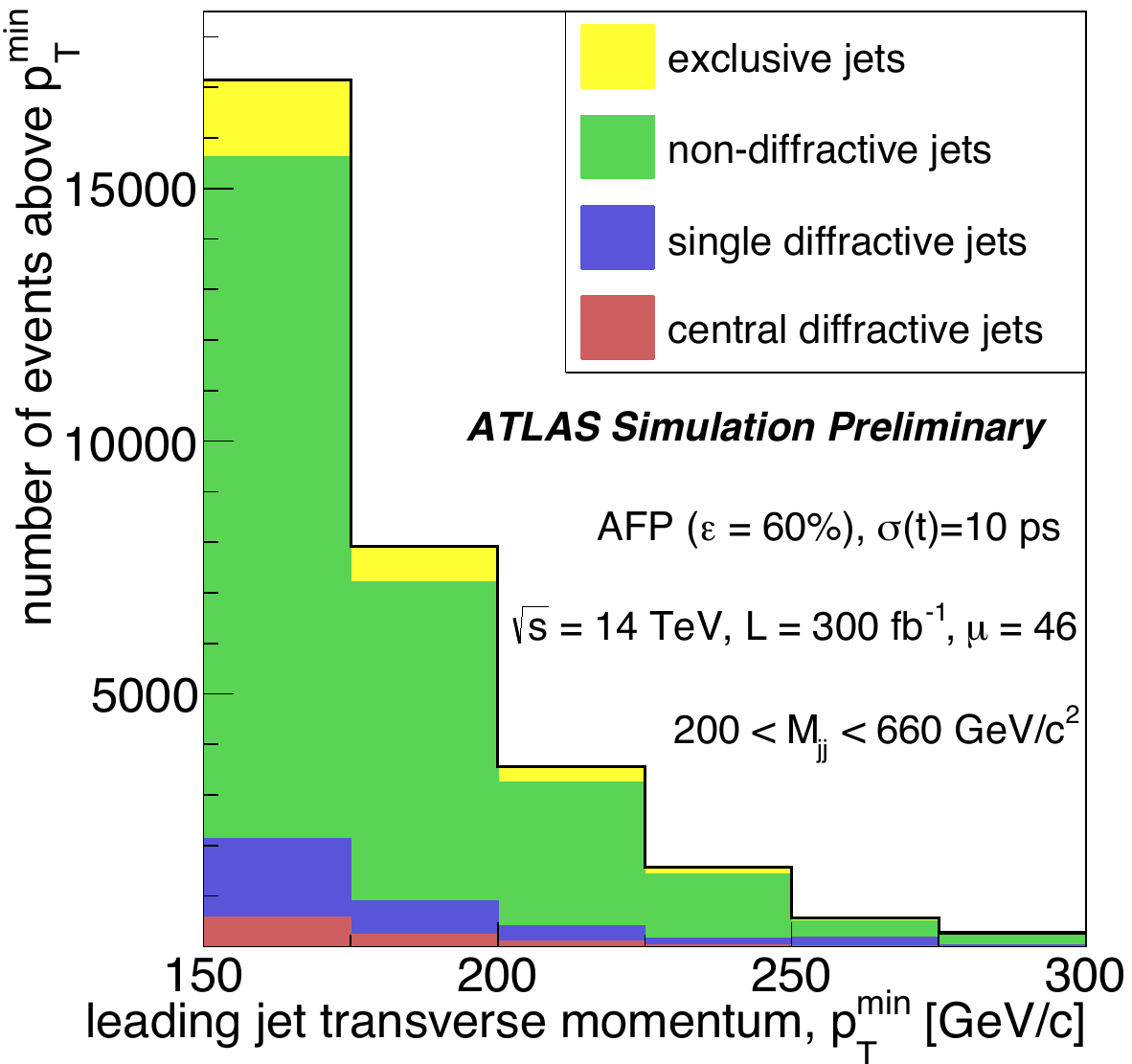}\qquad\qquad
\includegraphics[width=.4\textwidth]{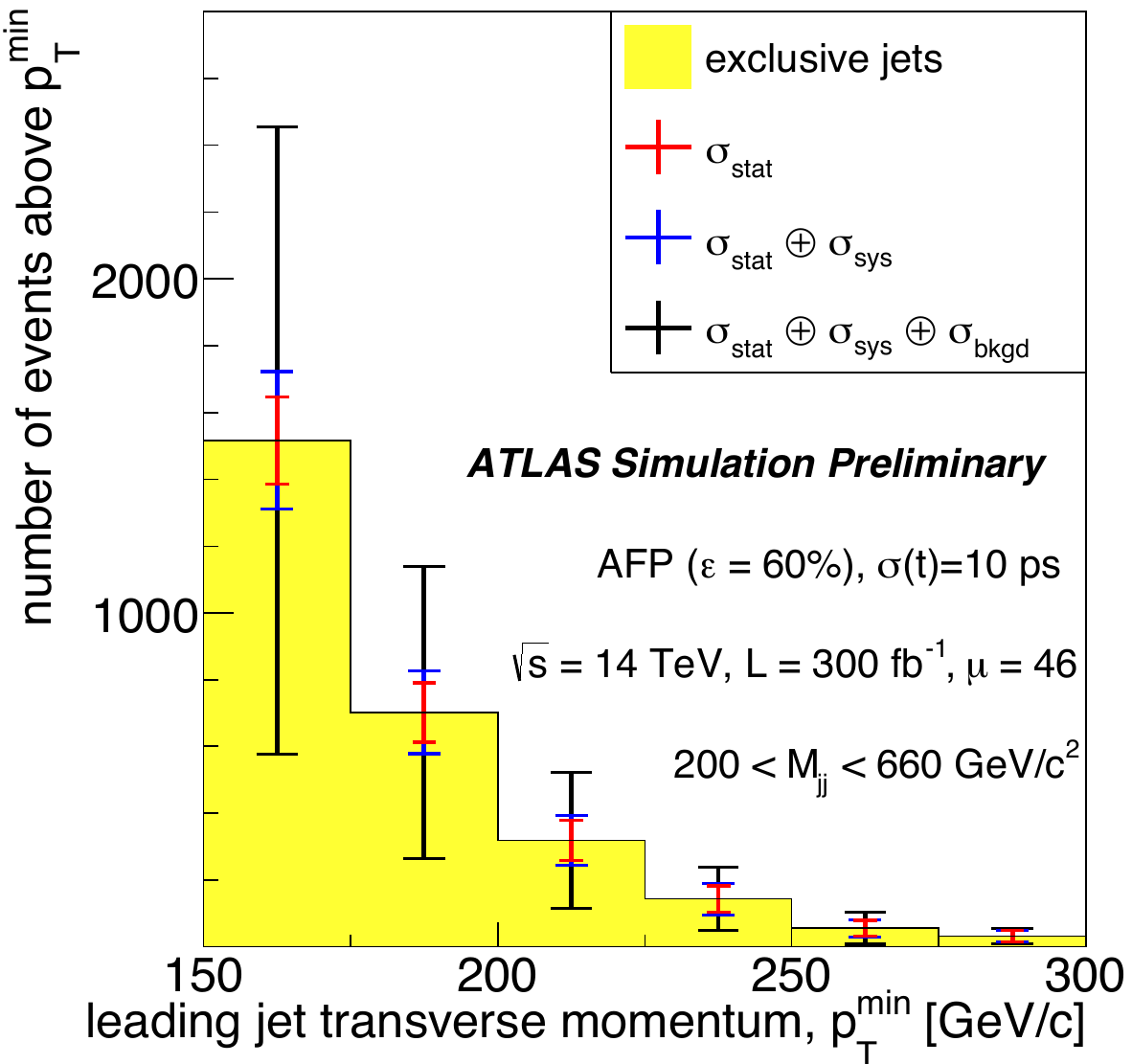}
\caption{\textbf{Left:} Number of accepted events as a function of the leading jet $p_T$ threshold. The background consists of the production of double Pomeron exchange (red), single diffractive (blue) and non-diffractive (green) jets. \textbf{Right:} Number of signal events, marked as yellow bar, with statistical ($\sigma_{stat}$), systematic ($\sigma_{sys}$) and background ($\sigma_{bkgd}$) uncertainties. The $\oplus$ sign indicates that the corresponding errors are added in quadrature. The upper (lower) plots correspond to an integrated luminosity of  $L = 40\,(300)$ fb$^{-1}$ and average number of interactions of $\mu = 23\,(46)$}
\label{fig_final_23_jet_pT_min}
\end{figure}

The lack of both underlying event activity and proton/Pomeron remnants provides another handle with which to improve the signal purity. A track with $\eta>0$ is considered to be outside the jet system if $\eta_{trk} > \eta_{jet} + 0.7$ and $\eta_{trk} > \eta_{jet} + w_{jet} + 0.2$, where $w_{jet}$ is the reconstructed jet width~\cite{Cacciari:2008gp:ch5}, with a similar condition if $\eta<0$. A track is considered as perpendicular in $\phi$ to the leading jet if $\frac{\pi}{3} < \Delta \phi < \frac{2 \pi}{3}$ or $\frac{4 \pi}{3} < \Delta \phi < \frac{5 \pi}{3}$, where $\Delta \phi$ is the azimuthal angle between the track and the leading jet. Then, by cutting on the number of tracks perpendicular in $\phi$ to the leading jet and outside the jet system in $\eta$ (required to be less than or equal to 2 and 5, respectively), the background can be reduced by almost two orders of magnitude.

In addition, three further selection criteria were applied in this analysis: the azimuthal angle between the two leading jets must satisfy $2.9 < \Delta\phi < 3.3$ -- exclusive jets are expected to be produced back-to-back; the missing mass must satisfy $m_{X} < 550$ GeV -- the $\xi$ distribution falls much more steeply for exclusive jets production; the mass fraction is required to be $0.9 < m_{jj}/m_{X}$ $<1.3$ -- this variable was proposed in~\cite{Khoze:2006iw} to reduce the effects of hard state radiation and is strongly correlated with mass ratio and rapidity difference requirements. The change in the fiducial cross section for signal and background processes after each selection requirement for an integrated luminosity of 40 fb$^{-1}$ and average number of interactions of $\mu = 23$ is shown in Fig.~\ref{fig_cut_summary_mu23}. The dominant background is due to non-diffractive jet production overlaid with protons from pile-up interactions. After all selection requirements the signal to background ratio is increased from $10^{-6}$ to $\sim 0.57$. 

The analysis was repeated for the average number of interactions of $\mu = 46$ and an integrated luminosity of 300 fb$^{-1}$. The selection criteria for these conditions are similar to the ones used in the $\mu = 23$ analysis. The dominant background from ND dijet events overlaid with protons from minimum bias events increases with respect to the signal, as the probability of producing this combinatorial background increases quadratically with the number of interactions per beam bunch crossing.

The leading jet transverse momentum above a given threshold for an integrated luminosity $L = 40\, (300)$ fb$^{-1}$ and an average number of interactions of $\mu = 23\, (46)$ are presented in the upper (lower) panels of Fig. \ref{fig_final_23_jet_pT_min}. Although the statistical significance is roughly the same in both scenarios, the impact of statistical and background uncertainties is much larger in the latter situation. Improvements in the AFP timing detector resolution and/or the analysis method are needed in order to control the background modelling uncertainties. For example, if the background is measured in control regions to an accuracy of $\sim 1\%$ then the accuracy of the cross section measurement would be similar to that for $\mu = 23$.

In summary, while the initial signal to background ratio for exclusive jet production in AFP is about $10^{-6}$, after dedicated signal selection cuts have been applied, this reduces to about 0.57 (0.16) for $\mu = 23$ (46). In both cases the statistical errors are considerably smaller. The biggest uncertainty is associated with the modelling of the background from ND dijet events overlapping with two protons from pile-up events. The impact of the ND background on the measurement ultimately depends on the success of data--driven methods using dedicated control regions. In case of $L = 40$ fb$^{-1}$ and $\mu = 23$, the measurement will be challenging, but potentially feasible. On the other hand, in order to make such a measurement in a higher pile-up environment, much better control of systematic effects is needed.

\subsection{Jet production: CT--PPS feasibility study}

In this section, a detailed study of the measurement possibilities for exclusive jet production with the CT-PPS detectors, based on the experimental techniques developed in Refs.~\cite{Aaltonen:2007hs,Chatrchyan:2012vc:ch5}, is summarised.

Events are selected by requiring a time coincidence in both arms of the CT-PPS. 
Leading protons are required to be in the CT-PPS fiducial region, 
and the arrival time difference at the CT-PPS location
depends on the z-vertex position, $\mathrm{z}_\mathrm{PPS}$, 
and must be consistent with the vertex position of the central di-jet system, $\mathrm{z}_\mathrm{vertex}$.
An expected time resolution of 10~ps (30~ps) is assumed.
Two jets with reconstructed transverse momenta $p_\mathrm{T}>100(150)$~GeV in the central ($|\eta|<2.0$) detector are required.
Jets are reconstructed using the anti-$k_\mathrm{T}$ jet clustering algorithm with a distance parameter of 0.5~\cite{Cacciari:2008gp:ch5}.
Finally, the instrumental background in the CT--PPS from additional sources is accounted for, as discussed in Chapter~\ref{chap:bema}.
The main physics backgrounds are from minimum bias events --including SD and DPE events--
in coincidence with either two jets in the central detector or another leading proton within the PPS detector acceptance. A cut on the time-of-flight difference $\Delta t$ that varies  according to the z-vertex position, which keeps approximately 60\%~(50\%) of signal events
while reducing the inclusive di-jet background by a factor 33~(18), for a 10~(30)~ps timing resolution, is chosen.
 
\begin{figure*}[htb]
\centering
\includegraphics[width=0.4\textwidth]{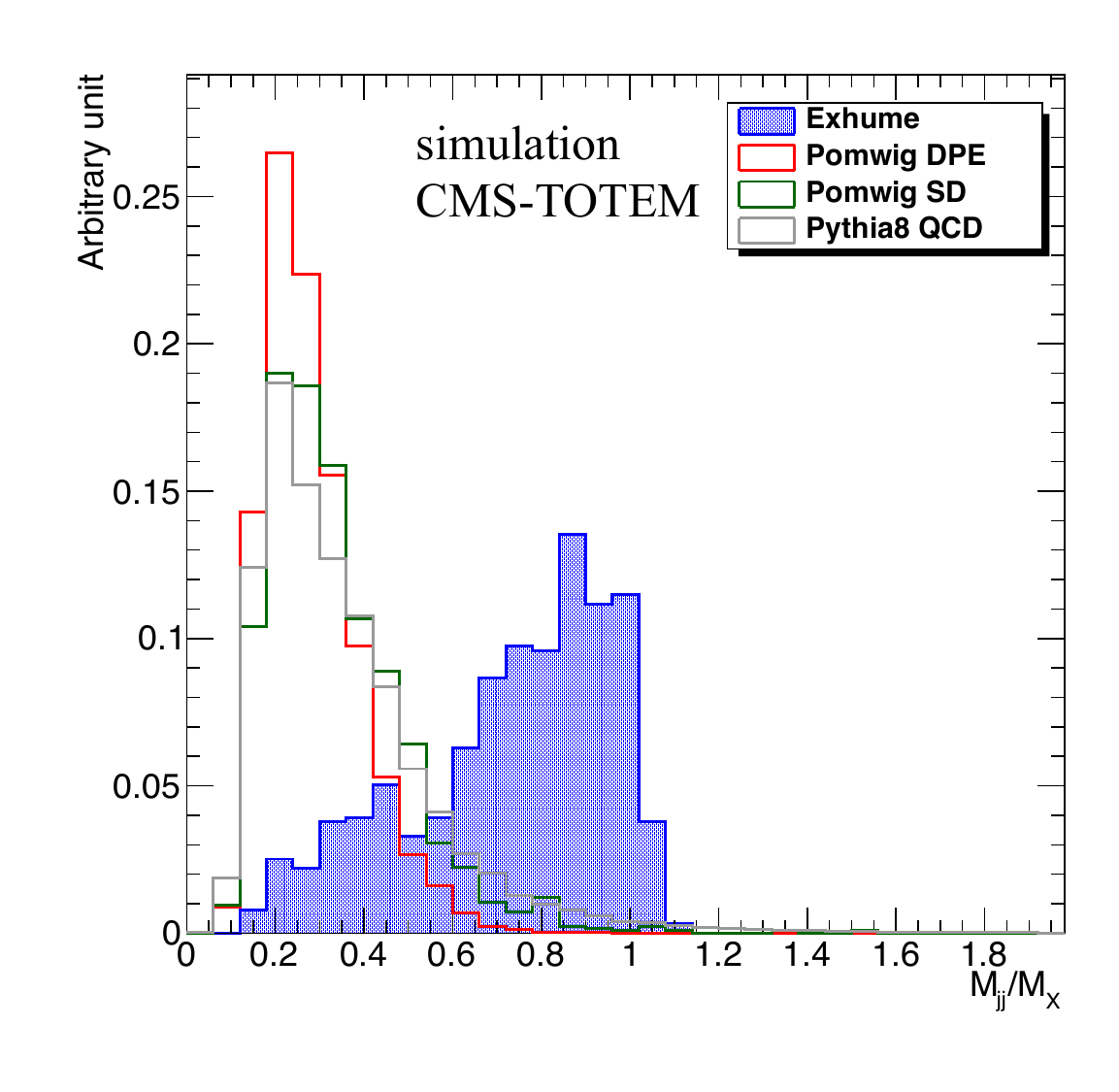}\qquad \qquad
\includegraphics[width=0.4\textwidth]{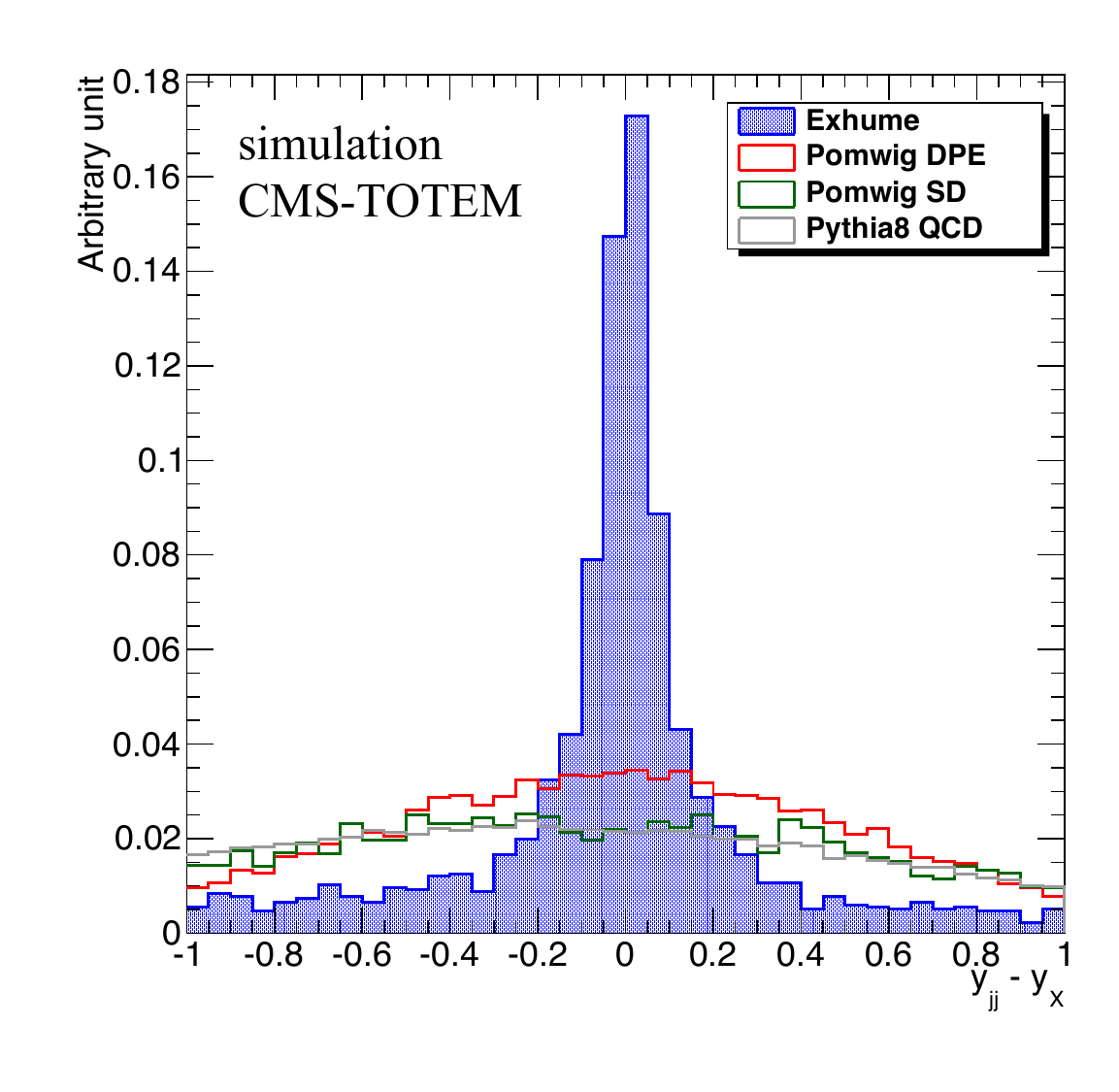}
\caption{
{\it Left:} Dijet mass fraction $R_\mathrm{jj}=M_\mathrm{jj}/M_\mathrm{X}$ (left). 
{\it Right:} Rapidity difference of jet ($y_\mathrm{jj}$) and proton ($y_\mathrm{X}$) systems. 
Distributions are shown
for exclusive di-jet signal (ExHuME) and for background (SD, DPE, inclusive dijets) events
and are normalized to unit area.
}
\label{fig:kinematics_jj2}
\end{figure*}

Fig.~\ref{fig:kinematics_jj2} (left) shows 
the di-jet mass fraction, $R_\mathrm{jj}=M_\mathrm{jj}/M_\mathrm{X}$, and
the rapidity difference (right) of the jet system ($y_\mathrm{jj}$) and the proton system, $y_\mathrm{X}=0.5\cdot \ln (\xi_1/\xi_2)$. 
Consistency is required between the values of the jet mass system measured in the central detector ($M_\mathrm{jj}$) and in the CT--PPS ($M_\mathrm{X}$), 
and the requirement $0.70<R_\mathrm{jj}<1.15$ is applied.
A selection cut of $|(y_\mathrm{jj}-y_\mathrm{X})|<0.1$ is also required.

\begin{figure*}[htb]
\centering
\includegraphics[width=0.4\textwidth]{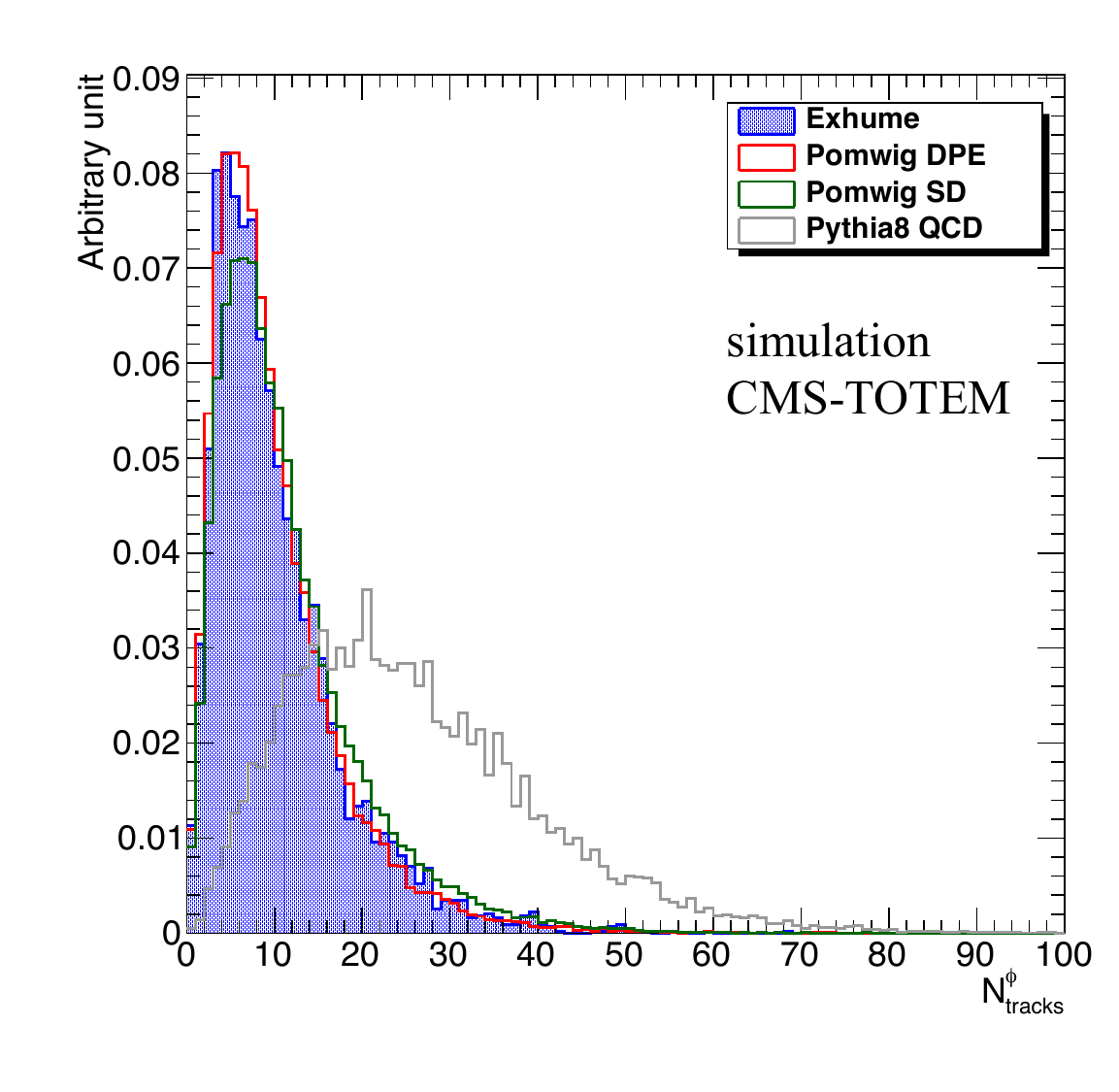}\qquad \qquad
\includegraphics[width=0.4\textwidth]{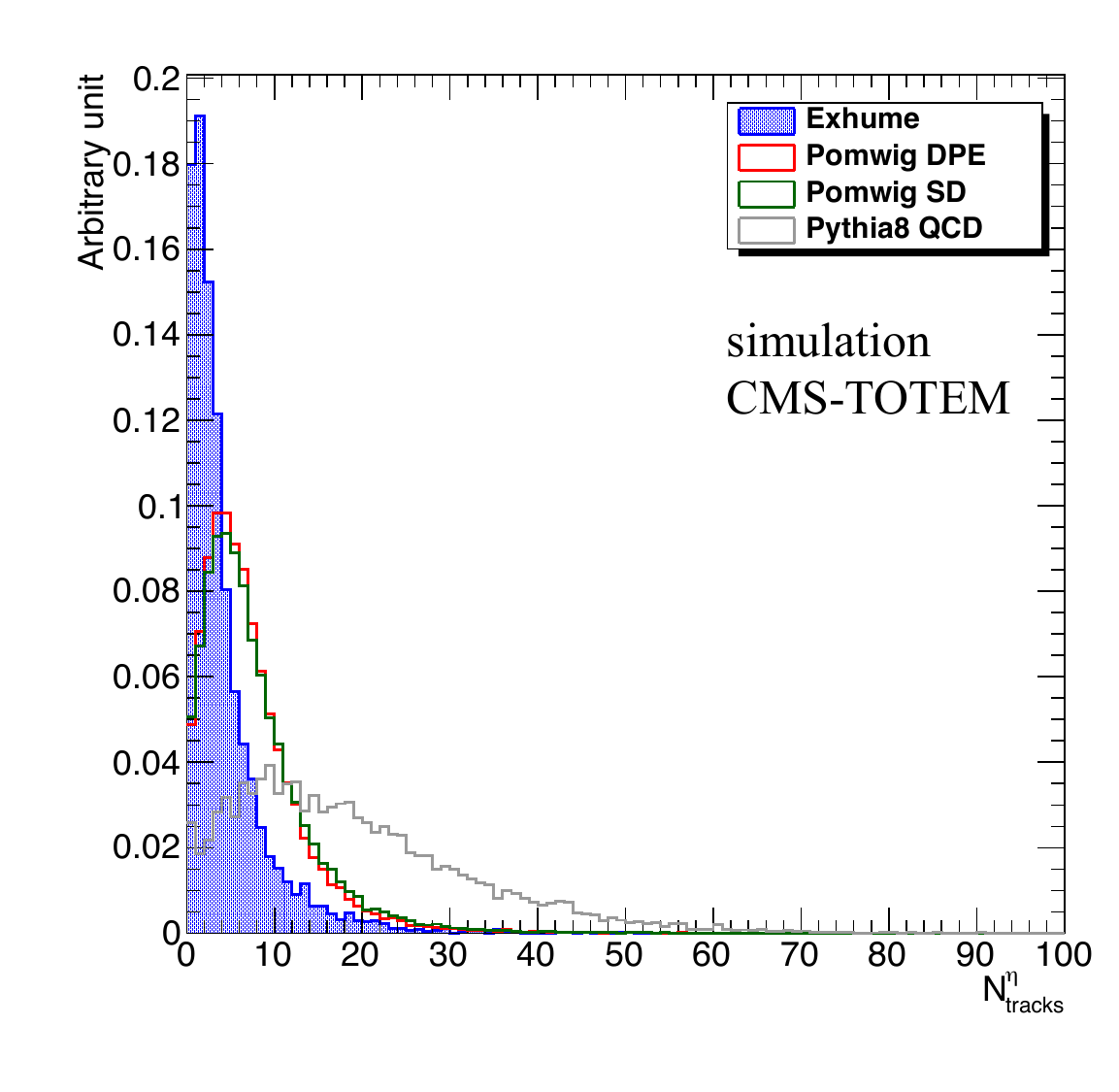}
\caption{
Number of extra tracks outside the jet system in $\phi$ (left) and in $\eta$ (right), associated to the di-jet vertex for exclusive dijet and background processes. 
Distributions are shown after the leading proton time-of-flight correlation requirements (with a 10~ps resolution).
Event yields are normalized to unit area.
}
\label{fig:trackmultiplicity_jj}
\end{figure*}

\begin{figure*}[htb!]
\centering
\includegraphics[trim=0 1.5cm 0 0, width=0.7\textwidth]{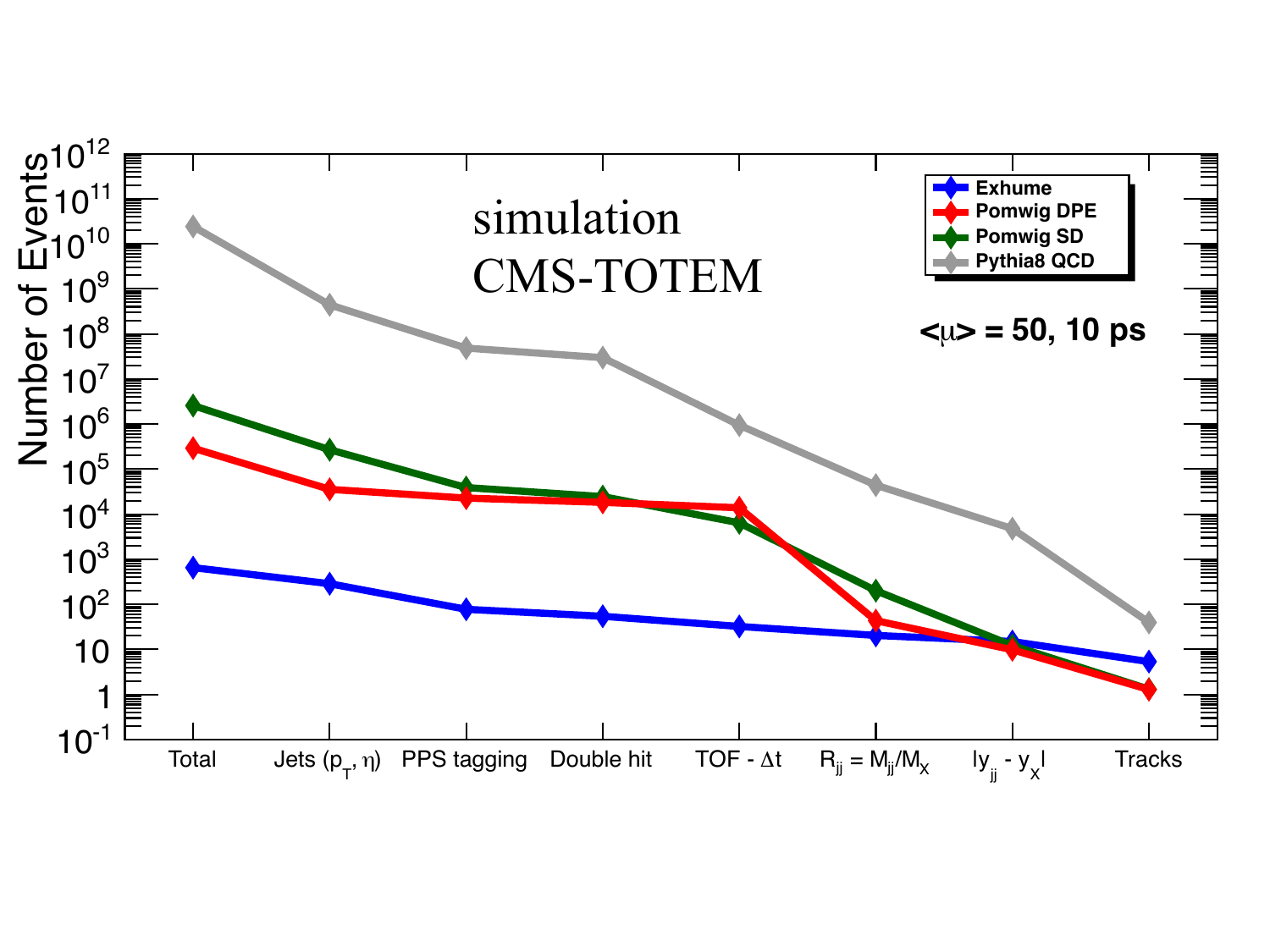}
\includegraphics[trim=0 2cm 0 2cm, width=0.7\textwidth]{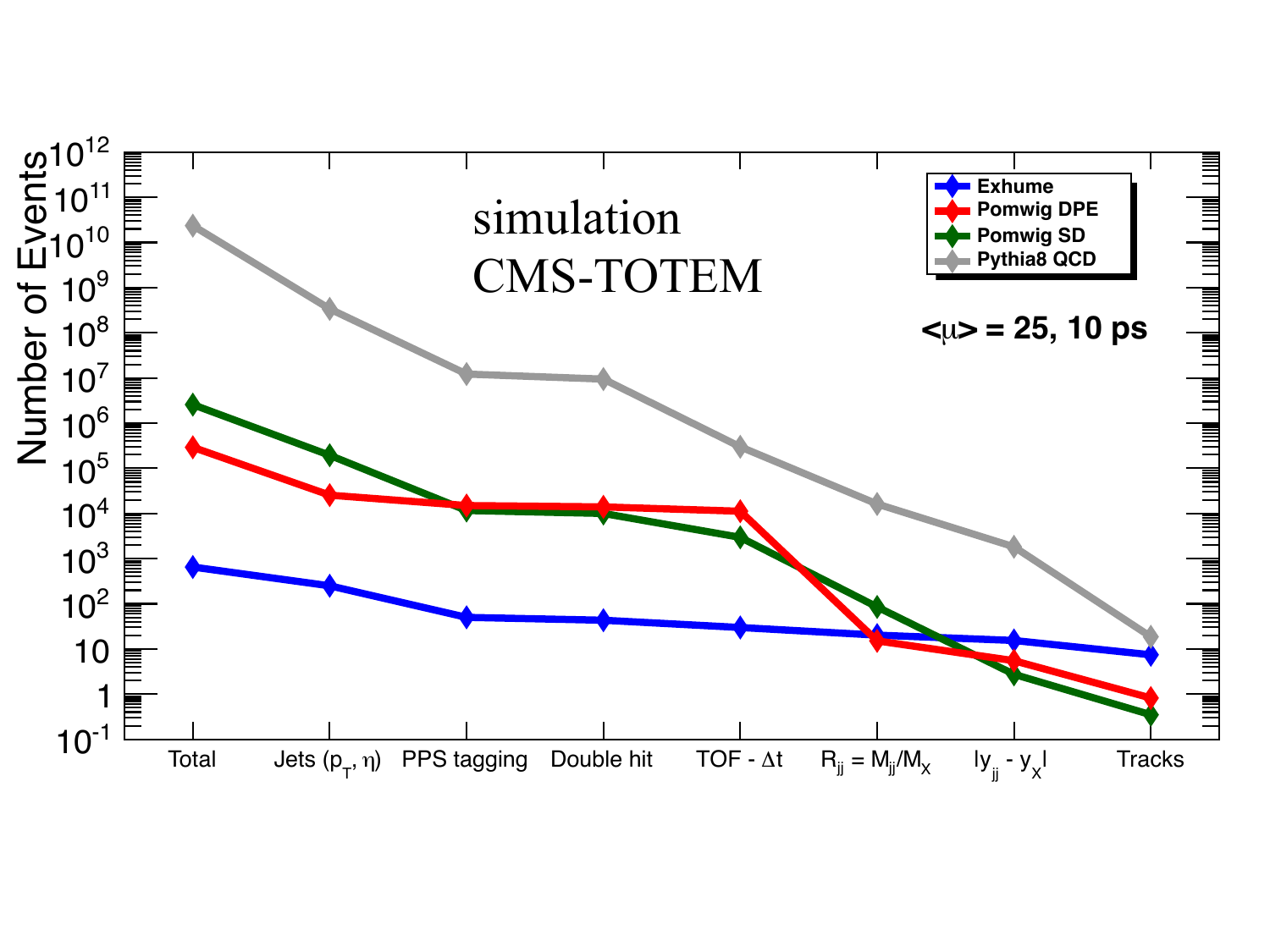}
\caption{
Graphical illustration of the event yields for signal and background processes as a function of the cuts applied.
A time resolution of 10~ps is assumed, and an average pile up multiplicity of $\mu=50$ (25) is taken in the left (right) figure.
Yields are normalized to an integrated luminosity of 1~fb$^{-1}$.
}
\label{fig:yields_pu50}
\end{figure*}

The track multiplicity associated to the di-jet vertex is used to discriminate exclusive signal events against backgrounds. 
In particular, techniques developed in~\cite{atlas_ntrack} are exploited.
Two variables that account for the ``exclusivity'' of the event by counting the number of extra tracks between the jets, both in $\phi$ and $\eta$, are built, denoted by $N_\mathrm{tracks}^{\phi}$ and $N_\mathrm{tracks}^{\eta}$. All tracks from the primary vertex are considered and the area of $-1.0$~($+1.0$) away from the minimum (maximum) jet $\eta$ coordinates is defined, 
$\eta_\mathrm{min}$ and $\eta_\mathrm{max}$.
Then,  the number of extra tracks that are below (above) the $\eta_\mathrm{min}$ ($\eta_\mathrm{max}$) position are counted.
Similarly, the $N_\mathrm{tracks}^{\phi}$ variable is built. In this case, the number of tracks that are perpendicular to the plane formed by the two-jet system are counted, in the region $0.54<\phi<2.60$. The track multiplicity variables after the timing selection cuts are shown in Fig.~\ref{fig:trackmultiplicity_jj}.
Exclusive signal events tend to have significantly lower track multiplicity than inclusive di-jet events in either $\phi$ and $\eta$. 
The tracking multiplicity separation in $\eta$ helps in further rejecting SD and DPE events.
Events are kept if the conditions $N_\mathrm{tracks}^{\phi}\leq 9$ and $N_\mathrm{tracks}^{\eta}\leq2$ are satisfied.
Fig.~\ref{fig:yields_pu50} illustrates the evolution of the event yields as a function of the cuts applied for a time resolution of 10~ps.
The cross section for signal events (given by ExHuME) is multiplied by a factor of 5/3 to simulate 
a gap survival probability of 5\% (i.e. the same used for DPE di-jet event processes in \textsc{pomwig}),
instead of the gap survival probability given by ExHuME of 3\%.

\begin{table*}[htp!]
\caption{
Number of expected signal and background (SD, DPE, and inclusive dijets) events (after the $N_\mathrm{tracks}$ cut), for separate bins of missing mass $M_\mathrm{X}$.
Yields normalized to an integrated luminosity of 1~fb$^{-1}$ are shown for
average pile up multiplicities of $\mu=25$ and $\mu=50$. 
Statistical uncertainties are shown.
A timing resolution of 10~ps is assumed.}
\label{tab:yields_jj_mx}
\begin{center}
\begin{tabular}{l|c|c|c|c|c} 
\hline\hline
									& Exclusive di-jets 	& DPE 			& SD 		& Inclusive dijets 	& S:B \\
\hline
\multicolumn{6}{l}{pile up $\mu=25$} \\
\hline
$M_\mathrm{X}\leq 500$~GeV				& 4.0$\pm$0.2		& 0.2$\pm$0.1		& 0$\pm$1	& $1\pm 1$		& 3:1\\
$500<M_\mathrm{X}\leq 800$~GeV			& 3.1$\pm$0.2		& 0.3$\pm$0.1		& 0$\pm$1	& $15\pm 1$		& 1:5\\
$M_\mathrm{X}>800$~GeV				& 0.3$\pm$0.1		& 0.3$\pm$0.1		& 1$\pm$1	& $4 \pm 1$		& 1:18 \\
\hline 
\multicolumn{6}{l}{pile up $\mu=50$} \\
\hline
$M_\mathrm{X}\leq 500$~GeV				& 2.8$\pm$0.2		& 0.6$\pm$0.2		& 0$\pm$1	& $5 \pm 1$		& 1:2\\
$500<M_\mathrm{X}\leq 800$~GeV			& 2.3$\pm$0.2		& 0.7$\pm$0.3		& 1.3$\pm$1.0	& $26\pm 1$		& 1:12\\
$M_\mathrm{X}>800$~GeV				& 0.3$\pm$0.1		& 0$\pm$1		& 0$\pm$1	& $9\pm 1$		& 1:30 \\
\hline\hline
\end{tabular}
\end{center}
\end{table*}

The study demonstrates the feasibility of measuring exclusive di-jet production in Run~2 with CT--PPS with an average pile up of $\mu=50$. 
A signal-to-background ratio of S:B$\simeq$1:8 can be achieved (after the ``$N_\mathrm{tracks}$'' cut), 
with the signal events appearing as an enhancement of the $M_{jj}/M_X$ distribution around $M_{jj}/M_X=1$.
 The case of a lower average number of pile up interactions, $\mu=25$, has also been studied. 
A signal-to-background ratio of S:B$\simeq$1:3 can be achieved in the less harsh condition with pile up of $\mu=25$ (after the ``$N_\mathrm{tracks}$'' cut).
Fig.~\ref{fig:yields_pu50} (right) illustrates the evolution of the event yields as a function of the cuts applied for a time resolution of 10~ps. Table~\ref{tab:yields_jj_mx} summarises the yields of signal and background events (after the $N_\mathrm{tracks}$ cut)
estimated in bins of separate missing mass regions, 
$M_\mathrm{X}<500$~GeV (where most of the signal is expected), 
$500<M_\mathrm{X}<800$~GeV, and $M_\mathrm{X}>800$~GeV.
Yields normalized to an integrated luminosity of 1~fb$^{-1}$ are shown for
average pile up multiplicities of $\mu=25$ and $\mu=50$, and a timing resolution of 10~ps is assumed.

  \section{Photon--induced and photoproduction processes}\label{CEP:phoex}

 In this Section, theoretical discussion of two--photon induced and photoproduction processes, and motivations for future measurements, are presented.

  \subsection{Introduction}

High energy charged particles are a source of a flux of Weizs\"acker-Williams (WW) photons~\cite{vonWeizsacker:1934sx:ch5,Williams:1934ad:ch5}. At the LHC, this opens the possibility to study photon--hadron interactions at unprecedented energies.  Such reactions may be observed in ultraperipheral heavy ions collisions, where the WW flux ($\propto Z^2$) is enhanced by the large charge $Z$ of the ion, as well as proton-proton (and proton-antiproton)  collisions. 

In this section the two--photon collision and photoproduction processes described in the introduction are discussed. The former proceeds via the $\gamma \gamma \to X$ subprocess and is a theoretically well--understood purely QED process, up to small corrections due to additional soft proton--proton interactions, while in the latter case one proton interacts strongly while the other interacts via photon exchange (in the language of Regge theory, via photon--pomeron fusion, $\gamma I\!\!P \to X$).  

Due to the well understood production mechanism in the initial state, two--photon collision processes provide a valuable handle on the electroweak sector, can in principle serve as an effective luminosity monitor, and as a tool for BSM particle discovery (see Section~\ref{sec:exploratory}), while studies of such processes with dissociating outgoing protons can provide information about the soft survival factors introduced in Section~\ref{CEP:gluonex:intro}.

The photoproduction of vector mesons, $pp \to p + V + p$, has been the focus of much theoretical study, see for example~\cite{Klein:2003vd,Schafer:2007mm,Khoze:2008cx,Motyka:2008ac,Jones:2013pga,Guzey:2014axa}.  The virtuality of equivalent photons is controlled by the electromagnetic form factors of the proton, for which quasi--real photon exchanges are dominant, so that the dominant momentum transfers are deeply in the non--perturbative region. On the other hand, a hard scale necessary for the application of perturbative QCD may be supplied by the quark mass. Therefore, among the possible final states, heavy $c\overline{c}$, $b\overline{b}$ quarkonium for which a pQCD approach may be considered, stand out as being of special interest; for heavy vector mesons ($J/\psi,\Upsilon$) it serves as a probe of the small--$x$ (unintegrated) gluon distribution, see for example~\cite{Jones:2013pga}, as well as of models of gluon saturation. In addition, 
it has long been speculated that besides the $C$-even predominantly imaginary Pomeron exchange, the Regge-phenomenology of strong interactions at high energies would feature a $C$--odd, dominantly real, trajectory, known as  the Odderon (see for example~\cite{Barone:2002cv,Donnachie:2002en,Ewerz:2003xi}).  
This exchange is experimentally very elusive, and as of yet no firm evidence for it exists, but it should in general contribute to the vector meson CEP cross section (via $O I\!\!P \to X$); thus measurements of exclusive vector meson production could constitute the first clear experimental evidence for the Odderon.
 A further process that may be studied is the photoproduction of $Z$ boson: this would represent a completely new observation and can provide a test of the pQCD model of the production mechanism. Further details of all of the topics discussed above can be found in the following sections.

   \subsection{Forward proton tagging: phenomenological insight and advantages}\label{sec:tagcepmotphot}

One process discussed in e.g.~\cite{Khoze:2000db}, for which proton tagging is particularly beneficial, is the exclusive production of lepton pairs. This purely QED cross section is known theoretically to very high, sub $1\%$ level, precision, and so represents a potential luminosity monitor. Achieving this level of theoretical precision, and in particular the high insensitivity to the effect of additional proton--proton soft rescatterings, relies crucially on the fact that the interaction is truly exclusive, that is the protons remain intact after the collision. Proton tagging is the only way to select such purely exclusive events. On the other hand, achieving this level of precision in the experimental measurement may be challenging due to the effect of systematic errors, and the relatively low cross sections within the invariant mass acceptance of the tagging detectors during high luminosity running.

 Proton tagging can also provide an additional handle regarding the Odderon, the C--odd partner of the Pomeron, discussed above. It has been shown~\cite{Motyka:2008kh} that any contribution of the Odderon to the vector meson photoproduction cross section suffers from important uncertainties and may be hard to disentangle from the photon--exchange contribution. On the other hand, a firm prediction is that the distribution with respect to the proton (or meson) transverse momentum $p_\perp$ should be much harder in the Odderon case, and so a measurement of this distribution, in particular at larger $p_\perp$ values, could provide evidence for the Odderon. However, at large $p_\perp$ the probability that the interacting proton will dissociate rapidly increases, and so such a measurement would be very challenging using rapidity gap based selection techniques. By tagging the protons, this dissociative contribution can be effectively eliminated, and a clean probe of the Odderon provided. In addition, as discussed in~\cite{Armesto:2014sma}, by measuring the proton $p_\perp$ distribution it may be possible to distinguish between models with and without gluon saturation effects. In particular, as a result of unitarity features of the colour dipole amplitude in the saturation regime, there is expected to be a pronounced dip (or multiple dips) in the higher $p_\perp$ region. The observation of these dips, for which proton tagging is clearly essential, would represent clear evidence in favour of such models.
 
\subsection{Two-photon collisions}

\subsubsection*{Motivation and theory}

Two--photon production in $pp$ collisions is described in the framework of Equivalent Photon Approximation 
(EPA)~\cite{vonWeizsacker:1934sx:ch5,Williams:1934ad:ch5,Budnev:1974de}. The almost real photons (with low photon virtuality $Q^2=-q^2$) are 
emitted by the incoming protons, producing an object $X$ in the $pp\rightarrow pXp$ process, 
through two--photon exchange $\gamma\gamma\rightarrow X$.
The photon spectrum of virtuality $Q^2$ and
energy $E_{\gamma}$ is proportional to the Sommerfeld fine--structure constant $\alpha$  and reads (in the lab frame)
\begin{equation}
\d N = \frac{\alpha}{\pi}\frac{\d E_{\gamma}}{E}\frac{\d Q^2}{Q^2}
 	 \left[ \left(1-\frac{E_{\gamma}}{E}\right)\left(1-\frac{Q^2_{min}}{Q^2}\right)F_E +
	         \frac{E_{\gamma}^2}{2E^2}F_M\right]\;,
\label{flux}
\end{equation}
where $E$ is the energy of the incoming proton of mass $m_p$, $Q^2_{min}\equiv
m^2_p E^2_{\gamma}/[E(E-E_{\gamma})]$ is the photon minimum virtuality allowed by
kinematics and $F_E$ and $F_M$ are functions of the electric and magnetic form factors. They read
in the dipole approximation~\cite{Budnev:1974de}
\begin{equation}
F_M=G^2_M  \qquad F_E=(4m_p^2G^2_E+Q^2G^2_M)/(4m_p^2+Q^2)\qquad G^2_E=G^2_M/\mu_p^2=(1+Q^2/Q^2_0)^{-4}\;.
\label{newera:elmagform}
\end{equation}
The magnetic moment of the proton is $\mu_p^2=7.78$ and the fitted scale $Q^2_0=0.71\,\GeV^2$.
The electromagnetic form factors are steeply falling  functions of $Q^2$: for this reason the two--photon cross section can be factorized into a sub--matrix element and  two photon fluxes. 

The theoretical framework described above is firmly established, and in principle allows pure QED two--photon collision processes to be described to a high degree ($\lesssim 0.1\%$) of accuracy. As discussed in Section~\ref{sec:tagcepmotphot}, this presents the possibility of using exclusive di-lepton production, which proceeds via this mechanism, as a luminosity monitor at the LHC if the systematics are low enough. 
Alternatively, this accurate knowledge of the two--photon initial state allows such processes to be used as effective probes of BSM physics. In particular, as discussed in Section~\ref{sec:excbos}, this can provide by far the most stringent probe that is currently possible of triple and quartic anomalous gauge boson couplings, as well as providing a complementary measurement of SM gauge boson pair production.

Finally, as discussed in~\cite{HarlandLang:2012qz}, while, due to the peripheral nature of the interaction, the probability of additional soft proton--proton interactions, which generate the soft survival factor, is very small, this is only the case provided proton dissociation can be effectively eliminated. Conversely, if proton dissociation is allowed in the event selection then such processes can provide a differential probe of the soft survival factor, for example by measuring the region of higher $p_\perp(l^+ l^-)$ in quasi--exclusive lepton pair production. Here, the deviation of the observed cross section from simulations including proton dissociation, but which do not include soft survival effects, can be evaluated. Such deviation has already been seen in the CMS measurement~\cite{Chatrchyan:2013foa} of $\gamma\gamma \rightarrow \mu\mu$, performed in the context of a $\gamma\gamma \rightarrow WW$
analysis.

  \subsubsection*{Experimental results and outlook}

\begin{figure}[ht]
{\includegraphics[width=0.49\linewidth]{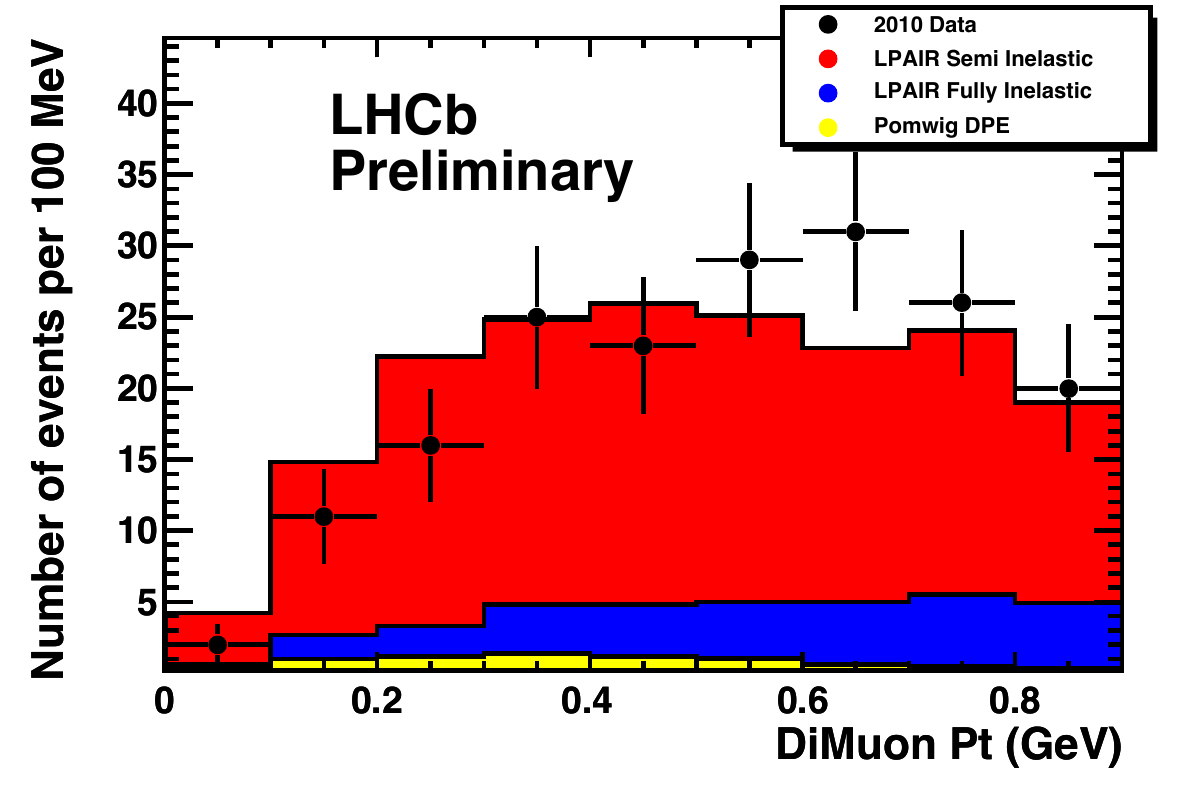}}
{\includegraphics[width=0.49\linewidth]{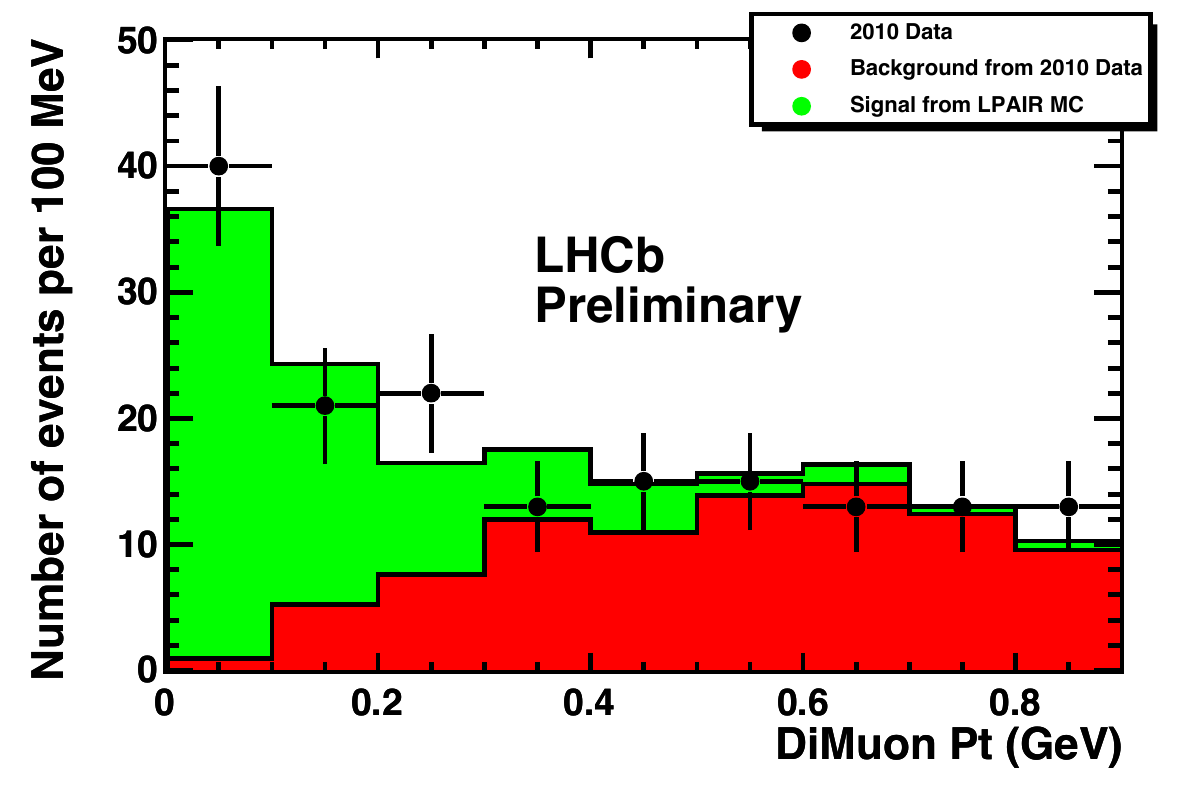}}
\caption{
Transverse momentum of di-muons that have an invariant mass above 2.5 GeV and are not 
consistent with vector meson production.  The plot on the left shows events with more
than two tracks compared to expectations for inelastic di-muon production.  The plot on
the right shows events with exactly two tracks and no other activity inside the LHCb detector.
The shape of the signal is taken from LPAIR.  The background shape is taken from the
data in the left-hand plot.
\label{fig:ggpt}}
\end{figure}

A preliminary measurement of the cross section for
muon pairs produced through two-photon fusion has been made 
by the LHCb collaboration~\cite{LHCb:2011dra}
using the small 2010 data sample of $37 {\rm\ pb}^{-1}$.
The selection is as described in Sec.~\ref{sec:photooutlook} and the candidate events are
those in Fig.~\ref{fig:mu2mass},
with masses above 2.5 GeV but outside mass windows around the vector meson
resonances.
To determine the elastic CEP component, a fit to the transverse momentum distribution is made,
using a template shape from the LPAIR simulation~\cite{Vermaseren:1982cz,Baranov:1991yq} to describe the elastic signal events
and using data, (low multiplicity di-muon candidates that have additional tracks) 
to describe the background.
A comparison of this data-driven background estimate to the simulation
of inelastic di-muon production, where one or both protons dissociate, 
shows good agreement (see the left plot in Fig.~\ref{fig:ggpt},)
albeit with rather large uncertainties
due to the limited statistics.
The fit to the signal candidates in the right plot of Fig.~\ref{fig:ggpt} also
shows good agreement and an almost
pure sample of di-muons from elastic di-photon fusion is obtained 
when requiring the $p_T$ of the pair to be below 100 MeV.
A cross section times branching fraction estimate of $67\pm 19$~pb for both muons produced inside the LHCb
acceptance is in agreement with the theoretical prediction of \mbox{42 pb}, but is a long way from 
the aim of a few-percent measurement.

\begin{figure}[ht]
\centerline{\includegraphics[width=0.49\linewidth]{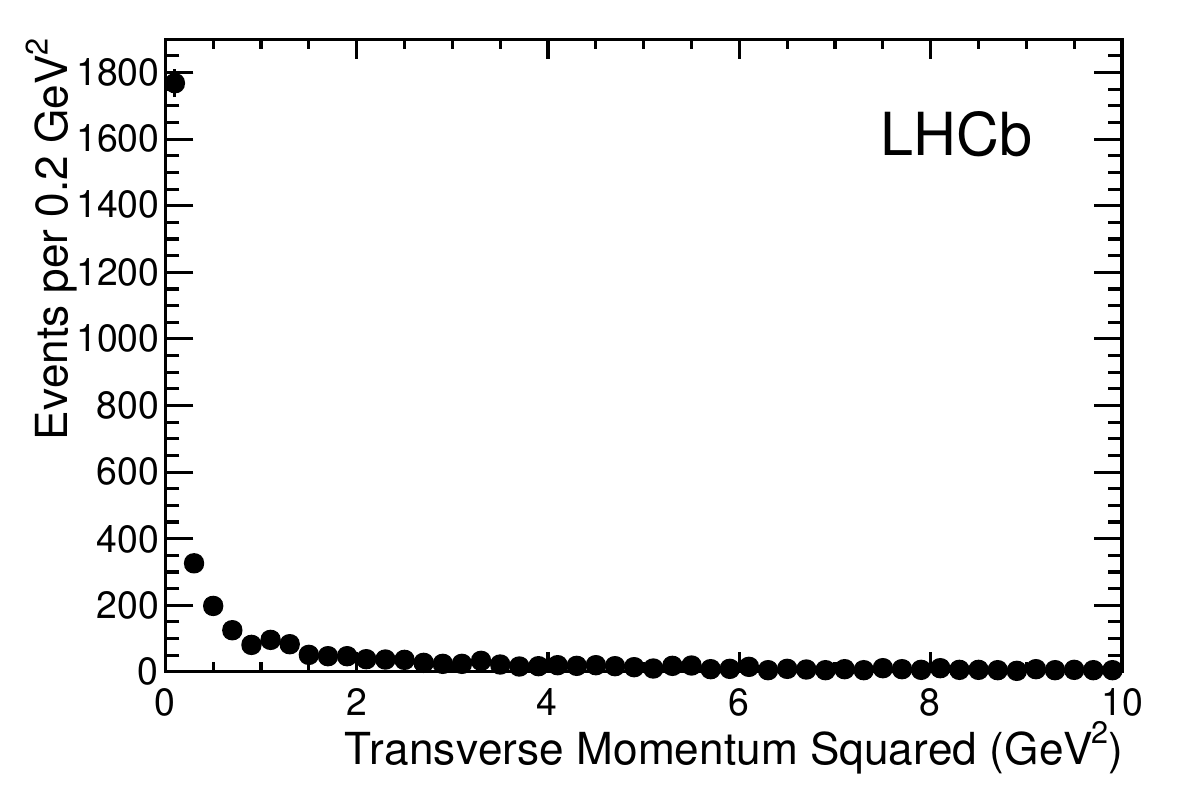}}
\caption{
Transverse momentum squared of di-muons 
with an invariant mass between 6 and 9 GeV.
\label{fig:mu2pt}}
\end{figure}

There are only 40 candidates with $p_T<100{\rm \ MeV}$ in the analysis of $37 {\rm\ pb^{-1}}$ of data, but improvements
to the trigger and a reduction in the average number of proton-proton interactions per beam
crossing suggest that about 10,000 candidates are available in the 
$3 {\rm\ fb^{-1}}$ of data taken in 2011 and 2012, sufficient for a 1\% statistical measurement.
Fig.~\ref{fig:mu2pt} shows the $p_T^2$ distribution for 
di-muon candidates with an invariant mass between 6 and 9 GeV in about
$3 {\rm\ fb^{-1}}$ of data taken at $\sqrt{s}=7$ and 8 TeV.
The strong peak below $0.2 {\rm \ GeV}^2$ is characteristic of the QED process.
Note that the distribution falls off much more rapidly than the $\approx \exp(-6p_T^2)$
dependence for the $J/\psi$ in Fig.~\ref{fig:pt2}.
An estimate of how much signal there is in the first bin requires
a complete description of the spectrum.

Measurements of $\gamma\gamma \rightarrow \mu\mu$~\cite{Chatrchyan:2011ci} and $\gamma\gamma \rightarrow ee$~\cite{Chatrchyan:2012tv} 
production at $\sqrt{s} = 7$~TeV were also performed by CMS using 40 pb$^{-1}$ and 36 pb$^{-1}$ of data,
respectively\footnote{In the final stages of the preparation of this report, ATLAS have reported the measurement of exclusive $e^+e^-$ and $\mu^+\mu^-$ production at $\sqrt{s}=7$ TeV, where the muons (electrons) are required to have  $p_\perp> 10(12)$ GeV, and in both cases $|\eta^l|<2.4$. See~\cite{Aad:2015bwa} for details.} In the $e^{+}e^{-}$ channel events were selected by requiring two
electrons with $E_{T} > 5.5$~GeV, no additional charged tracks, and no unassociated
activity in the calorimeters above the noise threshold. A dedicated trigger was
employed to maintain a low electron threshold throughout the data-taking period.
The background, extrapolated from the sideband region in the calorimeter tower
and track multiplicities, was determined to be $0.85 \pm 0.28$ events.
In the data 17 events were observed, in agreement with the LPAIR prediction
of $16.3 \pm 1.3$ events from the sum of elastic and proton dissociation
production. In the $\mu^{+}\mu^{-}$ channel events were selected by requiring two
muons with $p_{T} > 4$~GeV, $|\eta| < 2.1$, and $m(\mu\mu)>11.5$~GeV, with
no other charged tracks associated to the dimuon vertex. This selection
method allowed for a much higher efficiency in the presence of pile up, compared
to the CMS $e^{+}e^{-}$ analysis. A template fit to the $p_{T}(\mu\mu)$ distribution
was then performed, to extract the elastic component of the cross section. The resulting fiducial
cross section was $3.38^{+0.58}_{-0.55} \mathrm{(stat.)} \pm {0.16}\mathrm{(syst.)} \pm {0.14}\mathrm{(lumi.)}~\mathrm{pb}$,
consistent at the $\sim 1\sigma$ level with the LPAIR prediction. In both the $e^{+}e^{-}$ and
$\mu^{+}\mu^{-}$ channels, the shapes of the single lepton and pair distributions were observed to 
be in good agreement with the predictions. In the context of the CMS $\gamma\gamma \rightarrow WW$
analysis, high-mass $\gamma\gamma \rightarrow \mu\mu$ events were also analyzed using a
much larger sample of 5 fb$^{-1}$ in order to study the proton dissociation contribution, as 
well as the effects of pile up on the selection~\cite{Chatrchyan:2013foa}.

Finally, one should note that exclusive processes in the $\gamma \gamma$ channel are very promising to reveal the elusive Odderon discussed above. Indeed, consider
the exclusive production of two $\pi^+ \pi^-$ pairs.
Since the $C-$parity of the amplitude describing this process is not fixed, both the Odderon and the Pomeron exchanges contributes. Considering charge asymmetries, one can therefore build an observable which involves the interference of Odderon and Pomeron, and thus linear in the tiny Odderon contribution~\cite{Pire:R1}.

    \subsection{Diffractive photoproduction $\gamma p \to V p$}\label{CEP:phoex:theory}

\subsubsection*{Motivation and theory: available models}
%%%%%%%%
\begin{figure}[ht] % Figure 1
\begin{center}
\includegraphics[height=4.0cm]{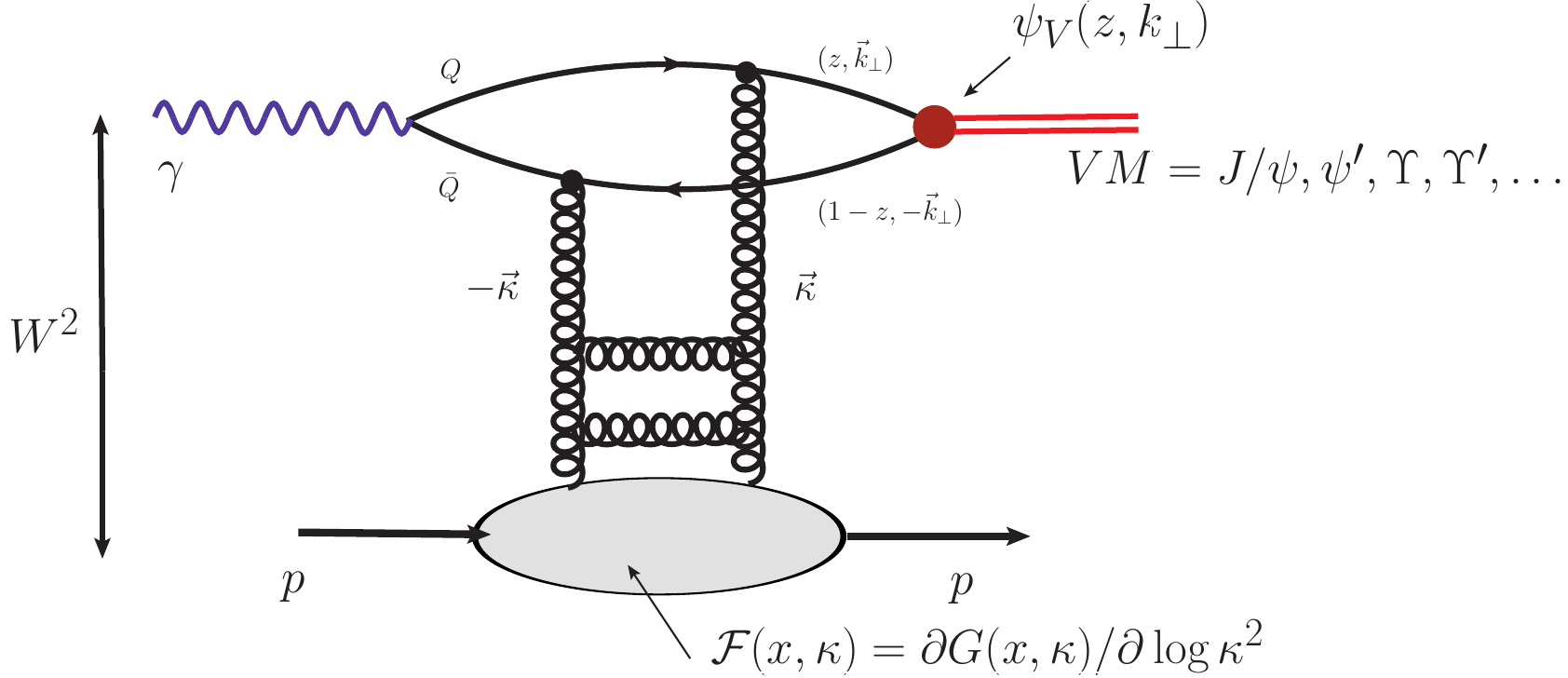}
\caption[*]{
Representative diagram for the exclusive $\gamma p \to Vp$ production of a vector meson $V$.
\label{fig:cep1}
}
\end{center}
\end{figure}
%%%%%%%%%
Two largely equivalent approaches to modelling the exclusive photoproduction of a vector meson
of mass $M_V$ at a $\gamma p$ center-of-mass energy $W$, applicable at small values of $x = M_V^2/W^2$,
are the color-dipole approach and $k_T$-factorization. Within the color-dipole framework (see e.g.~\cite{Motyka:2008ac,Cox:2009ag,Goncalves:2011vf} and references therein), the amplitude depicted in Fig.~\ref{fig:cep1} takes the form
\begin{equation} \label{eq:exclamp}
  A^{\gamma p\rightarrow E+p}_{T}(x,\vec{\Delta}) = \mathrm{i}\,\int\!\dif^2\vec{r}\int_0^1\!\frac{\dif{z}}{4\pi}\int\!\dif^2\vec{b}\;(\Psi_{E}^{*}\Psi_\gamma)_{T}\;\mathrm{e}^{-\mathrm{i}[\vec{b}-(1-z)\vec{r}]\cdot\vec{\Delta}}\;\frac{\dif\sigma_{q\bar q}}{\dif^2\vec{b}}\;\sqrt{(1+\beta^2)},
\end{equation}
where $z$ $(1-z)$ is the
longitudinal momentum fraction of the quark (anti--quark),  $\Delta$ denotes the transverse 
momentum lost by the outgoing proton ($t = - \Delta^2$) and $x$ is the Bjorken variable. 
The variable $\rb$ is the transverse distance from the center of the target to the center of mass of the 
$q \bar{q}$  dipole and the factor  in the exponential  arises when one takes into account 
non-forward corrections to the wave functions \cite{Bartels:2003yj}. The factor of $\sqrt{1+\beta^2}$ in \eqref{eq:exclamp} is a correction to account for the real part of the $S$-matrix element for dipole--proton scattering. A common ansatz for the  differential dipole cross section for the $q\bar{q}$ pair to scatter elastically off the proton is given by~\cite{Kowalski:2006hc}
\begin{equation} \label{eq:dsigmad2b}
  \frac{\dif\sigma_{q\bar{q}}}{\dif^2\vec{b}} = 2\left[1-\exp\left(-\frac{\pi^2}{2N_c}r^2\alpha_S(\mu^2)\,R_g\,xg(x,\mu^2)\,T(b)\right)\right],
\end{equation}
where the factor $R_g$ relates the generalized gluon PDF (the same object introduced in (\ref{bt})) that is relevant in this situation to the standard diagonal gluon PDF, see~\cite{Shuvaev:1999ce,Harland-Lang:2013xba}. The scale $\mu^2$ is related to the dipole size $r$ by $\mu^2=4/r^2+\mu_0^2$. In the case of exclusive production in $pp, pA$ or $AA$ collisions the photoproduction regime $Q^2\approx 0$ prevails, so that for example for $J/\psi$ photoproduction the hard scale is $\sim 2.4 \, \mathrm{GeV}^2$. It is worth noting that such a scale is quite close to what one may expect for a saturation scale, e.g. in the case of a heavy nucleus. These saturation effects manifest themselves in the small-$x$ behaviour of the (unintegrated) gluon and therefore mainly affect the energy dependence of the photoproduction cross section. Finally, other approaches to modelling the dipole cross section exist in the literature, see e.g.~\cite{Goncalves:2011vf} for phenomenological studies.

A related approach is given by the $k_T$-factorization representation of the forward amplitude, see~\cite{Ivanov:2004ax} for a detailed discussion and references . The imaginary part of the amplitude for the $\gamma p \to V p$ process,
for vanishing transverse momentum transfer $\bDelta=0$,
can then be written as a convolution of an impact factor for the $\gamma \to V$ transition
and the unintegrated gluon distribution of the target:
\begin{eqnarray}
\Im m \; {\cal M}_{\lambda_{\gamma},\lambda_V}(W,\bDelta^2=0) =
W^2 \frac{c_\Upsilon \sqrt{4 \pi \alpha_{em}}}{4 \pi^2}
\int \frac{d^2\bkappa}{\kappa^4} \alpha_S(q^2)  
f_g(x_1,x_2,\bkappa) 
\nonumber \\
\times
\int \frac{dz d^2 \bk}{z (1-z)}  \psi_V(z,\bk) \, 
I_{\lambda_{\gamma}, \lambda_V}(z,\bk,\bkappa) \; ,
\label{full_imaginary}
\end{eqnarray}
Here, the unintegrated gluon distribution $f_g(x_1,x_2,\bkappa_1)$ is again the same off-diagonal (``skewed'') object introduced in (\ref{bt}), which as above can be reconstructed from the diagonal one.
The explicit expressions for $I_{\lambda_{\gamma}, \lambda_V}$ 
can be found in~\cite{Ivanov:2004ax}. For heavy vector mesons, helicity--flip 
transitions may be neglected, so that one can safely take 
$\lambda_\gamma = \lambda_V$. 

Besides the unintegrated gluon distribution the second important
non--perturbative input, in both the colour dipole and $k_\perp$ factorisation approaches, is the (``radial'') light-cone wave function
$\psi_V(z,\bk)$ of the vector meson. The relativistic vertex  $V \to Q \bar Q$ for an $S$--wave vector meson is~\cite{Jones:2013pga,Jones:2013eda,Ivanov:1999pb}:
\begin{eqnarray}
\varepsilon_\mu \, \bar u(p_Q) \Gamma^\mu v(p_{\bar Q}) = [M^2 - M_V^2] \, \psi_V(z,k^2) \, 
\bar u(p_Q) \Big( \gamma^\mu - {p_Q^\mu - p_{\bar Q}^\mu \over M + 2 m_Q} \Big)
v(p_{\bar Q}) \, \varepsilon_\mu \, ,
\end{eqnarray}
where $\varepsilon_\mu$ is the polarization vector of the vector meson.
and $p_{Q, \bar Q}^\mu$ are the on-shell four--momenta of the $Q,\bar Q$ quarks, 
$p_{Q, \bar Q}^2 = m_Q^2$.  The invariant mass of the $Q \bar Q$ pair is given in terms
of light-cone variables as
\begin{equation}
M^2 = {k^2 + m_Q^2 \over z(1-z)} \, .
\end{equation}
The radial wave--function $\psi_V(z,k^2)$ can be regarded as a function
not of $z$ and $\bk$ independently, but rather of the three--momentum $\vec{p}$
of, say, the quark in the rest frame of the $Q \bar Q$ system of invariant mass $M$,
$ \vec{p} = (\bk , (2z-1) M/2) $.
Then,
\begin{eqnarray}
\psi_V(z,k^2) \to \psi_V(p^2) \, , \, {dz d^2 \bk \over z(1-z)} \to {4 \, d^3\vec{p} \over M} 
\, , p^2 = {M^2 - 4 m_Q^2 \over 4} \, .
\end{eqnarray}
It is assumed that the Fock--space components of the $V$--states
are exhausted by the two-body $Q \bar Q$ components. In the absence of first-principles calculations of the light-cone wave function, different phenomenological models are available~\cite{Ivanov:2004ax}. For example, the Gaussian, harmonic--oscillator--like wave functions:
\begin{equation}
\psi_{1S}(p^2) = C_1 
\exp\left( - \frac{p^2 a_1^2}{2} \right) \, , \, 
\psi_{2S}(p^2) = C_2 (\xi_0 - p^2 a_2^2) 
\exp\left( - \frac{p^2 a_2^2}{2} \right) \, , 
\label{harmonic_oscillator_WF}
\end{equation}
and the Coulomb--like wave functions, with a slowly decaying
power--like tail:
\begin{equation}
\psi_{1S}(p^2) = {C_1 \over \sqrt{M}} \, {1 \over (1 + a_1^2 p^2)^2} \, , \, 
\psi_{2S}(p^2) = {C_2 \over \sqrt{M}} \, {\xi_0 - a_2^2 p^2 \over (1 + a_2^2 p^2)^3} \, .
\end{equation}
The parameters $a_i^2$ are obtained from fitting  to the $e^+ e^-$ decay widths, whereas $\xi_0$, and therefore
the position of the node of the $2S$ wave function, is obtained
from the orthogonality of the $2S$ and $1S$ states.

While (\ref{full_imaginary}) is written for the case of zero momentum transfer $\Delta$, the full amplitude, at finite momentum transfer within the diffraction cone, is given by
\begin{eqnarray}
{\cal M}(W,\Delta^2) = (i + \beta) \, \Im m {\cal M}(W,\Delta^2=0) \, \exp(-B(W) \Delta^2/2 ) \, ,
\end{eqnarray}
where $\beta$ accounts for the real part of the scattering amplitude, as in (\ref{eq:exclamp}), while $B(W)$ is
the energy--dependent slope parameter:
\begin{equation}
B(W) = B_0 + 2 \alpha'_{eff} \log \Big( {W^2 \over W^2_0} \Big) \, ,
\end{equation}
where $B_0$ depends on the vector meson in question, whereas $\alpha'_{eff}$ can be taken as universal
in the present applications.

\begin{figure}[ht] % Figure 3
\begin{center}
\includegraphics[height=5.0cm]{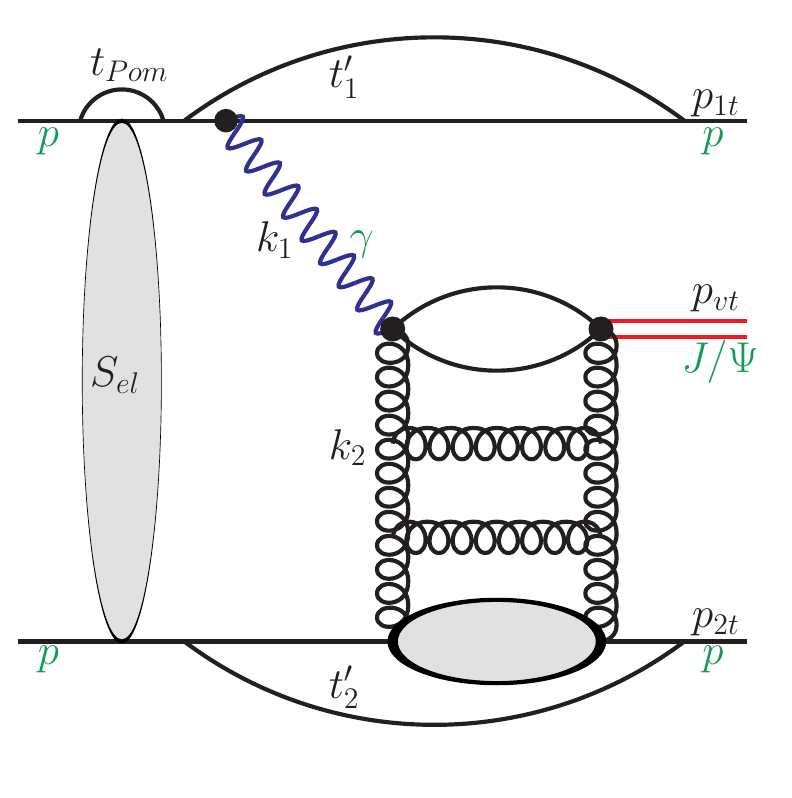}
\includegraphics[height=5.0cm]{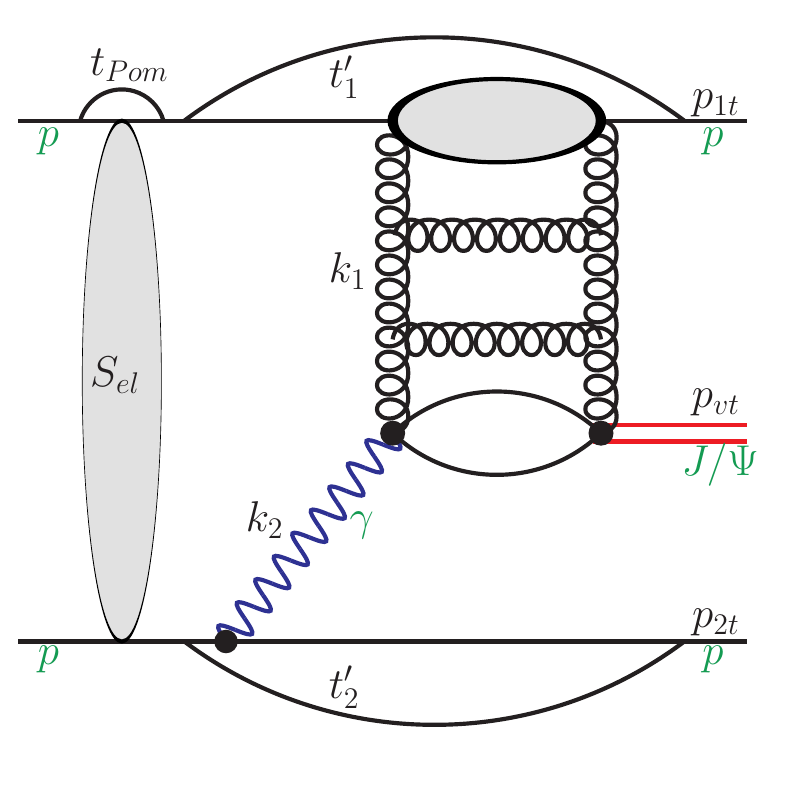}
\caption{
Diagrams representing the two interfering amplitudes for the $p p \to p p J/\psi$ process,
including absorptive corrections.
\label{fig:2:ch5}
}
\end{center}
\end{figure}
In going from $\gamma p$ to $pp$ collisions (see Fig. \ref{fig:2:ch5}), there are two complications. Firstly, either of the colliding protons can emit the photon, and the amplitudes for these two processes interfere. This interference mainly affects the transverse momentum distributions. In the Born approximation, the interference contribution will cancel after azimuthal integration but in the presence of absorption, a small effect of the interference can remain. Secondly, as for all central exclusive diffractive processes, the fact that protons may also interact strongly must be taken into account, i.e. the gap survival probability~\cite{Bjorken:1992er,Ryskin:2009tk,Ostapchenko:2010gt,Khoze:2014aca,Gotsman:2014pwa} discussed in Section~\ref{CEP:gluonex:intro}. The effects of gap survival are however expected to be much weaker than for the purely QCD process, due to the peripheral nature of the photon exchange. See~\cite{Khoze:2001xm,Schafer:2007mm,Khoze:2008cx} for further discussion of these points. 

\subsubsection*{Motivation and theory: predictions}

A selection of results for the rapidity spectrum of exclusive vector mesons at Tevatron and LHC energies are now presented. At Tevatron energies and at central rapidities, the subprocess energies for $\gamma p \to Vp$ or $\gamma \bar p \to V \bar p$ cover the known HERA domain. At LHC energies, it is possible to extend the energy range beyond the one already studied at HERA. For example, for $J/\psi$ production at central rapidity $y =0$, we have $W_{\gamma p} \sim 140 \, \rm{GeV}$, whereas at $y=4$ this extends to $W_{\gamma p} \sim 1 \, \rm{TeV}$.

\begin{figure}[ht]
\begin{center}
\includegraphics[width=6cm]{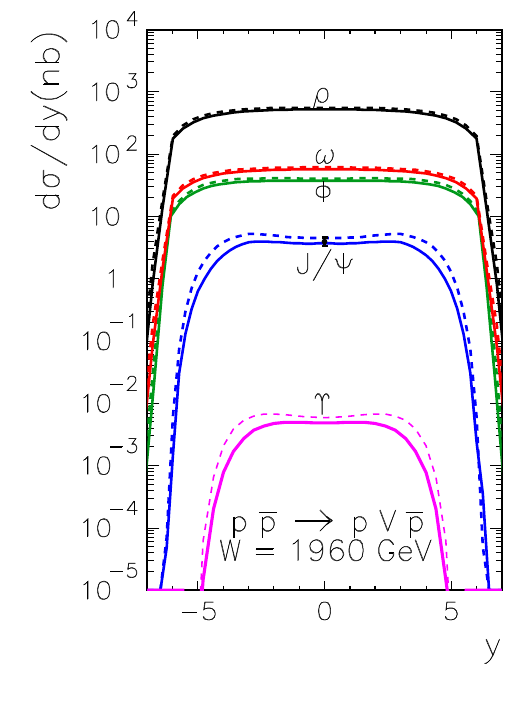}
\includegraphics[width=6cm]{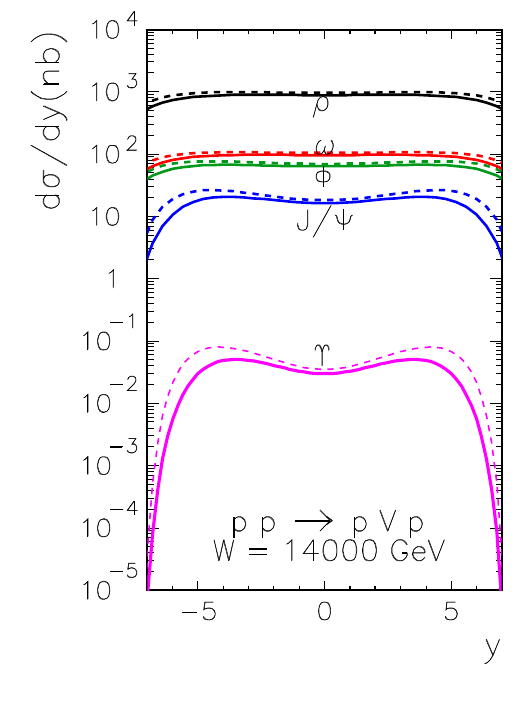}
\caption{Rapidity dependence of central exclusive vector meson production in proton-(anti)proton
collisions. Left panel: proton-antiproton collisions at $\sqrt{s}=$W=1960 GeV. Right panel: 
proton-proton collisions at $\sqrt{s}=$W=14 TeV. The left panel also shows
the data point from the CDF collaboration. Dashed lines are without absorptive corrections,
while solid lines include them.}
\label{fig:3:ch5}       % Give a unique label
\end{center}
\end{figure}
%%%

\begin{figure}[!htb] % Figure 2
\begin{center}
\includegraphics[width=5.0cm]{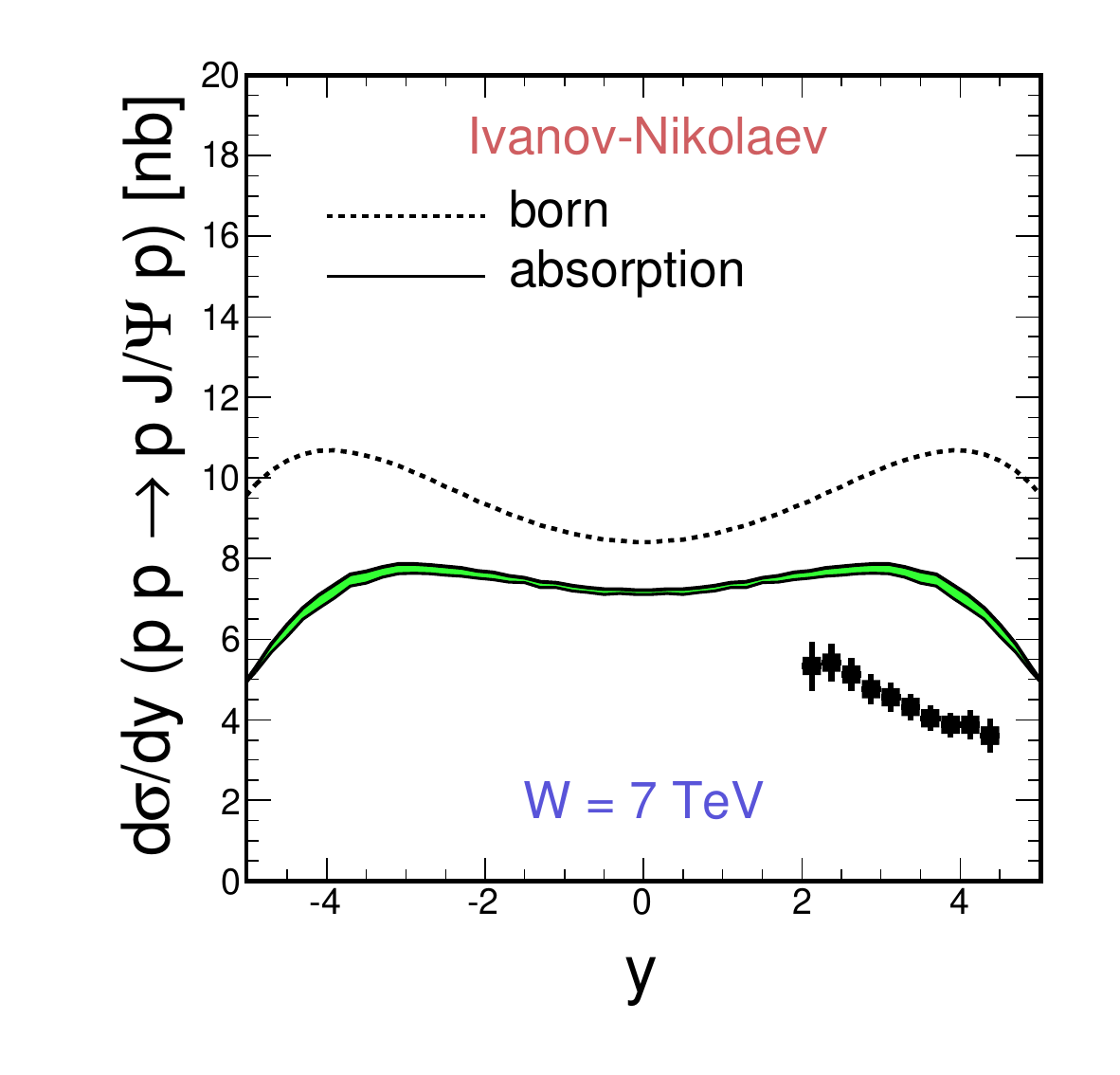}
\includegraphics[width=5.0cm]{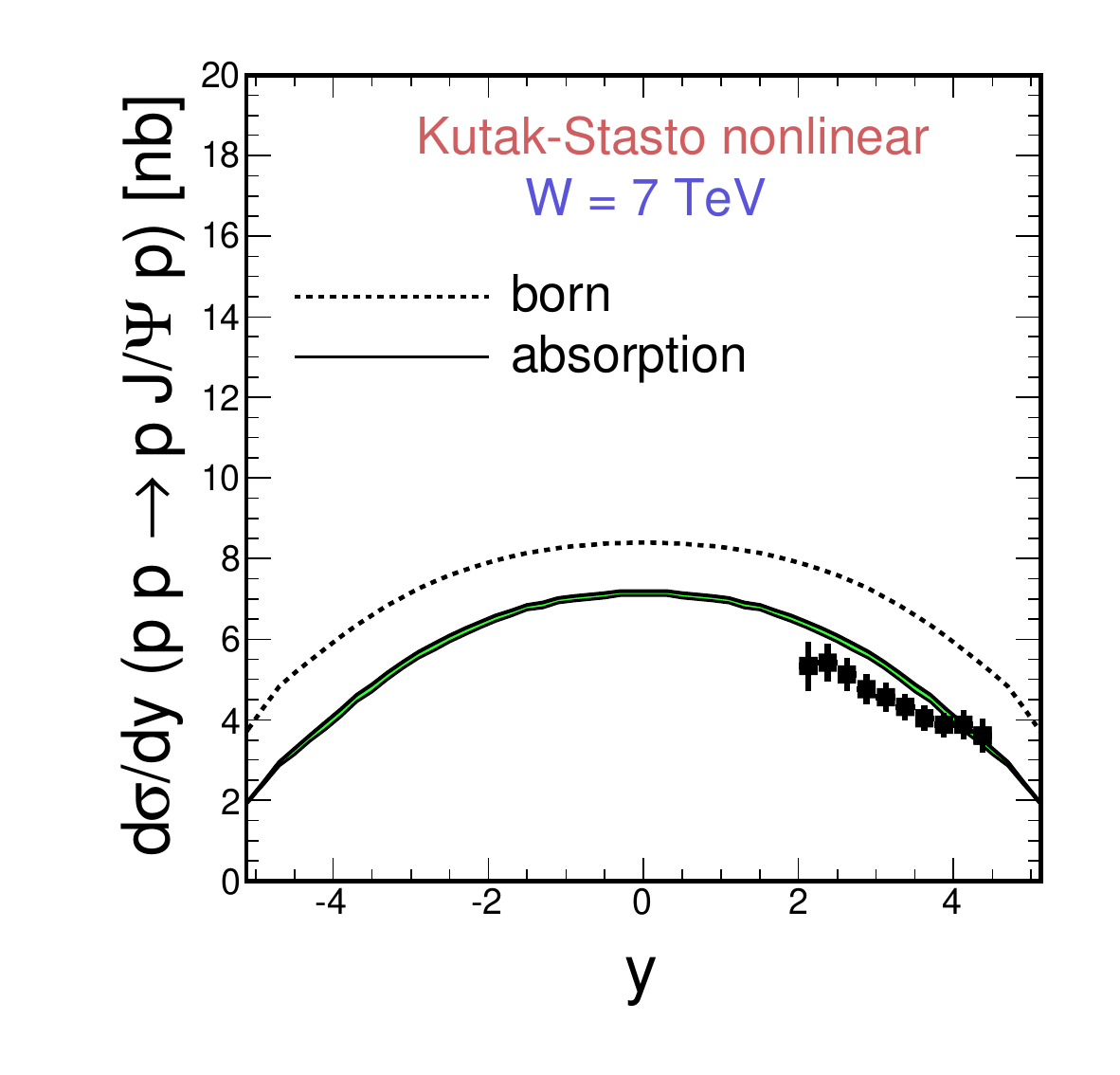}
\includegraphics[width=5.0cm]{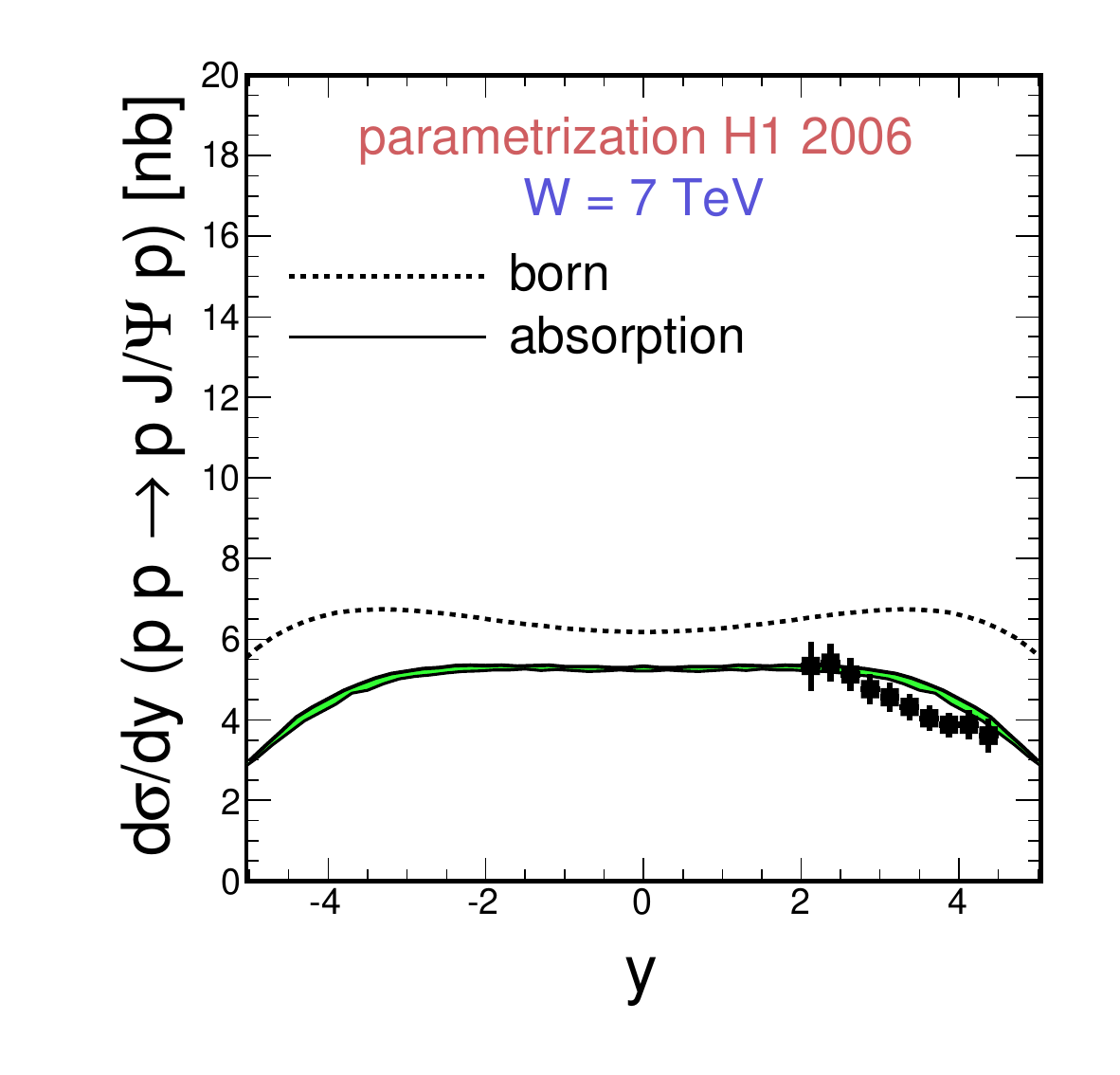}
\caption[*]{ 
$J/\psi$ rapidity distribution calculated with the inclusion of 
absorption effects (solid line), compared to the Born result 
(dashed line) for $\sqrt{s}$ = 7 TeV. 
The LHCb data points from~\cite{Aaij:2014iea} are shown for comparison.
\label{fig:4:ch5}
}
\end{center}
\end{figure}

%-------------------------------------------------------------------------
\begin{figure}[!htb] % Figure 3
\begin{center}
\includegraphics[width=5.0cm]{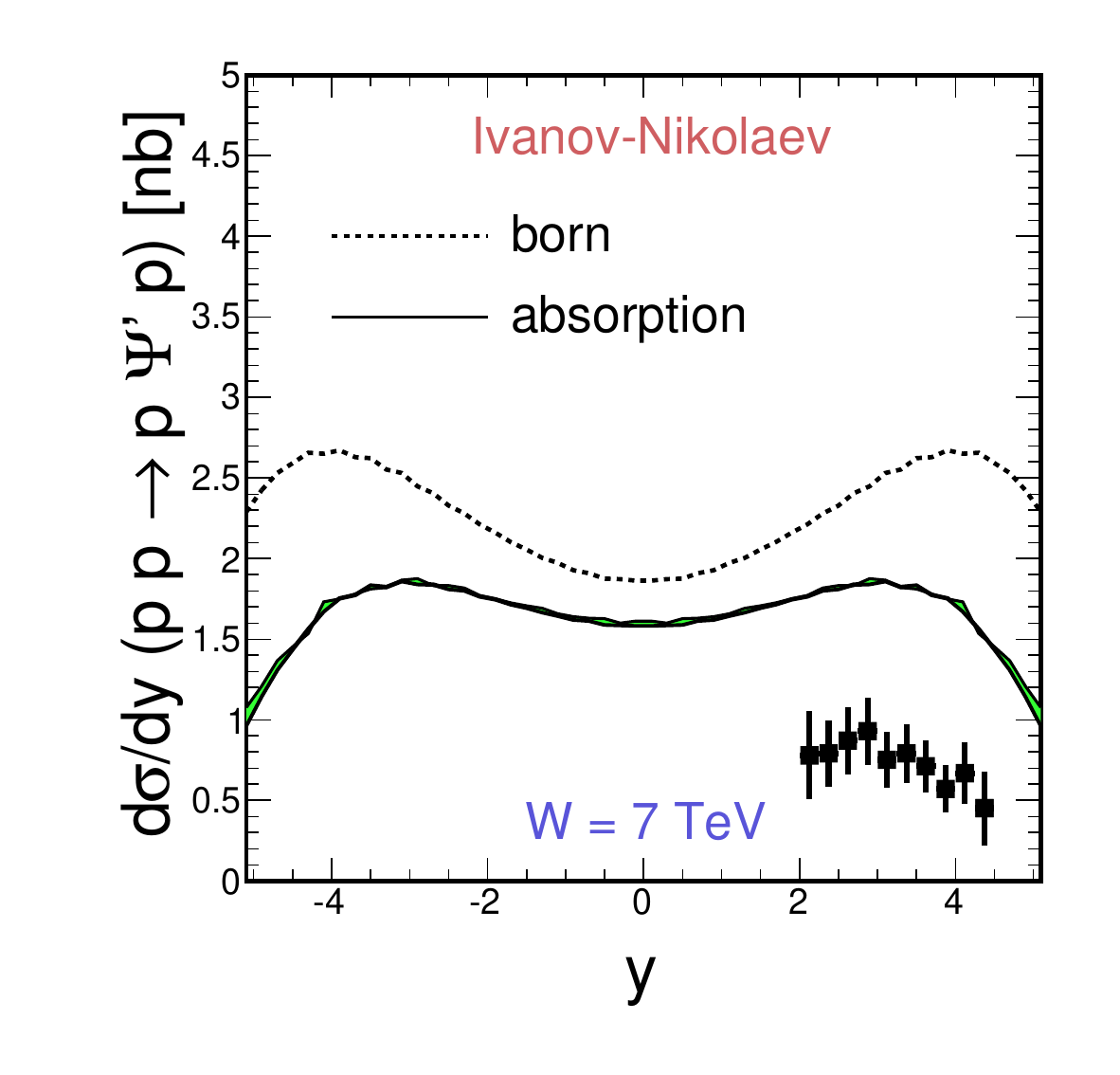}
\includegraphics[width=5.0cm]{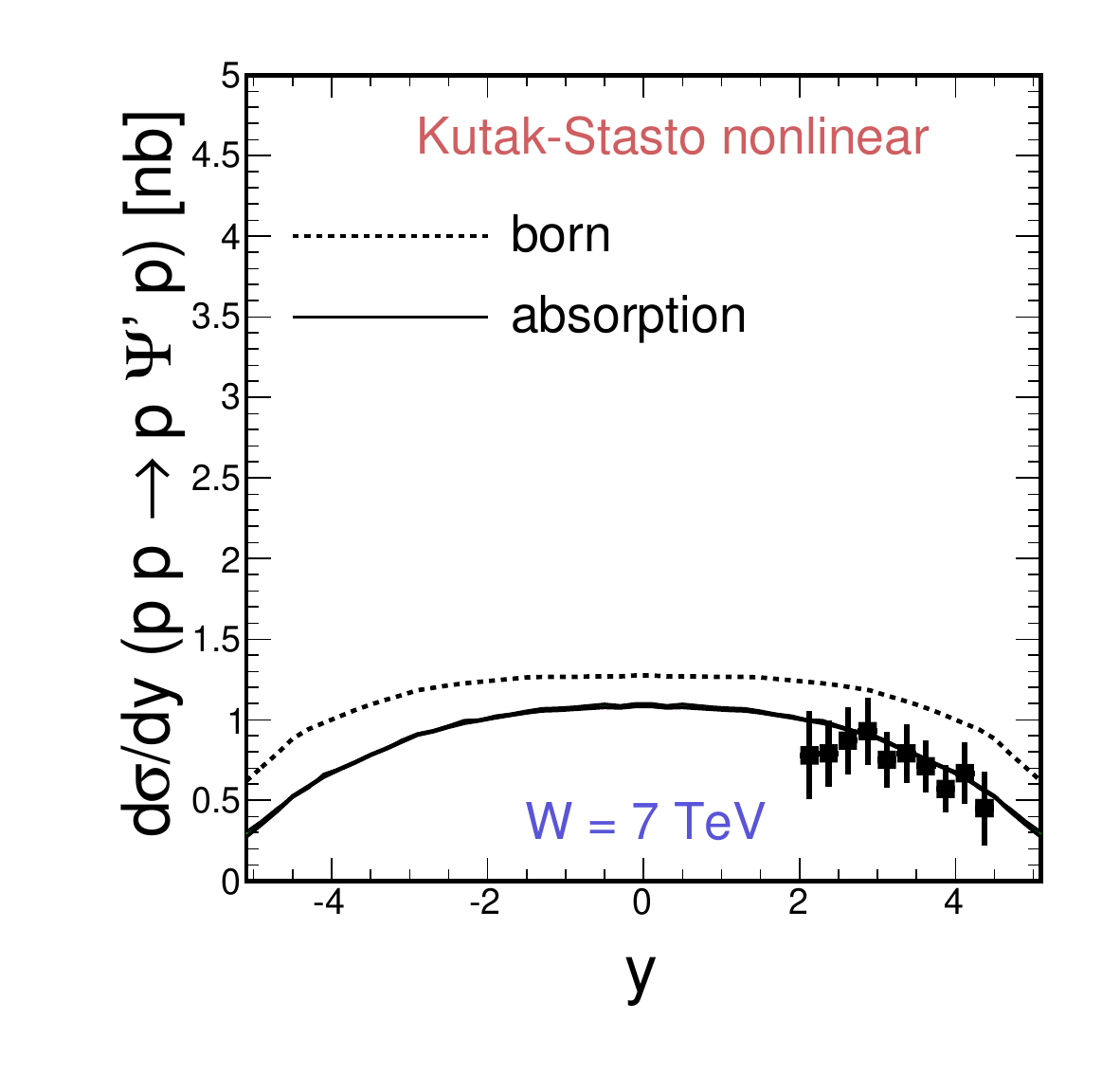}

\caption[*]{ 
$\psi'$ rapidity distribution calculated including
absorption effects (solid line), compared with the result when 
absorption effects are ignored (dotted line) for $\sqrt{s}$ = 7 TeV.
The LHCb data points from~\cite{Aaij:2014iea} are shown for comparison.
\label{fig:5}
}
\end{center}
\end{figure}
%-------------------------------------

%%%

Considering first numerical results from a recent work~\cite{Rybarska:2008pk,Cisek:2010jk,Cisek:2011vt,Cisek:2014ala} based on the $k_T$-factorization formalism,
 in Fig.~\ref{fig:3:ch5}  the predicted vector meson rapidity distributions at the Tevatron and LHC are shown.
The first measurement of exclusive $J/\psi$ production was made
by the CDF collaboration at the Tevatron~\cite{Aaltonen:2009kg}, and this data point is also shown.
At the LHC, data for exclusive $J/\psi, \psi'(2S)$ have been obtained by the 
LHCb collaboration~\cite{Aaij:2013jxj,Aaij:2014iea}. In Fig.~\ref{fig:4:ch5},  the predicted rapidity distributions of $J/\psi$ compared to these data~\cite{Aaij:2014iea} are shown. 
In the first two panels from the left, the results for two different unintegrated gluon distributions
 are shown. In the first panel, the gluon from~\cite{Ivanov:2000cm} is used 
while in the second panel one of the distributions from~\cite{Kutak:2004ym} is used. 
Both these distributions describe the HERA data for $F_2$ well, but the latter is obtained from a nonlinear evolution equation which accounts for the physics of gluon saturation
at small $x$. The former, Ivanov-Nikolaev, gluon does not include explicit saturation effects. 
We observe that the data appear to prefer the gluon distribution including saturation effects.
On the other hand in the rightmost panel, a calculation using
a fit to the vector meson production amplitude by the H1-collaboration, used in~\cite{Schafer:2007mm},
and simply extrapolated to LHC energies, is shown. It describes the data quite well,
and being an effective Pomeron-pole approximation casts some doubt on the saturation 
interpretation.

In Fig.~\ref{fig:5} the predicted rapidity distribution for the $\psi'(2S)$ production is shown.
Again it is seen that the gluon from~\cite{Kutak:2004ym} gives a very good description
of the data. It should be added that these calculations use the Gaussian-type wave function of 
the vector meson, which is strongly favoured by the $\psi'(2S)$ data.
It should be emphasised that only the $k_T$-factorization or color dipole approaches make
reliable predictions for the production of excited vector meson states. 
Such predictions cannot be obtained in the collinear approach, and it is worth
stressing that the convolution of impact factor with unintegrated gluon cannot be simply reduced
to a choice of the scale in the collinear gluon distribution.

In the future it will be important to include proton dissociative processes in theoretical models, which generally concentrate on the exclusive case. On the
$\gamma$-side one can proceed analogously to~\cite{daSilveira:2014jla} (the $p \to \gamma \Delta^+$
transition has recently been included in~\cite{Guzey:2014axa}). A treatment of the Pomeron side
requires a better understanding of the quasi-diffractive $ \gamma p \to J/\psi X$ 
process, as well as a coupled channel treatment of absorptive corrections.

\begin{figure}
\begin{tabular}{cc}
\includegraphics[scale=0.29]{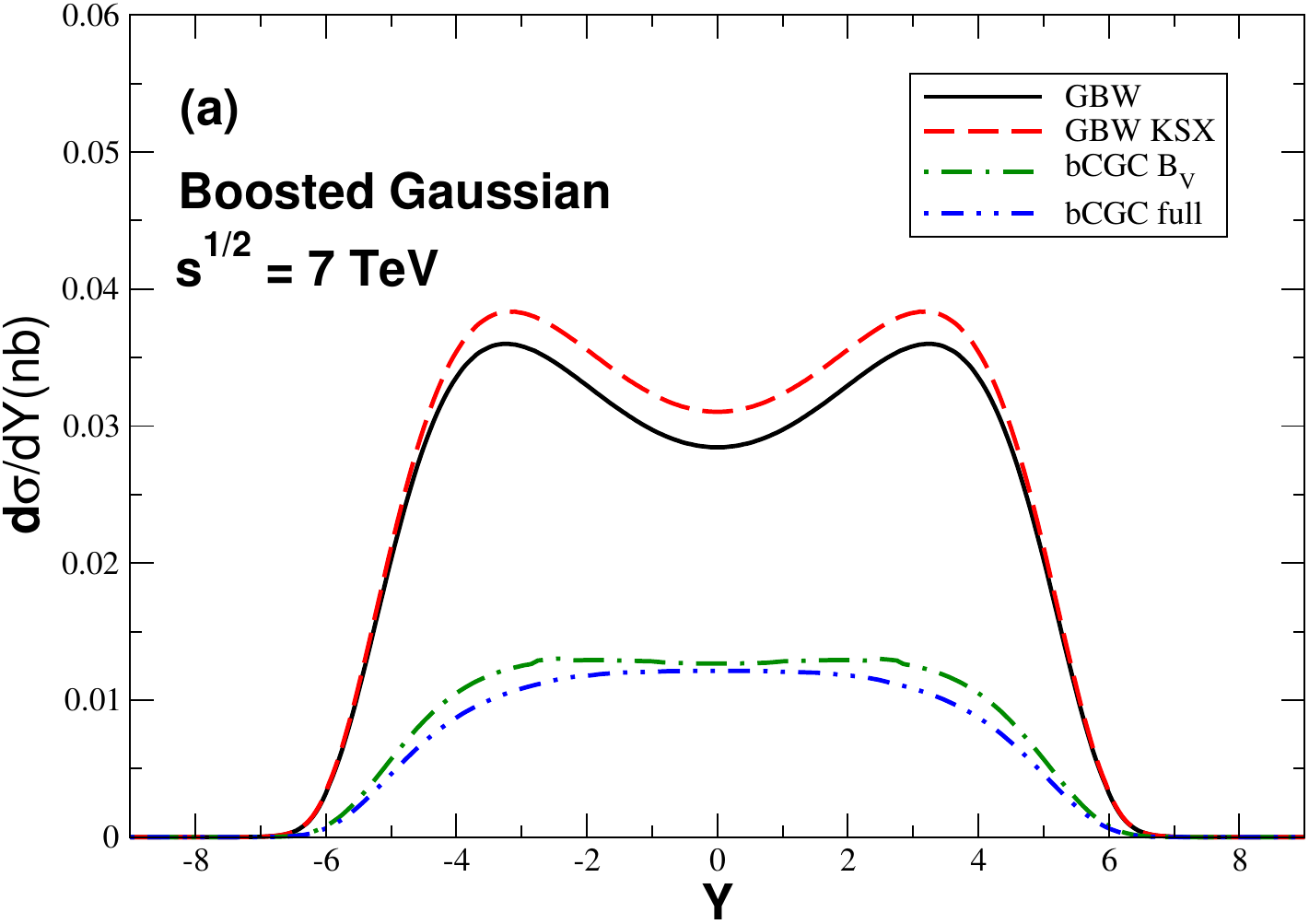} & 
\includegraphics[scale=0.3]{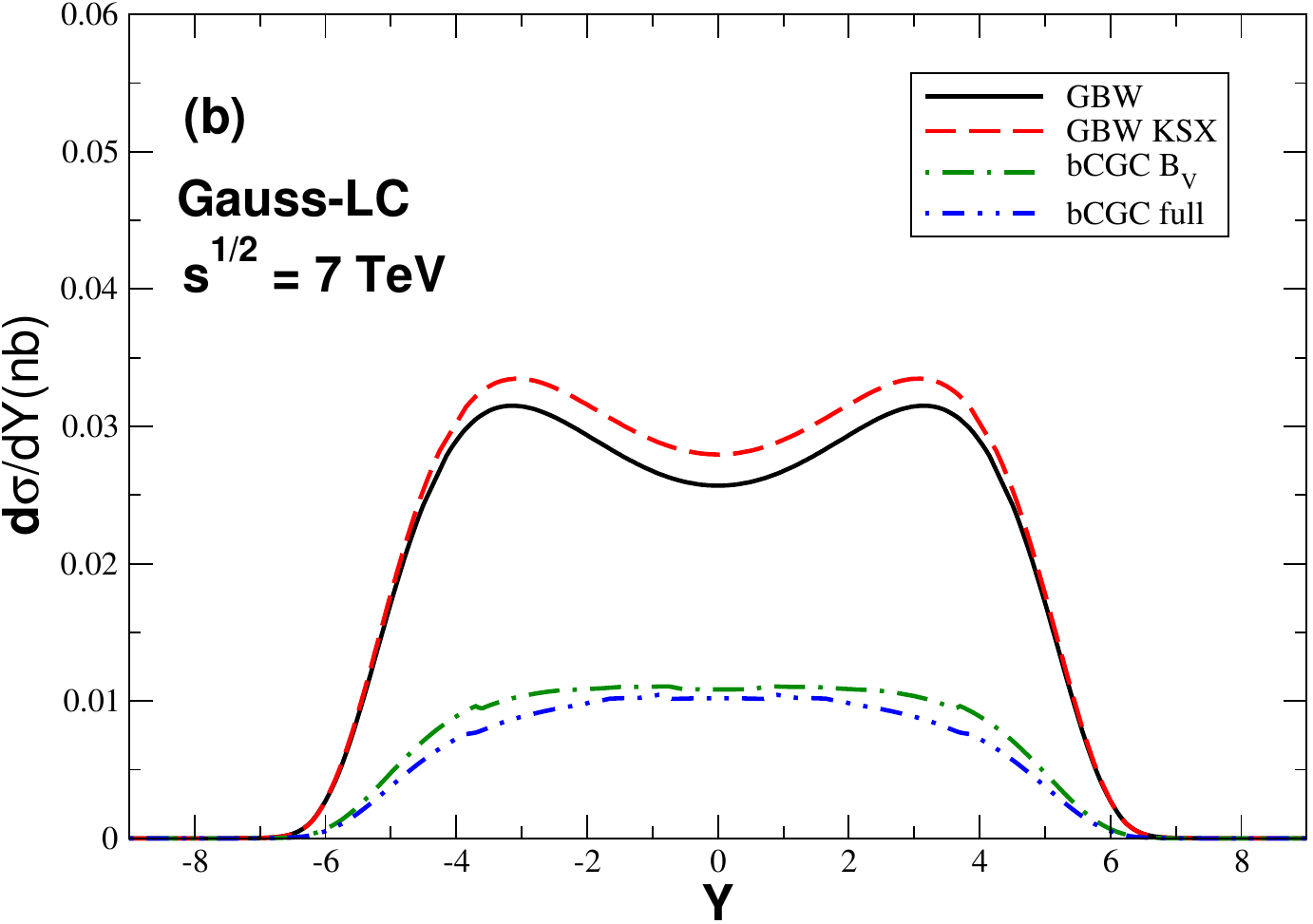}
\end{tabular}
\caption{(Color online) Exclusive $\Upsilon$ photoproduction in $pp$ collisions at $\sqrt{s} = 7$ TeV. }
\label{fig3}
\end{figure}

\begin{figure}
\begin{tabular}{cc}
\includegraphics[scale=0.29]{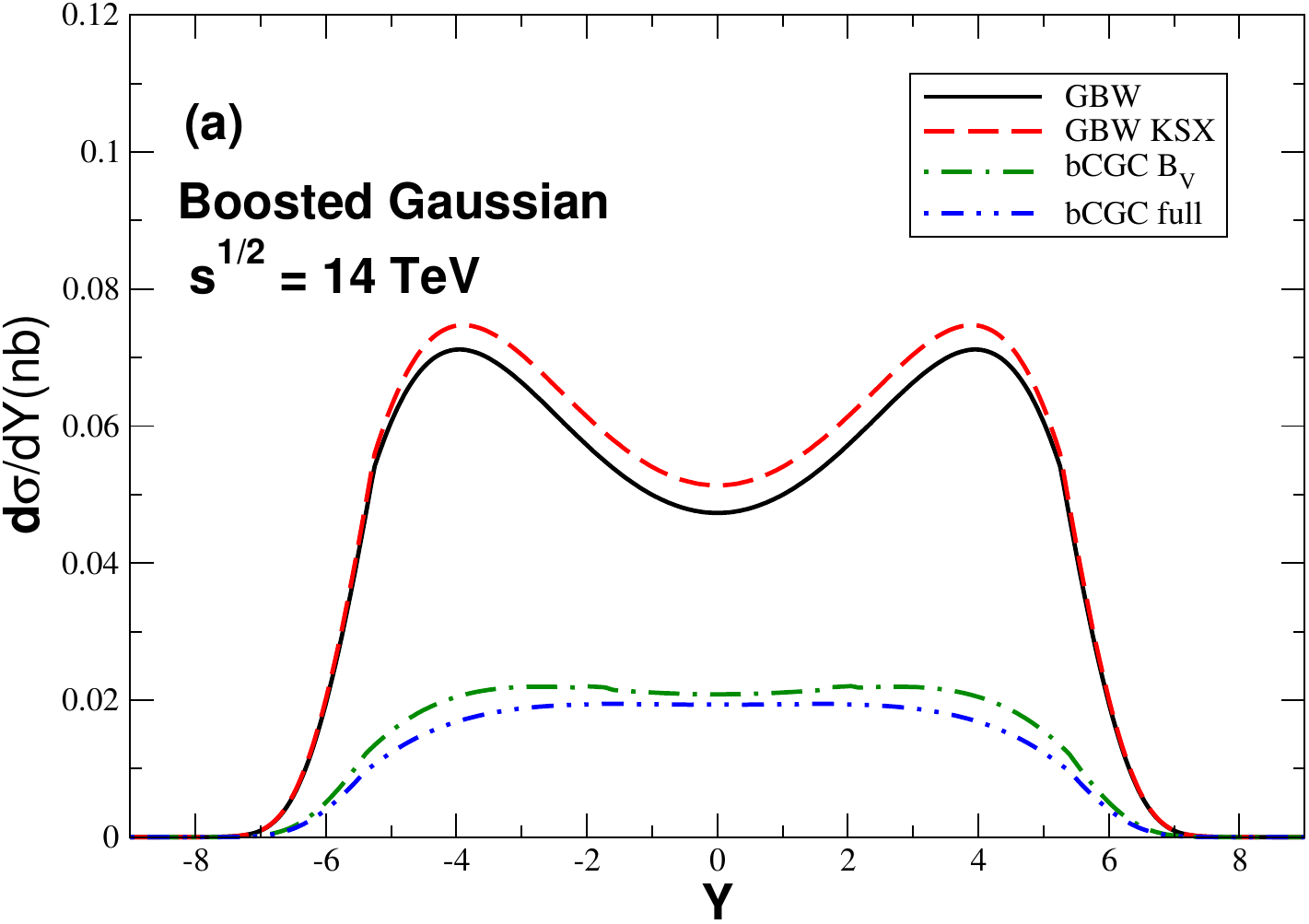} & 
\includegraphics[scale=0.3]{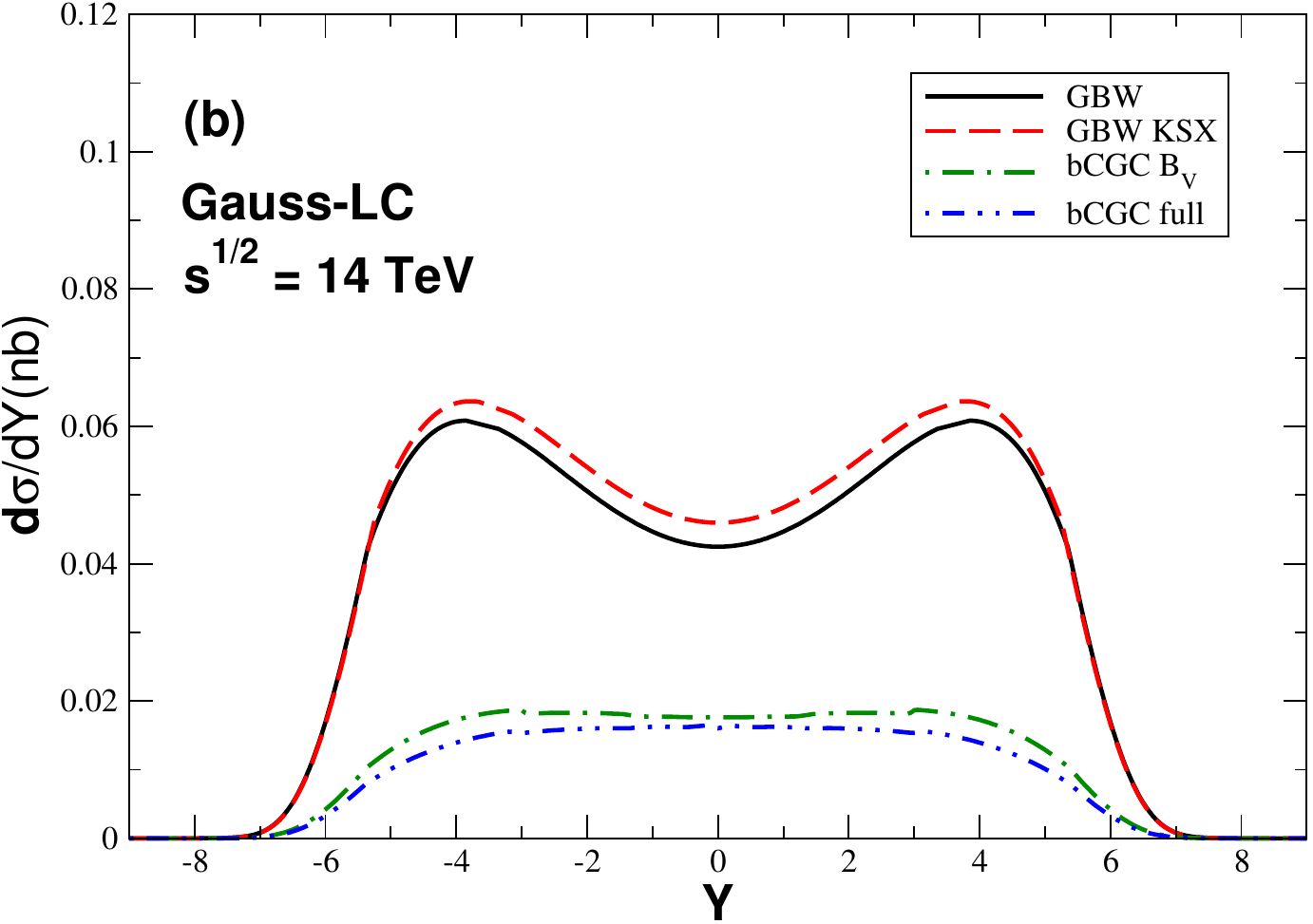}
\end{tabular}
\caption{(Color online) Exclusive $\Upsilon$ photoproduction in $pp$ collisions at $\sqrt{s} = 14$ TeV. }
\label{fig4}
\end{figure}

As previously noted, the results presented above are from the recent work of~\cite{Rybarska:2008pk,Cisek:2010jk,Cisek:2011vt,Cisek:2014ala}, based on the $k_T$-factorization formalism.  However, it is also possible to model the photoproduction process using the color dipole approach, see for example~\cite{Motyka:2008ac,Goncalves:2011vf,Ducati:2013tva}.
In Figs.~\ref{fig3} and \ref{fig4} the predicted rapidity distributions for exclusive $\Upsilon$ photoproduction at $\sqrt{s}=7$ TeV and  $\sqrt{s}=14$ TeV, respectively, are shown using the colour dipole approach (see~\cite{Goncalves:2014swa}). Here `bCGC' and `GBW' correspond to two alternative phenomenological models for the dipole cross section introduced in (\ref{eq:exclamp}). In addition the results from approximate and more precise evaluations of the $t$--dependence of the cross section are given, and for two different forms of the meson wave functions (`Boosted Gauss' and `Gauss-LC'), defined in~\cite{Goncalves:2011vf}. While there is some difference between the choices of wave functions, the variation between the different models of the dipole cross section are dramatic. The `bCGC' is mildly disfavoured by (low statistics) HERA measurements but LHC data on $\Upsilon$ photoproduction can greatly clarify this.

\begin{figure}
\begin{center}
\includegraphics[width=0.45\textwidth,angle=-90]{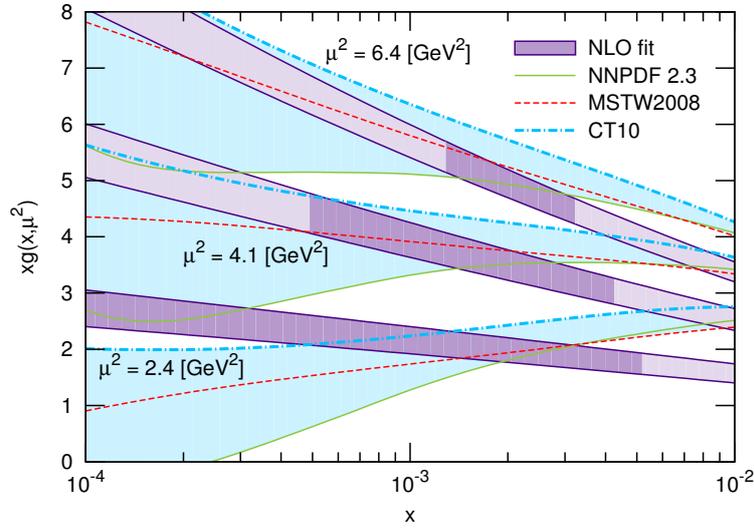}
\caption{NLO gluon resulting from a fit to the available HERA and LHCb data on $J/\psi$ photoproduction, taken from~\cite{Jones:2013pga}.} 
\label{fig:gluonfit}
\end{center}
\end{figure}

A complementary approach, using the same $k_\perp$ factorisation formalism described in the previous sections, but differing in some elements, is described in~\cite{Jones:2013pga,Jones:2013eda}.  Simple parametric forms for the low--$x$ gluon PDF fitted to the existing data from HERA and LHCb, and both a LO and a NLO--type fit are considered; in the latter case this is not the result of a full NLO calculation, but rather the form of the gluon is chosen so as to reproduce the effect of NLO DGLAP evolution. Such an approach is motivated by the large uncertainty in the low--$x$, low--$Q^2$ gluon that results from global parton analyses, and conversely highlights the way in which such processes can help reduce this uncertainty, as is shown in Fig.~\ref{fig:gluonfit}. This fit then allows predictions for the cross sections and rapidity distributions at higher energies, and for other processes such as exclusive $\Upsilon$ and $\psi(2S)$ photoproduction, to be given. In all cases a full treatment of soft survival effects, and in particular the rapidity (i.e. $W$) dependence of the survival factor is given, which as seen above, must be included for a precise comparison to data in hadronic collisions. Finally, it should be noted that a proper inclusion of the `skewedness', i.e. the relation between the off-diagonal gluon in~\ref{full_imaginary} and the standard gluon PDF, for the case that the gluon is unintegrated over the parton transverse momentum, as in the $k_\perp$--factorization approach, may be of some importance in future high--precision theoretical work~\cite{Harland-Lang:2013xba}. It is worth emphasising that measurements of the ratios of these cross sections at different $\sqrt{s}$ values would be particularly sensitive to the gluon PDF in this low$-x$ and $Q^2$ region, with other uncertainties due to e.g. soft survival effects, largely cancelling in this ratio.

Finally, in the case of photoproduction processes with the momentum transfer square playing the role of the hard scale, one can perform a systematic analysis of this process based on the twist expansion of meson distribution amplitude in an analogous way as it was done in ~\cite{RREF1, RREF2} for meson hard electroproduction at HERA. One should note that such a twist expansion can be done in the color dipole picture~\cite{RREF3}, offering a natural way for including saturation effects~\cite{RREF4}.

In addition to these lighter vector mesons states, exclusive $Z$-boson photoproduction is also accessible at the LHC, but only in high luminosity running. This proceeds via an interesting ``vertex", actually a loop diagram, with electromagnetic, strong and weak lines together. The process $\gamma\gamma \rightarrow Z$ is forbidden by the Landau-Yang theorem, and so this photoproduction process naturally dominates. The central $e^+e^-, \mu^+\mu^-$ or $\tau^+\tau^-$ pairs have $p_T \lesssim$ 2 GeV and so ($\Delta\phi(\ell^+\ell^-) - 180^\circ) < 1^\circ$, with no other tracks associated with the vertex. 

 The $Z$ photoproduction cross section at the LHC in both $pp$ and heavy ion collisions has been calculated in~\cite{Goncalves:2007vi,Motyka:2008ac}. The virtual photon fluctuates into a $q\bar{q}$ pair (a colour dipole), which scatters off the proton diffractively by two-gluon exchange and recombines as a $Z$. The wave functions and the dipole-proton cross sections are reasonably well known from HERA photoproduction data, and consequently $Z$-photoproduction can be considered as a good test of pQCD (due to the high $Z$-mass scale). The largest uncertainty is the gap survival probability; such measurements would therefore be sensitive to such soft survival effects. The  prediction of~\cite{Motyka:2008ac} is, for $Z$ production at the LHC, $\frac{d\sigma}{dy}(y=2.5)$ = 1.7 fb, and 1.4 fb at $y$ = 0. The prediction of~\cite{Goncalves:2007vi} is in agreement; for $\sqrt{s}$ = 14 TeV it is 11 fb for all $y(Z)$, and $\frac{d\sigma}{dy}$ peaks at $|y|$ = 3.

Measuring the process with both protons detected requires forward proton detectors at 420 m, but with CT-PPS Stage 1 or AFP at $z \sim$ 220 m, and a $Z$ boosted to $2 \lesssim |y_Z| \lesssim 3$ one proton can be measured. If the event is really exclusive the proton momentum is well known from $p_z(Z)$, even if the other proton dissociates. However, allowing for the $Z$ branching fractions of only 3.63\% for each of the $e^+e^-$ and $\mu^+\mu^-$ channels, an observation of this process is challenging, but could be possible in a large enough data sample. For example, assuming an efficient trigger, which should include a proton tag, the prediction gives 24 $\times A \times f$ events in 100 fb$^{-1}$ of integrated luminosity, where $A$ is the acceptance and efficiency, and $f \sim 1.5$ -- $2$ is a factor allowing the other proton to dissociate. A control of the background is provided by considering $W \rightarrow e/\mu + E_T \!\!\!\!\!\!/$ candidates, which cannot occur exclusively.

\subsubsection*{Experimental results and outlook}\label{sec:photooutlook}

\begin{figure}[ht]
\centerline{\includegraphics[scale=0.5]{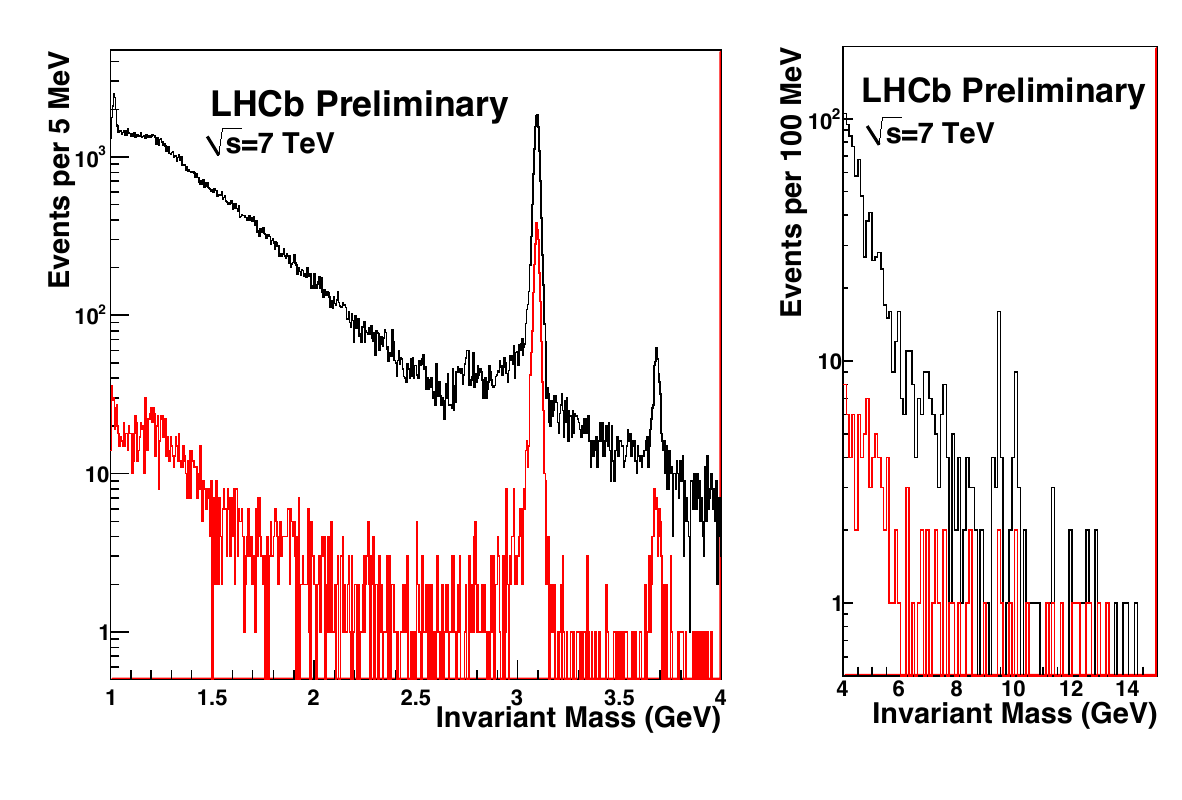}}
\caption{
Invariant mass of di-muons in $37 {\rm\ pb}^{-1}$ of data
after the low-multiplicity di-muon trigger (black)
and after requiring no other activity inside LHCb (red).
The discontinuity at 2.5 GeV is due to a trigger threshold.
\label{fig:mu2mass}}
\end{figure}

\begin{figure}[ht]
\centerline{\includegraphics[height=5cm]{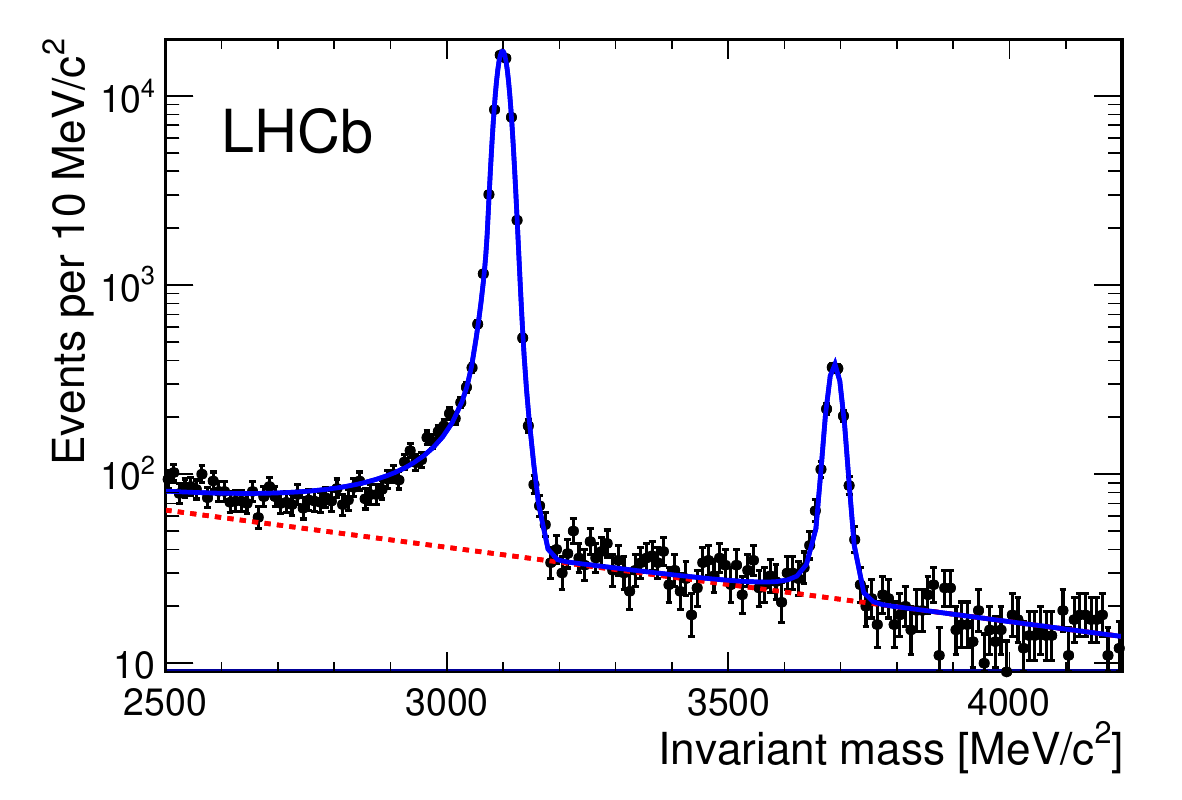}}
\caption{
Invariant mass of selected di-muon candidates
in $930 {\rm\ pb}^{-1}$ of data.
\label{fig:jmass}}
\end{figure}

The LHCb collaboration has made two measurements of $J/\psi$ and $\psi(2S)$
production at $\sqrt{s}=7$ TeV, 
one with an integrated luminosity of 
$37{\rm \ pb}^{-1}$ (2010 data)~\cite{Aaij:2013jxj}, and one
with $930{\rm \ pb}^{-1}$ (2011 data)~\cite{Aaij:2014iea}\footnote{In the final stages of the preparation of this report, LHCb have reported a measurement of exclusive $\Upsilon\to\mu+\mu^-$ production at $\sqrt{s}=7$ and 8 TeV, see~\cite{Aaij:2015kea} for details.}.
The selection starts by triggering on low multiplicity events
containing two muons.  The events are then selected as exclusive inside the
LHCb acceptance by requiring no additional charged tracks or neutral deposits in 
the detector.
The invariant mass of the two muons after the trigger and after the selection
is shown in Fig.~\ref{fig:mu2mass} for the $37{\rm \ pb}^{-1}$ sample.
Within a falling continuum, there are clear signals after the trigger requirements
for several vector mesons: $\phi,J/\psi,\psi(2S),\Upsilon(1S),\Upsilon(2S)$.
With the additional exclusivity requirements, only charmonia signals remain 
visible in this limited data sample.
Candidate events for $J/\psi$ and $\psi(2S)$ mesons in the larger $930{\rm \ pb}^{-1}$
sample can be seen in Fig.~\ref{fig:jmass}.  

Three backgrounds are considered in extracting the elastic signal: non-resonant di-muon production, feed-down from other mesons 
and inelastic $J/\psi$ production.
The non-resonant background
is determined from the fit shown in Fig.~\ref{fig:jmass}.
Feed-down is only considered for the $J/\psi$ selection and can come from 
$\chi_{c0},\chi_{c1},\chi_{c2}$ or $\psi(2S)$ decays,
with the other decay products being below the threshold for detection or outside
the LHCb acceptance.  Feed-down from $\chi_c\rightarrow J/\psi\gamma$
is evaluated to be $(7.6\pm0.9)\%$ by selecting events in which the photon is seen and using the simulation
to estimate the number of events in which it would be undetected.  Feed-down from the
decays $\psi(2S)\rightarrow J/\psi X$
is estimated from the simulation, which has been normalised to the
observed number of events from the decay $\psi(2S)\rightarrow\mu\mu$,
and contributes $(2.5\pm 0.2)\%$ of the $J/\psi$ sample.

\begin{figure}[ht]
{\includegraphics[width=0.49\linewidth]{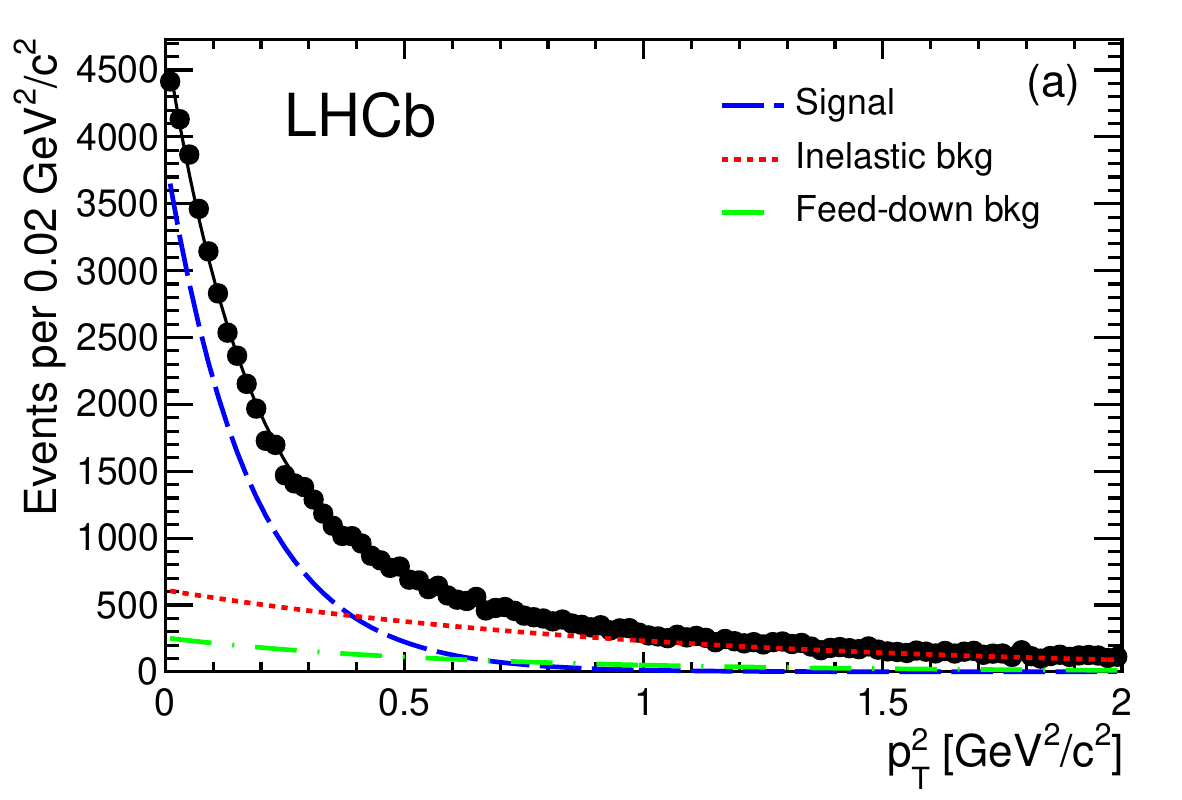}}
{\includegraphics[width=0.49\linewidth]{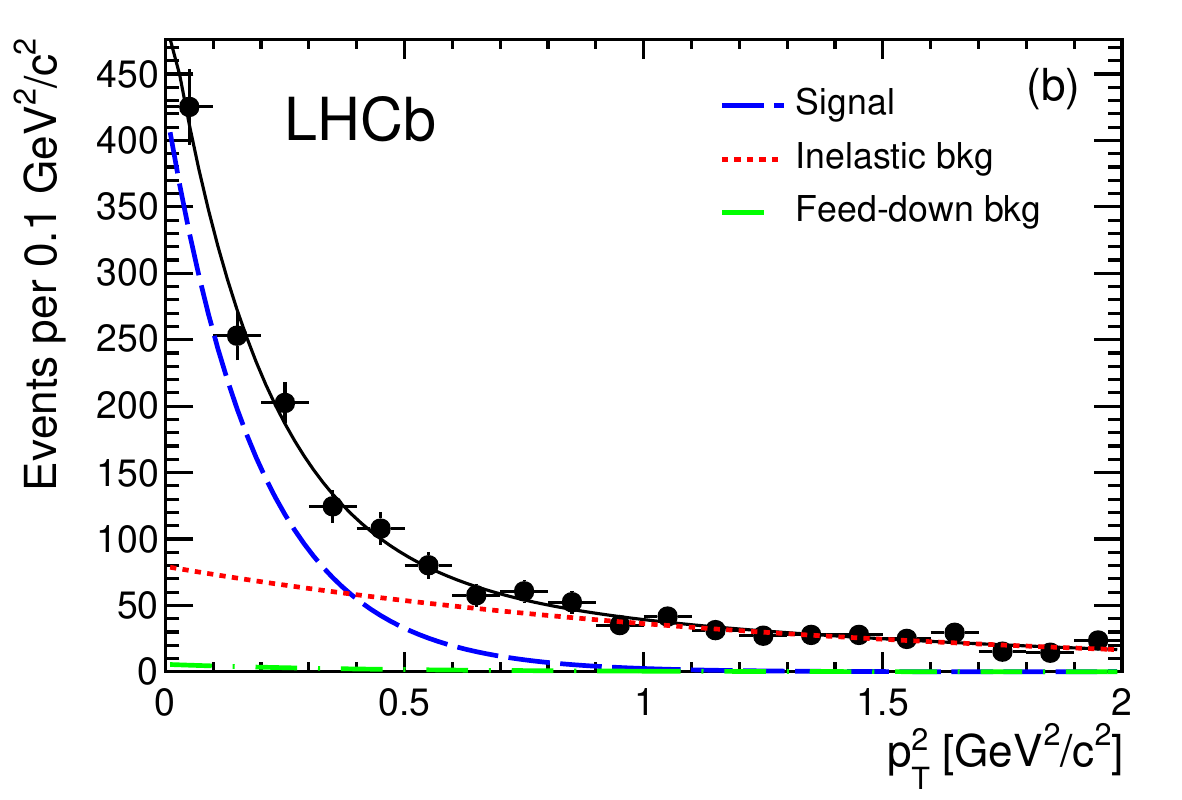}}
\caption{
Transverse momentum squared of (a) $J/\psi$ and (b) $\psi(2S)$ candidates.
The fitted contributions from the CEP signal, the inelastic and feed-down backgrounds
are indicated in the legend.
\label{fig:pt2}}
\end{figure}

The third background source is  
the largest and is also the most poorly determined for this analysis and all other analyses of the same kind
that LHCb has performed.
This consists of 
centrally produced $J/\psi$ or $\psi(2S)$ mesons that appear exclusive inside
the LHCb acceptance, but have activity outside of the active area of the detector, originating either
from additional gluon radiations or low mass diffractive dissociation of one or both protons.
Assuming that the $p_T^2$ distribution for both the elastic and inelastic components can
be described by exponential functions, $\exp(-bp_T^2)$, a fit to the data is performed to
determine the $b$ values and the relative sizes of both components.
The results are shown in Fig.~\ref{fig:pt2} and an overall purity of $0.592\pm 0.012$ is obtained
for the $J/\psi$ sample and $0.52\pm 0.07$ for the $\psi(2S)$ sample.
It is also worth noting that the fitted $b$ values  are consistent with the photoproduction results
from the H1 collaboration~\cite{Alexa:2013xxa}, once the difference in the centre-of-mass of 
the photon-proton system has been taken into account. 

The ALICE collaboration have reported in~\cite{TheALICE:2014dwa} a measurement of coherent $J/\psi$ photoproduction in ultra peripheral $p$--$Pb$ collisions at $\sqrt{s_{NN}}=5.02$ TeV. Results are selected with a dimuon pair produced either in the rapidity interval, in the laboratory frame, $2.5 < y < 4$ (p--Pb) or $-3.6 < y < -2.6$
(Pb--p), and with no other particles observed in the ALICE acceptance. The data were found to be consistent with a power law dependence of the $J/\psi$ photoproduction cross section in $\gamma p$ energies from $W_{\gamma p} \sim 20$ -- 700 GeV, and with results that are consistent with the LHCb measurement.

\begin{figure}[ht]
{\includegraphics[width=0.49\linewidth]{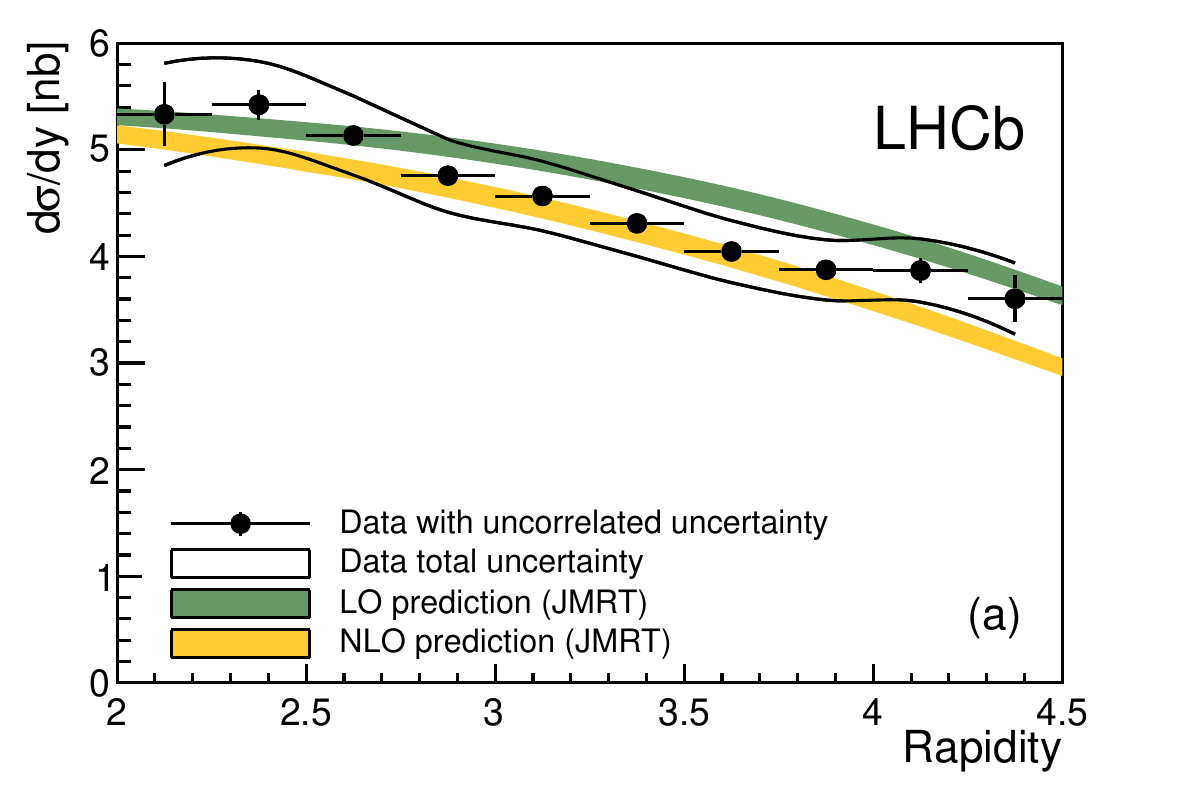}}
{\includegraphics[width=0.49\linewidth]{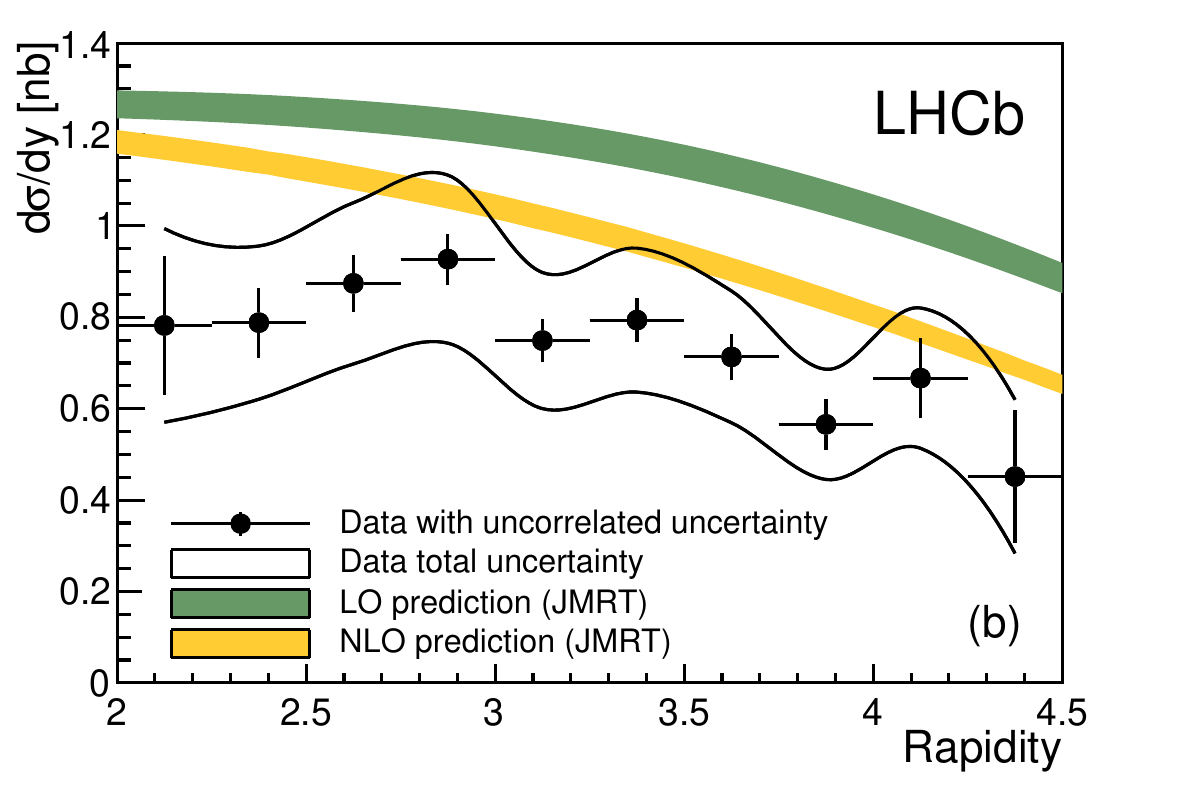}}
\caption{
Differential cross section for 
(a) $J/\psi$ and (b) $\psi(2S)$ 
compared to LO and NLO predictions.
\label{fig:cs}}
\end{figure}

After correcting for the LHCb data for detector efficiency and acceptance, 
the differential cross section as a function of rapidity, $y$, is calculated and is shown
in Fig.~\ref{fig:cs} compared to predictions at LO and `NLO' from Refs.~\cite{Jones:2013pga,Jones:2013eda}, see also Fig.~\ref{fig:gluonfit}.  
The `NLO' calculation is not a
full next-to-leading-order calculation but includes the dominant effects.
The experimental points are plotted with their statistical and total uncertainties.  
Most of the total uncertainty is correlated bin-to-bin and so the overall shape is well determined
by the data, which fits the `NLO' predictions better than LO, 
for both the $J/\psi$ and $\psi(2S)$ mesons. 

 The difference between the `NLO' and `LO' predictions becomes more pronounced in $\Upsilon$ production~\cite{Jones:2013pga}.
Analyses are ongoing at LHCb, using about 600 pb$^{-1}$ of pile-up free Run--I data,
to measure the CEP of $\Upsilon$
as well as other vector mesons such as $\phi$ and $\omega$.
Measurements of the CEP of $J/\psi$ and $\Upsilon$ mesons will be repeated at $\sqrt{s}=13$ TeV
and the ratio of these to measurements at  $\sqrt{s}=7$ TeV will provide important constraints
on the gluon PDF at parton fractional momenta around $x\sim$10$^{-5}$.

 In addition to this rapidity gap based analysis there are also possibilities for future LHC measurements with tagged forward protons.
It is worth emphasising that all of the existing measurements in hadronic collisions suffer from proton dissociation backgrounds; such a background can only be fully eliminated by tagging the outgoing protons. The possibility of such a measurement with CMS-TOTEM joint data taking is being actively pursued (the ATLAS--ALFA detector shows similar potential). To overcome efficiency losses in the muon reconstruction and triggering at low muon $p_T$'s, the exclusive J/$\psi$ analysis will be done on two charged-particle-only final states using the double arm RP triggered event sample without any muon identification requirement. In 5 pb$^{-1}$ 
of integrated luminosity at high $\beta^{*}$, more than 1000 J/$\psi$ candidates are expected in the $\mu^+\mu^-$ decay mode with little background. This will allow a detailed study of the azimuthal angular difference $\phi$ of the outgoing protons and for the $p_T$ spectrum of the produced J/$\psi$ meson to be determined, essentially without background,
even at larger $p_T$'s, where proton dissociation events dominate the existing measurements. As discussed in Section~\ref{sec:tagcepmotphot}, the Odderon, the C-odd partner of the Pomeron, is predicted to significantly modify the large $p_T$ part of the spectrum \cite{Motyka:2008kh}, which a CMS-TOTEM or ATLAS-ALFA measurement could test. Up to now there is no compelling experimental evidence for the existence of Odderon exchange, despite it being predicted by QCD. 

  \section{Exploratory physics}\label{sec:exploratory}

The study of BSM signatures in the CEP channel, which usually have very low cross sections and signal to background ratios, can be very competitive with and complementary compared to standard LHC searches. In this section, some examples of such processes are given. 

 \subsection{Search for invisible objects via the missing mass and momentum methods}

\subsubsection*{Motivation and theory}

CEP processes provide a possibility for simultaneous and precise measurements of the initial and final state kinematics, which can be used to search for events with missing 
mass or missing momentum signatures, see e.g.~\cite{Khoze:2010ba}. This opens up ways to search for new physics that might have escaped the searches of the general purpose 
detectors, CMS and ATLAS, e.g. in scenarios where the new physics couples dominantly or only to gluons.

\subsubsection*{Experimental results and outlook}

A preliminary analysis has been performed on the data of the common CMS-TOTEM $\beta^*$ = 90 m run at $\sqrt{s}$ = 8 TeV in July 2012, with a search for missing mass or missing momentum events performed on the existing data samples of double arm RP triggered and jet triggered events \cite{Hubert_seminar}. Only CEP events 
with a central mass, $M_{central}$, $\lesssim M_X$ are examined to avoid contamination from pile up events.  $M_{central}$ is reconstructed 
from the sum of the CMS particle flow objects and the missing momentum, $P\!\!\!\!/ \:$, is reconstructed from the difference of the sum of 
the proton momenta and the sum of the momenta of the particle flow objects. The rapidity gaps, $\Delta \eta = - $ln $ \xi$, predicted by the 
proton $\xi$ measurements (momentum loss fraction) are verified using the T2 detector with a rapidity coverage of $5.3 < |\eta| < 6.5$. To probe $O$(pb) cross sections 
for the two signal topologies described below, a statistics of double-arm RP-triggered and of jet-triggered events corresponding to an 
integrated luminosity of $\sim$ 100 pb$^{-1}$ is needed. 

\begin{figure}[htbp]
      \centering
      \includegraphics[width=0.7\textwidth]{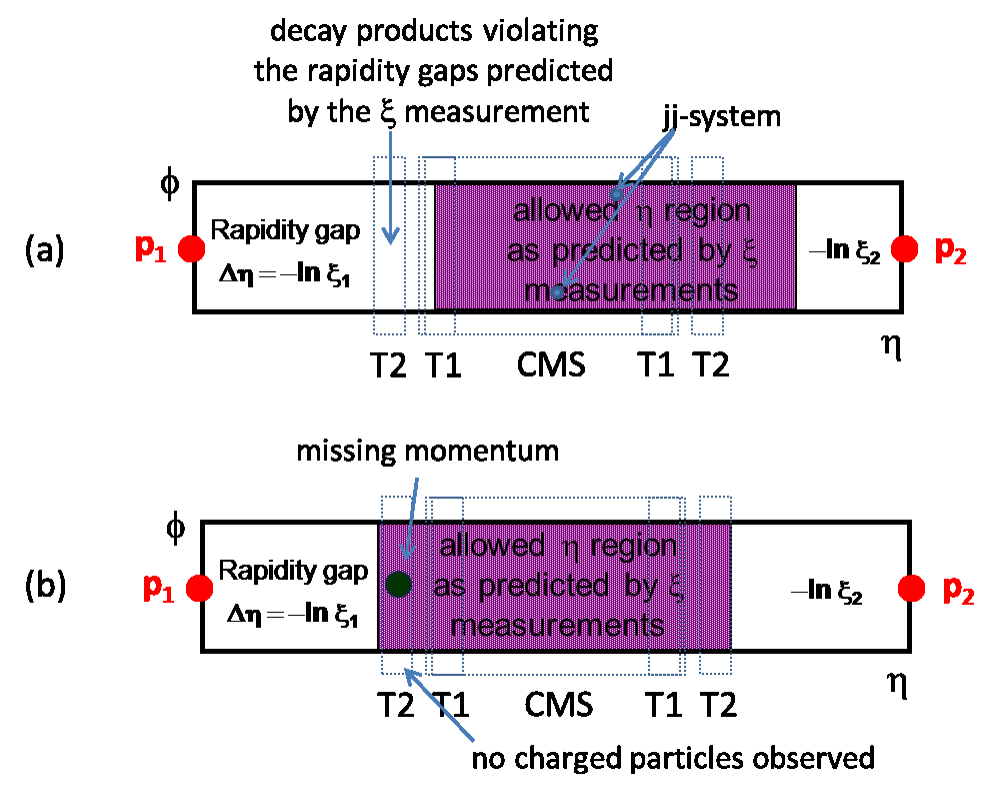}
      \caption{Schematical drawings of the two event topologies used in the search for missing mass and momentum 
signatures in CEP events. (a) Events with charged particles in T2, violating the rapidity gaps predicted by the $\xi$ measurement, 
from e.g. the decay of a CEP-produced particle.  (b) Events with high missing momentum, pointing towards an $\eta$ region with 
good CMS-TOTEM instrumentation and no charged particles or energy deposits observed in the $\eta$ region where the missing 
momentum points. T2 $\eta$ region is given as example; this could be due to a CEP-produced particle escaping undetected.}
      \label{fig:missing_mass}
\end{figure}
   
To verify the performance and the search methodology, control samples of events were selected both in double arm RP triggered and in the jet triggered 
samples with the following requirements: the presence of charged particles in T2, when allowed by the rapidity gaps predicted by the $\xi$ measurements, and no charged 
particles in T2, when not allowed by the rapidity gaps predicted by the $\xi$ measurements. Many such events, corresponding to standard CEP events, were 
found in both the double arm RP triggered and jet triggered data samples and these will be used for a determination of the inclusive CEP event and 
CEP jet cross sections, respectively. One such candidate in the jet sample with $M_{central} \approx M_X$ is shown in Fig. \ref{fig:trijet}. 

A first signal topology, depicted schematically in Fig.~\ref{fig:missing_mass}(a), are events with charged particles in 
T2 violating the $\xi$-predicted rapidity gaps. This could happen if a particle is created in the CEP reaction and some of its decay products go into 
the T2 $\eta$ acceptance region. Such events would be used  to search
for the production of new particles by studying the $M_X$ (and $M_X - M_{central}$) 
distributions. No candidate events were found in the available jet sample. Remaining single diffractive pile up and beam 
halo background makes the double-arm RP triggered sample unusable for such searches. 

An even more striking signature is 
events with high missing momentum pointing towards the region with good CMS-TOTEM instrumentation 
($|\eta|<$ 6.5) and no charged particles or energy deposits in the $\eta$ region close to where the missing momentum points. 
Fig.~\ref{fig:missing_mass}(b) depicts the case where the missing momentum points towards T2. This could happen if a particle is created 
in the CEP reaction and escapes the experimental apparatus, undetected in the T2 acceptance region. Events are rejected if more forward 
rapidity gaps than T2 ($|\eta|>$ 6.5) would be allowed by the $\xi$ measurements. This confines the search to the mass region between 
a few times the combined resolution of the $M_{central}$ and $M_X$ measurements and the maximal central mass allowed by the T2 acceptance.
For $\sqrt{s}$ = 13 TeV, this implies a 150-600 GeV mass range. Events with missing mass up to 400 GeV
were found in both the double arm RP  and jet triggered data set at  $\sqrt{s}$ = 8 TeV with background events expected from neutral 
particles escaping 
detection in the T2 acceptance region, due to ``acceptance gaps'' between the forward detectors as well as from  $p + p \rightarrow N^* \,+\, X  \,+\,  p$ 
or $p  \,+\,  X  \,+\,  N^*$ reactions. In the latter case, one of the observed protons would come from a decay of a nucleon resonance, $N^*$, and the other 
decay products of the $N^*$ would escape detection. With increased statistics, it is expected that these backgrounds will be modelled sufficiently well.

 \subsection{Searching for magnetic monopoles with forward proton detectors}
 
 \subsubsection*{Motivation and theory}
 
One theoretical possibility which could produce the signal described in the previous section is the exclusive production of magnetic monopoles. The existence of magnetic monopoles has been discussed since the discovery of
magnetism, although it is well established that all magnetic and
electromagnetic phenomena surrounding us can be explained with electric
charges and electric currents. However, while the existence of magnetic monopoles is
not required, it is also not excluded.
Our current understanding of electromagnetism is formulated in terms of
Maxwell's equations. These are in fact a special case of more general relations, which contain additional terms connected to magnetic charges ($\rho_{\mathrm
m}$) and currents ($\mathbf{j}_{\mathrm m}$):

\vspace{1mm}
\begin{minipage}{0.4\textwidth}
    %Gauss's law for electricity
  \[ \nabla \cdot \mathbf{E} = 4 \pi \rho_{\mathrm e}\;,\] 
    %Faraday's law of induction  
  \[ -\nabla \times \mathbf{E} = \frac{1}{c}\frac{\partial \mathbf{B}} {\partial t} 
  + \textcolor{blue}{\frac{4 \pi}{c}\mathbf{j}_{\mathrm m}}\;, \]
\end{minipage}
\begin{minipage}{0.2\textwidth}
\end{minipage}
\begin{minipage}{0.4\textwidth}
    %Gauss's law for magnetism   
  \[ \nabla \cdot \mathbf{B} = \textcolor{blue}{4 \pi \rho_{\mathrm m}}\;, \]
    %Ampère's law
  \[ \nabla \times \mathbf{B} = \frac{1}{c}\frac{\partial \mathbf{E}}{\partial t}
  + \frac{4 \pi}{c} \mathbf{j}_{\mathrm e} \;.\]
\end{minipage}
\vspace{2mm}

\noindent The existence of magnetic monopoles are therefore not mathematically inconsistent with Maxwell's equations; it is simply that in their usual form they assume their absence in Nature.
In the theory of quantum mechanics, the electromagnetic field is described in
terms of the vector potential. It has been shown by Dirac~\cite{Dirac1931} that
it is possible to incorporate magnetic monopoles into this description: a
vector potential singular along an infinite line starting at some point $\vec
x$ describes the field of a magnetic monopole placed at $\vec x$. This is
equivalent to a solenoid of an infinitely small radius starting at $\vec x$ and
ending at infinity. Such a description is called the Dirac string.
Naturally, such a picture can only describe a particle (the monopole) if the
string is unobservable. This leads to a constraint on the possible values of
the magnetic charge, known as Dirac's quantisation:
\[eg = \frac{n}{2} \hbar c, \quad n = 0, \pm 1, \pm 2, \dots \]
where $e$ is the elementary electric charge and $g$ is a possible value of
magnetic charge. This result has a significant implication -- if magnetic
monopoles exist, electric charge is quantised; the existence of magnetic monopoles would therefore predicts electric charge quantisation, the origin of which is one of the biggest questions in
particle physics. In addition, the relation between the electric and magnetic
elementary charges allows the former to be calculated as
\[ \quad g = \frac{ne}{2\alpha} \approx e \cdot 68.5 n\;,\]
where $\alpha$ is the fine-structure constant. This high value of the magnetic
charge means that a magnetic monopole would interact with electromagnetic field
like a heavy nucleus.

\begin{figure}[htpb]
  \begin{minipage}[t]{0.75\textwidth}
    \includegraphics[width=\textwidth,clip, trim = 0 0 0 40mm]{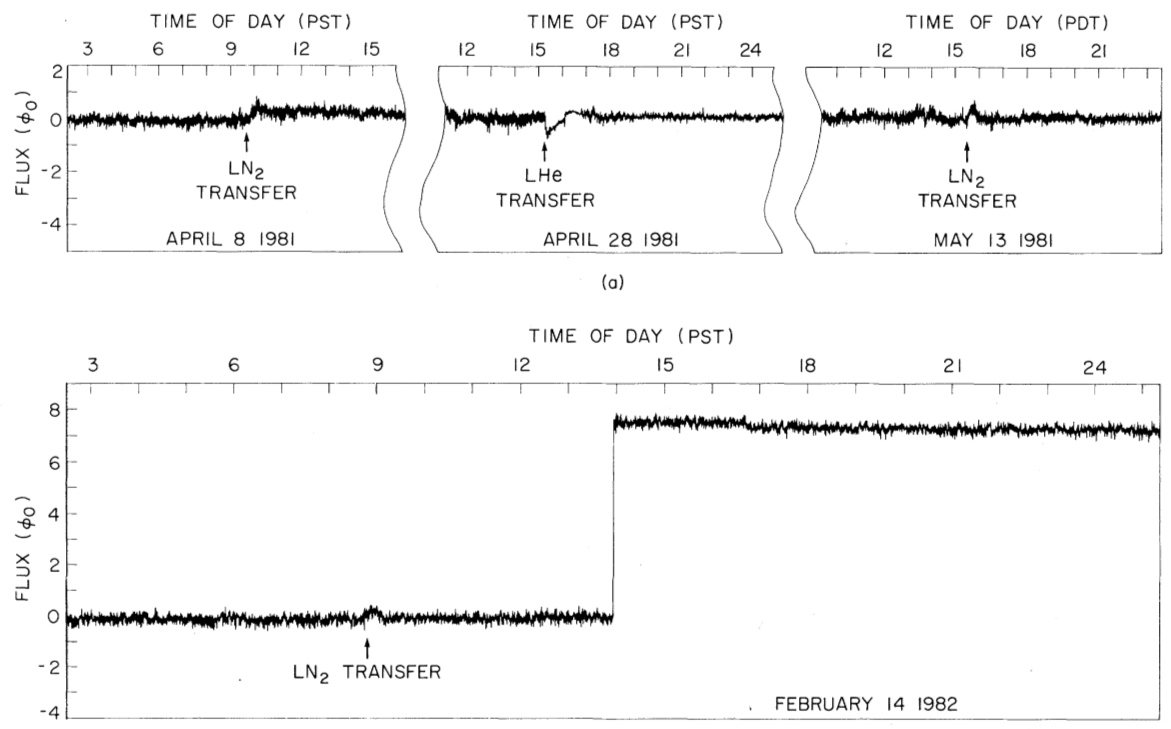}
  \end{minipage}\hfill
  \begin{minipage}[t]{0.21\textwidth}
    \includegraphics[width=\textwidth]{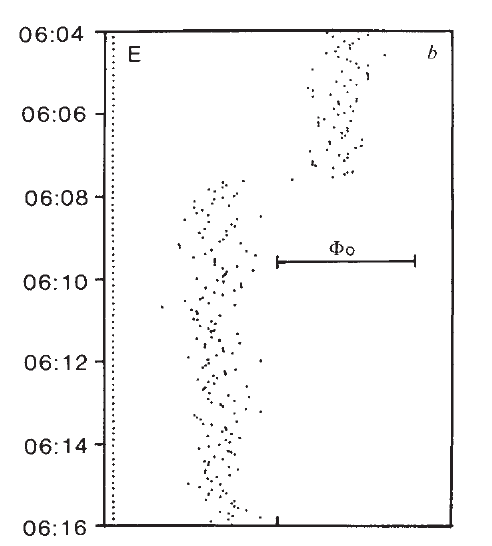}
  \end{minipage}
  \begin{minipage}[t]{0.75\textwidth}\centering(a)\end{minipage}\hfill
  \begin{minipage}[t]{0.21\textwidth}\centering(b)\end{minipage}
  \caption{Two magnetic monopole candidate events: (a)
    from~\cite{Cabrera:1982gz}, (b) from~\cite{Caplin:1986kw}.
  }
  \label{fig:monopole_observations}
\end{figure}

Many searches for magnetic monopoles have been performed over the years~\cite{Milton:2006cp}. Two of them~\cite{Cabrera:1982gz, Caplin:1986kw} observed signals consistent with a magnetic
monopole passing through superconducting coils (manifesting as a sudden change of the magnetic flux, see
Fig.~\ref{fig:monopole_observations}). While no alternative explanations for these signals have been
found, each of the experiments only observed a single such event; moreover, each pointed towards a different value of the magnetic charge. 
Magnetic monopoles have also been searched for at particle colliders. Typical
methods include searches for highly ionizing particles, see for example the ATLAS result~\cite{Aad:2012qi}. 

However, it is possible that magnetic
monopoles would not be observed by the main LHC detectors, even if they are
produced in collisions. For example, if monopoles are heavy, their velocity
will be low and they can miss the trigger window. Another possibility is that
the monopoles could be trapped inside dead material, such as the beam pipe.
This leads to another widely used search method -- scanning the exposed beam
pipe with very sensitive magnetometers~\cite{Kalbfleisch:2003yt,Aktas:2004qd}.
MoEDAL~\cite{MoEDAL,
Acharya:2014nyr}  is a dedicated experiment at the LHC, devoted to magnetic
monopole searches. Its aim is to address the typical drawbacks of general purpose
detectors, and it consists of trapping detectors, which can capture
magnetic monopoles for further study with magnetometers, and plastic tracking
detectors, which are sensitive to highly ionising particles and do not suffer
from the trigger issue described above.

\begin{figure}[htbp]
  \centering
    \includegraphics[width=0.3\textwidth]{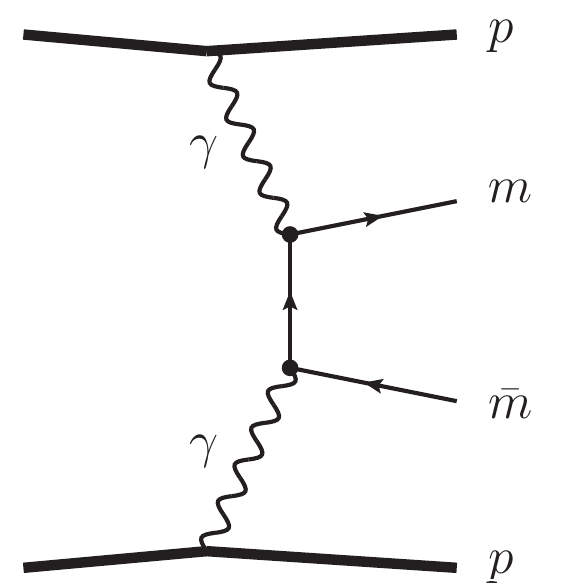}
  \caption{Exclusive magnetic monopole production mechanism.}
  \label{fig:monopole_production}
\end{figure}

\subsubsection*{Experimental results and outlook}
 
 In assessing the measurement feasibility for magnetic monopole searches with tagged forward protons, one of the main issues is the lack of a reliable model for magnetic monopole production.
Although one can as usual draw a Feynman diagram for the production process, the very high value
of coupling (magnetic charge) invalidates perturbative calculations. It is
therefore not possible to predict the expected cross section value using current calculation tools. On the other hand, it is possible to consider the potential sources of
backgrounds to assess the cross section sensitivity. Most simply, an analysis could be performed in a low
pile-up environment, in which case the signature will consist of two forward protons
tagged in the forward detectors and an empty central detector.
 
 Two types of background can be considered. First, the two measured
protons may not directly originate from the same collision, i.e. they can be produced by two separate
soft single diffractive events, or they can be two halo particles (or a combination
of these two cases). For diffractive protons, a lack of
particles in the central detector means that the proton could have lost only a
tiny part of its momentum. Therefore, these types of background will be
important for measurements with high-$\beta^\ast$ optics. On the other hand,
the background from two halo particles is relevant for all optics scenarios.
Moreover, such a background cannot be accurately predicted and can change
from run to run, or even during the run.  However, since the two particles are
independently produced, the probability of such an event should be equal to the product of
probabilities of having protons on each side.  In addition, halo protons are
expected to be close to the beam in low $\beta^{*}$ optics, so they should have a sizable impact on the sensitivity of the search mainly in the low mass monopole region.

For high mass monopoles searches, where the cross sections are expected to be lower, low $\beta^\ast$ optics are necessary.  Here, the main background is expected to originate from soft double
diffractive processes, where the central detector is empty, and the forward protons can then be present in the dissociated state as a simple consequence of baryon number conservation (i.e. the dissociated state will always contain either a proton or a neutron). Usually, the energy of such protons is too small to reach the forward detectors, however the high cross section of the process means that the resulting background is non-negligible.

\begin{figure}[htbp]
  \centering
  \includegraphics[width=0.6\textwidth]{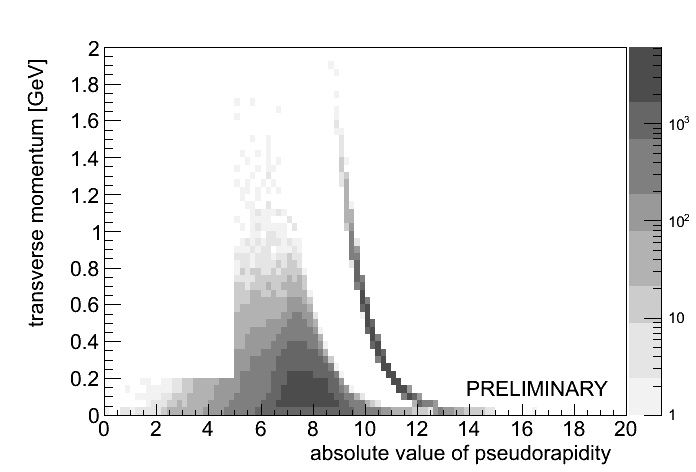}
  \caption{Particle flow for double diffractive events consistent with the signature of
    invisible particle production (empty central detectors and two forward
  protons with $0.02<\xi<0.12$).}
  \label{fig:particle_flow}
\end{figure}

 In Fig.~\ref{fig:particle_flow} the particle flow for
double diffractive events with the signature of invisible particles production is presented, i.e.
empty central detectors and two forward protons with $0.02<\xi<0.12$
(measurements with low-$\beta^\ast$ optics). It has been assumed that the central
detector can measure particles with $|\eta|<5$ and $p_T>200$ MeV (both charged
and neutral), and Pythia 8 has been used here and in the results which follow.  The very thin parabolic shape at $10<|\eta|<14$ corresponds to the forward protons. One can clearly see that the majority of particles have
$6 < |\eta| < 9$, and thus vetoing on activity in this regions will suppress a
significant part of the background.

\begin{figure}[htbp]
  \centering
  \includegraphics[width=0.5\textwidth]{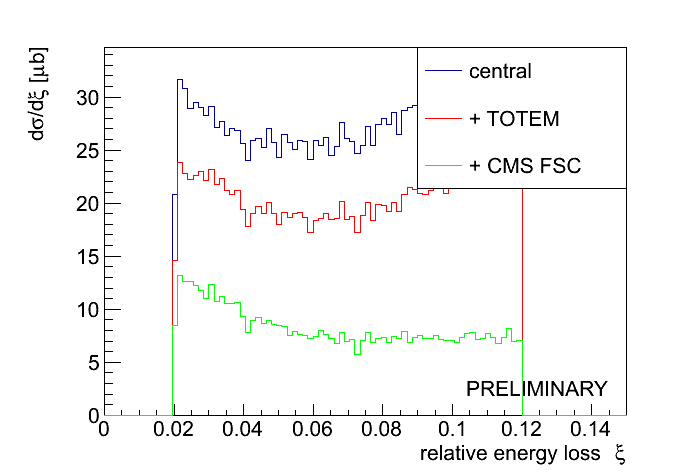}%
  \includegraphics[width=0.5\textwidth]{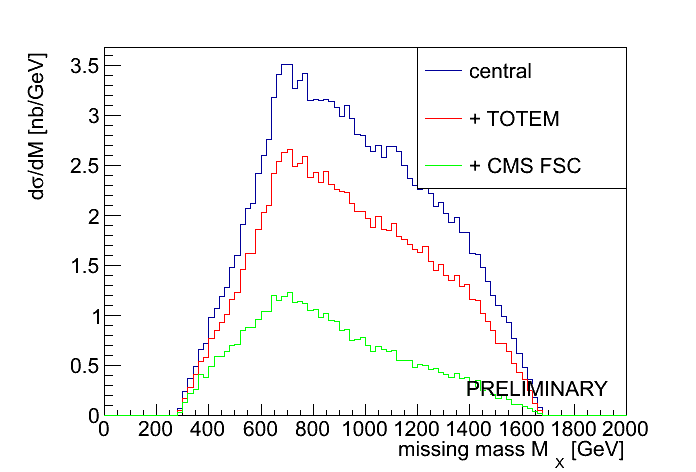}
  \begin{minipage}{0.5\textwidth}\centering (a) \end{minipage}%
  \begin{minipage}{0.5\textwidth}\centering (b) \end{minipage}
  \caption{Distributions of (a) relative momentum loss and (b) missing mass
  for events with the signature of invisible particle production (empty central
detectors and two forward protons with $0.02<\xi<0.12$). The effect of an additional
veto from the TOTEM tracker and CMS FSCs is presented.}
  \label{fig:xi_M_distributions}
\end{figure}

It is interesting to consider the relative momentum loss and missing mass
distributions for these background events. These are presented
in Figs.~\ref{fig:xi_M_distributions} (a) and \ref{fig:xi_M_distributions} (b),
respectively, where in addition the effects of a veto with the TOTEM T2
detectors and CMS FSCs are presented.

Even with the most stringent selection, the background cross section remains at the mb level. However, one must remember that for the case of magnetic
monopoles, the electromagnetic coupling is very large, and the signal cross section
may therefore be significant. Although it is not possible to reliably predict  the expected
cross section, one can compare the obtained value to the most recent limits.
ATLAS put limits on the cross section of 145 -- 16 fb for monopole
masses of 200 -- 1200 GeV. It is therefore clear that the forward proton method described above will not be competitive with these general monopole searches. However, it
will remain useful for scenarios in which the monopoles are not visible in the
standard ways, in which case these may be missed by standard searches.

    \subsection{Standard Model exclusive production of $\gamma \gamma$, $WW$ and $ZZ$ via photon induced processes}\label{sec:excbos}
    
    \subsubsection*{Motivation and theory}

        \begin{figure}
    \begin{center}
    \includegraphics[width=0.35\textwidth]{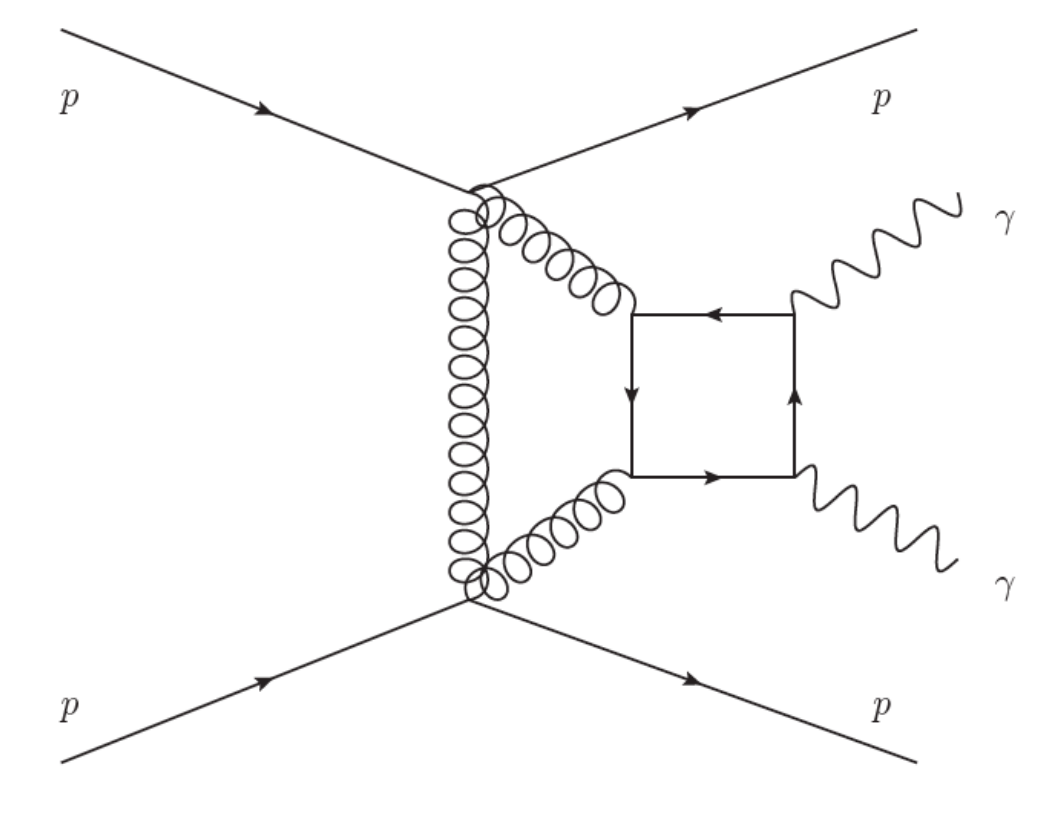}
    \includegraphics[width=0.35\textwidth]{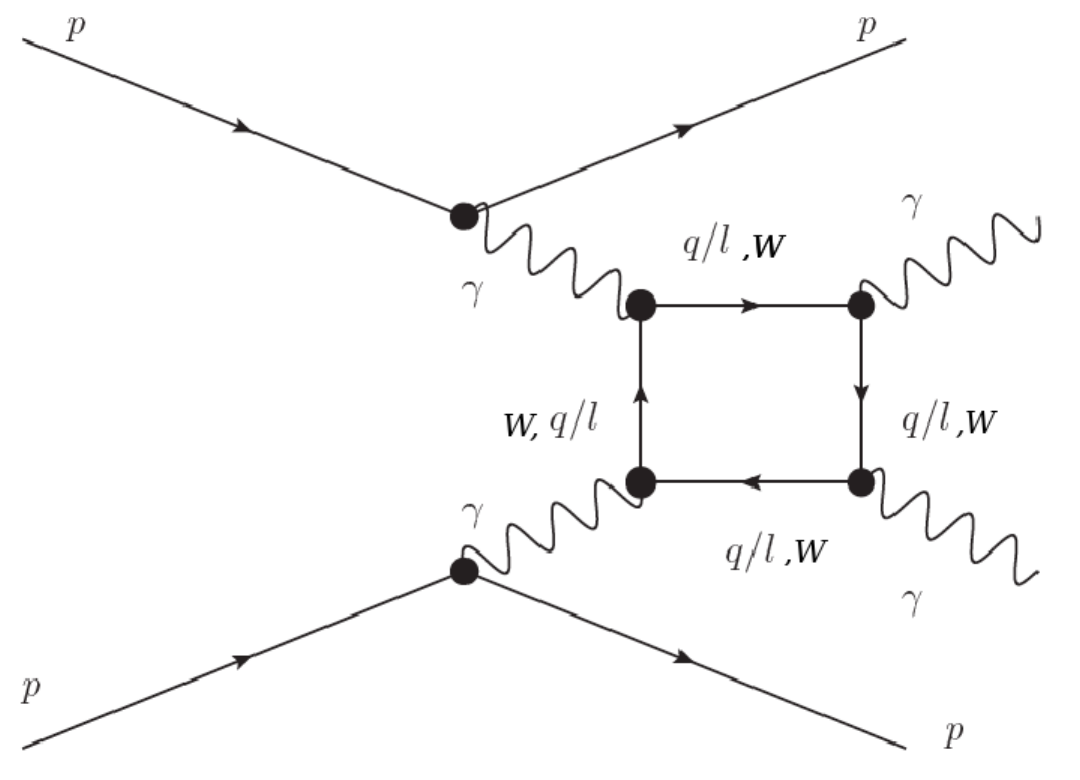}
    \caption{Di-photon exclusive Standard Model production via QCD (left) and photon induced (right) processes at the lowest order of perturbation theory.}
    \label{fig:pCp-fpmc}
    \end{center}
    \end{figure}

    \begin{figure}
    \begin{center}
    \includegraphics[width=0.49\textwidth]{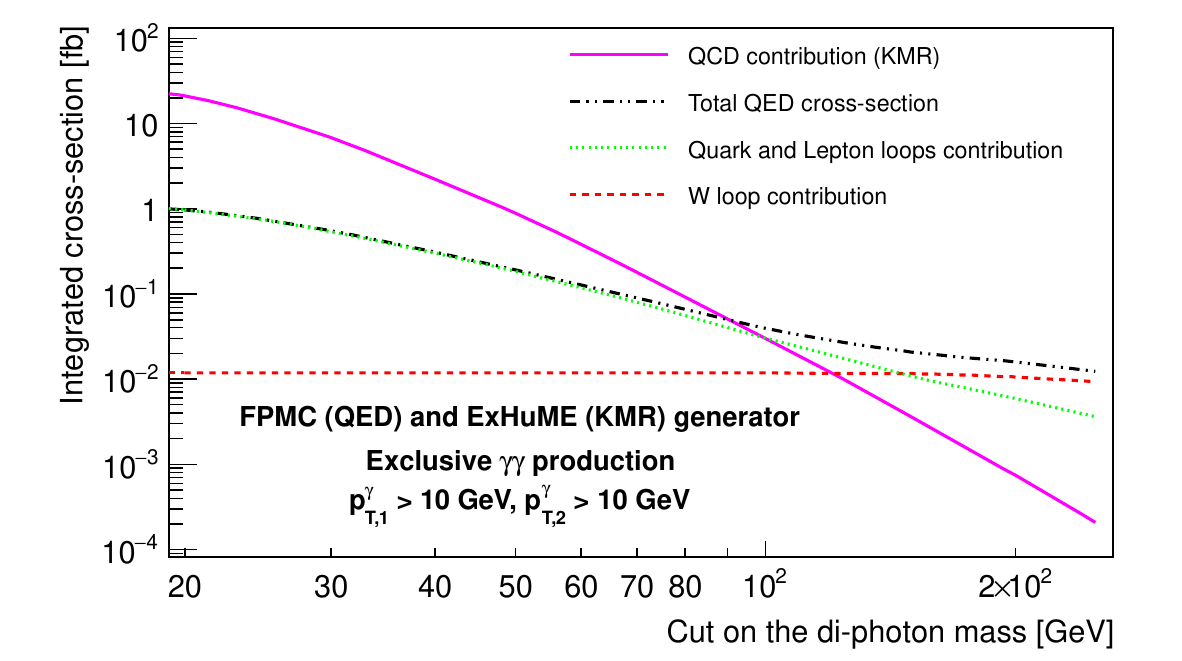}
    \caption{Integrated di-photon production cross section as a function of the minimum di-photon mass
    requirement. In addition, both photons must have 
    a transverse momentum $p_\perp>$ 10 GeV. The QCD exclusive  (Durham)  processes, solid line,
    dominates at low masses while QED di-photon production dominates at higher masses
    (dashed lines). The QED production corresponds to di-photon production via
    lepton/fermion loops (dotted line) and $W$ boson loops (dashed-dotted line).}
    \label{matthias}
    \end{center}
    \end{figure}

    In the SM, the couplings of fermions and 
    gauge bosons are constrained by the gauge symmetries of the Lagrangian.
    Therefore, the measurement of $W$, $Z$ boson and $\gamma$ pair production via the exchange of
    two photons  
    can directly provide stringent tests
    of one of the most important and least understood
    mechanism in particle physics, namely 
    electroweak symmetry breaking.

    Considering first $W$ pair production
    induced by the 
    exchange of two photons~\cite{deFavereaudeJeneret:2009db}, this is a pure electroweak process
    in which the decay products of the $W$ bosons are measured in the central 
    detector and the scattered protons may remain intact.
    This process, as well as the different diffractive backgrounds, are all
    implemented in the FPMC Monte Carlo~\cite{FPMC:ch5} and Herwig++. After simple cuts to select exclusive $W$ pairs decaying into leptons, such
    as a cut on the proton momentum loss ($0.0015<\xi<0.15$) ---
    assuming that the protons are tagged in AFP or CT-PPS at 210 and 420 m ---
    on the transverse momentum of the leading and second leading leptons at 25 and
    10 GeV respectively, on $\met>20$ GeV, $\Delta \phi>2.7$ between leading
    leptons, and $160<M_{X}<500$ GeV, the diffractive mass reconstructed using the
    forward detectors, the background is found to be less than 1.7 event for 30
    fb$^{-1}$ for a SM signal of 51 events~\cite{Kepka:2008yx}. 

\begin{table*}
\begin{center}
\small
\begin{tabular}{|c|c|c|c|}
\hline
Cut / Process & QCD Exclusive (KMR) & QED Fermion loop & $W$ loop \\
\hline
\hline
$m_{\gamma\gamma}>10$~GeV,$p_{T1,2}>5$~GeV       & 372.1  fb    & 5.5   fb   & 0.01  fb  \\
\hline
\hline
$m_{\gamma\gamma}>20$~GeV,$p_{T1,2}>10$~GeV      & 20.4   fb    & 1.   fb   & 0.012  fb  \\
$m_{\gamma\gamma}>50$~GeV,$p_{T1,2}>10$~GeV      & 0.87   fb    & 0.18  fb   & 0.012  fb  \\
$m_{\gamma\gamma}>100$~GeV,$p_{T1,2}>10$~GeV     & 0.030  fb    & 0.03  fb   & 0.012  fb  \\
$m_{\gamma\gamma}>200$~GeV,$p_{T1,2}>10$~GeV     & 7.4$\cdot$10$^{-4}$ fb    & 5.$\cdot$10$^{-3}$ fb   & 0.010 fb  \\
$m_{\gamma\gamma}>500$~GeV,$p_{T1,2}>10$~GeV     & 3.2$\cdot$10$^{-6}$ fb    & 3.$\cdot$10$^{-4}$ fb   & 0.004 fb  \\
\hline
\end{tabular}
\end{center}
\caption{Integrated cross sections of the different SM exclusive di-photon production processes at the LHC at $\sqrt{s} = 14$ TeV for various requirements on the di-photon mass (m$_{\gamma\gamma}$) and photon transverse momenta (p$_{T1,2}$).}
\label{tab_smprod}
\end{table*}

   Considering the $\gamma\gamma$ final state, in Fig.~\ref{fig:pCp-fpmc}, the leading processes that produce two photons and two 
    intact protons
    in the final state are shown.
    Fig.~\ref{fig:pCp-fpmc} (left) corresponds to exclusive QCD di-photon
    production, while
    Fig.~\ref{fig:pCp-fpmc} (right) correspond to photon--induced production, which is of interest here. It is worth noticing that quark, lepton and $W$ loops
    need to be considered in order to get the correct SM cross section for di-photon
    production, as shown in Fig~\ref{matthias}. The QCD induced process, discussed in Section~\ref{sec:gamgam}, is dominant at low masses whereas
    the photon induced process dominates at higher di-photon 
    masses~\cite{Khoze:2001xm}.
    It should be emphasised that the $W$ loop contribution dominates at high di-photon 
    masses~\cite{Fichet:2013gsa, Fichet:2014uka, d'Enterria:2013yra,Sun:2014qba}; all of these terms have been implemented for the first time in a single generator in FPMC~\cite{FPMC:ch5}. In Table~\ref{tab_smprod}, some selected values of the cross sections from Fig.~\ref{matthias} discussed above are shown. The threshold where the $W$-loop contribution becomes dominant is for a di-photon mass slightly above 100 GeV. 

While, as  can be seen in Table~\ref{tab_smprod}, and discussed further in Section~\ref{sec:gamgam}, the expected cross sections for the QCD production mechanism are sufficient for measurements at the LHC, including during special luminosity runs, there appears to be limited sensitivity to the  SM photon--induced mechanism. On the other hand, it may be possible to
study di-photon production via quark, lepton and even $W$ loops at the LHC in
the heavy ion mode~\cite{d'Enterria:2013yra}.

\subsubsection*{Experimental results and outlook}

    An initial study of $\gamma\gamma \rightarrow WW$~\cite{Chatrchyan:2013foa} was performed at CMS
    using 5 fb$^{-1}$ of data collected at $\sqrt{s} = 7$~TeV, based on
    the final state consisting of an electron, a muon, and undetected neutrinos.
    Events were selected by requiring the presence of a $\mu^{\pm}e^{\mp}$
    vertex with zero additional charged tracks associated, and
    $p_{T}(\mu^{\pm}e^{\mp})>30$~GeV. The first requirement was used to
    suppress inclusive backgrounds, while the $p_{T}(\mu^{\pm}e^{\mp})$
    requirement also suppressed backgrounds from
    $\gamma\gamma \rightarrow \tau\tau$. As the outgoing protons could not
    be tagged, the selected sample also contained a large fraction 
    of proton dissociation, which was estimated from data using
    control samples of high-mass $\gamma\gamma \rightarrow \mu\mu$ events.
    The backgrounds were estimated using simulation and control regions in
    the data. In the signal region two events were observed in the data,
    compared to an expectation of $2.2 \pm 0.4$ signal events and
    $0.84 \pm 0.15$ background. The event properties such as the
    $\mu^{\pm}e^{\mp}$ invariant mass and acoplanarity, and the missing
    transverse energy, were compatible with Standard Model expectations.
    
   \subsection{Anomalous gauge couplings: $\gamma \gamma  \gamma\gamma$} 
    
    \subsubsection*{Motivation and theory}
    
     Assuming that the new physics mass scale $\Lambda$ is heavier than the
    experimentally accessible 
    energy $E$, all new physics manifestations can be described using 
    an effective Lagrangian valid for  $\Lambda\gg E$.
    Among these operators, the pure photon dimension-eight operators
    \begin{eqnarray}
    {\cal L}_{4\gamma}= %\frac{\zeta_1^\gamma}{\Lambda^4} 
    \zeta_1^\gamma F_{\mu\nu}F^{\mu\nu}F_{\rho\sigma}F^{\rho\sigma}
    +\zeta_2^\gamma F_{\mu\nu}F^{\nu\rho}F_{\rho\lambda}F^{\lambda\mu}
    \label{zetas}
    \end{eqnarray}
    can induce the $\gamma \gamma \gamma \gamma$ process, highly 
    suppressed in the SM~\cite{Fichet:2013gsa, Fichet:2013ola}.
  Different new physics processes can contribute to 
    $\zeta_{1,2}^\gamma$.
    For example, loops of heavy charged particles contribute to the $4\gamma$
    couplings~\cite{Fichet:2013gsa, Fichet:2013ola} as 
    $\zeta_i^\gamma=\alpha^2_{\rm em} Q^4\,m^{-4}\, N\,c_{i,s}$, where
    $c_{1,s}$ is related to the spin of 
    the heavy particle of mass $m$ running in the loop and $Q$ 
    its electric charge. 
    The factor $N$  counts all additional multiplicities such as color or flavor.
    These couplings scale as $\sim Q^4$ and are enhanced 
    in the presence of  particles with large charges. For example, certain light 
    composite fermions, characteristic of composite Higgs models, have typical
    electric charges of several units. For a 500 GeV 
    vector (fermion) resonance with  
    $Q=3\, (4)$,
    large couplings $\zeta_i^{\gamma}$ of the order of
   $10^{-13}-10^{-14}$ GeV$^{-4}$ can be reached. The difference of sensitivity between vector and fermions comes from the $c_{i,s}$ factors. 
    Beyond perturbative contributions to $\zeta_i^\gamma$ from  
    charged particles,  non-renormalizable interactions of neutral particles are 
    also present in common extensions of the SM.  Such theories can contain 
    scalar, pseudo-scalar and spin-2 resonances that couple  to the photon
    and generate  the $4\gamma$ couplings by tree-level exchange as 
    $\zeta_i^\gamma=(f_{s}\, m)^{-2}\,d_{i, s}$, where
    $d_{1,s}$ is related to the spin of the particle.
    Strongly-coupled conformal extensions of the SM contain a scalar particle 
    $(s=0^+)$, 
    the dilaton. 
    Even a dilaton of mass 2 TeV can produce a sizeable effective 
    photon interaction, $\zeta_1^\gamma\sim 10^{-13}$\gev$^{-4}$.
    These features are reproduced for a large number of colors by the 
    gauge-gravity correspondence in warped extra dimensions.
    Warped-extra dimensions also feature Kaluza-Klein (KK) 
    gravitons~\cite{Randall:1999ee}, that can induce anomalous couplings~\cite{Fichet:2013ola}
     \begin{equation}
    \zeta_i^\gamma= \frac{\kappa^2}{8 \tilde k^4} \,d_{i, 2}
     \end{equation}
    where $\tilde k$ is the IR scale that determines the first KK graviton mass 
    and $\kappa$ is a parameter that can be taken 
    $\mathcal O(1)$. 
    For $\kappa\sim 1$, and $m_{2}\lesssim 6$ TeV, the photon vertex can easily 
    exceed $\zeta_2^\gamma\sim 10^{-14}$ \gev$^{-4}$.

    \begin{figure}
    \begin{center}
    \includegraphics[scale=0.45]{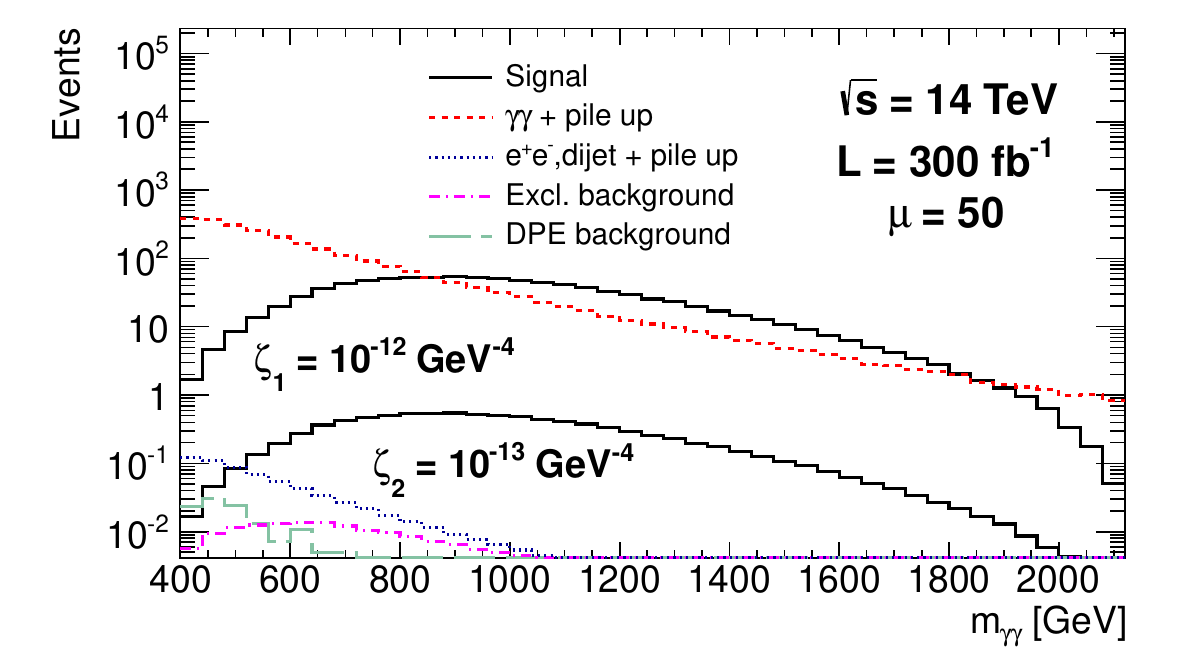}
    \caption{ Di-photon invariant
    mass distribution for the signal ($\zeta_{1} = 10^{-12},~10^{-13}$ GeV$^{-4}$, 
    see Eq.~\ref{zetas}) and for the backgrounds 
    (dominated by $\gamma\gamma$ with protons from pile up), requesting
    two protons in the forward detectors and two photons of $p_T >$ 200 (100)~GeV for the leading (subleading) photon with 
    at least one converted photon in the central detector, for a luminosity of 
    300 fb$^{-1}$ and an average pile up of $\mu = 50$.}
    \label{fig:mass}
    \end{center}
    \end{figure}
    \begin{table}[h]
    \begin{center}
    \small
    \caption{Number of signal (for a baseline coupling of 2.10$^{-13}$ GeV$^{-4}$) and background 
    events after 
    various selections for an integrated
    luminosity of 300 fb$^{-1}$\ and $\mu=50$ at $\sqrt{s}=14$ TeV. At 
    least one converted photon is required. The standard cuts correspond to the AFP
    or CT-PPS
    acceptance ($0.015<\xi<0.15$) and the request of the
    photon $p_T$ to be above 50~GeV}
    \begin{tabular}{|c||c||c|c|c|c|}
    \hline
    Cut / Process & Signal & Excl. & DPE & e$^{+}$e$^{-}$, di-jet  
    & $\gamma\gamma$ +\\
    & & & & + pile up & pile up \\
    \hline
    \hline
    %\vspace{1mm}
    standard                  & 20.8  & 3.7 & 48.2 & $2.8\cdot10^{4}$ & $1.0\cdot10^{5}$ \\
    $p_{\mathrm{T}1}>200$GeV, $p_{\mathrm{T}2}>100$~GeV                    & 17.6  & 0.2 & 0.2  & 1.6          & 2968         \\
    $m_{\gamma\gamma}>600$~GeV                          & 16.6  & 0.2 &  0    & 0.2          & 1023         \\
    $p_{\mathrm{T2}}/p_{\mathrm{T1}}>0.95$, $|\Delta\phi|>\pi-0.01$      & 16.2  & 0.1 & 0   & 0          & 80.2         \\
    $\sqrt{\xi_{1}\xi_{2}s} = m_{\gamma\gamma} \pm 3\%$                                      & 15.7  & 0.1 & 0   & 0          & 2.8          \\
    $|y_{\gamma\gamma}-y_{pp}|<0.03$                   & 15.1  & 0.1 & 0   & 0            & 0            \\
    \hline

    \end{tabular}
    \label{tab:event}
    \end{center}
    \end{table}

	\begin{center}
    \begin{figure*}
    \includegraphics[width=0.48\textwidth]{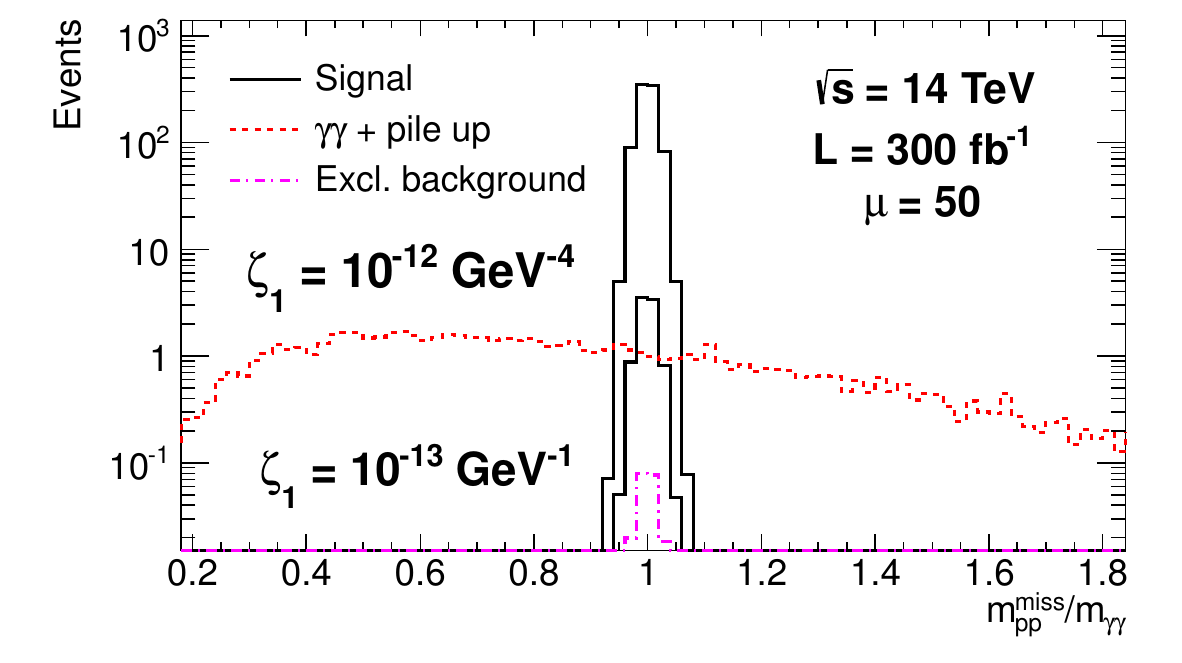}
    \includegraphics[width=0.53\textwidth]{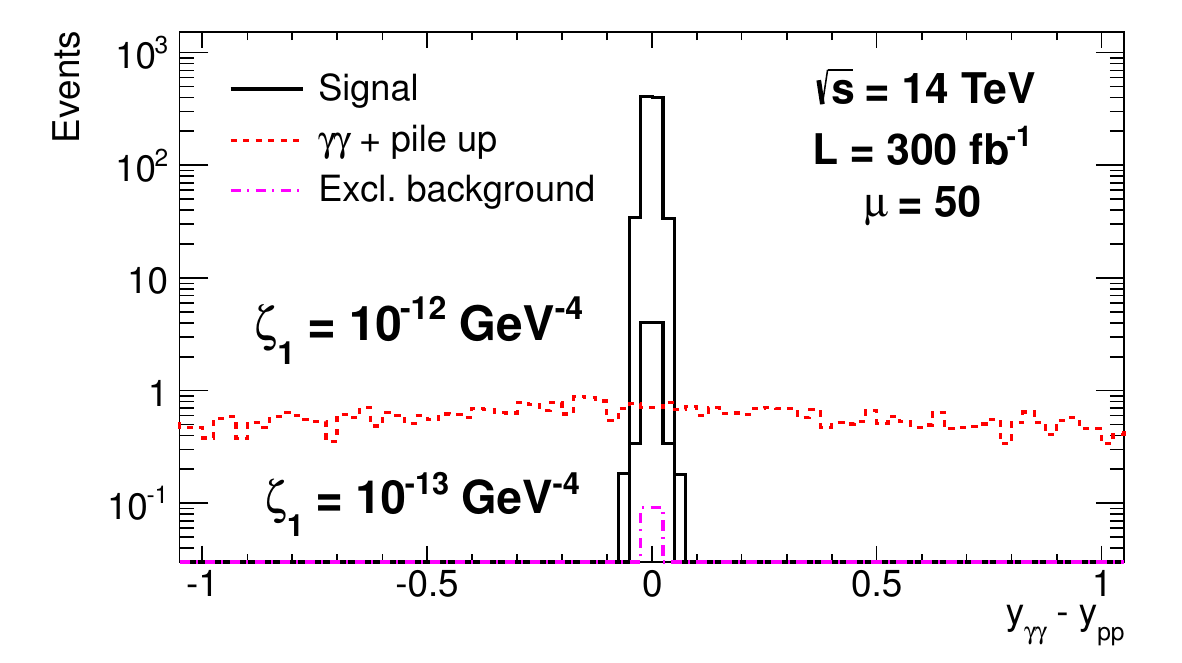}
    \caption{\label{fig:massratio} Di-photon to missing proton mass ratio (left) and rapidity difference (right)
    distributions for signal considering two 
    different coupling values ($10^{-12}$ and $10^{-13}$\gev$^{-4}$, 
    see Eq.~\ref{zetas}) and for 
    backgrounds after requirements on photon $p_T$, di-photon invariant mass, $p_T$ ratio between the two photons and on the angle between the two photons. At least one converted photon is required. The integrated luminosity 
    is 300~fb$^{-1}$ and the average pile up is $\mu=50$.}
    \end{figure*}
    \end{center}
    
    \subsubsection*{Experimental results and outlook}
    
    The $\gamma \gamma \gamma \gamma$ process (Fig.~\ref{fig:pCp-fpmc}) can be probed 
    via the detection of 
    two intact protons in the forward proton detectors and two 
    energetic photons in the corresponding electromagnetic 
    calorimeters.     
    The SM cross section for exclusive di-photon production  is dominated by the
    QED process at high di-photon mass --- and not by gluon exchanges --- and is 
    thus very well known. 
    The photon identification efficiency is 
    expected to be around 75\% for $p_T > 100$~GeV, with jet rejection 
    factors exceeding 4000 even at high pile up ($>$100)~\cite{Aad:2008zzm:ch5}. In addition, about 1\% of the electrons are 
    mis-identified as photons. These numbers are used in the phenomenological
    study presented below.
    As in the previous studies, the anomalous $\gamma \gamma \gamma \gamma$ 
    process has been 
    implemented in the FPMC generator~\cite{FPMC:ch5}.
    The FPMC generator was also used to simulate the background 
    processes giving rise to two intact protons accompanied by two photons, 
    electrons or jets that can mimic the photon signal. These include exclusive 
    SM production of $\gamma \gamma \gamma \gamma$ via lepton and quark boxes and 
    $\gamma\gamma\rightarrow e^{+}e^{-}$. The CEP of $\gamma\gamma$ via two-gluon exchange, not present in FPMC,
    was simulated using ExHuME~\cite{Monk:2005ji}. More details on those Monte Carlo generators can be found in Chapter~\ref{chap:montecarlo}. This series of backgrounds is 
    called
    ``Exclusive" in Table~\ref{tab:event} and Figs.~\ref{fig:mass} and~\ref{fig:massratio}.
    FPMC was also used to produce $\gamma\gamma$, 
    Higgs to $\gamma\gamma$ and di-jet productions via double pomeron exchange 
    (`DPE' in Table~\ref{tab:event} and Fig.~\ref{fig:mass}).
    Such backgrounds tend to be softer than the signal and can be suppressed with 
    requirements on the transverse momenta of the photons and the di-photon invariant 
    mass. 
    In addition, the final-state photons for the signal are typically back-to-back 
    and have 
    about the same transverse momenta. Requiring a large azimuthal 
    angle $|\Delta \phi| > \pi -0.01$ between the two photons and a 
    ratio $p_{T,2} / p_{T,1} > 0.95$ greatly reduces the contribution of 
    non-exclusive processes.
    Additional background processes include the quark and gluon-initiated 
    production of two photons, two jets and  Drell-Yan processes leading to two 
    electrons. The two intact 
    protons arise from pile up interactions (these backgrounds are called 
    $\gamma\gamma$ + pile up and e$^{+}$e$^{-}$, di-jet + pile up in 
    Table~\ref{tab:event}). These events were produced using HERWIG~\cite{Corcella:2002jc} and
    PYTHIA~\cite{Sjostrand:2007gs:ch5}. 
    The pile up background is further suppressed by requiring the proton 
    missing mass to match the di-photon invariant mass within 
    the expected 
    resolution and the di-photon
    system rapidity and the rapidity of the two protons to be similar.

     \begin{table}

\begin{center}
\begin{tabular}{|c||c|c||c|c||c|}
\hline
Luminosity & 300~\fbi & 300~\fbi & 300~\fbi & 300~\fbi & 3000~\fbi \\
\hline
 pile up ($\mu$) & 50 & 50 & 50 & 50 & 200 \\
\hline
\hline
coupling & $\ge$~1 conv. $\gamma$ & $\ge$~1 conv. $\gamma$ & all $\gamma$ & all $\gamma$  & all $\gamma$\\
(GeV$^{-4}$) & 5 $\sigma$ & 95\% CL & 5 $\sigma$ & 95\% CL & 95\% CL \\

\hline
$\zeta_1$~f.f.   &  $8\cdot10^{-14}$   & $5\cdot10^{-14}$   & $4.5\cdot 10^{-14}$ & $3\cdot 10^{-14}$ & $2.5\cdot10^{-14}$ \\
$\zeta_1$~no f.f.&  $2.5\cdot10^{-14}$ & $1.5\cdot10^{-14}$ & $1.5\cdot10^{-14}$ & $9\cdot10^{-15}$  & $7\cdot10^{-15}$\\
\hline
$\zeta_2$~f.f.   &  $2\cdot10^{-13}$   & $1\cdot10^{-13}$   & $9\cdot10^{-14}$ & $6\cdot10^{-14}$  & $4.5\cdot10^{-14}$ \\
$\zeta_2$~no f.f.&  $5\cdot10^{-14}$   & $4\cdot10^{-14}$   & $3\cdot10^{-14}$& $2\cdot10^{-14}$  & $1.5\cdot10^{-14}$ \\
\hline

\end{tabular}
\end{center}

\caption{5\,$\sigma$ discovery and 95\% CL exclusion limits on $\zeta_1$ and $\zeta_2$
couplings in~\gev$^{-4}$ (see Eq.~\ref{zetas}) with
and without form factor (f.f.), requesting at least one converted photon ($\ge$~1 conv. $\gamma$) or
not (all $\gamma$). All sensitivities are given for 300 fb$^{-1}$
and $\mu=50$  pile up events (medium luminosity LHC) except for the numbers of the last column which are given for 3000
fb$^{-1}$ and $\mu=200$  pile up events (high luminosity LHC). }
\label{sensitivities}

\end{table}
     
    The number of expected signal and background events passing respective 
    selections is shown in 
    Table~\ref{tab:event} for an integrated luminosity of 300 fb$^{-1}$\ 
    for a center-of-mass energy of 14\,TeV.
    Exploiting the full event kinematics with the forward proton detectors 
    allows the background to be suppressed with a signal selection 
    efficiency after the acceptance cuts exceeding 70\%. Tagging the protons
    is essential to suppress the $\gamma \gamma$ + pile up events.
    Further background reduction is possible by requiring the photons 
    and the protons to originate from the same vertex, providing an additional 
    rejection 
    factor of 40 for 50 pile up interactions.
    A similar study 
    at a higher pile up of 200 was performed 
    and led to a very small background ($<$ 5 expected background events for 300 fb$^{-1}$ without
re-optimizing the event selection)
.
    The sensitivities
    on photon quartic anomalous couplings are given in Table~\ref{sensitivities}.
    The sensitivity extends up to $7\cdot10^{-15}$ GeV$^{-4}$, allowing  the models of new
    physics described above to be probed further. 
    
\begin{figure}
\begin{center}
\includegraphics[width=0.49\linewidth]{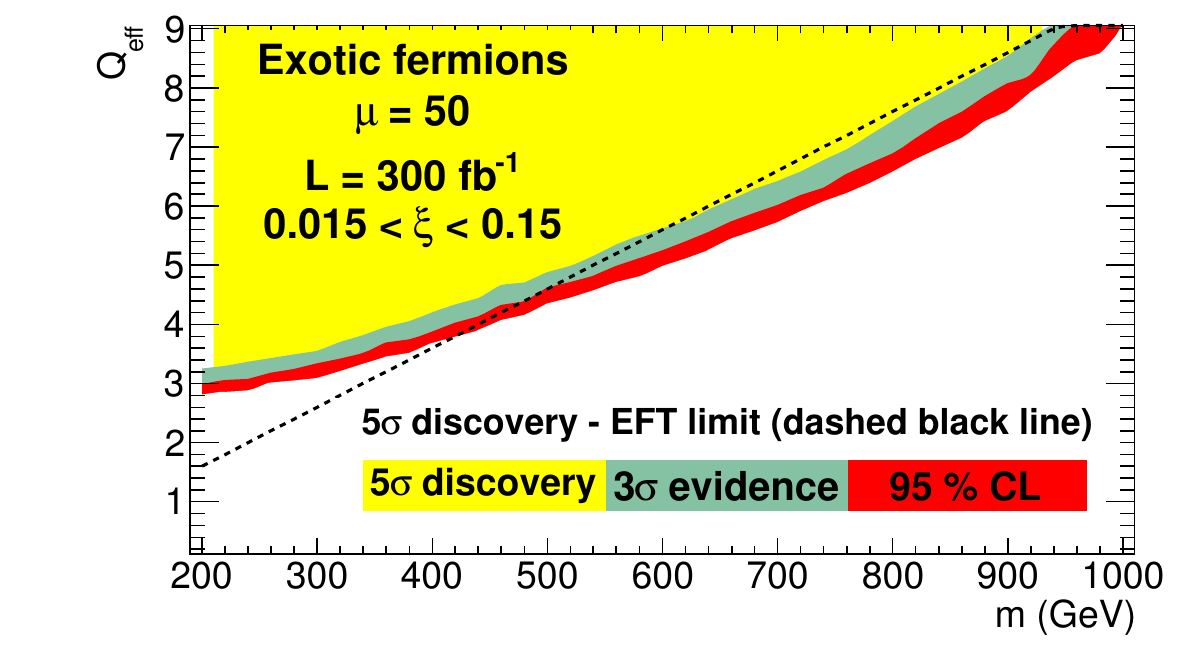}
\includegraphics[width=0.49\linewidth]{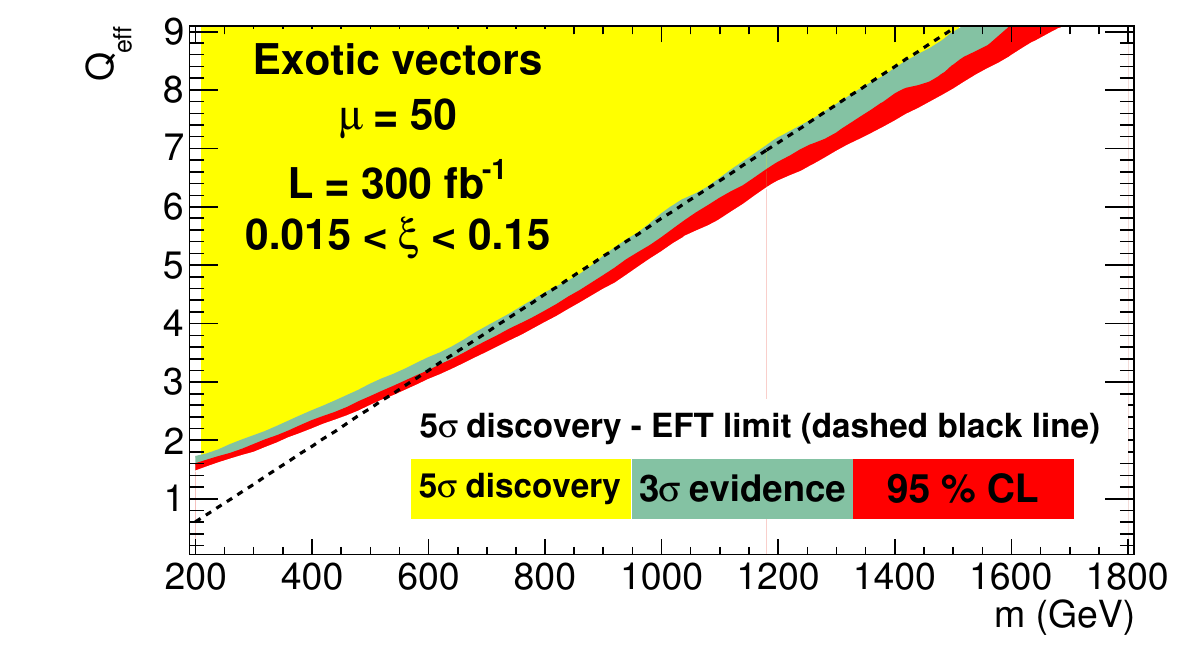}
\end{center}
\caption{Exclusion plane in terms of mass and effective charge of generic fermions and vectors. in the case of no requirement of photon conversion at the analysis stage and full integrated luminosity of the LHC (300~\fbi, $\mu=50$).}
\label{fig:mqplane}
\end{figure}

    A more recent study has been performed using a full amplitude calculation in the case of BSM contributions from generic new heavy charged particles~\cite{Fichet:2014uka}. Here, the cross section is fully determined by the particle spin (vectors or fermions), mass, charge and multiplicity, which allows an effective charge of the new particles, $Q_{\rm eff}=Q\cdot N^{1/4}$, to be defined. In Fig.~\ref{fig:mqplane} the mass-effective charge exclusion plane with 300 fb$^{-1}$ at the 14 TeV LHC with a comparison with the effective field theory study results is shown, while Tab.~\ref{fullamp_values} gives the 5\,$\sigma$ discovery limits for various mass scenarios. More details can be found in~\cite{Fichet:2014uka}. 

\begin{table}

\begin{center}
\begin{tabular}{|c||c|c|c|c|c|}
\hline
Mass (GeV) & 300 & 600 & 900 & 1200 & 1500 \\
\hline
$Q_{\rm eff}$ (vector)  & 2.2 & 3.4 & 4.9 & 7.2 & 8.9 \\
\hline
$Q_{\rm eff}$ (fermion) & 3.6 & 5.7 & 8.6 & - & - \\
\hline
\end{tabular}
\end{center}

\caption{5\,$\sigma$ discovery limits on the effective charge of new generic charged fermions and vectors for various masses scenarios in the case of no requirement of photon conversion at the analysis stage and full integrated luminosity at the LHC (300~\fbi, $\mu=50$).}
\label{fullamp_values}

\end{table}

\subsection{Anomalous gauge couplings: $\gamma \gamma WW$ and $\gamma \gamma  ZZ$}\label{sec:anomwwzz}
\subsubsection*{Motivation and theory}

    \begin{figure}
    \begin{center}
    \includegraphics[width=0.49\textwidth]{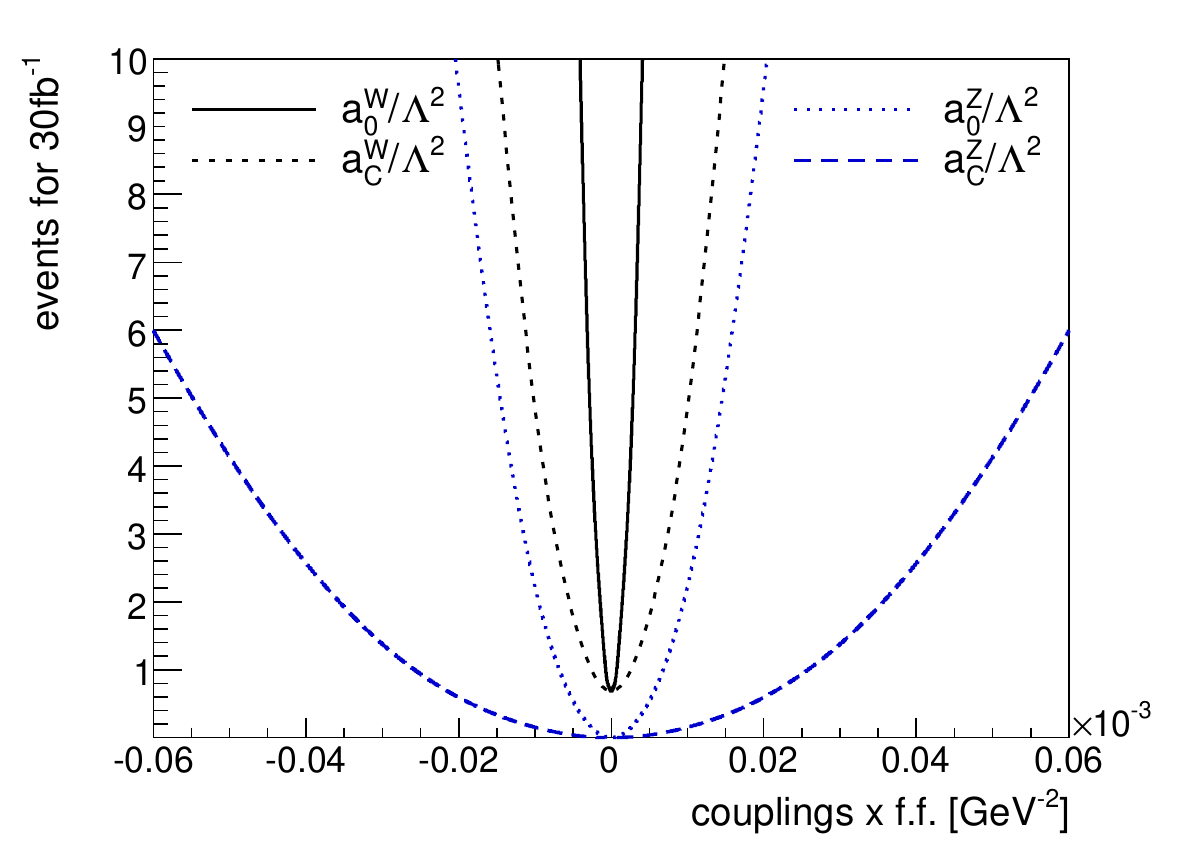}
    \includegraphics[width=0.49\textwidth]{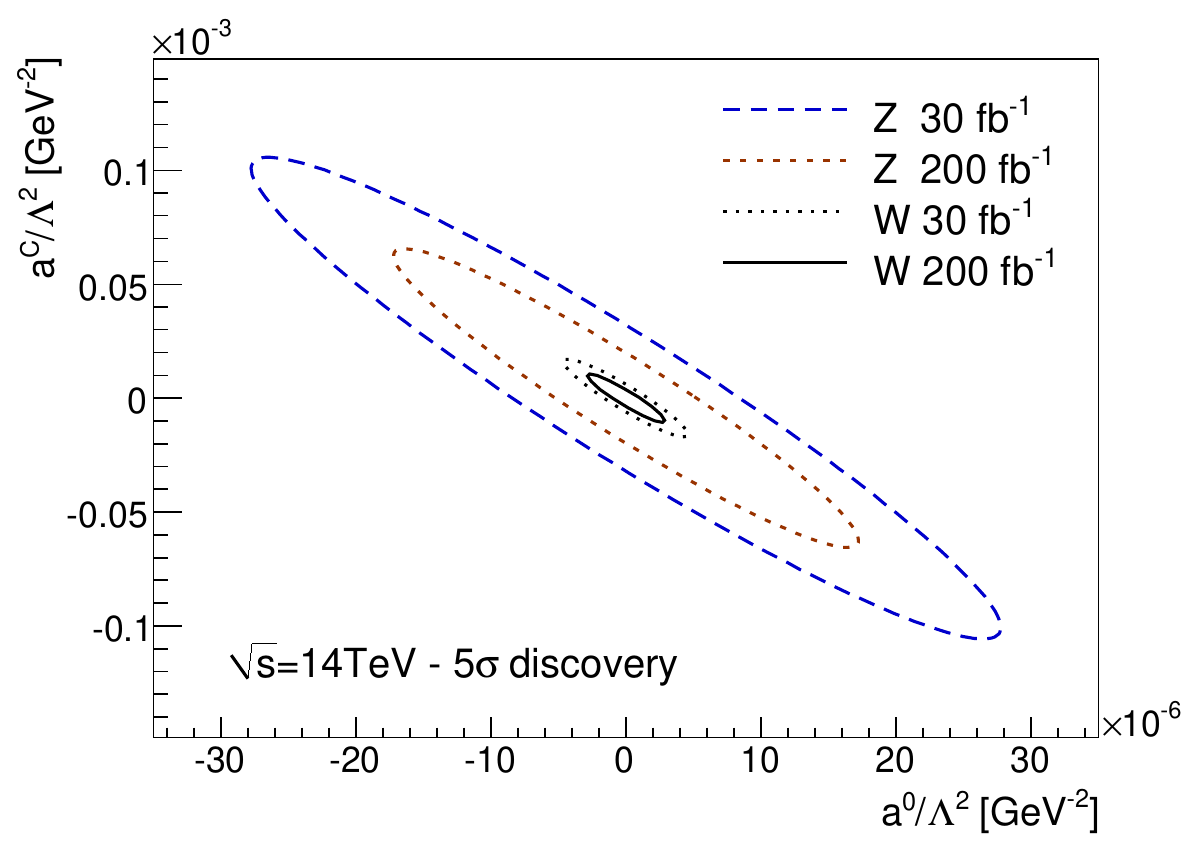}
    \caption{(left) Number of events for signal due to different values of 
    anomalous couplings after all cuts (see text) for a luminosity of
    30 fb$^{-1}$ and (right) $5\sigma$ discovery contours for all the $WW$ and $ZZ$ quartic 
    couplings at $\sqrt{s}=14$ TeV for a luminosity of 30 fb$^{-1}$ and 200
    fb$^{-1}$.}
    \label{fig:tgc}
    \end{center}
    \end{figure}

As well as altering the $4\gamma$ coupling from SM expectations, BSM physics such as the scenarios described in the previous section can also effect the couplings to electroweak bosons, namely $\gamma\gamma WW$ and $\gamma\gamma ZZ$.
    The parameterization of~\cite{Belanger:1992qh} can be adopted for the quartic anomalous gauge couplings. 
    The cuts to select the anomalous $WW$ events are similar to the
    ones mentioned in the Section~\ref{sec:excbos}, namely $0.0015<\xi<0.15$ for the
    tagged protons corresponding to the AFP or CT-PPS detector at 210 and 420 m, $\met>$ 20 GeV, 
    $\Delta \phi<3.13$ between the two leptons. In
    addition, a cut on the $p_T$ of the leading lepton $p_T>160$ GeV and on the
    diffractive mass $W>800$ GeV are requested since anomalous coupling events
    appear at high mass.  
    After these requirements, about 0.7 background
    events for a signal of 17 events are expected, for a luminosity of 30 fb$^{-1}$, if the anomalous coupling is about
    four orders of magnitude lower than the present LEP limit~\cite{Abbiendi:2004bf} ($|a_0^W / \Lambda^2| =
    5.4\cdot10^{-6}$)  or two orders of magnitude lower with respect to the D0 and CDF
    limits~\cite{Chatrchyan:2013foa,Abazov:2013opa}, or CMS measurement at $\sqrt{s}=7$ TeV~\cite{Chatrchyan:2013foa} mentioned above. The strategy to select anomalous 
    coupling $ZZ$ events is similar and the presence of three leptons or two like sign leptons are 
    requested. 
    Table \ref{tab:quartic_anom_coupling} summarizes the reach on anomalous couplings at the LHC for
    luminosities of 30 and 200 fb$^{-1}$ compared to the present OPAL limits from
    the LEP accelerator~\cite{Abbiendi:2004bf}.
    Fig.~\ref{fig:tgc} shows the expected number of signal events as a function
    of the anomalous coupling value (left) and the 5$\sigma$ discovery contours for all $WW$ and $ZZ$
    anomalous couplings for 30 and 200 fb$^{-1}$ (right); sensitivity to values expected in 
    extra dimension models~\cite{Fichet:2013gsa,Fichet:2014uka} is demonstrated. Proton tagging is the only method at present to test 
    quartic anomalous couplings down to such small values.

Finally, we note that the LHC sensitivity to triple gauge anomalous couplings at the LHC has been studied in~\cite{Kepka:2008yx}, however in this case the limits obtained in the context of proton tagging based analyses are comparable to the inclusive ones, and therefore do not appear to be particularly competitive. 

    \begin{table}[h]
    \begin{center}
       \begin{tabular}{|c||c|c|c|}
        \hline
        %\raisebox{-1.5ex}[0pt][0pt] 
        Couplings & 
        OPAL limits & 
        \multicolumn{2}{c|}{Sensitivity @ $\mathcal{L} = 30$ (200) fb$^{-1}$} \\
        &  \small[GeV$^{-2}$] & 5$\sigma$ & 95\% CL \\ 
        \hline
        $a_0^W/\Lambda^2$ & [-0.020, 0.020] & $5.4\cdot10^{-6}$ & $2.6\cdot 10^{-6}$\\
                          &                 & ($2.7\cdot 10^{-6}$) & ($1.4\cdot 10^{-6}$)\\ \hline               
        $a_C^W/\Lambda^2$ & [-0.052, 0.037] & $2.0\cdot 10^{-5}$ & $9.4\cdot 10^{-6}$\\
                          &                 & ($9.6\cdot 10^{-6}$) & ($5.2\cdot 10^{-6}$)\\ \hline               
        $a_0^Z/\Lambda^2$ & [-0.007, 0.023] & $1.4\cdot 10^{-5}$ & $6.4\cdot 10^{-6}$\\
                          &                 & ($5.5 \cdot10^{-6}$) & ($2.5 \cdot10^{-6}$)\\ \hline               
        $a_C^Z/\Lambda^2$ & [-0.029, 0.029] & $5.2\cdot 10^{-5}$ & $2.4  \cdot10^{-5}$\\
                          &                 & ($2.0\cdot 10^{-5}$) & ($9.2\cdot 10^{-6}$)\\ \hline               
        \hline
    \end{tabular}
    \end{center}
    \caption{Reach on anomalous couplings obtained in $\gamma$ induced processes
    after tagging the protons in AFP or CT-PPS compared to the present OPAL limits. The $5\sigma$ discovery and 95\%
    C.L. limits are given for a luminosity of 30 and 200 fb$^{-1}$~\cite{deFavereaudeJeneret:2009db}.} 
    \label{tab:quartic_anom_coupling}
    \end{table}
    
    \subsubsection*{Experimental results and outlook}
%\label{sec:physicsProc}

Measurements of two-photon production of 
W boson pairs, in the process $pp\rightarrow p \mathrm{W^+W^-} p$, were performed 
in the $\mu^\pm \mathrm{e}^\mp$ final state, 
using 5.05~fb$^{-1}$ of data collected in proton-proton collisions at $\sqrt{s}=7$~TeV 
with the CMS detector at the LHC in 2011, but without proton tagging~\cite{Chatrchyan:2013foa}.
Model-independent upper limits were extracted and compared to predictions involving anomalous quartic gauge couplings (AQGCs).
This resulted in limits of $|a^{W}_{0}/\Lambda^{2}| < 0.00015~\mathrm{GeV}^{-2}$ and $|a^{W}_{C}/\Lambda^{2}| < 0.0005~\mathrm{GeV}^{-2}$ at 95$\%$ CL 
on the dimension-six AQGC operators, including a dipole form factor with $\Lambda_{\mathrm{cutoff}} = 500$~GeV to preserve unitarity.

The prospects of sensitivities to quartic $\gamma\gamma WW$ couplings, using a full detector simulation, have been studied in detail within the context of the AFP and CT-PPS detectors, following the phenomenological study presented above. As these studies and anomalous coupling searches represent an important part of the high--luminosity AFP and CT--PPS programs, these are described in more detail in Section~\ref{sec:wwzzdetailed}.

\subsection{Anomalous $\gamma\gamma WW$ couplings: detailed studies}
\label{sec:wwzzdetailed}

As discussed in Section~\ref{sec:anomwwzz}, the prospects of sensitivities to quartic $\gamma\gamma WW$ couplings, using a full detector simulation, have been studied in detail within the context of AFP and CT-PPS. The first study was presented in the Letter of Intent of Phase-I ATLAS upgrade~\cite{ATLAS_LOI_PhaseI}, confirming that 
results from phenomenological studies can be obtained with realistic detector setup. 
As these studies and anomalous coupling searches represent an important part of the high--luminosity AFP and CT--PPS programs, these are described in some detail below.
This ATLAS study is summarized first, followed by a more detailed discussion of the more recent CMS analysis,
performed for this report, and based on the experimental techniques developed in~\cite{Chatrchyan:2013foa}. While these studies differ in selections, they give 
similar overall obtainable sensitivity. Before this, some more general aspects of the studies and processes under consideration are described.

With the integrated luminosity expected to be collected during Run~II and with the AFP and CT-PPS detectors,
the experimental reach on the anomalous quartic coupling $\gamma\gamma WW$ can be extended by several orders of magnitude with respect to the best limits obtained so far.
In the process $pp\rightarrow p \mathrm{WW} p$,
both forward-scattered protons are detected in the CT-PPS, depending on the acceptance on the mass of the WW central system produced.
The AFP and CT-PPS detectors were assumed to be operating at 10 and 15$~\sigma$ from the beam, respectively.
The process is characterized by a primary vertex from the two leptons $\ell^\pm\ell '^\mp$ (where $\ell=e,\mu$) from the W boson pair decays, no other tracks,
a large transverse momentum of the dilepton system, $p_\mathrm{T}(\ell^\pm\ell '^\mp)$, and a large invariant mass, $M(\ell^\pm\ell '^\mp$).
A simulated sample of
SM exclusive $pp\rightarrow p \mathrm{WW} p$ signal events is used,
in conjunction with samples in which anomalous quartic gauge couplings (AQGC) are assumed. 
   
\begin{figure*}[htbp]
\centering
\includegraphics[width=0.48\textwidth]{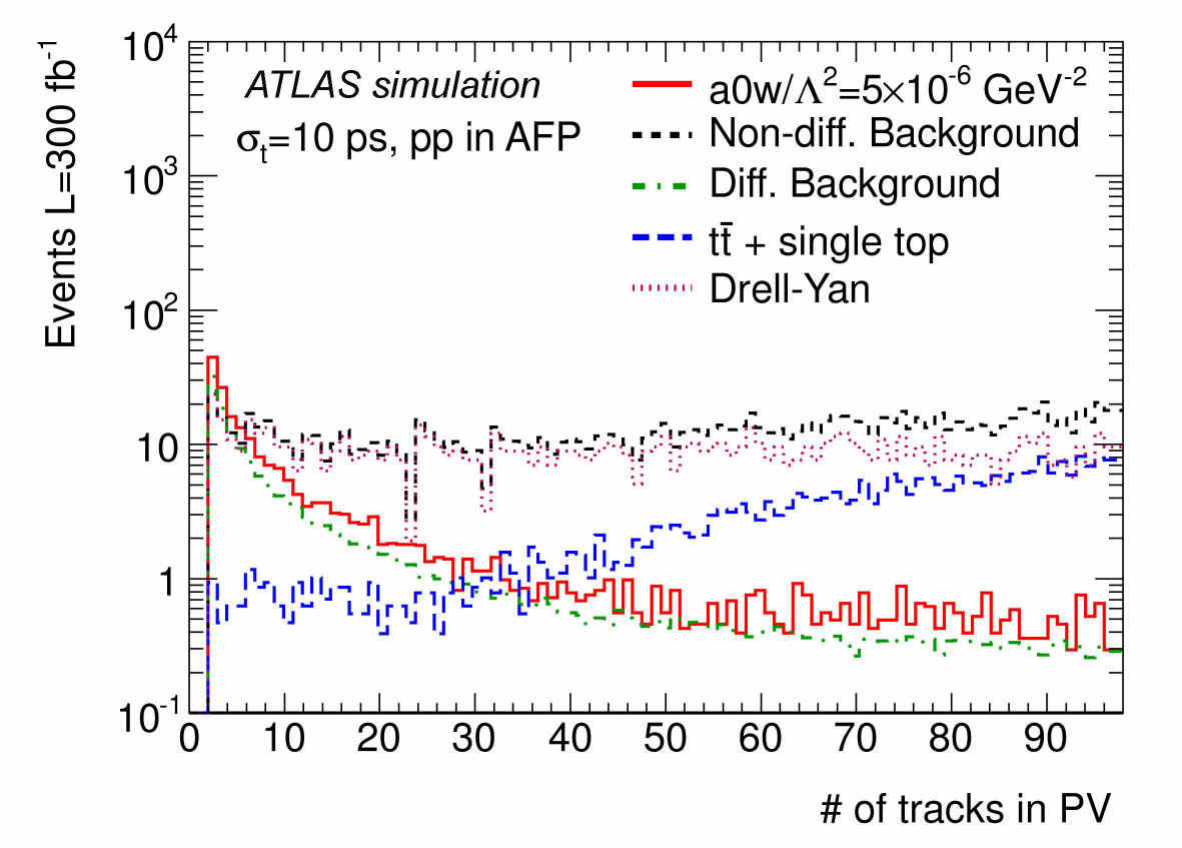}
\includegraphics[width=0.48\textwidth]{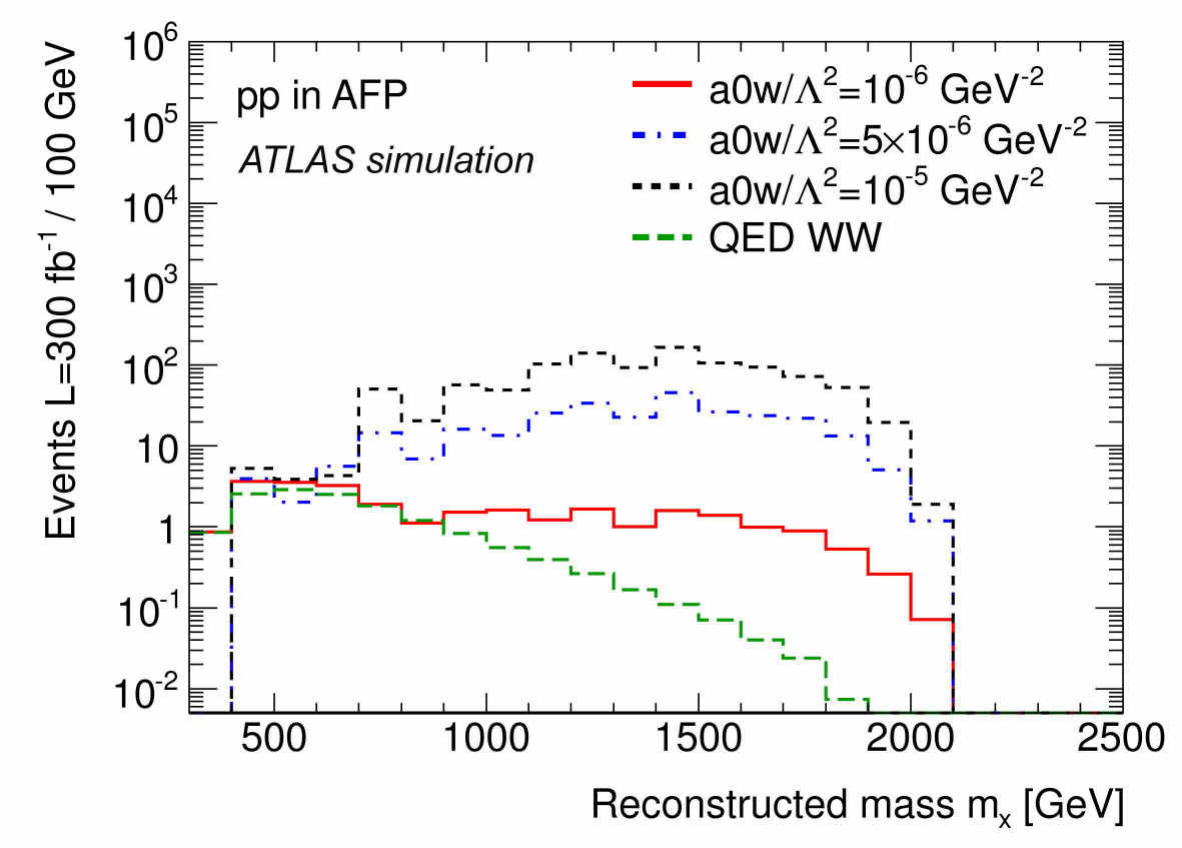}
\caption{Distribution of the number of tracks fitted to the primary vertex (left) and of the reconstructed missing mass in the forward detectors (right) for signal and the different backgrounds. Those results were obtained using a full simulation of the ATLAS detector.}
\label{fig:afp_fullsim_aww}
\end{figure*}

 Considering first the ATLAS study, the full list of background processes 
    used for the ATLAS measurement of Standard Model $WW$ cross section was
    simulated, namely $t \bar{t}$, $WW$, $WZ$, $ZZ$, $W+$jets, Drell-Yan and 
    single top events. In addition, processes having two forward protons denoted as `diffractive backgrounds' were simulated. They 
    include two-photon and double pomeron exchange production of dileptons and $WW$. In addition, single diffractive production of dileptons, $W$ and $WW$
    was simulated. Inclusive and single diffractive events have zero or  one forward proton in the final state, however, due to significant amount of multiple
    proton-proton interaction rate in Run--II, similar final state signatures may emerge as for the signal due to coincidences with soft diffractive events. 
    The requirement of the presence of at least one
    proton on each side of AFP  within a time window of 10 ps allows  the 
    background to be reduced by a factor of about 200 (50) for $\mu =$ 23 (46), by matching the vertex position reconstructed inside ATLAS with that calculated from time arrival of the protons.
    The $p_T$ of the leading 
    lepton originating from the
    leptonic decay of the $W$ bosons is required to be 
    $p_T >$ 150 GeV, and that of the next-to-leading
    lepton $p_T>$ 20 GeV. An additional requirement of the dilepton mass to 
    be above 300 GeV allows
    most of the diboson events to be removed. Since only leptonic decays of the 
    W bosons are considered, in addition less than 3 tracks associated 
    to the primary vertex are required, which allows  a large fraction of the
    non-diffractive backgrounds (e.g. $t \bar{t}$, diboson
    productions, $W+$jet, etc.) to be rejected. This is illustrated in Fig.~\ref{fig:afp_fullsim_aww} (left) where signal events clearly peak at low track multiplicities. 
    The remaining Drell-Yan and
    QED backgrounds are suppressed by requiring the difference in azimuthal angle between the
    two leptons to satisfy $\Delta \phi <$ 3.1. 
    
Fig.~\ref{fig:afp_fullsim_aww} (right) displays the reconstructed missing mass $m_{X}$ in AFP for the
irreducible QED WW background and for signal events produced with
three values of anomalous coupling. Anomalous coupling events are seen
to appear at high $W$-pair invariant masses. The requirement that $m_{X}>800$\,GeV,
reconstructed using the two scattered protons
in the AFP, allows the backgrounds to be rejected by an additional factor of four.

    With the above event selection, a similar
    sensitivity with respect to fast simulation without pile up was obtained. The 95\% C.L. limits are $a^{W}_0/\Lambda^2 = 2.4\cdot10^{-6} (1.3\cdot10^{-6}) {\rm GeV}^{-2}$
    for the assumed collected integrated luminosity of 40 (300)\,fb$^{-1}$ with $\mu=$23(46) mean number of interactions per bunch crossing.

Considering now the CMS-TOTEM study, experience from the CMS measurement~\cite{Chatrchyan:2013foa} discussed above was used and only the significant backgrounds have been considered. 
The dominant SM backgrounds come from inclusive $\mathrm{W^+W^-}$ and exclusive $\tau^+\tau^-$ pair production.
Drell-Yan (DY) production of $\tau^+\tau^-$, where one $\tau$ decays via the electron channel and
the other one via the muon channel, in combination with pile up protons, may lead to a similar event configuration
as the exclusive $WW$ signal.
A selection cut on the transverse momentum of the dilepton pair $p_T(\ell^\pm\ell '^\mp)>30$~GeV 
was used in~\cite{Chatrchyan:2013foa} to reject the DY background almost completely. 
A further rejection of the other SM backgrounds, and a reduction of signal events by approximately 20\% is expected.
In this study, only the e$\mu$ final state is selected. 
The simulated samples for the signal process are compared to the SM background expectations, and 
the tails of the $M_X$ distribution ($M_X=\sqrt{s\cdot \xi_1\cdot \xi_2} >$ 1~TeV), 
where the SM $\gamma\gamma\rightarrow \mathrm{W^+W^-}$ contribution 
is expected to be small, are investigated to look for AQGCs.
Events are selected by requiring two central ($|\eta|<2.4$) leptons
with a minimum transverse momentum $p_\mathrm{T}>20$~GeV.
In order to reduce the contamination from the W+jet (or other non-prompt lepton) background, 
``tight'' lepton identification criteria are imposed (as in~\cite{Chatrchyan:2013foa}).
Leptons are also required to have charges of opposite sign and to come from the same primary vertex. 
For signal events (either SM or AQGC), the scattered protons are in the region covered by CT-PPS, and the presence 
of hits in both tracking and timing detectors is therefore required: a large background suppression is expected.
A source of inefficiency comes from high detector occupancy, as hits are required not to overlap in the timing detector cells.

\begin{figure*}[htbp]
\centering
\includegraphics[width=0.4\textwidth]{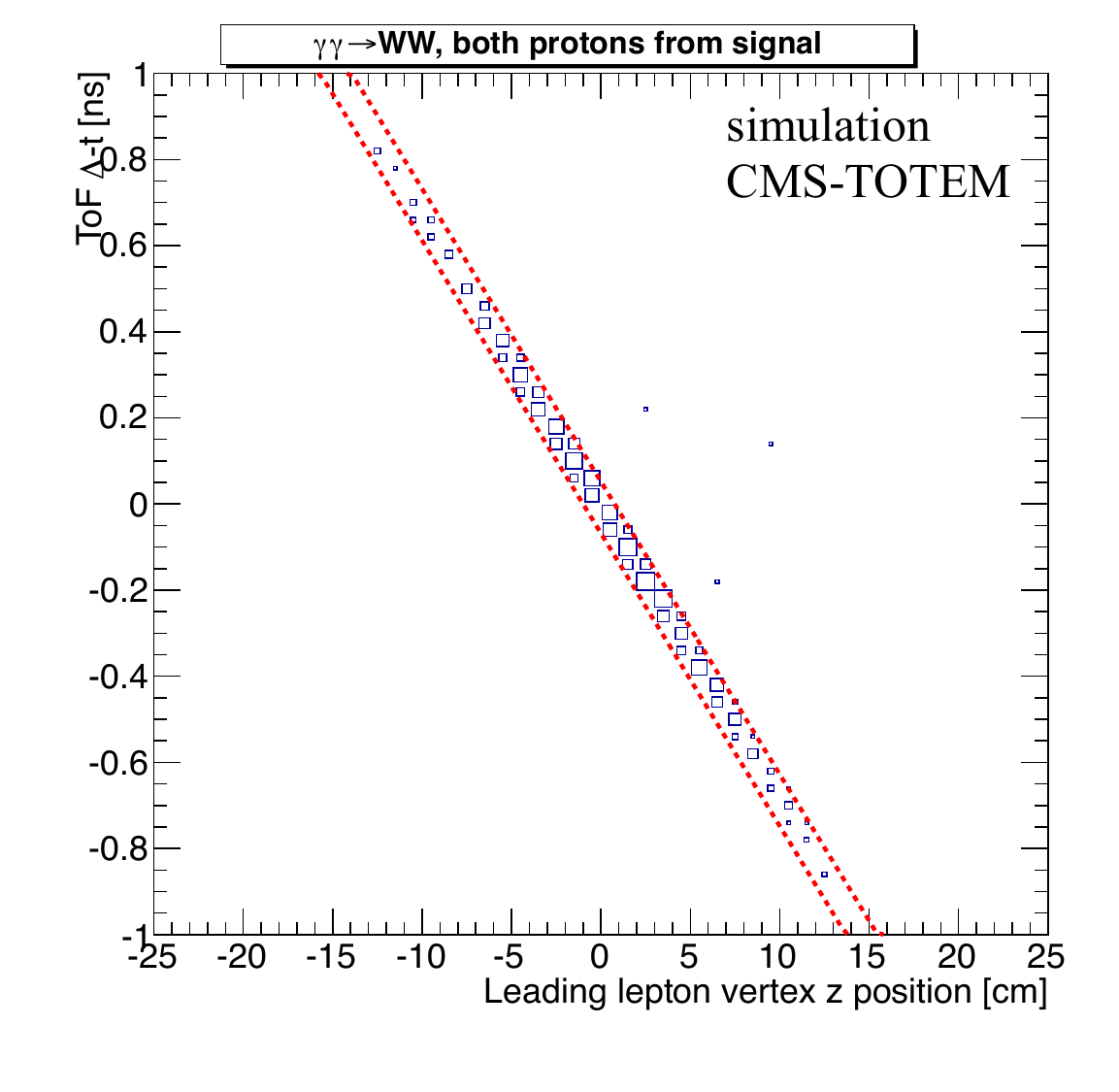}\qquad\qquad
\includegraphics[width=0.4\textwidth]{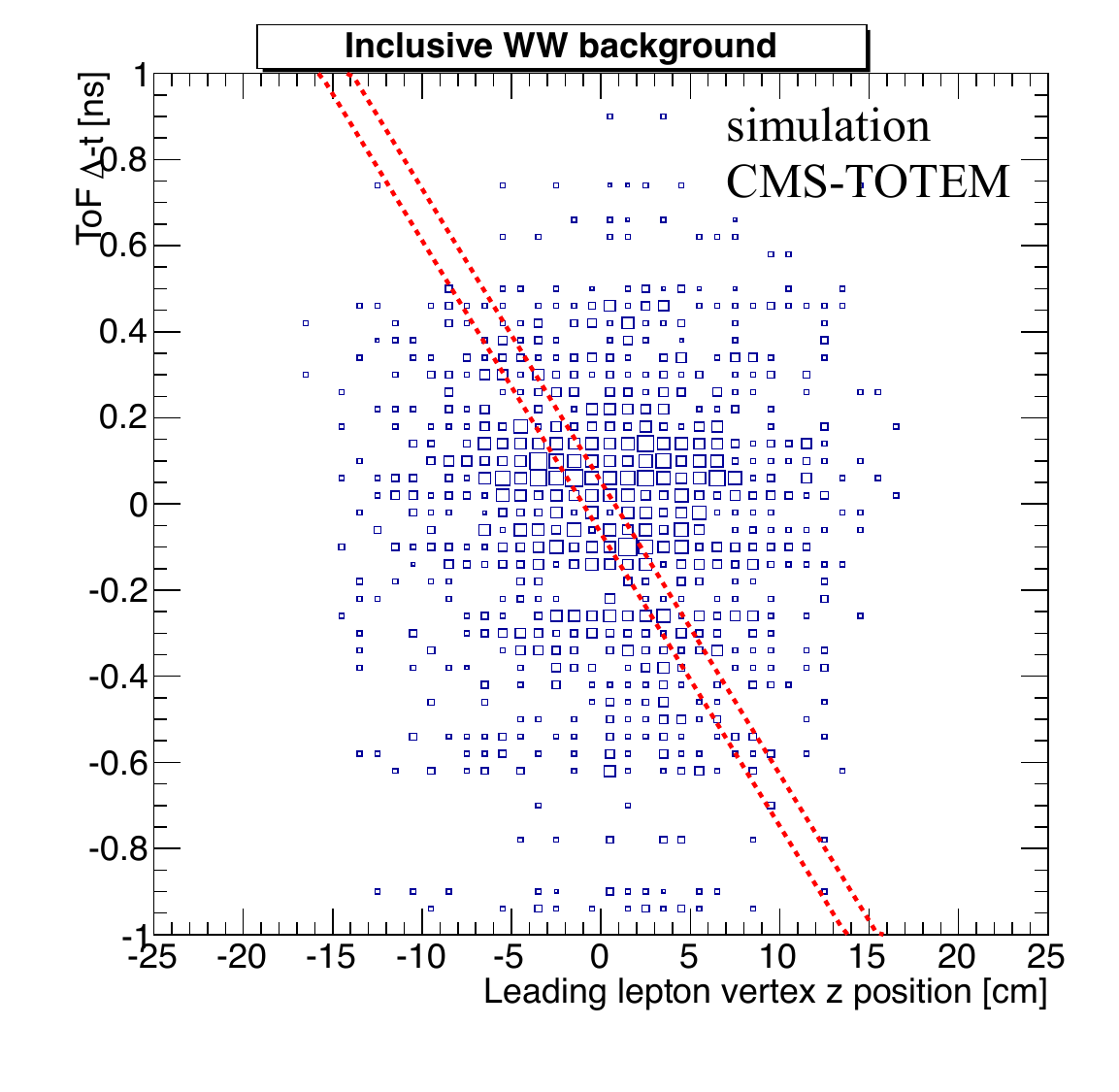}
\caption{
Time-of-flight difference between the two leading protons arriving at the CT-PPS detector location on opposite sides of the IP
as a function of the z-vertex position of the leading central lepton for exclusive signal (left) and background inclusive WW (right) events. 
A timing resolutions of 10~ps is assumed.
Distributions are shown for events where both leading protons are within the CT-PPS detector acceptance
(after selecting the closest match of the vertices of the dilepton system and of the leading protons), 
and before the time-of-flight difference requirement.
The dotted lines show an ideal window retaining close to 100\% of signal events.
An arbitrary normalization is used in the distributions.
}
\label{fig:timing}
\end{figure*}

Signal events from SM exclusive WW events are correlated in time with the leading protons detected in the CT-PPS detectors, 
whereas inclusive WW events -superimposed with additional pile up events- are not.
Therefore, the information of the proton time-of-flight arrival at the CT-PPS detector location can be used as an additional background rejection.
After requiring the coincidence of hits in both tracking and timing detector stations, the time-of-flight 
difference between the two leading protons arriving at the CT-PPS detector location on opposite sides of the IP
is shown in Fig.~\ref{fig:timing} as a function of the z-vertex position of the leading central lepton, for signal (left) and background (right) events.
For each event, if there is more than one $pp$ combination in the CT-PPS detector, only the one with the closest match 
between the time-of-flight $\Delta t$ and the lepton vertex position in $z$ is kept (``vertex matching'' in Table~\ref{tab:yields_ww_xsec}). 
The background is shown for inclusive WW events in coincidence with pile up events.
Distributions corresponding to a timing resolution of 10~ps are shown for signal and background events.

\begin{figure*}[ht!]
\centering
\includegraphics[width=0.4\textwidth]{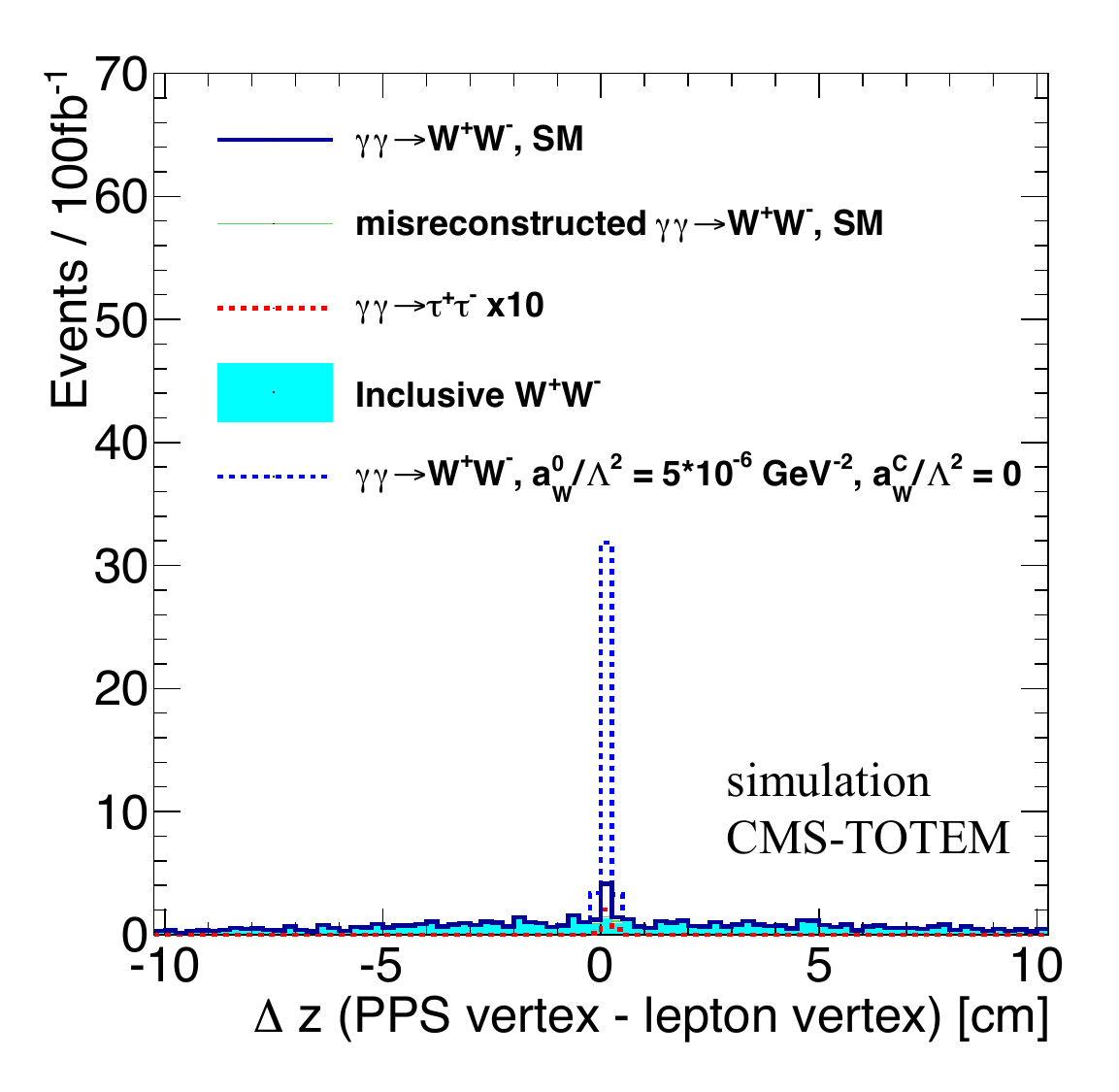}\qquad \qquad
\includegraphics[width=0.4\textwidth]{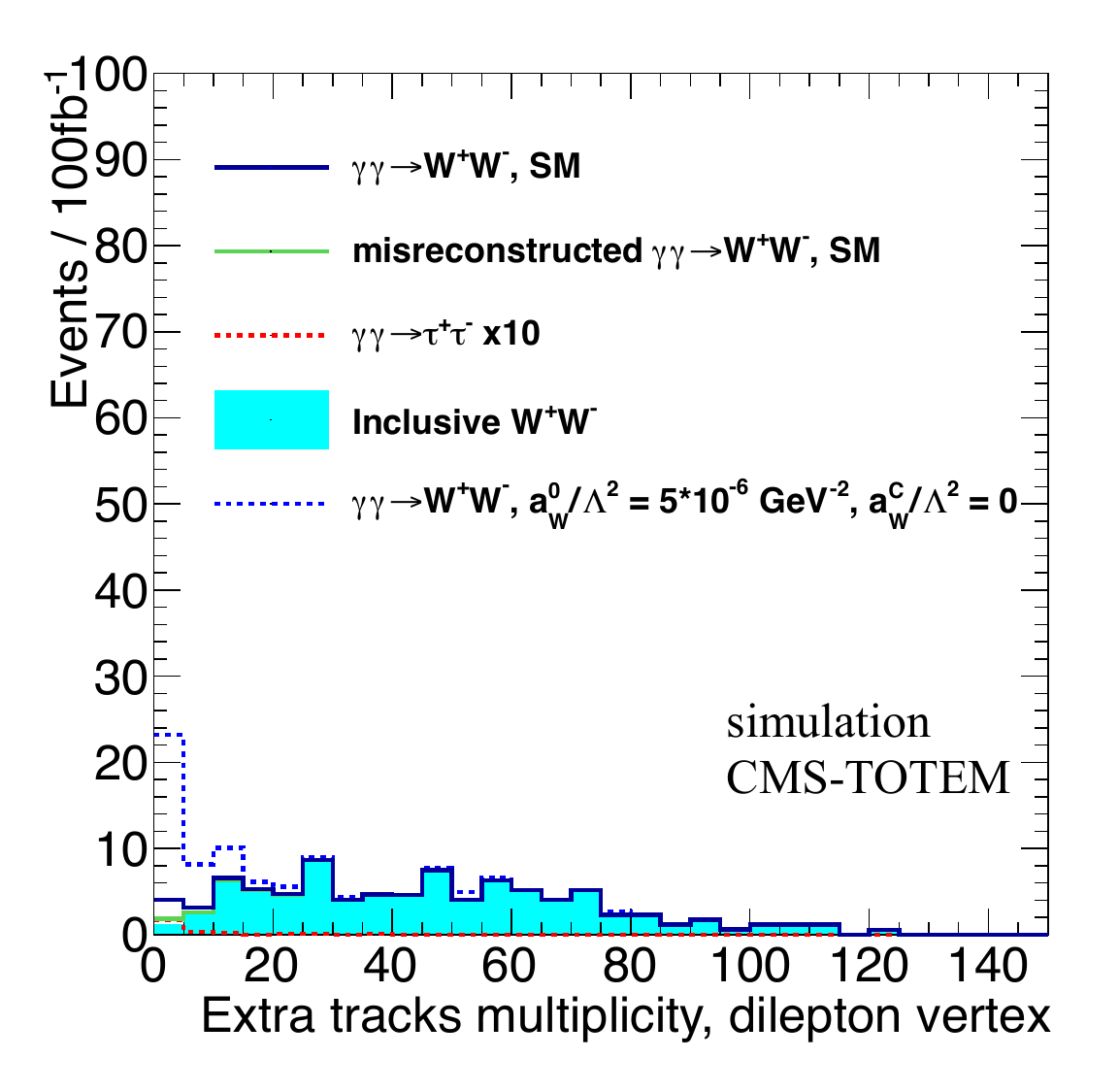}
\caption{
{\it Left:} Distance along the z-axis (in cm) of the vertex positions as measured from the CT-PPS timing detectors and from the leading lepton in the central detector,
$\Delta z= z_\mathrm{PPS} - z_\mathrm{lead~lepton}$. Distribution is shown after all cuts, except for the time-of-flight information requirement.
{\it Right:}
Number of extra tracks associated to the dilepton vertex for exclusive (WW and $\tau\tau$) and inclusive (WW) events.
Distribution is shown after all cuts, except for the track multiplicity cut.
Distributions are shown for SM exclusive WW/$\tau\tau$, inclusive WW events, and AQGC exclusive WW events, and a timing resolutions of 10~ps is assumed.
Event yields are normalized to an integrated luminosity of 100fb$^{-1}$ and a timing resolution of 10 ps. 
Histograms are stacked, except for that of the exclusive $\tau\tau$ background, which is not stacked and is multiplied by a factor of 10.
}
\label{fig:timing_distrib}
\end{figure*}

The distance (in z) of the vertex positions measured from the CT-PPS timing detectors and from the leading lepton in the central detector,
$\Delta z= z_\mathrm{PPS} - z_\mathrm{lead~lep}$,
is shown in Fig.~\ref{fig:timing_distrib} (left)
for SM exclusive WW/$\tau\tau$ and inclusive WW events, and for AQGC exclusive WW events,
after all cuts, except for the time-of-flight information requirement.
Time-of-flight requirements may help reduce the inclusive WW background by a factor of 10~(5), for a timing resolution of 10~ps (30~ps).
The track multiplicity associated to the dilepton vertex after the timing selection cuts is shown in Fig.~\ref{fig:timing_distrib} (right)
for SM signal and backgrounds, after all cuts, except for the track multiplicity cut.
The number of extra tracks associated to the dilepton vertex is significantly larger for inclusive WW events, 
and a selection of $N_\mathrm{tracks}<10$ is expected to suppress the inclusive background by 90\%, 
while retaining approximately 90\% of the exclusive events.
A signal-to-background ratio of 1:1 can be achieved for the standard model production after applying a cut on the maximum number of reconstructed tracks 
in the central detector (except the two selected leptons).

\begin{figure*}[htp!]
\centering
\includegraphics[width=0.4\textwidth]{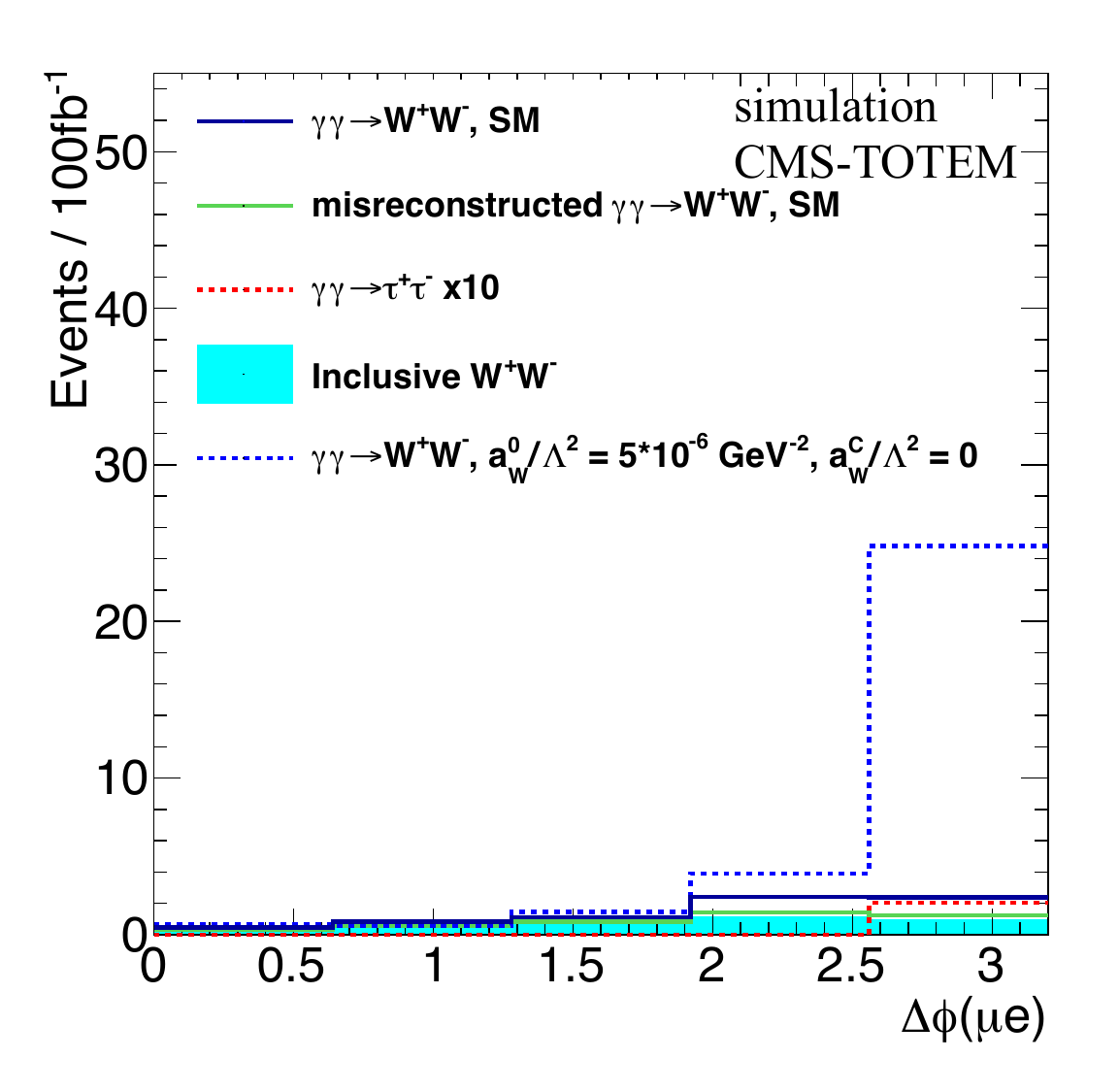} \qquad \qquad
\includegraphics[width=0.4\textwidth]{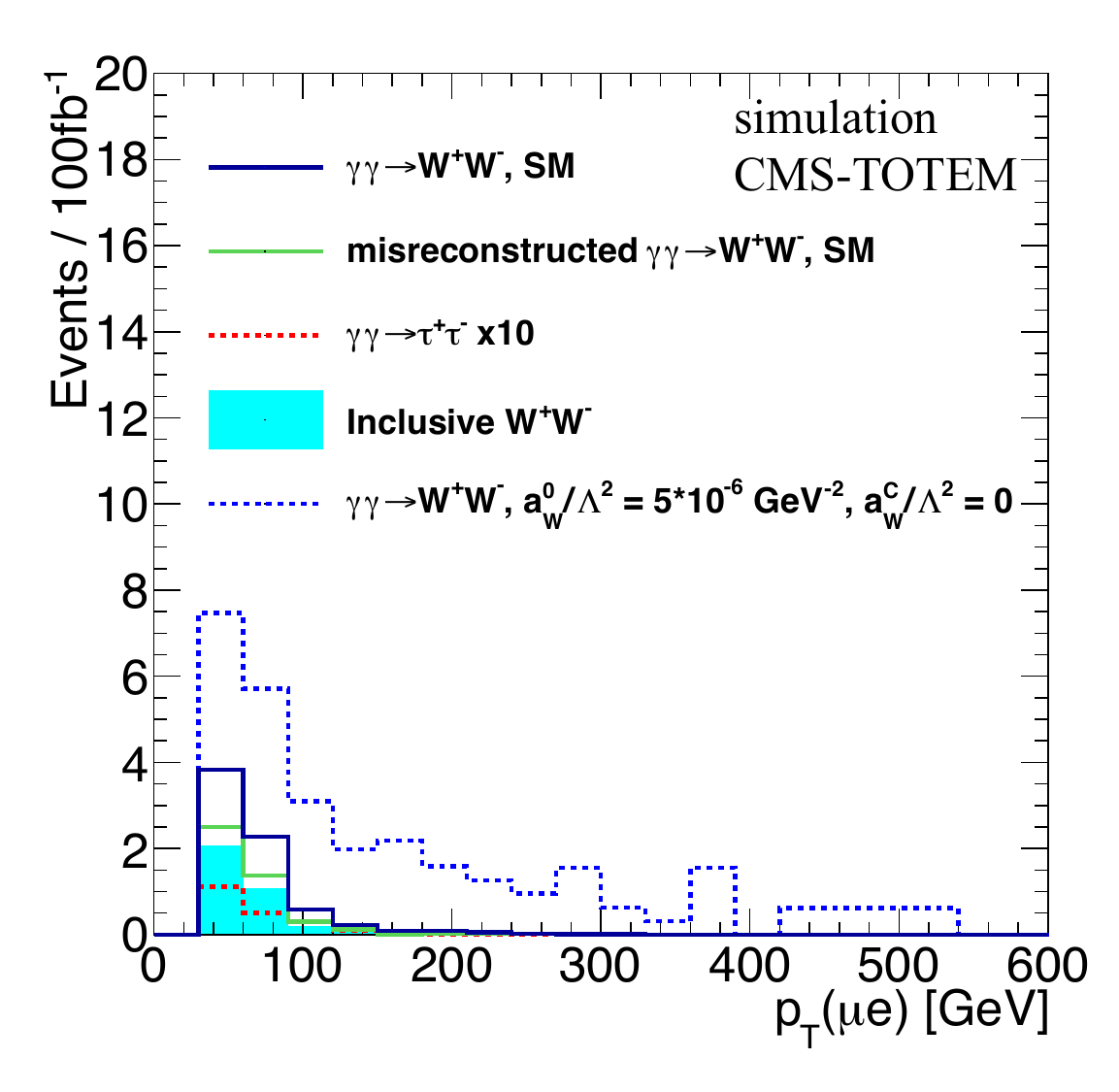}\\
\includegraphics[width=0.4\textwidth]{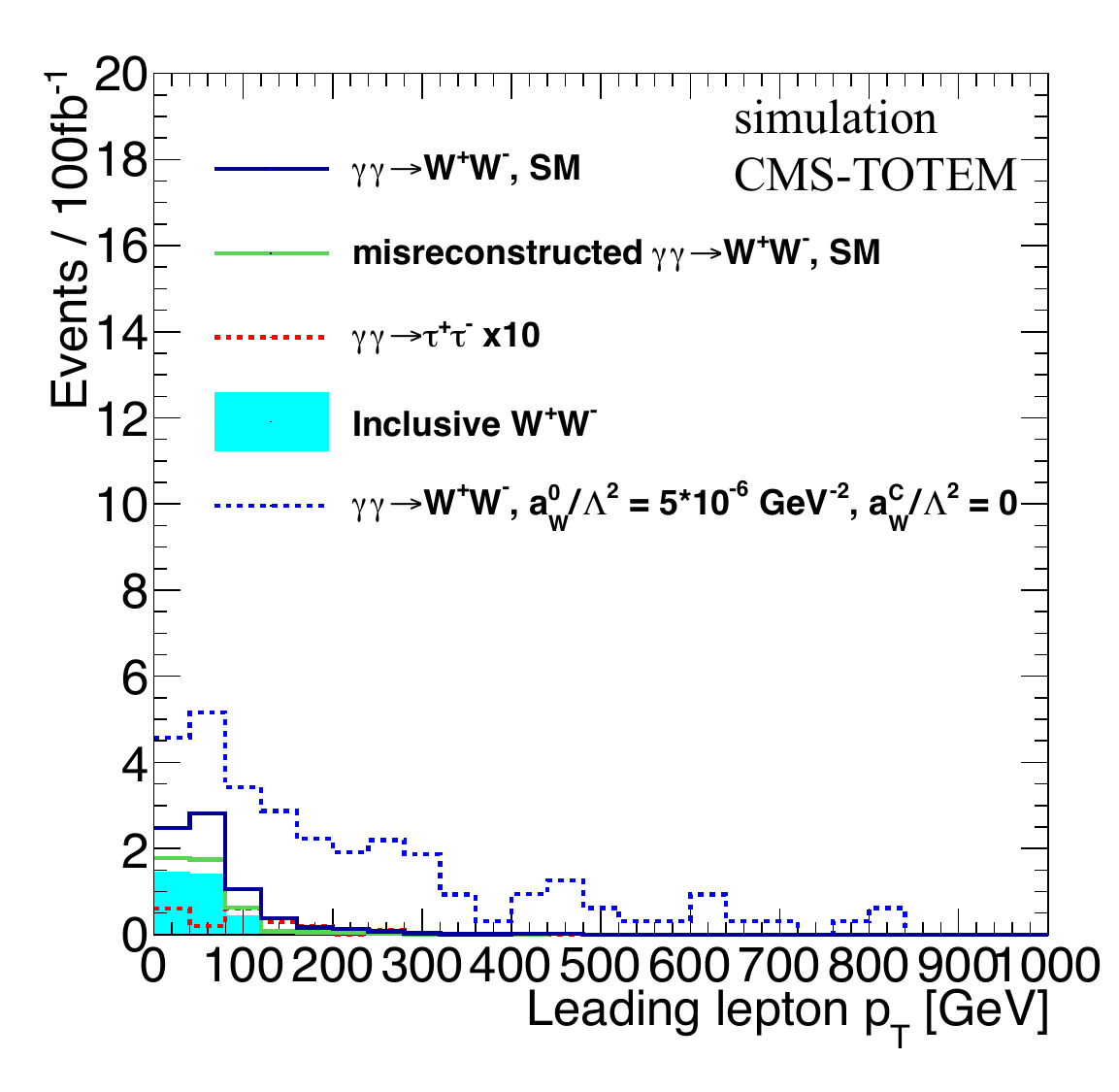} \qquad \qquad
\includegraphics[width=0.4\textwidth]{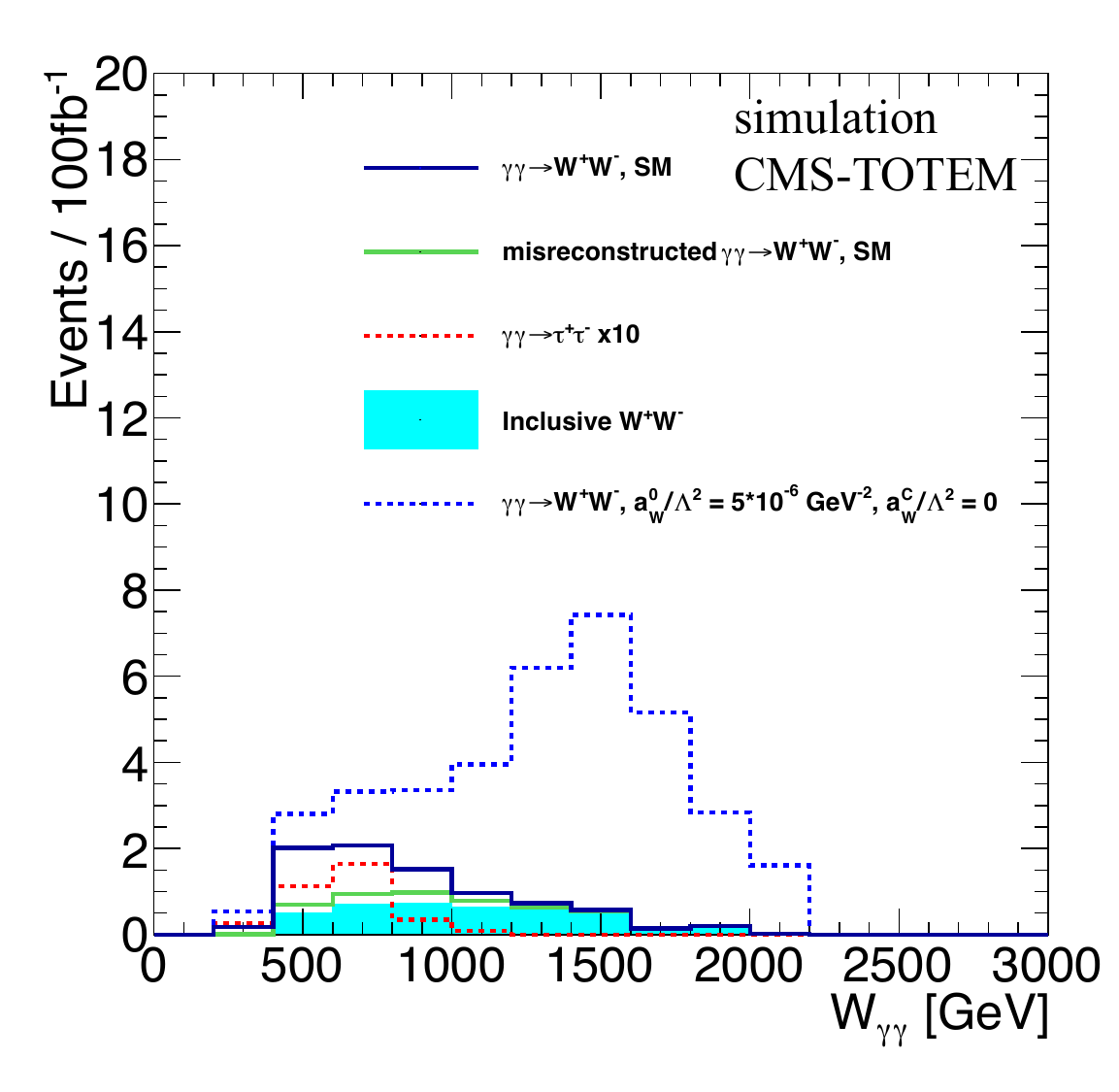}
\caption{
Azimuthal angle difference between the two leading leptons (top, left), transverse momentum of the dilepton pair (top, right), 
leading lepton transverse momentum (bottom, left), and  missing mass (bottom, right) for signal and background events, 
and in the presence of exclusive WW events due to AQGC processes.
Distributions are shown after the full event selection, for an integrated luminosity of 100~fb$^{-1}$.
Histograms are stacked, except for that of the exclusive $\tau\tau$ background, which is not stacked and is multiplied by a factor of 10.
}
\label{fig:kinematics_ww}
\end{figure*}

Kinematic distributions after the full event selections are shown in Fig.~\ref{fig:kinematics_ww}.
The transverse momentum of the dilepton system, the azimuthal angle difference between the two leading muons, and the missing mass distributions 
are shown for signal and for the background exclusive $\tau\tau$ event yields.
The missing mass $M_X=\sqrt{s\cdot \xi_1\cdot \xi_2}$ (also indicated as $W_{\gamma\gamma}$ in Fig.~\ref{fig:kinematics_ww})
is estimated from the reconstructed values of the two leading protons, $\xi_1$ and $\xi_2$.
The yields of exclusive $\tau\tau$ background events are multiplied by a factor of 10, in order to allow comparison of the shapes. 

Table~\ref{tab:yields_ww_xsec} summarizes the cross sections (in fb) 
after each selection cut, while Table~\ref{tab:yields_aqc} summarizes the cross sections (in fb) for the expected exclusive WW events due to AQGC
for two different values of the coupling parameters, $a_0^W$ and $a_C^W$. 
Cross sections include the branching fractions and 
are given for the dominant SM processes within the geometrical and detector acceptance. 
A small contribution from incorrectly reconstructed exclusive WW signal events, 
where at least one of the leading protons comes from a pile up or SD/DPE event, is also estimated separately.

\begin{table}
\tiny{
\caption{
Cross section (in fb)
for the expected SM processes, exclusive and inclusive WW, and exclusive $\tau\tau$ events,
after each selection cut (for a timing resolution of 10~ps). 
In case of different values, numbers in parentheses are for a timing resolution of 30~ps.
Only the e$\mu$ final state is considered. Statistical uncertainties are shown.}
\label{tab:yields_ww_xsec}
\begin{center}
\hspace*{-1.5cm}
\begin{tabular}{l|c|c|c|c} 
\hline\hline
Selection &  \multicolumn{4}{c}{Cross section (fb)}  \\ \hline
 & exclusive WW & exclusive WW & inclusive WW & exclusive $\tau\tau$\\
 & & (incorrectly reconstructed) & & \\
\hline 
generated $\sigma\times{\cal B}(WW\rightarrow e\mu ~\nu\bar{\nu})$	
											& 0.86$\pm$0.01  			& N/A 					& 2537 				& 1.78$\pm$0.01 \\
$\ge 2$ leptons ($p_\mathrm{T}>20$~GeV, $\eta<2.4$)	& 0.47$\pm$0.01			& N/A 					& 1140$\pm$3			& 0.087$\pm$0.003 \\
opposite sign leptons, ``tight'' ID					& 0.33$\pm$0.01			& N/A 					& 776$\pm$2			& 0.060$\pm$0.002 \\
dilepton pair $p_\mathrm{T}>30$~GeV				& 0.25$\pm$0.01			& N/A					& 534$\pm$2			& 0.018$\pm$0.001
\\
protons in both PPS arms (ToF and TRK)				& 0.055~(0.054)$\pm$0.002	& 0.044~(0.085)$\pm$0.003	& 11~(22)$\pm$0.3		& 0.004$\pm$0.001 \\
no overlapping hits in ToF + vertex matching 			& 0.033~(0.030)$\pm$0.002	& 0.022~(0.043)$\pm$0.002 	& 8~(16)$\pm$0.2 		& 0.003~(0.002)$\pm$0.001 \\
ToF difference, $\Delta t=(t_1-t_2)$					& 0.033~(0.029)$\pm$0.002	& 0.011~(0.024)$\pm$0.001	& 0.9~(3.3)$\pm$0.1		& 0.003~(0.002)$\pm$0.001 \\
$N_\mathrm{tracks}<10$							& 0.028~(0.025)$\pm$0.002	& 0.009~(0.020)$\pm$0.001 	& 0.03~(0.14)$\pm$0.01	& 0.002$\pm$0.001 \\
\hline\hline
\end{tabular}
\end{center}
}
\end{table}
\begin{table*}[htp!]
\caption{
Cross section (in fb)
for the expected exclusive WW events due to anomalous quartic gauge couplings, 
for different values of anomalous coupling parameters ($a_0^W$ and $a_C^W$)
after each selection cut (for a timing resolution of 10~ps). 
In case of different values, numbers in parentheses are for a timing resolution of 30~ps.
Only the e$\mu$ final state is considered. Statistical uncertainties are shown.}
\label{tab:yields_aqc}
\begin{center}
\begin{tabular}{l|c|c} 
\hline\hline
Selection &  \multicolumn{2}{c}{Cross section (fb)}  \\ \hline
 & $a_0^W/\Lambda^2=5\cdot10^{-6}$GeV$^{-2}$  & $a_C^W/\Lambda^2=5\times10^{-6}$GeV$^{-2}$  \\
 & ($a_C^W=0$) & ($a_0^W=0$) \\
\hline 
generated $\sigma\times{\cal B}(WW\rightarrow e\mu ~\nu\bar{\nu})$	& 3.10$\pm$0.14		& 1.53$\pm$0.07 \\
$\ge 2$ leptons ($p_\mathrm{T}>20$~GeV, $\eta<2.4$)				& 2.33$\pm$0.08		& 1.00$\pm$0.04 \\
opposite sign leptons, ``tight'' ID								& 1.82$\pm$0.08		& 0.78$\pm$0.03 \\
dilepton pair $p_\mathrm{T}>30$~GeV							& 1.69$\pm$0.07		& 0.68$\pm$0.03 \\
protons in both PPS arms (ToF and TRK)							& 0.52~(0.50)$\pm$0.04	& 0.18~(0.17)$\pm$0.02\\
no overlapping hits in ToF detectors								& 0.35~(0.32)$\pm$0.03	& 0.12~(0.11)$\pm$0.01\\
ToF difference, $\Delta t=(t_1-t_2)$								& 0.35~(0.32)$\pm$0.03	& 0.12~(0.11)$\pm$0.01\\
$N_\mathrm{tracks}<10$										& 0.27~(0.24)$\pm$0.03	& 0.11~(0.10)$\pm$0.01\\
\hline\hline
\end{tabular}
\end{center}
\end{table*}

\begin{figure*}[htp!]
\centering
\includegraphics[width=0.4\textwidth]{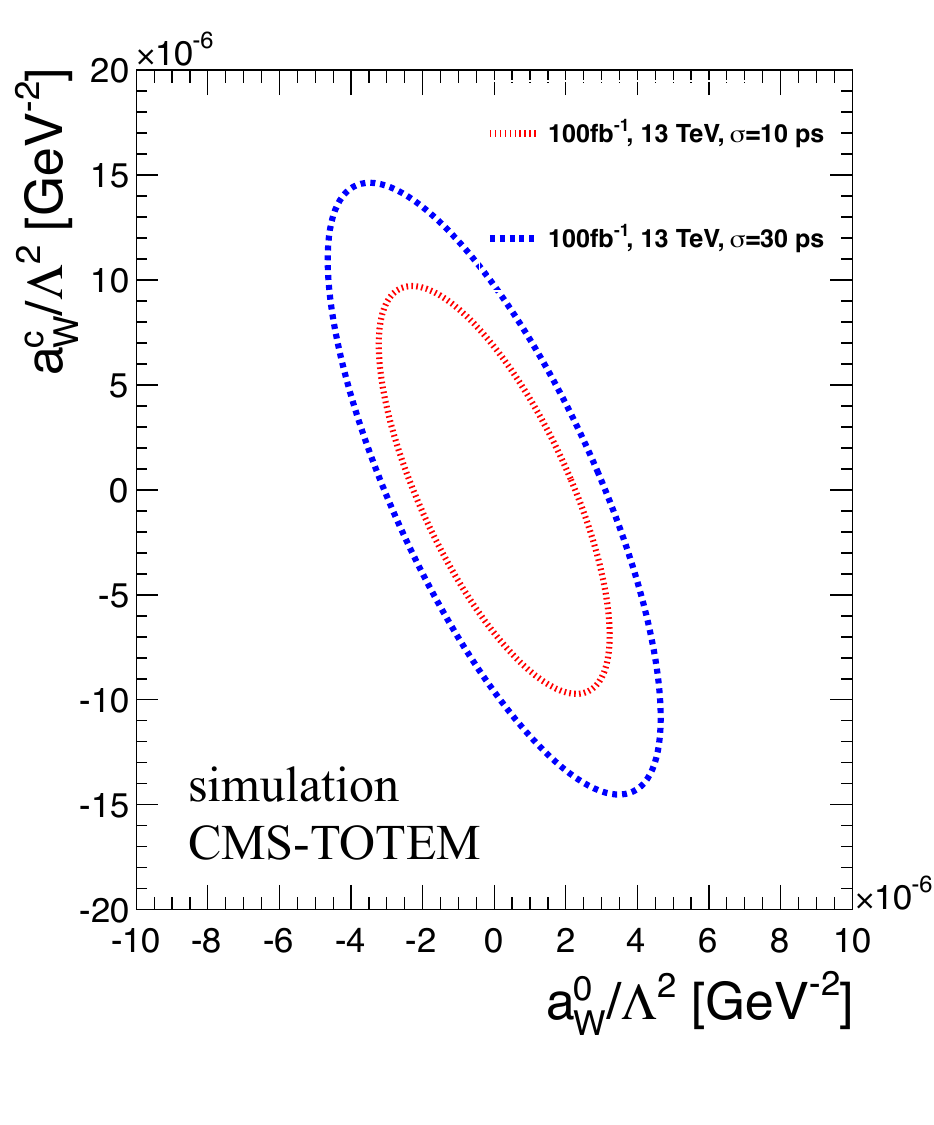} \qquad \qquad
\includegraphics[width=0.4\textwidth]{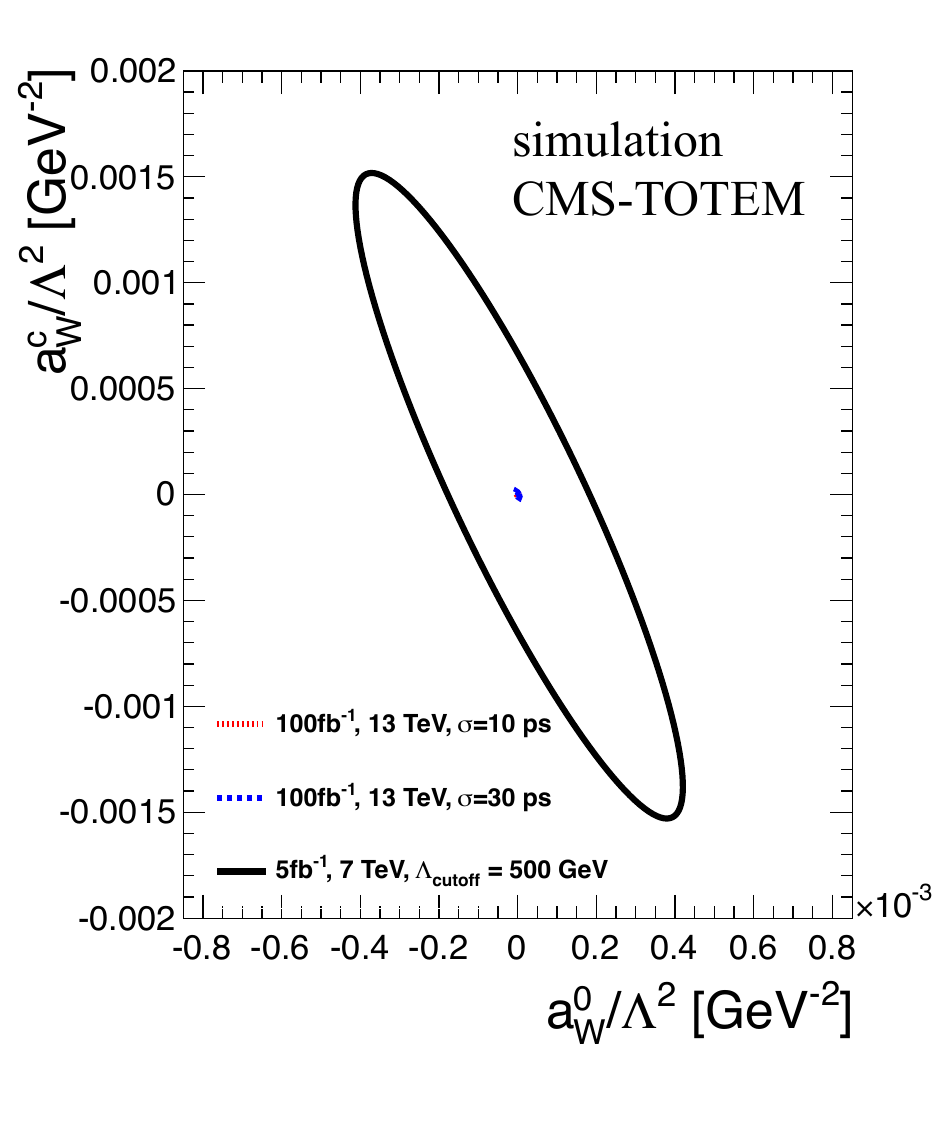}
\caption{
Excluded values of the anomalous coupling parameters $a_0^W/\Lambda^2$ and $a_C^W/\Lambda^2$.
The areas outside the contours are excluded at 95\% CL.
Approximate limits expected with 10~ps and 30~ps timing resolutions (left)
compared to the current 2011 CMS limits from exclusive WW events (right).
}
\label{fig:limits_ww}
\end{figure*}

The cross section times the acceptance for SM exclusive WW events is already sizeable at small reconstructed 
values of the missing mass ($M_X\simeq 300\div 400$~GeV), and a close approach to the beam can provide a rapid increase of SM signal event yield.
The variation of the acceptance as a function of the detector distance from the beam has been evaluated. 
In particular, the visible cross section for signal exclusive WW events 
increases by a factor of two when the detector distance from the beam decreases from $15~\sigma$ to $10~\sigma$~\cite{pps_tdr}. 
The selected events are used to set limits on the AQGC parameters, $a_0^W/\Lambda^2$ and $a_C^W/\Lambda^2$.
The resulting limit values are of the order of
$a_0^W/\Lambda^2=2\times10^{-6}$ ($3\times10^{-6}$), and $a_C^W/\Lambda^2=7\times10^{-6}$ ($10\times10^{-6}$),
in case of a 10~ps (30~ps) time resolution.
Approximate 95\% CL limits expected with 10~ps and 30~ps timing resolutions (Fig.~\ref{fig:limits_ww}, left)
are compared to the 2011 CMS results~\cite{Chatrchyan:2013foa} from exclusive WW events (Fig.~\ref{fig:limits_ww}, right).
Expected limits for Run~2 are estimated for an integrated luminosity of 100~fb$^{-1}$.
The areas outside the contours are excluded at 95\% CL.

The study demonstrates the feasibility of measuring exclusive WW production in Run~2. 
With an integrated luminosity of 100~fb$^{-1}$, approximately 3~SM exclusive WW signal events are expected and a similar number of background events, 
even when looking at the e$\mu$ channel alone.
Anomalous quartic gauge couplings would produce a striking, very visible signal. Approximately 30~(10)~events would be visible in the presence of AQGC, 
with coupling parameters $a_0^W (a_C^W) /\Lambda^2=5\cdot10^{-6}$GeV$^{-2}$.

\subsection{New strong dynamics in exclusive processes}

After the recent discovery of the Higgs boson \cite{Aad:2012tfa,Chatrchyan:2013lba} 
at the LHC and follow-up precision studies of its interactions with SM particles, 
a rough picture of consistency with the SM has begun to emerge. This consistency, however,
does not yet mean that the nature of the Higgs boson and electroweak
symmetry breaking (EWSB) is completely understood~\cite{Ellis:2015dha}. An
immediate question that challenges our current understanding of
symmetries in Nature is what initiates EWSB in the SM.

One of the major questions to be answered in the ongoing search for
New Physics at the LHC is whether a fine structure of the Higgs-like
signal exists in the low invariant mass interval $110\ -\ 140$ GeV,
predominantly in $\gamma\gamma$, $W\gamma$ and $Z\gamma$ channels,
or not. There exists a possibility that yet unknown resonances which
decay into two photons could be very difficult to identify in 
inclusive measurements. As indicated by for example the CMS data~\cite{Chatrchyan:2013lba} on Higgs boson production, such a fine structure is not yet completely excluded, and this is being explored theoretically 
in various BSM scenarios. It is therefore interesting
to study such a structure in the $\gamma\gamma$ and $Z\gamma$
decay channels, both in QCD and VBF-initiated exclusive production
mechanisms. An exclusive measurement has the advantage 
that $\gamma\gamma$-resonance signals could be
enhanced relative to the two-photon background. This
offers important advantages compared to searches of new
$\gamma\gamma$-resonances in inclusive reactions.

New strongly-coupled dynamics at the TeV energy scale is one possible cause for EWSB in the SM~\cite{Weinberg:1975gm,Susskind:1978ms}. This initiates EW symmetry breaking dynamically by means of confined
techniquark condensation at low energy scales. Such new dynamics 
unavoidably predicts a variety of new states; most importantly, 
composite Higgs-like particles \cite{Barducci:2013wjc} whose properties depend on
the group-theoretical structure of underlying theory and its
ultraviolet (UV) completion. The discovery of such a family of new
(pseudo)scalar states with invariant masses not exceeding $200$ GeV
in these channels is of high priority for strongly-coupled dynamics
searches at the LHC.

A number of realisations of such new dynamics at the TeV
scale, known as ``Technicolor'' (TC) or ``compositeness''
scenarios, have been proposed in the literature  
(for a review, see e.g.~\cite{Hill:2002ap,Sannino:2009za}).
However, these have been strongly restricted by
electroweak (EW) precision tests and recent SM Higgs-like
particle observations. At present, among the most appealing scenarios 
for dynamical EWSB consistent with current constraints is a class 
of TC models with vector-like (Dirac) UV completion -- 
vector-like Technicolor (VLTC). The simplest
realisation of  a VLTC scenario with two vector-like or Dirac techniflavors and a
SM-like Higgs boson has been studied for the first time in~\cite{Pasechnik:2013bxa,Lebiedowicz:2013fta} and 
very recently has emerged in composite Higgs scenarios with 
confined $SU(2)_{\rm TC}$ \cite{Cacciapaglia:2014uja,Hietanen:2014xca}.

Here an important case of light exotic resonances is considered, namely the pseudo-Goldstone
T-pions, commonly predicted by Technicolor extensions of the
Standard Model. Since T-pions in the consistent VLTC scenario 
do not couple directly to SM fermions 
and gluons, the only way to produce them is in vector-boson
($\gamma\gamma,\,\gamma Z,\,ZZ$) fusion.
At Born level, the pseudoscalar T-pions can only be pair-produced in $\gamma\gamma$ and VBF reactions. 
At one-loop level, T-pions are coupled to photon and vector bosons via either a
T-quark triangle or box diagrams, depending on the number of T-colors. 
In a QCD-like scenario with $N_{\rm TC}=3$ and a degenerate T-quark doublet, 
the T-pion decays into two gauge bosons $V_1$ and $V_2$, while 
in the case of $N_{\rm TC}=2$ the T-pion can only decay into three 
gauge bosons via a T-quark box diagram. 
Thus, in the former case one expects single
T-pion $\tilde \pi^0$ production, predominantly, in $\gamma\gamma$-fusion 
via the T-quark triangle, whereas in the latter case a single T-pion will be
produced in $V_1V_2$-fusion in association with an extra gauge boson 
$V_3$. Then, each produced T-pion should further decay either into two or 
three gauge bosons, depending on $N_{\rm TC}$, or into a pair of Higgs 
bosons $\tilde \pi\to hh$.
%-------------------------------------------------------------
\begin{figure*}[!tbh]
\centering
\begin{minipage}{0.35\textwidth}
 \centerline{\includegraphics[width=1.0\textwidth]{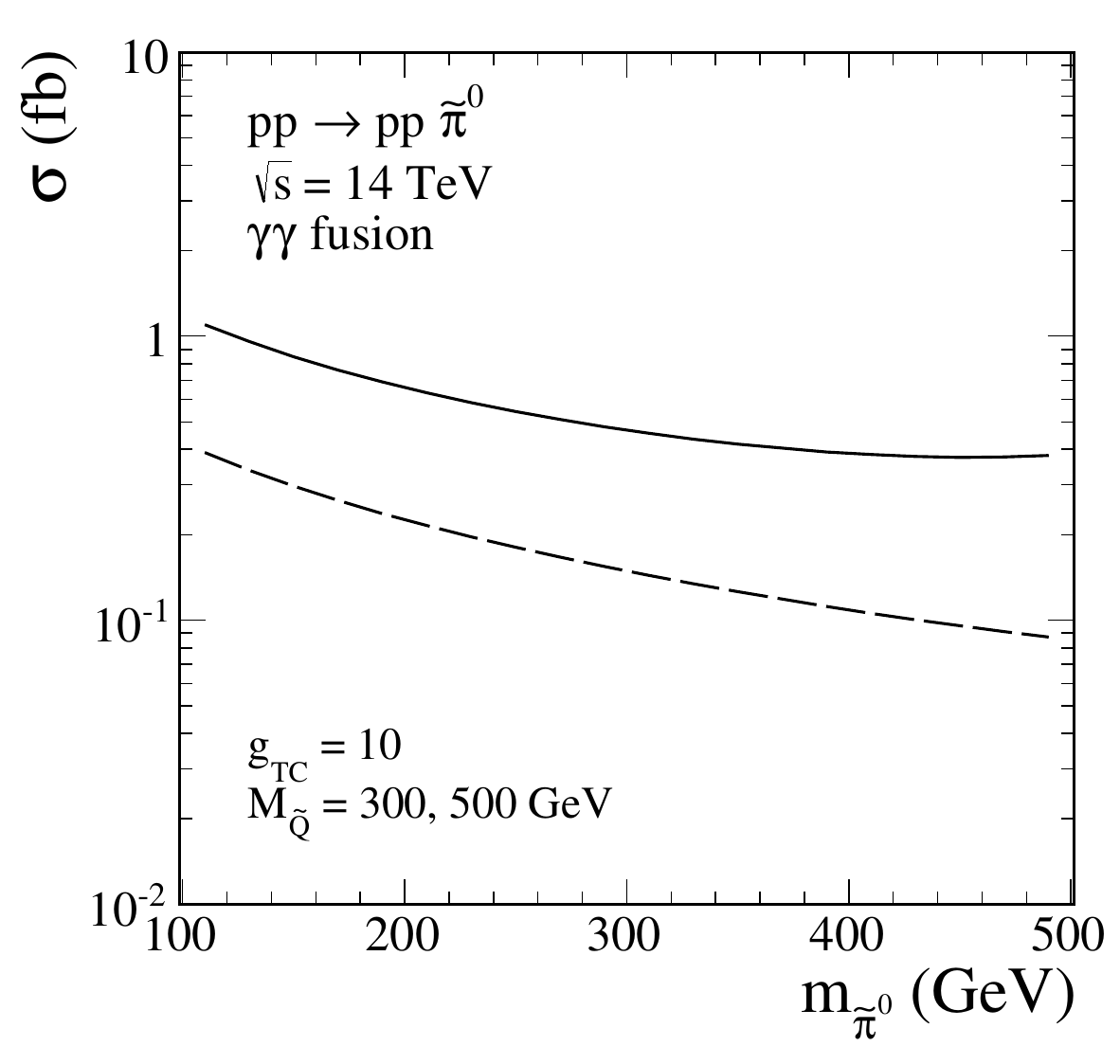}}
\end{minipage}
\begin{minipage}{0.35\textwidth}
 \centerline{\includegraphics[width=1.0\textwidth]{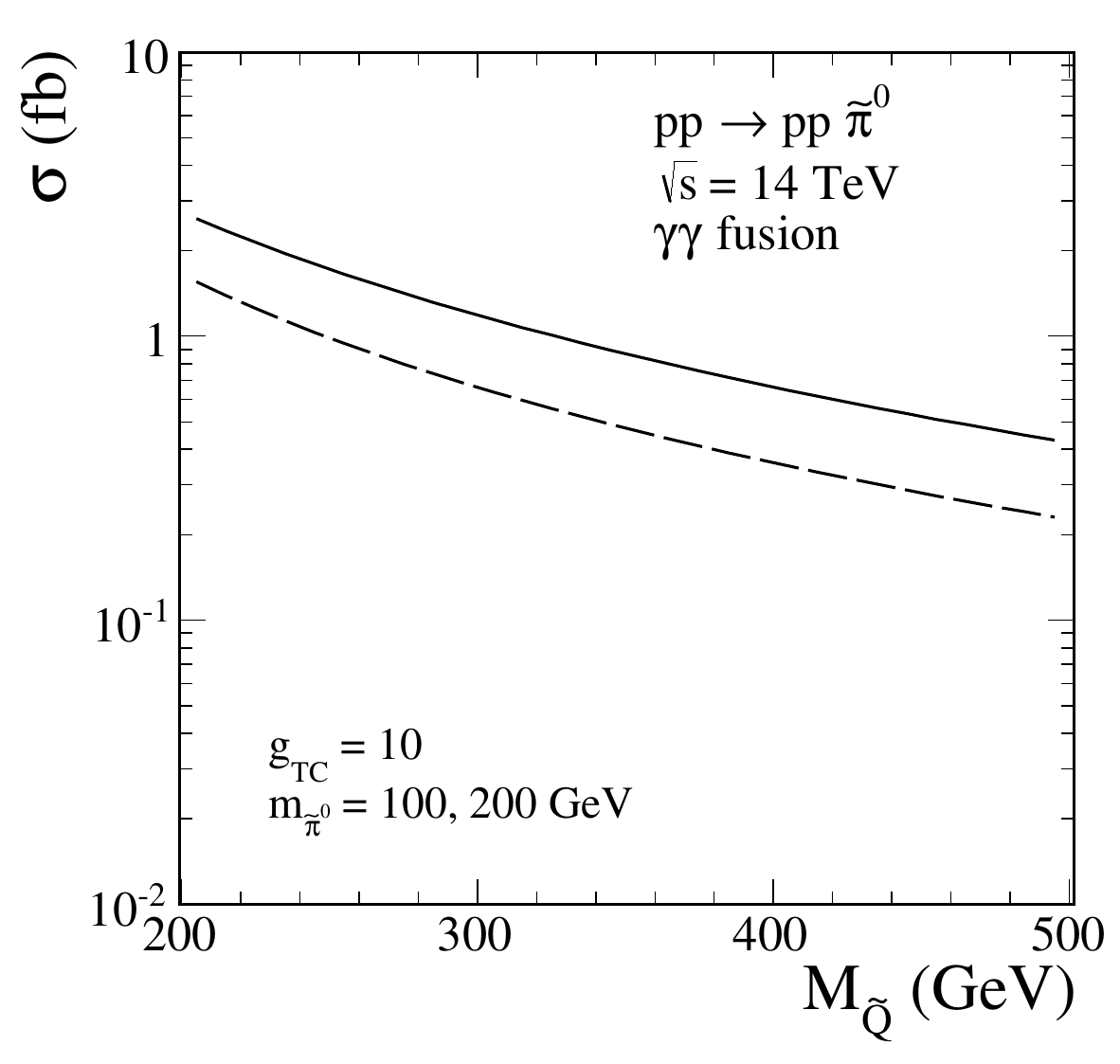}}
\end{minipage}
   \caption{
\small Integrated cross section of central exclusive T-pion 
as a function of T-pion mass (left) and T-quark mass (right) for fixed
model parameters. For more details, see~\cite{Lebiedowicz:2013fta}.}
 \label{fig:parameters}
\end{figure*}
%-------------------------------------------------------------

The central exclusive $p p \to p \tilde{\pi}^0 p$
reaction dominated by the $\gamma^* \gamma^* \to \tilde{\pi}^0\to \gamma\gamma$ 
hard subprocess is of special interest for the $N_{\rm TC}=3$ scenario (the
$N_{\rm TC}=2$ case is much more challenging). 
As can be seen in Fig.~\ref{fig:parameters}, the predicted central exclusive
${\tilde \pi}^0$ production cross section in the $\gamma\gamma$ channel 
can be of the same order or even exceed the traditional Higgs boson 
CEP cross section, making the considered proposal particularly relevant for the forward physics program at the LHC
\cite{Khoze:2001xm,Albrow:2010yb}; see also~\cite{Goncalves:2015oua} for a study of the CEP of a BSM dilaton, which is for example predicted within a range of techicolour scenarios.

In order to estimate the feasibility of such an exclusive T-pion measurement 
the exclusive $\gamma\gamma$ background must be considered. 
There are two basic non-resonant leading order box-induced contributions -- the QCD
mechanism via $gg\to \gamma\gamma$ and the QED (light-by-light)
scattering mechanism ($\gamma\gamma\to\gamma\gamma$),
depicted in Fig.~\ref{matthias}.
At relatively low masses, the QCD mechanism dominates, 
however above $M_{\gamma \gamma} >$ $O$(100) GeV,
the photon-photon mechanism takes over. The latter is therefore the
most important potential background for the T-pion signal in the $\gamma\gamma$ channel. After inclusion 
of the ATLAS detector resolution, the
$S/B$ ratio for the T-pion CEP is significantly better than for
the inclusive case as well as for Higgs boson CEP in the $b
\bar b$ channel. This analysis demonstrates that 
the exclusive reaction $p p \to p p \gamma\gamma$ is probably the best 
suited in searches for T-pions at the LHC; see~\cite{Lebiedowicz:2013fta} for further details.

In general, this indicates that the exclusive production of two photons in $pp$ collisions 
can be an especially attractive channel for the search of new heavy resonances that
predominantly couple to photons. This process is interesting by itself 
and rather unique due to relatively well understood QED mechanism. Any deviation 
from the Standard Model prediction here may 
be a signal of New Physics contributions. This motivates the search for both continuum and resonance $\gamma\gamma$ signals of New Physics in exclusive processes, via the ATLAS and CMS forward proton tagging detectors.

\section{Conclusion}

In this chapter, the central exclusive production (CEP) process has been considered, where a system $X$ is produced in the central detector while the outgoing protons remain intact after the collision. This may be mediated purely by the strong interaction, in the language of Regge theory by double pomeron exchange, as well as by two--photon and photon--Pomeron collisions (that is, in photoproduction reactions). Theoretically, these require the development of a framework which is sensitive to both the hard and soft QCD regimes, and lead to predictions and effects which are not seen in the more common inclusive modes. Experimentally, CEP represents a very clean signal (in the absence of pile--up), while the outgoing protons can be measured by proton tagging detectors situated far from the interaction point. This is the aim of the installed CMS--TOTEM, CT--PPS and ALFA and planned AFP detectors at the LHC. This chapter has discussed the motivation and possibilities for performing exclusive measurements both with and without tagged protons at low to medium luminosity, as part of special high $\beta^*$ runs with ATLAS and CMS, or during general LHCb and ALICE running, and at higher luminosity with tagged protons, where tools such as precision timing detectors will be fundamental to control pile up effects. A wide range of final--states has been considered, and it has been shown how the exclusive mode can provide additional insight and information about physics both within and beyond the SM, which is complementary to and can extend beyond the possibilities with more conventional inclusive measurements.
%Measurements of CEP with tagged protons are well motivated by the additional information that these can provide about the physics of the production process, as well as the benefit in terms of selecting such events, while many interesting studies can also be performed without tagged protons.  

\clearpage

\section*{Acknowledgments}
    
Pieces of this chapter has been supported in part by Polish Nation Science Centre grants number: 2013/08/M/ST2/00320, DEC-2011/01/B/ST2/04535 and 2012/05/B/ST2/02480. Some of the results presented here come from a fruitful collaboration with Emilien Chapon,
Sylvain Fichet, Gero von Gersdorff, Old\v{r}ich Kepka, Bruno Lenzi.

%% file: cosmic/cosmic.tex
In this chapter, measurements needed for the simulation of cosmic ray air showers are reported. A general overview of the air-shower measurements is presented in Section \ref{sec7:Introduction}, followed by a brief overview of the air-shower simulations in Section \ref{sec7:LHCandEAS}. Sections \ref{sec7:EnergyFlow}-\ref{sec7:Spectra} give an overview of the past and future planned measurements that are relevant to understand and fine-tune hadronic models used for air shower simulations. Finally, a proposal for a dedicated beam conditions for min-bias analyses in Run-2 is outlined in Section \ref{sec7:Beam}.

\section{Introduction}
    \label{sec7:Introduction}
    \input{cosmic/Introduction}

\section{LHC and air showers}
    \label{sec7:LHCandEAS}
    \input{cosmic/LHC_and_EAS}

\section{Energy Flow}
    \label{sec7:EnergyFlow}
    \input{cosmic/EnergyFlow}
\section{Particle multiplicities}
    \label{sec7:Multiplicity}
    \input{cosmic/Multiplicity}

\section{Spectra}
   \label{sec7:Spectra}
    \input{cosmic/Spectra}

\section{Beam}
   \label{sec7:Beam}
   \input{cosmic/Beam}

%% file: cosmic/Introduction.tex
Understanding the sources and the propagation of cosmic rays are central questions of
astroparticle physics. While there is increasing evidence that supernova remnants accelerate 
cosmic rays up to energies of $\sim Z\times 10^{14}$\UeV\ (with $Z$ being the charge of
the cosmic ray nucleus), the sources of the particles of energies up to $10^{20}$\UeV\ are not
known~\cite{Kotera:2011cp}. Accelerating particles to such energies requires exceptional
astrophysical objects~\cite{Hillas:1985is}. Using
the LHC technology of superconducting magnets one would have to build the ring of an LHC-like
accelerator as big as the orbit of the planet Mercury to be able to accelerate protons up to
$10^{20}$\UeV. Ultra-high energy cosmic rays are not only messengers from the extreme Universe,
they also allow us to study the laws of physics under extreme conditions (for example, to search for 
violation of Lorentz invariance or extra dimensions) and provide a window to particle physics 
at energies up to $\sqrt{s} \sim 400$\UTeV.
Therefore it is not surprising that the physics of ultra-high energy cosmic rays is
subject of very intensive research.

The study of cosmic rays of energies higher than about $10^{14}$\UeV\ is hampered by the low flux of the 
particles arriving at Earth. With the flux being too low for direct particle detection one has to resort to
the observation of extensive air showers produced by cosmic rays entering the atmosphere~\cite{Bluemer:2009zf}.
Although being very efficient for large aperture measurements, the drawback of this approach is the need
of detailed air shower simulations for deriving the primary particle energy, and more importantly,
the particle type and its mass number from the measured air shower observables.

\begin{figure}[htb!]
  \begin{center}
     \includegraphics[width=0.8\textwidth]{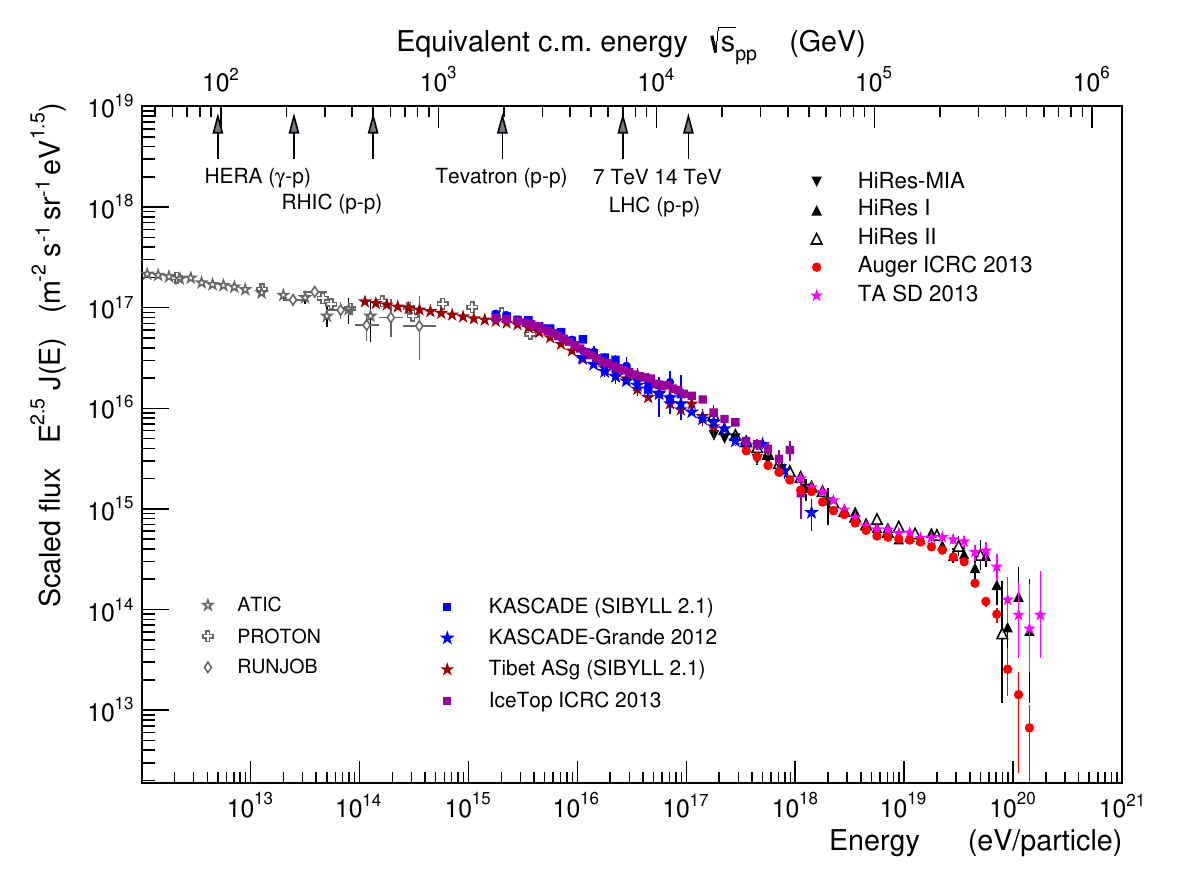}
  \end{center}
  \caption{Compilation of representative measurements of the flux of cosmic rays at
  Earth (from~\cite{Engel:2011zzb}, updated). All data sets shown in color are derived from air
  shower measurements. The gray (open) symbols show direct measurements covering the energy range
  below the knee. The equivalent energies of LHC proton-proton collisions at different c.m.s.\ energies
  are shown at the upper axis.
  \label{fig7:intro:flux}
  }
\end{figure}

An overview of recent measurements of the flux of cosmic rays is given
in Fig.~\ref{fig7:intro:flux}~\cite{Engel:2011zzb}. Both the lab.\ energy measured in cosmic ray experiments
and the equivalent c.m.s.\ energy, assuming the primary particles are protons, are shown. LHC measurements
allow us to access, for the first time, energies beyond the {\it knee} in the cosmic ray spectrum, a break of
the power law of the flux at about $3\times 10^{15}$\UeV, which is not yet understood.
In a number of models this break has been attributed
to an unexpected change of particle physics above $\sqrt{s}\sim 2$\UTeV, which could be ruled out by the
first LHC measurements (see~\cite{dEnterria:2011kw} and references therein). Furthermore
the cosmic ray flux exhibits an ankle at about $3\times 10^{18}$\UeV\ 
(probably related to the transition between galactic and extragalactic cosmic rays~\cite{Berezinsky:2002nc,Allard:2005ha}) and a very strong suppression
at energies above $7\times 10^{19}$\UeV.

To understand the sources of cosmic particles at energies
above $10^{15}$\UeV\ and the astrophysical origin of the striking features in the cosmic ray flux
it is needed to derive the mass composition of cosmic rays. A good example is the flux
suppression at the highest energies, which was expected due to energy loss effects in the CMB
first described by Greisen~\cite{Greisen:1966jv}
and Zatsepin and Kuzmin~\cite{Zatsepin66e} (GZK). However, recent data of the Auger Observatory on the mass composition
indicate that
the upper end of the energy spectrum is more likely related to the maximum injection energy of the particles at the
sources~\cite{Abraham:2010yv,Allard:2011aa,Abreu:2013env}.

Being able to derive reliably the mass composition of cosmic rays from air shower measurements is of
fundamental importance and is currently hampered mainly by the large uncertainties in predicting hadronic
multiparticle production at high energy~\cite{Engel:2011zzb,Kampert:2012mx}. Due to not being
able to calculate corresponding predictions within QCD, performing further measurements at accelerators
is the only way to make progress.

The relation between the characteristics of hadronic interactions at high energy and air shower observables has
been reviewed in~\cite{Engel:2011zzb} and recent numerical studies
can be found in~\cite{Drescher:2007hc,Ulrich:2010rg,Pierog:2006qv}. In each
hadronic interaction a number of $\pi^0$ are produced that decay immediately, feeding the electromagnetic
shower component with high-energy photons. Already after less than $5$ generations of hadronic interactions more than
$80$\% of the primary particle energy is transferred to the em.\ shower component. In contrast, the production
of muons, which are mainly coming from the decay of low energy pions, takes place only after $8-12$ consecutive hadronic interactions. Only then the energy of the produced
charged pions is low enough ($E_{\pi^\pm} \sim 30$\UGeV) that they decay instead of interacting again~\cite{Meurer:2005dt}.
While muons are most directly linked to the hadronic shower component, interactions of a very wide
range of energies are important for understanding the properties of the muonic shower component.

Understanding the energy transfer from the hadronic to the electromagnetic and muonic shower components
is of key importance for reliably predicting shower observables. This means that both secondary particle
multiplicities as well as the energy given to different particle types are of direct relevance to air 
shower physics. In particular, knowing the number of baryons, charged and neutral pions as well as kaons at large
Feynman-$x$ are of outstanding importance. In addition, the interaction cross sections of the different 
particles are needed to estimate the depth at which different stages of showers develop. On the other hand,
transverse momentum distributions of particles are only of secondary importance for cosmic ray interactions. Due to
the large Lorentz boost needed to transfer $\sqrt{s} \sim 14$\UTeV\ collisions to the lab.\ frame,
in which these interactions take place in an air shower, even rather large transverse momenta lead
to very small angles of the particles relative to the shower axis.

The interaction models used in cosmic ray physics have been tuned to describe not only particular data sets
at certain collider energies but also the energy dependence of multiparticle production.  While cosmic ray models were developed before LHC was turned on, their predictions bracket many distributions measured in minimum-bias mode at LHC,
%The model predictions derived before LHC data became available bracket the measured distributions,
as shown section~\ref{sec7:models} and in~\cite{dEnterria:2011kw}. This is a great success of the
phenomenology developed for soft hadronic interactions. On the other hand, the predictions can be improved 
considerably by tuning the models to match the LHC data, as it will be illustrated
in section~\ref{sec7:eas}.
In the following we will discuss LHC measurements that can further improve the understanding of hadronic 
multiparticle production with direct relevance to cosmic ray physics and air shower simulations.

%% file: cosmic/LHC_and_EAS.tex
% The core of air shower being driven by hadronic interactions, Monte Carlo 
% generators are the main source of uncertainties in air shower simulations.
% To build and test hadronic interaction models high quality data at various
% energies and for different phase space regions are required. Since 2009 LHC experiments
% provide a wide range of new data which can be used to constrain and tune
% the interaction models used in air shower simulations.

There are two categories of LHC measurements that are important for improving
predictions for extensive air showers. First of all, the measurement of the 
multiplicities and energy fractions given to the different secondary particles, and their production
cross sections can be used directly in shower simulations. However, due to the limited 
phase space covered by collider experiments such data can cover only a small part
of the relevant phase space and will always be restricted to certain interaction
energies. The second category of measurements is closely related to theoretical
and phenomenological concepts implemented in interaction models. While these measurements
can be limited to phase space regions that are, in general, not of direct relevance to
air showers, they test fundamental concepts of the models and, thanks to 
energy-momentum and quantum number conservation as well as inherent correlations,
allow us to make predictions on particle production in phase space regions or at
energies not accessible. 

\begin{figure}
  \includegraphics[angle=0,width=.49\textwidth]{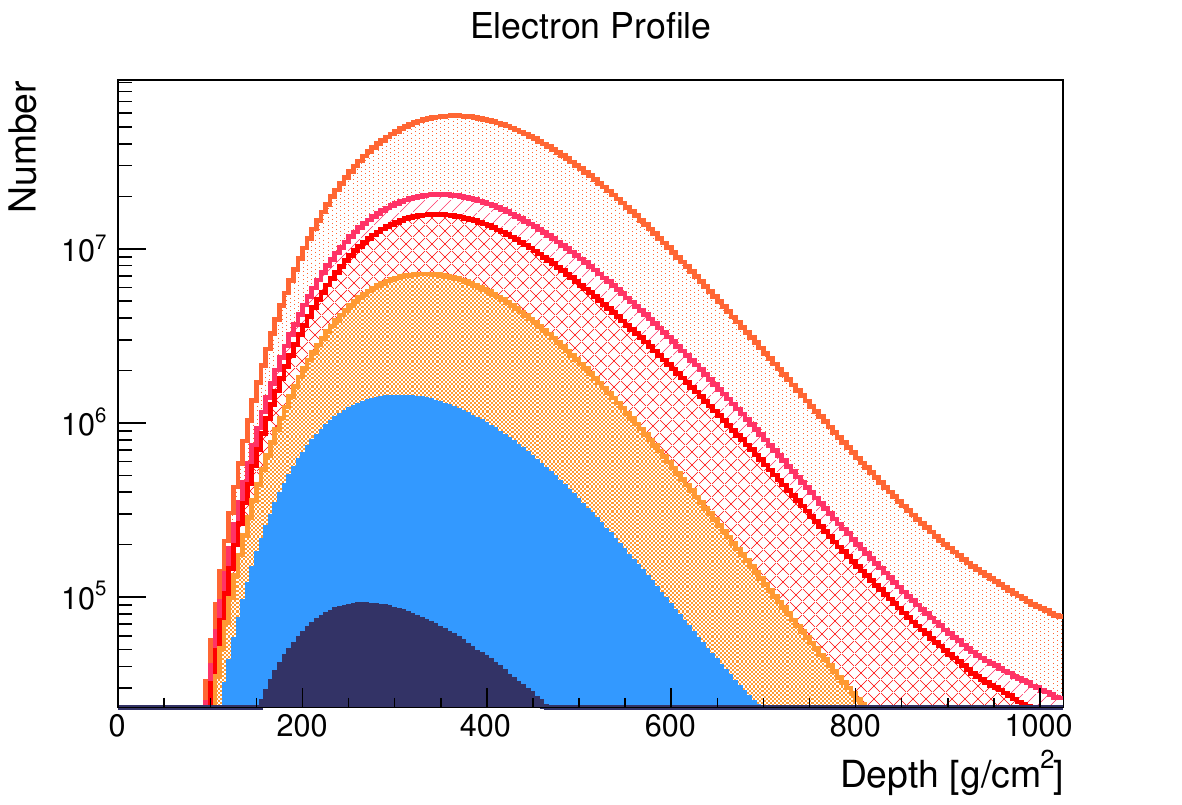}
  \includegraphics[angle=0,width=.49\textwidth]{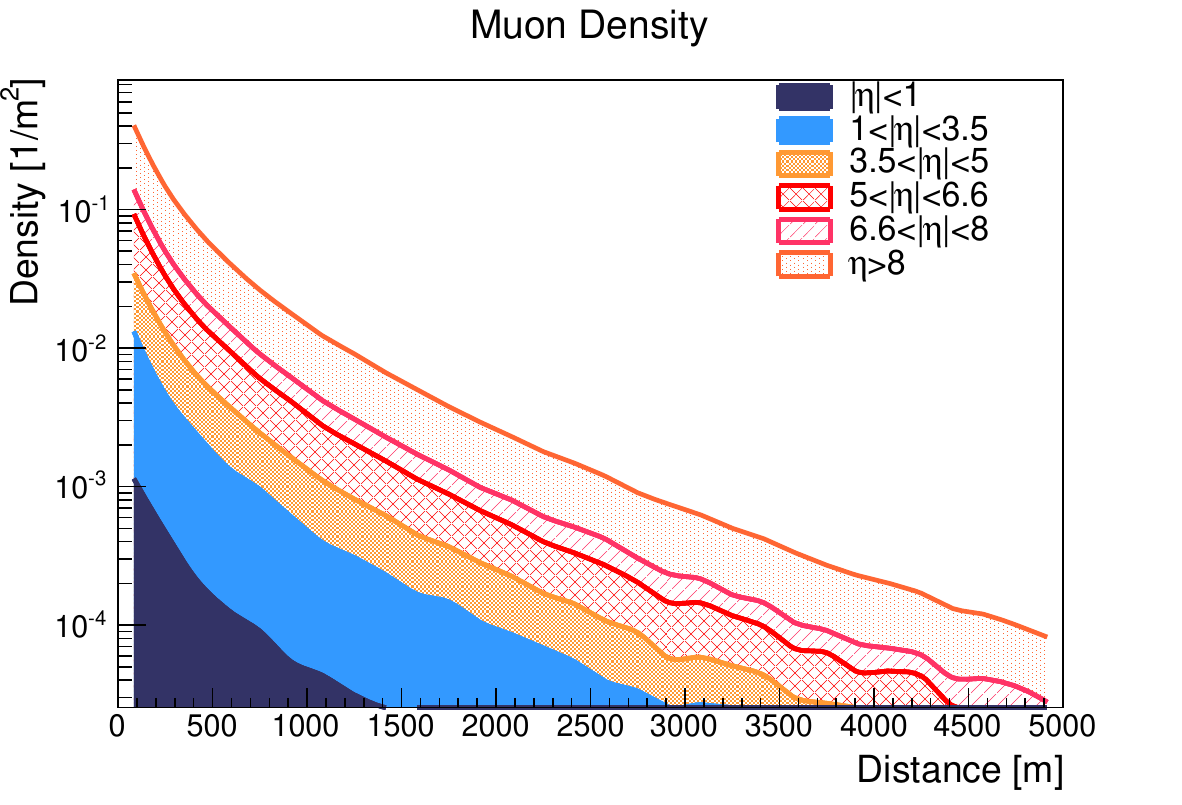}
  \caption{Fraction of the air shower development for one proton induced air 
shower at $10^{17}$eV primary energy, which is determined by the 
hadronic particle production in the initial inelastic p-air collision in 
different acceptance regions for electrons in longitudinal profile (left hand-side and muons in lateral distribution at ground (right hand-side). The acceptance is calculated in the 
center-of-mass frame of the collision, and the shown values are related 
to typical LHC detectors. The major part of the air shower is determined 
by particle production in the forward region.}
  \label{fig7:eas:longldflhc}
\end{figure}

However, the particle production, which carries the majority of the primary energy, is what drives
the development of extensive air showers and, thus, determines all the relevant features necessary
for the analysis of cosmic ray data. It is a problem for the reliability of hadronic interaction models
in extensive air showers, when they are tuned to central particle production ($|\eta|<3$) only, since this
only explains about 5$\%$ of the resulting observed air shower particles as shown on Fig.~\ref{fig7:eas:longldflhc} (for a primary proton at
E$_{0} = 10^{17}$\UeV\ the corresponding energy of the primary interaction is $\sqrt{s_{NN}} = 14$\UTeV). 
Crucial observables, as the location of the air shower
maximum or also the muon fraction at ground level cannot be reliably predicted in this situation.
Considering forward  acceptance up to  $\eta \sim 7$ allow to reach about $50\%$ of
the observed particles in such extensive air showers.

Of course,  the better the forward direction is covered in the measurements the smaller are the model 
extrapolations. But it will always be necessary to use hadronic interaction models as tool
to link LHC data, and accelerator measurements in general, to air shower physics. 

The LHC data on total, elastic and diffractive cross sections and other measurements related to soft diffraction (rapidity gaps, energy loss, ...) are examples of
the first category, while mean particle multiplicities, multiplicity
distributions, jet cross sections at low $p_\perp$, particle spectra and correlations between observables
belong to the second one.

    \subsection{LHC data and hadronic interaction models}
    \label{sec7:models}

\begin{figure}
  \includegraphics[angle=-90,width=.49\textwidth]{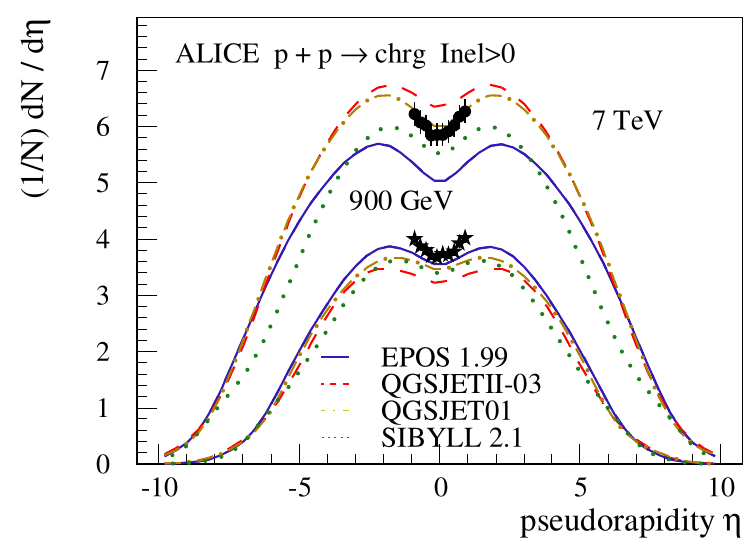}
  \includegraphics[angle=-90,width=.49\textwidth]{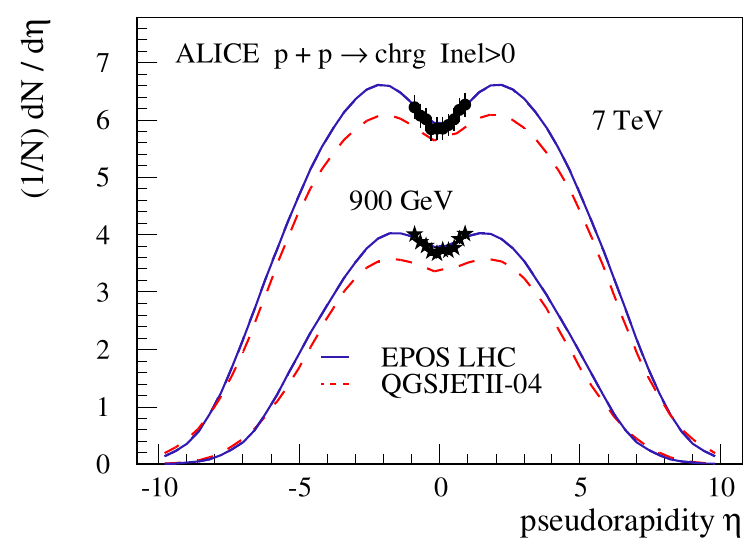}
  \caption{Pseudorapidity distribution $dN/d\eta$ of charged particles for events with at least one charged particle with $|\eta|<1$ for {\it p-p} interactions at 900\UGeV\ and 7\UTeV. Simulations with \epos~1.99 (solid line), \texttt{QGSJETII}-03 (dashed line), \texttt{QGSJET01} (dash-dotted line) and \texttt{SIBYLL}~2.1 (dotted line) on left panel, and \epos~\texttt{LHC} (solid line) and \texttt{QGSJETII}-04 (dashed line) on right panel, are compared to data points from ALICE experiment~\protect\cite{Aamodt:2010pp}.}
  \label{fig7:eas:dndetapp}
\end{figure}

%Tim Martin for ATLAS
For instance, measurement of the pseudorapidity dependence of the transverse energy flow and charged particle multiplicity distributions in proton-proton collisions are sensitive to the modeling of soft fragmentation effects, MPI and diffractive interactions. As well as allowing for a deeper understanding of these effects in their own right, the tuning of MC models yields more accurate simulations of the ``underlying event'' - comprising MPI and additional soft interactions between the primary partons in events with a hard perturbative scatter. The dynamics of soft interactions are also important to understand at the LHC due to the large number of soft interactions (pile-up) which occur during every event.
An example of how models can be retuned using these data is shown on Fig.~\ref{fig7:eas:dndetapp}.
On the left-hand side, predictions of pre-LHC models used for air shower simulations (\epos~1.99~\cite{Werner:2005jf,Pierog:2009zt} (solid line), \texttt{QGSJETII}-03~\cite{Ostapchenko:2005nj,Ostapchenko:2006vr} (dashed line), \texttt{QGSJET01}~\cite{Kalmykov:1997te,Kalmykov:1993qe} (dash-dotted line) and \texttt{SIBYLL}~2.1~\cite{Engel:1992vf,Fletcher:1994bd,Ahn:2009wx} (dotted line)) are compared to ALICE data~\cite{Aamodt:2010pp}, while on the right-hand side results are presented for the two
models (\epos~\texttt{LHC}~\cite{Pierog:2013ria} (solid line) and \texttt{QGSJETII}-04~\cite{Ostapchenko:2010vb} (dashed line)) which where retuned using first LHC data.              

By requiring a forward proton to be tagged in a LHC Roman pot based detector, a subset of inelastic interactions are probed which will allow diffraction to be investigated in more detail. This in turn will lead to more accurate tunes and possibly highlight areas of tension where the current phenomenological models are unable to describe the data and would therefore need revisiting. Such samples are especially sensitive to the modeling of the forward regions and will be of use to constrain cosmic-ray air shower physics. 

The CASTOR (CMS) calorimeter provides the unique possibility to minimize the gap in the forward
coverage of detectors at LHC. While other forward charged particle detectors reach up to $|\eta|<5$,
this is extended by CASTOR up to 6.6. For the physics in extensive air showers this is very
important, since the vast majority of primary energy is directed in the very forward phase space. Even more 
forward, the LHCf experiment can measure the neutral particle spectra for the highest pseudorapidities.

Other measurements like the various cross-sections by TOTEM~\cite{Csorgo:2012dm} or the rapidity 
gap distributions~\cite{Aad:2012pw} and many other distributions from CMS, ATLAS, ALICE or LHCb experiments
were taken into account to improve the models used for air shower simulations (see \cite{Pierog:2013ria} 
for \epos~\texttt{LHC}).

    \subsection{Hadronic interaction models and air showers}
    \label{sec7:eas}

Since min-bias measurement of antiproton-proton interactions at Tevatron had large uncertainties,
LHC data provided the first high quality data useable for cosmic ray MC 
since the RHIC measurements at 510\UGeV, thus a gain by about a factor 15 in center-of-mass energy. As
a consequence the modifications of the models due to LHC data have a strong impact on air shower
observables.

\begin{figure}
  \includegraphics[width=.49\textwidth]{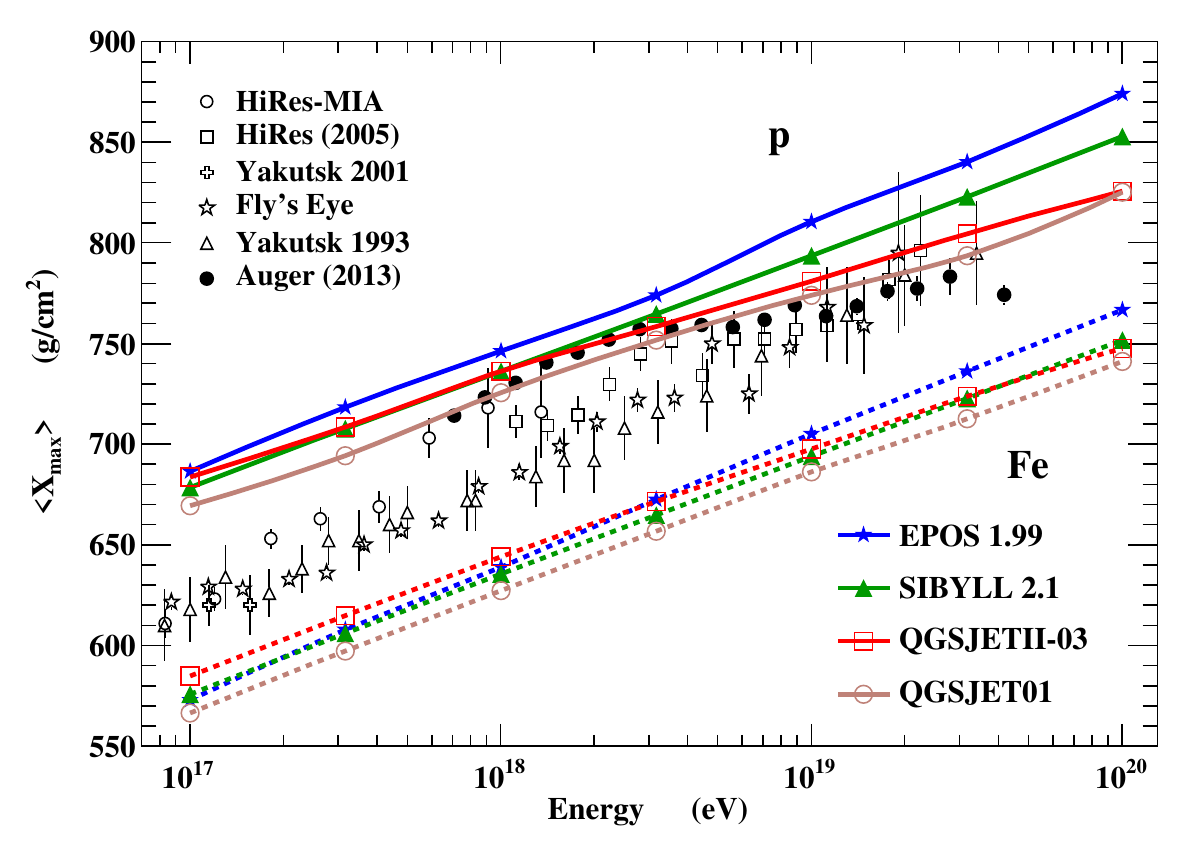}
  \includegraphics[width=.49\textwidth]{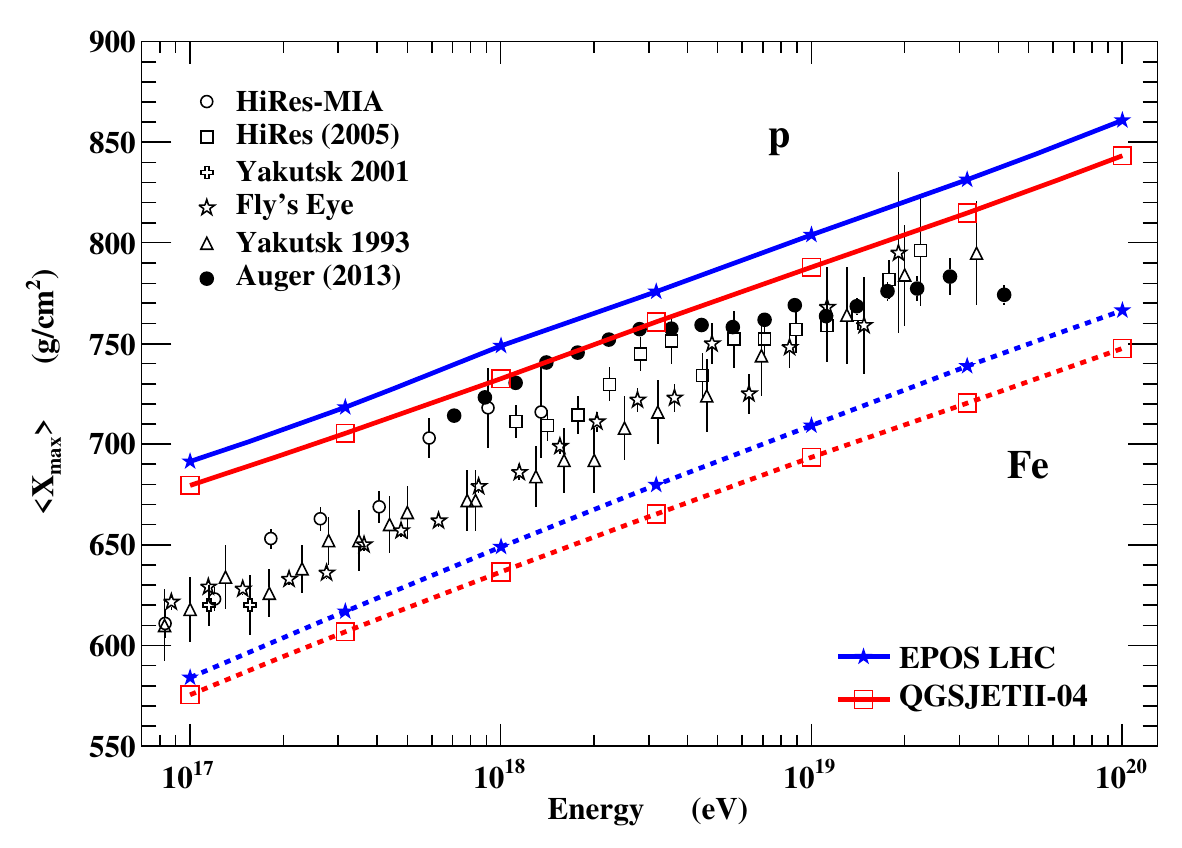}
  \caption{Mean X$_{\rm max}$ for proton and iron induced showers as a 
  function of the primary energy. Predictions of different high-energy 
  hadronic interaction models, full lines for proton and dashed lines for iron 
  with full stars for \epos~1.99, open squares for
  \texttt{QGSJETII}-03, open circles for QGSJET01, and full triangles for 
  \texttt{SIBYLL}~2.1 on top panel and full stars for \epos~\texttt{LHC}, open squares for
  \texttt{QGSJETII}-04 on bottom panel, are compared to data. Refs. to the data 
  can be found in \protect\cite{Bluemer:2009zf} and \protect\cite{Aab:2013ika}.}
  \label{fig7:eas:Xmax}
\end{figure}

One of the most important measurement of air shower property is the depth of shower maximum which
is sensitive to the mass composition of primary cosmic rays. As shown on Fig.~\ref{fig7:eas:Xmax}
left-hand side, the simulations before LHC were such that at the highest energy the difference
between model predictions was almost as large as the maximum range expected between the lightest 
(proton) and the heaviest (iron) element, making any mass composition measurement very difficult.
Furthermore the slope of the mean $X_{\rm max}$ as function of the primary energy, the elongation
rate, was very different between the models. So even a change in composition (change in the slope)
could be interpreted very differently depending on the simulation used.

Using the post-LHC models \epos~\texttt{LHC} and \texttt{QGSJETII}-04, Fig.~\ref{fig7:eas:Xmax} right-hand side,
there is still some remaining difference of about 20\Ug/cm$^{-2}$ (same order than the experimental
systematic error) due to the different predictions of the models for nuclear and pion interaction
(see section~\ref{sec7:pO}) but the elongation rates are now very similar leading to the 
same interpretation of the break in the slope of the mean X$_{\rm max}$.

\begin{figure}
  \includegraphics[width=.49\textwidth]{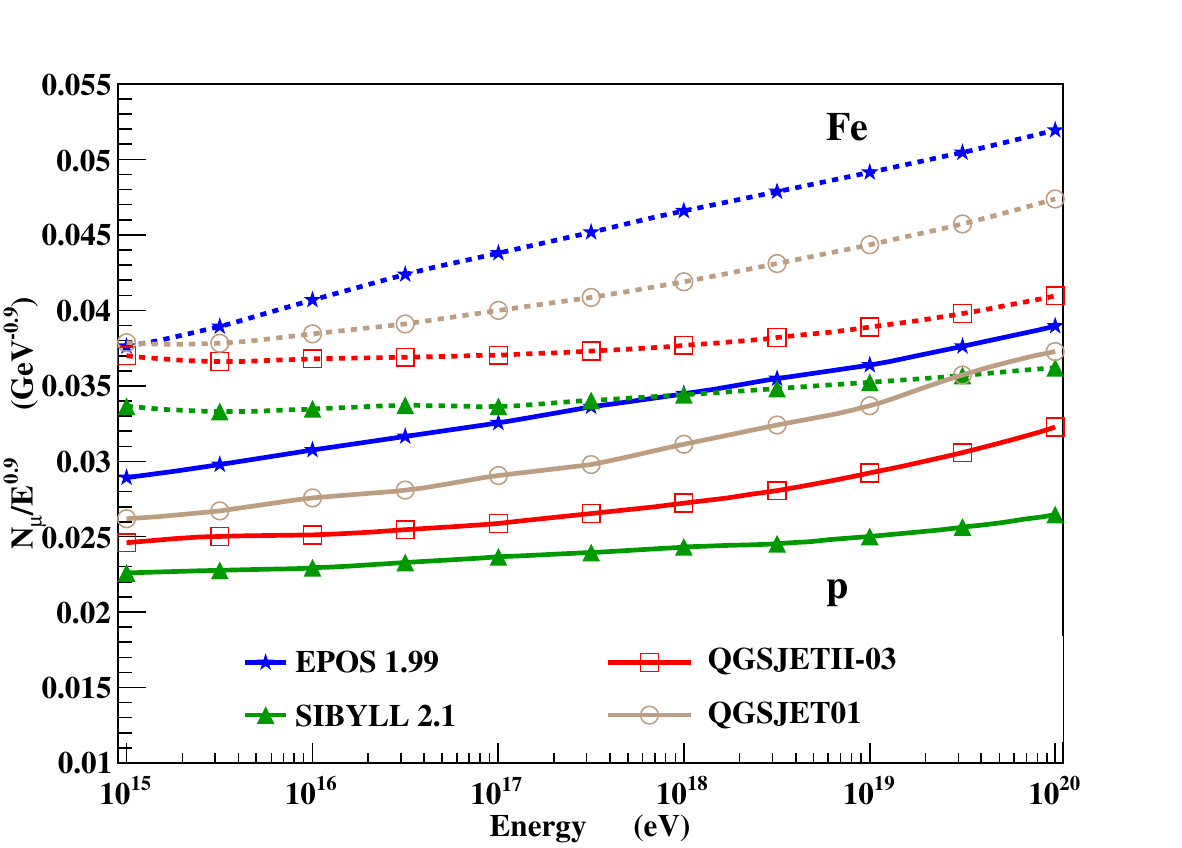}
  \includegraphics[width=.49\textwidth]{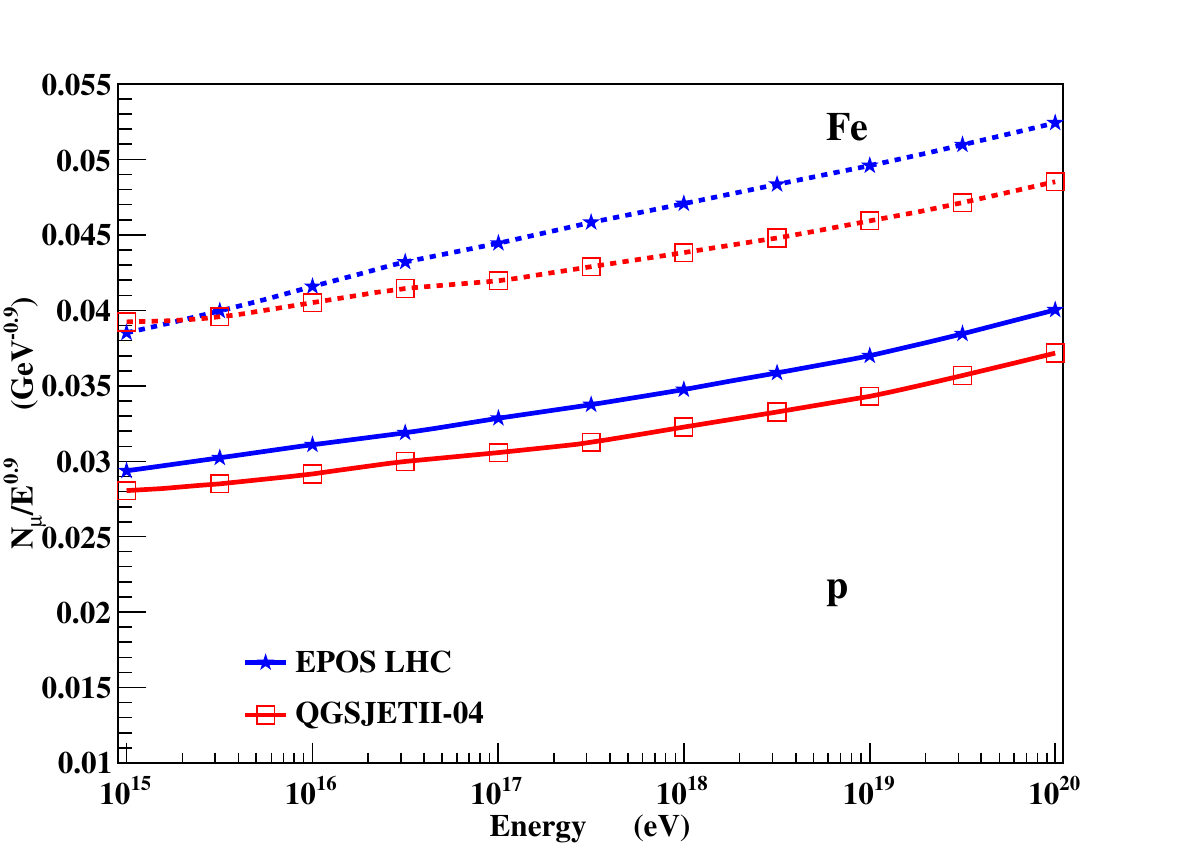}
  \caption{Mean number of muons at ground divided by the primary energy to the power 0.9 for proton and iron induced showers as a 
  function of the primary energy. Predictions of different high-energy 
  hadronic interaction models are presented: full lines for proton and dashed lines for iron 
  with full stars for \epos~1.99, open squares for
  \texttt{QGSJETII}-03, open circles for QGSJET01, and full triangles for 
  \texttt{SIBYLL}~2.1 on left panel and full stars for \epos~\texttt{LHC}, open squares for
  \texttt{QGSJETII}-04 on right panel.}
  \label{fig7:eas:Nmuon}
\end{figure}

Looking at the muon production at ground on Fig.~\ref{fig7:eas:Nmuon}, the situation really
improved a lot from a confusing overlap between proton prediction from \epos~\texttt{LHC} and \texttt{SIBYLL}~2.1
for instance to an almost constant shift of about 7$\%$ between the two post LHC models.

    \subsection{Need for measuring proton-oxygen interactions}
    \label{sec7:pO}

In air showers, all hadronic interactions are with nuclei of air as target particles.
While there are many measurements available and planned for proton-proton interactions, only a few fixed-target
measurements exist for interactions of protons or pion/kaons with light nuclei~\cite{Engel:1999zq}.
This means that, even if we had an arbitrary good understanding of p-p interactions, still a
model-dependent extrapolation is needed to apply this knowledge in simulations of cosmic ray
interactions in the atmosphere.

The data on interactions of heavier nuclei (p-Pb and Pb-Pb)
can be used to improve and tune hadronic interaction models. However, there are a number of collective effects
that are of central importance in heavy ion collisions and much less of relevance
for proton interactions with light ions, in which typically only 3 nucleons participate on average.
This means that the modeling of most of the heavy ion data involves additional effects that will not
help much to understand cosmic ray interactions. The most promising way could be, perhaps, to select peripheral
p-Pb collisions with the same mean number of interaction nucleons as expected for air. This would allow us
to compare inclusive cross sections or other quantities that do not depend on event-by-event fluctuations.
The key point of such a measurement would be that the number of interacting nucleons should not be
determined by any quantity related to secondary particle production
(such as transverse energy, for example) but by the number of spectator nucleons.

Given the lack of heavy ion data selected by the number of spectator nucleons 
and its limited applicability to average quantities, the most
promising way to reduce the uncertainties of describing interactions with light nuclei is 
the direct measurement of p-O interactions at LHC. (O is preferred over N only because it
is used already as carrier ion for accelerating Pb and it is hoped it can be injected without
too high an effort of re-tuning the accelerator.)

    % \subsubsubsection{Model predictions and uncertainties}
    % \label{sec7:predictions}

\begin{figure}[htb!]
  \includegraphics[angle=-90,width=.49\textwidth]{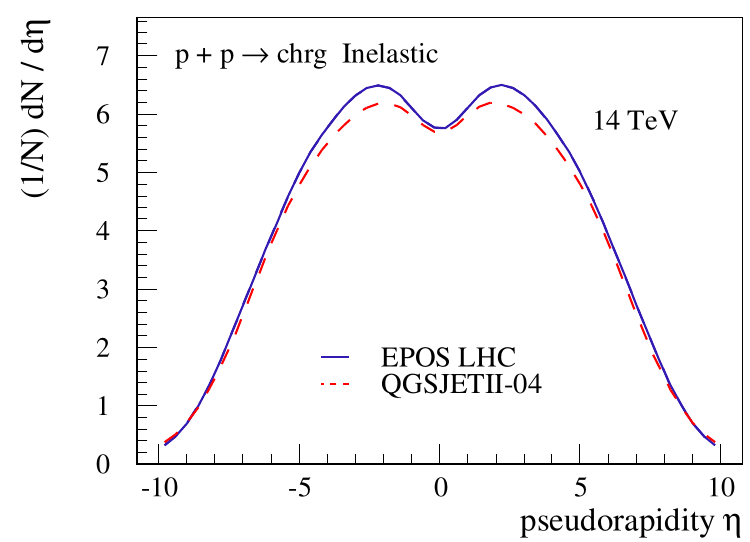}
  \includegraphics[angle=-90,width=.49\textwidth]{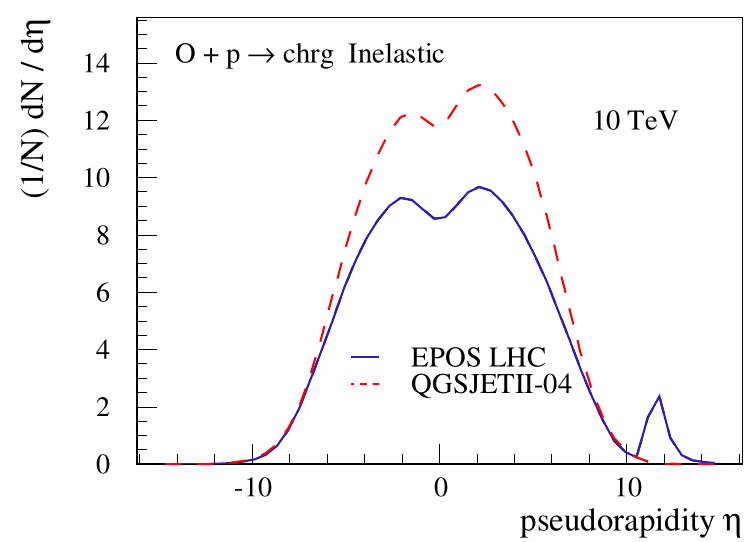}
  \caption{Pseudorapidity distribution $dN/d\eta$ of charged particles for inelastic events for {\it p-p} interactions at 14\UTeV\ on left panel and  {\it O-p} interactions at 10\UTeV\ on right panel. Predictions are from \epos~\texttt{LHC} (solid line) and \texttt{QGSJETII}-04 (dashed line). The difference at very large pseudorapidity is related to spectator nucleons that are not treated in \texttt{QGSJET}.}
  \label{fig7:eas:dndetapO}
\end{figure}

To illustrate the potential gain in accuracy of air shower predictions we first compare the
pseudorapidity distributions of charged particles in p-p interactions with that
predicted for p-O collisions, see Fig.~\ref{fig7:eas:dndetapO}. Only models already tuned to
the new p-p data from LHC are shown. Still there is a difference of $20-30$\% 
between the predictions for, for example, the charged particle multiplicity of p-O interactions.

    % \subsubsubsection{Impact on air showers}
    % \label{sec7:airshowers}

\begin{figure}[htb!]
\begin{centering}
  \includegraphics[width=.6\textwidth]{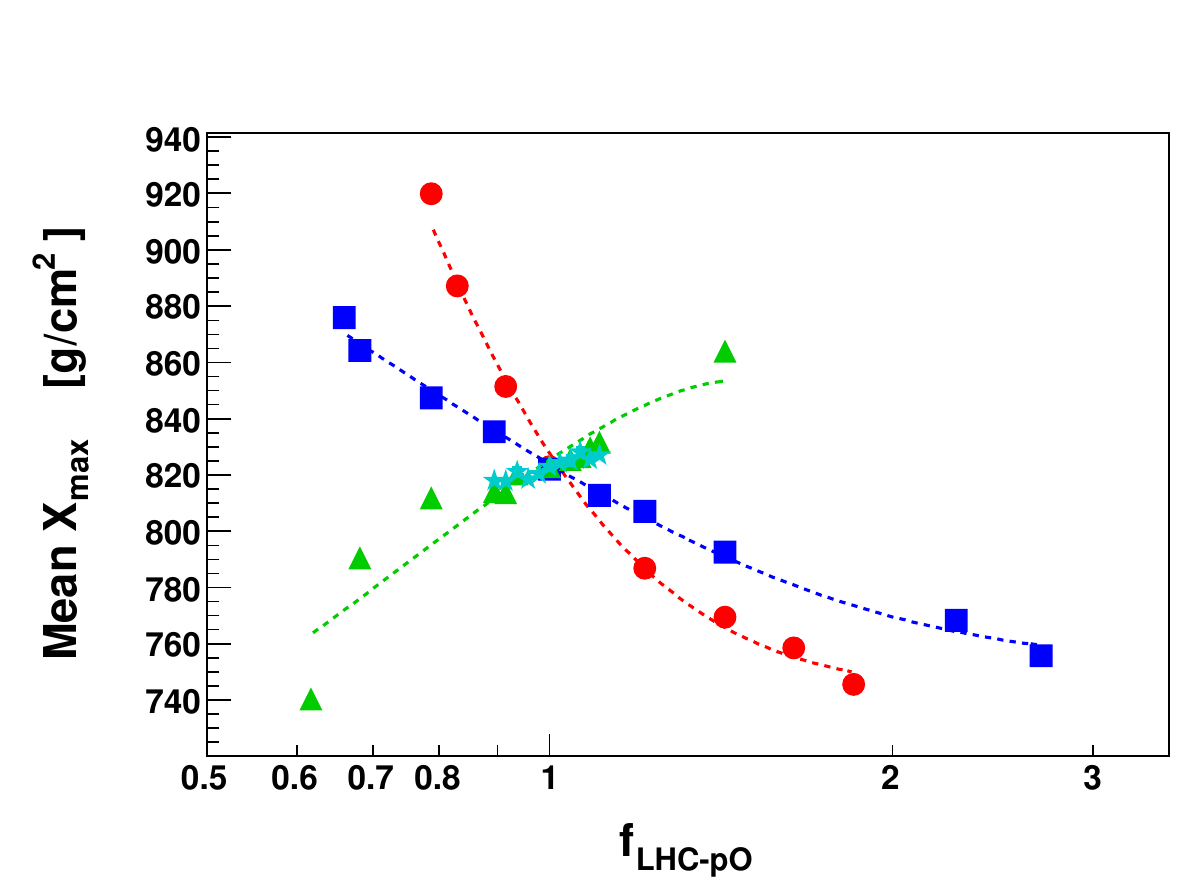}

  \caption{Change of the predicted mean depth of shower maximum, $X_{\rm max}$, as a function of the relative difference between
  the expected and measured quantity for p-O at LHC, f$_{\rm LHC-pO}$. The observables are
  particle production cross section -- red circles,
  charged particle multiplicity -- blue squares,
  energy fraction carried by the most energetic secondary particle (elasticity) -- green triangles,
  and ratio between charged and neutral particles -- green stars.
  The curves have been obtained in the same way as those in~\cite{Ulrich:2010rg}.
  \label{fig7:eas:xmaxmean}
  }
\end{centering}
\end{figure}

These differences in modeling p-O interactions lead to corresponding uncertainties in the predictions
of important air shower observables. This is shown in Fig.~\ref{fig7:eas:xmaxmean} considering the mean depth of
shower maximum, one of the observables typically used to determine the mass composition of cosmic rays.
The curves give the change in the predicted $X_{\rm max}$ as function of the relative difference
between the nominal model prediction at LHC energy and a potential measurement. The largest dependence is found for
the p-O cross section for particle production. If this cross section were $10$\% smaller than current expectations
the predicted depth of shower maximum would shift by $\sim 30\,{\rm g/cm^2}$, more than the difference between
proton and helium primaries. Similarly deviation of the particle multiplicity by $30$\% would correspond to a
change equivalent to going from proton to helium as primary cosmic ray particles.

Whereas the LHC data from Run-1 have made it possible to distinguish more reliably between proton and iron primaries,
see Figs.~\ref{fig7:eas:Xmax} and \ref{fig7:eas:Nmuon}, the direct measurement of p-O interactions could improve
the separation power to the level of the proton-helium difference.

%% file: cosmic/EnergyFlow.tex
The energy flow of hadronic interactions is one of the most important component of air
shower simulations. The whole cascade development depends on the way the energy is transferred from
one generation to the other. For air showers the total energy flow is important
while in some cases only the transverse energy flow has been measured by LHC
experiment. Nevertheless this measurement is important to understand the energy
transfer in the hadronic interaction models which are used for air shower 
simulations.

\subsection{Past Measurements of Energy Flow}
\label{sec7:EnergyFlow:past}

Measurements of the transverse energy flow in minimum bias
interactions have been performed by the ATLAS~\cite{Aad:2012mfa},
CMS~\cite{Chatrchyan:2011wm}, and LHCb~\cite{Aaij:2012pda}
experiments, in all cases without the use of forward proton tags.
Instead, based on event triggers requiring minimal activity in the
rather central regions covered by the detectors, calorimeters are used
to measure the energy flow in inelastic proton interactions over a
large angular range. The contributions from neutral particles are in
all cases included.

% can't we be consistent here? Mention trigger requirements for both
% ATLAS and CMS, and mention either selection criteria like CMS
% (tracks and vertices) or particle level criteria like ATLAS

The ATLAS measurement at $\sqrt{s} = 7 \TeV$ used an event
selection based on the number of charged particles $\Nch \geq 2$ with
$\pt > 200 \MeV$ and $|\eta| < 2.5$. The measured differential
transverse energy flow was corrected to a particle level definition
based on the transverse energy of all stable charged (neutral) final
state particles with $p > 500 (200) \MeV$ and $|\eta| < 4.8$.

The CMS measurement was performed on events with at least four
tracks ($\pt > 75 \MeV $) associated to a primary vertex and with
signals in the forward and backward BSC ($3.9 < |\eta| < 4.4$).
The data were corrected to a particle level definition of all charged
and neutral stable final state particles within $3.15 < |\eta| < 4.9$,
excluding muons and neutrinos. Additionally one or more charged
particles were required within the forward and backward acceptance of
the BSC to replicate the detector-level definition.

% transverse momentum or momentum range covered?
%The LHCb measurement selected events with one or more
%reconstructed particles in the range $1.9 < \eta < 4.9$. Data were
%corrected to a particle level definition of all stable charged and
%neutral particles within this range.

The LHC beauty experiment (LHCb) has measured the energy flow
in the pseudorapidity range $1.9 < \eta < 4.9$ with data collected
by the LHCb experiment in pp collisions at $\sqrt{s}=$ 7\UTeV\ for 
inclusive minimum-bias interactions, hard scattering processes
and events with enhanced or suppressed diffractive 
contribution~\cite{Aaij:2012pda}.
In this study, the primary measurement is the energy flow carried by 
charged particles. For the measurement of the total energy flow,
a data-constrained MC estimate of the neutral component is used.
The energy flow is found to increase with the momentum transfer in an
underlying pp inelastic interaction. The evolution of the energy flow
as a function of pseudorapidity is reasonably well reproduced by the 
\texttt{PYTHIA}-based and cosmic-ray interaction models. Nevertheless, the 
majority of the \texttt{PYTHIA} tunes underestimate the measurements at large 
pseudorapidity, while most of the cosmic-ray interaction models 
overestimate them, except for diffractive enriched interactions. For 
inclusive and non-diffractive enriched events, the best description of 
the data at large $\eta$ is given by the \texttt{SIBYLL}~2.1 and \texttt{PYTHIA}~8.135 
generators. The latter also provides a good description of the energy 
flow measured with diffractive enriched events, especially at large 
$\eta$. The comparison shows that the absence of hard diffractive 
processes moderates the amount of the forward energy flow meaning that 
their inclusion is vital for a more precise description of partonic 
interactions. It also demonstrates that higher-order QCD effects as 
contained in the Pomeron phenomenology play an important role in the 
forward region. None of the event generators used in this analysis
are able to describe the energy flow measurements for all event
classes that have been studied.

All experiments applied subsequently harder selection criteria, such
as additionally requiring a high-\pt\ particle or di-jet event. This allow a
better understanding of the underlying physics in the models.

\subsection{Future Measurements of Energy Flow}
\label{sec7:EnergyFlow:future}

Future measurements of the energy flow in ATLAS will include a forward proton
tag. MC predictions for the pseudorapidity differential
transverse energy density at $\sqrt{s} = 14 \TeV$ are presented for
the ATLAS inclusive selection ($\Nch \geq 2$ with $\pt > 200
\MeV$, $|\eta| < 2.5$) and for a sample of events where the only event
selection is exactly one tagged forward proton in either the AFP
or ALFA detectors (see Section~\ref{sec:timm:protontagging} for
forward proton selection and MC details). In both cases, the event
averaged transverse energy density is calculated from all
charged~(neutral) final state particles with $p > 500 (200) \MeV$. In
order to preserve any asymmetry which may be modeled in the energy
flow, the tagged proton in the AFP and ALFA triggered
samples is required to be at $+z$. Events where the tag is at $-z$ are
therefore inverted along the $z$ axis.

\begin{figure}
\centering
\begin{subfigure}[b]{0.47\textwidth}
\includegraphics[width=\textwidth]{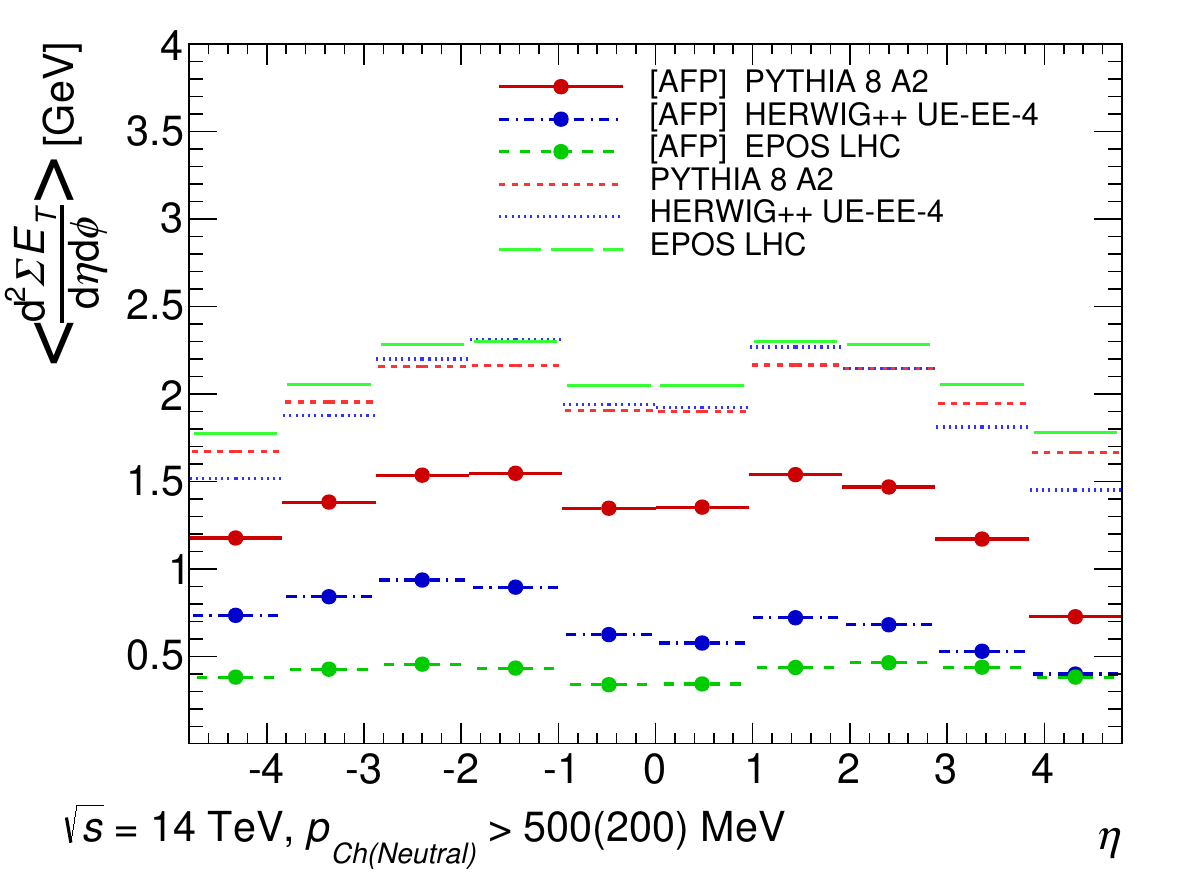}
\caption{}
\label{fig:forwardEnergy_1:a}
\end{subfigure}
\hspace{5mm}
\begin{subfigure}[b]{0.47\textwidth}
\includegraphics[width=\textwidth]{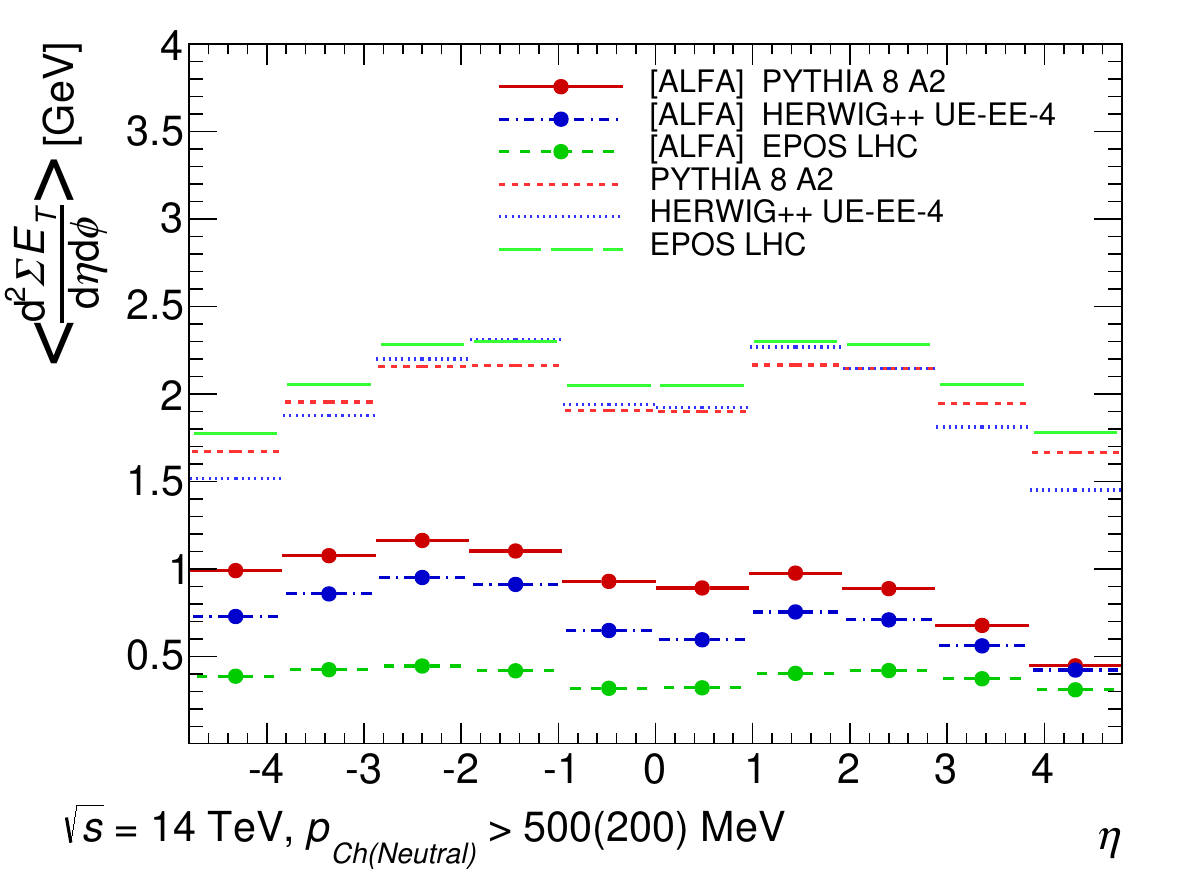}
\caption{}
\label{fig:forwardEnergy_3:b}
\end{subfigure}
\caption{Transverse energy density in the range $|\eta| < 4.8$
  compared between MC models, (a) for the inclusive and AFP
  selections and (b) for the inclusive and ALFA selections (see
  text). }
\label{fig:forwardEnergy_1}
\end{figure}

In Fig.~\ref{fig:forwardEnergy_1}, the transverse energy density from
the ATLAS inclusive selection is compared to an event selection
requiring exactly one tagged forward proton in either AFP or
ALFA. The model spread for ATLAS central event selection is
shown to be relatively small, 15\% at central-pseudorapidity and 20\%
at forward-pseudorapidity. The dip at central-pseudorapidity is
predominantly driven by the exclusion of very low momentum particles
which make up a larger fraction of the energy flow in this region
where $p \approx \pt$. \epos consistently predicts the largest average
energy density while \hpp predicts the lowest.

Upon requiring a forward proton tag, the situation changes
dramatically. For an AFP tag, the predicted energy density from
all models is significantly lower, with \epos now predicting the
lowest average energy density, \pyeight predicting the highest - and
differing from \epos by a factor of 4.2. The distributions are also of
interest for their asymmetry -- it was explored above
(Section~\ref{sec:timm:forwardprotonsources}) that a fraction of
these proton tags from \pyeight originate from low mass double
dissociation giving an independent probe of the dissociation of the
proton traveling in the $-z$ direction over a wide range of possible
diffractive masses. Such topologies are expected to deposit more
energy at negative pseudorapidity due to the presence of the
pseudorapidity gap between the two diffractive systems. In \pyeight
this results in the transverse energy density in the $-4.80 < \eta <
-3.84$ region being 70\% greater than the opposite $3.84 < \eta <
4.80$ region. Although with less overall activity, \hpp also predicts
a 75\% increase in the transverse energy density between the two
forward regions while \epos predicts zero asymmetry in events with an
AFP tag, probably because in \epos the independent remnant 
scheme~\cite{Liu:2002gw} imply a different proton spectrum and the 
AFP tag might select different kind of diffractive events (like
central diffraction which is symmetric).

When the proton tag is required to be within ALFA acceptance,
both \hpp and \epos change relatively little - with \epos developing a
small asymmetry of around a 30\% increase in the $-z$ direction (\hpp
remains the same at a 75\% increase). The predicted activity from
\pyeight is substantially reduced with the average central transverse
energy density falling from 1.35 to 0.4 \UGeV. The predicted asymmetry
is also increased to a factor of 2.4 difference between the edge bins.

\begin{figure}
\centering
\begin{subfigure}[b]{0.46\textwidth}
\includegraphics[width=\textwidth]{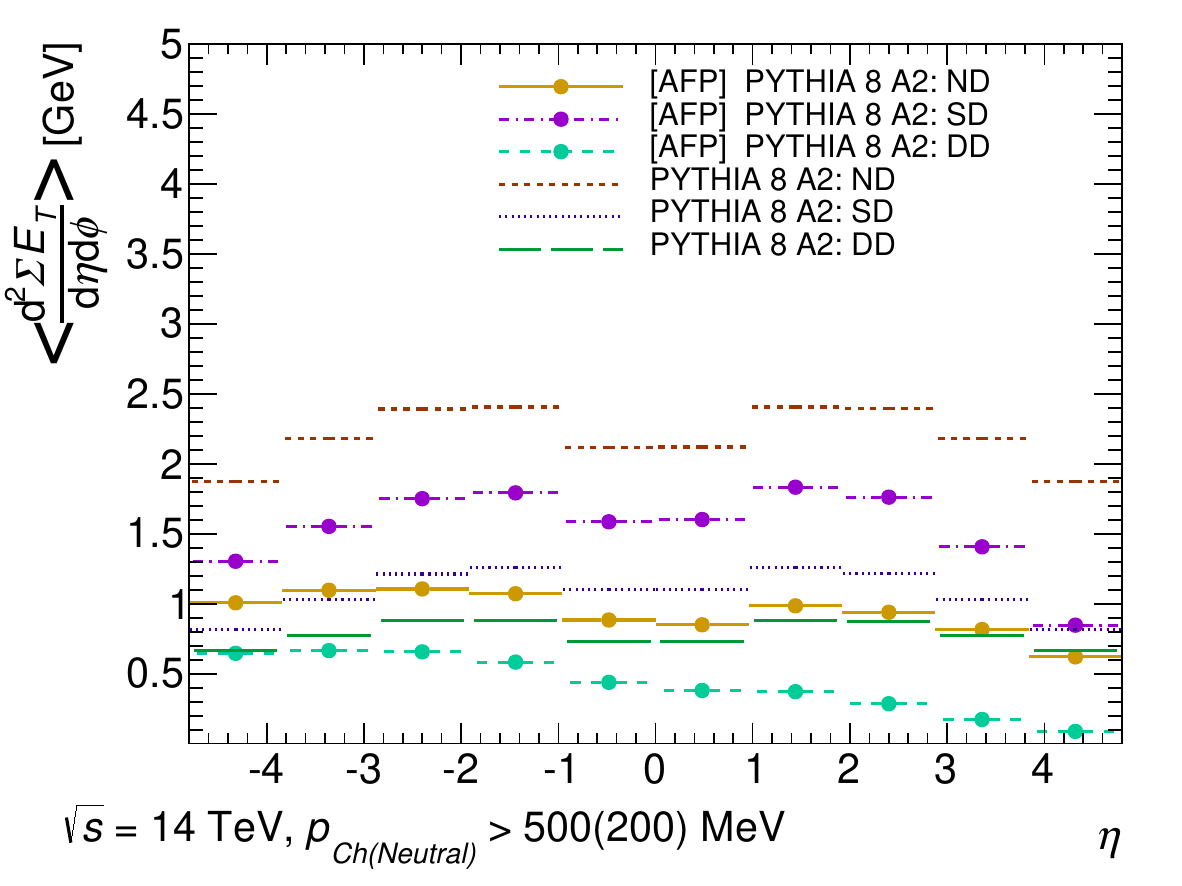}
\caption{}
\label{fig:forwardEnergy_2:a}
\end{subfigure}
\hspace{5mm}
\begin{subfigure}[b]{0.46\textwidth}
\includegraphics[width=\textwidth]{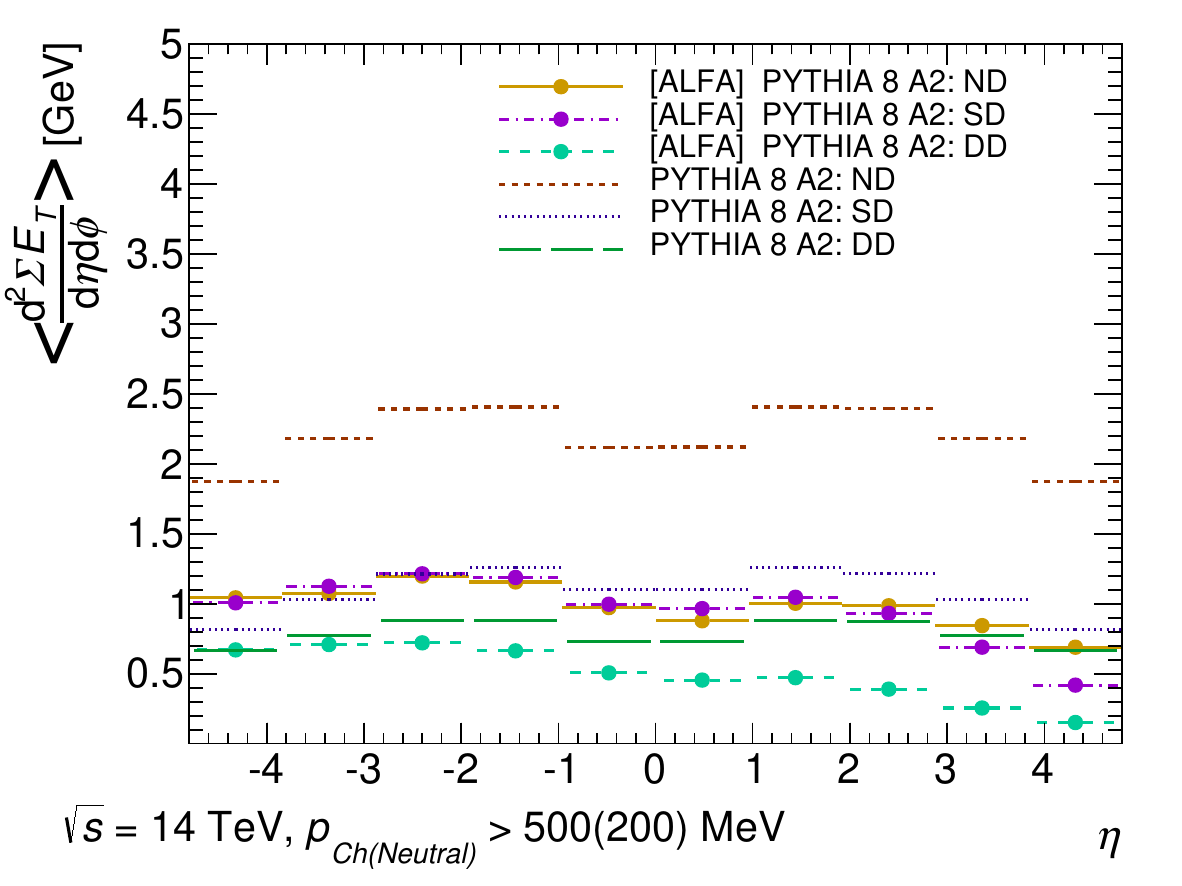}
\caption{}
\label{fig:forwardEnergy_4:b}
\end{subfigure}
\caption{Transverse energy density in the range $|\eta| < 4.8$
  compared between single-, double- and non-diffractive components of
  the inelastic cross section from \pyeight, (a) for the inclusive and
  AFP selections and (b) for the inclusive and ALFA
  selections (see text). }
\label{fig:forwardEnergy_2}
\end{figure}

\pyeight generates samples of pure single-, double- and
non-diffractive interactions. Using these, the generator's proposed
mechanisms which drive these changes are explored in more
detail. Figure~\ref{fig:forwardEnergy_2} illustrates that the dramatic
overall drop in the predicted transverse energy density is driven by
the single diffractive component. AFP has good acceptance for
high mass single diffractive events whereas in ALFA all masses
are accessible but are suppressed by the small but non-zero
requirement on $p_{\textrm{T,}p}$. In both cases it is indeed the
double-diffractive events which display the largest asymmetry, with a
factor of 7 increase in the transverse energy density at $-z$ compared
to $+z$. It must be noted however that double dissociation measured to
make up only around 5\% (see TOTEM references
\cite{Antchev:2013any, Antchev:2013iaa}) of the inelastic cross
section and hence the asymmetries observed in single dissociation
(increase of 50\%) and in non-diffractive events (increase of 60\%)
are also of importance when comparing the inclusive distributions.

Having larger differences between models thanks to the proton tagging means
that it is easier to identify the sources of discrepancy between them 
and hence to improve our understanding of the underlying physics. As a 
consequence, the predictive power of such model can be really improved and
reduce uncertainties in air shower simulations.

\begin{figure}
  \begin{center}
    \includegraphics[width=.525\linewidth]{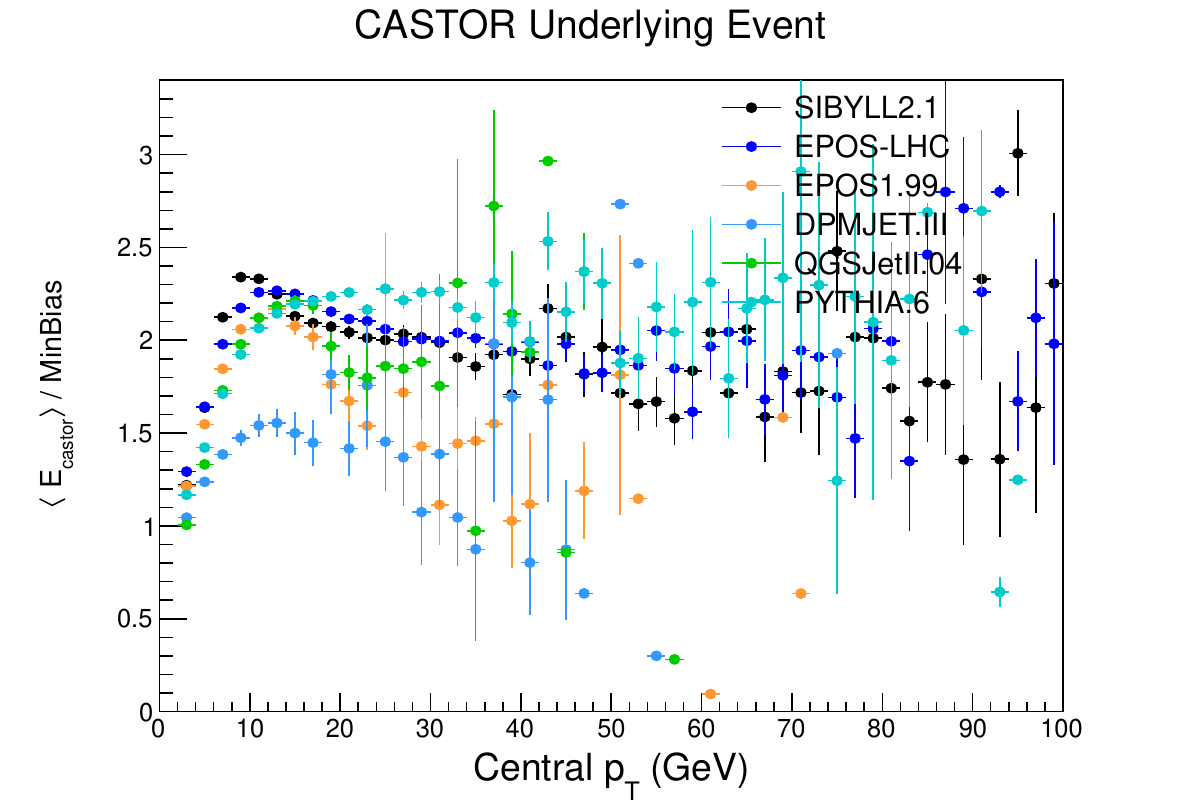}
    \includegraphics[width=.375\linewidth]{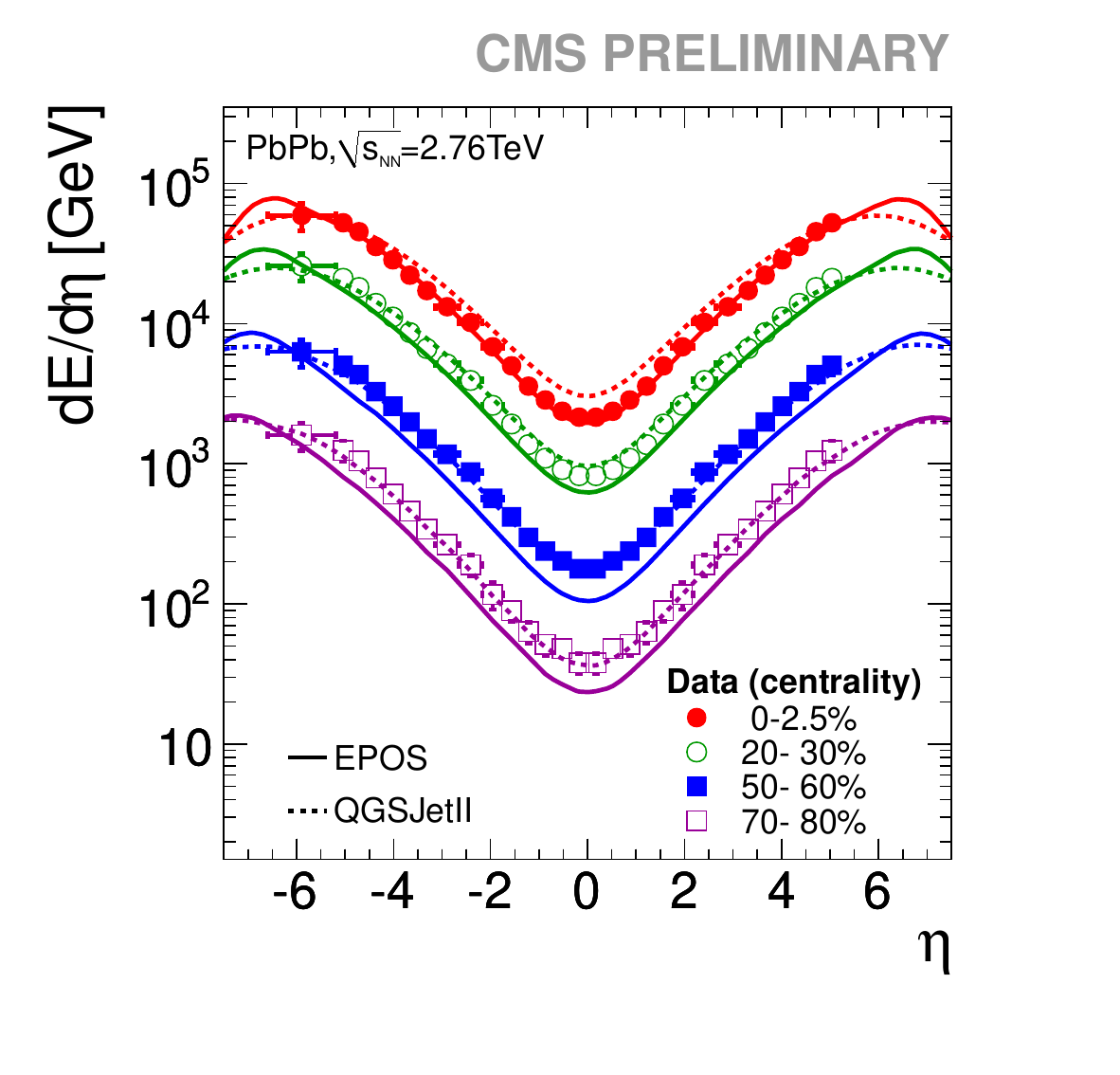}
    \caption{ Left panel: Expected CASTOR underlying event measurement at 13\UTeV.
Shown is the average CASTOR response normalized to the inclusive minimum bias response as a
function of central jet $p_T$. Right panel: Measured
dE/d$\eta$ over the full CMS phase space, from -6.6 to +5 in PbPb data at $\sqrt{s_{NN}}$ of 2.76\UTeV
\protect\cite{Wohrmann:2013nta}}
    \label{fig7:ulrich:ue}
  \end{center}
\end{figure}

For CASTOR experiment, one of the first measurements of Run-2 may be the underlying event in very forward direction
\cite{Chatrchyan:2013gfi}. The model predictions for this measurement at 13\UTeV\ are shown in Fig.~\ref{fig7:ulrich:ue} (left). For this analysis data on the order of 2\Unb$^{-1}$\ at a pileup of
$\mu$<0.05 are needed. Furthermore, after a full high statistics halo muon intercalibration of all
channels of CASTOR, which can be done using the scraping runs before the first technical stop of
LHC (TS1), also an absolute energy measurement will be possible and dE/d$\eta$ over the full
acceptance of CMS $-6.6<\eta<+5$, can be performed, see an example of this in Fig.~\ref{fig7:ulrich:ue} (right).

\begin{figure}
\centering
\includegraphics[width=0.5\textwidth]{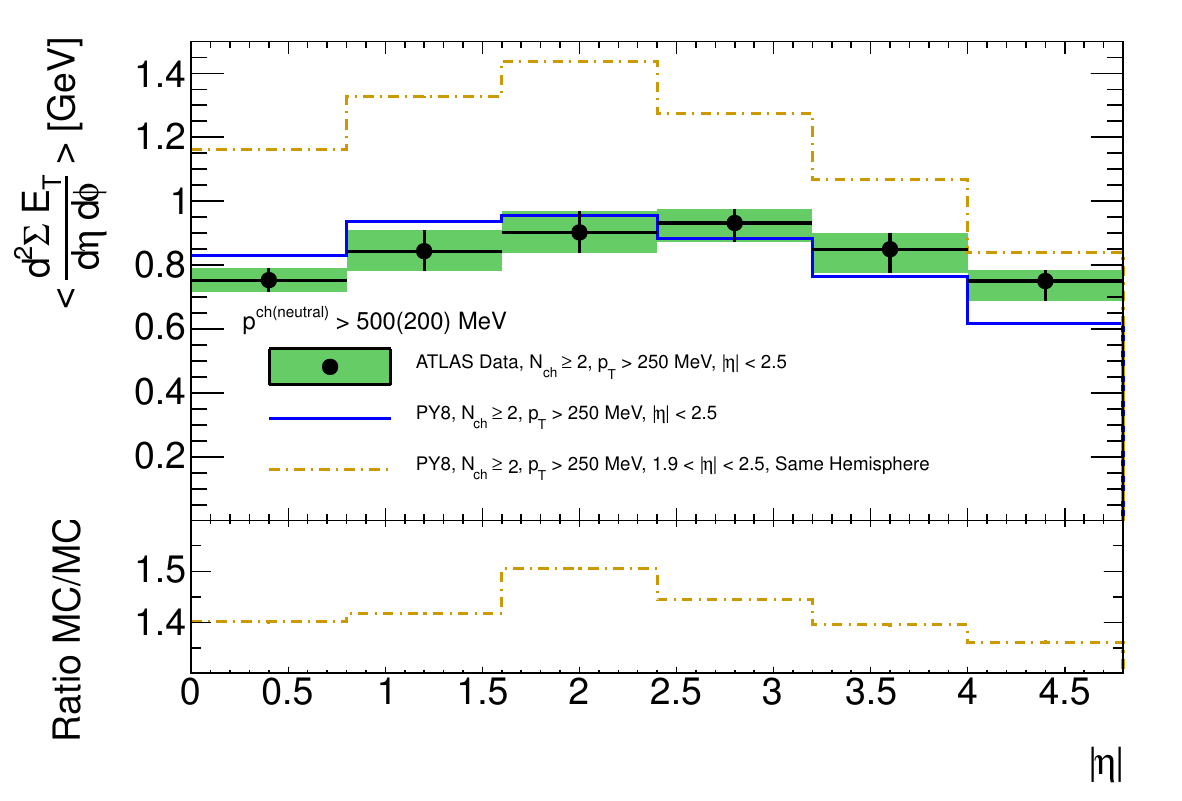}
\caption{The differential transverse energy flow as measured by
  ATLAS at $\sqrt{s} = 7 \TeV$ compared to \pyeight tune 4C. The
  dashed line denotes the prediction of \pyeight 4C using the new
  common selection as proposed in the text.}
\label{fig:eflow_common}
\end{figure}

The full dE/d$\eta$ distribution measured in p-p interaction at 13\UTeV will
be of primary importance to test and improve models used for air shower 
simulation.
\paragraph*{Common Fiducial Definition for Energy Flow Measurement}
The current $\sqrt{s} = 7 \TeV$ measurement of the differential energy
flow performed by LHC experiments are not directly comparable due
to the different choices in event and kinematic selections detailed
above. A common definition accessible to ATLAS, CMS and
LHCb for use in future analyzes in addition to the experiment's
preferred selection will allow for direct comparisons in the regions
of overlapping pseudorapidity between the experiments. The proposed
common selection is as follows:
\begin{itemize}
\item For each event, treat the $+z$ and $-z$ hemispheres separately
  (does not apply to LHCb). 
\item Per hemisphere, require $\Nch \geq 2$ with $\pt > 250 \MeV$ and
  $\pm1.9 < \eta < \pm2.5$.
\item Measure hemisphere transverse energy flow differential in
  pseudorapidity.
\item Unfold measured transverse energy flow to a hadron level
  definition of all charged~(neutral) stable particles with $p >
  500 (200) \MeV$.
\end{itemize}
An example of this new definition is presented in
Fig.~\ref{fig:eflow_common}, here ATLAS experimental data from
reference~\cite{Aad:2012mfa} and the corresponding prediction from
\pyeight 4C are shown along side the \pyeight 4C prediction for this
proposed common selection. The mean energy flow for $|\eta| < 4.8$ is
predicted to be approximately 40\% larger by MC, peaking in the
region of the new common event selection requirement. With this new selection,
events with a very central activity and low transverse energy flow are excluded.

%% file: cosmic/Multiplicity.tex
After the energy flow, the particle multiplicity is a very important ingredient of
air shower development since it impacts directly the speed at which the cascade
grows in the atmosphere having a direct influence on the position of the shower maximum. To improve the model prediction not only the measurement at higher energy is important but the separation between diffractive and non-diffractive events with an extended $\eta$ coverage is fundamental to reduce significantly the
model uncertainties.

\subsection{Past measurements of charged particle multiplicities}
The LHC collaborations have published charged particle
multiplicity spectra at various center of mass energies spanning the
range $\sqrt{s} = $ 900--7000\UGeV\ \cite{Aamodt:2010pp, Aad:2010ac,
  Khachatryan:2010nk, Aaij:2011yj, Aspell:2012ux}. Complimentary
acceptances of the different LHC experiments allow for the
charged particle multiplicity to be investigated over the ranges
$|\eta| < 2.5$ (ALICE (partial), ATLAS and CMS), $2.0 <
\eta < 4.5$ (LHCb), $3.1 < |\eta| < 4.7$ and $5.3 < |\eta| < 6.5$
(TOTEM, telescopes $T1$ and $T2$). In addition, the definition of
common selection requirements decided by the minimum bias and
underlying event LPCC working group have allowed direct comparisons
between the experiments.

\begin{figure}[tbp]
\centering % \begin{center}/\end{center} takes some additional vertical space
\includegraphics[width=0.53\textwidth]{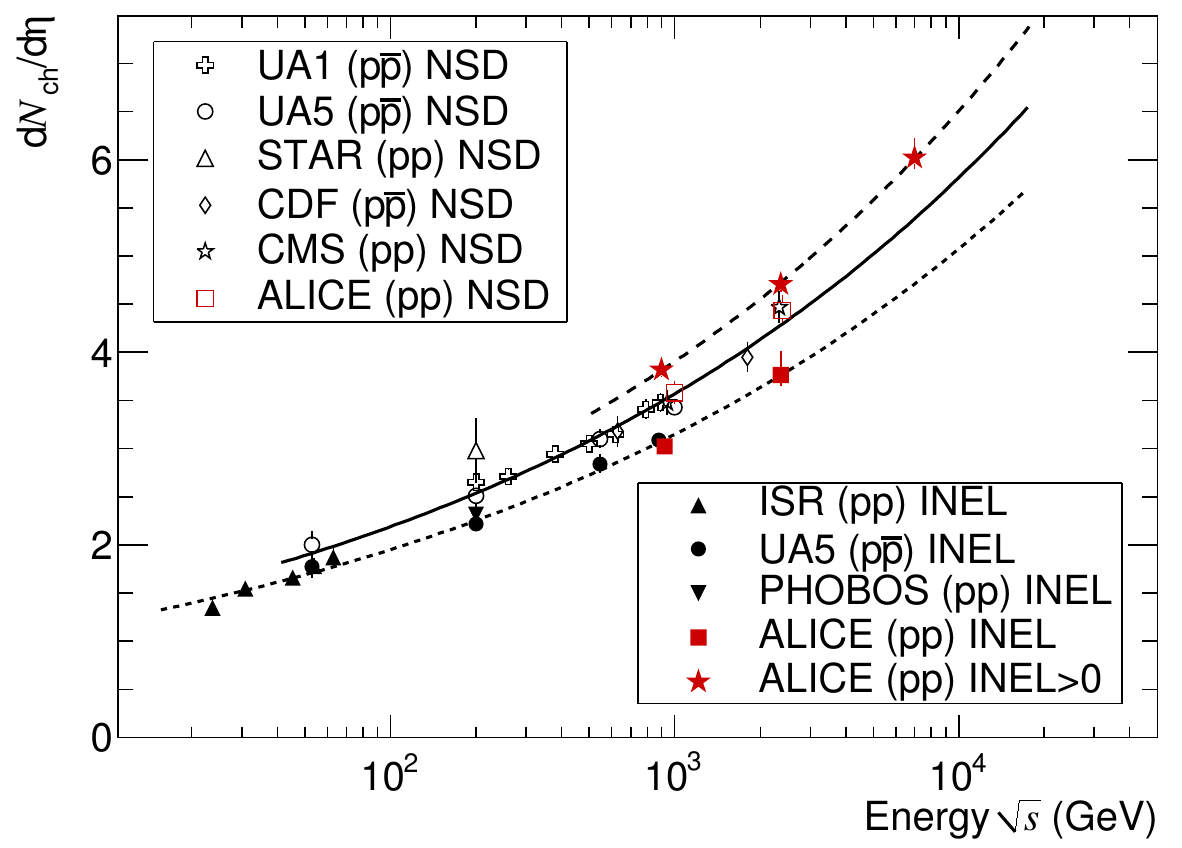}
\includegraphics[width=0.42\textwidth]{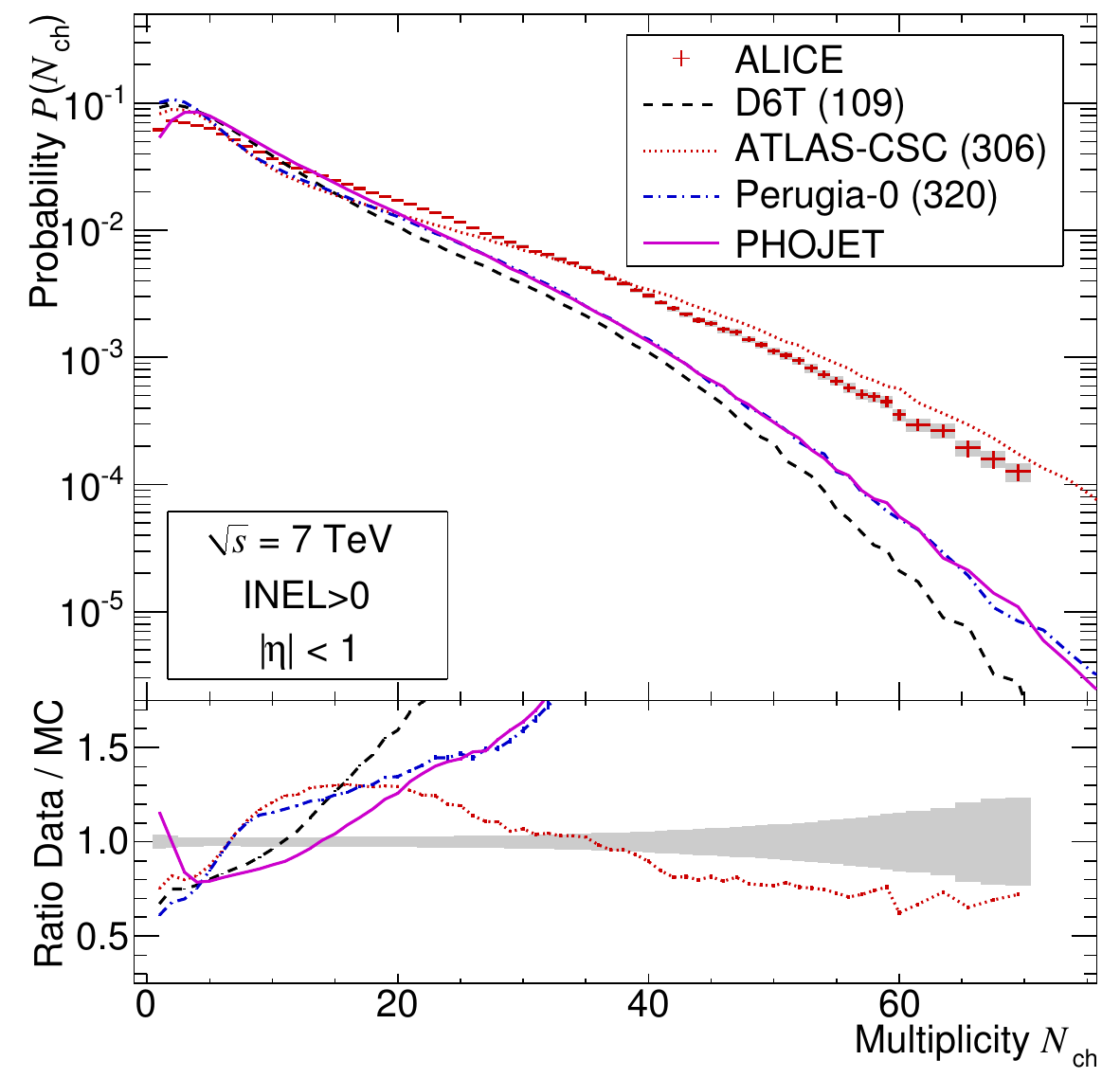}
\caption{\label{fig:ALICEmult} Left: The charged-particle density as a function of center-of-mass energy as measured by ALICE is compared to results from other experiments. Right: The multiplicity distribution of charged particles at mid-rapidity measured by ALICE at $\sqrt{s}=7$\UTeV\ is compared to several MC models. Figures from \cite{Aamodt:2010pp} }
\end{figure}

The ALICE Collaboration has measured the density of charged particles at mid-rapidity and the multiplicity distribution in pp collisions at 
0.9\UTeV, 2.36\UTeV~\cite{Aamodt:2009aa,Aamodt:2010ft} and 7\UTeV~\cite{Aamodt:2010pp}. The results are presented for non-single diffractive and for inelastic processes. The charged-particle density is presented as a function of energy in the left panel of Figure \ref{fig:ALICEmult}. ALICE measurements for the inelastic case with at least one charged particle in the pseudo rapidity range $|\eta|<1$ are compared to those of several other collaborations. The measurements are well described by a power-law increase with energy and reach up to 6 charged particles per unit of pseudo rapidity at mid rapidity for the highest energies. The left panel of the same figure shows the  multiplicity distribution  measured at 7\UTeV\ for inelastic events and compares it to predictions of different MC models. Data are reasonably well described by a Negative Binomial Distribution, which slightly under(over) estimates data at small (large) multiplicities. 

The CMS and TOTEM experiments have measured the charged particle multiplicity as
function of $|\eta|$ at a center-of-mass energy of $\sqrt{s}=8$\UTeV\ in three different event samples
based on the T2 event topology: an inclusive pp interaction sample that includes more than 
90 \% of all inelastic pp collisions,  one sample depleted and one sample enhanced in single diffractive 
pp interactions~\cite{Chatrchyan:2014qka,Antchev:2014lez}. The measurement is performed  in 
$|\eta|<2.2$ for $p_T>100$\UMeV/$c$ and in $5.3<|\eta|<6.4$ for $p_T>40$\UMeV/$c$ with at least one charged particle per event
required in $5.3<|\eta|<6.5$ range and then corrected down to $p_T=0$\UMeV/$c$. The measurement is compared to several models used to describe 
high-energy hadronic interactions. None of the models considered provide a consistent description 
of the measured distributions.

By using the full spectrometer information, the previous
LHCb results~\cite{Aaij:2011yj} have been extended to include momentum
dependent measurements. The LHCb experiment has measured charged particle multiplicities
and mean particle densities as functions of $p_{\rm T}$ and $\eta$
in inclusive pp interactions at a center-of-mass energy of
$\sqrt{s}=7$\UTeV~\cite{Aaij:2014pza}. The measurement is performed
in the kinematic range $p>2$\UGeV/$c$, $p_{\rm T}>0.2$\UGeV/$c$ and 
$2.0<\eta<4.8$, in which at least one charged particle per event
is required. 
The comparison of data with predictions from \texttt{PYTHIA}-based and 
\texttt{Herwig}-based event generators shows that predictions from recent 
generators, tuned to LHC measurements in the central rapidity region, 
are in better agreement than predictions from older generators.
While the phenomenology in some kinematic regions is well described by 
recent \texttt{PYTHIA} and \hpp simulations, the data in the higher
$p_{\rm T}$ and small $\eta$ ranges of the probed kinematic region are 
still underestimated. None of the event generators considered in this 
study are able to describe the entire range of measurements.

Such data on the charged particle multiplicity spectrum and
pseudorapidity dependence, charged particle \pt\ spectrum, and
correlation between the average charged particle \pt\ and the charged
particle multiplicity continue to be used in the tuning of soft
inelastic MC models.

\subsection{Future measurements of charged particle multiplicities}
The effect of requiring a
forward proton tag on the charged particle multiplicity spectra  is investigated in $\sqrt{s} = 14 \TeV$\  MC for an
ATLAS inclusive selection ($\Nch \geq 2$, $\pt > 100 \MeV$,
$|\eta| < 2.5$) and for an event selection requiring exactly one
forward proton tag in either AFP or ALFA (see
Section~\ref{sec:timm:protontagging} for proton tagging and MC
details).

\begin{figure}
\centering
\begin{subfigure}[b]{0.47\textwidth}
\includegraphics[width=\textwidth]{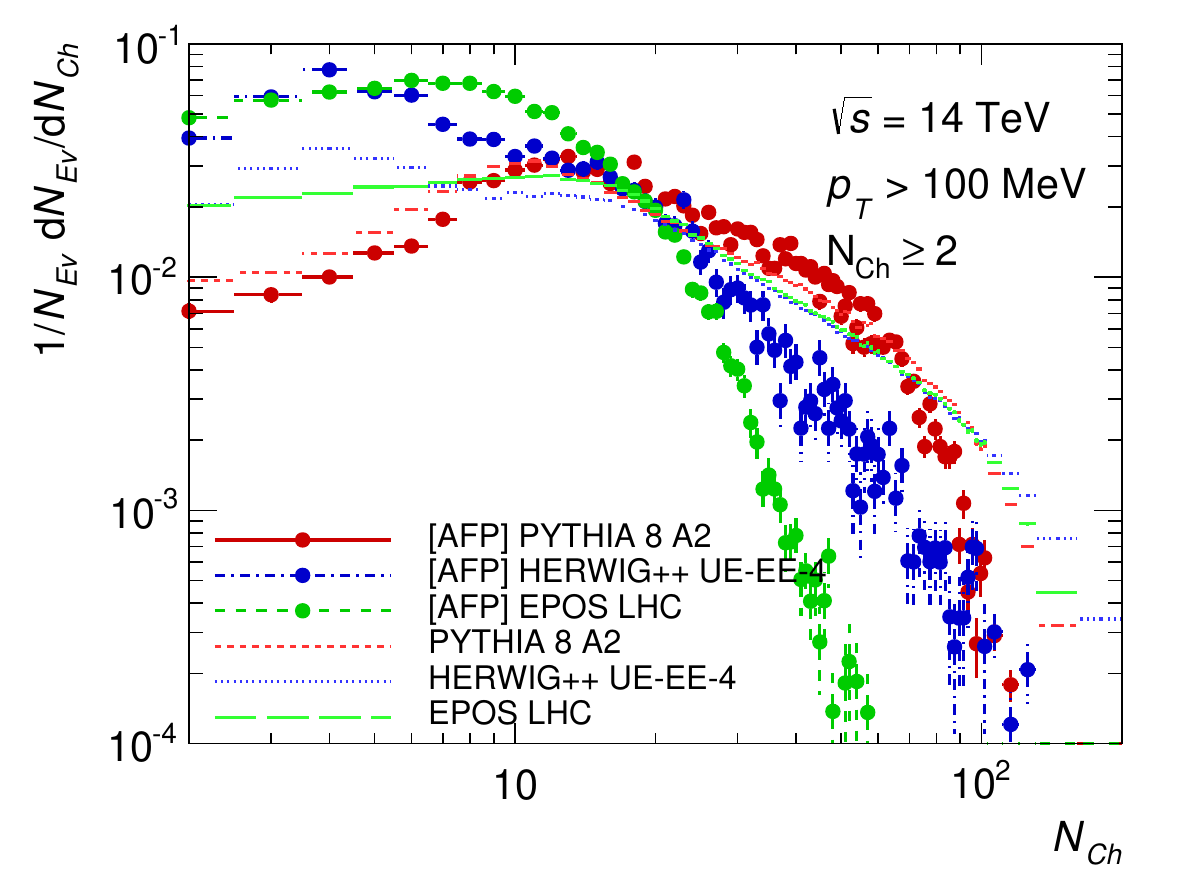}
\caption{}
\label{fig:ChargedParticle_3:a}
\end{subfigure}
\hspace{5mm}
\begin{subfigure}[b]{0.47\textwidth}
\includegraphics[width=\textwidth]{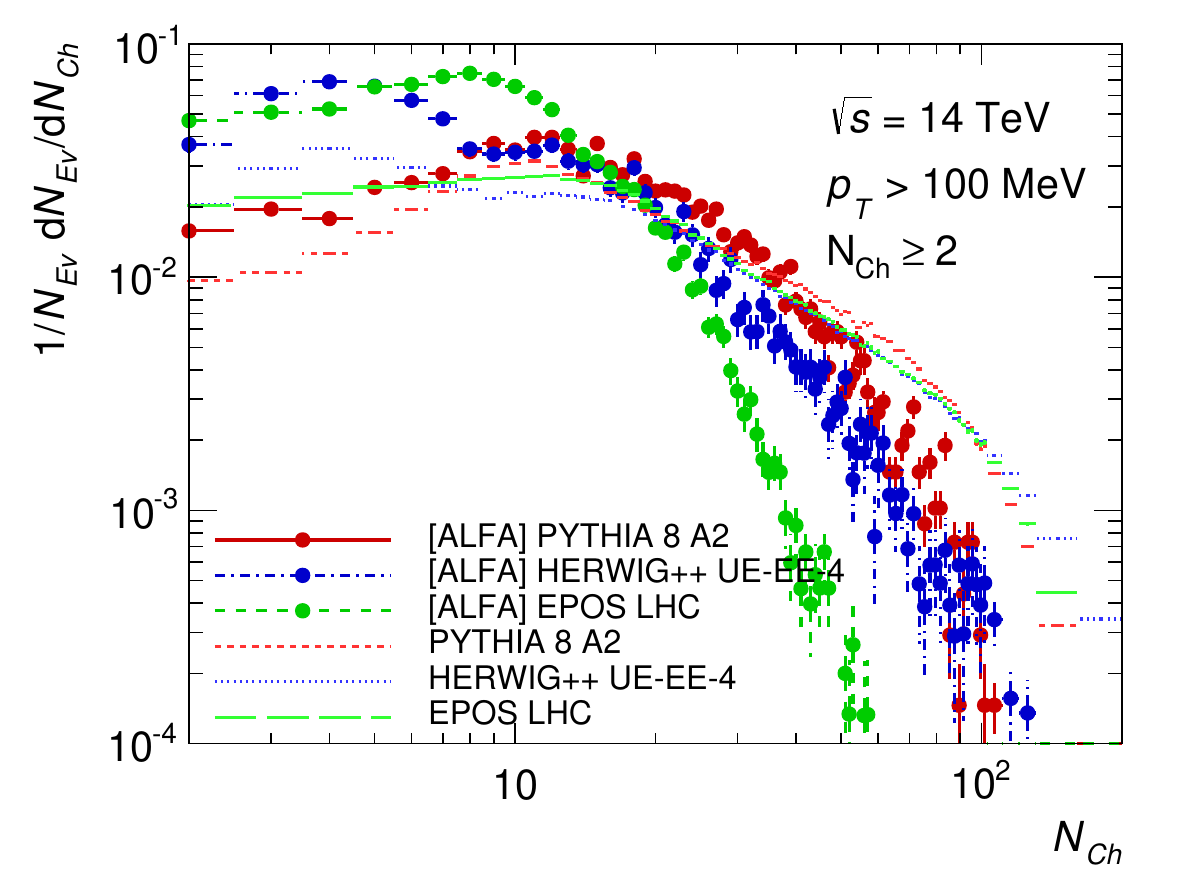}
\label{fig:ChargedParticle_3:b}
\caption{}
\end{subfigure}
\caption{Event normalized charged particle multiplicities compared
  between MC models for inclusive and AFP selections (a) and
  inclusive and ALFA selections (b). }
\label{fig:ChargedParticle_3}
\end{figure}

The event normalized charged particle spectra is plotted in
Fig.~\ref{fig:ChargedParticle_3}. For larger \Nch, the tails of the
distributions are observed to fall more sharply with AFP or
ALFA tags than for the inclusive selection. \epos in particular
has a very low probability of generating events with $\Nch > 50$ and a
forward proton while the tail in \hpp and \pyeight forward-tagged
events extends to $\Nch = 100$ for the same number of events.

\begin{figure}
\centering
\begin{subfigure}[b]{0.47\textwidth}
\includegraphics[width=\textwidth]{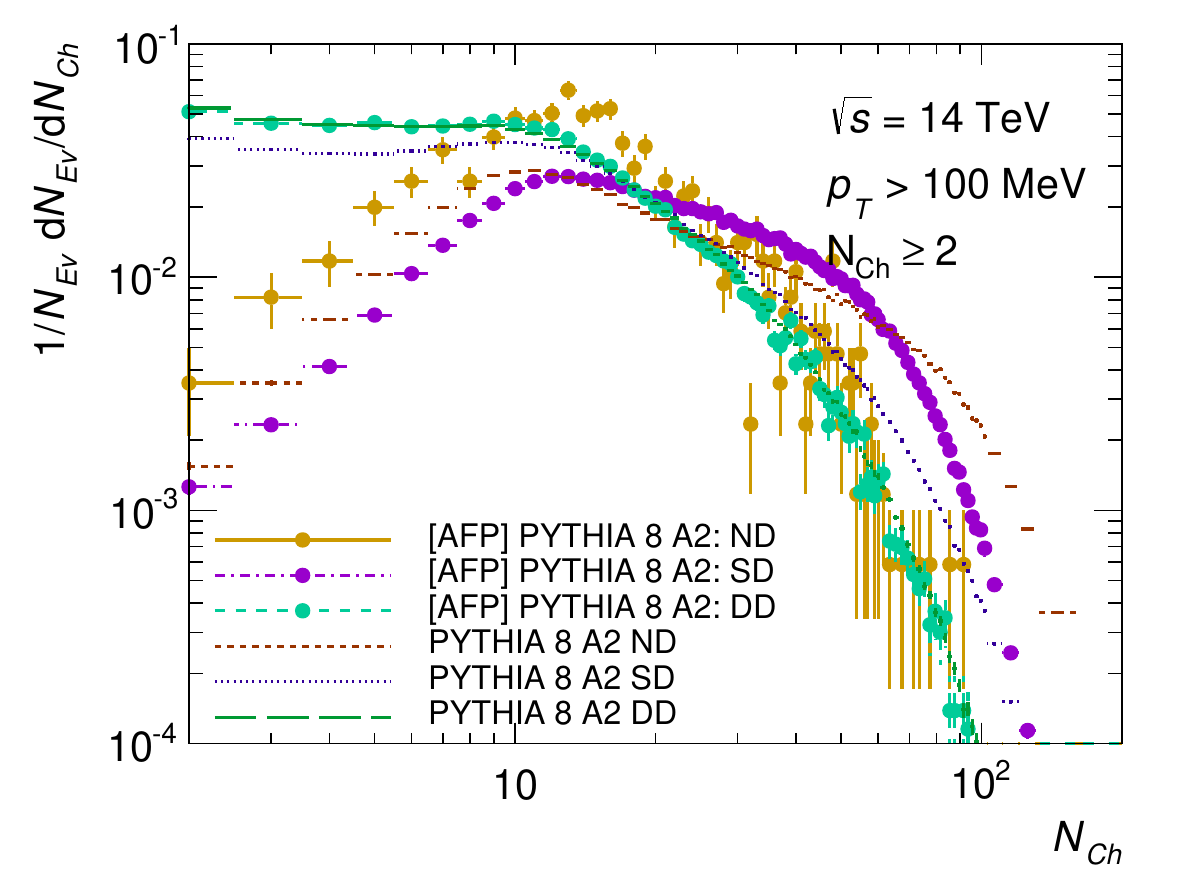}
\caption{}
\label{fig:ChargedParticle_7:a}
\end{subfigure}
\hspace{5mm}
\begin{subfigure}[b]{0.47\textwidth}
\includegraphics[width=\textwidth]{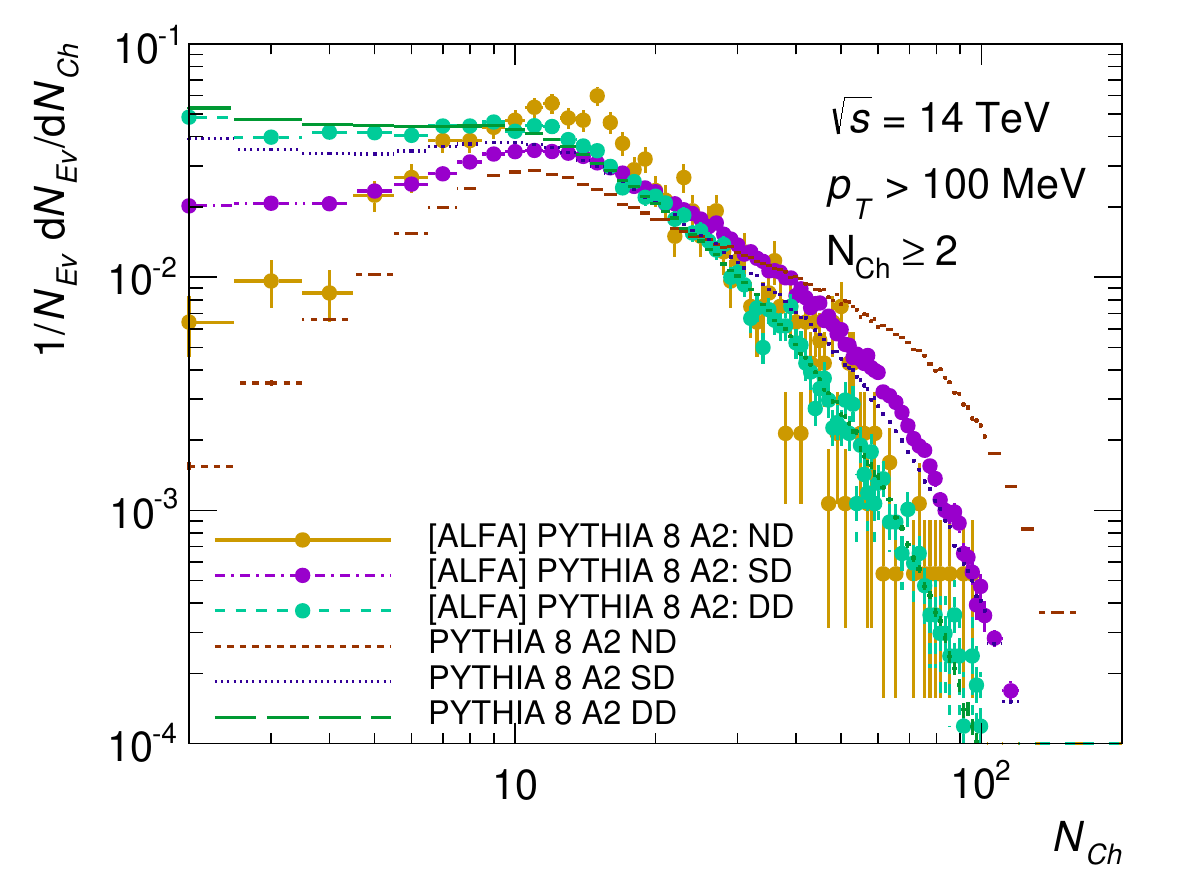}
\caption{}
\label{fig:ChargedParticle_7:b}
\end{subfigure}
\caption{Event normalized charged particle multiplicities compared
  with single-, double- and non-diffractive components of the
  inelastic cross section from \pyeight for inclusive and AFP
  selections (a) and inclusive and ALFA selections (b). }
\label{fig:ChargedParticle_7}
\end{figure}

For the region at low multiplicity, where diffractive signatures may
be expected due to lower particle production in rapidity gap events, a
large spread of predictions is observed. \epos and \hpp both support
this hypothesis and predict an enhancement for $2 < \Nch < 20$ with
regards to their inclusive distributions, as does \pyeight for the
ALFA selection. For the AFP selection however, the opposite
is true and fewer events are expected in this region. This is
understood by examining the breakdown of \pyeight into its diffractive
and non-diffractive components in
Fig.~\ref{fig:ChargedParticle_7}. 
%The fraction of the tagged sample
%originating from high-$\xi$ single diffractive events within the
%AFP acceptance (whose mass is sufficiently large that no gap is
%expected within $|\eta| < 2.5$) biases the selection to a harder
%spectrum than for the inclusive distribution in the region $20 < \Nch
%< 70$. 
The AFP selection is biased to a harder
spectrum than for the inclusive distribution in the region $20 < \Nch
< 70$ by high-mass single diffractive events within AFP acceptance, 
the final state particles from the dissociated proton in such events 
fully span the $|\eta| < 2.5$ acceptance of the central tracking detectors.
The tail is still observed to fall faster than for the
inclusive case as very high multiplicity events ($\Nch > 100$) are
predicted to be dominated by low impact-parameter non-diffractive
interactions which are suppressed by requiring an AFP or
ALFA tag. For low \Nch, only the double-diffractive sample (which
as discussed in Section~\ref{sec:timm:forwardprotonsources} may cause
large rapidity gaps) retains the enhancement in the tagged sample, but
with a smaller fractional cross section, see
Table~\ref{tab:timm:crosssections}.

\begin{figure}
\centering
\begin{subfigure}[b]{0.47\textwidth}
\includegraphics[width=\textwidth]{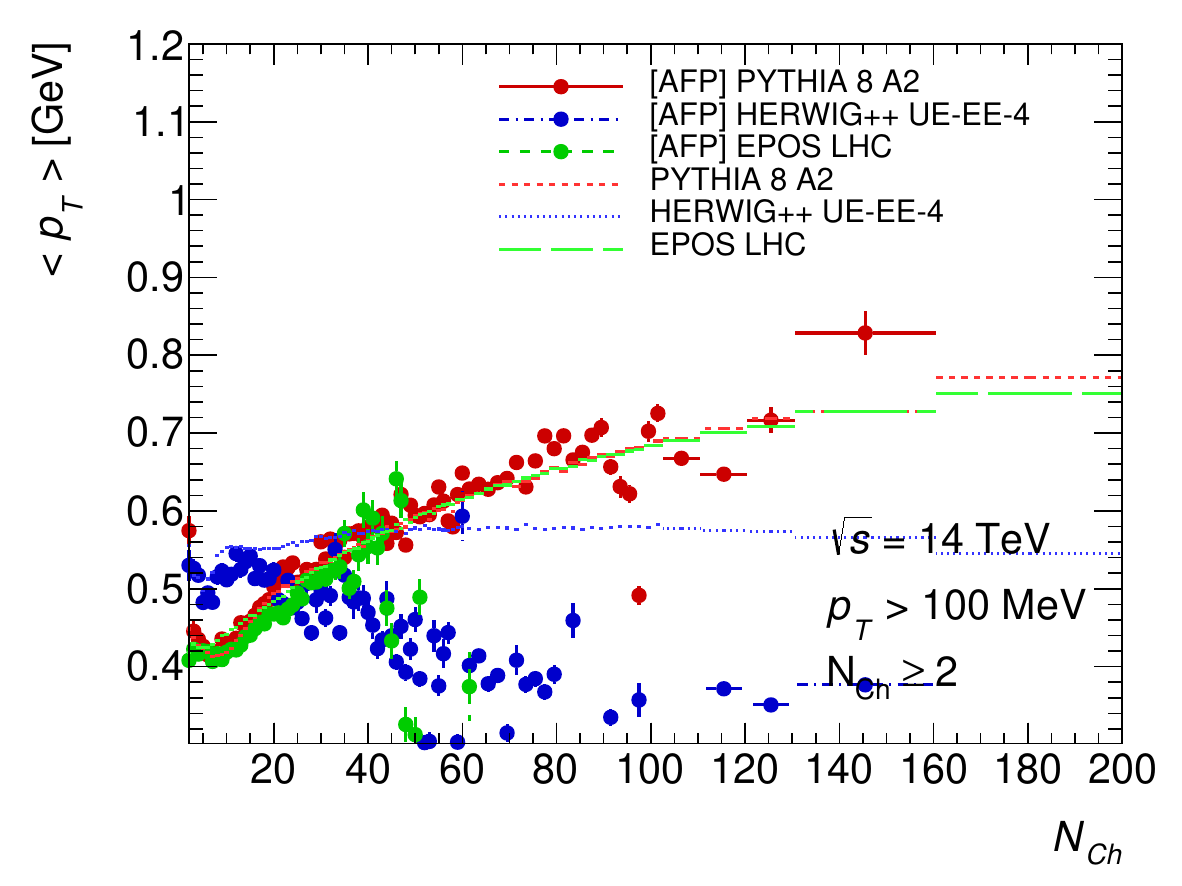}
\caption{}
\label{fig:ChargedParticle_2:a}
\end{subfigure}
\hspace{5mm}
\begin{subfigure}[b]{0.47\textwidth}
\includegraphics[width=\textwidth]{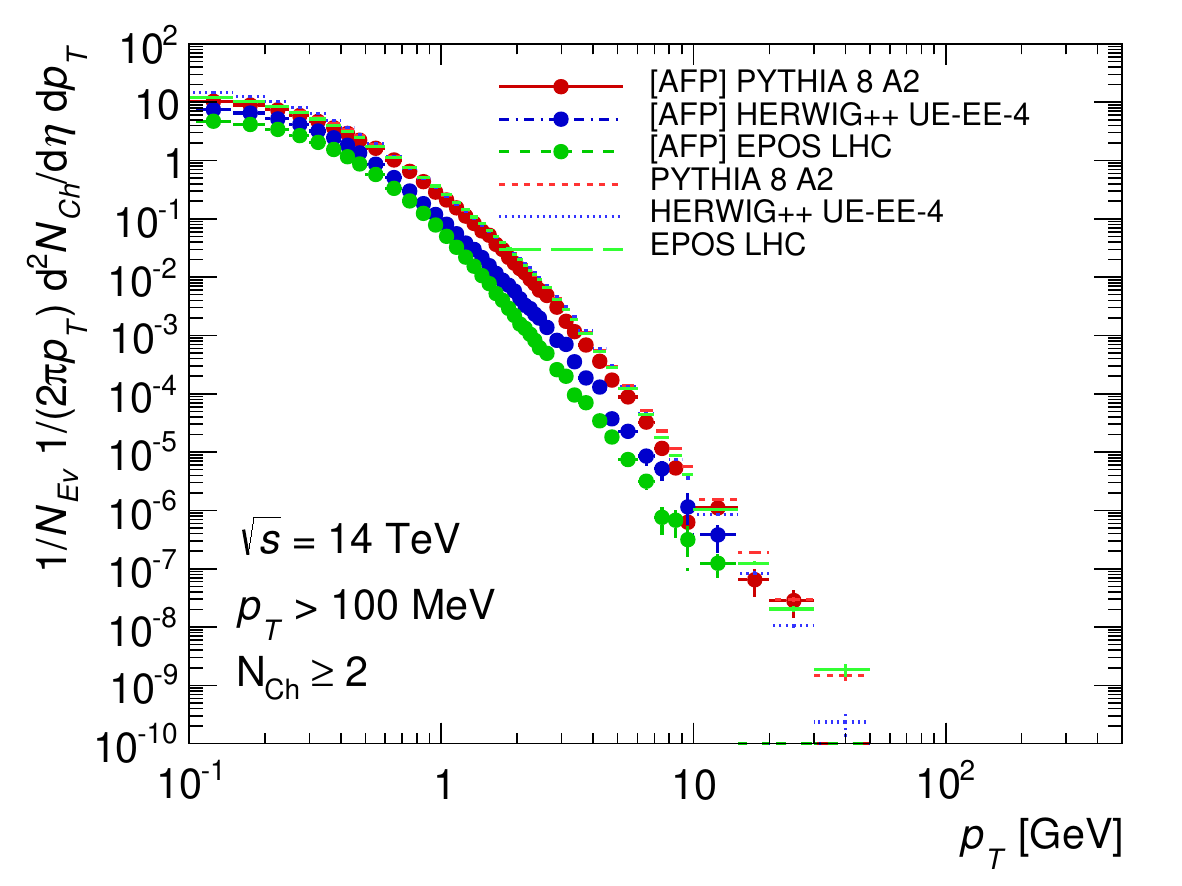}
\caption{}
\label{fig:ChargedParticle_2:b}
\end{subfigure}
\caption{Event normalized charged particle $\eta$ and \pt\
  distributions compared between MC models for inclusive and
  AFP selections using log-log axes.}
\label{fig:ChargedParticle_2}
\end{figure}

In addition to the multiplicity spectrum, the charged particle \pt\
spectra are investigated for just the inclusive and AFP
selections in Fig.~\ref{fig:ChargedParticle_2}. The corresponding
ALFA distributions were noted to be very similar to the AFP
ones for these selections (see section \ref{sec:timm:protontagging}) 
are hence not reproduced here. In
Fig.~\ref{fig:ChargedParticle_2:a}, the charged particle multiplicity
is plotted versus the mean particle \pt\ summed over all tracks in all
events of a given multiplicity. All the AFP tagged distributions
suffer from statistical fluctuations in their respective tails from
Fig.~\ref{fig:ChargedParticle_3}. The \pyeight prediction is very
similar for the inclusive and AFP tagged distributions over the
range which is comparable with the available statistics. With \epos,
the shape of the distributions agree, however the particle spectrum is
on average softer for the tagged distribution. Unlike the other MCs,
\hpp predicts a $<\pt>$ distribution which is roughly invariant in
\Nch for the inclusive distribution. The tagged \hpp sample becomes
increasingly soft at higher multiplicities indicating that for this
class of event, there are kinematic constraints imposed by the
generator limiting the energy available to perform the cluster
hadronization of the final state.

Figure \ref{fig:ChargedParticle_2:b} shows the charged particle \pt\
spectra. The above observations are conformed
in the normalizations of the AFP tagged samples. Some differences
in slope are also visible, most notably for \hpp.

The large $\eta$ coverage ($|\eta|$ < 6.5) of the combined CMS and TOTEM tracking will enable interesting charged multiplicity measurement at $\sqrt{s}$ = 13 and 14\UTeV. In addition to straight forward multiplicity measurements are also charged multiplicity correlation (e.g. forward vs backward, forward vs central) measurements envisaged. The capability to deplete or enhance the event sample in single diffractive interactions either based on the forward event topology (using T2) or the detection of the diffractive proton in the Roman Pots even further enhances the sensitivity of these measurements.
 
Having new type of correlations, in particular with a proton tagging, greatly 
enhance the possibility to distinguish between diffractive and non-diffractive 
events and hence increase the visible difference between models. 
It will first of all allow a better understanding of diffractive event
as such, but furthermore the tuning of the hadronic interaction models will be 
improve being based on almost pure samples of diffractive and/or non-diffractive
 while until now it was always a mixing.

%% file: cosmic/Spectra.tex
Identified particle spectra are a key component of model tuning for air shower simulations
since the muon production depends on the relative abundance of $\pi^{0}$, strangeness and baryon
production in the total multiplicity. With the new LHC run not only larger energies will be reached but special triggers can be developed to better understand
the particle production mechanisms and then improve model extrapolations.

\subsection{Measurement of identified charged particle spectra in pp and p-Pb collisions with ALICE}

      Recent measurements of identified particle spectra in proton-lead and lead-lead 
      collisions show relatively large discrepancies with standard hadronic model predictions in particular for strange baryons. Presenting the results according to event multiplicity and centrality, respectively, have shown 
      interesting signs of hadrochemistry changes and of modifications to transverse momentum 
distributions of identified particles. Even if these analysis are based on rare particles and at midrapidity which are not so relevant for direct air shower physics, knowing the correct hadronization scheme is of primary importance for the modelisation of hadronic interactions in particular for air shower simulations which require large extrapolations to unexplored energies and phase space. An interesting goal to constrain hadronic interaction models would be thus to 
      perform extensive identified particle spectra measurements according to multiplicity also 
      in proton-proton collisions. 
      Existing data allows for these studies only for the more abundantly produced particles, 
      but the more exotic hyperons such as the $\Omega$ essentially require very large statistics that are 
      currently unavailable. Performing such a measurement is particularly important for  
      hyperons, as this may help in understanding the mechanisms behind the
      strangeness enhancement already observed in nuclear collisions \cite{ABELEV:2013zaa}. In particular, this may 
      further constrain or discriminate some aspects of thermal models such as non-equilibrium, 
      the use of $\gamma_{S}$ and canonical strangeness production. Furthermore, the high mass of the 		
hyperons makes them particularly susceptible to the presence of collective effects in high multiplicity 	
proton-proton collisions, further underlining the need for such a measurement. 

 \subsubsection{Past measurements of charged particle spectra}

The ALICE collaboration has systematically measured the production of identified particle spectra in pp collisions at mid-rapidity ($|\eta|<0.5$)for a variety 
of light flavour particle species, ranging from the more abundant $\pi^{\pm}$, $K^{\pm}$, p and $\bar{p}$, the multi-strange $\Xi$ and $\Omega$ and resonances such as $K^*(892)^0$, $\phi(1020)$, $\Sigma(1385)^{\pm}$ and $\Xi(1530)^{0}$. 

The more abundant of the light flavour species, $\pi^{\pm}$, $K^{\pm}$, p and $\bar{p}$, have been studied at energies of 900\UGeV, 2760\UGeV\ and
7\UTeV~\cite{Aamodt:2011zj,Abelev:2014laa, Chojnacki:2011pv}. These analyses employ several different particle identification techniques, such as specific energy loss in
both Time Projection Chamber and Inner Silicon tracker as well as time of flight measurements, to cover a broad range of transverse momentum. The 
measured spectra have been compared to QCD-based models such as PYTHIA in various tunes as well as PHOJET, and none of the event generators 
have been able to reproduce all spectra simultaneously with a precision better than 30\%. These serve
as important reference data for modelling efforts as well as benchmarks for comparisons to other systems such as p-Pb and Pb-Pb. 

\begin{figure}[tbp]
\centering % \begin{center}/\end{center} takes some additional vertical space
\includegraphics[width=0.49\textwidth]{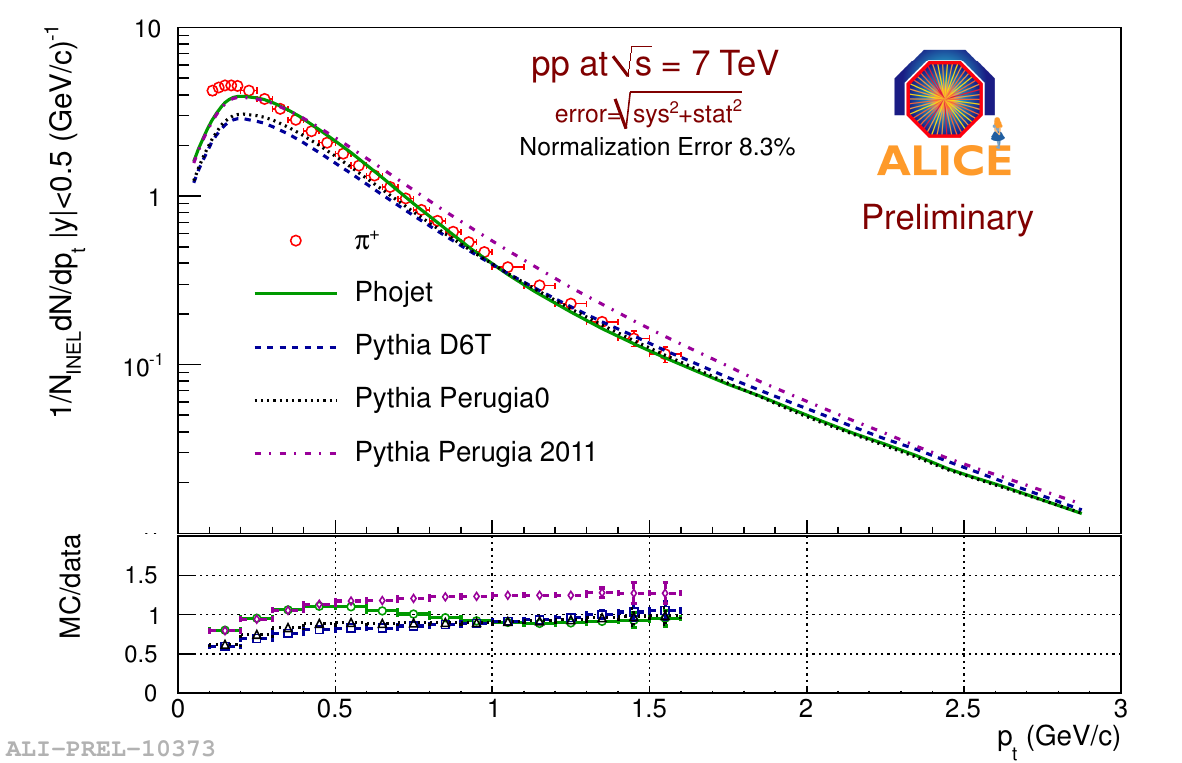}
\includegraphics[width=0.49\textwidth]{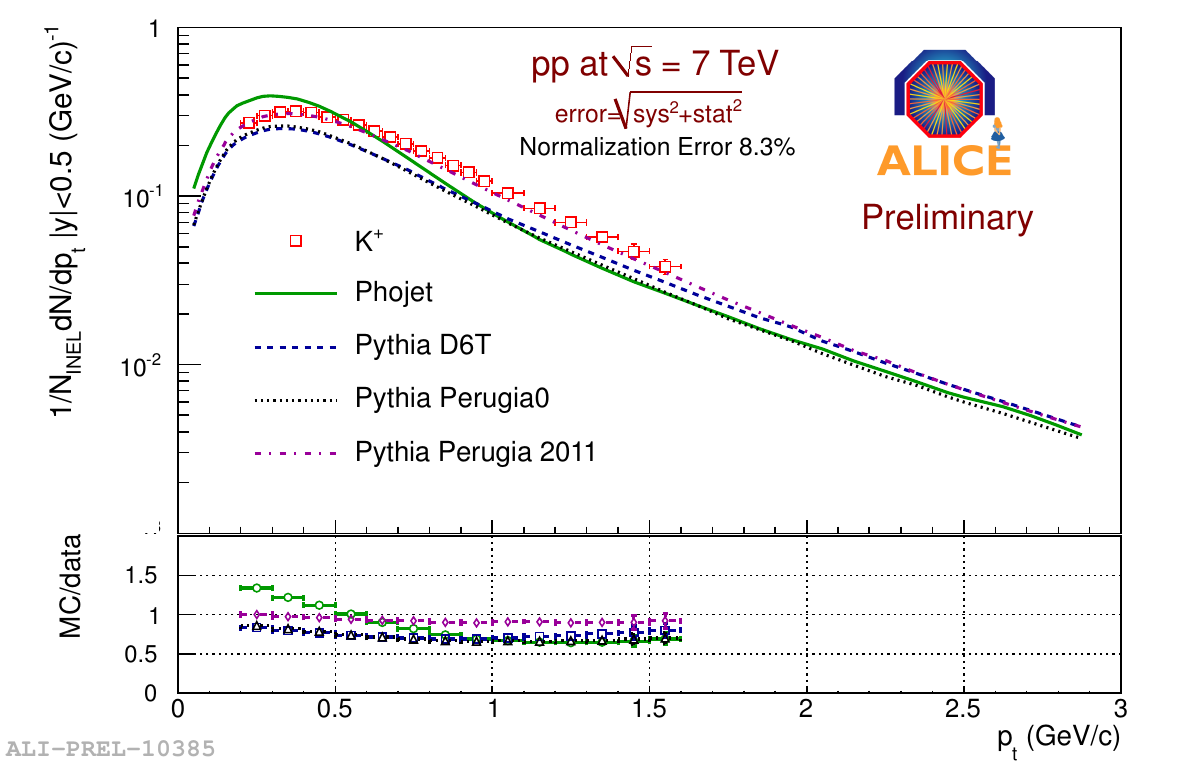}
\includegraphics[width=0.49\textwidth]{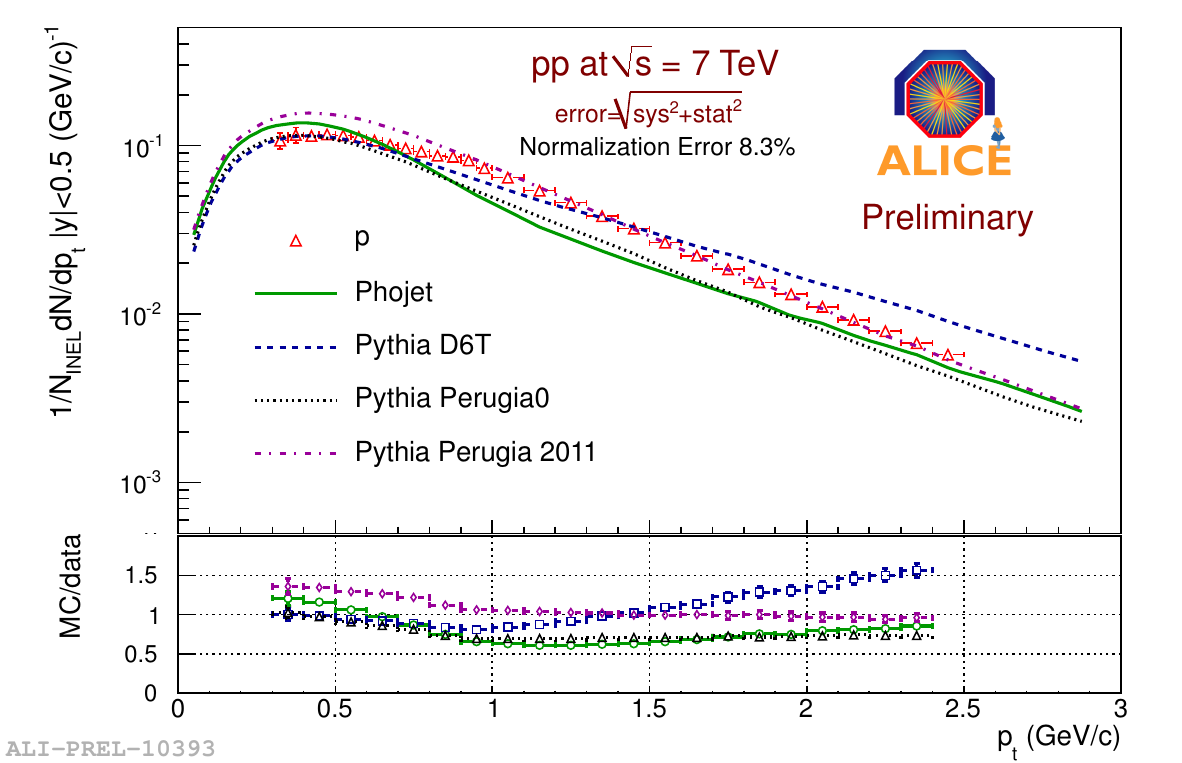}
\caption{\label{fig:ALICEpiKp} $\pi^{+}$ (top left), $K^{+}$ (top right) and $p$ (bottom) spectra in pp collisions at $\sqrt{s}=7$\UTeV. The corresponding
antiparticles are not drawn since they are identical within uncertainties. Figures from \cite{Chojnacki:2011pv}. }
\end{figure}

Furthermore, multi-strange baryon measurements in pp collisions at 7\UTeV \cite{Abelev:2012jp} over a wide range in momentum are made possible by 
exploiting the weak decay topology and employing particle identification via energy loss in the Time Projection Chamber on the daughter tracks. The
resulting measurements can be seen in Figure \ref{fig:ALICEmult} together with predictions from the PYTHIA6 
event generator using its Perugia-2011 tune. The Monte Carlo simulations are unable to reproduce the measured yields, underpredicting the 
$\Xi$ spectra by up to a factor 2 and the $\Omega$ spectra by up to a factor 4-5, although better agreement can be observed for $\Xi$ for 
larger momenta of around 6-7\UGeV/c. 

\begin{figure}[tbp]
\centering % \begin{center}/\end{center} takes some additional vertical space
\includegraphics[width=0.53\textwidth]{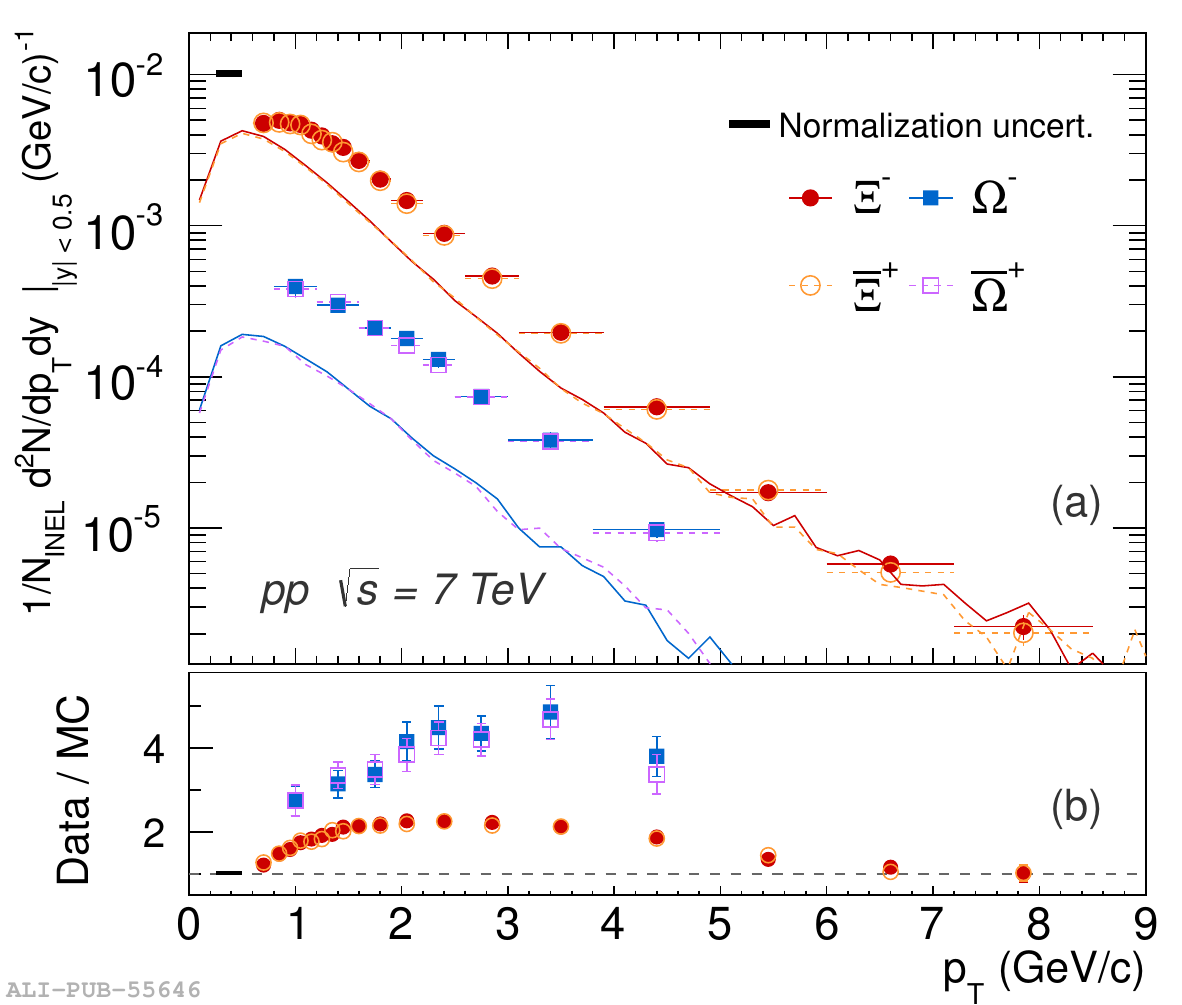}
\caption{\label{fig:ALICEcascades} $\Xi^{-}$, $\bar{\Xi}^{+}$, $\Omega^{-}$ and $\bar{\Omega}^{+}$ transverse momentum spectra in pp collisions 
at $\sqrt{s}=7$\UTeV\ measured by ALICE compared with predictions from PYTHIA6 using tune Perugia-2011. Figure from \cite{Abelev:2012jp}. }
\end{figure}

Further identified particle spectra measured by ALICE in pp collisions include the $K^*(892)^0$ and $\phi(1020)$ \cite{Abelev:2012hy} as well as the $\Sigma(1385)^{\pm}$ and $\Xi(1530)^{0}$ resonances \cite{Abelev:2014qqa}. Since for those resonances the measured decay daughters originate from 
the primary vertex, one cannot exploit the decay topology to isolate their signals. Instead, methods such as event mixing and like-sign background estimation 
have to be used in order to compute the combinatorial background to be subtracted from the measured invariant mass distributions, which then 
results in invariant mass peaks that are suitable for signal extraction. 

\begin{figure}[tbp]
\centering % \begin{center}/\end{center} takes some additional vertical space
\includegraphics[width=0.45\textwidth]{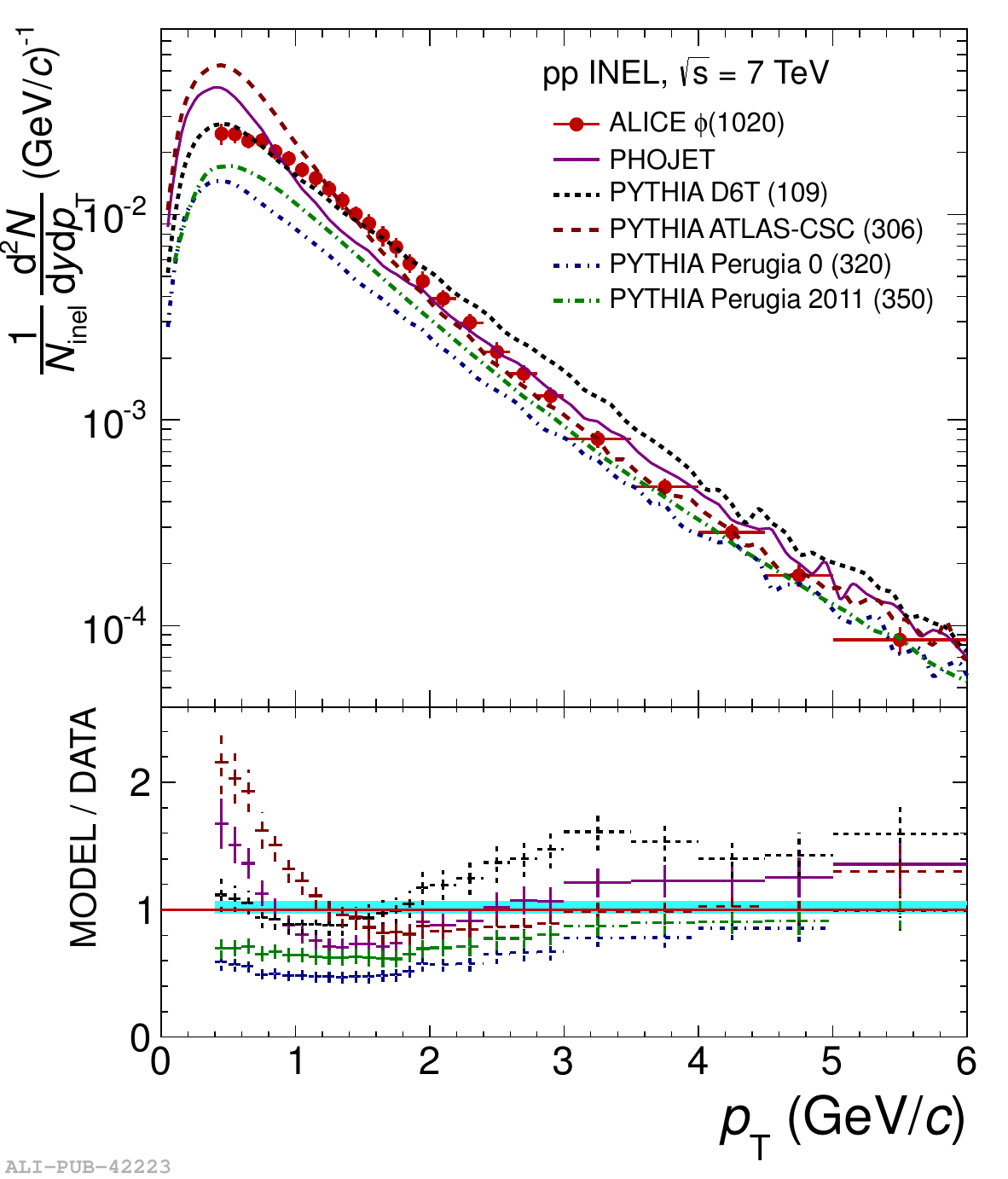}
\includegraphics[width=0.45\textwidth]{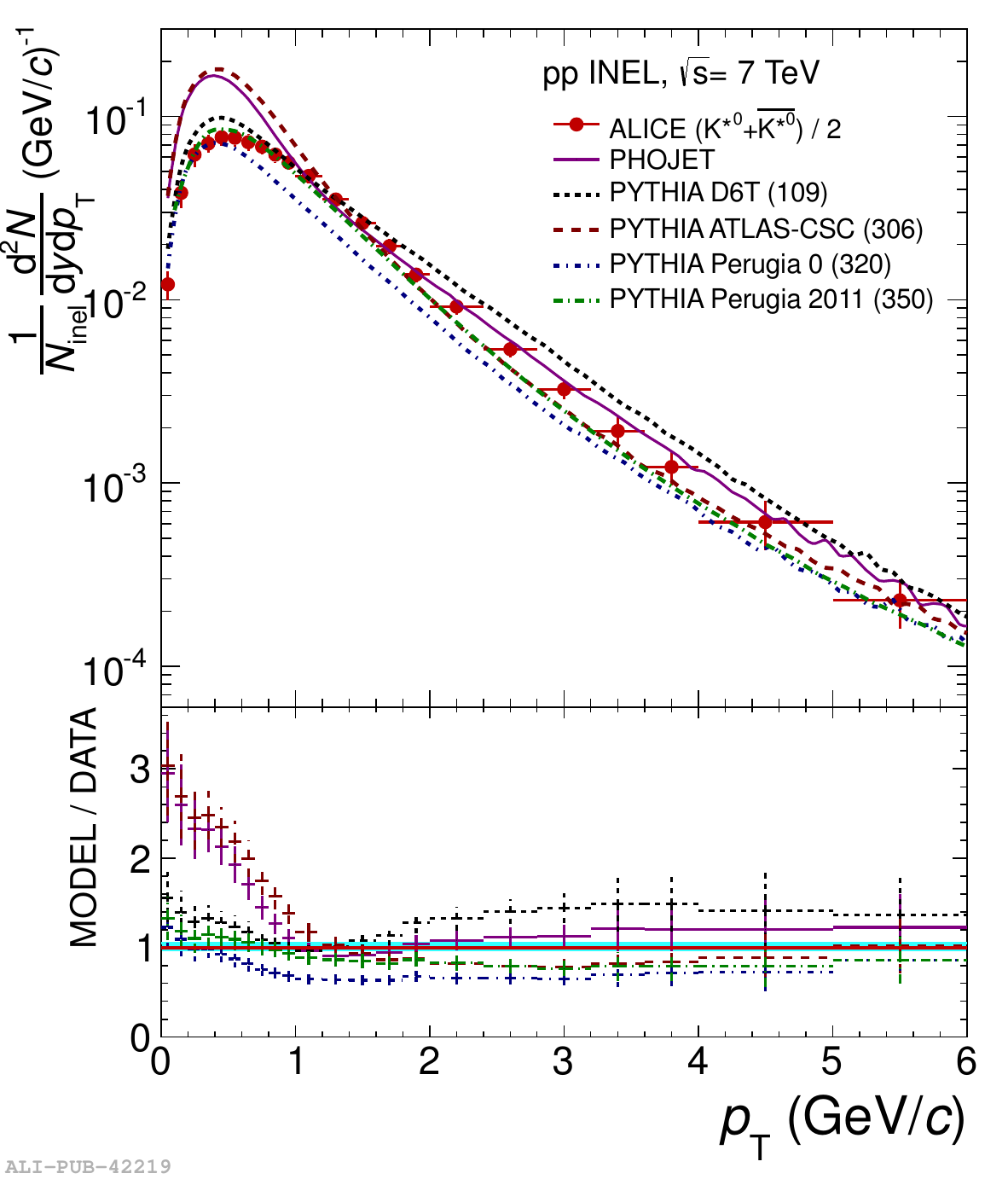}
\caption{\label{fig:ALICEresonances1} $\phi(1020)$ (left) and $K^*(892)^0$ (right) transverse momentum spectra at mid-rapidity in pp collisions at $\sqrt{s}=5.02$\UTeV. Figures from \cite{Abelev:2012hy}. }
\end{figure}

\begin{figure}[tbp]
\centering % \begin{center}/\end{center} takes some additional vertical space
\includegraphics[width=0.45\textwidth]{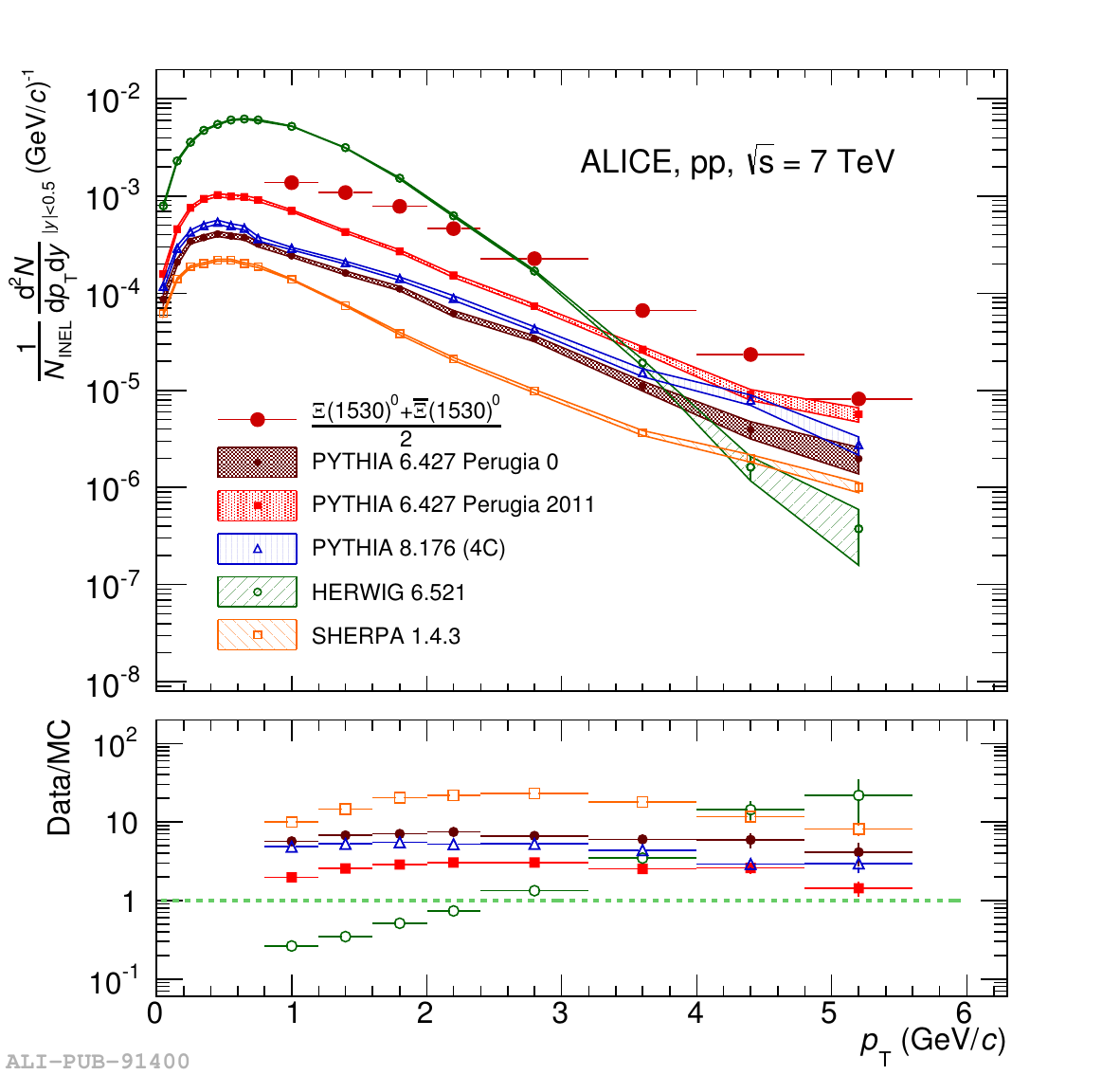}
\includegraphics[width=0.45\textwidth]{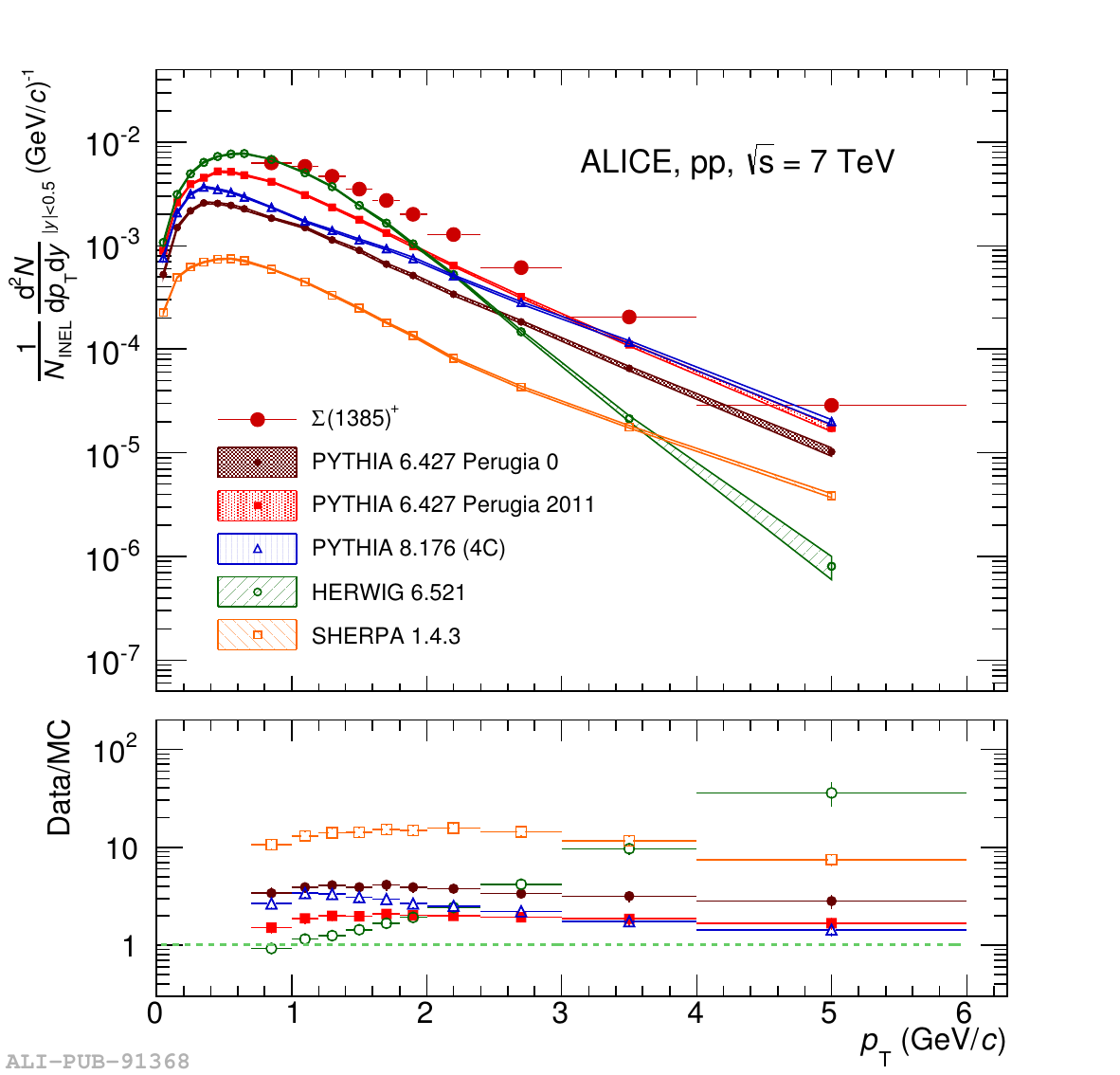}
\caption{\label{fig:ALICEresonances2} $\Sigma(1385)^{\pm}$ (left) and $\Xi(1530)^{0}$ (right)   transverse momentum spectra at mid-rapidity in pp collisions at $\sqrt{s}=5.02$\UTeV. Figures from \cite{Abelev:2014qqa}. }
\end{figure}

The transverse momentum spectra obtained in these analyses can be seen in Figures \ref{fig:ALICEresonances1} and \ref{fig:ALICEresonances2}. Also in those cases, PYTHIA fails to correctly reproduce the measured yields by up to a factor 2. Other models, such as PHOJET, have also been tested but also fail at reproducing
both integrated yields as well as spectral shapes. These various measurements serve as reference for better 
understanding of particle production mechanisms in pp collisions and introduce further constraints on future modelling efforts. 

\begin{figure}[tbp]
\centering % \begin{center}/\end{center} takes some additional vertical space
\includegraphics[width=0.45\textwidth]{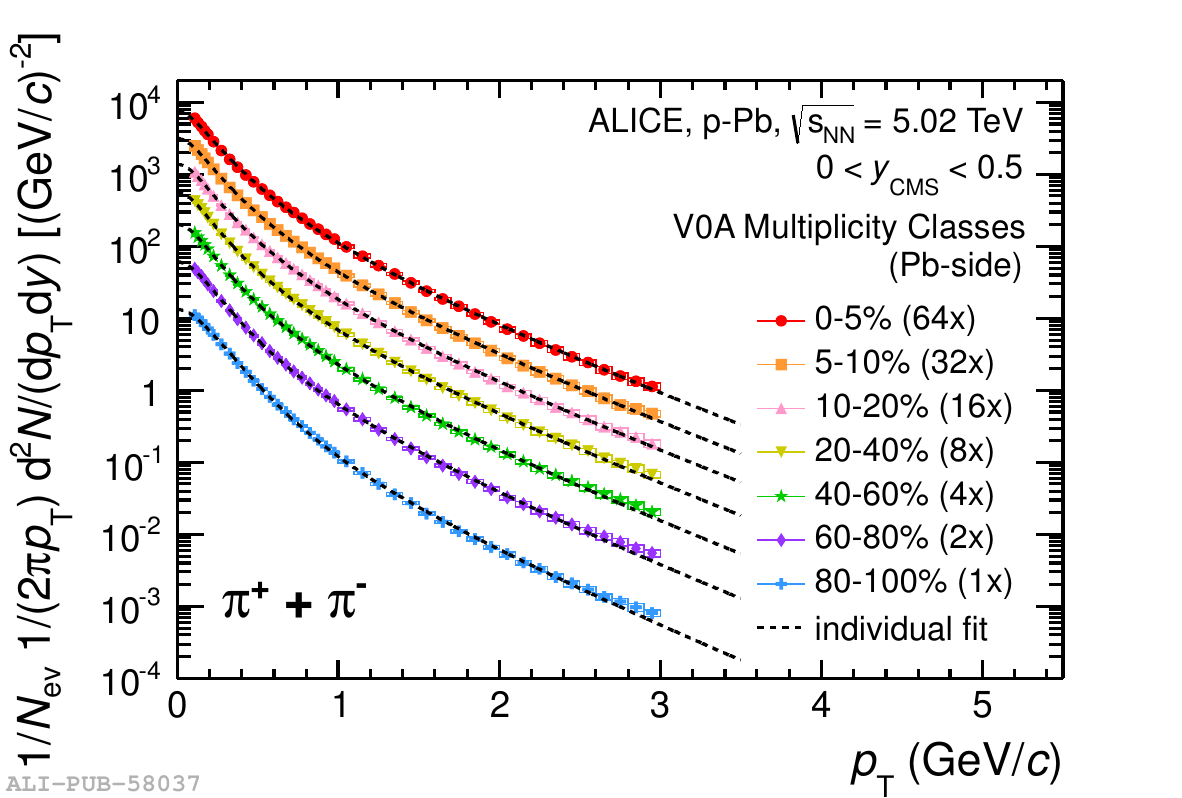}
\includegraphics[width=0.45\textwidth]{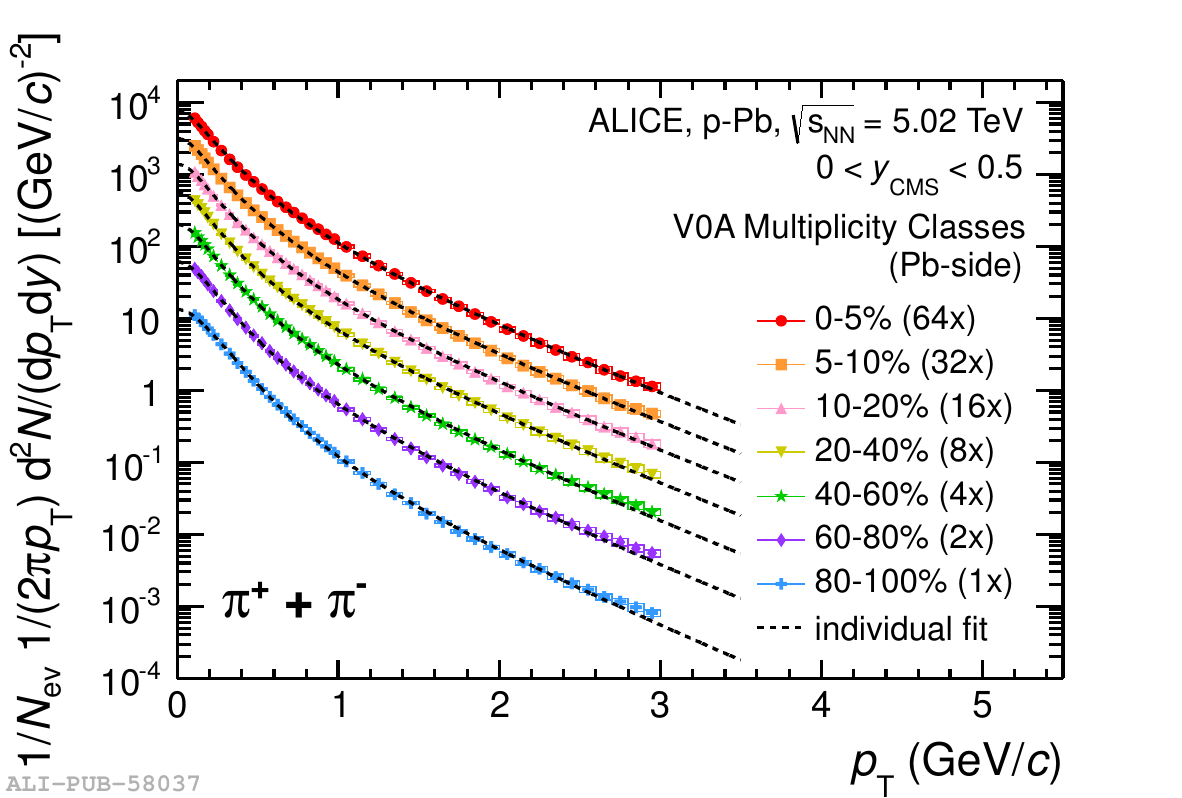}
\includegraphics[width=0.45\textwidth]{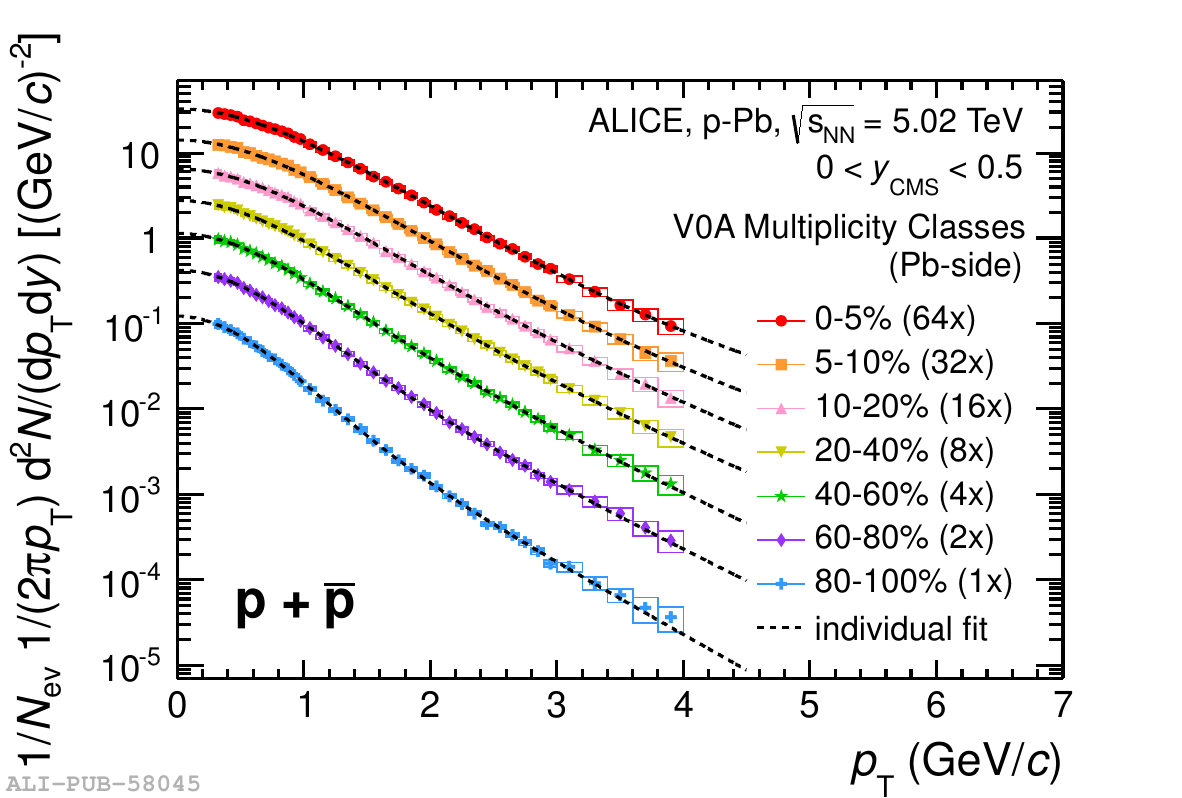}
\includegraphics[width=0.45\textwidth]{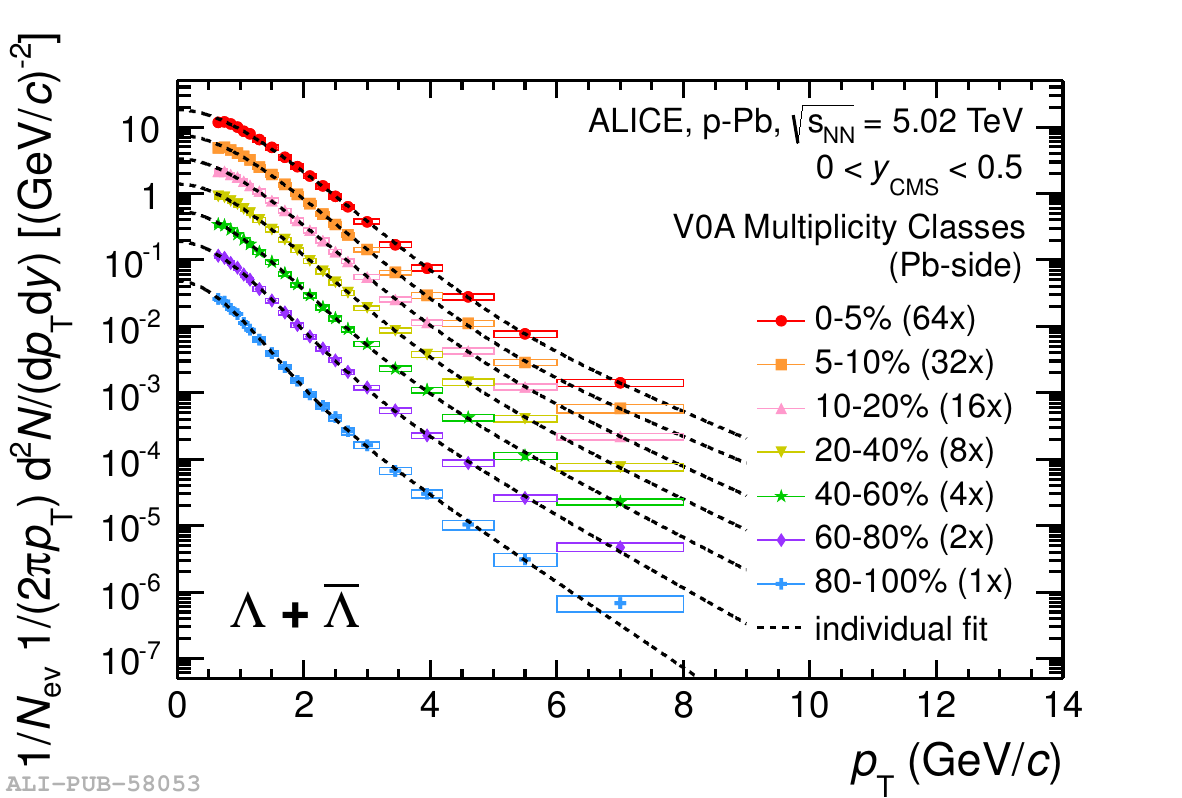}
\caption{\label{fig:ALICEpPbPIDpiKp} $\pi^{\pm}$ (top left), $K^{\pm}$ (top right), $p+\bar{p}$ (bottom left) and $\Lambda+\bar{\Lambda}$ (bottom right) transverse momentum spectra as a function of event multiplicity in p-Pb collisions at $\sqrt{s}=5.02$\UTeV. Figures from \cite{Abelev:2013haa} }
\end{figure}

In addition to the more elementary pp collisions, ALICE has been performing systematic measurements of identified particle spectra also 
in the p-Pb colliding system. Measurements of $p_{T}$-differential yields of $\pi^{\pm}$, $K^{\pm}$, $K^{0}_{S}$, $p$, $\bar{p}$, $\Lambda$ 
and $\bar{\Lambda}$ have been performed as a function of charged particle multiplicity \cite{Abelev:2013haa}. These measurements also employ the same
techniques as used for the pp observations, such as energy loss in the Time Projection Chamber, time of flight measurements and reconstruction by
decay topology for weakly decaying hadrons. The observed spectra show a clear 
evolution with multiplicity, as can be seen in Figure \ref{fig:ALICEpPbPIDpiKp}, similar to what occurs in high energy pp and Pb-Pb collisions. They have been compared to predictions from DPMJet, a QCD-inspired event generator based on the Gribov-Glauber approach which is not able to reproduce spectral shapes. Other models, such as EPOS which include a phase of collective hadronization, are significantly better at describing the measured spectra shapes. 

\subparagraph*{Identified particle spectra according to multiplicity}

    \begin{figure}
    \centering\includegraphics[width=.825\linewidth]{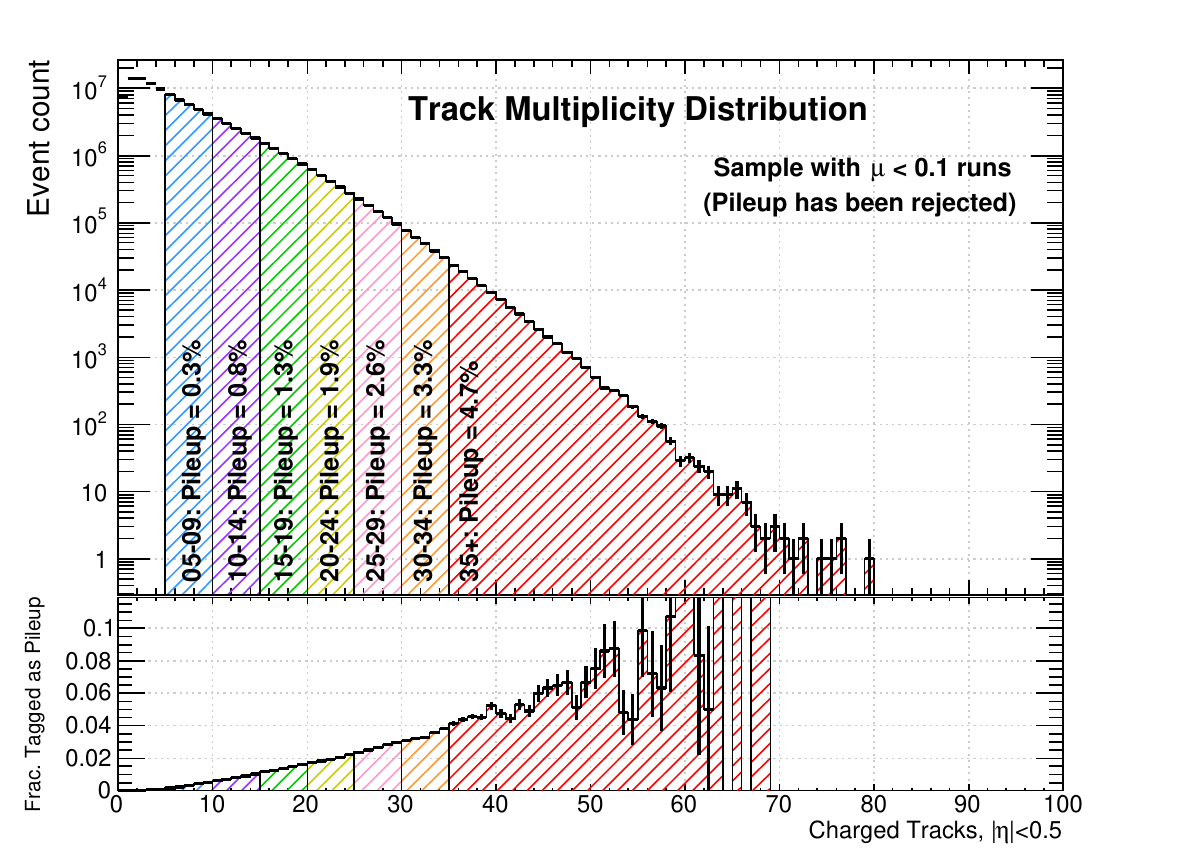}
    \caption{Reconstructed charged particle density distribution in proton-proton collisions at 7\UTeV\ (top pad) and 
             fraction of events thereof that have been tagged as pileup (bottom plot).}
    \label{fig:pileupvsmult}
    \end{figure}

      Currently existing data enables particle spectra measurements according to multiplicity for $\pi^{\pm}$, $K^{\pm}$, 
      $p$, $\bar{p}$ \cite{Andrei:2014vaa}, $\Lambda$, $\bar{\Lambda}$ and even $\Xi^{-}$ and $\Xi^{+}$, albeit with limited momentum and
      multiplicity reach for the $\Xi$ baryons. An ongoing analysis has shown that the spectra for all particles except the $\Xi$ can be computed according to mid-rapidity track counts reaching a charged particle
      density of up to approximately a factor 6 higher than the minimum-bias value using a sample of approximately
      $150\times10^{6}$ minimum-bias proton-proton collisions at 7\UTeV. For the $\Xi$, the multiplicity distribution is such that another 		
factor $10$ in statistics would be required to reach the same multiplicity. 
These analyzes used runs for which operational conditions
      were such that  the interaction probability $\mu<0.1$, and standard inner tracking system (ITS) pileup tagging methods revealed no more than an 	 
average $5\%$ pileup rate for the      
      largest multiplicity bin\footnote{though note that these methods have been 
      estimated to only successfully tag half of the piled up events, and so the real pileup rates could be as much as twice these values.}, as can be seen in Fig.~\ref{fig:pileupvsmult} . This would     
      represent an acceptable rate of pileup for which
      systematic uncertainties would still be under control, and it is an important consideration since simply collecting 
      data with large luminosities will incur in prohibitively large pileup rates for the high-multiplicity event classes. 

      Assuming that one wishes to perform a similar measurement for the $\Omega$ baryon under the same conditions and having the 
      same reach in multiples of minimum-bias charged particle density, two components need to be considered: 
      
    \begin{itemize}
    \item The $\Omega/\Xi$ ratio, which, while being one of the goals of such a measurement, can be estimated to be of 
      in the range of the published minimum-bias $\Omega/\Xi$ ratio or more and is of approximately 0.085 (see Ref.~\cite{Abelev:2012jp});
    \item The fact that $\Omega$ has a lower reconstruction efficiency than $\Xi$, which is mostly a consequence of the fact
      that the measured $\Omega$ channel has a lower branching ratio than the measured $\Xi$ channel, in addition to different
      decay kinematics. The resulting difference is of about a factor 2. 
    \end{itemize}
  
      Having these considerations in mind, the number of minimum-bias events required for the $\Omega$ analysis according
      to charged particle multiplicity would be of approximately $5.0\times10^{10}$ detected events. 
      Assuming a worst-case scenario trigger efficiency and analysis selection event loss, this would require a 
      delivered integrated luminosity of at least $1.0~pb^{-1}$ or more and ensuring $\mu<0.1$. Operationally, it will
      also be important to consider high-multiplicity triggering during such data taking. 

 \subsubsection{Future measurements of charged particle spectra : dedicated triggering}
      
If selecting large multiplicity events, a much smaller sample of triggered events would be sufficient to
      perform an $\Omega$ analysis according to multiplicity. The triggering can be performed based on 
      a charged particle density estimate acquired either in mid-pseudorapidity, such as a counting tracks in $|\eta|<0.5$ or 
      $|\eta|<0.8$, where ALICE has full central barrel detector coverage, or in forward pseudorapidity, as would 
      be the case if using the amplitudes measured by the VZEROA and VZEROC scintillators, located in
      $2.8<\eta<5.1$ and $-3.7<\eta<-1.7$, respectively. It would be of interest to study the possibility to trigger on either one, since it is 		      
	known that using mid-rapidity multiplicity estimators adds biases
      to the measurements of primary charged particle yields with respect to weakly decaying particles such
      as the V0 and cascade decays. Having two trigger strategies will provide a tool to study such biases. 

      Furthermore, in order to study the multiplicity dependence of identified particle production, it will be important to also set up different
      multiplicity triggers with different downscalings to compensate for the steeply falling multiplicity distribution. In any case, this 
       special triggering would have to be accompanied by a short min-bias data taking period to calibrate and
      determine proper normalization.
\begin{itemize}
\item Triggering on mid-rapidity track density: One of the simplest possibilities to study high multiplicity proton-proton events is to classify events based on the 
number of charged particle tracks at mid-pseudorapidity as reconstructed by the ALICE ITS and TPC. By doing so, one can 
reach relatively high multiplicities, as can be seen Fig.~\ref{fig:pileupvsmult}, where the highest multiplicity class 
corresponds to $0.1\%$ of the measured cross-section. Triggering using a similar criterion would thus ensure
a recorded data reduction of a factor $10^{3}$. 
\item Triggering on VZERO amplitudes:
    \begin{figure}
    \centering\includegraphics[width=.7\linewidth]{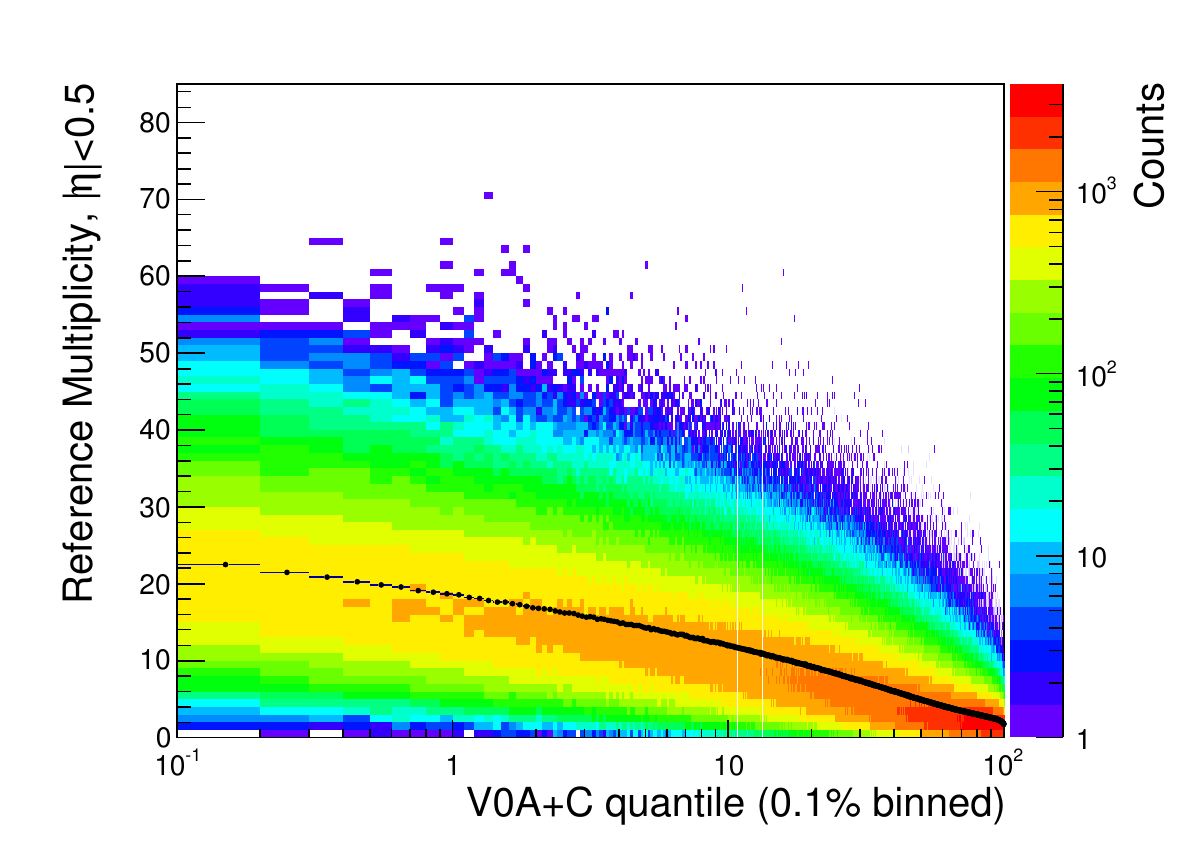}
    \caption{Mid-rapidity raw multiplicity estimator according to VZERO cross-section quantiles.}
    \label{fig:quantiles}
    \end{figure}
      The effects of selecting on the VZERO amplitudes for reaching high mid-rapidity charged particle
      densities can be seen in Fig.~\ref{fig:quantiles}, where the mid-rapidity raw multiplicity is drawn as a function of the
      VZERO quantile. It is also noteworthy that, for the minimum-bias data sample, the mid-rapidity
      multiplicity estimator returns an average multiplicity of approximately 6. Thus, for instance, selecting 
      on the $1\%$ of events with highest VZERO amplitudes would yield a sample with approximately
      $4\times(dN_{ch}/d\eta)_{MB}$\ and one would need only $5\times10^8$ recorded events in that case. 
    \end{itemize}

\subsection{Neutral particle spectra}

The ALICE Collaboration has measured the production of the neutral pion and the $\eta$ meson in 
pp collisions at $\sqrt{s}=0.9$\UTeV\ and  $\sqrt{s}=7$\UTeV~\cite{Abelev:2012cn}. The production of the
neutral pion has also been measured at $\sqrt{s}=2.76$\UTeV~\cite{Abelev:2014ypa}. The measurements were performed using the two-photon decay channel and cover the mid-rapidity region and a large range in transverse momentum. The spectra are compared to NLO calculations using a variety of fragmentation functions. Data is well described at the lowest energy, while the predictions are not so successful at the largest energies. Nonetheless, the ratio $\eta/\pi^0$ is well described over the full energy range.

The LHCf detector~\cite{Adriani:2008zz} is purpose made for measurements
in the very forward direction. It is installed on both sides of the
ATLAS interaction point behind the inner beam separation dipoles that
sweep the charged particles from the collisions aside. Therefore only
neutral particles reach the detector and are measured by LHCf.

LHCf has so far operated at the LHC 900\UGeV\ and 7\UTeV\ p-p collisions
in 2009--2010 \cite{ADRIANI:2013ira} and 5.02\UTeV\ p-Pb collisions and
2.76\UTeV\ p-p collisions in 2013.  During the 2013 operation, only the
Arm2 detector was installed.  To operate at the 13\UTeV\ p-p collisions
in 2015 under a high radiation dose, the LHCf detectors were upgraded
by replacing plastic scintillators to Gd$_{2}$SiO$_{5}$ (GSO)
scintillators \cite{Kawade:2011zza,Suzuki:2013dva}.

\subsubsection{Past measurements of neutral particle spectra}

\begin{figure}
  \begin{center}
    \includegraphics[width=.9\linewidth]{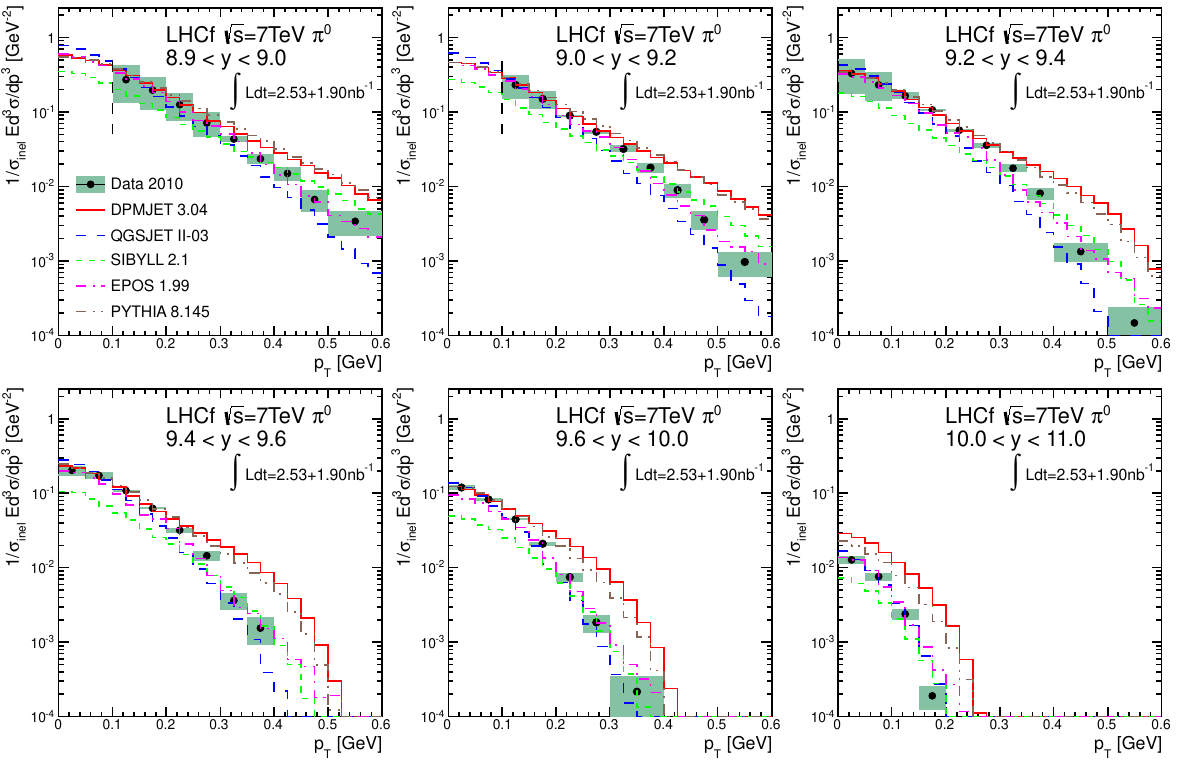}
    \caption{
              Transverse momentum spectra of $\pi^{0}$ observed at LHC 7\UTeV\ p-p collision.
              Experimental results are presented in six rapidity ranges together with the model predictions.
             }
    \label{fig:sako:pi0}
  \end{center}
\end{figure}

\begin{figure}
  \begin{center}
    \includegraphics[width=.9\linewidth]{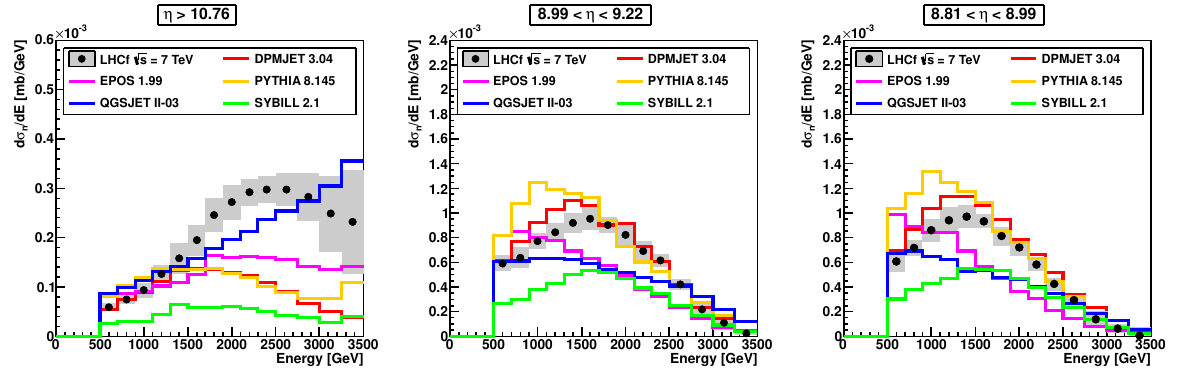}
    \caption{
               Forward neutron spectra measured at $\eta>$10.76, 8.99$<\eta<$9.22 and 8.81$<\eta<$8.99 in 
               LHC 7\UTeV\ p-p collision.
               Spectra are corrected for the detector efficiency and resolution, then compared with the model predictions~\cite{Adriani:2015nwa}.
             }
    \label{fig:sako:neutron}
  \end{center}
\end{figure}

\subparagraph*{Single photon spectra from 900\UGeV\ and 7\UTeV\ p-p collisions}
LHCf published photon spectra at 8.81$<\eta<$8.99 and 10.94$<\eta<\infty$ from
the 7\UTeV\ p-p collision data \cite{Adriani:2011nf}.  Similar
analysis was also performed from the 900\UGeV\ data for 8.77$<\eta<$9.46
and 10.15$<\eta<\infty$ \cite{Adriani:2012fz}.  Experimental results
were compared with the model predictions, but no model could perfectly
reproduce the experimental results.  On the other hand, the
experimental results were well between the variation of model
predictions.
\subparagraph*{$\pi^{0}$ spectra from 7\UTeV\ p-p collisions} Transverse
momentum ($p_{T}$) spectra of $\pi^{0}$ for 6 different rapidity ($y$)
ranges were obtained from the 7\UTeV\ p-p collision data as shown in Fig.\ref{fig:sako:pi0} 
\cite{Adriani:2012ap}.  Again no model could explain the experimental data
but \epos~1.99 \cite{Werner:2005jf,Pierog:2009zt} had a better agreement than the other
models.  Mean $p_{T}$ measurements, $<p_{T}>$, were obtained from 6 rapidity bins
by fitting the observed spectra using empirical functions.  When
plotting $<p_{T}>$ as a function of y$_{beam}$--y together with the
UA7 results \cite{Pare:1989mr} from Sp\={p}S 630\UGeV\ p-\={p} collisions, data
points were found to be smoothly connected.
\subparagraph*{$\pi^{0}$ spectra from 5\UTeV\ p-Pb collisions and nuclear
  effect} p$_{T}$ spectra of $\pi^{0}$ were also obtained from the
$\sqrt{s_{NN}}$=5.02\UTeV\ p-Pb collision data \cite{Adriani:2014mfa} at the
direction of the proton beam.  A characteristic shoulder structure was
observed and it was identified to be a result of Ultra-Peripheral
Collisions between a proton and electro-magnetic field around the Pb
nuclei.  After subtracting this UPC effect based on a calculation,
$\pi^{0}$ from QCD interaction were extracted.  The $\pi^{0}$ spectra
in p-p collisions at the corresponding collision energy were estimated
by interpolating the p-p data at 2.76\UTeV\ and 7\UTeV.  Dividing the
$\pi^{0}$ spectra in p-Pb collisions by those in p-p collisions and
theoretical number of binary collisions, the nuclear modification
factors were calculated.  Large factors around 0.1 were observed and
these values were explained by existing models.
\subparagraph*{Neutron spectra from 7\UTeV\ p-p collisions} Although the LHCf
calorimeters were optimized for the electro-magnetic shower
measurements, they have a sensitivity to the hadronic showers.
Hadronic shower events, predominantly neutrons, were analyzed from the
7\UTeV\ p-p collision data \cite{Adriani:2015nwa}.  In both folded and
unfolded spectra, energy spectra were compared between the data and MC
predictions. The unfolded spectra obtained for $\eta>$10.76, 8.99$<\eta<$9.22 and 8.81$<\eta<$8.99 are shown in Fig.\ref{fig:sako:neutron}.
  At the most forward direction including the zero degree,
a very hard spectrum was obtained and that was similar to the
prediction by the \texttt{QGSJETII}-03 model \cite{Ostapchenko:2004qz,Ostapchenko:2005nj} both in the shape and
absolute cross section.  At the smaller rapidities, the data were
close to the models predicting a high neutron yield.  When comparing
the number of neutrons over number of photons with energy larger than 100\UGeV,
 the experimental data showed a larger neutron yield than the
 models.

\subsubsection{Future measurements of neutral particle spectra}
\subparagraph{13\UTeV\ p-p in 2015} The original goal of the LHCf was to take
data at the highest accelerator energy possible.  The 14 (13)\UTeV\ 
p-p collisions to be achieved at the LHC correspond to 1.0
(0.9)$\times$10$^{17}$\UeV\ in the laboratory frame.  This is an
important energy in the cosmic-ray physics because the transition from
the galactic to the extra-galactic cosmic ray is expected at this
energy.  Because of a high radiation dose at the TAN location, LHCf
cannot survive for a long time even though using radiation-hard GSO
scintillators.  LHCf will take the highest energy data soon after the
start of RUN2 and the target collisions energy is 13\UTeV. 

\subsection{Heavy flavor particle spectra}

The measurement of inclusive very forward electrons in proton-proton collisions is a challenging
opportunity with a potential to provide unique insights into the very low-x quark structure of
hadrons which is very important for the forward physics leading air shower development. This specifically includes the heavy quark content which are also of particular interest because it is a background for astrophysical neutrino detection.

\subsubsection{Past measurements of heavy flavor particle spectra}
 
The ALICE Collaboration has measured the production of heavy flavour in pp collisions at 2.76 and 7\UTeV\ both at mid and at forward rapidity using semi-leptonic decays~\cite{Abelev:2012pi,Abelev:2012xe,Abelev:2012qh,Abelev:2012sca,Abelev:2014gla,Abelev:2014hla}. The measurements at forward rapidity are performed with the MUON spectrometer, which covers the rapidity range $2.5<y<4$ and transverse momenta from around 2 to 15 GeV/c. The dependence on these two variables is well described by theoretical models. In particular,  FONLL pQCD is in good agreement with data within experimental and theoretical uncertainties, although the data are close to the upper limit of the model calculations.

The results in the mid-rapidity region are obtained measuring the electrons from the semi-leptonic decays of hadrons containing heavy quarks. These analyses use the particle-identification capabilities of ALICE. The electrons are identified using their energy loss when traversing the Time Projection Chamber (TPC) complemented with the information from the Time-Of-Flight system and the Transition Radiation Detector. Another complementary technique is to use the TPC in conjunction with the Electromagnetic Calorimeter. The spectra are measured for the rapidity interval $|y|<0.5$ and the transverse momentum is measured from 0.5 to 8\UGeV/c.
To separate the electrons originating from the decay of beauty hadrons, the impact parameter of the lepton tracks with respect to the main interaction vertex is used. With this technique the contribution from charm and from beauty can be separated and measured independently. In both cases perturbative QCD calculations agree with the measured cross section within the experimental and theoretical uncertainties.

\begin{figure}[tbp]
\centering % \begin{center}/\end{center} takes some additional vertical space
\includegraphics[width=0.488\textwidth]{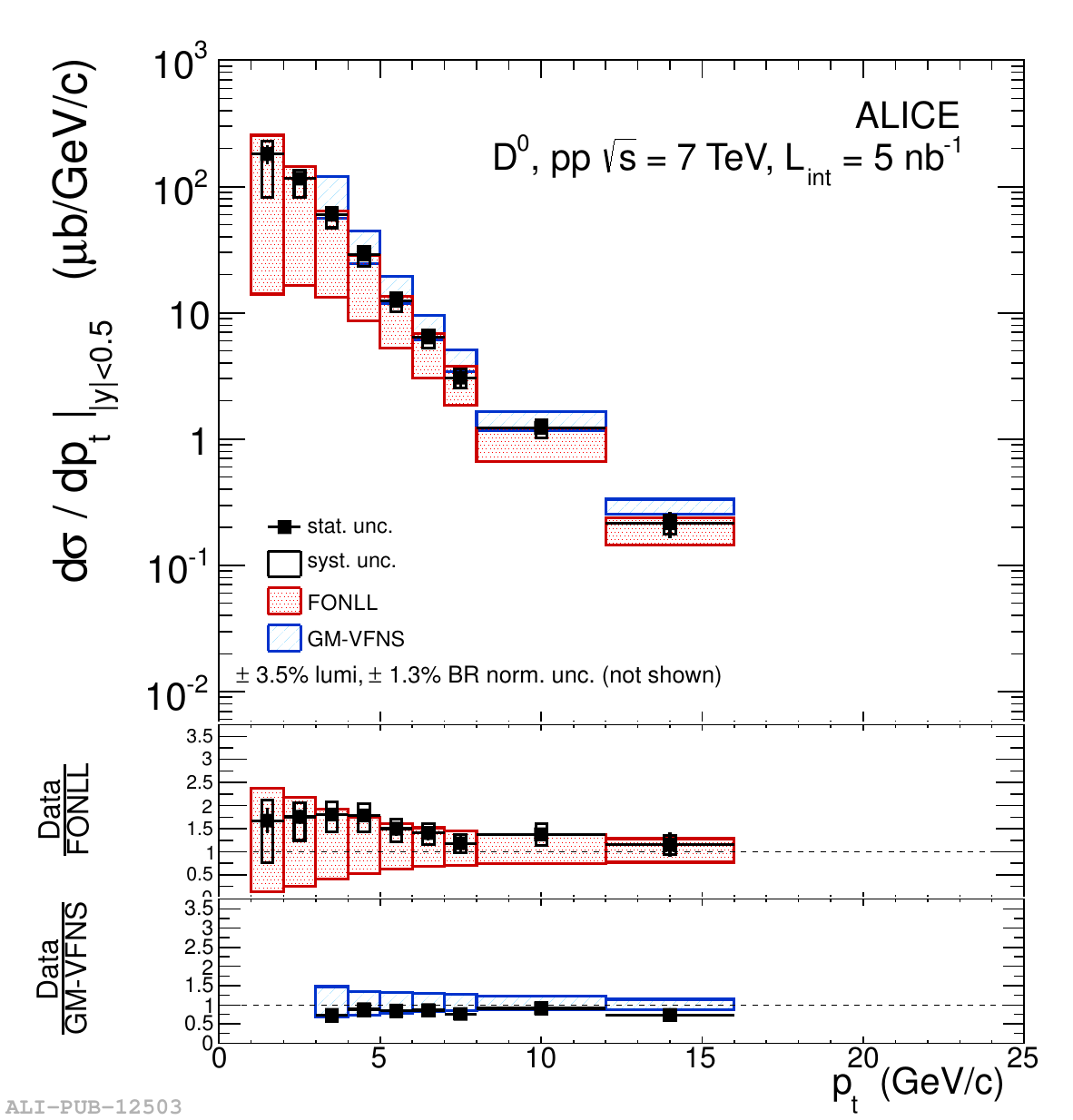}
\includegraphics[width=0.488\textwidth]{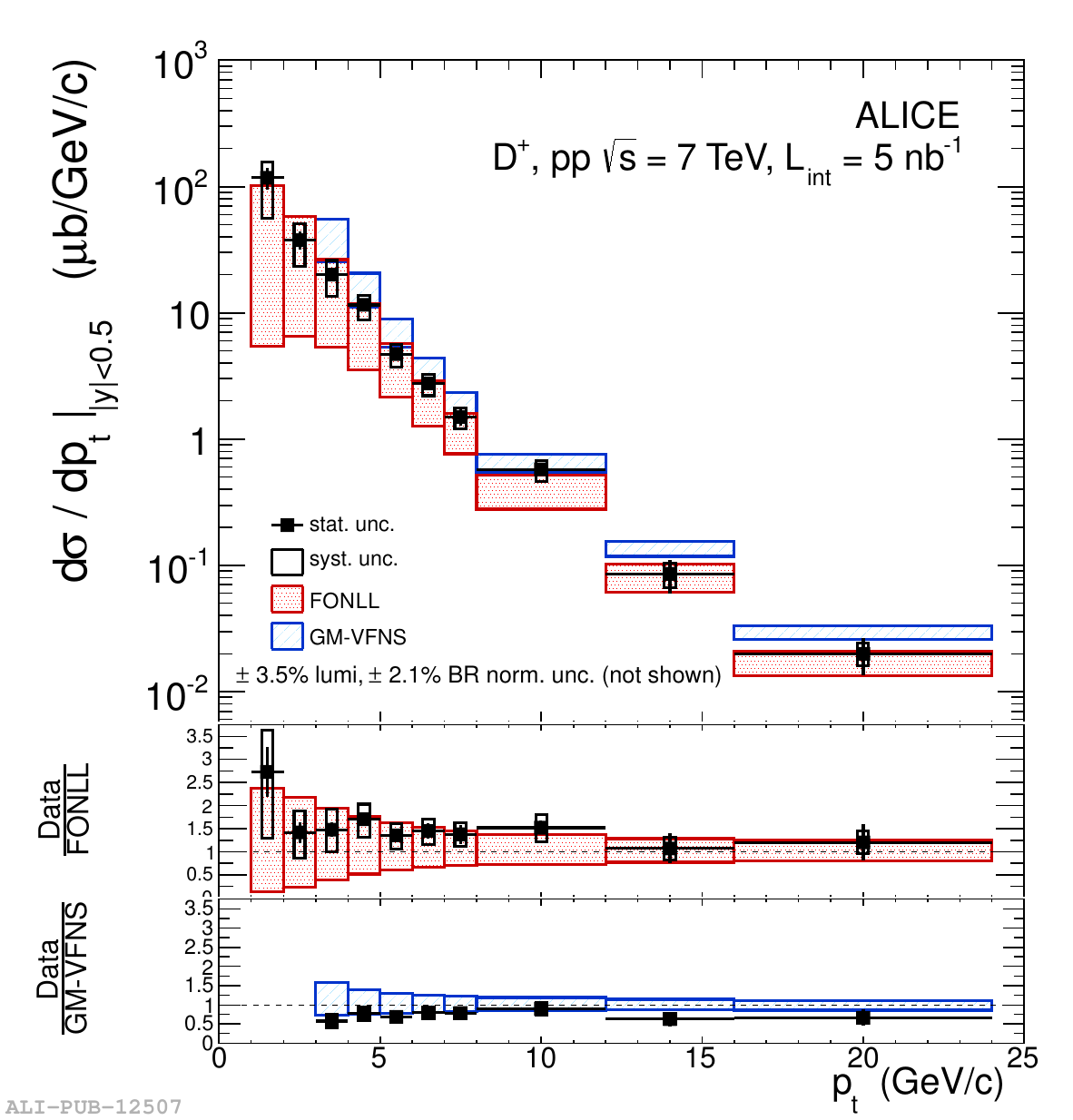}
\includegraphics[width=0.488\textwidth]{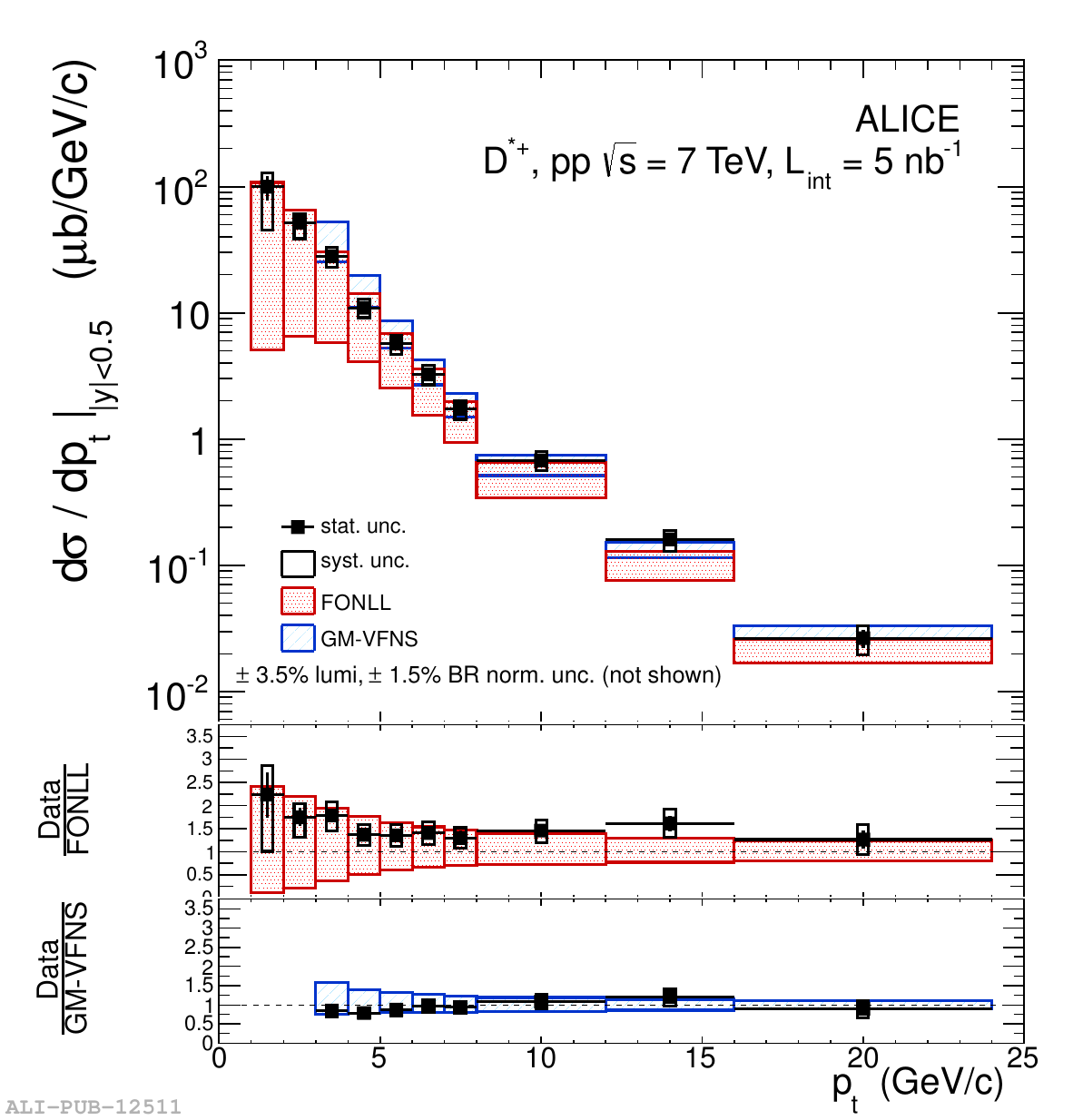}
\caption{\label{fig:ALICEdmesons} $D^{0}$ (top left), $D^{+}$ (top right) and $D^{*0}$ (bottom) inclusive $p_{T}$-differential cross-sections in pp collisions at 7~TeV. Figure from \cite{ALICE:2011aa} }
\end{figure}

In addition the ALICE collaboration has measured the production of charmed mesons such as the $D^{0}$, $D^{+}$ and $D^{*+}$ in pp collisions at 7\UTeV~\cite{ALICE:2011aa} at mid-rapidity ($|\eta|<0.5$). The weakly decaying $D^{0}$ and $D^{+}$ have decay vertices which 
are typically displaced by a few hundreds of $\mu m$ from the primary vertex. This displacement is such that, given the high resolution of ALICE tracking, topological selections are able to discern between decay daughters of these mesons and primary particles. Furthermore, the strongly decaying 
$D^{*+}$ will have decay positions indistinguishable from the primary vertex but can measured in its $D^{0}\pi^{+}$ decay channel, whereas topological selections
can again be used for the $D^{0}$ daughters. The resulting measurements are shown in Figure \ref{fig:ALICEdmesons} and are reproduced within
uncertainties by theoretical calculations based on QCD such as FONLL and GM-VFNS. 

\subsubsection{Future measurements of heavy flavor particle spectra}

As explained in section \ref{sec7:EnergyFlow:future}, the CASTOR experiment is ideally placed to study
forward energy flow relevant for air shower physics. However the complexity of the very
forward detectors is much reduced with respect to the central detectors, firstly, due to a more
extensive amount of dead material in front as well as other geometrical limitations, and, secondly,
because of the missing magnetic bending power to precisely measure particle momenta and particle
identification.

The combination of the TOTEM T2 tracking station with the CMS CASTOR calorimeter can to
some extend overcome some of these limitations. While T2 can precisely measure and tag charged
particles, CASTOR can identify electromagnetic from hadronic particles and also perform energy
measurements. For the example of electrons this allows a full reconstruction, since the particle
identification can be performed very reliably and, thus, also the particle four-vector can be
measured precisely. For other particles further assumptions are necessary, which is in general in the
very forward phase space not a major limitation. In particular the measurement of very forward
scattered partons, which fragmented into jets, may provide unique insight into the low-x structure of
gluons in hadrons. The measurement of inclusive very forward electrons will provide information
on the Drell-Yan process, and thus the quark content at low-x, but also on the heavy quark content at
very low-x via the decay of heavy hadrons into electrons. Forward Charm production, being very 
poorly known until now, could provide a very interesting input for high energy muon production
in air showers.

\begin{figure}
  \begin{center}
    \includegraphics[width=.45\linewidth]{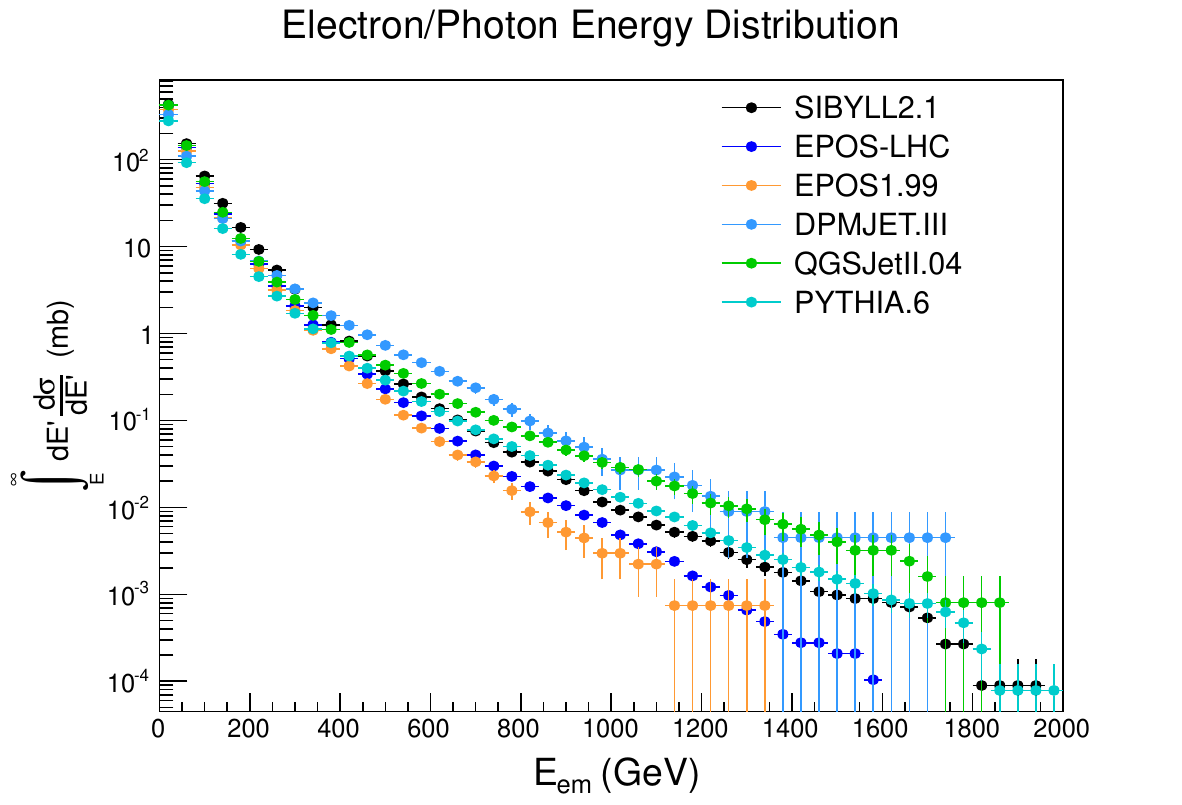}
    \includegraphics[width=.45\linewidth]{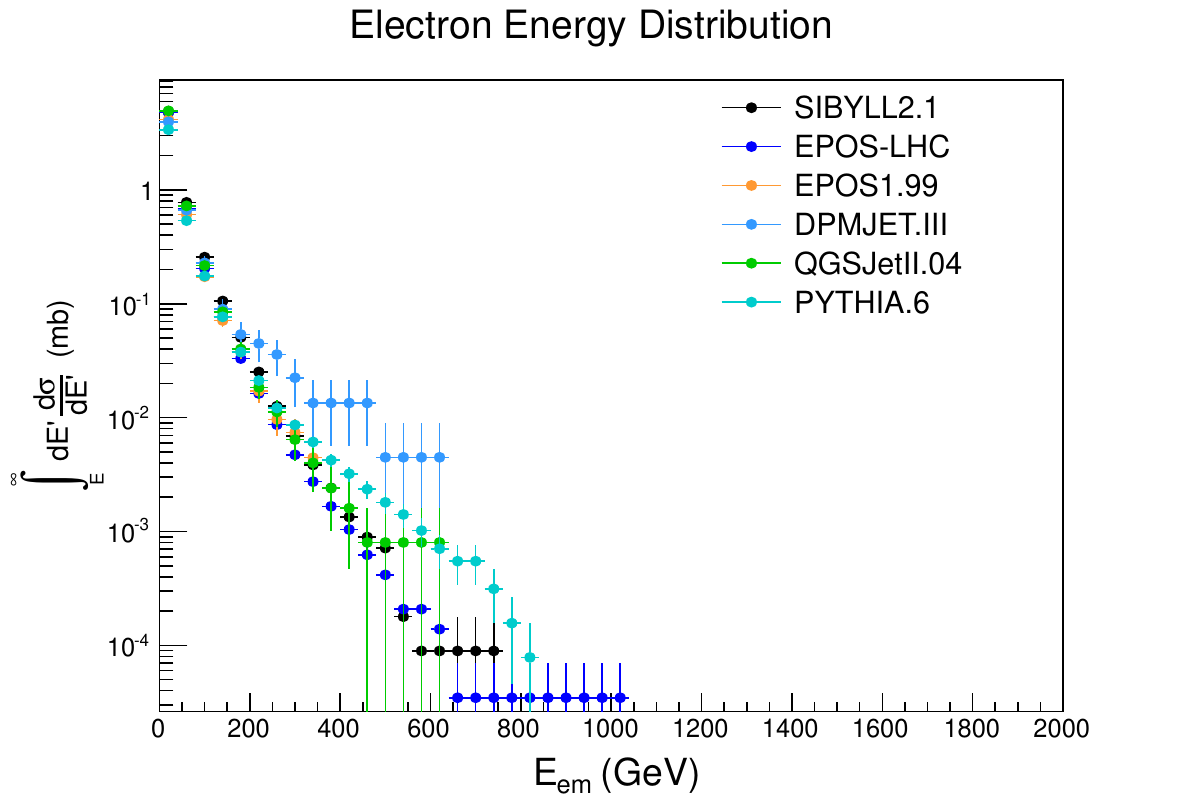}
    \caption{Left panel: Integrated inclusive production of very forward
electron and photons in 13\UTeV\ pp collisions. Right panel: same only
for very forward electrons. The cross sections drops rapidly as a function of the energy.}
    \label{fig7:ulrich:hcasem}
  \end{center}
\end{figure}

In order to reach to very high energy
electrons in the CMS experiment, and thus very low-x physics, an integrated luminosity of about 100\Unb$^{-1}$\ is necessary and 1\Upb$^{-1}$\ will be useful, where the average pileup should be small since at $\mu\sim0.07$ the average
underlying event contribution per event is on the level of $10\%$ for a 500\UGeV\ electron, which is the
case for $\mu \sim 0.15$ for 1\UTeV\ electrons. The best performance of this data analysis will be
achieved, when a dedicated trigger menu with an isolated very forward electron trigger is used.
However, it is not possible to perform a geometric matching of TOTEM T2 tracks with
electromagnetic clusters in proton-proton collisions, thus, photons cannot be distinguished from
electrons at trigger level with CASTOR. Fig.~\ref{fig7:ulrich:hcasem} indicates that only about $1\%$ of all electromagnetic
particles are in fact electrons. This situation requires to record the triggered events without any
prescale and do an offline event separation of electrons from photons in a dedicated data analysis.
At the same time this will set the allowed lower energy threshold for the trigger very high, in order
to keep the trigger rates low. The details depend on the luminosity and the pileup of the data taking.

%% file: cosmic/Beam.tex
\subsection{Proton-proton collisions} 
%T. Martin for ATLAS
For inclusive inelastic studies, a data sample of 10 million events is
required to sufficiently populate the tails at large charged particle
multiplicity and \pt. Previous experience from ATLAS at $\sqrt{s}
= 8 \TeV$ during LHC fill number 2470 indicates that such a
sample is obtainable with approximately three hours data taking with
an output rate of 1 kHz from ten colliding bunches and an average
number of interactions per bunch-crossing $<\mu> = 0.003$. Low $<\mu>$
($\sim 1\%$) is important to allow for the correlation of the central
and forward detectors and to preserve any large rapidity gaps.

MC predictions (Table~\ref{tab:timm:crosssections}) indicate
that a forward tag rate of the order $1\%$ should be expected. Therefore
three hours with $\sim$1000 bunches or 30 hours with $\sim$100 bunches
is sufficient to collate a forward-tagged data sample of equivalent
size for a given optics arrangement.
%T.Sako for LHCf
 To avoid
collision pileup, an operation at low luminosity,
L$\sim$10$^{29}$\,cm$^{-2}$s$^{-1}$, is required.

For LHCf experiment, in addition %At the same time,
an angle of individual particle $\beta^{*}$ larger than
10\,m is necessary.  To cope with a slow data acquisition, number of
collision bunches below 40 are also required.  Because these requests
can be shared with those of the van der Meer scan, the dedicated
operation of LHCf will be carried out with the vdM scan campaign in May 2015.  Beam parameters for this
operation are summarized as follows.
\begin{itemize}
\item Instantaneous luminosity; 1.5$\times$10$^{29}$\Ucm$^{-2}$s$^{-1}$
\item Integrated luminosity 15\Unb$^{-1}$
\item Number of collision bunch pair; 40, a few non-collision pairs
\item Maximum crossing angle (half angle 145\Uurad)
\item $\beta^{*}$=19\Um\ as defined from vdM scan
\end{itemize}

These beam conditions will allow most of the analysis needed to tune the hadronic
interactions models for air shower simulations presented in this chapter.

LHCf is also planning to take data with ATLAS. LHCf will send its
trigger signal to ATLAS and ATLAS will record the data after
prescaling this trigger. By tagging the forward events using the
central detector, classification of diffractive and non-diffractive
events at the event-by-event basis will be available.

Models for extensive air-showers constrained by the collider data rely on extreme extrapolations to high energies. These extrapolations depend on the underlying assumptions in the models, related to particular types of interactions and their properties. The LHCf experiment has provided useful information on the forward energy flow from neutral particles. However, without knowing the true nature of the pp interaction, the production mechanism cannot be fully understood. Therefore, a combined LHCf+ATLAS pp run would help to classify the interactions and further tune the air-shower models. We propose the following two measurements:

\begin{itemize}
\item Measure the energy flow of photons, $\pi^0$ and neutrons in LHCf in the case of an activity in the central ATLAS detector (non-diffractive events) and in the case of no activity in central ATLAS detector (low mass single diffractive or double-diffractive events).
\item Measure $dN/d\eta$, multiplicity, $dN/dp_t$ and <$p_t$> vs. multiplicity in ATLAS for the events triggered by LHCf.
\end{itemize}

Such measurements would complement possible outcomes of the analysis of the p+Pb data taken in early 2013 where a common LHCf trigger was defined in the ATLAS trigger menu. Nuclear effects need to be taken into account in order to understand diffractive p+Pb interactions and pp can be used as a reference here. For the cosmic air-shower models, it is the p-Air processes that matter in the end.

\subsection{Light ion collisions} 

Due to its SMOG system, which allows injection of noble gases into
the interaction point, the LHCb detector is able to perform fixed
target physics with beam-gas interactions~\cite{LHCb-CONF-2012-034}.
In particular, during the 2012 proton-ion pilot run, the neon gas
has been injected into the LHCb interaction region that has
increased the beam-gas interaction rate by two orders of magnitude.
It allows accurate measurements of the beam profile for
a precise determination of the absolute luminosity.
In this particular case the nucleon-nucleon center-of-mass energy was 
$\sqrt{s_{NN}}=87$\UGeV\ and the whole
system was boosted by $\Delta y\approx 4.5$\/ units in the direction
of the proton beam.
The rate of beam-gas interactions between the proton beam
and the injected gas was sufficient to measure light quark and
strangeness production. In particular, clear signals have been
observed for different strange hadrons~\cite{LHCb-CONF-2012-034}.
Studies of light quark and strangeness production in these collisions
provide unique input to cosmic-ray interaction models and
will be conducted by the LHCb experiment although the center-of-mass energy
is still very low compared to nominal LHC data.

As a consequence, to get full advantage of the LHC beam and reach a center-of-mass energy
per nucleon of $\sqrt{s_{NN}}=10$\UTeV,
LHCf is discussing a future
possibility of light-ion collisions at LHC.  This is ultimate goal for
the cosmic-ray physics to simulate CR-atmosphere interactions as explained in~\ref{sec7:pO}.  A
technical feasibility was presented in \cite{django}.  From the
detector point of view, even with light ion A-A collisions a
multiplicity around zero degree is significantly higher than the case
of p-p collisions.  Even using a LHCf-like small calorimeter, more
than one particle enter a single calorimeter in $80\%$ of events.  To
reconstruct the energies of these `multi-hit' particles, energy
measurement with a higher granularity is required.  Silicon-pad
detectors having a pad size of $\sim$mm are thought to fit this
request.  (Note that the Moli\`{e}re unit of Tungsten is 9\Umm\ and
any very fine structure is not useful.)  A basic R\&D to design a
future light-ion collision experiment is started.

Such a light ion beam could be used by all other experiments for measurements
presented in this section and will be of great help to further constrain the MC generator
necessary for air shower simulation and to reduce uncertainties in cosmic ray measurements as explained in \ref{sec7:pO}.

%% file: heavyion/heavyion.tex
\section{Introduction}
\label{hi:intro}
The LHC is not only the most powerful collider for proton-proton and heavy-ion collisions, but also for photon-photon and photon-hadron ($\gamma \rm{p}$ and $\gamma \rm{Pb}$) interactions, offering a unique opportunity to study fundamental aspects of QED and QCD via photon-induced processes. The protons and ions which are accelerated by the LHC themselves carry an electromagnetic field, which can be viewed as a source of photons~\cite{Fermi:1924tc,Fermi:1925fq,Williams:1934ad,vonWeizsacker:1934sx,Bertulani:2014dsa}. That is, a photon generated by one of these hadrons can interact with another photon (or with a hadron) producing a wide variety of particles. 

In recent years there has been an increasing interest in these physics processes that can be studied in ultra-peripheral collisions (UPC) in hadronic colliders~\cite{Baltz:2007kq,Bertulani:2005ru}. The relevant collisions typically occur at impact parameters of several tens (or even hundreds) of femtometres --cases when the incoming ions barely overlap, and well beyond the range of the strong force. This is because these reactions occur when the protons or ions pass by each other with impact parameters larger than the sum of their radii and are mediated by the exchange of virtual photons between the nuclei. The number of photons scales as the square of the nuclear charge while typical photon energies scale with the Lorentz contraction of the nuclei and so increase with beam energy. 

The beam energies at the LHC are high enough to make the LHC the most energetic photon source ever built. In Pb-Pb collisions the LHC can reach $W_{\gamma \rm{Pb}}$ energies up to 500 GeV, while in p--Pb collisions it can reach $W_{\gamma \rm{p}}$ up to about 1500 GeV in $\gamma \rm{p}$ interactions. Photon-induced processes have by far the largest cross sections in PbPb collisions at the LHC. The total cross section for breaking up one of the nuclei through a photonuclear process is over 200 barns. In most of these reactions the nucleus just breaks up without any particle production. However, the cross section for having at least one photoproduced charged particle at mid-rapidity is still substantial, about 4 b. But both these numbers are dwarfed by the total cross section for producing an $e^{+}e^{-}$ pair from an interaction between two photons. This cross section is about 3 million times larger than that for normal hadronic pp collisions.

We discuss recent results on ultra-peripheral heavy-ion collisions measured at the LHC. Special emphasis is given to the measurements of exclusive photonuclear processes, as well as the prospects for future photon-induced measurements.

\section{Exclusive photonuclear processes}

A photonuclear interaction that has attracted a lot of interest is exclusive vector meson production. That is, a reaction where only a vector meson is produced in the final state, and nothing else. The large cross section of this process can be understood from what is known as Vector Meson Dominance. This means that the photon may fluctuate into a quark-anti-quark pair and, since the photon has spin 1 and negative parity, the fluctuation will most likely be to a vector meson. 

Exclusive vector meson production in heavy-ion collisions provides a way to probe the nuclear gluon density for which there is a considerable uncertainty at low values of Bjorken-$x$, where $x$ is the fractional parton momentum $x = p_{\rm{parton}}/p_{\rm{hadron}}$. For example, a J/$\psi$ produced at rapidity $y$ is sensitive to the gluon distribution at $x=\frac{M_{J/\psi}}{\sqrt{s}}e^{\pm y}$ at hard scales $Q^2$ $\sim \frac{M_{J/\psi}^2}{4}$. 

\begin{figure}[htbp]
\centering
%[width=0.8\columnwidth]{
\includegraphics[width=0.4\columnwidth]{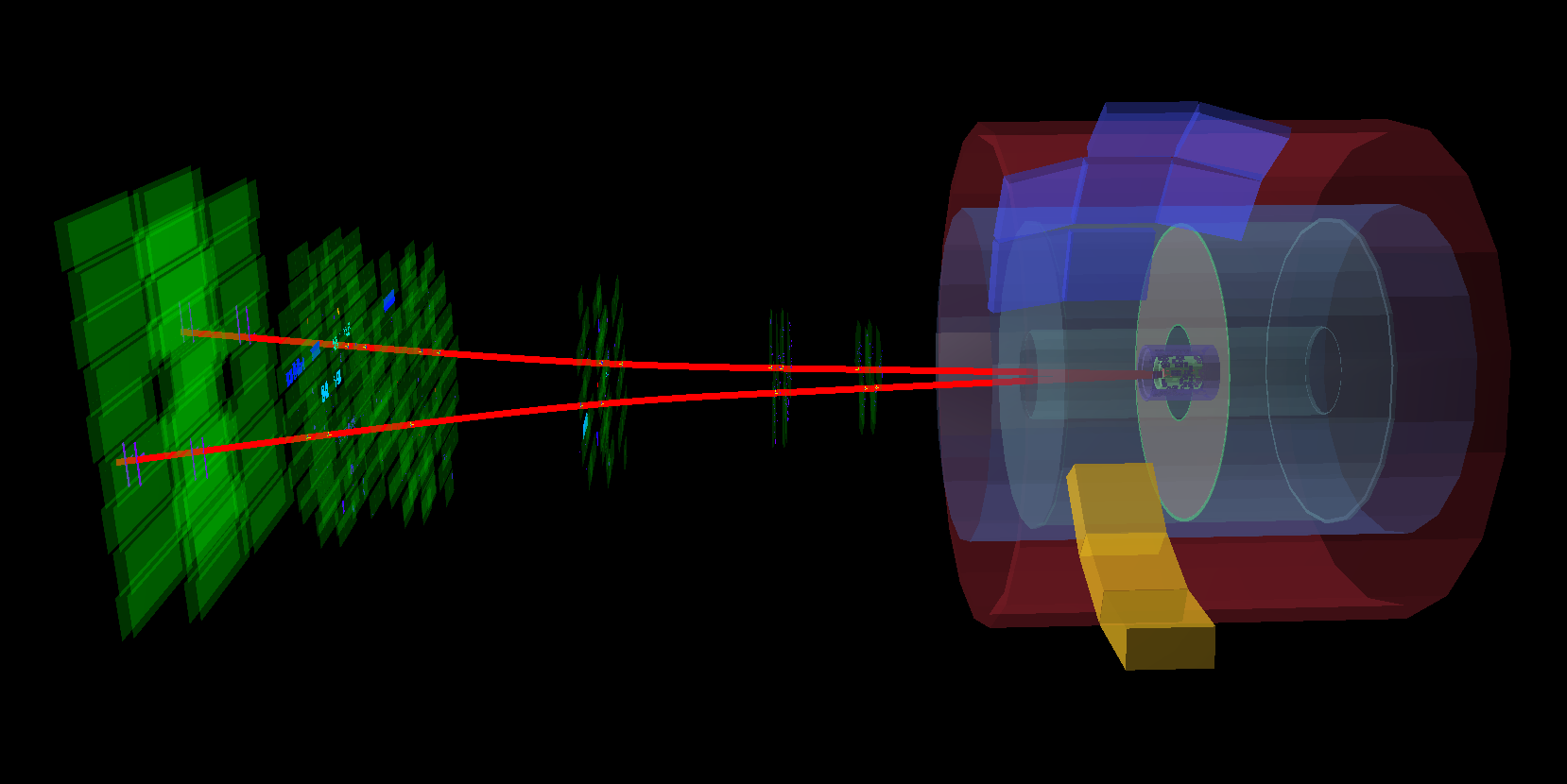}
\caption{Event display of $J/\psi$ candidate produced in an ultra-peripheral Pb--Pb collision at  $\sqrt{s_{NN}} = 2.76$ TeV.}
\label{Fig:JpsiEventDisplay}
\end{figure}

\begin{figure}[t]
\begin{center}$
\begin{array}{cc}
(a)\includegraphics[width=0.45\textwidth]{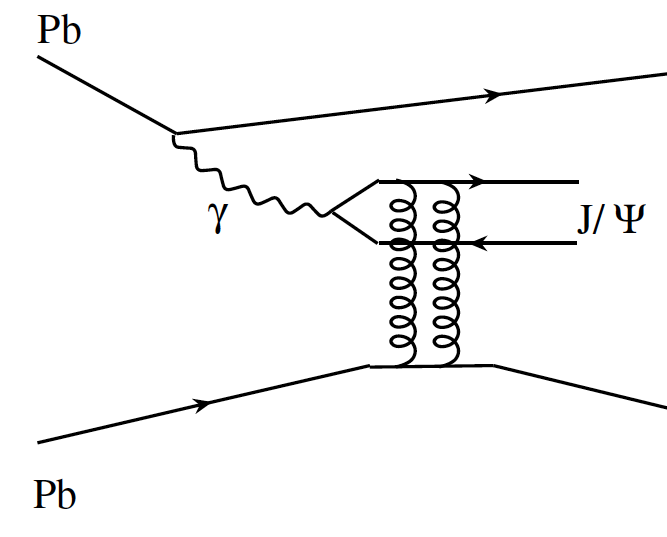} & 
(b)\includegraphics[width=0.45\textwidth]{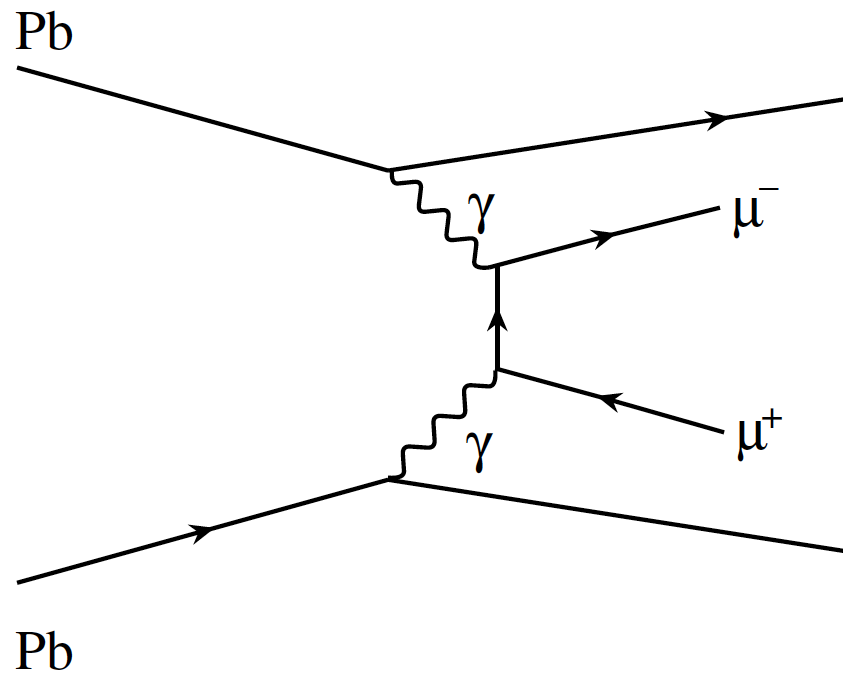} \\
\end{array} $
\end{center}
   \caption{Feynman diagrams for the photoproduction of vector messons (panel a) and the two-photon process (panel b) in ultra-peripheral Pb--Pb collisions.}
\label{fig:FeynDiag}
\end{figure}

Up to now, two types of  UPC processes have bean measured at the LHC: (1) The photoproduction of a vector meson in photo-nuclear interactions, where the vector meson is reconstructed from its decay products, and (2) the two-photon process decaying to a di-lepton pair ($\gamma\gamma \rightarrow l^{+}l^{-}$), where $l = e,\mu$. Studies to $\tau$ decays have not been performed so far. The experimental signature of these events is characterized by their very low transverse momenta. Apart from two tracks in the final state the detector is otherwise empty. Figure~\ref{Fig:JpsiEventDisplay} shows an event display for a J/$\psi$ candidate produced in an ultra-peripheral Pb--Pb collision at  $\sqrt{s_{NN}} = 2.76$ TeV with CMS. The Feynman diagrams for these proceses are shown in Figure~\ref{fig:FeynDiag}. These processes are further classified into the following classes of events

\begin{itemize}
\item Photoproduction off nuclei in ultra-peripheral Pb--Pb collisions
\begin{itemize}
\item[$\circ$] {\bf Coherent production:} The photon interacts coherently with the whole nucleus. The coherence condition, both in the emission of the photon and in the interaction with the nuclear target, constraints the transverse momentum of the produced di-lepton or vector meson to be of the order of $1/2R_{\rm Pb}$ -- where $R_{\rm Pb}$ is the radius of the lead nucleus -- which corresponds to a $p_{T}\sim$ 60 MeV/$c$. 
\item[$\circ$] {\bf Coherent production with nuclear break up:} Owing to the intense electromagnetic fields of the lead nuclei it is possible to have independent electromagnetic interactions between the nuclei. These additional interactions may excite at least one of the nuclei, resulting in the emission of at least one neutron in the same direction to that of the emitting nucleus. 
%The experimental signature is similar as the one in the previous case, but with the addition of at least a neutron which can be detected by zero degrees neutron calorimeters.
%
\item[$\circ$] {\bf Incoherent production}. In this case the photon interacts not with the whole nucleus, but rather with a single nucleon. There are two main differences with respect to the coherent case. First, as the radius of the nucleon is smaller than that of the nuclei, the transverse momentum of the produced system is larger, around 300 MeV/$c$. Second, in the incoherent case the interaction makes the nuclei to break up producing, in almost all cases, forward neutrons. 
\end{itemize}
\item  Photoproduction off protons in ultra-peripheral p--Pb collisions
\begin{itemize}
\item[$\circ$] {\bf Exclusive production:} The photon interacts with the proton without breaking it. The transverse momentum of the produced system is of the order of 300 MeV/$c$.
\item[$\circ$] {\bf Dissociative production:} The proton is excited by the interaction and dissociates. The transverse momenta of the produced di-lepton or vector meson extends to well above 1 GeV/$c$.
\end{itemize}
\end{itemize}

The cross section for the vector meson photoproduction in ultra-peripheral Pb--Pb collisions is given by

\begin{equation}
\frac{d\sigma_{\rm PbPb}(y)}{dy}= N_{\gamma/\rm Pb}(y,M)\sigma_{\gamma \rm Pb}(y) + N_{\gamma/\rm Pb}(-y,M)\sigma_{\gamma \rm Pb}(-y),
\label{eq:SigPbPb}
\end{equation}
where $M$ is the mass of the produced vector meson state, and $y$ is the rapidity given by $y = \ln(2k/M)$, where $k$ is the photon energy and  $\sigma_{\gamma \rm Pb}(y)$ is the corresponding photoproduction cross section. $N_{\gamma/\rm Pb}$ is the photon flux. There are two terms because each of the incoming lead nuclei may act as the photon source. 

A similar formula can be written for the production of vector mesons in ultra-peripheral p--Pb collisions. In this case, the term involving the photon emission by the proton is very small and the dominant contribution is given by
 
\begin{equation}
\frac{d\sigma_{\rm pPb}(y,M)}{dy} \approx N_{\gamma/\rm Pb}(y,M)\sigma_{\gamma \rm p}(y) 
\end{equation}

In a similar way,  the two-photon cross section can be calculated. In this case two photon fluxes are convoluted with the corresponding photon-photon cross section.

The photon flux per unit area in the semi-classical description is given by (see for example \cite{Baur:2001jj})

\begin{equation}
n(k,\vec{b}) = \frac{\alpha Z^2}{\pi^2 b^2}x^2\left[K^2_1(x)+\frac{1}{\gamma}K^2_0(x)\right],
\label{eq:FluxPerArea}
\end{equation}

where $k$ is the photon energy in the nucleus frame with Lorentz factor $\gamma$, $Z$ is the electric charge of the emitting heavy nuclei, $K_{0}$ and $K_1$ are Bessel functions and $x=kb/\gamma$.
This formula is a good approximation for heavy nuclei and at impact parameters $b$ larger than $b_{\rm min}$, the sum of the radii of the interacting particles.  In this case, the photon flux $n(k) = \int d^2\vec{b}\ n(k,\vec{b})$ is given by

\begin{equation}
n(k) = \frac{2\alpha Z^2}{\pi}\left[ \xi K_0(\xi)K_1(\xi)-\frac{\xi^2}{2}\left(K^2_1(\xi)-K^2_0(\xi)\right)\right],
\label{eq:FluxHS}
\end{equation}

where $\xi=kb_{\rm min}/\gamma$. 

The photon flux from a lead nucleus is then obtained using the corresponding values of $Z$ and $\gamma$ and using rapidity instead of photon energy as the relevant variable:

\begin{equation}
N_{\gamma/\rm Pb}(y,M)\equiv \left. k\frac{dn(k)}{dk}\right|_{\rm Pb}. 
\end{equation}

%t [tbh!f] 
\begin{figure}[tbh!] 
\begin{center}$
\begin{array}{c}
\includegraphics[width=1.\columnwidth]{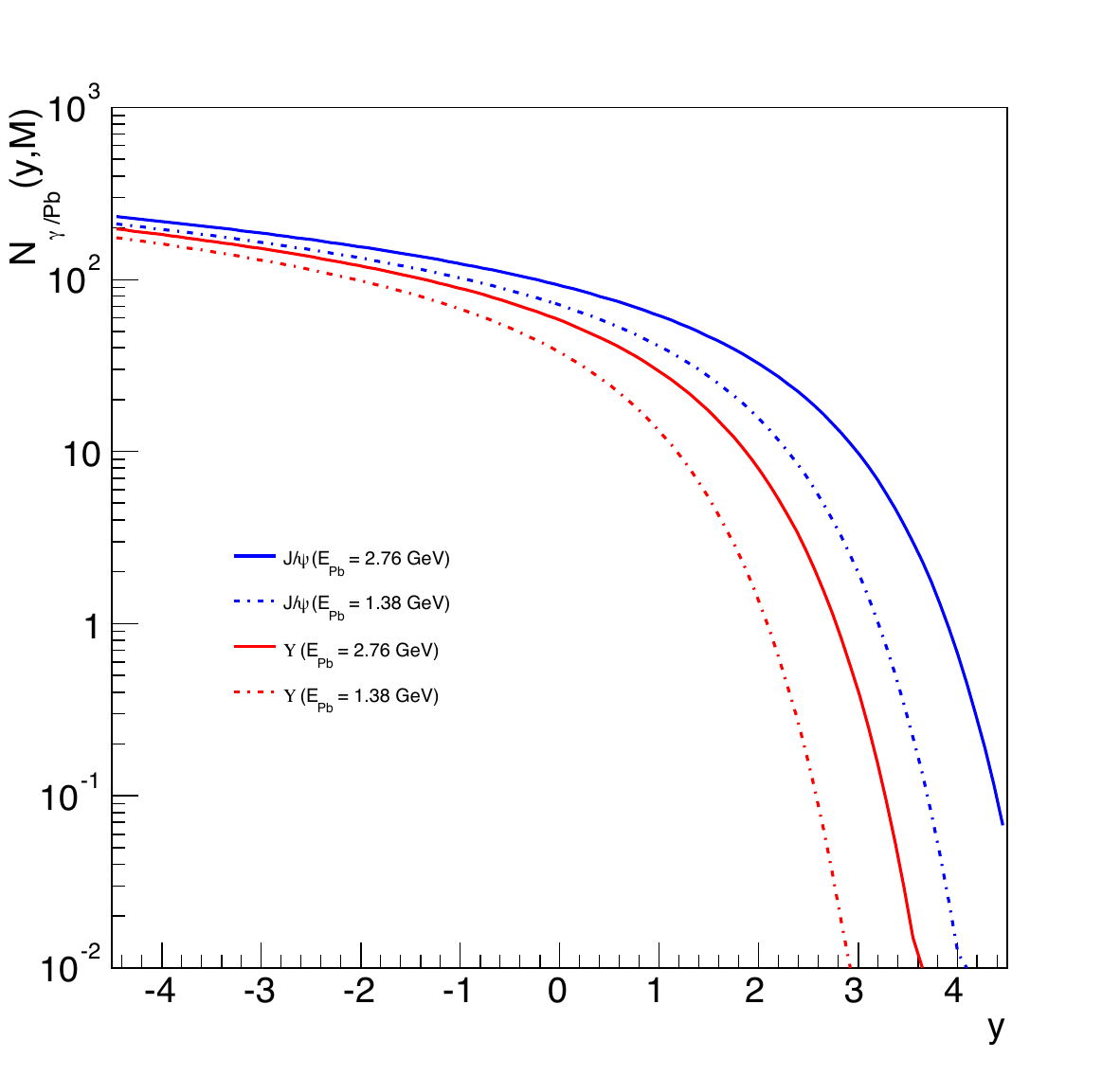} \\
\end{array} $
\end{center}
   \caption{Photon flux emitted by a lead nucleus at two different LHC energies  as a function of rapidity for photon energies corresponding to the production of a  $\jpsi$ and a $\Upsilon$ vector meson at the corresponding rapidity. }
\label{fig:Flux}
\end{figure}	

Figure \ref{fig:Flux} presents  $N_{\gamma/\rm Pb}(y,M)$ as a function of rapidity for the masses of the $\jpsi$ and the $\Upsilon$ for two different energies of the lead beam: 1.38 GeV --used during Run 1 in the Pb--Pb data taking periods of 2010 and 2011-- and 2.76 TeV which will be possible in Run 2. The drop of the flux at large rapidities is given by the behaviour of the Bessel functions at large photon energies. Comparing the Run 1 and Run 2 scenarios, there is a significant increase in the acceptance at forward rapidities. According to STARLIGHT (see section on Monte Carlo generators for more details), this represents a 60\% increase in statistics for $\jpsi$, 70\% increase for $\psi$(2S) and a 200\% increase for $\Upsilon$(1S). This would allow differential studies using Run 2 data. The colliding system for nucleus-nucleus collisions to be taken from Run 3 onwards will be decided depending on the physics outcome from Run 2. Possibilities include not only p--Pb and Pb--Pb collisions, but also light ion colliding systems such as Ar--Ar. 

%%%%%%%%%%%%%%%%%%%
%  M O D E L S                               %
%%%%%%%%%%%%%%%%%%%

\section{Models for photonuclear production}
\label{sec:theory}

The following models have predicted cross sections for photonuclear production at LHC energies will be discuss in this section:
\begin{itemize}
\item[] {\bf AB-AN}: Model by Adeluyi and Bertulani \cite{Adeluyi:2012ph} and Adeluyi and Nguyen \cite{Adeluyi:2013tuu};
\item[] {\bf CSS}: Model by Cisek, Sch\"afer and Szczurek \cite{Cisek:2012yt};
\item[] {\bf KN}: Model by Klein and Nystrand implemented in the STARLIGHT Monte Carlo program  
\cite{Klein:1999qj,Baltz:2002pp,Klein:2003vd};
\item[] {\bf LM}: Model by Lappi and Mantysaari \cite{Lappi:2010dd,Lappi:2013am};
\item[] {\bf GM-GDGM}: Model by Goncalves and Machado \cite{Goncalves:2011vf} and by Gay-Ducati, Griep and Machado \cite{Ducati:2013bya};
\item[] {\bf RSZ}: Model by Rebyakova,  Strikman and Zhalov \cite{Rebyakova:2011vf}, and 
\item [] {\bf IKS}: Model by Ivanov, Kopeliovich and Schmidt~\cite{Ivanov:2007ms}. 
\end{itemize}

All models are based on Equation (\ref{eq:SigPbPb}) which has two ingredients: the photon flux and the photonuclear cross section. The first difference among the models is that some of them (CSS, LM, GM-GDGM) use the hard sphere approximation of the photon flux; i.e., equation (\ref{eq:FluxHS}), and other models (AB-AN, KN, RSZ-GZ) integrate the convolution of  equation (\ref{eq:FluxPerArea}) with the probability of no hadronic interaction. 

As for the photonuclear cross section the models contain the following ingredients: ($i$)  the models have to assume a nuclear distribution in the transverse plane, ($ii$) the models also include implicitly or explicitly a prescription for the wave function of the vector meson and finally ($iii$)
all models fix some of the parameters using data on exclusive photoproduction of charmonium off the proton and thus have to include a prescription to link the photoproduction off protons with that of the photonuclear interaction. For these reasons, the models can be grouped into three different groups: models based on the generalized vector dominance model (KN), on LO pQCD (AB-AN, RSZ) and on the color dipole model (CSS, LM, GM-GDGM).

\subsection{Models based on the vector dominance model}

The only model in this class is the KN model. The main ingredients of this model are three. The vector dominance model (VDM) relates both the $\gamma \mathrm{+Pb}\to\mathrm{Pb+V}$ and the $\gamma \mathrm{+p}\to\mathrm{p+V}$ processes to $\mathrm{Pb+V}\to\mathrm{Pb+V}$ and $\mathrm{p+V}\to\mathrm{p+V}$, respectively. Here $V$ represents a vector meson. The optical theorem relates these last processes to the total cross section. Finally a classical Glauber model relates the total cross section on protons to that on nuclei. This can be expressed in the following way 

\begin{equation}
\sigma_{\gamma \rm Pb}(y) \equiv\sigma(\gamma{\rm+Pb}\to{\rm V+Pb}) =
\left. \frac{d\sigma(\gamma{\rm+Pb}\to{\rm V+Pb}) }{dt} \right|_{t=0}
\int^\infty_{t_{min}} dt |F(t)|^2,
\label{eq:Fwd}
\end{equation}
where  $F(t)$ is the nuclear form factor and $t$ the momentum transferred to the nucleus. 
Using VDM and the optical theorem yields
\begin{equation}
\left. \frac{d\sigma(\gamma{\rm+Pb}\to{\rm V+Pb}) }{dt} \right|_{t=0}
= \frac{\alpha \sigma^2_{\rm TOT}({\rm Pb+V})}{4f^2_V},
\end{equation}
where $f_V$ is the vector meson photon coupling. A classical Glauber model produces
\begin{equation}
\sigma_{\rm TOT}({\rm Pb+V}) =\int d^2\vec{b}\left(
1-\exp\left[-\sigma_{\rm TOT}({\rm p+V}) T_{\rm Pb}(\vec{b}) \right]
\right),
\end{equation}
where $T_{\rm Pb}$ is the nuclear thickness function and $\sigma_{\rm TOT}({\rm p+V})$ is obtained from  the optical theorem, now applied at the nucleon level
\begin{equation}
\sigma^2_{\rm TOT}({\rm p+V}) = 16\pi
\left. \frac{d\sigma({\rm V+p}\to{\rm V+p}) }{dt} \right|_{t=0}.
\end{equation}
Using VDM leads to
\begin{equation}
\left. \frac{d\sigma({\rm V+p}\to{\rm V+p}) }{dt} \right|_{t=0} = \frac{f^2_V}{4\pi\alpha}
\left. \frac{d\sigma(\gamma{\rm+p}\to{\rm V+p}) }{dt} \right|_{t=0},
\end{equation}
where  the elementary cross section
\begin{equation}
\left. \frac{d\sigma(\gamma{\rm+p}\to{\rm V+p}) }{dt} \right|_{t=0}
= b_V\left( XW^\epsilon_{\gamma\rm p} + YW^{-\eta}_{\gamma\rm p}\right)
\end{equation}
is fitted to experimental data to obtain the values for the 
$X$, $Y$, $\epsilon$, $\eta$ and $b_V$  parameters.
 
 \subsection{Models based on LO pQCD}

These models start from Equation \ref{eq:Fwd} and use the LO pQCD calculation 
\cite{Ryskin:1992ui,Brodsky:1994kf} for the forward cross section
\begin{equation}
\left. \frac{d\sigma(\gamma{\rm+Pb}\to{\rm V+Pb}) }{dt} \right|_{t=0}=
\frac{16\pi^3\alpha^2_s \Gamma_{ee}}{3\alpha M^5}
\left[ xG_A(x,Q^2)
\right]^2,
\label{eq:LOpQCD}
\end{equation}
where $\Gamma_{ee}$ is the decay width to electrons and $G_A$ is the nuclear gluon density distribution at a scale $Q^2$, which for the models described below was chosen to be $Q^2 = M^2/4$, although other options are possible and may describe better the experimental data \cite{Guzey:2013qza}. It is important to note that this equation contains implicitly a model for the wave function of the vector meson, but in the final result the only trace of it is the presence of $\Gamma_{ee}$. 

The AB-AN model modifies equation \ref{eq:LOpQCD} by adding a normalization parameter to the right side, which should take into account effects missing in the approximation. This factor is then fitted to reproduce HERA data using the same type of equation applied to the $\gamma \mathrm{+p}\to\mathrm{p}+\jpsi$ case. Nuclear effects are modelled as $G_A(x,Q^2) = g_p(x,Q^2) R^A_g(x,Q^2)$, where $g_p$ is the gluon distribution in the proton and $R^A_g$ is the nuclear modification factor of the gluon distribution.
MSTW08 \cite{Martin:2009iq} is used for the gluon distribution in the proton, while several different choices are made for $R^A_g$ to estimate nuclear effects: EPS08 \cite{Eskola:2008ca}, EPS09 \cite{Eskola:2009uj}, HKN07 \cite{Hirai:2004wq} and  $R^A_g = 1$ to model the absence of nuclear effects.

The RSZ model computes  $R^A_g$ in the leading twist approach to nuclear shadowing 
\cite{Frankfurt:2011cs}. The main ingredients are the factorization theorem for hard diffraction  and the theory of inelastic shadowing by Gribov. The evolution  is done using DGLAP equations. The experimental input to fix the parameters of the model is given by inclusive diffractive parton distribution functions of nucleons as measured at HERA.
For the gluon distribution in the proton they use the LO distribution from \cite{Martin:2007sb}.

\subsection{Models based on the colour dipole approach}

The basic idea of this formalism is that long before the interaction, the photon splits into a quark-antiquark pair, which forms a colour dipole. Long time afterwards, this dipole interacts with the target and after another long time the dipole creates a vector meson. The cross section in this formalism is given by
\begin{equation}
\frac{d\sigma(\gamma{\rm+Pb}\to\jpsi + {\rm Pb}) }{dt} =
\frac{R^2_g(1+\beta^2)}{16\pi}\left|
A(x,Q^2,\vec{\Delta})\right|^2,
\end{equation}
where the so called skewdness correction $R^2_g$ compensate for the fact that only one value of $x$ is used, even thought the two gluons participating in the interaction have different $x$ \cite{Shuvaev:1999ce}, while $(1+\beta^2)$ is the correction that takes into account the contribution from the real part of the amplitude. The amplitude is given by
\begin{equation}
A(x,Q^2,\vec{\Delta})= i\int dz d^2\vec{r}d^2\vec{b} e^{-i(\vec{b}-(1-z)\vec{r})\cdot\vec{\Delta}}
\left[\Psi^*_{\jpsi}\Psi\right] 2\left[
1-\exp\left\{
-\frac{1}{2}\sigma_{\rm dip}T_{\rm Pb}(b)
\right\}
\right],
\end{equation}

where the integration variable $\vec{r}$ represents the distance between the quark and the antiquark in the plane transverse to the collision, $z$ quantifies the fraction of the photon momentum carried by the quark and $b$ is the distance between the centres of the  target and the dipole; $\vec{\Delta}$ is the transverse momentum transferred to the nucleus; the virtuality of the incoming photon is denoted by $Q^2$ and for the case of photoproduction discussed here is zero; $\Psi$ describes the splitting of the photon into the dipole and $\Psi_{\jpsi}$ is the wave function of the $\jpsi$;  the term $i(1-z)\vec{r}\cdot\vec{\Delta}$ in the exponential is a third correction to take into account non-forward contributions to the wave function $\Psi_{\jpsi}$, which is modelled for the forward case \cite{Bartels:2003yj}; and finally $\sigma_{\rm dip}$ is the universal cross section for the interaction of a colour dipole with a nuclear target. The models differ in the functional form of $\Psi_{\jpsi}$, in corrections they consider and in their formulation of the universal dipole cross section. 

In the case of LM they do not consider the non-forward correction to the wave function. They use two different prescriptions for the wave function:  the Gauss-LC \cite{Kowalski:2003hm} and the boosted Gaussian \cite{Nemchik:1994fp,Nemchik:1996cw}. They write $\sigma_{\rm dip}$ in terms of the cross section of a dipole and a proton, $\sigma^p_{\rm dip}$; assuming a Gaussian profile in impact parameter for the proton, $\exp(-b^2/(2B_p))$ they arrive at 
\begin{equation}
\frac{1}{2}\sigma_{\rm dip} = 2\pi B_p A N(r,x),
\end{equation}
where $N(r,x)$ is the dipole target amplitude. They use two different models for $N(r,x)$: The IIM model \cite{Iancu:2003ge} which is a parameterisation of the expected behaviour of the solution to the BK equation \cite{Balitsky:1995ub,Kovchegov:1999yj,Kovchegov:1999ua} which includes a non-linear term for the evolution of  $N(r,x)$; and the IPsat model \cite{Kowalski:2003hm,Kowalski:2006hc} which uses DGLAP equations to evolved an eikonalized gluon distribution.

The GM-GDGM model uses the boosted Gaussian prescription for the wave function. The dipole cross section is given by $\sigma_{\rm dip} = R^A_g(x,Q^2)\sigma^{\rm p}_{\rm dip}$, where $\sigma^{\rm p}_{\rm dip}$ is given according to the IIM model and the leading twist approximation is used for $R^A_g(x,Q^2)$.

\section{Experimental results on exclusive photonuclear processes}

\subsection{Exclusive J/$\psi$ photoproduction off protons in ultra-peripheral p--Pb collisions}

Exclusive J/$\psi$ photoproduction has been studied in previous colliders at HERA~\cite{Chekanov:2002xi,Aktas:2005xu,Alexa:2013xxa} and the Tevatron~\cite{Aaltonen:2009kg}. Recently, both ALICE~\cite{TheALICE:2014dwa} and LHCb~\cite{Aaij:2014iea} reported results on exclusive J/$\psi$ photoproduction. By studying photon-proton collisions one can get insights about one of the most interesting QCD discoveries from the last decade: the density of gluons carrying a small fraction of the momentum of hadrons grows extremely rapidly. The growth of the probability density function (PDF) for small-$x$ gluons cannot continue forever. Gluon saturation~\cite{Gribov:1984tu,Mueller:1989st} is the most straightforward mechanism to slow down the growth of the gluon PDF at small-$x$, and it would have important implications in the early stages of ultra-relativistic heavy-ion collisions at RHIC and LHC. Consequently, finding evidence for gluon saturation has become a central task for present experiments and for future projects that aim to study QCD. Although gluon saturation regime should manifest in terms of new physics in the strongly interacting sector, only hints for this QCD phenomena at HERA, RHIC, CEBAF and LHC have been found so far.

ALICE results on exclusive J/$\psi$ in p--Pb collisions~\cite{TheALICE:2014dwa} provide a unique opportunity to study the proton gluon distribution over an unprecedented range in Bjorken-$x$, from $2\times 10^{-2}$ to $\times 10^{-5}$, and do not suffer from the ambiguities and assumptions that have to be made when studying symmetric systems such as in pp collisions. These results indicate no significant change in the behavior of the gluon density from HERA to LHC energies, extending by a factor five the $x$ values previously explored. These findings thus substantially advance our understanding of the proton structure and set important constraints on gluon saturation. 

The measurement of exclusive J/$\psi$ production in ultra-peripheral heavy-ion collisions~\cite{Klein:1999qj} is one of the key measurements for the future electron-ion collider such as the LHeC electron-proton and electron-ion collider~\cite{AbelleiraFernandez:2012ty}. Similar studies have been highlighted in the U.S. electron-ion collider design study reports~\cite{Accardi:2012qut,Aschenauer:2014cki}.

\begin{figure*}[htbp]
\centering
\includegraphics[width=1.\columnwidth]{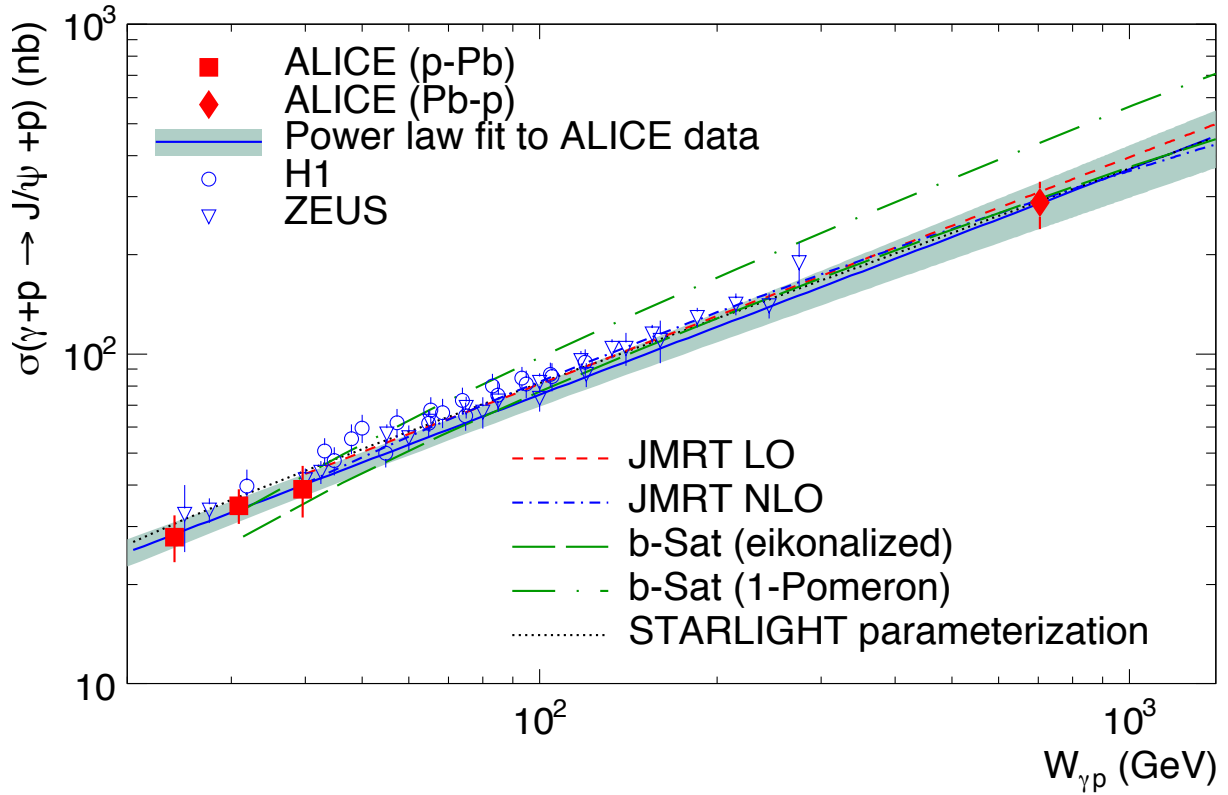}
\caption{Exclusive J/$\psi$ photoproduction cross section off protons measured by ALICE and compared to HERA data~\cite{TheALICE:2014dwa}. Comparisons to STARLIGHT, JMRT and the b-Sat models are shown. The power law fit to ALICE data is also shown.
\label{ALICE_UPC2}}
\end{figure*}

\subsection{Coherent and incoherent J/$\psi$ photoproduction from ultra-peripheral Pb--Pb collisions}

One of the most important questions in relativistic heavy-ion physics is the nature of the initial state produced in high energy heavy-ion collisions. A common denominator in heavy-ion physics analyses is the need to distinguish between final states effects expected from the QGP from those inherent to the nuclei themselves. The competing Glauber and Color Glass Models are two well studied alternatives but both may be wrong. Uncertainty over the initial state is an impediment to measuring fundamental properties of the QGP such as viscosity. The photon-nuclear measurements at the LHC are putting important constraints on the initial state by measuring the nuclear effects on the parton distribution functions (PDFs). The modifications due to nuclear effects in the gluon PDF are called nuclear shadowing. This is related to the fact that nuclear parton distributions at small-$x$ are suppressed compared to the case of a free proton. The degree of gluon shadowing effects for a $x<$ 0.01 is poorly known. The study of photo-nuclear reactions at the LHC allows us to put important constraints on theoretical models that predict nuclear gluon shadowing. Although similar studies were performed by PHENIX~\cite{Afanasiev:2009hy}, there studies were limited by their small sample size. 

\begin{figure*}[htbp]
\centering
\includegraphics[width=1.\columnwidth]{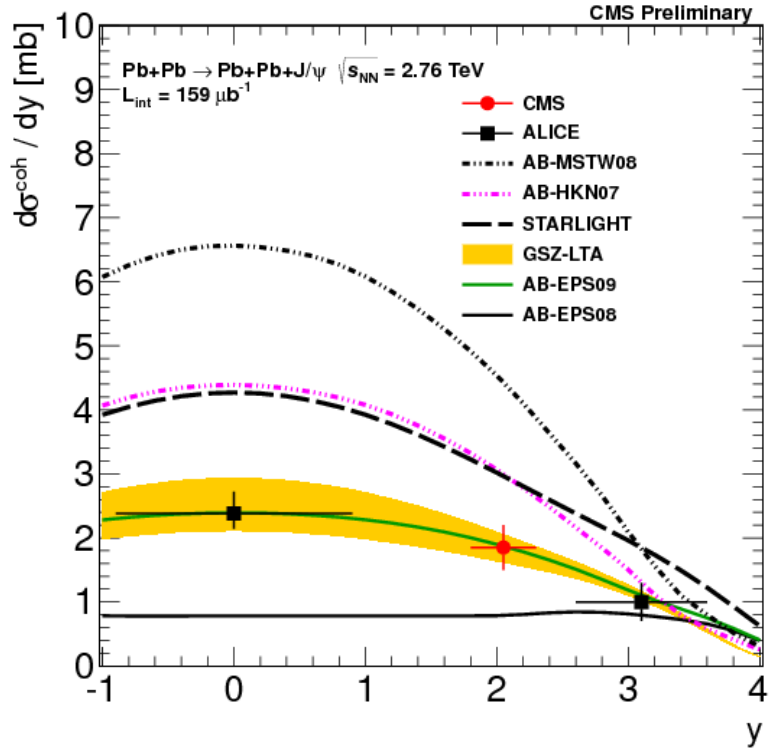}
\caption{Coherent J/$\psi$ photoproduction cross section in ultra-peripheral Pb--Pb collisions at $\sqrt{s_{NN}}$ = 2.76 TeV measured by ALICE and CMS~\cite{Abelev:2012ba,Abbas:2013oua,CMS:2014ies}. CMS results corresponds to preliminary results and have not been corrected for feed-down contributions from $\psi$(2S) decays. Data is compared to model calculations.
\label{CMS_UPC}}
\end{figure*}

The coherent J/$\psi$ measured by ALICE is detected through its dimuon decay in the muon spectrometer of the ALICE detector, which also provides the trigger for these events, or in its dielectron or dimuon decay in the central barrel. At the rapidities~\cite{Abelev:2012ba} (y around 3) studied in the muon arm, J/$\psi$ photoproduction is sensitive mainly to the gluon distribution at values of Bjorken-$x$ of about 10$^{-2}$, whereas at mid-rapidity on probes x $\sim$ 10$^{-3}$~\cite{Abbas:2013oua}. Preliminary CMS results on the measured cross section for coherent J/$\psi$ were presented for the first time in Fall 2014~\cite{CMS:2014ies}. Figure~\ref{CMS_UPC} shows the comparison between the coherent J/$\psi$ cross section in ultra-peripheral PbPb collisions at $\sqrt{s_{NN}}$ = 2.76 TeV~\cite{Abelev:2012ba,Abbas:2013oua} and theoretical predictions~\cite{Santos:2013lpa,Goncalves:2013ixa,Guzey:2013taa,Guzey:2013qza,Ducati:2013tva,Ducati:2013bya,Adeluyi:2013tuu,Lappi:2013am,Cisek:2012yt,Guzey:2013xba}. Models which do not include nuclear gluon shadowing are inconsistent with the measured ALICE results~\cite{Adeluyi:2012ph,Klein:1999qj}. Best agreement is found for models that incorporate the EPS09 shadowing parameterization~\cite{Adeluyi:2012ph}. 

%\subsubsection{Determining the nuclear parton distribution}
The ALICE measurements have provided the first direct experimental evidence~\cite{Abelev:2012ba,Abbas:2013oua} for nuclear gluon shadowing at small values of Bjorken-$x$~\cite{Guzey:2013xba}. In addition, the ALICE measurements have shown that certain models can be rejected~\cite{Vogt:2014pta}. The next step is to use these measurements to determine the nuclear parton distribution. 

Incoherent production of $\jpsi$ in Pb--Pb UPC  has  been measured at mid rapidities~\cite{Abbas:2013oua} using the same trigger and detectors as for the coherent case. The incoherent contribution was obtained from the distribution of transverse momentum. The centre of mass energy in the $\gamma$-Pb system are the same as for the coherent case. The measured cross section is 0.98 $^{+0.19}_{-0.17}$ (stat+sys) mb.

\subsection{Coherent $\psi$(2S) photoproduction from ultra-peripheral Pb--Pb collisions}

Preliminary results on coherent production of $\psip$ in ultra-peripheral Pb--Pb collisions at mid-rapidity have been reported. The $\psip$ has been identified in the following channels: to $l^+l^-$ and to $\jpsi\rightarrow \pi^+\pi^-$, with $\jpsi\to l^+l^-$, where $l=e,\mu$. The measurement correspondeds to $W_{\gamma\rm{Pb}}\approx 100$ GeV ($x\approx 1.3\cdot10^{-3}$). The coherent $\psi$(2S) is expected to be sensitive to nuclear gluon shadowing as for the J/$\psi$ case. Despite the small sample size and the considerable uncertainty in the underlying $\gamma + \rm{p}\rightarrow \rm{V} + \rm{p}$ cross section, the ALICE measurement concludes that models with no nuclear effects or with a strong nuclear gluon shadowing are disfavored. 

\begin{figure*}[htbp]
\centering
\includegraphics[width=1.\columnwidth]{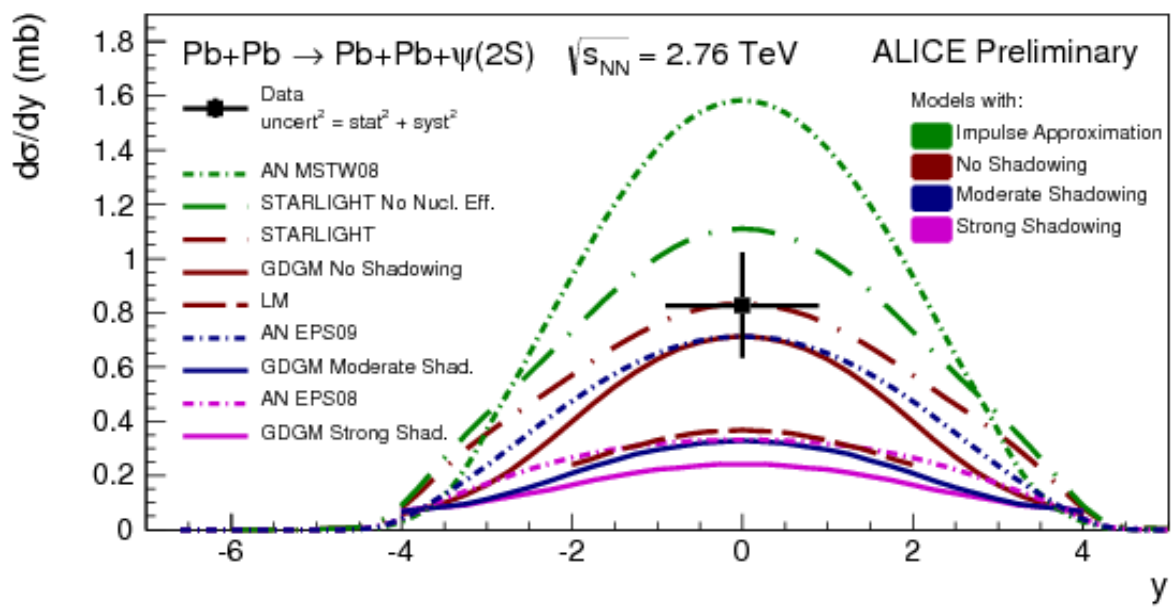}
\caption{Coherent $\psi$(2S) photoproduction cross section measured by the ALICE experiment (preliminary results) and compared to model calculations.
\label{ALICE_PSI2S}}
\end{figure*}

\subsection{Coherent $\rho^{0}$ photoproduction from ultra-peripheral Pb--Pb collisions}

ALICE has recently reported preliminary results on coherent $\rho^{0}$ photoproduction from ultra-peripheral Pb--Pb collisions at $\sqrt{s_{NN}}$ = 2.76. The measured cross section was found to be in agreement with both STARLIGHT~\cite{Klein:1999qj} and the Goncalves and Machado (GM) model. The prediction by Glauber-Donnachie-Landshoff (GDL) is about a factor 2 larger than in data. This confirms the STAR findings~\cite{Abelev:2007nb}. However, it is surprising that the measured cross section agrees with the STARLIGHT that does not include the elastic component of the total cross section. It would be important to understand why the scaling of the $\gamma \rm{p}$ cross section using the Glauber model~\cite{Frankfurt:2002wc,Rebyakova:2011vf} overpredicts the measured cross section.

\begin{figure*}[htbp]
\centering
\includegraphics[width=1.\columnwidth]{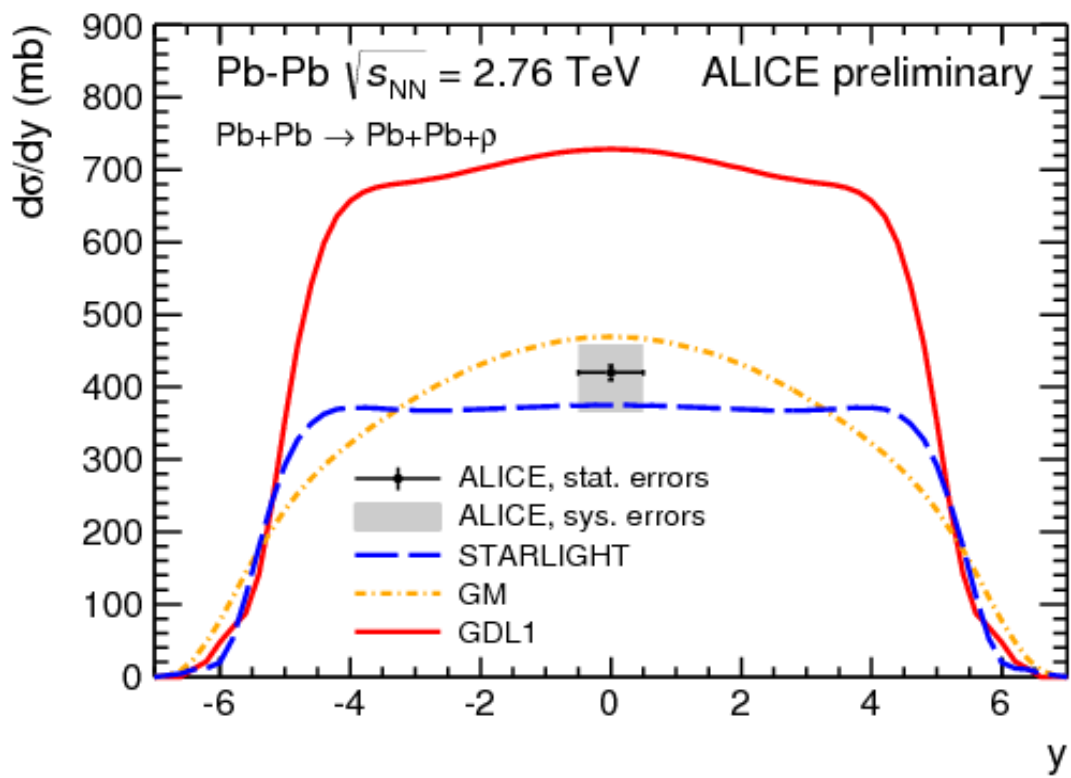}
\caption{Coherent $\rho^{0}$ photoproduction cross section in ultra-peripheral Pb--Pb collisions at $\sqrt{s_{NN}}$ = 2.76 TeV measured by the ALICE experiment and compared to model predictions.
\label{ALICE_RHO0}}
\end{figure*}

\subsection{Four-pion production in ultra-peripheral Pb--Pb collisions}

It is interesting to look for excited states of photo-produced $\rho^{0}$ mesons. It is not clear how many excited states exist or their quantum numbers (see special PDG review~\cite{Beringer:1900zz}). STAR reported on four-pion production in UPC~\cite{Abelev:2009aa}, albeit at much lower centre-of-mass energies. No HERA publications on photo-production of a $\rho^{0}$ excited states exist. Figure 4 in~\cite{Takaki:2014vva} shows the four-pion $p_{T}$ distribution, where a clear coherent peak can be seen at low transverse momenta. This corresponds to data collected during the 2011 Pb-Pb run, where we have 10 times more statistics than those published by STAR. One of the current research interesest is understanding the possible production mechanics~\cite{Klusek-Gawenda:2013dka}.

\section{Two-photon physics}

The two-photon process is governed by QED. Here, the coupling between the photon and the emitting nucleus is enhanced by a factor $Z$. Thus, higher order terms might be important. However, recent ALICE results on $\gamma \gamma \rightarrow e^{+}e^{-}$ are well described by STARLIGHT which only includes the leading QED order terms~\cite{Abbas:2013oua}. The published analysis~\cite{Abbas:2013oua} was carried out using data from the 2011 Pb--Pb run, a data that was recored using a topology trigger. This restricted the analysis to invariant masses $M_{e^{+}e^{-}}>$ 2.2 GeV/$c^{2}$. A preliminary analysis of the 2010 Pb--Pb run was presented at Quark Matter and ICHEP 2014~\cite{Nystrand:2014vra,TapiaTakaki:2014vra}. The 2011 data does not include a topology trigger and so the analysis can be performed to go down to $M_{e^{+}e^{-}}>$ 0.6 GeV/$c^{2}$. 

\begin{figure*}[htbp]
\centering
\includegraphics[width=0.8\columnwidth]{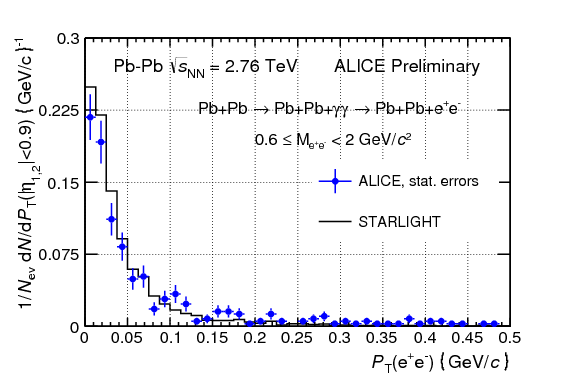}
\includegraphics[width=0.8\columnwidth]{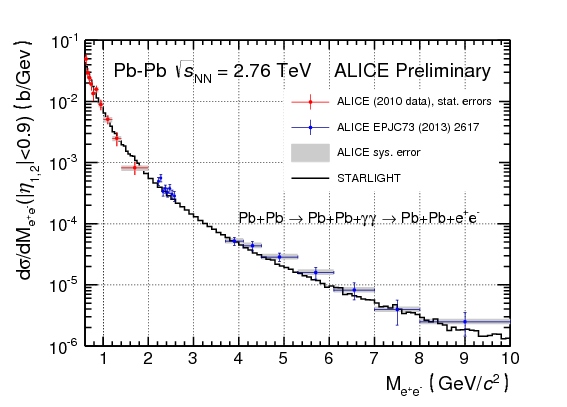}
\caption{Transverse momentum distribution of exclusive $e^{+}e^{-}$ pairs measured by ALICE and compared to STARLIGHT (upper panel). Measured cross section for the exclusive two-photon process measured by ALICE and compared to STARLIGHT~\cite{Abbas:2013oua,Nystrand:2014vra}. 
\label{ALICE_GG}}
\end{figure*}

\section{UPC lessons from LHC Run 1}

The experimental challenge for such measurements consists in having dedicated UPC triggers that are often orthogonal to the general trigger strategy of the experiments. Moreover, validating an exclusive analysis requires a good understanding of the trigger efficiency for the exclusivity conditions imposed at the online and offline levels, for which control triggers are usually required. The main experimental challenges during Run 2 are threefold

\begin{itemize}
\item {\bf Background suppression}. The detectors do not have a complete acceptance in rapidity. It is possible that some of the tracks of low multiplicity non-UPC processes are not detected and thus the events are included in the UPC sample. This has been dealt with data driven approaches to describe and subtract the remaining background. The situation will improved be in Run 2

\begin{itemize}
\item[$\circ$] There will be new detectors~\cite{Albrow:2014jxd,HerreraCorral:2011zz} which will increase the capabilities to tag, and thus to veto the background events. 
\item[$\circ$] CMS and TOTEM will work together, which will allow, for some special beam optics, to tag the elastically scattered proton in p--Pb collisions and thus tag directly the exclusive photoproduction of vector mesons.
\end{itemize}

\item {\bf Triggering}. The vector mesons or low mass di-leptons produced in UPC produce few particles each with low transverse momenta. It is very difficult to trigger on these configurations, because the detectors where optimized to study  hard processes and/or very high multiplicities.  Nonetheless the experience gained during Run 1 will help improving both the purity and the efficiency of the triggers. Also the new detectors mentioned above will be key in this respect. 

\item {\bf Statistics.} The low statistics collected in Run 2 is due to a combination of the problems related with triggering on these events and to the small production cross sections for some of these UPC processes. For Run 2, one expect the following improvements:

\begin{itemize}
\item[$\circ$] It is expected that the triggers will be more efficient as mentioned previously;
\item[$\circ$] It is expected that the luminosity will be a factor 50 larger in Pb--Pb, and 
\item[$\circ$] It is expected that the energy will be larger in Pb--Pb, and the case of p--Pb is under discussion. For the UPC program a larger energy in p--Pb would allow to reach even lower values of Bjorken-$x$ which could be crucial for finding the onset of gluon saturation in exclusive vector meson photoproduction. 
\end{itemize}

\end{itemize}

\section{Experimental prospects}

Apart from expanding the program on exclusive photonuclear processes, the next LHC runs will open the possibility to study inclusive photonuclear processes. For example, the charm photoproduction through photon-gluon fusion~\cite{Klein:2002wm}. Most of the attention has been put on small-$x$ physics and in the determination of the nuclear parton distribution. However, it would be possible to study physics signatures sensitive to physics beyond the standard model~\cite{Baltz:2007kq}.  

During Run 1 the ALICE analyses were performed on a integrated luminosity of about 20 $\mu \rm{b}^{-1}$ for the mid-rapidity measurement and about 50 $\mu \rm{b}^{-1}$ for the muon arm analysis for 2.76 TeV Pb--Pb collisions. Note that the trigger efficiency for the muon analysis during the 2011 Pb--Pb was low. During Run 2, the plan is to collect about 1 $\rm{nb}^{-1}$ at 5.5 TeV Pb--Pb collisions. The factor 50 increase in luminosity represent a significant increase in the number of reconstructed events even assuming if the detector conditions will remain the same as for the 2011 Pb--Pb run. 

The increase of statistics will allow us to perform differential studies in $p_{T}$ and $y$ for coherent and incoherent J/$\psi$ and $\psi$(2S). Fourier transforms of the $p_{T}$ distribution is thought to be sensitive to the spacial distribution of glue in the nucleus~\cite{Armesto:2014sma,Toll:2012mb}. The $t$ distribution is expected to present a pronounced diffractive peak which can then be used to clearly discriminate between saturation and non-saturation models. Light vector mesons like the $\phi$ would be more sensitive to gluon saturation compared to the J/$\psi$. It would be very interesting to have UPC $\phi \rightarrow K^{+}K^{-}$ and $\omega \rightarrow \pi^{+}\pi^{-}\pi^{0}$ measurements at the LHC during Run 2, which might be possible with the ALICE detector.  

By tagging the neutron in UPC J/$\psi$ measurements one can disentangle the coherent and incoherent processes~\cite{Strikman:2005ze,Guzey:2013jaa}. This would also allow us to identify the low and high $W_{\gamma \rm{Pb}}$ contributions to the total cross section. This is very interesting as both contributions are sensitive to different values of Bjorken-$x$.

It would be possible to study exclusive J/$\psi$ in p--Pb collisions during Run 2. The UPC program would benefit by having the largest possible energy. During Run 1, ALICE reported measurements at 5.02 TeV that reached Bjorken-$x$ values of about 10$^{-5}$, while by colliding at 8.16 TeV we would be able to reach up to $10^{-6}$. 

Measurement the $\Upsilon$(1S) would be performed for the first time during Run 2. CMS and ATLAS would be in an ideal position to collect these events which can be used to constraint nuclear shadowing models, although there is a significant large uncertainty on the corresponding $\gamma \rm{p}$ cross section at present. Given the expected integrated luminosity of 1 nb$^{-1}$ for the 5.1 TeV Pb--Pb run, CMS and ATLAS should be able to measure UPC dijets from $\gamma$Pb interactions over a wide rapidity and $p_T$ range~\cite{Baltz:2007kq,Strikman:2005yv}. Together with the UPC quarkonia measurements, these data can then be used to map the gluon distribution over a wide range of Bjorken-$x$.  Apart from the study of four-pion photoproduction, other exotic quarkonium states can also be studied~\cite{Baltz:2007kq}. The UPC program at RHIC have also explored other measurements that can be studied in the future~\cite{Klein:2015qna}, for example, the study of low transverse momentum vector mesons and interference\cite{Abelev:2008ew}.

The two-photon physics program would also benefit from the experimental conditions expected during Run 2. In the future it would be interested to study whether there is any asymmetries between the positive and negative lepton as this has been proposed as a possible signature for higher order terms~\cite{Brodsky:1968rd}.  It would be possible to study various $\gamma\gamma$ decay channels, namely, $\eta_{c}$, $\chi_{c0}$(2P) and $\chi_{c2}$(2P) in the decay channels $\pi^{+}\pi^{-}\pi^{+}\pi^{-}$, $\rm{K}^{+}\pi^{-}\rm{K}^{-}\pi^{+}$ and $\rm{K}^{+}\rm{K}^{-}\rm{K}^{+}\rm{K}^{-}$. It would also be possible to perform first studies on light-by-light scattering~\cite{d'Enterria:2013yra,Fichet:2014uka}. Furthermore, the study of bound-free pair production $e^{+}e^{-}$ originating from an electron bound to one of the ions~\cite{Klein:2000ba,Bruce:2007mx} will be possible. This study will provide useful information about the LHC accelerator performance for high luminosity heavy-ion collisions at the LHC. 

\section{Experimental Summary}

In summary, we have reviewed recent experimental results on ultra-peripheral heavy-ion collisions at the LHC. The coherent J/$\psi$ results provide the first direct experimental evidence for nuclear gluon shadowing. Results from coherent $\rho^{0}$ are also very interesting. The quantum Glauber calculation which works for low fixed-target energies does not work for RHIC and LHC energies. The two-photon QED process has also been measured and does not favour higher order terms. 

Given the significant increase in energy and luminosity for Run 2 and the increase capabilities LHC detectors it is important to be open to unexpected phenomena. For example the production of open $b$ or $c$ quarks in ultra-peripheral collisions may reveal new insight into the initial state produced in high energy heavy-ion collisions. Photon-induced processes are also sensitive to physics beyond the standard model. To prepare for the unexpected it is important to spend considerable effort on developing clean and efficient UPC triggers since these will be the key for capturing new rare physics. Furthermore, the studies of photon-nuclear processes would allow us to gain insights that will be important as the electron-ion collider facilities are developed.

\section{Theoretical proposals}

Here we will focus on several promising directions of the UPC studies which emerged in the last few years after the review\cite{Baltz:2007kq} was published.

\subsection{Tracking fast small color dipoles through strong gluon fields at the LHC}
It was discussed in a number of papers that the process of photon - proton scattering with production of a leading meson at $p_t \ge \mbox{few GeV}$ with a rapidity gap
\begin{equation}
\gamma + p \to J/\psi (\rho) + \mbox{rapidity gap } +Y,
\end{equation}
is dominated by the scattering of the photon in a small transverse size configuration. The process is mediated by the elastic  scattering of $q\bar q$ pair off  a gluon (quark) of the target (with $x_g$ expressed through the value of the rapidity gap interval). In the discussed kinematics one may expect that the elastic scattering is  mediated by the  perturbative gluon ladder.  Hence such process provides a unique method  to studying  the BFKL dynamics by measuring  the cross section at fixed $-t=p_t^2$ as a function of the rapidity gap for fixed $t, M_Y^2$:
\begin{equation}
  {d\sigma^{\gamma + p\to J/\psi + \mbox{gap + Y}}(t, y)\over dt}\propto \exp \left[2(\alpha_{I\!\!P}(t)-1)\Delta y\right],
  \end{equation}
see~\cite{Blok:2010vj} and references therein. Such studies should be feasible in the $pp$ scattering at the LHC in the low lumi runs.  They may require detection of the leading proton to suppress the contributions of non UPC processes.

The analogous process in $AA$ collisions provides a unique opportunity for studying interaction of small color dipoles propagating through the nuclear media~\cite{Frankfurt:2008et}. In particular,  it allows to study the  onset of the  novel perturbative QCD regime of strong absorption for the interaction of small dipoles at the collider energies. Such a study would be clearly complementary to the study of suppression of the forward hadron production discussed in Sec.~\ref{leading}. It appear that it would be feasible to probe in the  forthcoming $AA$ run  interaction of $q\bar q$ dipoles of sizes $\sim$ 0.2 fm with nuclear media down to $x = (m_V^2-t)/s \sim 10^{-5}$. Two possible mechanisms of the deviation of the yield from linear in $A$ regime are multiple interactions of the dipole with the media and the gluon  leading twist  shadowing. Since the first effect is primarily a function of  $W_{\gamma N}$  while the second one is  a function of $x_g$ (the rapidity gap) these two effects can be easily separated for example by comparing  $x_g \ge 0.01$ and $x_g \sim 10^{-3} $ cross sections. Study of the structure of hadronic system $Y$ would provide an additional information on the mechanism of absorption and in particular dependence of the cross section on the centrality.
\subsection{Studies of the color fluctuation phenomena}
Dominance of large longitudinal distances in the hadron (photon) nucleus interactions at high energies implies that the quark- gluon configurations in which projectile interacts with the nucleus can be considered as frozen during the passage of the nucleus. At the same time different configurations in the projectile, $h$, are expected to interact with different strength - in particular due to the different area occupied by color in different configurations.  This phenomenon is referred to as the color fluctuation phenomenon, for the recent discussion and references see \cite{Alvioli:2014sba}. It is characterized by the distribution over the strength of the interaction $P_h(\sigma)$ which satisfies normalization conditions: $\int P(\sigma) d\sigma=1,  \int \sigma P(\sigma) d\sigma=\sigma_{tot}$. Information about variance of $P(\sigma)$  - $\omega_\sigma= \int (\sigma/\sigma_{tot} - 1) P(\sigma) d\sigma$ can be extracted using Miettinen - Pumplin relation from diffractive data \cite{Miettinen:1978jb}. Color fluctuations lead to a significant broadening of the distribution over the number of the wounded nucleons, $\nu$, as compared to  the conventional Glauber model. Evidence for such an effect was reported in ref.\cite{ATLAS}. Also, it was demonstrated recently \cite{Alvioli:2014eda}  that the pattern of violation of the Glauber picture for centrality dependence of the rate of forward jet production observed in $pA$ collisions at the LHC \cite{ATLAS,Chatrchyan:2014hqa} provides evidence that configurations in the proton containing a parton with large $x_p$ interact with a signicantly smaller than average cross section and have smaller than average size.

UPC in $AA$ collisions provide a  complementary tool as compared to the t $pA$ scattering for studies of the  color fluctuations. Indeed, one expects a much broader distribution over $\sigma$ in the  case of the photon projectile. Such a pattern already in the vector dominance model approximation as the variance of the distribution over $\sigma$  is larger in the   the case of meson - nucleon interaction \cite{Blaettel:1993rd}. An additional broadening comes from the enhanced contribution of small size configurations which lead to $P_\gamma (\sigma\to 0) \propto 1/\sigma$~\cite{Frankfurt:1997zk} as compared to $P_N (\sigma\to 0) \propto \sigma$.

In the UPC it would be possible to large extent to regulate the transverse size of the configuration by selecting  final states with leading pions ($\rho, \omega $ - like  configurations), kaons ($s\bar s$ configurations in the photon), and $D$-mesons (small size, $d\propto 1/m_c$, configurations). Another way to select small size dipoles would be triggering on the leading pion production  and studying distribution over the number of wounded nucleons as a function of $p_t$. Moreover it would be possible to study effects of color fluctuations as a function of the collision energy $ W_{\gamma N}$ in a wide interval of $W$  

It is worth emphasizing that in the kinematics where the small  dipole nucleon interaction is far from the BDR, but $x= 4p_t^2/s < 10^{-2}$ is  small enough, we expect a significant suppression of the forward back to back pion production due to the leading twist gluon shadowing. For example for the pion transverse momenta $\sim $ 1 GeV/c we expect shadowing comparable to that for the amplitude of the coherent $J/\psi$ elastic photoproduction corresponding to the suppression of the yield by a factor of $\sim$ 0.6 (0.5)  for $x\sim 10^{-3} (10^{-4})$ and average number of wounded nucleons $\nu = 1.6 (2)$ \cite{Frankfurt:2011cs}. A much larger suppression is expected for the energy range where BDR is reached.

\subsection{Multiparton interactions in the direct photon kinematics}
In addition to the studies of the leading twist hard processes we mentioned in sec \ref{hi:intro},  UPC can be used to study multiparton interactions (MPI).
 A promising kinematics is the one  dominated by the contribution of the  direct photon mechanism. In this mechanism   photon splits into $q\bar q$ pair  with comparable light-cone fractions and which carries practically all momentum of the photon. Next,   quark and antiquark experience hard collisions with a parton of the target with sufficiently larger rapidity gap between the quark (antiquark) and the balancing jet. In this kinematics one can suppress  the leading twist mechanism by both choosing back-to-back kinematics and optimal distances in rapidity. The 
 cleanest channel is the one where photon splits into charm -- anticharm pair since in this case the photon wave function is purely perturbative
 with the final state being  two dijets each containing charm (anticharm) quarks and carrying $x_1,x_2>0.2$ fractions of the photon momentum and two balancing jets. It was demonstrated in \cite{Blok:2014rza}  that in this case cross section is more directly related to the nucleon (nucleus) generalized parton distribution than in the case of double parton interactions in the proton--proton (nucleus) collisions. A significant number of such double parton interactions should be produced in $p - Pb$ and $Pb- Pb$ collisions at the LHC in a $10^6$ sec run: $\sim 6\cdot 10^4$ for $Pb-Pb$, and $\sim  7 \cdot 10^3$ for p--Pb  collisions for the same transverse momentum cutoff. 
\section{Propagation of partons through nucleons and nuclei at ultrahigh energies}
\label{leading}
\subsection{Introduction}
In perturbative QCD interaction of a small color dipole of size $d$  with  target grows rapidly with energy. In the leading log approximation the  probability  for a $q\bar q$ dipole to interact is at a given impact parameter is 
\begin{equation}
\Gamma_{inel}(d,x,b)= {\pi^2\over 3} \alpha_s(Q^2_{eff}) d^2
   \left[x G_T(x,Q^2_{eff},b) + \frac{2}{3} x S_N(x, Q^2_{eff},b)\right],  
   \label{sigma}
    \end{equation}
where $Q^2_{eff} = \lambda/d^2, \lambda= 4 \div 10, x=Q^2_{eff}/s$ and $G_T, S_T$ are generalized parton distributions in the target. In the case of a color octet dipole the gluon term is larger by a factor of 9/4 while the second term is missing.  One can see from Eq. \ref{sigma} that $\Gamma_{inel}(d,x,b)$ rather rapidly grows with increase of energy (this explains for example the energy dependence of the $J/\psi$ exclusive photoproduction) and starts to approach  black disk (unitarity limit) regime (BDR)  of $\Gamma_{inel}(d,x,b)=1$  for small $b$ at sufficiently small $x$.

It can be argued that   the pattern of interaction  of a quark  (gluon)  of virtuality $\mu^2$ through the target is similar to that of the $q\bar q$ ($gg$)  dipole with $d^2\propto 1/\mu^2$. Hence measurements in  the very forward kinematics  at the  LHC allow to probe  theBDR down to  $x_p$, $x_A \sim  10^{-6}$  for virtualities of the order of few GeV$^2$.

In the BDR   partons with virtuality $< \mu^2$ should   interact with probability of the order one, acquire transverse momenta on the scale $\mu$ and also split into partons with virtualities $\sim \mu^2$. The latter effect is manifested as  effective {\it fractional} energy losses which explicitly breaks the QCD factorization for hard processes \cite{Frankfurt:2001nt,Frankfurt:2007rn}. 

As a result, one expects that for collisions at small impact parameters in $pp$ and especially $pA$ collisions   the coherence of quarks in the proton should be completely destroyed. As a result in this limit   all leading partons should fragment independently (if $z\ge 0.1 \div 0.2 $($z$ is the fraction of the projectile momentum carried by produced hadron)  at the LHC) leading to  (a)  a strong suppression of the leading hadron spectrum \cite{Berera:1996ku,Dumitru:2002wd}  (see e.g. fig.1 from \cite{Dumitru:2002wd}):
\begin{equation}
d\sigma(p+ T \to h +X)/d x_F \propto  (1-x_F)^{n(p_t)}, n_N(p_t)  \sim 5\div 6,
\end{equation}
with $n_h(p_t)$ decreasing with increase of $p_t$, (b)  $\left<p_t\right>$ broadening of the spectrum: $p_t\sim \mu$~\cite{Dumitru:2002wd,Gelis:2006tb},  and dominance of the leading meson production as compared to that of baryons: $d\sigma(p+ T \to \pi  +X)/d z \gg d\sigma(p+ T \to N  +X)/d z$ for $z\ge 0.5$.  Since for peripheral collisions  distribution of nucleons in $z$  is known to be practically flat for $z\le 0.8$ and average $ p_t \sim \mbox{ .5 GeV/c}$  we expect a gross change of the spectrum in the central $pA $ collisions.
 \begin{figure}[htp]
  \vskip -3.0cm
  %\centerline{\hspace{0cm}
  \centerline{\includegraphics[width=8.0cm]{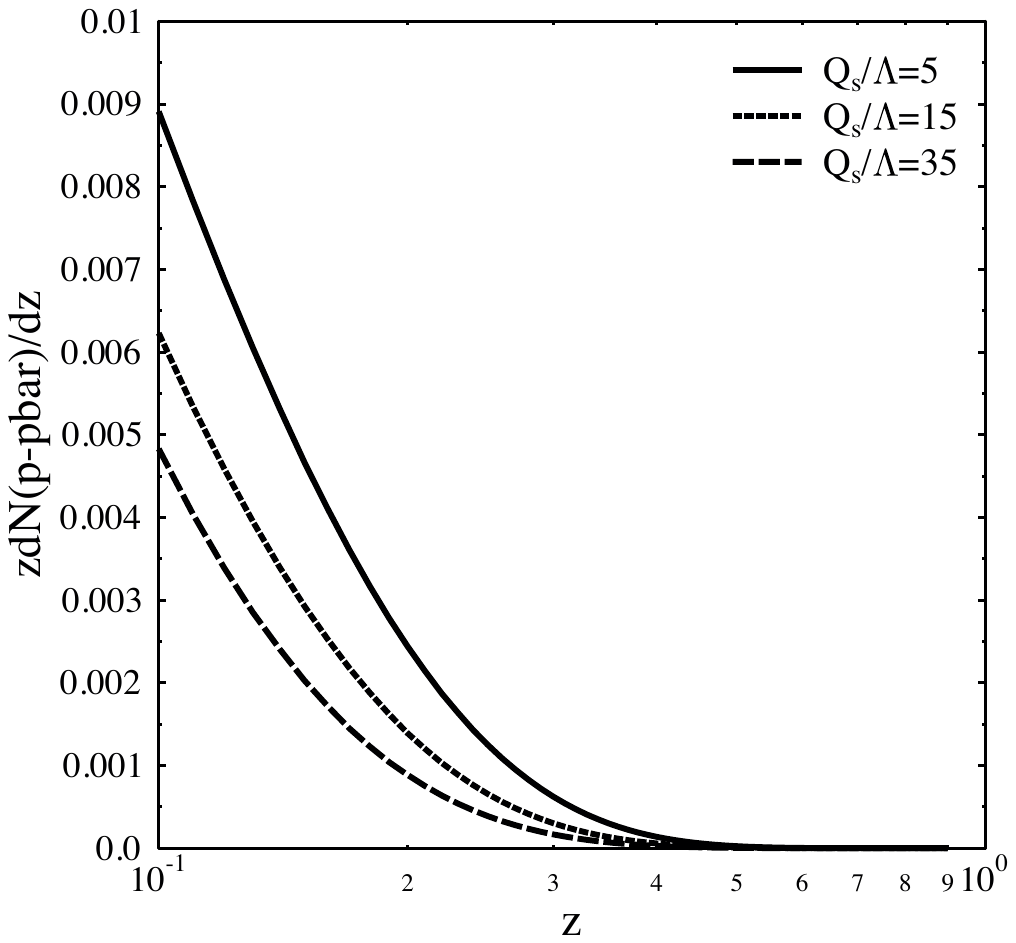}}
\caption{Distribution of protons over $z$ in central $pA$ collisions  for different $p_t$ ranges of dominance of black disk dynamics. The neutron yields for large $z$ are a factor of two smaller. }
  \label{fig1}
\end{figure}
These predictions are based on very generic features of the   interaction of leading quarks in  the  black disk regime. Hence the study of the leading hadron production would provide a critical test of the expectation that the BDR is reached at the LHC in the interactions with nuclei and nucleons for $p_t\le \mbox{few GeV/c}$.
 
Note here that the current calculations neglect effective fractional energy losses which further amplify the discussed effects. Measurement of these losses at the LHC would require observations of hard interactions with hard scale $\le $5~GeV. Observation of the dijet production is hardly possible   in this case. One would have to focus on the study of the leading hadron production. At  RHIC a strong suppression of the pion spectrum was observed for $z\ge 0.3$\cite {Arsene:2004ux}, and the observed regularities of the process (in particular the forward -- central correlation)~\cite {Adams:2006uz} are consistent with the effective energy loss scenario \cite {Frankfurt:2007rn}.  It would be desirable to study at the LHC centrality dependence of the leading pion (nucleon) spectrum in  similar kinematics since for $x_F\le 0.2$  effective fractional energy losses lead to a small reduction of the cross section which maybe compensated by $p_t$ broadening. 

In the case of large $z$ measurements, of say  the leading $\pi^0$, it would be desirable  to study  also the centrality dependence of the  production of the recoil hadrons   in the $x_A \ge 0.01$ range where nuclear PDFs are close to the nucleon ones. 

{\it A warning: } in the measurements one has to put a requirement that there is  some activity at negative rapidities (towards the  nucleus fragmentation region) - otherwise the processes of  photon-proton interaction  would give the dominant contribution.

Similar questions can be asked for $pp$ collisions. One can study how the differential multiplicity of leading baryons  depends on the centrality - for example one can investigate whether  it is strongly suppressed for the high multiplicity events - say for $N/\left<N\right> \ge 3$ where collisions are dominated by scattering 
at small impact parameters b$\le$  0.5\, fm \cite{Azarkin:2014cja} . One can also investigate  modification  of the baryon  spectrum at $x_F\ge 0.3$ for a moderate increase of the centrality  using as a trigger  $p_t$ of a hadron / jet  at central  (or negative  rapidities)   -- for sufficiently large $p_t$ ($\ge 10 GeV/c $) as  this selects significantly smaller  $b$ than average (0.7 fm vs 1.2 fm) \cite{Frankfurt:2010ea}. Overall one expects that for such a trigger the neutron distribution would be softer but not change significantly with $p_t$ of the trigger (similar to pattern observed for the underlying events for central rapidities). Modeling of the dependence of the nucleon spectrum on centrality was performed in~\cite{Drescher:2008zz} where it was suggested  that combining several centrality triggers (including veto on the neutron production) may allow to probe high gluon densities in $pp$ collisions at the LHC which are comparable to those present in the central $pA$ collisions.

Use of two ZDC's in $pp$  scattering would allow to study correlation between  production of hadrons at the huge rapidity interval up to $\Delta y \sim 20$. If a forward neutron is detected it is likely to enhance the selection of large impact parameters and hence enhance the probability to have a neutron in the opposite ZDC. Calculations of~\cite{Drescher:2008zz}  indicate that there should be also  a correlation (though rather mild)  between  absences of  signal in the ZDCs. However the model was pretty crude and it is easy to imagine models with a stronger correlation. An important issue  here is how much the correlation is diluted by the acceptance of the ZDCs. Current versions of PYTHIA allow to look for such effects.

\section*{Acknowledgements}M.S.'s research was supported by the US Department of Energy Office of Science, Office of Nuclear Physics under Award No.
 DE-FG02-93ER40771.

%% file: forward/forward.tex
\section{Introduction}

\begin{figure}[htb]
\centerline{
\includegraphics[height=4.5cm]{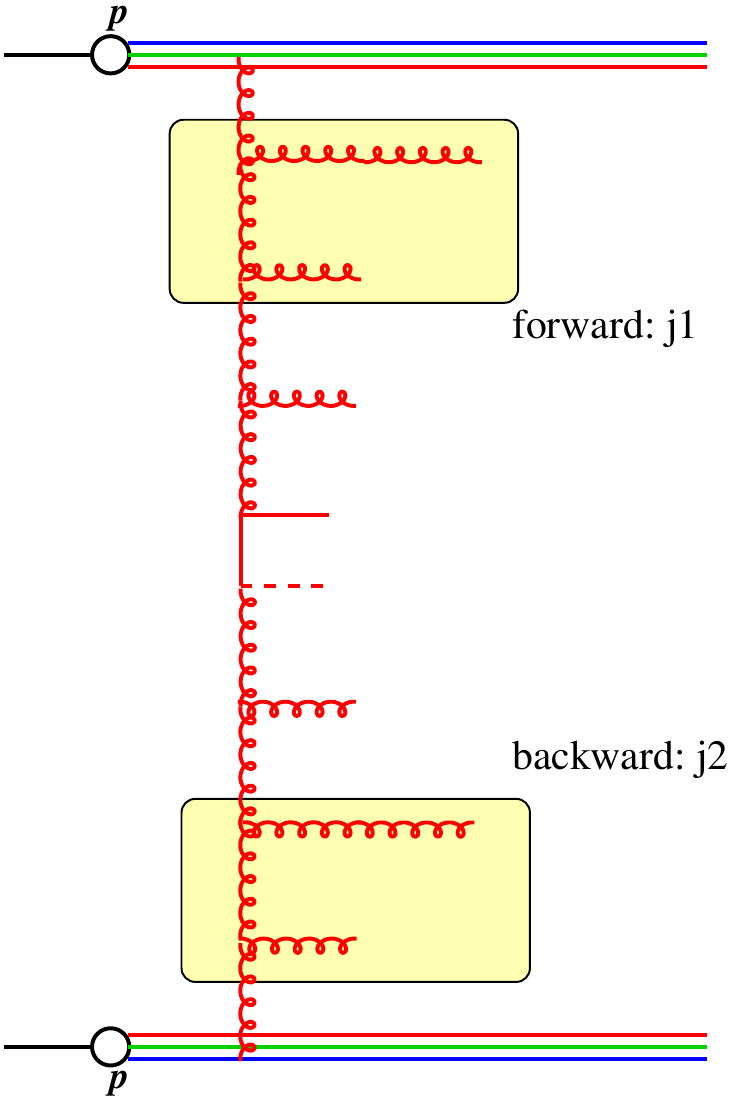}\hskip .5cm
\includegraphics[height=4.5cm]{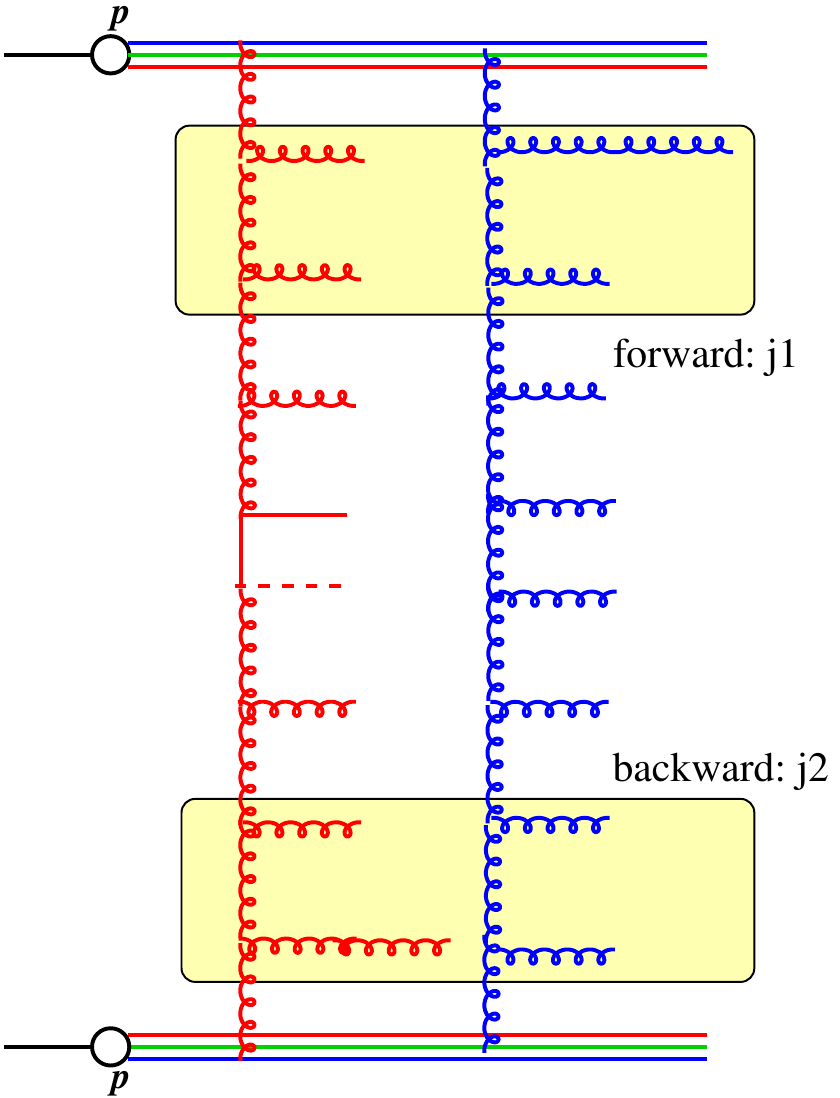}\hskip 1.5 cm
\includegraphics[height=4.5cm]{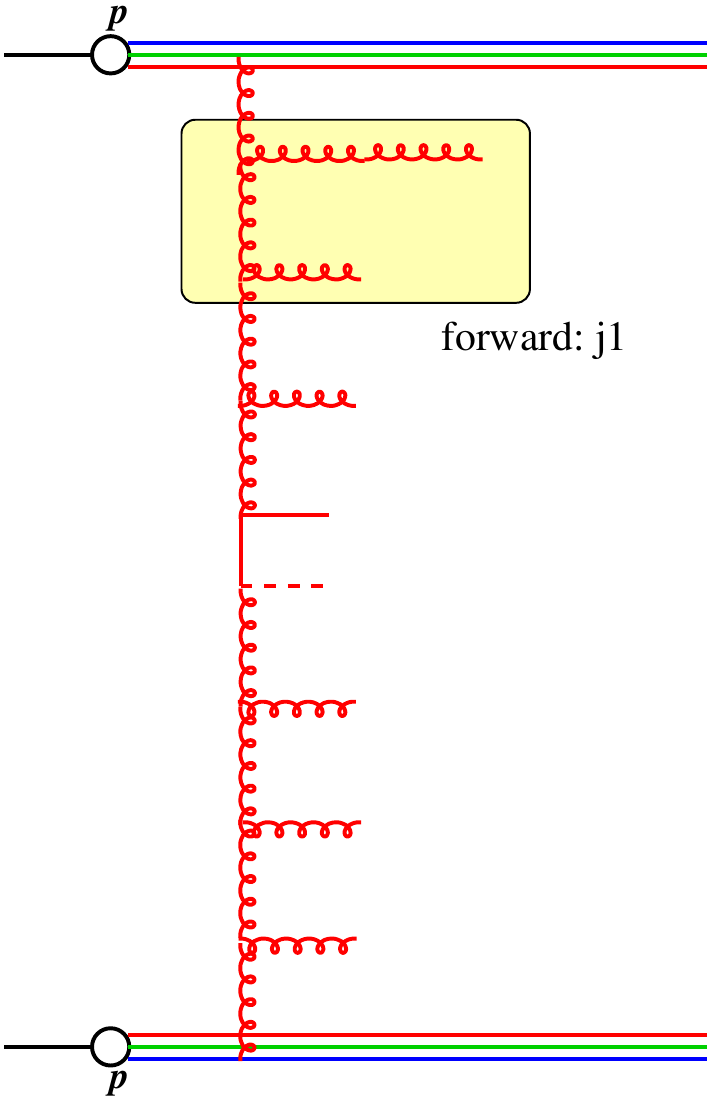} \hskip 0.5 cm
\includegraphics[height=4.5cm]{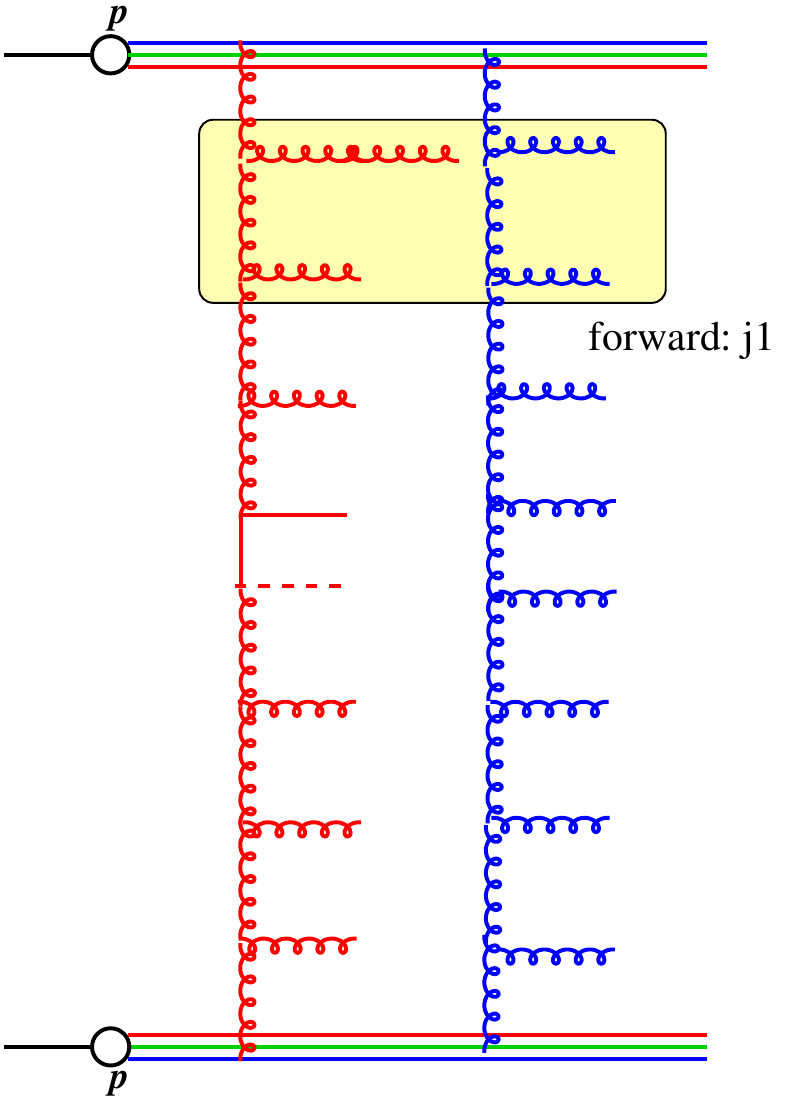}}
\hskip 3.5cm (a) \hskip 8cm (b) 
\caption{Schematic view of (a)  forward-backward jets and (b) inclusive forward jets}
\label{fig:fwdjet}
\end{figure}

\begin{figure}[htb]
\centerline{
\includegraphics[height=4.5cm]{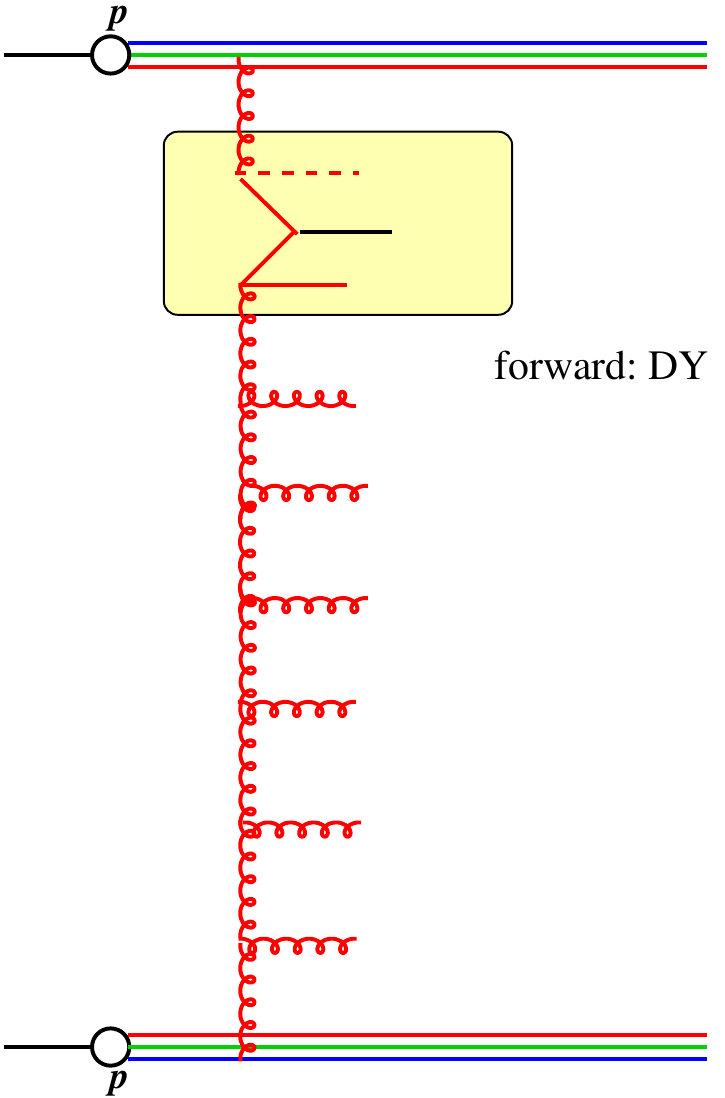}\hskip .5cm
\includegraphics[height=4.5cm]{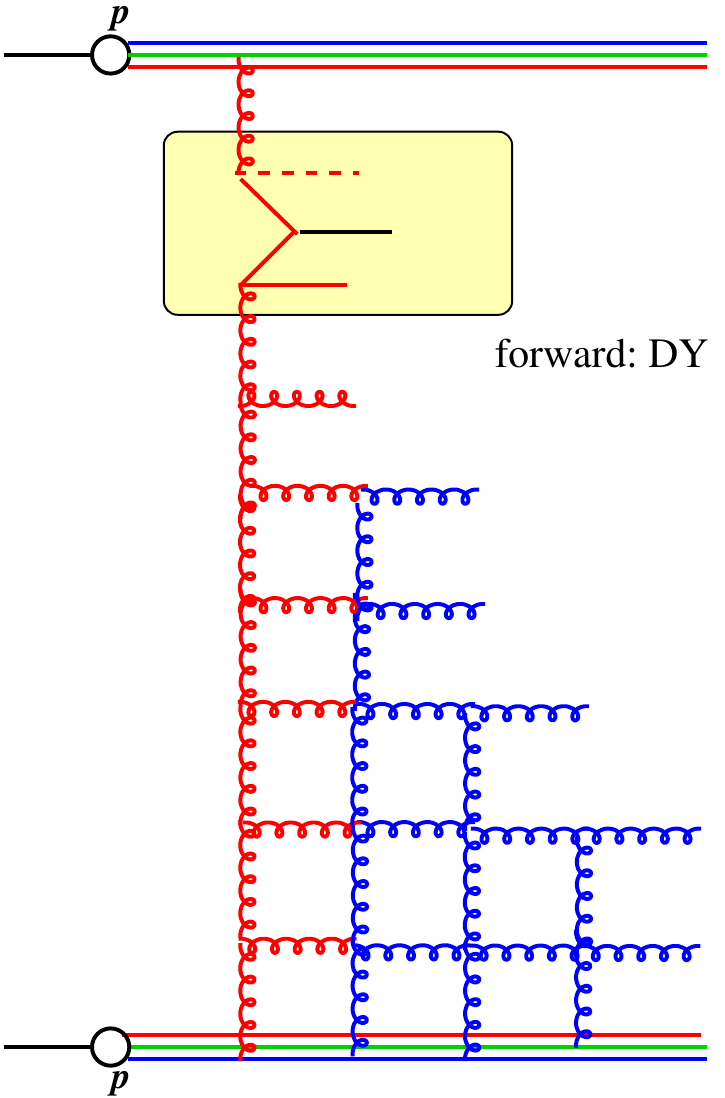} \hskip 1.5cm
\includegraphics[height=4.5cm]{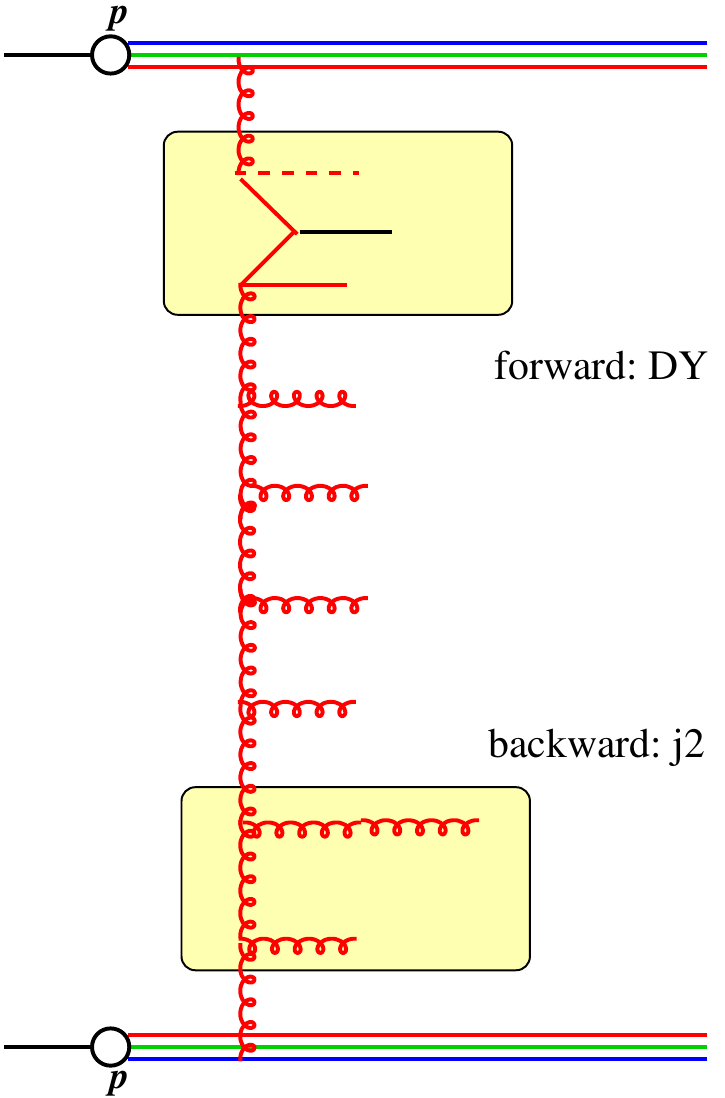} \hskip 1.5cm
\includegraphics[height=4.5cm]{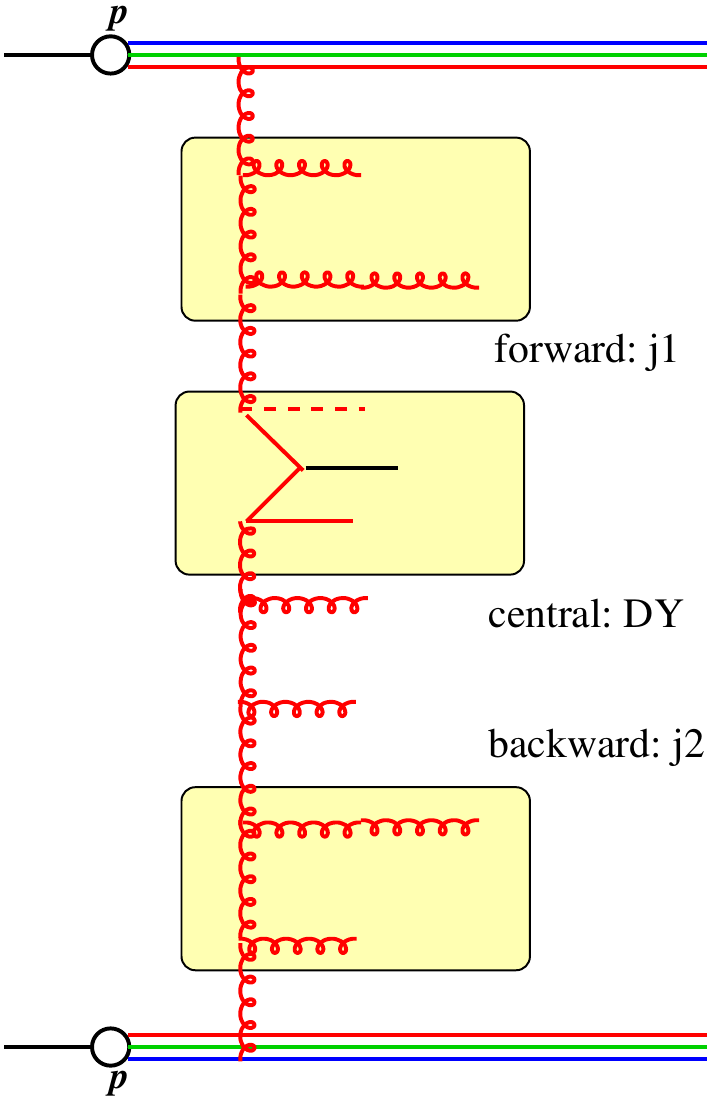}}
\hskip 3.cm (a) \hskip 6 cm (b) \hskip 4 cm (c) 
\caption{Schematic view of (a) forward Drell-Yan production,  (b) forward Drell-Yan + jet, 
(c) central DY + forward - backward jets}
\label{fig:fwdcent}
\end{figure}

In this section we discuss jets in the forward region, as well as the production of Drell-Yan lepton pairs close to the beam axis (large $\eta$ values). Such measurements will allow to analyze QCD beyond the collinear factorization, in particular BFKL dynamics\cite{Fadin:1975cb,Kuraev:1976ge,Kuraev:1977fs,Balitsky:1978ic}, multiparton interactions, and saturation. In more detail:    
\begin{itemize}
\item the cross section for di-jets with a large rapidity separation (forward-backward jets, Fig.~\ref{fig:fwdjet} a, left) includes multi-parton (gluon) radiation in the rapidity interval between the jets and thus 
is sensitive to the BFKL Pomeron\cite{Mueller:1986ey}: "Mueller-Navelet"-jets. For large rapidity separations and not too high transverse momenta of the jets multiparton interactions can play a significant role (Fig~\ref{fig:fwdjet} a, right).
\item the cross section of inclusive forward jets is sensitive to very small and very large values of the parton momentum fraction $x$ and thus can be used to constrain the parton densities at very small or very large $x$ (Fig~\ref{fig:fwdjet} b, left). In particular, the measurement of multijets at very large rapidities allows to address unintegrated gluons densities, saturation, and contributions from multiparton interactions (MPI, see Fig~\ref{fig:fwdjet} b, right).
\item the observation of Drell-Yan lepton pairs close to the beam directions at large rapidities 
represents the outstanding possibility to study small-$x$ dynamics at the LHC. In particular,
inclusive forward DY production can be sensitive to saturation effects (Fig.~\ref{fig:fwdcent} a)
The study of Drell-Yan pairs near the forward direction, together with a jet in the backward direction is sensitive to multiparton radiation (higher order radiation, similar to the "Mueller Navelet" jets): see Fig.~\ref{fig:fwdcent} b. In an alternative measurement, central DY production in association with jets in the forward or backward region, small-$x$ resummation can be probed in a different region of phase space (Fig.~\ref{fig:fwdjet} c).   
\end{itemize}

While the basic idea to investigate forward-backward jets (Fig~\ref{fig:fwdjet} b) as a probe for BFKL dynamics is old (the first proposal was made by Mueller and Navelet \cite{Mueller:1986ey} in 1986), the actual measurement and the interpretation is still very challenging, and there 
are questions and measurements which have not yet been addressed. 
Measurements performed until now required a lower $p_t$ threshold for the jets  and have been  found to be compatible (within uncertainties) with predictions coming from simulations based on collinear parton distribution functions supplemented by parton showers and multi-parton interactions. In contrast, differential measurements requiring a $p_t$ window for the forward-backward jets should be better suited for discriminating between predictions based on collinear factorisation and BFKL-based predictions. As to 'clean' BFKL signals, most attention has been 
given to azimuthal decorrelations. Different machine energies at the LHC provide the unique 
opportunity of looking also for the BFKL intercept: keeping the momentum fractions of the two parton densities fixed, a change of the machine energy directly translates into a variation of the size of the rapidity gap between the forward-backward jets. A new aspect is also the potential importance of multiparton contributions: first quantitative estimates indicates that, for highest LHC energies and moderate $p_t$ of the jets, multiparton contributions can become sizable and hence should be taken into account when 
searching for BFKL signals.
            
In addition to the search for such 'clean' BFKL signals which result from the infinite sum of gluon emissions 
between the forward and backward jets, it will be useful to test BFKL predictions also in another way. With the state-of-the art calculations for multiple jet production allowing to access matrix elements with high parton multiplicity matched to parton showers,  one is able to calculate to sufficiently high order in perturbation theory jet cross sections with full matrix-elements which include a large part of what is expected from a BFKL calculation in the high energy limit. However, a fixed order calculation cannot predict a stabilisation of the cross section in the high energy limit, which comes eventually only from a resummation to all orders. The search for BFKL thus is to identify regions of phase space, where a fixed order calculation (even at high orders) will be not sufficient and resummation effects become important. In this case BFKL provides an effective method to calculate the resummation. In this context it is extremely helpful to have a dedicated, purely BFKL-based Monte Carlo event generator (BFKLex): this allows to investigate final states with fixed numbers of jets (partons) with large rapidity separations. 

Jet production at large rapidities probes the region of small and large $x$ momentum fractions of the partons involved in the hard scattering. While jet production at central rapidities is rather well described by calculations based on collinear factorization and next-to-leading order (NLO) partonic calculations, jet production at forward rapidities is 
well-suited to study unintegrated parton densities and is also sensitive to multiple parton radiation beyond NLO, which show up in systematic kinematic shifts (see for example \cite{Dooling:2012uw}). Such effects come from multiple hard parton radiation, which becomes significant, when $x$ is small, and eventually are sensitive to small $x$ resummation treated in BFKL. The large-x region attracts interest since it allows to look for 
the "intrinsic" heavy quark content of the proton.   

%Saturation lies at the interface between hard scattering and nonperturbative strong interaction physics,
% and it is of particular importance for understanding the initial state in heavy ion collisions. In $pp$ scattering 
% at the LHC, a particularly promising place is the Drell Yan production of lepton pairs very close to the forward 
% direction. This kinematic region probes the gluon density at very small x and low lepton mass and thus could be sensitive to saturation effects. Several studies compare predictions based upon saturation models with the 
% collinear approach and find significant differences. 

Saturation lies at the interface between hard scattering and nonperturbative strong interaction physics, and it is of particular importance for understanding
the initial state in heavy ion collisions. In  $pp$ scattering at the LHC, a particularly promising observable is the Drell-Yan production of lepton pairs of low mass
in the forward direction. This kinematic region probes the gluon density at very small x and thus could be sensitive to saturation effects.
Several studies compare predictions based upon saturation models with the collinear
approach and find significant differences.

Unfortunately, in pA collisions which are very interesting for saturation studies due to the higher gluon density, 
the measurement of Drell-Yan may not be feasible due to luminosity constraints.
Here the measurement of real photons might provide a valuable alternative probe, 
which is more abundantly produced. However, such a
measurement would require upgraded detectors at forward rapidity.
In the following we discuss these questions in more detail.
%=======
In the following we discuss these questions in more detail.
%=======
\section{Forward backward jet production in $p\bar{p}$ and $pp$: the BFKL program}
We discuss first new theoretical developments in forward-backward jet production, followed by a survey of measurements from Tevatron and LHC run1. 

\subsection{Theoretical remarks}

We start with a brief theoretical introduction and a description of observables where the BFKL formalism is expected to apply. The BFKL approach identifies and calculates, for scattering amplitudes in the Regge limit, 
at each order of perturbation theory those contributions which have powers of $\ln s$ and thus grow 
with the center-of-mass energy. These terms are associated to final states extending over large rapidity 
intervals. Compared to other high energy calculations (e.g. hard scattering processes within the collinear 
factorization), the BFKL formalism exhibits new degrees of freedom:  reggeized gluons and quarks. 
These reggeons are universal in the sense that they drive a large variety of cross sections with very different initial and final state configurations, and they are present in lepton-lepton, lepton-hadron and hadron-hadron induced scattering processes. Most prominent is the Pomeron which, in QCD, appears as a bound state of two 
reggeized gluons. This reggeization picture~\cite{Fadin:1998fv} in QCD has now been tested  at leading (LL)~\cite{Fadin:1975cb,Kuraev:1976ge,Kuraev:1977fs,Lipatov:1976zz, Balitsky:1978ic,Lipatov:1985uk} and next-to-leading (NLL)~\cite{Fadin:1998py,Ciafaloni:1998gs} accuracy, and it has been cross-checked by many different methods, including sophisticated techniques in string theory~\cite{Bajnok:2008qj}. 
It is important to mention that reggeization and the existence of a Pomeron state also holds for electroweak ~\cite{Bartels:2006kr}, and gravitational interactions at high energies ~\cite{Bartels:2012ra}. In particular, they have become of central interest in AdS/CFT duality and in string theory  (integrability). 

All this provides a very strong motivation to search for experimental evidence of the QCD Pomeron in high 
energy scattering processes. The LHC, because of the large center-of-mass energy, offers the unique 
possibility of testing this part of strong interaction dynamics.     
 
\subsection{Signals based upon inclusive all-order summation}  

\begin{figure}[htbp]
  \centering
  \includegraphics[width=12.cm,angle=0]{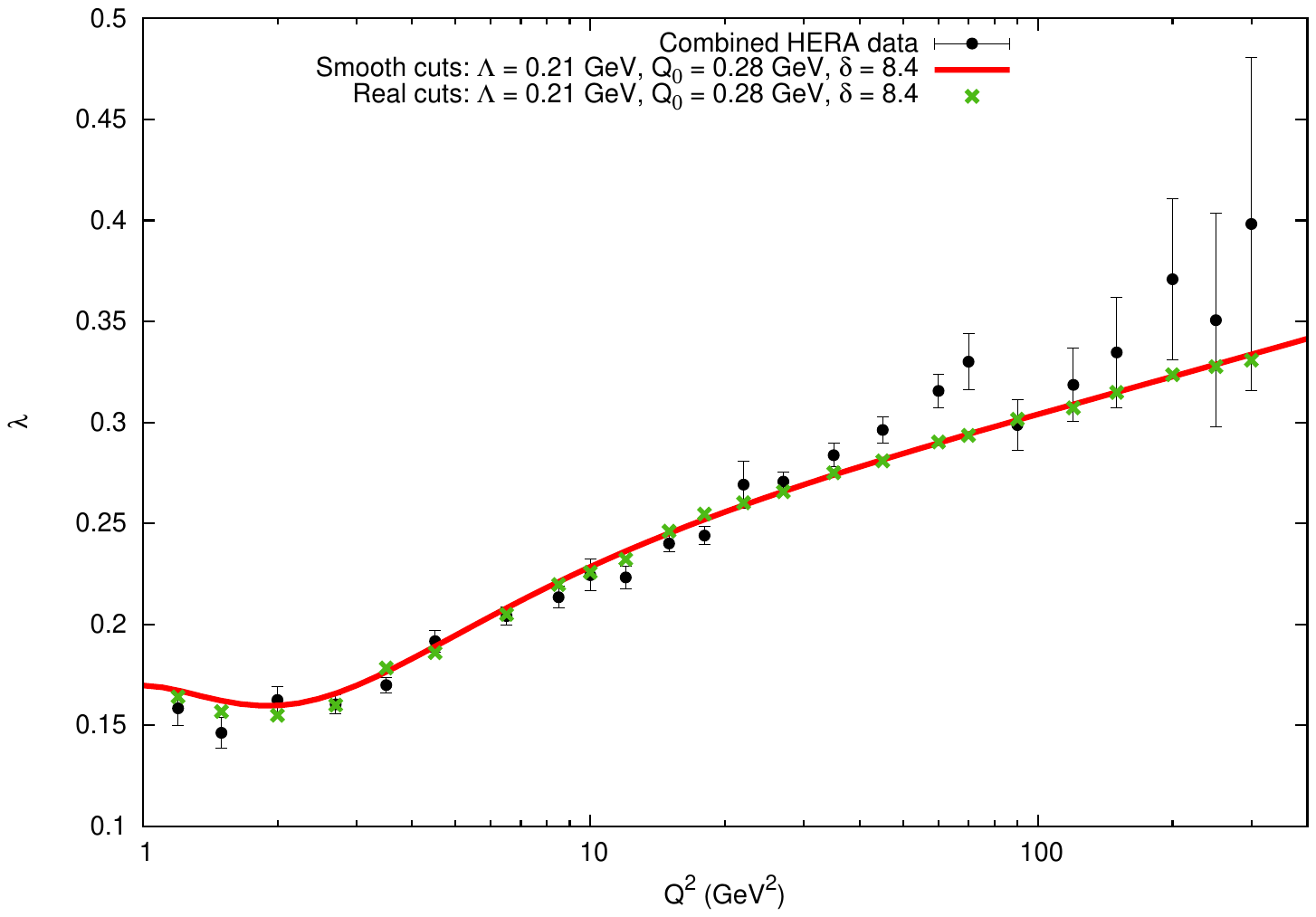}
  \caption{{\small BFKL calculation for $\lambda(Q^2)$ in the parametrization of the structure $F_2 (x,Q^2) = c (Q^2) x^{-\lambda(Q^2)}$ at NLL with collinear improvements and a model for the proton impact factor of the form $\simeq (p^2/Q_0^2)^\delta e^{-p^2/Q_0^2}$}, for values of $x< 0.01$. $\Lambda$ enters in a regularization of the infrared behavior of the running coupling with the form $\alpha_s (\mu^2) = (4 \pi / \beta_0)\left[1/ \ln(\mu^2/\Lambda^2) + 125 (1+4 \mu^2/\Lambda^2) /((1-\mu^2/\Lambda^2)(4+ \mu^2/\Lambda^2)^4)\right] $.}
  \label{F2FL}
\end{figure}

From a phenomenological point of view, it is a pressing matter to find windows of applicability of the BFKL formalism.  In the traditional approach, one searches for BFKL effects in a scattering process initiated by two hard partons with equal momentum scales: in $pp$ scattering, this idea  is realized in the Mueller-Navelet jets,  and the momentum scales are related to  the transverse momenta of the forward and backward jets (in $ep$ scattering the analogous final state are the forward jet configuration). Whether the cross section for this process is described by BFKL or by the standard DGLAP \cite{Gribov:1972ri,Lipatov:1974qm,Altarelli:1977zs,Dokshitzer:1977sg} evolution depends upon the size of the 
rapidity separation of the two jets and upon the ratio of the momentum scales of the two jets: BFKL applies 
to configurations with large rapidity separations and (almost) equal momentum scales: for configurations with very different momenta it is DGLAP which provides the correct description. In a realistic scenario the situation may be not so clear: a characteristic example is the description of the growth of the proton structure functions at low values of Bjorken $x$ where the two momentum scales are given by the photon virtuality and 
by the (lower) momentum scale of the proton. Indeed it is possible to get a good fit of the most recent combined HERA data for $F_2$ and $F_L$ with a NLL BFKL calculation, see Fig.~\ref{F2FL}, taken from \cite{Hentschinski:2012kr,Hentschinski:2013id}. 
However, it is equally possible to fit these data with other approaches. Moreover, there is quite some model dependence in the BFKL calculation, which includes three free parameters for the structure of the coupling of the gluon ladder to the hadron, a model with a frozen behavior of the coupling in the infrared and the use of a collinear resummation together with a particular renormalization scheme which allows us to reach very low values of $Q^2$ in the fit. 

All this indicates that particular attention has to be given to the ratio of the scales. As it has already been said,
a clean BFKL test requires similar transverse momenta of both jets. However, such a configuration becomes unstable~\cite{Andersen:2001kta,Fontannaz:2001nq}, in particular when a comparison at fixed order is performed. Thus, it would be very important phenomenologically to have also experimental data in a slightly asymmetric configuration with different transverse momenta of the tagged jets. This may allow to observe the transition from BFKL-like dynamics to collinear descriptions.   
This is not an easy task since the cross-section is strongly peaked near $|k_{\perp1}| \sim |k_{\perp2}|$. However, this could be offset by the larger cross-section at $\sqrt{s}=13$ TeV compared to $\sqrt{s}=7$ TeV. In addition, lowering the minimal value of the transverse momenta of the tagged jets would be a very motivating experimental challenge since it would further increase the cross-section and reduce the statistical uncertainties. 
A low transverse momentum of the jets also increases the contribution of multiparton 
corrections (see below). In general it may be safer to define 'windows' of transverse momenta rather than 
'lower limits'. In the latter case, asymmetric configurations are included which may spoil BFKL signals.   

\subsubsection{Energy dependence: the BFKL intercept} 
\begin{figure}[h!]
\centering
\includegraphics[scale=0.5]{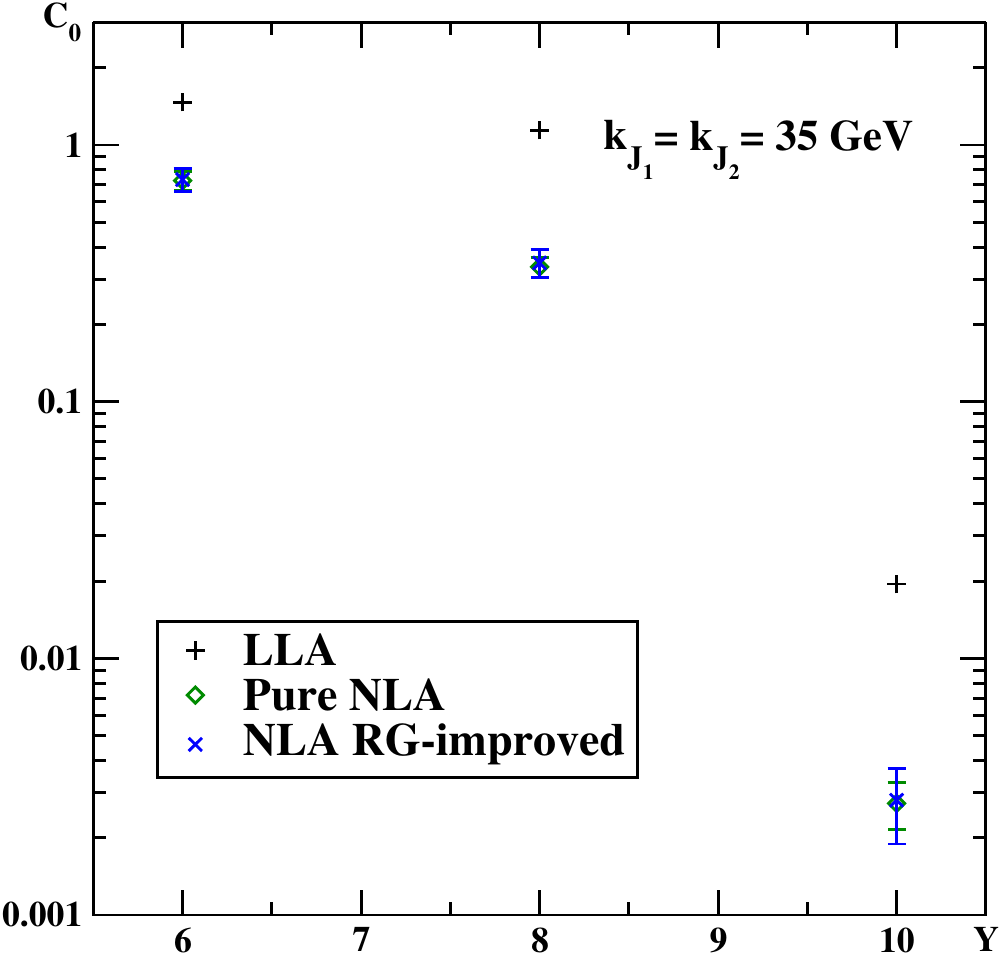}
\caption[]{$Y$ dependence of Mueller-Navelet cross section $C_0$ for $|\vec k_{J_1}|=|\vec k_{J_2}|=35$ GeV at $\sqrt s=14$ TeV.}
\label{MNC0}
\end{figure}

Let us now look in more detail into possible measurements at the LHC.   A natural test of BFKL dynamics is the 
growth of the cross section with increasing rapidity separation of the two jets.    
There is, however, a problem which is related to the fact that the BFKL-like growth due to the emission of gluons between the jets is contaminated by the $x$ dependence of the parton densities of the two protons.  
The further the two jets are separated in rapidity, the deeper we are driven into the $x \to 1$ limit of the collinear parton distribution functions. This region is characterized by a drop of the cross section due to energy momentum conservation. This hides the BFKL effect inside the growth of the cross section, as it can be seen in Fig.~\ref{MNC0} (obtained in Ref.~\cite{Caporale:2013uva}). Here the rapidity dependence of the cross section for Mueller-Navelet jets is investigated with the same values of the transverse momenta of the tagged jets. The calculations are performed for an LHC run at $\sqrt{s} =  14$ TeV, and we can see how the cross section decreases as $Y$ increases. 
As a possible way out of this difficulty, one should make use of different machine energies offered by the LHC and measure  the cross-section at different center-of-mass energies $s$. By keeping the longitudinal momentum fractions of the jets, which are given by $x_{J,i}=\frac{k_{\perp i}}{\sqrt{s}}e^{y_{J,i}}$, fixed, the change in the overall 
energy $s$ directly translates into a change of the difference of the  rapidities $y_{J,1}$ and $y_{J,2}$ of the tagged jets. In this way, at least in the LO description where the jet vertex imposes $x_i = x_{J,i}$, one should get access to the rapidity dependence of the partonic subprocess itself. In NLO the jet vertex contains an integral over $x_i$, and thus makes the theoretical prediction less straightforward.     
    
\subsubsection{Angular decorrelation}

A very important signal for BFKL dynamics is contained in the angular dependence of the forward and backward jets, in particular in their decorrelation at large separation in rapidity ~\cite{DelDuca:1993mn,Stirling:1994zs}.  In practice, one can study the coefficients
\begin{equation}
\label{def:dsigma}
  \langle\cos(n\varphi)\rangle \equiv \langle\cos\big(n(\phi_{J,1}-\phi_{J,2}-\pi)\big)\rangle\,.
\end{equation}
These coefficients are related to the azimuthal distribution of the jets $\frac{1}{{\sigma}}\frac{d{\sigma}}{d \varphi}$
through
\begin{equation}
\label{exp_cos}
 \frac{1}{{\sigma}}\frac{d{\sigma}}{d \varphi}
  ~=~ \frac{1}{2\pi}
  \left\{1+2 \sum_{n=1}^\infty \cos{\left(n \varphi\right)}
  \left<\cos{\left( n \varphi \right)}\right>\right\}.
\end{equation}
In a pure leading order (LO) collinear treatment, the two jets should be emitted back to back: $\varphi=0$ and $k_{\perp1}$=$k_{\perp2}$, since there is no phase space for (untagged) emission between them. This simple picture should of course be corrected by radiative corrections. 
For large rapidities between the jets, the multiple emission of semi-hard gluons between these two jets is expected to modify dramatically this picture. This should lead to enhanced cross-sections as well as to strong decorrelation effects, i.e. a decrease of   $< \hspace{-.12cm}\cos{(n \phi)} \hspace{-.12cm}>$. 

Several comments are in place. First, it is known that passing from a leading logarithmic (LL) to a next-to-leading-logarithmic (NLL) treatment in the BFKL framework can modify significantly this picture.   
A complete NLL BFKL analysis of Mueller-Navelet jets (for more details, see refs.~\cite{Ducloue:2013hia,Ducloue:2013bva}), including the NLL corrections both to the Green's function~\cite{Fadin:1998py,Ciafaloni:1998gs} and to the jet vertex~\cite{Ciafaloni:1998kx,Ciafaloni:1998hu,Bartels:2001ge,Bartels:2002yj,Caporale:2011cc,
Hentschinski:2011tz,Chachamis:2012cc}, showed that the NLL corrections to the jet vertex have a very large effect, of the same order of magnitude as the NLL corrections to the Green's function~\cite{Vera:2007kn,Marquet:2007xx}, leading to a lower cross-section and a much larger azimuthal correlation~\cite{Colferai:2010wu}. However, these results are very dependent on the choice
of the scales, especially the renormalization scale $\mu_R$ and the
factorization scale $\mu_F$, in particular in the case of realistic kinematical cuts for LHC experiments~\cite{Ducloue:2013hia}. This dependency can be reduced by including a set of higher order contributions, according to the Brodsky-Lepage-Mackenzie (BLM) prescription~\cite{Brodsky:1982gc} adapted to the resummed perturbation theory \`a la BFKL~\cite{Brodsky:1998kn,Brodsky:2002ka}. Such a full NLL BFKL analysis supplemented by the BLM scale fixing procedure has been performed~\cite{Ducloue:2013bva},
leading to a very satisfactory description of
the most recent LHC data extracted by the CMS collaboration
for the azimuthal correlations of Mueller-Navelet jets at a center-of-mass energy $\sqrt{s}=7$ TeV~\cite{CMS-PAS-FSQ-12-002}.   

\begin{figure}[hbtp]
  \begin{center}
    \includegraphics[width=0.4\textwidth]{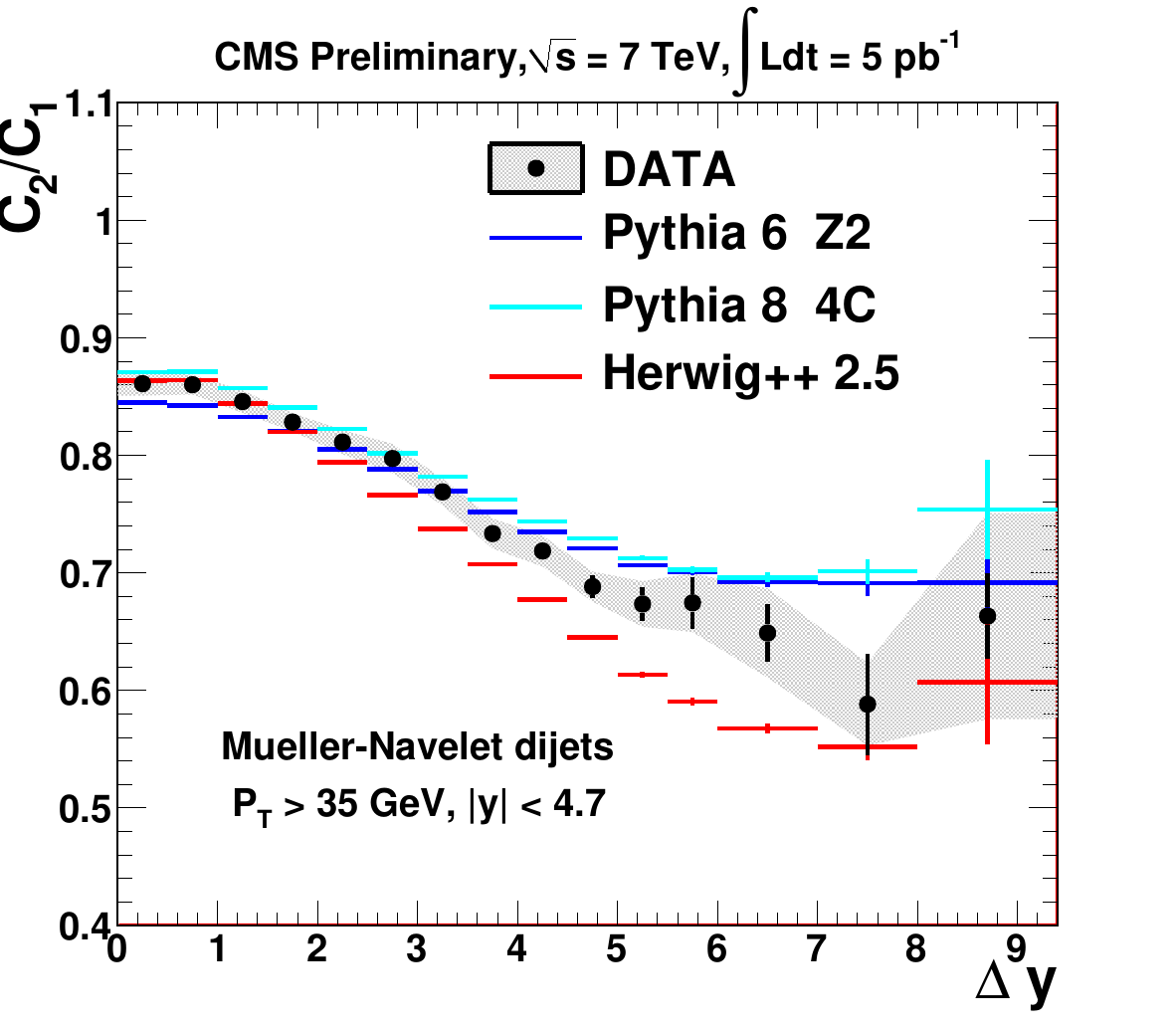}     \includegraphics[width=0.4\textwidth]{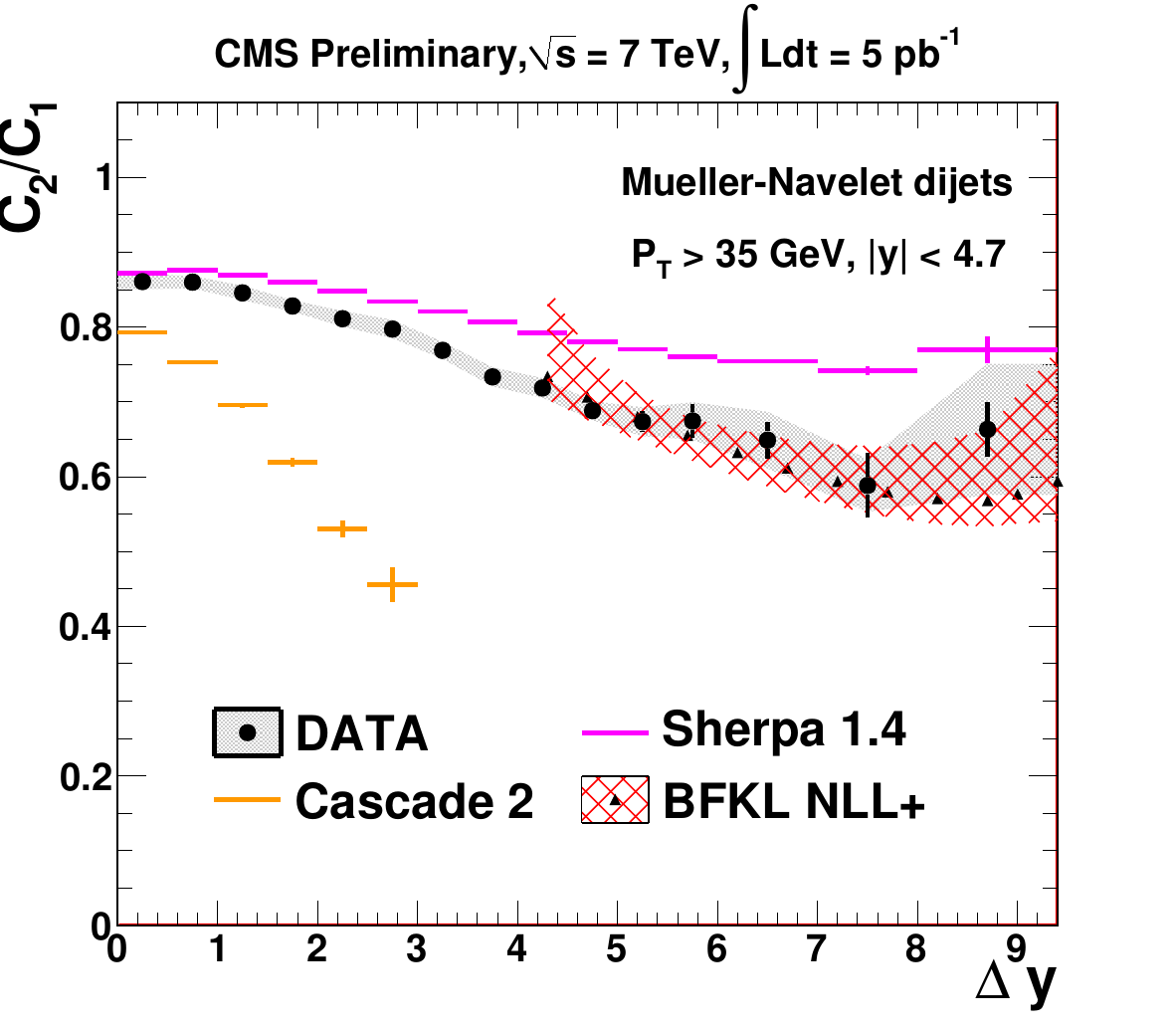} 
    \caption{Ratio $C_2/C_1$ as a function of $\Delta y$ compared to various theory predictions. }
    \label{fig:crat}
  \end{center}
\end{figure}

Second, in ~\cite{Vera:2006un,Vera:2007kn} it has been noted that these differential distributions still suffer from a large influence of the collinear region. This is due to the fact that $< \hspace{-.12cm}\cos{(n \phi)} \hspace{-.12cm}> \simeq \exp{(\alpha_s Y (\chi_n (1/2) - \chi_0(1/2)))}$, where 
$\chi_n(\gamma)$ is, in Mellin space, the $n$-th Fourier component in azimuthal angle of the BFKL kernel where the region $\gamma \simeq 1/2$ gives the largest contribution to the cross section at high energies. It turns out that the $n=0$ component is very sensitive to collinear dynamics well  beyond the original multi-Regge kinematics. It is possible to take care of these collinear contributions in an approximated way by including their leading all-orders expression in a resummation ``on top" of the BFKL original calculation. 
Nevertheless, at present  it may be more important to fix the real region of applicability of the original BFKL formalism at NLL by using observables which are far less sensitive to this collinear ``contamination". It is only following this philosophy that one will be able to find ``distinct" BFKL observables. An important step in this direction was taken in Ref.~\cite{Vera:2006un,Vera:2007kn} where instead of $< \hspace{-.12cm}\cos{(n \phi)} \hspace{-.12cm}>$ it was proposed to remove the $n=0$ dependence by studying ``conformal ratios"\footnote{We call them conformal because they capture the $SL(2,C)$ nature of the effective theory of QCD at high energies. When the same ratios are calculated in the $N=4$ supersymmetric Yang-Mills model, which enjoys four-dimensional conformal invariance, they are in agreement with those obtained in QCD using the Brodsky-Lepage-Mackenzie (BLM) scale-fixing procedure in momentum-subtraction (MOM) renormalization scheme (see Ref.~\cite{Angioni:2011wj}).} 
${\cal C}_{m,n} = < \hspace{-.12cm}\cos{(m \phi)} \hspace{-.12cm}> / < \hspace{-.12cm}\cos{(n \phi)} \hspace{-.12cm}>$ which 
behave like $\simeq \exp{(\alpha_s Y (\chi_m (1/2) - \chi_n (1/2)))}$. It is important to note that the BFKL kernel for $n \neq 0$ is insensitive to collinear regions, as it was proven in Ref.~\cite{Vera:2006un,Vera:2007kn}. In that work it was shown that these new ratios are very stable under radiative corrections with the LO result (including running of the coupling) giving very similar results to the full NLL calculations. After the arrival of LHC data in recent years it has been seen  that the NLL predictions, including NLO forward jet vertices, are in agreement with the experimental results. Furthermore, these observables are so fine tuned  to the multi-Regge limit that it is difficult for other approaches to fit them with accuracy.
This can be clearly seen in recent studies as those presented in Ref.~\cite{Ciesielski:2014dfa} 
( Fig.~\ref{fig:crat}), where a BFKL analysis at NLL is able to fit the large $Y$ tail of the Mueller-Navelet ``conformal ratios" proposed in Ref.~\cite{Vera:2006un,Vera:2007kn} whereas DGLAP Monte Carlos clearly deviate.

Last, a general weakness of the BFKL approach is the fact that it does not respect strict energy momentum conservation. While such kinematic constraints are in principle subleading in the BFKL approach, numerically their effect could be sizable, in particular in the context of Mueller-Navelet jets at LHC. There have been many attempts to estimate
these effects of energymomentum non-conservation \cite{REF1, REF2, REF3}. An effective rapidity interval $Y_{\rm eff}$ to study energy
momentum conservation effects in this process was introduced and studied in the BFKL LL approximation \cite{REF1}. Later, an estimation with
LO vertices and NLL Greens function was performed in \cite{Marquet:2007xx}. Recently, a detailed study of the $2 \gamma \to 2 \gamma$ process  at order $O (\alpha_s),$ treated either exactly or based on a NLL BFKL approximation has shown that
one obtains a very significant improvement of energymomentum
conservation when including NLO vertex corrections. This is true in the region where the two outgoing
jets are in a slightly asymmetric configuration, the most important domain phenomenologically in view of a comparison with fixed-order approaches, as discussed at the beginning of section 8.2.2. We thus believe that
energy-momentum non-conservation in NLL BFKL should not be a
major issue in future phenomenological studies.

\subsection{Multiple parton interactions (MPI) vs BFKL contribution} 

\begin{figure}[h!]
\centerline{\includegraphics[width=12cm]{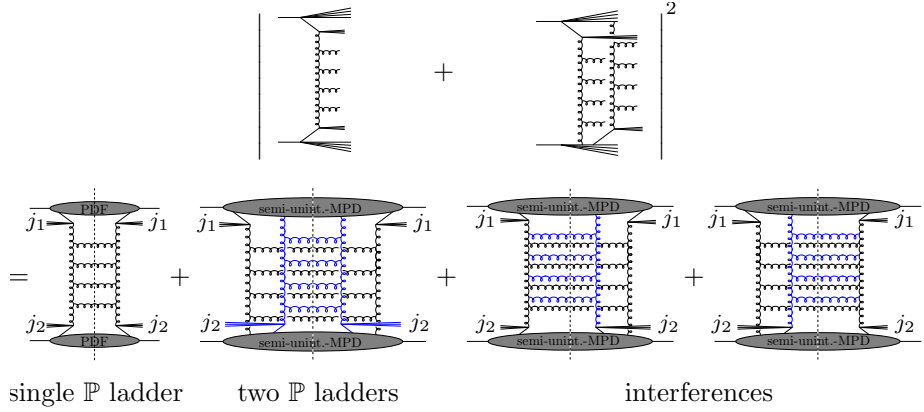}}
\caption{The Mueller-Navelet squared amplitude, which involves a single chain \`a la BFKL (first line, left) added to a multipartonic contribution with two BFKL chains (first line, right). This squared amplitude can be expanded as the sum of a single BFKL ladder (second line, first term, scaling as $s^{\alpha_\mathbb{P}}$), of two BFKL ladders (second line, second term, scaling as $s^{2\alpha_\mathbb{P}}$) and of interference contributions (last terms of the second line, with an unknown scaling).}
\label{fig:MPI}
\end{figure}

So far our discussion of BFKL signals has been restricted to the simplest set of diagrams with gluon emissions, the  single chain process illustrated in Fig.~\ref {fig:fwdjet}a. There exists, however, another 
contribution to the Mueller-Navelet cross section which may become sizable for not too large transverse momenta of the jets, the two-ladder diagram illustrated in Fig-\ref{fig:MPI}. 
This correction contains the product of two gluon densities. Since at small $x$ each gluon density becomes large, this double parton contribution could be sizable, in particular when dealing with small transverse momenta of the tagged jets (which hopefully will become accessible at CMS and ATLAS ). Note that as long as we do not integrate over the transverse momenta of the jets, the usual twist suppression is not applicable ~\cite{Diehl:2011yj}. Thus it may happen that this MPI correction interferes with the BFKL signal. On the other hand, as an MPI contribution, it is of interest also in its own rights. 

\begin{figure}[!h]
\begin{center}
\includegraphics[width=6cm]{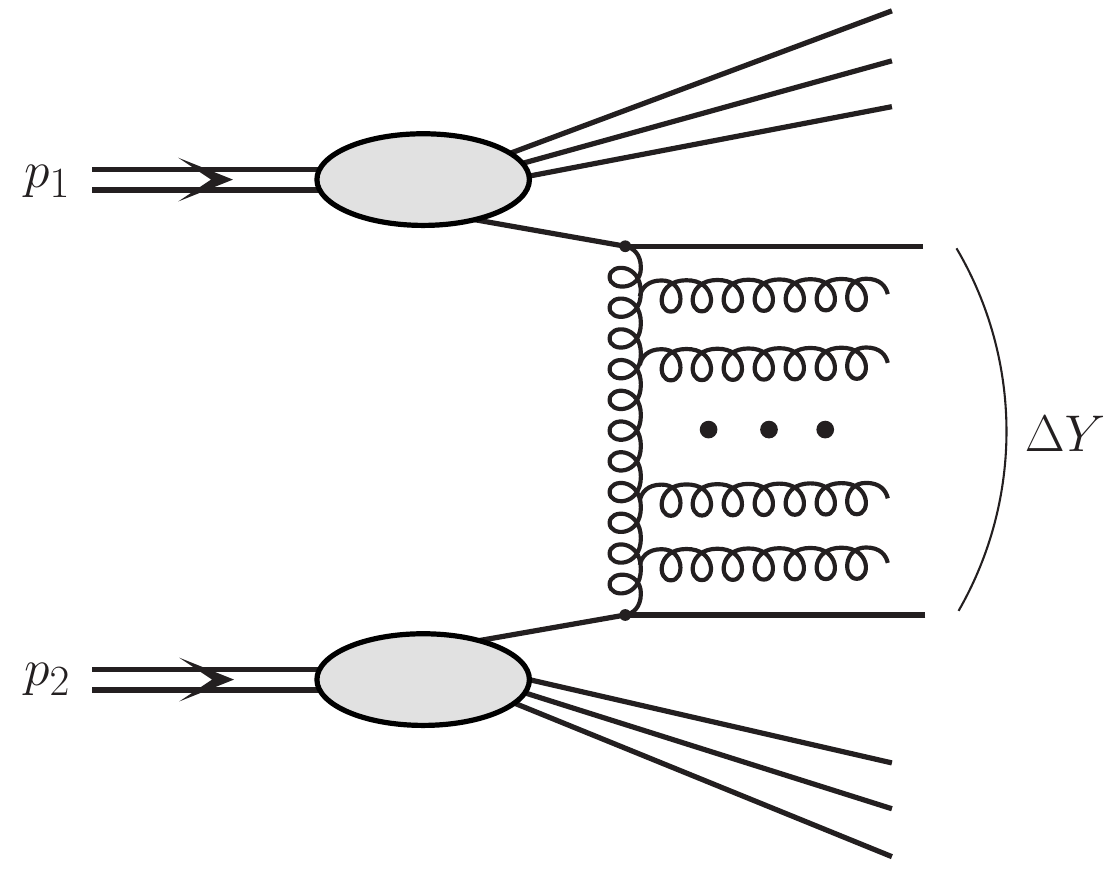}
\includegraphics[width=6cm]{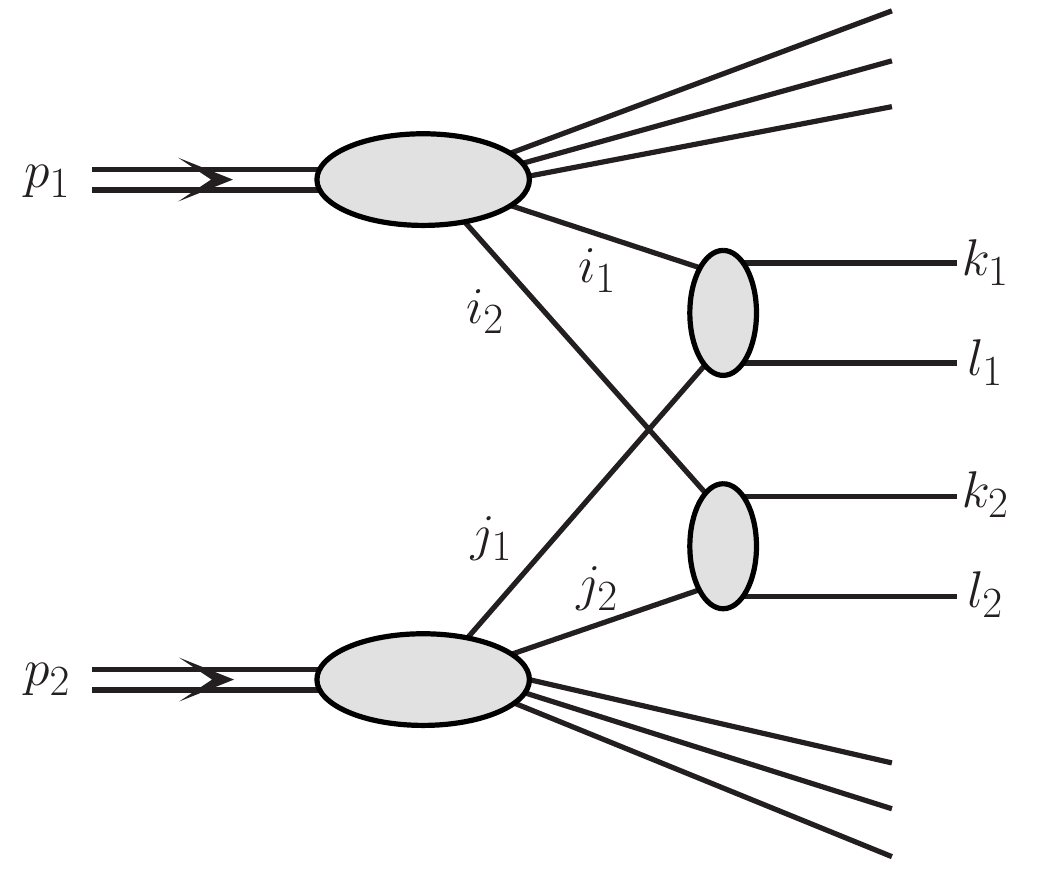}
\end{center}
   \caption{
\small The production of jets widely separated in rapidity via
Mueller-Navelet mechanism (left) and within double-parton scattering 
mechanism (right).
}
 \label{fig:diagrams}
\end{figure}

Let us report on a specific calculation \cite{Szczurek:2014zfa}. We start with a sketch of the formalism. We considers the production of two pairs of jets, one in the forward, the other in the backward direction. We illustrate this process  in Fig.~\ref{fig:diagrams}.
 For a single pair of jets we write:      
\begin{equation}
\frac{d \sigma(i j \to k l)}{d y_1 d y_2 d^2p_t} 
= \frac{1}{16 \pi^2 {\hat s}^2}
\sum_{i,j} x_1 f_i(x_1,\mu^2) \; x_2 f_j(x_2,\mu^2) \;
\overline{|\mathcal{M}_{i j \to k l}|^2} \;.
\label{LO_SPS}
\end{equation}
The calculations include only leading-order $i j \to k l$ partonic 
subprocesses.
The $K$-factor for dijet production is rather small, of the order of 
$1.1 - 1.3$ (see e.g. \cite{K-factor1,K-factor2}, 
but it can be easily incorporated. Below we shall show that
already the leading-order approach gives results in sufficiently reasonable 
agreement with recent ATLAS \cite{ATLASjets} and CMS \cite{CMSjets} data.
With this the cross section for dijet production can be written as:
\begin{equation}
\frac{d \sigma^{DPS}(p p \to \textrm{4jets} \; X)}{d y_1 d y_2 d^2p_{1t}
  d y_3 d y_4 d^2p_{2t}} 
= \sum_{i_1,j_1,k_1,l_1;i_2,j_2,k_2,l_2} \; 
\frac{\mathcal{C}}{\sigma_{eff}} \;
\frac{d \sigma(i_1 j_1 \to k_1 l_1)}{d y_1 d y_2 d^2p_{1t}} \; 
\frac{d \sigma(i_2 j_2 \to k_2 l_2)}{d y_3 d y_4 d^2p_{2t}}\;, 
\label{DPS}
\end{equation}
where
$\mathcal{C} = \left\{ \begin{array}{ll}
\frac{1}{2}\;\; & \textrm{if} \;\;i_1 j_1 = i_2 j_2 \wedge k_1 l_1 = k_2 l_2\\
1\;\;           & \textrm{if} \;\;i_1 j_1 \neq i_2 j_2 \vee k_1 l_1 \neq k_2 l_2
\end{array} \right\} $, and the summation extends over the partons species $j,k,l,m = g, u, d, s, \bar u, \bar d, \bar s$. 
The combinatorial factors take care of the identity of the two subprocesses.
Each step of DPS is calculated in the leading-order approach 
(see Eq.(\ref{LO_SPS})). 
Experimental data from Tevatron \cite{Tevatron} and 
LHC \cite{Aad:2013bjm} 
provide an estimate of $\sigma_{eff}$ in the denominator of formula 
(\ref{DPS}). In the calculations, whose main results are presented here, 
we have taken $\sigma_{eff}$ = 15 mb.
As to the comparison with BFKL calculations, we consider 
the cross section integrated over $\phi$ (relative azimuthal angle between
jets). Then the cross section can be simplified by including only
one term $n$=0 in the sum over conformal spins.
For comparison we also calculate correlations between
jets obtained in the $k_t$-factorization approach. The calculations use Kimber-Martin-Ryskin unintegrated parton distributions \cite{KMR}. 
The formulae for the off-shell matrix elements were obtained e.g. 
in \cite{NSS2013}, and corresponding formulae can be used in our calculations.

\begin{figure}[!h]
\begin{center}
\includegraphics[width=6cm]{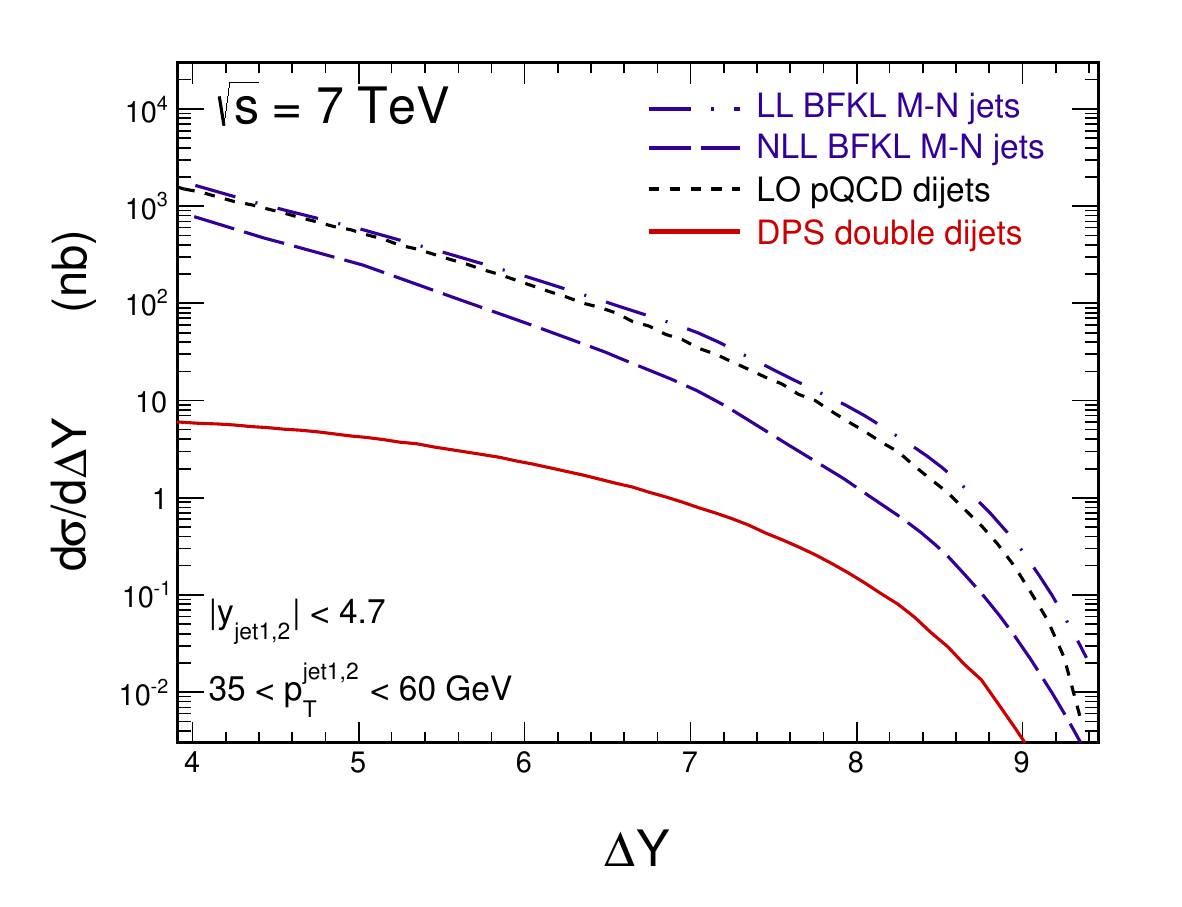}
\includegraphics[width=6cm]{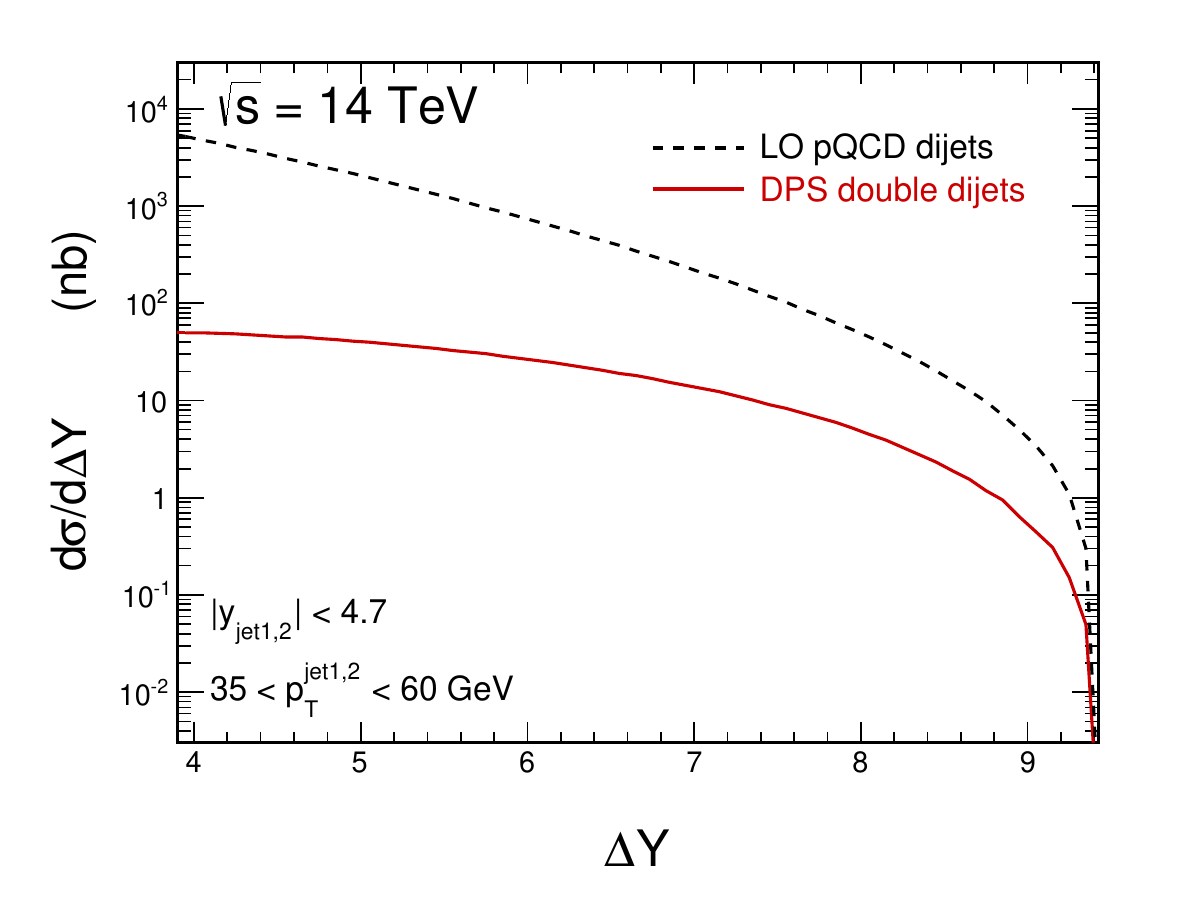}
\end{center}
   \caption{
\small Distribution in rapidity distance between the jet 
(35 GeV $< p_t <$ 60 GeV) with maximal (the most positive) and minimal 
(the most negative) rapidities. 
The collinear pQCD result is shown by the short-dashed line
and the DPS result by the solid line for $\sqrt{s}$ = 7 TeV (left panel)
and $\sqrt{s}$ = 14 TeV (right panel). For comparison we also show
results for the BFKL Mueller-Navelet jets in leading-logarithm 
and next-to-leading-order logarithm approaches from 
Ref.~\cite{Ducloue:2013hia}.
}
 \label{fig:Deltay1}
\end{figure}

\begin{figure}[!h]
\begin{center}
\includegraphics[width=5.5cm]{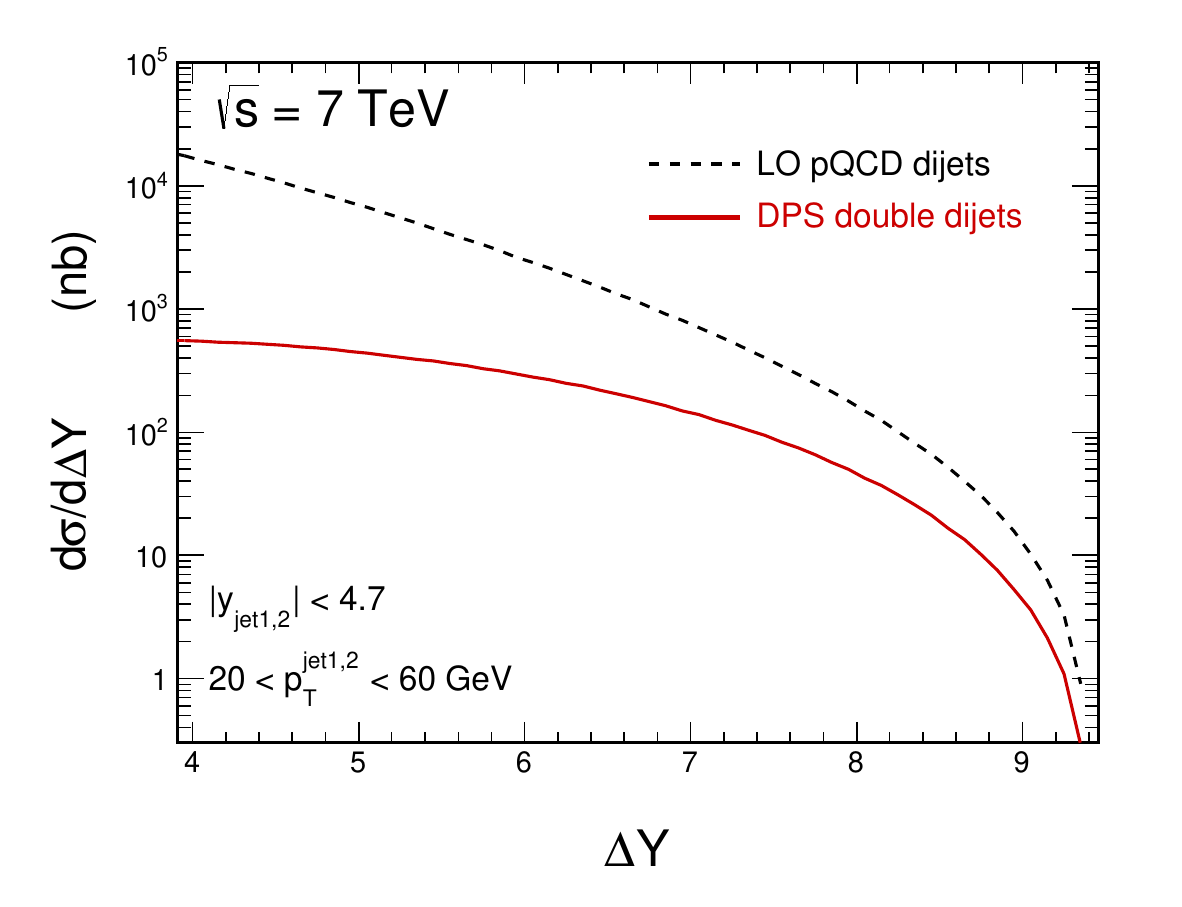}
\includegraphics[width=5.5cm]{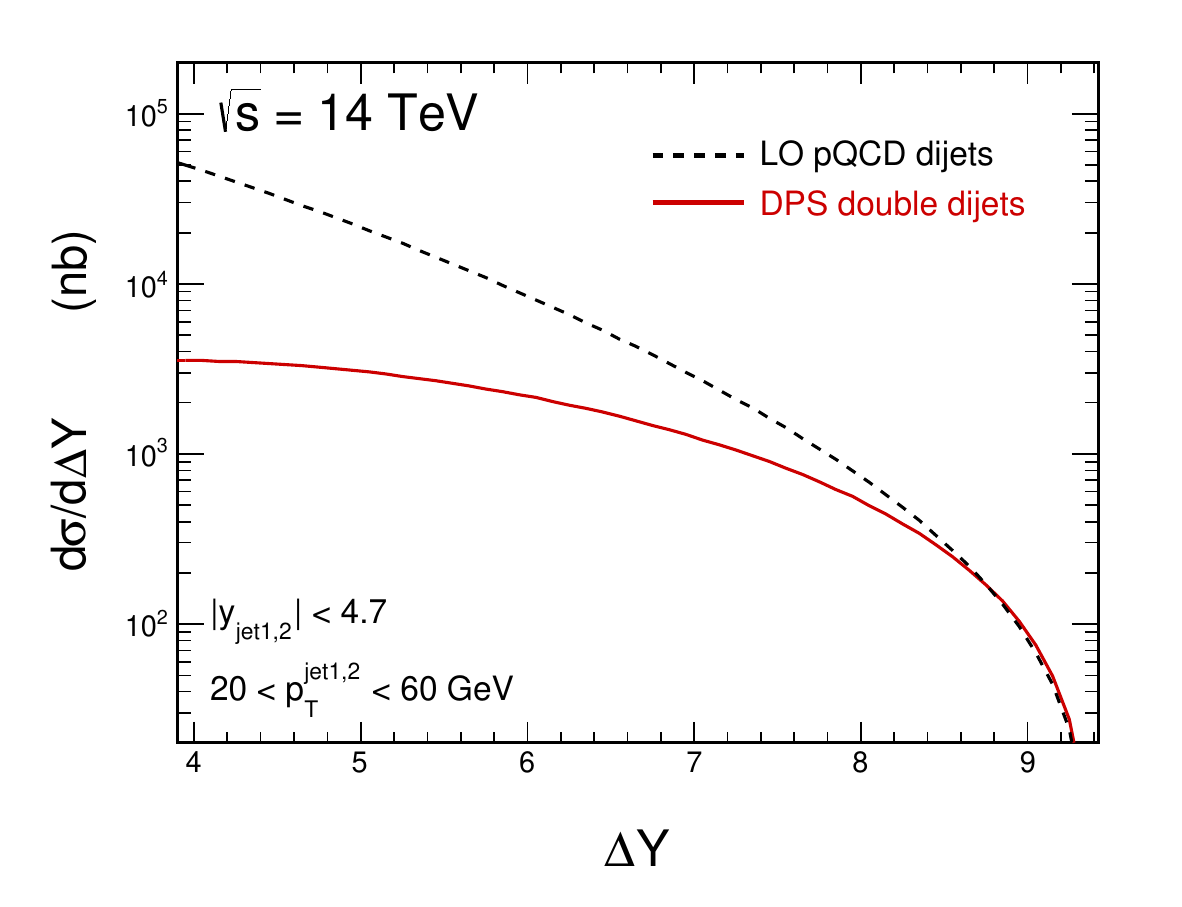} \\
\includegraphics[width=5.5cm]{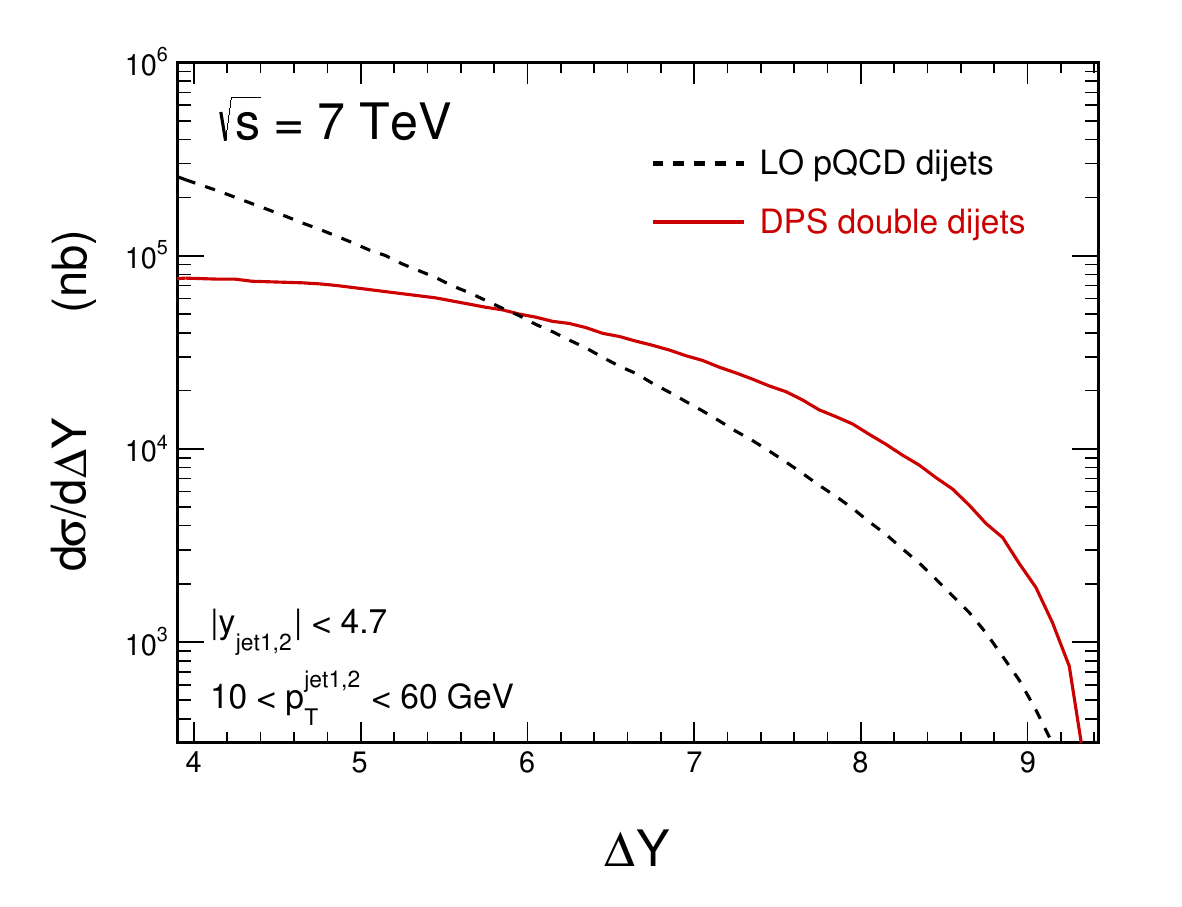}
\includegraphics[width=5.5cm]{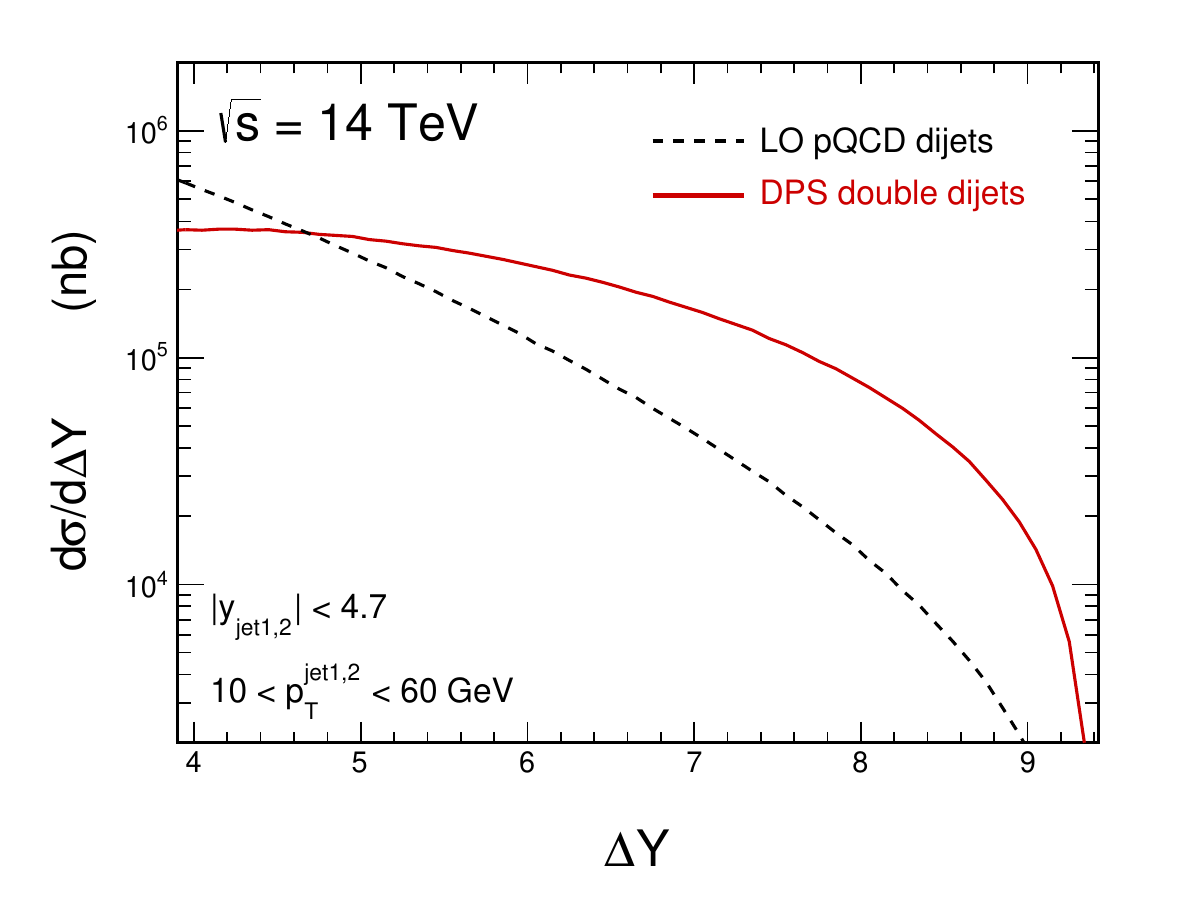}
\end{center}
   \caption{
\small The same as in the previous figure but now for somewhat smaller lower
cut on minijet transverse momentum.
}
 \label{fig:Deltay-2}
\end{figure}

Let us present a few numerical results. In Fig.~\ref{fig:Deltay1} we show distributions in the rapidity 
distance between two jets in leading-order collinear calculation
and between the most distant jets in rapidity in the case of four double parton scattering (DPS) jets.
In this calculation we have included cuts characteristic for the
CMS experiments \cite{CMS_private}.
For comparison we show also results for BFKL calculation from
Ref.~\cite{Ducloue:2013hia}. For this kinematics the DPS jets
give sizeable contribution only at large rapidity distance.
However, the four-jet (DPS) and dijet (LO SPS) final state can 
be easily distinguished and in principle one can concentrate on the
DPS contribution which is interesting by itself.
The NLL BFKL cross section (long-dashed line) is sizeably lower than 
that for the LO collinear approach (short-dashed line). For higher LHC energies the suppression of the 
DPS-cross section becomes weaker.
Fig.~\ref{fig:Deltay-2} shows results for smaller transverse momenta.
Now the double parton scattering (DPS)  contribution may even exceed the standard 
single parton scattering (SPS) 
dijet contribution. A measurement of such minijets may be, however, difficult.  One could measure, for instance, correlations of 
semihard ($p_t \sim$ 10 GeV) neutral pions with the help of 
so-called zero-degree calorimeters (ZDC). This will be discussed elsewhere.

\begin{figure}
\begin{center}
\includegraphics[width=5cm]{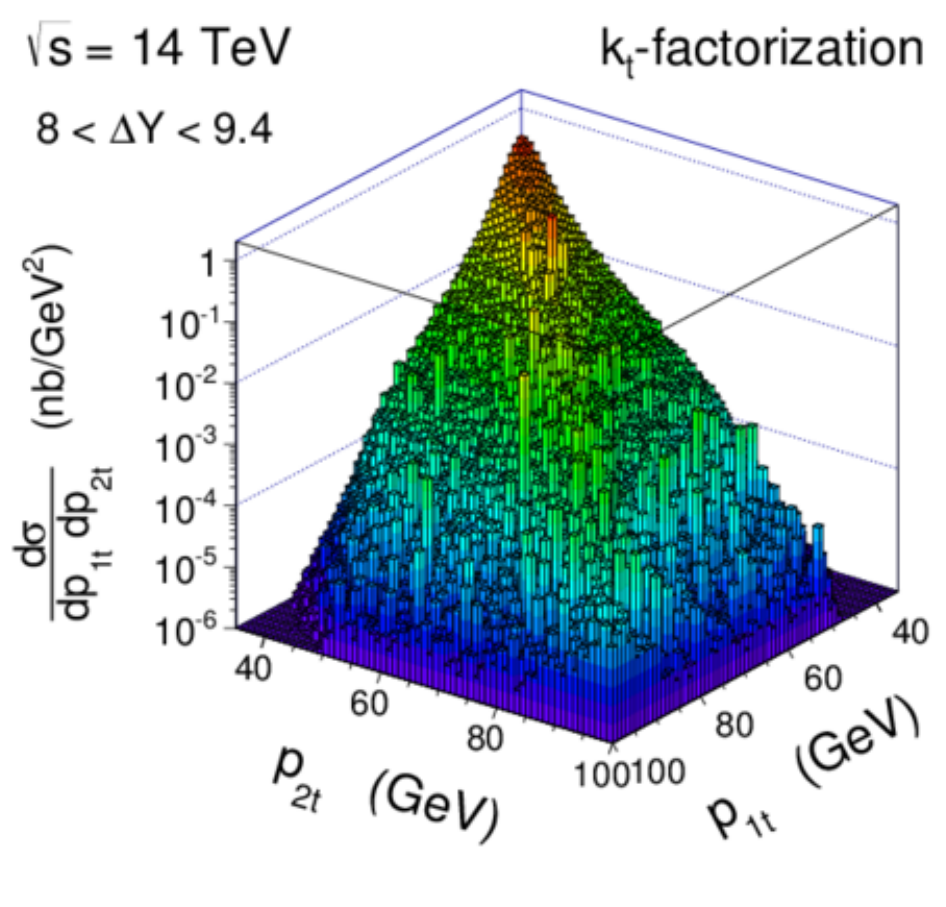}
\includegraphics[width=5cm]{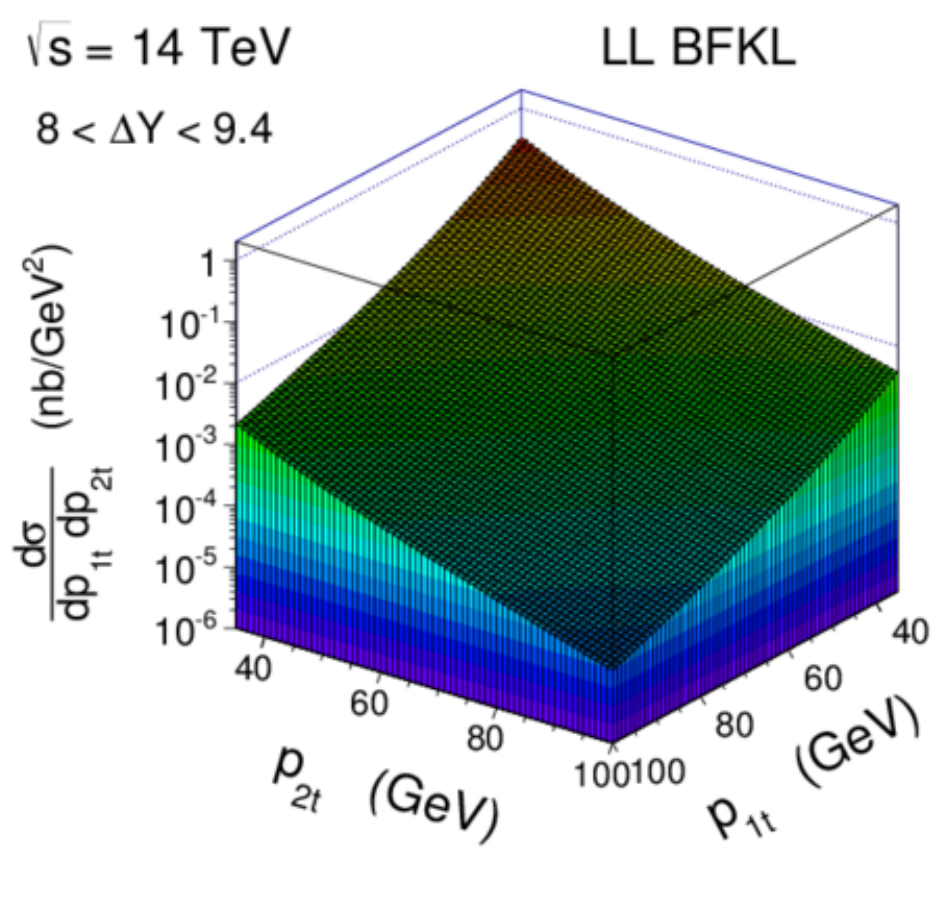}
\includegraphics[width=5cm]{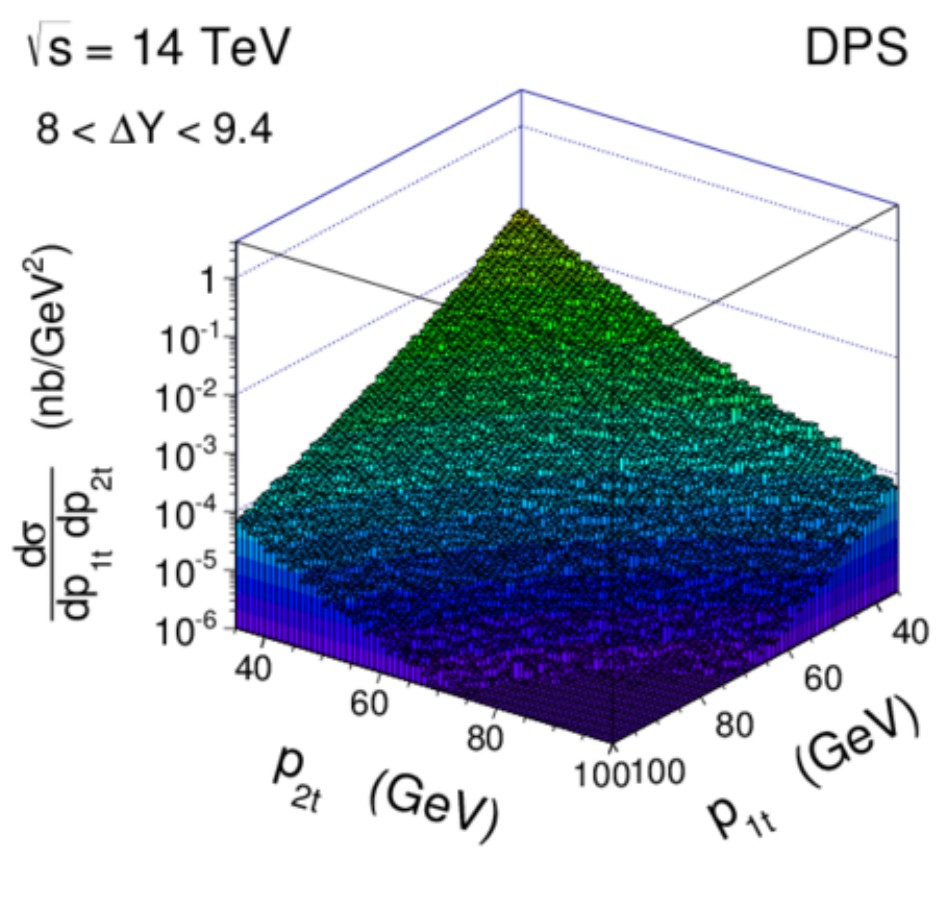}
\end{center}
\caption
{\small Two-dimensional distributions in jets transverse momenta
for large rapidity distance 8 $< y <$ 9.4 (CMS apparatus for jets)
for $k_t$-factorization approach (left), LL BFKL (middle)
and DPS (right).}
\label{fig:maps2dim}
\end{figure}

Let us compare results of the different mechanisms still in another form.
In Fig.~\ref{fig:maps2dim} we show two-dimensional 
distributions in transverse momenta of dijets widely separated
in rapidity (8 $< y <$ 9.4)) at $\sqrt{s}$ = 14 TeV (Run II).
Again we show results for $k_t$-factorization approach with KMR UPDF (left panel),
the LL BFKL approach (middle panel) and the DPS mechanism
calculated here in LO collinear approach (right panel).
All mechanisms lead to cross sections of the same order of magnitude.
We obtain integrated cross section for $p_{1t},p_{2t} >$ 35 GeV
$\sigma$ = 18.7 nb ($k_t$-factorization, KMR PDF), 40.1 nb (LL BFKL), 
2.8 nb (DPS, collinear LO), respectively. Such cross section could 
be measured in Run II both by the CMS collaboration and by ATLAS. 
The distribution in the BFKL approach and that corresponding to 
the DPS are rather similar as far as the shape is considered. 
The LL BFKL cross section is probably overestimated. 
NLL cross sections are much smaller than LL ones \cite{Colferai:2010wu}. 
Corrections for energy-momentum conservation (violated in analytic BFKL)
are expected to lower the cross section in addition.
The distribution in the $k_t$-factorization approach is also interesting 
and shows a ridge along the diagonal $p_{1t} = p_{2t}$, which corresponds  to
back-to-back jets in the collinear LO approach.
Therefore to understand which mechanism dominates one should study such 
distributions. This would require rather good statistics.
The presence of the ridge along the diagonal would clearly signal 
that the BFKL approach is insufficient in this context.
Distribution in the azimuthal angle (to be discussed in \cite{CMS2014}) 
would be another useful possibility to pin down the underlying mechanism.
Many more interesting distributions will be discussed in Ref.\cite{CMS2014}.

As our main result,  we have shown that\\
- the contribution of the DPS mechanism increases with increasing distance
in rapidity between jets,\\
- the relative effect of DPS increases if one lowers the transverse momenta of jets 
(although such measurements may be difficult) .\\
The DPS effects therefore cannot be neglected completely when searching for BFKL signals. 
Also, double parton scattering contributions are interesting not only in the context how they 
contribute to distribution in rapidity distance between the jets 
but also per se.  One could also make use of other observables like correlations in jet transverse momenta or jet transverse-momentum
imbalance (see \cite{MS2014}) to enhance the relative contribution of DPS.

\subsection{Exclusive radiation patterns: \\towards a new class of BFKL observables}

In our opinion, it is mandatory to continue, in the coming years of analysis of LHC data, this line of searches  for BFKL signals which are obtained from  the all-order summation of gluon emissions. 
However, it is also needed to propose new quantities sensitive to multi-Regge kinematics and avoiding the influence of the, otherwise widely dominant, collinear regions of phase space. 
These analysis are more complicated from the theoretical point of view, since we need two key ingredients: new NLO impact factors and Monte Carlo techniques to control the gluon Green function and to fully extract its physical content in the most differential form. The former are mandatory to fairly test the theory and correctly control the dependence on the scales appearing in the calculations (running of the coupling and energy scale separating the universal gluon radiation in central regions of rapidity and that stemming from the fragmentation regions). The latter are needed in order to effectively generate differential distributions which are otherwise complicated to be calculated analytically. 

\begin{figure}[htbp]
  \centering
  \includegraphics[width=6.5cm]{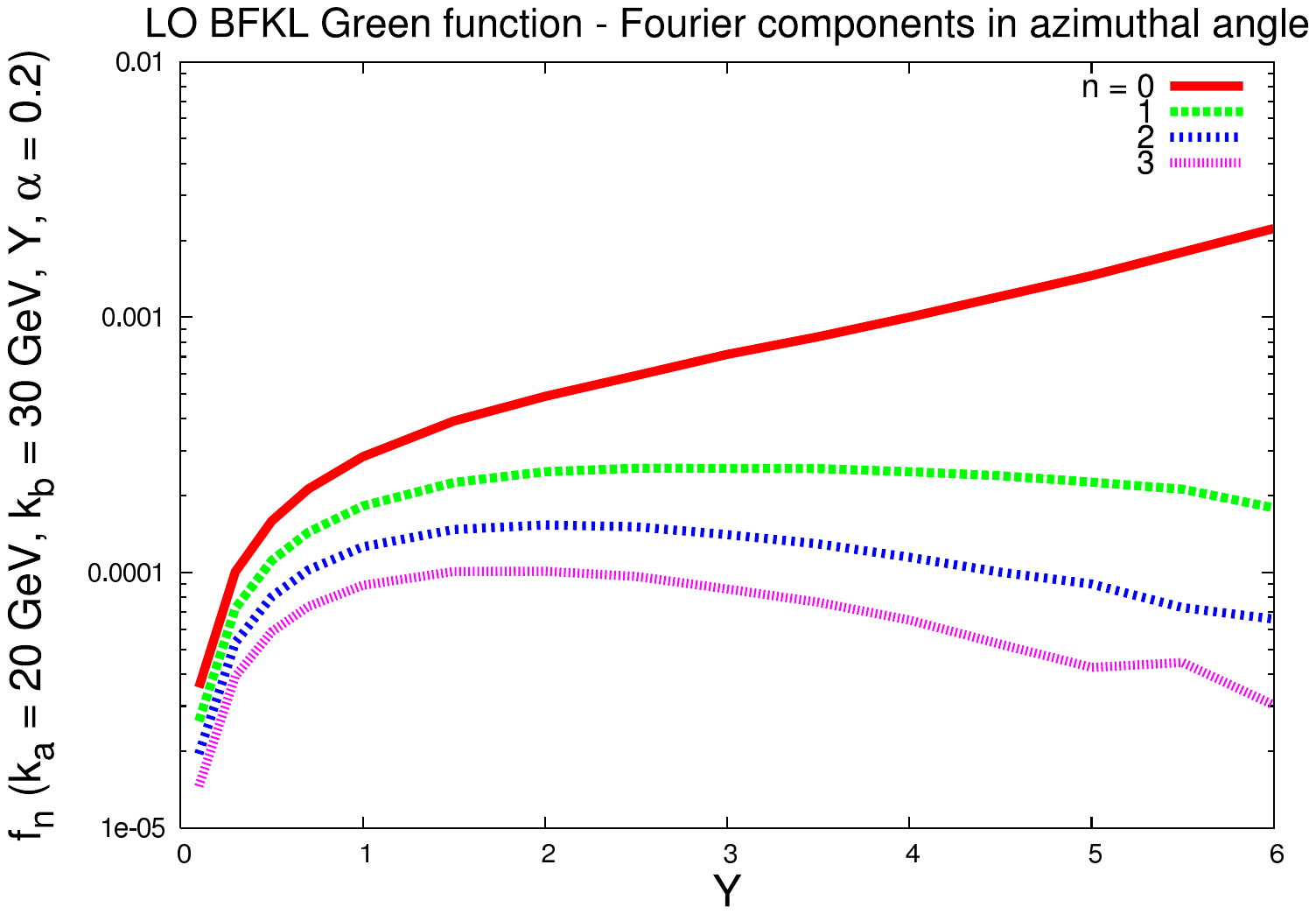}
   \includegraphics[width=8cm]{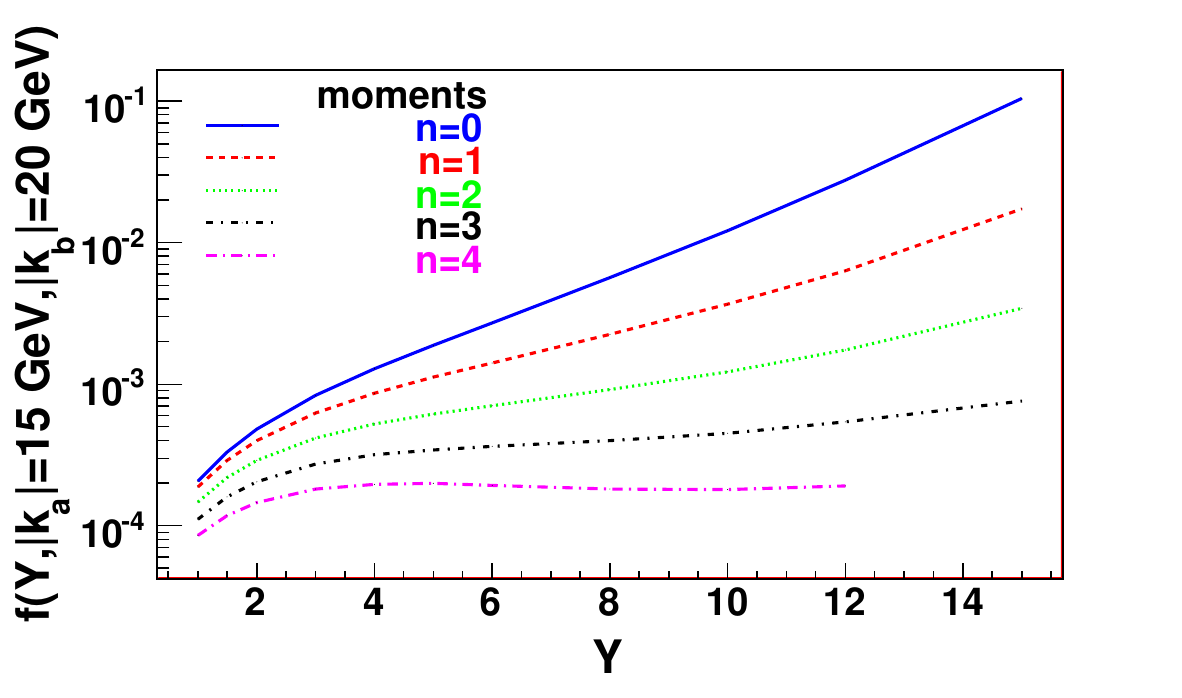}
  \caption{Variation with rapidity of the different components of the Fourier expansion on the azimuthal angle of the BFKL (left) and CCFM (right) gluon Green function.}
  \label{BFKLvsCCFMFourier}
\end{figure}

In order to illustrate how details of the radiation pattern may influence some of the quantities we have discussed before, we return to the azimuthal angle projections and show that it is a good idea to distinguish among different approaches. 
Let us compare the above mentioned $n$-moments of the BFKL cross section with those obtained from the Catani-Ciafaloni-Fiorani-Marchesini (CCFM)~\cite{Catani:1989yc,Marchesini:1994wr,Ciafaloni:1987ur,Catani:1990gu}, which,  in principle, interpolates the large $x$ and small $x$ limits of unintegrated gluon densities (we will see that this is only true for the $n=0$ projection). This has been studied in Ref.~\cite{Chachamis:2011rw} and we illustrate it here in Fig.~\ref{BFKLvsCCFMFourier}. In the left plot we see the BFKL result, and in the right panel  we show the CCFM analysis. We can see that the $n=0$ component has a similar behavior in both cases since it grows with  $Y$. But this is not the case for the other $n>0$ components which in the BFKL case decrease with $Y$ and for CCFM increase with $Y$. This is a fundamental difference between the Regge limit and approaches based on QCD coherence which should be exploited further in order to disentangle BFKL from other dynamics. 

These two points were confirmed in another evaluation of the DPS contribution for this process~\cite{Ducloue:2015jba}, performed in a
slightly different framework. Similarly to the results shown above, these predictions are based on the assumption that the DPS cross
section can be factorized into the product of two SPS subprocesses. However, here each subprocess consists in the emission of a single
jet in the forward (or backward) direction from one on-shell parton and one off-shell gluon. The former is described by a usual collinear
PDF while the latter is described by an unintegrated gluon distribution (UGD), which is the source of the jet's transverse momentum. This
is in the spirit of the second contribution shown in the first line of Fig.~\ref{fig:MPI}. The uncertainty on the DPS cross section,
coming from the choice of the UGD parametrization and the variation of $\sigma_{eff}$ in a range compatible with experimental
determinations of this quantity, is rather large. Nevertheless, in this approach the DPS cross section is always smaller than the SPS one
computed at NLL accuracy in the BFKL framework if one considers kinematics similar to those studied in Ref.~\cite{MS2014}. When looking
at the angular correlation between the jets, the effect of including the DPS contribution is always small, except for transverse momenta
of the order of 10 GeV at $\sqrt{s}=14$ TeV. For such small transverse momenta, double parton scattering could lead to a significantly
smaller correlation than predicted by a SPS calculation, but the large uncertainty on the DPS cross section makes it impossible to draw
firm conclusions in these kinematics at the moment.

There are other distributions which relate the average $p_T$ of the mini jet radiation to its position in rapidity space, the so-called ``diffusion plots" which can be connected to, {\it e.g.}, energy-energy correlations or with the distribution of mini jet multiplicity in different regions of rapidity in the detectors. With these tools at hand it is needed to be imaginative and try to tag different particles or jets in the final state which can serve as ``projecting operators" on multi-Regge kinematics.  It is clear that tagging different jets in the final state and allowing for associated mini jet radiation, together with stringent cuts on the $p_T$ and rapidity separation of the tagged jets, will allow for a much more precise identification of the multi-Regge region. 

\begin{figure}[htbp]
\begin{center}
\includegraphics[width=7.5cm,angle=0]{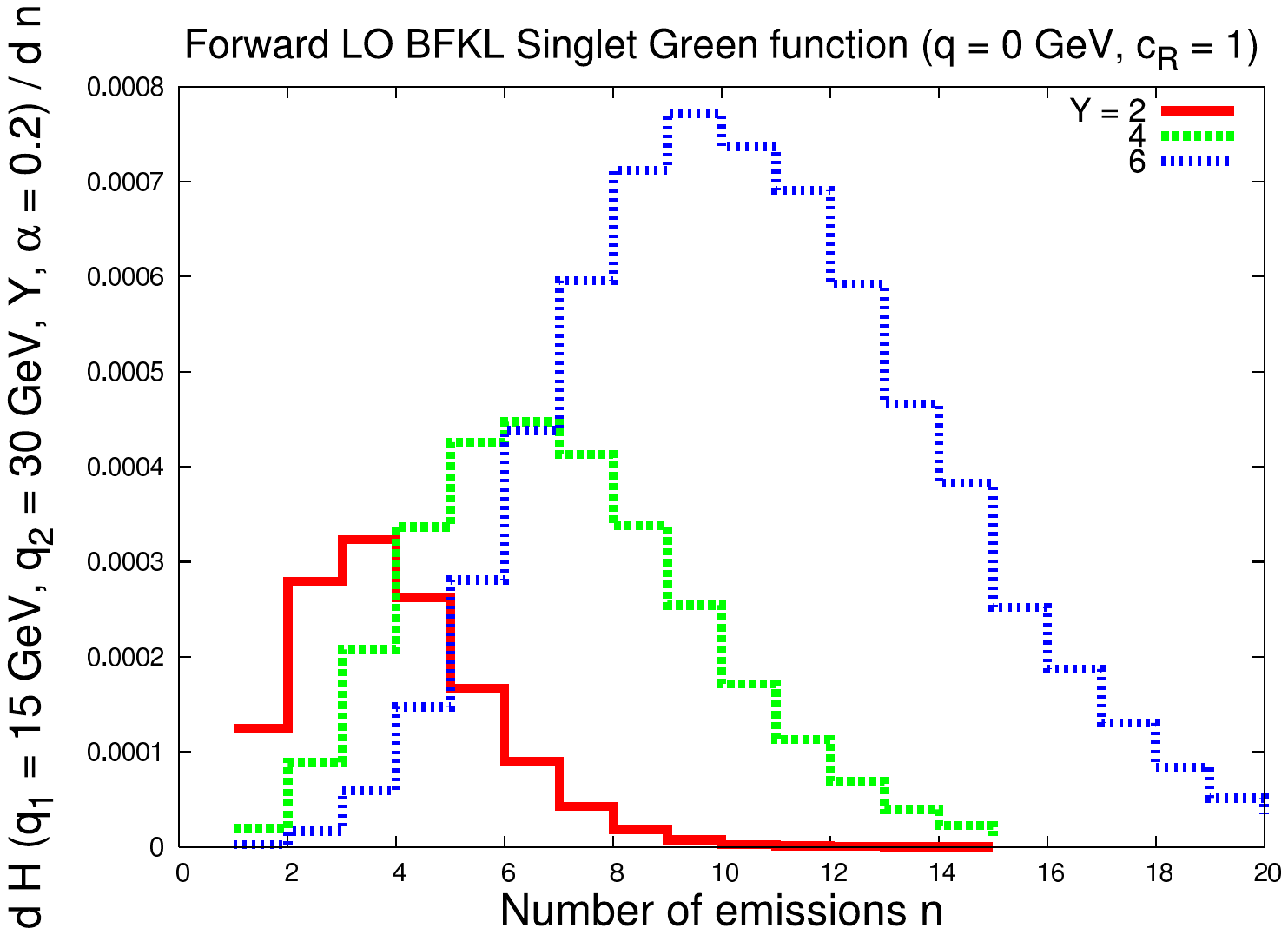}
\includegraphics[width=7.5cm,angle=0]{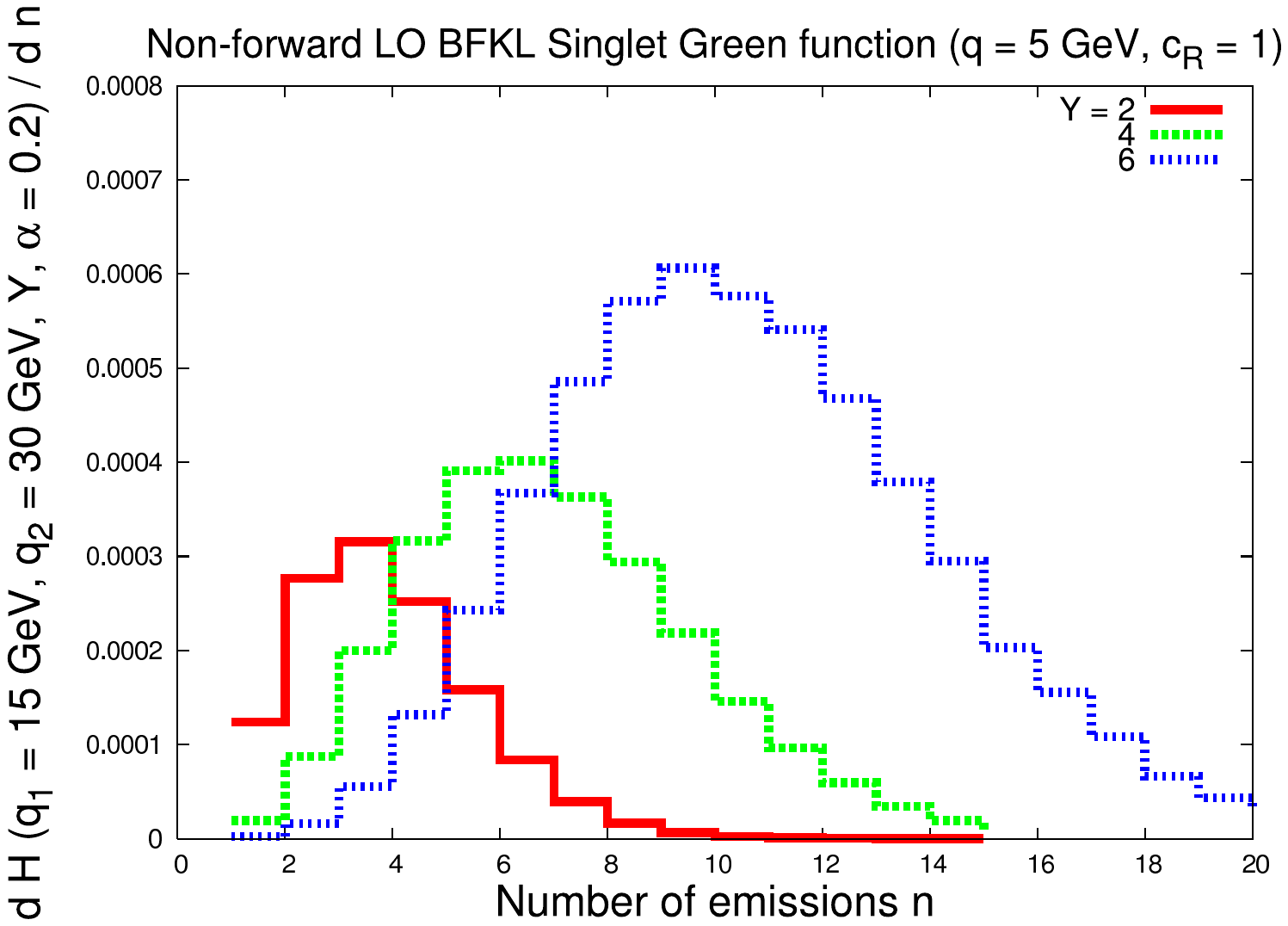}
    \caption{Distribution in the contributions to the BFKL gluon Green function (left is the forward and right the non-forward case)
    with a fixed number of iterations of the kernel, plotted for different values of $Y$ and ${\bar \alpha}_s = 0.2$. }
  \label{NumberofRungs}
\end{center}
\end{figure}

When using Monte Carlo methods it is possible to access very exclusive information not only related to multiple particle production, but also of the internal structure of the ``pomeron ladder", which is relevant in diffractive events. As an example we show Fig.~\ref{NumberofRungs}, from Ref.~\cite{Chachamis:2011nz}, where we can see the distribution in the number of minijets in the case of a total cross section (left) and the distribution in the number of rungs present in the pomeron exchange in a high energy elastic scattering with nonzero momentum transfer (right). The former includes final state radiation with any $p_T$, but a similar analysis can be performed for different bins in $p_T$ at LL and NLL. These Poissonian shapes are characteristic of BFKL radiation. 
From these few examples it should be clear that the use of a BFKL-based Monte Carlo,
in particular BFKLex described in \cite{Chachamis:2013rca}  allows to address 
many more comparisons of exclusive final states with theoretical predictions.

Let us also mention a specific final state which allows to address BFKL physics: the jet-gap-jet configuration 
in which the jet plus proton remnants are attached to a color singlet exchange ~\cite{Hentschinski:2014lma,Hentschinski:2014bra,Hentschinski:2014esa} without any activity inside the gap.    
As to the calculation of impact factors, the authors of   
~\cite{Chachamis:2013hma,Chachamis:2013qca,Chachamis:2013oga} have developed a procedure based on Lipatov's high energy effective action which is very powerful and useful in this regard. In ~\cite{Hentschinski:2011tz,Chachamis:2012gh,Chachamis:2012mw,Chachamis:2012cc} the
NLO forward jet vertex coupled to an octet has been reproduced which is related to forward jet production with associated mini jet radiation, and  more recently  ~\cite{Hentschinski:2014lma,Hentschinski:2014bra,Hentschinski:2014esa} the impact factor through which  the jet plus proton remnants are attached to a color singlet exchange. Further processes at full NLO, like forward production of electroweak bosons, will also require the calculation of NLO impact factors, a problem which can be suitably addressed with the high energy effective action approach. Turning to the color singlet exchange, because of the hard scales of the jets the color singlet exchange  should be described by the exchange of a BFKL Pomeron: this makes this process 
an interesting candidate for BFKL searches.  However, for these diffractive events we have the "traditional" problem of the gap survival probability which cannot be calculated yet in a reliable manner.
As a step forward towards circumventing this problem, one should investigate exclusive distributions related to ratios of cross sections which somehow remove this gap factor and the influence of parton distribution functions.

Finally, it may also be useful to think about 'BFKL signals' in a quite different way. Most of the searches 
discussed so far look for signals which result from the inclusive radiation  of a large (in most cases even 'infinite') number of gluons. Modern calculation techniques now allow for the computation of 
fixed order matrix elements of hard subprocesses with many partons (gluons) in the final state.
At large rapidities, these matrix elements generate some of those logarithms which the BFKL approach sums to all orders (in LO or NLO accuracy). From this point of view BFKL can be seen as the most efficient tool 
for performing an all-order summation. A search for BFKL dynamics, therefore, could start from a  close 
inspection of fixed-order calculations and a comparison with the corresponding BFKL predictions.

\subsection{Previous measurements and experimental aspects} 

After this theoretical part let us now turn to experimental aspects. Some of the measurements discussed 
before have already been started at the Tevatron and in the previous LHC run. It may therefore be useful to briefly review these measurements and then say a few words about RUNII expectations and future experimental techniques.  

\subsubsection{Measurements from D0}

Early measurements of forward-backward jet production in proton-(anti)proton collisions were performed by the D0 experiment. In\cite{Abachi:1996et}, published in year 1996, azimuthal decorrelations between jets with large rapidity separation are presented. Jets above 20 GeV with $\vert \eta \vert < 3.0$ with largest rapidity separation in the event were selected from the data sample of 83 nb$^{-1}$ collected at collision energy of 1800 GeV. If one of these jets was above 50 GeV the pair was selected for the analysis. An average cosine of $\Delta \phi$ between jets was measured as a function of rapidity separation, $\Delta y$, in the range $\Delta y < 5$. The measurements were compared with the \textsc{herwig} Monte Carlo (MC) generator\cite{Hw6} based on leading order matrix elements supplemented with parton shower, a NLO calculation based on \textsc{jetrad} \cite{Giele:1993dj, Giele:1994gf} as well as LL BFKL calculations performed in \cite{DelDuca:1994ng, DelDuca:1993mn}. \textsc{herwig} has demonstrated the best agreement with the data. The LL BFKL  calculation was above the measurements  while the NLO calculations were below. 

In \cite{Abbott:1999ai} the ratio of dijet production at different energies was measured. Data samples of $0.7$ nb$^{-1}$ and 31.8 nb$^{-1}$ obtained at collision energies of 630 GeV and 1800 GeV were used. The ratio was originally thought to demonstrate the strong growth of partonic cross-section with collision energy as predicted by BFKL.  The measurements were compared to predictions of the \textsc{herwig} MC generator, fixed order calculation in LO pQCD and LL BFKL calculation \cite{Orr:1998hc}. The measurement showed considerably larger growth of cross-section with energy than all theory predictions.   For a detailed discussion on the interpretation of this measurement see \cite{Andersen:2001kta}.

\subsubsection{Forward backward jet production -  measurements from CMS}

CMS has performed two measurements of jets with large rapidity separation based on LHC data taken in 2010.

\subsubsection{\bf Inclusive to exclusive dijet production ratio}

In this subsection the measurement \cite{Chatrchyan:2012pb} performed in collisions at $\sqrt{s} = 7$ TeV taken in 2010 is discussed.
The measurement was performed with the effective luminosity of 5 pb$^{-1}$. 
The ratio of inclusive to exclusive dijet production was proposed in \cite{Kim:1995zu} as an observable sensitive to higher order QCD radiation and a potential manifestation of BFKL effects. 
Jets with transverse momenta above $p_{\mathrm{T,min}} = 35$ GeV are considered. Events with at least one pair of jets  are denoted as ``inclusive". Events with exactly one pair of jets within the acceptance are called ``exclusive".
In the inclusive case, the rapidity separation is evaluated for each pairwise combination of jets above threshold. 
The ratio of inclusive to exclusive dijet production is measured in bins of rapidity separation $\Delta y$, covering a range up  to $\Delta y = 9.2$. 

\begin{figure}[hbtp]
  \begin{center}
    \includegraphics[width=0.35\textwidth]{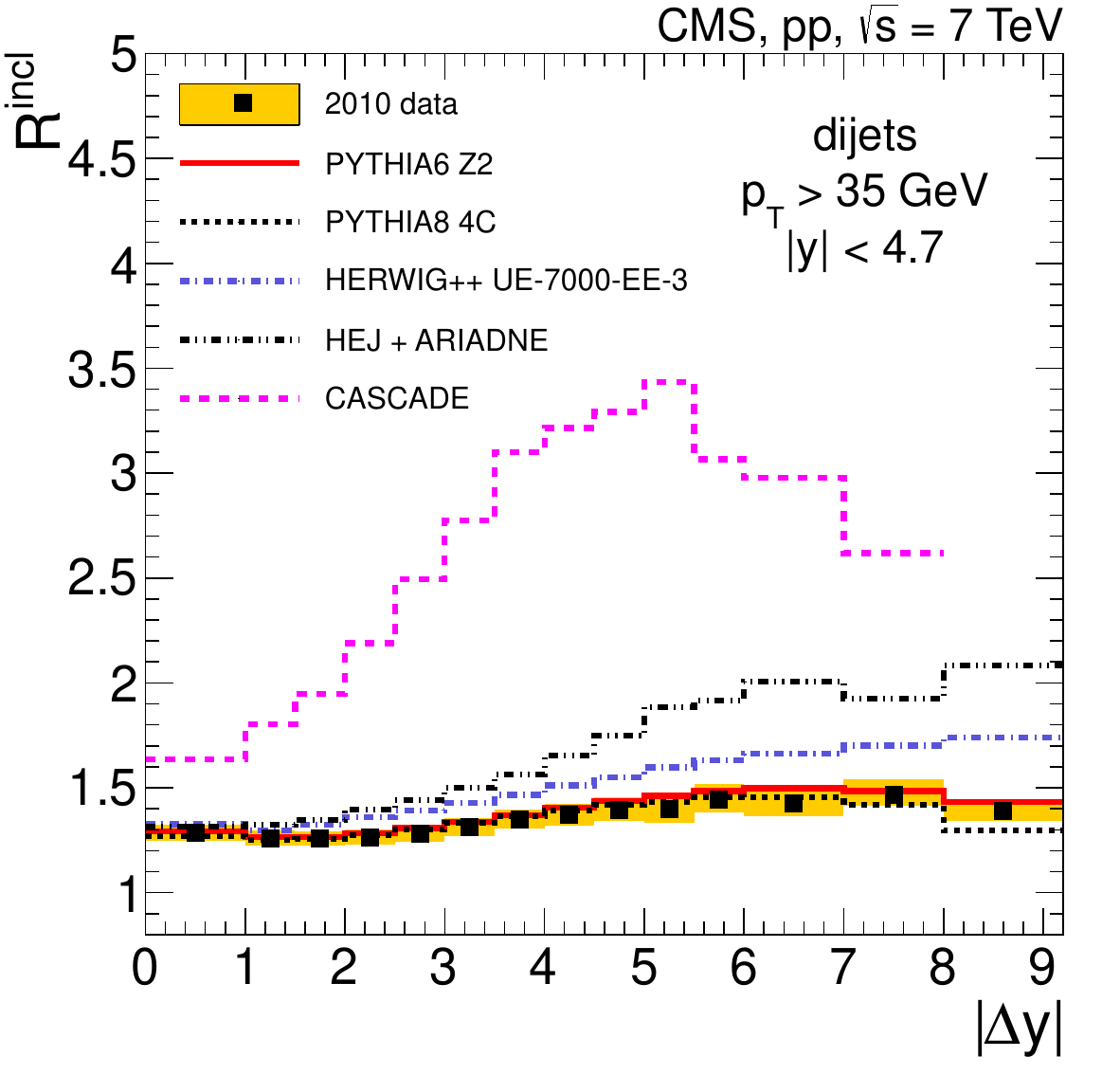}     \includegraphics[width=0.36\textwidth]{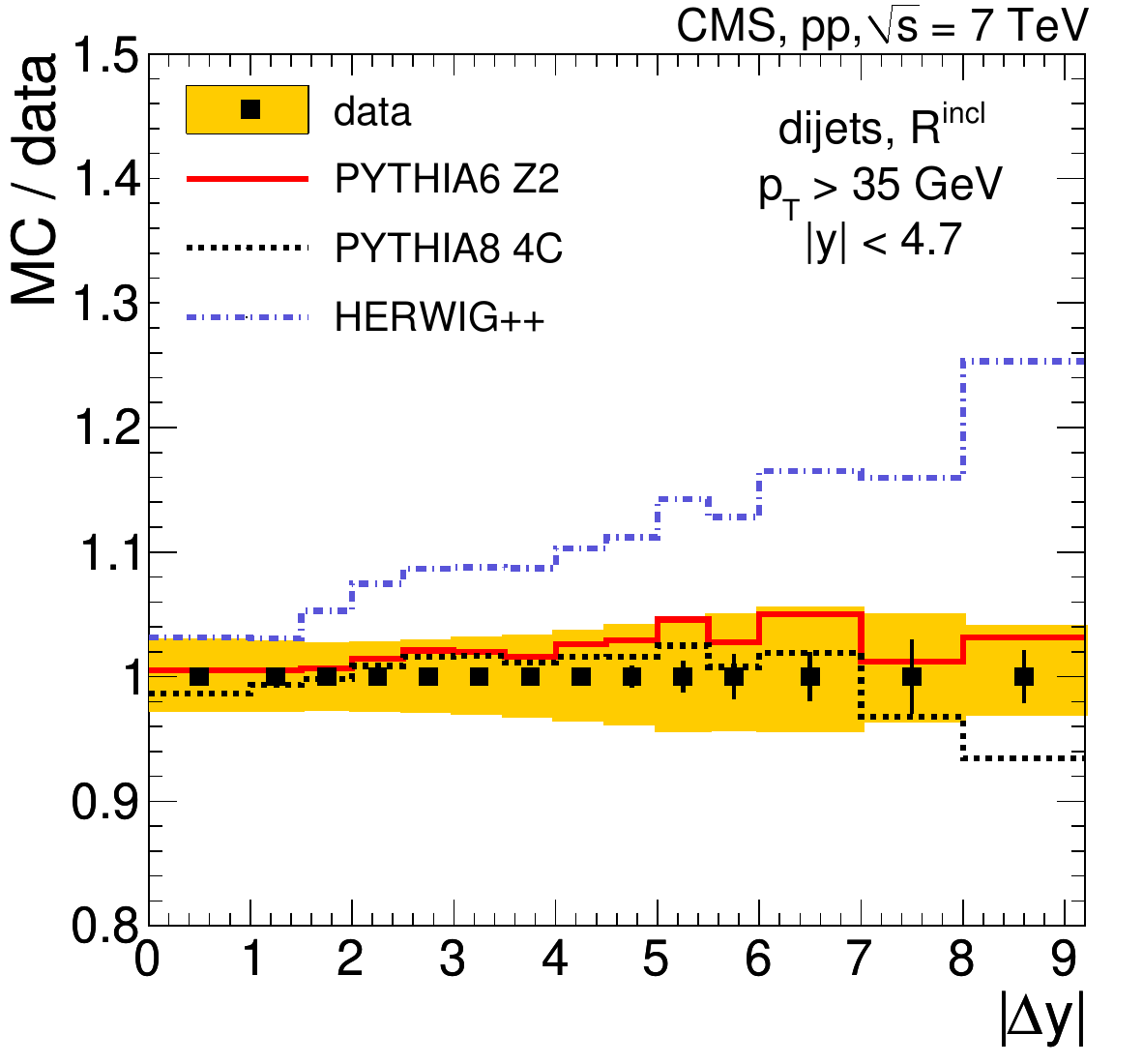}     \caption{Inclusive to exclusive dijet production ratio compared to predictions of DGLAP-based MC \textsc{pythia}6, \textsc{pythia}8, \textsc{herwig}++ and MC with elements of LL BFKL - \textsc{hej} and \textsc{cascade} (left). Comparison to DGLAP-inspired MCs is presented as data/MC ratio on the right. }
    \label{fig:kf}
  \end{center}
\end{figure}

The measurement of a ratio of cross sections  allows to reduce many theoretical and experimental uncertainties with 
the jet energy scale calibration and model dependence of unfolding procedure being the most important systematic uncertainties. Total experimental uncertainty goes up to 5\%.

The results of the measurement are presented at Fig.~\ref{fig:kf}. The measurement is compared to predictions of DGLAP-based Monte Carlo generators \textsc{pythia}6 \cite{Pyt}, \textsc{pythia}8 \cite{Pyt8} and \textsc{herwig++} \cite{Hw++} and to generators which include BFKL effects:  \textsc{hej} \cite{HEJ} and \textsc{cascade} \cite{Jung:2010si}. The  \textsc{hej} MC generator accounts for wide-angle gluon emission to all orders of perturbation theory using LL BFKL formalism. \textsc{cascade} uses the CCFM evolution equation for initial state parton cascade. 
\textsc{pythia}8 and \textsc{pythia}6 predictions agree with the data within the experimental uncertainties. \textsc{herwig}++ shows stronger rise of the ratio with the rapidity separation increase. \textsc{hej} and \textsc{cascade} predict a too strong rise. 
However, \textsc{cascade} uses only gluon induced processes and thus the exclusive cross section at large rapidity separations  becomes very small, leading to and increased ratio. This effect has been confirmed~\cite{Daniela2014} by a simulation using gluon induced processes in \textsc{pythia}6.

\subsubsection{\bf Azimuthal de-correlation of forward backward jets}

The measurement \cite{azimu} of azimuthal de-correlation for jets separated by a  large rapidity interval was performed using  5 pb$^{-1}$ of pp collisions at $\sqrt{s} = 7$ TeV in 2010.

Perturbative QCD at leading order predicts exactly two outgoing jets in a parton-parton interaction, which are back-to-back in azimuthal angle $\phi$. In higher order calculations the jets can become de-correlated. 
In calculations based on BFKL, a significant de-correlation in azimuthal angle is predicted, however also higher order matrix element calculations  and calculations supplemented with parton showers and multi-parton interactions predict significant de-correlation effects.
Thus the study of azimuthal de-correlation of jets as a function of rapidity separation may give further insights into the production mechanism.  

\begin{figure}[hbtp]
  \begin{center}
    \includegraphics[width=0.4\textwidth]{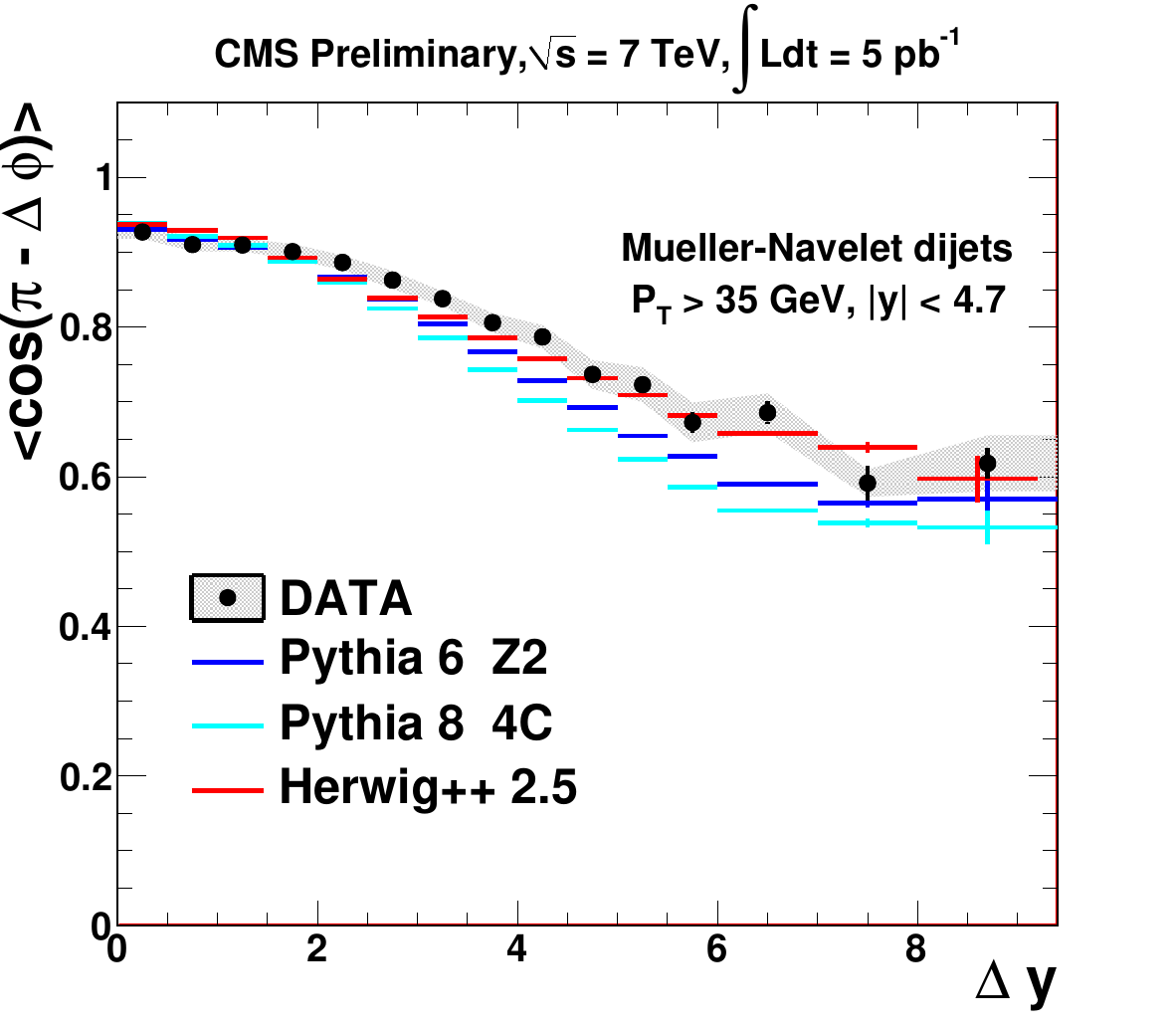}     \includegraphics[width=0.4\textwidth]{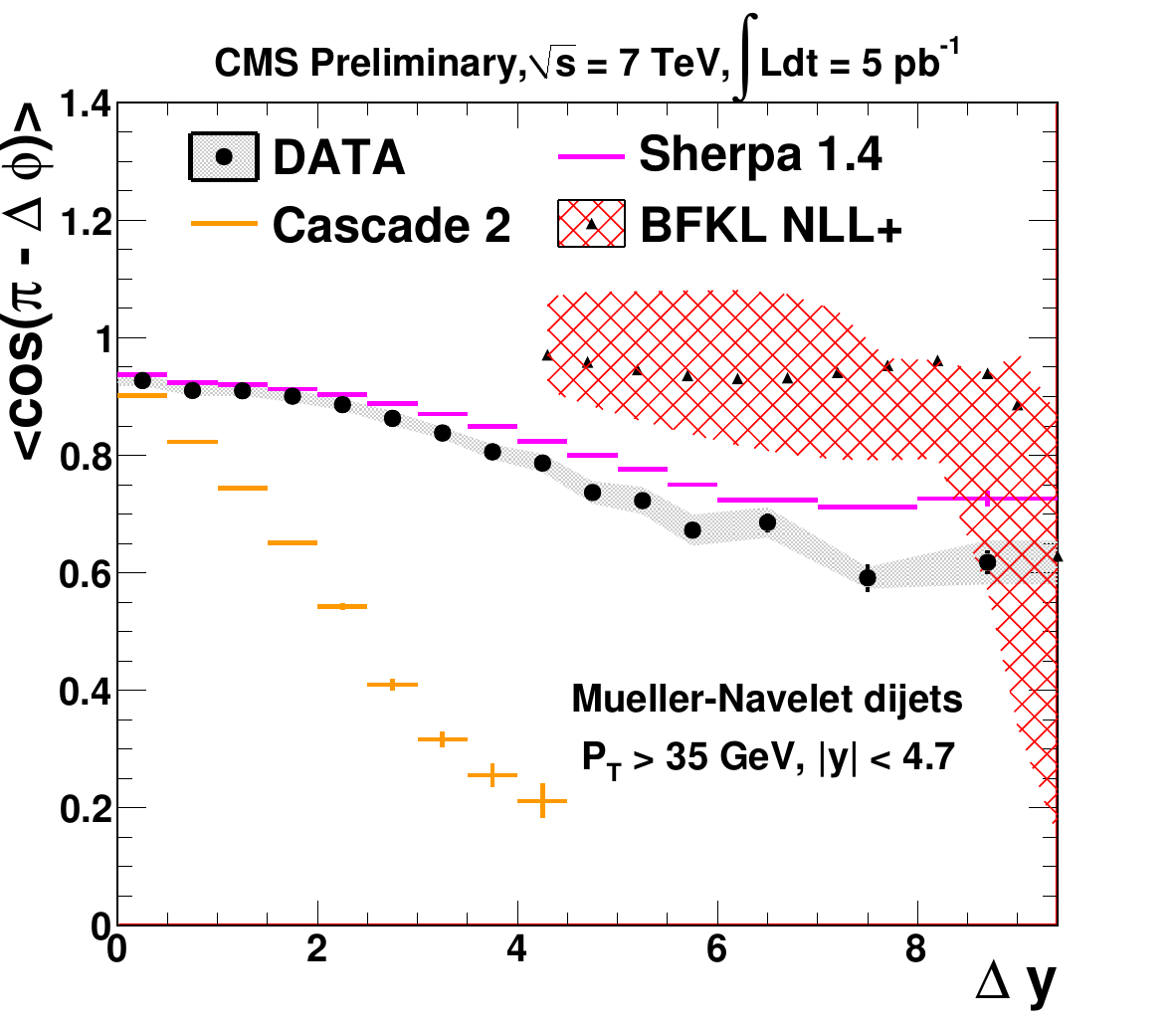}     \caption{$C_1$ as a function of $\Delta y$ compared to various theory predictions. }
    \label{fig:cos}
  \end{center}
\end{figure}

The dijet production cross section as a function of azimuthal angle difference can be written as a Fourier series:
\begin{equation}
\frac{1}{\sigma} \frac{d \sigma}{d (\Delta \phi) } (\Delta y, p_{\mathrm{T min}})   = 
\frac{1}{2 \pi} \biggl[ 1 + 2 \sum_{n=1}^{\infty}  C_n (\Delta y, p_{\mathrm{T min}}) \cdot \, cos(n (\pi - \Delta \phi)) \biggr]. 
\end{equation}

The Fourier coefficients $C_n (\Delta y, p_{\mathrm{T min}})$  are equal to the average cosines of the de-correlation angle: 
$C_n (\Delta y, p_{\mathrm{T min}}) = \langle cos(n ( \pi - \Delta \phi))\rangle$, where $ \Delta \phi = \phi_1 - \phi_2 $ 
is the difference between the azimuthal angles $\phi_1$ and $\phi_2$ of the jets most forward and backward in rapidity.

In  \cite{Vera:2007kn} the ratio $C_2/C_1$ and $C_3/C_2$ was proposed as observables, which are particularly sensitive to higher order corrections and to BFKL effects.
CMS has measured both $C_n$ and the cosine ratios. The average cosines of the azimuthal angle and their ratios were measured in bins of the rapidity separation between the jets, $\Delta y$. Jets with $p_{\mathrm{T}} > $35 GeV and $\vert y \vert < 4.7$ covering a rapidity separation of $\Delta y < 9.4$ were investigated. The measurements are compared to theoretical predictions in Fig.~\ref{fig:cos}. The measurement of the ratio $C_2/C_1$ is shown in Fig.~\ref{fig:crat}. The measurements are corrected for detector effects. The leading source of the experimental uncertainty is the jet energy scale of up to 24\% depending on the observable and  the rapidity separation range. The total experimental uncertainty, however,  does not exceed 25\%.

The next-to-leading logarithmic approximation (NLL) BFKL calculations \cite{Ducloue:2013hia} provides a good description of $C_2/C_1$. It should be noted that improved NLL BFKL calculations \cite{Ducloue:2013bva,Caporale:2014gpa} were released later than the present measurement was published, in mentioned work comparison with the CMS data can also be found. 
However, also calculations supplemented with parton showers and multiparton interactions, \textsc{pythia6}, \textsc{herwig++}, provide a reasonable good description of the measurement over the full range in  $\Delta y$. \textsc{cascade} predicts too a large de-correlation.

\subsubsection{Measurements from ATLAS}
 ATLAS detector  \cite{Aad:2008zzm} calorimeter system covers pseudorapidity range $\vert \eta \vert < 4.9$ allowing jet reconstruction up to $ \vert y \vert < 4.4$. Tracking system extends to $\vert \eta \vert < 2.5$. Fine-segmented calibrated calorimeter energy deposits are combined in dedicated manner and clustered with anti-kT algorithm with distance parameter $R=0.6$ (for measurements presented here). Jets are calibrated using various in-situ techniques \cite{Aad:2014bia, Aad:2012ag}. 

In \cite{Aad:2011jz} the measurement of dijet as a function of rapidity separation for different requirements on additional jets is described,
either considering the two leading $p_{\mathrm{T}}$ jets or the most-forward and most-backward jets using sets above $Q_{0} = 20$ GeV. Two different cuts (veto) were applied on the  additional jets between  the jets defining the dijet system. In the first scenario, no additional jets above $Q_{0}$ are allowed, while in the second scenario no additional jets above $\overline{p_{\mathrm{T}}}$ were allowed. Measurements were compared to the predictions from NLO MC \textsc{powheg} \cite{Nason:2004rx} interfaced to parton shower with \textsc{pythia6} or \textsc{herwig++} and MC generator \textsc{hej}. Predictions for \textsc{hej} were obtained at parton level. The rapidity separation range covered by the measurement extends to $\Delta y = 6$.
It was observed that for the veto $Q_{0} = 20$ all predictions give smaller cross section than observed 
(Fig.~7~in~\cite{Aad:2011jz}), while for veto scenario $Q_{0} = \overline{p_{\mathrm{T}}}$ \textsc{powheg}-based predictions show a good agreement with the data while \textsc{hej} overestimated the measurement (Fig.~8~in~\cite{Aad:2011jz}). This is in qualitative agreement with the CMS dijet production ratio measurement (see Fig \ref{fig:kf}).

\begin{figure}[hbtp]
  \begin{center}
    \includegraphics[width=0.4\textwidth]{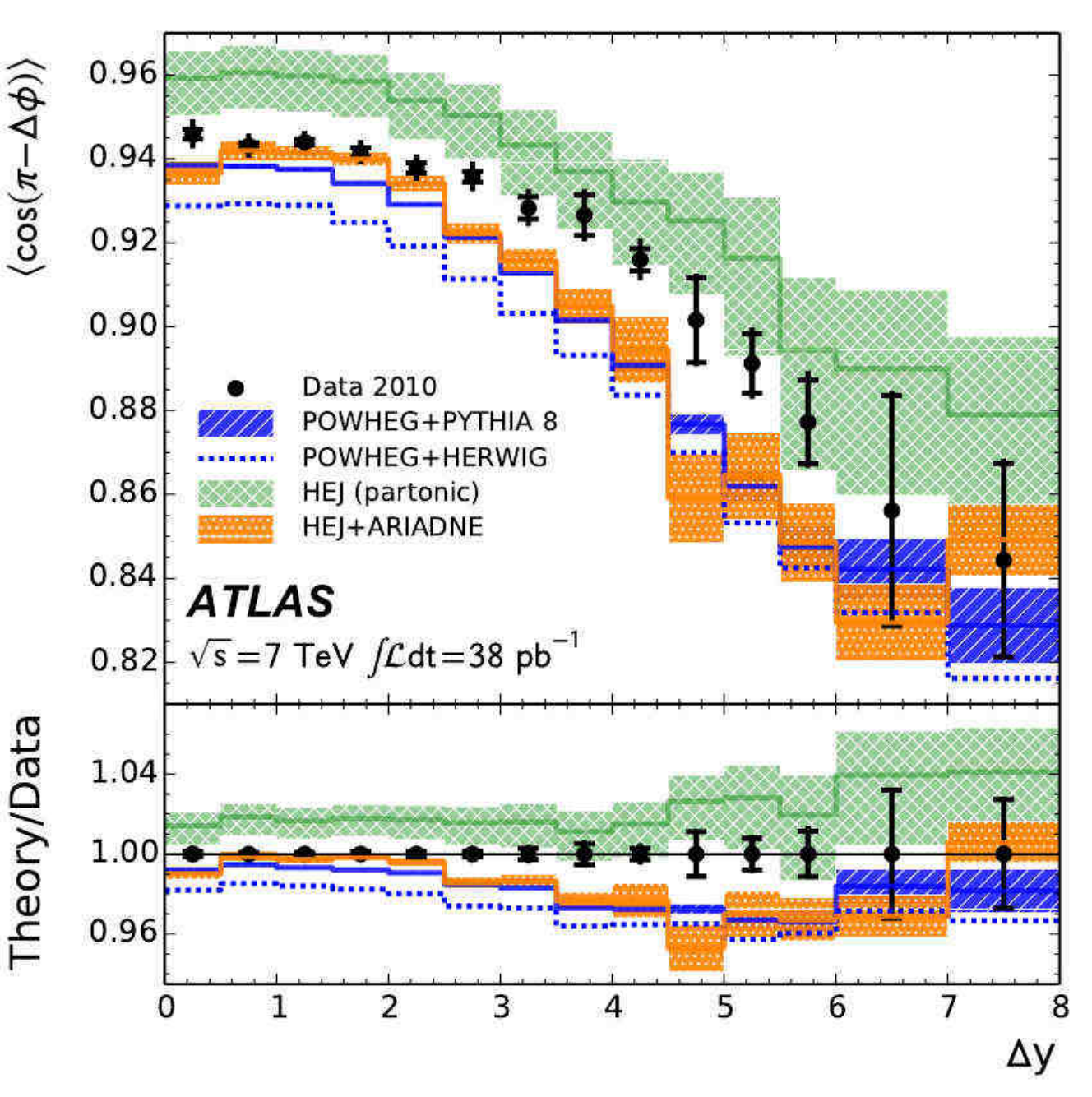}     \includegraphics[width=0.4\textwidth]{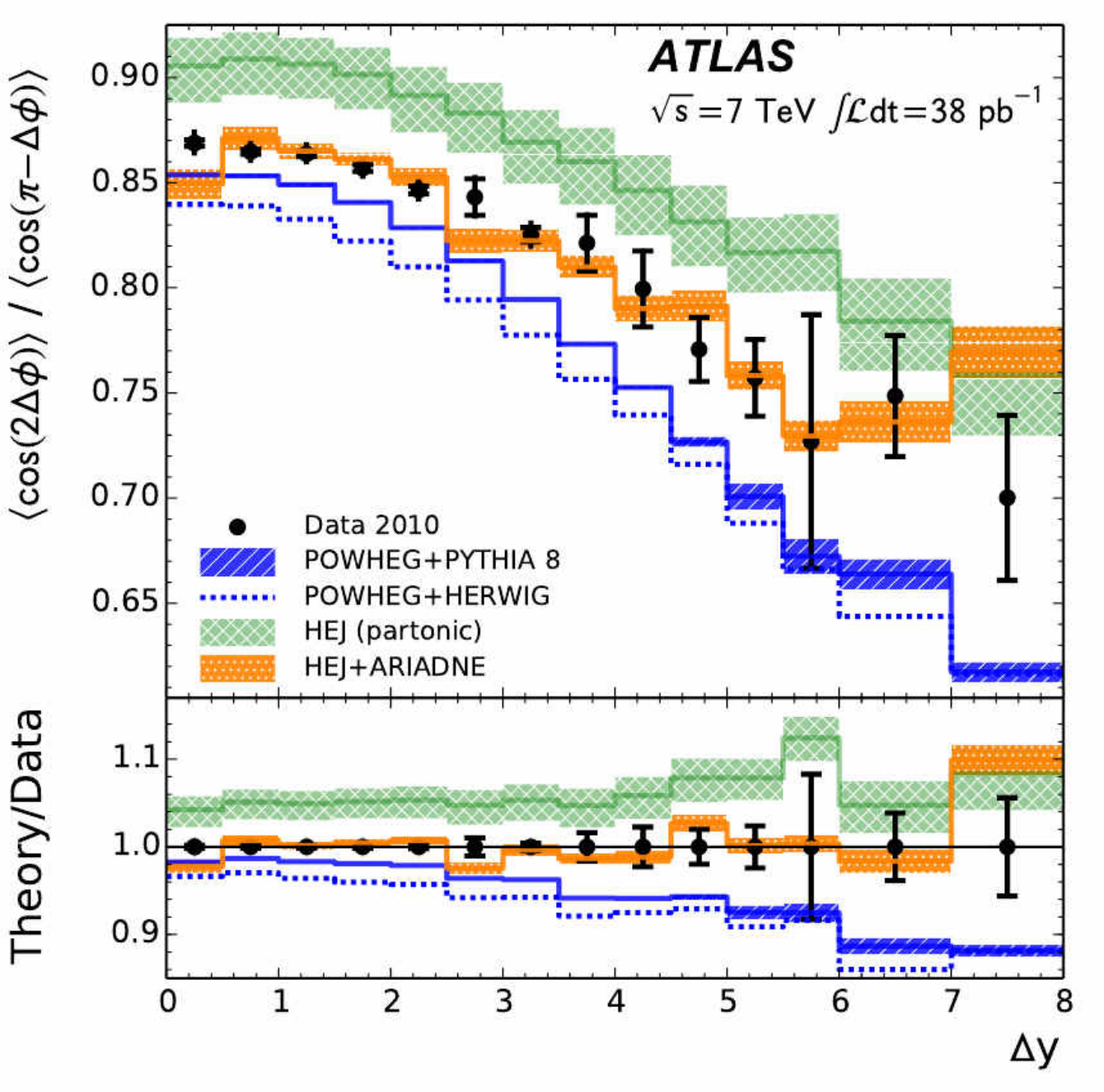}      \caption{ATLAS results on azimuthal decorrelation measurement, plots are taken from Fig. 5 in \cite{Aad:2014pua}. On the left average cosine as a function of rapidity separation is shown, on the right ratio $C_2/C_1$ is presented. }
    \label{fig:atlasAz}
  \end{center}
\end{figure}

The azimuthal de-correlation between jets as a function of rapidity separation has been measured\cite{Aad:2014pua} for jets above $Q_{0}=20 (30)$ GeV for data taken in 2010 (2011). The two leading jets were selected, and events were rejected if $p_{\mathrm{T}}$ of leading jet is below 60 or $p_{\mathrm{T}}$ of subleading jet is below 50 GeV. Events were also rejected if jets with transverse momenta of 20 (30) GeV were found between the leading jets. Jets were considered in the acceptance region of $\vert \eta \vert < 4.4$ for 2010 data. For data taken in 2011 the jets were restricted to the tracker acceptance $\vert \eta \vert < 2.4$ as this allows to identify jets belonging to the same interaction vertex in high-pileup environment. 

The angular moments $C_{\mathrm{1}}$ and $C_{\mathrm{2}}$ as well as the ratio $C_{\mathrm{2}}/C_{\mathrm{1}}$ were measured. Below we will focus on measurement of $C_{\mathrm{1}}$ and $C_{\mathrm{2}}/C_{\mathrm{1}}$ as functions of rapidity separation.
The data are compared to predictions of NLO MC \textsc{powheg} supplemented with parton showers generated by \textsc{pythia} or \textsc{herwig++} and to \textsc{hej}.  The
measurements of the azimuthal de-correlations are shown at Fig.~\ref{fig:atlasAz}. The measurements extend to $\Delta y = 8$. The value of the average cosine obtained by ATLAS is bigger than the CMS result (~0.8 for ATLAS and ~0.6 for CMS in $7 < \Delta y < 8$ bin) which can be attributed to $p_{\mathrm{T}}$ ordering of jets and higher $p_{\mathrm{T}}$ of jets in case of ATLAS. \textsc{powheg} interfaced to \textsc{pythia} as well as \textsc{herwig} overestimate the de-correlation. \textsc{hej+ariadne} package also overestimates the de-correlations. The ratio $C_{\mathrm{2}}/C_{\mathrm{1}}$ is well-described by \textsc{hej+ariadne} package while \textsc{powheg} + parton shower underestimates the ratio.

To summarize, measurements of forward-backward dijet production performed by different experiments do not allow to make clear conclusions concerning observation of BFKL signal. Although a good description of the azimuthal de-correlation by the NLL BFKL calculations is obtained, the DGLAP MC parton shower calculations give a surprisingly good description of the measurement over the full acceptance region.

\subsection{RunII expectations}

BFKL resummation is performed for ``pQCD high energy limit'' asymptotic region or when the collision energy is much larger than the momentum transfer: $s \gg Q^{\mathrm{2}}$. Obviously the validity of a given relation will increase for jets measured at the same $p_{\mathrm{T}}$'s and increased collision energy (we imply here transition from 7 to 13 TeV). 
%Thus one may expect BFKL-like signatures in hard parton radiation. An amount of data equal to 100 pb$^{-1}$ is needed to perform careful analysis of dijet production at large rapidity intervals which is comparable to what was available in 2010. 
For further investigations, a dataset of $\sim$100 pb$^{-1}$ is needed to perform careful measurements of dijet production at large rapidity intervals at pp collision energy of 13 TeV. An average pileup of 1 can be tolerated by analysis techniques used in LHC RunI.

%{\bf Notes on experimental techniques}

The ability to measure jets with low $p_{\mathrm{T}}$ is important for approaching the high energy limit and revealing BFKL contributions into production of jets. There are two obstacles making this task rather challenging. The first one is the jet trigger efficiency. For example in CMS experiment for 2010 RunI data 99\% efficiency was reached at around $p_{\mathrm{T}}=35$ GeV. Another obstacle is the jet energy scale calibration. Due to nonlinear effects in the jet composition and calorimeter response, the jet energy resolution worsens for lower $p_{\mathrm{T}}$ which leads to larger uncertainties on the jet energy scale calibration. In measurements described here the energy scale was valid for jets above $p_{\mathrm{T}}$ of 20 GeV which can be considered as a realistic offline threshold for imposing vetoes or for Mueller-Navelet jet selection. Though it should be noted that the particle flow technique used for jet reconstruction allows to reconstruct jets starting from 10 GeV \cite{JME-10-011pub}. This technique uses all detector components to form particle candidates which are clustered into jets and that allows to improve jet energy resolution significantly. 
 
For the measurement of Mueller-Navelet jets, a large rapidity acceptance of the detector is required. CMS calorimeter system covers the pseudorapidity range up to 5 allowing jet reconstruction within $\vert \eta \vert < 4.7$ range. In presence of additional interaction in the beam-crossing (pile-up) it is important to identify the pair with largest rapidity separation belonging to the same hard scattering. This may be possible in case the acceptance region is instrumented with a tracking detector. This is not the case for the forward region of CMS. The tracking system of CMS covers a pseudorapidity range of $\vert \eta \vert < 2.4$. Thus measurements of Mueller-Navelet jets across the full detector acceptance requires low-pileup collision events. The same is applicable to ATLAS detector where the calorimeter system allows to measure jets up to 4.4 in pseudorapidity while the tracking system extends to $\eta=2.5$. Combinatorial background from pileup also saturates the bandwidth allocated to dijet triggers widely used for jet energy scale calibration in forward region or for selecting Mueller-Navelet jets. The pile-up scenario for data-taking should not have an average pileup of $\langle \mu \rangle > 1$. Note that both CMS and ATLAS analyses applied requirement of exactly one primary vertex per event to reduce or remove the combinatorial background in 2010 data. This requirement leaves $\sim 37$\% of events in case of mentioned scenario.

\subsection{Summary}

It is clear that we are entering a very exciting period where both rich data and new powerful calculations  will be available allowing us to pin down in a precise fashion what is the underlying dynamics governing the high energy limit of QCD. Some of the BFKL searches presented  in this section have been discussed earlier and experimentally investigated before  (HERA, Tevatron, low energy runs of the LHC). However, we have also listed new questions and proposed novel measurements which have not been addressed yet. In combination with the unprecedented high energy, this defines a really promising program for the coming years.      

This program is important since there are uncertainties in the BFKL approach itself which need to be fixed. As an example, it is needed to find the dependence of each proposed observable on the renormalization schemes (the above mentioned conformal ratios were shown to be independent of these choices), but also the correct treatment of the running of the coupling must be addressed in an accurate way. Only a fair comparison to experimental data can solve many of these theoretical questions. 
Once we control a class of observables where only BFKL can fit them then we can introduce corrections to the original calculations in the form of hadronization, non-linearities, collinear radiation or even the possible connection with soft-collinear effective theories, in order to extend their applicability beyond the multi-Regge kinematics. Particularly interesting ``deformations" of the original theory are those where non-perturbative effects (of confinement type, non related to high parton densities) are included~\cite{Kowalski:2010ue,Kowalski:2012ur,Kowalski:2014iqa}. But all of these can  be studied only after we have a clear idea of the phenomenological window of applicability of the perturbative linear BFKL program.

Experimentally, with the large $\sqrt{s}$ reachable in run2, a study of the
energy dependence of Mueller-Nevelet jets can be performed, by comparison
with measurements from run1. Important are also dedicated measurements,
where the transverse momenta of the forward and backward jets are in a $p_T$
window as advocated in the introduction.
Besides dedicated searches for BFKL effects, multi-jet measurements over the largest
rapidity range are essential, as those measurements might not be well described by
fixed higher order calculations (even supplemented with parton showers) and they could
show the need for small $x$ resummation to all orders.
Experimentally challenging will be the high pileup scenario in run2 and dedicated methods
for pileup identification and subtraction, in a region where there is no tracking, are desperately needed.

\section{Inclusive forward di-jet production in $pp$
}

Let us now turn to our next topic, inclusive forward jets (Fig~\ref{fig:fwdjet} b, left).
Questions of highest interest include unintegrated gluon densities, saturation, and multiple 
interactions.  

Unintegrated Gluon Densities (UGDs) are crucial ingredients of $k_T$-factorization-based approaches which once convoluted with off shell matrix elements allow to provide predictions for observables at low $x$. UGDs are in their nature more exclusive than collinear parton densities, since they depend not only on longitudinal degrees of freedom, but also on 
a transversal momentum of a gluon. However, the properties of UGD are still unexplored in wide kinematical regime \cite{Hautmann:2013tba}. A natural tool to access UGDs at low $x$ are various configurations of forward jets.
In particular, 
configurations with dijet or trijet system in a forward rapidity region and forward-central jet 
configurations are of great interest. 
They offer a possibility to perform a scan of UGDs in a large kinematical domain, in particular to access
a kinematic region where the saturation phenomenon 
\cite{Gribov:1984tu,Albacete:2010pg}
 eventually emerges.

One of the frameworks which allows to access the observables sensitive to UGDs is so-called High Energy Factorization (HEF)
\cite{Catani:1990eg} (see also \cite{Gribov:1984tu}, \cite{Collins:1991ty}). It relies on off-shell gauge invariant matrix elements 
and UGDs convolved together in longitudinal and transversal degrees of freedom. For the
configurations of final states populating (at least partially) the forward rapidity region, the following
HEF formula may be used
\begin{equation}
d\sigma_{AB\rightarrow X}=\int \frac{d^{2}k_{TA}}{\pi}\int\frac{dx_{A}}{x_{A}}\,\int dx_{B}\,
\sum_{b}\mathcal{F}_{g^{*}/A}\left(x_{A},k_{TA}\right)\, f_{b/B}\left(x_{B}\right)\,
 d\hat{\sigma}_{g^{*}b\rightarrow X}\left(x_{A},x_{B},k_{TA}\right),\label{eq:HENfact_2}
\end{equation}
where $\mathcal{F}_{g^{*}/A}$ is the UGD, $f_{b/B}$ are the collinear parton distribution functions and $b$ runs over gluon and all the quarks that can contribute
to the production of multiparticle state $X$ (see also \cite{Deak:2009xt}). The off-shell gauge invariant matrix elements for multiple final states 
(residing in $d\hat{\sigma}_{g^{*}b\rightarrow X}$) can be calculated
along the lines of Refs. \cite{vanHameren:2012uj,vanHameren:2012if}. The restriction that the multiparticle state $X$ populates the forward rapidity
region follows from the fact, that Eq. (\ref{eq:HENfact_2}) is valid when $x_{B}\gg x_{A}$, i.e. when the events are highly asymmetric.

Recently, some new forward jet observables were calculated within this framework for existing experimental setup
\cite{Kutak:2012rf,vanHameren:2013fla,vanHameren:2014lna}
and some UGDs relevant for large $p_t$'s \cite{Kutak:2004ym} accounting for saturation phenomenon in both proton and lead. 
Below we briefly discuss the relevant observables and outline the main observations. We also give new predictions for a possible extension of CASTOR
detector allowing for jet reconstruction.

\subsection{Dijet production at forward and very-forward rapidities}

Forward dijets offer practical
advantages over forward-central dijets since the nearness in rapidity lowers the phase space for an emission of further jets. Moreover, the events are highly asymmetric as 
required in view of HEF formula (\ref{eq:HENfact_2}). Depending on the cuts applied we probe $x_A$ between $10^{-4}$ and $10^{-6}$ and $x_B$ around unity.

An observable of particular interest in p+p and p+Pb collisions are angular decorrelations, i.e. the cross section as a function of the azimuthal angle between the two jets. Such a cross section reflects a pattern of gluon radiation as summed by the evolution equations. For instance when the jets are nearly back-to-back the transverse momentum of the incoming off-shell gluon is small and possibly affected by the saturation. In the other limit when the gluons are observed in a similar direction in the transversal plane the gluon density is probed at the large momentum and is subject to large sub-leading effects of higher orders. Those corrections come from non-singular pieces of DGLAP splitting function at low $x$  and from energy conservation.

\begin{figure}[t]
  \begin{center}
    \includegraphics[width=0.3\textwidth]{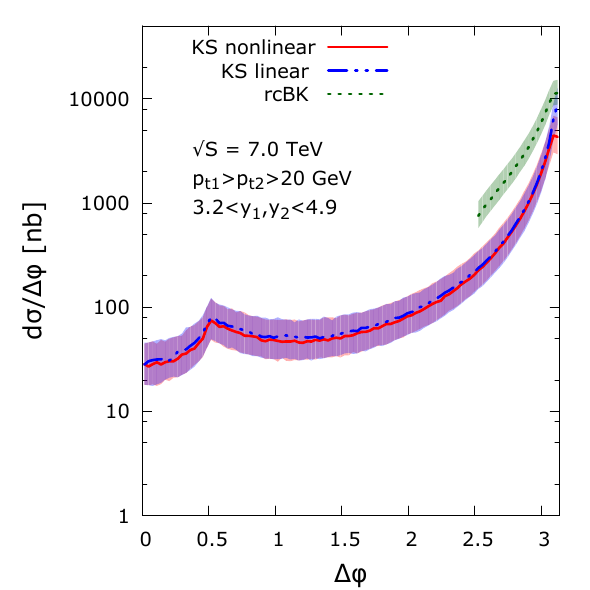}
    \includegraphics[width=0.3\textwidth]{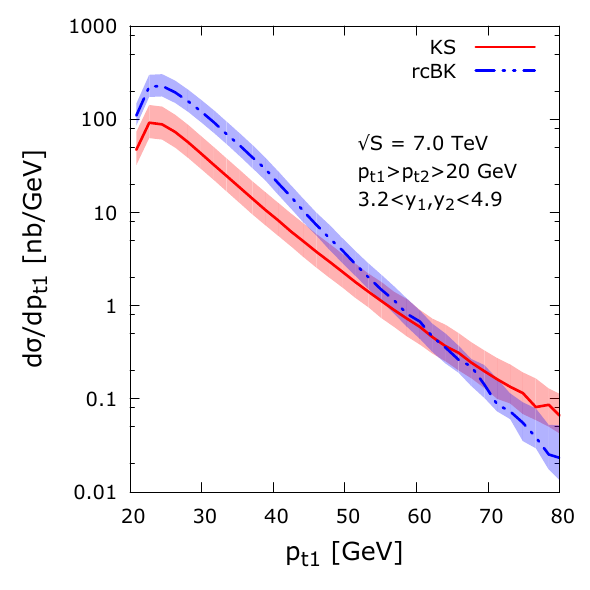}
    \includegraphics[width=0.3\textwidth]{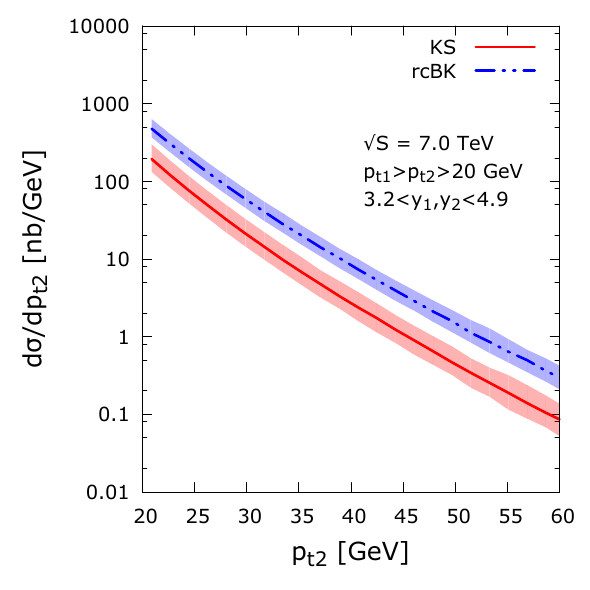} \\
    \includegraphics[width=0.3\textwidth]{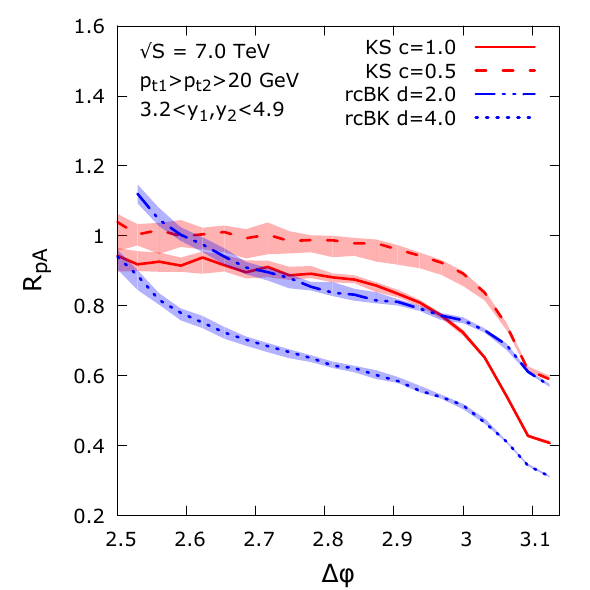}
    \includegraphics[width=0.3\textwidth]{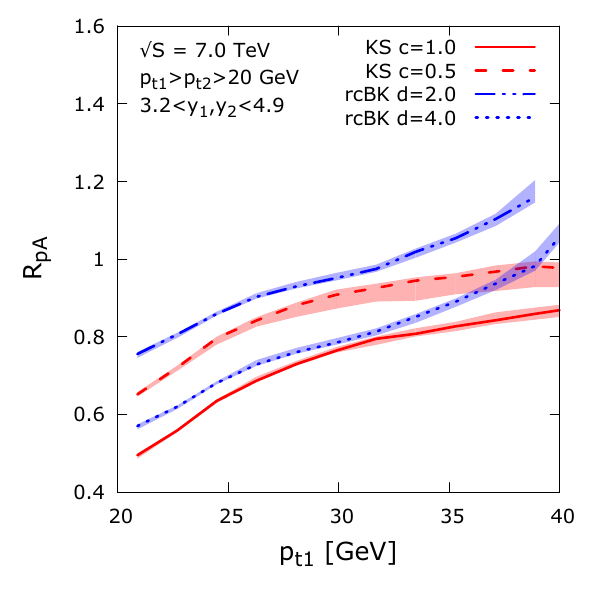}
    \includegraphics[width=0.3\textwidth]{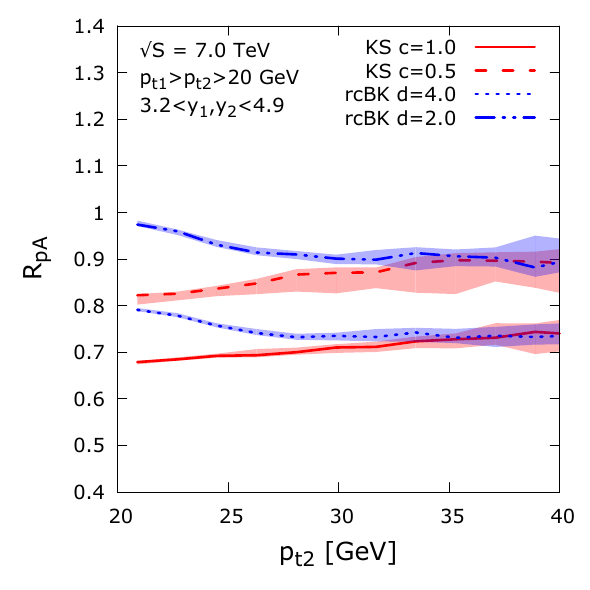} 
  \end{center}
  \caption{
  \small
The results for forward-forward dijet production within HEF framework with two models for UGDs described in the text. The uncertainty bands come from
the scale choice uncertainty. We refer to \cite{vanHameren:2014lna} for further details.
  }
  \label{fig:fw2j}
\end{figure}

In Fig.~\ref{fig:fw2j}  recent results \cite{vanHameren:2014lna} are presented for dijet system in forward rapidity region $3.2<y<4.9$. The potential singularities in matrix elements were cut by using the anti-$k_T$ jet algorithm with $R=0.5$ and the $p_T$ cut of $20\,\mathrm{GeV}$. We show both the absolute predictions and nuclear modification ratios $R_{pA}$ defined as the ratio of the p+A cross section to p+p cross section normalized to the number of nuclei. The last observable is sensitive to the saturation effects. The calculations were made using two frameworks for UGDs: KS \cite{Kutak:2012rf} and rcBK \cite{Balitsky:2008zza,Albacete:2009fh}. Both of the approaches are extensions of the Balitsky-Kovchegov (BK) equation \cite{Balitsky:1995ub,Kovchegov:1999yj}. The KS is formulated in the momentum space and includes corrections of higher order coming from energy conservation, non-singular  pieces of splitting functions  at low $x$ and running coupling  \cite{Kutak:2004ym}. It also assumes a homogeneous target. The rcBK is an extension of the BK equation to running coupling case and similarly to KS it assumes homogeneous target.  Applying these densities to considered observables one clearly sees the sensitivity of the results to a particular evolution scenario (potentially this might be an  effect of different initial conditions but the essential difference comes from different effects incorporated in the evolution kernel).

A particularly interesting observable is $R_{pA}$ as a function of the $p_T$ of the sub-leading jet. The striking feature is that the $R_{pA}$ stays constant and is significantly lower than one in wide range of $p_T$. This suggests that the sub-leading jets are more affected by the saturation scale than the leading jet.

\begin{figure}[t]
  \begin{center}
    \includegraphics[width=0.3\textwidth]{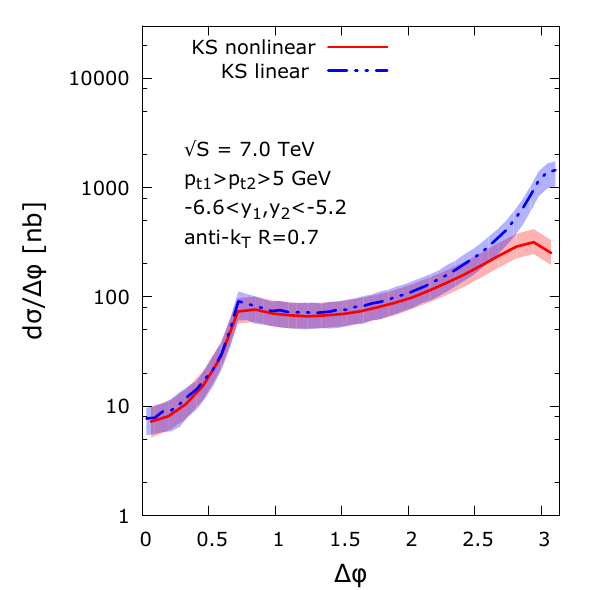}
    \includegraphics[width=0.3\textwidth]{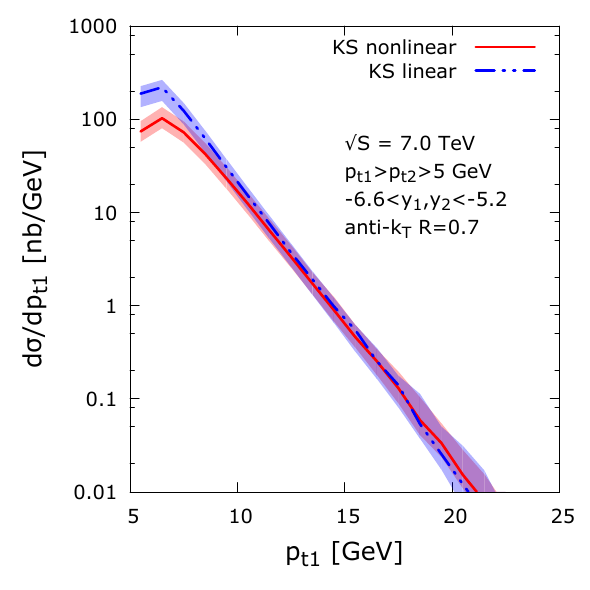}
    \includegraphics[width=0.3\textwidth]{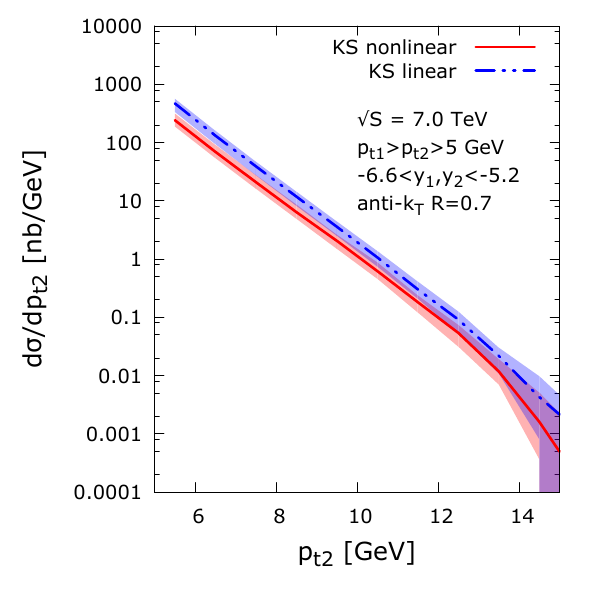} \\
    \includegraphics[width=0.3\textwidth]{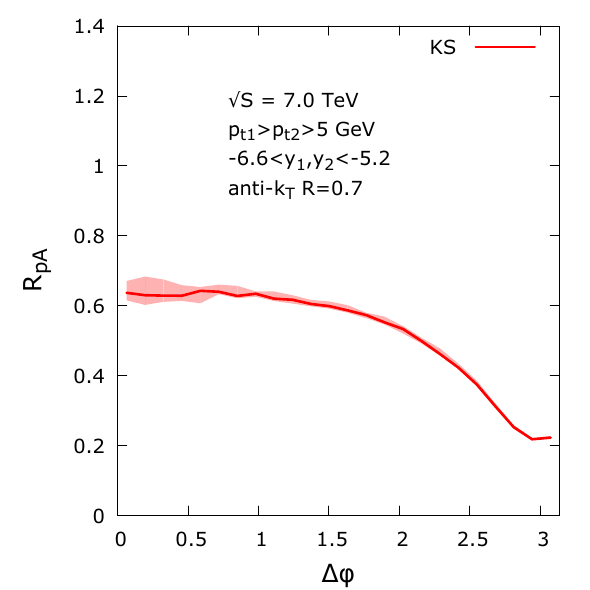}
    \includegraphics[width=0.3\textwidth]{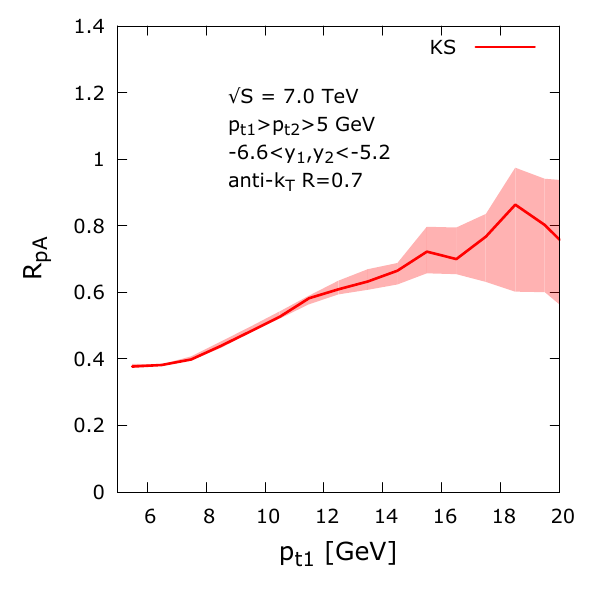}
    \includegraphics[width=0.3\textwidth]{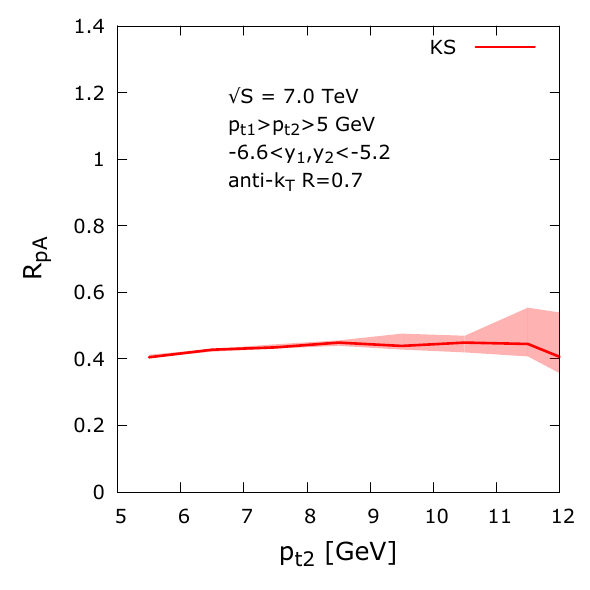} 
  \end{center}
  \caption{
  \small
The results for very-forward dijet production in a potential extension of the CASTOR detector \cite{Pierre_talk:2014}. The uncertainty bands come from the energy scale uncertainty
(the scale enters the HEF factorization formula (\ref{eq:HENfact_2})).
  }
  \label{fig:vvfw2j}
\end{figure}

In order to see an impact of possible upgrades of the LHC on our observables, and for the special purposes of this note, the calculation for a potential extension of the CASTOR detector, which would  allow for jet reconstruction \cite{Pierre_talk:2014}, is extended. We have assumed that the reconstruction of the jets is possible for $5\,\mathrm{GeV}<p_T<30\,\mathrm{GeV}$ using anti-$k_T$ algorithm with $R=0.7$. We refer to this scenario as the very-forward case. The calculations were made using $\mathtt{LxJet}$ program \cite{LxJet:2013} with the KS UGDs  for p+p and p+A collisions. We present the results in Fig.~\ref{fig:vvfw2j}. We see significant difference between
the non-linear evolution of UGDs and the scenario where the non-linear term is removed from the equation. Saturation effects are very strong, as is also evident from the
nuclear modification ratios. The values of $x$ probed here lie between $10^{-5}$ and $10^{-6}$.

\begin{figure}[t]
  \begin{center}
    \includegraphics[width=0.3\textwidth]{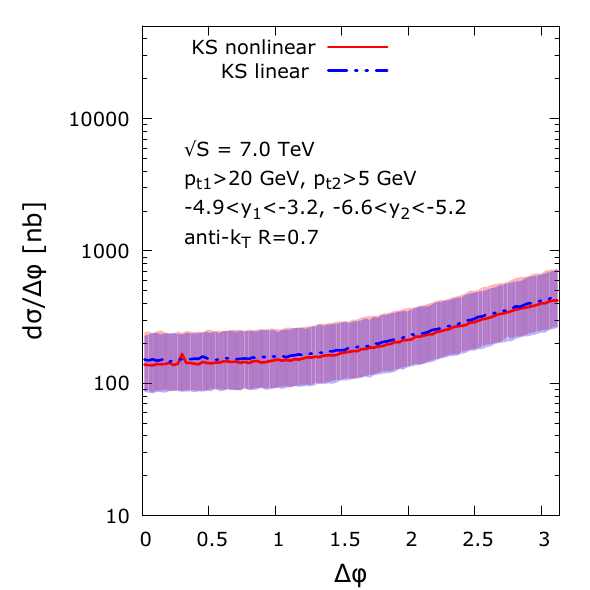}
    \includegraphics[width=0.3\textwidth]{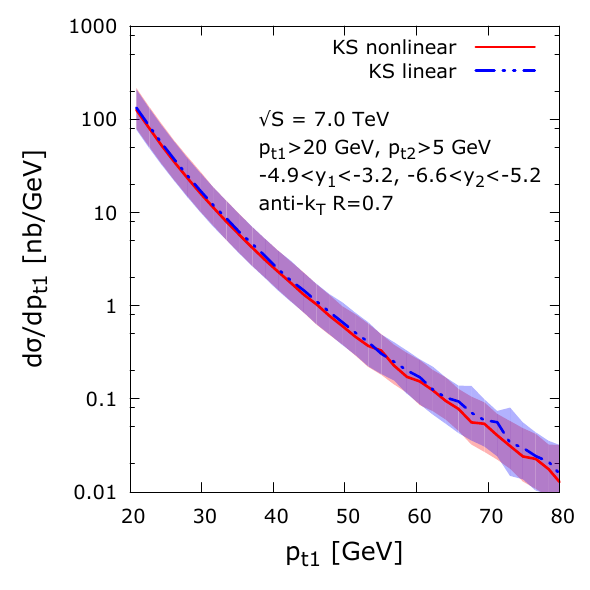}
    \includegraphics[width=0.3\textwidth]{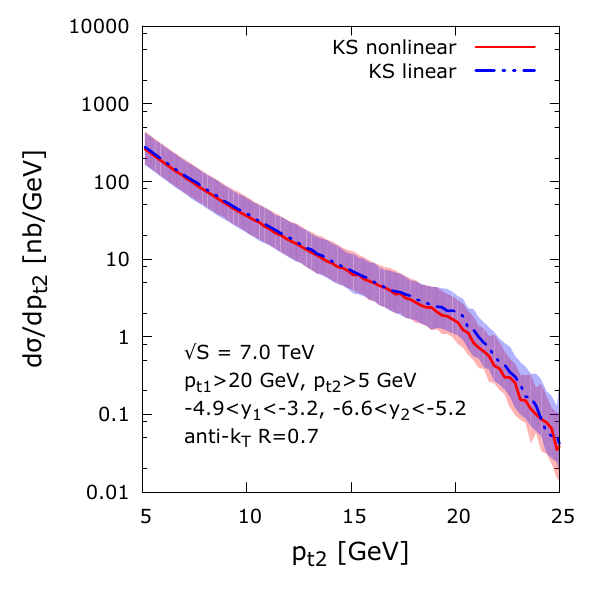} \\
    \includegraphics[width=0.3\textwidth]{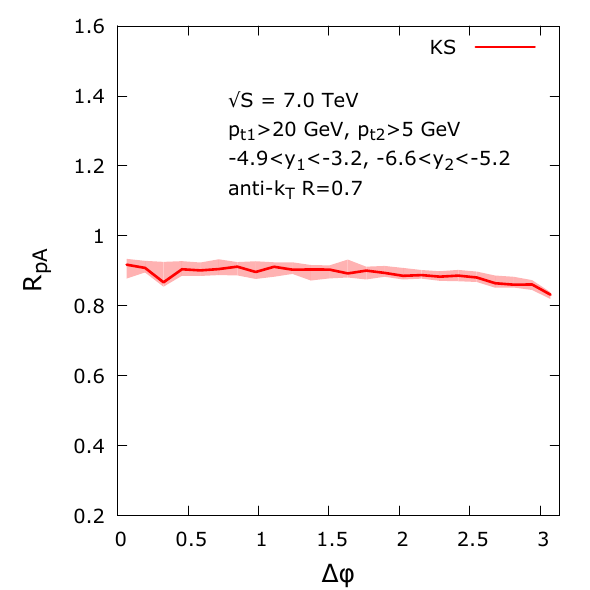}
    \includegraphics[width=0.3\textwidth]{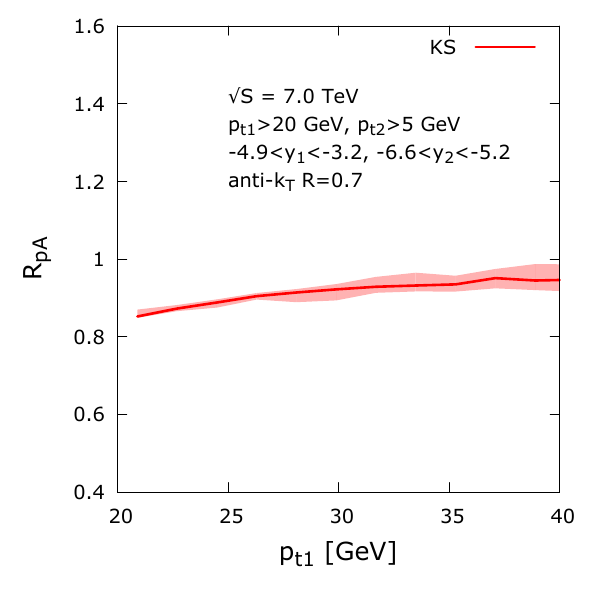}
    \includegraphics[width=0.3\textwidth]{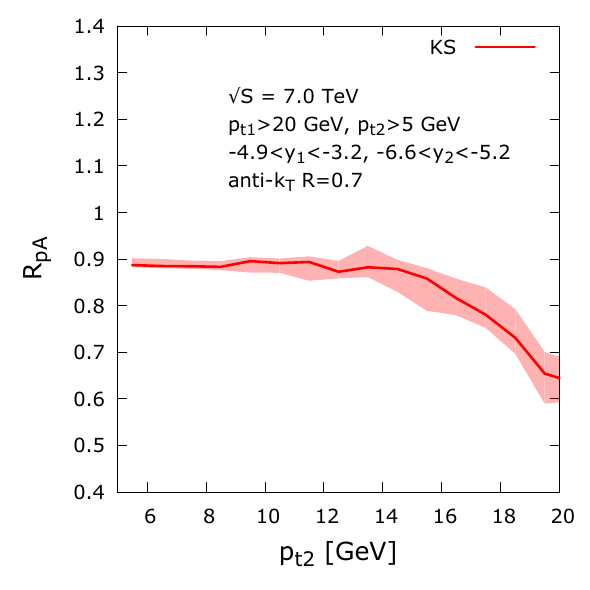} 
  \end{center}
  \caption{
  \small
The result for forward-very-forward scenario, i.e. the softest jet of the dijet system is in the CASTOR detector, while the harder is around HF detector.
The uncertainty bands come from the scale choice uncertainty.
  }
  \label{fig:vfw2j}
\end{figure}

There is another interesting scenario with an extension of the CASTOR detector.  
We may look also at the case where
there is one (leading) jet with $p_{T1}>20\,\mathrm{GeV}$ within rapidity interval $-4.9<y_1<-3.2$ and the second (soft) jet with 
$p_{T2}>5\,\mathrm{GeV}$ is within $-6.6<y_2<-5.2$. This gives an opportunity to study UGDs at low $x$ for large gluon off-shellness. The results
prepared are shown in Fig.~\ref{fig:vfw2j}. An interesting feature is a relative flatness of the decorrelation distribution, meaning 
that one probes large transversal momenta of gluon density. A similar scan is also possible using three jet events, as explained below.

\subsection{Trijet production at forward-central and purely forward rapidities}

Three jets observables may give an additional insight into the UGDs due to possible additional cuts one may apply to scan certain regions of the phase space.
For instance, one may consider the case where one of the jets is in the forward rapidity region while the two hardest jets are in the central rapidity region. In addition, we may restrict the two central jets to balance each other on the transverse plane within a cut on the sum of the two transverse momenta, $D_{\mathrm{cut}}$. This allows to access the UGD at large transverse momentum almost directly by the third forward jet.

\begin{figure}[t]
  \begin{center}
    \includegraphics[width=0.45\textwidth]{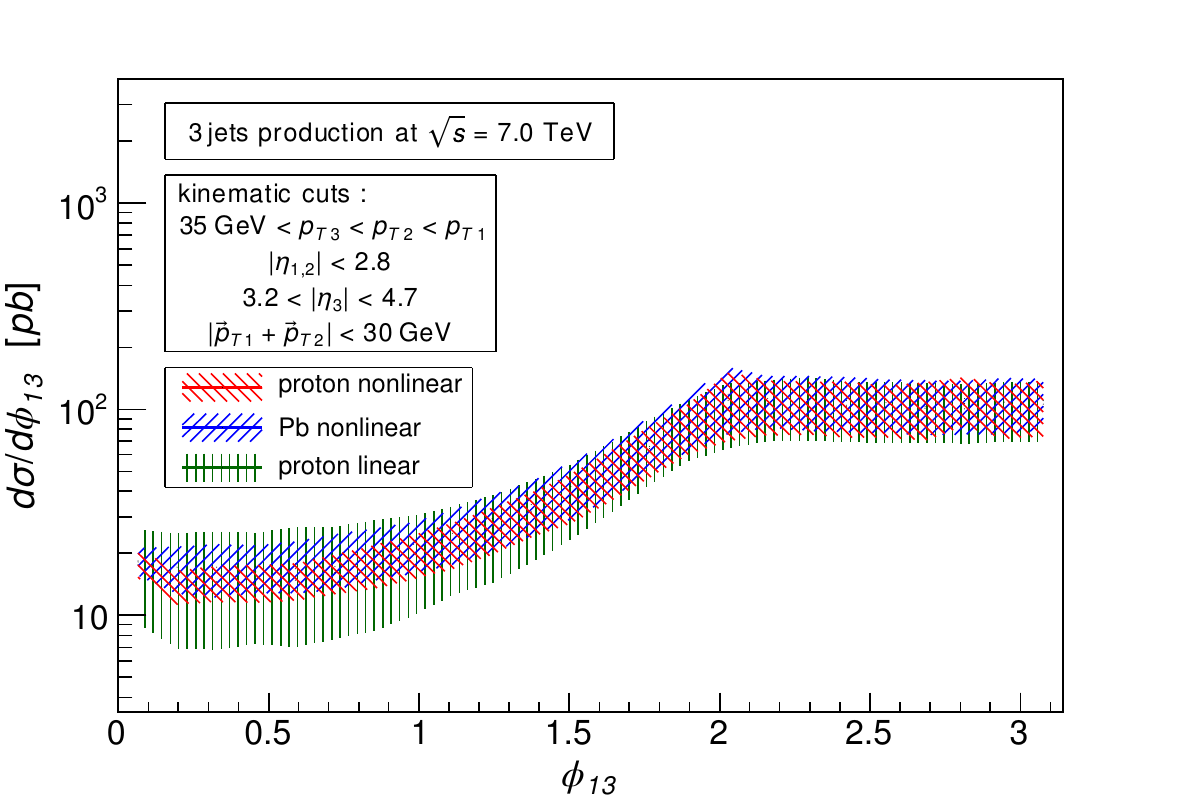}
    \includegraphics[width=0.45\textwidth]{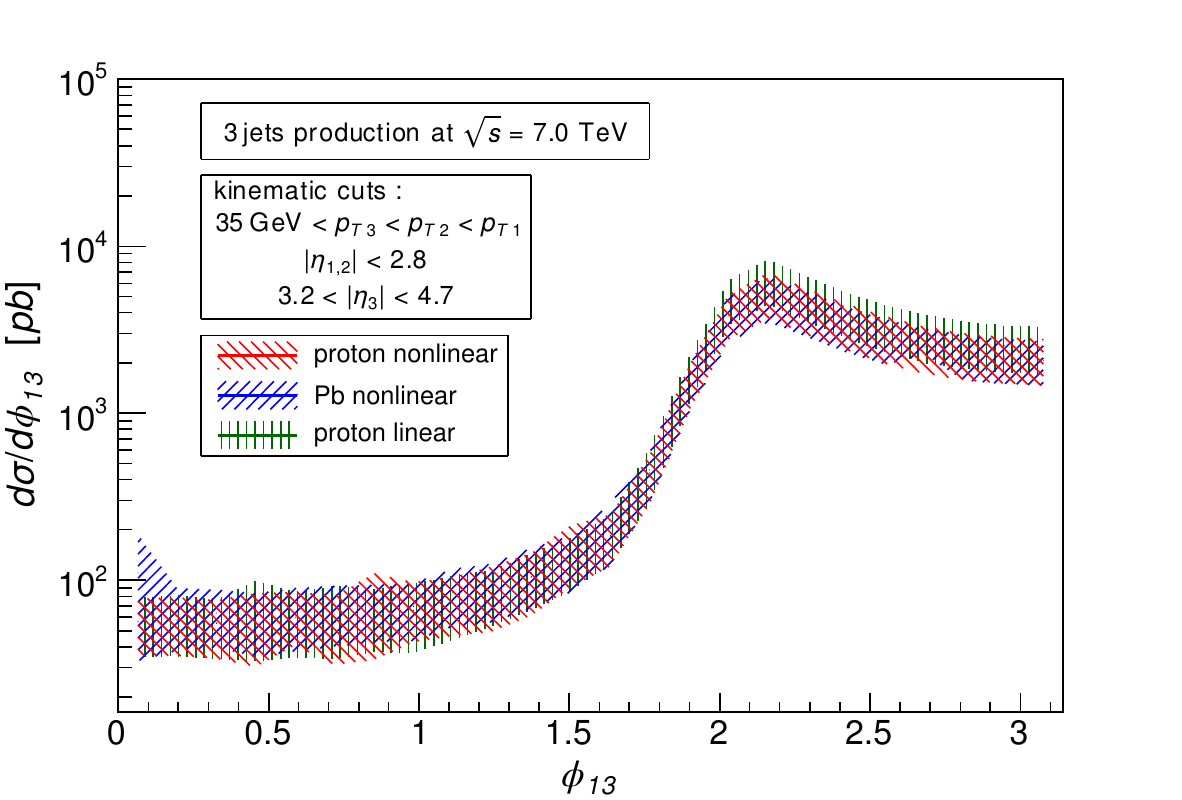}
  \end{center}
  \caption{
  \small
Azimuthal decorrelations for forward-central three jet production. Two hardest jets are in the central detector, while the softest jet is in the forward region. For the left plot 
an additional cut is applied on the central jets, namely we require that they should almost balance each other. This flattens the distribution (left) comparing to
the case without this cut (right) and makes it sensitive to UGDs for large transverse momenta.  For more details refer to \cite{vanHameren:2013fla}.
  }
  \label{fig:btb3j}
\end{figure}

In Fig.~\ref{fig:btb3j} (left) the decorrelations are shown between the hardest and the softest jet \cite{vanHameren:2013fla} for the LHC setup available at present. The calculations were made for KS UGDs and for p+p and p+A collisions. The cuts applied are listed in the plots. An important feature of the result is a relative flatness of the distribution compared to calculations without the back-to-back cut on the central jets (right of Fig.~\ref{fig:btb3j}). Such a distribution is very sensitive to the transversal degrees of freedom of UGDs.

\begin{figure}[t]
  \begin{center}
    \includegraphics[width=0.45\textwidth]{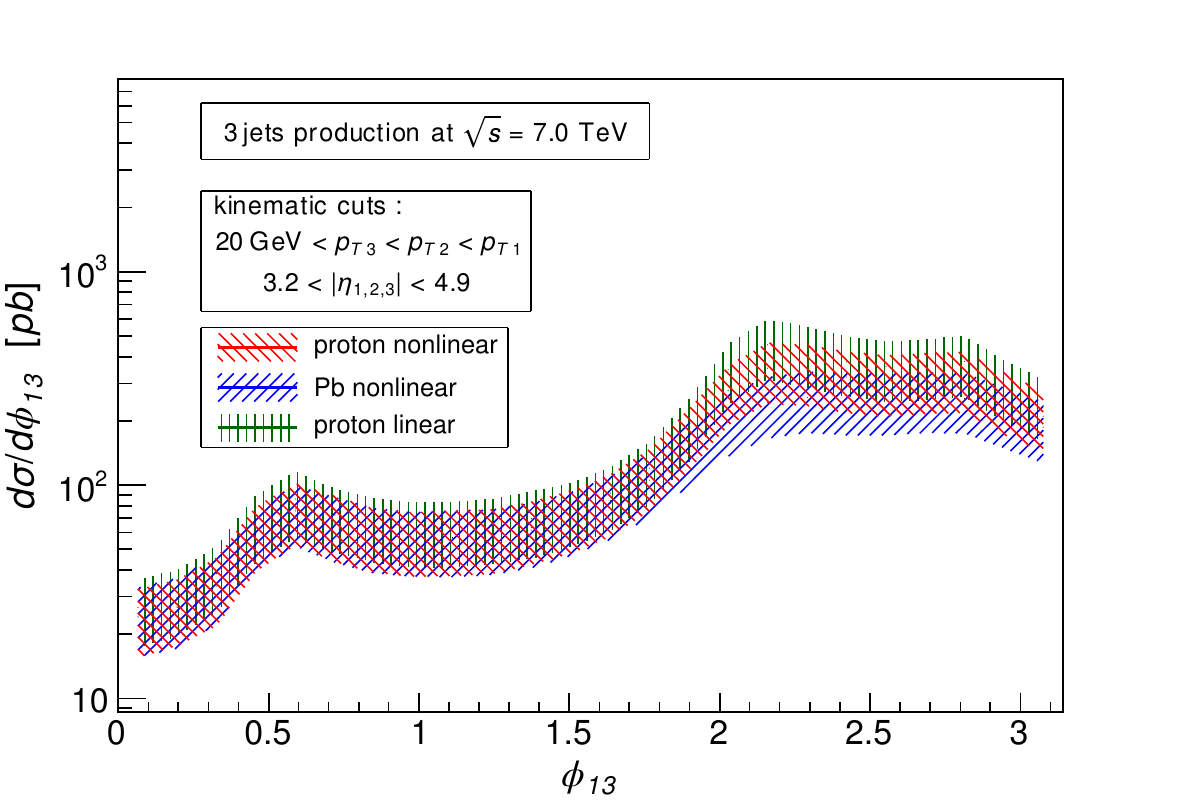}
    \includegraphics[width=0.45\textwidth]{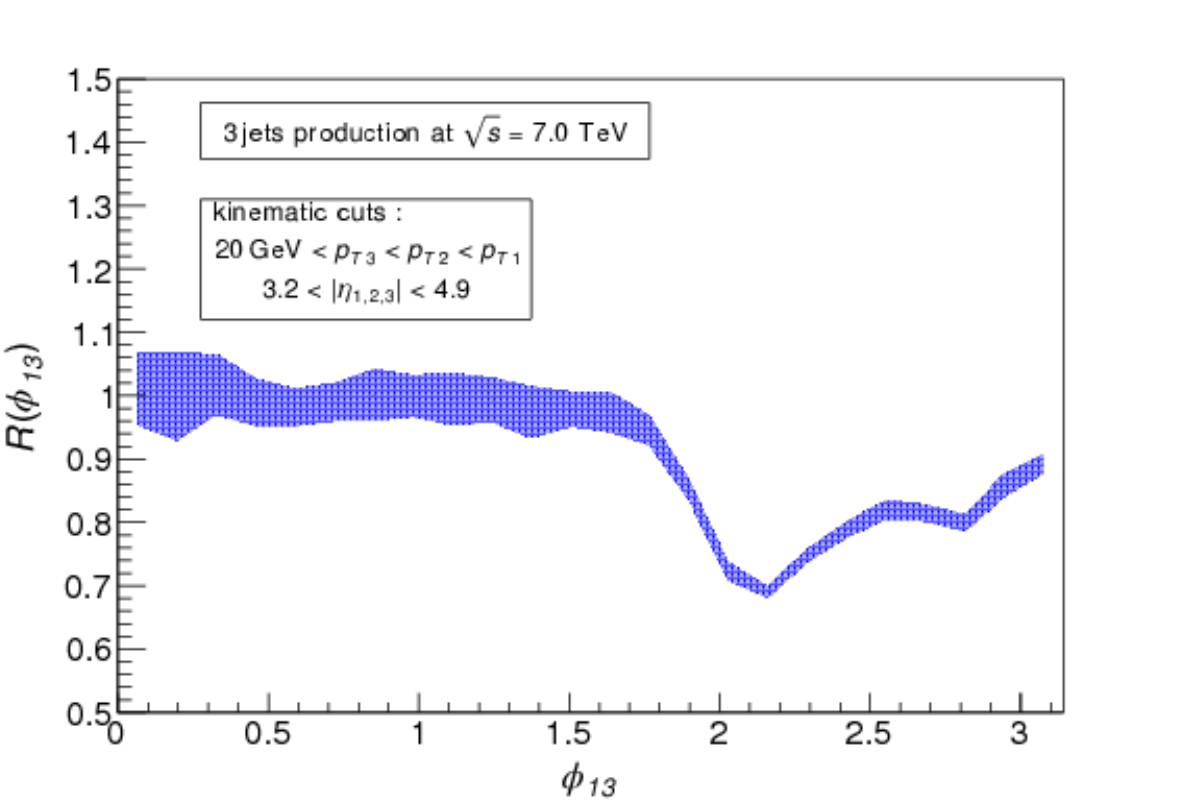}
  \end{center}
  \caption{
  \small
Azimuthal decorrelations for forward-forward-forward three jet production. For large values of the azimuthal angle between the softest and the hardest jets
we see significant differences between different evolution scenarios. For more details refer to \cite{vanHameren:2013fla}.
  }
  \label{fig:ffw3j}
\end{figure}

Another interesting three jets observable is conveyed by configuration where all three jets are produced in the forward direction. This kinematical setup is similar to forward dijet case, i.e. the events are highly asymmetric, but the allowed phase space is larger. Particularly interesting is again the cross section for decorrelations measured as a dependence of the cross section on an angle between softest jet and hardest jet $\phi_{13}$. We see on Fig.~\ref{fig:ffw3j} (left) that this quantity  is sensitive for large $\phi_{13}$ to gluon saturation, since we observe a depletion of the plateau as we move from KS linear to KS non-linear and KS for lead. This is also reflected in the nuclear modification ratio \ref{fig:ffw3j} (right).

\subsection{Forward jet production  -  measurements at very large rapidities}

The very forward calorimeter of CMS (CASTOR) in the acceptance region of $-6.6<\eta<-5.2$ allows to access very small and very large $x$ values in jet measurements (see Fig \ref{fig:fwdjet}c). 
CASTOR has  a 16-fold $\phi$ segmentation and a 14-fold longitudinal segmentation. While the longitudinal energy deposits are grouped together to yield a tower response, it was shown that an anti-kt algorithm with a radius parameter of 0.7 can reconstruct jets in CASTOR. Due to the missing $\eta$ segmentation, all jets are reconstructed at a fixed pseudorapidity of ~6.0. The $p_T$ of these jets is within the lowest accessible at LHC and reaches up to only 6 GeV.

\begin{figure}[h]
\centerline{
\includegraphics[scale=0.5]{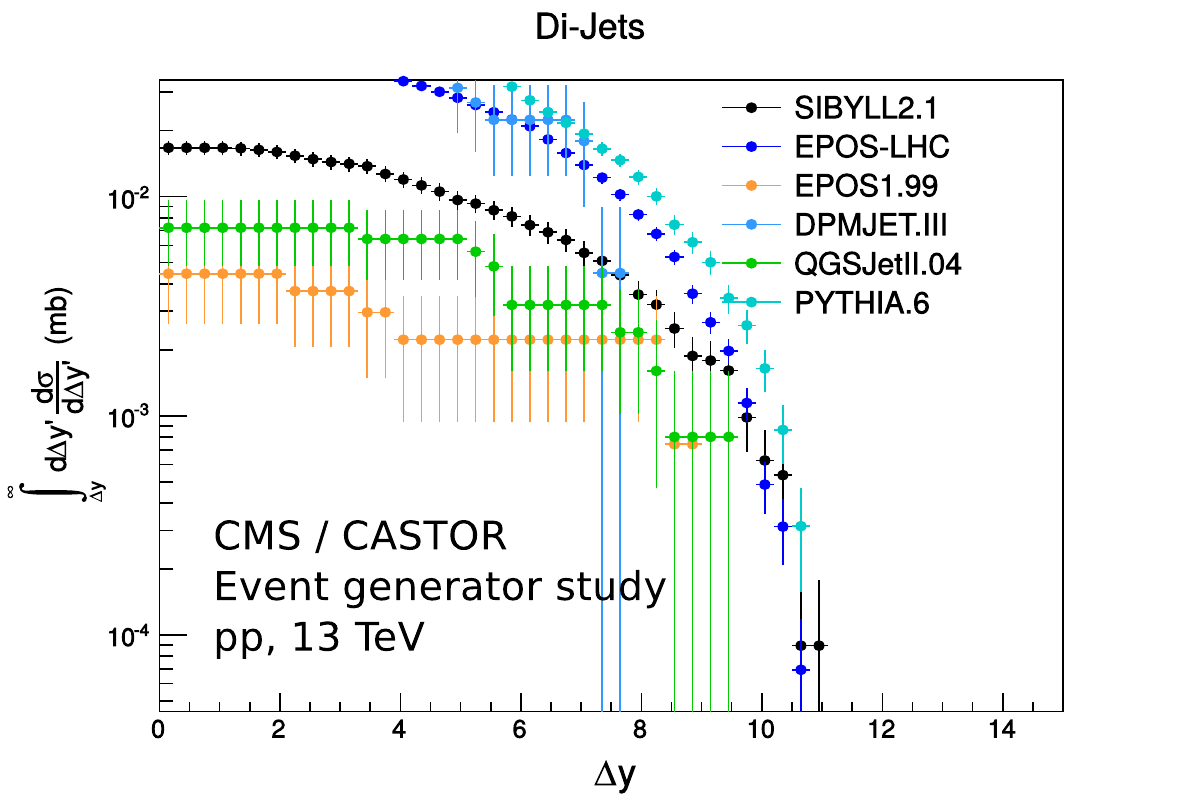}  
}
\caption{Integrated cross sections for di-jets pairs with one jet reconstructed in CASTOR with E > 500 GeV and the other in the central part of CMS with $p_T$ > 25 GeV. Very large model differences are visible as a function of di-jet opening angle Delta y. 
}
\label{fig:ralf2}
\end{figure}

In Fig.~\ref{fig:ralf2} the cross section for di-jets with one jet with $p_T > 25$~GeV in the central region and another jet with $E> 500$~GeV in the CASTOR acceptance region is shown as predicted by different calculations. The predictions vary by more than factors of 10, so  a measurement is needed to constrain the calculations.

Due to its 14-fold longitudinal segmentation, CASTOR is a very good detector to distinguish electromagnetic from hadronic energy deposits. Furthermore, when CASTOR is combined with the T2 tracking detector, charged electrons can be distinguished from neutral photons. First studies of isolated electrons with CASTOR and T2 have shown some of the potential behind this \cite{woehrmann}. Given improved techniques to handle the underlying event subtraction in pp runs at 13 TeV, the detection of very forward electrons at $-6.6<\eta<-5.2$ with fully resolved position (T2) and energy (CASTOR) has the potential to enhance the reach of Drell Yan and $Z$-production studies towards much smaller values of $x$ compared to what is possible to study so far at LHC. For this purpose we implement a dedicated isolated electron trigger in CASTOR on L1 hardware trigger level, which can be used on HLT level to select the relevant event topologies. Given this trigger it will require $1\;nb^{-1}$ in order to study very forward electrons up to ~500 GeV and $100 \;  nb^{-1}$ to extend this range towards ~1 TeV. Since the current proposal assumes to analyse isolated electrons, small values of average pileup are needed. This is also mandatory if a dedicated isolated electron trigger should be used. The pileup should not exceed one for this reason.

The first measurement of jets in the CASTOR calorimeter based on a sample of $pp$ collisions at $\sqrt{s} = 7$ TeV is presented in \cite{castor-jet-dpn}.
Events are selected with at least one track-jet in the central CMS detector, with transverse momentum
$p_T > 1$ GeV and $|\eta| < 2$. Track-jets are reconstructed with the anti-$k_T$ jet clustering algorithm~\cite{Cacciari:2008gp} with a distance parameter $R = 0.5$.  The number of selected events in the minimum-bias sample is 4.6 million. The event selection is similar to that described in \cite{Chatrchyan:2013gfi}.

\begin{figure}[htb]
\includegraphics[scale=0.42]{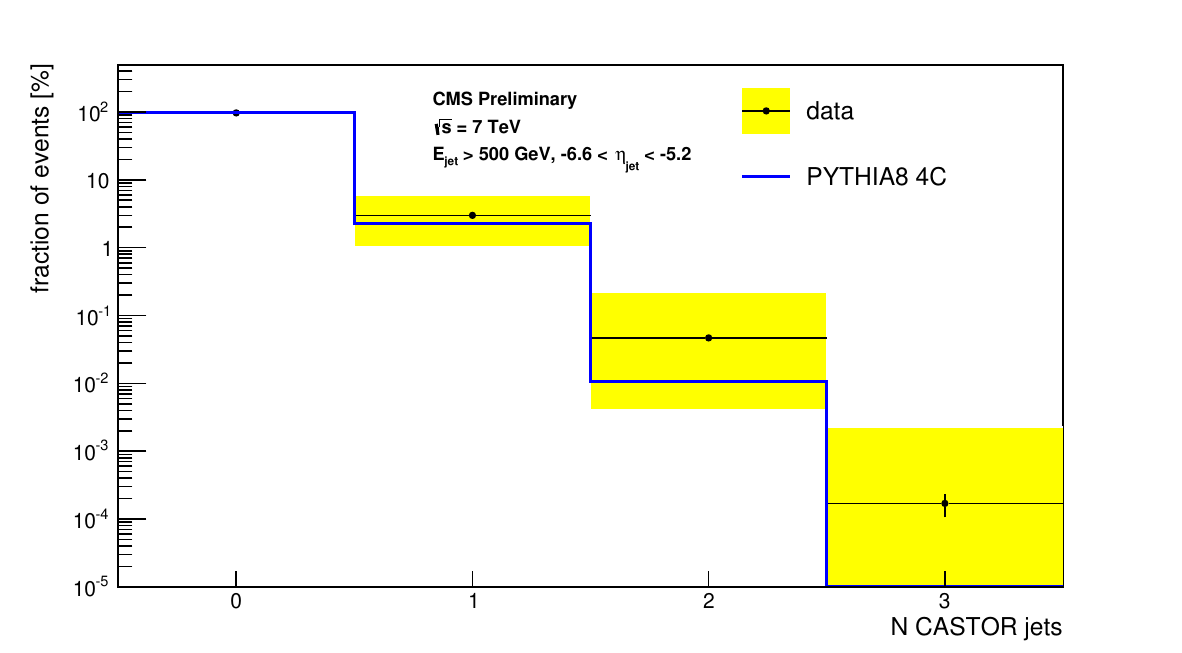}
\includegraphics[scale=0.42]{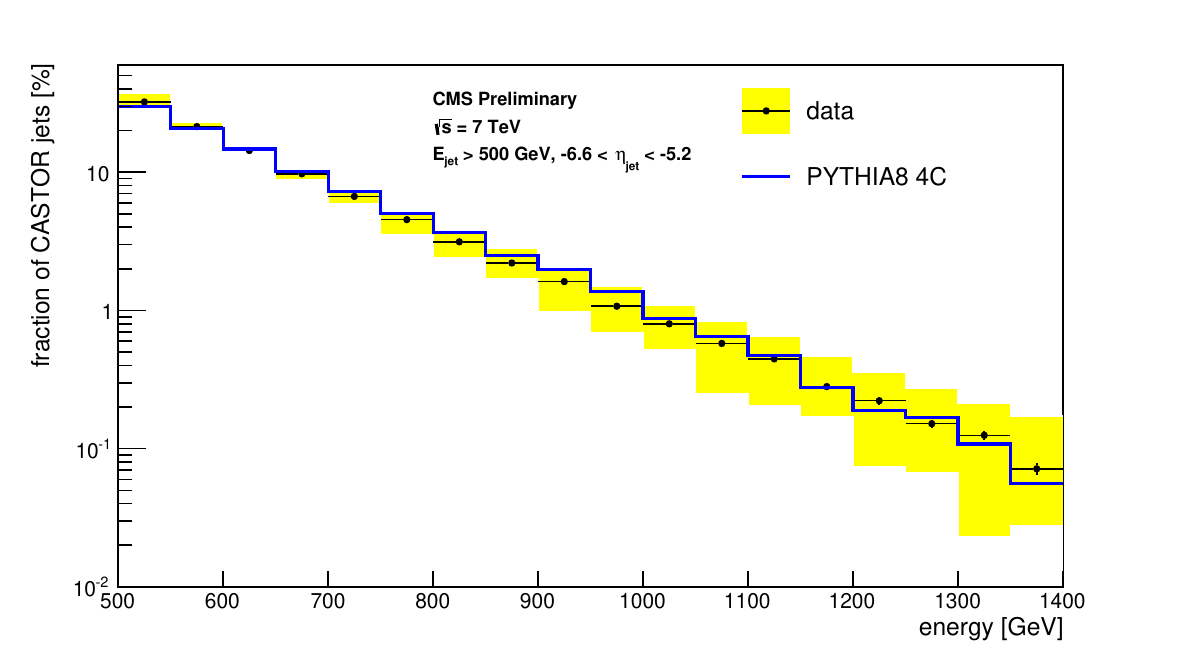}
\caption{Normalised distributions of the jet multiplicity and jet energy spectrum for jets reconstructed in CASTOR with an energy $E > 500$ GeV. The data are compared to the predictions of the MC 
event generator {\sc PYTHIA}8-4C. The error band represents the $22 \%$ uncertainty on the CASTOR calorimeter absolute energy scale.}
\label{fig:multi-energy}
\end{figure}

The normalised distributions of the jet multiplicity and jet energy spectrum are shown in figure \ref{fig:multi-energy} for jets reconstructed in 
$-6.6 < \eta < -5.2$ and $E > 500$ GeV. The data are compared to the predictions of the Monte Carlo (MC) event generator {\sc PYTHIA}8~\cite{Sjostrand:2007gs} with tune 4C~\cite{Corke:2010yf}.
The error band represents the $22 \%$ uncertainty on the absolute energy scale of the CASTOR calorimeter. 

\begin{figure}[ht]
\centering
\includegraphics[scale=0.4]{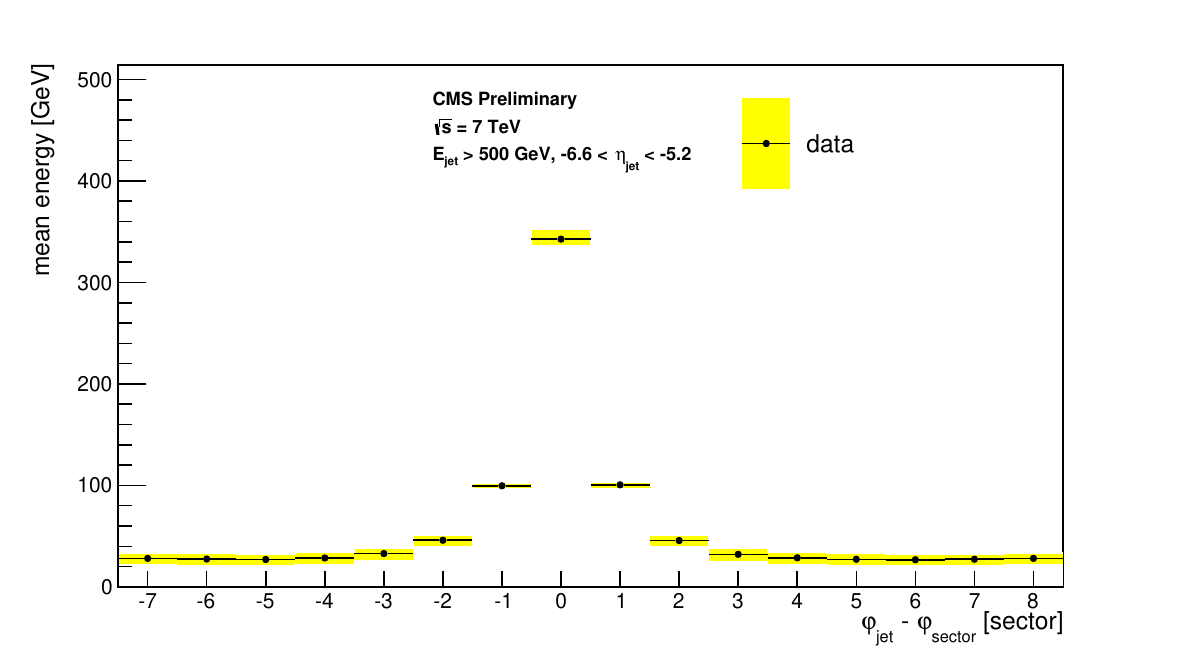}
\caption{Energy weighted azimuthal $\phi$ profile for jets reconstructed in CASTOR with an energy $E > 500$ GeV. The error band represents the $22 \%$ uncertainty on the CASTOR calorimeter absolute 
energy scale.}
\label{fig:profile}
\end{figure}

The energy weighted azimuthal $\phi$ profile \cite{castor-jet-dpn} for jets reconstructed in 
CASTOR with an energy $E > 500$ GeV is presented in figure~\ref{fig:profile}. The peak around the jet axis as well as the width of the distribution are found to reproduce the characteristics 
of the jet profile as determined in a generator level study.

\section{Saturation physics in p+p and p+A collisions}

\subsection{Introductory remarks}

Saturation has attracted much interest, in connection with the behavior of the gluon 
density of the proton at small $x$ and low $Q^2$, and for the understanding of the initial states in heavy ion collisions. First evidence for saturation  has 
been found in electron-proton scattering at HERA  (e.g. geometrical scaling),  and in heavy ion collisions at RHIC and in the first run of LHC (e.g. two-particle correlations, nuclear modification factor). In $pp$ collisions at the LHC, the best place to look for the small-$x$ behavior of parton densities are Drell-Yan production processes in the forward direction; therefore, this also one of the 
most promising regions where saturation can be looked for.          

\subsection{Forward Drell-Yan production -  collinear vs small-x approach}
% ({\it Emilia Lewandowska$\,\,^{\dagger}$})}\let\thefootnote\relax\footnote{$^{\dagger}$ \it Institute of Nuclear Physics Polish Academy of Sciences, Cracow, Poland}
\subsubsection{Drell-Yan cross section in the collinear approach}

\begin{figure}[t]
\begin{center}
\vskip 0.5cm
\includegraphics[height=3cm]{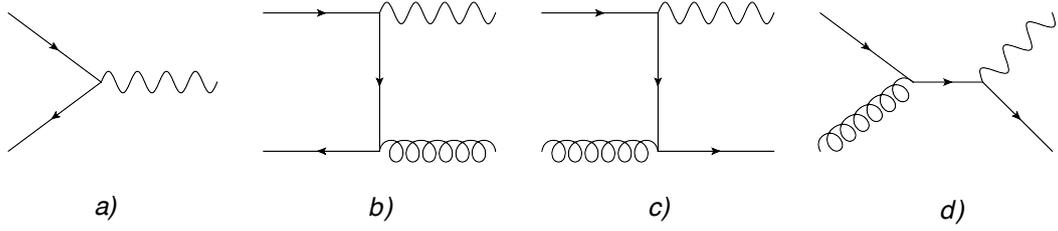}
\caption{The Drell-Yan production in the leading (a) and next-no-leading (b-d) order approximation. The diagrams c) and d) are enhanced in the small-$x$ limit due to a strongly rising gluon distribution.}
\label{dy1}
\end{center}
\end{figure}

The Drell-Yan production is a unique process  which offers high sensitivity to the parton distribution functions in hadrons. In the leading order (LO) approximation, the Drell-Yan lepton pair of invariant mass $(M > 1~{\rm GeV})$ is produced by annihilation of two quarks from the colliding hadrons, see the diagram (a) in Fig.~\ref{dy1}: 
\begin{eqnarray}
\nonumber
q_{f}\bar{q}_{f} \rightarrow \gamma^{*} \rightarrow l^{+}l^{-}.
\end{eqnarray}
The cross section in this approximation is given by the quark/antiquark distributions in the colliding hadrons taken at the scale ${M^2}$:\\
\begin{eqnarray}
\label{eq1}
\frac{d^2\sigma^{LO}}{dM^2 dx_F}=\frac{4\pi\alpha^2_{em}}{3N_cM^4} \frac{x_1x_2}{x_1+x_2} \sum_f e_f^2
 \left\{q_f(x_1,M^2)\,\bar q_f(x_2,M^2)+\bar q_f(x_1,M^2)\,q_f(x_2,M^2)\right\},
\end{eqnarray}
where ${\alpha_{em}}$ is the fine structure coupling constant, ${N_c}$ is the number of quark colors and ${q_f,\bar{q}_f}$ are quark/antiquark distributions.
The quark momentum fractions, ${x_1}$ and ${x_2}$, are determined by the lepton pair kinematics:
\begin{eqnarray}
x_{1}=\frac{1}{2}\left(\sqrt{x^{2}_{F}+4\frac{M^2}{s}}+x_F\right), ~~~~~~~  x_{2}=\frac{1}{2}\left(\sqrt{x^{2}_{F}+4\frac{M^2}{s}}-x_F\right),
\end{eqnarray}
where ${x_F=x_1-x_2}$ is the Feynman variable of the lepton pair and ${s}$ is the hadronic center-of-mass energy squared.

%%%%%%%%%%%%%%%%%%%%%%%%%%%%%%%%%%%%%%%%%%%%%%%%%%%%%%%%%%%%

In the next-to-leading (NLO) approximation additional emissions of a parton into the final state has to be taken into account, see the diagrams (b-d) in Fig.~\ref{dy1}.
Because of the emission, one of the quarks entering the photon vertex 
carries a~fraction ${z < 1}$ of the 
incoming parton momentum.
Thus, the incoming parton momentum fractions take now the form:
\begin{eqnarray}
x_{1}=\frac{1}{2}\left(\sqrt{x^{2}_{F}+4\frac{M^2}{{z}\/s}}+x_F\right),~~~~~~~  
x_{2}=\frac{1}{2}\left(\sqrt{x^{2}_{F}+4\frac{M^2}{{z}\/s}}-x_F\right).
\end{eqnarray}
The NLO corrections to the Drell-Yan cross section 
are proportional to the strong coupling ${\alpha_s}$ and are given by \cite{Altarelli:1978id,Altarelli:1979ub,KubarAndre:1978uy}\\
\begin{eqnarray}
\label{eqDY1}
\nonumber
\frac{d^2\sigma^{NLO}}{dM^2 dx_F}&=&\frac{4\pi\alpha^2_{em}}{3N_cM^4}  
{\frac{\alpha_s(M^2)}{2 \pi}}
\int^{1}_{z_{min}} dz \frac{x_1x_2}{x_1+x_2}  \sum_f e_f^2 \Big\{q_f(x_1,M^2)\,\bar q_f(x_2,M^2){D_{q}(z)}+g(x_1,M^2) 
\\ 
&\times& [q_f(x_1,M^2)+\bar q_f(x_2,M^2)]{D_{g}(z)}+(x_{1} \leftrightarrow x_{2})\Big\}\,,
\end{eqnarray}
where the coefficient functions ${D_{q,g}}$ are calculated perturbatively and ${g}$ is a gluon distribution. 
Thus, up to the order $\alpha_s$, the Drell-Yan cross section in the collinear approach is the sum of the leading and next-to-leading contributions:
\begin{eqnarray}
\label{eq:collfinal}
 \frac{d^2\sigma^{col}}{dM^2dx_F} =\frac{d^2\sigma^{LO}}{dM^2dx_F}+\frac{d^2\sigma^{NLO}}{dM^2dx_F}.
\end{eqnarray}

%%%%%%%%%%%%%%%%%%%%%%%%%%%%%%%%%%%%%%%%%%%%%%%%%%%%%%%%%%%%
\subsubsection{Drell-Yan process in the small-$x$ limit}
In  the small-$x$ limit, the dilepton mass is much smaller than the center-of-mass energy of the colliding  hadrons, ${M_{\ll} <<\sqrt{s}}$. In this case,  a momentum fraction of one of the incoming partons is very small, e.g.:
\begin{eqnarray}
x_1\sim 1,   ~~~~~~~~~~~~~~~x_2=\frac{M^2_{\ll}}{s\,x_1}.
\end{eqnarray}
If the small momentum fraction is carried by a gluon,  the fast incoming quark 
probes high gluon density system in which saturation effects may occur.

\begin{figure}[t]
\begin{center}
\includegraphics[height=2cm]{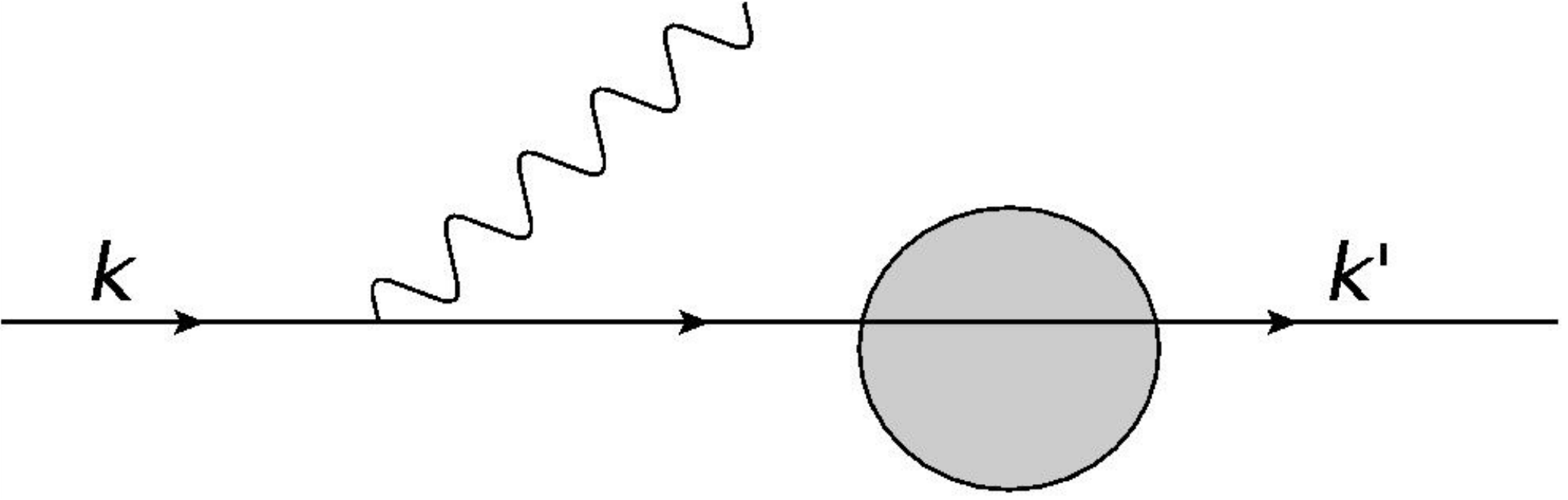}~~~~\includegraphics[height=2cm]{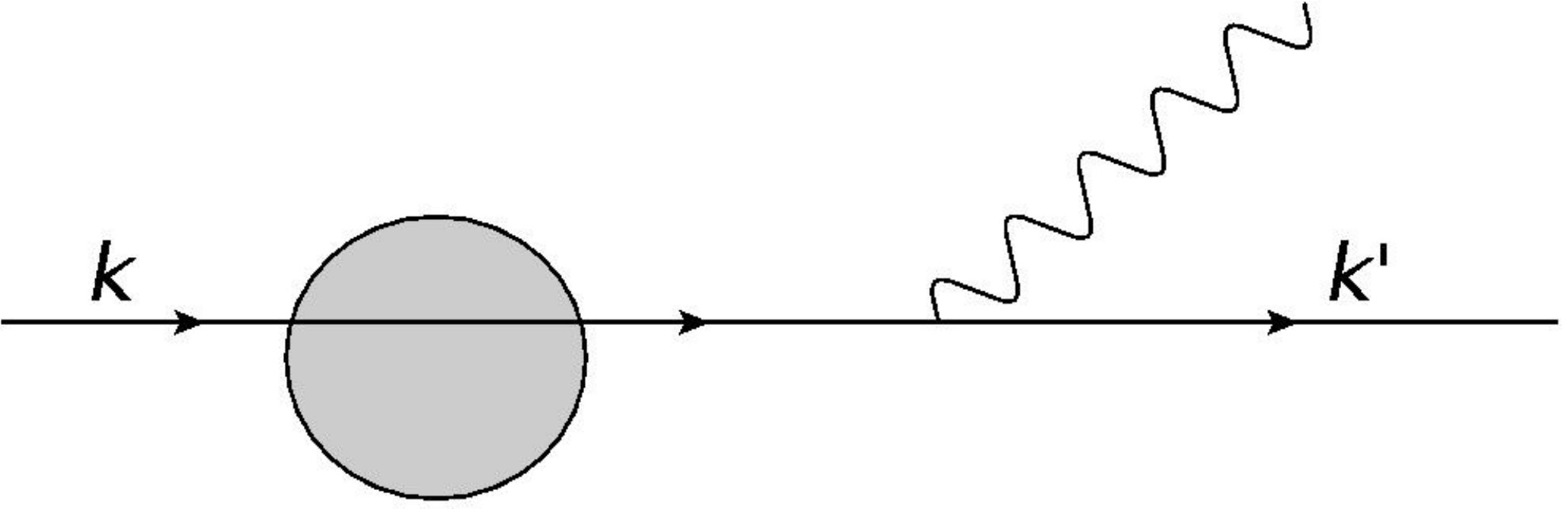}
\caption{The Drell-Yan process in the target rest frame point of view.}
\label{3}
\end{center}
\end{figure}

The target rest frame point of view is particularly attractive for physical interpretation
of these effects. In this frame, the  fast incoming  quark interacts with the target gluon field before or after scattering, emitting a virtual photon. This is shown by diagrams in 
Figs.~\ref{3}.
The photon then decays producing a lepton pair which moves into the region of forward rapidity.

The cross section for radiation of a  photon 
with the momentum fraction ${z}$ of the fast quark is given by \cite{Kopeliovich:2000fb}
\begin{eqnarray}
\sigma(q p\rightarrow \gamma^*X)= \int d^2 r \,W(z,r,M^2)\, {\sigma_{qq}(x_2,zr)},
\end{eqnarray}
where ${r}$ is the photon-quark transverse separation  and ${W}$ is
the photon wave function squared, computed perturbatively in \cite{Brodsky:1996nj,Betemps:2001he}.
The dipole cross section ${\sigma_{qq}}$ \cite{Nikolaev:1990ja} is known from DIS scattering at small Bjorken-${x}$ and describes the interaction of the incoming quark with strong gluon fields of the target hadron. The dipole form comes form the interference of the two shown amplitudes
in the formula for the cross section.
The final form of the Drell-Yan cross section in the dipole framework is given by the equation
\begin{eqnarray}
\label{eq2}
\frac{d^2\sigma^{DY}}{dM^2\,dx_F}&=&\frac{\alpha_{em}}{6\pi M^2}\frac{1}{x_1+x_2} 
\int_{x_1}^1\frac{dz}{z}\,F_2\!\left(\frac{x_1}{z},M^2\right) \sigma(qp\to\gamma^*X)\,,
\end{eqnarray}
where ${F_2}$ is the proton structure function. We will compare predictions given by this formula
with those given by the collinear factorization approach (\ref{eq:collfinal}). Before presenting our results, we will describe the dipole cross sections used in the analysis.

The following three models of the dipole cross sections ${\sigma_{qq}}$ with gluon saturation effects have been used in the calculations:
\begin{itemize}
\item {Golec-Biernat--W\"{u}sthoff}  (GBW) \cite{Golec-Biernat:1998js,Golec-Biernat:1999qd}
\begin{eqnarray}
\label{gbw}
\sigma_{qq}(x,r)=\sigma_0\{1-\exp(-r^2Q_s^2(x)/4) \},
\end{eqnarray}
\item {Bartels--Golec--Kowalski--Sapeta} (BGKS) \cite{Bartels:2002cj,GolecBiernat:2006ba}
  \begin{eqnarray}
\label{gs}
\sigma_{qq}(x,r)=\sigma_0\Big\{1- \exp(-\pi^2 r^2 \alpha_s(\mu^2) xg(x, \mu^2)/3 \sigma_0)\Big\},
\end{eqnarray} 
\item {Color Glass Condensate} (CGC) \cite{Iancu:2003ge,Soyez:2007kg}
\begin{eqnarray}
\label{cgc}
\sigma_{qq}(x,r)=\sigma_0 \times \Bigg\{ 
\begin{array}{ll}
 N_0 \left(\frac{rQ_s}{2}\right)^{2(\gamma_s + \frac{1}{\kappa \lambda y} \ln \frac{2}{rQ_s})} &  :\,\,\,\,  rQ_s \leq 2  
\\ 
1-e^{-A \ln^2 (B rQ_s)} &  :\,\,\,\,  rQ_s >2 
\end{array}
\end{eqnarray}
\end{itemize}
In the formulas, $Q_s$ is the saturation scale: $Q_s=Q_0\, x^{-\lambda}$. The  parameters in the above formulas are determined from the analysis of the HERA data on deep inelastic scattering.

%%%%%%%%%%%%%%%%%%%%%%%%%%%%%%%%%%%%%%%%%%%%%%%%%%%%%%%%%%%%
\subsubsection{Results}

\begin{figure}[t]
\begin{center}
\includegraphics[width=7cm]{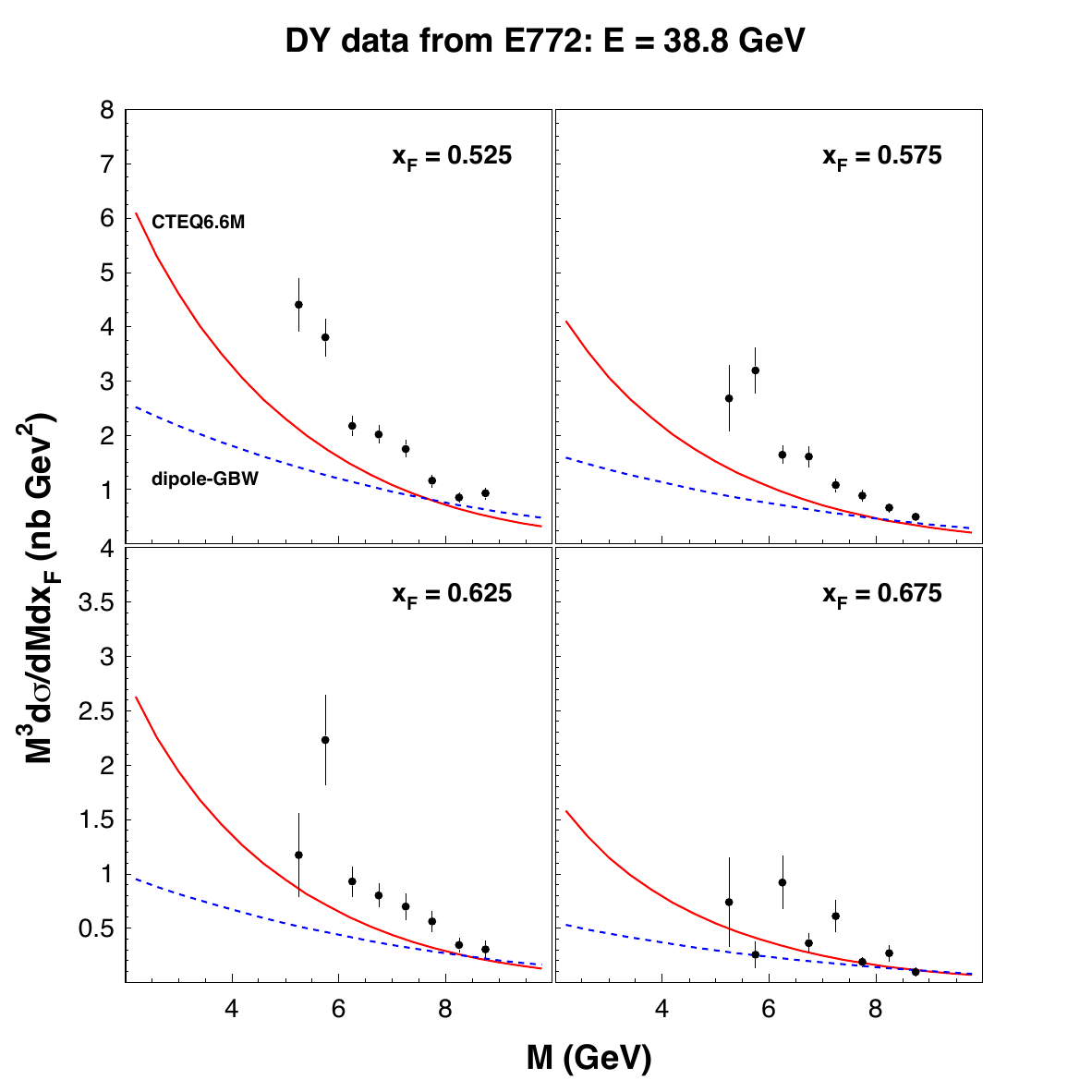}
\includegraphics[width=7cm]{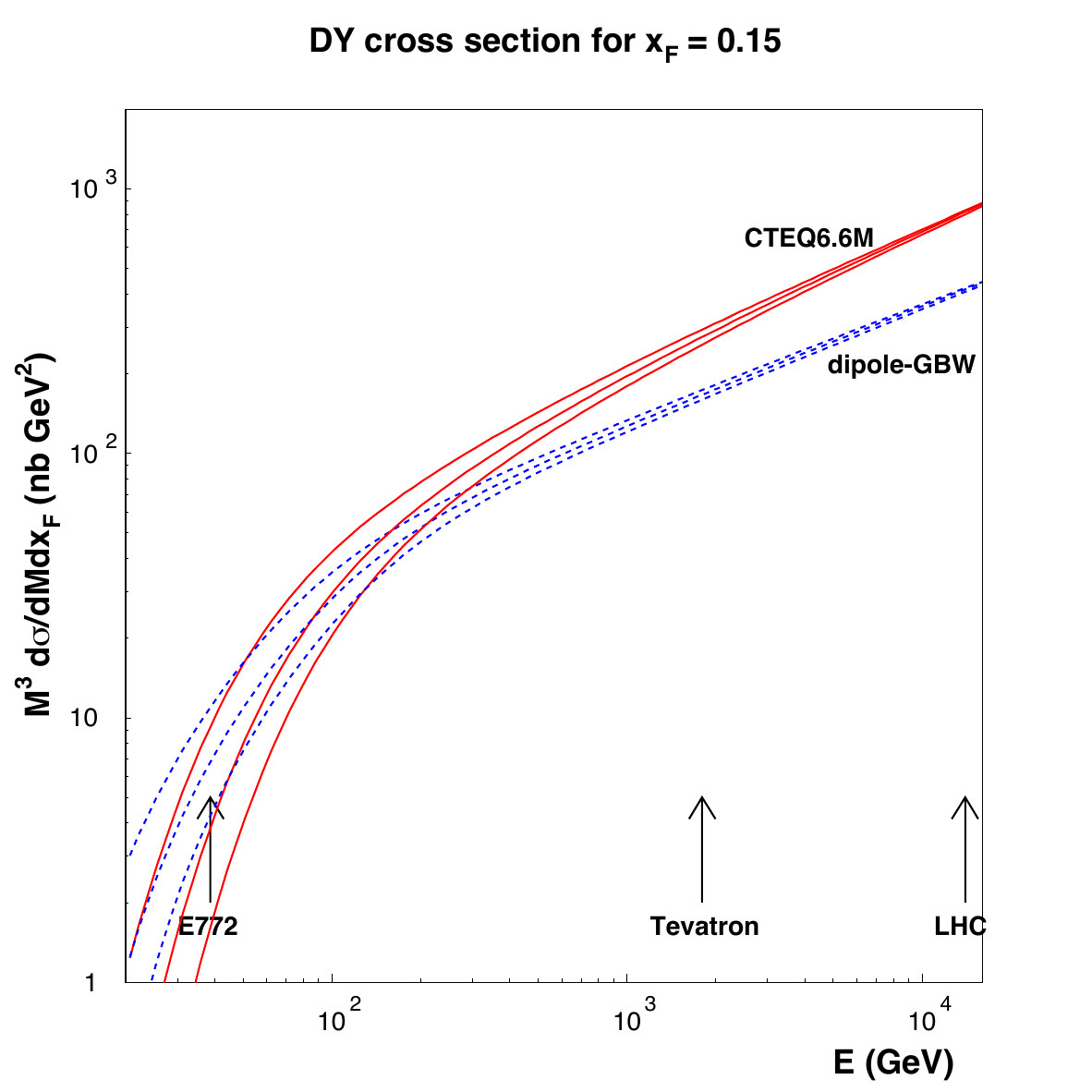}
\caption{Left: The Drell-Yan cross section in the collinear and dipole formulas against the E772 Collaboration data. Right: predictions for the LHC energies and three values of the lepton pair mass ${M=6,8,10}$ GeV.}
\label{dy3}
\end{center}
\end{figure}

In Fig.~\ref{dy3}-left we present a comparison of the results from the collinear factorization 
(\ref{eq:collfinal}) and the dipole approach (\ref{eq2}) formulas with the existing data from the Fermilab E772 Collaboration \cite{McGaughey:1994dx}. We use the NLO CTEQ6.6M parton distributions \cite{Nadolsky:2008zw} for the collinear factorization and 
the  GBW parametrization  \cite{Golec-Biernat:1998js,Golec-Biernat:1999qd}
for the dipole cross section. It is clearly seen that for the different values of the Feynman variable ${x_F}$, the E772 data are above the results from both approaches.

\begin{figure}[t]
\begin{center}
\includegraphics[width=9.9cm]{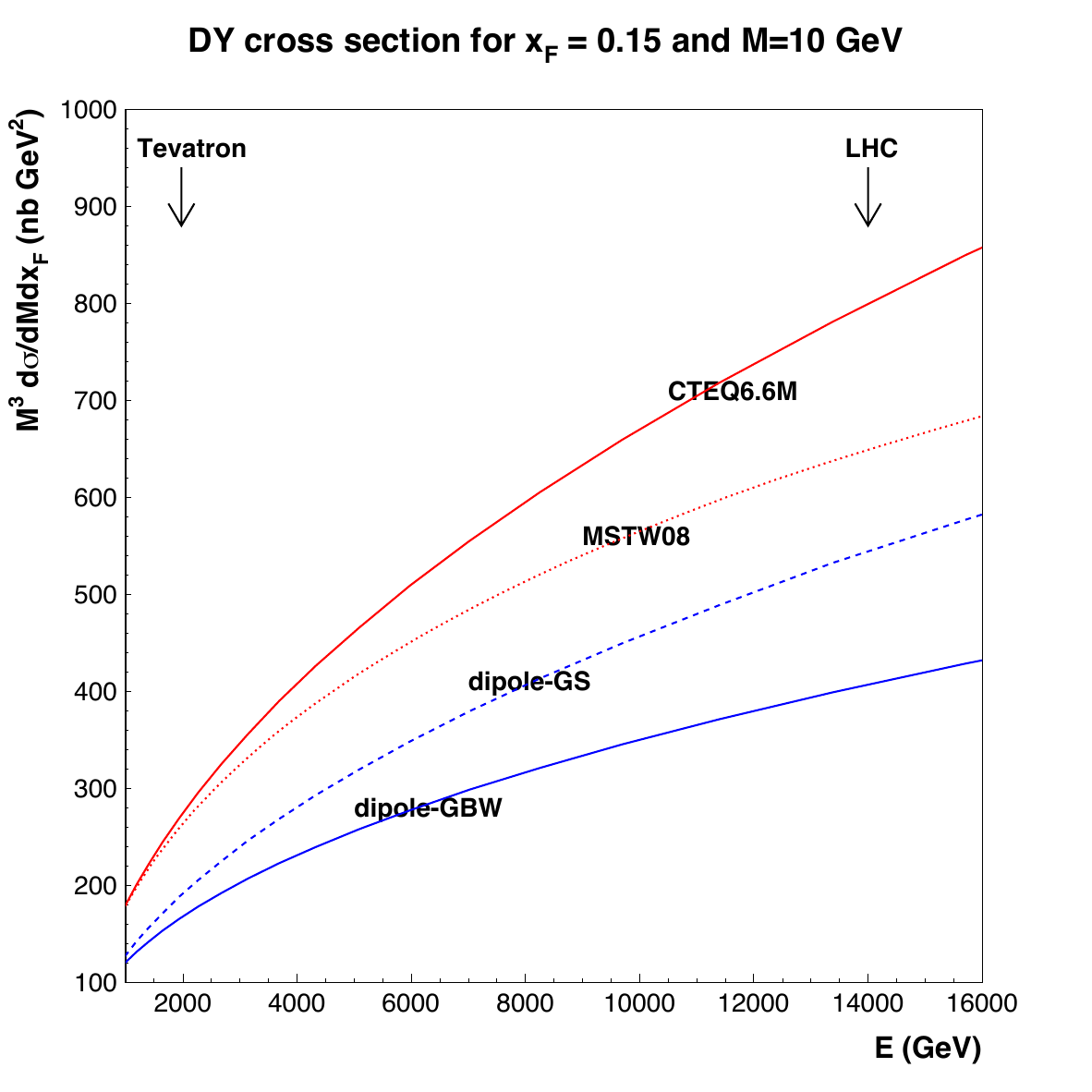}
\caption{The Drell-Yan cross section from the collinear and dipole approaches for fixed ${x_F=0.15}$ and lepton pair mass ${M=10}$ GeV.}
\label{dy5}
\end{center}
\end{figure}

Fig.~\ref{dy3}-right presents predictions for the Drell-Yan cross section as a function of the center-of-mass energy $\sqrt{s}$ at fixed ${x_F=0.15}$ for three values of the lepton pair mass ${M=6,8,10}$ GeV. At the LHC energy, saturation effects in the dipole model give results which are significantly below the collinear factorization predictions. 
The same results are shown using the linear scale in Fig.~\ref{dy5}.  The CTEQ6.6M and MSTW08 parton distributions,  and the GBW and BGKS \cite{Bartels:2002cj,GolecBiernat:2006ba} dipole models are used in these plots. The CGC model (\ref{cgc}) gives results which are very close to the GBW lines. 

Thus, the predictions from the dipole approach with gluon saturation give a significant suppression of the Drell-Yan production cross section  in comparison to the collinear factorization results.

\subsection{Forward Drell-Yan production - Further prospects in the collinear approach}
%  ({\it Emmanuel de Oliveira$\,\,^{\dagger}$})}

%\let\thefootnote\relax\footnote{$^{\dagger}$ \it Durham, UK}

Drell--Yan process is one of the standard observables used in parton distribution function determination. It has been measured in LHC at Atlas \cite{Aad:2011dm, Aad:2013iua, Aad:2014qja}, CMS \cite{CMS:2011aa, Chatrchyan:2011cm, Chatrchyan:2013tia, CMS:2014hga}, and LHCb~\cite{LHCb:2012fja, Aaij:2012vn, Aaij:2012mda}. With this new very high energy data, new information about the parton distribution can be obtained. In particular, Drell--Yan is most sensitive to quark and antiquark parton distributions. In fact, most of the major parton distribution function parametrizations are already taking steps to include these data in their analyses.

In collinear factorization, one can schematically write the Drell--Yan cross section as (cf.(\ref{eqDY1})):
\begin{equation}
\frac{d \sigma}{d^3p} = \int d x_1 d x_2 \, \text{PDF}(x_1, \mu_F) \,|\mathcal{M} (p; \mu_F, \mu_R)|^2 \, \text{PDF} (x_2,\mu_F). 
\end{equation}
In the above equation we have the factorization scale $\mu_F$, renormalization scale $\mu_R$, and a sum over the different flavours of PDF is implied.
The partons that take part in the process carry longitudinal momentum fraction given by:
\begin{equation}
x_{1,2} = \frac{m_\text{hard}}{\sqrt{s}} \exp (\pm y)
\end{equation}
where $m_\text{hard}$ is the subprocess mass and $y$ is its rapidity (at leading order $m_\text{hard} = M$, i.e., the dilepton mass). The centre of mass energy is given by $\sqrt{s}$.

At forward rapidities, say, $y=4$, for LHC compatible energy and dilepton mass of 5 GeV, one would probe the PDFs at $x_1 \approx 10^{-2} $ and $x_2 \approx 10^{-5}$, an uncharted region in $x_2$. Unfortunately, for such a small $x$ the parton distributions are very sensitive to the factorization scale. Another point is that one cannot neglect the possibility that some saturation and multiple parton interactions take part in the process.

The factorization scale uncertainty is related with many gluon emissions during the evolution. The $g\rightarrow g$ DGLAP splitting function:
\begin{equation}
P_{g\rightarrow g}(z<1) = 2 C_A \left[\frac{1-z}{z} + \frac{z}{1-z} + z (1-z) \right]
\end{equation}
has a divergence $1/z$. Therefore, for small $x$, when the factorization scale increases, many more gluons can be emitted in the DGLAP ladder, rapidly increasing the cross section. Ideally this would be compensated by the matrix element, however at next-to-leading order the matrix element can compensate only one gluon emission. In the view of this it is possible to understand the huge factorization scale uncertainty observed. 

In Ref.\ \cite{deOliveira:2012ji} a method to reduce the factorization scale uncertainty was developed. The idea was to fix the factorization scale of the LO contribution based on the known NLO matrix element, determining the optimal scale at which all NLO contribution is already included in the DGLAP ladder at small $x$. The result of such calculation pointed that $\mu_F = 1.4 M$ is the optimal scale. This choice was shown to greatly reduce the factorization scale dependence of the cross section, contributing for a better convergence of the perturbative series. Given that, observations of this process at the LHC can make a direct measurement of parton distribution functions (PDFs) in the low x region, $x < 10^{-4}$.

Unfortunately, LHC experiments do not have very good precision for such a small dilepton mass. Therefore one would like to probe a little larger dilepton mass ($M \approx 20$ GeV) at forward rapidities ($y > 3$). To do so while still probing PDFs in the important low-scale, low-$x$ domain, a cutoff in the dilepton transverse momentum can be introduced~\cite{deOliveira:2012mj}. The act of introducing a cutoff in the matrix element has to be matched by the inclusion of corresponding Sudakov form factors in the parton distribution evolution, as detailed in Ref.~\cite{deOliveira:2012mj}. Taken into account that, the calculation of the optimal scale to reduce factorization scale uncertainty was redone, now with the cutoff. Therefore, a LHC measurement of such distribution is a direct measurement of an uncharted region of the PDFs.

%-------------------------------------------------------------------------------------
While successful, collinear factorization for the Drell--Yan process only takes into account the leading twist terms. Using the dipole formulation in Ref.~\cite{GolecBiernat:2010de}, it was possible to include higher twist effects, as well as saturation effects. Saturation is expected to happen at sufficiently high energies, however the exact line where it becomes indispensable is not known. In Ref.~\cite{GolecBiernat:2010de} it was shown that, at forward rapidities, leading twist is a good approximation as long as the dilepton masses are larger than $M\approx 6$ GeV. For lower masses, the full twist resummation is necessary. Therefore, the observation of such low mass dileptons could test the boundaries of saturation and higher twist effects.

\begin{figure}[t]
\begin{center}
\includegraphics[width=9.9cm]{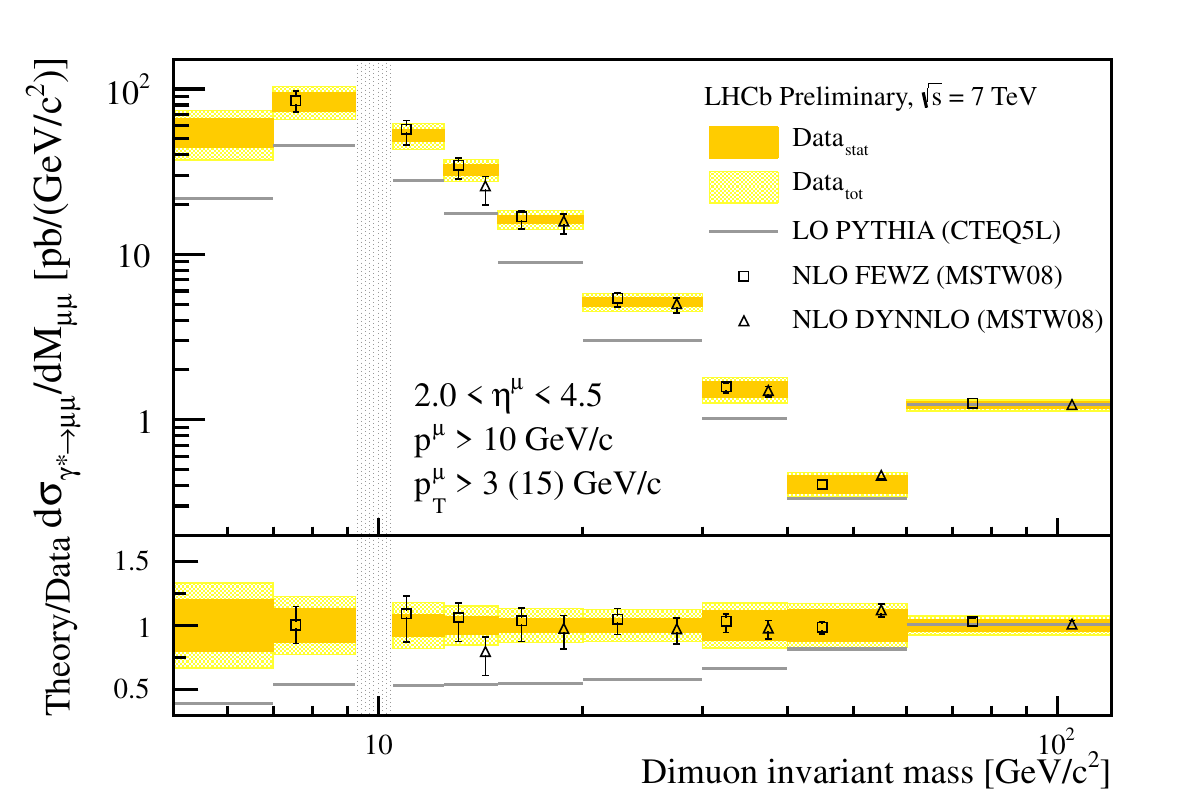}
\caption{Differential cross-section for $\gamma' \to \mu\mu$ as a function of $M_{\mu\mu}$. The dark shaded (orange) bands correspond to the statistical uncertainties, the light shaded (yellow) band to the statistical and systematic uncertainties added in quadrature. Superimposed are the PYTHIA predictions and the NLO predictions from FEWZ and DYNNLO; they are displaced horizontally for presentation. The shaded vertical band corresponds to the mass region of the Υ which is not included in the measurement. The uncertainties of the NLO predictions contain the PDF uncertainties evaluated at the 68\% confidence level and the theoretical errors added in quadrature. The two bins with $M_{\mu\mu}$ > 40 GeV/c have a cut of $p_{\mu T}$ > 15 GeV/c for the data and the predictions. The lower plot shows the ratio of the predictions or the uncertainties to the data.}
\label{LHCb-8}
\end{center}
\end{figure}

LHCb ~\cite{LHCb} has measured the Drell Yan cross-section at $\sqrt{s}=7$ TeV
differentially as functions of rapidity and mass down to 5 GeV with a small
data sample corresponding to an integrated luminosity of just 37pb-1. 
Within the limited statistical precision consistency is seen with both FEWZ
and Pythia predictions (Fig.~\ref{LHCb-8}).  An analysis of the full Run 1 dataset of 3fb-1 will allow significant improvements to the measurement precision and could allow
discrimination between theoretical models that include higher order twist or
saturation effects.

%-------------------------------------------------------------------------------------
From the theoretical point of view, everything that was done for $pp$ collisions can be done as well for heavy ion collisions. In particular, for $pA$ collisions there is a good possibility of analyzing nuclear parton distributions when one can disregard final state effects. One could study the same problems as before, except that saturation, multiple parton scattering and higher twist effects are much more important in the $pA$ setup. In this case, backward and forward rapidities are different, but both are interesting. For backward rapidities, one is probing the nucleus at high $x$, but higher twist effects from the interaction between multiple nucleons should be present. For forward rapidities, one has small $x$ in the nucleus and saturation effects should be important, on top of the fact that these nuclear distributions are now well known in this region. Higher twist effects should play a role as well, since for small $x$ the effect of nuclear shadowing is related to multiple nucleon interactions.

%-------------------------------------------------------------------------------------
As discussed above, the LHC can contribute to the determination of the parton distribution functions in the low $x$ and low scale region. Of course one could imagine an upgrade in the detectors to fully account for small mass dileptons in a run with low luminosity. However, it would be much easier to keep the experiments as they are and just have some analysis for more exclusive observables. Instead of only working with the integrated dilepton distribution, one could do the necessary cuts to guarantee that the partons are probed at small $x$. In this context, LHCb has a very promising potential with its geometry that covers forward rapidities.

\subsection{Forward Drell-Yan production - Further prospects in the small-x approach}
% ({\it M.B. Gay Ducati, M.T. Griep and M.V.T. Machado$\,\,^{\dagger}$})}

%\let\thefootnote\relax\footnote{$^{\dagger}$ \it High Energy Physics Phenomenology Group, GFPAE  IF-UFRGS, Caixa Postal 15051, CEP 91501-970, Porto Alegre, RS, Brazil}

The current LHC detector configurations can explore small-$x$ hard phenomena with nuclei and nucleons at photon-nucleon
center-of-mass energies TeV scale, extending the $x$ range of HERA by a factor of ten \cite{baltz}. The LHC is in the kinematic range
where nonlinear effects are several times larger than at HERA.
In these regions,
dileptons production in hadronic collisions (Drell-Yan process) can be used to investigate the limit of high partonic density,
 since this process probes the gluon distribution through QCD Compton process. In particular, the Drell-Yan transverse
momentum ($p_{T}$) distribution can be extended to be sensitive to saturation effects.\
The Drell-Yan (DY) process in the kinematical region where the dilepton mass M is small compared to the center of mass energy $\sqrt{s}$
is of similar theoretical interest as deep-inelastic scattering (DIS) at low Bjorken-$x$. Both processes probe the target at high gluon
density where one expects to find new physics. In contrast to DIS, where only the total cross section can be measured, there is a variety
of observables which can be measured in the DY process, such as the transverse momentum distribution or the angular distribution of the
lepton pair.\

The Drell-Yan (DY) process cross sections have been proven to still
 fulfill the factorization property and are finite to first  orders in perturbation theory at 
sufficiently large transverse momenta, $p_T$. The conventional factorization approaches 
to the DY process give divergent results at $p_T\rightarrow 0$, but the low $p_T$ regions
are treated in an extensive program of research. (see Ref. \cite{Kang} and references therein).
The differential cross section in the region 
$p_T\geq M_{\ell\ell}/2$ is driven by subprocesses initiated by incident gluons, and
 massive lepton-pair differential cross sections can be used to constrain the gluon 
density \cite{Klasen1}.
The DY cross section with large values of dileptons transverse
 momentum is related to deep-inelastic-lepton scattering (DIS), prompt 
photon production and jet production as an important probe of
short-distance hadron dynamics.
The production of dileptons in DY process in nuclear targets can help to constrain the parton
distribution functions (PDFs) in the nucleons and are colorless probes of the dynamics 
of quarks and gluons \cite{Ayala}.
Namely, they escape through
the colored medium of the high-energy collision. The dileptons interact with the medium only electromagnetically, thus they can be a powerful probe of
 the initial state of matter created in heavy ion collisions.
Refs. \cite{betemps,jamal_tam} show that those electromagnetic 
probes are crucial to determine the dominant physics in the forward region at RHIC and at the LHC. 
\\
Direct (prompt) photon production and Drell-Yan 
dilepton pair production processes can be described within the same color dipole approach without 
any free parameters \cite{Kope_gamma}.
Such a formalism, developed in \cite{bb} for the case of the total and 
diffractive cross sections, can be also applied to
radiation \cite{hir,brodsky}. In the target rest frame, the DY process looks like a 
bremsstrahlung \cite{Kopeliovich:2000fb}  of a massive photon from an incoming quark.
 The photons can be emitted before or after a quark to be scattered on a proton.
The cross section can be expressed through the more elementary cross section
$\sigma_{dip}$ of the interaction of a $Q\bar{Q}$ dipole \cite{Kopeliovich:2000fb},
 although  no real quark dipole participates in the process of 
electromagnetic bremsstrahlung by a quark.
The relation
 between this formalism and the usual collinear pQCD factorization has been studied in details 
in Ref. \cite{Raufeisen:2002zp}.  The dipole formalism offers an easy way to calculate the
 transverse momentum distribution in DY processes even in the low-$p_T$ region.
 This contribution investigates the low mass DY cross section at the LHC energies using color 
dipole approach, discussing several phenomenological aspects. The main focus is at forward 
rapidities at the energy available at the LHC.

\begin{figure}[t]
\begin{center}
\includegraphics[scale=0.4]{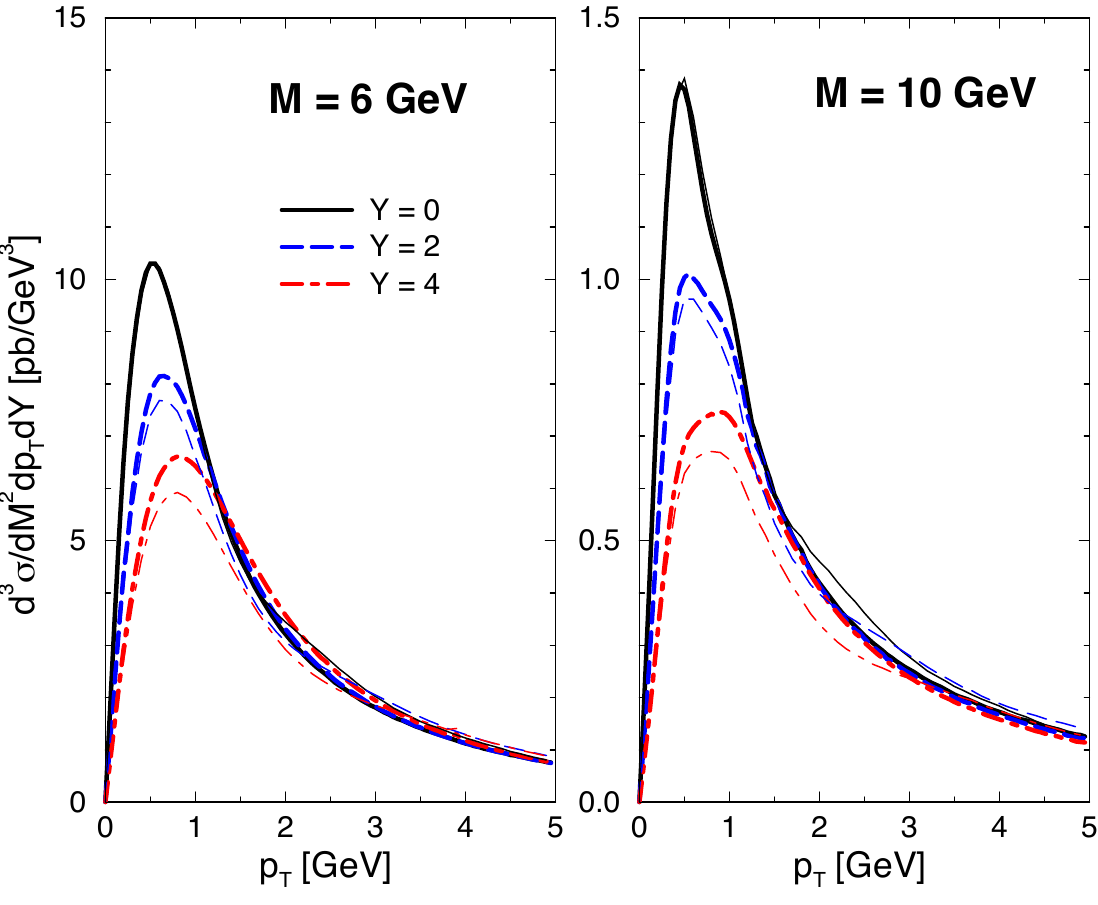}
\end{center}
\caption{Low mass DY differential cross sections, $d^3\sigma /dM^2dydp_T$, as a function of dilepton transverse momentum, $p_T$, at energy of $\sqrt{s}=7$ TeV. The plots are shown for fixed dilepton mass ($M=6$ and $10$ GeV) and distinct lepton pair rapidities ($y=0,2,4$). The results are presented using the GBW dipole cross section (bold curves) and the CGC dipole cross section (thin curves).}
\label{fig:1}
\end{figure}

 The Fig.~\ref{fig:1}, presents the results for the differential cross section, 
 $d^3\sigma /dM^2dydp_T$ (in units of pb), as a function of the dilepton transverse momentum $p_T$.
The bold curves present the predictions using the GBW dipole cross section and the thin curves
present the predictions using the CGC dipole cross section. The hard scale considered is
 $\mu^2 = (1-x_1)M^2+p_T^2$. The $p_T$-spectrum is quite sensitive to the particular model of dipole cross section 
(specially at large transverse momentum) as it depends on the behavior of effective anomalous
 dimension as discussed in the previous section.
 The left panel shows the case for fixed
 invariant mass $M=6$ GeV and for sample values of dilepton rapidity including central and forward
 rapidities, i.e. $y=0,2$ and 4, respectively. The same notations hold for the right panel,
 where now the invariant mass is $M=10$ GeV. As expected, the large rapidity cases give smaller 
cross sections and the peak on the distributions is shifted to larger values on transverse momentum.
 In the kinematical situation investigated here the peak is located at momentum around $p_T\approx 1$
 GeV. 

\begin{figure}[t]
\begin{center}
\includegraphics[scale=0.4]{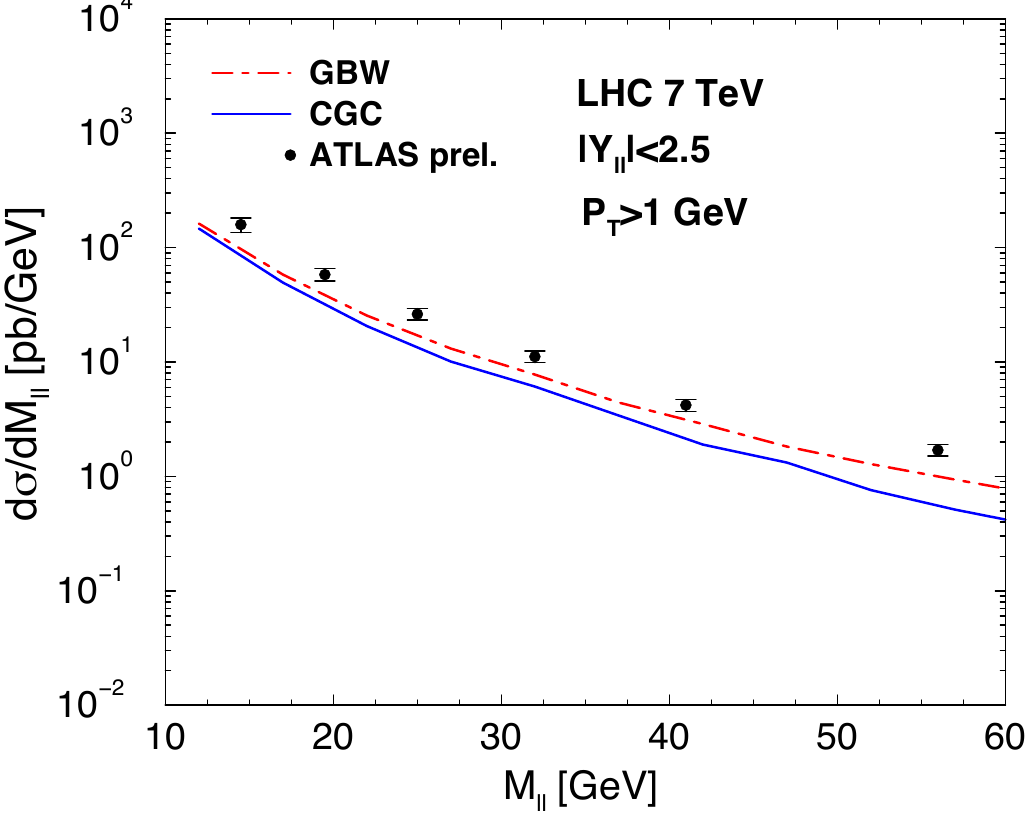}
\end{center}
\caption{Invariant mass distribution in the range $12< M_{\ell\ell}< 60$ GeV. The imposed cuts at energy of $\sqrt{s}=7$ TeV are lepton pair rapidities $|y_{\ell\ell}|<2.5$ and dilepton transverse momentum $p_T\geq 1$ GeV. Preliminary ATLAS data  \cite{ATLAS} are shown for sake of comparison.}
\label{fig:2}
\end{figure}
 
The Fig.~\ref{fig:2} presents the invariant mass distribution at midrapidities considering the GBW 
model (dot-dashed line) and  the  phenomenological saturation model,
 labeled here CGC (solid line), which involves a running anomalous dimension. 
In the large $p_T$ region occurs the main deviation between these two models,
which gives distinct overall normalizations for the dilepton invariant mass distribution.
The considered cuts are 
 presented by the A\-TLAS analysis \cite{ATLAS} for low mass Drell-Yan di-muon process. 
The selection cuts on that analysis at energy of $\sqrt{s}=7$ TeV and integrated luminosity of 
36 pb$^{-1}$ were low muon transverse momentum, $p^{\mu}_T>6$ GeV and low di-muon mass region
 $12< M_{\ell\ell}< 66$ GeV. Here, we consider the integration over the boson rapidity in the range
 $|y_{\ell\ell}|<2.5$ and dilepton transverse momentum $p_T\geq 1$ GeV. Distinct $p_T$ cuts will
 lead to a different overall normalization for the invariant mass distribution. At this stage we did
 not impose the selected cuts on individual muons as done by ATLAS analysis.
 The results presented here are somewhat
 consistent with the extrapolated Born level differential cross section using the symmetric analysis.
 For sake of comparison, we include the preliminary data \cite{ATLAS} in Fig. \ref{fig:2} 
(filled circles).

\begin{figure}[t]
\begin{center}
\includegraphics[scale=0.4]{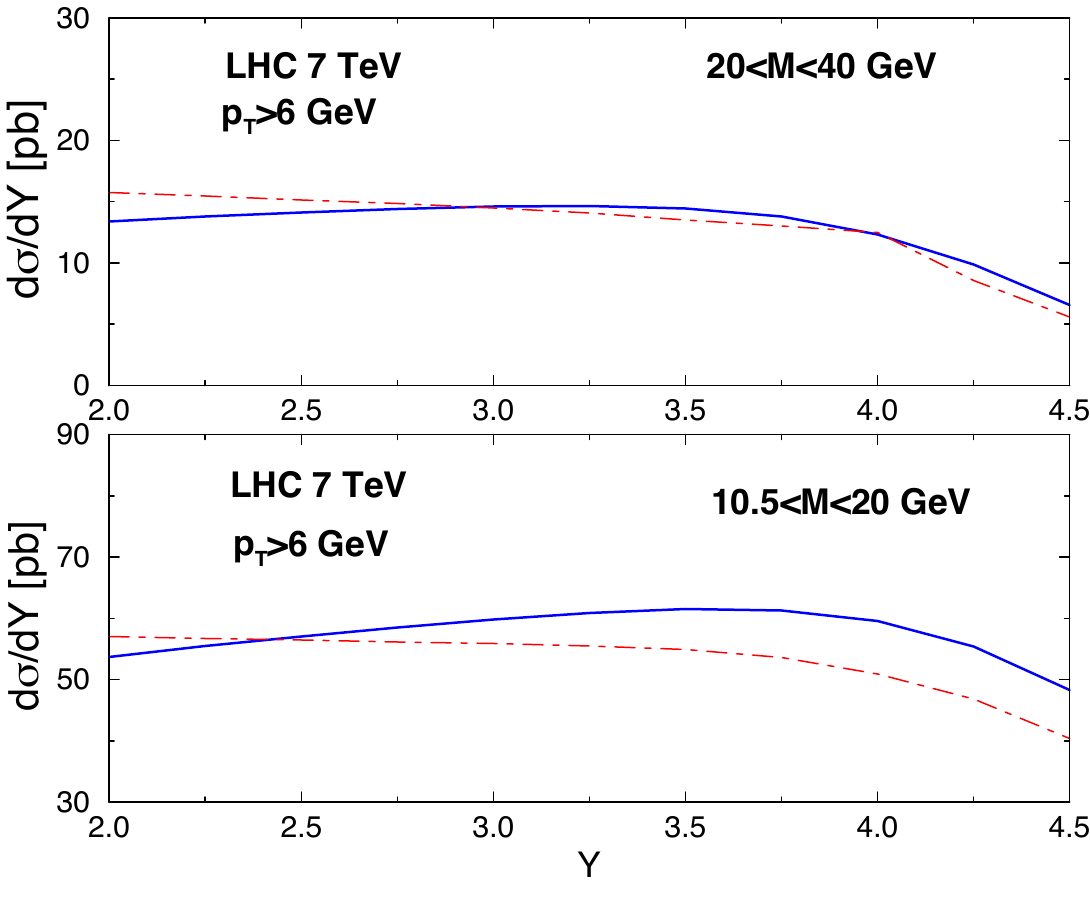}
\end{center}
\caption{The dilepton rapidity distribution at $\sqrt{s}=7$ TeV imposing the cut on dimuon transverse momentum $p_T>6$ GeV and two invariant mass regions: (upper plot)  $20\leq M_{\ell\ell}\leq 40$ GeV and (lower plot) $10.5\leq M_{\ell\ell}\leq 20$ GeV.}
\label{fig:3}
\end{figure}

\begin{figure}[h]
\begin{center}
\includegraphics[scale=0.35]{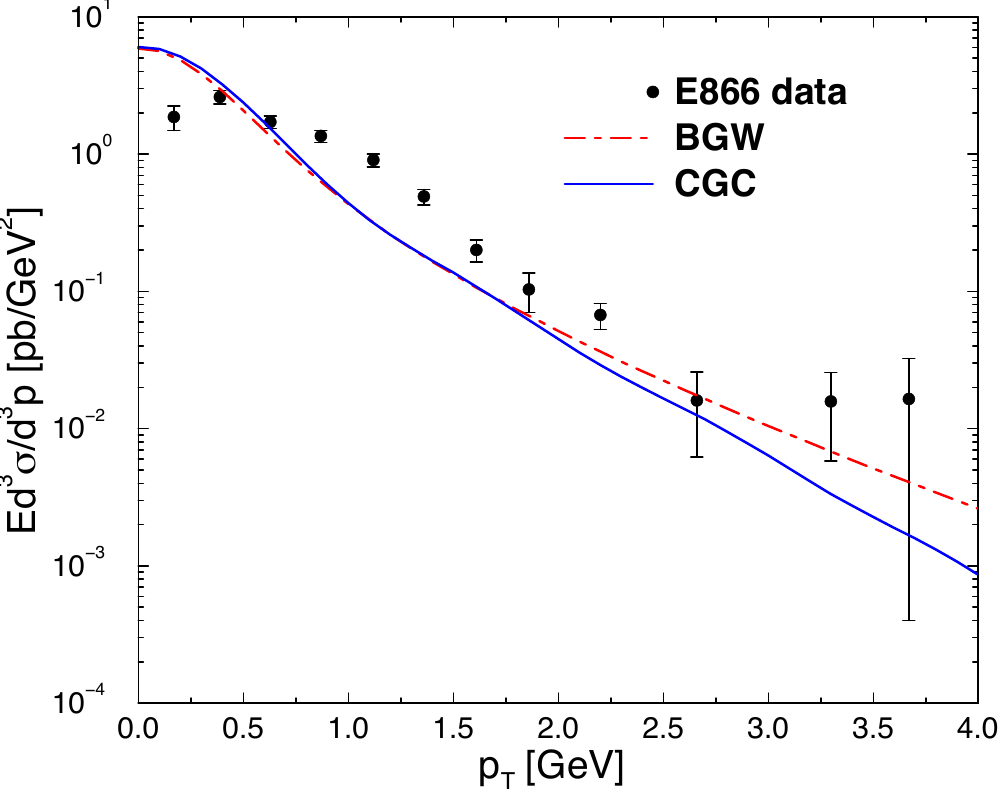}  \includegraphics[scale=0.35]{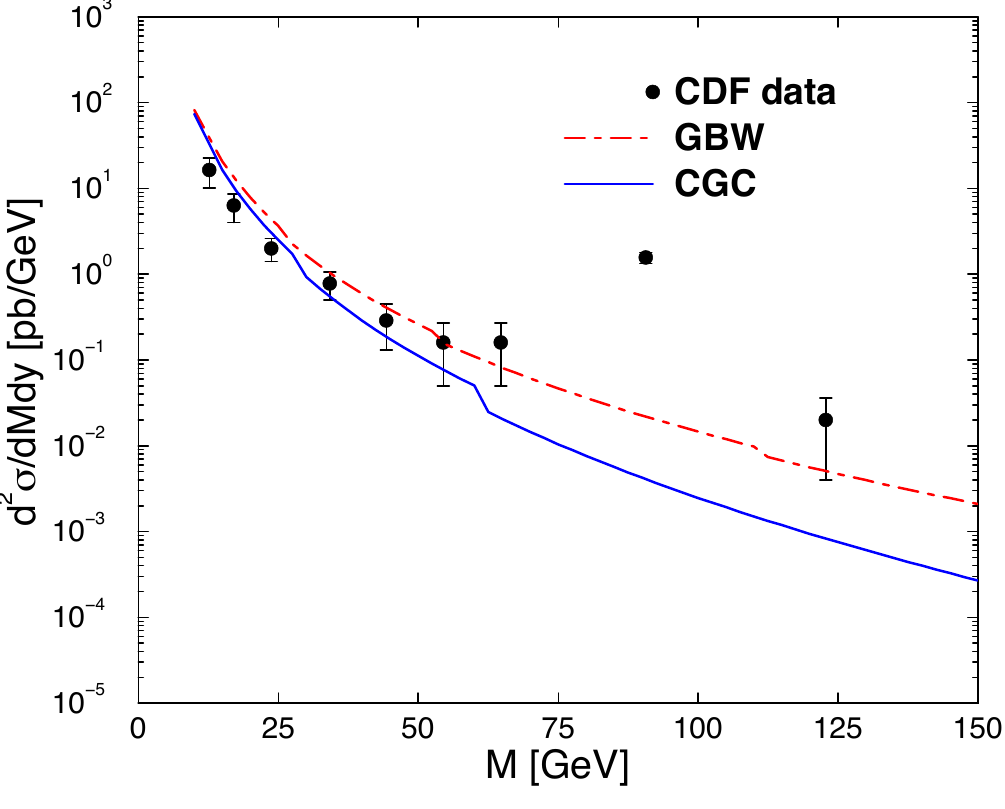}
\end{center}
\caption{The DY invariant cross section (left panel) at $\sqrt{s}= 39$ GeV as a function of $p_T$ compared to the E866 Collaboration data \cite{E866}. In the left panel, the differential cross section $d^2\sigma /dMdy$ at $\sqrt{s}=1800$ GeV as a function of invariant mass is presented and compared to the CDF Collaboration data \cite{CDFDY}.}
\label{fig:4}
\end{figure}

The rapidity distribution, 
$d\sigma/dy$, is computed for the interval $2<y<4.5$ in Fig.~\ref{fig:3}. The
 phenomenological models considered here are the same as the previous plot and the same notation 
was used. The hard scale, in this case, is
 $\mu^2 = \frac{1}{2}[(1-x_1)M^2+p_T^2]$. 
The distinct anomalous dimension in the models causes the deviations between the predictions using
distinct models.
%since the rapidity distribution is driven that.
The cut imposed for the dilepton transverse momentum is $p_T>6$ GeV and two distinct intervals of invariant
 mass are considered. In the upper plot one has  $20\leq M_{\ell\ell}\leq 40$ GeV whereas in the lower 
plot one has $10.5\leq M_{\ell\ell}\leq 20$ GeV. The cut motivation is the recent
 LHCb collaboration \cite{LHCb} measurement of low mass DY cross section. The measurement collected 
with an integrated luminosity of 37 pb$^{-1}$ are for the di-muon final state having muons within 
pseudorapidities of 2 to 4.5, muon transverse momentum $p^{\mu}_T>3$ GeV  ($p^{\mu}_T>15$ GeV for
 higher masses) in two distinct mass regions. In the forward rapidities considered here,
 the saturation scale is in the interval
 $0.6 \leq \langle Q_{\mathrm{sat}}^2 \rangle \leq 1.2$ GeV$^2$  for
  $\langle M_{\ell\ell} \rangle \simeq 15.25$ GeV. Slightly lower  values are found also for higher
 mass $\langle M_{\ell\ell} \rangle \simeq 30$ GeV.  Fig.\ref{fig:4} (upper panel) shows the invariant cross section as a function of
 $p_T$ at energy $\sqrt{s}= 39$ GeV. The experimental results from the E866 Collaboration 
 \cite{E866} are also presented
 ($\langle x_F \rangle \simeq 0.63 $ and $4.2 \leq M_{\mu^+\mu^-}\leq 5.2$ GeV). 
The bottom panel shows the differential cross section $d^2\sigma / dMdy$ (for $|y|<1$) for
 the energy $\sqrt{s}=1800$ GeV as a function of dilepton invariant mass. The data from CDF
 Collaboration \cite{CDFDY} are included in the plot, considering also the large invariant masses
 data points. The solid curves refer to CGC and dot-dashed curves to GBW dipole cross section,
 respectively. The color dipole picture reasonably describes the cross section from low to high
 energies in the kinematical regions where it is expected to be valid. The approach is also somewhat
 consistent with calculations carried out in next-to-leading order QCD at both fixed target and 
collider energies \cite{Klasen1}. 

As a summary, the main physics motivation for DY studies at the LHC are: extraction of PDFs in
extended kinematics regions (high sensitivity to PDFs) and the saturation region study. This is due
to the rather large cross sections (statistics) expected, the clear experimental signature and no
uncertainties from fragmentation function.
 The low mass DY  production can be addressed in the color dipole picture
without any free parameters by using the dipole cross sections determined from current  phenomenology
in DIS. 
It has been shown before \cite{Mariotto} that saturation physics is not directly relevant
for RHIC at midrapidity, by considering in the kinematic
range of data the saturation scale from the GBW model 
$Q^{2}_{sat}=(x_{0}/x_{2})^{\lambda}=(x_{0}\sqrt{s}e^{y}p_{T})^{\lambda}$, getting $0.1\leq Q^{2}_{sat}\leq 0.5\,\, GeV^{2}$, which is very small compared to the transverse momenta
$4\leq p^{2}_{T} \leq 100\,\,GeV^{2}$. Therefore, saturation effects do not play an important 
role at RHIC midrapidity.
The same statements about the role played by saturation effects remain valid for Tevatron at
midrapidity, where the saturation scale is in the range $0.2\leq Q^{2}_{sat}\leq 0.8\,\, GeV^{2}$.
In the forward rapidities considered here, the saturation scale is in the interval 
$0.6\leq Q^{2}_{sat}\leq 1.2\,\, GeV^{2}$ for $\langle M_{ll}\rangle \simeq 15.25\,\,GeV $.
Thus, in the LHC this situation can be changed even at midrapidities as the saturation scale
is enhanced by a sizable factor. So, the suppression of the DY cross section due to saturation effects can be large. 
The LHC opens a new kinematic regime at high energies, where QCD evolution leads
to the fast growth of the gluon density. At these high densities it is possible that the novel
phenomena related to the nonlinear dynamics of the gluon fields will occur, and DY production
offers high sensitivity to the parton distribution in the hadron. 
Therefore, the LHC experiments will be important to study the production of low-mass Drell Yan
with the goal to achieve higher PDF's precision.

\subsection{Forward photon production and gluon saturation - theoretical overview and measurement proposal}
%  ({\it Thomas Peitzmann$\,\,^{\dagger}$})}\let\thefootnote\relax\footnote{$^{\dagger}$ \it Utrecht University, Utrecht, Netherlands}
\subsubsection{Gluon saturation and photon production}

Gluon saturation should affect the total multiplicity of produced particles in high-energy collisions. Furthermore, the predicted scaling properties of the momentum scales should lead to a behaviour called geometric scaling. Both the multiplicities and the scaling properties of particle distributions have been measured and compared to saturation models. The results appear to be consistent with expectations from the models, but unfortunately these observables are not specific enough to provide a proof for gluon saturation.

More discriminative power may come from more detailed studies of transverse momentum distributions and from two-particle angular correlations. The main interest lies here in the comparison of particle production in pp and p--A collisions, as saturation effects should be stronger for the higher gluon density in nuclei. Specifically, one expects that
\begin{itemize}
\item the nuclear modification factor 
\begin{equation}
R_{pA} \equiv \frac{\slfrac{dN}{dp_T} (p+A)}{\left\langle N_{\mathrm{coll}} \right\rangle \slfrac{dN}{dp_T} (pp)}
% R_{pA} \equiv \frac{\nicefrac{dN}{dp_T} (pA)}{\left\langle N_{\mathrm{coll}} \right\rangle \nicefrac{dN}{dp_T} (pp)}
\end{equation}
should show a suppression of particle production $R_{pA} < 1$ in a characteristic $p_T$ range, and
\item the jet-like peak at $\Delta \phi = 0$ usually observed in two-particle correlations in pp collisions should be modified (weakened and/or broadened) in nuclear collisions.
\end{itemize}
The interesting kinematic range is defined by the values of Bjorken $x$ and the corresponding saturation scale $Q_s$. The relevant production processes will be affected by saturation when $Q < Q_s$, which calls for small to intermediate momentum probes. However, as one would like to use a calculable probe as a reference, this excludes too low momentum transfers -- ideally one would want to study momentum ranges, where perturbative QCD should be applicable. To access small $x$, which for leading-order processes on the parton level can be approximated as $x \approx \slfrac{2 p_T  \exp(-y)}{\sqrt{s}}$, particle production at large (i.e. \textit{forward}) rapidities should be studied.

Transverse momentum spectra and angular correlations for neutral pions at forward rapidity have been studied in pp and d--A collisions at RHIC, and a suppression has been observed in the nuclear modification factor  \cite{Ada06,Ars04} and a suppression and broadening in the angular correlation \cite{Bra10,Ada11}. However, the transverse momenta studied are still very small (on the order of 1~GeV or only slightly higher), a momentum range where particle production is anyhow not well understood. In addition, the relation between the kinematic variables in the final state are only weakly correlated to the parton kinematic parameters due to fragmentation and possibly other final state modifications of particle production.

Similar studies at LHC should allow to study both higher $Q^2$  and smaller $x$ contributions, thus should be able to use well-defined particle production processes, while still being sensitive to saturation as the saturation scale should be much larger at the lower $x$ values.
First measurements of hadron production at forward rapidities have been performed at the LHC, but results are not conclusive. The production of $\phi$ mesons shows a strong suppression in p+A collisions compared to pp \cite{Ura13}. There is also a suppression of $J/\psi$ production \cite{LHCb-jpsi,Abe14}, which is consistent with calculations using shadowing and final state energy loss. A CGC calculation predicts a stronger suppression than seen in the data, however this calculation has a number of uncertainties related to the coupling of the $J/\psi$ to the gluon field, and it does not use a state-of-the-art CGC implementation. In general, also at LHC, hadron production will most likely not provide an ideal probe because of final state modifications and their uncertainties, which will obscure the kinematics.
 
\begin{figure}[htb]
\begin{center}
\includegraphics[width=0.7\textwidth]{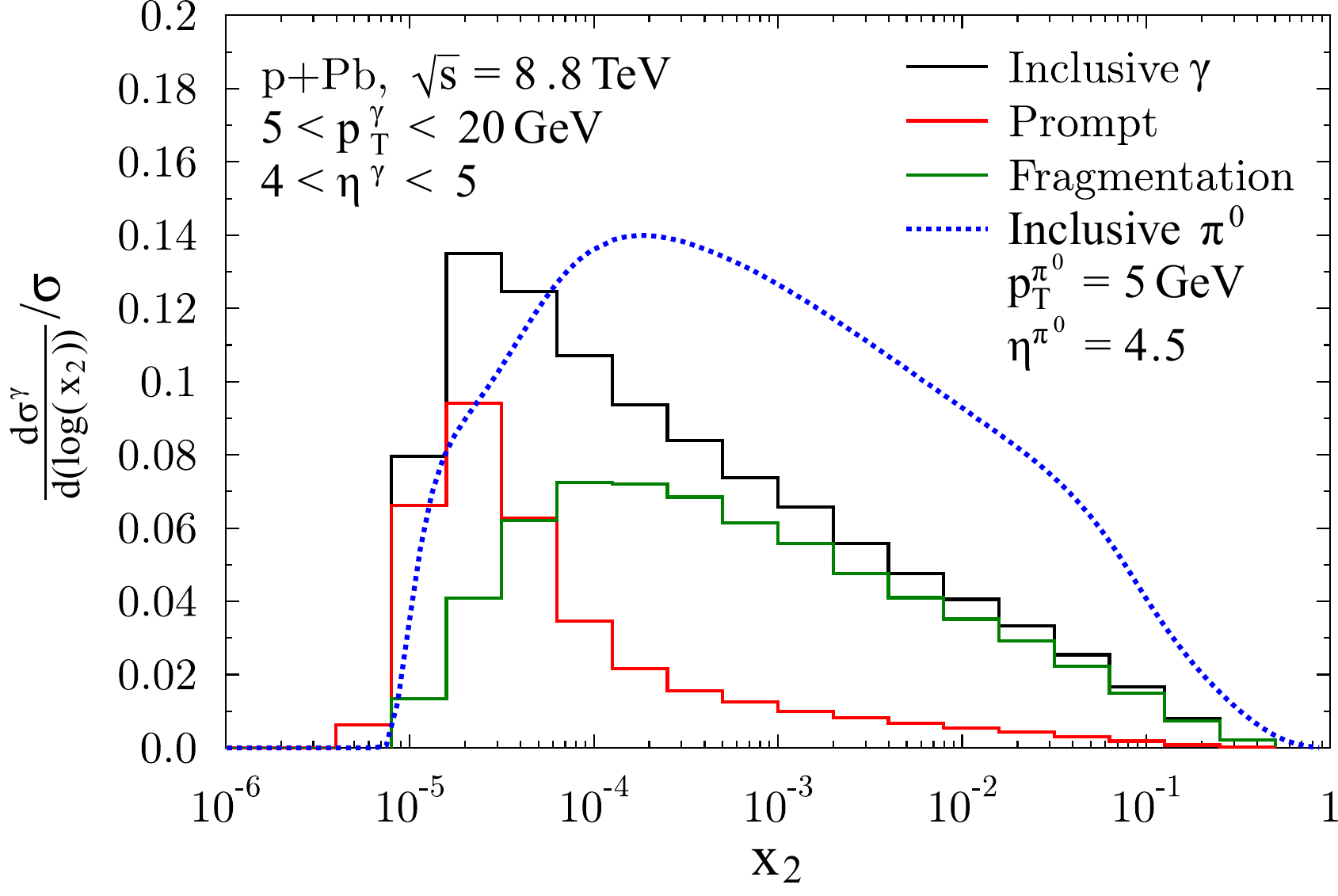}
\caption{Distribution of $x_2$ (momentum fraction of parton from the nucleus) probed in direct photon production at forward rapidity in p+Pb collisions at 8.8 TeV as calculated in JEPHOX using EPS09 structure functions in Ref.~\cite{helenius2014}. The different components of photon production are also shown separately. For comparison, the $x_2$ distribution for pion production of similar kinematics are included.}
\label{fig:xdist}
\end{center}
\end{figure}

\begin{figure}[htb]
\begin{center}
\includegraphics[width=0.7\textwidth]{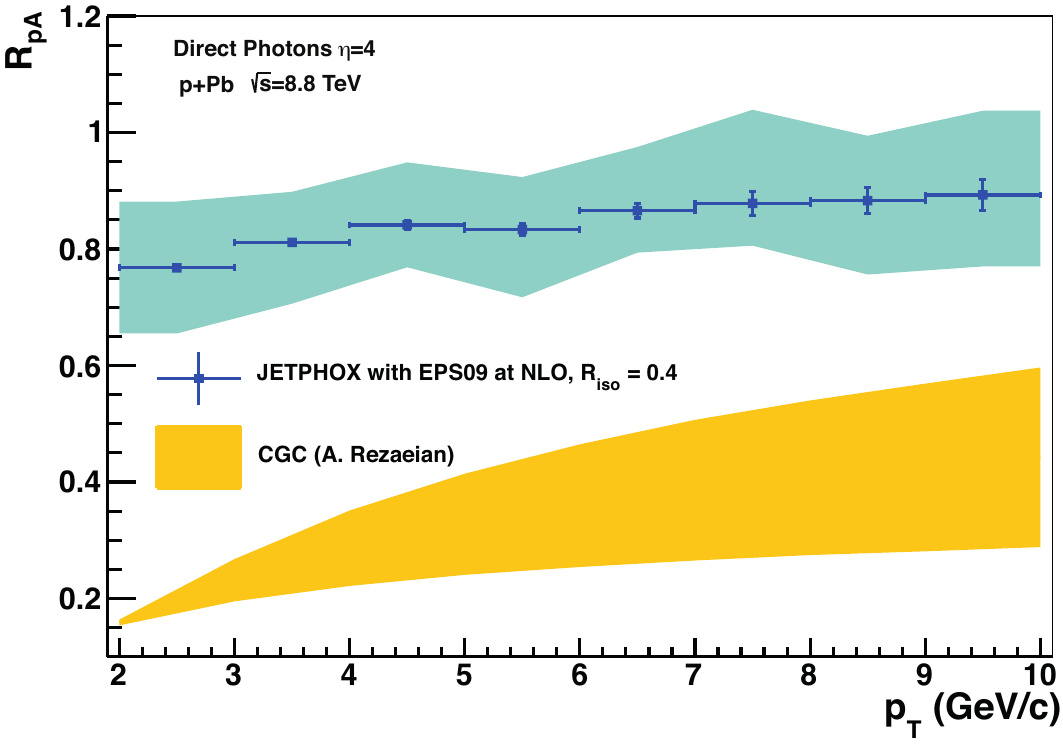}
\caption{Nuclear modification factor $R_{pA}$ as a function of $p_T$ for forward direct photon production. Shown are results of CGC calculations from Ref.~\cite{rezaeian2012} (orange) and from NLO pQCD calculations with JETPHOX (blue). The shaded bands show the systematic error estimates.}
\label{fig:rpa}
\end{center}
\end{figure}

Probes, which directly access the parton kinematics, would be strongly preferable, which points to direct photons as an ideal probe. Direct photons have a number of advantages compared to other, in particular hadronic probes:
\begin{enumerate}
\item The production processes of direct photons are well understood theoretically.
\item The leading order process (q-g Compton scattering) is directly sensitive to the gluon density.
\item Fragmentation contributions, though significant at LHC, are less important than for hadrons, and can be suppressed by isolation cuts.
\item No other strong final state nuclear modification (like e.g. energy loss) is expected.
\end{enumerate}
The advantage of photons with respect to the sensitivity to parton kinematics can be illustrated with the distribution of momentum fractions $x_2$ probed in the nucleus in p+A reactions at the LHC as displayed in Figure~\ref{fig:xdist}. The $x_2$ distribution for photon production with $4 < y < 5$ and $5 < p_T < 20 \, \mathrm{GeV}/c$ is peaked between $10^{-5}$ and $10^{-4}$, while the maximum contribution for pion production is generated from partons with $x_2$ about an order of magnitude larger. Thus, already the inclusive direct photon distribution has a clear sensitivity advantage, which can be further enhanced by applying isolation cuts on the photons, although the studies in Ref.~\cite{helenius2014} show that isolation may not be as effective as previously thought.
Another advantage of photon production is that the theoretical description also in the context of models of gluon saturation is very well understood. State-of-the-art calculations have shown a clear sensitivity to gluon saturation effects, as demonstrated in Figure~\ref{fig:rpa}, which shows the nuclear modifications factor $R_{pA}$ of direct photons from the CGC calculation in Ref.~\cite{rezaeian2012} and from a pQCD calculation at NLO with EPS09 PDFs using JETPHOX. While the pQCD prediction shows only a slight reduction of reaching $R_{pA} \approx 0.8$ at low $p_T$ related to nuclear shadowing, the CGC calculation shows a strong suppression to $R_{pA} < 0.4$ for $p_T < 5 \, \mathrm{GeV}/c$.

In principle the measurement of Drell-Yan production -- i.e. virtual direct photons -- would provide an alternative means of assessing low-$x$ parton distributions with similar advantages, i.e. no final state modifications. The measurement is not as directly accessing the gluon distribution, however, as gluons play a role only via second order diagrams, or in the quark/antiquark PDFs via DGLAP evolution. It may be an additional complication for the interpretation to rely heavily on DGLAP evolution in theoretical predictions -- finally a search for gluon saturation effects should \textit{challenge} DGLAP evolution. Still this is likely not a major argument against using Drell-Yan. 

The major disadvantage of Drell-Yan measurements is the very low cross section compared to real photon production. Ref.~\cite{LHCb-DY} shows a measurement of forward Drell-Yan muon pairs in pp collisions at 7~TeV from LHCb. For the low mass range relevant for this discussion ($5 < M < 7 \, \mathrm{GeV}$) they quote a statistical error of $\approx 20 \% $ for the rapidity-integrated measurement in a sample of 37~pb$^{-1}$. A low-mass rapidity-differential measurement would not be possible from this sample. The situation is considerably worse for p+Pb collisions, where an integrated luminosity of 50~nb$^{-1}$ (corresponding to a nucleon-nucleon-equivalent luminosity of $\approx$10~pb$^{-1}$) is considered reasonable. Measurements of Drell-Yan production will therefore not be competitive to those of real photon production.

\subsubsection{A new detector for measurements of forward direct photons in ALICE}

A detector upgrade with a calorimeter at forward rapidities (FoCal) to measure forward direct photon production is currently being discussed in the ALICE collaboration \cite{Pei13}. This detector would be intended to measure direct photons, electrons/positrons and jets for rapidities $\eta > 3$. Such a detector would offer a wealth of physics possibilities, but its main focus is on measurements related to the structure of nucleons and nuclei at very low Bjorken-$x$ and possible effects of gluon saturation.

FoCal would consist of an electromagnetic calorimeter most likely positioned at a distance from the IP of $z \approx 7 \,\mathrm{m}$ covering $3.2 < \eta < 5.3$ backed by a standard hadronic calorimeter. A distance of $z = 3.6 \,\mathrm{m}$, which corresponds to a maximum reachable pseudorapidity of $\eta = 4.5$, has also been studied in simulations. Both positions are equivalent in terms of measurement conditions such as the particle density, such that it is sufficient at this stage to not explicitly perform all studies for both positions. 
The main challenge of an electromagnetic calorimeter in this region of phase space is the requirement to discriminate decay photons from direct photons at very high energy, which will require extremely high granularity.

The design option currently under study is a SiW sandwich construction. It consists of 20 layers of a 3.5~mm W plate ($\approx 1 X_0$) interleaved with active layers with Si sensors. The active layers use two different sensor technologies: low granularity layers (LGL), which consist of sensors with 1 cm$^2$ pads summed longitudinally in segments and equipped with analog readout, and high granularity layers (HGL) based on CMOS monolithic active pixel sensors (MAPS). The MAPS will have a pixel size of a few $10 \times 10 \, \mu\mathrm{m}^2$ with internal binary readout.\footnote{The current model for MC simulations uses a pixel size of  $100 \times 100 \, \mu\mathrm{m}^2$.} On-chip processing will convert the pixel count in a \textit{macro-pixel} of $1 \, \mathrm{mm}^2$ to a pseudo-analog value.

\begin{figure}[htb]
 \begin{center}
    \includegraphics[width=0.45\textwidth]{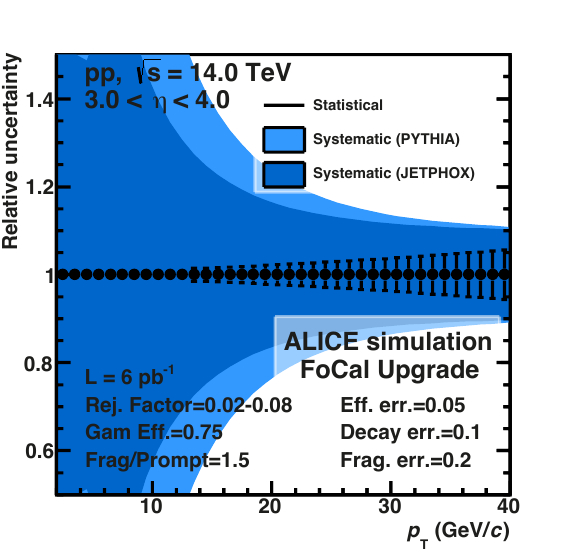}
    \includegraphics[width=0.45\textwidth]{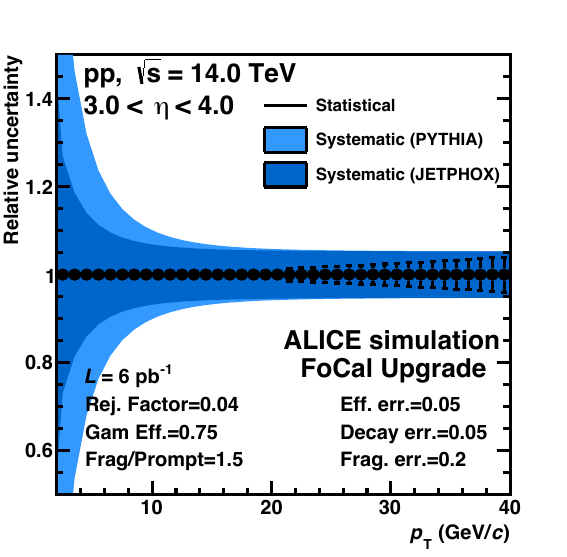}
   \caption{Estimated relative uncertainties on
  the cross section measurement for direct photon production in p+p collisions at $\sqrt{s}=14$
  TeV, based on direct photon spectra from JETPHOX (dark band)
  and PYTHIA (light band), and background spectra from PYTHIA events. Statistical uncertainties are shown
  as error bars and the systematic uncertainty is shown as a band. }
   \label{fig-errors}
 \end{center}
\end{figure}

The HGL are crucial for  $\gamma/\pi^0$ discrimination. The LGL have an effective tower width of the order of the Moli\`ere radius. Their two-shower separation power is similar to existing standard electromagnetic calorimeters.\footnote{Those conditions are in fact very similar to the ones of the electromagnetic calorimeter of the LHCb experiment.} However, the shape of electromagnetic showers allows us to make use of much finer granularity for shower separation and additional shower shape analysis for very high energy $\pi^0$, when the two photons can no longer be resolved. The impact of the granularity is shown in Fig.~\ref{fig-errors}, which shows uncertainty estimates for a direct photon measurement in pp collisions at 14~TeV. The panel on the left hand side shows the expected performance using only the LGL, while the right panel shows the performance of the full detector. A low-granularity detector would only determine the photon yield with a much larger systematic error, mainly due to the merging of $\pi^0$-decay photons. Only the high-granularity option has a good  sensitivity for such a photon measurement. While FoCal would offer coverage towards higher rapidities than other LHC experiments, it is in particular the superior granularity at these large rapidities that would give FoCal a unique advantage.

\subsubsection{Required beam times and luminosities}

\begin{figure}[htb]
 \begin{center}
    \includegraphics[width=0.45\textwidth]{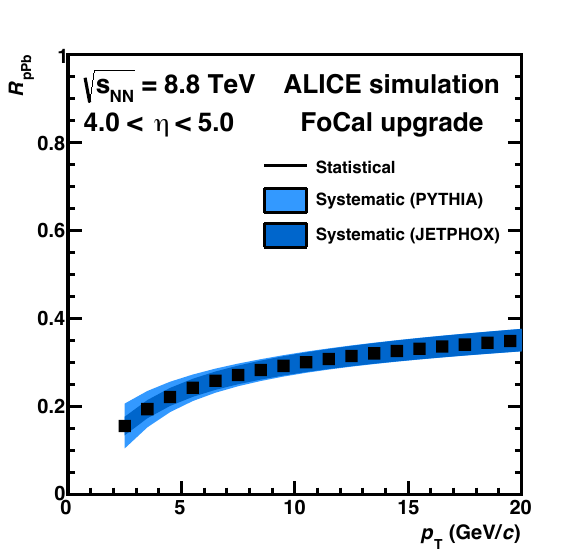}
   \caption{Estimated relative uncertainties on
 measurement of the nuclear modification factor $R_{pA}$ for direct photon production at $\sqrt{s}=8.8$
~TeV, based on direct photon spectra from JETPHOX (dark band)
  and PYTHIA (light band), and background spectra from PYTHIA events. Statistical uncertainties are shown
  as error bars and the systematic uncertainty is shown as a band. This simulation assumes a location of the detector at $z = 3.6 \, \mathrm{m}$. Results are shown using only low granularity layers ({\it left}) and for the full detector including high granularity layers ({\it right}).
         }
   \label{fig-rpa}
 \end{center}
\end{figure}

The detector upgrade would be installed in the LHC long shutdown 3 (~2024), and measurements would be performed together with the full ALICE setup. Beam intensity conditions should thus be similar to the standard requirements of the upgraded ALICE experiment \cite{Abe14a}. The main signal requires measurements of pp and p+Pb collisions -- minimising systematic uncertainties requires running at the same $\sqrt{s}$ for both systems. The FoCal detector would also participate in additional Pb+Pb running of ALICE. The estimated requirements are summarised in table~\ref{tab:lumi}. Figure~\ref{fig-rpa} shows an estimate of the measurement uncertainty for the nuclear modification factor for these conditions.

\begin{table}[h]
\begin{center}
\caption{Collision systems, beam conditions and integrated luminosities required for gluon saturation studies with FoCal in ALICE.}
\label{tab:lumi}
\begin{tabular}{p{3cm}rrr}
\hline\hline
\textbf{system}             & \textbf{luminosity}
                                                & \textbf{max. event rate} & \textbf{int. luminosity}\\
\hline
pp     & $3 \times 10^{30} \, \mathrm{cm}^{-2}\mathrm{s}^{-1}$   & 200~kHz & $\approx 6 \, \mathrm{pb}^{-1}$ \\
p--Pb     & $10^{29} \, \mathrm{cm}^{-2}\mathrm{s}^{-1}$   & 200~kHz & $50 \, \mathrm{nb}^{-1}$ \\
Pb--Pb$^{a}$     & $7 \times 10^{27} \, \mathrm{cm}^{-2}\mathrm{s}^{-1}$   & 50~kHz &  \\
\hline\hline
\multicolumn{3}{l}{$^{a}$ \footnotesize Not required for this measurement.}
\end{tabular}
\end{center}
\end{table}

\subsection{Summary}

Forward Drell Yan production is a promising place for studying gluon densities at small-$x$ and for finding saturation. For low lepton pair masses, dipole model predictions 
are significantly below the collinear predictions. Preliminary LHCb data indicate that, in fact,
it will be possible to discriminate between different models. Distributions in transverse momentum of the Drell-Yan pair at low $p_T$ should provide further hints for deviation from collinear factorization.  Forward direct photon production is another final state sensitive to saturation. A detector upgrade at ALICE (FoCal) would offer a wide spectrum of physics possibilities.

\section{Large-x physics in p+p and p+A collisions}

%({\it Stan Brodsky$\,\,^{\dagger}$})}

%\let\thefootnote\relax\footnote{$^{\dagger}$ \it SLAC, USA}

One of the most important reasons to extend the capabilities of LHC experiments to high rapidities (high $x_F$) is to access the physics of heavy quarks and heavy hadrons~\cite{Brodsky:2015fna}.

The original discovery of the $\Lambda_c(cud)$  and $\Lambda_b(bud)$~\cite{Bari:1991ty}  at the ISR was possible because of the ``split-field" magnet which allowed the measurement of  ``leading hadrons" at high $x_F$. 
The hadroproduction of the $J/\psi$  and even double $J/\psi$ was originally observed by NA3~\cite{Badier:1982ae} in fixed-target experiments at  high $x_F$.
The production of the double-charm baryons  $ccd$ and $ccu$ at forward rapidities  has been reported by SELEX~\cite{Engelfried:2007at}. 

High $x_F$ processes in proton collisions follow from the fact the heavy quarks in the five-quark  $|uudQ\bar Q > $ light-front Fock state wave-functions of the proton appear at high light-front fraction $x$  due to multi-connected diagrams -- the ``intrinsic heavy quark" distributions ~\cite{Brodsky:1980pb}. 
The high- $x$ component is in addition to the usual low-$x$ contribution from gluon splitting or DGLAP evolution. The intrinsic amplitude is maximal at minimal off-shellness;  i.e., equal rapidity of the constituents and thus  $x_i \propto m_{\perp i} = \sqrt {m^2_i +k^2_{\perp i}} $
and  the heavy quarks have the highest momentum fraction. The probability for intrinsic heavy quarks falls as $1/m^2_Q$ in non-Abelian QCD. 
These features are rigorous properties of QCD and the operator product expansion~\cite{Brodsky:1984nx,Franz:2000ee}. 

The intrinsic charm distribution was measured at high $x $ by the EMC collaboration~\cite{Aubert:1982tt}. 
The intrinsic contribution dominates at high $x$ despite the  $1/m^2_Q$ suppression since the DGLAP contribution falls rapidly   as $(1- x)^5.$

The leading hadrons are formed in a collision when the quarks in the proton light-front wave-function coalesce at nearly equal rapidity. 
For example, in $pp \to  \Lambda_c X $  the $cud $ coalesce at  equal rapidity. 
The momentum of the $\Lambda_c$ is the sum of the three momenta of the $cud$ and thus is dominated by the intrinsic high-$x$ charm quark.

There are enormous implications for the LHC~\cite{Brodsky:2014hia}: 
One can create a vast array of heavy-quark hadrons at high $x_F$. 
For example, the $B_c(b\bar c)$ meson will be formed at high $x_F$  in $ pp \to B_c X $ collisions from the coalescence of the heavy quarks from the $7$-quark 
$|uudccb \bar b > $ Fock state proton.   One can make many other exotic heavy quark meson and and baryons, e. g., $ |bcu > ,|ccc >, $ etc. 
One can also create tetraquark states at high $x_F$, such as the $Z_c(cc\bar u \bar d) $ and $Z_b(bb \bar u \bar d). $.

The existence of intrinsic heavy quarks in the proton leads to a novel mechanism for the production of the Higgs at high $x_F$.~\cite{Brodsky:2006wb,Brodsky:2007yz}
For example, the Higgs can couple to the $b$ and $ \bar b$ in the protonÕs five-quark Fock state  $|uud b\bar b >. $
Thus the Higgs will be produced in $pp \to H X $ at high  $x_F = x_b +x_{\bar b}$  -- at $x_F$ as large as $x_F  \simeq 0.9.$ 
Similar contributions can arise from each of the $|uudQ\bar Q >$  Fock states, since the Higgs coupling compensates for the  $1/ m^2_Q$ fall-off of the intrinsic heavy-quark Fock state probability. 
The high $x_F$ Higgs tends to travel down the LHC beam pipe. 
Thus the decay channels $\gamma \gamma$, and  $\mu^+ \mu^-  \mu^+ \mu^-$ could be particularly advantageous channels for the detection of high-$x_F$ very forward Higgs events.

%%%%%%%%%%%%%%%%%%%%%%%%%%%%%%%%%%%%%%%%%%%%%%%%%%%%%%%%%%%%%%%%%%%%%%%%% 

%% file: detectors/detectors.tex
%\chapter{Forward Detector Systems at LHC}
\label{sec:forward_detector_systems}

The following paragraphs describe the detector systems of the different LHC experiments optimized for the forward physics program as outlined in previous chapters of this report.
Already existing detectors are described shortly, or a reference to the detailed articles~\cite{jinst-global} is given. The newly developed components that have been integrated during the long 
shutdown 1 (LS 1) and the ongoing R\&D activities are described in more detail with possible reference to the corresponding technical design reports (TDRs) or upgrade proposals. 
The forward detector systems can in general be separated in two different types, the movable beam inserts with detectors and the standard detectors that are integrated in the central detector or 
installed in the LHC tunnel. The following table provides information on the coverage of LHC detectors with $|\eta| >$ 5.

\begin{table}[ht]
\begin{center}
\begin{tabular}{|c|c|c|c|c|}
\hline
\hline
Experiment  & Detector & Hardware  &  Acceptance    &  Comment\\
            &          & Tecnology &           &    \\
\hline
ATLAS & LUCID  &  Gas Cherenkov tubes  & 5.9 $<|\eta|<$ 6.0   & \\
  & ALFA   & Scintillating fibres  &    & Forward proton tracking   \\
  &     &    &    & $0<\xi<0.2$ high-$\beta^*$   \\
  & ZDC    & Quartz rods, tungsten    &$|\eta|>$ 8.3  &   Neutrals only\\
  & AFP    &  Tracking: Silicon  &  &   Forward proton tracking \\
  &   &      Timing:Quartz Cherenkov  &   & $\approx 0.03<\xi<0.2$  low-$\beta^*$\\
\hline
 ALICE & ADA    & Scintillator   &-7.0 $<\eta<$ -4.8     &  \\
  & ADC    & Scintillator   &4.7 $<\eta<$ 6.3    &  \\
\hline
 CMS   & CASTOR & Quartz plates, tungsten & -6.8 <\eta< -5.2    &  \\
   &  FSC   & Scintillator   &6 $<|\eta|<$ 8  &    \\
    & ZDC    &  Quartz fibres, tungsten  & $|\eta|>$ 8  & Neutrals only\\
 TOTEM & T1     & Cathode Strip Chambers   &3.1 $<|\eta|<$ 4.7    & \\
  & T2     & GEM Chambers  & 5.3 $<|\eta|<$ 6.5    & \\
  & RP &  Silicon  &  &    Forward proton tracking   \\
  &  &    &    & $0<\xi<0.2$ high-$\beta^*$   \\
 CMS+TOTEM & CT-PPS  & Tracking: Silicon  &    & Forward proton tracking \\
           &           &Timing: Quartz Cherenkov     &    &  $\approx 0.03<\xi<0.2$  low-$\beta^*$\\
\hline
 LHCb & HERSCHEL & Scintillator  &  5 $<|\eta|<$ 8   & \\
 
   \hline
 \end{tabular}
 \\[4.0mm]
 Table 1 : Summary of very forward detector coverage of LHC experiments \\
  in Run II. Several numbers are approximate. \\ 
 \end{center}
  \label{sigmas}
\end{table}

\paragraph*{Movable beam inserts with detector sensors}

The detector carriers which are integrated in the LHC beam tube and enter into the LHC beam vacuum are generically called ``movable beam inserts''. 
These beam inserts carry the detector sensors for tracking or timing and allow the approach of the  sensor to a distance down to a millimetre from the beam center. 
The movable beam inserts have to comply with LHC requirements in view of impedance, ultra high vacuum and safety. The material budget plays an important role, 
as a fraction of the secondary particles which are generated by  interaction of high energy particles  with the material of the movable beam inserts are scattered 
in the material of the supra conducting LHC magnets close by.
The beam insert design is universal and can be used in the different corresponding locations of the LHC machine.
  
The  specific LHC optics generates a specific particle occupancy pattern in the sensors integrated in the movable beam inserts, depending on both, the distance of 
the sensor relative to the corresponding interaction point and the location in the tunnel (experiment). For each experiment and location, the size and  pixelization of 
the sensor is specific and optimized accordingly and differs strongly for the high and low $\beta^{*}$ optics. The insertion of the beam inserts is a complex procedure 
requiring the close collaboration with the LHC collimator experts and the operators in the LHCC control room. In this sense the movable beam inserts become an integral element of the LHC machine.
  
The movable beam inserts are also present in the parking position during the standard runs and interfere with the LHC machine mostly due to their impedance. 
The beam-induced heating of the movable beam device can lead to local vacuum degradation and thus create perturbations of the machine operation during the energy ramp-up phase and the 
later stable operation of the LHC machine.

\paragraph*{Standard detectors}

The standard detector for LHC forward physics can be either integrated in the central detector of an LHC experiment, or in the tunnel of LHC, outside of the vacuum beam pipe. 
The operation of these detectors is in the autonomy of the experiment.

\paragraph*{Considerations for the Design and Operation of Forward Detectors at LHC with High and Low $\mathbf{\beta^{*}}$ Optics}

Certain physics observables require that the forward detector information is combined with that of the corresponding central detector. In this case the synchronisation of the specific  
forward detector systems with the central detector needs to be considered in the design of the trigger and data acquisition. The distance of up to 250\,m from the forward detectors to 
the central detectors requires fast signal transmission and precision clock distribution systems. 
An important impact on the detector design and general layout is given by the machine settings. 
The physics programme outlined previous chapters of this report specifies the different LHC beam optics, beam parameters and instantaneous luminosity for the measurement of 
a given physics variable. The requirements for the operation of detectors and movable beam inserts at low and high $\beta^{*}$ are quite different.

The LHC beam intensity setting and instantaneous luminosity for the special runs (high $\beta^{*}$) will be defined by the forward physics community in agreement with the LHC operation group, 
while the integral time for these special runs is subject to negotiation with the LHC scientific management of CERN. 

For high $\beta^{*}$ runs, the integral luminosity per run and the integral beam time per year is much lower with respect to the standard runs, leading to significantly lower instantaneous and 
integral radiation exposure to ionizing and non-ionizing radiation of the dedicated detectors and carriers. However, forward tracking detectors optimized in view of radiation hardness and 
multi-track resolution combined with timing detectors synchronized with precision clock distribution systems allow background elimination and vertex separation in the central detectors. 
Such optimized forward detector systems allow higher instantaneous luminosities with the advantage of reaching the physics goals with significantly shorter beam time. 

The forward physics programme at low $\beta^{*}$ in contrary needs to cope with the beam parameters determined by the mainstream LHC physics community, and all detector components to 
be used under these beam conditions are forced to be adopted accordingly. The detector components installed for this purpose can be exposed to very high radiation levels depending on 
the final location of the detector the tunnel or in the central detector.  
In particular, the movable beam inserts need to cope with these machine boundary conditions during insertion under standard run conditions. The design of the beam inserts in view of 
impedance and material budget is one key issue to avoid any impact on the machine stability. To assure the compliance of the beam inserts with the machine requirements, the design and 
production of the beam inserts undergoes strict quality control in close collaboration with the LHC machine safety experts. The requirements for movable beam inserts and their corresponding 
detectors differ for the usage in high and low $\beta^{*}$ beams: the following sections describe the detector systems of the different experiments for the different optics settings.

\input{detectors/alpha_yr.tex}

\section{TOTEM Experiment}
\label{sec:totemdetector}

TOTEM~\cite{totem-tdr,jinst-totem} is a  LHC experiment with two detectors embedded in the CMS experiment (T1 and T2 telescopes) and  Roman Pot units  
integrated in the LHC beam lines on both sides of the interaction point IP5/CMS. Since the start of LHC in 2010 the T1, T2 detectors and the  
24 Roman Pot installed at $\pm 147\,$m and at $\pm 220\,$m from IP5 were operated successfully to perform the physics program of TOTEM during Run-I.
The detectors of TOTEM, designed and optimized for special runs of  low luminosity, were used for measurements  
at low- and high-$\beta^{*}$ optics and fundamental results like the total p-p cross section at the LHC energy of  $\sqrt s$ = 7 TeV were measured and published in 2011.
In 2012,  TOTEM developed and proposed an upgrade strategy~\cite{totem-upgrade-proposal,ecr-consolidation,ecr-upgrade,totem-upgrade-tdr:ch9} with the goal to operate the Roman Pots at high 
luminosities in special and standard runs of LHC after LS1, allowing the reach of new physics observables.
The ``TOTEM upgrade proposal"  which was approved by the LHCC in September 2013,
comprises  the relocation of the Roman Pot from $\pm 147\,$m  in  the $\pm 210\,$m  region at IP5, the upgrade of four 
existing horizontal Roman Pots with a Radio Frequency shield, the installation of two newly developed low impedance 
Roman Pots in the $\pm 220\,$m region and  the installation of additional collimators (TCL4 and TCL6)  in the LHC beam line in the region of IP5.
This newly proposed  forward physics spectrometer at IP5  combining  the existing Roman Pots with the upgraded and newly 
developed Roman Pots was going  beyond the original goals and scope of the TOTEM collaboration and overlapped with the future CMS forward physics program.
Therefore a  new collaboration between CMS and TOTEM was created in the year 2014, to develop the 
CMS-TOTEM Precision Proton Spectrometer (CT-PPS) related to the physics goals reachable with low-$\beta^{*}$  optics (standard LHC luminosity settings), 
using as baseline carriers the four standard horizontal Roman Pots (box shape) with RF shield to house pixel tracking detectors and the new 
horizontal circular Roman Pot to house timing detectors. The TOTEM collaboration will  independently continue with its enlarged physics 
program related to special runs and started  the development of timing  detectors optimized for the  vertical Roman Pots allowing the vertex separation at high luminosity, high-$\beta^{*}$ runs.

\begin{figure}[h]
  \centering
\includegraphics[width=\textwidth]{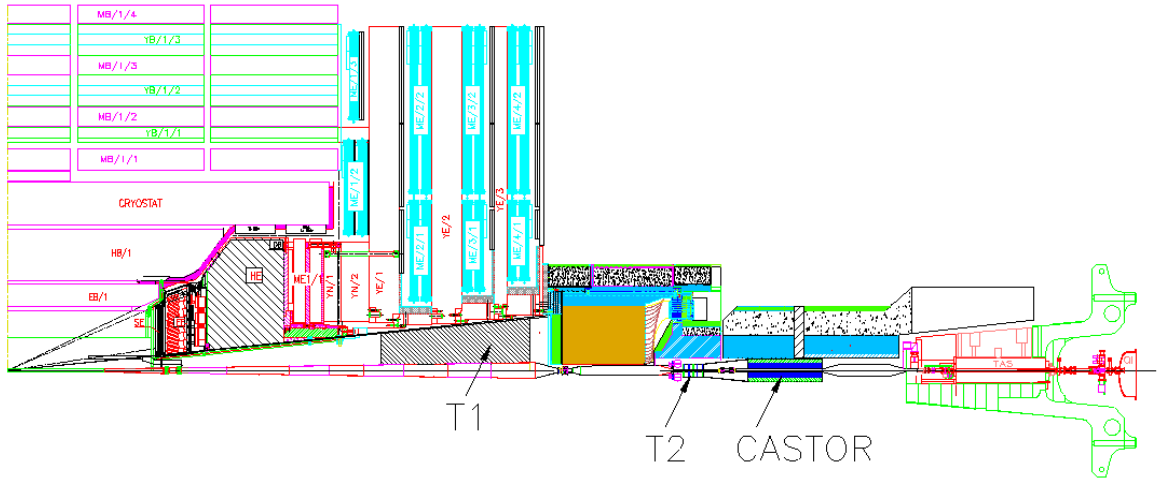} 
  \caption{The TOTEM forward trackers T1 and T2 embedded in the CMS detector}
  \label{fig:totem_layout}
\end{figure}

\begin{figure}[h]
  \centering
\includegraphics[width=0.9\textwidth]{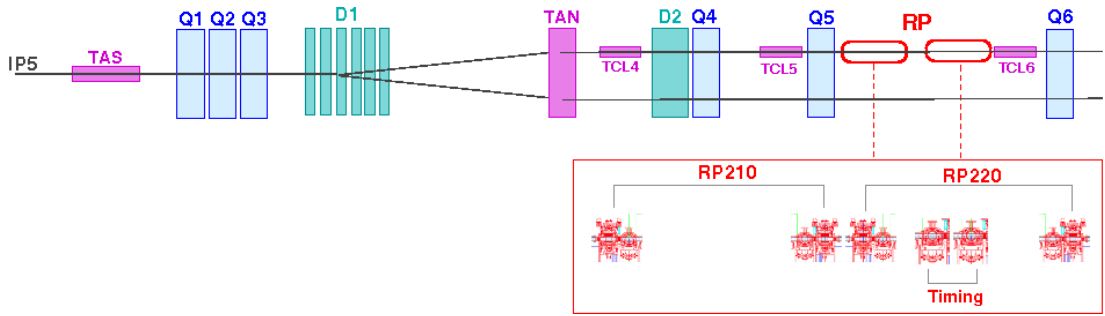}  
  \caption{The LHC beam line on one side of interaction point IP5 after LS1: the TOTEM Roman
Pots are installed at distances of 210-220\,m.}
  \label{fig:totem_layout_rp}
\end{figure}

In a common effort of CMS and TOTEM the spectrometer was installed during the LS1 period, while the development of the detectors (timing and tracking) is ongoing.
%Figures~\ref{fig:totem_layout},~\ref{fig:totem_layout_rp} show the layout of the spectrometer installed in the LHC tunnel at ip5 before and after LS1. 
The layout of the TOTEM experiment before LS1 can be found in all detail in~\cite{jinst-totem} for both the telescopes (T1,T2) and the RPs. 
The following section puts the main focus on the new RP layout 
as it was realized during LSl.
Figures~\ref{fig:totem_layout},~\ref{fig:totem_layout_rp} shows the TOTEM experiment with the detectors T1 and T2 embedded in the CMS central detector and the RPs on both sides of CMS after LS1. 

The following sections 
are based on different documents~\cite{totem-upgrade-proposal,totem-upgrade-tdr:ch9,ctpps-tdr:ch9}.

\subsection{Standard TOTEM Detectors Operated during Run-I at High $\beta^{*}$}

\begin{figure}[h]
  \centering
  \includegraphics[width=0.5\textwidth]{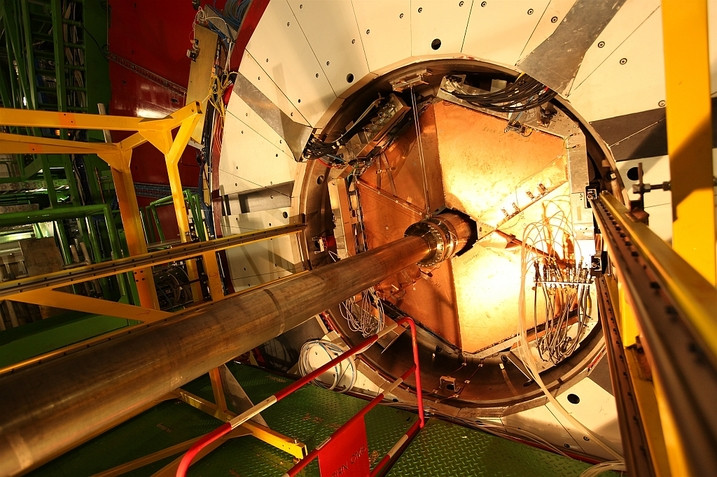} 
  \includegraphics[trim=0mm 55mm 0mm 10mm,clip,width=0.49\textwidth]{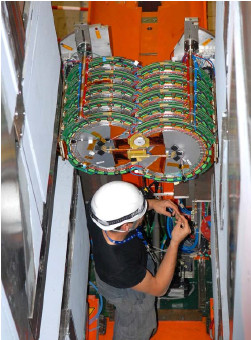} 
  \caption{The T1 (left) and T2 (right) telescopes integrated in CMS central detector.}
  \label{fig:t1+t2}
\end{figure}

\paragraph*{The Telescopes T1 and T2}
Figure~\ref{fig:t1+t2} shows the telescopes T1 and T2 and their integration in CMS.
The T1 telescope consists of two arms, on either side of the IP5, fully integrated in the high-$\eta$ cone of the CMS end-cap at a distance between 7.5 and 10.5\,m from the IP. T1 
detects charged particles in the pseudorapidity range 
$3.1 < |\eta| < 4.7$. Each arm is composed of 5 planes of Cathode Strip Chambers, with six chambers per plane~\cite{jinst-totem}. 
The T2 telescope, based on ``Gas Electron Multiplier'' (GEM) technology~\cite{gem_reference}, allows charged track reconstruction in the pseudorapidity range 
$5.3 < |\eta| < 6.5$. Each T2 telescope arm located at $\sim13.5$\,m on either side of IP5 is composed of 20 semi-circular GEM planes - with overlapping regions - interleaved on both 
sides of the beam vacuum chamber to form ten detector planes of full azimuthal coverage~\cite{jinst-totem}. This novel triple GEM detector technology with the combined pad and strip readout 
was optimized to cope with the specific event topology and high radiation load close to the beam tube. In the triple GEM structure, a 3\,mm drift space is followed by two 2\,mm deep charge 
transfer regions (Transfer 1 and Transfer 2) and a 2\,mm charge induction space. Each GEM readout board contains 256 concentric circular strips for the radial coordinate (80\,$\mu$m wide, 
pitch of 400\,$\mu$m) allowing a track resolution of about 100\,$\mu$m, and a matrix of 1560 pads (with varying size from $2\times2\,\rm mm^{2}$ to $7\times7\,\rm mm^{2}$) for azimuthal 
coordinate reconstruction and for the T2 local trigger. The T2 GEM detector is operated with the gas mixture Ar/CO$_{2}$ (70\%/30\%). 

\paragraph*{Roman Pot System}

\begin{figure}[h]
  \centering
  \includegraphics[width=0.8\textwidth]{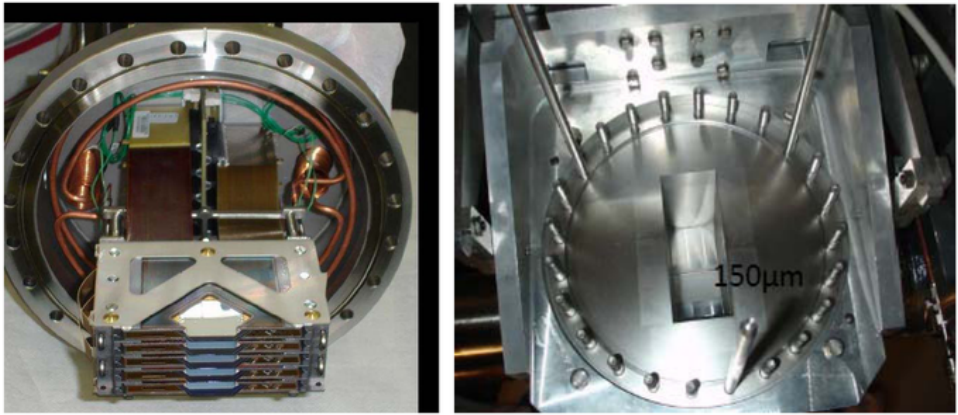} 
  %\vspace*{5cm}
  \caption{Si detector package composed of 10 Si planes (left). Roman Pot, to
separate the Si detector from the LHC vacuum via a thin window of 150$\mu$m 
thickness (right).
  }
  \label{fig:romanpots}
\end{figure}

\begin{figure}[h]
  \centering
\includegraphics[width=0.6\textwidth]{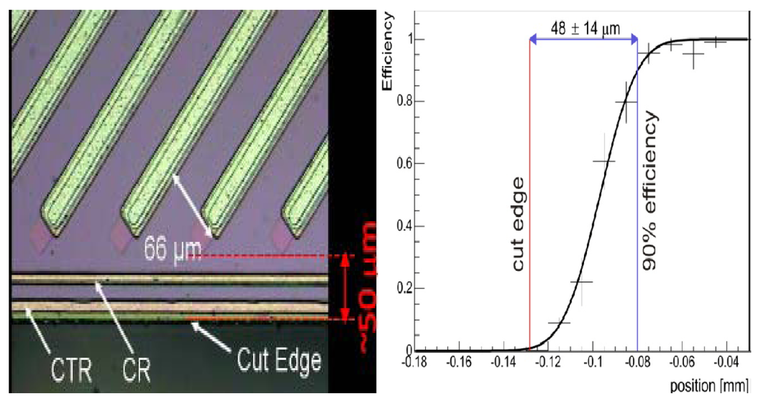} 
%  \includegraphics[width=0.35\textwidth]{figs/detector/} 
  %\vspace*{5cm}
  \caption{(left) Edgeless Si strip sensor with Current Terminating Structure. The
detector reaches full efficiency at 50$\mu$m distance from the cutting edge
(right)}
  \label{fig:cts_detectors}
\end{figure}

The movable beam-pipe insertions called Roman Pots~\cite{jinst-totem} are optimized to detect the leading protons scattered at very small angles. Each RP station is composed of 
two units (near and far) separated by a distance of $\sim 5$\,m, each consisting of two vertical pots and one horizontal pot. On each side of the interaction point IP5, 2 RP stations are installed, 
resulting in a total of 24 single RP detectors, not counting the two new horizontal RPs for timing, added during LS1. Each RP detector is equipped with a stack of 10 edgeless Si strip planes fixed 
in a frame with a high mechanical precision of better than 50\,$\mu$m 
(Figure~\ref{fig:romanpots}, left). The 512 strips with a pitch of 66\,$\mu$m of a single Si-plane are oriented at an angle of $+45^{\circ}$ (u-planes) or $-45^{\circ}$ (v-planes) 
relative to the detector edge facing the beam. Each stack is composed of 5 pairs of u- and v-planes mounted back-to-back and centered inside the RP 
(Figure~\ref{fig:romanpots}, right), that separates the sensors and the associated frontend electronics from the LHC vacuum via a thin window of 150\,$\mu$m thickness. 
The Si detectors with the frontend electronics are operated at $-30^{\circ}$C by means of evaporative cooling. The pressure in the pot is kept between 10\,mbar and 40\,mbar absolute to avoid 
condensation. In case of vacuum problems or increase of the absolute pressure above 50\,mbar absolute, the cooling system is automatically switching to the `warm mode', stabilizing the 
temperature at $+12^{\circ}$C.  To optimize the acceptance for protons scattered at the smallest angles, the TOTEM experiment has developed planar edgeless silicon detectors~\cite{cts_ref1} 
with a Current Terminating Structure (CTS) to reduce the insensitive area at the edge facing the beam down to 50\,$\rm\mu$m 
(Figure~\ref{fig:cts_detectors}, left). The edgeless Si sensors are processed on very high resistivity n-type silicon wafers ($> 10\,\rm k\Omega\,cm$), with a thickness of 300\,$\mu$m. 
The silicon detector hybrid carries the sensor with the 512 strips wire-bonded to the input channels of 4 VFAT readout chips. Beam tests have shown that the full detection efficiency 
is achieved at a distance of only 50\,$\mu$m from the cut edge 
(Figure~\ref{fig:cts_detectors}, right).

\subsection{Detector Upgrade for Vertical Roman Pots}
To reach the physics goals at high $\beta^{*}$, TOTEM stand-alone and combined runs of the CMS detector with the TOTEM RPs are envisaged. 
The operations of the RPs during Run-I  have shown that the RP carriers and the Si strip detectors worked fully satisfactorily both in stand-alone mode and in combination with the CMS 
detector.
%tracker. 

\begin{itemize}

\item
Within the consolidation project performed during LS1, the RPs at 147\,m have been relocated to the 210\,m region upstream of the existing RPs at 220\,m to improve the resolution of this 
proton spectrometer. The Roman Pot unit at 210\,m far was rotated by 8 degrees around the LHC beam axis 
(Figure~\ref{fig:rp-rotation}) relative to the 220\,m station, to increase the multiple-track resolution of the silicon strip detector system~\cite{totem-upgrade-proposal}. 
Furthermore, all RPs of TOTEM were equipped with new ferrites~\cite{ferrites} of higher Curie temperature as it was requested by the LHC machine committee with the aim to improve the beam vacuum.

\begin{figure}[h]
  \centering
  \includegraphics[width=0.9\textwidth]{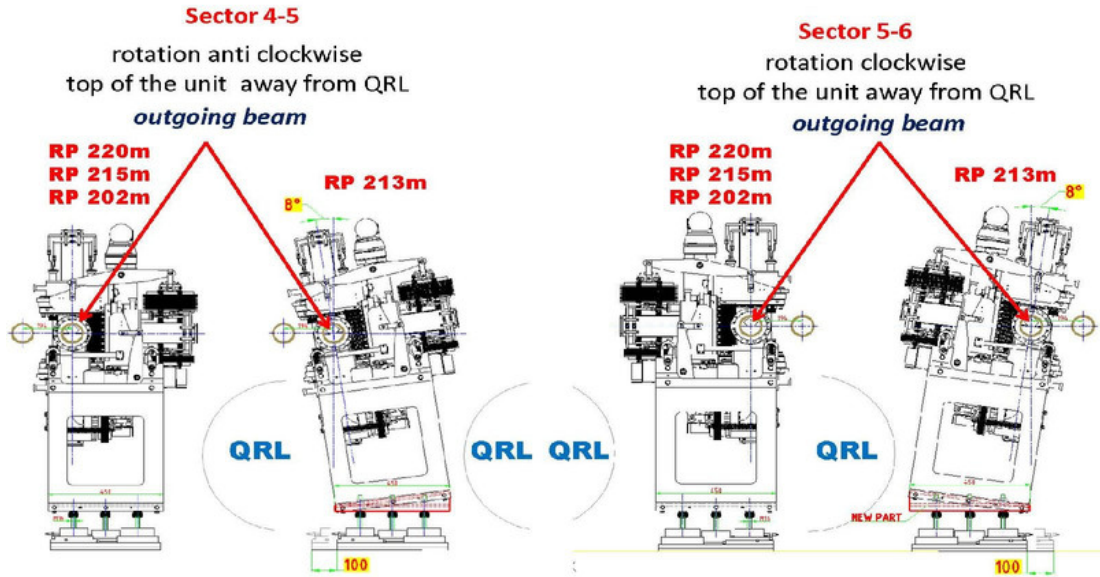} 
  \caption{Drawings of the rotated RP units. The view is directed along the outgoing beam.
  }
  \label{fig:rp-rotation}
\end{figure}

\item
As outlined in the physics chapter, high luminosity runs are needed to reach the integrated luminosity within reasonable beam time for special runs of the machine.
Analysis of special Run-I  data in combination with simulations have shown that the integration of timing detectors with a timing resolution of 50\,ps can improve the vertex reconstruction 
significantly up to a pileup of $\mu\sim 1$. 

Timing detectors with a resolution of 50\,ps in combination with analogue and digital front-end  electronics components are the main elements of the TOTEM upgrade programme as described 
in detail in~\cite{totem-upgrade-tdr:ch9}. 
The pixel size and occupancy of the timing detector sensor and the necessary rate capability of the sensor in combination with the full electronics chain were determined by detailed 
simulation studies. As part of the timing detector infrastructure, a precision clock distribution to synchronize the readout electronics of the timing detectors in the two arms of the $\pm$ 210\,m stations 
will be integrated in the LHC tunnel.
While the timing detector infrastructure will be  already installed during LS1, the detector components -- still under development -- can be integrated, thanks 
to the design of the RPs, in short term technical stops.

\item
The space available to house the timing detectors in the  Vertical RPs  is limited by the dimensions of the carrier box allowing to stack detectors up to 5\,cm. The 
Cerenkov detectors as proposed~\cite{Albrow-ARX-2012} for movable beam inserts cannot be integrated in the vertical RPs as the radiator bar requires at least 12\,cm. 
However, the new horizontal RPs of the CT-PPS project, installed during LS1 with a cylindrical shape and a diameter of 12.5\,cm, can house them. This technology will be 
explained in Section~\ref{sec:cylindricalpot} %on upgrade detectors for low $\beta^{*}$ runs later.

To maintain a low occupancy for each detector channel, its segmentation must be properly tuned.
A simple increase of the granularity reducing the pixel size would lead to an impractical growth of the number of channels, which in turn would reflect on the readout, 
requiring for example the development of custom ASICs.

These considerations led to study a design with pixels of different sizes since the track density due to diffractive events and overlapping background is not uniform as 
can be seen in the RP data recorded during Run-I . The pixel size is defined in view of having the same track occupancy in all pixels.
The simulations to study the occupancy of a single pixel and the minimization of the number of channels required in each detector plane suggested that the minimal number of 
pixels of different sizes needed for a good efficiency at higher luminosity is 10 per plane with areas ranging from a few mm$^2$, for the pixels where the track density is larger, 
to several hundreds of mm$^2$ in the peripheral parts of the detection plane. 
The simulation is explained in detail in~\cite{totem-upgrade-tdr:ch9} (Section~4.2).
\end{itemize}

% Pixels of different sizes and the required timing resolution determine the selection of the detector technology:  the choice is to use diamond detectors in order to have to deal with
% a variation of the timing response with the pixel size small enough not to degrade the timing resolution.
% In fact for a diamond detector the pixel size barely affects the time response of the signal due to the extremely high impedance of the material.
\subsubsection*{Diamond Detector as Baseline Technology for Timing Sensor} 
The selection of the detector technology has to take into account the required timing resolution and the variable size of the pixels. %The best choice is 
In the proposed diamond sensor, 
the pixel size minimally affects the time response of the signal due to the extremely high impedance of the material, guaranteeing the same resolution all over the detector plane. 
However the charge released from a diamond sensor is small in absolute terms ($\approx\,15\,000~e$ for a thickness of $500\,{\rm \mu m}$, or $\approx 3\,$fC/MIP), and a low noise amplifier 
is needed to keep the S/N ratio large enough. 
Since the diamond resistivity is very high the main noise source is the first stage of the amplifier. 
It is also easy to implement a pattern with pixels of different sizes by means of a simple metallisation on the diamond crystal. 
The front end electronics design will be then a compromise between speed and low noise.

\begin{figure}[htb!]
 \begin{center}
 \includegraphics[trim=50mm 1mm 60mm 20mm, clip,width=0.6\textwidth]{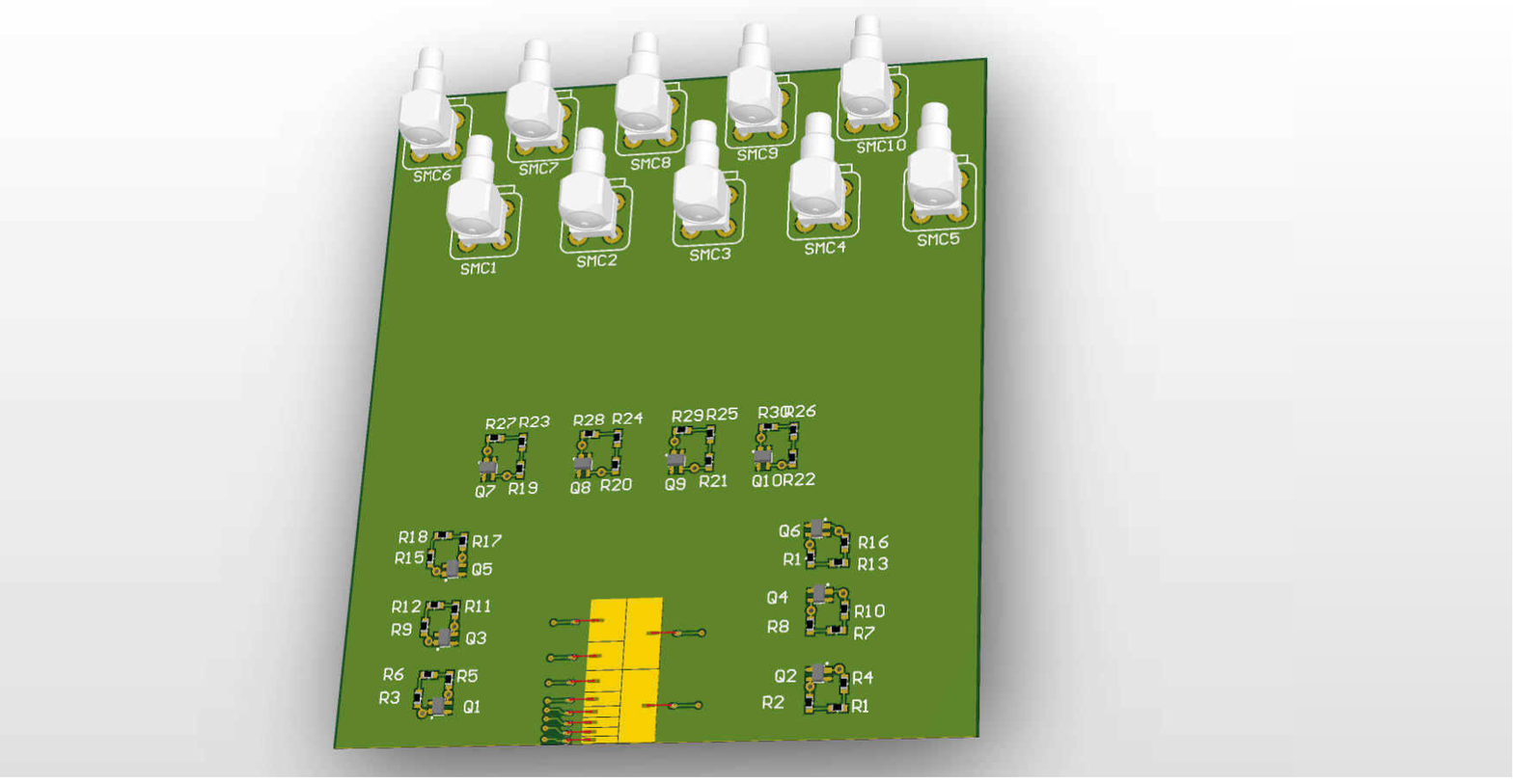}
 \caption{The layout of the board showing the pixel position for one diamond detector plane.}
  \label{fig:diamond-plane}
 \end{center} 
\end{figure}

The present R\&D considers
ten channels per plane (as shown by simulation in ~\cite{totem-upgrade-tdr:ch9}) and the preamplifier stage located near the detector itself. 
A single plane is a $10\times20\,\rm mm^{2}$ diamond sensor.
A board built with controlled impedance material (Roger) will be the mechanical support for the detectors, glued with the smallest pixel sizes located near the edge closer to the beam, 
and for the preamplifier electronics in order to reduce to the minimum the input capacitance 
(see Figure~\ref{fig:diamond-plane}).
A package of 4 detector planes, with thickness up to 500\,$\mu$m, will fit in a vertical RP. Among the commercially available diamond substrates it is possible to choose detectors
with resolution of the order of 100-150 ps, as the multiple measurements allow to reduce the overall time resolution down to the required $\sim$ 50\,ps.
The readout board will be located as close as possible to the detector . 

\subsubsection*{Electronics for Timing Detectors}
\label{sec:Readoutelectronics}
Given the small number of electronics channels required for the readout, TOTEM is developing a discrete component amplifiers.
This  single channel preamplifier is made up of two stages, i) the first is a simple CE transimpedance amplifier, with low amplification power and high bandwidth  that allows fine tuning of 
the input impedance (Silicon-Germanium transistors from Infineon are under test). 
The controlled output signal has 50\,$\Omega$ impedance. ii)  The next stage amplifies the signal to an output voltage range of 0-1\,V to match the readout requirements. 
The detector hybrid will contain only the first stage of the amplification chain and the signals are sent through coaxial cables to the second stage amplification board.
The Hybrid is in the secondary vacuum and the cooling is performed passively through the metallic layers of the board itself.

Two possible ways of adding a time stamp to the recorded protons are: a TDC connected with single or multiple threshold discriminator or a high bandwidth signal sampler. 
The two possibilities considered have slightly different performances: the first gives a better trigger capability, while the second has a better time resolution.

{\underline {Discriminator and  time-over-threshold measurement with a TDC}}: each pixel is equipped with a wide bandwidth transconductance preamplifier and the output voltage is 
proportional to the input current generated from the collected charge that discharges on the input resistance. 
A single threshold discriminator detects the edge of the signal.
The time walk of a single threshold discriminators, consequence of charge fluctuations, can be corrected  measuring the time over threshold for each signal. 
In this case the signal rise time is limited by the bandwidth and with the present electronics is possible to obtain rise time down to $\sim180\,\rm ps$.
The criticality of this approach is that for a MIP the signal to noise ratio is lower than 2.
A way to improve the S/N is either to add coherently with one preamplifier the signal produced in two (or more) diamond planes connected in parallel or to increase the input 
resistance of the amplifier as discussed in~\cite{osipenko2013comparison}. 
However the understanding of both solutions requires a certain amount of simulations and tests.
 \begin{figure}
 \begin{center}
 \includegraphics[width=.5\linewidth]{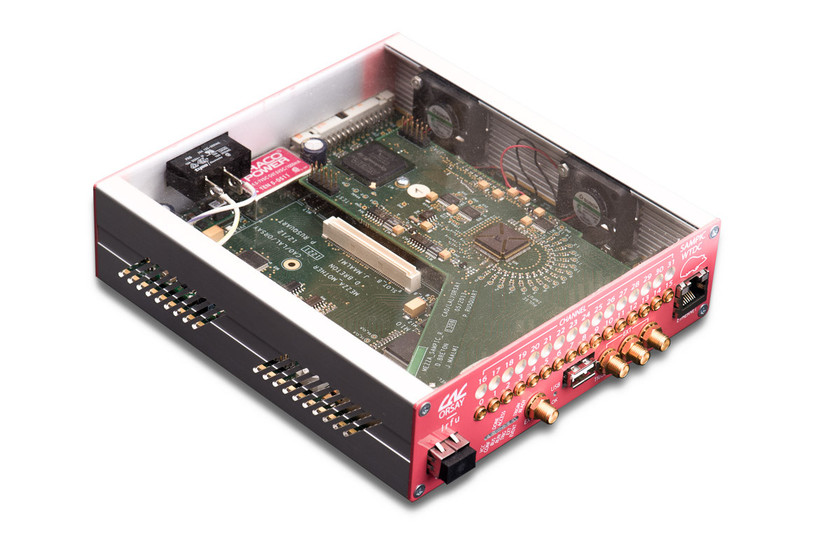}
 \caption{The board with the SAMPIC chip used for the first tests.}
  \label{fig:sampic-board}
 \end{center} 
\end{figure}

The NINO chip~\cite{anghinolfi2004nino} provides this possibility, the output signal length being proportional to the time over threshold of the input analog signal. 
The maximum acquisition rate of this device is around 30\,MHz which in turn implies an average rate for each pixel of less than 10\,MHz, well below the maximum rate expected in the experiment.
The front-end board for one plane will have 10 LVDS output signals each one providing both the pixel information of time of arrival with the leading edge and the charge released with the signal length.

{\underline {Digital sampling}}: the signal from each pixel is integrated and then sampled with a high bandwidth signal sampler.
To extract the timing information from the output of a preamplifier with a known transfer function, an appropriate algorithm reconstructs the original signal.

The sampling can be performed with the SAMPIC chip developed in Saclay and Orsay~\cite{GrabasPhD}. 
The chip has 16 input channels with a  sampling rate up to 10\,Gs/s which provides a good signal reconstruction, due to the fact that the preamplifier has a rise time of 2\,ns 
(see Figure~\ref{fig:sampic-board}).

\begin{figure}[htb]
 \begin{center}
  \includegraphics[width=.7\linewidth]{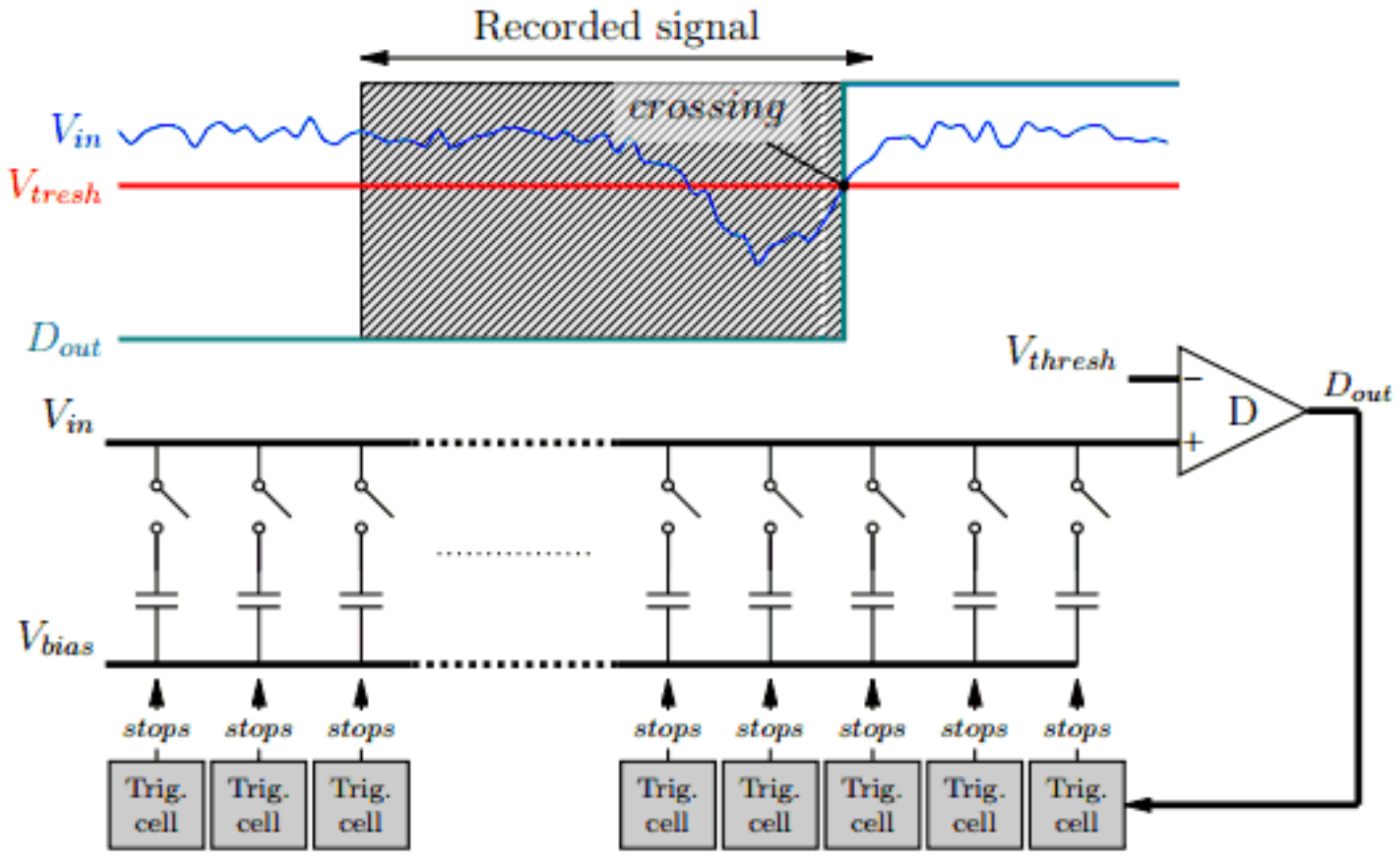}
  \caption{Inside the SAMPIC, the input signal is continuously sampled in a ring analog buffer. In internal trigger mode, the signal is compared to a programmable threshold to stop the sampling 
   and start the ADC conversion.}
  \label{fig:SAMPIC_Principle}
 \end{center} 
\end{figure}

For each channel a circular buffer of capacitors continuously samples the input signal.
Digitization  of the buffer using a 11 bit Wilkinson ADC starts either  when an external trigger is provided or when the input signals goes above a programmable threshold 
(see Figure~\ref{fig:SAMPIC_Principle} for a diagram of the chip).

A future version of the chip will allow a minimum 50\,ns dead time on each independent  channel by using  a faster interleaved readout between two or four channels and a function to control the 
internal trigger, for instance start conversion only when the internal trigger fires in coincidence with the bunch crossing, and the possibility of a fast read-out of the internal trigger time-stamp 
to be complemented with a more precise timing information after the digital analysis of the sampled signal will be completed.

\begin{figure}[htb!]
 \begin{center}
  \includegraphics[width=.7\linewidth]{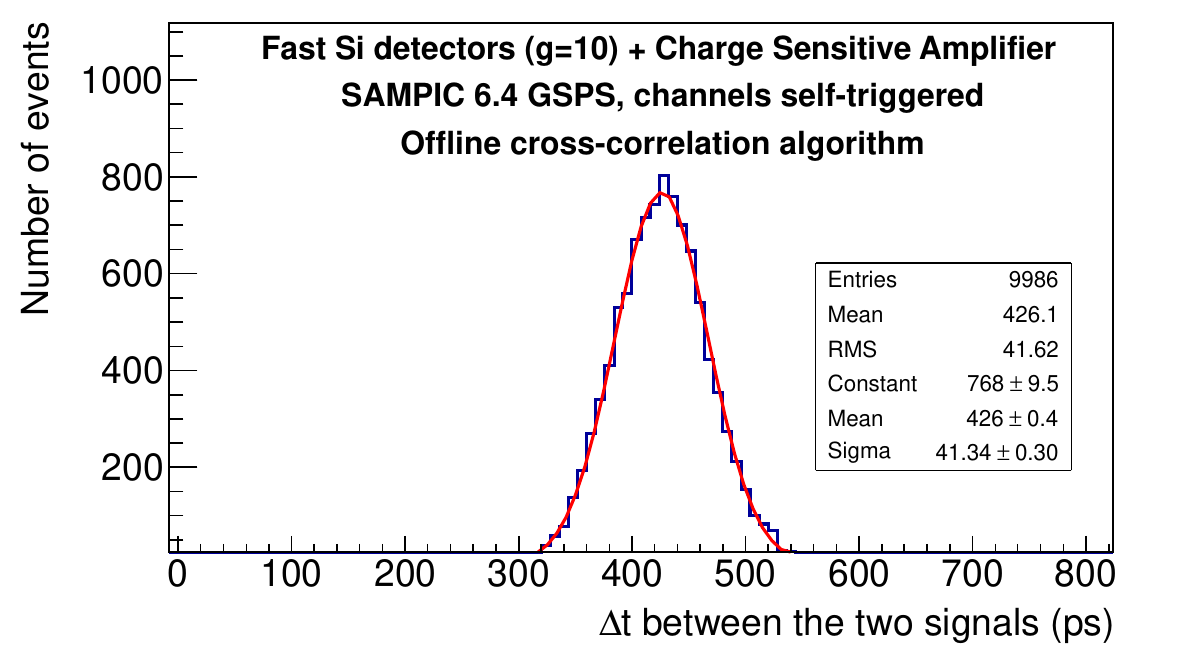}
  \caption{Distribution of the time difference between two Ultra-Fast Silicon detectors pulsed with the same laser and read with fast CSA and SAMPIC, using an advanced off-line algorithm. } 
  \label{fig:Si_CSA_6-4_CC}
 \end{center} 
\end{figure}

The SAMPIC with a CSA preamplifier has been tested with a pair of ``Ultra-Fast'' Silicon detectors~\cite{cartiglia2013performance}.
Figure~\ref{fig:Si_CSA_6-4_CC} shows the time difference measured between the two detectors pulsed with the same laser via an optical splitter and using an off-line algorithm. 
The resolution achieved on the timing difference is of $\sim 40 \,$ps, which indicates a resolution of $\sim 30\,$ps for a single measurement. More studies will follow with diamond detectors in a real 
test beam or cosmic rays.

\begin{figure}[htb!]
 \begin{center}
\includegraphics[width=.48\linewidth]{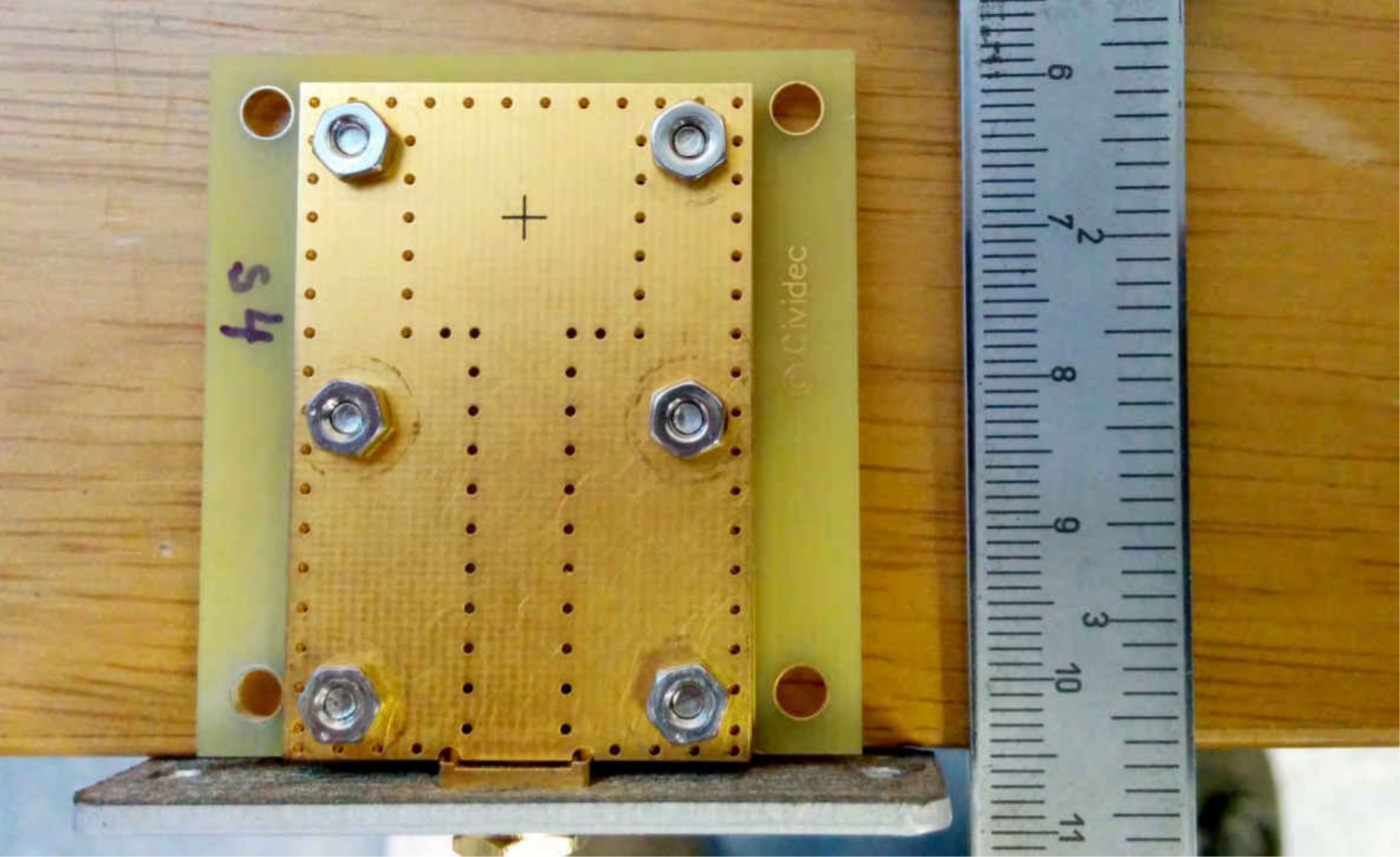}  
\includegraphics[width=.48\linewidth]{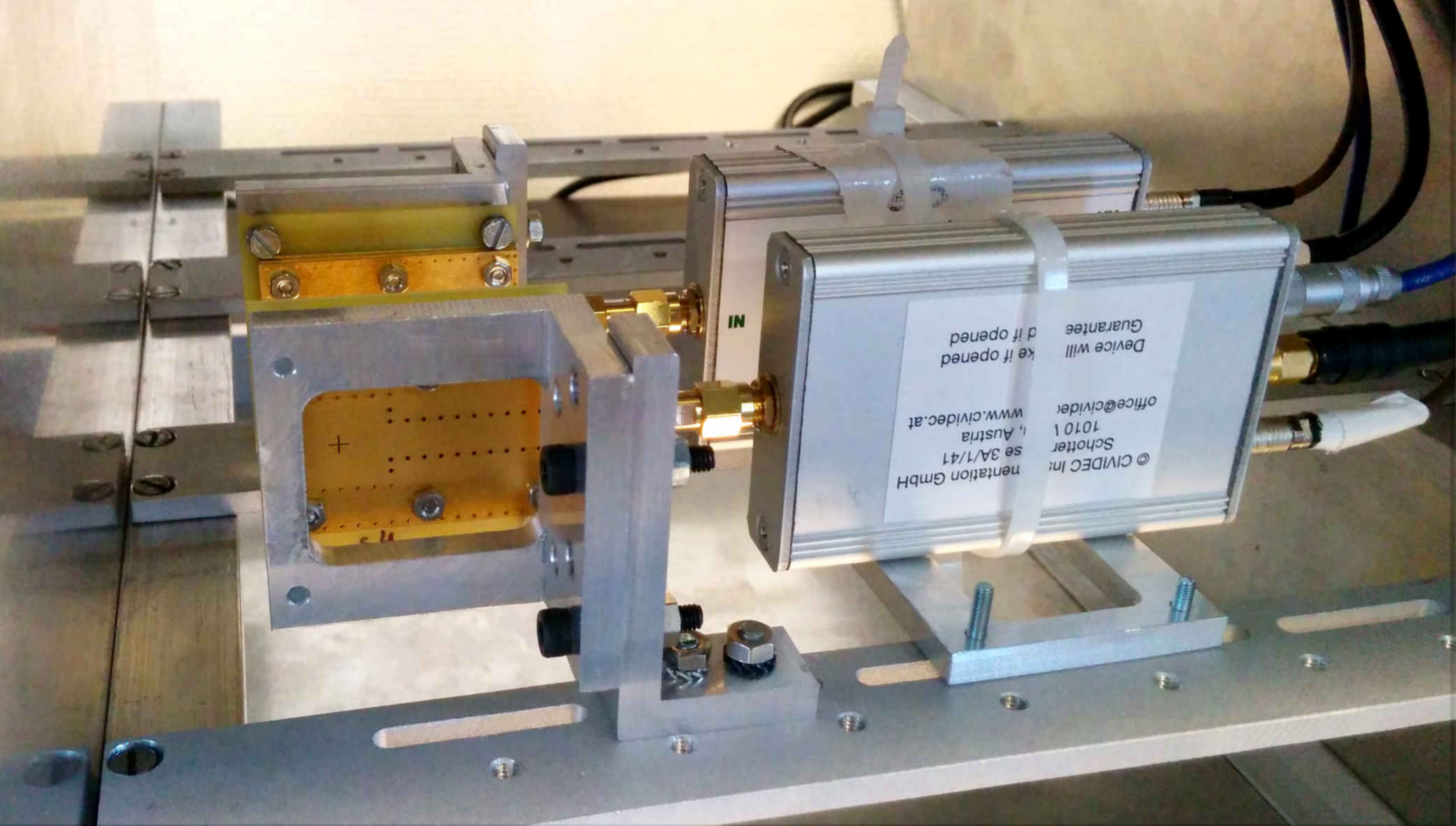}
  \caption{Prototype of the diamond detector from Cividec Instrumentation (left) and the assembly of  the test  telescope (right).}
  \label{fig:DiamondDet}
 \end{center} 
\end{figure}

\begin{figure}[htb!]
 \begin{center}
  \includegraphics[width=.7\linewidth]{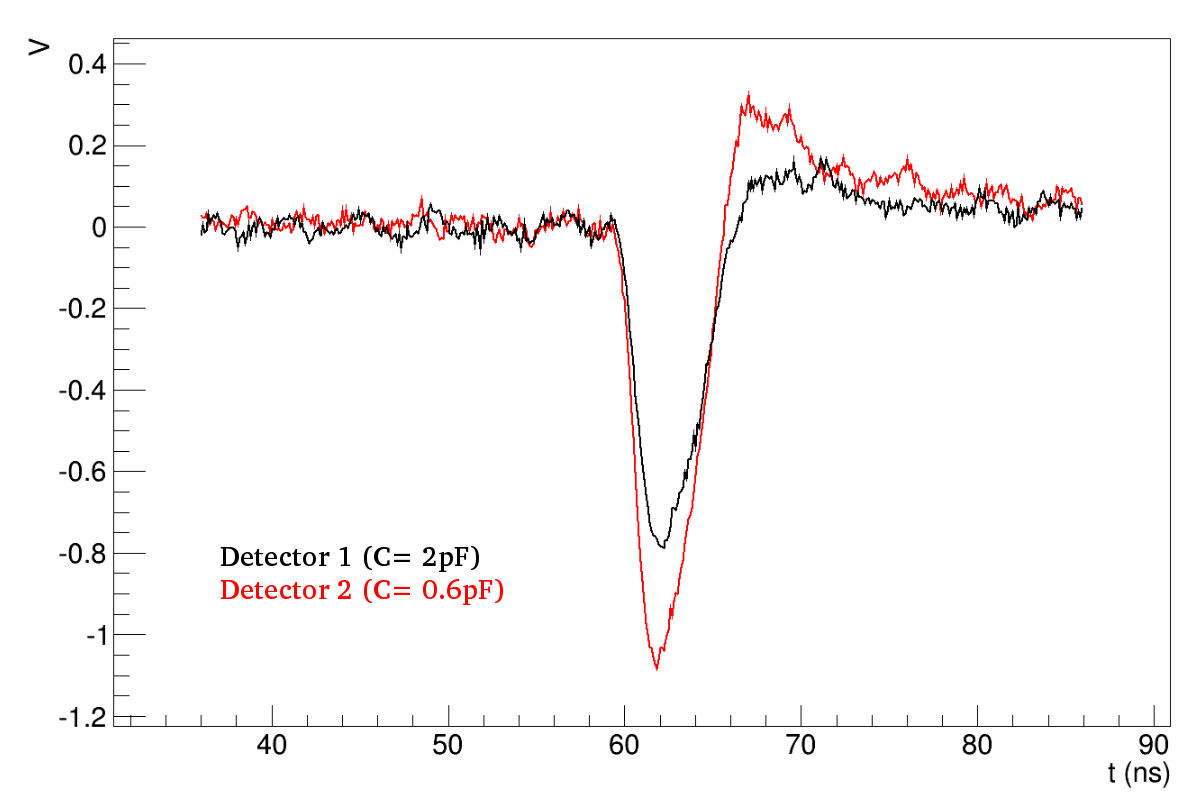}
  \caption{  
  Signals of two diamond detectors of different capacity recorded by the newly developed TOTEM Fast Amplifier in a test with 4\,GeV electrons.}
  \label{fig:DiamondSig}
 \end{center} 
\end{figure}

{\underline {Preliminary Tests:}}
Two diamond detectors, $0.5\times0.5\,\rm mm^{2}$ in size and 500\,$\mu$m in thickness, have been purchased from Cividec Instrumentation~\cite{cividec} together with state of the art CSA and wide-band 
amplifiers with specs optimized to  our request, and assembled in a telescope 
(Figure~\ref{fig:DiamondDet}) for measurements with particles on a test beam (Figure~\ref{fig:DiamondSig}).
Moreover new transimpedance preamplifiers have been developed in house in order to study and optimise the input impedance of the circuits.
Three beam test have been performed in PSI and Cern PS, with different configurations. The detectors where connected to the preamplifiers with SMA connectors. The input capacitance, of about 0.5\,pF from 
the detector, was dominated by the connectors (5-10 pF). The rise time of the signals is strongly affected by this parameter and therefore a reduction in performance is expected. The resolution obtained 
is around 190 ps, well in line with the expected value, and an improvement is expected with a design of the hybrid that removes the connector. 
The final step is therefore to design and bring to a test-beam a hybrid with first-stage preamplifiers bonded as close as possible to the detector, in order to keep the capacitance below
$\sim 1$\,pF. A similar design has already been used successfully elsewhere~\cite{pietraszko:2009}. 
Construction of the new hybrid has already started.

{\underline {Other Technical Considerations:}}
A cooling system will be provided for the electronics only, since the diamond detectors do not dissipate any power from the polarization power supply.

All the electronics that need to be as close as possible to the actual detector and that makes use of FPGAs (as the control/transmission board for the SAMPIC or, in alternative, the TDC board) can operate 
only in regions with reasonable radiation levels.
Studies performed on Altera Cyclone FPGA with ion and neutron beams  show that even in the surrounding area of the beam pipe we could expect for high luminosity runs a SEU (single event upset) rate of 
1 every 3 hours, which is already orders of magnitude higher than what was experienced in the special high beta optics runs. 
For this reason space close to the Roman Pot station located in the floor of the tunnel will be available  to keep the electronics as far possible from the beam pipes. 
In case of SEU a Resync request will be needed only for  the TDC board.


TDCs  with a time resolution of $\sim$\,10\,ps inside an FPGA are in an advanced stage of development by TOTEM and under evaluation. 
The time reconstruction algorithm measures the  crossing time for a single threshold and the time over threshold and a correction matrix. 
In case we will use the SAMPIC chip, the data have to be fitted in a simple FPGA board.
The advantage of using on board FPGAs is that the Trigger and DAQ information will be formatted on the same board.
The event is formatted for DAQ with a header, a start of frame patterns and counters, the list of TOA (Time Of Arrival) and TOT (Time Over Threshold) for each fired pixel, and a footer. 
The information is transmitted without on-line corrections.

The Trigger algorithm instead will perform an on-line rejection based on the number of tracks (see Section~4.2.2 in ~\cite{totem-upgrade-tdr:ch9}). 
In order to filter out noisy channels that could contaminate the time measurement a trigger signal (a track road) is generated only if the signals from aligned pixels from adjacent planes satisfy a 
majority-AND condition. 
Track counting is done locally and, if the event is accepted, the time of arrival is formatted into 4 words and sent to the central trigger unit (TOTEM LONEG board). 

\subsubsection*{Clock Distribution for Timing Detectors at $\pm\,$220\,m from IP5}
The challenge of combining measurements with picosecond range precision for Timing signals generated in locations separated by large distances (order of 220\,m) requires a clock distribution system 
capable of the highest precision and of the utmost time stability. 

The following pages aim at describing the Clock Distribution system for the TOTEM Timing Upgrade.
The system is adapted from the \textit{Universal Picosecond Timing System}~\cite{Bousonville:2009}, developed for FAIR (Facility for Antiproton and Ion Research), the new, unique international 
accelerator facility for research with antiprotons and ions presently under construction at GSI (D), where a Bunch phase Timing System (BuTiS) based on this 
concept has been implemented~\cite{Moritz:2011zz}.

The optical clock distribution network will use a Dense Wavelength Division Multiplex (DWDM) technique that makes it possible to transmit multiple signals of different wave-lengths 
over a common single mode fibers. This will allow to use standard telecommunication modules compliant to ITU (International Telecommunications Union) international standards.

The experiment requires two very stable clocks for the precise timing reference of the measurement and the bunch identification. 
These reference clock signals are sent from the counting room to a set of receivers positioned near the timing detectors in various location of the LHC tunnel on both sides of IP5. 
A third signal added  on the same optical fiber will be simply reflected back to be used to continuously measure the time delay of each optical transmission line: these delay measurements 
are necessary to correct the time information generated at the detector location for fiber delay variations (thermal and mechanical instabilities). 

\begin{figure}[ht!]
    \begin{center}
    \includegraphics[width=0.95\textwidth]{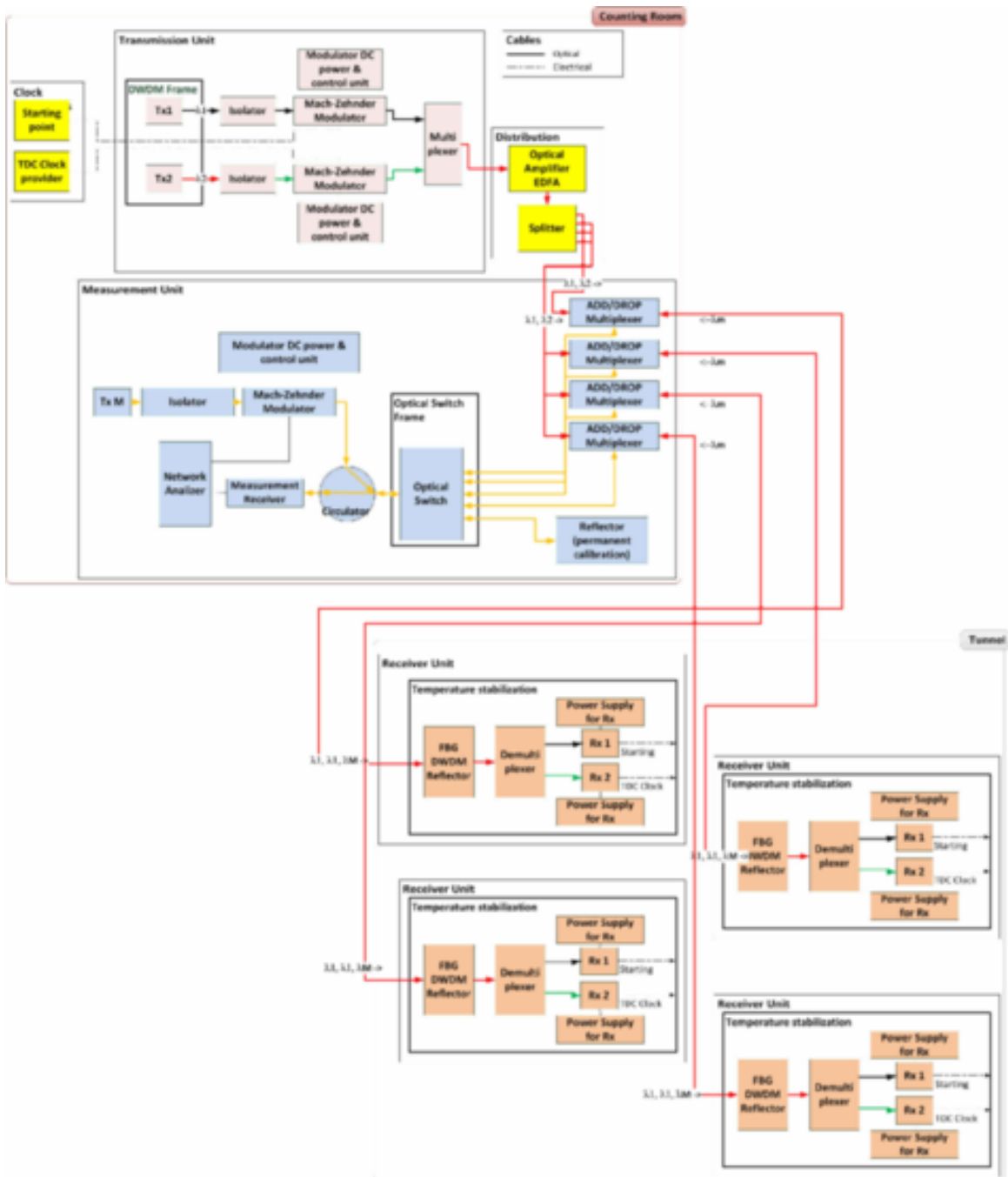}
    \caption{The clock distribution system (see text).}
    \label{fig:CD-fig1-fullpage}
    \end{center}
\end{figure}

The system can be logically subdivided in four major blocks: the Transmission Unit, the Distribution Unit, the Measurement Unit and the Receiving Unit. 
One Receiving Unit must be installed very near each Roman Pot location, the Transmission, Distribution and Measurement Units will be located in the TOTEM racks in the IP5 counting room.
A block diagram of the entire system is reproduced in Figure~\ref{fig:CD-fig1-fullpage}. 

Measurements performed with the prototype of the ``BuTiS'' system show that the influence of the transmission system on the signal jitter, is of the order of 0.4\,ps~\cite{Moritz:2011zz}, mainly 
dependent on the quality of the clock source signal, the noise added by the optical components and the bandwidth of the transmission system itself. 
Using a transmission system based on this concept, the total jitter of the TOTEM clock transmission system will also be due mainly to the  inherent jitter of clock sources and the end user electronics.

\begin{itemize}
\item Transmission unit\\
The Transmission Unit  optically modulates  the two reference clocks in signals with different wavelength  $\lambda_1$ and $\lambda_2$.
Via DWDM these optical signal are multiplexed into a single fiber 
and re-transmitted at a specific wavelength using a 1\,550\,nm band laser to the Distribution Unit. 
The signal is amplified with an erbium-doped fiber amplifier (EDFA) to compensate the attenuation due to further splitting and the multiplexers. 

A Thorlabs PRO8000~\cite{thorlabs} 
platform has been chosen to generate the two DWDM wavelengths on channels ITU 32 and ITU 34.
 This complete platform is designed to operate and control electrical and optical modules for telecommunication testing and application developments from a broad family of interchangeable modular 
 devices and can be controlled by an external computer using industrial control protocols.
The modulation of the optical signals is performed by two military grade Mach-Zehnder modulators with a 20\,GHz bandwidth.
This unit can be rack mounted and is suitable for use in the experiment control room harsh environment. 

\item Distribution Unit\\
The DWDM optical signal, as generated by the Transmission Unit, needs to be split in order to be transmitted to the four Receiving Units.
Moreover a third DWDM modulated optical signal of wavelength $\lambda_M$ is needed to measure the transmission delays over each fiber and is added to the other two clock signals. 
 
The JDSU~\cite{jdsu} Multiple Application Platform (MAP-200) has been chosen for the Distribution Unit optical amplification, optical signal splitting and switching. 
This platform is a highly configurable, scalable and industrially controlled system that can host several optical modules with a wide range of functions.
The EDFA amplifier developed for this platform will be used for the signal optical amplification.

\item Measurement Unit\\
The signals' delays are measured in this unit.  
A reference signal is generated, optically modulated using the wavelength $\lambda_M$ and sent via an optical switch to every Receiving Unit and to a reflector, which will be used for calibration.

Add/drop multiplexers combine this reference signal to each of  the  4 DWDM optical signals generated in the Transmission Unit and split in the Distribution Unit. 
The multiplexed signal, that now contains the three modulated optical wavelengths: $\lambda_1$, $\lambda_2$ and $\lambda_M$, is transmitted to the Receiving Units located in the tunnel at $\pm220$m and $\pm210$m.

The  $\lambda_M$ optical reference signal once at  the Receiving Unit is reflected back to the Measurement Unit, where the optical add/drop multiplexer separates the $\lambda_M$ optical signal 
coming back from the RUs and pass it back to the optical switch.  
 A ``circulator'', placed between the DWDM modulator and the switch, distributes the reflected signals to a measurement instrument without interrupting the transmission from the generation module to the switch. 
A phase comparison of the reflected signal with the reference one is performed, using a vector network analyzer. 
The phase differences obtained by this measurement determine the delay of each clock signal distribution channel.

As for the Transmission Unit, a Thorlab PRO8000 module, and the same Mach-Zender modulator, will be used.  
The DWDM wavelengths proper of channel ITU 36, will be used to modulate the reference signal.

\item Receiving Unit\\
The Receiving Unit separates the multi-wavelength optical signal at the RP stations into individual signals.

The signal from the single mode fiber (SMF) encounters first a  Bragg grating (FBG) DWDM reflector and reflects back the signal component of $\lambda_M$ wavelength. 
The other components of the signal are routed to a DWDM demultiplexer that separates the two wavelengths, $\lambda_1$ and $\lambda_2$, and outputs them on separate fibers for conversion to 
electrical signals and delivered to the front end electronics and DAQ cards eventually.

This unit should be located as close as possible to every Roman Pot location.
A temperature stabilization of this unit, depending on the temperature characterization of the installation point in the tunnel, may be needed to reduce the long term shift of the measured delay. 
\end{itemize}

\input{detectors/FSC.tex}

\section{CMS-TOTEM Proton Spectrometer (CT-PPS)}
\label{sec:ctppsdetector}

\begin{figure}[h!]
\begin{center}
\includegraphics[width=0.95\linewidth]{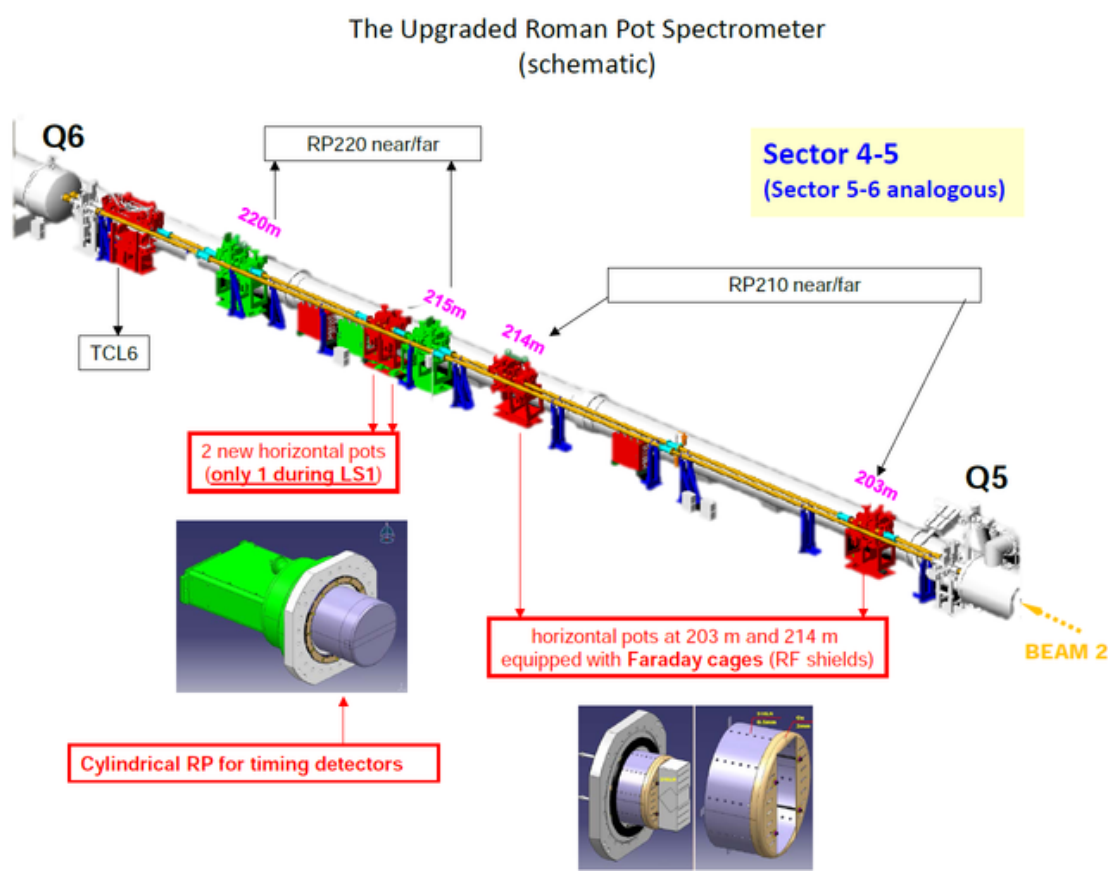}
%\vspace*{5cm}
\caption{The layout of the beam line in the 200 m region after LS1}
\label{fig:layout}
\end{center}
\end{figure}

A new collaboration between CMS and 
TOTEM was created to develop the 
CMS-TOTEM Precision Proton Spectrometer~\cite{ctpps-tdr:ch9} related to the physics goals reachable with low $\beta^{*}$ optics (standard LHC settings). The baseline  carriers of 
this spectrometer are the newly developed low-impedance RPs integrated in the beam line at $\pm (203\div214)$\,m from IP5, designed to host future pixel and picosecond-timing detectors (Figure~\ref{fig:layout}). 
To optimize the beam quality and to protect the LHC magnets near the RPs against collision debris, new collimators were installed upstream (TCL4) and downstream (TCL6) of the RPs during LS1. 
The TCL4 collimators were installed in the former locations of the RP147 stations, and TCL6 was installed between the most downstream unit of RP220 and the supra-conducting magnet Q6.
Comprehensive simulation studies were performed and it could be shown that optimized collimator settings can result in a satisfactory machine protection without limitation of the physics goals. 
Additional beam loss monitors (BLMs) were installed in that region to monitor beam losses caused by RP and TCL6 insertions in that critical region.  
The installed spectrometer will consist of a total of 24 RPs (for tracking) and 2 RPs (for timing) that can be inserted selectively.

%\subsubsection{Movable Beam Inserts, Collimators and Detectors for Low $\beta^{*}$} 
%
Within the CT-PPS collaboration new movable beam inserts, tracking and timing detectors for the low $\beta^{*}$ forward physics were developed for the LHC straight line beam region of $\pm$\,210\,m from the interaction point IP5.
The requirements on beam inserts and detectors operated at low $\beta^{*}$ are quite different from those at high $\beta^{*}$, as already outlined in Chapter~\ref{chap:bema}. 
The underlying strategy foresees to integrate where possible the 
newly developed  detectors in existing carrier systems of TOTEM and the use of existing infrastructure in the LHC tunnel. This concept lead to an inter weaved project of the TOTEM consolidation and upgrade program with the CT-PPS project. 
Timing detectors with a resolution better than 10\,ps are required to achieve  the necessary vertex separation in the CMS central detector with a O(mm) precision, for pileup of 25-50. The double hit probability per bunch crossing and the acceptable single proton rate of a 
single timing cell  determine the pixel size.   
The detector systems with this time resolution and pixels size, combined with the high rate requirements, represent the forefront of present detector developments. With Cerenkov detectors, that can be integrated in the newly developed Roman Pots,  
the time resolutions in the range of 10\,ps could already be achieved in test beams, however the high material budget and limited pixel size of this technology might turn out to be a limiting factor in the final operation under LHC conditions. 
Therefore R\&D of different solid state detector have started to obtain the required time resolution with a lower material budget and smaller pixel cell size.
The edgeless Si strip tracking detectors of TOTEM can not be used for the low $\beta^{*}$ operation due to the intrinsic limited  multi hit resolution. New slim edge radiation hard  pixel detectors that cope with the required space resolution and 
optimized for the use in movable beam inserts are under development. 

The upgrade of the present Roman Pots and the development of new cylindrical Roman Pots was mandatory for the ambitious goal to insert the detectors under standard LHC conditions close to the beam. 
At this point it is emphasized that alternative beam inserts like the movable beam pipe with low impedance (Hamburg Beam Pipe) have been studied for many years and are still under development~\cite{hamburg-beampipe}. The movable beam pipe concept, primarily 
developed for locations at LHC where the operation of horizontal Roman Pots is almost impossible due to space constraints of the LHC beam line, was as well proposed as alternative in that location, where the Roman Pots 
are at present integrated in the beam line of $\pm$220\,m from IP5. However, till today  was neither reached a final design nor the construction of a full size prototype, that complies with all requirements imposed on a beam insert for LHC. The possible usage of 
these type of beam inserts requires further R\&D and the successful construction and test of a full size prototype.

In the absence of feasible alternatives to the concept of the Roman Pot at the time of LS1, the TOTEM collaboration decided for time and cost reasons to study and develop in close collaboration with the 
LHC Beams Impedance and engineering groups, the low 
impedance Roman Pot concepts, based on the experience made during three years of operation during Run-I of LHC. These developments lead over a prototype phase, including mechanical \& vacuum tests  and  RF measurements simulating the EM radiation field of 
the LHC beam close to the Roman Pot to a final design, that was approved for serial production. Even though the simulation and lab test have shown significant improvement of all critical parameters that determine the successful and safe operation of these 
beam inserts,  a final prove of the expected performance can only be made under realistic conditions of LHC. In this sense the newly installed Roman Pots within the CT-PPS project can be considered as R\&D for new beam inserts designs. 
     
\begin{figure}[h!]
\begin{center}
\includegraphics[width=0.95\linewidth]{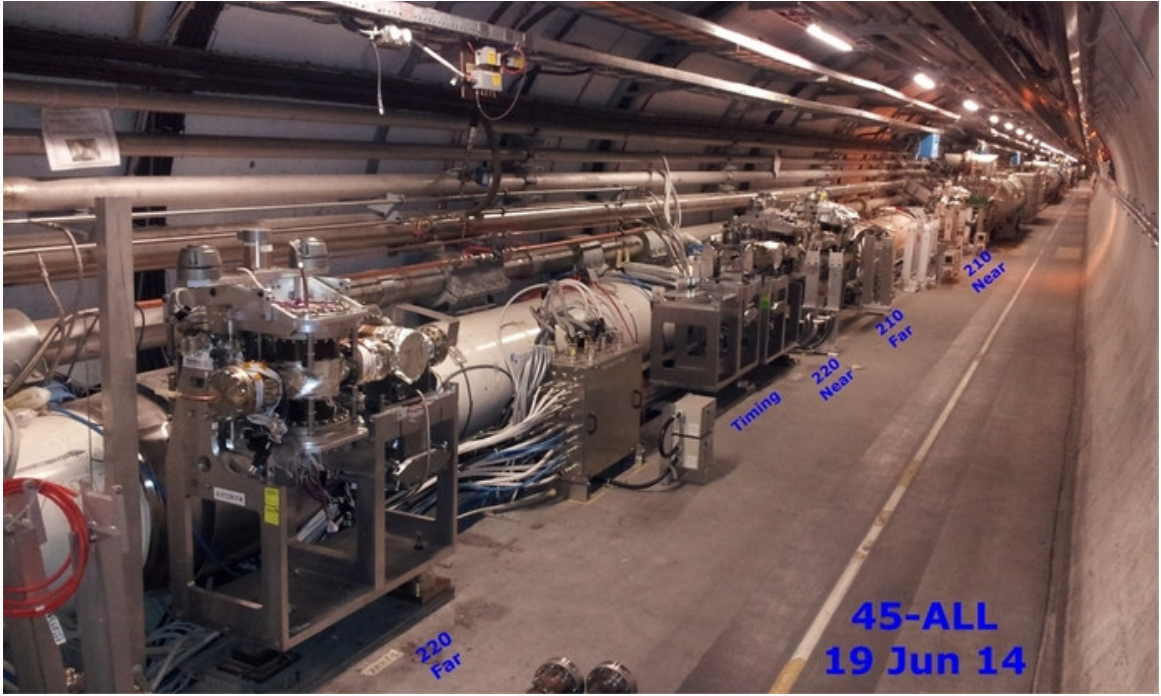}
%\vspace*{5cm}
\caption{The beamline in Sector 4-5 with the RPs installed}
\label{fig:layoutphoto}
\end{center}
\end{figure}

Test insertions of horizontal Roman Pots by TOTEM during Run-I  have shown, that the generated secondary particle production rate impinging on the magnets downstream of the 220\,m far Roman Pot stations can exceed the acceptable level. 
For that reason a new collimator system (TCL4 and TCL6) was proposed for IP1 and IP5, and installed during LS1~\cite{ecrcollim}. 
Figures~\ref{fig:layout} and~\ref{fig:layoutphoto} show the beamline in Sector 4-5 with the RPs installed.

%\iffalse 
 
% \subsubsection{The Roman Pot System and Collimator in the 200\,m Region of IP5}
%The TOTEM consolidation and CT-PPS project combines the relocated Roman Pots (box shape) to house new pixel tracking detectors (RP210 near-horizontal and RP210 far-horizontal) with the
%newly fabricated cylindrical Roman Pots to house timing detectors. The requirements for both the beam inserts and the detectors are very different for the operation at standard LHC conditions (low beta*)   
% Within the TOTEM R\&D project a cylindrical RP 
%has been newly designed to house timing detectors and was optimized in view of material budget and RF interaction with the LHC beam. Furthermore, 
%the horizontal RP210 destined to house the new tracking pixel detectors,  have been optimized to reduce the RF interaction with LHC by keeping the original box design of the standard TOTEM RP. 
%        
%TCL4 and TCL6 collimators have been installed~\cite{ecrcollim} in front of the Q4 and Q6 quadrupoles, respectively.
%Figure~\ref{fig:layoutphoto} shows the beamline in Sector 4-5 with the RPs installed.

%\fi
 
\subsection{Development of Low-Impedance Roman Pots}
\label{sec:cylindricalpot}

The new RPs installed between the existing units 
220-N and 220-F are intended to host timing detectors. Hence their design was
subject to the following main requirements:
\begin{enumerate}
\item Among several potential detector technologies for the timing
measurements (see Section~\ref{sec:timing_cherenkov}), \v{C}erenkov counters~\cite{Albrow-ARX-2012,atlas_upgrade}
are already at well advanced development stage.
For the full timing resolution of $\sim\,15$\,ps a total 
length of 24\,cm of quartz is needed. Distributing this length over the two 
new pots requires each pot to accomodate two slabs of 6\,cm length, too big
for the space provided by the traditional TOTEM pots.
If at a later stage thinner timing detectors (e.g. diamond detectors) become 
available, tracking and timing functionality may be combined in the same pots,
reducing the number of pots to be inserted.
\item The RPs housing the timing detectors will have to operate in  
high luminosity running scenarios. They will have to approach very intense 
beams to the same distance as the tracking RPs, i.e. down to about a mm from
the centre. At that distance beam-coupling impedance effects, machine vacuum
compatibility in terms of outgassing, and particle 
shower development have to be taken into account in the geometrical design and
in the choice of materials (Section~\ref{sec:beamenvironment2012}).
\end{enumerate}
\begin{figure}[ht!]
\begin{center}
\includegraphics[height=5cm]{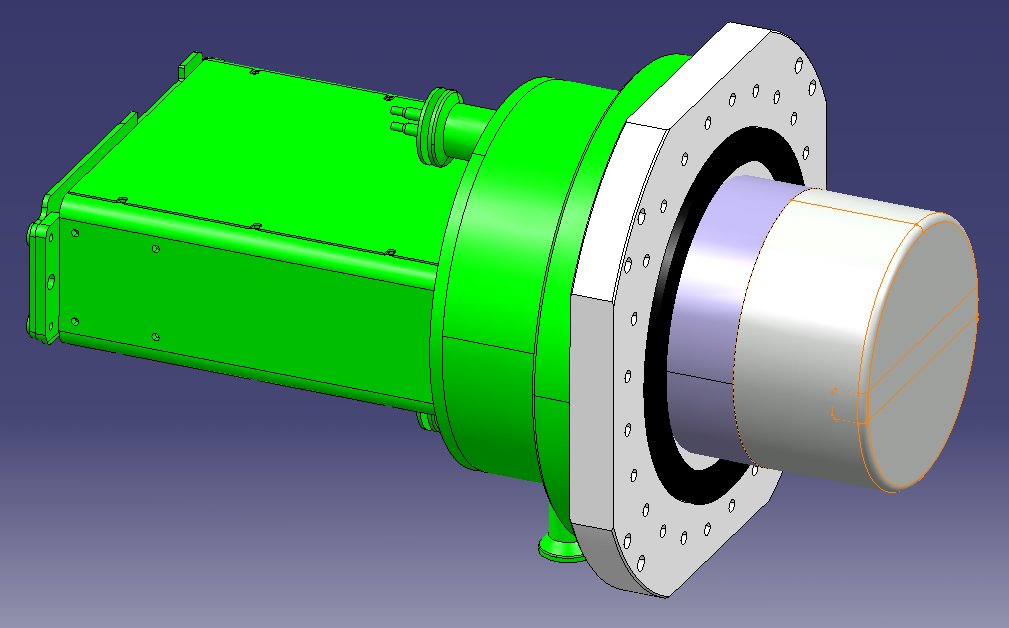}
\includegraphics[height=5cm]{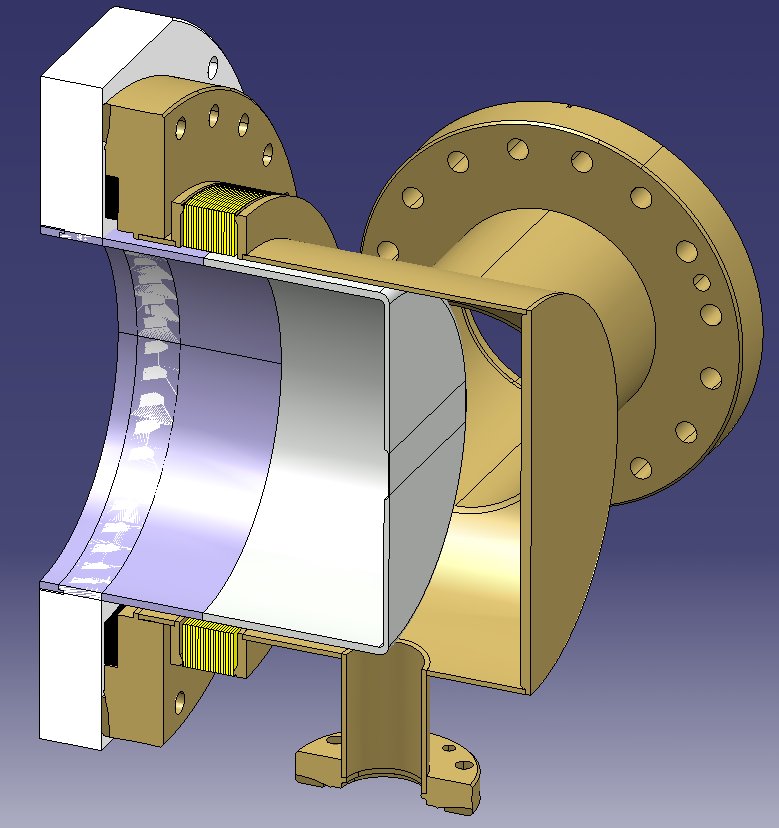}
\includegraphics[height=5cm]{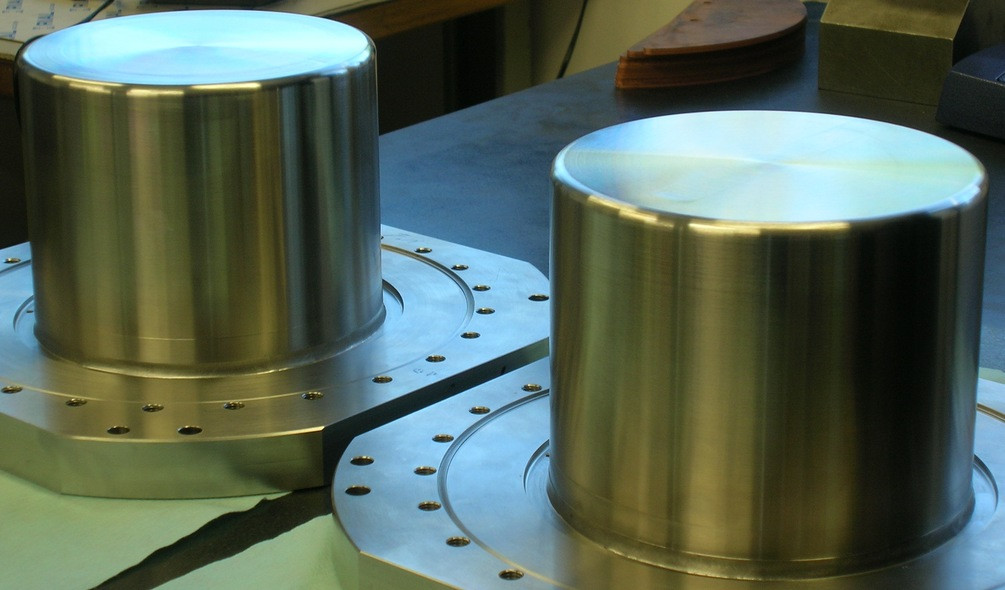}
\caption{Top: drawings of the cylindrical detector housing for the new RPs designed 
to accomodate timing detectors. Bottom: the manufactured pots.
}
\label{fig:cylindricalpot}
\end{center}
\end{figure}
\begin{figure}[h!]
\centering
\includegraphics[width=0.8\linewidth]{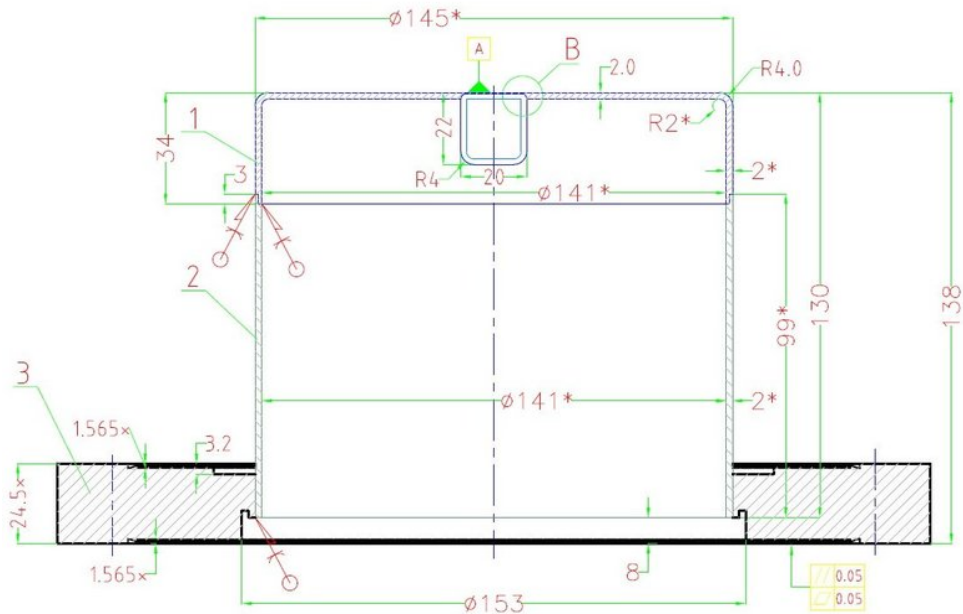}
%%\vspace*{5cm}
\caption{Dimensions of the cylindrical RP.}
\label{fig:cylindricalpot_dimensions}
\end{figure}
After considering various options and after an iterative optimisation, the 
following design has been adopted for the new RPs 
(Figures~\ref{fig:cylindricalpot} and~\ref{fig:cylindricalpot_dimensions})~\cite{JBEDMS}.
The volume housing the detectors will have a cylindrical rather than rectangular
box shape. This choice provides the necessary space for all potential 
technologies of timing detectors and at the same time reduces the beam 
coupling impedance by minimising resonant cavities.
The ferrite in this design will be integrated in the 
(stationary) flange rather than mounted on the moving detector housing. It will have
a ring geometry (inner diameter = 150\,mm, radial 
width = 15\,mm, thickness = 5\,mm).
Furthermore, all vacuum-side surfaces of the RP stations are foreseen to
receive a $2\,\mu$m thick Non-Evaporative Getter (NEG) coating.

\subsubsection{The Mechanical Tests of New Roman Pot Cylinders}
The new cylindrical Roman Pots with the thin window of 300\,$\mu$m thickness have been produced (see Figure~\ref{fig:cylindricalpot}) in a collaboration of CERN with industries. 
After the production of the first prototype in fall 2013 a series of tests have been performed at CERN in collaboration with different support groups, 
to approve the compatibility  of this new RP design with the LHC requirements.
The deflection of the thin window was measured as function of the applied air overpressure
simulating the possible pressure difference seen by the RP when the LHC beam tube is under vacuum and the inner side of the RP is under atmospheric pressure.
Such pressure difference will occur during the installation of the detector components or a failure of the vacuum system and a leak of the 
feedthrough integrated in the flange separating the atmosphere from the inner side of the RP~\cite{edms_rpwind}.  
Furthermore the He transmission was measured in a special setup and the compatibility of this design with the LHC leak rate requirements was shown~\cite{edms_rpleak}.

\subsubsection{The RF Shield for the Box-Shaped Horizontal Roman Pots}
\label{sec:rfshield}
\begin{figure}[h!]
\begin{center}
\includegraphics[height=4.6cm]{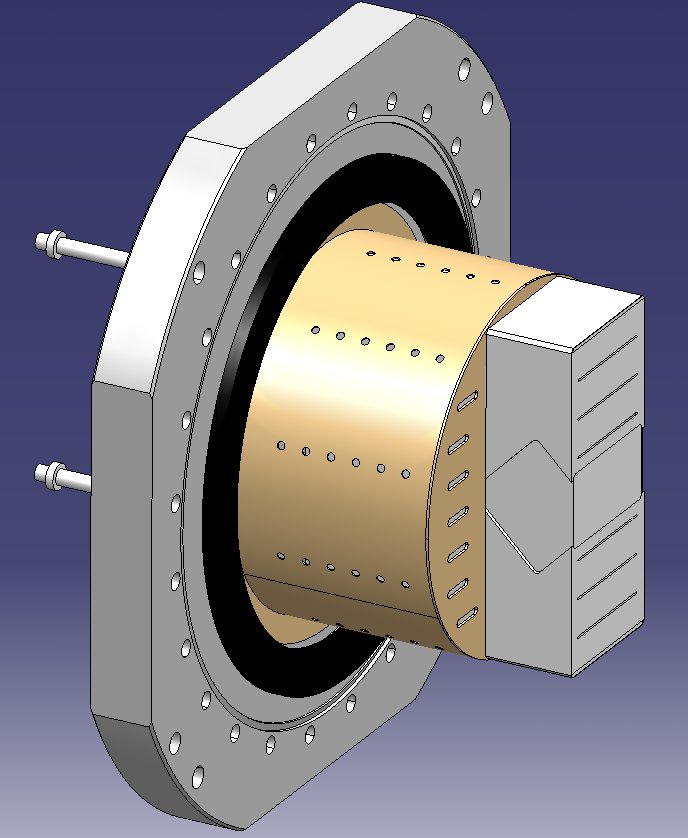}
\includegraphics[height=4.6cm]{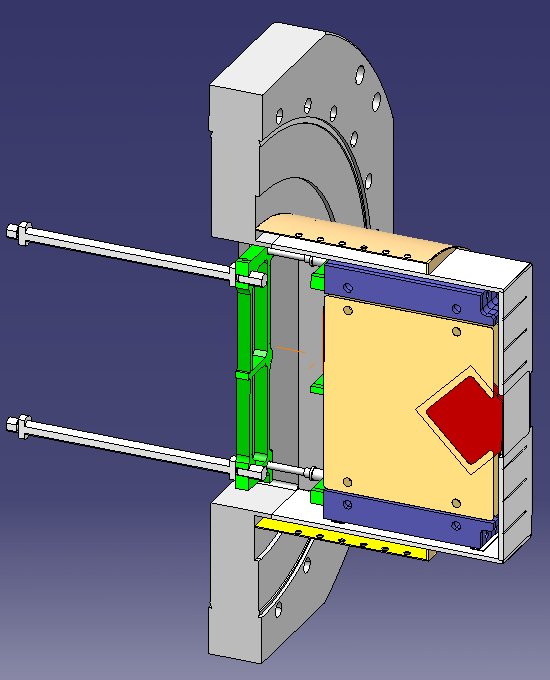}
\includegraphics[height=4.6cm]{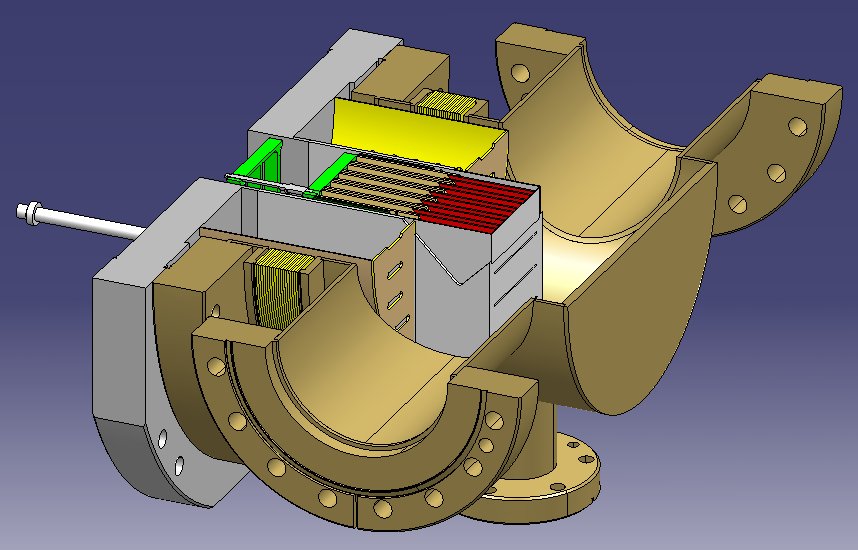}
\includegraphics[width=6cm]{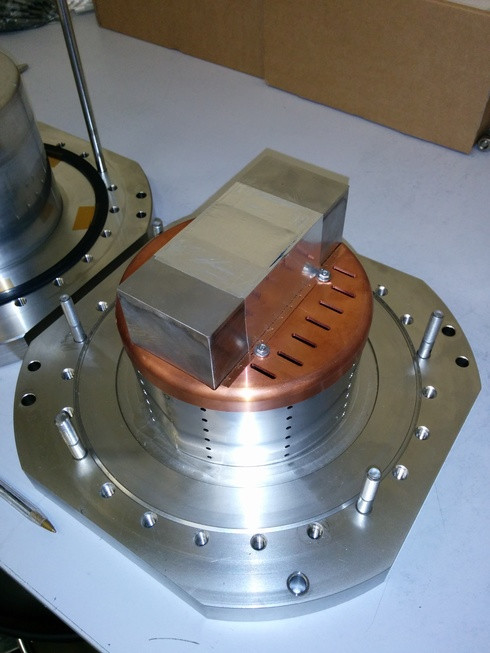}
\caption{Top: Drawings of the cylindrical RF shield for the box-shaped horizontal RPs. Bottom: the manufactured shield.}
\label{fig:cylindricalshield}
\end{center}
\end{figure}

Given that the existing horizontal pots housing tracking detectors will have to 
cope with the same high luminosity conditions as the new timing RPs, some 
adaptations will be made:
\begin{itemize}
\item To reach the same impedance reduction as for the new cylindrical pots 
(see Section~\ref{sec:impedance}),
the rectangular detector boxes will be successively 
equipped with 1\,mm thick cylindrical 
copper RF shields (Figure~\ref{fig:cylindricalshield}). Holes in the shield allow 
for the gas flow
necessary to establish a vacuum equilibrium inside and outside the shield.
The number and dimensions of these holes have been defined in cooperation with the 
LHC vacuum group: In the lateral, cylindrical wall there are 3 rows of 15 circular 
holes each with a diameter of 1\,cm; the wall facing the beam has 8
slits of $3 \times 12 \rm mm^{2}$ with rounded corners (2\,mm radius).
The shield is retracted by 30\,mm from the box window facing the beam, in 
order not to intercept any signal protons with the shield material. In the first
step, during LS1, the horizontal pots of the RP210 station will 
receive the RF shields, in order to gain experience without touching the RP220
station.
\item 
The horizontal RP210 stations will be equipped with new vacuum bellows and modified flanges that allow to integrate 
the same ferrite geometry as used in the new RPs. 
The ferrites of the vertical RPs of the RP210 stations will be exchanged with the new TT2 material as in the RP220 stations.
\end{itemize}

\subsubsection{Interaction of the Roman Pot with the Beam Environment}
\label{sec:beamenvironment2012}
\begin{figure}[h!]
\begin{center}
\includegraphics[height=5.7cm]{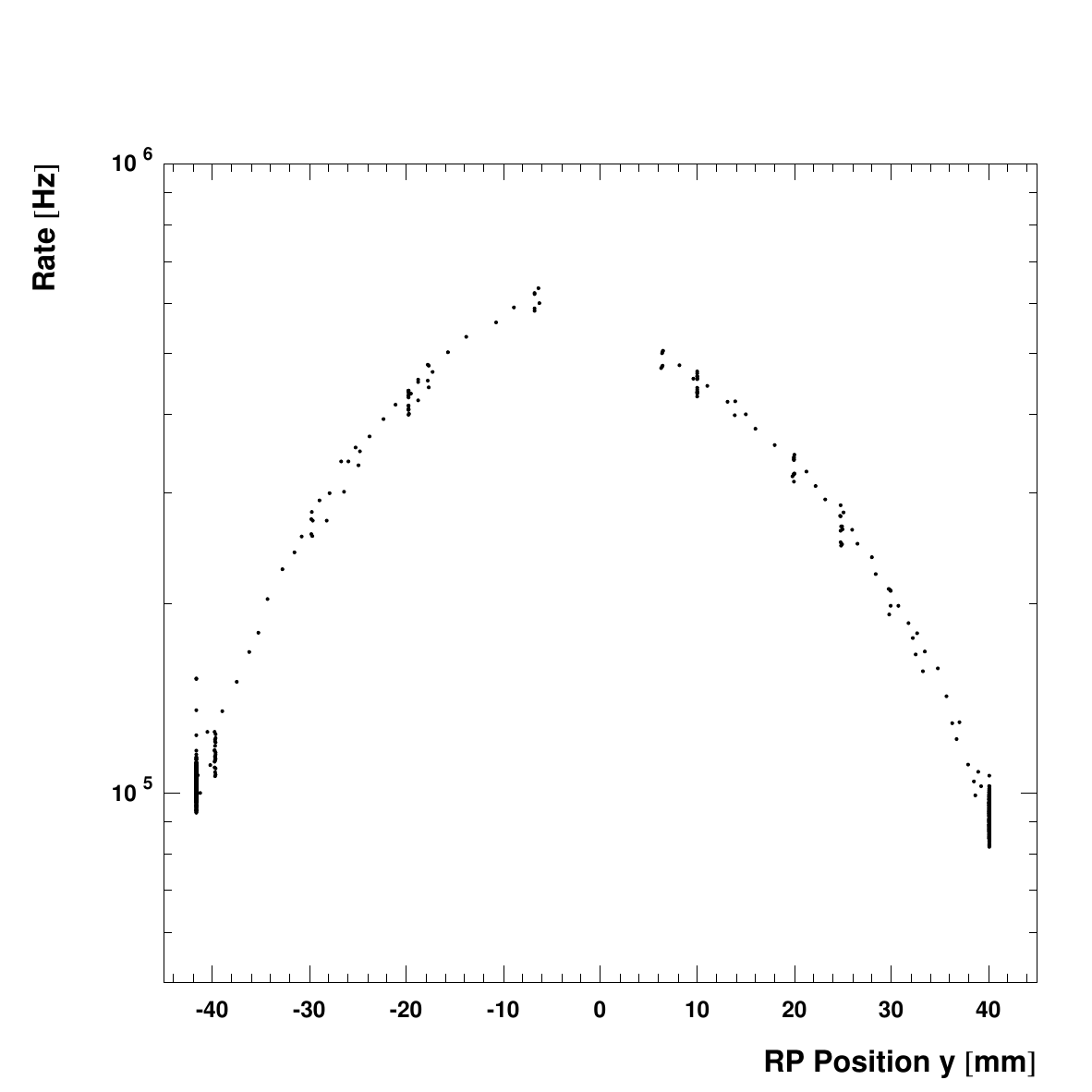}
\includegraphics[height=5.4cm]{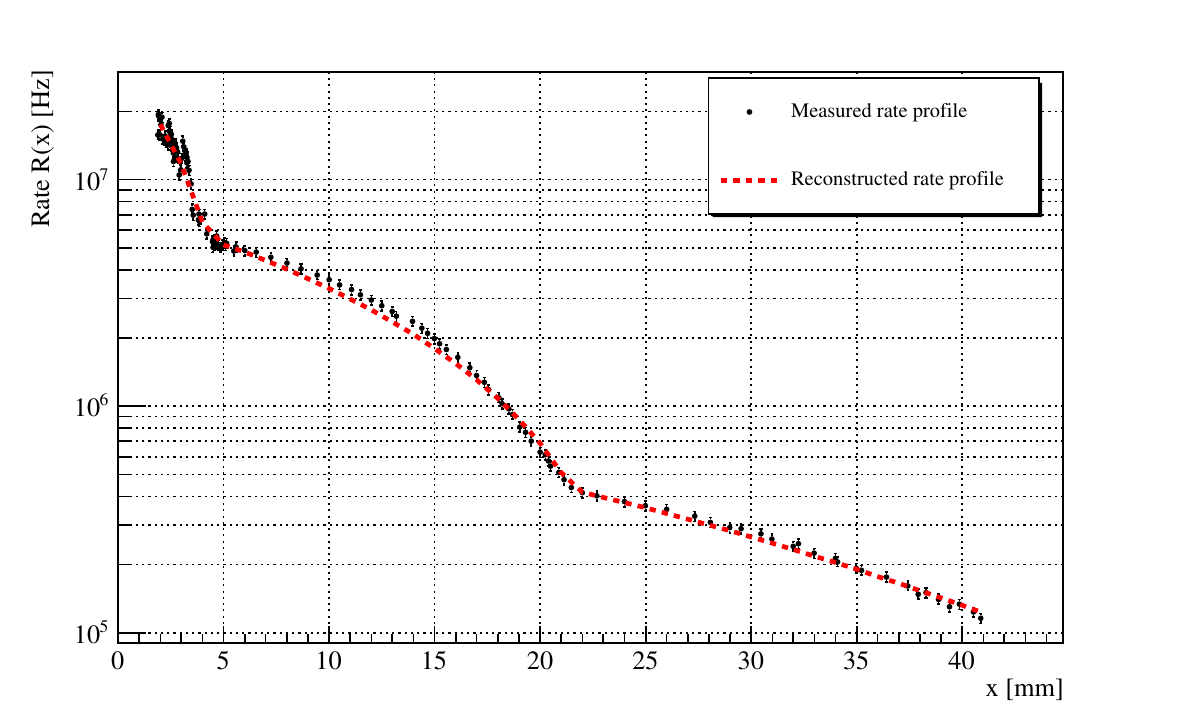}
%%\vspace*{5cm}
\caption{Left: vertical beam profile measured via the trigger rates in the 
detectors of the top and bottom pots of Sector 56, 220-N. Right: horizontal beam profile
measured in Sector 45, 220-N; the luminosity reduction by beam separation has been 
corrected for. The reconstructed curve is the result of a convolution fit 
discussed in~\cite{frici_note}.
}
\label{fig:beamprofiles}
\end{center}
\end{figure}

In October and November 2012 several test insertions of the RPs in 
normal high-luminosity fills at $\sqrt{s}=8\,$TeV with $\beta^{*}=0.6\,$m were 
performed. While the vertical pots had no problems to reach the target 
distance of 12\,$\,\sigma$ from the beam centre, 
the horizontal pots encountered a very intense collision 
debris halo, and repeatedly the beam was dumped by 
showers hitting the Beam Loss Monitors at a pot position of about 30\,$\sigma$.
Separating the beams in IP5 finally reduced the luminosity -- hence the debris
halo -- by a factor 22.7, enabling the approach to the horizontal target distance
of 14$\,\sigma = 1.6\,$mm from the beam containing 1368 bunches of -- at 
RP insertion time --
$1.1\times10^{11}$ protons or a total charge of $1.45\times10^{14}$ protons. 
The beam profiles (Figure~\ref{fig:beamprofiles}) 
measured during these insertions can be used to benchmark shower simulations 
(Section~\ref{sec:showerMC}).
The first lesson for the upgrade from this exercise is that 
a horizontal RP approach to physics-relevant positions of $10\div15\,\sigma$
will require to absorb the showers produced by the RPs in order to protect the
quadrupole Q6. The solution is the addition of the new collimators
TCL6 between the RR220 station and Q6 (see Section~\ref{sec:collimation}).

While the horizontal pots were stationary at 14$\,\sigma$ from the beam centre,
i.e. for about 30 minutes, the temperature sensors on the detector hybrid 
boards in those RPs registered a temperature increase by about 4$^{\circ}$C, 
despite the active cooling of the detector packages. This effect is explained
by impedance heating of the ferrite collar mounted around the box-shaped 
housing on the beam vacuum side. A direct temperature measurement 
near the ferrite was not available, but given the long thermal conduction 
path from
the heat source to the detector package, and the absence of convection inside
the pot due to the secondary vacuum, the temperature of the ferrites may have
reached values well above 100$^{\circ}$C, the Curie temperature of the ferrites
(material 4S60 from Ferroxcube) above which they are ineffective.
Another piece of evidence for substantial heating of the ferrites was given by
the vacuum deterioration observed after the very close insertion of the 
horizontal RPs. First laboratory tests have shown that the ferrite material 
installed around the pots shows substantial outgassing at high temperatures.

Triggered by the problems and observations described above, a programme of 
simulations, extended laboratory tests, and design optimisations
was defined; it is discussed in the following section.

\subsubsection{Impedance}
\label{sec:impedance}

As mentioned in the previous section, during a RP insertion to 1.6\,mm 
from a high-intensity beam (1368 bunches of $1.1\times10^{11}$ protons) a 
temperature increase was observed on the detector hybrid boards. This 
effect can probably be attributed to impedance heating. It is 
hypothesised that the 4S60 ferrite mounted around the RP box reached a temperature above
100$^{\circ}$C, the Curie temperature, which resulted in the loss of ferrite 
effectiveness and hence even stronger heating by the now non-damped cavity resonance 
near 550\,MHz~\cite{oldrfnote}. Note, however, that no other impedance effects were 
observed, in particular, no beam instabilities.

The aim of the work presented here~\cite{nicolaminafra} is the optimisation of the RP design to minimise the 
beam-coupling impedance, in particular at very close distances to the beam, in view of
more regular and extended RP insertions in the future.

The impedance seen by a beam of particles has contributions from the shape of the vacuum chamber (geometrical impedance) and from the finite conductivity of the material used for its construction. 
The remainder of this section focusses on the dominant geometrical contribution of three RP designs: the standard box-shaped RP, the new cylindrical RP, and the
improved box-shaped RP with shield (introduced in Section~\ref{sec:rfshield}).

The study was performed by simulating the passage of a charge distribution (source charge) through a cavity, in this case through a RP, and computing the wake field felt by a longitudinally or transversely displaced second charge (test charge).
The potential felt by the test charge is then used to compute the longitudinal or transverse impedance using Fourier Transforms.

Three impedance effects have to be addressed:
\begin{itemize}
\item \textbf{Beam-induced heating}, i.e. the transfer of power from the beam to the 
lossy wall of a cavity, is determined by the frequency-dependent real part of the 
longitudinal impedance in conjunction with the power spectrum of the beam

%:
%\begin{equation}
%P_{\rm loss} = 2 \, I^2 \sum_{p=0}^{\infty} \emph{PS} (p M\rq{} f_{\rm rev}) \Re [Z_{\rm long} (p \, M\rq{} \, f_{\rm rev})] \: ,
%\label{eqn:ploss}
%\end{equation}
%where
%\begin{itemize}
%	\item $\emph{PS}(f)$ is the power spectrum;
%	\item $f_{\rm rev}$ is the revolution frequency, 11.245\,kHz for the LHC;
%	\item $M\rq{}$ is the number of buckets, 1782 for the LHC with a bunch spacing of 50\,ns;
%	\item $I = M e N_B f_{\rm rev}$ is the beam current, with $M$ number of bunches, $e$ charge of the proton, $N_B$ number of protons per bunch;
%	\item $Z_{\rm long}$ is the longitudinal impedance.
%\end{itemize}

The main contribution to the heating comes from resonances below 1.5\,GHz; at higher
frequencies the beam power spectrum is attenuated by more than $\sim30$\,dB relative to its value at $f = 0$~\cite{baudrenghien2011lhc}.
For all power calculations a current of 0.6~A (corresponding to $M$=2808 and $N_B$= 1.2 $\cdot$ 10$^{11}$ protons) was used.
\item \textbf{Longitudinal instabilities} are related to the effective longitudinal impedance.
The effective impedance is the impedance actually felt by the beam: it is given by the impedance convoluted with a
weighting function $\sigma(f)$ which is determined by the bunch profile.

%\begin{equation}
%\label{efflongdef}
%	Z^{\rm eff} =\frac{\sum\limits_{f} Z(f) \sigma(f)}{\sum\limits_{f} \sigma(f)}
%\end{equation}
%
A conservative estimation of the effective longitudinal 
impedance is the slope of the imaginary part of the longitudinal impedance
at low frequencies $Z_{\rm long}^0/n$
%:
%\begin{equation}
%\label{efflong}
%	\Im Z_{\rm long}^0/n = \lim_{f \to 0}  \, f_{\rm rev} \,\frac{ 
%{\rm d} \Im Z_{\rm long}}{{\rm d}f}
%\end{equation}
where $n=f/f_{\rm rev}$ is the harmonic number.
It is possible to show~\cite{nicolaminafra} that
$\Im Z_{\rm long}^0/n < (\Im Z_{\rm long}/n)^{\rm eff}$.

The simulated value for $\Im Z^{0}_{\rm long}/n$ will be compared with the measured value for $ (\Im Z_{\rm long}/n)^{\rm eff}_{\rm LHC} = \rm 90\,m\Omega$~\cite{Chapochnikova}.
\item \textbf{Transverse instabilities} have, analogously, their origin in the 
low-frequency behaviour of the transverse impedance. Following the same approach for the effective transverse 
impedance, it is possible to compute the \emph{driving} (or \emph{dipolar}) impedance and relate it to the transverse impedance~\cite{EMetral}, $\Im Z_{\rm t}^{\rm driving}$.

%:
%\begin{equation}
%\label{efftrans}
%\Im Z_{\rm t}^{\rm driving} = \frac{\partial \Im Z_{\rm t}}{\partial t_{\rm source}} 
%\end{equation}
%where $t = x, y$ and $t_{\rm source}$ represents a small transverse displacement of the source charge, which creates
%the wake field, from the nominal position.
%The value is usually constant at low frequency ($< 500$\,MHz).
A normalisation with
the ratio of the beta function value at the equipment under study, 
$\beta_t$, and the 
average over the ring, $\langle\beta_t\rangle = 70\,$m, facilitates the 
comparison with other equipments at different positions in the 
machine:
\begin{equation}
\label{efftransnorm}
\overline{\Im Z_{\rm t}^{\rm driving}} = \Im Z_{\rm t}^{\rm driving} \frac{\beta_t}{\langle\beta_t\rangle} 
\end{equation}
The new RP will be horizontal ($t=x$); moreover, among all the RPs the highest value (worst case) of $\beta_{x} = 98\,$m is reached at the unit 210-N. 
This value can be compared with 25 $\rm M\Omega /m$, a conservative value of the value expected for the full machine~\cite{EMetralTransMeas}.
\end{itemize}

For the new cylindrical pots, simulation results indicate that no low frequency resonances 
($<~1.4\,$GHz) are present if the gap between the detector housing and the flange is 
completely closed, which of course prevents any RP movement.
Mechanical constraints require at least 2.5\,mm gap between the housing and the flange.
With this gap a resonance at 470\,MHz appears, % as shown in Figure~\ref{fig:CylWOF_25mm}; 
however its impedance is smaller than for the standard box-shaped RP.
%
%\begin{figure}[h!]
%\centering
%\includegraphics[height=4.5cm]{figs/detector/CylWOF_25mm.jpg}
%%\vspace*{5cm}
%\caption{Simulated $\Re [Z_{\rm long}]$ of the cylindrical RP without ferrite. The resonance at 470\,MHz is due to the cavity between the flange and the detector housing. 
%The darkened part of the graph has negligible impact on the heating due to the 
%strongly attenuated LHC power spectrum at such high frequencies.}
%\label{fig:CylWOF_25mm}
%\end{figure}
The position and the dimensions of the ferrite has been optimised through various iterations considering also vacuum and mechanical construction.
The final design consists of a ferrite ring (inner diameter = 150\,mm, radial 
width = 15\,mm, thickness = 5\,mm) integrated into the 
flange, as far as possible from the beam %(Figure~\ref{fig:cyl_ferrite}).
%This design is feasible and can be easily integrated in the existing design.

\begin{figure}[h!]
\begin{center}
\includegraphics[width=0.495\textwidth]{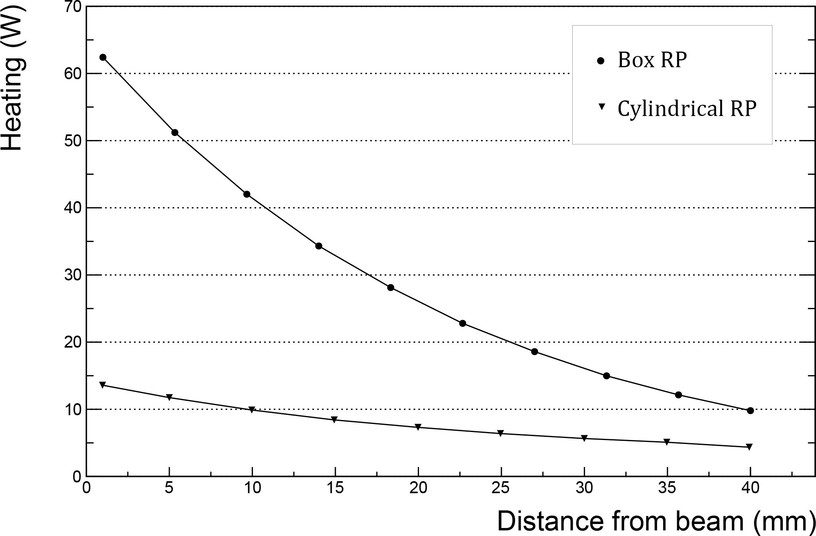}
\includegraphics[width=0.495\textwidth]{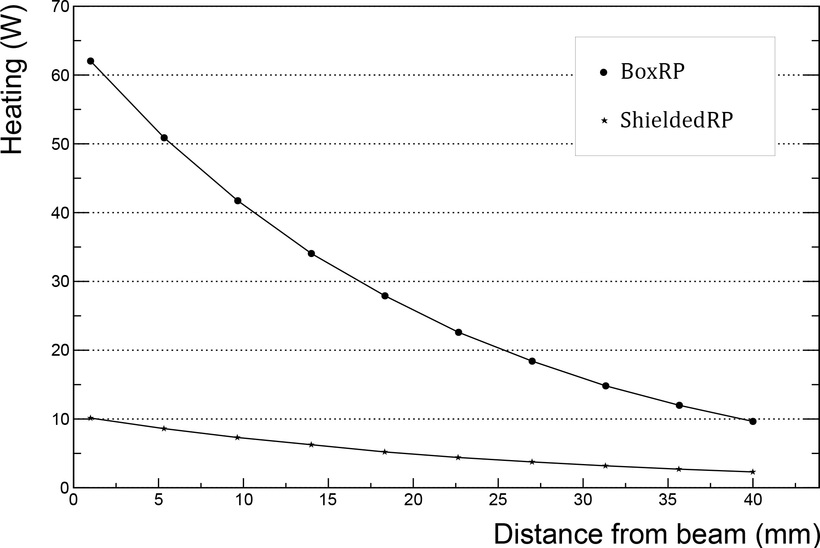}
%%\vspace*{5cm}
\caption{Power lost by the beam passing through the RP, for the three RP designs ($I=0.6$\,A).
}
\label{fig:Ploss_vs_d}
\end{center}
\end{figure}

\begin{figure}[h!]
\begin{center}
\includegraphics[width=0.495\textwidth]{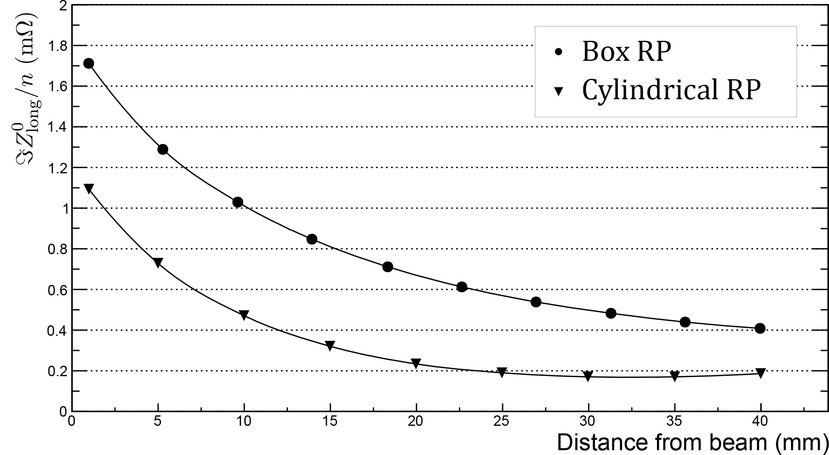}
\includegraphics[width=0.495\textwidth]{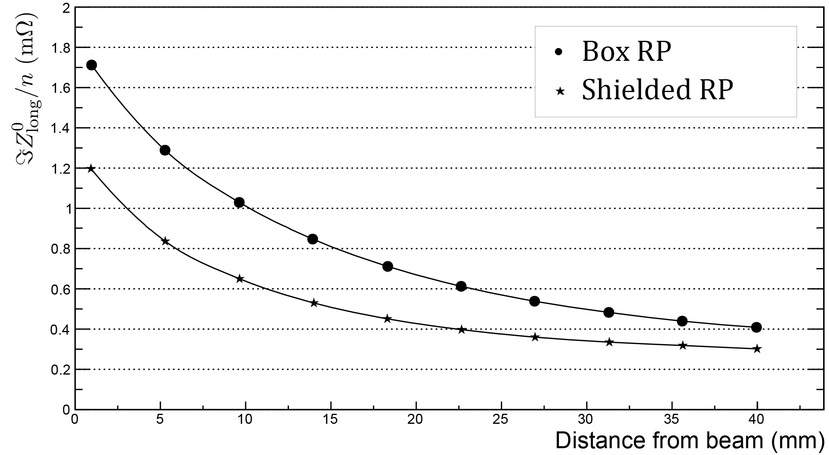}
%%\vspace*{5cm}
\caption{Effective longitudinal impedance as a function of 
the RP distance from the beam, for the three RP designs.
}
\label{fig:Zlong_vs_d}
\end{center}
\end{figure}

%
%\begin{figure}[h!]
%\begin{center}
%\includegraphics[height=6cm]{figs/detector/Ferrite5x15.jpg}
%%\vspace*{5cm}
%\caption{Detail of the geometrical model used in the impedance simulations. It shows the
%new ferrite ring to be integrated in the flanges of the cylindrical pots and of the
%improved pots with shields. 
%}
%\label{fig:cyl_ferrite}
%\end{center}
%\end{figure}
%
%Figure~\ref{fig:Zlong} shows the real and imaginary parts of the longitudinal impedances
%of the three RP designs with ferrites. 
In all cases, the 470\,MHz resonance is damped and
smeared beyond recognition. At low frequencies, the cylindrical and shielded RPs have 
a smaller $\Re [Z_{\rm long}]$ than the standard RP. This is also reflected by the 
reduced heating for the new designs (Figure~\ref{fig:Ploss_vs_d}).
%
%\begin{figure}[h!]
%\begin{center}
%\includegraphics[width=0.495\textwidth]{figs/detector/ZBox.jpg}
%\includegraphics[width=0.495\textwidth]{figs/detector/ZCyl.jpg}\\
%\includegraphics[width=0.495\textwidth]{figs/detector/ZShield.jpg}
%%\vspace*{5cm}
%\caption{Simulated longitudinal impedance for the standard box-shaped RP (top left), 
%the cylindrical RP (top right) and the shielded RP (bottom). 
%Both real and imaginary part are shown.
%}
%\label{fig:Zlong}
%\end{center}
%\end{figure}
%
%
Figure~\ref{fig:Zlong_vs_d} shows the effective longitudinal impedance as a function of 
the RP distance from the beam. Also here, the new designs have led to a significant 
reduction.

These results are numerically summarised in Table~\ref{table:impedances}.
\begin{table}[!ht]
\footnotesize
\centering
\renewcommand{\arraystretch}{1.5}
\begin{tabular}{|p{0.14\textwidth}|p{0.13\textwidth}|p{0.09\textwidth}|p{0.11\textwidth}|p{0.09\textwidth}|p{0.11\textwidth}|p{0.09\textwidth}|}
\cline{1-7}
		&	Distance from the\newline beam [mm]	&	$\frac{\Im Z^{0}_{\rm long}}{n}$ [m$\Omega$] &	fraction of $(\frac{\Im Z_{\rm long}}{n})^{\rm eff}_{\rm LHC}$ (90\,m$\Omega$)	&	$\overline{\Im Z_{\rm trans}^{\rm driving}}$ [M$\Omega$/m]
		& fraction of $\Im (Z_{\rm x})^{\rm eff}_{\rm LHC}$ (25\,M$\Omega$/m)		&	Heating [W] I=0.6~A\\
	\cline{1-7}
	\multirow{3}{1cm}{\centering \mbox{Box RP}}			&	1		&	1.7					&	$< 1.9$\,\%				&	0.15							&	$< 0.6$\,\%				&	62 \\
								&	5					&	1.3					&	$< 1.4$\,\%				&								&							&	52 \\
								&	40 (garage)				&	 0.41					&	$< 0.45$\,\%	 			&								&							&	10 \\
	\cline{1-7}
	\multirow{3}{1cm}{\centering \mbox{Cylindrical RP}}		&	1			&	1.1		 			&	$< 1.2$\,\%				&	0.11							&	$< 0.5$\,\%				&	13 \\
								&	5					&	0.73					&	$< 0.81$\,\%				&								&							&	11 \\
								&	40 (garage)				&	0.18					&	$< 0.20$\,\%				&								&							&	4 \\
	\hline
	\multirow{2}{1.5cm}{Shielded RP}		&	1					&	1.2					&	$< 1.3$\,\%				&	0.2							&	$< 0.8$\,\%				&	10 \\
								&	40 (garage)				&	0.30					&	$< 0.33$\,\%				&								&							&	2 \\
	\cline{1-7}
	\end{tabular}
\caption{Main results of the simulation of the present box RP (Box RP), the cylindrical RP (Cylindrical RP), and the Box RP with Shield. Longitudinal and transverse impedances are compared with the total values estimated for the present LHC effective impedances.}
\label{table:impedances}
\end{table}

\subsubsection{The RF Test in the Lab}
The new cylindrical RP and the RF shield in combination with ferrites were developed to reduce the 
RF interaction with the LHC beam. 
Prior the serial production of components a new RP cross with flanges and bellow
was manufactured. The prototype of the cylindrical RP and the RF shield in 
combination with the box-shaped RP have been integrated in this new  RP cross for RF measurements. 
In February 2014 measurements were performed by TOTEM and the LHC impedance group in the TIF lab of CERN. In detailed measurements with and 
without the new ferrites and at different insertion positions of the RP, the RF characteristics of this new geometry could be determined and good agreement 
with the predictions, based on simulations could be found~\cite{note_rprf}.

\subsubsection{Impact of Ferrite outgassing on LHC Vacuum}
The vacuum degradations observed in 2012 after very close horizontal RP 
insertions to high-intensity beams triggered the following consolidation
activities for improving the vacuum compatibility of the RP system:
\begin{itemize}
\item Ferrite material improvements:\\
The 4S60 ferrites used in the RP system before LS1 are now (but not in 2006) 
known to show high outgassing rates unless they are baked out at 
1000$^{\circ}$C~\cite{gregory_test1}.
In the TOTEM RPs these ferrites were installed as received from the 
manufacturer and then baked out \textit{in situ} at about 200$^{\circ}$C like all
other beam-pipe components, which turned out not to be sufficient.
Since the 4S60 ferrites have in addition a low Curie temperature of only  
100$^{\circ}$C, alternative ferrite materials are being investigated instead of 
only baking out the 4S60 material at 1000$^{\circ}$C.
The material used for the TOTEM RPs is TT2-111R from 
TransTech with a Curie temperature of $375^{\circ}$C and an acceptable 
outgassing after bake-out at 1000$^{\circ}$C~\cite{gregory_test2}.
An alternative material 
for possible future use is
4E2 (Ferroxcube) with a Curie temperature
of about $400^{\circ}$C; its outgassing behaviour remains to be tested.
\item The new geometrical ferrite configuration reduces the ferrite surfaces exposed
to the vacuum by an order of magnitude from 220\,cm$^{2}$ per standard RP to 23\,cm$^{2}$
per cylindrical RP.
\item All components exposed to the primary beam vacuum have been proposed
by the vacuum group to be coated with NEG, as far as technically possible.
\end{itemize}

\subsubsection{Generation of Particle Showers}
\label{sec:showerMC}
To assess the generation of particle showers by RPs interacting with the 
intense debris halo (see 2012 experience discussed in 
Section~\ref{sec:beamenvironment2012}), Geant4 simulations implementing 
detailed models of the rectangular and the new cylindrical RPs 
have been carried out~\cite{frici_note}. 

The first goal of the study was to identify the contributions from the 
different structural elements of a RP to the shower creation.

As expected, the number of secondary particles is mostly determined by the 
amount of material traversed. The key observations are:
\begin{itemize}
\item 
In both RP designs, the bottom foil produces by far the highest number 
of secondary particles, followed by the thick body walls with 2 to 3.5 orders 
of magnitude lower rates. The orthogonally traversed thin front and back windows
produce the lowest numbers of secondaries.
\item 
The bottom foil of the cylindrical pots produces more than 10 times more 
secondaries than the much shorter foil of the standard pots. In the other 
elements the shower production is similar in the two designs.
\item
The showers from the bottom foil of the cylindrical pot are 3 times 
wider than the ones
from the standard pots: 99\% contained in 0.6\,rad$\,=34^{\circ}$  rather than in 0.2\,rad$\,=11^{\circ}$.
\item
The two projections, horizontal and vertical, are almost identical.
\end{itemize}

The simulated secondary 
particle distribution in a scoring plane 6\,m
downstream of a standard horizontal RP inserted to 2\,mm from the beam centre
shows that
at the entrance point of TCL6,
25\,\% of the secondary particles, carrying 90\,\% of the energy, are contained within the beam-pipe radius and thus 
intercepted by TCL6. How much of this flow leaks through the TCL6 aperture and
thus hits Q6 will be the subject of the FLUKA study discussed in 
Section~\ref{sec:collimation}.

%----------------------------------------------------------------------------

\subsubsection{Interplay between Roman Pots and Collimators}
\label{sec:collimation}
The modified RP system with relocated and additional units will be 
embedded in an upgraded collimation system. This section discusses the 
performance of the new combined layout in terms of physics acceptance and 
machine protection.

\paragraph*{The New Collimators TCL4 and TCL6}

LHC operation at highest luminosities may require additional protection of 
the quadrupoles Q5 and Q6 against collision debris from IP5~\cite{stefano}.

To protect Q5, new collimators, TCL4, have been installed on the outgoing beams
in the old location of the RP147 station, and the already existing collimators
TCL5 may be partially closed. Since both TCL4 and TCL5 are located upstream of
the RP stations and can intercept diffractive protons if too tightly 
closed, the aperture settings of these collimators will be the
result of an optimisation study maximising the physics acceptance as far as 
compatible with the necessary magnet protection.

Downstream of the last RP unit, 220-F, another collimator, TCL6, has been
installed~\cite{ecrcollim} on the outgoing beam to protect the quadrupole Q6 against 
debris from IP5, thus taking over a part of the original role of TCL5 
which cannot be too tightly closed without intercepting all the signal protons 
to be measured by the RP system. Another beneficial effect of this new 
collimator is its capability to absorb showers created by the insertions of 
the horizontal RPs close to the beam. RP operations at low $\beta^{*}$
and high luminosities in 2012 have demonstrated that without any absorber 
behind the RP stations, insertions were limited to distances greater than 
30\,$\sigma$, because the showers caused by the pots' interaction with the 
debris halo brought the dose rates measured by the Beam Loss Monitors above 
the beam dump thresholds. The improvement by the addition of TCL6 will be 
the subject of a FLUKA study.

\paragraph*{Optimisation of Roman Pot and Collimator Settings}

This section discusses the strategy for defining an optimal combined set of jaw 
positions for the RPs and the collimators TCL4, TCL5 and TCL6. Given that 
the TCLs are only required at highest luminosities, only the low-$\beta^{*}$
running scenarios are relevant for these considerations.
The TCL collimators are designed to protect the quadrupoles Q5 and Q6 against
debris from collisions at IP5. Their jaws approach the beam 
horizontally and potentially intercept diffractively scattered protons, thus
interfering with the physics measurements in the RPs.
Therefore, the aim of the optimisation is to find:
\begin{enumerate}
\item jaw positions for TCL4 and TCL5 that leave the aperture as widely open as 
allowed by the protection needs of Q5, i.e. the dose rate received by Q5 has
to stay well below the magnet quench threshold;
\item RP positions as close to the beam as allowed by the protection
capacity of TCL6 to prevent Q6 from quenching.
\end{enumerate}
The upper limit $\xi_{\rm max}$ for accepted momentum losses of diffractive 
protons is given by minimum value of the ratio $d_{x}/D_{x}$ between horizontal 
aperture and dispersion along the path from the interaction point to the RP.
Table~\ref{tab:apertures} gives the values of $10\,\sigma_{x}$ beam width and
the dispersion $D_{x}$ in all TCL collimators and in some RP locations for the 
$\beta^{*} = 0.55\,$m optics at $\sqrt{s}=14\,$TeV.

\begin{table}[h!]
\begin{tabular}{|l|c|c|c|c|}
\hline
Beam Element & Position $s$ [m] from IP5 & $10\,\sigma_{x}(s)$ [mm] & $D_{x}(s)$ [mm] & $|\xi(10\,\sigma)|$\\
\hline
TCL4         & 149                       & 5.2       & -66   & 0.079 \\
TCL5         & 185                       & 2.8       & -83   & 0.034 \\
RP 210-N     & 202                       & 2.2       & -90   & 0.024 \\
RP 220-F     & 220                       & 0.90      & -80   & 0.011 \\
TCL6         & 221                       & 0.89      & -80   & 0.011 \\
\hline
\end{tabular}
\caption{Horizontal beam envelope ($10\,\sigma$) and dispersion at the TCL 
collimators and at the first and last RP unit. The last column gives the 
$\xi$-value at $10\,\sigma$ from the beam centre (for $t=0$).
}
\label{tab:apertures}
\end{table}
The most stringent impact on diffractive proton acceptance is made by TCL5
which at its nominal jaw position of $10\,\sigma$ from the beam centre would 
intercept all protons with $\xi>0.034$ while for the physics programme a 
cut-off greater than 0.1 would be desirable. 
Therefore the collimation group developed an alternative
scheme for fills with RP operation where TCL5 would be fully open and both 
TCL4 and TCL6 closed to $10\,\sigma$. However, in that scheme TCL4 would 
be the bottleneck producing a cut at 0.079. A new study presently carried out
by the FLUKA team investigates the possibility to open TCL4 to $15\,\sigma$
and complement its protection by closing TCL5 to $35\,\sigma$. In this way, 
both collimators would lead to the same upper $\xi$ cut-off at 0.11.

Once the optimal settings for TCL4 and TCL5 will be fixed, another FLUKA study
will focus on the impact of RP insertions on Q6 and its mitigation by closing
TCL6 to $10\,\sigma$. It is expected that a horizontal RP approach to 
a minimum distance between 11 and $14\,\sigma$ should be possible, corresponding
to minimum accepted $\xi$-values between 0.012 and 0.016.

\subsection{Requirements on the Timing Detectors and Strategy}
\label{sec:timing_cherenkov}
 
The need for precise timing detectors measuring the time difference of the
protons in the two arms of the spectrometer is justified 
%in Chapter \ref{sec:??????} 
as an effective way to reduce the pileup background. 
A baseline time resolution of $\sigma(t)=10$\,ps, corresponding to
a vertex resolution $\sigma(z_{pp})=2.1$\,mm, is set as an ambitious target of the CT-PPS project.
%A backup scenario where the time resolution is limited to a conservative value
%of 30\,ps is also evaluated and justified in the physics performance studies
%presented in Chapter~\ref{sec:?????????}.

The required detector area is small ($\leq4\rm cm^2$). In addition to a good time resolution, 
the use of a detector with a small dead region is a key requirement.  On the
side adjacent to the beam, the dead region should be at the level of $ \sim 200\,\mu$m or below, matching that of the tracking detectors. The distance between
the active area and the vacuum includes in addition the bottom of the RP
($0.3^{+0.02}_{-0.10}$\,mm). The scattered protons are deflected out of the 
beam by the LHC magnets, but at the $z$ position 
of the detectors they are displaced by only a few mm, so any inactive area (on the inner edge) 
causes a loss in acceptance at low masses.

The detectors should be radiation hard. Close to the beam where the detectors
are located, we expect a proton flux of about $5 \times 10^{15}\rm\,cm^{-2}$ per
100\,fb$^{-1}$. The expected thermal neutron flux extrapolated from TOTEM measurements is about
$10^{12}\rm\,cm^{-2}$ per 100\,fb$^{-1}$. In the case of Cherenkov detectors the
photodetectors will be farther from the beam, where radiation field is reduced to the neutron component. 
Replacing the photodetectors or solid-state timing detectors approximately once a year is feasible, 
as they are accessible and relatively inexpensive.

As there is often more than one proton in the acceptance from the same bunch crossing, 
a fine segmentation is also required. The detectors should have the capability
of measuring the times of two or more particles from the same bunch crossing, and of being read out every 25\,ns, 
with no significant remnant signals from earlier crossings. This implies segmentation. 

A detector based on Cherenkov technology is developed for precise timing as the
baseline proposal. 

%Similar technology was studied earlier in the context of the 
%FP420 project R\&D project~\cite{fp420}. 
%These included detectors based on quartz (or sapphire) radiators as well as gas
%Cherenkov detectors. Several prototypes have been built and evaluated in test beams.
%Recently there has been growing interest in timing measurements, which has led
%to several on-going R\&D efforts on the use of solid state detectors for timing, 
%notably in the frame of the LHC Phase 2 Upgrade programme. Several CT-PPS groups are 
%participating to these efforts and are willing to explore the option of using such detectors in PPS.

Presently, the state-of-the-art of time resolution with minimum ionizing
particles in a single detector layer is the following:
1) gas Cherenkov $\sim$\,15\,ps~\cite{gastofnim}; quartz Cherenkov $\sim$\,30\,ps
\cite{clermont}; diamond sensors $\sim$\,100\,ps \cite{diamond,diamond2};
silicon sensors $\sim$\,100\,ps \cite{silicontiming}. Complete systems with several detector
layers allow for improved performance.

While Cherenkov based detectors have intrinsically better time resolution and are more mature timing 
technologies, they have some important drawbacks. In the existing prototype implementations, the quartz 
detector is segmented in elements of $3\times3$ mm$^2$, which implies a large
rate of double hits in the same bar per bunch crossing, approaching 50\% in the
sensors close to the beam. We assume that two hits in the same channel cannot be resolved. 
%This source of inefficiency is taken into account in the physics performance simulations in Chapter\ref{sec:??????????}.  
Finer granularity near the beam, where it is most
needed, may be possible but needs further development.

The amount of material introduced by the quartz detector itself is not negligible. In the foreseen 
configuration, the probability that a proton has a nuclear interaction in one
detector is between 7.2\% and 14.6\% (depending on the proton position in the
detector). As the timing detector is located downstream of the pixel detectors  these 
interactions do not affect the track measurement, however they may smear or fully corrupt 
the time measurement introducing another source of inefficiency.

%The gas Cherenkov detector is not affected by this issue, but on the other hand its granularity 
%is only indirect, in the sense that a multi anode Micro Channel Plate
%photomultiplier (MCP-PMT) could possibly allow to disentangle Cherenkov photons
%from different protons. Indeed the reconstruction of the time of two or more protons in 
%the detector is affected by the uncertainty in associating MCP-PMT channel
%information to individual protons.
%This question is investigated with GEANT simulation (\cite{ctpps-tdr:ch9}, Section~5.3).
%The MCP-PMT devices have also a limited life time in terms of integrated charge,
%which requires using the device at reduced gain, preventing the optimal time
%resolution.

Solid state timing detectors have the important advantage of being very thin and allowing for fine 
granularity. Typically solid state detectors are a few hundred microns thick
making it possible to stack ten or more detectors, which to a first
approximation improves the time resolution by the square root of the number of layers. A 
resolution of 30\,ps per detector, possibly achieved as a result 
of the current R\&D effort, would allow for a timing system with the baseline 10\,ps resolution. On the other hand the possibility of defining small size pixels permits 
reducing the rate per channel, 
which improves 
the time measurement and reduces significantly the inefficiency due to double hits. Of course the 
detectors should be able to sustain the high radiation doses involved in this
application, which requires still considerable development.  

Taking into account the previous considerations, we have chosen as
the baseline timing detector the L-bar Quartic (Quartz Timing
Cherenkov) design with $5\times4 = 20$ independent
channels of $3\times3\rm\,mm^2$ area. The SiPM photodetectors are relatively far from the beam, in 
a region where the neutron flux is $\sim 10^{12}$\,neq/cm$^2$ per 100\,fb$^{-1}$.
SiPM devices that tolerate this radiation level are available, as found in
the framework of the HCAL Upgrade project, however a
increase of the leakage current is observed~\cite{hcal_upgrade}. The SiPMs will
probably require replacement after 100\,fb$^{-1}$, which is feasible given the small number of
devices involved. We will also consider the possibility of using GaInP
photosensors, under development for the upgrade of the CMS endcap calorimeter,
given its potentially better tolerance to radiation. Two Quartic detectors
fit inside a cylindrical Roman Pot, providing a combined resolution of the order of 20\,ps. 
The Quartic baseline is presented in Section~\ref{sec:quartic}.

The relatively high fraction of nuclear interactions in the quartz bars prevents
the use of more than two Quartic detectors per spectrometer arm. In order to reduce the amount of the dense
material, we explore the possibility of complementing the Quartic measurement
by using a short ($\sim 10$\,cm) Gas Cherenkov Time-of-Flight detector (GasToF)
inside a second, upstream, cylindrical RP.
The GasToF detector with a multi-anode MCP-PMT may be able to time individual
photoelectrons to achieve multi-proton capability. Combined with the Quartic measurements, this additional 
detector could allow 
to approach the 10\,ps time resolution. While GasToF prototypes have been built
and validated, there are not yet test beam results confirming the multi-hit performance predicted 
from simulation. The possible use of GasToF 
in the experiment is therefore dependent on successful test beam results with final prototypes. The 
GasToF option is described in~\cite{ctpps-tdr:ch9} (Section~5.3).

Both the Quartic and GasToF detectors have a relatively small number of channels
and produce electrical pulses with similar characteristics. Therefore the proposed readout system, based on two well 
known integrated circuits (the amplifier-discriminator NINO and the High
Performance time-to-digital converter HPTDC), can be used by
both detectors. This solution offers a potential for possible future upgrades
as new improved versions of the HPTDC and of the NINO chips are already in the
pipeline.
A reference clock system, complementary to the CMS timing system, provides time synchronization 
with less than 1\,ps jitter between the detectors in opposite arms.  
%Sections~\ref{sec:timing_electronics} and \ref{sec:?????????} describe the
%Cherenkov readout system and the reference timing system, respectively.

In parallel we intend to pursue the R\&D on solid state options for timing, in
particular diamond sensors and silicon sensors with avalanche gain. There are still many challenges to overcome 
before any of these options become a viable timing detector for CT-PPS. This includes the improvement 
of the intrinsic detector resolution, the demonstration of resistance to radiation, and the development 
of suitable low noise and fast electronics. Prototypes will be built and evaluated in test beams. 
The small area, and therefore cost, of the timing detectors allows to foresee the replacement of the 
CT-PPS timing baseline when a better solution is available. 
The solid state options and respective R\&D plans are described in Sections~5.5 and~5.6 in~\cite{ctpps-tdr:ch9}.

\subsubsection{The Cherenkov Quartic Detector as Baseline for Timing}
\label{sec:quartic}
Cherenkov light is prompt and therefore ideal for fast timing, although the amount of light is small
compared to that in scintillator. Radiators need to be transparent, i.e. with a long absorption length
$L_{abs}(\lambda)$, where $\lambda$ is the optical wavelength, preferably into the
ultraviolet, $\lambda \approx$ 200\,nm, where most photons are generated.  The
number of Cherenkov photons radiated is proportional to $1-1/n^2(\lambda)$; more completely
(for charge $Q = 1$, and $\beta = 1$):
\begin{equation}
\frac{d^2N}{dxd\lambda}=\frac{2\pi\alpha}{\lambda^2}\left(1-\frac{1}{n^2(\lambda)}\right) \: ,
\end{equation}
where $\alpha$ is the fine structure constant.

The approximate rule for the number of photoelectrons in a typical detector is:
   \[N_{pe} \sim 90 \: \mathrm{cm}^{-1} \cdot \mathrm{L(cm)\: sin}^2 \theta_{ch} \sim 50\; \mathrm{cm}^{-1},\]
    which gives about 200 photoelectrons for a quartz detector of length 40 mm, to be scaled by a factor for the acceptance of the photons.

The light is emitted along the particle's path in a cone
with half angle (Cherenkov radiation angle) $\theta_{ch}$ given by cos$(\theta_{ch}) = 1/n(\lambda)$.
We have developed fast detectors with both gas and solid radiators.
%The gas Cherenkov detector, or GasToF, is described in~\cite{ctpps-tdr:ch9} (Section~5.3).
%The Cherenkov light is much smaller than in a solid radiator but is very collimated. It is detected with a 
%MCP-PMT,
%and by timing individual photoelectrons one can measure the time of more than one proton 
%from the same bunch crossing. 

 Among solid radiators, fused silica, SiO$_2$, or quartz
(ultraviolet grade, UVT) is commonly used, and it is chosen as our baseline material. As $n(\lambda) \sim 1.48$, there 
is much more light per cm than in a
gas, but since $\theta_{ch} \sim 48^\circ$, the light does not arrive as
promptly, and fine segmentation, in our case with quartz bars, is limited. 
%The refractive index of quartz varies from 1.455 at $\lambda = 700$\,nm to 1.475 at $\lambda = 350$\,nm. The corresponding Cherenkov angles 
%are $46.6^\circ$ and $47.3^\circ$. Over this wavelength range the optical absorption length of quartz is $> 110$\,cm. 
%The density of quartz is 2.20\,g/cm$^{3}$, the radiation length is 12.3\,cm and the interaction length $\lambda_I = 44.5$\,cm. 

\subsubsection{Quartic Design for Roman Pots, with L-Bar Geometry}

We have developed detectors \cite{Albrow-ARX-2012} with quartz bars of 
$3 \times 3\rm\,mm^2$ cross section in the form of an `L', called L-bar Quartic
(QUARtz TIming Cherenkov), the light being detected with SiPMs. 

This configuration allows segmentation in both $x$ and $y$. 
  The photodetectors are located at $\sim 8$\,cm from the beam in the horizontal plane and can be 
  partially shielded to reduce their radiation dose.

\begin{figure*}[t]
\centering
\includegraphics[width=70mm]{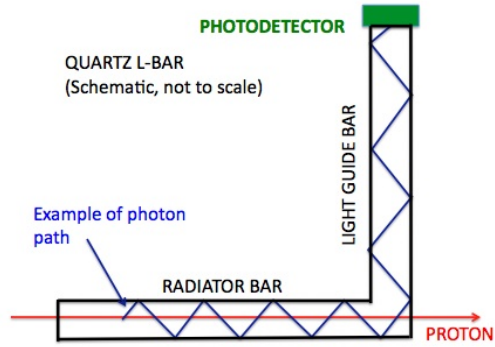}
%\vspace*{5cm}
\caption{Cherenkov light rays in the radiator and light guide bar, for $n$ = 1.48, in the plane of the `L'
 ($\phi_{ch} = 90^\circ$).}
 \label{lbarsketch}
\end{figure*}

In the Quartic design 
there is an array of $3 \times3\rm\,mm^2$ ``radiator bars'', R, parallel to the beam. 
The Cherenkov angle is the complement of the critical angle for Total
Internal Reflection (TIR) on the bar sides, and as the proton paths are almost exactly parallel to the bars all the Cherenkov
light is internally reflected to the back end of the radiator, as shown schematically in Figure~\ref{lbarsketch}. 
Most (about 2/3) of the light is transmitted to the SiPM along 
the light-guide (LG) bar, also with total internal reflection. The remaining $\sim 1/3$ is
reflected back to the entrance of the radiator bar, where there will be a black absorbing surface. 
The LG bars end in a (vertical in the horizontal RP) plane, 73\,mm from the beam pipe wall. 
This distance is a compromise between being away from the
beam for radiation issues, and keeping the LG bars short to minimise the absorption, 
number of reflections, and optical dispersion.
The spread of the travel time of the photoelectrons is mainly caused by the length of the 
radiator bar. The component of the speed of the light along
the radiator bar is $dz/dt = c/n^2 \sim 0.140$\,mm/ps. 

%Sapphire, Al$_2$O$_3$, has a higher refractive index, $n \sim 1.78$ and hence about 27\% more light per cm, however it is denser and has a 
%shorter nuclear interaction length ($\lambda_I = 24.8$\,cm cf. 44.5\,cm). To minimize interaction issues we choose quartz.
%Our baseline is to have two Quartic modules per arm, in one RP. 
% The second RP foreseen for timing detectors can be used for housing a GasToF or solid-state timing
%detectors. In both cases, the alternative timing detectors will be installed upstream of the Quartics, 
%to provide complementary time measurements. 
Detailed studies of the Quartic and other timing detectors can be found in  Refs.\cite{clermont,Albrow-ARX-2012,mazzillo,ronzhin,vavra}.

\begin{figure}
\centering
\includegraphics[width=120mm]{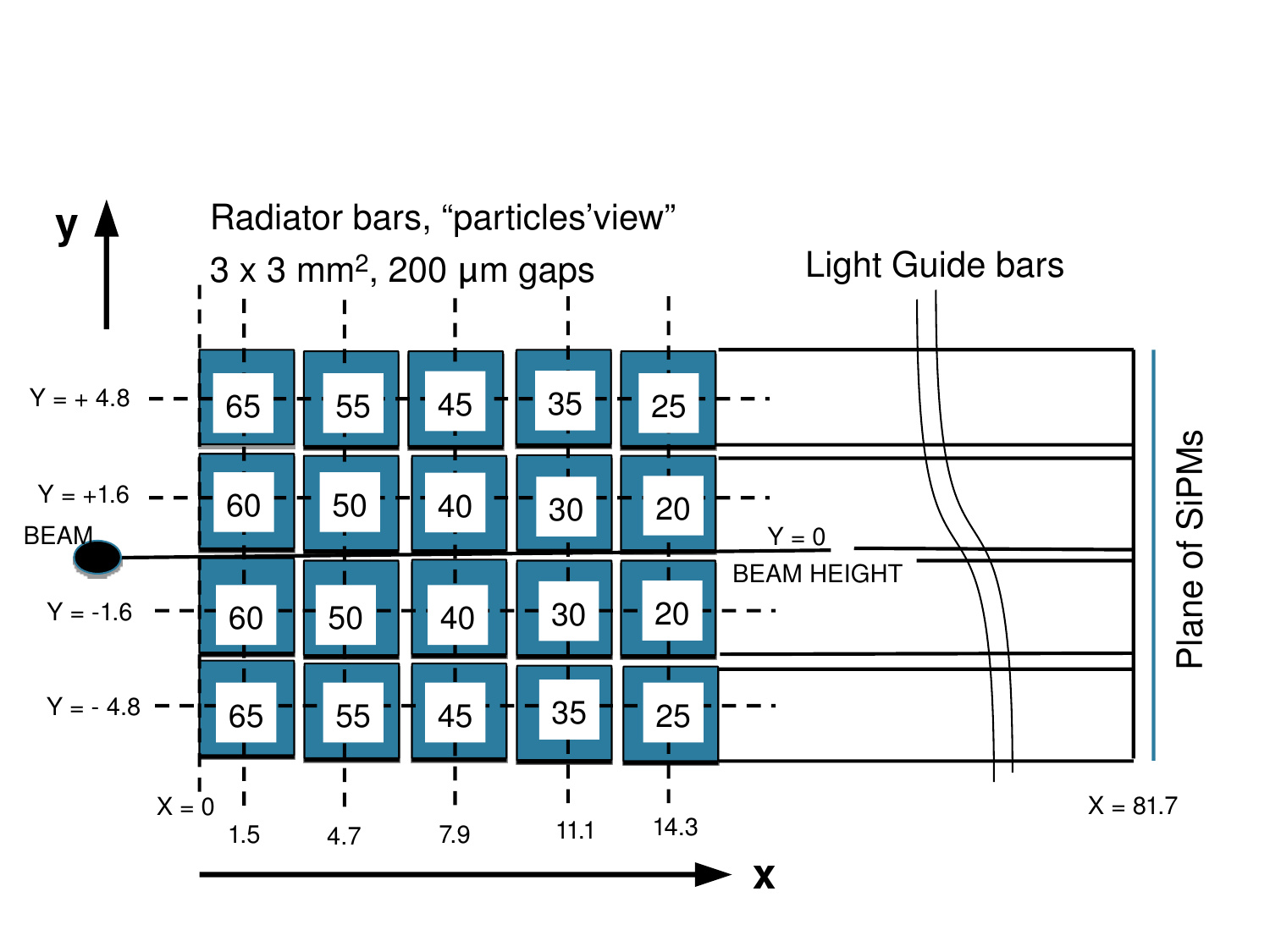}
%\vspace*{5cm}
\caption{Schematic layout of quartz bars looking in the direction of the protons. 
Numbers on the $3 \times 3\rm\,mm^2$ radiator bars are their lengths in mm, and
coordinates (mm) are the centers of the bars. The light guide bar lengths are chosen to all end in a common 
plane 81.7\,mm from the edge closest to the beam.}
  \label{fig:barl}
\end{figure}

The CT-PPS Quartic ``module'' 
is a light tight box with a very thin ($\sim 100\,\mu$m) side wall on the beam side, and blackened interior. 
This side wall may be removed on insertion in the pot as the interior will be dark; it is to protect the inside prior 
to insertion and to optically close the
box for beam tests. One module consists of ($4 \times 5 = 20$) independent $3 \times 3\rm\,mm^2$
bar elements. This allows a time measurement of two or more protons from the same bunch crossing (which has a 
time spread $\sigma_t \sim 150$\,ps) if they are in different elements. The active area is 12.6\,mm
(vertically, $y$) $\times$ 15.8\,mm (horizontally, $x$).  
This includes 200\,$\mu$m spacers (a wire grid) 
to separate the bars, allowing total internal reflection, and avoiding light leakage.
The dimensions of the bars are given in Figure~\ref{fig:barl}.
    The ends of the light guide bars arrive at an array of SiPMs, coupled with a thin 
    silicon ``cookie'' for good optical coupling. 
    We use SiPMs Hamamatsu MPPC Type S12572-050 mounted in a flat plate
    holder. The SiPMs fit in
   rectangular holes in the plate
   and as they are not fixed to the read-out board they can be very simply
   replaced. 
    The SiPMs are connected to the read-out board through
   an anisotropic conducting sheet (embedded very short wires give an electrical connection through 
   the sheet but not in the plane).  
  
 The bars pass through circular holes for locating the
bars against the SiPMs, with better than 25\,$\mu$m accuracy. These holes are countersunk, as the 20 bars all have to be inserted together.
%The U-clamp is then tightened and the temporary clamp removed. 
The complete bar assembly is then inserted in the box on precision grooves; the front
window is placed in position later, after position and optical checks are done. 
%Figures~\ref{twoinpot2} and~\ref{twoinpot} 
Figure~\ref{twoinpot} shows the assembly of two modules in one RP. 
We have the option of displacing one module in $(x,y)$ with respect to the other by 250\,$\mu$m (e.g.) to avoid any dead regions between
the bars. Alignment of the radiator bars parallel to the protons (at the level of $\lesssim 10$\,mrad) 
will be needed to maximise light collection and avoid light leakage into neighbouring bars.

\subsubsection{Integration with Roman Pots}

\begin{figure*}[ht]
\centering
\includegraphics[width=0.47\textwidth]{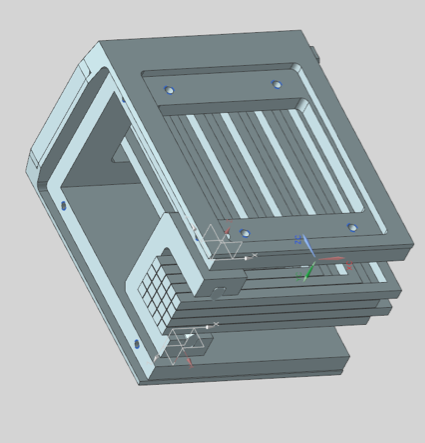}
%%\vspace*{5cm}
%\caption{Design (B. Ellison, FNAL) of module for insertion in Roman Pot (two in one pot).}
%\label{box1}
%\end{figure*}
%\begin{figure*}[ht]
%\centering
%\includegraphics[width=78mm]{figs/detector/twoinpot3.png}
%%\vspace*{5cm}
%\caption{Assembly of two Quartic modules in Roman Pot.}
%  \label{twoinpot2}
%\end{figure*}
%\begin{figure*}[ht]
%\centering
\includegraphics[width=0.49\textwidth]{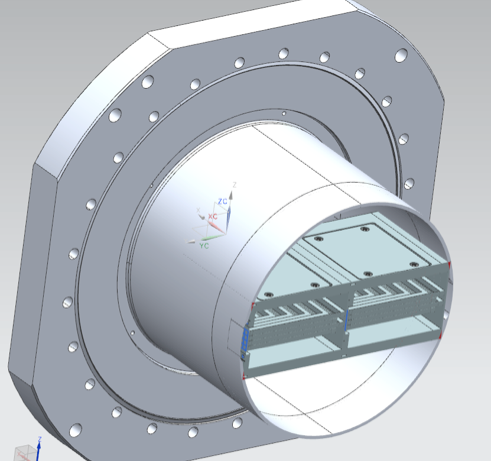}
%%\vspace*{5cm}
\caption{Left: Design of module for insertion in Roman Pot (two in one pot). Right: Assembly of two Quartic modules in Roman Pot. The beam comes from the left.}
  \label{twoinpot}
\end{figure*}

The longitudinal space in the RPs is approximately 140\,mm. The L-bar geometry 
allows installation of Quartic two modules in one RP (Figure~\ref{twoinpot}) %(Figures~\ref{twoinpot2} and~\ref{twoinpot}). 
The two modules in a pot will be fitted together precisely using dowel pins. 

\begin{figure*}[ht]
\centering
\includegraphics[width=80mm]{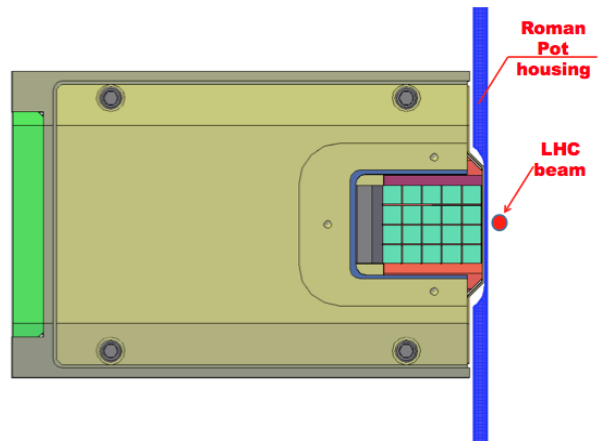}
%%\vspace*{5cm}
\caption{``Particle eye view'' of bar array in pot. This is a modified design reducing the material close to the beam.}
  \label{pev}
\end{figure*}

%Figures~\ref{linediag} and~\ref{box1} 
Figure~\ref{twoinpot} 
shows the design of the module for insertion into the horizontal RP. 
A slightly modified version (Figure~\ref{pev}) reduces the material close to the beam.
The protons enter through a thin
window (nominally 100\,$\mu$m aluminium, but it has no mechanical purpose, it is for absorption of reflected light and light exclusion). 
The array of radiator bars
is clamped on the three sides away from the beam with a plastic U-clamp, touching each bar only along a fine line. 
For assembly of the bar array a 4-sided clamp is placed temporarily 
around the front of the array. 
The SiPM plate and read-out board are precisely positioned by dowels with respect 
to the L-bar positioning plate.

\subsubsection{Photodetectors: Silicon Photomultipliers}
     Silicon photomultipliers, SiPMs, are solid state photon counters comprised of a large number of avalanche 
     photodiodes (APDs) or ``pixels" of order 20\,$\mu$m
     dimensions, 
     with a high gain (up to $10^6$) in Geiger mode, with an applied voltage just above the breakdown 
     voltage (about 30\,V to 70\,V depending on the type).
     Each discharged pixel has a recovery time of $\sim 50$\,ns, but with e.g.
     100 photoelectrons per event and thousands of pixels per mm$^2$ this is acceptable with 25\,ns bunch-crossing time. For the SiPMs, the single photon detector efficiency is the product of the quantum efficiency 
     and the fractional area coverage of the APDs. SiPMs are rugged, simple to use and relatively
     cheap per unit, but at present are only available
     commercially with effective active areas from $1\times1\rm\,mm^2$ to $3.5\times 3.5\rm\,mm^2$. Smaller SiPMs have less capacitance and are
     intrinsically faster. 
 
\begin{figure}[ht]
\centering
\includegraphics[width=80mm]{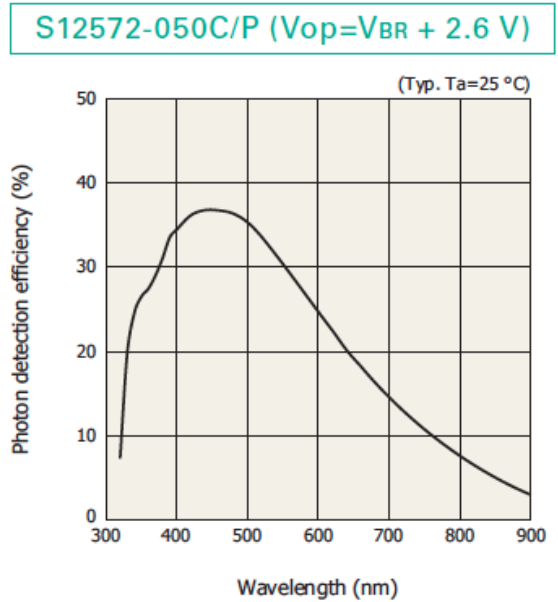}
%\vspace*{5cm}
\caption{Photon detection efficiency of Hamamatsu S12572 SiPM. This is the product of the quantum efficiency
and the fill factor (50\,$\mu$m pixels).}
\label{sipmpde}
\end{figure}

 The SiPMs, Hamamtsu MPPC type S12572-050,  
 %(Figure~\ref{s12572}), 
 operate at $\sim 72$\,V, 
 just above the breakdown voltage (they operate in Geiger mode, 
discharging one or two pixels per detected photon). These have 3600 pixels of 50\,$\mu$m diameter. 
The single photon detection efficiency and the wavelength-dependence of
the response is shown in Figure~\ref{sipmpde}. Improved efficiency in the UV is being investigated.
Individual HV values can be applied to each SiPM, and their leakage currents monitored
through a high-resistance (6.4\,M$\Omega$) bleed resistor to ground. 
While the gains of SiPMs are very sensitive to temperature, the time resolution is not. Air cooling is 
expected to be sufficient; an alternative is to use the same cooling system as
for the tracking detectors. The temperature of the SiPM boards will be monitored with thermistors. 
The SiPM board receives an individually controllable bias voltage $\sim 72$\,V from a local programmable supply. 
The signals are read out with miniature coaxial cable with SMA connectors.

%\begin{figure}
%\centering
%\includegraphics[width=100mm]{figs/detector/S12572.png}
%\vspace*{5cm}
%\caption{Details of Hamamatsu MPPC (SiPM) S12572-050.}
%\label{s12572}
%\end{figure}

\subsubsection{Monitoring, Alignment, and in situ Calibration}
\label{sec:time_align}
The rates in each bar are monitored both online and offline. The
rates are expected to be up-down symmetric about the beam height (as can also be determined from the tracker), 
column-by-column. This gives a measure of the centre of
the beam in $y$, assuming the backgrounds are relatively small. For a given row in $y$ the rates will
fall with $x$ with two components: protons from collisions and beam halo background. The former will be compared 
with predicted $t$-distributions, and it is valuable to
measure the latter. 
When the RPs are inserted at the beginning of stores the rates will be carefully monitored, e.g. 
partially withdrawing in the event
of an abnormal increase.

Initial relative alignment of the Quartic bars and tracker will be achieved by matching the individual bar edges in $x$ 
and $y$ with the track distributions.

A pair of light-emitting diodes will be mounted inside the module in such a way that some light is captured in each bar. 
This will be pulsed when there is no beam as a control
of all the SiPMs and their readout. 

Occupancy of the bars will be monitored in real time, as functions of the instantaneous luminosity 
and background conditions.
Geometrical matching between the
bar array and the trackers will also be measured; the 200\,$\mu$m gaps between the bars can also provide a check. 

A calibration of the absolute time difference between the protons, $\Delta t_{pp}$, can be derived by matching $z_{pp} = z_X$
using real events of the type $p+p \rightarrow p+X+p$, where $X$ is a set of particles measured in the central
   detector. After kinematic matching of $X$ to the protons (four-momentum conservation) the 2D plot of $z_{pp}$ vs $z_X$ will show a ridge which 
   calibrates both $z_{pp} = 0$ as well as checking the time scale. 
   
   While the time difference between the ``left'' and ``right'' protons, $t_L-t_R$, gives $z_{pp}$, the time sum, or $(t_L+t_R)/2$ (minus a constant), 
   would provide another, orthogonal, variable for pileup
   rejection if the actual event time were known much better than the spread in collision times, $\sim$150\,ps.

\subsubsection{Beam Tests}
\label{sec:beamtest}
\begin{figure*}[ht]
\centering
\includegraphics[width=80mm]{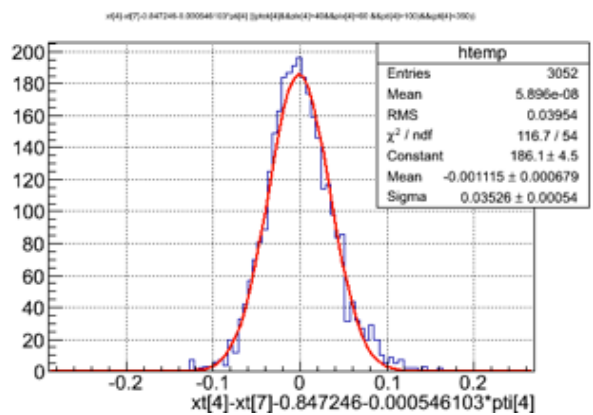}
%\vspace*{5cm}
\caption{The time difference between one L-bar (30\,mm radiator, 40\,mm light guide, 
  Hamamatsu MPPC type S10362-330050C) and the
  reference time signal (PMT240 in beam). It shows $\sigma = 35.3$\,ps.}
  \label{deltat}
\end{figure*}

A prototype Quartic with L-bars was tested in the Fermilab test beam, with 120\,GeV protons, using a mechanical design that does not include
 some features needed for the CT-PPS version. 
 Figure~\ref{deltat} shows one example, with $\sigma(\Delta t) = 34.9$\,ps, showing no background or inefficiency.
The  single-photon time resolution of the SiPM we used is quoted by Hamamatsu to be $\sim 300$\,ps;
giving an expected time resolution for 100 p.e. of 300\,ps$/\sqrt{100} = 30$\,ps, in reasonable agrement with the observations. 
 See~\cite{Albrow-ARX-2012} for more details of the test procedures and results.

\subsubsection{Future Developments}
We have described the baseline Quartic detectors, which will be ready for installation in 2015
when the RPs have been installed and commissioned.  Several potential improvements will be investigated. 
These include finer segmentation, especially near the beam, a more compact SiPM array (``buttless'' type), and the use of a multi-anode MCP-PMT 
in place of SiPMs. The latter have had photocathode lifetime problems limiting their use in such a high-rate environment, but developments 
to mitigate that have been made.

Also, faster SiPMs with higher photon detection efficiency and possibly more
sensitivity in the UV may become available, potentially allowing easy replacement.

SiPMs from STMicroelectronics (STM) with new p-on-n structure (rather than N-on-P SiPMs, also from STM) 
show significantly better timing properties~\cite{mazzillo}~\footnote{We thank STMicroelectronics for providing samples.}. 
Tests with a PiLas (Picosecond
Injection Laser) showed the photon
detection efficiency at $\lambda = 405$\,nm, 5\,V above breakdown voltage (28\,V), to be  43\% higher (31.1\% cf. 21.7\%). Also the 
SPTR, is 174\,ps cf. 231\,ps, i.e. smaller by 25\% than for STM n-on-p detectors. 
Together these improvements lead one to expect that the single bar resolution can be improved from the measured 32\,ps to $\sim 20$\,ps.

Unlike the large central CMS detectors, due to the small number of channels,
upgrades to the timing detectors can be considered on a yearly timescale, and we expect to continue the 
necessary R\&D. Also, if timing detectors based on other principles, e.g. solid state detectors, demonstrate
good performance and acceptable properties (e.g. radiation hardness) they can supplement, or possibly 
replace, the Quartic detectors.

%\subsubsection{Summary of Quartic Detectors}

%The L-bar Quartic design satisfies the requirements of edgelessness (within about 100\,$\mu$m), 
%sufficient radiation hardness, ability to measure
%several protons within a bunch crossing (time spread $\sigma = 150$\,ps) if in separate bars, and to be 
%active every 25\,ns (the bunch separation). 
%We plan to construct and install a pair of L-bar Quartic modules, with SiPM photodetectors, in a 
%RP on each arm, and use them in 2015
%to assess their performance in realistic conditions, as well as for physics in the low pileup runs. 
%The modules will be constructed at Fermilab 
%and the signals from the
%SiPM board sent to a local DAQ.% (which will be the responsibity of the Lisbon Group).

%Further development will continue to improve their time resolution and possibly the spatial segmentation. 
%Therefore R\&D will continue, and improved versions could be ready for installation at the end of 2015.  

\subsubsection*{Readout System of the Cherenkov Detectors}
\label{sec:timing_electronics}

The task of the readout system is to provide time and amplitude measurements of
the pulses generated by the photosensors associated to the Cherenkov detectors, and to transmit the digitized data to 
the data acquisition system. 
%Given the similar characteristics of the Quartic
%and Gastof detectors, the readout system is expected to be suitable for both
%detectors.

The Cherenkov timing detectors are composed of a number of modules installed in
one or more RP. The readout system follows the same
modularity, and is composed of independent units (readout module) interfacing to the CT-PPS
DAQ/Trigger system. Each readout module has $64$ channels suitable for use with
the baseline Quartic module ($20$ channels), but also with possible Quartic modules with finer granularity.
% and with the Gastof $8\times8$ anode MCP-PMT.

%Each electronic channel must be able to handle very fast pulses (rise time $\sim 1$ ns) from
%silicon photomultipliers or micro-channel plates. The photon yield of the
%Cherenkov signals per channel is %range from $1$ photon in the case of Gastof MCP to
%about $100$ photons in the case of Quartic.
%The SiPM gain is of the order of $10^{6}$ while the MCP-PMT will be operated at
%about 10 times lower gain to limit the total charge collected (see~\cite{ctpps-tdr:ch9}, Section~5.3).
%Therefore the dynamic range of the input signals is in the range of 20\,fC to 20\,pC.

%The timing precision of the readout system is required to be 20\,ps, which,
%combined with a similar precision of the Quartic detector, would yield 30\,ps
%precision per module, and global precision (two modules) of about 20\,ps. When
%associated to one Gastof detector, assuming 7 photoelectrons per proton and a
%precision of 30\,ps per MCP-PMT channel, we may estimate that the readout system would
%provide a measurement with 15\,ps precision.
 
The readout system is also required to provide the measurement of the input pulse amplitude. 
While this measurement is not directly used in the reconstruction of
the collision vertex, the knowledge of the amplitude is mandatory for detector
calibration, time corrections (e.g. ``time-walk'') and pulse pile-up rejection.

The readout system is required to have double hit resolution better than 25\,ns, suitable for operation with 25\,ns bunch separation without loss of
efficiency, and to sustain a maximum rate of 6\,MHz
per channel, corresponding to a maximum channel occupancy of 20\% at 25\,ns
LHC beam operation, averaged over all channels. While the average occupancy of the Quartic channels ($3\times3\rm\,mm^{2}$ quartz bars) is 20\% for average pile-up of
50 events, the highly non-uniform occupancy of the detector (the
occupancy of the innermost channels reach 70\%) induces a
significant readout inefficiency for a number of pile-up events larger than 25.
%To sustain twice this value, as foreseen in Run 2, a fine-granularity Quartic
%version will be required. In the case of the Gastof detector, a channel occupancy 
%smaller than 20\% is expected.

The readout system is required to provide on-detector L1 trigger matching, allowing extraction 
of the detector data in a time window around the L1 time, local event building
and data transmission to the DAQ system. 
The data rate of a readout module, assuming readout 
of 10 channels after zero suppression, 3 bunch crossings time window, 32 bit
event data per channel, and 100\,kHz L1 rate, is estimated to be 100\,Mb/s.

The timing detector readout system is expected to provide hit information and
time measurement at the bunch crossing rate to be used by the trigger system. By
combining the information from the two PPS arms, the L1 can estimate the
z-vertex coordinate, allowing to select events in the tails of the z-vertex
distribution where the pileup density is smaller. This capability would provide
a reduction of the L1 rate of high cross-section processes so that it fits
within the L1 rate constraints, selecting at the same time the events less affected by pileup. 
The trigger requirement implies the use of a low latency TDC delivering
conversion data at the bunch crossing rate.

In order to achieve the desired time resolution, the front-end timing
electronics must be located in the RPs or a nearby region (1-2\,m
distance). This raises issues of radiation tolerance since the
radiation levels in the RPs, in the region 200\,m from IP, are expected to be 100\,Gy and $10^{12}$ neq/cm$^2$ for 100\,fb$^{-1}$ of integrated luminosity.
%Together with machine
%experts, we have explored the feasibility of digging a hole in the
%tunnel floor concrete below the timing RPs, which would provide a substantial
%reduction of the radiation exposure. While this possibility is technically feasible, the
%proposal could not be accommodated in the present LHC LS1 schedule but would be
%possible if necessary in the winter technical stop 2015-16.

\paragraph*{System Design}

The main guideline in the design of the timing detector readout system was to
reuse well known components with adequate performance, allowing to streamline the design and implementation of the system 
so that it may be possible to evaluate the timing detectors in the LHC beam in 2015.

We have therefore decided to base our system on the amplifier-discriminator
NINO and the time-to-digital converter HPTDC. Both chips have been developed by
CERN's microelectronics group for the LHC experiments. The HPTDC chip is used in the CMS muon system, while the NINO 
and HPTDC chips are associated in the time-of-flight detector of the ALICE experiment. 
These chips are now widely used in many applications, including PET Time-of-Flight. 

\begin{figure}
\centering
\includegraphics[width=100mm]{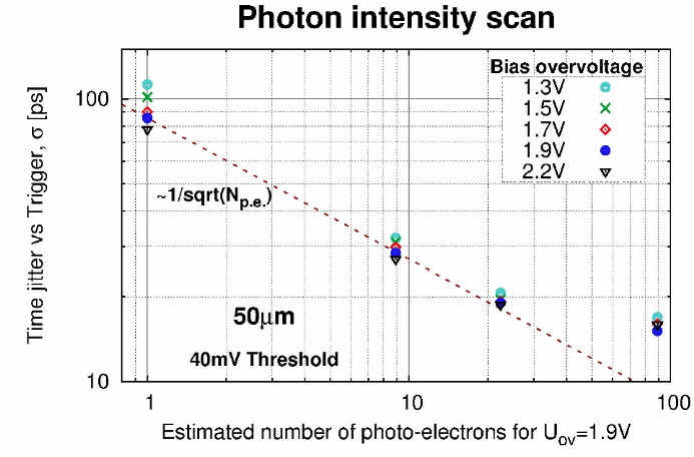}
%\vspace*{5cm}
\caption{Time resolution of laser pulses as a function of the number of
photo-electrons measured with MPPC and the NINO-HPTDC electronics chain.}
\label{fig:nino_hptdc_resolution}
\end{figure}

The EndoTOFPET-US collaboration\footnote{This 
project  have  been  funded  by  the  European  Union  7th  Framework  Program  (FP7 / 2007-2013) under Grant Agreement No. 256984 (EndoTOFPET-US)} has studied the
time resolution of the NINO-HPTDC readout chain, using laser pulses detected by $3\times3$\,mm$^{2}$ SiPMs (Hamamatsu MPPC) with
SPAD size of $50\,\mu$m~\cite{Gundacker}.
The measured time resolution as a function of the number of photo-electrons is
shown in Figure~\ref{fig:nino_hptdc_resolution}.
For 100 photoelectrons a resolution better than 20\,ps is achieved.

The NINO amplifier-discriminator is implemented in a 8 channel ASIC. 
%A newer version with 32 channels developed by an external group is also suitable.
The time binning and number of channels of the HPTDC ASIC is configurable. In our case we use 
the HPTDC high-resolution mode, which provides 8 channels with 25\,ps time
binning.
The timing readout system is designed to be integrated in the common CT-PPS DAQ and Control 
system, based on the CMS Pixel FED and FEC boards. The FED board provides input
to 400\,Mb/s optical links transmitting detector data, builds event packages and
transmits them to the central DAQ.
The FEC board transmits fast controls (LHC clock, L1 and fast signals) to the detector, as 
well as front-end configuration data. We plan to use the same components used in the Pixel 
detector to implement the on-detector interface to the FED and FEC board, namely the  
transmission and reception optical hybrids and the CCU control chips.

Physically, the timing readout system is implemented in two electronics boards:
%(see Figure~\ref{fig:readoutsystemdiagram}):
1) the frontend board, housing 8 NINO chips and installed in the RP, 
which receives the SiPM signals transmits the LVDS output on the feed-through
connector; 2) the digital board, which receives the LVDS output of the discriminators and
houses 8 HPTDC chips. The digital board integrates a radiation resistant FPGA to
serve as readout controller of the HPTDC chips, and provides on-board connectors to the opto-hybrid mezzanines (PoH and DoH). 
If required, the 
digital board could be installed a few meters away of the RPs in a radiation protected place.

%\begin{figure}
%\centering
%\includegraphics[width=150mm]{figs/detector/readoutsystemdiagram.pdf}
%%\vspace*{5cm}
%\caption{Diagram of the timing readout system.}
%\label{fig:readoutsystemdiagram}
%\end{figure}

%--------------------------
%\input{ctpps_el.tex}     
%-------------------------- 

\subsection{Pixel Tracking System}
\label{sec:silicon}
\begin{figure}[h!]
\begin{center}
\includegraphics[width=0.6\linewidth]{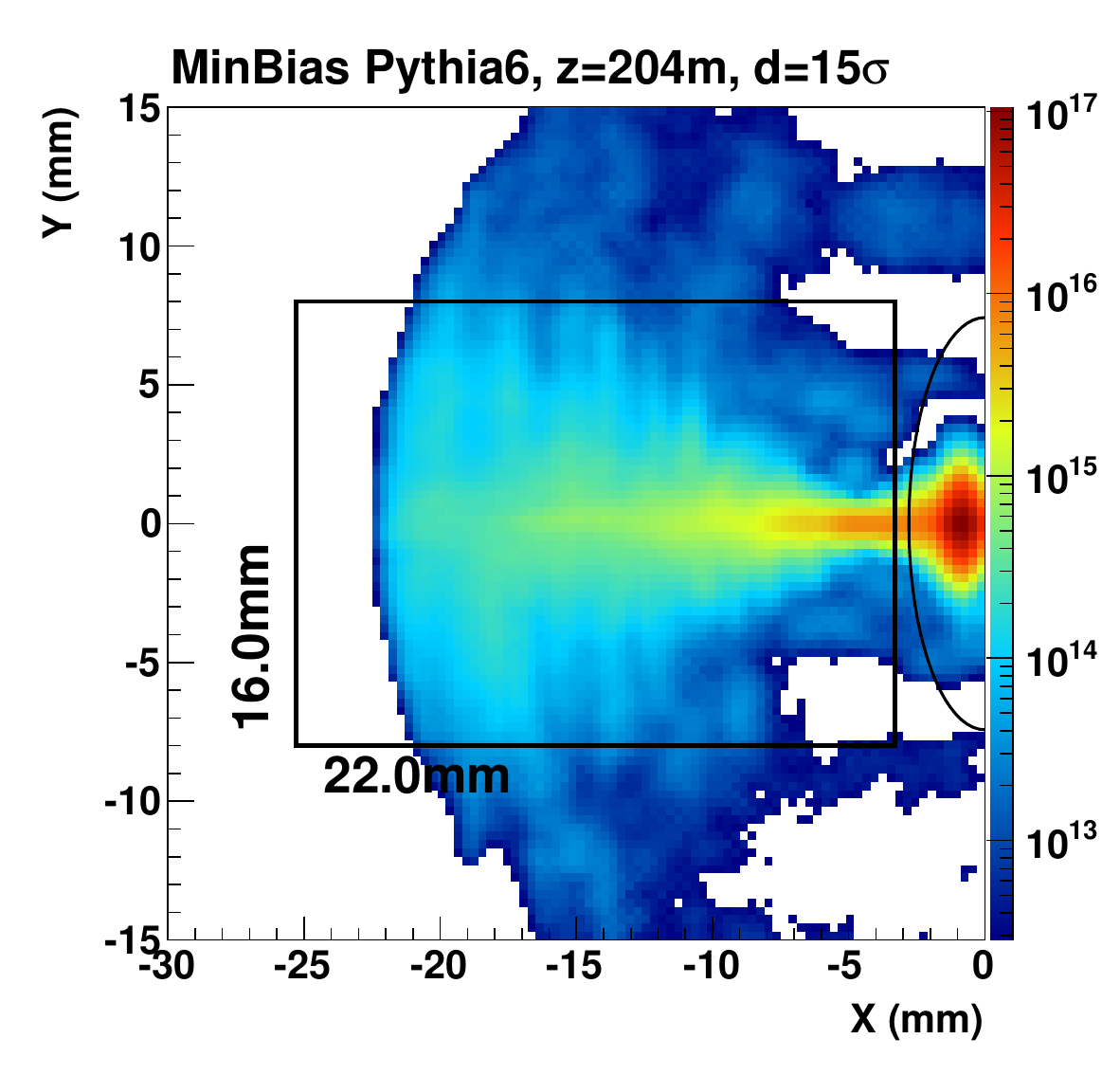}
%\vspace*{5cm}
\caption{Simulated proton fluence in the tracking station at 204~m from 
the IP for the integrated luminosity of 100\,fb$^{-1}$. The rectangle 
indicates the detector surface transverse to the beam assuming a detector 
tilt angle of 20$^{\circ}$. The ellipse shows the 15 sigma beam contour.
In the detector edge a value of the order of $5\times 10^{15}$\,p/cm$^2$ 
is obtained. This value is compatible with the extrapolation from TOTEM data.}
\label{fig:ProtonFluence}
\end{center}
\end{figure}

The key requirements for the CT-PPS tracking system are:
\vspace{-0.5\baselineskip}
\begin{itemize}
\item Efficient pixel based tracking as close as possible 
to the sensors physical edge, providing hit resolution better than 
30~$\mu$m.
\item Radiation hardness: 
a design figure of $5\cdot 10^{15}$~protons/cm$^2$ for 100~fb$^{-1}$ of
integrated luminosity is required (Figure \ref{fig:ProtonFluence}).
\item Reliable operation at the highest LHC luminosity.
\end{itemize}
\vspace{-0.5\baselineskip}

Since the construction of the original pixel tracking systems for 
the LHC experiments, there has been considerable progress in 
silicon sensor technology. Both CMS and ATLAS have pursued improved 
pixel designs for the high-luminosity upgrades of the LHC. 
These ongoing R\&D efforts have already achieved two proven sensor 
designs that meet the needs 
of CT-PPS and can be produced by industry: 3D and planar slim-edge 
silicon pixel sensors. Both types of sensors are being installed in 
the new Insertable Barrel Layer (IBL) of the ATLAS vertex 
detector~\cite{ATLAS_IBL}.

\begin{figure}
\centering
\begin{center}
\includegraphics[height=1.5in]{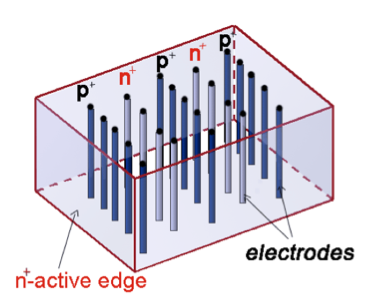}
%\vspace*{5cm}
\end{center}
\caption{Sketch of a 3D sensor.}
\label{fig:silicon-3D}
\end{figure}

3D sensors~\cite{parker-3d} consist of an array of columnar electrodes (radius 
$\sim$5\,$\mu$m) of both doping types that penetrate through the 
silicon bulk perpendicularly to the surface, as shown in 
Figure~\ref{fig:silicon-3D}. The bulk is usually 
of type p. Junction n-type electrodes are read out on the 
front side of the sensor while ohmic p-type electrodes are 
connected on the back side for applying the bias voltage. 
This structure decouples the inter-electrode distance from the 
sensor substrate thickness, allowing to reduce the drift path 
of the charge carriers without decreasing the total generated charge. \\
The close electrode spacing provides several advantages compared to 
the planar sensor design:
%\vspace{-0.5\baselineskip}
\begin{itemize}
\item low full depletion voltage ($\sim$10\,V),
\item fast charge collection time, 
\item reduced charge-trapping probability and therefore high radiation hardness. 
\end{itemize}
%\vspace{-0.5\baselineskip}

%As a result, 3D detectors are emerging as one of the most 
%promising technologies for the innermost layers of tracking devices for 
%the foreseen upgrades of the LHC.
%A very interesting feature of 3D technology is the possibility of 
%realizing the so-called ``active edge''~\cite{active-edge}, 
%i.e. a deep trench all around 
%the sensor that reduces the dead region to a few microns from the 
%physical edge of the device, compared to the hundreds of $\mu$m for 
%standard planar detectors.

The baseline CT-PPS tracking system is based on  
3D pixel sensors, produced either by FBK (Trento, Italy)~\cite{FBK} or CNM (Barcelona, 
Spain)~\cite{CNM}, which we think provide the best performance in terms 
of active region and radiation hardness. 
These companies have already produced 200~$\mu$m slim-edge 
3D sensors for the IBL project with satisfactory yield. 
However, for CT-PPS we would like to pursue the option with 
a 100~$\mu$m, or better, slim-edge design, where the active region 
of the sensor is as close as 100~$\mu$m to the edge. 
Since such slim edges have not yet been produced, we mitigated 
our schedule risk by designing the CT-PPS tracking system 
to allow rapid installation, or replacement, 
of the unit during a LHC technical stop.

%As a backup solution, we have identified a second sensor concept, 
%a slim-edge planar design 
%produced by CIS (tested by ATLAS) and SINTEF(tested by CMS), based 
%on a more mature fabrication process. We have designed 
%the CT-PPS sensor readout and packaging to accept either sensors 
%as a drop-in replacement. 

The chosen configuration for the tracking system consists of two 
detector units in each arm, for a total of four detector units. 
These are the horizontal RPs located at $\sim$210\,m.
Each station will contain one stack of silicon tracking detectors. 
Each stack will consist of six planes, where each plane contains 
a 1.6$\times$2.4\,cm$^2$ pixel sensor read out by six PSI46dig readout chips
ROCs~\cite{CMS_digROC}. Each ROC reads 52$\times$80 pixels with 
dimensions 150$\times$100\,$\mu$m$^2$. Given the small area of the detector,
covered by a small number of individual sensors, we have chosen a number of
planes that provide confortable redundancy making the system resilient
to possible failures.  The design of 
the front-end electronics and of the DAQ is based on that developed for the 
Phase 1 upgrade of the CMS silicon pixel 
detectors~\cite{Ph1-TDR}.

The resolution of the x-coordinate is determined by the sharing of charge in
the pixel clusters, which depends on the detector tilt angle in the x-z plane.  
While this parameter is not yet defined, test beam results with similar sensors 
indicate that for an angle of $<$20 degrees the two-pixel clusters have resolution
of the order of 10\,$\mu$m.  Since there is no tilt in the y-z plane, the
resolution of the y-coordinate is of the order of 30\,$\mu$m.

\subsubsection*{The Readout System}

The readout of the pixel detector is largely based on components developed in the framework of the CMS Pixel 
Upgrade Phase 1 readout project. 
%This readout chain has been proven to work efficiently. 
Most components
are currently at an advanced test stage and/or in final production. 
%In this section, an overview of the readout system of the pixel detector is given shortly with special 
%emphasis on CT-PPS specific points. 
Extensive documentation of each component can be obtained elsewhere~\cite{Ph1-TDR}.

%The complete readout chain of one detector package is shown in Figure~\ref{silicon:readout}. 
By convention a detector package is a set of six modules, one for each sensor present in 
a RP. There are in total four detector packages.
The specific CT-PPS elements in the readout system are the six RPix Modules (Roman Pot PIXel Modules) 
and the RPix Portcard. 
The low voltage system and high voltage system are part of the low voltage and high
voltage system of TOTEM, already in place.
The backend DAQ electronics, to be installed in the service cavern, is being actively developed  
by CMS as part of the Pixel Upgrade Phase 1 project.

%\begin{figure}[htb!]
%\centering
%\includegraphics[height=4.2in]{figs/detector/readout.pdf}
%%\vspace*{5cm}
%\caption{Detector readout and control strings. All components up to PP1 are in the LHC tunnel, while
%         FEDs and FECs are in the service cavern.}
%\label{silicon:readout}
%\end{figure}

Each RPix Module consists of a flexible hybrid circuit hosting: a silicon sensor, six ROCs ``bump-bonded'' to the silicon sensor, and one Token Bit Manager  (TBM) chip. 
The ROC is responsible for charge collection, charge discrimination and data sparsification. 
%Two slightly different versions of the chip, named V2.1 and V2.1b, are suitable: 
%which version to use is currently under evaluation.

The TBM is responsible for reading out the six ROCs in the module (it can manage up to 16) using a token ring protocol
and serialising the data over a single output line.
The TBM %versions suitable for this application, named TBM08a and TBM08b, 
manages two token bit ring protocols 
at the same time, multiplexing the two data streams on the same output line before encoding the data stream.
% using a 4B/5B NRZI schema.
The net result is an uplink data stream running at 400 Mb/s. 
Dispatching these high speed signals on the flexible hybrid without degrading them
is one of the major design requirements of the RPix Module, together with the need to
deliver the high voltage to the sensor safely. 

The RPix Portcard accomplishes a variety of tasks. The board receives the output data from six RPix
modules  and retransmits them on six optical fibres towards DAQ modules, using a POH7 opto-electrical 
converter mezzanine card~\cite{POH}.
This board also receives fast configuration commands from the Pixel FEC via optical fibres, translates these 
signals using detector optical receivers (DOH) and dispatches them to the modules. 
These functionality are similar to those of the Forward Pixel project portcard developed by Fermilab.
Moreover the RPix Portcard integrates other components such as the radiation sensors, part of the TOTEM DCS 
radiation monitoring system, and the CMS Tracker Optical Control Link components which are
capable of receiving and decoding the commands sent from the Tracker FEC. 
Finally, the newly developed DC/DC converters~\cite{DCDC1}, developed by CMS, are installed on the board, 
in order to generate the different voltages needed by the portcard itself and by the attached modules.

For the backend DAQ system, the plan is to use the new uTCA crates and boards developed for the Pixel Upgrade 
Phase 1 project. A fallback solution, using VME electronics, is available in case the baseline solution could suffer 
long delays.

It should be noted that the tracking front-end based on the
CMS pixel readout chip PSI46dig chip does not have trigger outputs and
therefore can not be integrated in the L1 Trigger.

\input{detectors/afp_yr_review.tex}

\input{detectors/yellow_lhcb.tex}

\input{detectors/adYellowReport-ed.tex}

\input{detectors/LHCfdetector-yrep-sako.tex}

%% file: detectors/alpha_yr.tex
\section{ATLAS-ALFA Experiment}
\label{sec:exp}
% ============================================================
ATLAS is a multi-purpose detector designed to study elementary processes in 
proton--proton interactions at the TeV energy scale. It consists of an inner 
tracking system, calorimeters and a muon spectrometer surrounding the interaction 
point of the colliding beams. 
% The inner tracking system is embedded in a 2$T$ axial magnetic field, the muon 
% system is completed by three air-core toroids.
The tracking system covers the pseudorapidity range $|\eta| < 2.5$ and the calorimetric 
measurements range to $|\eta| = 4.9$.\footnote{ATLAS uses a right-handed coordinate system 
  with its origin at the nominal 
  interaction point in the centre of the detector and the $z$-axis along the beam pipe. 
  The $x$-axis points from the interaction point to the centre of the LHC ring and the 
  $y$-axis points upwards.
  The pseudorapidity $\eta$ is defined in terms of the polar
  angle $\theta$ as $\eta = -\text{ln tan}(\theta/2)$.}
To improve the coverage in the forward 
direction three smaller detectors with specialized tasks are installed at large 
distance from the interaction point. The most forward detector, ALFA, is sensitive to 
particles in the range $|\eta| > 8.5$, while the two others have acceptance 
windows at $|\eta| \approx 5.8$ (LUCID) and $|\eta| \approx 8.2$ (ZDC). 
A detailed description of the ATLAS detector can be found in Ref.~\cite{atlas1}.

The ALFA detector (Absolute Luminosity For ATLAS) 
is designed to measure small-angle proton scattering. 
% For the measurement of small-angle proton scattering ATLAS is complemented by the 
% ALFA detector. 
Two tracking stations are placed on each side of the central ATLAS 
detector at distances of 237 m and 245 m from the interaction point. 
The tracking detectors are housed in  Roman Pots which can be moved close 
to the circulating proton beams. Combined with special beam optics,
% the large distance to the interaction point, 
this allows the detection of protons at scattering angles down to 
 $\mu$rad. 
% For a good resolution at very low $t$ a small angular divergence at 
% the interaction point is required. For this reason the data taking is performed in 
% special runs with a high $\betastar$ beam optics \cite{HBLumiday11}.

\begin{figure}[h]
  \centering
  \includegraphics[width=80mm]{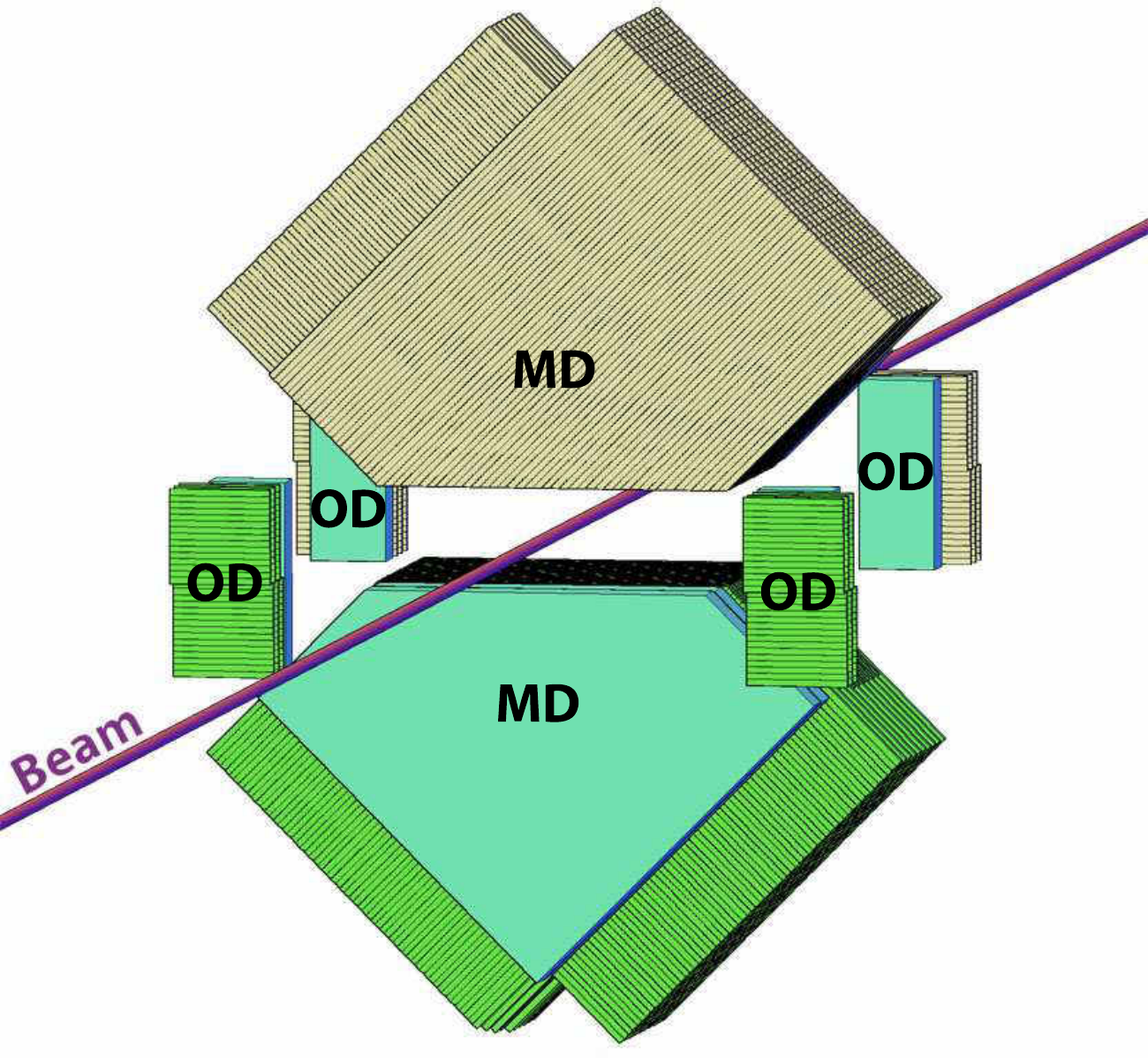}  
%  \vspace*{5cm}
  \caption{A schematic view of a pair of ALFA tracking detectors in the upper and lower RPs.
           Although not shown, the ODs on either side of each MD are mechanically 
           attached to them. 
           The orientation of the scintillating fibres is indicated by dashed lines. 
           The plain objects visible in front of the lower MD and ODs are the 
           trigger counters. For upper MD and the lower ODs they are hidden at the opposite 
           side of the fibre structures.}   
%           The distance between the two MDs is measured using particles which pass through both ODs 
%           on either side of each MD. Although not shown in the figure, the ODs are mechanically 
%           attached to the MDs.}
  \label{fig:ALFA_front_view}
\end{figure}

Each station carries an upper and lower RP connected by flexible bellows 
to the primary LHC vacuum. The RPs are made of stainless steel with thin windows 
of 0.2 mm and 0.5 mm thickness at the bottom and front sides to reduce the interactions 
of traversing protons.
Elastically scattered protons are detected in the main detectors (MDs) while dedicated overlap 
detectors (ODs) measure the distance between upper and lower MDs. 
The arrangement of the upper and lower MDs and ODs with respect to the beam 
is illustrated in Fig.~\ref{fig:ALFA_front_view}. 

Each MD consists of 2 times 10 layers of 64 
% Kuraray SCSF-78 ==> taken out by Edboard
square scintillating fibres with 0.5 mm side length 
%\footnote{KURARAY CO., LDT., Tokyo, Japan, SCSF-78, S-type}
glued on titanium plates. The fibres on the front and back sides of each titanium plate 
are orthogonally arranged at angles of $\pm$45$^\circ$ with respect to the $y$-axis.
% to the lower detector edge. 
The projections perpendicular to the fibre axes define the   
$u$ and $v$ coordinates which are used in the track reconstruction.
To minimize optical cross-talk, each fibre is coated with a thin aluminium film. The 
individual fibre layers are staggered by multiples of 1/10 of the fibre size 
to improve the position resolution. 
The theoretical resolution of 14.4 $\mu$m per $u$ or $v$ coordinate is degraded  
due to imperfect staggering, cross-talk, noise and inefficient fibre channels.
To reduce the impact of imperfect staggering on
the detector resolution, all fibre positions were measured by microscope.
% and fed into the track reconstruction.  
In a test beam~\cite{testbeam1,testbeam2} with 120\,GeV hadrons, the position resolution
was measured to be between 30 $\mu$m and 35 $\mu$m. The efficiency to detect a traversing 
proton in a single fibre layer is typically 93$\%$, with layer-to-layer variations of 
about 1$\%$. 
% This value includes 4$\%$ efficiency loss due to the 10$\mu$m non-scintillating cladding 
% material surrounding the fibre core. 
%
The overlap detectors (OD) consist of three layers of 30 scintillating fibres per layer measuring 
the vertical coordinate of traversing beam-halo particles or shower fragments.\footnote{Halo 
particles originate from beam 
particles which left the bunch structure of the beam but still circulate in the beam pipe.}
Two independent ODs are attached at each side of both MDs,
as sketched in Fig.~\ref{fig:ALFA_front_view}.
The alignment of the ODs with respect to the coordinate system of the MDs was performed 
by test-beam measurements using a silicon pixel telescope. 
% \cite{EUDET}. out since no real message
A staggering by 1/3 of the fibre size results in a single-track 
resolution of about 50 $\mu$m. 
% Large samples of halo particles traversing the ODs are used for the measurement of
% the distance between upper and lower MDs which is the main ingredient for the vertical 
% alignment described in Section~\ref{sec:alignment}.   
% With a few thousand halo-tracks and a MD-to-OD alignment of 8$\mu$m precision 
% an uncertainty of 20$\mu$m was achieved for the distance value.
The signals from both types of tracking detectors are amplified by 
64-channel 
% Hamamatsu R7600 
multi-anode photomultipliers (MAPMTs). 
% \footnote{HAMAMATSU PHOTONICS K.K., R7600-00-M64}. 
The scintillating fibres are directly coupled to the MAPMT photocathode.  
Altogether, 23 MAPMTs are used to read out each MD and its two adjacent ODs.

Both tracking detectors are completed by trigger counters 
% to select elastic scattering and halo or shower events for the distance measurement. 
% These counters 
which consist of 3 mm thick scintillator plates covering the active areas of MDs and ODs. 
% The two MD trigger counters reduce noise contributions by demanding a coincidence of 
% signals. 
Each MD is equipped with two trigger counters and their signals are used in coincidence to
reduce noise contributions. The ODs are 
covered by single trigger counters and each signal is recorded. Clear-fibre bundles are 
used to guide all scintillation signals from the trigger counters to 
% Hamamatsu R7400 
single-channel photomultipliers. 

Before data taking, precision motors move the RPs vertically in 5 $\mu$m steps towards 
the beam. The position measurement is realized by inductive 
displacement sensors (LVDT) 
% (LVDT HCA 2000) 
% \footnote{MEASUREMENT SPECIALITIES TM, Hampton VA, USA, AC LVDT, HCA 2000} 
% \footnote{Linear Variable Differential Transformer AC HCA 2000} 
calibrated by a laser survey in the LHC tunnel. The internal precision of these 
sensors is 10 $\mu$m. In addition, the motor steps are used to cross-check the 
LVDT values. 
% Each station is equipped with a system 
% of micro-switches and electrical stoppers to define innermost and outermost positions. 
% The outermost electrical stoppers serve as reference points to follow the
% stability of the LVDT values. Since upper and lower RPs can in principle be moved 
% across the beam an anti-collision switch protects them against accidental collision. 

The compact front-end electronics is assembled in a three-layer structure attached to the 
back side of each MAPMT. The three layers comprise a high-voltage 
divider board, a passive board for signal routing and an active board for signal amplification, 
discrimination and buffering using the MAROC2 chip~\cite{MAROC1,MAROC2}. 
% It comprises three functionalities: a passive connector board for the 64 
% MAPMT channels, a high-voltage divider board, and an active board with the readout chip. 
% The MAROC chip amplifies the signal and stores them in a readout buffer. 
The buffers of all 23 MAPMT readout chips of a complete detector are serially transmitted 
by five kapton cables to the mother-board.
% The trigger signals are amplified in a separate board. 
% with two MAROC chips. 
All digital signals are transmitted via a fibre optical link to the central ATLAS 
data acquisition system. The analogue trigger signals are sent by fast air-core cables to 
the ATLAS central trigger processor.   

\begin{figure}[hb!]
  \centering
  \includegraphics[width=\textwidth]{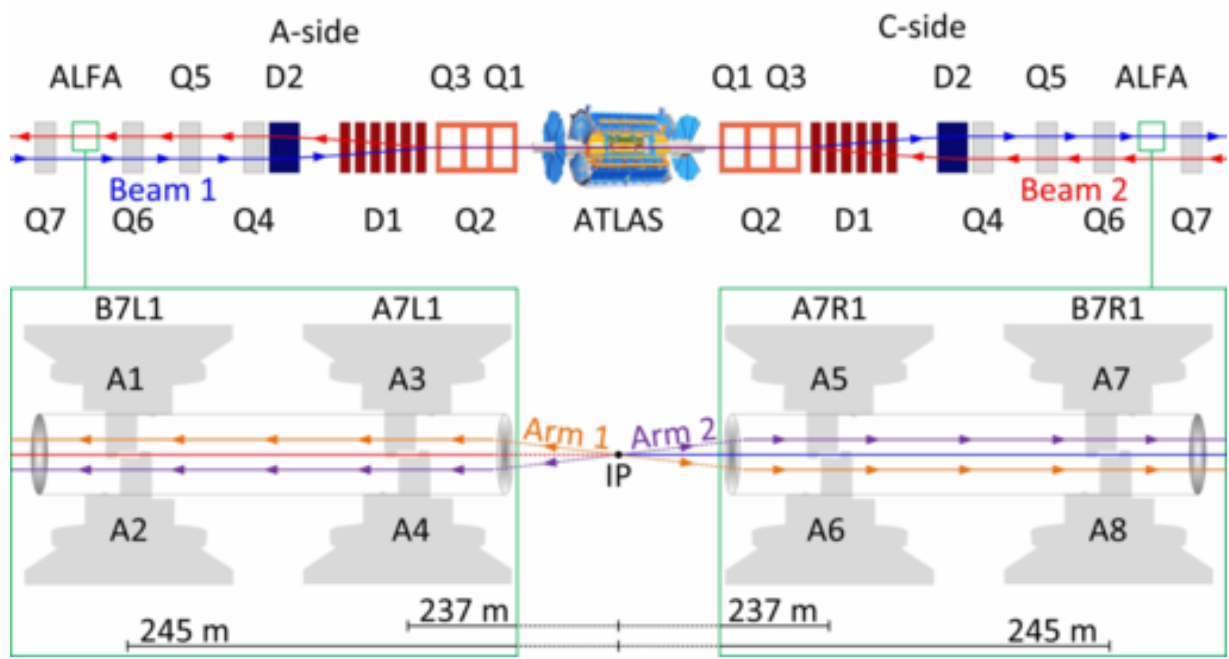} 
%  \vspace*{5cm}
  \caption{A sketch of the experimental set-up, not to scale, showing the
   positions of the ALFA Roman Pot stations in the outgoing LHC beams, and the quadrupole
   (Q1--Q6) and dipole (D1--D2) magnets situated between the interaction point and ALFA.
   The ALFA detectors are numbered A1--A8, and are combined into inner stations A7R1 and A7L1, 
   which are closer to the interaction point, and outer stations B7R1 and B7L1.The positions of the outer stations correspond to the new positions as defined for Run-II. 
   The arrows indicate in the top panel the beam directions and in the bottom panel the scattered 
   proton directions.}
  \label{fig:alfa_layout_new}
\end{figure}

The station and detector naming scheme is depicted in Fig.~\ref{fig:alfa_layout_new}.
The stations A7R1 and B7R1 are positioned at $z$ = $-$237 m and $z$ = $-$245 m respectively 
in the outgoing beam 1 (C side), while the stations A7L1 and B7L1  are situated 
symmetrically in the outgoing beam 2 (A side). 
The detectors A1--A8 are inserted in 
increasing order in stations B7L1, A7L1, A7R1 and B7R1 with even-numbered detectors in the 
lower RPs. Two spectrometer arms for elastic-scattering event topologies are defined by the following 
detector series: arm 1 comprising detectors A1, A3, A6, A8, and arm 2 comprising 
detectors A2, A4, A5, A7.
The sequence of dipoles and quadrupoles between the interaction point and ALFA is also shown in 
Fig.~\ref{fig:alfa_layout_new}. Among them, the inner triplet Q1--Q3 is most important for the 
high-$\beta^{*}$ beam optics.

The ALFA detectors were operated in 2011 and 2012 under various beam conditions, with the nominal LHC collision optics
but mostly at a dedicated high-$\beta^{*}$ beam optics of 90 m. The latter setting was optimized to measure the
total $pp$ cross-section at the LHC at $\sqrt{s}=7$\,TeV (see Ref.~\cite{alfa1}) and $\sqrt{s}=8$\,TeV. Data were also taken at a $\beta^{*}$ value of 
1 km to give access to even lower $t$, the momentum transfer, values. In addition to these low intensity $pp$ runs, diffractive data were taken
with about 110 bunches with the detectors being as close as 7 mm to the beam as well as data in proton-lead collisions.

During operation of the ATLAS/ALFA system in Run-I a systematic increase of the temperature at the level of the detectors was noticed,
starting at injection of the beam and reaching the maximum typically 3 hours after the energy ramp; the higher the densities of the bunches the higher the 
temperature increase. An increase of up to 20 degrees Celsius was observed for a total beam intensity of 2$\times$10$^{14}$protons. Therefore 
absolute temperatures close to 40-45 degrees Celsius were reached, putting the gluing of fibers under stress and risking to put in danger the safe operation of the detector.
The increase in temperature was traced back to be the consequence of RF losses in the cavity near the detector. Simulation work confirmed the hypothesis
with an estimated power deposit of typically 10 W, which translate into the 20 degrees Celsius. Extrapolations to Run-II conditions
gave values of up to 80 W in power dissipated; this would have damaged definitely the ALFA detectors. 

\begin{figure}[ht!]
  \centering
  \includegraphics[width=\textwidth]{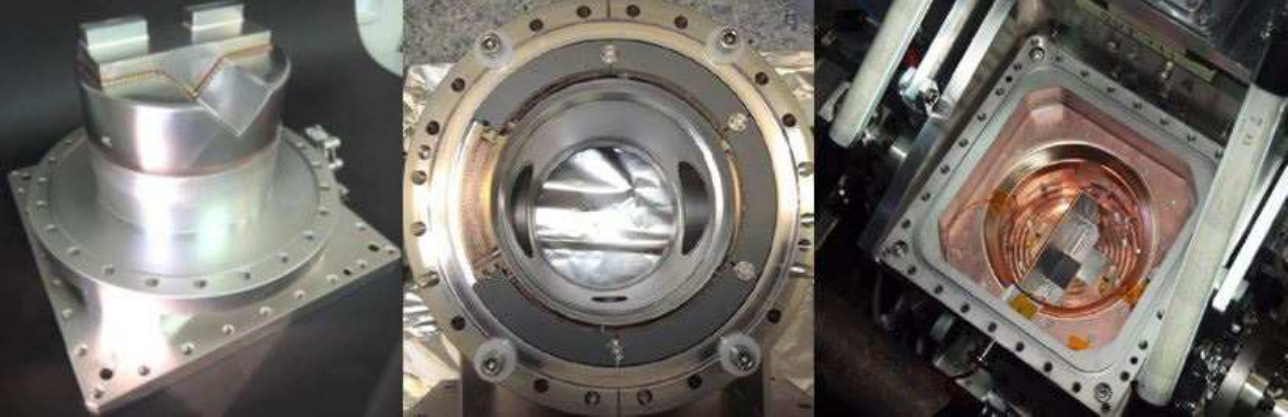} 
%  \vspace*{5cm}
  \caption{Three of the main modifications implemented to minimise the effect of RF.
  From left to right: the RP filler with the copper contacts for proper grounding, the ferrites distributed as a ring on the flange 
  of the station and the HDS (heat distribution system) in copper with the temperature probes attached to it.}
  \label{fig:alfa_upgrade_new}
\end{figure}

During LS1 it was therefore decided to revise the design of the ALFA RP and stations to minimise the RF losses. Four main actions were taken.
The first was to reduce the cavity to its minimum by extending the RP by a RP filler; the second was to move the ferrites to a more appropriate 
position to absorb more efficiently the wake field; the third was to add a heat distibution system in copper to extract more easily the heat if it gets to the 
detector and finally the fourth was to implemented an air cooling circuit. The three first options are illustrated on Fig.~\ref{fig:alfa_upgrade_new}. After implementation of the 
changes and extensive testing, the four ATLAS/ALFA stations are back in operation. With all the measures taken a reduction of close to a factor 50 is expected in the 
power deposited on the detectors.

Two other important upgrades were undertaken during LS1; moving stations apart by 4 m (see Figure~\ref{fig:alfa_layout_new}) and changing the Trigger system. 
Moving the B7L1 and B7R1 stations 4 m downstream from their original positions will improve the local angle resolution by factor of 2. The new trigger Back End boards will
%significantly 
reduce the latency budget, making it possible for ATLAS to use the ALFA triggers with readout of the full detector system.
The trigger Front End electronics of the ALFA detectors where upgraded to reduce dead time from about 550 ns to 87.5 ns. This makes
efficient triggering possible with bunches separation down to 100 ns, corresponding to up to about 700 colliding bunches in LHC.

Before the forthcoming data taking period, Run-II, quite some activity will go into recommissioning of the system, in particular on the 
new Trigger system. The ALFA approved physics programme, total cross-section measurement and independant luminosity measurement, will be the 
main focus during Run-II. Data at a $\beta^{*}$ of 90 m at $\sqrt{s}=13\,\rm TeV$ is the plan for 2015; during the shutdown 2015-2016 a new set of 
cables will be installed to power separately Q4 and Q7 allowing higher values in $\beta^{*}$ and therefore giving access to lower $|t|$ values. 
If supported by ATLAS, ALFA will also participate in runs aiming at diffractive physics; in particular ALFA would have the 
capability to participate in high intensity runs with up to 700 bunches.

%% file: detectors/FSC.tex
\section{Forward Shower Counters in CMS}	

The true rapidities of the proton beams are 
$|y_{beam}| <$ 8.92 (9.94) at $\sqrt{s}$
= 7 TeV (13 TeV) respectively. Neutrons can be measured up to the beam rapidity
with the Zero Degree Calorimeter, ZDC, which also measures photons, mostly from $\pi^0$'s, at $\theta$ = 0 ($\eta = \infty$). 
 All charged particles have been swept by upstream magnets out of the ZDC acceptance. 
Up to the TAN at $z$ = 140 m both incoming and outgoing beams are in a common vacuum pipe.

\begin{figure}[!ht]
\centering
 \includegraphics[width=0.80\textwidth]{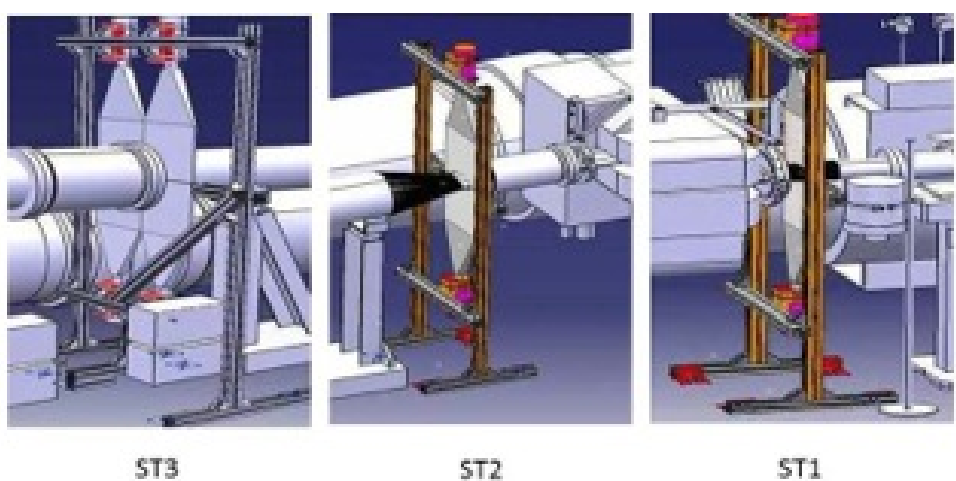}    
\caption{The layout of three FSC stations in CMS. The fourth station is identical to ST3.}
\label{fig:st123}
\end{figure}

Apart from quasi-elastically scattered protons with $y \sim y_{beam}$ detected in Roman pots in TOTEM in high-$\beta^*$ runs, and in future with CMS-TOTEM
Precision Proton Spectrometers (CT-PPS) there are rapidity regions $+5 < \eta < +9$ and $-9 < \eta < -6.6$ not instrumented
for \emph{direct} particle detection. The asymmetry is because CASTOR  has $-6.6 > \eta > -5.2$ but only on the negative-$\eta$ side. 
Sets of scintillation counters, called Forward Shower Counters, FSC \cite{fsc,fscijmpa}, have been installed surrounding the outgoing beam 
pipes at $ z = \pm 59$ m, $\pm$85 m and $\pm$114 m, see Figure~\ref{fig:st123}.
%Particles in the region $6 \lesssim |\eta| \lesssim 8$ hit the beam pipes and surrounding material and may interact and cause showers of charged and neutral
%particles that will be detected in the FSC. The $\eta$-coverage is not well defined. Magnets upstream (quadrupoles and the beam separation dipoles, MBX) deflect
%charged particles; the counters are rectangular with elliptical or circular cut-outs for the beam pipe, and the upsteam material is not simple.
%For Run-II fourth sets of (4) counters at $z = \pm$140 m just upsteam of the TAN is being prepared, and the readout will be upgraded.

These scintillation counters are fully efficient for minimum ionising particles, 
but their efficiency for primary particles is determined by the material of the beam pipes and surroundings 
and is a function not just of $\eta$ but also of $p_T, \phi$ and charge $Q$, as well as the machine optics $\beta^*$. 
Particles in the region $6 \lesssim |\eta| \lesssim 8$ hit the beam pipes and surrounding material and may interact and cause 
showers of charged and neutral particles that will be detected in the FSC. Then $\eta$-coverage is not well defined.

\begin{figure}[!ht]
\centering
\includegraphics[width=100mm]{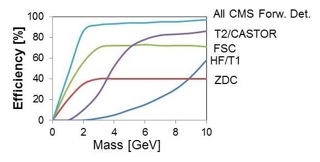}
\caption{Detection efficiency for single diffractive events as a function of diffractive mass, for different CMS
detectors.}
\label{fig:sde}
\end{figure}

An FSC gap by itself is not enough to select diffractive events efficiently it can be combined with either a proton in the TOTEM Roman pot in the same direction in the high-$\beta^*$ runs, or gaps in other forward detectors: ZDC, CASTOR, T2 and also
sometimes HF ($ 3 < \eta < 5$).
\textsc{geant} was used \cite{fsc} to simulate forward particle production, transport through the beam lines and
showering in the materials.
Fig. \ref{fig:sde} shows estimated detection efficiencies for low mass diffraction, as
a function of diffractive mass, for different combinations of forward detectors. At least five hits in any FSC counter were required, or a track or signal in the $|\eta|$ region
covered by T1, T2, HF, CASTOR or the ZDC. Approximately 25\% of the single diffractive cross section is for masses below 10 GeV/c$^2$.

CMS-TOTEM measurements at $\sqrt{s}$ = 8 TeV \cite{dndeta} show $dN_{ch}/d\eta \sim$ 3 at $\eta$ = 6.  
Thus over two units of $\eta \sim 6 - 8$ most non-diffractive interactions have several charged particles plus neutrals 
(photons from
$\pi^0$, $K^0_L$ and neutrons) which may interact in the beam pipe or other material and make showers. 
Therefore the FSC counters are only useful for low pileup running, in particular for bunch crossings with only one inelastic collision.
Using FSC as rapidity gap detectors one can also require the adjacent T2 to be empty, and when
using them to detect proton-dissociation products they can be used to extend the mass range.
 
During the mean pileup $\mu \sim$ 0.05 TOTEM run with $\beta^*$ = 90 m in July 2012 the FSC counters on the positive-$\eta$ side were operational and correlations between
their activity (hits or no-hits) and the direction of a proton, as well as with the mean-$\eta$ of associated central jets, were observed.
Their noise levels are low enough for them to be used as rapidity gap detectors.

%During the mean pileup $\mu \sim$ 0.05 TOTEM run with $\beta^*$ = 90 m in July 2012 the FSC counters on the positive-$\eta$ side were operational and correlations between
%their activity (hits or no-hits) and the direction of a proton, as well as with the mean-$\eta$ of associated central jets, were observed. 
%Their noise levels are low enough for them to be used as rapidity gap detectors. 
%On the other hand simulations show that their efficiency for detecting particles
%in $6 < |\eta| < 8$ is less that 100\%, typically $\sim$80\% depending on the modeling of the forward particle distributions.
%The acceptance $\times$ efficiency is a function of the primary particle's $p_T, p_z, \phi$ and 
%charge $Q$ (neutral as well as positive and negative) and a detailed model of the material along the beam pipes is also needed.

In addition to their use as ``gap detectors", they extend the rapidity coverage close to $\Delta\Omega = 4 \pi$, which minimises the
uncertainty in extrapolating the total inelastic cross section $\sigma_{inel}$ to low diffractive masses.
Events (in low pileup running) in which all the CMS detectors covering $ -5 < \eta +5$ are empty (in noise levels) but which have signals in the FSC counters
(on one or both sides) from low-mass diffraction events can be measured and included. Furthermore the patterns of hits in the three stations (which cover different
$\eta$-ranges) can be tested against models of low-mass diffraction, e.g. mass-dependent $ p \rightarrow p \pi^+\pi^-$ or $n\pi^+$.


%An FSC gap by itself is not enough to select diffractive events. However it can be combined with either a proton in the TOTEM Roman pot in the same
%direction in the high-$\beta^*$ runs, or gaps in other forward detectors: ZDC, CASTOR (when installed), T2 and also 
%sometimes HF ($ 3 < \eta < 5$). 
%\textsc{geant} was used \cite{fsc} to simulate forward particle production, transport through the beam lines and 
%showering in the materials.
%Fig. \ref{fig:sde} shows estimated detection efficiencies for low mass diffraction, as
%a function of diffractive mass, for different combinations of forward detectors. At least five hits in any FSC counter were required, or a track or signal in the $|\eta|$ region 
%covered by T1, T2, HF, CASTOR or the ZDC. Approximately 25\% of the single diffractive cross section is for masses below 10 GeV/c$^2$. 

%% file: detectors/afp_yr_review.tex
\section{The AFP Detector}
\label{sec:afp}

The ATLAS Forward Proton (AFP) program aims to intercept and measure protons emitted in the very forward directions from the ATLAS interaction point (IP). 
Forward protons are characterized by their energy fractional loss $\xi=(E - E_{beam})/E_{beam}$ and by the four-momentum transfer 
squared $t=(p - p_{beam})^2 \simeq (p\theta)^2$, where $E$ and $p$ characterize the forward proton, and $\theta$ is its scattering angle. 
The program and its detectors is described in detail in the AFP Technical Design report.\cite{bib:AFP-TDR2015}

A varied physics program using forward protons becomes available with AFP: single and double diffraction measurements, Pomeron structure functions, 
rapidity gap survival probability, and double Pomeron exchange (DPE) and double photon exchange (DPhE) processes. The latter give access to anomalous 
quartic coupling measurements of interest to beyond-Standard-Model Physics signals.

The AFP detectors consist of two forward arms, with two detector station per arm located at 206~m and 214~m from the ATLAS IP. The detectors are housed 
inside so-called Roman Pots, stainless steel pots that are able to move inside the beam pipe aperture after stable collisions are established. The pots 
have thin 300~$\mu$m windows facing the beam, to minimize interactions and to enable the detectors to approach the beam as close as possible to 
intercept protons with energy losses as low as $\xi\simeq 1.5$\%. The upper limit to the $\xi$-acceptance is about 15\% and determined by the LHC optics 
and the upstream beam apertures.

Although the non-exhaustive list of AFP physics processes covers a wide range of cross sections, some of the more interesting processes are rare and 
require running at the highest luminosity available. Characterizing the instantaneous luminosity by the average number of interactions, \textmu, 
occurring at a bunch crossing, \textmu\ is expected to be in excess of 25 for the upcoming LHC run period and may well reach 50 or more interactions per 
bunch crossing.

\subsection{Beam Interface}

The AFP Beam interface of choice is the Roman Pot (RP) because of its proven service record in the ALFA and TOTEM experiments, 
and the acceptable additional impedance that the RP presents to the circulationg LHC beams: less than 1\%\ of the total LHC transverse impedance. 
This was extensively simulated by the LHC Impedance team.

\begin{figure}[ht]
\centerline{
\includegraphics[width=.45\columnwidth]{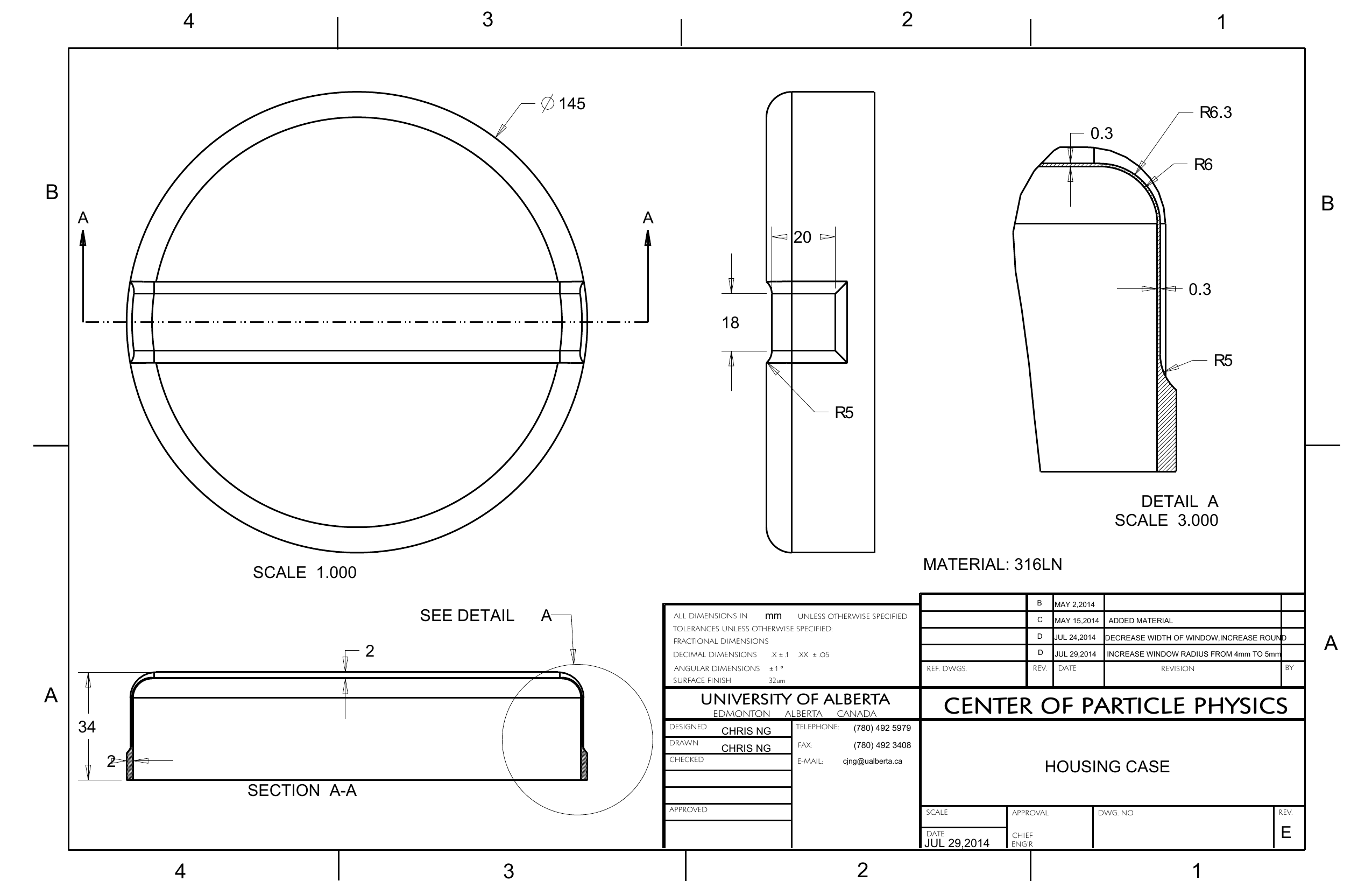}\hspace{0.5cm}
\includegraphics[width=.45\columnwidth]{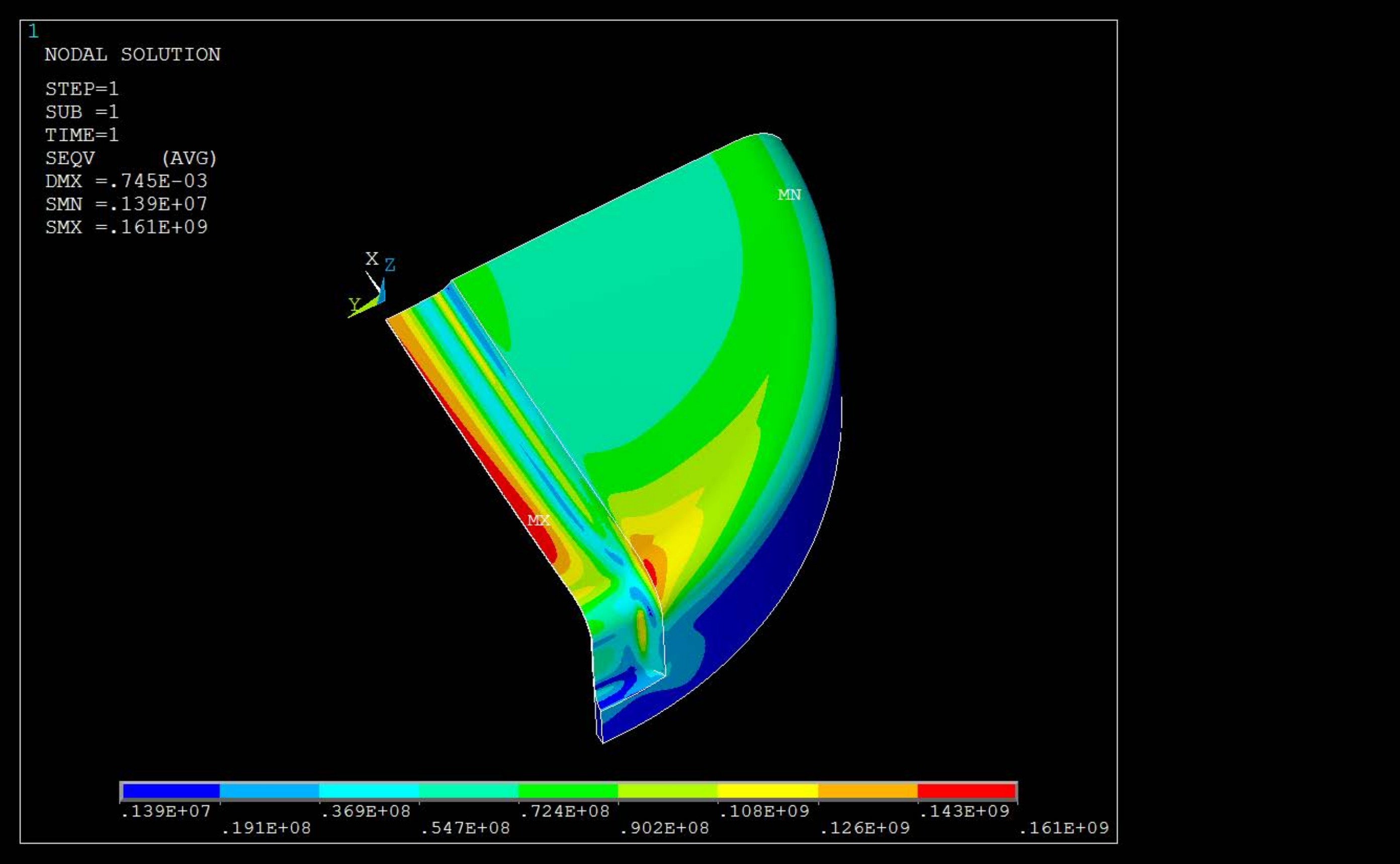}}
\caption{Left: the design of the AFP Roman Pot with 300~\textmu{m} windows. Note the exterior `groove' forming the window facing the beam.
 Right: the calculated stresses in the material} 
\label{fig:AFP-RP}
\end{figure}

The TOTEM collaboration has developed a full design of a cylindrical horizontal RP station based on the existing horizontal and vertical 
RP stations of TOTEM and ALFA, see the relevant chapters in this report. The AFP RP station is a single-sided horizontal cylindrical RP station 
virtually identical to the TOTEM design presented in this report. Differences concern the support table, which is specific to the AFP locations near 
the ATLAS Interaction Point 1 (IP1). Also, the RP itself has a slightly different design: whereas the TOTEM RP design has a beam window of 300~\textmu{m} 
thick formed by a `groove' of 1.7~mm deep and 18~mm wide on the inside of the RP bottom, the AFP RP is instead flat on the inside of the RP, 
with a similar `groove' facing the circulating beam. Fig.~\ref{fig:AFP-RP} shows the AFP RP design. In normal operation the AFP RP will have a 
secondary vacuum inside the pot, and only during the installation and removal of the detectors will the inside RP pressure go to 1 bar. 
The AFP RP design was simulated for mechanical stability in two extreme cases: inside pressure of 1 bar, and outside vacuum, and the reverse. 
The maximum stress of 0.14~GPa, see Fig.~ref{fig:AFP-RP-Stresses}, is 70\% of the assumed 0.20~GPa yield strength of the 316LN Stainless Steel 
and occurs near the outside corners of the thin window facing the beam. The maximum displacement occurs at the center of the thin window and is 
approximately 0.8 mm.

\begin{figure}[ht]
\centering
\includegraphics[width=.45\columnwidth]{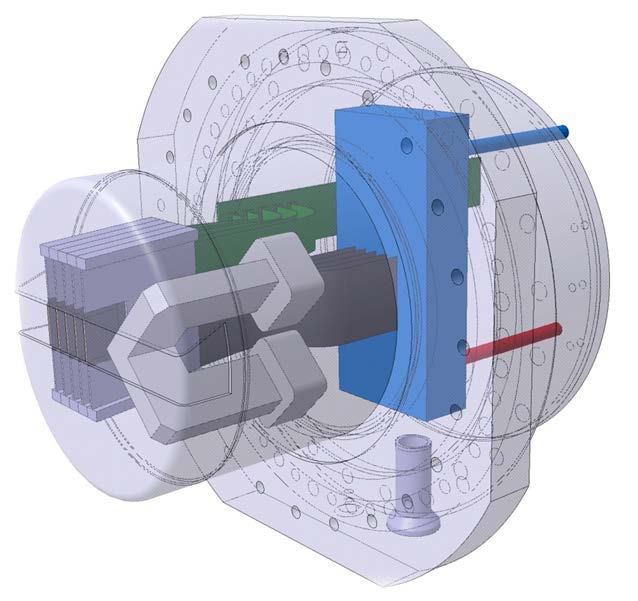}
\caption{The design of the AFP Roman Pot with a 5-plane Silicon 3D Pixel tracker and a double LQbar Time-of-Flight detector. 
Note the cold-air heat exchanger near the top flange (dark blue) which is using cold air from a vortex `aircooler' apparatus.} \label{fig:AFP-RP_SiT-ToF}
\end{figure}

The upstream pot, closest to the ATLAS IP, will contain a Silicon Tracker (SiT) using 3D pixel sensors of the ATLAS IBL design. 
The upstream pot, at 214~m, will contain both a SiT and a Time-of-Flight detector (ToF) using L-shaped Quartz radiator bars (LQbar). 
A draft design is shown in Fig.~\ref{fig:AFP-RP_SiT-ToF}. The detectors are described in the following sections.

\subsection{Silicon Tracker}

The AFP design foresees a high resolution pixelated silicon tracking system
placed at 210~m from the ATLAS interaction point (IP). Combined with the magnet
systems of the LHC accelerator, the AFP tracker will provide the momentum
measurement of the scattered protons. The full AFP tracker will consist of four
units, each composed of four to five pixel sensor layers, which will be placed in Roman Pots, two on each side of the ATLAS IP.  

To ensure good momentum resolution, the AFP tracker is required 
to provide high spatial resolution (about 10~\textmu{m}) in the short
pixel direction. 
Furthermore, it is vital for the physics program to measure very small 
scattering angles. To this end the detectors will be placed
almost perpendicular to the beam (under a small tilt of 15$^{\circ}$) with 
one side only 2-3~mm away from it. This leads another two
critical requirements for the pixel detectors:
\begin{enumerate}
\item The inactive region of the detector side facing the beam 
has to be minimized to about 100-200~\textmu{m}.
\item Due to the proximity to the beam, the detectors have to withstand a highly non-uniform
irradiation profile with a high maximum fluence in the area closest to the beam and 
several orders of magnitude lower away from it. The magnitude of the maximum fluence depends on the
run scenario: about 5$\times$10$^{12}$~p/cm$^2$ are expected for initial low-luminosity 
runs and about 5$\times$10$^{15}$~p/cm$^2$ for a possible later high-luminosity scenario.
\end{enumerate}
 
\subsubsection{Modules}

The most critical component of the AFP tracking system is the pixel module.  It
consists of a 3D pixel sensor bump-bonded (connected) to a front-end chip which
in turn is glued and wire-bonded to a flexible printed circuit (flex).  The
flex provides clock and command signals and routes the data output. The AFP
module will consist of a single FE-I4B front end chip, which will provide an
active area of $1.68\times2.00$~cm$^2$. The AFP single-chip 3D pixel modules are similar to the ones used in the 
ATLAS Insertable B-Layer (IBL) detector. However, some
important modifications have to be implemented to meet the specific requirements
for AFP. 

\paragraph*{3D Sensors}

In 3D pixel sensors, n- and p-type column-like electrodes
penetrate the substrate defining the pixel configuration.
Though the fabrication process is complex,
the technology is less demanding in terms of bias voltage
and cooling than the standard planar approach, and
the reduced drift path makes 3D devices more radiation
hard. In recent years significant progress has been made
in the development of 3D sensors, which culminated in
the sensor production for the ATLAS IBL~\cite{bib:IBL-TDR2010}.
The AFP pixel detectors will be based on the 3D double
sided sensors developed by CNM (Barcelona) and FBK
(Trento) for the IBL. 

\begin{figure}[htbp]
\centering
\includegraphics[width=\textwidth]{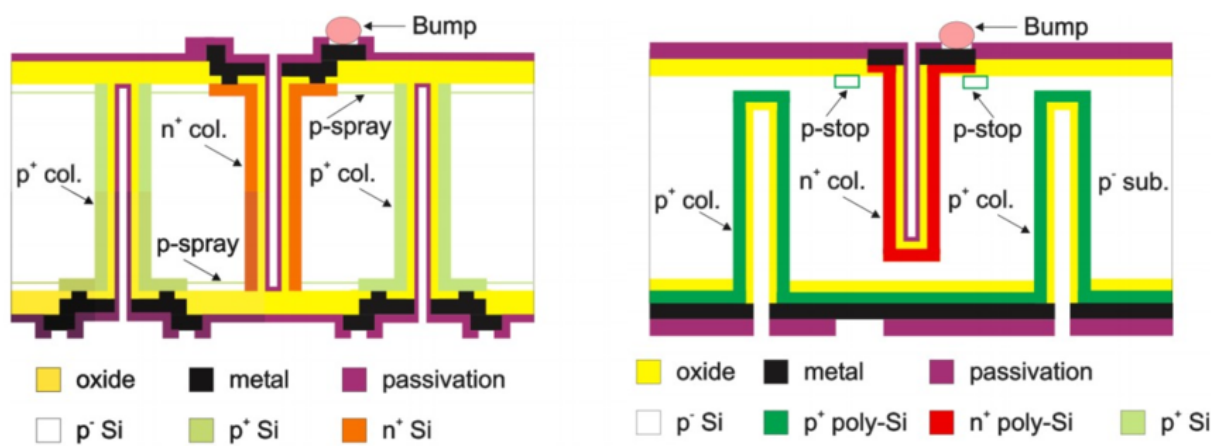}
\caption{\textbf{Design of the columns of the FBK (left) and CNM (right) 3D sensors.}}
\label{fig:3d_design}
\end{figure}

The AFP 3D sensors were already fabricated 
at CNM on Float Zone, p-type, 100~mm diameter, wafers, with $<100>$ crystal
orientation, 230~\textmu{m} thickness, and a very high resistivity (10 to 30
k$\Omega$~cm). Columnar electrodes, 12~\textmu{m} wide, were obtained by Deep
Reactive Ion Etching (DRIE) and dopant diffusion from both wafer sides
(n+ columns from the front side, p+ columns from the back side), without
the presence of a support wafer. By doing so, the substrate
bias can be applied directly on the back side. 
The sensor design features an array
of $336\times 80$ pixels with a pixel size of $50\times 250$ \textmu{m}$^2$. 
Each pixel consists of 2 n+-junction columns and 6 surrounding p+-ohmic 
columns.
Figure~\ref{fig:3d_design} shows details of the 3D sensor layout.

The CNM production for AFP concluded in July 2014. Unfortunately, due 
to a machine problem, a large portion of the $13$~wafers of the run were
damaged and only 6 are expected to work. Each wafer provides $8$
3D sensors, so this is not a problem for the first phase of the 
AFP program.

\paragraph*{Front-end Electronics}

The pixel readout electronics will be the FE-I4B~\cite{bib:FE-I4b_doc}. The
sensors are DC coupled to the chip with negative charge
collection. Each readout channel contains an independent
amplification stage with adjustable shaping, followed
by a discriminator with independently adjustable
threshold. The chip operates with a 40 MHz externally
supplied clock. The time over threshold (ToT) with 4-
bit resolution together with the firing time are stored for
a latency interval until a trigger decision is taken. The
FE-I4 chip can also send a trigger signal via the HitOr line. This HitOr will be used for the first-phase AFP trigger, 
when there is no installed Time-of-Flight detector.

The FEI4 chip has been extensively tested for the ATLAS IBL detector. The
radiation hardness has been well established to fluencies of 250~Mrad and beyond,
surpassing the AFP requirements. The trigger capabilities have also been proven,
as the chip is used to trigger the ATLAS Diamond Beam Monitor detector.

\subsection{Time-of-Flight Detector}

Because single diffraction, which produces a single forward proton, are relatively common (about 10\% of 
the total cross section), an interaction of high interest, like Double Pomeron or Photon Exchange, which yields two forward protons, may easily be faked by the 
occurrence of two single-diffraction interactions in the same bunch crossing. The only possible rejection of this background %to DP(h)E processes 
is to 
measure the arrival time difference of the forward protons in the two arms with pico-second accuracy. For a genuine %DP(h)E 
two-proton event, the arrival 
time difference of the forward protons $\Delta{t}=t_{Left arm} - t_{Right arm}$ is directly related to the interaction vertex location $z_{vertex}$ ($z$ 
measured along the beam from zero at the ATLAS IP and positive toward the 'Right' arm) as: $z_{vertex} = c\Delta{t}/2$. Thus, a $\sigma_t=10$~ps time-of-flight 
resolution translates into a $\sigma_z=2.1$~mm vertex resolution. The vertex location derived from fast timing is compared to the location measured 
from the ATLAS inner detector tracking; if the two locations differ, the protons stem from unrelated background events. Extensive simulations have shown 
that a 10~ps Time-of-Flight measurement provides a background rejection factor around 20.

\begin{figure}
\centering
\includegraphics[width=.34\columnwidth]{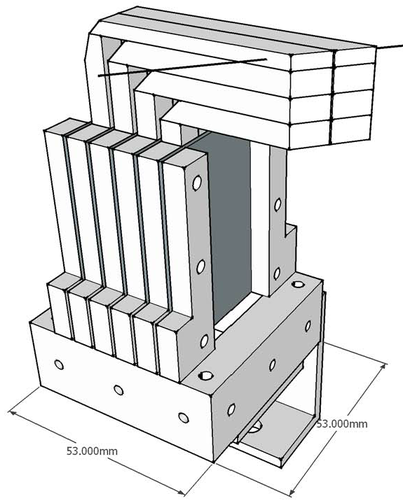}
\includegraphics[width=.22\columnwidth]{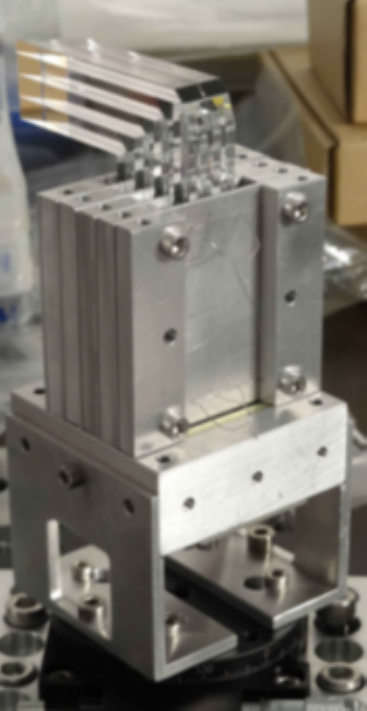}
\includegraphics[width=.40\columnwidth]{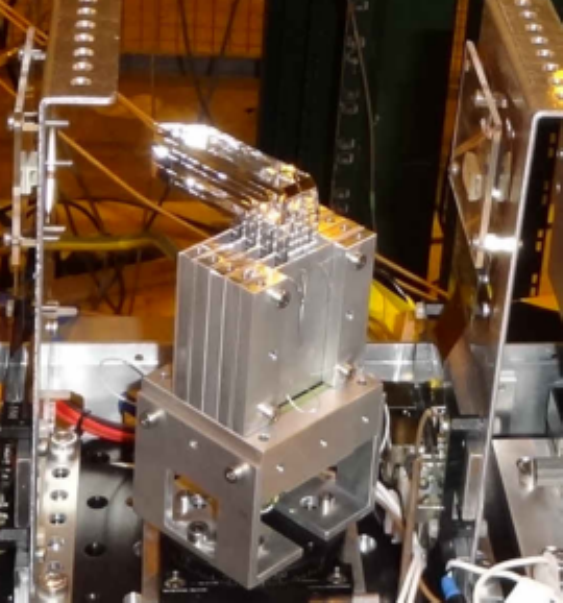}
\caption{Left: the design of the AFP Time-of-Flight Detector (LQbar). The straight line is an example of a diffractive proton entering one of the 
quartz radiator bars from the right. Middle: The LQbar detector before test beam installation in November 2014. Note the quartz radiator bars 
with the 45$^{\circ}$ Aluminized mirrors. Right: the LQbar ToF installed in the beam test. Two 3D Silicon tracker planes are also visible.} \label{fig:LQbarToF}
\end{figure}

The proposed AFP Time-of-Flight detectors consist of Quartz bars positioned at the Cerenkov angle with respect to the proton directions. 
Because of the constraints imposed by the Roman Pot housing, the quartz bars must be bent out of the $z$(beam) - $y$(vertical) plane into the $x$(horizontal) 
direction; a structure that was named `LQbar'. A picture of an 8-channel LQbar detector, and its protype implemented for the November 2014 
AFP Beam Test is shown in Fig.~\ref{fig:LQbarToF}. The Cerenkov light travels up the bars and is converted to a signal by a specialized $4\times 4$-pixel 
Multi-Channel-Plate Photomultiplier tube by Photonis.\cite{bib:PhotonisPMT}

A possible upgrade of this detector allowing a better pixelisation is under study, it could benefit from other technologies such as fast Si or damonds as for CMS-TOTEM.

The PMT output signal is approximately Gaussian with a 700~ps full width at half maximum (rms$\simeq 300$~ps). Photon statistics (the mean number of 
photo-electrons is about 10) affect the signal amplitude but preserve the shape precisely. The goal of the electronics is to preserve the signal shape 
information and derive the best possible timing of the signal, independent of the signal amplitude. 

\begin{figure}
\centerline{\includegraphics[width=.9\columnwidth]{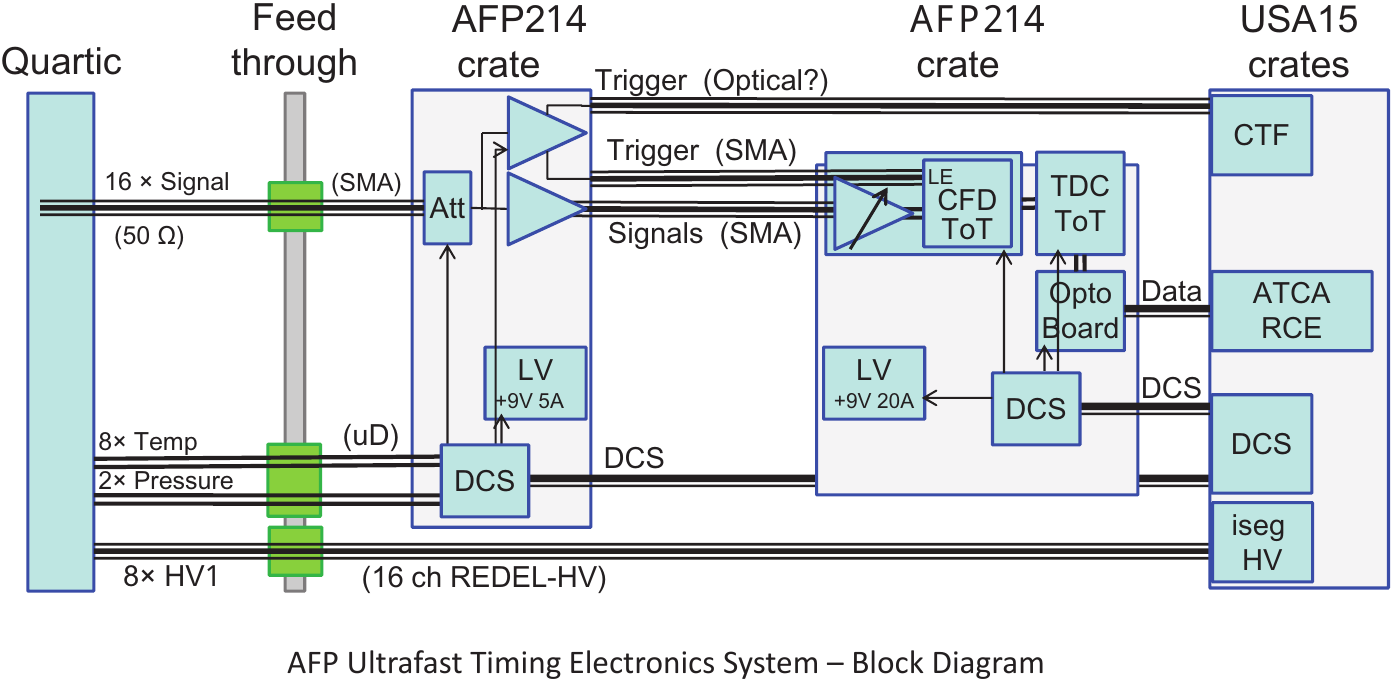}}
\caption{A schematic diagram of the components of the fast timing electronics chain described in the text, together with their physical locations in the LHC tunnel. } \label{fig:Electronics-Chain}
\end{figure}

The approach chosen by the AFP timing group is low-noise amplification followed by constant-fraction discrimination (CFD) and high-precision time 
digitization (HPTDC) and readout, see Fig.~\ref{fig:Electronics-Chain} where the various components and their locations are depicted.\cite{bib:Rijssenbeek2014} (with a possible upgrade using SAMPIC).

Other approaches are possible, as discussed for example by E.~Delagnes, E.~Breton, and S.~Ritt in Ref.~\cite{bib:PicosecondWkshp2014}. 
The sampling methods described by these authors are best performing, and the cost reduced compared to CFD, and the SAMPIC chip is considered as an upgrade for the readout electronics~\cite{bib:JFGenat2014}.

Beam tests (Fermilab 2012, CERN 2013) have shown that the single-channel resolution of PMT, Preamplifiers and CFD is 20~ps, limited by the PMT signal shape, 
statistics, and noise. Two or more sequential measurements (two or more successive quartz bars) will reduce the proton ToF resolution accordingly; four 
sequential measurements will provide 10~ps resolution
(although this configuration has to be tested due to the material budget).

The beauty of the system is its modularity: the resolution can be tuned by the changing the number of quartz bars in succession, while the resolution 
requirement per-channel is somewhat relaxed. Somewhat arbitrarily, in order to preserve the per-channel timing resolution, we require the time jitter of 
the electronics to be 5~ps or less.

Adjacent sequences of quartz bars cover intervals in proton energy loss $\xi$. It is assumed that the $\xi$ acceptance is subdivided into 4 intervals, 
each individual $\xi$ interval covered by a sequence of 2-4 quartz bars. Because the central missing mass $MM$ measured from the two protons is directly 
related to their $\xi$, $MM^2=\xi_1\xi_2(2E_{beam})^2$, a selection for events in which a large missing mass is produced can be formed already at the trigger level.

\subsubsection{Preamplifiers}
The PMT is used at a low gain of about $5\times10^4$ to maximize the lifetime of the tube in the high-rate LHC environment close to the circulating beam. 
The typical PMT output signal at this gain is about 8 mV for 10 photoelectrons. The AFP CFD used has a dynamic range from 250 - 1200~mV (a new design is 
in the works with a further improvement anticipated in dynamic range). 

\begin{figure}
\centerline{\includegraphics[width=.9\columnwidth]{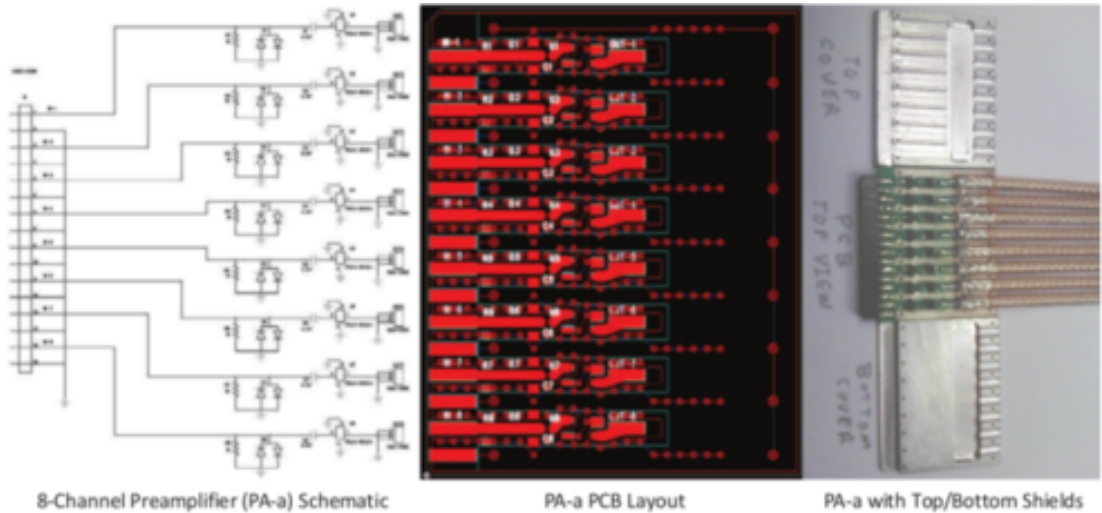}}
\caption{The schematic diagram, layout, and photo of the low noise pre-amplifier, to be located on the photodetector base in the secondary vacuum.}
\label{fig:PA-a}
\end{figure}

In order to match the CFD dynamic range, and to provide for gain variations as function of PMT pixel and ageing, the preamplification is 
done in two 20~dB stages. The first stage PA-a is located directly on the base of the PMT. The 8-channel preamplifier PCB is based on the 
PSA4-5043+ low-noise (NF=0.7~dB) InGaP E-PHEMT MMIC gain block (gain 18.6 dB at 1 GHz) from MiniCircuits.com, see Fig.~\ref{fig:PA-a}.
The PA-a has been tested under power and demonstrated to be radiation tolerant to at least 9~kGy (LANSCE, February 2014 run with 800 MeV protons, 
$2.2\times10^{13}$~p/cm$^2$); this dose corresponds to the dose expected at the preamplifier location for 300~fb$^{-1}$ or three years at a 
luminosity of $10^{34}$~cm$^{-2}$s$^{-1}$.

The first preamplifier stage is connected by coaxial cable to the second preamplification stage (PA-b) located at floor level below the detector, 
where the high-energy proton flux is expected to be a factor 20 lower (the low energy neutron flux is a factor 10 lower there). 

\begin{figure}
\centerline{\includegraphics[width=.9\columnwidth]{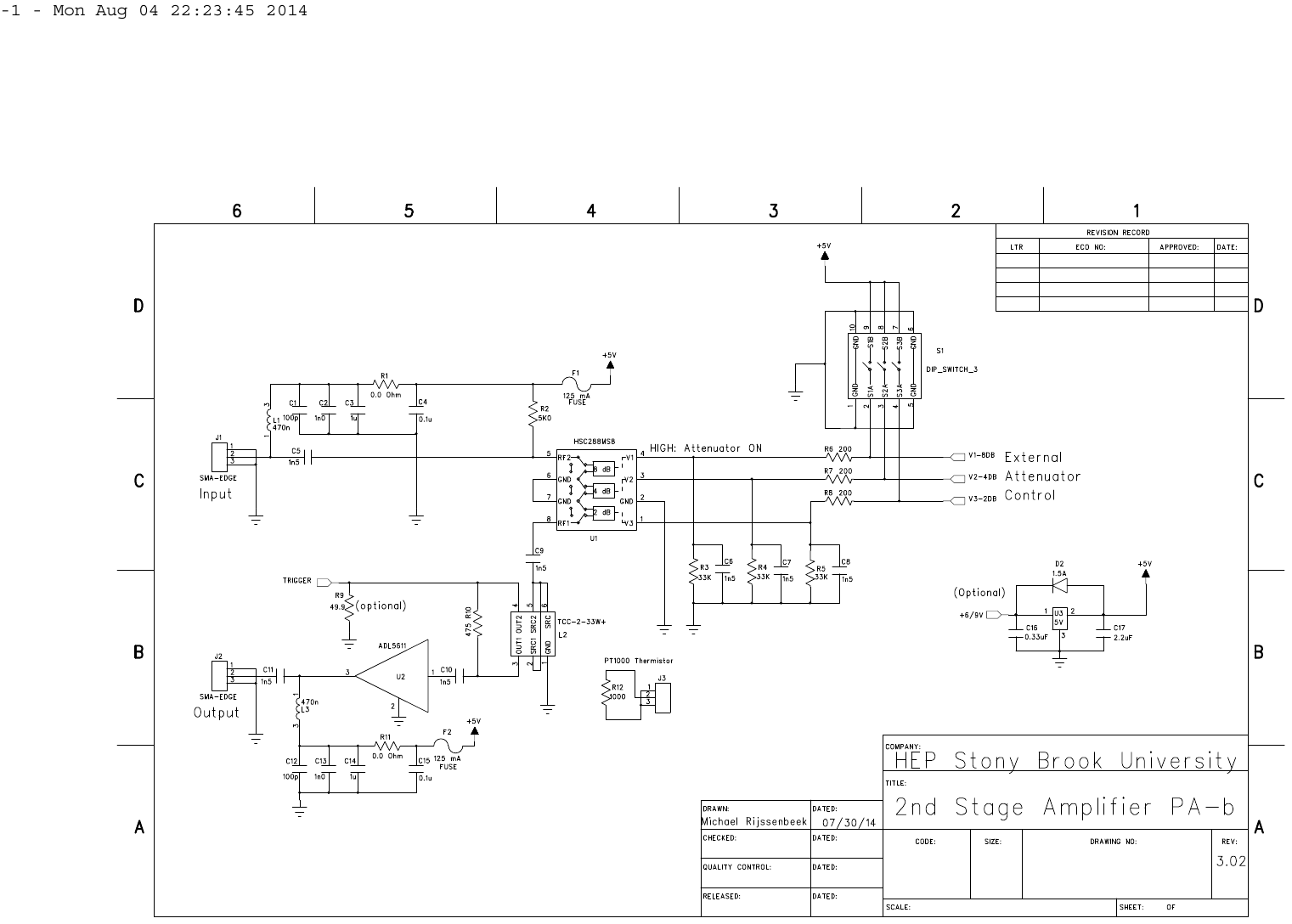}}
\caption{The schematic diagram of the new second-stage variable-gain amplifier PA-b, to be located below the detector at floor level.}
\label{fig:PA-b}
\end{figure}

The PA-b provides DC power (5 V) to the PA-a via the coaxial connection. The PA-b further  includes (in order): a programmable 3-bit attenuator 
(Hittite HMC288MS8 2 dB LSB GaAs MMIC, range 1~dB - 15~dB ), a 2 Way-$0^o$ splitter (MiniCircuits TCP-2-33W+, $-4$~dB insertion loss) providing a 
trigger pick-off, and a ADL5611 gain block (Analog Devices, gain 22(20)~dB at 1(4) GHz), see Fig.~\ref{fig:PA-b}. The PA-b has successfully survived the 
same irradiation runs and doses as PA-a.

\subsubsection{Trigger}
A trigger board has been designed and will be produced in the near future. The design is based on the 8-channel GaAs Discriminator MMIC `NINO' 
(developed and produced by CERN.\cite{bib:NINO}), followed by programmable majority circuitry to form a `N out of M' type trigger combination on two 
LVPECL outputs. The option to include a (properly timed) bunch crossing gate (LVPECL) is implemented. The PCB has been laid out but not yet been produced. Let us notice that the upgrade SAMPIC solution will provide a trigger directly.

The trigger signals from several adjacent quartz bar sequences are combined into a bit stream and sent over fast air-core coax cables to the ATLAS 
Central Trigger Processor. The trigger information from the two AFP arms can be used to form a large proton-proton `missing mass' trigger and can be 
combined with various central ATLAS trigger terms.

\subsubsection{Constant Fraction Discriminator}
The Constant Fraction Discriminator principle has long be used to correct for time walk in cases where the signal fluctuates in amplitude but is constant 
in shape. The AFP design was initially developed for FP420 by Luc Bonnet of the Universit\'{e} Catholique de Louvain, and was further developed for 
AFP by the HEP group (J.~Pinfold, S-L.~Liu) at the University of Alberta at Edmonton (Alberta). The measured time-walk is 5~ps or less over the 
range 250 - 1200~mV. The design is currently revisited to obtain a larger dynamic range and to implement a time-over-threshold functionality, 
which will allow off-line timing corrections if so required. Moreover, the new CFD design includes an optional bunch crossing gate to reduce output rates.

A single CFD channel is implemented on a small $28\times70$~mm$^2$ daughter board, with RF I/O connectors for signal in and signal NIM out, and 
differential LVPECL outputs.

\subsubsection{High Precision Time Digitizer}
The High Precision Time Digitizer board, HPTDC, was developed by Alberta. The 12-channel board uses 4 HPTDC ASICs developed and produced by 
CERN in 0.25~$\mu$m CMOS technology (HPTDC, J.~Christiansen et al., http://tdc.web.cern.ch/tdc/hptdc/hptdc.htm). The four ASICs are controlled 
by an on-board FPGA which also handles the flow of data and controls. This and previous versions of the HPTDC board have been used successfully at 
various beam tests. The HPTDC and new developments were presented in Ref.~\cite{bib:JPinfold2014}.

The intrinsic resolution of the current HPTDC is 16~ps, which is a significant contributor to the per-channel resolution. However, new HPTDC ASIC 
development with smaller feature size are ongoing at CERN and may lead to significant improvements in the near future. Note that the 16~ps 
resolution of the HPTDC is per channel and that the contribution for a system of four quartz bars in sequence would only be 8~ps. 

The radiation tolerance of the HPTDC is not guaranteed. The HPTDC ASIC is expected to be radiation tolerant to a degree sufficient for it 
to be located on the tunnel floor, near the the detectors. The FPGA firmware must be re-designed to provide the appropriate checking of HPTDC 
registers for upsets. Moreover, the FPGA itself has to be radiation tolerant, which can be done by choosing a radiation-hard part (expensive!) 
or going to a fuse-programmable part. Alternatively, the FPGA can be programmed to do self-checking and organized with majority decisions 
in critical paths. It is the latter choice that will be pursued.

\subsubsection{Reference Clock}
A major component of any time-of-flight system using two widely separated detector arms (424~m apart measured along the beam line), is a 
synchronizing Reference Clock. As for other components, the requirement is that the two local detector clocks are synchronized to well within 5~ps.

The University of Texas group (A.~Brandt, V.~Shah, et al.) has developed a prototype Reference Clock based on a design originally by SLAC. 
Every Daughter Clock sends its signal to a central Reference Clock which produces a DC phase error signal that is read (on the same cable) 
by the Daughter Clock. The Daughter Clock adjusts its phase until the phase error signal is zero. 

The design is not fully complete at this time but initial tests indicate the desired performance can be reached.

In addition to the synchronized local clock, clock fanouts at the local detectors are required. We intend to implement these with high 
performant LVPECL Clock FanOut buffers from Micrel.

\subsection{Data Acquisition}
The Data Acquisition system currently foreseen is based on the Reconfigurable Cluster Element (RCE) computer daughterboards in the ATCA 
telecom standard.\cite{bib:ATCA-RCE} This system has successfully been employed for the testing of the ATLAS Intermediate B-Layer 
Silicon pixel detectors. Because the same Silicon sensors are used for the AFP tracker, the RCE-based DAQ system can be used essentially 
without any new development. However, the HPTDC board has to be interfaced to the RCE readout. This requires new FPGA firmware 
(also required for radiation tolerance!) as well as additions and modifications to the RCE software.

Recently, the Time-of-Flight time digitizer HPTDC board was successfully interfaced to the RCE DAQ with an interfacing very similar 
to the interfacing with the FE-I4 silicon tracker front-end chip. Both tracker and Time-of-Flight detectors can therefore be read with 
the same RCE-based DAQ. The IBL Optoboard, the optical data and command interface board for the ATLAS Intermediate B-Layer detector, 
will be used to interface between the optical data cable and the copper lines to the front-ends in the same way as for the ATLAS IBL. 

Because the RCE hardware is located in a low radiation and accessible area near the ATLAS detector, radiation tolerance is not an issue for the DAQ.

%% file: detectors/yellow_lhcb.tex
.

\newpage

\section{LHCb Experiment}

As described in Chapter~\ref{chap:cep} LHCb is well suited for central exclusive production (CEP) physics \cite{Aaij}. 
It has excellent acceptance, tracking and particle identification in
the forward region, good sensitivity to low $\rm p_T$ particles, and is able to reject activity in the backward
region using tracks reconstructed in the VELO.   The experiment runs
at lower luminosity using $\beta^*$ settings and offset levelling
techniques, and so benefits from low pile-up conditions and is able to
select CEP events with no additional interactions.  Of the 3.2 $\rm
fb^{-1}$ accumulated in Run I about $21\%$ is useful for 
studying exclusive events with no additional pile-up 
activity.  Run II will extend this potential; after the move to a 25 ns filling scheme, LHCb plans to accumulate more than 
$5 \: \rm fb^{-1}$, of which the useful fraction for CEP studies with
no pile-up will rise to $37 \%$.  A further advantage of LHCb is the
availability of the low level trigger operating at 40 MHz, which has
access to information about backwards activity and has been used since
2012 to enhance the event yield for CEP hadronic final states.

Previously published LHCb results have shown that with the current coverage there is still a 
significant irreducible background to the central exclusive signal consisting of events where the proton has dissociated 
in the forward direction outside the experimental acceptance.   For
this reason LHCb is currently installing a new system of scintillator
detectors in the LHC tunnel to detect showers from high rapidity
particles interacting with beam-pipe elements.
The concept is modelled on previous successfully operating systems, in
particular at CDF at the Tevatron~\cite{CDF} and CMS for low pile up LHC running.
The absence of activity in the scintillators can be used to confirm the existence of a rapidity gap extending beyond the spectrometer acceptance, 
and to reduce backgrounds to CEP candidates and allow the study of
many interesting central states with lower masses which are currently
systematics limited.   The HERSCHEL (High Rapidity Shower Counters for
LHCb) system will also act as a general rapidity 
gap detector, identifying very forward showers in low mass diffractive
excitation.   The LHCb readout offers the potential to incorporate 
scintillator signals into the low level trigger.  In this way a highly
efficient trigger can be developed which will allow LHCb to exploit
dedicated low luminosity running for forward physics purposes.  
In addition to their value for studying rapidity gaps without pile-up, the FSC counters can provide real-time monitoring of beam 
conditions and may extend the usefulness of the Beam Gas Imaging
luminosity measurement by complementing existing coverage in the
backwards direction.

\subsection{HERSCHEL configuration}
\label{sec:detecor:lhcb:herschel}

\begin{figure}[hbt]
\centering
\includegraphics[width=0.9\textwidth]{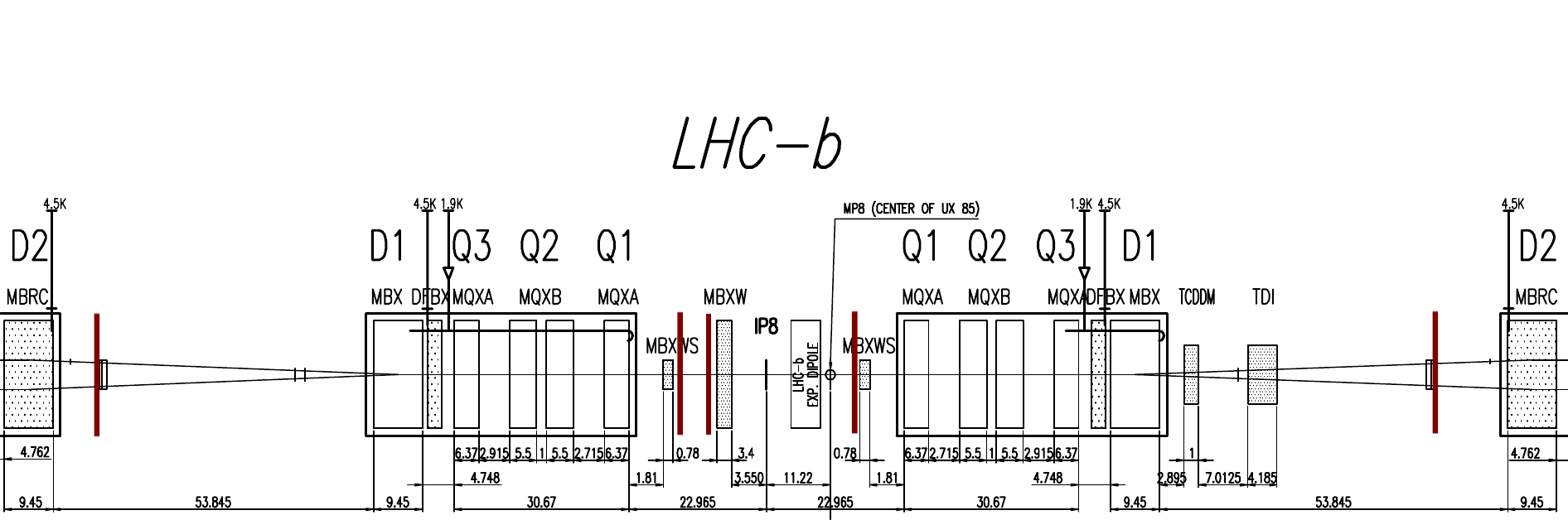}
\caption{Layout of IR8. The positions of the shower counter stations are indicated by the red arrows.}
\label{fig:i8beamline}
\end{figure}

\begin{figure}[hbt]
  \centering
  \includegraphics[width=0.95\textwidth]{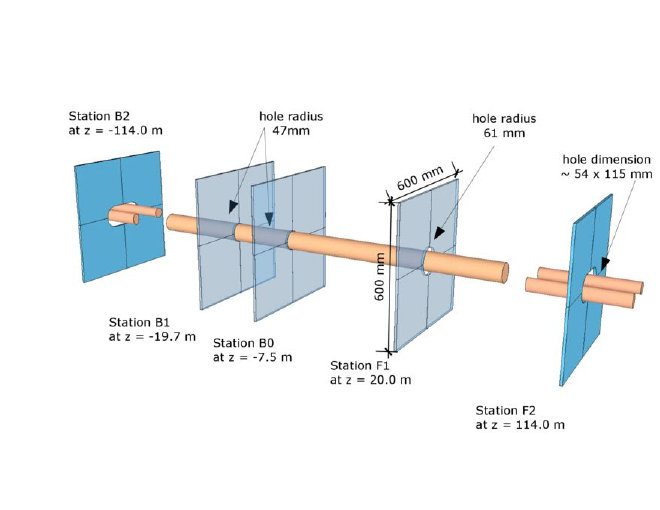}
  \caption{Schematic overview of the scintillator configuration.  The
  $z$ scale has been compressed and the long section of the beam pipe
  between the B/F1 and B/F2 stations omitted.}
  \label{fig:configurations1}
\end{figure}

\begin{figure}
  \centering
  \includegraphics[width=0.45\textwidth]{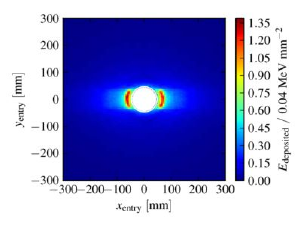}
  \includegraphics[width=0.45\textwidth]{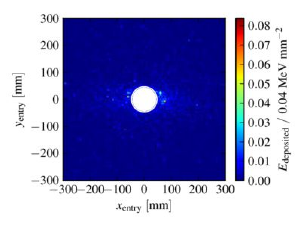}
  \caption{Simulated energy deposited in the innermost backward Herschel counter
    (B0) for inelastic collisions (left) and for events mimicking a CEP-like
    physics event with a disassociated proton (right)}
  \label{fig:configurations2}
\end{figure}

As illustrated in Fig.~\ref{fig:i8beamline}, the Herschel system 
comprises three stations in the RB84 section of the LHC 
on the left side of the LHCb interaction region IP8, and two stations
in the RB86 section on the right side of IP8.  The left-side stations,
which are upstream of the VELO, are labelled B (backwards) and the
right side stations downstream of the muon chambers are labelled F (forwards).
The $z$-positions of the stations with respect to IP8 are largely
defined by the available space, and are placed as symmetrically as
possible.  The outermost stations (B1,B2,F1 and F2) are at
$\pm~ $20~m and 114~m, and on the backward side an additional station
(B0) is placed at -7.5~m.    Due to differences in the vacuum chamber layout at the 
proposed locations, the innermost cut-out of the scintillator panels
are adapted appropriately.  The smallest achievable radii are at B0
and B1, where the vacuum chamber has a circular cross-section with an
outer diameter of 84\,mm.  At station F1 the scintillator plates must
fit around the vacuum bellow, and the far stations are situated in the
region where each beam has an individual chamber, and are enlarged
accordingly.  Each station is equipped with four scintillator plates,
with outer quadrant dimensions of $300\times300$ mm$^2$. 
The layout is indicated schematically in Fig~\ref{fig:configurations1}~\ref{fig:configurations2}~\ref{fig:configurations3}.
The stations start to be efficient for primary particles with a
pseudorapidity of about 7.5 or less, due to the presence of beam
elements where showering can be initiated giving signals in the scintillators.

The shower counter environment has been fully simulated with a
description of the beam material and magnetic fields, and the
scintillator occupancy has been studied with single particles and
full LHCb events.
A typical inelastic collision will give rise to showers that will produce a high level of activity in most or 
all of the scintillator stations.  In these events the number of charged particles that will give light is 
expected to be $\sim 100$ per quadrant in stations B0, B1 and F1, and
up to ten times this value in B2.  
The photomultipliers must be able to withstand the light flux produced from such activity occurring in the majority of 
beam-crossings. At the same time the system must be sensitive to the activity from classes of interaction 
that constitute background to the CEP analyses, for example a single diffractive interaction that gives 
rise to a low level of activity in the spectrometer acceptance.   Simulation studies indicate that the hit 
multiplicity in such interactions is about five times lower than in other classes of inelastic 
events. 

The radiation environment is most severe for B2 and F2.   Here
simulation indicates that the dose in the 
innermost region of the scintillator plane will approach 
1~MRad/${\rm fb^{-1}}$, and to be an order of magnitude lower for those regions more than  5~cm from the 
closest acceptance. The other stations will experience a lower dose, following the reduced multiplicity 
in these planes.  Irradiation measurements for calorimeter modules exposed in the tunnel during Run~1 
close to the location of B0 give results that are compatible with
these estimates.
In order to protect the scintillators from accumulating irradiation
while not in use, each station is equipped with a remoted controlled
pneumatic moving system
allowing the planes to be retracted and rotated away from the beamline.
The stations are not located in a high magnetic field, the largest being  $\sim$ 30~G  at F1.

\subsection{Detector Design and Installation}

The scintillators are manufactured from EJ-200\footnote{Eljen Technology, Sweetwater, Texas 79556, United States (\tt{http://www.eljentechnology.com}} 
plastic scintillator material, which features a rise time of 0.9\,ns, a decay time of 2.1\,ns
and a light yield of 10000 photons/1~MeV~e$^{-}$. 
Light guides in the shape of a fish tail are glued to the top
or bottom of the scintillator plates.  For calibration purposes, two
``Mega Bright Blue'' LEDs~\footnote{Manufactured by Multicomp} are included with each scintillator, close to and far away from
the light guide.  The scintillators and light guides are 
covered by thin aluminium sheets appropriate for the LHC tunnel environment.

The main challenge for the photomultiplier tube is the high photon
flux, which even when the tube is operated with reduced gain will lead
to a high anode current. The Hamamatsu\footnote{Hamamatsu Photonics
  K. K., Hamamatsu City, Japan (\tt{www.hamamatsu.com})} R1828-01
$2^{''}$ diameter PMT is used, which allows a maximum average anode
current of $200\,{\rm \mu A}$.  It has a relatively fast signal
response of 1.3~ns rise time.  This 12-stage PMT also has a large
range of gain adjustment which is suitable for both a low gain
operation ($\sim 10^3-10^4$)  in the experimental environment of the
LHC, and a high gain operation ($\sim 10^6 - 10^7$) required for the
calibration of the counters with cosmic muons.  In order to cope with
the rates a special resistive divider design is used which features
Zener diodes to stabilise the dynodes and a bias extra-current to
allow sufficient current in the vicinity of the anode.  Such a design
has already been operated successfully in the LHCb Beam Loss
Scintillator (BLS) system in similar radiation conditions.   Each
photodetector is mounted in a standard steel housing, including a shielding tube that together provides protection against magnetic fields up to $\sim 1$~kG.  

\begin{figure}
  \centering
  \includegraphics[width=0.95\textwidth]{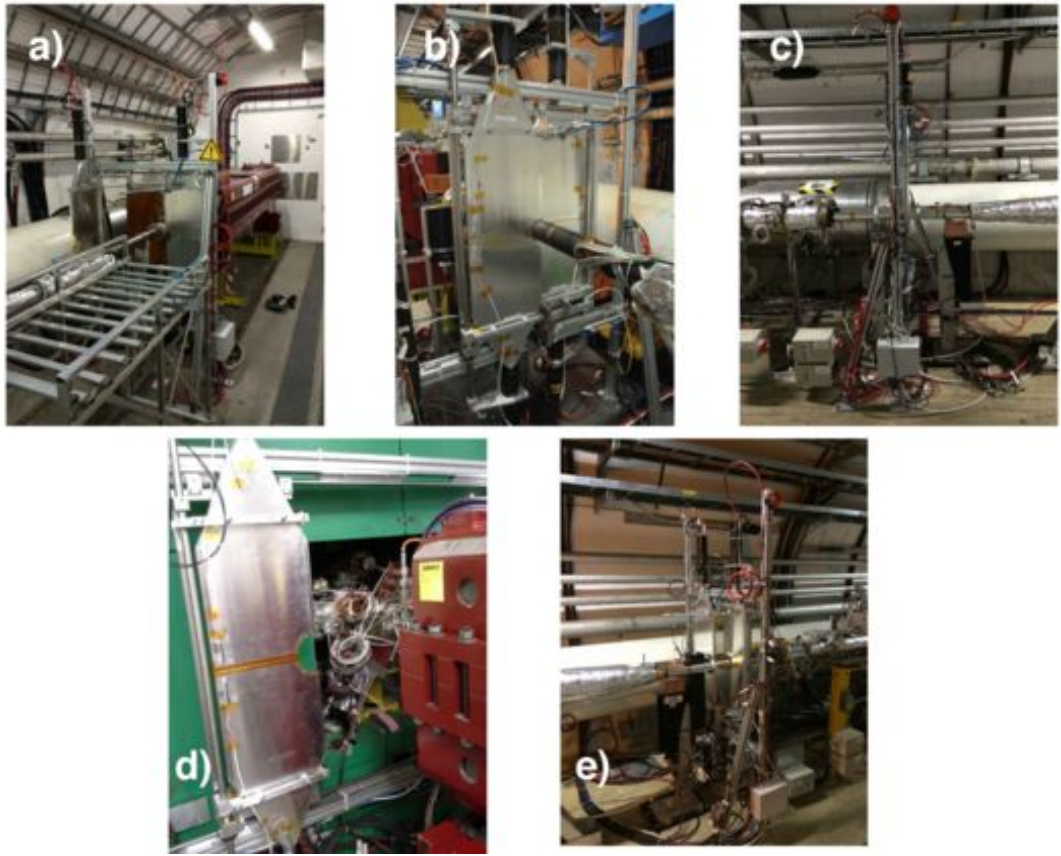}
  \caption{Photographs of the installed Herschel scintillators in the
    LHC tunnel.  a) shows station B0, 7.5 m upstream of LHCb, b) shows
  station B1 at 19.7m upstream and c) shows station B2 at 114 m
  upstream.  On the downstream side, d) shows station F1, here in the
  open position such that the scintillator shape can be clearly seen,
  and e) shows station F2.}
  \label{fig:configurations3}
\end{figure}

All scintillators were calibrated before installation using cosmic and
LED signals.  The light yield is estimated to be $~200$ electrons per
mip and the cosmic signals show a clear separation from the pedestal.
A clipping scheme is used to contain the signals within a 25 ns
window.  The scintillators and cables were installed during available
access times in the LHC tunnel over a six month period towards the end
of 2014 and the start of 2015.  Photographs of the complete set of
stations can be seen in figure~\ref{fig:configurations3}.

\begin{figure}
%  \centering
  \includegraphics[width=0.95\textwidth]{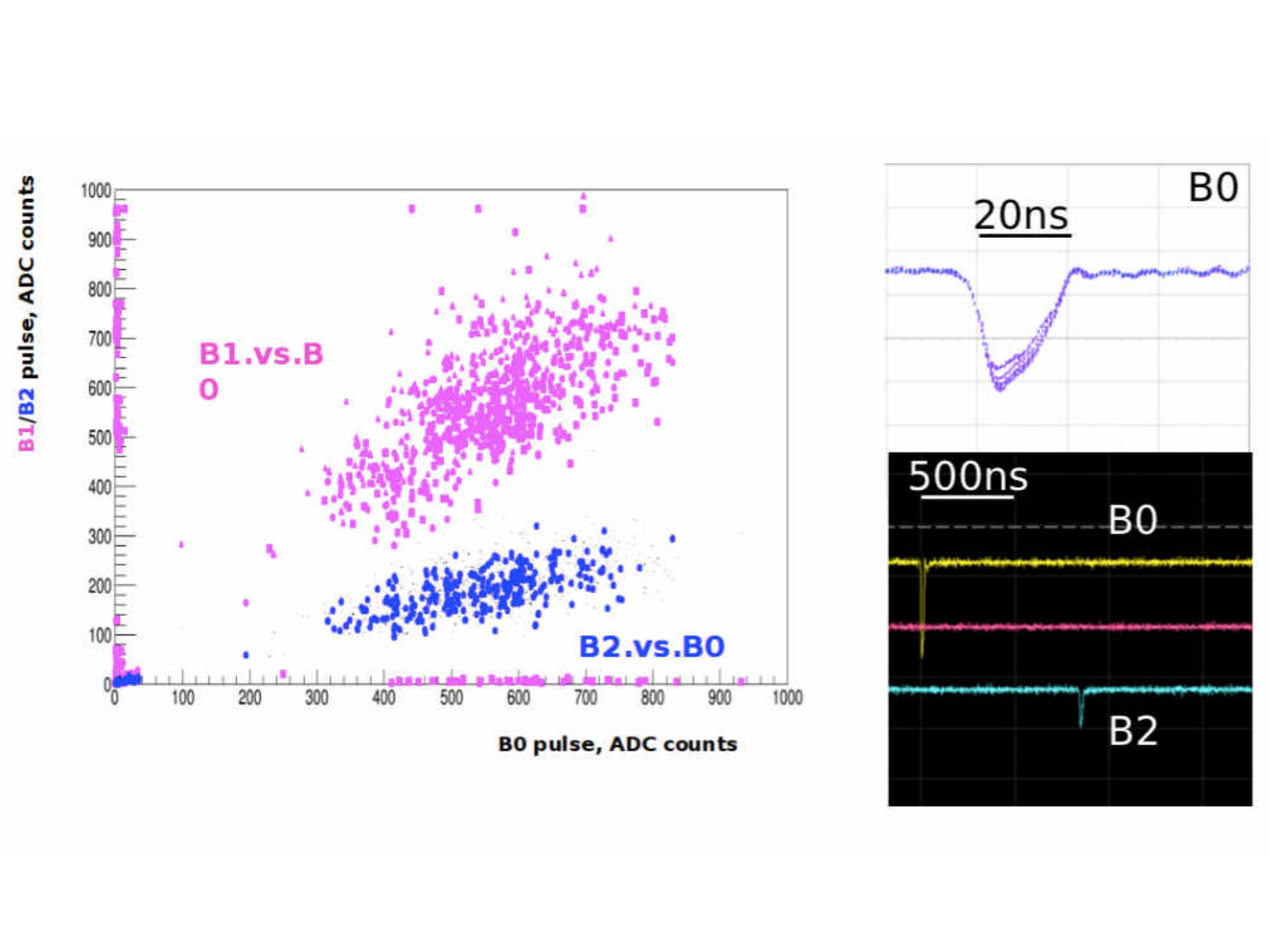}
  \caption{Preliminary plots giving a qualitative indication of the
    Herschel data quality from the November 2014 ``TED'' run.  The
    scatter plot shows the correlation between the three downstream
    stations.  As expected, B2 is less populated than B0 or B1.  The
    right figures show the characteristics of the signals in the
    channels connected to the oscilloscope. The signal is contained
    within 25 ns.}
  \label{fig:tedherschel}
\end{figure}

The Herschel system is required to identify events with low signals,
which typically occur after a minimum bias event with large energy
deposition in the scintillators, with the event occupancy being very
close to 100$\%$.   For this reason careful attention was paid
to the signal speed and choice of electronics.   The readout system
uses components from the readout system of the LHCb
preshower%~\cite{preshower} 
detector.  Each channel is treated by
alternately by two integrators, which are each reset after 50~ns.  The
electronics for the backwards and forwards stations are installed in
independent crates on either side of the LHCb cavern, before routing
to the barracks and combination with the other LHCb sub detectors.
The commissioning of the system started in November 2014 and March
2015, when data was provided in LHCb during the so called ``TED''
shots during LHC sector tests, where particles emerge from the beam
stopper downstream of LHCb.  This provided an opportunity to check the
synchronisation of Herschel relative to the rest of LHCb, tune the
time alignment, and evaluate
the readout chain.  For the purposes of this test some Herschel
channels were connected directly to an oscilloscope.  The first
performance indications showed that the scintillators were working
well, that the signal is contained within 25 ns, and that correlations
in signals are seen between the stations.  Data from the November 2014
TED run is shown in figure~\ref{fig:tedherschel} and a more detailed
analysis is ongoing.

It is also of great interest to include Herschel information in the
LHCb L0 trigger, which is the lowest level trigger of the experiment
running at 40 MHz.  By vetoeing events with activity in the
scintillator, it should prove possible to extend the kinematic range
of exclusive hadronic channels by relaxing the multiplicity and
transverse momentum cuts which are currently imposed to suppress the
rate.  In addition it should be possible to exploit Herschel to
enhance the L0 trigger for single diffractive physics events.  The
LHCb luminosity monitoring also needs to suppress background pp
interactions where there are particles in both directions but may
mimic beam-gas interactions due to the lack of angular coverage of
LHCb and the Herschel trigger is expected to play a role.  The
commissioning of Herschel at L0 requires additional hardware
interventions and is scheduled for the first technical stop after a
few months running at Run II.  

\subsection{Conclusions}

The addition of the Herschel scintillation counters around the beam
pipes on both sides of LHCb increases their rapidity coverage and
enables studies of diffractive processes, 
both as gap detectors and as detectors of propton diffractive
dissociation.  The system is currently being installed and
commissioned and is expected to operate fully during Run II.

%% file: detectors/adYellowReport-ed.tex
\section{AD: The Alice Diffractive Detector }
\label{sec:TheAliceDiffractiveDetector}

%\color{green}
%In 2012, two station composed by scintillator paddles on both sides of the interaction point of ALICE were installed. They were successfully used as a beam radiation monitor \cite{ref1}. The system was capable of detecting minimum ionizing particles. It was used to measure the beam gas background and provided an online determination of luminosity. The asynchronous read out of the charge deposited in the detector worked rather well.
%\color{red}\
%{\it Given that we do not have much space, we should not divert attention to a detector that reads out asynchronously. I propose this paragraph be dropped.}
%\color{black}\

There are only a few %forward
sub-detectors in ALICE that are used to measure charged particle multiplicity at large pseudo-rapidities $\eta$, ($|\eta|>2$). The VZERO detector, made of scintillator plastic, is used as a level zero trigger and it is suitable for multiplicity measurements, the Forward Multiplicity Detector (FMD) made of silicon detectors and the Photon Multiplicity Detector (PMD) at moderately large forward rapidity, the Zero Degree Calorimeter (ZDC), which can be used to tag neutrons and protons from the nuclear break-up,  are some of the devices used for diffractive physics studies.

%ALICE performed a dedicated study of diffractive events, selecting experimental observables sensitive to the diffractive content of the selected data sample. An adjustment of the Monte Carlo event generators (PYTHIA 6 and PHOJET) was required. As a result, the ratio of the single (double) diffraction cross section to the total inelastic cross section was measured, allowing a robust measurement of the p-p inelastic cross section. The details of these results  can be found in the ALICE publication \cite{ref2}.

%ATLAS and CMS have also made proposals for installing forward detectors for diffractive studies \cite{ref3}. On the other hand,
ALICE has excellent particle identification capabilities in the central rapidity region and can resolve low transverse momentum tracks, ($p_T >150$ MeV),  i.e. ALICE is in a position to study soft and hard diffractive events at the LHC. In order to extend the rapidity coverage of ALICE and enhance the efficiency for detecting events with rapidity gaps,  during LS1  a small detector was installed made of scintillation counters with optical fibre readout (AD, the ALICE Diffractive detector).

\begin{figure}[t]
\centering
\includegraphics[scale=0.6]{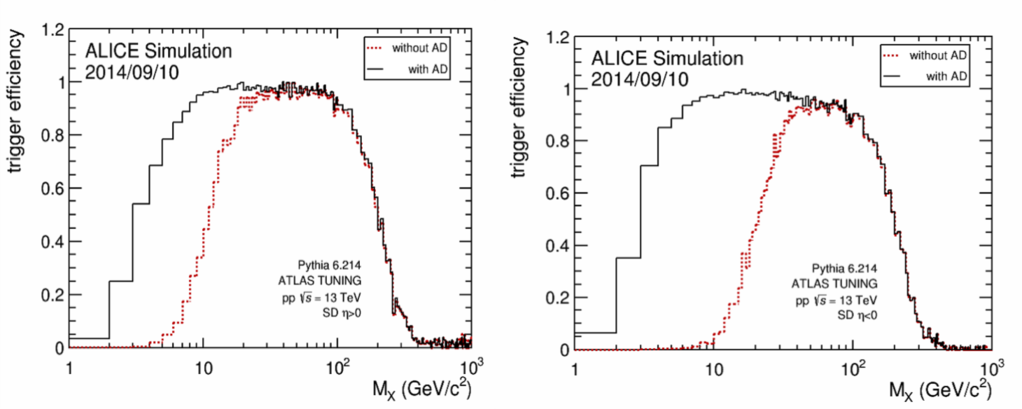}
\caption{Trigger efficiency for single diffractive events (SD) for $\eta > 0$ (right) and $\eta < 0$ (left). In both cases the efficiency of the trigger given by AD increases considerably at low diffractive masses}
\label{fig:fig1}       % Give a unique label
\end{figure}

The AD detector will increase the sensitivity to diffractive masses close to  threshold ($m_p+m_\pi$) and also partially compensate for the loss of trigger efficiency for Minimum Bias events and diffractive events when reaching the design LHC energies (see figure \ref{fig:fig1}). This detector will provide a level zero trigger signal which will be useful for diffractive cross section measurements. It will extend the pseudorapidity gap trigger, crucial in the study of central diffraction, where the physics reach is limited by statistics.
In addition, the possibility of triggering on the charge deposited in the AD scintillator modules  will provide an extended centrality trigger in both Pb--Pb and pPb collisions studies.

\subsection{Design of AD}

\begin{figure}[t]
\centering
\includegraphics[scale=0.5]{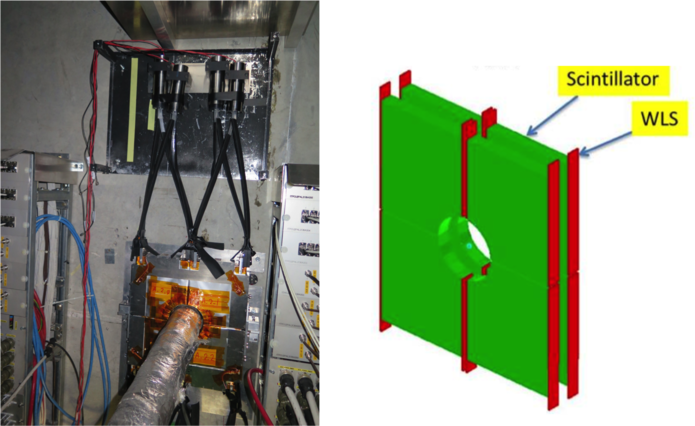}
\caption{Picture of one AD station installed inside the ALICE' cavern (left). Drawing of one AD station (right) showing eight scintillator cells (green) and two WLS bars per cell (red)}
\label{fig:fig2}       % Give a unique label
\end{figure}

\begin{figure}[h]
\centering
\includegraphics[scale=0.3]{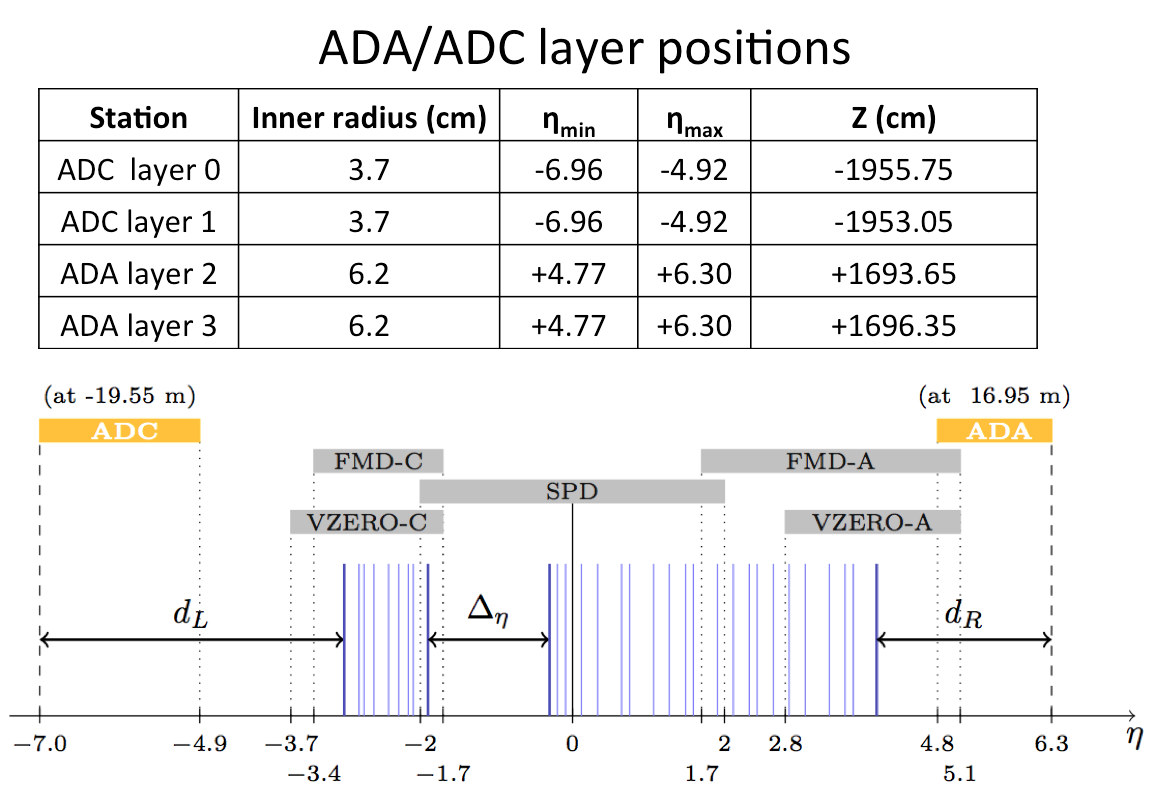}
\caption{Pseudorapidity coverage of the AD system. The diffractive trigger will be generated by AD, VZERO and SPD systems.}
\label{fig:fig3}       % Give a unique label
\end{figure}

The AD detector consists of two stations of scintillator pads (see figure \ref{fig:fig2}), one on each side of the interaction point (see figure \ref{fig:fig3}), ADA on the cavern A-side, ADC on the  tunnel C-side ).
Each station has two layers, each with four BC404 scintillator pads of dimension $181\times216\times 25 $ $mm^3$, arranged around the beam pipe. In each station, a coincidence between corresponding pads is required to reduce background and electronic noise.
On the C-side the two layers are placed inside the LHC tunnel next to the compensator dipole magnet, a position that avoids synchronization with beam-gas when LHC runs with 25 ns bunch spacing.

\subsection{Commissioning of AD}

The light produced by BC404 plastic scintillator material is collected by two Wave Length Shifting bars (WLS) attached, but not glued, on each side of the pads. Each WLS bar transfer the collected light to a bundle of 96 transparent optical fibres, which conducts the light to the PMTs (inside the ALICE cavern).  The light is converted into an electric pulse by a fine mesh PMT from Hamamatsu R5946 (hybrid assembly H6153-70). The signal from the PMT is sent to a preamplifier card which delivers two signals: one, amplified by a factor of 10 and clamped at about 300 mV which is used for timing measurement, the second, direct unmodified signal, is used for charge integration. The preamplifiers are installed close to the Front End Readout electronics.
The Front End Electronics provides signals for the level 0 trigger of ALICE. It is of the same kind as that presently used in the VZERO detector \cite{ref4}. The trigger signals of the AD counters will expand the acceptance of the Minimum Bias trigger.  Moreover it will be possible to trigger on charge deposition in the two AD detectors providing an extended centrality trigger in both Pb-Pb and proton-Pb collision studies.

\begin{figure}[h]
\centering
\includegraphics[scale=0.6]{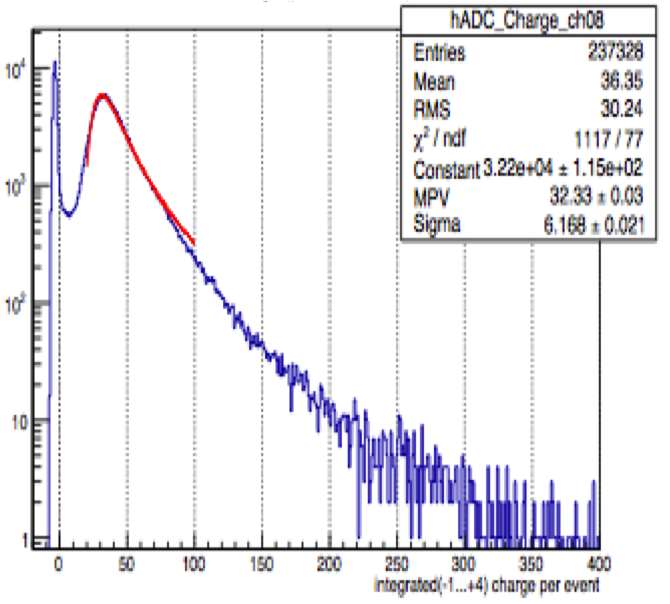}
\caption{Charge distribution of one channel of AD. The pedestal is well separated from the signal.}
\label{fig:fig4}       % Give a unique label
\end{figure}

The photomultipliers and scintillators were calibrated with cosmic ray data  and LED signals in the laboratory.
The measured time resolution is about 0.8 ns and a clear separation obtained between the signal and the pedestals as shown in figure \ref{fig:fig4} for cosmic muons.

\subsection*{Conclusion}

Diffraction is an important part of  non-perturbative QCD studies. It is also important in the tuning of Glauber models used to simulate Pb-Pb and pPb collisions.
For this purpose, during LS1 the ALICE collaboration installed the AD detector, which will increase the trigger efficiency for diffractive events.
The system is fully installed and currently is being commissioned so as to be completely operational during Run-II.

%% file: detectors/LHCfdetector-yrep-sako.tex
 \section{LHCf Detectors}

The Large Hadron Collider forward (LHCf) experiment installed two independent detectors at either sides of IP1 (ATLAS) \cite{LHCf-TDR} .
The detectors are installed in the instrumentation slots of the TAN absorbers 140\,m away from the IP.
The detector at the IP8 (LHCb) side is called Arm1 and the other at the IP2 (ALICE) side is called Arm2.
This location allows the detection of neutral particles with $\eta>8.4$.
Photons, predominantly decay products of $\pi^{0}$, and neutrons are dominant particles arriving at the detectors.
The detectors to be used in Run-II are essentially the same used in Run-I except some upgrades described below.
 
Each of the LHCf detectors consists of two small calorimeters with a double tower structure \cite{LHCf-JINST}.
The dimensions of the towers transverse to the beam direction are 20\,mm$\times$20\,mm and 40\,mm$\times$40\,mm 
for Arm1 and  25\,mm$\times$25\,mm and 32\,mm$\times$32\,mm for Arm2.
The longitudinal structure of the towers is a stack of 44 radiation lengths of Tungsten interleaved with 16 sampling
scintillators.
Plastic scintillators were used during Run-I but they have been replaced with Gd$_{2}$SiO$_{5}$ scintillators for Run-II  
\cite{GSO-JINST} to make the calorimeters radiation harder.
Four X-Y pairs of strip sensors, SciFi in Run-I \cite{SciFi-mizuishi} and GSO-bar bandles in Run-II \cite{GSObar-JINST} for Arm1  
and Silicon strip sensors for Arm2 \cite{Silicon-JINST}, 
are inserted to measure the lateral distribution of the showers.
The longitudinal locations and the readout circuit of the Silicon strip sensors have also been updated in Run-II to optimize 
the energy determination ability using the Silicon sensors. 
Thanks to the double tower structure and position sensitivity, invariant mass of photon pairs hitting each calorimeter 
can be estimated.
By selecting $\pi^{0}\rightarrow\gamma\gamma$ events, momenta of $\pi^{0}$ are obtained \cite{menjo-APP} \cite{LHCf-pi0}.

The performances of the LHCf detectors have been carefully studied by using data from SPS beam test and LHC Run-I,
as well as  through detailed MC simulations, and are summarized in  \cite{mase-NIM}  \cite{kawade-JINST} \cite{LHCf-IJMPA}.

\begin{figure}
  \begin{center}
    \includegraphics[width=120mm]{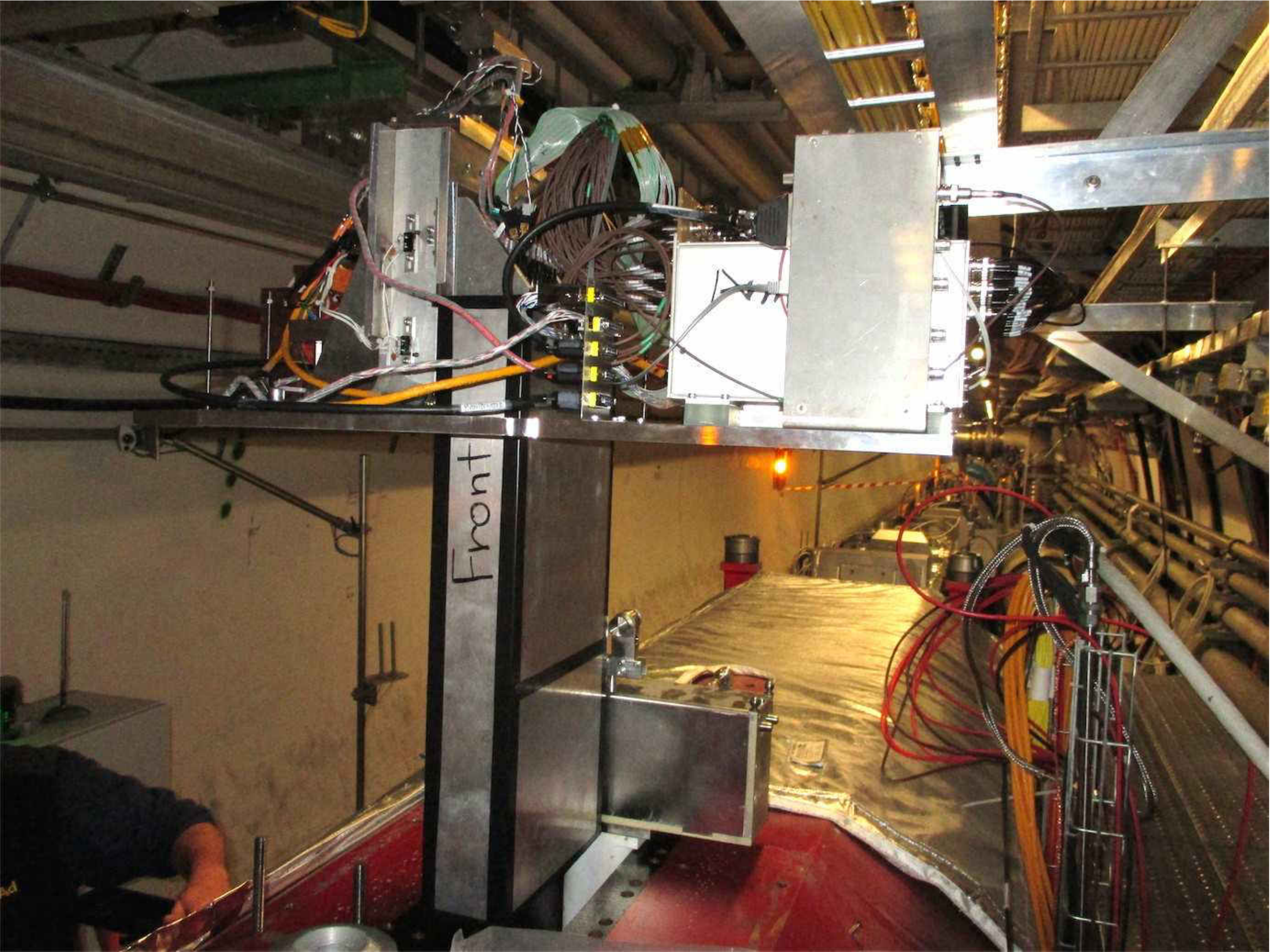}
    \caption{LHCf Arm1 detector and its front-end electronics being installed into the TAN instrumentation slot.}
    \label{fig-lhcf-install}
  \end{center}
\end{figure}

%% file: summary/summary.tex
This yellow report presents a summary of our current understanding in the field of high-energy forward particles physics and indicates a path for future studies.

Forward physics in the next few years has the potential to provide strong new contributions to the understanding of perturbative and non perturbative QCD, to be instrumental in the understanding of processes involving forward jets, and to open a new window on Beyond Standard Model searches.

The activities that have lead to this document have lasted about 18 months, with regular meetings at CERN and in several countries. These meetings have been characterized by a strong participation from many experiments at LHC and in cosmic rays and from the theory community, bringing together the experts working in different fields and theory divisions.

High cross studies, mostly centered around the physics of diffractive processes and particle multiplicities, will be carried out first while central exclusive production, with its unique capabilities of studying events completely contained in the detector, and saturation processes require more luminosity.

The forward physics program, as outlined in the report, relies strongly on new detectors added to current LHC experiments to gain access to part of the final state phase space currently not reachable.  These new detectors, combined with both dedicated and standard LHC running conditions will enable in the next few years to explore the physics program here detailed. 

%% file: acknowledgements/acknowledgements.tex
We are very grateful to the LPCC for all their help in creating
an excellent scientific environment while organizing the Forward Working
Group meetings at CERN. V.A. Khoze thanks the Leverhulme Trust for an emeritus fellowship.
This work was also supported in part by the Polish National Science Centre grant
UMO-2012/05/B/ST2/02480.